\documentclass[12pt,twoside,a4paper,normalheadings,headsepline]{scrbook}
\usepackage{palatino,cite,epsfig,amsmath,amssymb}
\usepackage{graphics}
\usepackage[small]{subfigure}
\usepackage[small]{caption2}
\usepackage{epsf}
\usepackage{rotating}
\usepackage{psfrag}
\usepackage{dcolumn}
\usepackage{multirow}

\input axodraw.sty
\input rotate

\allowdisplaybreaks

\def\abstract{\list{}{\listparindent 0pt
    \setlength{\leftmargin}{2cm}
    \itemindent\listparindent
    \rightmargin\leftmargin }\item[]{\bf Abstract}\\}

\begin{document}

\setcounter{chapter}{0}
\setcounter{secnumdepth}{3}
\setcounter{tocdepth}{2}


\thispagestyle{empty}

\begin{flushright}
hep-ph/0410364
\end{flushright}

\vspace{.4em}

\begin{center}


{\Large\sc \bf Physics Interplay of the LHC and the ILC}

\vspace*{.7cm}

%
%
%
%
%

The LHC / LC Study Group

\vspace{1.5em}

Editors:\\[.5em]

{\sc 
G.~Weiglein$^{1}$,
T.~Barklow$^{2}$, 
E.~Boos$^3$, 
A.~De Roeck$^4$,
K.~Desch$^5$,
F.~Gianotti$^4$, 
R.~Godbole$^{6}$,
J.F.~Gunion$^7$, 
H.E.~Haber$^{8}$,
S.~Heinemeyer$^{4}$, 
J.L.~Hewett$^{2}$,
K.~Kawagoe$^{9}$, 
K.~M\"onig$^{10}$,
M.M.~Nojiri$^{11}$,
G.~Polesello$^{12,4}$, 
F.~Richard$^{13}$,
S.~Riemann$^{10}$,
W.J.~Stirling$^{1}$
}

\vspace{1.5em}

Working group members who have contributed to this report:\\[.5em]

{\sc 
A.G.~Akeroyd$^{14}$,
B.C.~Allanach$^{15}$, 
D.~Asner$^{16}$,
S.~Asztalos$^{17}$,
H.~Baer$^{18}$,
T.~Barklow$^{2}$, 
M.~Battaglia$^{19}$,
U.~Baur$^{20}$, 
P.~Bechtle$^{5}$,
G.~B\'elanger$^{21}$
% ...
}

\clearpage


{\sl
$^1$Institute for Particle Physics Phenomenology, University of Durham,\\
Durham DH1~3LE, UK\\
$^{2}$Stanford Linear Accelerator Center, 
Menlo Park, CA 94025, USA,\\
$^3$Skobeltsyn Institute of Nuclear Physics,
Moscow State University,\\
   119992 Moscow, Russia\\
$^4$CERN, CH-1211 Geneva 23, Switzerland\\
$^5$Universit\"at Hamburg, Institut f\"ur Experimentalphysik, Luruper
Chaussee,\\
   D-22761 Hamburg, Germany\\
$^{6}$Centre for Theoretical Studies, Indian Institute of Science, 
Bangalore, 560012, India\\
$^7$Davis Institute for HEP, Univ.~of California, Davis, CA 95616, USA\\
$^8$Santa Cruz Institute for Particle Physics, UCSC, Santa Cruz CA 95064, USA\\
$^{9}$Kobe University, Japan\\
$^{10}$DESY, Deutsches Elektronen-Synchrotron, D-15738  Zeuthen,
Germany\\
$^{11}$YITP, Kyoto University, Japan\\
$^{12}$INFN, Sezione di Pavia, Via Bassi 6, Pavia 27100, Italy\\
$^{13}$LAL-Orsay, France\\
$^{14}$KEK Theory Group, Tsukuba, Japan 305-0801\\
$^{15}$DAMTP, CMS, Wilberforce Road, Cambridge CB3 0WA, UK\\
$^{16}$University of Pittsburgh, Pittsburgh, Pennsylvania, USA\\
$^{17}$LLNL, Livermore, Livermore, California, USA\\
$^{18}$Physics Department, Florida State University,
Tallahassee, FL 32306, USA\\
$^{19}$Univ.~of California, Berkeley, USA\\
$^{20}$Physics Department, State University of New York,
Buffalo, NY  14260, USA\\
$^{21}$LAPTH, 9 Chemin de Bellevue, BP 110, F-74941 Annecy-Le-Vieux, France\\
% ...
%
}

\end{center}

\clearpage

\noindent
ANL-HEP-PR-04-108, CERN-PH-TH/2004-214, DCPT/04/134, DESY 04-206,\\
IFIC/04-59, IISc/CHEP/13/04, IPPP/04/67, SLAC-PUB-10764,
UB-ECM-PF-04/31,\\
UCD-04-28, UCI-TR-2004-37\\[4em]

\begin{abstract}
Physics at the Large Hadron Collider (LHC) and the International 
$e^+ e^-$ Linear Collider (ILC) 
will be complementary in many respects, as has been demonstrated at previous
generations of hadron and lepton colliders. This report addresses the 
possible interplay between the LHC and ILC in testing the
Standard Model and in 
discovering and determining the origin of new physics.
Mutual benefits for the physics programme at both machines can
occur both at the level of a combined interpretation of Hadron Collider
and Linear Collider data and at the level of combined analyses of the
data, where results obtained at one machine can directly influence the
way analyses are carried out at the other machine. Topics under study
comprise the physics of weak and strong electroweak symmetry breaking,
supersymmetric models, new gauge theories, models with extra dimensions,
and electroweak and QCD precision physics. 
The
status of the work that has been carried out within the LHC~/~LC Study Group 
so far is summarised in this report. 
Possible topics for future studies are outlined.
\end{abstract}

\vfill


\clearpage
\pagestyle{empty}
\tableofcontents
\pagestyle{headings}







\chapter*{Executive Summary}
\setcounter{page}{1}
\label{chapter:execsumm}

\addcontentsline{toc}{chapter}{Executive Summary}

\newcommand{\lsim}{\buildrel<\over{_\sim}}
\newcommand{\gsim}{\buildrel>\over{_\sim}}

The present level of understanding of the fundamental interactions of nature 
and of the structure of matter, space and time will enormously 
be boosted by the experiments under construction at the 
Large Hadron Collider (LHC)~\cite{execsumm_lhcexp}
and those planned for the International Linear Collider 
(ILC)~\cite{execsumm_lctdrs}. 
The LHC, which will collide protons with protons, is currently under
construction and is scheduled to go into operation in 2007. The ILC, which
will bring the electron to collision with its antiparticle, the positron,
has been agreed in a world-wide consensus to be the next large experimental 
facility in high-energy physics. The concept of the ILC has been proved
to be technologically feasible and mature, allowing a timely realisation
leading to a start of data taking by the middle of the next decade.

One of the fundamental questions that the LHC and the ILC will most 
likely answer is what gives particles the property of mass.
Furthermore, 
the results of the LHC and ILC are expected to be decisive in the quest for 
the ultimate unification of forces. This will provide
insight, for instance, about the possible extension of space and time by
new supersymmetric coordinates. The LHC and ILC could reveal the nature of
Dark Matter, which forms a large but as yet undisclosed part of all the matter
occurring in the Universe, and could advance our understanding of the origin
of the dominance of matter over antimatter in the Universe. At the
energy scales probed at the LHC and the ILC new space--time dimensions
might manifest themselves. Thus, results from LHC and ILC could dramatically 
change our current picture of the structure of space and time.

The way the LHC and ILC will probe the above-mentioned questions will be very
different, as a consequence of the distinct experimental conditions of
the two machines. The LHC, due to its high collision energy, in particular 
has a large mass reach for direct discoveries. 
Striking features of the ILC are its clean experimental environment,
polarised beams, and known collision energy,
enabling precision measurements and therefore detailed studies of
directly accessible new particles as well as a high sensitivity to
indirect effects of new physics.
The need for instruments that are optimised in different
ways is typical for all branches of natural sciences, for example 
earth- and space-based telescopes in astronomy. 
The results obtained at the LHC and ILC will complement and supplement
each other in many ways. Both of them will be necessary in order 
to disentangle the underlying structure of the new physics that lies ahead 
of us. The synergy between the LHC and ILC will likely be very similar to 
that demonstrated at previous
generations of electron--positron and proton--\mbox{(anti-)proton}
colliders running concurrently, where the interplay between the two
kinds of machines has
proved to be highly successful. There are many examples in the past where 
a new particle has been discovered at one machine, and its properties have
been studied in detail with measurements at the other. Similarly,
experimental results obtained at one machine have often given rise to
predictions that have led the searches at the other
machine, resulting in ground-breaking discoveries. 

The synergy from the interplay of the LHC and ILC can occur in
different ways.
The combined interpretation of the LHC and ILC data will lead to a much
clearer picture of the underlying physics than the results of both
colliders taken separately. Furthermore, in combined analyses of the data 
during concurrent running of both machines
the results obtained at one machine can directly influence the way
analyses are carried out at the other machine, leading to optimised
experimental strategies and dedicated searches. 

An important example is the physics of the Higgs
boson, which, if it exists, will be the key to understanding the
mechanism of generating masses of the elementary particles.
The combination of the highly precise measurements possible at the ILC
and the large mass and high-energy coverage of the LHC will be crucial
to completely decipher the properties of the Higgs boson (or several
Higgs bosons) and 
thus to disentangle the mechanism of mass generation. The discovery of 
particles predicted by supersymmetric theories 
would be a breakthrough in our understanding of matter, space and time. It is
likely in this case that the LHC and the ILC will be able to access
different parts of the spectrum of supersymmetric particles. Using
results from the ILC as input for analyses at the LHC will
significantly improve and extend the scope of the measurements carried
out at the LHC. The information from both the LHC and ILC will be crucial
in order to reliably determine the underlying structure of the
supersymmetric theory, which should open the path to the ultimate
unification of forces and give access to the structure of nature at
scales far beyond the energy reach of any foreseeable future
accelerator.
Another possible
extension of the currently known spectrum of elementary particles are 
heavier copies of the $W$ and $Z$ bosons, the mediators of the weak
interaction. The LHC has good prospects for discovering heavy states of
this kind, while the ILC has sensitivity exceeding the
direct search reach of the LHC through virtual effects of the new
particles. If the mass of the heavy state is known
from the LHC, its properties can be determined with high precision at
the ILC. A detailed study of the properties of these heavy states will be
of utmost importance, since they could arise from very different
underlying physics scenarios, among them the existence of so far
undetected extra dimensions of space.
Thus, the intricate
interplay between the LHC and ILC during concurrent running of the two
machines will allow to make optimal use of the capabilities of both
machines.

The present report contains the
results obtained within the LHC / LC Study Group since this working
group formed in spring 2002 as a collaborative effort of the hadron
collider and linear collider experimental 
communities and theorists. The aim of the 
report is to 
summarise the present status and to guide the way towards further studies. 
Many different scenarios have been investigated, and significant
synergistic effects benefiting the two collider programmes have been
demonstrated.
For scenarios where detailed experimental
simulations of the possible measurements and the achievable accuracies are 
available both for the LHC and ILC, the LHC / ILC interplay has been 
investigated 
in a quantitative manner. In other scenarios the most striking synergy
effects arising from the LHC / ILC interplay have been discussed in a 
qualitative
way. These studies can be supplemented with more detailed analyses in
the future, when further experimental simulations from the LHC and ILC
physics groups are available.



\chapter{Introduction and Overview}
\label{chapter:introoverview}

\section{Introduction}

\subsection{The role of LHC and LC in revealing the nature of matter, space 
and time}

The goal of elementary particle physics is to reveal the innermost
building blocks of matter and to understand the fundamental forces
acting between them. The physics of elementary particles and their
interactions played a key role in the evolution of the Universe from the
Big Bang to its present appearance in terms of galaxies, stars, black
holes, chemical elements and biological systems. Research in elementary
particle physics thus addresses some of the most elementary human questions:
what are we made of, what is the origin and what is the fate of the
Universe?

The past century was characterised by an enormous
progress towards an understanding of the innermost secrets of the
Universe, which became possible through a cross-fertilisation of
breakthroughs on the experimental and theoretical side.
The results obtained in particle physics have revealed a complex
microphysical world, which however seems to obey surprisingly simple
mathematical descriptions, governed by symmetry principles. We now
believe that there are four fundamental forces in nature, the strong,
electromagnetic, weak and gravitational forces. The seemingly disparate
electromagnetic and weak interactions have been found to emerge from the
unified electroweak interaction. We have been able to
formulate a quantum theory of elementary particles based on the strong
and electroweak interactions, which will stand as one of the
lasting achievements of the twentieth century. The quantum nature of the
interactions means in particular that they arise from the interchange of 
particles, namely the massless photon, massive $W$ and $Z$ bosons for the 
electroweak interaction, and the massless gluon for the strong interaction.

According to our current understanding there seem to be indications 
pointing towards a unification of the strong and electroweak forces,
and it appears to be conceivable that also gravity, with the graviton as
mediator of the interaction, may be incorporated into a unified framework. We
know, however, that our picture of the observed forces and particles ---
the known particles comprise the constituents of matter, the quarks and 
leptons, and the mediators of the interactions --- 
is incomplete. There needs to be another ingredient, being related to
the origin of mass and the breaking of the symmetry governing the
electroweak interaction. Its effects will manifest itself at the energy
scales that can be probed at the next generation of colliders. The favourite 
candidate for this ingredient is the
Higgs field, a scalar field that spreads out in all space. Its
field quantum is the Higgs particle. If no fundamental Higgs boson exists
in nature, electroweak symmetry breaking can occur, for instance,
via a new kind of strong interaction. 

The Higgs boson is the last missing ingredient of the ``Standard Model''
(SM) of 
particle physics, which was proposed more than three decades ago. The SM
has provided an extremely successful description of the
phenomena of the electroweak and strong interaction, having passed
hundreds of experimental tests at high precision. The direct search for
the Higgs boson has excluded a SM Higgs boson with a mass below about
114~GeV~\cite{sec0_Barate:2003sz}, which is about 120 times the mass of the
proton. The precision tests of the SM, based on the interplay of
experimental information obtained at different accelerators, allow one to set 
an indirect upper bound for a SM Higgs of about 250~GeV~\cite{ewfits}. 

However, even if a
SM-like Higgs boson is found, the SM cannot be the ultimate theory,
which is obvious already from the fact that it does not contain gravity.
There are indications that new physics beyond the SM should manifest
itself below an energy scale of about 1~TeV ($\equiv 10^3$~GeV). A 
particular shortcoming of
the SM is its instability against the huge hierarchy of vastly different
scales relevant in particle physics. We know of at least two such
scales, the electroweak scale at a few hundred GeV and the Planck scale
at about $10^{19}$~GeV, where the strengths of gravity and the other
interactions become comparable. The Higgs-, $W$- and $Z$-boson masses are all
unstable to quantum fluctuations and would naturally be pushed to the Planck
scale without the onset of new physics at the scale of few hundred GeV.

There are also direct experimental indications for physics
beyond the SM. While in the SM the neutrinos are assumed to be massless, 
there is now overwhelming experimental evidence that the neutrinos
possess non-zero (but very small) masses~\cite{neutrinomass_exp}. A
neutrino mass scale in agreement with the experimental observations
emerges naturally if there is new physics at a scale of about
$10^{16}$~GeV. We furthermore know that ``ordinary'' matter, i.e.\
quarks and leptons, contributes only a small fraction of the matter density of
the Universe~\cite{darkmatter}. There is clear evidence, in particular, for
a different kind of ``cold Dark Matter'', for which the SM does not offer an
explanation. The known properties of Dark Matter could arise in particular 
if new weakly interacting massive particles exist, which requires an
extension of the SM.

A very attractive possibility of new physics that stabilises the
hierarchy between the electroweak and the Planck scale is supersymmetry
(SUSY), i.e.\ the extension of space and time by new supersymmetric 
coordinates. Supersymmetric models predict the existence of partner
particles with the same properties as the SM particles except that their
``spin'', i.e.\ their internal angular momentum, differs by half a unit.
Other ideas to solve the hierarchy problem postulate extra spatial
dimensions beyond the three that we observe in our every-day life, or
new particles at the several TeV scale.

Supersymmetric theories allow the unification of the strong,
electromagnetic and weak interactions at a scale of about $10^{16}$~GeV.
In such a ``grand unified theory'', 
the strong, electromagnetic and weak interactions can be understood as being 
just three different manifestations of a single fundamental interaction. 
(In contrast, in the absence of supersymmetry, the three interactions
fail to unify in the SM.)
It should be noted that the possible scale of grand unification is
approximately the same as the one that would give rise to neutrino masses 
consistent with the experimental observations. In the minimal
supersymmetric extension of the SM the lightest SUSY particle (LSP) 
is stable. The LSP has emerged as our best candidate for cold Dark
Matter in the Universe.

\bigskip
The current understanding of the innermost structure of the Universe
will be boosted by a wealth of new experimental information which we
expect to obtain in the near future within a coherent programme of
very different experimental approaches. They range from astrophysical
observations, physics with particles from cosmic rays, neutrino physics
(from space, the atmosphere, from reactors and accelerators), precision
experiments with low-energy high-intensity particle beams to experiments 
with colliding beams at the highest energies. The latter play a central
role because new fundamental particles can be discovered and studied
under controllable experimental conditions and a multitude of observables
is accessible in one experiment.

While the discovery of new particles often requires access to the
highest possible energies, disentangling the underlying
structure calls for highest possible precision of the measurements.
Quantum corrections are influenced by the whole structure of the model.
Thus, the fingerprints of new physics often only manifest themselves in tiny 
deviations.
These two requirements --- high energy and high precision --- cannot
normally be 
obtained within the same experimental approach. While in hadron
collisions (collisions of protons with protons or protons with
antiprotons) it is technically feasible to reach the highest 
centre-of-mass energies, in lepton collisions (in particular collisions
of the electron and its antiparticle, the positron) the highest
precision of measurements can be achieved. This complementarity has
often led to
a concurrent operation of hadron and lepton colliders 
in the past and has undoubtedly created a high degree of synergy of the
physics programmes of the two colliders.
As an example, the $Z$~boson, a mediator of the weak interactions,
has been discovered at a proton--antiproton collider, i.e.\ by colliding
strongly interacting particles. Its detailed properties, on the other
hand, have only been measured with high precision at electron--positron
colliders. These measurements were crucial for establishing the SM. 
Contrarily, the gluon, the mediator of the strong interactions, 
has been discovered at an electron--positron collider rather than at a
proton collider where the strong interaction dominates.

Within the last decade, the results obtained at the electron--positron
colliders LEP and SLC had a significant impact on the physics programme
of the Tevatron proton--antiproton collider and vice versa. The
electroweak precision measurements at LEP and SLC gave rise to an
indirect prediction of the top-quark mass. The top quark was
subsequently discovered at the Tevatron with a mass in 
agreement with the indirect prediction. The measurement of the top-quark
mass at the Tevatron, on the other hand, was crucial for deriving
indirect constraints from LEP/SLC data
on the Higgs-boson mass in the SM, while 
experimental bounds
from the direct search were established at LEP. The experimental results
obtained at LEP have been important for the physics programme of the
currently ongoing Run~II of the Tevatron. 
There are further examples of this type of synergy between different
colliders in the recent past.
Following an observed excess of events with high momentum squared
at HERA in 1997, and
their possible interpretation as leptoquark production, dedicated leptoquark
searches at the concurrently running Tevatron were subsequently carried
out. These Tevatron searches provided strong constraints on the leptoquark
model, information that was in turn fed back to the HERA analyses.
The most recent example of the
interplay of lepton and hadron colliders is the discovery of the state 
X(3872) at BELLE~\cite{Choi:2003ue}, which gave rise to a dedicated search
at the Tevatron, leading to an independent confirmation of the new
state~\cite{Acosta:2003zx,Stark:2004td}.

The enormous advance of accelerator science over the last decades has put
us in a situation where both the next generation of hadron and 
electron--positron colliders are technologically feasible and mature. 
The Large Hadron Collider (LHC)~\cite{lhcexp} is under construction at CERN
and is scheduled to start taking data in 2007. It will collide protons
with an energy of 14~TeV. Since the proton
is a composite particle, the actual ``hard'' scattering process takes
place between quarks and gluons at a fraction of the total energy.

The Linear Collider (LC)%
\footnote{The shorthands LC and ILC are used
synonymously in this report.}
will bring the electron and the 
positron to
collision with an energy of up to approximately 1~TeV and high
luminosity~\cite{lctdrs}. 
The LC has been agreed in a world-wide
consensus to be the next large experimental facility in high-energy
physics. Designs for this machine have been developed in a world-wide
effort, 
and it has been demonstrated that a LC can be built and reliably
operated. 
The technology for the accelerating cavities has recently been
chosen, and the development of an internationally federated design has
been endorsed~\cite{ITRPrecomm}.

Ground-breaking discoveries are expected at the LHC and LC.
The information obtained from these machines will be
indispensable, in particular, for deciphering
the mechanism giving rise to the breaking of the electroweak
symmetry, and thus establishing the origin of the masses of particles. 
Furthermore it is very likely that we will 
be able to
determine the new physics responsible for stabilising the hierarchy 
between the electroweak and the Planck scale, which may eventually lead us 
to an understanding of the ultimate unification of forces. We expect new
insights into the physics of flavour and into the origin for the
violation of the charge conjugation and parity (CP) symmetry. This could
lead to a more fundamental understanding of the observed matter--antimatter 
asymmetry in the Universe.

Thus, the physics case is well established for both the LHC and the LC.
While the physics
programme at each of the machines individually is very rich, further important
synergistic effects can be expected from an intimate interplay of the
results from the two accelerators, in particular during concurrent
running. This will lead to mutual benefits for the
physics programme of both machines. 
In this way the physics return for the investment 
made in both machines will be maximised.

\subsection{Objectives of the study}


The goal of the studies contained in this document is to
delineate through detailed examples the complementarity
of the LHC and LC programs and the enormous synergy that
will result if the two machines have a very substantial overlap
of concurrent operation.

One of the great assets of the LHC is its large mass reach for direct
discoveries, which extends
up to typically $\sim$6--7~TeV for singly-produced particles with
QCD-like couplings (e.g.\ excited quarks) and $\sim$2--3~TeV for
pair-produced strongly interacting particles. The reach for singly
produced electroweak resonances (e.g.\ a heavy partner of the $Z$~boson)
is about 5~TeV. The
hadronic environment at the LHC, on the other hand, will be
experimentally challenging. 
Kinematic reconstructions are normally restricted to the transverse
direction. Since the initial-state particles carry colour charge, QCD
cross sections at the LHC are huge, giving rise to backgrounds which are
many orders of magnitude larger than important signal processes of
electroweak nature. Furthermore, operation at high luminosity
entails an experimentally difficult environment such as pile-up events.

The envisaged LC in the energy range of
$\sim$0.5--1~TeV provides a much cleaner experimental environment that
is well suited for high-precision physics. It has a well-defined initial
state which can be prepared to enhance or suppress certain processes
with the help of beam polarisation. The better knowledge of the momenta
of the interacting particles gives rise to kinematic constraints which
allow reconstruction of the final state in detail. The signal--to--background
ratios at the LC are in general much better than at the LHC. Direct
discoveries at the LC are possible up to the kinematic limit of the 
available energy. In many cases the indirect sensitivity to effects of 
new physics via precision measurements greatly exceeds the kinematic
limit, typically reaching up to 10~TeV.

While the complementarity between the LHC and LC is qualitatively obvious, 
more quantitative analyses of the possible interplay between the LHC and LC 
have been lacking until recently. They require detailed case studies,
involving input about various experimental aspects at both the LHC and LC.
In order to achieve this, a close collaboration is necessary between 
experimentalists from the LHC and LC and theorists.

The LHC / LC Study Group has formed as a collaborative effort of
the hadron collider and linear collider communities. This world-wide 
working group investigates how analyses carried
out at the LHC could profit from results obtained at the LC and vice
versa. In order to be able to carry out analyses of this kind, it is 
necessary to assess in detail the capabilities of the LHC and LC in different 
scenarios of physics within and beyond the Standard Model. Based on these 
results, the LHC~/~LC Study Group investigates the possible synergy
of a concurrent running of the LHC and LC. This synergy can arise from a
simultaneous interpretation of LHC and LC data, leading to a clearer
physics picture. Furthermore, results from one collider can be directly
fed into the analyses of the other collider, so that experimental
strategies making use of input from both colliders can be established.
The LC results can in this context directly influence the running
strategy at the LHC. In particular, the LC could predict properties of
new particles, leading to a dedicated search at the LHC. This could
involve the implementation of optimised selection criteria or
modifications of the trigger algorithms. LC results could also guide
decisions on required running time and sharpen the goals for a
subsequent phase of LHC running.


In general, the untriggered operation of the LC has the potential to
reveal new physics that gives rise to signatures that do not pass the
LHC triggers. Such a situation occurred in the past, for instance, at
ISR where the discovery of the $J/\psi$ at the electron--positron collider 
SPEAR (and independently at AGS) lead to a
modification of the trigger, and the signal could subsequently be
confirmed at ISR. In the TeV regime one cannot exclude the possibility
that unexpected new physics manifests itself in events which will not be
selected by the very general trigger strategies adopted by the LHC
experiments.
Insight from the LC could help in such a case to 
optimise the search strategy at the LHC.


While the LHC is scheduled
to take first data in 2007, the LC could go into operation in about the
middle of the next decade. This would allow a substantial period of
overlapping running of both machines, since it seems reasonable to
expect that the LHC (including upgrades) will run for about 15 years
(similarly to the case of the Tevatron, whose physics programme
started more than 20 years ago).
During simultaneous running of both machines there is obviously the highest
flexibility for adapting analyses carried out at one machine according
to the results obtained at the other machine. 

The results obtained in the framework of the LHC / LC Study
Group are documented in this first working group report. The report
should be viewed as a first step that summarises the present status and
guides the way towards further studies. Many different scenarios were
investigated, and important synergistic effects have been established.
Topics under study comprise the physics of weak and strong
electroweak symmetry breaking, electroweak and QCD precision physics, the 
phenomenology of supersymmetric models, new gauge theories and models with 
extra dimensions. For scenarios where detailed experimental simulations
of the possible measurements and the achievable accuracies are available
both for the LHC and LC, the LHC / LC interplay could be investigated in a
quantitative manner. In other scenarios the assessment of the current
situation has revealed the need for further experimental
simulations at the LHC and LC as input for studying the interplay
between the two machines. Thus, the present work of the LHC and LC physics
groups will serve as an important input for
future LHC / LC analyses. 


\section{Overview of the LHC / LC Study}

In the following, a brief overview of the
results obtained in this working group report is given.

\subsection{Electroweak symmetry breaking}

Revealing the mechanism of electroweak symmetry breaking will be the
central issue for the LHC and LC. 
Within the SM, the mass of the Higgs boson is a free parameter.
The comparison of the SM predictions with the
electroweak precision data point towards a light Higgs
boson with $m_h \lsim 250$~GeV~\cite{ewfits}. In the Minimal Supersymmetric
extension of the Standard Model (MSSM) the mass of the lightest CP-even
Higgs boson can be directly predicted from the other parameters of the
model, yielding an upper bound of 
$m_h \lsim 140$~GeV~\cite{execsumm_mhboundMSSM}.
The MSSM
predicts four other fundamental Higgs bosons, $H$, $A$ and $H^{\pm}$. The
Higgs sectors of the SM and the MSSM are the most commonly studied
realisations of electroweak symmetry breaking. 

However,
the electroweak precision data do not exclude the possibility
of a Higgs sector with unconventional properties. In particular, new
physics contributions to electroweak precision observables can in
principle compensate the effects of a heavy Higgs boson, mimicking in
this way the contribution from a light SM-like Higgs boson. While a
light Higgs boson is also required
in extensions of the MSSM, the
properties of such a light Higgs boson (and also the other states in the Higgs 
sector) can significantly differ from the MSSM.

Furthermore, the possibility that no fundamental Higgs particle exists
has to be investigated. This necessitates, in particular, the study of 
scenarios
where electroweak symmetry breaking occurs as a consequence of a new strong
interaction.

In the following, four different scenarios of electroweak symmetry
breaking will be discussed and the impact of the LHC / LC interplay
will be highlighted.

\subsubsection{Scenarios with a light SM-like or MSSM-like Higgs boson}


If a state resembling a Higgs boson is detected, it is crucial to
experimentally test its nature as a Higgs boson. To this 
end the couplings of the new state to as many particles as possible must 
be precisely determined, which requires observation of the candidate
Higgs boson in several different production and decay channels.
Furthermore the spin and the other CP-properties of the new state need to 
be measured, and it must be clarified whether there is more than one Higgs 
state. The LHC will be able to address some of these questions, but in
order to make further progress a comprehensive programme of precision
Higgs measurements at the LC will be necessary. The significance of the
precision Higgs programme is particularly evident from the fact that many
extended Higgs theories over a wide part of their parameter space have a
lightest Higgs scalar with nearly identical properties to those of the
SM Higgs boson. In this so-called decoupling limit additional states of
the Higgs sector are heavy and may be difficult to detect both at the LHC and
LC. Thus, precision measurements are crucial in order to distinguish the
SM Higgs sector from a more complicated scalar sector. In this way the
verification of small deviations from the SM may be the path to
decipher the physics of electroweak symmetry breaking. 

While the LC will
provide a wealth of precise experimental information on a light Higgs
boson, the LHC may be able to detect heavy Higgs bosons which lie
outside the kinematic reach of the LC (it is also possible, however,
that the LC will detect a heavy Higgs boson that is not experimentally
observable at the LHC due to overwhelming backgrounds). 
Even in the case where only one
scalar state is accessible at both colliders important synergistic effects 
arise from the
interplay of LHC and LC. This has been demonstrated for the example 
of the Yukawa coupling of the Higgs boson to a pair of top quarks. 
At a 500~GeV LC
the $t \bar th$ coupling can only be measured with limited precision
for a light Higgs boson $h$. The LHC will provide a measurement
of the $t \bar th$ production cross section times the decay branching ratio 
(for $h \to b \bar b$ or $h \to W^+W^-$). The LC, on the other hand, will
perform precision measurements of the decay branching ratios.
Combining LHC and LC information will thus allow one to extract the top
Yukawa coupling.
A similar situation occurs for the determination
of the Higgs-boson self-coupling, which is
a crucial ingredient for the reconstruction of the Higgs potential. The
measurement of the Higgs self-coupling at the LHC will require precise
experimental information on the top Yukawa coupling, the $hWW$ coupling
and the total Higgs width, which will be available with the help of the
LC.
An important synergy between LHC and LC results would even exist if
nature had chosen the (very unlikely) scenario of just a SM-type Higgs boson
and no other new physics up to a very high scale.
The precision measurements
at the LC of the Higgs-boson properties as well as of electroweak
precision observables, the top sector, etc.\ (see Sec.~\ref{sec:SMgauge}),
together with the exclusion bounds from the direct searches at the LHC
would be crucial to verify that the observed particles are sufficient for
a consistent description of the experimental results.

The LHC and LC can successfully work together in determining the CP
properties of the Higgs bosons. In an extended Higgs-sector with
CP-violation there is a non-trivial mixing between all neutral Higgs states.
Different measurements at the LHC and
the LC (both for the electron--positron and the photon--photon collider
option) have sensitivity to different coupling parameters. In the
decoupling limit, the lightest Higgs boson is an almost pure CP-even
state, while the heavier Higgs states may contain large CP-even and
CP-odd components.
Also in this case, high-precision measurements of the properties of
the light Higgs boson at the LC may reveal small deviations from the SM
case, while the heavy Higgs bosons might only be accessible at the LHC.
In many scenarios, for instance the MSSM, CP-violating effects are
induced via loop corrections. The CP properties therefore depend on the
particle spectrum. The interplay of precision measurements in the Higgs
sector from the LC and information on, the SUSY spectrum
from the LHC can therefore be important for revealing the CP structure.
As an example, if CP-violating effects in the Higgs sector in a SUSY
scenario are established at the LC, one would expect CP-violating
couplings in the scalar top and bottom sector. The experimental strategy at the
LHC could therefore focus on the CP properties of scalar tops and bottoms.


In supersymmetric theories Higgs boson masses can directly be predicted
from other parameters of the model, leading, for instance, to the upper
bound of $m_h \lsim 140$~GeV~\cite{execsumm_mhboundMSSM} for the mass of the
lightest CP-even Higgs boson of the MSSM. A precise determination of the
Higgs masses and couplings therefore gives important information about
the parameter space of the model. If at the LHC the $h \to \gamma\gamma$
decay mode is accessible, the LHC will be able to perform a first
precision measurement in the Higgs sector by determining the Higgs-boson 
mass with an accuracy of about 
$\Delta m_h^{\rm exp} \approx 200$~MeV~\cite{lhcexp}. As a consequence
of large radiative corrections from the top and scalar top sector of
supersymmetric theories, the
prediction for $m_h$ sensitively depends on the precise value of the
top-quark mass. For the lightest CP-even Higgs-bosons of the MSSM, an
experimental error of 1~GeV in $m_t$ translates into a theory
uncertainty in the prediction of $m_h$ of also about 1~GeV. As a
consequence, the experimental accuracy of the top-mass measurement
achievable at the LHC, $\Delta m_t^{\rm LHC} \approx
1$--2~GeV~\cite{mtexpLHC}, will not be sufficient to exploit the high
precision of the LHC measurement of $m_h$. Thus, in order to match the
experimental precision of $m_h$ at the LHC with the accuracy of the
theoretical prediction, the precise measurement of the top-quark mass at the
LC, $\Delta m_t^{\rm LC} \lsim 0.1$~GeV~\cite{mtexpLC}, will be
mandatory. 

If the uncertainty in the predictions for observables in the Higgs
sector arising from the experimental error of the top-quark mass 
is under control (and the theoretical predictions are sophisticated
enough so that uncertainties from unknown higher-order corrections are
sufficiently small), one can make use of precision measurements in the
Higgs sector to obtain constraints on the masses and couplings of the
SUSY particles that enter in the radiative corrections to the Higgs
sector observables. The results can be compared with the direct
information on the SUSY spectrum. In general, the LHC and LC are sensitive to
different aspects of the SUSY spectrum, and both machines will provide
crucial input data for the theoretical interpretation of the precision
Higgs programme.
In a scenario where the LHC and LC only detect one light Higgs boson,
precision measurements of its properties at the LC allow to set indirect
limits on the mass scale of the heavy Higgs bosons, provided that
combined information from the LHC and LC on the SUSY spectrum is available.
If the heavy Higgs bosons are directly accessible at the LHC, the
combination of the information on the heavy Higgs states at the LHC with
the LC measurements of the mass and branching ratios of the light Higgs
will allow one to obtain information on the scalar quark sector of the
theory. In particular it was demonstrated that the trilinear coupling $A_t$ 
of the Higgs bosons to the scalar top quarks
can be determined in this way. 

A promising possibility for the detection
of heavy Higgs states at the LHC is from the decay of heavy Higgs bosons
into SUSY particles, for instance a pair of next-to-lightest
neutralinos. The next-to-lightest neutralino will decay into leptons and
the lightest neutralino, which will escape undetected in most SUSY
scenarios. For this case it was demonstrated that the reconstruction of the 
mass of
the heavy Higgs bosons at the LHC requires as input the precision
measurement of the mass of the lightest neutralino at the LC.

A fundamental parameter in models with two Higgs doublets (e.g., the
MSSM) is $\tan\beta$, the ratio of the vacuum expectation
values of the two Higgs doublets. It not only governs the Higgs sector, but
is also important in many other sectors of the theory. A precise 
experimental determination of this parameter will be difficult, and it
seems very unlikely that it will be possible to extract $\tan\beta$ from
a single observable. Instead, a variety of measurements at the LHC and
LC will be needed to reliably determine $\tan\beta$. The measurements involve
observables in the Higgs sector, 
the gaugino sector and the scalar tau sector as well as information on
the SUSY spectrum.

\subsubsection{Higgs sector with non-standard properties}

While the most studied Higgs boson models are the SM and the MSSM, more
exotic realisations of the Higgs sector cannot be ruled out. Thus, it is
important to explore the extent to which search strategies need to be
altered in such a case.

A possible scenario giving rise to non-standard properties of the Higgs
sector is the presence of large extra dimensions, motivated for instance 
by a ``fine-tuning'' and ``little hierarchy'' problem of the MSSM.
A popular class of such models comprise those
in which some or all of the SM particles live on 3-branes in the extra
dimensions. Such models inevitably require the existence of a radion
(the quantum degree associated with fluctuations of the distance
between the three branes or the size of the extra dimension(s)).

The radion has the same quantum numbers as the Higgs boson
and in general the two will mix.  Since the radion has couplings that
are very different from those of the SM Higgs boson, the two physical
eigenstates will have  unusual properties corresponding to
a mixture of the Higgs and radion
properties; the prospects for detecting them
at the LHC and LC must be carefully analysed. 
One finds that there are significant portions
of the parameter space for which it will
not be possible to observe the Higgs-like $h$ state at the LHC.
For most of this region, the radion-like $\phi$ state will be
observable in the process $gg\to \phi \to ZZ^* \to 4\ell$, leading
thus to a situation where one scalar will be detected at the LHC.
Disentangling the nature of this scalar state will be a very important
but experimentally challenging task.

For instance, if the LHC observes a scalar state with a non-SM-like
production or decay rate, it will be unclear from LHC data alone
whether this is due
to mixing with a radion from extra dimensions or due to the presence
of an extended Higgs sector, such as that predicted by the MSSM
or its most attractive extension, the Next-to-Minimal Supersymmetric
Model (NMSSM), which has two more neutral Higgs bosons.
The difficulty in interpreting the LHC experiments
would also be severe if an intermediate-mass scalar, with a mass above
the SM bound from electroweak precision tests (e.g.\ $m\sim 400~$GeV),
is observed alone. It will then be very challenging to determine
whether the observed particle is the radion (with the Higgs particle
left undetected), a heavy Higgs boson within a multi-doublet Higgs sector 
(with additional contributions to precision electroweak observables
that compensate for the non-standard properties of the observed
scalar) or something else.

In the above scenarios, the LC can observe both the Higgs boson and the
radion, and covers most of the parameter space where detection of either
state at the LHC is difficult.
The Higgs--radion mixing effect would give rise
to the same shift in the Higgs couplings $g_{hWW}$, $g_{hZZ}$ and
$g_{h\bar f f}$. Thus, ratios of couplings would remain unperturbed and
correspond to those expected in the SM. Since the LHC will measure
mostly ratios of couplings, the Higgs--radion mixing could easily be
missed. The LC, on the other hand, has the capability to measure the
absolute values of the couplings to fermions and gauge bosons with
high precision. Furthermore, an accurate determination of the total
Higgs width will be possible at the LC. These capabilities are crucial
in the described scenario, since there would be enough measurements and
sufficient accuracy to experimentally establish the
Higgs--radion mixing effects. It has been demonstrated in
a detailed analysis that the parameter regions for which the Higgs
significance is below $5 \sigma$ at the LHC are covered by the regions where 
precision measurements of Higgs couplings at the LC establish the Higgs--radion
mixing effect. The LHC, on the other hand,
will easily observe the distinctive signature of 
Kaluza--Klein graviton excitation production in these scenarios
over a substantial range of the radion vacuum expectation value,
$\Lambda_\phi$.

A case where the LHC detects a heavy (500~GeV--1~TeV) SM-like Higgs
boson rather than a light CP-even Higgs boson as apparently needed to
satisfy precision electroweak constraints can also occur in a general
two-Higgs-doublet model. The source of the extra contributions mimicking
the effect of a light Higgs boson in the electroweak precision tests
may remain obscure in this case. The significant improvement in the
accuracy of the electroweak precision observables obtainable at the LC
running in the GigaZ mode and at the $WW$ threshold will be crucial to
narrow down the possible scenarios.

If nature has chosen a scenario far from the decoupling limit,
even in the case of the MSSM, electroweak symmetry breaking dynamics
produces no state that closely resembles the SM Higgs boson. Within the
MSSM a situation is possible where the neutral Higgs
bosons are almost mass-degenerate and $\tan\beta$ is large. In this
case,
detection of the individual Higgs boson peaks is very challenging at the
LHC, whereas the different Higgs boson signals can more easily be
separated at the LC. The measured characteristics at the LC will then
allow to determine further Higgs-boson properties at the LHC.

Another situation which was investigated in this report is the case of a
fermiophobic Higgs boson ($h_f$)
decaying to two photons with a larger branching ratio than in the
SM. In this case the standard Higgs
production mechanisms are very much suppressed
for moderate to large $\tan\beta$, both at the LC and the LHC.
It has been shown that the search for $pp\to H^\pm h_f$ should
substantially benefit from a previous signal at a LC in the channel
$e^+e^-\to A h_f$, and would provide important confirmation of any LC
signal for $h_f$.

There are many other scenarios where Higgs detection at the LHC can be
difficult, or the Higgs signal, while visible, would be hard to interpret. 
If no clear Higgs signal has been established at the LHC, it
will be crucial to investigate with the possibilities of the LC whether
the Higgs boson has not been missed at the LHC because of its non-standard
properties. This will be even more the case if the gauge sector doesn't
show indications of strong electroweak symmetry breaking dynamics. The
information obtained from the LC can therefore be crucial for
understanding the physics of mass generation and for guiding the future 
experimental programme in high-energy physics.
The particular power of the LC is its ability to look for $e^+e^- \to ZH$
in the inclusive $e^+e^- \to ZX$ missing-mass, $M_X$, distribution
recoiling against the $Z$ boson. Even if the Higgs boson decays completely
invisibly or different Higgs signals overlap in a complicated way, the
recoil mass distribution  will reveal the Higgs boson mass spectrum
of the model.


An example studied in this context is a scenario where the Higgs boson 
decays primarily into hadronic jets, possibly without definite 
flavor content. Such a situation could be realised for instance in the
MSSM if the scalar bottom quark turns out to be very light. A light
Higgs boson decaying into jets, undetected
at the LHC, could thus lead one to conclude erroneously that the
Higgs sector has a more exotic structure than in the MSSM.
Such a state could be discovered at the LC and 
its properties measured with high accuracy.
The LHC, on the other hand, should be able to produce, discover, and 
study in great detail possible new physics at the weak scale 
(the superpartners in the SUSY example).
In order to truly understand electroweak symmetry breaking and
the solution of the hierarchy problem, the synergy of the LHC and the LC
is crucial. As a
further possibility, one might produce superpartners at the LHC that
decay through light Higgs bosons as intermediate states into jets and 
not realise the identity of the 
intermediate states. In such a situation, it might even
be impossible to identify the parent superparticles, despite their
having rather ordinary properties from the point of view of the MSSM.
The analysis and understanding of data from concurrent operation of
the LHC and LC would very likely prove crucial.

New Higgs boson decay modes can also open up in extensions of the MSSM,
for instance the NMSSM.
A case in which Higgs detection may be difficult occurs if
there is a light (CP-even) Higgs boson
which dominantly decays into two light CP-odd Higgs
bosons, $h\to aa$. Confirmation of the nature
of a possible LHC signal at the LC would be vital. For example,
the $WW\to h\to aa$ signal, as well
as the usual $e^+e^-\to Z h\to Z aa$ signal,
will be highly visible at the
LC due to its cleaner environment and high luminosity. The LC will
furthermore be able to measure important properties of the CP-odd
scalar.

Another challenging NMSSM scenario is a singlet dominated light Higgs.
While this state has reasonably large production cross sections at the
LHC, it would be difficult to detect as it mainly decays hadronically. 
Such a state could be discovered at the LC. From the measurement of its
properties, the masses of the heavier Higgs bosons could be predicted,
guiding in this way the searches at the LHC. For a very heavy singlet
dominated Higgs state, on the other hand, the kinematic reach of the LHC
will be crucial in order to verify that a non-minimal Higgs sector is
realised. Thus, input from both the LHC and the LC will be needed in
order to provide complete coverage over the NMSSM  parameter space.

Another very difficult scenario for Higgs boson detection would be the 
case of a ``continuum'' Higgs model, i.e.\ a large number of doublet and/or
singlet fields with complicated self interactions. This could result in
a very significant diminution of all the standard LHC signals. The
missing-mass signal from the LC will be crucial in this case to guide
the search strategy at the LHC. In all cases with Higgs properties such
that the Higgs boson remains undetected at the LHC, experimental
information from the LC 
will be crucial in order
to identify the phenomenology responsible for making the Higgs boson
``invisible'' at the LHC.

\subsubsection{The Little Higgs scenario}

New approaches to electroweak symmetry breaking dynamics have led to
pheno\-meno\-logies that may be quite different from the conventional
expectations of weakly coupled multi-Higgs models.

Little Higgs models revive an old idea to keep the Higgs boson naturally
light: they make the Higgs particle a pseudo-Nambu-Goldstone boson of a
broken global symmetry. The new ingredient of little Higgs models is
that they are constructed in such a way that at least two
interactions are needed to explicitly break all of the global symmetries
that protect the Higgs mass.  Consequently, the dangerous 
quadratic divergences in the
Higgs mass are forbidden at one-loop order. In this way a cutoff scale 
$\Lambda \approx 10$~TeV could be naturally accommodated, solving the
``little hierarchy'' problem. 

The phenomenology of Little Higgs models is normally very rich, giving
rise to new weakly coupled fermions, gauge bosons and scalars at the TeV
scale. The LHC has good prospects in such a scenario to discover new heavy 
gauge bosons and the vector-like partner of the top quark.
The LC has a high sensitivity to deviations in the
precision electroweak observables and in the triple gauge boson couplings,
and to loop effects of the new heavy particles on the Higgs boson
coupling to photon pairs. Therefore, both the LHC direct observations
and the LC indirect measurements will be important to clarify
the underlying new physics. If only part of
the new states are detectable at the LHC, the high-precision
measurements at the LC may allow to set indirect constraints on the
masses of new states, indicating in this way a possible route for
upgrades in luminosity or even energy for a subsequent phase of LHC
running.

\subsubsection{No Higgs scenarios}

If no light Higgs boson exists, quasi-elastic scattering processes of
$W$ and $Z$ bosons at high energies provide a direct probe of the dynamics
of electroweak symmetry breaking. The amplitudes can be measured in 
6-fermion processes both at the LHC and the LC. The two colliders are 
sensitive to different scattering channels and yield complementary 
information.  

The combination of LHC and LC data will considerably increase the LHC
resolving power. In the low-energy range it will be possible to
measure anomalous triple gauge couplings down to the natural value of
$1/16\pi^2$. The high-energy region where resonances may appear can be
accessed at the LHC only. The LC, on the other hand, has an indirect
sensitivity to the effects of heavy resonances even in excess of the
direct search reach of the LHC. Detailed measurements of cross
sections and angular distributions at the LC will be crucial for
making full use of the LHC data. In particular, the direct
sensitivity of the LHC to resonances in the range above 1~TeV can be
fully exploited if LC data on the cross section rise in the region
below 1~TeV are available. In this case the LHC measures the mass of the
new resonances and the LC measures their couplings.
Furthermore, the electroweak precision measurements (in particular from
GigaZ running) at the LC will be crucial to resolve the conspiracy that 
mimics a light Higgs in the electroweak precision tests. 
Thus, a thorough understanding of the 
data of the LC and the LHC combined will be essential for 
disentangling the new states and identifying the underlying physics. 

Besides the mechanism of strong electroweak symmetry breaking, recently 
Higgs\-less models have been proposed in the context of higher-dimensional
theories. In such a scenario boundary conditions on a brane in a warped 5th 
dimension are responsible for electroweak symmetry breaking.
The unitarity of $WW$ scattering is maintained so
long as the KK excitations of the $W$ and $Z$ are
not much above the TeV scale and therefore
accessible to direct production at the LHC. Experimental information
from the LC, in particular electroweak precision measurements, will be 
important in this case in order to correctly identify the underlying
physics scenario.

\subsection{Supersymmetric models}

Experimental information on the masses and couplings of the largest
possible set of supersymmetric particles is the most important input to the
reconstruction of a supersymmetric theory, in particular of the SUSY 
breaking mechanism. The lightest supersymmetric particle (LSP) is an
attractive candidate for cold Dark Matter in the Universe.
A precise knowledge about the SUSY spectrum and the 
properties of the SUSY particles will be indispensable in order to
predict the Dark Matter relic density arising from the LSP (see
Sec.~\ref{sec:lhclcintro_dm} below). 

The production of
supersymmetric particles at the LHC will be dominated by the production
of coloured particles, i.e.\ gluinos and squarks.
Searches for the signature of jets and missing energy at the LHC will cover
gluino and squark masses of up to 2--3~TeV~\cite{lhcexp}. The main
handle to detect uncoloured particles will be
from cascade decays of heavy gluinos and squarks, since in most 
SUSY scenarios the uncoloured particles are lighter
than the coloured ones. An example of a possible decay chain is
$\tilde g \to \bar q \tilde q \to \bar q q \tilde\chi_2^0 \to
\bar q q \tilde\tau \tau \to \bar q q \tau\tau \tilde\chi_1^0$,
where $\tilde\chi_1^0$ is assumed to be the LSP.
Thus, fairly long decay chains giving rise to the production of several
supersymmetric particles in the same event and leading to rather
complicated final states can be expected to be a typical feature of SUSY 
production at the LHC. In fact, the main background for measuring SUSY
processes at the LHC will be SUSY itself.

The LC, on the other hand, has good prospects for the production of
the light 
uncoloured particles. The clean signatures and small backgrounds at the
LC as well as the possibility to adjust the energy of the collider to
the thresholds at which SUSY particles are produced will allow a precise
determination of the mass and spin of supersymmetric particles and of
mixing angles and complex phases~\cite{lctdrs}. 

In order to establish SUSY experimentally, it will be necessary to
demonstrate that every particle has a superpartner, that their spins
differ by $1/2$, that their gauge quantum numbers are the same, that
their couplings are identical and that certain mass relations hold.
This will require a large amount of experimental information, in
particular precise measurements of masses, branching ratios,
cross sections, angular distributions, etc. A precise knowledge of as
many SUSY parameters as possible will be necessary to disentangle the
underlying pattern of SUSY breaking. In order to carry out this physics
programme, experimental information from both the LHC and the LC will be
crucial.

\subsubsection{Measurement of supersymmetric particle masses, mixings and
couplings at LHC and LC}

As mentioned above, at the LHC the dominant production mechanism is pair 
production of
gluinos or squarks and associated production of a gluino and a squark. 
For these processes, SUSY particle masses have to be determined from the 
reconstruction of long decay chains which end in the production of the
LSP. The invariant
mass distributions of the observed decay products exhibit thresholds and 
end-point structures. The kinematic structures can in turn be expressed
as a function of the masses of the involved supersymmetric 
particles. The LHC is
sensitive in this way mainly to mass {\em differences}, resulting in a
strong correlation between the extracted particle masses. In particular,
the LSP mass is only weakly constrained. This uncertainty propagates
into the experimental errors of the heavier SUSY particle masses.

At the LC, the colour--neutral part of the SUSY particle spectrum
can be reconstructed with high precision if it is kinematically accessible. 
It has been demonstrated that
experimental information on properties of colour--neutral particles from
the LC can significantly improve the analysis of cascade decays at the
LHC. In particular, the precise measurement of the LSP mass at the LC 
eliminates a large source of uncertainty in the LHC analyses. This leads
to a substantial improvement in
the accuracy of the reconstructed masses of the particles in the decay
chain.

In general LC input will help to significantly reduce the model
dependence of the LHC analyses. Intermediate states that appear in the
decay chains detected at the LHC can be produced directly and
individually at the LC. Since their spin and other properties can be
precisely determined at the LC, it will be possible to unambiguously
identify the nature of these states as part of the SUSY spectrum. In
this way it will be possible to verify the kind of decay chain observed at the
LHC. The importance of this has been demonstrated, for instance, in a
scenario with sizable flavour-changing decays of the squarks.
Once the particles in the lower parts of the decay cascades have
been clearly identified, one can include the MSSM predictions for their
branching ratios into a constrained fit. This will be helpful in order
to determine the couplings of particles higher up in the decay chain.

Most of the studies of the LHC / LC interplay in the reconstruction of
SUSY particle masses carried out so far have been done for one
particular MSSM benchmark scenario, the SPS~1a benchmark
point~\cite{intro_sps}, since only for this benchmark point detailed
experimental simulations are available both for the LHC and LC. 
As an example, the scalar top and bottom mixing angles can be extracted from
the reconstructed scalar bottom masses from cascade decays and the measurement
of ratios of branching ratios at the LHC in the SPS~1a scenario,
provided that precise
information on the parameters in the neutralino and chargino mass
matrices is available from the LC.

A detailed study of important synergistic effects between LHC and LC has
been carried out in the gaugino sector within the SPS~1a scenario. In this 
analysis the 
measurements of the masses of the two lightest neutralinos, the lighter
chargino, the selectrons and the sneutrino at the LC were used to predict
the properties of the heavier neutralinos. It was demonstrated that this
input makes it possible to identify the heaviest neutralino at the LHC
and to measure its mass with high precision. Feeding this information back
into the LC analysis significantly improves the determination of the 
fundamental SUSY parameters from the neutralino and chargino sector at 
the LC. 

The described analysis is a typical example of LHC / LC synergy.
If a statistically not very pronounced (or even marginal)
signal is detected at the LHC, input from the LC can be crucial in order
to identify its nature. In fact, the mere existence of a LC prediction
as input for the LHC searches increases the statistical sensitivity of
the LHC analysis. This happens since a specific hypothesis is tested,
rather than performing a search over a wide parameter space. In the
latter case, a small excess {\em somewhere} in the parameter space is
statistically much less significant, since one has to take into account
that a statistical fluctuation is more likely to occur in the
simultaneous test of many mass hypotheses. On the other hand, if the LHC
does not see a signal which is predicted within the MSSM from LC input,
this would be an important hint that the observed particles cannot be
consistently described within the minimal model. 

Beyond the enhancement of the statistical sensitivity, predictions based
on LC input can also give important guidance for dedicated searches at
the LHC. This could lead to an LHC analysis with optimised cuts or even
improved triggers. LC input might also play an important role in the
decision for upgrades at later stages of LHC running. For instance, the 
prediction of states being produced at the LHC with very small rate
could lead to a call for an LHC upgrade with higher luminosity.

In order to establish SUSY experimentally and to determine the
SUSY-breaking patterns, it is necessary to accurately measure as many
Lagrangian parameters as possible. Since most observables depend on a variety 
of parameters, one will have to perform a global fit of the SUSY model
to a large number of experimental observables. As the measurements at the 
LHC and the LC in general probe different sectors of the MSSM Lagrangian, 
the combination of LHC and LC data will be crucial in order to obtain
comprehensive information on the underlying structure of the model. For
the studied cases, it has
been demonstrated that only the combination of measurements of both the
LHC and the LC offers a complete picture of the MSSM model parameters in
a reasonably model independent framework. In fact, it turned out that
attempts to fit only individual sectors of the theory are unsuccessful,
and a converging fit is only obtained from the combination of LHC and LC
data.

As mentioned above, 
most of the studies of the LHC / LC interplay in the reconstruction of
SUSY particle masses carried out so far have been done for the SPS~1a
benchmark scenario~\cite{intro_sps}. The SPS~1a
benchmark point is a favourable scenario both for the LHC and LC. The
interplay between the LHC and LC could be qualitatively rather different in 
different regions of the MSSM parameter space. It seems plausible that
synergistic effects from the LHC / LC interplay will be even more important
in parameter regions which are more challenging for both colliders. 
In order to allow a quantitative assessment of the LHC / LC
interplay also for other parameter regions, more experimental
simulations for the LHC and LC are required.


\subsubsection{Distinction between different SUSY-breaking scenarios and
extrapolation to physics at high scales}

The importance of a precise knowledge of the fundamental SUSY parameters
for discriminating between different SUSY-breaking scenarios has been
demonstrated for several examples. In general, a per-cent level accuracy
appears to be mandatory in order to have a suitable sensitivity to
discriminate between different scenarios. This will require detailed
experimental information from both the LHC and LC.

The interplay between the LHC and LC is also important 
for the determination of the nature of the lightest and 
next-to-lightest SUSY particle (LSP and NLSP). In a scenario where the
long-lived NLSP is a charged slepton, methods have been established to
discover a massive gravitino, and thereby supergravity, at the LHC and
LC. It is crucial to verify supergravity predictions for the NLSP
lifetime as well as angular and energy distributions in 3-body NLSP decays.
With the gravitino mass inferred from kinematics, the measurement of the
NLSP lifetime will test an unequivocal prediction of supergravity. 
It has been demonstrated that the characteristic couplings of
the gravitino, or goldstino, can be tested even for very small masses.

Combining the experimental results from the LHC and LC, 
stable extrapolations can be performed from the
electroweak scale to the grand unification scale, 
provided that the low-energy spectrum can be fully reconstructed.
This has been done by studying the evolution
of the three gauge couplings and of the soft supersymmetry breaking
parameters, which approach universal values at the GUT scale in 
minimal supergravity scenarios. For the example of the SPS~1a benchmark
point it has been shown that from LHC data alone no reliable
extrapolation to the GUT scale can be performed. The coherent analyses
based on combined information from the LHC and LC, in which the measurements
of SUSY particle
properties at the LHC and LC mutually improve each other, result in a
comprehensive and detailed picture
of the supersymmetric particle system. In particular, the gaugino sector
and the non-coloured scalar sector are under excellent control. 

Though minimal supergravity has been chosen as a specific example, the 
methodology can equally well be applied to more general supersymmetric 
theories.
High-precision high-energy experiments at the LHC
and LC, providing accuracies at the level of per-cent to per-mille,
allow a thorough analysis of the mechanism of supersymmetry breaking and
give access to the structure of nature
at scales where gravity is linked with particle physics.


\subsection{Gauge theories and precision physics}

\subsubsection{Standard Model gauge sector}

\label{sec:SMgauge}

A detailed analysis of the properties in the gauge sector is important
for determining the structure of the underlying physics and for
distinguishing between different models. Examples of extensions of the
SM that however have the same gauge sector as the SM are the MSSM or
more general two-Higgs-doublet models.  

Models possessing the same gauge sector can be distinguished via 
quantum corrections that are influenced by the whole structure of the
model. The LC provides precision data obtained from running at the 
$t \bar t$ threshold, from fermion pair production at high energies, from 
measurements in the Higgs sector, etc. Furthermore, running the LC in
the GigaZ mode yields extremely precise information on the effective weak 
mixing angle, the total $Z$-boson width, the $Z$ partial widths and the mass 
of the $W$~boson (the latter from running at the $WW$ threshold).
Comparing these measurements with the predictions of different models 
provides a very sensitive test of the theory, in the same way as many 
alternatives to the SM have been found to be in conflict with the electroweak
precision data in the past. In this way, the electroweak precision tests 
can give access to effects of heavy particles which are
beyond the direct reach of the LHC and the LC. 
However, in order to fully exploit the sensitivity to new physics, as
much information as possible is necessary about the part of the spectrum
that is directly accessible experimentally. This will most likely
require measurements from both the LHC and the LC.

Interesting LHC / LC synergy can also be expected in the
determination of the self-couplings of the gauge bosons. 
The couplings among the gauge bosons can be measured at
both colliders independently. Therefore the combination of the uncorrelated
LHC and LC measurements may lead to a significantly higher accuracy
than the individual measurements. 
Deviations from 
the prediction of a SM-like gauge sector could reveal the existence of new 
(and so far unknown) high mass scales.

The physics of the top quark plays an important role as a possible window
to new physics. The top quark is the heaviest elementary particle found so 
far. Since it decays much faster than the typical time for formation of top 
hadrons, it provides a clean source of fundamental information.
Accurate measurements of the top quark properties, such as its mass, 
couplings, in particular the couplings to gauge bosons and Higgs fields,
and branching ratios of rare decay modes, probe possible deviations from 
the SM predictions in a sensitive way. Since there are many
possibilities of anomalous couplings of the top quark to gauge bosons,
one will greatly benefit from the combined results of the LHC and LC. 
The analysis of spin correlations between production and decay of the
top quarks is of particular interest in this context.
For example, the top-quark pair production at the LHC involves
the strong coupling of gluons to the top quark. In order to probe deviations 
from 
the SM structure for this coupling via spin correlations one needs to take 
into account the information on the $Wtb$ coupling structure which occurs 
in the top-quark decay. 
The latter can be accurately measured at the LC. Such an 
information from the LC could also be useful for measuring the $b$-quark 
structure function via single top-quark production at the LHC.

An interesting interplay between the LHC and LC can also occur in QCD
analyses. For instance, the clean experimental environment at the LC
will allow important measurements relevant for determining fragmentation 
functions. This information can lead to an improvement of the
understanding of two-photon events at the LHC.

\subsubsection{New gauge theories}

Many kinds of extensions of the SM lead to an enlarged gauge-boson
sector. Determining the nature of the new gauge bosons 
will require a variety of detailed experimental results
that can be provided by the interplay of the LHC and LC. The LHC has a
large mass reach for direct detection of new gauge bosons, while the LC
has a large indirect reach arising from virtual effects of the new
states that result in deviations from the SM predictions. The indirect
search reach of the LC is substantially larger than the direct discovery
reach achievable at the LHC.

The LC is sensitive to $Z$--$Z'$ interference effects through the
fermion pair-production process, $e^+e^- \to f \bar f$, running at its highest
energy. If the mass of the $Z'$ is known from the LHC, the LC information
on the ratio of the $Z'f\bar f$ couplings and the $Z'$ mass can be used to
determine the $Z'$ couplings with high precision.
Furthermore, the measurements of the electroweak precision
observables in the GigaZ mode of the LC
yield important information
for distinguishing different models of new physics. This
input can be helpful for optimising the search strategies at the LHC.

Careful analyses are required to distinguish a $Z'$ from
other possible manifestations of new physics, which can have a somewhat
similar phenomenology, but a completely different physical origin. An
example is the study of the lightest Kaluza--Klein (KK) 
excitations of the SM electroweak 
gauge bosons, which arise in models with large extra dimensions. 
Little Higgs models provide 
a class of models with an extended gauge sector.
Detailed experimental information is necessary in order to 
determine the structure of Little Higgs models from the properties of
the observed new particle states.
In all these cases, combined information 
from the LHC and LC can be crucial.

\subsection{Models with extra dimensions}

Collider signatures for the presence of extra spatial dimensions
are wide and varied, depending on the geometry of the additional
dimensions.  The basic signal is the observation of a 
KK tower of states corresponding to a particle propagating in the
higher dimensional space-time.  The measurement of the properties
of the KK states determines the size and geometry of the extra
dimensions.

In the scenario of large extra dimensions, where gravity alone
can propagate in the bulk, the indirect effects and direct
production of KK gravitons are both available at the LHC and at
the LC.  For the indirect effects of KK gravitons, the search
reach of the LC exceeds that of the LHC for an LC centre of mass energy
of $\sqrt s \gsim 800$~GeV.  
Measurement of the moments of the resulting angular
distributions at the LC can identify the spin-2 nature of the
graviton exchange. If positron polarisation is available, then 
azimuthal asymmetries can extend the search for graviton exchange 
by a factor of two, probing fundamental scales of gravity up to 21~TeV for 
$\sqrt s= 1$~TeV with 500 fb$^{-1}$ of integrated luminosity.
In the case of direct KK graviton production, the LHC and LC
have comparable search reaches.  However, the LHC is hampered
by theoretical ambiguities due to a break-down of the effective
theory when the parton-level centre of mass energy exceeds
the fundamental scale of gravity.
Measurements of direct graviton
production at two different centre of mass energies at the LC
can determine the number of extra dimensions, and the absolute
normalisation of the cross section can determine the fundamental
scale.  Simultaneous determination of all the model parameters
has been examined in a quantitative fashion with the result
that data from the LC and LHC analysed together substantially
improves the accuracy of this determination over the LHC data
taken alone.

Standard Model fields are allowed to propagate in extra dimensions
with size less than TeV$^{-1}$.  Signatures for the KK states of
the SM gauge fields mimic those for new heavy gauge bosons in
extended gauge theories.  The LHC may discover electroweak gauge
KK states via direct production in the mass range $M_c\simeq 4$--6
TeV (lower masses are excluded by LEP/SLC data), while indirect
detection at the LC is possible for $M_c\lsim 20$~TeV for
$\sqrt s=1$~TeV.  Indirect detection of electroweak gauge KK
states is also possible at the LHC for $M_c\lsim 12$~TeV via a
detailed study of the shape of the Drell-Yan lepton-pair
invariant mass distribution.  If discovered, 
the determination
of the mass of the first gauge KK excitation at the LHC, together
with indirect effects at the LC can be used to distinguish the
production of a KK gauge state from a new gauge field in
extended gauge sectors.

The possibility of universal extra dimensions, where all SM
fields are in the bulk, can be mistaken for the production
of supersymmetric states, since the KK spectrum and phenomenology
resembles that of supersymmetry.  In fact, the lightest KK
state is a Dark Matter candidate.  In this case, threshold
production of the new (s)particle at the LC can easily determine
its spin and distinguish universal extra dimensions from
supersymmetry.  Spin determination analyses are on-going for
the LHC.

Lastly, the presence of warped extra dimensions results in the
resonance production of spin-2 gravitons.  This produces a
spectacular signature at the LHC for the conventional construction
of the Randall-Sundrum model.  However, extensions to this model,
such as the embedding in a higher-dimensional manifold, or the
inclusion of kinetic brane terms, may result in reduced coupling strengths
and extremely narrow-width graviton KK states. Narrow resonances of this
kind, in particular if they are closely spaced, may be difficult to
disentangle in a general search at the LHC. This can be the case even
for very light KK states. Radiative return at the LC 
may pinpoint the existence of these states, which can then be
confirmed by a dedicated search at the LHC.

%


\subsection{The nature of Dark Matter}

\label{sec:lhclcintro_dm}

The properties of Dark Matter as understood today imply that
it should be stable, cold, 
and non-baryonic. This behaviour is not compatible with the particles and
interactions of the SM. Thus, the existence of Dark Matter, which
is strongly implied from cosmological observations, is unambiguous
evidence for new physics. 

It seems suggestive that the physics giving rise to Dark Matter is
related to the weak scale, possibly closely tied to the origin of
electroweak symmetry breaking. Thus, there is an intriguing possibility
at the LHC and LC that Dark Matter particles will be produced 
in the laboratory. 

Several of the new physics scenarios discussed above provide possible
Dark Matter candidates that give rise to an acceptable relic density. 
Supersymmetry offers the LSP as a very attractive candidate. In many
scenarios the lightest neutralino is the LSP. The cosmological 
implications of a
stable neutralino have been very thoroughly studied in the literature
for many years, and it has been shown that the constraints on the SUSY
parameter space from cosmology, direct particle searches, electroweak
precision tests and flavour physics can be simultaneously satisfied.
Besides a neutralino LSP, other manifestations of Dark Matter can 
occur in SUSY, such as the gravitino. Possible
explanations for Dark Matter have also emerged from models with extra 
dimensions, for instance `Kaluza--Klein Dark Matter', `warped Dark
Matter' or `branons' (see e.g.\ Ref.~\cite{Bertone:2004pz} for a recent
review).  

The collider signatures in different kinds of Dark Matter scenarios can
be rather similar, giving rise, for instance, to jets and missing energy
at the LHC. Measurements at both the LHC and the LC will be crucial in
order to identify the underlying
physics~\cite{Polesello:2004qy,Bambade:2004tq,Battaglia:2004mp,Janot:2004mw}. 
In particular it will be
important to determine the quantum numbers, the spin and the
interactions of the possible Dark Matter candidate. Precise measurements
of the properties of this particle are indispensable as input for
predicting the cold Dark Matter density, which is a decisive test of the 
hypothesis that a particular physics scenario is in fact the origin of
cold Dark Matter in the Universe. Studying decays of heavier particles into
the Dark Matter particle can provide a unique window to processes that
happened in the early Universe.

Thus, it is of utmost importance to obtain precise and 
comprehensive experimental
information on the Dark Matter candidate and its annihilation channels
in a model-independent way.
This experimental requirement is 
analogous to the experimental information necessary to disentangle the
mechanism of electroweak symmetry breaking and, within supersymmetric
models, the SUSY breaking mechanism. 

If the physics responsible for Dark Matter in the Universe is
accessible at the next generation of colliders, a major goal of these
colliders will be to
predict the Dark Matter density at the same level of accuracy as it can
be measured experimentally. The WMAP results~\cite{intro:wmap} have led
to a measurement of the Dark Matter content at the 10\% level. This
precision will further improve with the Planck satellite 
mission, scheduled for
2007, which aims at a 2\% measurement~\cite{intro:planck}. 

The impact of the LHC and LC for the precision of the predicted
Dark Matter density has mainly been studied for SUSY
scenarios so far. For the SPS~1a benchmark point, which as mentioned
above is a favourable scenario both for the LHC and LC, the prediction of
the Dark Matter density has been studied based on the experimental precision
achievable at the LHC. Under the
theoretical assumption that the minimal supergravity (mSUGRA)
scenario is realised in nature, 
an accuracy of about 3\% can be achieved for the full LHC-design
integrated luminosity~\cite{Polesello:2004qy,Janot:2004mw}. A
similar accuracy can be reached at the LC for the SPS~1a point 
{\em without} the mSUGRA assumption (and a significantly higher accuracy
if mSUGRA is assumed).
The prospects at the LHC appear to be much worse for less 
favourable scenarios~\cite{Janot:2004mw}. Possible experimental
strategies for these scenarios are currently assessed in detailed studies.

Precision measurements of the properties of the LSP at the LC in a
model-indepen\-dent way, in particular of its mass, its couplings and
its annihilation channels, will be crucial for a precise prediction of the
relic density that can be compared with the results from observational
cosmology. 
Even in the experimentally 
challenging co-annihilation scenario, a reliable prediction of the Dark
Matter density can be obtained at the LC, based on a model-independent
determination of the mass of the LSP and the lightest 
slepton~\cite{Bambade:2004tq}. An additional critical input from the LC is
the precise measurement of the top-quark mass. The dependence of the
relic density prediction on the precise value of $m_t$ can be very
pronounced if a particular SUSY-breaking scenario is assumed, since in
this case the top-quark mass enters via the renormalisation group
running from the high scale to the low-energy parameters.

In the case where rapid annihilation occurs through resonant
Higgs-boson exchange, 
the LC measurement of the LSP mass and combined LHC and LC information on
the Higgs-boson properties and the heavier neutralinos could be used to 
reconstruct the relevant SUSY parameters for predicting the Dark Matter
density~\cite{Allanach:2004xn}. 
In the mSUGRA region where the LSP has a significant Higgsino
fraction (focus point region), precise measurements of the neutralino
masses and of $\tan\beta$ will be crucial. While this scenario 
may be problematic for the LHC because of its very heavy sfermions, the LC 
sensitivity to the neutralino sector leads to a coverage up to very
large values of the common scalar mass parameter in the WMAP-allowed
region of the mSUGRA parameter space~\cite{Baer:2003ru}. 


Thus, the combined information from LHC and LC can be crucial in order
to disentangle possible collider signatures of Dark Matter and to
precisely determine its origin. This could be a breakthrough in the
quest to identify the fundamental composition of the Universe.



\chapter{Experimental Aspects of the LHC and LC}

{\it K.~Desch and F.~Gianotti}

\vspace{1em}


  The main experimental aspects of the
 operation at the LHC and at a 0.5-1~TeV Linear Collider are
 summarized in this Section.
Emphasis
 is given to the comparison between the two machines and their environments
leading to complementary experimental approaches to explore the new
 physics at the TeV scale.

 The LHC and LC have, to some extent, similar features as
previous hadron and electron colliders. 
  At the same time, they are much more powerful engines
 than their predecessors in terms of energy and luminosity, 
 which implies more difficult experimental environments than in the past
 and more challenging detectors.

The main asset of the LHC is its high
mass reach for direct discovery, which extends 
up to typically $\sim$6-7~TeV for singly-produced particles.
The high luminosity and the 
excellent expected performance of the experiments, in particular
their trigger capabilities, will allow a very large fraction
of New Physics signatures to be covered.
In most cases, the backgrounds to New Physics processes will
be measured by using the data itself. In particular, 
processes like e.g. W, Z, and top production will offer
high-statistics data samples to calibrate the detector and
understand standard physics (e.g. structure functions, 
higher-order QCD corrections).
In addition, it has been shown~\cite{Branson:2001ak,ATLAS,CMS} that 
several precise measurements of the new particles 
will be possible, which should provide first constraints 
on the underlying New Physics.

The main asset of the LC is two-fold: first, within the kinematically
accessible range determined by the center-of-mass energy, new particles can be directly
produced and their properties can be studied in great detail.
In particular new particles can be also dicovered if their cross
sections are fairly low
or if their decays are complicated, e.g.\ purely hadronic.
Second, high-precision measurements
exhibit sensitivity to new phenomena at
scales far above the center-of-mass energy due to virtual effects.
High precision, like at previous $e^+e^-$ machines, 
can be achieved thanks to the well known initial momenta,
which allows complete reconstruction of the final states
with high efficiency and resolution, and  to the absence
of event pile-up (see Section~\ref{sec:exp_environment}).
In contrast to LEP
a precise determination of the beamstrahlung spectrum is required as discussed below.
Both beams can be polarized longitudinally offering
the possibility to disentangle the helicity structure of SM and
New Physics processes.
The center-of-mass energy is tunable allowing for precise
mass and quantum number measurements from threshold scans.
Optional
high-luminosity running at the Z resonance (``Giga-Z'') and at the
$W^+W^-$ threshold, as well as $e^- e^-$, $\gamma\gamma$,
and $\gamma e^-$ collision modes, offer additional flexibility.

These very different experimental conditions and capabilities of LHC
and LC {\it together} allow for the discovery and structural understanding
of the new phenomena at the TeV energy scale and may even open up windows
to physics of Grand Unification of forces and Quantum Gravity.

The main experimental features of the LHC and a LC are discussed
and compared briefly below. More details can be found in
Refs.~\cite{Branson:2001ak,ATLAS,CMS,TESLA,NLCreport,ACFAreport}.

\section{The interaction rate and the environment}
\label{sec:exp_environment}
    The total inelastic proton-proton cross-section is about 
   80~mb at $\sqrt{s}$~=~14 TeV, therefore
   the event rate at the LHC is expected to be 10$^9$ interactions 
   per second when running at design luminosity (10$^{34}$ cm$^{-2}$ s$^{-1}$). 
   Since the bunch spacing is 25~ns, on average $\sim$25 soft 
   interactions (minimum-bias events) are expected to be produced
   at each bunch crossing. This ``pile-up" gives rise
   to e.g. $\sim$800 charged particles per crossing inside the detector 
   region used for tracking ($\vert\eta\vert~<~2.5$).
    
   The need of minimizing the impact of this huge pile-up on the physics 
   performance has had a major impact on the technological choices for and
   the design of ATLAS and CMS~\cite{ATLAS,CMS}, 
   leading to three main requirements: 
    
\begin{itemize}

\item  Fast detector response (typically 25-50~ns), 
    in order to integrate over two bunch-crossings at most and therefore
    to minimize the number of piled-up minimum-bias events. This implies
    novel-technology readout electronics. 
    
\item   Fine detector granularity, 
  in order to minimize the probability that pile-up particles 
  hit the same detector element as an interesting object 
  (e.g. a photon 
  coming from a $H\to\gamma\gamma$ decay). This implies a large 
  number of readout channels, and therefore a  challenging
  detector operation (e.g. in terms of calibration and monitoring). 
 
\item  High radiation resistance. The detectors are exposed to an intense
   flux of particles produced by the pp collisions at each bunch-crossing,
   which has to be integrated over at least ten years of operation. 
   For instance, in ten years the forward calorimeters will absorb neutron 
   fluences of up to $10^{17}$ neutrons/cm$^2$ and doses of up to 10$^7$~Gy. 
 
\end{itemize}

   In spite of these detector features, pile-up is expected
   to have some residual impact on the physics performance~\cite{ATLAS,CMS}. 
   For example, in the calorimeters pile-up fluctuations contribute 
  an additional (noise) term to the energy resolution. 
   The pile-up noise inside the volume 
  of electromagnetic calorimeter needed to contain an electromagnetic 
 shower has a typical r.m.s. of $E_T\simeq250$~MeV, giving a contribution of 
 $\sim$2.5\% ($\sim0.25$\%) to
 the energy resolution of electrons and photons of $E_T=10~(100)$~GeV. 
  The pile-up noise inside a calorimeter cone of size $\Delta R=0.4$ has
  an r.m.s. of $E_T\simeq7$~GeV, 
  giving a contribution of 7\% (2\%) to the energy resolution of jets with
  $E_T=100 (300)$~GeV.  
      
   In summary, the large pile-up renders the operation at the LHC
   more challenging than at previous hadron colliders, and
  the experimental environment more dirty, 
  and represents a high price to be payed for the huge machine 
  luminosity. 
  
   Another challenging issue at the LHC is the trigger. 
  Since the interaction rate is $10^9$ events/s (dominated
  by minimum-bias and QCD interactions, see Section~\ref{physcross}), 
   whereas the maximum affordable rate-to-storage is of
  the order of 100~Hz, a powerful and highly-selective 
  trigger system, providing a rate reduction of $10^{7}$ while
  preserving a high efficiency for the interesting physics processes, 
  is needed. The ATLAS and CMS trigger will be based on a multi-level
  selection, where the first-level trigger is provided
  by fast hardware signals from the
  calorimeters and the muon spectrometers, and the higher-level
  triggers by software algorithms using the information
  from all sub-detectors. 

    The situation is much simpler at a LC, in spite of a even
   higher peak luminosity than at the LHC, 
   in the range 2-6$\times$10$^{34}$ cm$^{-2}$ s$^{-1}$ depending on the exact 
   center-of-mass energy.
     The average bunch-crossing rate will be 15-30 kHz
   (with bunch trains whose structure depends on the chosen technology)    
     and    the interaction rate will be dominated by
   $\gamma\gamma$ interactions ($\sim$0.1 events per crossing). 
    Because of such a low
   rate, the experiment can be run in continual ``triggerless" mode,
   so that the acceptance for physics is maximized, allowing for
   an unbiased search for new phenomena. In this scenario
   the LC detector~\cite{TESLA}
  will have no hardware level-one trigger,  but only a software-based 
  relatively loose selection.
   
    Beam-related backgrounds, although smaller than the 
   LHC pile-up, are much more severe than at previous $e^+e^-$ colliders
   because of the high luminosity. 
   The main source of beam-related backgrounds are 
   beam-beam interactions. The
   high charge density of the colliding beams produces intense emission
   of  beamstrahlung photons (about $6\cdot10^{10}$ photons per crossing
   when running at $\sqrt{s}\simeq500$~GeV, carrying a total energy of
   about 3$\cdot10^{11}$~GeV). These photons, although they disappear
   in the beam pipe, have two main effects.
    First, they broaden the energy spectrum of the colliding beams
   towards lower energies, with typically 85-90\% of the luminosity
   being produced at energies higher than 95\% of the nominal center-of-mass
   energy. The energy loss due to beamstrahlung is roughly of the same size as
   initial state radiation.
    Second, beamstrahlung photons give rise to secondary particles, among
    which particularly dangerous are $e^+e^-$ pairs from photon 
    conversions in the interaction region. If a sufficiently high
    magnetic field (3-4~T) is used in the detector, as it
    is indeed planned (see below), most of these pairs are confined
    inside a
    cylinder of radius $<3$~cm around the beam line, thereby 
    affecting mainly  the first layer of the vertex detector.  
     However,  these pairs tend to move longitudinally in $z$ towards the 
    machine quadrupoles, where they create a large number of secondaries. 
    The latter are potentially a harmful source of background for the detectors,
     which must be shielded with suitable masks~\cite{TESLA}.

\section{Physics cross-sections and backgrounds\label{physcross}}

      In addition to the backgrounds  
    and challenges related to the environment
    discussed in the previous Section, other sources 
    of (physics) backgrounds need to be considered. 
   
    At the LHC, the physics cross-sections are dominated by QCD jet production, 
    which is many orders of magnitude larger than the production of
    the most interesting physics channels. The latter are usually 
   characterized by electroweak cross-sections, or are expected to
   yield low rates because they involve new massive  particles.  
   
\begin{figure}[h]
\begin{center} 
\includegraphics[width=0.485\textwidth]{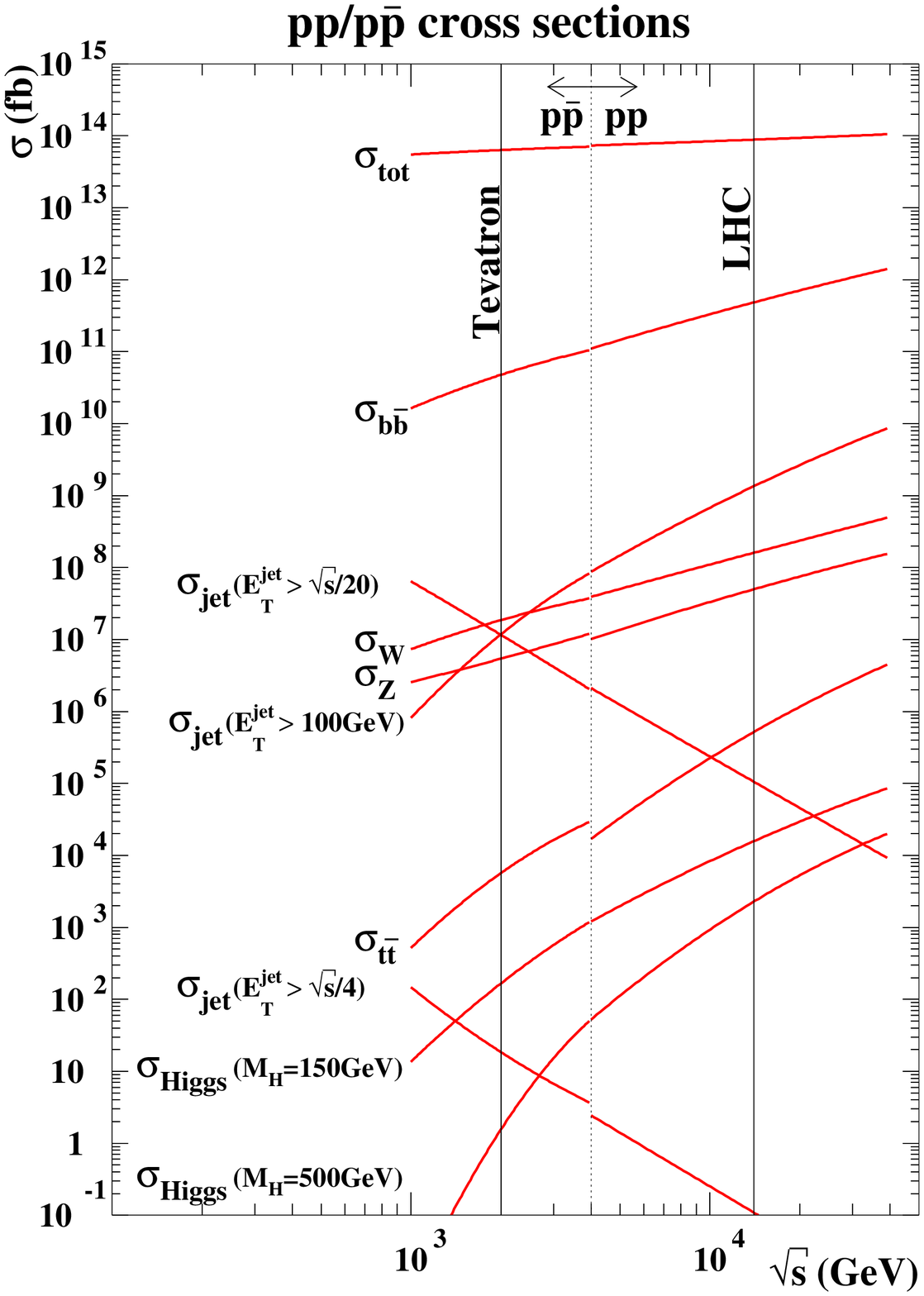}
\includegraphics[width=0.485\textwidth]{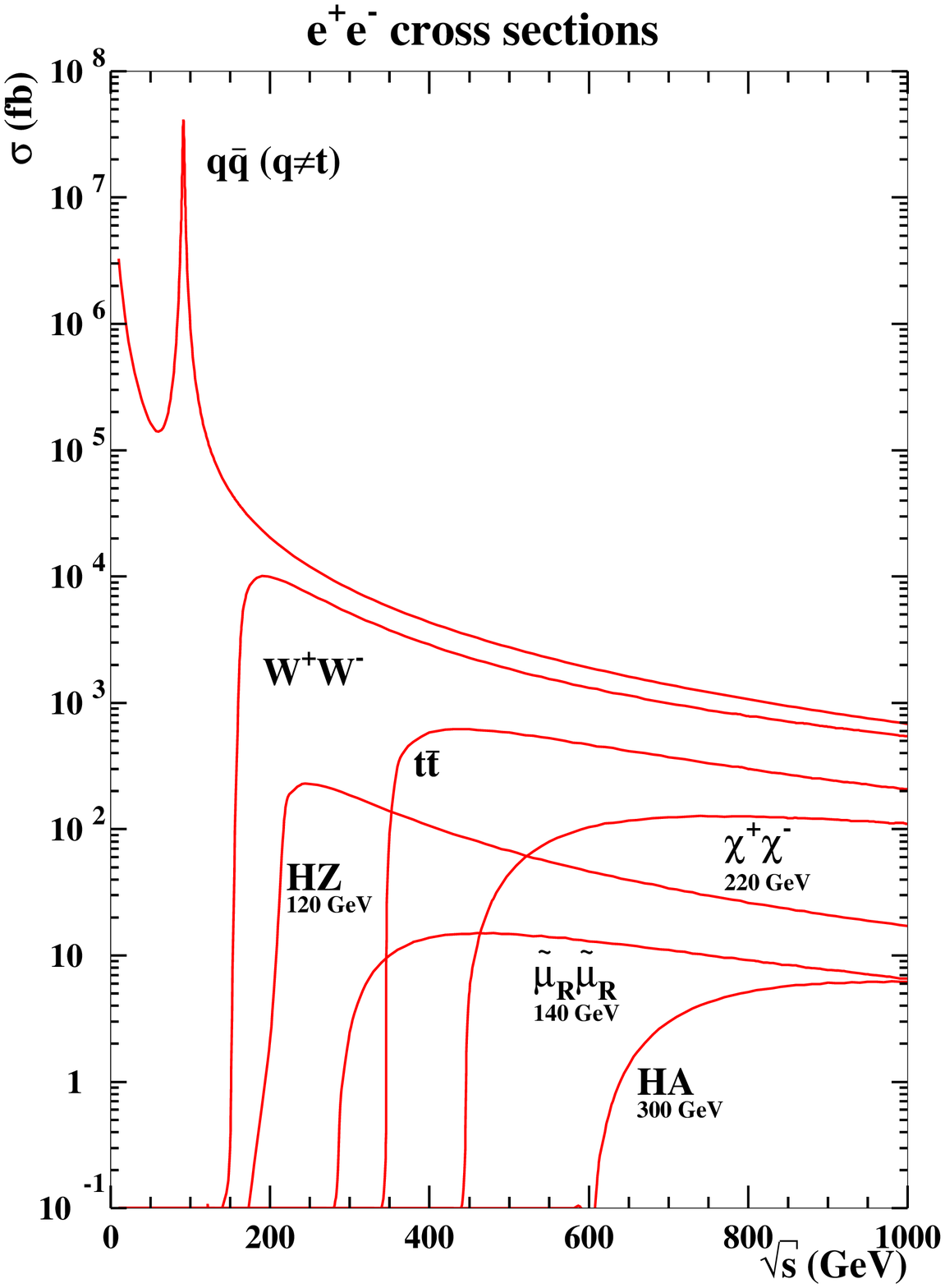} 
\caption{Production cross-sections for several representative processes 
 at hadron colliders (left) and $e^+e^-$ colliders (right), 
 as a function of the machine center-of-mass energy.}
\label{cross} 
\end{center}
\end{figure}

   Figure~\ref{cross} shows the production cross-sections for several 
   representative processes at hadron colliders, as a function of 
   the center-of-mass energy (left panel). 
    It can be seen, for instance, that at the LHC energies the 
   cross-section for jets with  $p_T~>~100$~GeV is five orders of magnitude 
   larger than the  cross-section for a Higgs boson of mass  150~GeV. 
    As a consequence,  
   there is no hope to detect a Higgs boson (or a $W$, or a $Z$ boson) 
    in the fully-hadronic decay 
   modes (unless it is produced in association with something 
    else), since such final states are swamped by the much higher,
     and to a large extent irreducible, jet background
    (hereafter referred to as ``QCD background").
     Decays into leptons and photons have to be used instead, and
     since they usually have 
     smaller branching ratios than decays into quarks, 
   a good part of the {\it a-priori} large
   production cross-section is {\it de facto} not used at the LHC. 

    Small signal-to-background ratios, due to the large QCD cross-sections
   compared to electroweak cross-sections,    
    are a general feature of hadron colliders. However, the situation 
    deteriorates with increasing center-of-mass energy, because 
    the small-$x$ region of the proton structure functions, where
    the gluon distribution has a huge enhancement, becomes more and more
    accessible. At the LHC, which is essentially a gluon-gluon collider, 
    the large contribution of gluon-gluon and gluon-quark 
    interactions enhances tremendously the QCD cross-sections compared
    to the ($qq$-dominated) electroweak cross-sections. This renders the
    signal-to-background ratio smaller than e.g. at the Tevatron. 
   As an example,  the ratio (e/jet) between the  
  inclusive rate of electrons (coming e.g. from $W$ and $Z$ decays) 
  and the inclusive rate of jets with $p_T~>~20$~GeV is 
  e/jet~$\simeq 10^{-3}$ at the Tevatron and $\simeq 10^{-5}$ 
    at the LHC. This implies
    that the 
  particle identification capabilities of the LHC detectors
   must be two orders
   of magnitude better than those of the Tevatron experiments. 

  Signal-to-background ratios are much more favorable at 
  $e^+e^-$ colliders, as shown in 
  the right panel of Fig.~\ref{cross}.  For instance, in the case of
  a light Higgs boson of mass $\sim$120~GeV, the signal production 
  cross-section is  only two orders of magnitude smaller than that
  of the backgrounds (e.g. $WW$ and $qq$ production), which in addition
  are to a large extent reducible. 
  With high luminosity, many physics scenarios can be explored 
  in a few years of operation.
  For some physics processes, relatively low signal rates call for
  very high integrated luminosity,
  and therefore require several years of operation in order to achieve high accuracy
  in the precision measurements.

\section{The detector performance requirements}

   Numerous detector performance requirements emerge 
  from the above considerations and from the physics goals of
  the two machines~\cite{ATLAS,CMS,TESLA,NLCreport,ACFAreport}. 
   The main issues are listed below: 
    
\begin{itemize}

\item  Measurements of leptons and jets over unprecedented dynamic ranges, 
 extending from a few GeV up to several hundreds GeV (LC) or
 up to a few TeV (LHC). This is needed in order to detect
 light particles, like the soft leptons produced in B-hadron decays, 
 as well as the very energetic objects which may be produced in the 
 decay of new massive particles. 
 
\item  Detector hermeticity. At both machines,
 full coverage in $\phi$ and coverage down to $\sim1^{\circ}$ 
 from the beam axis in $\theta$ 
 are needed mainly for a reliable measurement of the
 event missing (transverse) energy, a signature of the 
 production of weakly-interacting particles which are expected in
 many New Physics processes (e.g. SUSY, Higgs production through
 $WW$-fusion at a LC, etc.). 
   At the LHC, calorimetric coverage over the above-mentioned
   angular range
  is also needed to detect the forward jets 
  produced in association with a Higgs boson in the $WW$-fusion process. 

\item  Excellent energy and momentum resolution. 
  At the LHC, a mass resolution of $\sim 1$\% for particles of masses 
  up to a few hundreds GeV decaying into photons, electrons or muons is
  needed, for instance  to extract a possible $H\to\gamma\gamma$ 
   signal on top of the irreducible $\gamma\gamma$ background.
   
   At a LC,  an excellent track momentum resolution is
   required in particular to measure the di-lepton mass, and hence
   the mass of the recoiling system,   
   in the $HZ$ process with $Z\to\ell\ell$. This should
   give access to the detection and study of Higgs production 
   independently of the Higgs decay modes. 
 The goal momentum
 resolution of  $\sigma(1/p_T)\leq5\cdot 10^{-5}$~(GeV/c)$^{-1}$, which
 is needed to suppress the combinatorial background, 
 calls for large tracking volumes and high magnetic fields ($\sim$4~T). 
 
  Accurate energy flow measurements is also a must at a LC. 
  Indeed, most signatures from new
 physics involve final states with many jets, coming e.g. from 
 top-quark or multiple W and Z production and decays. These jets must
 be efficiently and precisely reconstructed in order to reduce
 the backgrounds. In addition, enhanced beamstrahlung, as
 compared to previous $e^+e^-$ colliders,
 render the kinematic constraints from the knowledge of the initial
 state weaker than in the past, which puts
 more weight on energy measurements provided by the detector.
  The goal energy-flow resolution for hadronic event
 is $\sigma/E\sim30\%/\sqrt{E}$, which is necessary e.g.~to separate 
 hadronic W and Z decays.
 This in turn requires a fine 3-dimensional detector granularity, 
 a coil located outside the calorimetry in order to minimize the amount
 of upstream material, and excellent detector
 energy resolution and particle identification capabilities.

\item  Particle identification. 
   Several stringent requirements on the identification of electrons, 
  photons, b-jets, taus, etc.  must be satisfied at the LHC in order
  to reject the huge QCD backgrounds.  
   As an example, excellent 
  electron/jet and photon/jet separation capabilities are needed. 
  Jets faking photons must be rejected by a factor of $\sim10^3$, 
   for a photon efficiency of $\sim$80\%,  
  in order to observe a possible $H\to\gamma\gamma$ signal on top of the 
  background.  As already mentioned, an unprecedented suppression factor 
  against  jets faking electrons of $10^{5}-10^{6}$ is needed to extract
   an inclusive clean electron signal. 
  
  At a LC, one of the strongest particle identification requirements 
  is flavor-tagging capabilities (vertexing), since 
  several channels from New Physics (e.g. SUSY) involve  $b$-quarks 
  or $\tau$-leptons. 
  Moreover, a detailed study of the Higgs sector, which requires
 the individual measurements
  of the Higgs decay branching ratios into $bb, cc, gg, \tau\tau$,
 is only possible with an excellent vertex detector, and an 
 innermost layer as close as possible to the beam pipe
 ($\leq$~2~cm).
  The  goal performance is to achieve 
 an impact parameter resolution of 
 about 5~$\mu$m in both $z$ and $R\phi$. 

\item At the LC, precision cross section measurements require an
  excellent luminosity measurement (10$^{-4}$), which has to be matched
by equally precise theoretical predictions. 
Luminosity measurement requires fast and highly granular forward calorimetry down
to angles of approximately 5 mrad. Beam polarization has to be
determined to a precision of at least 0.5\%.
A quasi-continuous monitoring of the differential luminosity spectrum
is also necessary.

\end{itemize}

\section{Summary of physics capabilities}

A detailed discussion of the physics goals of both machines and 
of their interplay is the subject of the next Sections of this document. 

Here only a few general (and fairly gross) conclusions are given:

\begin{itemize}

\item The LHC has the highest mass reach for the direct discovery
 of new particles. This reach extends up to masses of 
 $\sim$5-6~TeV for singly-produced electroweak
 particles (e.g. possible new gauge bosons $W'$ and $Z'$),
  up to $\sim$7~TeV for singly-produced strongly-interacting
  particles (e.g. possible excited quarks), and up to $\sim$~3~TeV
  for pair-produced SUSY particles with strong interactions. 
  
 The direct mass reach of a LC is much more modest, being 
 limited by the available center-of-mass energy
 to less than 1~TeV for the case considered in this document. 
 However, due to the cleaner environment, also direct signals
 of particles which are produced with low cross sections or
 which only decay into hadrons or
 which leave only a small amount of visible energy in the detector
 can be discovered.
  
\item  A Linear Collider has an indirect discovery 
 sensitivity to New Physics, 
 through precise measurements of known processes
 (e.g. two-fermion production) and the detection of 
  permil-level deviations from the Standard Model expectation, 
 which extends up to energies of $\sim$10~TeV. Therefore 
 a 1~TeV LC should be able to probe New Physics  lying
 at energy scales much higher than the machine $\sqrt{s}$
 through the measurements of the low-energy (quantum-level) 
 tails of the theory. 
   
   Because precise measurements are more difficult at hadron colliders
 for the reasons mentioned above, the LHC indirect sensitivity to New Physics   
 is more modest, except
 for some strongly-interacting  scenarios (like Compositeness) which
 are expected to manifest themselves through anomalous contributions
 to di-jet production. 

\item Precision measurements are the strongest asset of a LC.
  In general, all particles and processes 
 which are kinematically accessible and produced
  can be measured with
  typical precisions ranging from the permil to the percent level,
  irrespective of the precise physics scenario. 
Such a precision allows for the exploration of quantum-level effects
and yields the possibility to extrapolate the observations to energy
scales far above the center-of-mass energy in a model-independent way.
Ultimately, GUT or Planck scale physics could be probed.
 
   Several precise measurements should also be possible at
 the LHC, thanks mainly to the large available event statistics. 
   As an example, if SUSY
 exists the LHC experiments should be able to perform several
 measurements of the sparticle masses, and therefore to constrain
 the fundamental parameters of the underlying 
 theory to $\sim10\%$ or better (at
 least in minimal models). Therefore, in addition to being
 a very powerful and motivated discovery machine, the LHC should  
 also provide a first and possibly quite deep exploration
 of the structure of New Physics. 
   There are however two main limitations. First, in general the 
  extent and the precision of the measurements are poorer 
  than at a LC. For instance, the LHC can only measure some of
 the Higgs couplings, and with an accuracy ($10-20\%$) which is not competitive
 with that of a LC ($\sim1\%$). Second, a complete, model-independent 
 and conclusive study of the new theory 
 is {\it a priori} not granted and looks difficult in most cases.  
  
\end{itemize}

These features give rise to a nice complementarity between the
two machines, and lead to synergy effects which are
discussed in more detail in the rest of this document.

\chapter{Higgs Physics and Electroweak Symmetry Breaking}
\label{chapter:ewsymmbreak}

Editors: {\it A.~De Roeck, H.E.~Haber, R.~Godbole, J.~Gunion,
G.~Weiglein}

\vspace{1em}

The search for the fundamental dynamics that is responsible
for electroweak symmetry breaking 
is the central challenge for particle physics today.
This dynamics, whose fundamental origin is as yet unknown,
is ultimately responsible for the generation of the
masses of the quarks, charged leptons and the massive gauge bosons.
Two broad classes of electroweak
symmetry breaking mechanisms have been pursued theoretically.  In one
class of theories, electroweak symmetry breaking dynamics is
weakly-coupled, and in the second class of theories the dynamics is
strongly-coupled.

The electroweak symmetry breaking dynamics that is employed by the
Standard Model is governed by a
self-interacting complex doublet of scalar fields~\cite{hhg}.  
The Higgs potential is chosen so that
the neutral component of the scalar doublet
acquires a vacuum expectation value, $v=246$~GeV, which sets the 
mass scale of electroweak symmetry breaking.
Consequently, three massless Goldstone bosons are generated
which provide the longitudinal degrees of freedom for the $W^\pm$ and 
$Z^0$, while
the fourth scalar degree of freedom that remains in the physical spectrum
is the CP-even neutral Higgs boson.
It is further assumed in the Standard Model that the scalar doublet also
couples to fermions via the
Yukawa interactions.  After electroweak symmetry breaking, these interactions
are responsible for the generation of quark and charged
lepton masses.  The couplings of the Higgs boson to the Standard Model
particles is then fixed (and proportional to the corresponding
particle mass).  However the Higgs mass is proportional to 
the strength of the Higgs self-coupling, and is therefore not
(directly) fixed by present day observations.

Although the Higgs boson has not been directly observed, its virtual
effects (primarily via its contributions to the $W^\pm$ and $Z$ boson 
vacuum polarization)
can influence electroweak observables.  Consequently, one can obtain
constraints on the Higgs boson mass ($m_h$) through a
global Standard Model fit to the electroweak data.
The results of the LEP
Electroweak Working Group analysis 
yield \cite{lepewwg}:
$m_h=114^{+69}_{-45}~{\rm GeV}$, and provides a 
one-sided 95\% CL upper limit of
$m_h<260$~GeV.
These results reflect the logarithmic sensitivity to the Higgs mass via
the virtual Higgs loop contributions to the various electroweak
observables.  The 95\% CL upper limit is consistent with the
direct searches at LEP~\cite{lepsmhiggs}
that show no conclusive evidence for the Higgs
boson, and imply that $m_h> 114.4$~GeV at 95\%~CL.
This range of Higgs masses is consistent
with a weakly-coupled Higgs scalar that is the result of the
Standard Model scalar dynamics.

In the weakly-coupled approach to electroweak symmetry breaking, the
Standard Model is very likely embedded in a supersymmetric
theory~\cite{susyreview} in
order to stabilize the large energy gap between the electroweak and the Planck
scales in a natural way~\cite{susynatural}.
These theories predict a spectrum
of Higgs scalars~\cite{susyhiggs},
with an expected mass of the lightest CP-even Higgs
boson below 200~GeV~\cite{Espinosa:1998re}
(less than 135~GeV in the simplest supersymmetric
models~\cite{Degrassi:2002fi}), and a 
spectrum of additional neutral and charged Higgs bosons
with masses up to of order 1~TeV.
Moreover, over a significant fraction of the supersymmetric parameter
space, the properties of the lightest Higgs scalar closely resemble those
of the Standard Model (SM) Higgs boson.  

An alternative approach to weakly-coupled scalar dynamics posits that
electroweak symmetry breaking is driven by the dynamics of a new
strongly-interacting sector of particles~\cite{hill}.  
Initial models of this kind
introduced QCD-like strong interactions near the
TeV-scale~\cite{susskind}, with numerous variations subsequently explored. 
More recently, so-called ``little Higgs models'' have been proposed in
which the scale of the new strong interactions is pushed up above
10~TeV~\cite{littlehiggs}, and the lightest Higgs scalar resembles
the weakly-coupled SM Higgs boson.  These models typically contain
additional particles, such as
new gauge bosons and vector-like fermions, which 
populate the TeV mass region.
In a more speculative direction,
a new approach to electroweak symmetry breaking has
been explored in which extra space dimensions beyond
the usual $3+1$ dimensional spacetime are
introduced~\cite{extradim} with
characteristic sizes of order $(\rm TeV)^{-1}$.
In scenarios of this type, it is possible to devise a
mechanism for
electroweak symmetry breaking that is inherently
extra-dimensional~\cite{quiros}.
In such models,
the resulting phenomenology can be significantly
different~\cite{dimhiggs} 
from the dominant paradigm of the weakly-coupled electroweak
symmetry breaking sector.  
Typically, the spectrum of the TeV-scale
will look quite different from the standard weak-coupling approaches.
Kaluza-Klein excitations of the Standard Model particles can play a
significant role in the resulting phenomenology.   In some cases,
the light Higgs boson is completely absent from the low-energy
spectrum, in so-called higgsless models of
electroweak symmetry breaking~\cite{terning}.

Although there is as yet no direct evidence 
for the origin of
electroweak symmetry breaking dynamics, present data
can be used to discriminate among the different approaches.  
As noted above, the 
precision electroweak data, accumulated in the past decade at LEP, SLC,
the Tevatron and elsewhere, seem to be consistent with
the Standard Model (or its supersymmetric extension),
with a weakly-coupled
Higgs boson whose mass lies roughly between 100 and 250 GeV~\cite{lepewwg}.
Moreover, the contribution of new physics, which
can enter through $W^\pm$ and $Z$ boson vacuum polarization
corrections~\cite{tatsu}, is severely constrained.  This fact has already
served to rule out nearly all of the initially proposed
models of strongly-coupled electroweak symmetry
breaking dynamics, and provides strong constraints on any alternative
to the Standard Model and its supersymmetric extensions.

It is still possible that the Tevatron will yield the first hints of
a SM-like Higgs boson prior to the start of the LHC.
However, the most likely scenario is one where the LHC provides the
definitive initial discovery of the physics of the electroweak symmetry
breaking sector~\cite{higgsreview}.  
This will be either in the form of a candidate Higgs
boson, or evidence that the electroweak symmetry breaking dynamics is
driven by some mechanism that does not involve scalar fields in a
fundamental way.  Any
program of Higgs physics at future colliders must address a number
of important questions.  First, 
does the SM Higgs boson (or a Higgs scalar with similar properties)
exist?  If yes, 
how many physical Higgs states are associated with the scalar sector?
Moreover, how
can one prove that a newly discovered scalar is a Higgs boson?
To answer these questions, one must observe the Higgs boson in more
than one production and decay channel, and map out its
properties in detail.  
One must verify that the spin of the candidate Higgs boson
is consistent with spin-zero.  It is essential to measure a variety of Higgs
couplings and demonstrate that these do indeed scale in proportion to
the corresponding masses. 
The LHC will be able to address some of these questions
with a program of Higgs measurements that can determine the
Higgs couplings to the top-quark, tau-lepton, 
$W$ and $Z$ to an accuracy in the range 
10--30\%, assuming an integrated luminosity of 300
fb$^{-1}$~\cite{Duhrssen:2004cv}.  Note, however, that LHC measurements only
weakly constrain the Higgs coupling to $b\bar b$, even though the
latter is the dominant Higgs decay channel for Higgs masses below 135~GeV.

To make further progress requires a comprehensive program of precision
Higgs measurements.  Such measurements
with typical accuracies in some channels approaching the $1\%$ level,
are necessary to fully decipher the dynamics responsible
for electroweak symmetry breaking.  Such a program can only be
achieved at the LC.  The significance of the precision Higgs program
is especially evident in the
so-called decoupling limit~\cite{decoupling}, 
in which the properties of the lightest
Higgs scalar are nearly identical to those of the SM Higgs boson.  This
limiting case arises in many extended Higgs theories over a
significant fraction of the parameter space.  Moreover, 
additional scalars of the Higgs sector are heavy in the decoupling limit
and may not be so easily discovered at the next generation of colliders.
Thus precision measurements that
can distinguish the SM Higgs sector from a more
complicated scalar sector are especially important  
if only one scalar state is discovered.  In particular, small deviations
from the Standard Model encode the physics of electroweak symmetry
breaking, as well as being sensitive to 
new physics that lies beyond the Standard Model.

Additional information is required in
order to fully probe the underlying scalar dynamics responsible for
electroweak symmetry breaking.  Ideally, one aims to
reconstruct the Higgs potential and directly demonstrate the
mechanism of electroweak symmetry breaking.  This requires precision
measurements of Higgs self-couplings, which may only be possible at a
very high energy LC.  One would like to know whether there are
CP-violating phenomena associated with the Higgs sector.  Near the
decoupling limit, the couplings of the lightest Higgs boson are
CP-conserving to a very good approximation, whereas
CP-violating couplings among the heavier Higgs states can
be unsuppressed.  The former certainly requires the precision Higgs
program of the LC, whereas the latter may depend on the LHC if the
heavier Higgs bosons are too massive to be produced at the LC.

If the Higgs sector is weakly-coupled, one would be very
interested in testing the consistency with the constraints of
supersymmetry.  In this case, knowledge of the spectrum of
supersymmetric particles is especially significant for the precision
Higgs program.  The supersymmetric particle spectrum enters in a
crucial way in the radiative corrections to Higgs masses and
couplings.  The LHC and LC are sensitive to different aspects of the
supersymmetric spectrum, and both machines will provide crucial input data
for the theoretical interpretation of the precision Higgs program.

If nature chooses a scenario far from the decoupling limit, then
electroweak symmetry breaking dynamics produces no state that closely
resembles the SM Higgs boson.  In this case, it is likely that there
will exist many new light states (below a TeV in mass) with a rich
phenomenology.\footnote{The possibility that no light states exist
below 1~TeV seems remote given the standard interpretation of the
precision electroweak data.  Nevertheless, such possibilities cannot
yet be excluded with complete certainty.}
In particular, new approaches to electroweak symmetry breaking
dynamics have led to phenomenologies that may be quite different from
the conventional expectations of weakly coupled multi-Higgs
models (with or without supersymmetry).  It will be essential to
formulate strategies for using precision Higgs studies at future
colliders to distinguish among the many possibilities. 

This chapter describes a number of studies that exploit the
complementarity of the LHC and LC for exploring the origin of
electroweak symmetry breaking.  Section 3.1 focuses on precision
studies of the lightest CP-even Higgs boson, with the motivation of
determining how close its properties are to that of the SM Higgs
boson.  Section 3.2 studies the CP-properties of the Higgs bosons.
Section 3.3 focuses on Higgs physics in the minimal supersymmetric
extension of the Standard Model (MSSM), while section 3.4 examines
non-minimal approaches (both supersymmetric and non-supersymmetric) to
the extended Higgs sector.  In section 3.5, Higgs physics in the
context of extra-dimensional models are studied, and in section 3.6,
the consequences of the so-called littlest Higgs model is explored.
Finally, a number of miscellaneous topics are treated in section 3.7.

\section{\label{sec:Higgscoupl}
Higgs coupling measurements and flavour-independent Higgs searches}

\def\flipright#1{%
   \setbox0=\vbox{\epsfbox{#1}}
   \setbox1=\vbox{\rotr0}
   \centerline{\copy1}}                 
\def\flipleft#1{%
   \setbox0=\vbox{\epsfbox{#1}}
   \setbox1=\vbox{\rotl0}
   \centerline{\copy1}}                 
\newcommand{\ra}{\mbox{$\to$}}
\newcommand{\ee}{\mbox{${e}^+ {e}^-$}}
\newcommand{\bb}         {\mbox{${b}\bar{b}$}}
\newcommand{\ttbar}      {\mbox{${t}\bar{t}$}}
\newcommand{\glgl}       {\mbox{${gg}$}}
\newcommand{\mH}         {\mbox{$m_{H}$}}
\newcommand {\Ho}        {\mbox{${H}^{0}$}}
\newcommand {\Zo}        {\mbox{${Z}^{0}$}}
\newcommand{\WW}         {\mbox{${W}^+{W}^-$}}
\newcommand{\WWn}        {\mbox{${WW}$}}
\newcommand{\sqrts}      {\mbox{$\sqrt{s}$}}
\newcommand{\fb}         {\mbox{$\mathrm{fb}^{-1}$}}
                                                                                
\newcommand{\Ndash}{\nobreakdash\textendash}
                                                                                
                                                                                
\def\lsim{\mathrel{\raise.3ex\hbox{$<$\kern-.75em\lower1ex\hbox{$\sim$}}}}
\def\gsim{\mathrel{\raise.3ex\hbox{$>$\kern-.75em\lower1ex\hbox{$\sim$}}}}
\def\BR{{\rm BR}}
\newenvironment{Eqnarray}%
     {\arraycolsep 0.14em\begin{eqnarray}}{\end{eqnarray}}

In this section, we examine how complementary measurements from the
LHC and the LC can contribute to the precision Higgs program at future
colliders.  Here we shall assume that the lightest CP-even Higgs boson
of the scalar spectrum has a mass in the range of 100 to 200~GeV.
This assumption is consistent with the standard interpretation of the
LEP Higgs search and the implications of the global fit to the
precision electroweak data based on the Standard Model (SM).  The
former implies that $m_h> 114.4$~GeV at 95\% CL~\cite{lepsmhiggs} (in
the context of the MSSM, this limit is somewhat weaker, $m_h>91.0$~GeV
at 95\% CL~\cite{lepmssmhiggs}), whereas the latter implies that
$m_h<260$~GeV at $95\%$ CL~\cite{lepewwg}.

The LHC will provide the first opportunity for precision Higgs
measurements.   For example, using methods developed in 
refs.~\cite{zep,bere,due}, the
Higgs couplings to the top-quark, tau-lepton, $W$ and $Z$ 
can be determined at the LHC to an accuracy in the range 
10--30\%, assuming an integrated luminosity of 300
fb$^{-1}$~\cite{Duhrssen:2004cv}.
At the LC, the expectations for precision Higgs measurements are well
documented~\cite{lchiggsreview}. Significant improvements can be
obtained in many channels, approaching accuracies in the range of a
few percent or better in a number of cases.

However, there are two important Higgs observables for which the
expected accuracy of the LC running at $\sqrt{s}=500$~GeV
is not particularly impressive.  These are
the Higgs-top Yukawa coupling and the triple Higgs self-coupling.  At
the LC, the Higgs-top Yukawa coupling is obtained via measurements of
$e^+e^-\to t\bar t h$ production and the determination of the triple
Higgs coupling requires the observation of $e^+e^-\to Zhh$ and/or
$e^+e^-\to \nu\bar\nu W^*W^*\to\nu\bar\nu hh$ production.  At 
$\sqrt{s}=500$~GeV, the cross-sections for these
processes are quite small (due primarily to the phase space
suppression of the three and four body final states);
consequently, the LC
alone can only make crude measurements of these Higgs observables,
(assuming sufficient luminosity).  It is here
where the LHC can play a strong complementary role.  Given the large
LHC energy and luminosity, the main challenge for the LHC is to
suppress backgrounds efficiently enough in order to produce a Higgs
signal of significance from which the Higgs-top Yukawa coupling and
the triple Higgs coupling can be extracted.  

In this section, we present two studies of the determination of the
Higgs-top quark Yukawa coupling.  In the contribution of K.~Desch and
M.~Schumacher, an experimental method is proposed to determine the
Higgs-top quark Yukawa couplings in a model-independent way at the LHC
and LC.  By combining the results of the measurements at both
colliders, the most accurate determination of this coupling can be
achieved.  The contribution of S.~Dawson {\it et al.} focuses on the
theoretical uncertainties due to higher order QCD corrections that
arise in the computation of the cross section for $t\bar th$
associated production at the LHC and LC.  The detection of a number of
different Higgs decay channels is considered (including $b\bar b$,
$W^+W^-$, $\gamma\gamma$ and $\tau^+\tau^-$).  Finally, a brief
discussion of the Higgs-top Yukawa coupling determination in the MSSM
is given.  In the latter case, new channels enter if the non-minimal
Higgs states of the MSSM ($H$, $A$ and $H^\pm$) are not too heavy.
The complementarity of the LHC and LC for measuring the $t\bar t h$ coupling
becomes less compelling once
higher energies are available for the LC.  For example, for the LC with 
$\sqrt{s}=800$--1000~GeV and an integrated luminosity of 1~ab$^{-1}$, 
the Higgs-top-quark Yukawa coupling could be
determined with an accuracy that is significantly
more precise than the corresponding $t\bar t h$ coupling
determination at the LHC.

We also present one study that contrasts the capabilities of the LHC
and LC in the measurement of the triple-Higgs coupling.  Ultimately,
this is the first step required in a program to experimentally
determine the parameters of the Higgs boson potential.  Center-of-mass
energies at the LC ranging from 500~GeV to 1~TeV were considered.  
In fact, the lower center-of-mass energy provided the more accurate
measurement of the triple-Higgs coupling for Higgs masses below about
140~GeV.  The prospects for a significant LHC measurement in this mass
range are poor.  For Higgs masses in the range of 150--200~GeV, the
relevant Higgs cross-sections at the LC are becoming too small to
allow for a useful measurement, whereas LHC data can yield the more
accurate determination of the triple-Higgs coupling.  However, the
latter can be reliably accomplished only with the input of other precision
Higgs properties obtained from measurements at the LC.
Nevertheless, the initial accuracies for the triple-Higgs couplings
(for $m_h<200$~GeV) will be crude at best, and higher energy and/or
luminosity colliders will be needed to make significant improvements.

Most of the above discussion assumes a pattern of Higgs boson partial
widths that is close to the Standard Model expectations.  However, if
this assumption proves false, the Higgs search strategies (for both
discovery and precision measurements) will have to be reconsidered.
For example, flavor-independent techniques can be critical for Higgs
searches if the theory of the Higgs sector deviates from Standard
Model expectations.  In the contribution of E.L.~Berger, T.M.P.~Tait
and C.E.M.~Wagner, the phenomenology of a Higgs boson that decays
predominantly into jets of hadrons with no significant $b$-quark
flavor content is examined.  This condition may be realized, for
instance, in supersymmetric models in which a light bottom squark is
present in the spectrum and in models in which the dominant Higgs
decay is into a pair of of light CP-odd scalars.  Berger \textit{et
al.} emphasize that in these scenarios, the viability of the standard
Higgs discovery channels at the LHC (say, for $m_h=120$~GeV) is
significantly degraded.  In contrast, the Higgs discovery potential
and precision Higgs measurements at the LC are generally much less
sensitive to assumptions about the specific pattern of Higgs partial
widths.  In particular, the Higgs boson can be observed (and its mass
determined) independently of its final state decays in $hZ$ production
via the missing mass recoiling against the $Z$.

In probing the physics of the Higgs sector, it is essential to
elucidate the nature of the TeV-scale physics associated with the
mechanism of electroweak symmetry breaking.  The well known
naturalness arguments associated with understanding the origin of the
electroweak scale make plausible the existence of new TeV-scale
physics beyond the Standard Model.  In nearly all models considered,
the LHC, with its enormous energy and luminosity, is ideally suited to
discover the new TeV-scale phenomena, and provide the initial
opportunity for probing the underlying new dynamics.  Thus,
even if the observation of Higgs bosons at the LHC is problematical or
ambiguous, one anticipates a rich phenomenology of TeV-scale physics
that is accessible to the LHC.  At the LC, the success of the Higgs
program for $m_h\lsim 200$~GeV is guaranteed, independently of the
details of the Higgs model.  This illustrates another way in which the
complementarity of the LHC and the LC can be essential for providing a
broad understanding of the physics of electroweak symmetry breaking.

\subsection{Model independent determination of the top Yukawa coupling
from LHC and LC}

{\it K.~Desch and M.~Schumacher}

\vspace{1em}
\subsubsection{Motivation}

The Yukawa coupling of the Higgs boson to the heaviest quark, the top quark,
is of great interest for the study of the nature of electroweak symmetry
breaking and the generation of masses. While the Yukawa couplings to bottom
and charm quarks and to tau leptons and muons are in principle accessible
through the Higgs boson decay branching ratios, the Higgs boson decay into 
top quark pairs is kinematically forbidden for light Higgs bosons as they
are favoured by theory and electroweak precision data. The only 
Standard Model process that probes the top Yukawa coupling at tree 
level is the associated production of a \ttbar\ pair 
with a Higgs boson. This process occurs at the LHC (mainly 
\glgl\ra\ttbar\Ho ) as well as at the 
LC ( \ee\ra\ttbar\Ho ).
In the latter case the cross section is only significant at 
centre--of--mass energies in excess of 800~GeV. At the LHC, the final
states that have been investigated so far are \ttbar\bb\
\cite{lhcbb,Dai:1993gm,lhctop,cmstth} 
and \ttbar\WW \cite{lhcww,mrw}, the $\ttbar\tau^+\tau^-$ final state
is under study~\cite{eilam,ito}.
At tree level, their production rates are proportional to
$g_{ttH}^2\,\mathrm{BR}(\Ho\to\bb) $ and 
$g_{ttH}^2\,\mathrm{BR}(\Ho\to\WW) $, respectively. The
absolute values of $\mathrm{BR}(\Ho\to\bb)$ and $\mathrm{BR}(\Ho\to\WW)$ 
can be measured
accurately in a model independent way at the LC from the corresponding
decay branching ratios \cite{Aguilar-Saavedra:2001rg}. These can be
measured already at a first phase of the LC (\sqrts\ between 350 and
500 GeV). Thus, the combination of the measurements of both machines
can be used to determine the value of $g_{ttH}$ without model
assumptions and presumable before a second phase of the LC
($\sqrts\sim 1 $ TeV) would come into operation.

\subsubsection{Measurements at the LHC}

%
The results from the following ATLAS analyses of the \ttbar\Ho\ process
are used:

1. \ttbar\Ho\ with \ttbar\ra bbq$\ell\nu$ and \Ho\ra\bb\ ~\cite{lhcbb};

2. \ttbar\Ho\ with \Ho\ra\WW\ and two like-sign leptons~\cite{lhcww};

3. \ttbar\Ho\ with \Ho\ra\WW\ and three leptons~\cite{lhcww}.

The expected numbers of selected signal and background events in the
three channels for various Higgs masses and total integrated luminosities 
of 30~\fb\, and 300~\fb\, are
listed in Tables~\ref{tab:lhcbb} and \ref{tab:lhcww}.  The results
obtained in this sub-section are based on the anticipated data sample of
\textit{one} LHC detector, with the luminosity per detector quoted below.

\begin{table}[h!]
\begin{center}
\scalebox{0.9}{
\begin{tabular}{|c||c|c||c|c|} \hline
\mH & \multicolumn{2}{c||}{30\fb} & \multicolumn{2}{|c|}{300\fb} \\ \cline{2-5}
(GeV) & \ttbar\Ho\, \Ho\ra\bb\ & background & \ttbar\Ho\, \Ho\ra\bb\ &
      background \\ \hline
100   & 83.4 &  303.4 & 279.0 & 1101.3  \\ \hline
110   & 63.0 &  275.7 & 232.5 & 1140.6  \\ \hline
120   & 43.0 &  234.1 & 173.1 & 1054.2  \\ \hline
130   & 26.5 &  200.1 & 112.5 & 1015.8  \\ \hline
140   & 13.9 &  178.2 &  62.4 & 947.1  \\ \hline
\end{tabular}
}
\end{center}
\caption{ \label{tab:lhcbb} Expected number of signal and background events
for the \ttbar\Ho\ with \ttbar\ra bbq$\ell\nu$ and \Ho\ra\bb\ analysis
at LHC ~\cite{lhcbb}.}
\end{table}
\begin{table}[h!]
\begin{center}
\scalebox{0.9}{
\begin{tabular}{|c||c|c||c|c||c|c||c|c||} \hline
 & \multicolumn{4}{|c||}{30 \fb} & \multicolumn{4}{|c|}{300 \fb} \\ \cline{2-9}
\mH  & \multicolumn{2}{|c||}{\ttbar\Ho\,\Ho\ra\WWn (2$\ell$)}
      & \multicolumn{2}{|c||}{\ttbar\Ho\,\Ho\ra\WWn (3$\ell$)}
      & \multicolumn{2}{|c||}{\ttbar\Ho\,\Ho\ra\WWn (2$\ell$)}
      & \multicolumn{2}{|c||}{\ttbar\Ho\,\Ho\ra\WWn (3$\ell$)} \\ \cline{2-9}
(GeV)  & signal& bckgr  
  & signal& bckgr  
  & signal & bckgr  
  & signal & bckgr  \\ \hline
120 & 4.4  & 19.6 &  2.7 & 21.2 & 12.7 & 80.6 &  10.5 & 97.6 \\ \hline
140 & 15.0 & 19.6 &  8.7 & 21.2 & 50.0 & 80.6 &  33.7 & 97.6 \\ \hline
160 & 21.1 & 19.6 & 13.0 & 21.2 & 72.3 & 80.6 &  55.3 & 97.6 \\ \hline
180 & 17.3 & 19.6 & 10.3 & 21.2 & 60.9 & 80.6 &  41.7 & 97.6 \\ \hline
200 & 10.5 & 19.6 & 5.7  & 21.2 & 43.2 & 80.6 &  26.4 & 97.6 \\ \hline
\end{tabular}
}
\end{center}
\caption{ \label{tab:lhcww} Expected number of signal and background events
for the \ttbar\Ho\ with \Ho\ra\WW\ (two like-sign leptons and three
leptons, respectively) analyses at LHC~\cite{lhcww}.}
\end{table}

From the expected event numbers we first estimate the 
uncertainty (statistical and systematic) on the measured cross
section $\sigma_{tth}^{data}$. Further uncertainties arise when
$\sigma_{tth}^{data}$ is compared to the theoretical prediction
as a function of $g_{tth}$.

The uncertainty on the observed cross section $\sigma_{tth}^{data}$
is calculated as
\begin{eqnarray*}
(\Delta\sigma_{tth}^{data}/\sigma_{tth}^{data})^2 & = &
(S+B) / S^2 + 
 (\Delta B_{syst})^2 / S^2 + 
 (\Delta {\cal L})^2 / {\cal L}^2  +
 (\Delta {\epsilon})^2 / {\epsilon}^2.
\end{eqnarray*}

Here, $S (B)$ is number of signal (background) events. $\Delta
B_{syst}$ is the uncertainty on the background determination from
sideband data (10\% in the $h\to\bb$ channel at high luminosity, 5\% otherwise).
$\Delta {\cal L}$ is the error on the integrated luminosity
(5\%) and $\Delta \epsilon$ is the error on the determination of the 
efficiency. This error involves uncertainties on the tagging
efficiency for individual b-jets (3\%) and leptons (3\% from isolation
requirement and 2\% from reconstruction efficiency) and an overall
detector efficiency uncertainty of 2\% (following ~\cite{due}). The total value
of $\Delta \epsilon$ is then calculated for each channel individually depending 
on the number of leptons and b-jets.

The expeced error inclduing systematic uncertainties and 
taking into account only the statistical error of each channel is 
shown in Table~\ref{tab:lhc2}. For the \Ho\ra\WW decay mode the signal and background
from the two lepton and three
lepton channels are added together since their signal contributions are 
exclusive and the overlap in the background is small.
The obtained result is consistent with the study presented in~\cite{due}.

\begin{table}[t!]
\begin{center}
\begin{tabular}{|c||c|c||c|c|} \hline
\mH\ & \multicolumn{2}{|c||}{30 \fb} & \multicolumn{2}{|c|}{300
  \fb} \\ \cline{2-5}

(GeV) & \Ho\ra\bb & \Ho\ra\WWn\ &\Ho\ra\bb & \Ho\ra\WWn\   \\ \hline
100   & 0.398(0.236)    &              & 0.249(0.133)&  \\ \hline
110   & 0.476(0.292)    &              & 0.287(0.159)&  \\ \hline
120   & 0.598(0.387)    & 1.023(0.974) & 0.345(0.202)& 0.732(0.611) \\ \hline
130   & 0.840(0.568)    & 0.524(0.492) & 0.488(0.299)& 0.362(0.295) \\ \hline
140   & 1.444(0.997)    & 0.370(0.339) & 0.804(0.509)& 0.252(0.193) \\ \hline
160   &                 & 0.287(0.254) &             & 0.196(0.137) \\ \hline
180   &                 & 0.331(0.300) &             & 0.221(0.163) \\ \hline
200   &                 & 0.486(0.454) &             & 0.282(0.222) \\ \hline

\end{tabular}
\end{center}
\caption{ \label{tab:lhc2} Expected relative precision
on $\sigma_{ttH} \times BR(H\to X)$ for the various LHC \ttbar\Ho\
analyses including systemtatic uncertainties (statistical error only). 
For $\Ho\ra\WW$~ the expected signal and background in the two and
three lepton final state have been added.}
\end{table}

In the next step the uncertainty on $g_{ttH}^2 * BR(H\to\bb/WW)$ which
arises when the observed $\sigma_{ttH} * BR(H\to \bb/WW)$ is compared to its
theoretical prediction. These uncertainties consist of the
uncertainties in the proton structure functions (5\%~\cite{sf,pdf}) and
uncertainties in the calculation of the production cross section.
Recent full
NLO calculations estimate the uncertainty of the cross section prediction
to be approximately 15\% from a variation of the hard 
scale~\cite{ref1_nlo_tev,ref1_nlo_lhc,ref2_nlo_tev,ref2_nlo_lhc}.
The total theoretical uncertainty $\Delta\sigma_{ttH}^{theo}$ is obtained by
adding the above two sources in quadrature.

Finally, the total uncertainty  $\Delta(g_{ttH}^2 * BR(H\to\bb/\WW))$ is
obtained according to
\begin{eqnarray*}
& \Delta(g_{ttH}^2 * BR(H\to\bb/\WW))^2/(g_{ttH}^2 * BR(H\to\bb/\WW)^2
& =  \\
& (\Delta\sigma_{ttH}^{theo})^2/(\sigma_{ttH}^{theo})^2 + 
(\Delta\sigma_{ttH}^{data})^2/(\sigma_{ttH}^{data})^2. & 
\end{eqnarray*}

\subsubsection{Measurements at the LC}


At the LC, the decay branching ratios into b quark pairs and W boson pairs
can be measured at \sqrts = 350~GeV to the precision listed in 
Table~\ref{tab:lc}\cite{Aguilar-Saavedra:2001rg}.

\begin{table}[t!]
\begin{center}
\begin{tabular}{|c|c|c|} \hline
\mH~(GeV)     & $\Delta$BR(\bb)/BR(\bb)   & $\Delta$BR(WW)/BR(WW) \\ \hline
100     & 0.024                   &                      \\ \hline
120     & 0.024                   & 0.051                 \\ \hline
140     & 0.026                   & 0.025                 \\ \hline
160     & 0.065                   & 0.021                 \\ \hline
200     &                         & 0.021                 \\ \hline
\end{tabular}
\end{center}

\caption{\label{tab:lc} 
Relative precision on the branching ratio for $\Ho\to\bb$
and $\Ho\to\WW$ expected
for a LC running at \sqrts = 350~GeV with 500 fb$^{-1}$.}

\end{table}

\subsubsection{Results}

For an extraction of the top quark Yukawa coupling at each Higgs mass
we combine the LHC rate measurement of \ttbar\Ho with \Ho\ra\bb\ or
\Ho\ra\WW\ with the corresponding measurement of the branching ratio
at the LC.  We make the tree level assumption that the cross section
$\sigma_{ttH}$ is proportional to $g_{ttH}^2$. Thus, the relative
error on $g_{ttH}$ is simply given by $\Delta g_{tth} / g_{ttH} = 0.5
\Delta \sigma_{ttH} / \sigma_{ttH}$. The relative error on
$\sigma_{ttH}$ is obtained by adding in quadrature the statistical
and systematic uncertainties as
described above and the error of the LC branching ratio
measurement. The combination of the \bb\ and \WW\ final states is
performed by
\begin{equation*}
\left( \frac{\Delta g_{tth}} {g_{ttH}} \right)_{comb.}^{-2}= 
\left( \frac{\Delta g_{tth}} {g_{ttH}} \right)_{WW}^{-2} +
\left( \frac{\Delta g_{tth}} {g_{ttH}} \right)_{b\bar{b}}^{-2}. 
\end{equation*}

\begin{table}[b!]
\begin{center}
\begin{tabular}{|c||c|c|c||c|c|c|} \hline
\mH  & \multicolumn{3}{|c||}{30 \fb} & \multicolumn{3}{|c|}{300
  \fb} \\ \hline
(GeV) & bb & WW & bb+WW& bb & WW & bb + WW \\ \hline
100  &  0.22(0.12) &  & & 0.15(0.07) &  & \\ \hline                              
110  &  0.25(0.15) &  & & 0.17(0.08) &  & \\ \hline                              
120  &  0.31(0.19) &  0.52(0.49)  & 0.27(0.18) & 0.19(0.10) &0.38(0.31) & 0.17(0.10) \\ \hline
130  &  0.43(0.28) &  0.28(0.25)  & 0.23(0.19) & 0.26(0.15) &0.20(0.15) & 0.16(0.11) \\ \hline
140  &  0.72(0.50) &  0.20(0.17)  & 0.19(0.16))& 0.41(0.26) &0.15(0.10) & 0.14(0.09) \\ \hline 
150  &              & 0.18(0.14)  &              & 1.88(1.21) &0.14(0.08) & 0.14(0.08) \\ \hline
160  &              & 0.16(0.13)  &              &              &0.13(0.07) &  \\ \hline
170  &              & 0.17(0.13)  &              &              &0.13(0.07) &  \\ \hline
180  &              & 0.18(0.15)  &              &              &0.14(0.08) &  \\ \hline
190  &              & 0.22(0.19)  &              &              &0.15(0.10) &  \\ \hline
200  &              & 0.26(0.23)  &              &              &0.16(0.11) &  \\ \hline
\end{tabular}
\end{center}
\caption{\label{tab:result} Expected relative error on the top 
Yukawa coupling $g_{ttH}$ from the rate measurement including all
systematic uncertainties (statistic errors only) at the LHC  
and from the branching ratio measurement at the LC.}
\end{table}

\begin{figure}[th!]
\begin{center}
\resizebox{1.06\textwidth}{!}{
\includegraphics*{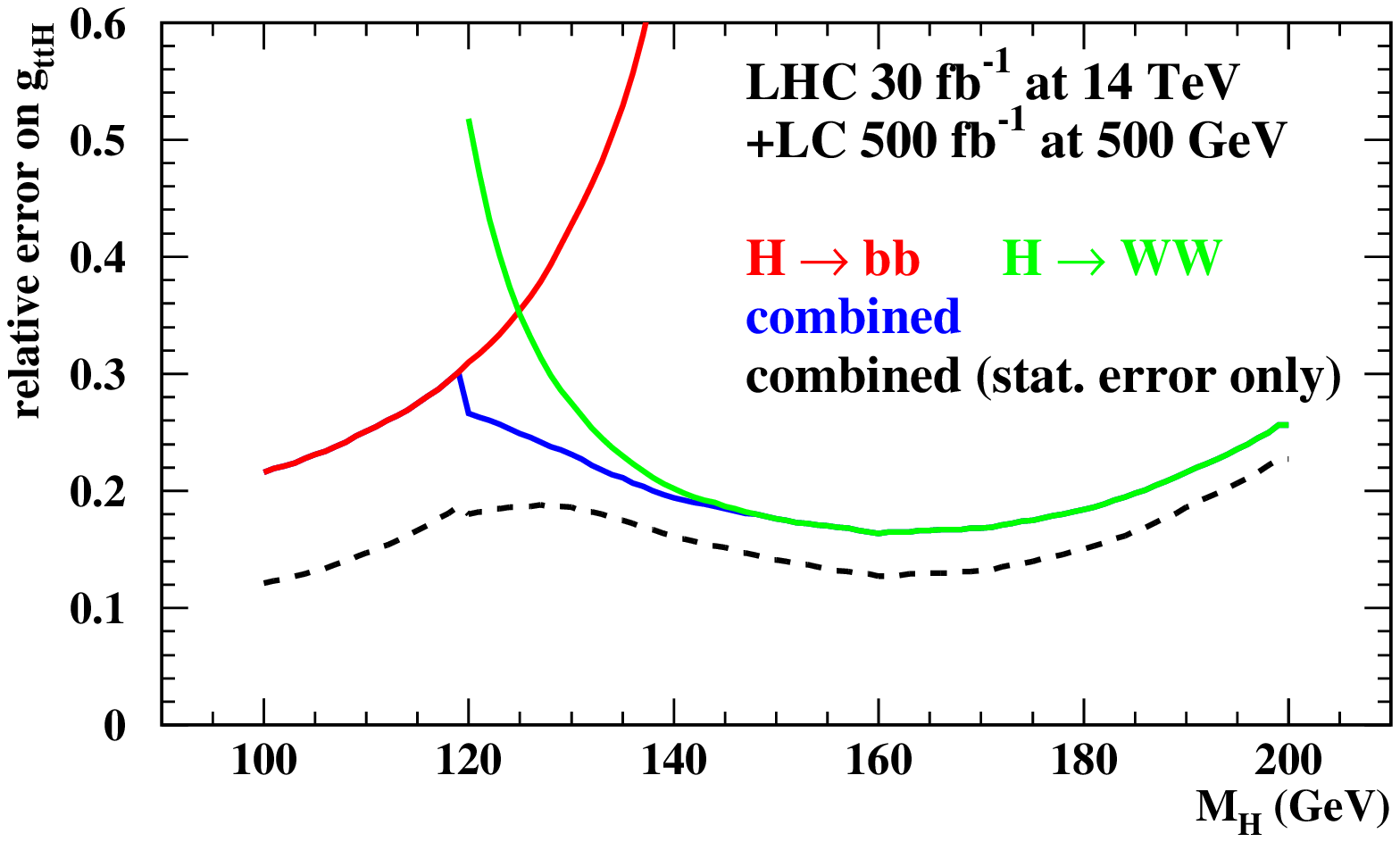}
\hspace*{3mm}
\includegraphics*{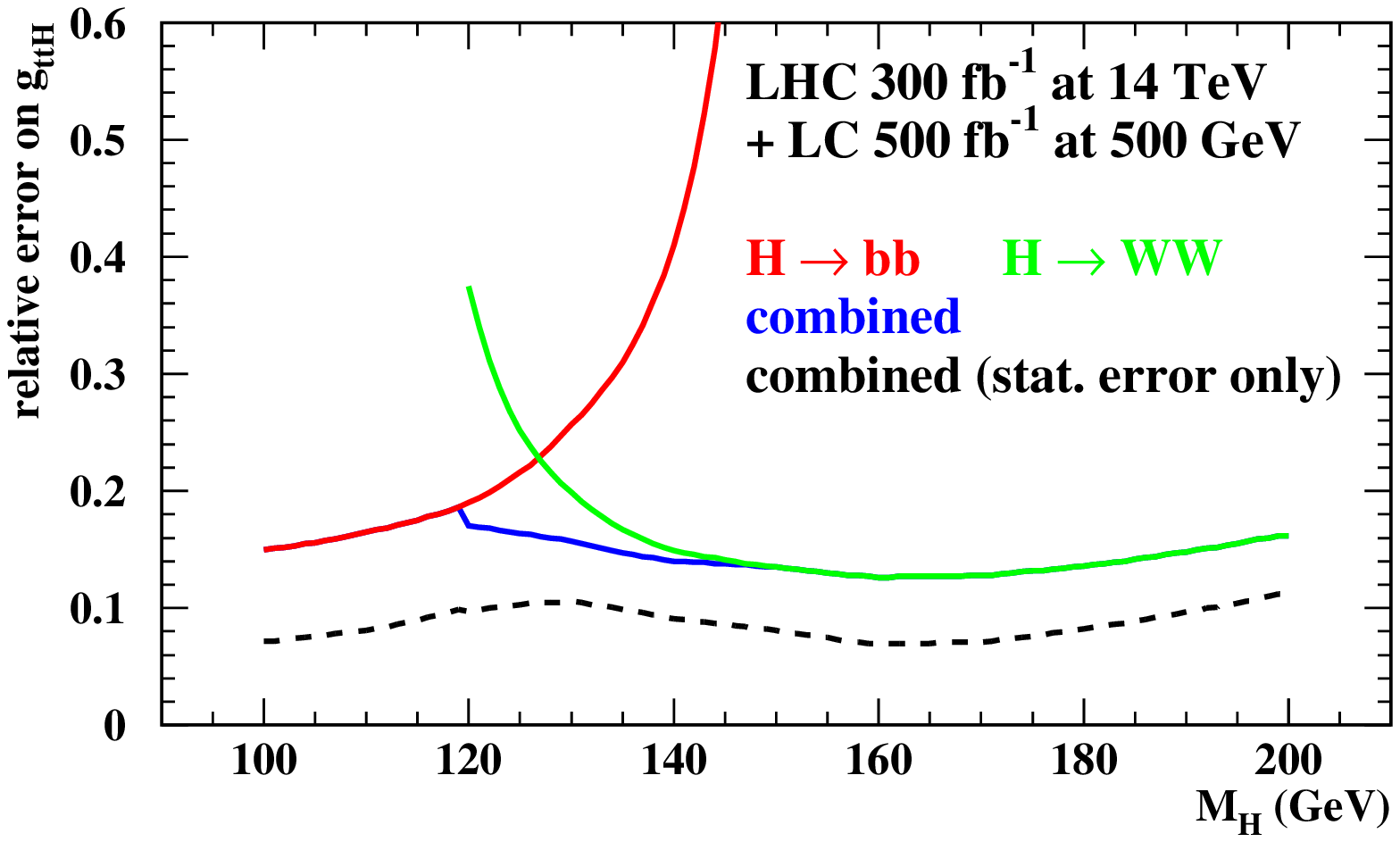}
}
\end{center}
\caption{\label{fig:result1} 
Achievable precision on the top Yukawa
coupling from 30~\fb at the LHC and 500~\fb at the LC (left),
and from 300~\fb at the LHC and 500~\fb at the LC
(right). The red curve shows the precision obtainable from the 
\Ho\ra\bb\ final state, the green from the \Ho\ra\WW\ final state and
the blue curve from the combination of the two. The dashed lines 
show the expected precision taking into account only statistical errors.}
\vspace{-0.2in}
\end{figure}

\begin{figure}[ht!]
\centerline{
\epsfig{file=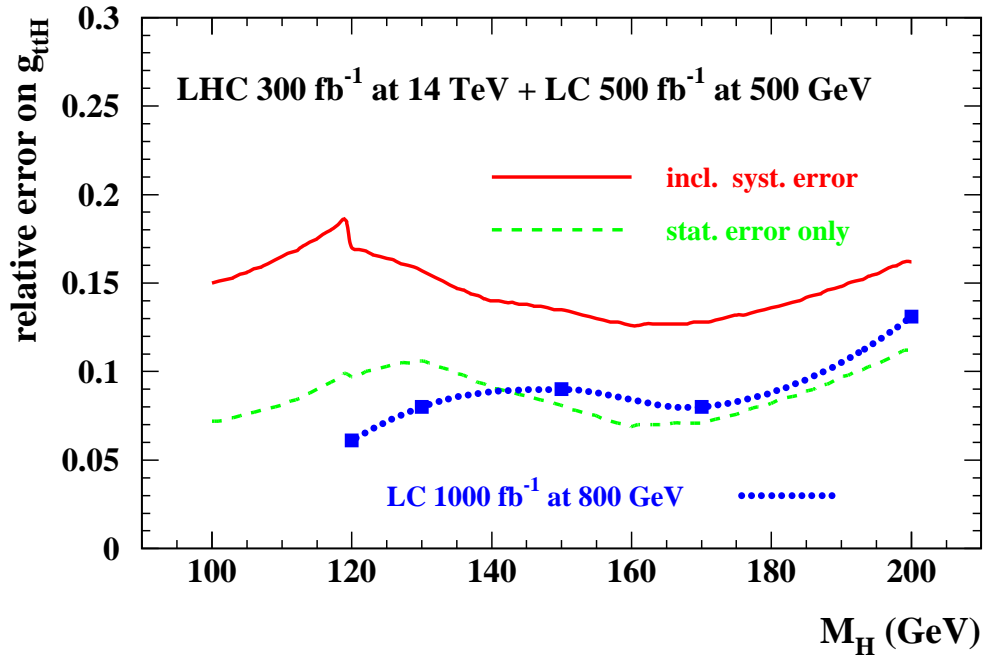,width=0.85\linewidth}  }
\caption{\label{fig:result2} Achievable precision on the top Yukawa
coupling from 300~\fb at the LHC and 500~\fb at the LC at 500 GeV
taking into account all systematic uncertainties (full red curve)  
and  using only statistical errors (dashed green curve) .
For comparison the expected precision from 1000~\fb at the LC at 800
GeV alone (dotted blue curve) is also shown.}
\end{figure}

The relative accuracies on the top quark Yukawa coupling 
that are achievable are
summarised in Table~\ref{tab:result}. 
In Fig.~\ref{fig:result1} the 
relative accuracy from the $H\to\bb$ and $H\to WW$ channels are
shown individually and combined both for low and high luminosity at 
the LHC. Also shown are the results that would be obtained if all
systematic errors were neglected. 
For 300~\fb at the LHC and 500~\fb at the LC,
the obtainable relative uncertainty is approximately 15\% for a Higgs
boson mass between 120 and 200 GeV. The purely statisical uncertainty
ranges from 7\% to 11\% as shown in Fig.~\ref{fig:result2}.

The size of the obtained uncertainties is comparable to those obtained
for the LHC alone~\cite{due} but in contrast to the latter no
model-dependent assumptions are made.
In Fig.~\ref{fig:result2} we also show the precision which can be achieved
at the LC alone if operated at 800~GeV center-of-mass energy~\cite{gay}
from the measurement of the \ee\ra\ttbar\Ho\ process with \Ho\ra\bb\ and
and \Ho\ra\WW\ combined.

%

\subsection{Associated $t\bar th$ production at the LC and LHC}

{\it S.~Dawson, A.~Juste, L.~Reina and D.~Wackeroth}

\vspace{1em}

\def\Ho{h_0}
\def\ttbar{t\overline{t}}
\def\bb{b\overline {b}}
\def\Ndash{-}

Once the Higgs boson has been discovered, it is
crucial to measure its couplings to fermions and gauge bosons.  The
couplings to the gauge bosons can be measured through the associated
production processes, $e^+e^-\to Zh$, $q\bar{q}^\prime\to W^\pm h$,
and $q\bar{q}\to Zh$, and through vector boson fusion, $W^+W^-\to h$
and $ZZ\to h$. The couplings of the Higgs boson to fermions are more
difficult to measure. We focus on prospects for measuring the top
quark Yukawa coupling.

At a Linear Collider (LC) with an energy $\sqrt{s}\!=\!500$~GeV, a
Higgs boson with mass less than around $200$~GeV will be copiously
produced in association with a $Z$ boson, $e^+e^-\to Zh$.  The
magnitude of the production rate probes the $ZZh$ coupling. Using
missing mass techniques, the LC can measure the Higgs branching ratios
(and hence couplings) for many decays\cite{desch,Brau:2001}.  The decay rates
measure the individual Yukawa couplings in a model independent fashion
and with small errors.  It is not possible, however, to probe the top
quark Yukawa coupling ($g_{tth}$) through the decay $h\to t\bar{t}$
for a light Standard Model (SM) Higgs boson ($M_h\!<\!200$~GeV), since
this decay is not kinematically accessible.

The LHC, on the other hand, will measure products of Higgs boson
production cross sections times Higgs boson branching fractions.
Ratios of various decay rates can then be extracted in a model
independent way, while measurements of individual Higgs boson
couplings to fermions and gauge bosons can be obtained under specific
assumptions~\cite{zep}.  The dominant source of Higgs bosons, gluon
fusion ($gg\to h$), proceeds through a top quark loop and in principle
is sensitive to the top quark Yukawa coupling.  However, heavy colored
particles beyond the Standard Model would contribute to the $gg\to h$
process and invalidate the interpretation of this production process
as a measurement of the top quark Yukawa coupling.

At both a linear and a hadron collider the top Yukawa coupling is only
directly accessible through the associated $t\bar{t}h$ production
mode. At a LC, the event rate for $e^+e^-\to t\bar{t}h$ is tiny for
$\sqrt{s}\!=\!500$~GeV and peaks for an energy scale between
$\sqrt{s}\!=\!700$\Ndash $800$~GeV for $M_h\!\sim\!120$\Ndash $130$~GeV.  
At the LHC,
$pp\to t\bar{t}h$ is an important discovery channel for a relatively
light Higgs boson ($M_{h}\!<\!130$~GeV).  Although the event rate is
small, the signature is quite distinctive.  The total cross sections
for $pp,p\bar{p},e^+e^-\to t\bar{t}h$ are known at next-to-leading
order (NLO) in QCD and we review the status of current theoretical
calculations.

For Higgs boson mass $M_h\!<\!135$~GeV the Higgs boson mainly decays
into $h\to b\bar{b}$, while above this threshold $h\to W^+W^-$
dominates. In the following discussion we will consider the decays
$h\to b\bar{b},\tau^+\tau^-$ and $W^+W^-$ and the corresponding
measurements of the top quark Yukawa coupling which can be obtained at
the LHC and a linear collider. A LC running at $\sqrt{s}\!=\!800$~GeV
will be able to determine $g_{tth}$ at the 4--5\% level
($M_h\!=\!120$~GeV, $L\!=\!1000~\mathrm{fb}^{-1}$), while at lower
energies the accuracy on this coupling deteriorates quickly. The LHC,
on the other hand, could, under some assumptions, measure $g_{tth}$ at
the 10-20\% level ($M_h\!=\!110$\Ndash $120$~GeV,
$L\!=\!100$\Ndash $300~\mathrm{fb}^{-1}$ per detector). However, by combining
the precisely measured branching ratios for the Higgs boson decay to
the lighter fermions and to gauge bosons, which could be obtained at a
$\sqrt{s}\!=\!500$~GeV LC, with a measurement of $t\bar{t}h$
production at the LHC, the LHC could extract a measurement of the top
quark Yukawa coupling which is free of theoretical uncertainties about
the other Higgs boson decay rates.

In the Standard Model, the couplings of the Higgs boson to fermions
are completely determined in terms of the corresponding fermion
masses, $g_{ffh}\!=\!{M_f}/{v}$, where $v\!=\!(\sqrt{2}G_F)^{-1/2}$ and so
the top quark-Higgs boson coupling is the largest Yukawa coupling.  In
extensions of the Standard Model, however, the Yukawa couplings can be
significantly different.  They are no longer strictly proportional to
the fermion masses, but depend on the parameters of the model.  The
minimal supersymmetric extension of the Standard
Model (MSSM) provides a useful benchmark for
comparison with the Standard Model.

In the following sections, we compare the $t {\overline t} h$
production rates in hadronic
and $e^+e^-$ collisions and examine the various Higgs decay channels.
We also discuss the possibility of probing the top-Higgs Yukawa coupling
by measuring the threshold energy dependence of the $e^+e^-\rightarrow
t {\overline t}$ cross section.
The current state of experimental studies of the $t\bar{t}h$ process
is surveyed, with an emphasis on the extraction of the top quark
Yukawa coupling. Unless stated otherwise, we concentrate
entirely on the Standard Model Higgs boson.
\subsubsection{Linear Collider}
%

\vspace{0.2cm}
\hspace{1.5em} $\bullet$ {\bf $t {\overline t} h$ Production Rates}
\vspace{0.2cm}
%

\begin{figure}[h!]
\begin{picture}(100,120)(-20,-25)
\SetScale{0.8}
\ArrowLine(0,100)(50,50)
\ArrowLine(0,0)(50,50)
\Photon(50,50)(100,50){3}{6}
\ArrowLine(100,50)(150,100) 
\ArrowLine(100,50)(150,0)
\DashLine(120,70)(170,70){5}
\put(180,35){$+$} 
\put(50,25){$\gamma, Z$} 
\put(0,85){$e^-$}
\put(0,-5){$e^+$}
\put(140,52){$h$}
\put(125,85){$t$}
\put(125,-5){${\overline t}$}
\SetScale{1}
\end{picture}
\begin{picture}(100,120)(-140,-25)
\SetScale{0.8}
\ArrowLine(0,100)(50,50)
\ArrowLine(0,0)(50,50)
\Photon(50,50)(100,50){3}{6}
\DashLine(75,50)(75,0){5}
\ArrowLine(100,50)(150,100) 
\ArrowLine(100,50)(150,0)
\put(180,35){$+ \cdots$} 
\put(45,25){$Z$} 
\put(0,85){$e^-$}
\put(0,-5){$e^+$}
\put(62,15){$h$}
\put(125,85){$t$}
\put(125,-5){${\overline t}$}
\SetScale{1}
\end{picture} 
\caption[]{Feynman diagrams contributing to the lowest
  order process, $e^+e^-\to t\bar{t}h$.}
\label{lofeyndiag}
\end{figure}
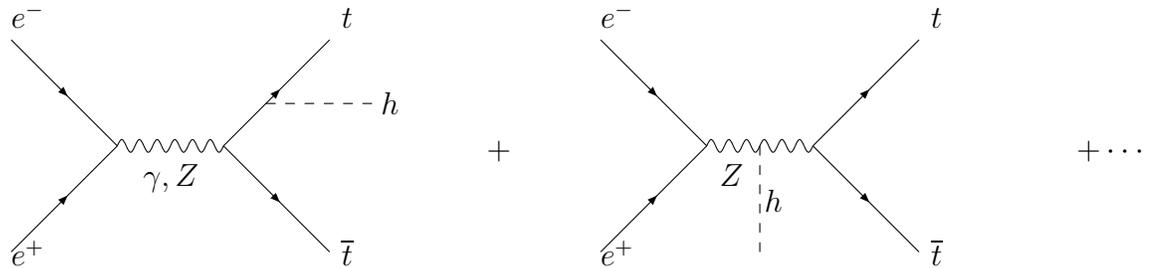

\noindent
The production of a $t\bar{t}h$ final state in $e^+e^-$ collisions
proceeds at the tree level through the Feynman diagrams shown in
Fig.~\ref{lofeyndiag}~\cite{goun}. 

\begin{figure}[t!]
\centering
\includegraphics[width=4.5in]{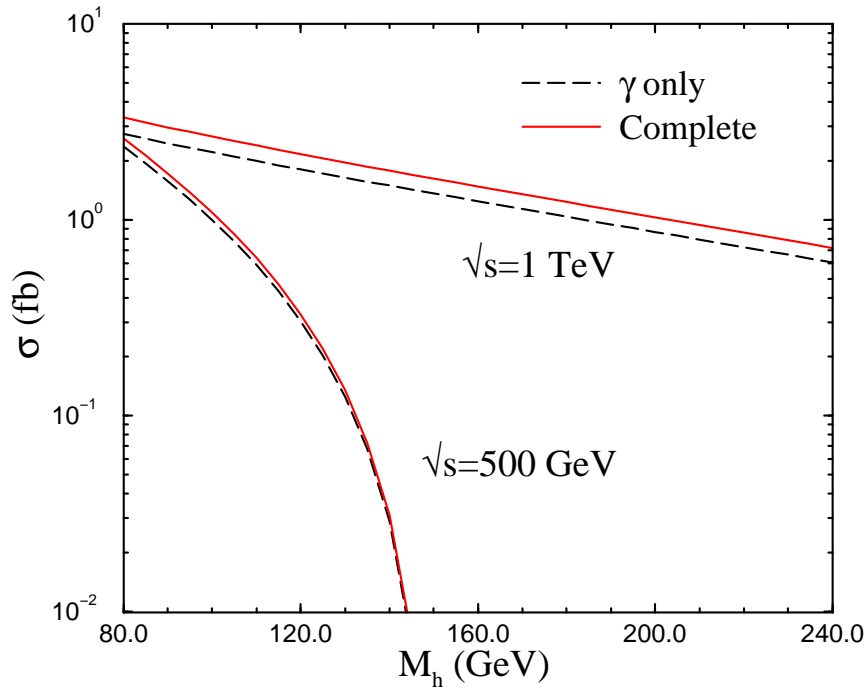}
\caption[]{Lowest order cross section for 
  $e^+e^-\to t\bar{t}h$ at $\sqrt{s}\!=\!500$~GeV and
  $\sqrt{s}\!=\!1$~TeV.  The curve labeled {\it complete} includes
  both $\gamma$ and $Z$ exchange, along with bremsstrahlung from the
  $Z$ boson~\cite{eenlo}.}
\label{tthlofig}
\end{figure}

To the extent to which the Higgs
boson radiation from the $s-$channel $Z$-exchange is negligible, the
cross section is directly proportional to $g_{tth}^2$.  In
Fig.~\ref{tthlofig}, the total rate from photon exchange only is
compared with the total contribution, showing that the effect of $Z$
exchange is only a few percent correction, in particular at low
center-of-mass energies.  Fig.~\ref{tthlofig} contains only the tree level
cross section:  at higher orders, the QCD and electroweak radiative
corrections could significantly change the relative 
contributions of the $\gamma$ and $Z$ exchange.
The process has an optimal energy since the
rate peaks around $\sqrt{s}\!\sim\!700$\Ndash $800$~GeV
($\sigma_{t\bar{t}h}\!\simeq\!2.5$~fb for $M_h\!\sim\!120$~GeV).  At
$\sqrt{s}\!=\!500$~GeV and for $M_h\!\sim\!120$~GeV, the cross section
is of order 0.5~fb.

QED corrections due to initial state radiation effects (bremsstrahlung
and beamtrahlung, the former being dominant) significantly distort the
$t\bar{t}h$ lineshape.  The main effect is to shift the maximum of the
cross section towards higher $\sqrt{s}$. As a result, the cross
section at $\sqrt{s}\!=\!500$~GeV is reduced by a factor $\simeq\!2$,
\textit{i.e.} $\sigma_{t\bar{t}h}\simeq\!0.2$ fb for 
$M_h\sim 120~$GeV\cite{juste,juste2}.

\begin{figure}[t!]
\centering
\includegraphics[width=4.5in]{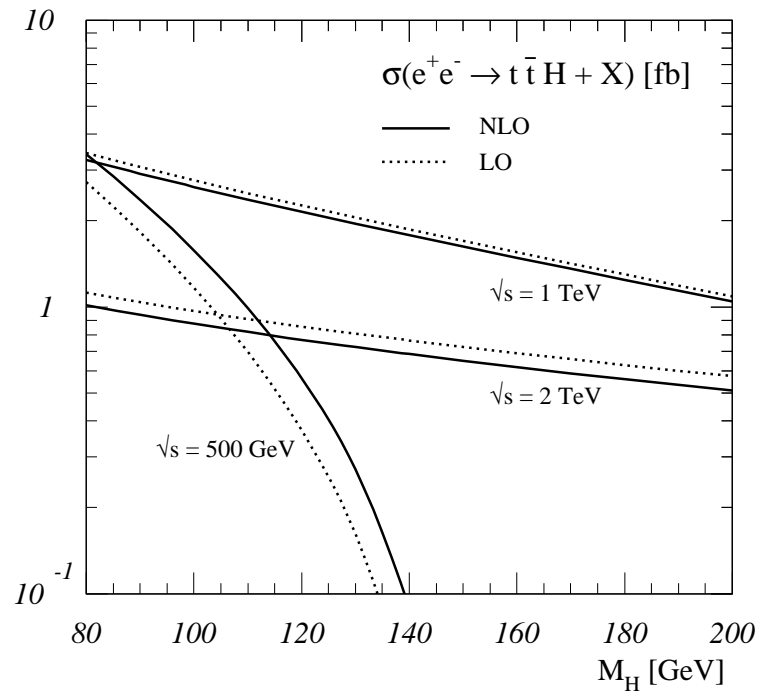}
\caption[]{LO and NLO cross sections
for $e^+e^-\to t\bar{t}h$ as a function of the Higgs mass $M_h$ for
$\sqrt{s}\!=\!500$~GeV, 1~TeV and 2~TeV.  The
renormalization scale is set to $\mu\!=\!\sqrt{s}$~\cite{eenlo_ditt}.}
\label{eenlo_fig}
\end{figure}

The first order QCD corrections to this process, which turn out to be
important near the $t\bar{t}$ threshold, have been computed including
only  $\gamma$ exchange  in
Ref.~\cite{eenlo} and with the complete $\gamma$ and
$Z$ contributions in Ref.~\cite{eenlo_ditt}.  The QCD corrected rate, as a function
of $M_h$ and for different center-of-mass energies, is compared to the
uncorrected one in Fig.~\ref{eenlo_fig}.  At $\sqrt{s}\!=\!500$~GeV,
the corrections are large and positive, with a strong dependence on
$M_h$ ($K_{NLO}\!\simeq\!1.5$, for $\mu\!=\!\sqrt{s}$ and
$M_h\!=\!120$~GeV).  At $\sqrt{s}\!=\!1$~TeV, they are small and
negative ($K_{NLO}\!\simeq\!0.9$, for $\mu\!=\!\sqrt{s}$), and
essentially independent of $M_h$.  The only scale dependence at NLO is
the running of the strong coupling constant, $\alpha_s(\mu)$.
Changing the scale parameter by a 
factor of 2 induces roughly a $10\%$ change in
$\sigma_{NLO}$, indicating that higher order QCD corrections may be
important to match the experimental precision of a high energy LC.

\begin{figure}[h!]
\centering
\epsfxsize=4.in
\leavevmode\epsffile{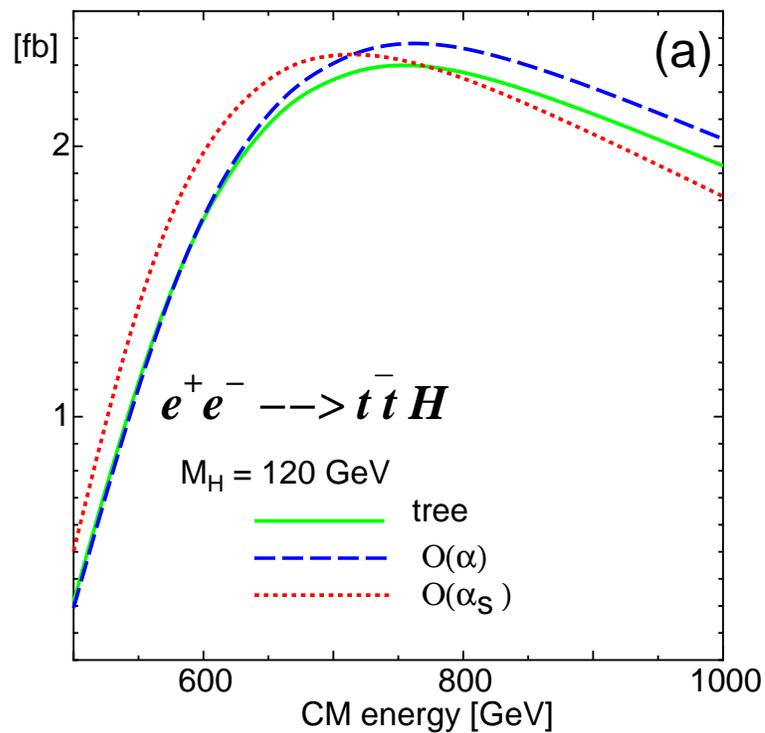}
\caption[]{LO and NLO cross sections
for $e^+e^-\to t\bar{t}h$ showing the effect of the QCD 
and electroweak corrections to the total
cross section for $M_h=120$~GeV.~\cite{grace}.}
\label{grace_fig}
\end{figure}

The electroweak corrections to $e^+e^-
\rightarrow t {\overline t} h$ are also important and
have been computed by three groups\cite{grace,ew2,ew2b}.
At $\sqrt{s}=500~$GeV, the QCD corrections are significantly larger than
the electroweak corrections, while at $\sqrt{s}=1~$TeV, the
electroweak corrections are roughly the same size as the QCD
corrections but opposite in sign, as shown in Fig.~\ref{grace_fig}.

\vspace{0.4cm}
$\bullet$ $\boldsymbol{h\to b\bar{b}}$
\vspace{0.2cm}

\noindent
The Higgs decays predominantly to $b\bar{b}$ pairs for
$M_h\!<\!135$~GeV.  The final state of interest is then
\begin{equation}
e^+e^-\to  t\bar{t}h\to t\bar{t}b\bar{b}\,\,\,.
\end{equation}
The possibility of fully reconstructing the two top quarks in the
final state allows efficient discrimination of the signal over the
background, and together with a good b-tagging efficiency is crucial
for extracting the signal and for increasing the precision with which
the top Yukawa coupling can be measured.  Both the semi-leptonic final
state
\begin{equation}
e^+e^-\to (bl\nu) + (bq\bar{q}^\prime) + (h\to b\bar{b})\,\,\,,
\end{equation}
and the fully hadronic final state
\begin{equation}
e^+e^-\to  (bq_1\bar{q}_1^\prime) + (bq_2\bar{q}_2^\prime)+ (h\to b\bar{b})
\end{equation}
can be observed.

In the semi-leptonic decay, the final state is 4 b-jets, 2 light quark
jets, a high $p_T$ isolated lepton and missing energy from the
neutrino.  The largest interfering background is the QCD background
from the gluon splitting process, $e^+e^-\to t\bar{t}g^*(g^*\to
b\bar{b})$. The $b\bar{b}$ pairs resulting from the gluon splitting,
however, tend to peak at low values of the $b\bar{b}$ invariant mass.
There is also an electroweak background, of which the dominant
contribution is $e^+e^-\to Zt\bar{t}$.  Although the electroweak
background is formally smaller than the QCD background, the $Z\to
b\bar{b}$ decay resonates in close proximity to the expected Higgs
signal. A parton level calculation of these interfering backgrounds
has been performed \cite{moretti} for the following final state:
\begin{equation}
e^+e^-\to b\bar{b} b\bar{b} l\nu q\bar{q}^\prime\,\,\,,
\end{equation}
using helicity amplitudes and no factorization of the production and
decay channels.

A feasibility study of the measurement of the top quark Yukawa
coupling was performed in Ref.~\cite{bdr}, including both semileptonic
and fully hadronic decay channels. This study considered the
production process $e^+e^-\to t\bar{t}b\bar{b}$, with QCD radiation,
hadronization and particles decays (including the top quark),
including a toy detector simulation, handled by ISAJET
\cite{Baer:1999sp}.  The analysis also included full reconstruction of
the top quark and $W$ boson masses.  The fully hadronic channel for
$e^+e^-\to t\bar{t}h$ has the advantage of initially higher rates than
the semi-leptonic channel due to the large $W$ boson hadronic
branching fraction.  However, in attempting mass reconstructions, a
greater combinatoric problem is presented.
The two studies \cite{moretti,bdr} found agreement on the size of the
backgrounds, lending validity to the use of the factorization
approximation for the decays.

At $\sqrt{s}\!=\!500$~GeV and $M_h\!=\!120$~GeV, a jet $b-$tagging
efficiency $\epsilon_b$, and an integrated luminosity $L$, the
precision on the top quark Yukawa coupling obtained by combining the
semi-leptonic channel with the hadronic channel is
predicted~\cite{bdr} to be:
\begin{equation}
\frac{\delta g_{tth}}{g_{tth}}\sim
21\%\sqrt{\frac{1000~\mathrm{fb}^{-1}}{L\epsilon_b^4}}\,\,\,.
\end{equation}
For $M_h\!=\!130$~GeV, the precision declines to
\begin{equation}
\frac{\delta g_{tth}}{g_{tth}}\sim
44\%\sqrt{\frac{1000~\mathrm{fb}^{-1}}{L \epsilon_b^4}}\,\,\,,
\end{equation}
primarily due to the suppression of the rate
for the heavier Higgs mass.

A more detailed preliminary study was performed for
$\sqrt{s}\!=\!500$~GeV and $L\!=\!1000~\mathrm{fb}^{-1}$, assuming
$M_{h}\!=\!120$~GeV, and considering so far only the semileptonic
channel~\cite{juste2}.  The analysis included both reducible and
irreducible backgrounds, realistic detector effects, and
reconstruction efficiencies. The main sources of efficiency loss are
from limitations of the jet-clustering algorithm and b-tagging
performances in a high jet multiplicity environment.  Signal and
backgrounds are discriminated by making use of a combination of highly
efficient preselections and multivariate techniques involving Neural
Networks.  Because of the large backgrounds, it is crucial that they
are well modeled both in normalization and shape. A conservative
estimate of $5\%$ in the overall background normalization was
included. The estimated total uncertainty in the top quark Yukawa
coupling is approximately $33\%$.  From the optimization of this
analysis (improved b-tagging performance, dedicated $\tau$ selections,
extensive use of kinematical information), together with the
combination with the hadronic channel and using NLO K-factors, an
ultimate accuracy of $\leq 20\%$ is expected.
The increased signal rate and reduced non-interfering backgrounds at
higher $\sqrt{s}$ allow for a more precise measurement of the top
quark Yukawa coupling.

A feasibility study~\cite{juste}, comparable in level of
sophistication to the previous one, but considering both semileptonic
and hadronic decay channels at $\sqrt{s}\!=\!800$~GeV and for
$M_{h}\!=\!120$~GeV, has been performed.  Assuming
$L\!=\!1000~\mathrm{fb}^{-1}$, the estimated total uncertainty in the
top-Higgs Yukawa coupling is:
\begin{equation}
\frac{\delta g_{tth}}{g_{tth}}\sim 5.5\%\,\,\,.
\end{equation}
Including only the statistical error, the precision becomes $4.2\%$.

\vspace{0.2cm}
$\bullet$ {$\boldsymbol{h\rightarrow WW^*}$}
\vspace{0.2cm}

\noindent
Within the SM and for $M_h\!\simeq\!135$~GeV, the branching ratio for
$h\to WW^*$ is comparable in size with that for $h\to b\bar{b}$. For
larger Higgs masses, $h\to WW^*$ becomes the dominant decay mode.
Therefore, it is important to explore the potential of this channel to
further increase the sensitivity to the top-Higgs decay Yukawa
coupling up to higher values of $M_h$.

The process under consideration contains 4 W bosons and 2 b-jets in
the final state:
\begin{equation}
e^+e^-\to t\bar{t}h \to W^+bW^-\bar{b}WW^*.
\end{equation}

The final state is fully determined by the W boson decay modes, thus
offering a wide variety of experimental signatures:

\begin{Eqnarray}
{\rm fully\:hadronic\:(10\:jets)}&:&\quad \BR\simeq 20.8\%\,\,\,;\nonumber\\ 
{\rm semileptonic\:(1\:lepton + 8\: jets)}&:&\quad \BR\simeq 40.0\% 
\,\,\,;\nonumber\\
{\rm 2\: opposite-sign\: leptons + 6\: jets}&:&\quad \BR\simeq 19.3\% 
\,\,\,;\nonumber\\ 
{\rm 2\: same-sign\: leptons + 6\: jets}&:&\quad \BR\simeq 9.6\% 
\,\,\,;\nonumber\\ 
{\rm 3\: leptons + 4\: jets}&:&\quad \BR\simeq 9.3\%\,\,\,; \nonumber \\ 
{\rm 4\: leptons + 2\: jets}&:&\quad \BR\simeq 1.0\%\,\,\,. \nonumber
\end{Eqnarray}

A feasibility study has been performed for $\sqrt{s}\!=\!800$~GeV in
the ``2 same-sign leptons+jets'' final state~\cite{gay}. This channel
is expected to have low backgrounds and therefore a relatively simple
topological selection appears to be sufficient.  A $5\%$ systematic
uncertainty in the overall background normalization has been included.
Assuming $L\!=\!1000~\mathrm{fb}^{-1}$, the estimated total
uncertainty in the top-Higgs Yukawa coupling is found to be $\simeq
15\%$ for $140\!\leq\!  M_h\!\leq\!180$~GeV.

In terms of statistics, the semileptonic channel is the most promising
one, although the backgrounds are expected to be much larger that for
the ``2 same-sign leptons+jets'' one.  This process could be analyzed
with a similar strategy to the one applied to the semileptonic channel
in $e^+e^-\to t\bar{t}h (h\to b\bar{b})$. Indeed, the increased jet
multiplicity in the final state makes it more challenging. Preliminary
studies~\cite{gay2,gay3} indicate a significant improvement in the
total uncertainty on $g_{tth}$ by combining it with the ``2 same-sign
leptons+jets'' channel.

Finally, the results of Refs.~\cite{gay2,gay3} suggest that the
combination of the $h\to b\bar{b}$ and $h\to WW^*$ decay channels
could yield an ultimate uncertainty of 
\begin{equation}
\frac{\delta g_{tth}}{g_{tth}}\!\leq\!15\%
\end{equation}
for $M_h\!\leq\!200$~GeV, assuming
$\sqrt{s}\!=\!800$~GeV and $L\!=\!1000~\mathrm{fb}^{-1}$.

\vspace{0.2cm}
$\bullet$ {\bf $t {\overline t}$ Threshold Scan}
\vspace{0.2cm}

\noindent
The total cross section for $e^+e^-\to t\bar{t}$ as a function of
center-of-mass energy is quite sensitive to the top quark mass,
$m_t$,~\cite{fujii}. It is also sensitive to a lesser degree to the
strong coupling constant $\alpha_s$, the total top quark decay width,
$\Gamma_t$, and the top quark Yukawa coupling, $g_{tth}$.  By
measuring the threshold energy dependence of the forward backward
asymmetry and the position of the peak of the top quark momentum
distribution, additional sensitivity to $\Gamma_t$ and $g_{tth}$ can
be obtained~\cite{fujii2,miquel}, although the sensitivity to
the top quark Yukawa coupling is quite small in all observables.

The total $e^+e^-\to t\bar{t}$ cross section at threshold
has been calculated 
including some of the next-to-next-to-leading 
logarithms, \cite{hoang1}. The complete set
of next-to-next-to-leading logarithmic contributions is not
yet complete, but the large size of the corrections
relative to the next-to-leading logarithmic terms\cite{hoang2}
 suggests
that the uncertainty on the cross section measurement will
be slightly larger than previously estimated,
$\delta\sigma_{tt}/\sigma_{tt}\sim \pm 6\%$.
 
In Ref.~\cite{miquel} an experimental study of the precision on
$g_{tth}$ has been performed which we will summarize in the following.
The study assumes an integrated luminosity of $300~\mathrm{fb}^{-1}$
and $M_h\!=\!120$~GeV. Theoretical predictions are based on the TOPPIK
program~\cite{kuhn}. By fixing all variables except $g_{tth}$ and
assuming a $3\%$ systematic uncertainty on the total cross section, a
measurement of
\begin{equation} 
\frac{\delta g_{tth}}{g_{tth}}=\mbox{}^{+0.18}_{-0.25} 
\end{equation} 
could be obtained from the threshold scan.  Of course, a more
realistic analysis would include the uncertainties on the other
variables. Leaving $m_t$ and $\alpha_s$ free with assumed 
errors
\begin{Eqnarray}
\Delta\alpha_s&=&0.001\,,\nonumber \\[4pt]
\Delta m_t&=& 27 {\rm MeV}\,\,, 
\end{Eqnarray}%
but fixing $\Gamma_t$ to the Standard Model value (and assuming a
systematic error of $\delta\sigma_{tt}/\sigma_{tt}\!=\!1\%$), the
precision would be reduced to
\begin{equation} 
\frac{\delta g_{tth}}{g_{tth}}=\mbox{}^{+0.33}_{-0.54}\,\,\,. 
\end{equation} 
Finally, when also leaving $\Gamma_t$ free,
\begin{equation} 
\frac{\delta g_{tth}}{g_{tth}}=\mbox{}^{+0.35}_{-0.65}\,\,\,. 
\end{equation} 
This precision is not competitive with that which
can be obtained through
$e^+e^-\to t\bar{t}h$ at $\sqrt{s}\!=\!500$~GeV.  In view of the
possibility of a systematic experimental error on the total $t\bar{t}$
threshold cross section of $1\%$, the effects of electroweak radiative
corrections\cite{ttew} also need to be included.
Interpreting the top quark threshold
measurements as a measurement of $g_{tth}$ requires the assumption
that there be no new physics in the $\gamma t\bar{t}$ 
and $Zt\bar{t}$ vertices.  Additionally, in the realistic
multi-parameter fit discussed above, the Yukawa coupling is $83\%$
correlated with the top quark mass, adding to the difficulty of the
interpretation of the threshold cross section as a measurement of the
Yukawa coupling.

\subsubsection{LHC} 
%

\vspace{0.2cm}
\hspace{1.5em} $\bullet$ {\bf Production Rates}
\vspace{0.2cm}

\noindent
The $pp\to t\bar{t}h$ channel can be used in the search for an
intermediate mass Higgs boson, $M_h\!<\!130$~GeV, at the LHC.  In this
region, the cross section for the associated production of a Higgs
boson with a pair of top quarks is still smaller than the leading
$gg\to h$ and $qq\to qqh$ cross sections, but the final state
$t\bar{t}h$ signatures are quite distinctive. At the Tevatron,
however, the Standard Model cross section for $p\bar{p}\to t\bar{t}h$
is probably too small to be observed\cite{ref1_nlo_tev,ref2_nlo_tev,goldstein}.

\begin{figure}[t!]
\centering
\epsfxsize=4.in
\leavevmode\epsffile{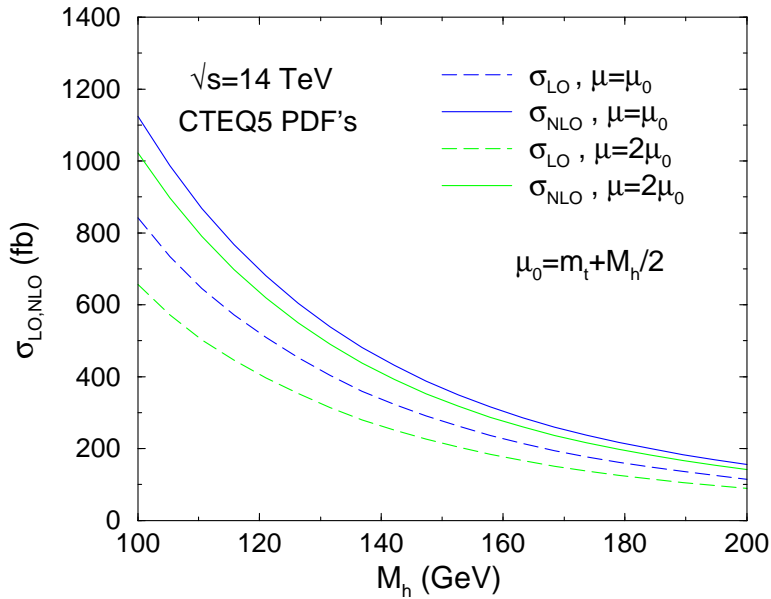}
\caption[]{Leading order (LO) and next-to-leading order (NLO)
  rates for $pp\to t\bar{t}h$ at the LHC, $\sqrt{s}\!=\!14$~TeV.  The
  rates are shown for two values of the unphysical
  factorization/renormalization scale, $\mu$\cite{ref2_nlo_lhc}.}
\label{fig_lhc_nlo}
\end{figure}

At the LHC energy, $\sqrt{s}=14$~TeV, the dominant subprocess for
$t\bar{t}h$ production is $gg\to t\bar{t}h$, but the other
subprocesses, $q\bar{q}\to t\bar{t}h$ and $g(q,\bar{q})\to
t\bar{t}h(q,\bar{q})$ are relevant and cannot be neglected. The cross
section for $pp\to t\bar{t}h$, at LO and NLO of QCD corrections, is
shown in Fig.~\ref{fig_lhc_nlo} as a function of the Higgs boson mass
$M_h$, for two different values of the renormalization/factorization
scale $\mu$. The major effect of the higher order corrections is to
reduce the unphysical scale dependence and to increase the rate from
the lowest order prediction by a factor of $1.2$--$1.6$, depending on the
value of the scale $\mu$ and on the set of parton distribution
functions (PDFs) used \cite{ref1_nlo_tev,ref1_nlo_lhc,ref2_nlo_lhc}.  The
overall theoretical uncertainty of the NLO cross section, including
the residual scale dependence, the error on $m_t$, and the uncertainty
in the PDFs, can be estimated around 15--20\%.

The cross section for $pp\to t\bar{t}h$ is directly proportional to
the top quark Yukawa coupling and can be instrumental to the
measurement of this coupling at the LHC. In spite of the fact that the
LHC does not allow for a completely model independent determination of
the Higgs boson couplings, strategies have been proposed to extract
individual couplings with reasonably good precision under some not too
restrictive assumptions \cite{zep,bere,due}. Typically both
$g_{WWh}/g_{ZZh}$ and $g_{bbh}/g_{\tau\tau h}$ ratios are assumed to
be SM like \cite{zep}, and the Higgs boson width is assumed to be
saturated by the allowed SM Higgs decay channels. In this picture, the
top quark Yukawa coupling could be determined from future LHC
measurement with a precision of about $10$\Ndash $15\%$ \cite{zep}.  Using the
$pp\to t\bar{t}h$ channel with both $h\to b\bar{b}$ and
$h\to\tau^+\tau^-$ allows one to measure the $g_{bbh}/g_{\tau\tau h}$
ratio in a model independent way\cite{bere,due}.
 Even in this more general scenario,
provided a good accuracy on $pp\to t\bar{t}h,h\to b\bar{b}$ and $pp\to
t\bar{t}h,h\to\tau^+\tau^-$ is confirmed by dedicated experimental
analyses, the precision on the top quark Yukawa coupling stays of the
order of $10\!-\!15\%$ \cite{bere} for Higgs boson masses below 130~GeV.

Moreover, both in the low and the high Higgs boson mass region, $pp\to
t\bar{t}h$, with $h\to\tau^+\tau^-$ in one case \cite{bere}
and $h\to W^+W^-$ in the other \cite{mrw}, will allow the measurement
of the ratio $g_{tth}/g_{ggh}$ in a model independent way. With enough
experimental accuracy, unambiguous evidence for contributions to $gg\to
h$ from exotic colored degrees of freedom could be provided.

Finally, in the most promising scenario, the LHC and a high energy LC
will complement each other. A LC with center of mass energy
$\sqrt{s}\!=\!500$~GeV will be able to measure all Higgs couplings,
except $g_{tth}$, at the few percent level.  We could therefore
imagine to use this knowledge of the $Br(h\to
b\bar{b},\tau^+\tau^-,W^+W^-,ZZ,\ldots)$ at the LHC to extract the top
quark Yukawa coupling from $pp\to t\bar{t}h$ with better precision .
This is indeed close to the philosophy adopted by most of the existing
experimental analyses of $pp\to t\bar{t}h$. In the following we
summarize some of the existing results.

\vspace{0.2cm}
$\bullet$ {$\boldsymbol{h\to b\bar{b}}$}
\vspace{0.5cm}

\noindent
One of the most important channels is the $t\bar{t}h,h\to b\bar{b}$
channel, which is the dominant channel for $M_h\!<\!135$~GeV
\cite{Richter-Was:sa,atlas_tdr,lhctop,cmstth}. The final state
consists of two $W$ bosons and four $b$-jets.  One $W$ boson is
required to decay leptonically in order to provide a trigger, while
the second $W$ is reconstructed from the decay to a $\bar{q}q^\prime$
pair.  This decay pattern yields eight fermions:
\begin{Eqnarray}
t\bar{t}h&\to& W^+bW^-\bar{b}\,b\bar{b}\nonumber\\[5pt]
&\rightarrow &
l\nu q\bar{q}^\prime b\bar{b}b\bar{b}\,\,\,.
\end{Eqnarray}

Both top quarks can be fully reconstructed, and this reduces most of
the $W+\mbox{jets}$ background.  The main backgrounds are the
irreducible continuum QCD $t\bar{t}b\bar{b}$ background, the
irreducible resonant $t\bar{t}Z$ background, and the reducible
backgrounds which contain jets misidentified as $b$-jets.  After the
reconstruction of the two top quarks, the most dangerous background is
$t\bar{t}b\bar{b}$.

\begin{figure}[b!]
\centering
\epsfxsize=4.in
\leavevmode\epsffile{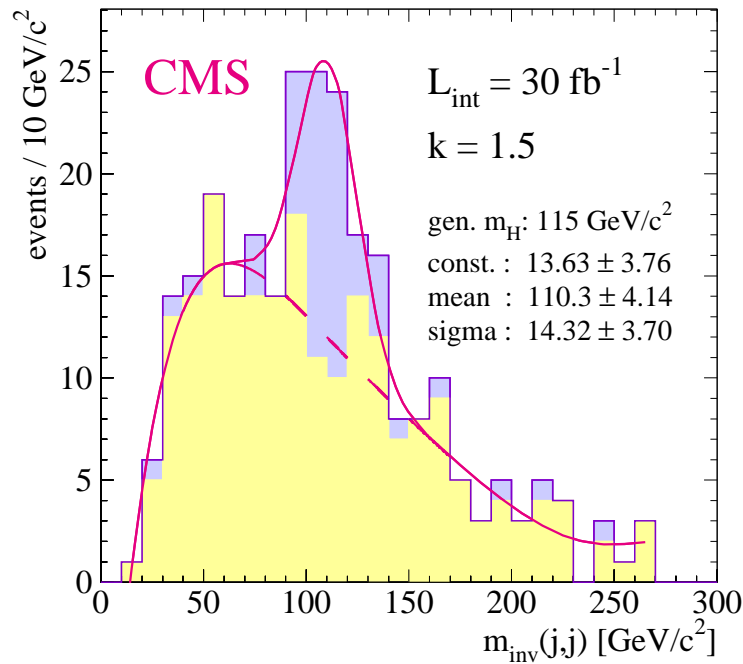}
\caption[]{Signal and background at the LHC for $pp\to
  t\bar{t}h\to l\nu q\bar{q}^\prime b\bar{b} b\bar{b}$ with the CMS
  detector and for $M_h\!=\!115$~GeV. (The NLO corrections are
  estimated using a $k$ factor of $1.5$ slightly higher than the
  actual value)\cite{cmstth}.}
\label{tthcms}
\end{figure}
 
Fig.~\ref{tthcms} shows the signal and background shapes from a CMS
simulation for
$M_h\!=\!115$~GeV and $30~\mathrm{fb}^{-1}$ of integrated luminosity
\cite{cmstth}.  This study obtains a value of $S/\sqrt{B}\!\sim\!
5$, yielding a value of
\begin{equation}
\frac{\delta g_{tth}}{g_{tth}}\sim 12\mbox{\textendash}15\%\,\,\,,
\end{equation}
assuming that the branching ratio for $h\rightarrow b {\overline b}$ 
is known with a negligible error.

Results for the $h\rightarrow
b {\overline b}$ channel have also been obtained by the ATLAS collaboration
\cite{atlas_tdr,lhcbb}. 
The ATLAS
analysis assumes a full reconstruction of the final state in order 
to eliminate combinatoric backgrounds.
 Assuming a $K$ factor of 1.0 and a
Higgs mass of $M_h=115$~GeV, ATLAS finds a value of $S/\sqrt{B}=2.4$
with $30$~fb$^{-1}$\cite{lhcbb}. 
  After adjusting for the 
different assumptions about efficiencies and the different cross
sections (due to differing choices of the scale factors and PDFs),
the ATLAS and CMS analyses are in agreement for the signal and
the reducible background.  The CMS analysis, however, finds a
significantly smaller reducible $t {\overline t} b {\overline b}$
background than does the ATLAS study.

\vspace{0.5cm}
$\bullet$ {$\boldsymbol{h\to\gamma\gamma}$}
\vspace{0.5cm}

\noindent
Because of the small rate, the $h\to\gamma \gamma$ channel is useful
only at high luminosity \cite{atlas_tdr,lhctop}.

\vspace{0.5cm}
$\bullet$ {$\boldsymbol{h\rightarrow \tau^+\tau^-}$}
\vspace{0.5cm}

\noindent
For $M_h\!<\!130$--$135$~GeV, the decay $h\rightarrow \tau^+\tau^-$ is
useful as part of a general strategy to determine Higgs boson
couplings with very few theoretical assumptions \cite{bere,due}.  It
offers the possibility of determining in a completely model
independent way the ratio $g_{bbh}/g_{\tau\tau h}$ by measuring the
ratio of $pp\to t\bar{t}h,h\to b\bar{b}$ to $pp\to
t\bar{t}h,h\to\tau^+\tau^-$.  Combining $pp\to t\bar{t}h$, $gg\to h$,
and $qq\to qqh$ channels, it also allows a model independent
determination of the $g_{tth}/g_{ggh}$ ratio.

First studies assumed that one of the top quarks decays leptonically,
while the other decays hadronically \cite{bere} .  The parton
level signature is $b\bar{b} l\nu q\bar{q}^\prime \tau^+\tau^-$, which
has the irreducible background $t\bar{t}\tau^+\tau^-$, where the
$\tau$ pair originates from a $Z$ boson or a photon.  The other
backgrounds are much smaller.  Ref. \cite{bere} considered only the
$\tau$ decays to $1$ or $3$ charged pions and found that with
$100~\mathrm{fb}^{-1}$ an accuracy of about $20\%$ on
$\delta\sigma/\sigma$ could be obtained for $M_h\!<\!120$~GeV,
declining to $50\%$ for $M_h\!=\!140$~GeV.  For larger Higgs masses
the rate becomes too small for this decay channel to be observed.

%

\vspace{0.5cm}
$\bullet$ {$\boldsymbol{h\rightarrow W^+W^-}$}
\vspace{0.5cm}

\noindent
For $M_h\!>\!135$~GeV, the dominant decay mode is $h\to W W^*$ and the
process $ pp\to t\bar{t}h,~h\to WW^*$ can be observed
despite the low signal and the lack of a reconstructed mass
peak.  By measuring
the ratio of
\begin{equation}
\frac{\sigma(pp\to t\bar{t}h,h\to b\bar{b})}
{\sigma(pp\to t\bar{t}h,h\to WW^*)}
\end{equation}
a measurement of $g_{bbh}/g_{WWh}$ can be made in which many of the
systematic uncertainties cancel \cite{mrw}. Combining $pp\to
t\bar{t}h,h\to WW^*$ with $gg\to h,h\to WW^*$, the ratio
$g_{tth}/g_{ggh}$ can be measured in a model independent way\cite{bere,due}.

\begin{figure}[t,b]
\centering
\includegraphics[width=4in]{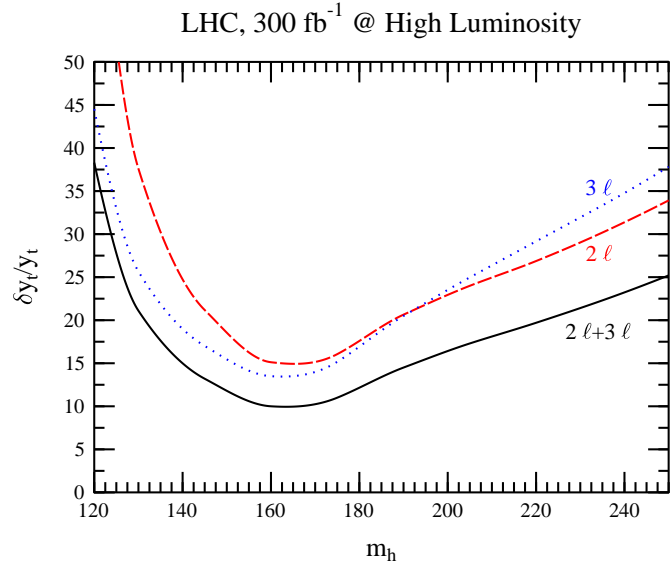}
\caption[]{Measurement of the top quark Yukawa coupling
  at the LHC in the $t\bar{t}h\,,$ $h\rightarrow W^+W^-$ channel
  \cite{mrw}.}
\label{tth_ww}
\end{figure}

The process $pp\to t\bar{t}h,h\to W^+W^-$ gives the final state
$W^+W^-W^+W^- b\bar{b}$.  The decays can then be classified according
to the number of leptonic $W$ decays\cite{lhcww}.  Ref.~\cite{mrw} performs a
parton model study where the background is reduced by requiring both
$b$ decays be tagged with an efficiency $\epsilon_b=0.6$.  The
reconstruction of the top quarks is not required and $p_T\!>\!20$~GeV
for one trigger lepton is required.  With $300~\mathrm{fb}^{-1}$ of
data, one can achieve an accuracy of
\begin{equation}
\frac{\delta g_{tth}}{g_{tth}}\sim 16\%\,,\,8\%\,,\,12\% 
\end{equation}
for $M_h\!=\!130$, 160 and 190~GeV respectively, as
illustrated in Fig.~\ref{tth_ww}\cite{mrw}.
The 
ATLAS collaboration \cite{lhcww} has found that in the most favorable
case, $M_h=160$~GeV, the cross section for
$pp\to
t\bar{t}h,h\to WW^*$, can be measured with $
\delta\sigma/\sigma \sim 26~\%~ (15\%)$ accuracy
with a luminosity of $30~fb^{-1} ~(300~fb^{-1})$  by combining several
final states.  

The $t{\overline t}h,~h\rightarrow b {\overline b}$ 
and $t {\overline t}h,~h\rightarrow
W^+W^-$ channels can be combined to get a measurement of the relative
decay widths $\Gamma(h\rightarrow b {\overline b})/\Gamma(h\rightarrow W^+
W^-)$ with a $50-60\%$ accuracy assuming a luminosity of 
$300~\mathrm{fb}^{-1}$ for $M_h \lsim 140$~GeV.\cite{due}
This measurement is free of theoretical assumptions.
With further assumptions about the $h\rightarrow W^+W^-$ width, a 
value for the $h b {\overline b}$ coupling can be extracted from this 
channel\cite{due}.

\subsubsection{SUSY Higgs Sector}
\label{sec:susy}

There are two important differences between the Higgs sector in the
Standard Model and in a supersymmetric model.  First, there are $5$
Higgs bosons, $A, h, H, H^\pm$, in the minimal SUSY model, so there
are more associated top quark-Higgs boson processes:
\begin{Eqnarray}
e^+e^-, pp &\to& t\bar{t} h\,,\,t\bar{t}H\,,\,t\bar{t}A\,\,\,,\nonumber\\[5pt]
e^+e^-, pp &\to& t\bar{b} H^- + b\bar{t} H^+\,\,\,.
\end{Eqnarray}%
At a linear collider, the additional Higgs bosons give important new
contributions, for example, $e^+e^-\rightarrow A H, A\rightarrow t
{\overline t}$.
Secondly, the Higgs boson couplings to the fermions are changed from
those of the Standard Model, and depend on the parameters of the
model, typically $\tan\beta$.

\begin{figure}[t!]
\centering
\includegraphics[width=3.8in,angle=-90]{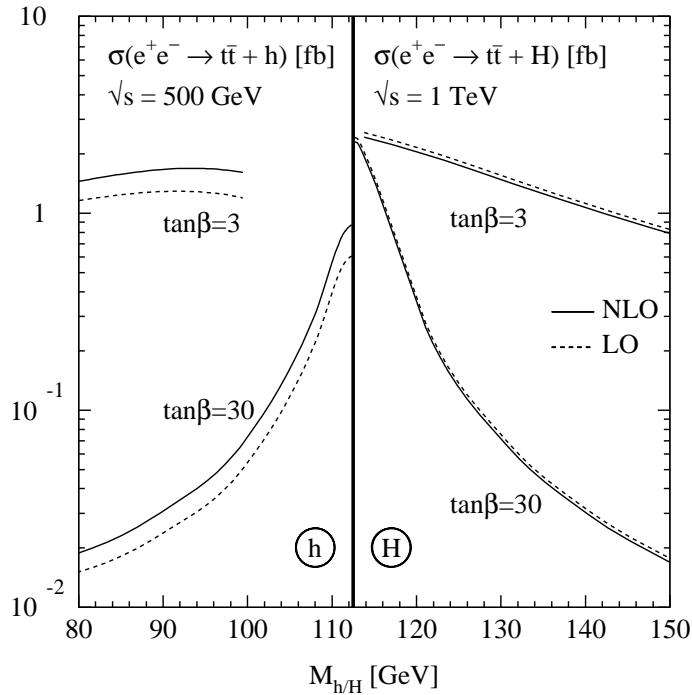}
\caption[]{The LO and NLO corrected rates for  
  $e^+e^-\to t\bar{t}h$ and $e^+e^-\to t\bar{t}H$ in the MSSM\cite{ditt_susy}.
}
\label{susyfig}
\vspace{-0.2in}
\end{figure}

\begin{figure}[t!]
\centering
\epsfxsize=4.in
\leavevmode
\epsffile{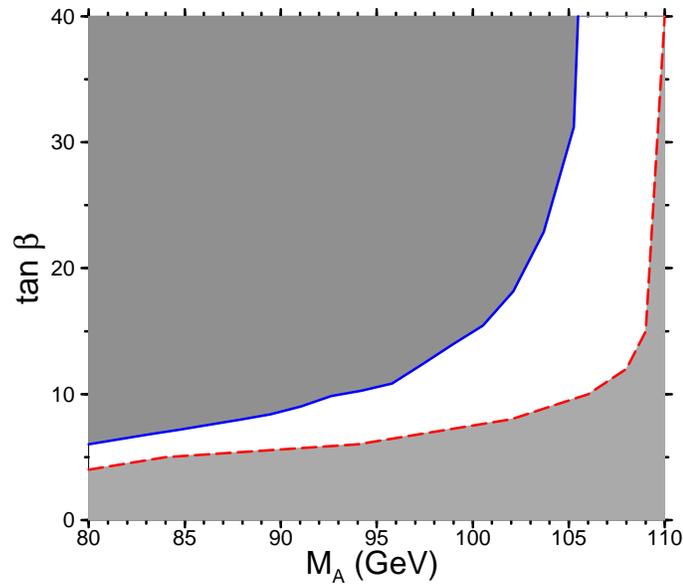}
\caption[]{Regions of parameter space where the
 QCD corrected NLO rate for $e^+e^-\to t\bar{t}h$ ($e^+e^-\to
  t\bar{t}H$) is larger than $0.75$~fb, bottom right (upper shading),
  at $\sqrt{s}\!=\!500$~GeV \cite{lr_susy}.}
\label{fig_susy}
\end{figure}

The NLO QCD corrections to $e^+e^-\to t\bar{t}h$ are
modest\cite{ditt_susy,lr_susy} and the QCD corrected rate for the
production of the neutral Higgs bosons is shown in Fig.~\ref{susyfig}.
For large $\tan\beta$, the associated production of the lighter Higgs
boson, $h$, is significantly suppressed due to the suppressed coupling
of the top quark to the lightest SUSY Higgs boson.  
The SUSY couplings
have the property that the $t\bar{t}h$ and $t\bar{t}H$ production
rates are complementary, as demonstrated in Fig.~\ref{fig_susy}.  The
associated production of the pseudoscalar Higgs boson, $e^+e^-\to
t\bar{t}A$, is significantly smaller than that of the scalars for all
values of the parameter space.

\begin{figure}[h!]
\centering
\epsfxsize=4.in
\leavevmode
\epsffile{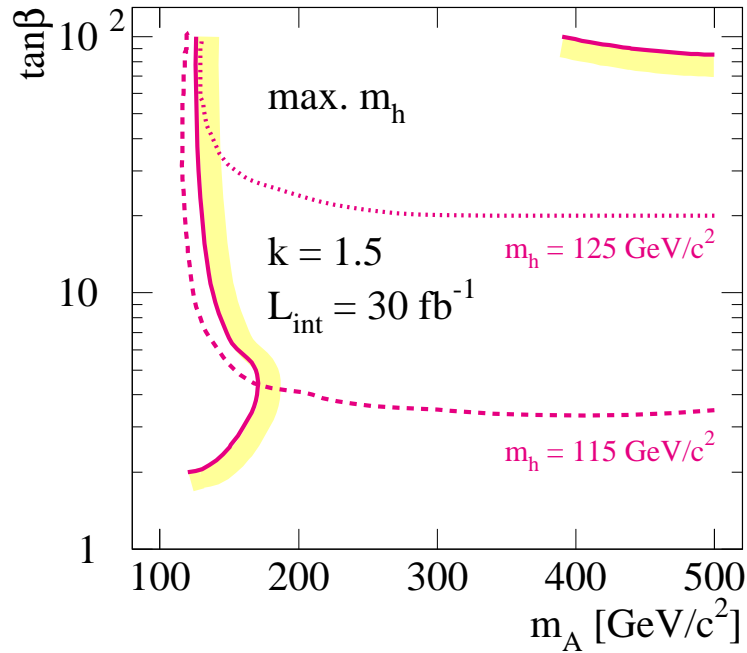}
\caption[]{$5\sigma$ discovery regions with $30~\mathrm{fb}^{-1}$ at 
  the LHC using the CMS detector and the process $pp\to t\bar{t}h,h\to
  b\bar{b}$ in the MSSM.  The plot is obtained by rescaling the
  Standard Model results by the appropriate SUSY couplings.  The
  discovery region is to the side of the shaded area \cite{cmstth}.
  The SUSY mixing parameters are chosen to give the maximum value of
  $M_h$ for a given $M_A$ and $\tan\beta$.}
\label{susycms}
\end{figure}

The LC can make precision measurements of the rates for
the $t\bar{t}h\,(H,A)$ production modes and use them to discriminate
between different models of supersymmetry breaking \cite{Dedes:2003cg}.  A
comparison of the rates in gauge mediated SUSY breaking models,
anomaly mediated SUSY breaking models, and the mSUGRA model suggest
that this channel is sensitive to the mechanism of SUSY breaking for
roughly $M_A\!<\!300$~GeV.

The NLO QCD corrections to $e^+e^-\to tbH^{\pm}$ have been computed in
Ref.~\cite{nlo_tbh}.  This process could be important in the energy
regime where it is not possible to pair produce the charged Higgs
boson through the process $e^+e^-\to H^+H^-$.  Unfortunately, at
$\sqrt{s}\!=\!500$~GeV, the rate is always small, $\sigma\!<\!0.03$~fb
for $M_H\!=\!260$~GeV and $\tan\beta\!=\!40$.

At the LHC, the primary effect in $t\bar{t}h$ production of
introducing a supersymmetric model is to change the Higgs couplings to
the top quark.  The rate for $pp\to t\bar{t}h,h\to b\bar{b}$ is
proportional to $g_{tth}g_{bbh}$ which has only a weak dependence on
$\tan\beta$.  For a wide choice of parameters and SUSY models, the
rate for $pp\to t\bar{t}h$ production is never suppressed by more than
$10\%$~\cite{Dedes:2003cg}. The $5\sigma$ discovery contours at the LHC for
$30~\mathrm{fb}^{-1}$ are shown in Fig.~\ref{susycms}. This curve is
obtained simply by rescaling the Standard Model results by the
appropriate SUSY couplings \cite{cmstth}. The discovery region is to
the side of the shaded area.

\subsubsection{Conclusion}

The associated production of a Higgs boson with a pair of $t\bar{t}$
quarks will be the only way of directly measuring the top quark Yukawa
coupling at both a high energy LC and the LHC.  An $e^+e^-$ collider
with $\sqrt{s}\!=\!500$~GeV can make a preliminary measurement of the
top quark Yukawa coupling.  However, the small rate at
$\sqrt{s}\!=\!500$~GeV implies that a precision measurement will
require a higher energy.  At $\sqrt{s}\!=\!800$\Ndash $1000$~GeV, the top
quark Yukawa coupling will be determined with a $4$\Ndash $5\%$ precision.

The LHC measures the product of the production cross section
multiplied by the Higgs branching ratios and will be 
instrumental in obtaining a first set of measurements of 
ratios of Higgs coupling constants.  In order to extract the top
quark Yukawa coupling, it is typically assumed that the Higgs decays
with Standard Model branching ratios. With this assumption,
the LHC will obtain absolute couplings in some channels.
This assumption could be removed
with precision measurements of the Higgs branching ratios at a Linear
Collider.

\subsection{Determining the parameters of the Higgs boson potential}

{\it U.~Baur, T.~Plehn and D.~Rainwater}

\vspace{1em}
The LHC is widely regarded as capable of directly observing the agent
responsible for electroweak symmetry breaking and fermion mass generation.
This is generally believed to be a light Higgs boson~\cite{lepewwg}. The LHC
will easily find a light SM Higgs boson with very moderate
luminosity~\cite{wbf_ww,wbf_ll} and have significant capability 
to determine many of its properties, such as its decay modes and
couplings~\cite{wbf_ll,zep,Yb,bere,cmstth,mrw}. An $e^+e^-$ linear 
collider could
significantly improve these preliminary measurements, in some cases by an
order of magnitude in precision~\cite{Aguilar-Saavedra:2001rg}.

Starting from the requirement that the Higgs boson has to restore unitarity of
weak boson scattering at high energies in the SM~\cite{unit}, perhaps the most
important measurement after the Higgs boson discovery is of the Higgs
potential itself, which requires measurement of the Higgs boson
self-couplings. These can be probed directly only by multiple Higgs boson
production. Several studies of Higgs boson pair production in $e^+e^-$
collisions have been conducted over the past few
years~\cite{LC_HH3,LC_HH1,LC_HH4}, deriving quantitative sensitivity
limits for the trilinear Higgs self-coupling for several proposed
linear colliders for $m_H\leq 140$~GeV. The potential of hadron
colliders has been examined only 
recently~\cite{SLHC,BPR,BPRrare}, investigating Higgs pair production via
gluon fusion and various subsequent decays. They established that future
hadron machines can probe the Higgs potential over a wide range of
Higgs mass.

To show the complementarity of hadron and lepton colliders we add to the
existing literature by looking at Higgs boson pair production for 
$m_H\leq 140$~GeV at hadron colliders, and estimate the prospects for 
probing the Higgs boson self-coupling at a linear collider if  
$m_H\geq 150$~GeV. Using this input we determine how well the Higgs 
potential could be reconstructed. We closely follow our argument in 
Ref.~\cite{orig}.\medskip

The trilinear and quartic Higgs boson couplings $\lambda$ and
$\tilde\lambda$ are defined through the potential
\begin{equation}
\label{eq:Hpot}
V(\eta_H) \, = \, 
\frac{1}{2}\,m_H^2\,\eta_H^2\,+\,\lambda\, v\,\eta_H^3\,+\,\frac{1}{4}\,
\tilde\lambda\,\eta_H^4 ,
\end{equation}
where $\eta_H$ is the physical Higgs field. In the SM,
$\tilde\lambda=\lambda=\lambda_{SM}=m_H^2/(2v^2)$. Regarding the SM as an
effective theory, the Higgs 
self-couplings $\lambda$ and $\tilde\lambda$
are {\it per se} free parameters. $S$-matrix unitarity constrains
$\tilde\lambda$ to $\tilde\lambda\leq 8\pi/3$~\cite{unit}. Since future
collider experiments likely cannot probe $\tilde\lambda$, we focus on
the trilinear coupling $\lambda$ 
in the following. 

\subsubsection{A low mass Higgs boson ($m_H\leq 140$~GeV)}
\label{sec:lowmass}

\vspace{0.2cm}
$\bullet$ LHC/SLHC
\vspace{0.2cm}

\noindent
At LHC energies, inclusive Higgs boson pair production is dominated by
gluon fusion, although weak boson fusion production and weak boson
or top quark pair associated production are also possible. Since Higgs pair
production is already rate limited, we consider only the gluon fusion process 
in the following.

For $m_H<140$~GeV, the dominant decay mode of the SM Higgs boson is
$H\to b\bar{b}$.  Unfortunately, the $4b$ channel (both Higgs bosons
decaying to $b\bar{b}$) is completely overwhelmed by the enormous QCD
background, which is larger by more than two orders of magnitude. For
completeness, we performed a calculation of the signal and background,
including the effects of NLO-QCD corrections to the signal via a
multiplicative factor. After including kinematic cuts and realistic
efficiency factors for the $b$ jets, a $\chi^2$ test on the $m_{vis}$
distribution for $m_H=120$~GeV we obtained meaningless $1\sigma$
bounds of $-6.8< \Delta\lambda_{HHH}<10.1$, where
$\Delta\lambda_{HHH}=\lambda/\lambda_{SM}-1$.

A more advantageous $S/B$ is conceivable if one of the Higgs bosons in
$gg\to HH$ decays into a $\tau$ pair. In this case, the main
contributions to the background arise from continuum
$b\bar{b}\tau^+\tau^-$ and $t\bar{t}$ production. The $\tau$-pair
invariant mass can be reconstructed with fairly good resolution. We
considered the leptonic-hadronic decay channel of the $\tau$ pair, to
satisfy detector trigger requirements. Unfortunately, no limit can be
extracted for the LHC. Even for a luminosity-upgraded LHC (SLHC)
achieving 3000~fb$^{-1}$, we expect a possible $1\sigma$ limit
extraction of only $-1.6<\Delta\lambda_{HHH}<3.1$, too weak to be
useful. We find even worse results for the final state
$b\bar{b}\mu^+\mu^-$: while background rejection is much better, this
channel suffers from simple lack of signal rate due to the extremely
small Higgs boson branching ratio to muon pairs - we could expect only
about 2 events at the SLHC.

The best strategy appears to be to consider the rare decay of one
Higgs boson to a photon pair, while allowing the other to decay to its
dominant channel, $b\bar{b}$~\cite{BPRrare}. The background consists
of QCD $b\bar{b}\gamma\gamma$ production and numerous similar QCD
processes where one or more of the $b$ jets or photons is a fake from
charm or light jets.  We also must include contributions from single
Higgs boson production in association with photons and jets, as well
as double parton scattering and fakes from multiple interactions. The
latter are of little concern at the LHC, so we may employ a single
$b$-tag strategy: for $\epsilon_b =50\%$, this retains three times the
number of signal events as double $b$-tagging. However, multiple
interactions is a serious problem at an SLHC, which we bring under
control with double $b$-tagging.

We impose kinematic cuts to enhance the signal relative to background,
and use conservative estimates for the efficiency to identify the
final state, $\epsilon_\gamma =80\%$ and $\epsilon_b =50\%$, as well
as the more pessimistic probabilities from ATLAS for charm and light
jets to fake the signal final state: $P_{c\to b}=1/13$, $P_{j\to
  b}=1/140(1/23)$ and $P_{j\to\gamma}=1/1600$ for the LHC(SLHC).
Under these conditions we expect to observe about 6(21) signal events
at the LHC(SLHC), on a background of 14(25) events. The $S/B$ ratio is
quite respectable, but the statistics are low at the LHC. A
$\chi^2$-test yields limits of $-1.1<\Delta\lambda_{HHH}<1.9$ for
600~fb$^{-1}$ at the LHC, and $-0.66<\Delta\lambda_{HHH}<0.82$ for
6000~fb$^{-1}$ at the SLHC. These limits improve by about $15\%$ if
the more optimistic ATLAS estimate of $P_{j\to\gamma}=1/2500$ can be
achieved. Another promising approach is to subtract the backgrounds
after measuring them in other kinematic regions, a technique already
used by CDF. In this case, the expected limits improve to
$-0.74(0.46)<\Delta\lambda_{HHH}<0.94(0.52)$ at the LHC(SLHC).

\vspace{0.2cm}
$\bullet$ LC
\vspace{0.2cm}

\noindent
We now turn our attention to Higgs boson pair production in $e^+e^-$
collisions.  A detailed study of how well the Higgs boson
self-coupling for $m_H=120$~GeV could be measured in $e^+e^-\to ZHH$
at $\sqrt{s}=500$~GeV can be found in Ref.~\cite{LC_HH4}.

Associated $ZHH$ production followed by $HH\to b+$jets is the dominant source
of $HH$ events in the SM if $m_H\leq 140$~GeV. The main backgrounds in this
channel are $\bar tt$ and $W^+W^-$ production. These are efficiently
suppressed by performing a NN analysis. Such an analysis, including a detailed
detector simulation, was presented in Ref.~\cite{LC_HH4} for $m_H=120$~GeV and
$\sqrt{s}=500$~GeV. It concluded that $\lambda$ could be determined with a
precision of about $23\%$ if an integrated luminosity of 1~ab$^{-1}$ could be
achieved. The limits achievable at hadron colliders for $m_H=120$~GeV are
significantly weaker. 

Since the $ZHH$ cross section and its sensitivity to $\lambda$
decrease with increasing collider energy, a linear collider operating
at 500~GeV offers the best chance for a precise measurement of
$\lambda$ for $m_H\leq 140$~GeV; the bounds we obtain for
$\sqrt{s}=500$~GeV are up to a factor~1.4 (1.9) more stringent than
those achievable for $\sqrt{s}=800$~GeV ($\sqrt{s}=1$~TeV). The
advantage of operating at 500~GeV gradually disappears with increasing
Higgs boson mass, due to the reduced phase space. The sensitivity
limits achievable weaken by a factor~1.8 (1.2) for $\sqrt{s}=500$~GeV
($\sqrt{s}=1$~TeV) if $m_H$ increases from 120~GeV to 140~GeV; for
$m_H=140$~GeV one will not be able to probe $\lambda$ with a precision
of better than $50\%$ for unpolarized beams.  If both the electron and
positron beams could be polarized, the bounds derived here improve by
a factor~1.3, assuming $80\%$ polarization for the electron beam and
$60\%$ for the positron beam, and the same integrated luminosity as
for unpolarized beams.

We conclude that a $0.5$--$1$~TeV 
linear collider offers a significantly better chance 
to probe $\lambda$ for the mass range from 120~GeV to 140~GeV, although an 
upgraded LHC could provide convincing proof of spontaneous symmetry breaking.

\subsubsection{A heavier Higgs boson ($m_H\geq 140$~GeV)}
\label{sec:highmass}

For $m_H\geq 140$~GeV, the principal Higgs boson decay is to $W$ pairs. This
completely changes the phenomenology of Higgs pair production.

\vspace{0.2cm}
$\bullet$ LHC/SLHC
\vspace{0.2cm}

We have previously examined this mass range for the LHC~\cite{BPR}.
Due to the huge QCD backgrounds in multijet final states, we can use
only multiplepton subsamples, where two same-sign $W$ bosons decay
leptonically and the other two give four jets, or where three $W$'s
decay leptonically and the events have only two extra jets.
Summarizing our previous results, the LHC could give $1\sigma$
constraints of $-0.4<\Delta\lambda_{HHH}<1.9$~\cite{BPR,LH03-Hcoup}
for $m_H=180$~GeV. This is sufficient to exclude the case of no
self-coupling to slightly better than $95\%$~c.l. for $150\lsim
m_H\lsim 200$~GeV, proving the existence of spontaneous symmetry
breaking, and is a fortuitous effect of the destructive interference
of the two diagrams in the gluon fusion process. An SLHC could instead
make a $20-30\%$ measurement (95\% CL) of the self-coupling over the
mass range $160<m_H<180$~GeV.

\vspace{0.2cm}
$\bullet$ LC
\vspace{0.2cm}

\noindent
If $m_H>140$~GeV, the channels yielding the largest event rates are $e^+e^-\to
ZHH$ with $Z\to jj$ and $HH\to 8$~jets or $\ell\nu+6$~jets.  Final states of
similar structure and complexity are encountered in $t\bar{t}H$ production.
The main background processes contributing both to $e^+e^-\to ZHH$ and
$e^+e^-\to t\bar{t}H$ are $WW+$~jets, $t\bar{t}+$jets and QCD multijet
production. Instead of using a NN analysis, we investigate how the sensitivity
bounds for $\lambda$ depend on the signal efficiencies and the signal to
background ratio. We then explore the prospects for determining the Higgs
boson self-coupling in $ZHH$ production for $m_H>140$~GeV using some 
rough guidelines.

Since the number of signal events is small, we combine the
$\ell\nu+8$~jet and 10~jet final states and use the total cross
section to derive sensitivity limits.  For the likely case that
efficiencies and $S/B$ are similar to the values obtained in the
$t\bar{t}H$ analysis, a first-generation LC could obtain only very
loose bounds on $\lambda_{HHH}$.  Using the values quoted in
Ref.~\cite{orig}, one finds $-4.1 < \Delta\lambda_{HHH} < 1.0$ at the
$1\sigma$ level for $m_H=180$~GeV, $\sqrt{s}=1$~TeV and 1~ab$^{-1}$.
The achievable limits improve by about a factor~1.3 for electron and
positron beam polarizations of $80\%$ and $60\%$, respectively, and if
the same integrated luminosity as in the unpolarized case could be
reached. We reach similar conclusions for $m_H=160$~GeV and
$\sqrt{s}=800$~GeV. For $m_H<160$~GeV and $m_H>180$~GeV, fewer than 5
signal events would be seen if efficiencies are smaller than 0.5,
disallowing bounds to be placed on $\lambda_{HHH}$.

At higher collision energies, one should consider $e^+e^-\to\nu\bar\nu HH$
production, which dominates over $ZHH$ above about $\sqrt{s}=1-1.5$~TeV,
depending on the Higgs mass.
We do not address this channel here, as it is under investigation 
elsewhere~\cite{dettaglia-private}.

\subsubsection{Reconstructing the Higgs Potential}
\label{sec:pot}

The results presented above can be used to compare the capabilities of future
lepton and hadron colliders to reconstruct the Higgs potential. In order to
translate bounds on $\Delta\lambda_{HHH}$ into constraints on the Higgs
potential which can be graphically displayed, it is convenient to consider the
scaled Higgs potential
\begin{equation}
\label{eq:scale_pot}
\frac{2}{v^2 m_H^2}\,V(x)=x^2+\lambda_{HHH}x^3+
\frac{1}{4}\,\tilde\lambda_{4H}x^4~ \, ,
\end{equation}
where $x=\eta_H/v$ and $\tilde\lambda_{4H}=\tilde\lambda/\lambda_{SM}$
is the four Higgs boson self-coupling normalized to the SM value,
$\eta_H$ is the physical Higgs field, and $v$ is the vacuum
expectation value. In the following we assume $\tilde\lambda_{4H}=1$.

It should be noted that the scaled Higgs potential of
Eq.~(\ref{eq:scale_pot}) is valid only in the vicinity of $x=0$. The
presence of a non-SM $HHH$ coupling requires higher dimensional terms
in an effective Lagrangian which would modify $\tilde\lambda_{4H}$ and
also create terms proportional to $x^n$ with $n>4$. These terms are
ignored in Eq.~(\ref{eq:scale_pot}).  Eq.~(\ref{eq:scale_pot}) with
$\tilde\lambda_{4H}=1$ thus represents a good approximation to the
true scaled Higgs potential only if the contributions of terms
proportional to $x^n$, $n\geq 4$, are much smaller than that of the
$x^3$ term.  This is guaranteed for sufficiently small values of $x$.
We restrict the range of $x$ for which we show the scaled Higgs
potential to $|x|\leq 0.2$. Provided that the coefficients of the
$x^n$, $n\geq 4$, terms are not much larger than $\lambda_{HHH}$, this
guarantees that Eq.~(\ref{eq:scale_pot}) is indeed a good
approximation of the true Higgs potential.

\begin{figure}[b!]
\begin{center}
\includegraphics[width=11.0cm]{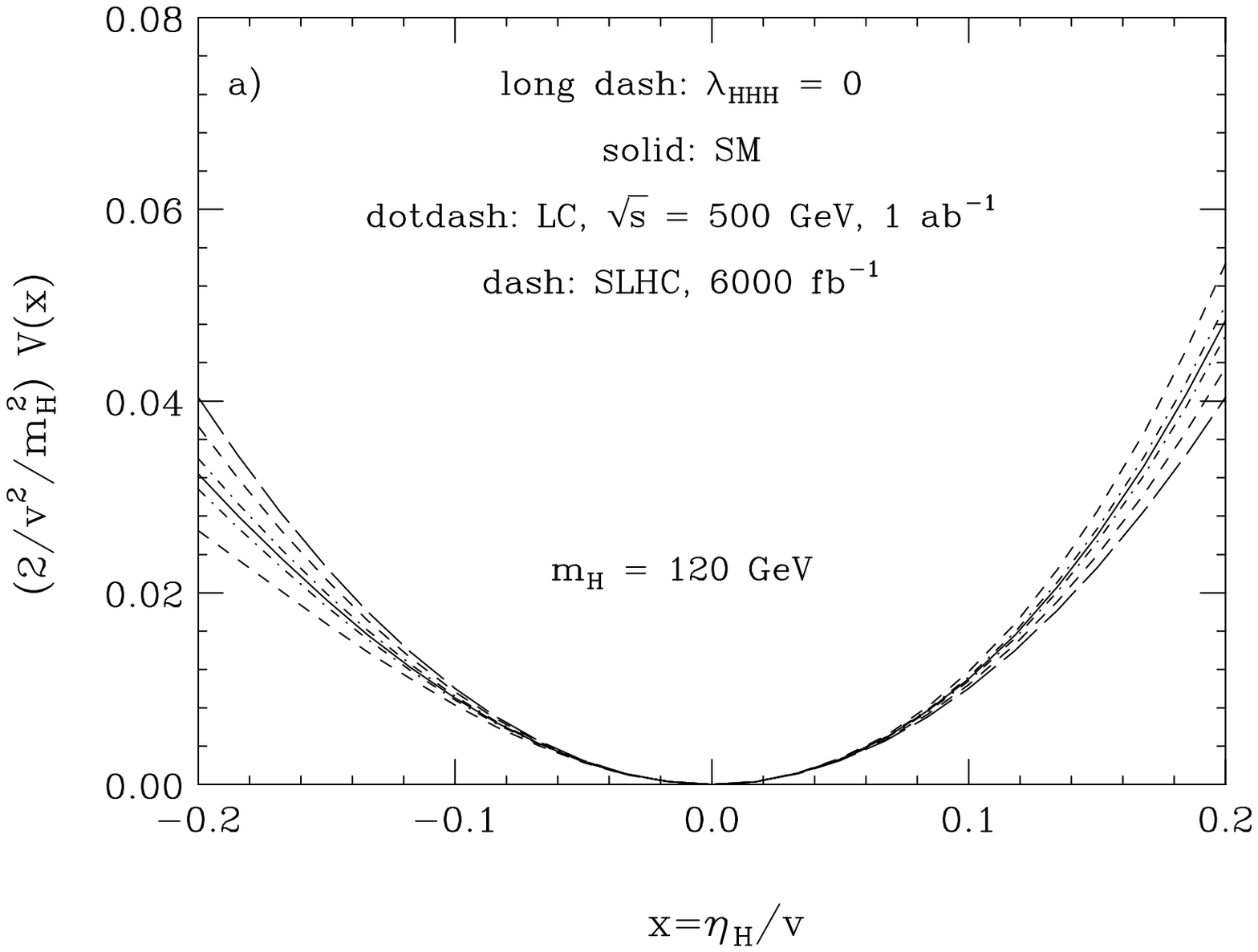}
\includegraphics[width=11.0cm]{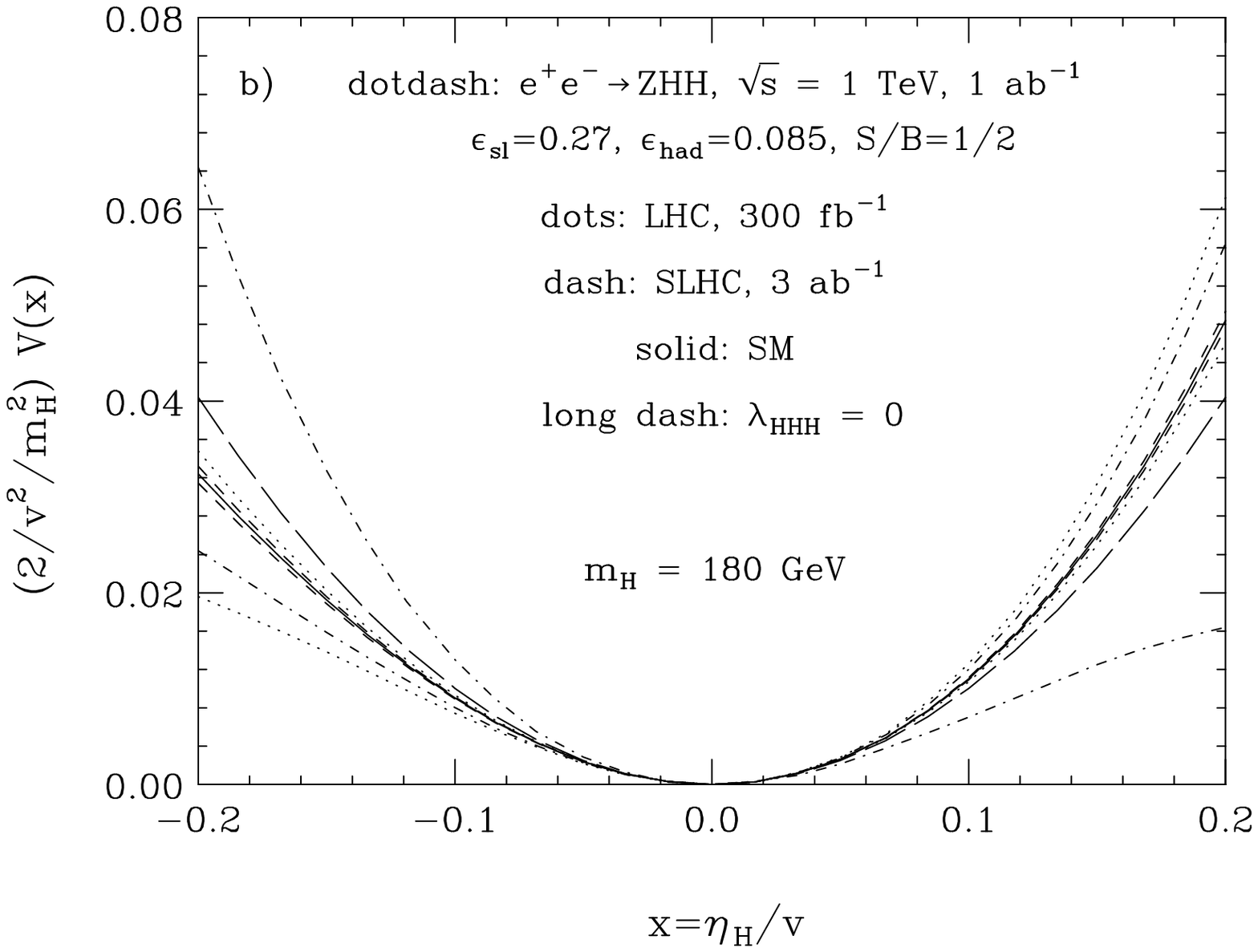}
\vspace*{2mm}
\caption[]{\label{fig:pot1} Constraints on the scaled Higgs potential
  for a) $m_H=120$~GeV and b) 180~GeV. The allowed region is between
  the two lines of equal texture. The solid line represents the SM Higgs
  potential, and the long-dashed line shows the result for a vanishing Higgs
  boson self-coupling.}
\end{center}
\vspace{-0.3in}
\end{figure}
In Fig.~\ref{fig:pot1} we show how well the scaled Higgs potential
could be reconstructed for a Higgs boson of mass $m_H=120$~GeV and
$m_H=180$~GeV.  Hadron colliders indeed have only very limited
capabilities to probe $\lambda$ if $m_H\leq 140$~GeV, although with
sufficient luminosity they can provide proof of spontaneous symmetry
breaking. The best channel appears to be the semi-rare decay to
$b\bar{b}\gamma\gamma$ final states. This channel should now be
studied with full detector simulation. In contrast, a 500~GeV linear
collider with an integrated luminosity of 1~ab$^{-1}$ could measure
$\Delta\lambda_{HHH}$ with a precision of about $20\%$~\cite{LC_HH4},
reconstructing the Higgs potential fairly accurately. We draw similar
conclusions for other Higgs boson masses in the range $120~{\rm
  GeV}<m_H<140$~GeV. The limits achievable for $\Delta\lambda_{HHH}$
both at lepton and hadron colliders gradually weaken by about a
factor~2 if $m_H$ is increased from 120~GeV to 140~GeV.

If the Higgs boson decays predominantly into a pair of $W$-bosons,
i.e. if $m_H\geq 150$~GeV, a completely different picture emerges.
While LHC experiments will only be able to put mild constraints on
$V(x)$, a luminosity upgrade of the LHC will make it possible to
reconstruct the Higgs potential quite precisely for this $m_H$ range.
Of course, to control systematic uncertainties associated with
knowledge of the Higgs branching ratios and top Yukawa coupling,
ideally one would have precision input from a linear collider. Thus,
having both collider data sets at hand would greatly increase our
understanding of the Higgs potential in this mass region.

At a linear collider with a center of mass energy in the $0.8$---1~TeV
range and an integrated luminosity of 1~ab$^{-1}$, the number of Higgs
boson pair events is very limited. The dominant $WW+$~jets and
$t\bar{t}\:+$~jets backgrounds are several orders of magnitude larger
than the signal. As a result, it will be difficult to constrain the
Higgs potential using linear collider data if $m_H\geq 150$~GeV. This
point is illustrated in Fig.~\ref{fig:pot1}b assuming the efficiencies
and the signal to background ratios to be equal to Ref.~\cite{juste}
for $e^+e^-\to t\bar{t}H$. We obtain results similar to those shown in
Fig.~\ref{fig:pot1} for $m_H=160$~GeV.

It should be noted that the prospects to determine the Higgs boson
self-coupling and to reconstruct the Higgs potential at an $e^+e^-$
collider for a Higgs boson with mass larger than 150~GeV improve
dramatically at larger energies.

\subsubsection{Conclusions}

Our results show that hadron colliders and $e^+e^-$ linear colliders
with $\sqrt{s}\leq 1$~TeV are complementary: for $m_H\leq 140$~GeV,
linear colliders offer significantly better prospects in measuring the
Higgs boson self-coupling, $\lambda$; for a Higgs boson in the range
$m_H\geq 150$~GeV, the opposite is true. However, to actually perform
a meaningful measurement at a hadron collider would demand precision
Higgs boson properties input for the top quark Yukawa coupling, the
$HWW$ coupling, and the total Higgs boson decay width.

While the LHC can obtain meaningful information on the Higgs total
width~\cite{due,LH03-Hcoup}, this is at best at the $15\%$ level for
$m_H\geq 160$~GeV, and about a factor of 2 worse in the lower mass
region. For individual couplings, the LHC would measure the $HWW$
coupling to about $10(25)\%$ in the upper(lower) mass region, and the
top Yukawa coupling at about the $25-30\%$ level.  These values
constitute large 
enough systematics to seriously complicate LHC $HH$ measurements. A LC
is vastly superior for the $HWW$ coupling and total Higgs width, with
a precision at the few percent level. For the top Yukawa coupling, a
LC with center of mass energy up to 1~TeV is superior in the lower
$m_H$ region, where it can improve the precision to about $15\%$ (see
Sec. 2.1.2). It is possibly competitive closer to the $WW$ threshold,
although the status of these studies is still not fully mature. A
laudable goal would be to remove many of the systematic uncertainty
assumptions both LHC and LC studies currently make, to provide more
accurate estimates of the ultimate precision achievable, as it would
greatly affect the interpretation of LHC results.




\subsection{Higgs boson decay into jets}

{\it E.L.\ Berger, T.M.P.\ Tait, C.E.M.\ Wagner}

\vspace{1em}
\subsubsection{Introduction}

Current strategies for the discovery and measurement of the properties the
neutral scalar Higgs particle $h$ with $m_h < 135$ GeV rely heavily on
the presumption that the principal branching fractions are close to
those predicted in the standard model~(SM) or in the usual minimal
supersymmetric standard model~(MSSM).  For masses in this range, the
decay width of the SM Higgs boson is dominated by its decay into a
pair of bottom quarks, $b \overline{b}$.  Since the SM bottom-quark
Yukawa coupling is small, these assumptions are not warranted in the
presence of non-standard light particles with order one couplings to
the Higgs boson. Searches can become particularly difficult at hadron
colliders if the Higgs boson decays predominantly into these new
particles and if these particles decay mainly into jets, with no
significant bottom or charm flavor content.  
Ref.~\cite{Berger:2002vs} shows explicitly that this situation can
arise in the MSSM in the presence of light bottom squarks
($\tilde{b}$'s), with mass smaller than about 10 GeV.  
In this case, $\tilde b$ is the lightest supersymmetric particle
(LSP).  The bottom
squarks are assumed to decay primarily into a pair of light quarks via
an R-parity-violating interaction.
The main results of our study, however, are applicable to a
general class of models in which the main modification to the Higgs
boson branching ratios is induced by the appearance of an additional
decay mode into hadron jets.

The possibility that the Higgs boson decays with substantial branching
ratio into jets is not unique to the light bottom squark scenario. It
results also in models in which there is a light CP-odd scalar in the
spectrum whose mass is less than $2m_b$.  
The Higgs boson then can decay into a pair of these scalars
(and in general, owing to the smallness of the bottom Yukawa coupling,
it will unless the coupling to the CP-odd scalars is very small).
This possibility was discussed in Ref.~\cite{Dobrescu:2000jt} in a
model in which the CP-odd scalar behaves like an axion.  It may arise
also in general two-Higgs-doublet models, and even in the MSSM with
explicit CP-violation~\cite{Carena:2002bb}.  Quite generally, if the
mass of the CP-odd scalar is above a few GeV, but below twice the
bottom quark mass, the CP-odd scalar decay into jets will overwhelm
the decay into two photons.  The Higgs boson would then decay in a way
that resembles the results in the light bottom squark scenario,
namely, it will decay predominantly into two scalars that subsequently
decay into jets. The exact phenomenology of the light CP-odd scalar
scenario will be different from that of the light bottom squark
scenario, and it should be analyzed in more detail. For instance, one
should determine the values of the branching ratio of the CP-odd
scalar into two photons that would permit observation of the Higgs
boson in the photon plus jet channel. For the aims of this work, we
assume that the branching ratio of the Higgs boson decay into photons
plus jets is sufficiently small, so that this search channel is not
suitable.

\subsubsection{Higgs decays in a model with a light bottom squark}

\begin{figure}[t!]
\centerline{\includegraphics[height=10.5cm]{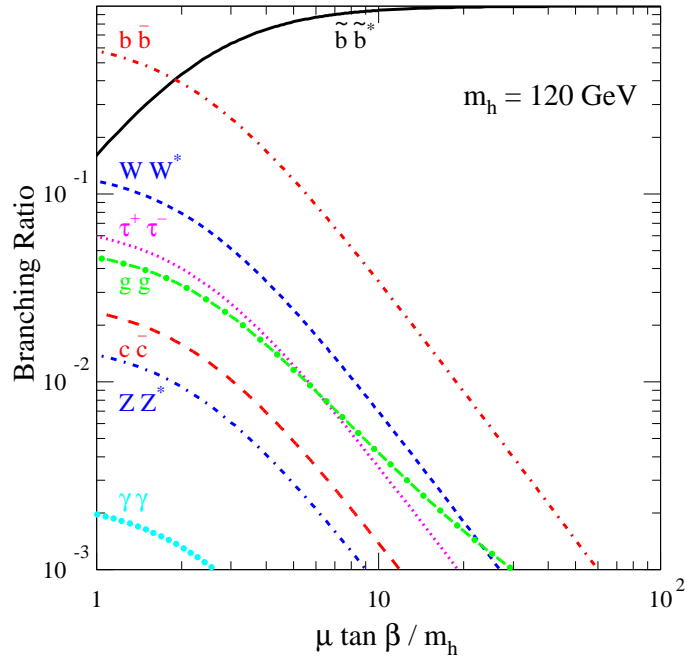}}
\caption[]{\it Branching fractions for various Higgs boson 
decay channels as a function of the ratio $\mu \tan \beta/m_h$, 
with $m_h = 120$ GeV and $m_{\tilde{b}} = 5$ GeV.  The partial widths
for Higgs decay into $WW^*$, $ZZ^*$, $b \overline{b}$, 
$c\overline{c}$ and $\tau^+\tau^-$ are assumed to be 
equal to their SM values.  The partial widths for Higgs decay into
$gg$ and $\gamma\gamma$ include the effects of the light bottom
squarks in the corresponding loop amplitudes.}
\label{fig:plot5}
\end{figure}

To exemplify quantitatively the modifications of the branching ratios
to SM particles, we summarize the case of Higgs boson decay to light
bottom squarks.  The possible existence of a light bottom squark
$\tilde{b}$ (and a light gluino $\tilde{g}$) is advanced in
Ref.~\cite{Berger:2000mp} to address the excess rate of bottom quark
production at hadron colliders.~\footnote{Various experimental
constraints and phenomenological implications are examined in
Ref.~\cite{Berger:2002kc}.}  We work in the decoupling limit in which
the mass of the CP-odd Higgs boson ($m_A$) is large compared to
$m_Z$, and we assume that the ratio of Higgs boson vacuum expectation
values $\tan \beta$ is large.  Under these conditions, 
$\Gamma_{\tilde b}/\Gamma_b\propto (\mu\tan\beta/m_h)^2$,
where $\mu$ is the Higgsino mass parameter.
In the case of light bottom squarks, the dominant
decay of $h$ is into a $\tilde b\tilde b^*$ 
pair\footnote{Here, $\tilde b^*$ is the anti-particle of 
$\tilde b$.}~\cite{Berger:2002vs,Carena:2000ka}, 
and the total decay
width of the $h$ may become several orders of magnitude larger than
the width for the decay into $b\bar b$.  
A plot of the Higgs branching fractions is
presented in Fig.~\ref{fig:plot5} as a function of $\mu
\tan\beta/m_h$.
Note that the partial Higgs widths into SM
decay modes (excluding the $gg$ and $\gamma\gamma$
modes\footnote{The partial 
Higgs widths into $gg$ and $\gamma\gamma$ (these decay modes 
are absent at tree-level and only arise at the loop level)
include the effects of the light bottom squarks in the 
corresponding loop amplitudes.})
are assumed to be given by their SM values (in particular,
we assume that radiative corrections to these widths due to
supersymmetric particle exchanges are negligible). 
Consequently, the Higgs branching fractions into SM decay
channels are reduced from their SM values by a factor proportional to
$\tan^{-2} \beta$ once the $\tilde b\tilde b^*$ decay mode is dominant.
At $m_h = 120$ GeV, the $b \overline{b}$ and $\tilde{b} \tilde{b}^*$
branching fractions cross each other for $\mu \tan\beta/m_h \simeq
1.9$, where each of the branching fractions is about 0.4.

\subsubsection{Higgs boson decay to jets at the LHC}

Having demonstrated the feasibility of the scenario under study, we
concentrate on searches for such a Higgs boson.  At the Large Hadron 
Collider (LHC), a SM-like Higgs boson of mass less than $\sim 135$ GeV 
is expected to be discovered through a variety of production processes
and decay modes~\cite{atlas_tdr,cms_tdr,higgsreview}.  
These standard searches look for Higgs boson decays into SM particles.  As 
indicated in Fig.~\ref{fig:plot5}, the presence of an additional, dominant
decay mode into hadron jets may
suppress the branching ratios of these decay modes by a factor of order of 
ten to several hundred, depending on the coupling of the Higgs boson to the new
particles.  This reduction raises serious questions as 
to the capability of experiments at the LHC to discover 
the Higgs boson.  The more standard decays are suppressed, and the principal 
decay mode into jets suffers from enormous QCD backgrounds. 

\begin{figure}[t!]
\centerline{\includegraphics[height=10.5cm]{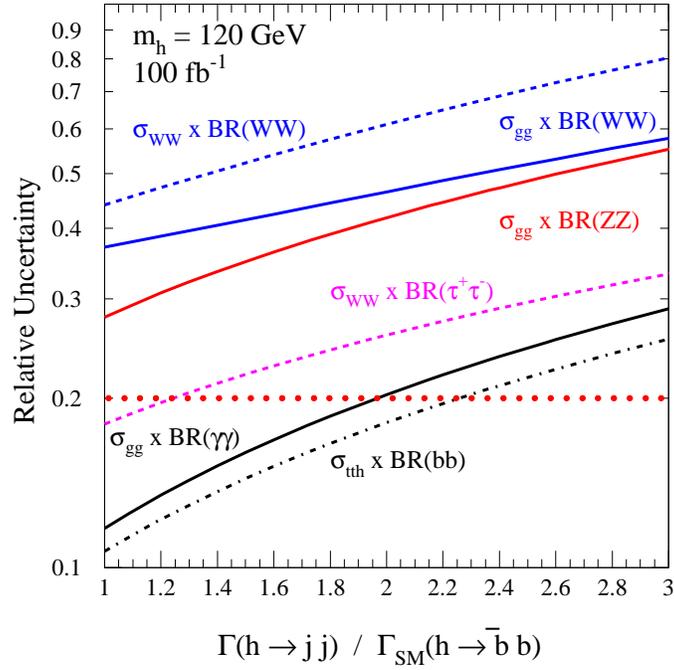}}
\caption[]{\it 
  Expected accuracy in LHC measurements of the product of production cross
  sections and branching ratios for the $WW$, $ZZ$, $b \overline{b}$, $\gamma
  \gamma$, and $\tau^+ \tau^-$ decay modes of a Higgs boson with mass 120 GeV
  as a function of the ratio of the jet-jet and the $b
  \overline{b}$ widths.  The horizontal dotted line at 0.2 indicates the
  $5\sigma$ discovery reach under the assumption $B \gg S$.  The partial widths
  for Higgs decay into $WW^*$, $ZZ^*$, $b \overline{b}$, 
  $c\overline{c}$, $gg$, $\gamma \gamma$, and $\tau^+
  \tau^-$ and the production cross sections are assumed to be 
   equal to their SM values.} 
\label{fig:plot6}
\end{figure}

At the LHC, it is difficult to obtain information about the Higgs
boson couplings in a model-independent way because it is impossible to
observe all possible decays in a single production mode.  One must be
content with measurements of cross sections times branching ratios and
cannot make definitive statements about the couplings themselves.  In
Fig.~\ref{fig:plot6}, for $m_h = 120$ GeV, we show the accuracies that
we expect could be achieved at the LHC (assuming a total 
integrated luminosity of $100~{\rm fb}^{-1}$)
for measurements of the rates
(cross sections times branching ratios) of gluon fusion into a Higgs
boson followed by $h \rightarrow \gamma \gamma$, $WW^*$, and $Z Z^*$,
and for weak boson fusion into a Higgs boson followed by the decays $h
\rightarrow WW^*$ and $h \rightarrow \tau^+ \tau^-$.  The accuracies are
shown as a function of the ratio of the jet-jet and the $b \overline
{b}$ widths.  By presenting the plot in this way, the results are more
general than the specific supersymmetric model discussed above, and
apply to any model of new physics in which the Higgs couplings to
light quarks and gluons are modified.  In particular,
the jet-jet width is the sum of the partial widths for
decay into the new hadron channel ($\tilde{b} \tilde{b}^*$ in the
supersymmetry example given above), plus $b \overline{b}$, $c
\overline{c}$, and $gg$.  The Higgs cross-sections\footnote{Details on
the Higgs cross-sections used in this analysis can be found in
in Ref.~\cite{Berger:2002vs}.}
and partial Higgs widths into SM
decay modes are assumed to be given by their SM values (that
is, radiative corrections to these widths due to
supersymmetric particle exchanges are neglected.)  
In the SM, $\Gamma (h \rightarrow
{\rm jets}) / \Gamma (h \rightarrow b \overline b) = 1.12$, primarily
due to the $WW^*$ (where the $W^*$ decays hadronically),
$gg$ and $c\bar c$ Higgs decay modes.\footnote{Note that the
left-most edge of the plot in Fig.~\ref{fig:plot6}, 
\textit{i.e.} $\Gamma (h \rightarrow
{\rm jets}) / \Gamma (h \rightarrow b \overline b) = 1$, would correspond
to a case in which the Higgs 
decay to bottom quarks is the only hadronic decay
mode of the Higgs boson.}  The relative uncertainties contain
statistical effects, $\sqrt{S + B} / S$, where we use estimates of the
backgrounds and SM signal rates presented in
Refs.~\cite{Cavalli:2002vs} and~\cite{zep}.  The 
production through weak boson fusion followed by decay into $W$ bosons
is based on leptonic $W$ decays into one electron and one muon, and includes
a presumed systematic uncertainty of $30\%$.

For values of the branching ratio ${\rm BR}(h \to {\rm jets})$
larger than two to five times that into bottom quarks, the large QCD jet 
backgrounds will make observation of the $h$ very difficult in Tevatron and 
LHC experiments.   In the particular example of light bottom squarks, however, 
experiments at these colliders may see 
evidence of low-energy supersymmetry and/or the heavy SUSY Higgs bosons
(note, however, the standard 
missing-energy signature of conventional SUSY is compromised 
if a charged $\tilde{b}$ is the LSP).  

\subsubsection{Higgs boson decay to jets at the LC}

\begin{figure}[t!]
\centerline{\includegraphics[height=10.5cm]{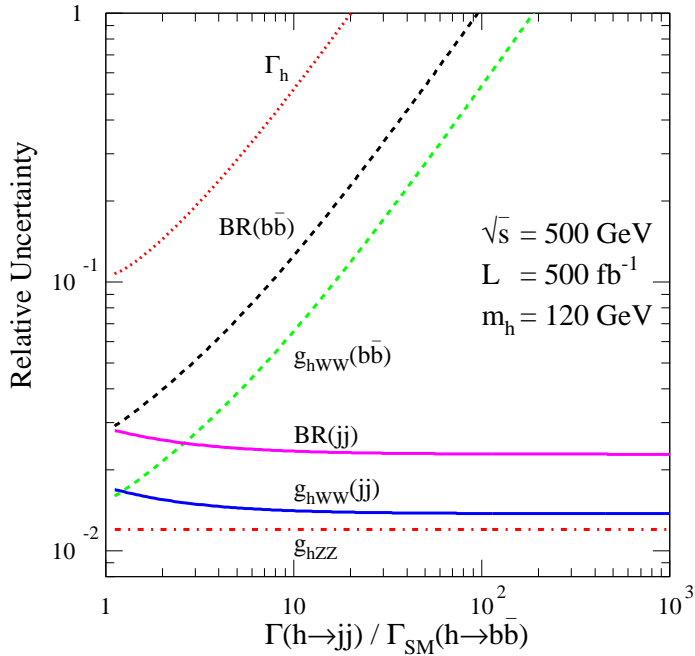}}
\caption[]{\it 
Expected accuracy in the measurements of the $b \overline{b}$ and jet-jet 
branching fractions, the $hZZ$ and $hWW$ coupling 
strengths, and the total width of the Higgs boson, as a function of the 
ratio of the jet-jet and the $b \overline {b}$ widths.  
The Higgs boson partial widths to $b \overline{b}$, $c\overline{c}$, $gg$,
$ZZ^*$, and $WW^*$ are taken to be equal to their SM values.}
\label{fig:plot7}
\end{figure}

We now consider a Higgs boson decaying into jets at a high energy $e^+
e^-$ linear collider.  For a light Higgs boson the dominant production
process is $e^+ e^- \rightarrow Z^0 h$ via an intermediate
$Z^0$~\cite{lcstudies}.  Once the $Z^0$ is identified, the Higgs boson
is discovered, independent of the Higgs boson decay modes, as a clean
enhancement in the distribution of mass recoiling from the
$Z^0$~\cite{lcstudies}, and the mass of the Higgs boson can be
measured.  The backgrounds from the $W$ fusion and $Z^0$ fusion
processes are small.  Because the $Z h$ cross section depends on the
$hZZ$ coupling strength, observation of the Higgs boson determines
this coupling with an expected accuracy of $\sim
1.2$\%~\cite{lcstudies} and can establish the Higgs boson as the
principal scalar responsible for electroweak symmetry breaking.  These
statements remain valid if the Higgs boson decays primarily into a
pair of jets since the method does not depend on the Higgs boson decay
products.

In the weak boson fusion process, the jet-jet Higgs boson decay
channel can also be used to determine the $hWW$ coupling at large $\mu
\tan \beta/m_h$ with significantly greater anticipated accuracy than
from the $b \overline{b}$ channel~\footnote{To obtain the uncertainty 
on the branching fraction
into a pair of jets, we must first estimate the number of signal and
background events in the jet-jet channel.  We begin with the numbers
presented in Ref.~\cite{Brau:2001} of simulated signal and background
events for $h \rightarrow b \overline{b}$ in the process $e^+ e^-
\rightarrow Z h$.  We remove the $b$-tag requirement, increasing both
signal and background by $(1/0.75)^2$.  The signal sample is then
multiplied by (1/0.69), the inverse of the SM $b \overline{b}$
branching fraction, and by the jet-jet branching fraction in our
model.  Since the primary background arises from $Z$ decay, the
background is increased by $1/R_b \simeq 1/0.21$, where $R_b$ is the
measured fraction of the hadronic width of the $Z$ into
$b\overline{b}$.}. 
In Fig.~\ref{fig:plot7}, we show the accuracies that we expect could be
achieved in the measurements of the $b \overline{b}$ branching
fraction, the $hZZ$ and $hWW$ coupling strengths, and the total width
of the Higgs boson, all as a function of the ratio of the jet-jet and
the $b \overline {b}$ widths.  We distinguish the accuracies to be
expected for the $hWW$ coupling strength depending upon whether the $b
\overline{b}$ or jet-jet decay mode of the Higgs boson is used.  In
this plot, the jet-jet width includes the partial widths into the new
non-standard particles ($\tilde{b} \tilde{b}^*$ in the MSSM example),
plus $gg$, $b \overline{b}$ and $c \overline{c}$ (the 
partial Higgs widths for the latter three channels
are fixed to their SM values).

Because they rely principally on the production process $e^+ e^-
\rightarrow h Z^0$, experiments at proposed $e^+ e^-$ linear colliders
remain fully viable for direct observation of the $h$ and measurement
of its mass and some of its branching fractions~\cite{Berger:2002vs}.
The possibility of measuring the Higgs boson width, however, is
diminished owing to the large suppression of the decay branching ratio
into the weak gauge bosons.  If the width exceeds about 2 GeV, a
direct measurement should be possible from the invariant mass
distribution in the jet-jet channel.

\subsubsection{Conclusions}

In the scenario considered here, the Higgs boson decays to a large
extent into hadronic jets, possibly without definite flavor content.
Measurements of various properties of the Higgs boson, such as its
full width and branching fractions, may therefore require a
substantial improvement in the experimental jet-jet invariant mass
resolution and a more thorough understanding of backgrounds in the
jet-jet channel.  Full event and reconstruction studies done for the
SM decay $h \rightarrow g g$ (where the SM branching fraction is
$\sim$ 5\% for $m_h =120$ GeV) should be pursued further to establish
the extent to which properties of the Higgs boson can be determined
solely from the jet-jet mode.

The existence of a light Higgs boson, observable at the LHC, is often
considered the hallmark of the {\em minimal} supersymmetric standard
model.  A light Higgs boson decaying into bottom squarks, undetected
at the LHC could thus lead one to conclude erroneously that the
underlying effective theory is more exotic than the minimal
supersymmetric extension.  On the other hand, a
Higgs boson that decays into jets can be studied quite well at a
linear collider (LC), a point that emphasizes the complementary nature
of the LHC and a LC in this scenario.  The LHC, with enormous energy
and luminosity should be able to produce, discover, and study in great
detail possible new physics at the weak scale (the super-partners in
the supersymmetry example).  The LC can discover and study the Higgs
boson.  In order to truly understand electroweak symmetry breaking and
the solution of the hierarchy problem, both aspects are crucial. As a
further possibility, one might produce super-partners at the LHC that
decay `through' light Higgs bosons into jets and not realize what the
intermediate state is.  In a situation such as this one, it might even
be impossible to identify the parent super-particles, despite their
having properties rather ordinary from the point of view of the MSSM.
The analysis and understanding of data from concurrent operation of
the LHC and a LC may prove crucial.



\section{Determination of CP~properties of Higgs bosons}

\def\ISAJET{Isajet\,7.64}
\def\SOFTSUSY{Softsusy\,1.71}
\def\SPHENO{Spheno\,2.01}
                                                                                
\renewcommand{\gsim}{\;\raisebox{-0.9ex}
           {$\textstyle\stackrel{\textstyle >}{\sim}$}\;}
\renewcommand{\lsim}{\;\raisebox{-0.9ex}{$\textstyle\stackrel{\textstyle<}
           {\sim}$}\;}
                                                                                

\hyphenation{Higgs-strah-lung}
                                                                                
In this section we briefly summarise the information on determination of
CP properties of Higgs bosons at different colliders and identify areas
of investigations to study the issue of LHC-LC synergy.

\subsection{CP studies of the Higgs sector}

{\it R.M.~Godbole, S.~Kraml, M.~Krawczyk, D.J.~Miller, P.~Nie\.zurawski
and A.F.~\.Zarnecki}

\vspace{1em}

\subsubsection{Introduction}

Studies of the CP properties of the Higgs sector, which will involve
establishing the CP eigenvalue(s) for the Higgs state(s) if CP is conserved,  
and measuring the mixing between the CP-even and CP-odd states if it is not, 
will certainly be part of the physics studies at future colliders 
\cite{Kinnunen:2002cr,Aguilar-Saavedra:2001rg,Badelek:2001xb}. 
CP violation in the Higgs sector~\cite{Weinberg:1976hu}, possible in 
multi-Higgs models, is indeed an interesting option to generate CP violation 
beyond that of the SM, possibly helping to explain the observed Baryon 
Asymmetry of the Universe~\cite{Dine:2003ax}.
  
In order to identify the CP nature of a Higgs boson, one must probe
the structure of its couplings to known particles, in either its
production or decay. At tree level, the couplings of a neutral Higgs 
boson $\phi$, which may or may not be a CP eigenstate,\footnote{For  
CP eigenstates, a pure scalar will be denoted by $H$ and a pure pseudoscalar 
by $A$. Otherwise we use the generic notation $\phi$.} 
to fermions and vector bosons can be written as 
\begin{equation}
   f\bar f\phi:~-\bar f(v_f+ia_f\gamma_5)f\,\frac{gm_f}{2m_W},\qquad
   VV\phi:~c_V\,\frac{gm_V^2}{m_W}\,g_{\mu\nu}\,
\label{eq:sec24-1}
\end{equation}
where $g$ is the usual electroweak coupling constant; $v_f$, $a_f$ give 
the Yukawa coupling strength relative to that of a SM Higgs boson, and 
$c_V$ ($V=W,\,Z$) are the corresponding relative couplings to gauge bosons.  
In the SM, for a CP-even Higgs $v_f=c_V=1$ and $a_f=0$. A purely CP-odd Higgs 
boson has $v_f=c_V=0$ and $a_f\not=0$, with the magnitude of $a_f$ depending 
on the model. 
In CP-violating models, $v_f$, $a_f$ and $c_V$ may all be non-zero at tree
level. In particular, in the case of a general 2HDM or the MSSM with CP 
violation, there are three neutral Higgs bosons $\phi_i$, $i=1,2,3$, which 
mix with each other and share out between them the couplings to the $Z$, $W$ 
and fermions; various sum rules are given in 
\cite{Gunion:1990kf,Gunion:1997aq,Ginzburg:2002wt}.
Due to this fact, limits on the MSSM (and 2HDM) Higgs sector implied
by LEP data are strongly affected by the presence of CP violation
\cite{Gunion:1997aq,Carena:2002bb,data2003}.

In most formulations of CP-violating Higgs sectors~\cite{Dedes:1999sj,
Pilaftsis:1999qt,Choi:2000wz,Carena:2002bb,Ginzburg:2002wt,Dubinin:2002nx} 
the amount of CP mixing is small, being generated at the loop level, 
with only one of the couplings to gauge bosons or fermions sizable. 
In most cases, the predicted CP mixing is also a function of the CP-conserving
parameters of the model, along with the CP-violating phases.%
\footnote{For the MSSM with explicit CP violation, computational tools 
for the Higgs sector are available\cite{Heinemeyer:2001qd,Lee:2003nt}.}
Thus observation and measurement of this mixing at the LC 
may give predictions for LHC physics; for instance for sparticle 
phenomenology in the MSSM.
Moreover, experiments at different colliders have different sensitivities 
to the various couplings of eq.~\ref{eq:sec24-1}. Hence a combination of
LHC, LC and photon collider (PLC) measurements of both CP-even and CP-odd 
variables may be necessary to completely determine the coupling structure 
of the Higgs sector. 
These are two ways in which the high potential of LHC-LC synergy 
for CP studies can be realized.

In what follows, we give an overview of the LHC, LC, and PLC potentials 
for CP studies in the Higgs sector. An example of  the LHC-LC synergy is 
presented as well.

\subsubsection{CP Studies at the LHC}

There are several ways to study the CP nature of a Higgs boson at the LHC. 
In the resonant s-channel process $gg \to\phi\to f\bar f$, the scalar or 
pseudoscalar nature of the Yukawa coupling gives rise to $f\bar f$ spin-spin 
correlations in the production plane \cite{Bernreuther:1997gs}.
A more recent study \cite{Khater:2003wq} looks at
this process in the context of a general 2HDM.

In the process $gg\to t\bar t\phi$, the large top-quark mass enhances the 
$v^2-a^2$ contribution, allowing a determination of the CP-odd and CP-even 
components of a light Higgs Boson~\cite{Gunion:1996xu,Field:2002gt}. 
While this method should provide a good test for verifying a pure scalar 
or pseudoscalar, examination of a mixed CP state would be far more 
challenging, requiring $600~{\rm fb}^{-1}$ to distinguish an equal 
CP-even/CP-odd mixture at $\sim 1.5\,\sigma$~\cite{Gunion:1996xu}.

Higgs decay into two real bosons, $\phi\to ZZ$, with $Z\to l^+l^-$,
~\cite{Choi:2002jk,Buszello:2002uu} can be used to rule out
a pseudoscalar state by examining the azimuthal or polar angle distributions 
between the decay lepton pairs. Below the threshold, $\phi \to Z^*Z$, 
extra information is provided by the threshold behavior of the virtual 
$Z$ boson invariant mass spectrum. 
This way,  one could rule out a pure $0^-$ state at $>5\sigma$
with $100~{\rm fb}^{-1}$ in the SM.  
An extension of these studies to scalar-pseudoscalar mixing is under progress.

In weak boson fusion, the Higgs boson is produced in association with two 
tagging jets, $qq \to W^+W^-qq \to \phi qq$. As with the decay to $ZZ$, 
the scalar and pseudoscalar couplings lead to very different azimuthal 
distributions between the two tagging jets~\cite{Plehn:2001nj}.
A similar idea may be employed in $\phi + 2 jets$ 
production~\cite{DelDuca:2001ad} in gluon fusion.  Higher order 
corrections~\cite{Odagiri:2002nd} may, however, reduce this correlation 
effect strongly.

Another approach uses the exclusive (inclusive) double diffractive process 
$pp\to p+\phi+p$ ($pp\to X+\phi+Y$)
\cite{Khoze:2001xm,Cox:2003xp,Khoze:2004rc} 
with large rapidity gaps between the $\phi$ and the (dissociated) protons. 
The azimuthal angular distribution between the tagged forward protons 
or the transverse energy flows in the fragmentation regions reflect 
the CP of the $\phi$ and can be used to probe CP mixing. 
This process is particularly promising for the region $m_\phi<60$~GeV, 
in which a Higgs signal may have been missed at LEP due to CP violation.

\subsubsection{CP Studies at an \boldmath $e^+e^-$ Linear Collider}

In $e^+e^-$ collisions, the main production mechanisms of neutral
Higgs bosons $\phi$ are 
(a) Higgsstrahlung $e^+e^-\to Z\phi$, 
(b) $WW$ fusion $e^+e^-\to \phi\,\nu\bar\nu$, 
(c) pair production $e^+e^-\to \phi_i\,\phi_j$ ($i \neq j$) and 
(d) associated production with heavy fermions, $e^+e^-\to f\bar f\phi$.  
Studies of CP at the Linear Collider aim at extracting the relevant
couplings mentioned in eq.~\ref{eq:sec24-1}.  Recall that a pure
pseudoscalar of the 2HDM or MSSM does not couple to vector bosons at tree
level.  The observation of all three $\phi_i$ $(i=1,2,3)$ in a given
process, e.g. $e^+e^-\to Z\phi_{1,2,3}$, therefore represents evidence
of CP violation
\cite{Mendez:1991gp,Grzadkowski:1999ye,Akeroyd:2001kt}.

In the Higgsstrahlung process, if $\phi$ is a pure scalar  the $Z$
boson is produced in a state of longitudinal polarization at high
energies \cite{Barger:1993wt,Hagiwara:1993sw}. For a pure pseudoscalar, 
the process proceeds via loops and the $Z$ boson in the final state is 
transversally polarized. The angular distribution of $e^+e^-\to ZH$ is 
thus $\propto\sin^2\theta_Z$, where $\theta_Z$ is the production angle 
of the $Z$ boson w.r.t.\ to the beam axis in the lab frame, while that 
of $e^+e^-\to ZA$ is $\propto(1+\cos^2\theta_Z)$. A forward-backward
asymmetry would be a clear signal of CP violation.  
Furthermore, angular correlations of the $Z\to f\bar f$ decay can be used 
to test the $J^{PC}$ quantum numbers of the Higgs boson(s). Measurements of
the threshold excitation curve can give useful additional information
\cite{Miller:2001bi,Dova:2003py}. A study in
\cite{Aguilar-Saavedra:2001rg} parametrized the  effect of CP violation
by adding a small $ZZA$ coupling with strength $\eta$ to 
the SM matrix element,  ${\cal M } = {\cal M}_{ZH} + i \eta  {\cal M}_{ZA}$,~
and showed that $\eta$ can be measured to an accuracy of $0.032$ with 
$500$ fb$^{-1}$.

Angular correlations of Higgs decays can also be used to determine the
CP nature of the Higgs boson(s), independent of the production
process; see 
\cite{Kramer:1993jn,Grzadkowski:1995rx,Gunion:1996vv} 
and references therein. 
The most promising channels are $\phi\to\tau^+\tau^-$ ($m_\phi<2m_W$) and 
$\phi\to t\bar t$ ($m_\phi>2m_t$) which in contrast to decays into $WW$ or 
$ZZ$ allow equal sensitivity to the CP-even and CP-odd components of $\phi$.  
  
A detailed simulation of $e^+e^-\to ZH$ followed by $H\to\tau^+\tau^-$ and 
$\tau^\pm\to\rho^\pm\bar\nu_\tau(\nu_\tau)$ 
\cite{Bower:2002zx,Desch:2003mw,Worek:2003zp} 
showed that CP of a 120~GeV SM-like Higgs boson can be measured to 
$\ge 95\%$ C.L. at a 500~GeV $e^+e^-$ LC with 500~fb$^{-1}$ of luminosity.
In case of CP violation, the mixing angle between the scalar and pseudoscalar 
states may be determined to about 6 degrees~\cite{Desch:2003rw}, the
limiting factor being statistics.

\subsubsection{CP Studies at a Photon Collider}

A unique feature of a PLC is that two photons can form a $J_z = 0$ state 
with both even and odd CP. As a result a PLC has a similar level of 
sensitivity for both the CP-odd and CP-even components of a CP-mixed state:
\begin{equation}
  {\rm CP\!-\!even:}
  \epsilon_1\cdot \epsilon_2 = -(1+\lambda_1\lambda_2)/2 , \quad
  {\rm CP\!-\!odd:}
  [\epsilon_1 \times \epsilon_2] \cdot k_{\gamma}
  =\omega_{\gamma} i \lambda_1(1+\lambda_1\lambda_2)/2,
\end{equation}
$\omega_i$ and $\lambda_i$ denoting the energies and  helicities of the 
two photons respectively; the helicity of the system is equal to 
$\lambda_1-\lambda_2$.
This contrasts the $e^+e^-$ case, where it is easy to discriminate between 
CP-even and CP-odd particles but may be difficult to detect small CP-violation 
effects for a dominantly CP-even Higgs boson~\cite{Hagiwara:2000bt}.  
For the PLC, one can form three polarization asymmetries in terms of helicity 
amplitudes which give a clear measure of CP mixing \cite{Grzadkowski:1992sa}. 
In addition, one can use information on the decay products of $WW$, $ZZ$, 
$t\bar t$ or $b\bar b$ coming from the Higgs decay.
Furthermore, with circular beam polarization almost mass degenerate 
(CP-odd) $A$ and (CP-even) $H$ of the MSSM may be separated 
\cite{Muhlleitner:2001kw, Niezurawski:2003ir,Asakawa:1999gz}.

A measurement of the spin and parity of the Higgs boson may also be
performed using the angular distributions of the final-state fermions
from the Z boson decay, which encode the helicities of Z's.   A
detailed study was performed for above and below  the ZZ threshold in
\cite{Choi:2002jk}. A realistic simulation based on this analysis was
made recently in~\cite{Niezurawski:2003ik}.

The same interference effects as mentioned above can be used in the process 
$\gamma \gamma \rightarrow \phi \rightarrow t \bar t$ 
\cite{Asakawa:2000jy,Godbole:2002qu} 
to determine the $t \bar t\phi$ and $\gamma \gamma \phi$ couplings 
for a $\phi$ with indefinite CP parity.

\subsubsection{Example of LHC-LC synergy}

As an example of the LHC-LC synergy, we consider the SM-like, type II 2HDM
with CP-violation~\cite{Niezurawski:2003ik,2HDM}.
%
We study production of  $\phi_2$ in the mass range 200 to 350~GeV, 
decaying to $VV$, $V = W/Z$, at the LHC, LC and PLC. In particular,
we investigate the interplay of different experiments for the determination 
of $\tan \beta$ and mixing angle $\Phi_{HA}$.

\begin{figure}[htb]
     \epsfig{figure=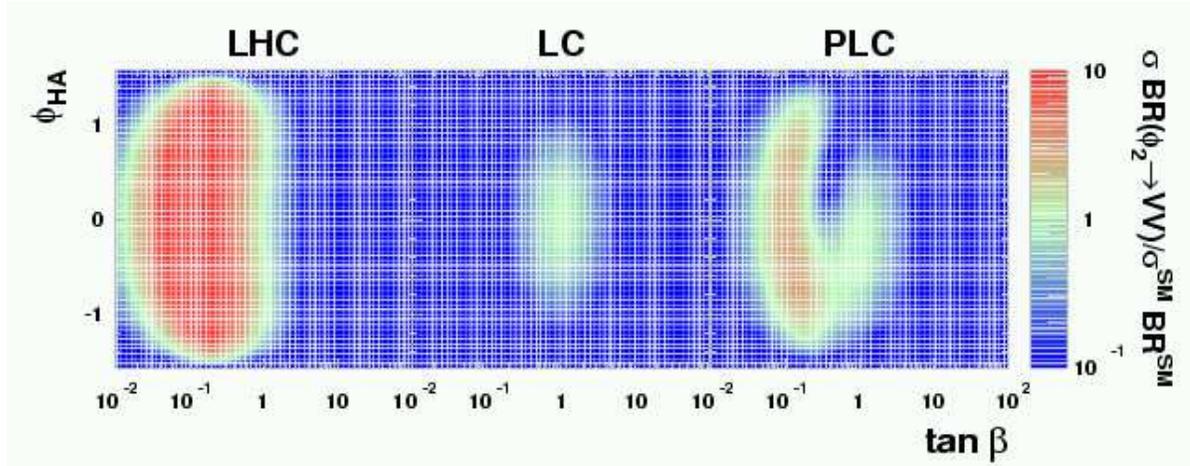,width=\textwidth,clip=}
\vspace{-1cm}
\caption{
$\sigma \times$ BR for $\phi_2 \rightarrow VV$ with $V = W/Z$, relative to the
SM expectation for the same for a mass of 250 GeV, as a function of
$\tan \beta$ and $\Phi_{HA}$ for  LHC, LC and PLC.}
 \label{nzk:cros2}
 \end{figure}

Figure~\ref{nzk:cros2} shows the expected rates for $\phi_2$ with 
$m_{\phi_2}=250$ GeV relative to the SM ones, as a function of 
$\tan\beta$ and  $\Phi_{HA}$.  For a SM Higgs boson, 
the expected precision on $\sigma \times BR(H \rightarrow VV)$ is 
$\sim 15$\% at the LHC \cite{cms_tn_95-018,cms_cr_2002-020} and 
better than  10\%  at LC and PLC \cite{lc-phsm-2003-066,nzk_wwzz}. 
A PLC will allow  to measure $\Gamma_{\gamma \gamma}$ with a precision 
of 3--8\%  and the phase of the $\phi \rightarrow \gamma \gamma$ 
amplitude, $\Phi_{ \gamma \gamma}$, to $40-120$~mrad \cite{nzk_wwzz}.   

\begin{figure}[htb]
  \begin{center}
     \epsfig{figure=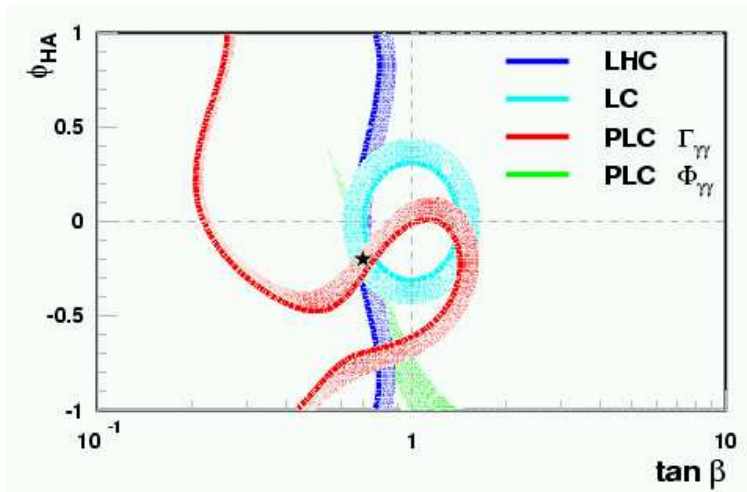,width=10cm,clip=}
  \end{center}
\vspace{-1cm}
\caption{ 
1-$\sigma$ bands for the determination of  $\tan \beta$ and  $\Phi_{HA}$ 
from  measurements at LHC, LC and PLC, for the case $\tan \beta = 0.7$ 
and $\Phi_{HA} = -0.2$. The assumed parameter values are  indicated by 
a star ($\star$).}
 \label{nzk:syn2} 
 \end{figure}

Figure~\ref{nzk:syn2} shows the $1\,\sigma$ bands for determination
of $\tan \beta$ and $\Phi_{HA}$, at the LHC, LC and PLC 
for a particular choice of parameters : $\tan \beta=0.7$ and 
$\Phi_{HA}=-0.2$. The chosen point is indicated by a star. For the PLC, 
information from $\Gamma_{\gamma\gamma}$ and $\Phi_{\gamma \gamma}$ 
is included. As can be seen, an accurate determination of both 
parameters requires combination of data from all three colliders.
%

\subsubsection{Summary}

The LHC, an $e^+e^-$ LC, and a LC in the photon collider option (PLC)
will be able to provide nontrivial information on the CP quantum numbers 
of the Higgs boson(s). 
We have summarized the potentials of the different colliders 
in this document and discussed the possible LHC-LC synergy.

In the MSSM, for instance, the size of CP-violating effects in the Higgs
sector depends in part on the sparticle spectrum.  
Observation and measurement of Higgs-sector CP mixing at the LC 
can hence give predictions for phenomenology at the LHC in the 
CP-conserving sector, thus providing a high potential of LHC-LC synergy.  
A detailed study of this issue is, however, still missing.

Moreover, experiments at different colliders have different sensitivities
to the various couplings of eq.~\ref{eq:sec24-1}. Hence a combination of
LHC and LC/PLC measurements of both CP-even and CP-odd variables may be
necessary to completely determine the coupling structure of the Higgs sector.
In this document we have presented a first analysis which exemplifies this
realization of LHC-LC synergy. While the example presented shows a high 
potential of the LHC-LC synergy for CP studies, detailed realistic simulations
still need to be performed.


Another possibility within CP-violating Higgs scenarios is the
following:
there are parameter choices (the CPX
model) for which rather light Higgs bosons would
not have been discovered at LEP
\cite{Carena:2002bb} and would also escape LHC
detection in the standard channels because of the
types of difficulties discussed earlier.  The
$WW\to h \to aa \to jj \tau^+\tau^-$ channel
(where $j=b$ if $m_a>2m_b$)
has not been examined in the context of this
model.  It can be anticipated (see also Sect.~\ref{sec251})
that an excess in the $2j2\tau$ mass distribution
might be observable, but that its confirmation as
a Higgs signal would require observation in the
$4b$ final state and probably also at the LC.




\section{SUSY Higgs physics}

The prediction of a firm upper bound on the mass of the lightest Higgs
boson is one of the most striking predictions of Supersymmetric theories
whose couplings stay in the perturbative regime up to a high energy
scale. Disentangling the structure of the Higgs sector and establishing
possible deviations from the SM will be one of the main goals at the next
generation of colliders.

In order to implement electroweak symmetry breaking consistently into
the MSSM, two Higgs doublets are needed. This results in eight degrees
of freedom, three of which are absorbed via the Higgs mechanism to give
masses to the $W^{\pm}$ and $Z$ bosons. The remaining five physical
states are the neutral CP-even Higgs bosons $h$ and $H$, the neutral
CP-odd state $A$, and the two charged Higgs bosons $H^{\pm}$. At lowest
order, the Higgs sector of the MSSM is described by only two parameters
in addition to the gauge couplings, conventionally chosen as $M_A$ and
$\tan\beta$, where the latter is the ratio of the vacuum expectation
values of the two Higgs doublets.

The tree-level upper bound on the mass of the lightest CP-even Higgs
boson, $m_h < M_Z$ in the MSSM, arising from the gauge structure of the 
theory, receives large radiative corrections from the Yukawa sector of the 
theory. Taking corrections up to two-loop order into account, it is
shifted by about 50\%~\cite{Degrassi:2002fi}. As a consequence, loop effects, in
particular from the top and scalar top sector and for large values of
$\tan\beta$ also from the bottom and scalar bottom sector, are very
important for SUSY Higgs phenomenology. 

While the Higgs sector of the MSSM is CP-conserving at tree level,
CP-violating effects can enter via loop corrections. The Higgs sector in
extensions of the MSSM contains further matter structure, for instance
additional Higgs singlets.

The LC will provide precision measurements of the properties of all
Higgs bosons that are within its kinematic reach. Provided that a
Higgs boson couples to the $Z$~boson, the LC will observe it independently of
its decay characteristics. At the LHC, on the other hand, Higgs boson
detection can occur in various channels. In many cases complementary 
information from more than one channel will be accessible at the LHC. In
particular, the LHC has a high potential for detecting heavy Higgs
states which might be beyond the kinematic reach of the LC.

In the following, the possible interplay between LHC and LC results in
SUSY Higgs physics is investigated for several examples.
In Section~\ref{sec:221} first a scenario is analysed where the LHC can detect
the heavy Higgs states of the MSSM, while the LC provides precise
information on the branching ratios of the light Higgs boson. This
allows to perform a sensitive consistency test of the MSSM and to obtain
indirect information on the mixing in the scalar top sector. 
Furthermore a scenario where LHC and LC only detect one light Higgs
boson is investigated and it is demonstrated how constraints on $M_A$
can be derived from combined LHC and LC data.
In Section~\ref{sec:222} the decay of the heavy Higgs bosons $H$ and $A$ 
into a pair of neutralinos is studied at the LHC. This decay can be used
to determine $M_A$, provided that a precise measurement of the mass of
the lightest neutralino from the LC is available.
Section~\ref{sec:223} investigates the situation where the neutral Higgs
bosons are almost mass-degenerate and $\tan\beta$ is large. In this case
detection of the individual Higgs boson peaks is very challenging at the
LHC, while the different Higgs boson signals can more easily be
separated at the LC. The measured characteristics at the LC will then
allow to determine further Higgs-boson properties at the LHC.
In Sections~\ref{sec:224} and \ref{sec:421} the determination of
$\tan\beta$ from combined LHC and LC information (including also LC
running in the $\gamma\gamma$ mode) is investigated.


\subsection{\label{sec:221} 
Consistency tests and parameter extraction from the combination of LHC
and LC results}

{\it K.~Desch, E.~Gross, S.~Heinemeyer, G.~Weiglein and
L.~\v{Z}ivkovi\'{c}}

\vspace{1em}
\renewcommand{\lsim}{\buildrel<\over{_\sim}}
\renewcommand{\gsim}{\buildrel>\over{_\sim}}
                                                                                
\noindent{\small
The interplay of prospective experimental information from both the 
LHC and the LC 
in the investigation of the MSSM Higgs sector is analyzed in
the SPS~1a and SPS~1b benchmark scenarios. Combining LHC 
information on the heavy Higgs states of the MSSM with precise
measurements of the mass and branching ratios of the lightest CP-even
Higgs boson at the LC provides a sensitive consistency
test of the MSSM. This allows to set bounds on the trilinear
coupling $A_t$. In a scenario where LHC and LC only detect one light Higgs
boson, the prospects for an indirect determination of $M_A$ are investigated. 
In particular,
the impact of the experimental errors of the other SUSY parameters is
analyzed in detail. We find that a precision of about 20\% (30\%) can be
achieved for $M_A = $ 600 (800) GeV.
}


\subsubsection{Introduction}

The Higgs sector of the MSSM is fully determined at lowest order by only
two parameters in addition to the gauge couplings, $M_A$ and $\tan\beta$. If
the heavy Higgs states $H$ and $A$ (and possibly the charged states
$H^{\pm}$) are accessible at the LHC, their detection will provide
experimental information on both these
parameters~\cite{Cavalli:2002vs}. Thus, in principle
the phenomenology of the light CP-even Higgs boson can be predicted if
experimental results on the heavy Higgs bosons are available. Comparing
these predictions with experimental results on the light CP-even Higgs
boson constitutes an important test of the MSSM. Deviations may
reveal physics beyond the MSSM.

A realistic analysis of such a scenario, however, needs to take into
account that the Higgs-boson sector of the MSSM is affected by large
radiative corrections, which arise in particular from the top/stop
sector (for large values of $\tan\beta$ also loops of scalar bottom
quarks
can be important). In this way additional parameters become relevant for
predicting the properties of the light CP-even Higgs boson. Experimental
information on the parameters entering via large radiative corrections
will therefore be crucial for SUSY Higgs phenomenology. This refers in
particular to a precise
knowledge of the top-quark mass, $m_t$, from the
LC~\cite{sec2_lctdrs,mtatLC,deltamt} and information about the SUSY spectrum
from both LHC and LC (see Sect.~\ref{chapter:susy} below).
An analysis within the Higgs sector thus becomes much more involved.
Furthermore,
while detection of the heavy Higgs states at the LHC will provide a
quite
accurate determination of $M_A$, the experimental information on
$\tan\beta$ will be rather limited~\cite{Cavalli:2002vs}.
On the other hand, the LHC will also be able to detect scalar top and
bottom quarks over a wide mass range, so that it can be expected that
additional experimental information on the scalar quark sector will be
available.

In the following, two examples of a possible interplay between LHC and
LC results in SUSY Higgs physics are investigated~\cite{eilipaper}. They
are based on the benchmark scenarios SPS~1a and SPS~1b~\cite{Allanach:2002nj}. 
In Section~\ref{sec:sps1b} a scenario is analyzed where the LHC can detect
the heavy Higgs states of the MSSM (see e.g.\ Ref.~\cite{Cavalli:2002vs}), 
providing experimental information on both tree-level parameters of
the MSSM Higgs sector, $M_A$ and $\tan\beta$.
The LC, on the other hand, provides precise 
information on the branching ratios of the light Higgs boson, which can
be compared with the theory prediction. 
This allows in particular to obtain
indirect information on the mixing in the scalar top sector, which is
very important for fits of the SUSY Lagrangian to (prospective)
experimental data~\cite{sfittino}.

In Section~\ref{sec:sps1a} another scenario is analyzed where no
heavy Higgs bosons can be detected at LHC and LC.
The combined information about the SUSY spectrum from the LHC and LC and
of Higgs-boson branching ratio measurements at the LC is used to obtain
bounds on the mass of the CP-odd Higgs boson, $M_A$, in the
unconstrained MSSM (for such analyses within mSUGRA-like scenarios, see
Refs.~\cite{Dedes:2003cg,Ellis:2002gp}).
Since a realistic analysis requires the inclusion of radiative
corrections, the achievable sensitivity to $M_A$ depends on the
experimental precision of the additional input parameters and the
theoretical uncertainties from unknown higher-order corrections. This
means in particular that observed deviations in the properties of the
light CP-even Higgs boson compared to the SM case cannot be attributed
to the single parameter $M_A$. We analyze in detail the impact of 
the experimental and theory errors
on the precision of the $M_A$ determination. Our analysis
considerably differs from existing studies of Higgs boson
branching ratios in the literature~\cite{MAdet}.
In these previous analyses, all parameters except for the
one under investigation (i.e.\ $M_A$) have been kept fixed and the effect
of an assumed deviation between the MSSM and the SM has solely been attributed
to this single free parameter. This would correspond to a situation with
a complete knowledge of all SUSY parameters without any experimental or
theoretical uncertainty, which obviously leads to an unrealistic
enhancement of the sensitivity to the investigated parameter. 


\subsubsection{Scenario where LHC information on heavy Higgs states is
available}
\label{sec:sps1b}

In this section we analyze a scenario where experimental results 
at the LHC are used as input for confronting the predictions for the branching
ratios of the light CP-even Higgs boson with precision measurements at
the LC. We consider the SPS~1b benchmark
scenario~\cite{Allanach:2002nj}, which is a 
`typical' mSUGRA scenario with a relatively large value of $\tan\beta$. 
In particular, this scenario yields an $M_A$ value of about 550~GeV,
$\tan\beta = 30$, and stop and sbottom masses in the range of
600--800~GeV. 
More details about the mass spectrum
can be found in Ref.~\cite{Allanach:2002nj}.

We assume the following experimental information from the LHC and the LC:
\begin{itemize}
\item
$\Delta M_A = 10\%$\\
This prospective accuracy on $M_A$ is rather conservative. The
assumption about the experimental accuracy on $M_A$ is not crucial in the
context of our analysis, however, since for $M_A \gg M_Z$ the
phenomenology of the light CP-even Higgs boson depends only weakly
on $M_A$.

\smallskip
\item
$\tan\beta > 15$\\
The observation of heavy Higgs states at the LHC in channels like 
$b \bar b H/A, H/A \to \tau^+\tau^-, \mu^+\mu^-$ will be possible in
the MSSM if  
$\tan\beta$ is relatively large~\cite{Cavalli:2002vs}. An attempted
determination of $\tan\beta$ 
from the comparison of the measured cross section with the theoretical
prediction will suffer from sizable QCD uncertainties, from the 
experimental errors of the SUSY parameters entering the theoretical
prediction, and from the experimental error of the measured cross
section.
Nevertheless, the detection of
heavy Higgs states at the LHC will at least allow to establish a lower
bound on $\tan\beta$. On the other hand, if $\tan\beta \lsim 10$ the LC
will provide a precise determination from measurements in the chargino
and neutralino sector. Thus, assuming a lower bound of $\tan\beta > 15$
seems to be reasonable in the scenario we are analyzing.

\item
$\Delta m_{\tilde t}, \Delta m_{\tilde b} = 5\%$\\
We assume that the LHC will measure the masses of the scalar top and
bottom quarks with $5\%$ accuracy. This could be possible if precise
measurements of parameters in the neutralino and chargino sector are
available from the LC, see Sec.~\ref{sec:41} below. On the 
other hand, the measurements
at the LHC (combined with LC input) will only loosely constrain the
mixing angles in the scalar top and bottom sectors.
Therefore we have not made any assumption about their values, but have
scanned over the whole possible parameter space (taking into account the
$SU(2)$~relation that connects the scalar top and bottom sector).
It should be noted that for the prospective accuracy on the scalar top
and bottom reconstruction at the LHC we have taken the results of
studies at lower values of $\tan\beta$ ($\tan\beta = 10$).
While the stop reconstruction should not suffer 
from the higher $\tan\beta$ value assumed in the present
study, sbottom reconstruction is more involved, see
Sec.~\ref{sec:41} below.
We assume that the reconstruction of hadronic $\tau$'s 
from the decay $\chi^0_2 \to \tau^+ \tau^- \chi^0_1$ will be possible and the 
di-tau mass spectrum can be used for a mass measurement of the scalar 
bottom quarks at the 5\% level. This still has to be verified by experimental 
simulation.

In the scenario we are studying here the
scalar top and bottom quarks are outside the kinematic limit of the LC.

\item
$\Delta m_h = 0.5$~GeV\\
At the LC the mass of the light Higgs boson can be measured with an
accuracy of 50~MeV. In order to account for theoretical uncertainties
from unknown higher-order corrections
we assume an accuracy of $\Delta m_h = 0.5$~GeV in this
study. This assumes that a considerable reduction of the present
uncertainty of about 3~GeV~\cite{Degrassi:2002fi} will be achieved until
the LC goes into operation. 

\item
$\Delta m_t = 0.1$~GeV\\
The top-quark mass (according to an appropriate short-distance mass
definition) can be determined from LC measurements at the $t \bar t$
threshold with an accuracy of 
$\Delta m_t \lsim 0.1$~GeV~\cite{sec2_lctdrs,mtatLC}.

\end{itemize}

The experimental information from the heavy Higgs and scalar quark
sectors that we have assumed above can be used to predict the branching
ratios of the light Higgs boson. Within the MSSM, the knowledge of these
experimental input quantities will significantly narrow down the
possible values of the light Higgs branching ratios. Comparing this
prediction with the precise measurements of the branching ratios carried
out at the LC will provide a very sensitive consistency test of the
MSSM.

\begin{figure}[htb!]
\begin{center}
\epsfig{figure=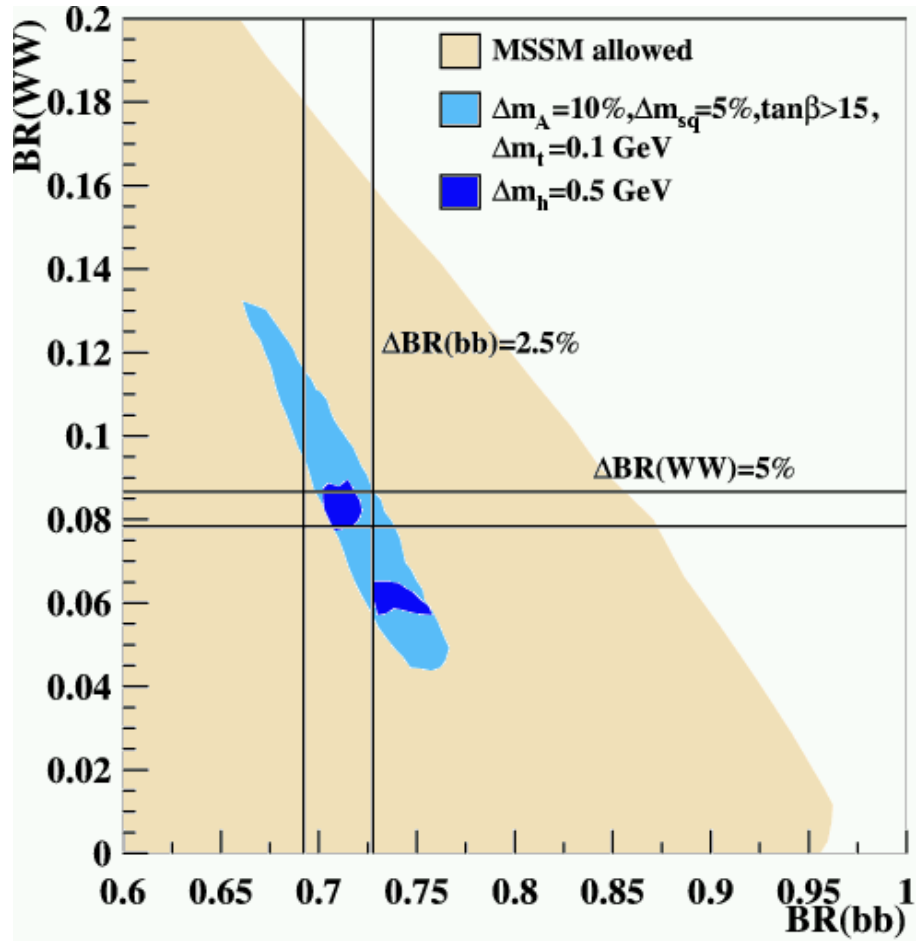, width=12cm}
\caption{
The experimental accuracies for the branching ratios BR($h \to b \bar b$) 
and BR($h \to WW^*$) at the LC of about 2.5\% and 5\%, indicated by a 
vertical and horizontal band, respectively, are
compared with the theoretical prediction in the MSSM. The light shaded
(yellow) 
region indicates the full allowed parameter space. The medium shaded
(light blue) region indicates the range of predictions in the MSSM
being compatible with the assumed experimental information from LHC
and LC,
$\Delta M_A = 10\%$, $\tan\beta > 15$,
$\Delta m_{\tilde t}, \Delta m_{\tilde b} = 5\%$,
$\Delta m_t = 0.1$~GeV.
The dark shaded (dark
blue) region arises if furthermore a measurement of the light CP-even
Higgs mass, including a theory uncertainty of $\Delta m_h = 0.5$~GeV,
is assumed. 
}
\label{fig:sec221plot}
\end{center}
\end{figure}

This is shown in Fig.~\ref{fig:sec221plot} for the branching ratios
BR($h \to b \bar b$) and BR($h \to WW^*$). The light shaded (yellow) region
indicates the full parameter space allowed for the two branching ratios
within the MSSM. The medium shaded (light blue) region corresponds to
the range of predictions in the MSSM being compatible with the assumed 
experimental information from the LHC as discussed above, i.e.\
$\Delta M_A = 10\%$, $\tan\beta > 15$, 
$\Delta m_{\tilde t}, \Delta m_{\tilde b} = 5\%$. The dark shaded (dark
blue) region arises if furthermore a measurement of the light CP-even
Higgs mass of $m_h = 116$~GeV, including a theory uncertainty of
$\Delta m_h = 0.5$~GeV, is assumed. 
The predictions are
compared with the prospective experimental accuracies for BR($h \to b
\bar b$) and BR($h \to WW^*$) at the LC of about 2.5\% and 5\%,
respectively~\cite{sec2_lctdrs,talkbrient}.

Agreement between the branching ratios measured at the LC and the
theoretical prediction
would constitute a highly non-trivial confirmation of the MSSM at the
quantum level. In
order to understand the physical significance of the two dark-shaded
regions in Fig.~\ref{fig:sec221plot} it is useful to investigate the
prediction for $m_h$ as a function of the trilinear coupling $A_t$ (see
also Ref.~\cite{deltamt}). If the masses of the scalar top and bottom quarks
have been measured at the LHC (using LC input), a precise measurement 
of $m_h$ will allow
an indirect determination of $A_t$ up to a sign ambiguity. It should be
noted that for this determination of $A_t$ the precise measurement of
$m_t$ at the LC is essential~\cite{deltamt}. It
also relies on a precise theoretical prediction for $m_h$, which
requires a considerable reduction of the theoretical uncertainties
from unknown higher-order corrections as compared to the present
situation~\cite{Degrassi:2002fi}, as discussed above. Making use of a 
prospective measurement of $m_h$ for predicting BR($h \to b \bar b$) and
BR($h \to WW^*$), on the other hand, is less critical in this respect,
since the kinematic effect of the Higgs mass in the prediction for the
branching ratios is not affected by the theoretical uncertainties.

\begin{figure}[htb!]
\begin{center}
\epsfig{figure=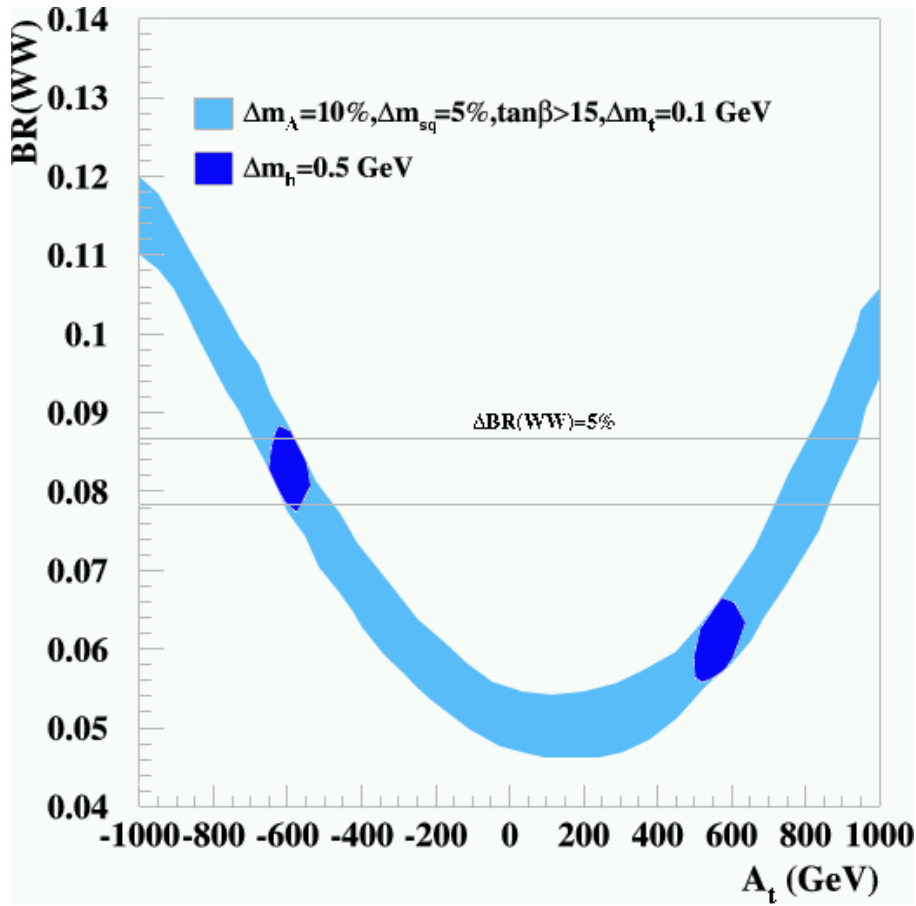, width=12cm}
\caption{
The branching ratio for $h \to WW^*$ is shown as a 
function of the trilinear coupling $A_t$. The light shaded (light blue)
region indicates the range of predictions in the MSSM
being compatible with the assumed experimental information,
$\Delta M_A = 10\%$, $\tan\beta > 15$,
$\Delta m_{\tilde t}, \Delta m_{\tilde b} = 5\%$,
$\Delta m_t = 0.1$~GeV. The dark shaded (dark
blue) region arises if furthermore a measurement of the light CP-even
Higgs mass, including a theory uncertainty of $\Delta m_h = 0.5$~GeV,
is assumed. The experimental
accuracy for BR($h \to WW^*$) at the LC of about 5\% is indicated by an
horizontal band.
}
\label{fig:sec221plotat}
\end{center}
\end{figure}

Fig.~\ref{fig:sec221plot} shows that the LC measurements of the
branching ratios of the light CP-even Higgs boson allow to discriminate
between the two dark-shaded regions. From the discussion above, these
two regions can be identified as corresponding to the two possible signs
of the parameter $A_t$. 
This is illustrated in Fig.~\ref{fig:sec221plotat},
where BR($h \to WW^*$) is shown as a function of $A_t$. It is demonstrated that
the sign ambiguity of $A_t$ can be resolved with the branching ratio
measurement. 
The determination of $A_t$ in this way will be crucial in 
global fits of the SUSY parameters to all available data~\cite{sfittino}.


\subsubsection{Indirect constraints on $M_A$ from LHC and LC\\
measurements}
\label{sec:sps1a}

In the following, we analyze an SPS~1a based scenario~\cite{Allanach:2002nj},
where we keep $M_A$ as a free parameter. We study in particular the
situation where the LHC only detects one light Higgs boson. For the
parameters of the SPS~1a scenario this corresponds to the region
$M_A \gsim 400$~GeV.

The precise measurements of Higgs branching ratios at the LC together
with accurate determinations of (parts of) the SUSY spectrum at the LHC
and the LC will allow in this
case to obtain indirect information on $M_A$ (for a discussion of
indirect constraints on $M_A$ from electroweak precision observables,
see Ref.~\cite{gigaz}). 
When investigating the sensitivity to $M_A$ it is crucial to take into
account realistic experimental errors of the other SUSY parameters that
enter the prediction of the Higgs branching ratios. 
Therefore we have varied all the SUSY
parameters according to error estimates for the measurements at
LHC and LC in this scenario. The sbottom masses and the gluino mass
can be obtained from mass reconstructions at the LHC with LC input,
see Sec.~\ref{sec:41} below. We have assumed a
precision of $\Delta m_{\tilde g} = \pm 8$~GeV and 
$\Delta m_{\tilde b_{1,2}} = \pm 7.5$~GeV.
We furthermore assume that the lighter stop (which in the SPS~1a
scenario has a mass 
of about 400~GeV, see Ref.~\cite{Allanach:2002nj}) will be accessible 
at the LC, leading to an accuracy
of about $\Delta m_{\tilde t_1} = \pm 2$~GeV. The impact of the LC information
on the stop mixing angle, $\theta_{\tilde t}$,
will be discussed below. For $\tan\beta$ we have 
used an uncertainty of $\Delta \tan\beta = 10\%$ (this accuracy can be 
expected from measurements at the LC in the gaugino sector for the SPS~1a 
value of $\tan\beta = 10$ \cite{Desch:2003vw}). We have
assumed a LC measurement of $m_h = 116$~GeV, but included a theory error from
unknown higher-order corrections of $\pm
0.5$~GeV~\cite{Degrassi:2002fi}.

\begin{figure}[htb!]
\begin{center}
\epsfig{figure=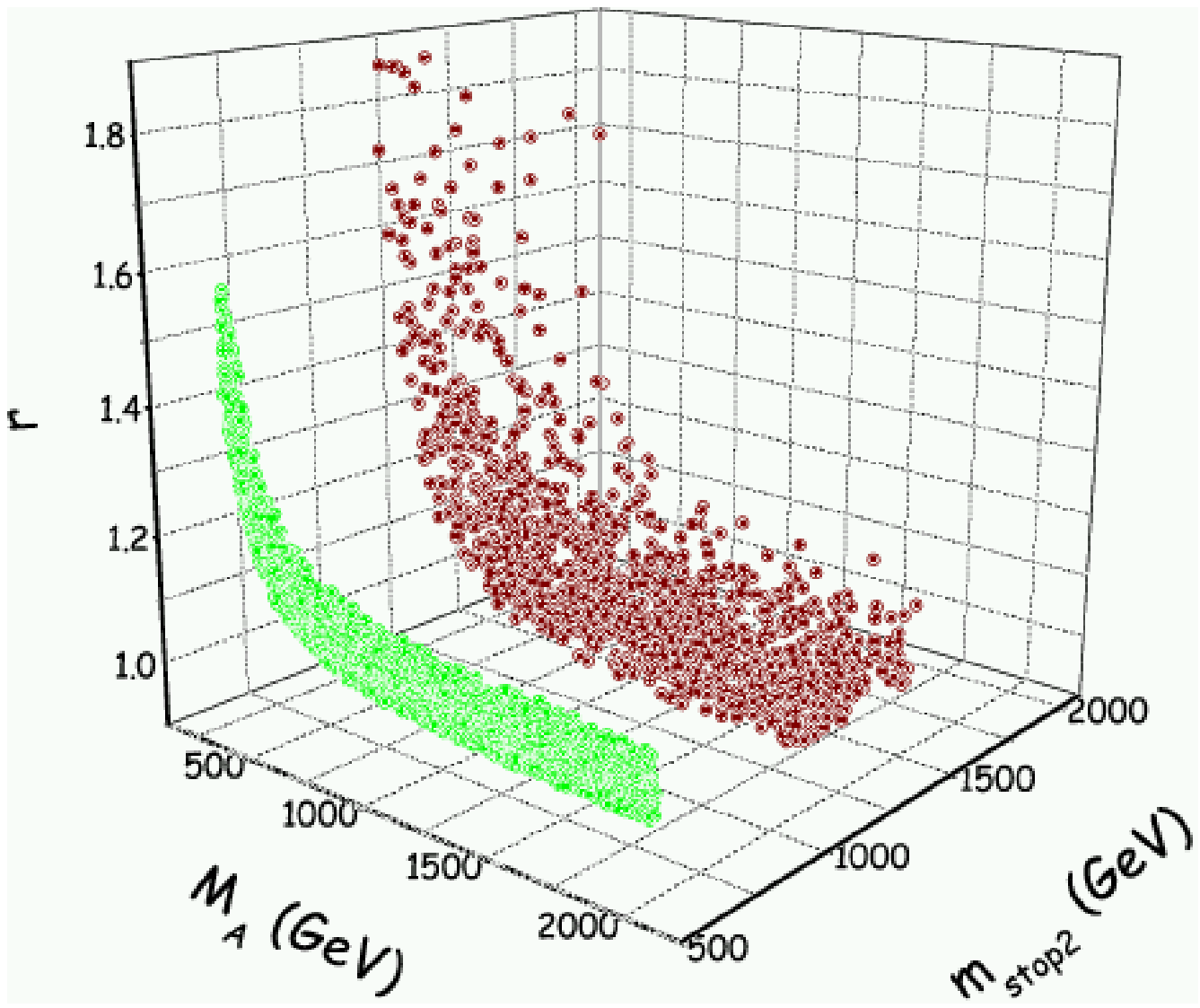, width=14cm,height=8.5cm}\\[1em]
\epsfig{figure=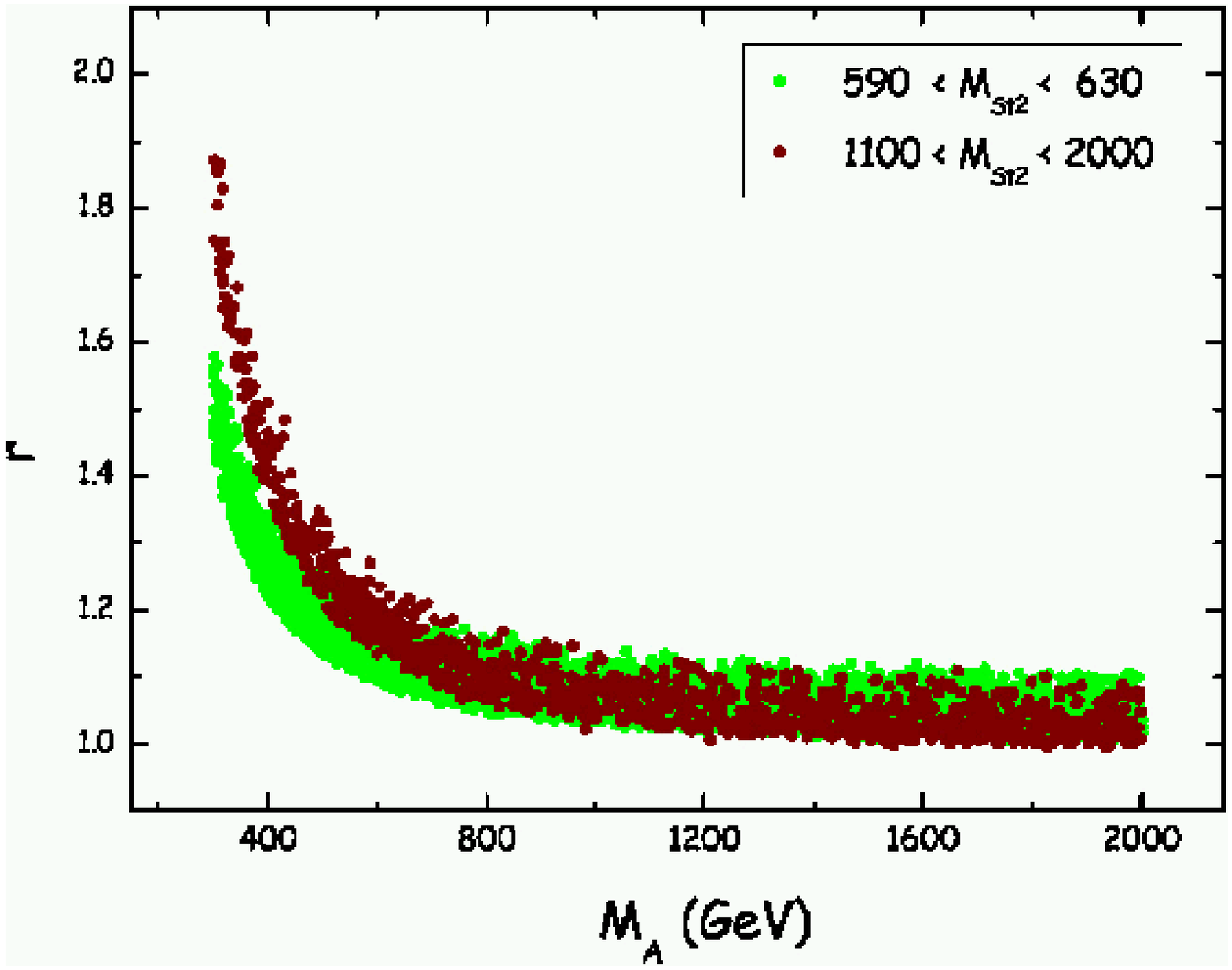, width=14cm,height=8.5cm}
\caption{
The ratio of branching ratios $r$, see eq.~(\ref{eq:sec221_r}),
is shown as a function of $M_A$ in the SPS~1a scenario. The
other SUSY parameters have been varied within the 3~$\sigma$ intervals
of their experimental errors (see text). The upper plot shows the 
three-dimensional $M_A$--$m_{\tilde t_2}$--$r$ parameter space, while
the lower plot shows the projection onto the $M_A$--$r$ plane.
}
\label{fig:sec221plotBR}
\end{center}
\end{figure}

\begin{figure}[htb!]
\begin{center}
\epsfig{figure=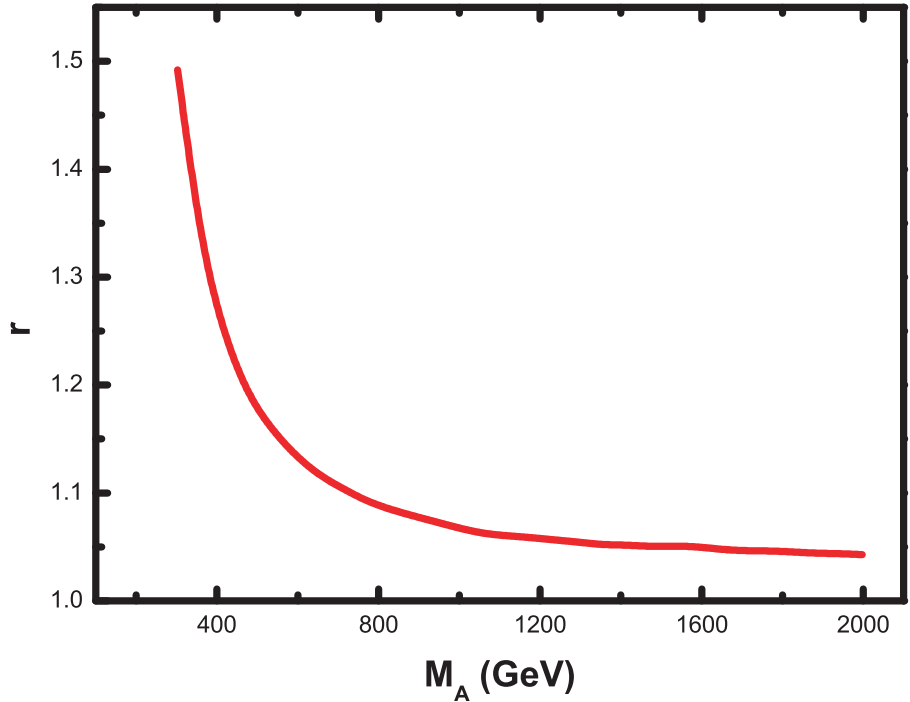, width=12cm}
\caption{The central value of $M_A$ corresponding to the central
  value of a prospective $r$~measurement is shown for the SPS~1a
  scenario. This relation between $r$ and $M_A$ would be obtained if all 
  experimental and theoretical uncertainties were negligible (see text).
}
\label{fig:rMA}
\end{center}
\end{figure}

In our analysis we compare the theoretical prediction~\cite{hff} for
the ratio of branching ratios 
\begin{equation}
r \equiv \frac{\left[{\rm BR}(h \to b \bar b)/
                     {\rm BR}(h \to WW^*)\right]_{\rm MSSM}}
              {\left[{\rm BR}(h \to b \bar b)/
                     {\rm BR}(h \to WW^*)\right]_{\rm SM~~~}\,}
\label{eq:sec221_r}
\end{equation}
with its prospective experimental measurement. 
Even though the experimental error on the ratio of the two BR's is
larger than that of the individual ones, the quantity $r$
has a stronger sensitivity
to $M_A$ than any single branching ratio.

In Fig.~\ref{fig:sec221plotBR} the theoretical prediction for~$r$
is shown as a function of $M_A$, where the scatter points result from
the variation of all relevant SUSY parameters within the 3~$\sigma$
ranges of their experimental errors. 
The constraint on the SUSY parameter 
space from the knowledge of $m_h$ is taken into account, where the
precision is limited by the theory uncertainty from unknown
higher-order corrections. The experimental information on $m_h$
gives rise in particular to indirect
constraints on the heavier stop mass and the stop mixing angle. Without
assuming any further experimental information, two distinct intervals
for the heavier stop mass (corresponding also to different intervals for
$\theta_{\tilde t}$) are allowed. This can be seen from the upper plot
of Fig.~\ref{fig:sec221plotBR}. The interval with lower values 
of $m_{\tilde t_2}$ corresponds to the SPS~1a scenario, while the
interval with higher $m_{\tilde t_2}$ values can only be realized in the
unconstrained MSSM. In the lower plot the projection onto
the $M_A$--$r$ plane is shown, giving rise to two bands with different
slopes. Since the lighter stop mass is accessible at the LC in this
scenario, it can be expected that the stop mixing angle will be
determined with sufficient accuracy to distinguish between the two
bands. This has an important impact on the indirect determination of
$M_A$. 

The central value of $r$ obtained from the band which is realized 
in the SPS~1a scenario is shown as a function of $M_A$ in
Fig.~\ref{fig:rMA}. The plot shows a non-decoupling behavior of $r$,
i.e.\ $r$ does not go to~$1$ for $M_A \to \infty$. This is due to the
fact that the SUSY masses are kept fixed in the SPS~1a scenario. 
In order to find complete decoupling, however, both $M_A$ and the mass
scale of the SUSY particles have to become large, see e.g.\
Ref.~\cite{decoupling2}. It should be noted that the sensitivity of $r$
to $M_A$ is not driven by this non-decoupling effect. In fact, for
larger values of the SUSY masses the slope of $r(M_A)$ even increases 
(one example being the second band depicted in
Fig.~\ref{fig:sec221plotBR}). Thus, even stronger indirect bounds on
$M_A$ could be obtained in this case. 

\begin{figure}[htb!]
\begin{center}
\epsfig{figure=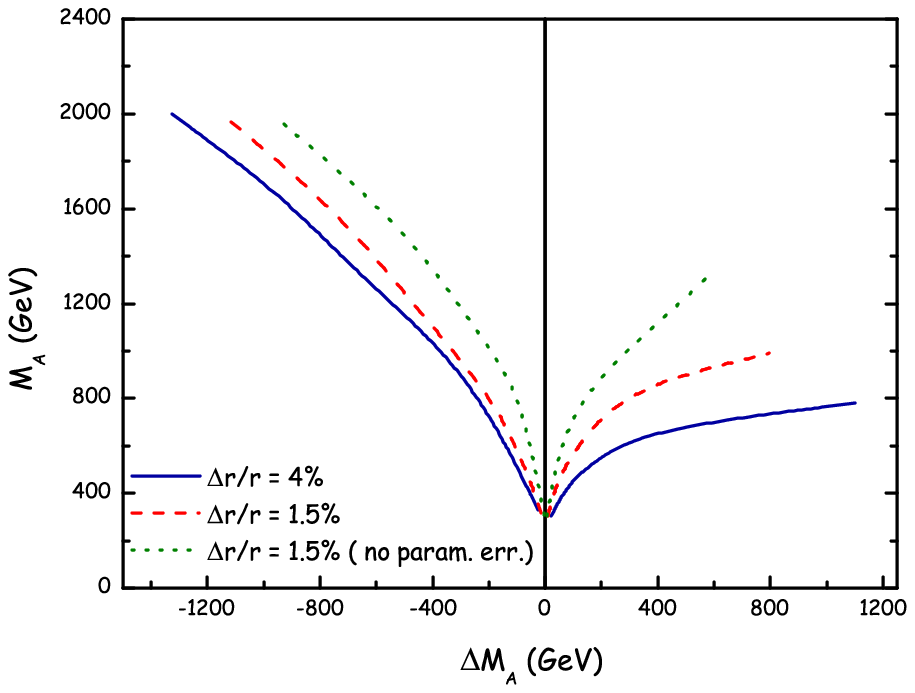, width=12cm}
\caption{The 1~$\sigma$ bound on $M_A$, $\Delta M_A$,
versus $M_A$ obtained from a
comparison of the precision measurement of $r$ (see text) at the LC 
with the MSSM prediction. The results for $\Delta M_A$ are shown
for a 4\% accuracy of $r$ (full line) and a 1.5\% accuracy of $r$
(dashed line). The parametric uncertainties in the
prediction of $r$ resulting from LHC/LC measurement errors on
$\tan\beta, m_{\tilde{b}_{1,2}}, m_{\tilde{t}_1}, m_{\tilde{g}},
m_h $, and $m_t$ are taken into account. Also shown is the accuracy on
$M_A$ which would be obtained if these uncertainties were neglected 
(dotted line).
}
\label{fig:sec221plotBlackHole}
\end{center}
\end{figure}

The comparison of the theoretical prediction for $r$
with the experimental
result at the LC allows to set indirect bounds on the heavy Higgs-boson
mass $M_A$. 
The relation between $r$ and $M_A$
shown in Fig.~\ref{fig:rMA} corresponds to an idealised situation where
the experimental errors of all input parameters in the prediction for
$r$ (besides $M_A$) and the uncertainties from unknown higher-order
corrections were negligibly small. 
Assuming a certain precision of $r$,
Fig.~\ref{fig:rMA} therefore allows to read off the best possible
indirect bounds on $M_A$ as a function of $M_A$, resulting from
neglecting all other sources of uncertainties. This idealised case is
compared with a more realistic situation based on the SPS~1a scenario in 
Fig.~\ref{fig:sec221plotBlackHole}.

For the experimental accuracy of $r$ we consider two
different values: a 4\% accuracy resulting from a first phase of LC
running with $\sqrt{s} \lsim 500$~GeV~\cite{sec2_lctdrs,talkbrient},
and a 1.5\% accuracy which can be achieved from LC running at 
$\sqrt{s} \approx 1$~TeV~\cite{barklow}.
In Fig.~\ref{fig:sec221plotBlackHole} the resulting 1~$\sigma$ 
bounds on $M_A$ are shown (the corresponding value of $r$ can be read off
from Fig.~\ref{fig:rMA}) for the experimental precisions of $r$ of 4\% 
and 1.5\%, respectively, where the estimated experimental errors on the
parameters 
$\tan\beta, m_{\tilde{b}_{1,2}}, m_{\tilde{t}_1}, m_{\tilde{g}},
m_h $, and $m_t$ based on the SPS~1a scenario 
are taken into account. Also shown is the 1~$\sigma$ error for 
$\Delta r/r = 1.5\%$ which would be obtained if all SUSY
parameters (except $M_A$) were precisely known, corresponding to the
idealised situation of Fig.~\ref{fig:rMA}. 

Fig.~\ref{fig:sec221plotBlackHole} shows that a 4\% accuracy on $r$ allows 
to establish an indirect upper bound on $M_A$ 
for $M_A$ values up to $M_A \lsim 800$~GeV (corresponding
to an $r$~measurement of $r \gsim 1.1$).
With an accuracy
of 1.5\%, on the other hand, a precision on $\Delta M_A / M_A$ of 
approximately 20\% (30\%) can be achieved for $M_A = $ 600 (800) GeV.
The indirect sensitivity extends to even higher values of $M_A$.
The comparison with the idealised situation where all SUSY parameters
(except $M_A$) were precisely known (as assumed in Ref.~\cite{MAdet})
illustrates the importance of taking 
into account the parametric errors as well as the
theory errors from unknown higher-order corrections. Detailed
experimental information on the SUSY spectrum and a precision
measurement of $m_t$ are clearly indispensable for exploiting the
experimental precision on $r$.


\subsubsection{Conclusions}

We have investigated indirect constraints on the MSSM Higgs and scalar
top sectors from
measurements at LHC and LC in the SPS~1a and SPS~1b benchmark scenarios.
In a situation where the LHC detects heavy Higgs bosons (SPS~1b) the
combination of the LHC information on the heavy Higgs states with
precise measurements of the mass and branching ratios of the lightest
CP-even Higgs boson at the LC gives rise to a sensitive consistency test
of the MSSM. In this way an indirect determination of the trilinear
coupling $A_t$ becomes possible. The measurement of $m_h$ alone allows
to determine $A_t$ up to a sign ambiguity, provided that a precise
measurement of the top-quark mass from the LC is available.
With the measurements of the branching ratios ${\rm BR}(h \to b \bar b)$ and
${\rm BR}(h \to WW^*)$ at the LC the sign ambiguity can be resolved and
the accuracy on $A_t$ can be further enhanced.

In a scenario where LHC and LC only detect one light Higgs
boson (SPS~1a, where $M_A$ is taken as a free parameter), 
indirect constraints on $M_A$ can be established from
combined LHC and LC data. Taking all experimental and theoretical
uncertainties into account, an indirect determination of $M_A$ with an
accuracy 
of about 20\% (30\%) seems to be feasible for $M_A = $ 600 (800) GeV.
In order to achieve this, a precise measurement of the branching ratios 
${\rm BR}(h \to b \bar b)$ and ${\rm BR}(h \to WW^*)$ at the LC and
information on the parameters of the scalar top and bottom sector from
combined LHC / LC analyses will be crucial.




\subsection{\label{sec:222} Importance of the $\tilde\chi_1^0$ mass measurement
for the $A^0,H^0 \rightarrow \tilde\chi_2^0\tilde\chi_2^0$ mass reconstruction}

{\it F.~Moortgat}

\vspace{1em}


\def\st{\scriptstyle}
\def\sst{\scriptscriptstyle}

\def\beq{\begin{equation}}
\def\eeq#1{\label{#1}\end{equation}}
\def\eeqn{\end{equation}}
\def\beqa{\begin{eqnarray}}
\def\eeqa#1{\label{#1}\end{eqnarray}}
\def\eeqan{\end{eqnarray}}
\def\CR{\nonumber \\ }
\def\leqn#1{(\ref{#1})}

\def\mco{\multicolumn}
\def\overbar#1{\overline{#1}}
\let\littlebar=\bar
\let\bar=\overbar
\def\sbar{\overline}
\def\stilde{\widetilde}
\def\ra{\rightarrow}
\def\Dslash{\not{\hbox{\kern-4pt $D$}}}
\def\dslash{\not{\hbox{\kern-2pt $\del$}}}
\def\half{\frac{1}{2}}
\def\thalf{\frac{3}{2}}
\def\D{{\cal D}}
\def\E{{\cal E}}
\def\F{{\cal F}}
\def\L{{\cal L}}
\def\M{{\cal M}}
\def\O{{\cal O}}
\def\W{{\cal W}}
\def\U{{\cal U}}
\def\tr{{\mbox{\rm tr}}}
\def\del{\partial}
\def\One{{\bf 1}}
\def\VEV#1{\left\langle{ #1} \right\rangle}
\def\bra#1{\left\langle{ #1} \right|}
\def\ket#1{\left| {#1} \right\rangle}
\def\vev#1{\langle #1 \rangle}
\def\eff{{\mbox{\rm eff}}}
\def\hc{{\mbox{\rm h.c.}}}
\def\Pl{{\mbox{\scriptsize Pl}}}
\def\eff{{\mbox{\scriptsize eff}}}
\def\SM{{\mbox{\scriptsize SM}}}
\def\Yuk{{\mbox{\scriptsize {\it Yukawa}}}}
\def\deltaeps{\delta_{\epsilon}}

\def\ee{e^+e^-}
\def\CM{{\mbox{\scriptsize CM}}}
\def\BR{\mbox{\rm BR}}
\def\FB{\mbox{\rm FB}}
\def\L{{\cal L}}
\def\M{{\cal M}}
\def\sstw{\sin^2\theta_w}
\def\cstw{\cos^2\theta_w}
\def\mz{m_Z}
\def\gz{\Gamma_Z}
\def\mw{m_W}
\def\mh{m_H}
\def\gh{\Gamma_H}
\def\mt{m_t}
\def\gt{\Gamma_t}
\def\gmu{G_\mu}
\def\GF{G_F}
\def\alphas{\alpha_s}
\def\msb{{\bar{\ssstyle M \kern -1pt S}}}
\def\lmsb{\Lambda_{\msb}}
\def\R{{\rm R}}
\def\ELER{e^-_Le^+_R}
\def\EREL{e^-_Re^+_L}
\def\ELEL{e^-_Le^+_L}
\def\ERER{e^-_Re^+_R}
\def\ch#1{\widetilde\chi^+_{#1}}
\def\chm#1{\widetilde\chi^-_{#1}}
\def\ne#1{\widetilde\chi^0_{#1}}

\def\etal{{\it et al.}}
\def\ie{{\it i.e.}}
\def\eg{{\it e.g.}}


\noindent{\small
Supersymmetric decay modes of the heavy MSSM Higgs bosons can be used to discover these
particles in the difficult low and intermediate $\tan\beta$ region of the MSSM parameter space.
In particular, the $A^0, H^0 \rightarrow \tilde{\chi}^0_2 \tilde{\chi}^0_2$ decay can lead to a 
cleanly observable $4\ell + E_T^{miss}$ final state, provided neutralinos and sleptons are 
sufficiently light.
In this contribution we investigate the importance of a precise measurement of the 
$\tilde{\chi}^0_1$ mass at a future LC for the Higgs mass reconstruction in this channel 
at the LHC.
We argue that the best way to determine the mass of the $A^0,H^0$ bosons is to fit the 4-lepton 
invariant mass $m_{4l}$, which is an estimator for the Higgs mass, to the Monte Carlo distribution.
However, the $m_{4l}$ estimator depends also strongly on the mass of 
the lightest neutralino $\tilde{\chi}^0_1$. 
Therefore, an accurate measurement of the $\tilde{\chi}^0_1$ mass with a precision better than 
about 1\% would be needed  
for an optimal measurement of $m_A$ through the neutralino channel at the LHC.
}


\subsubsection{Introduction}

In the MSSM, the couplings of the heavier neutral $A^0$ and $H^0$ bosons 
to down-type fermions are enhanced at high $\tan\beta$:
\begin{equation}
g_{A^0 b \bar{b}} \sim m_b \tan\beta   \> \qquad 
g_{H^0 b \bar{b}} \sim m_b \frac{\cos\alpha}{\cos\beta} 
\label{eq:fermion}
\end{equation}
at the expense of the couplings to up-type fermions.
This results in an important production mechanism for $H^0$ and $A^0$ Higgs bosons
in association with $b\bar{b}$ pairs at the LHC: $gg \ra H^0,A^0 b\bar{b}$.  
For the same reasons, the dominant decay modes of the heavy neutral Higgs bosons
are $H^0,A^0 \ra b\bar{b}$
and $H^0,A^0 \ra \tau \bar{\tau}$. The $b\bar{b}$ channel has large difficulties to
overcome the experimental trigger thresholds at the LHC and therefore 
does not show much potential; the $H^0,A^0 \ra \tau \bar{\tau}$ mode on the 
other hand, where both the hadronic and leptonic $\tau$ decaus can be exploited, 
allows to discover the heavy neutral Higgs bosons over a large part
of the parameter space, notably for high and intermediate 
values of $\tan\beta$.
Also the $H^0,A^0 \ra \mu \bar{\mu}$ channel shows some interesting potential, in spite of the very 
small braching ratio of Higgs into muons ${\cal{O}}(10^{-4})$, since muons are 
objects that can be experimentally measured with high precision and efficiency.\\
\\
In spite of the good coverage of the large $\tan\beta$ region, the low and 
intermediate $\tan\beta$ domain remains largely inaccessible for
the previously mentioned channels, and new channels need to be found.
Apart from the Higgs couplings to fermions (\ref{eq:fermion}), 
also some of the Higgs couplings to gauge bosons might be exploited:
\beqa
g_{H^0VV} \sim \cos(\beta-\alpha) & & g_{H^0A^0Z^0} \sim \sin(\beta-\alpha)\nonumber\\
g_{A^0VV} = 0 && g_{H^0H^0Z^0} = g_{A^0A^0Z^0} = 0 
\label{eq:boson}
\eeqan
where $V=W^{\pm},Z^0$. Since $\cos(\beta-\alpha)$ is very close to zero for $m_A > 200$
GeV, the only substantial coupling is $g_{H^0A^0Z^0}$.
In the Standard Model this coupling is not viable since the masses of $H^0$ and
$A^0$ differ with only a few GeV, however supersymmetry dictates that 
the coupling remains the same if two of the three
particles are replaced by their superpartners: $H^0\stilde{A}^0\stilde{Z}^0$.
We thus arrive at a sizeable coupling of the heavy neutral MSSM Higgs bosons to mixtures
of higgsinos and gauginos, i.e. neutralinos and charginos, whose masses can be
much below the Higgs mass.\\
The discovery reach of the Higgs decay modes to neutralinos and charginos at the LHC 
has been studied in \cite{filip1, filip2}. It was found that the most promising channel
is the decay of the heavy neutral Higgs bosons into two next-to-lightest neutralinos, 
with each of the neutralinos in turn decaying as $\ne2 \rightarrow l^+ l^- \ne1$, i.e.
into two (isolated) leptons + $E_T^{miss}$, thus leading to
\begin{equation}
H^0, A^0 \rightarrow  \ne2 \, \ne2 \rightarrow 4 \, l^{\pm}  \; + \; E_T^{miss}     \; \; \; \;
\, \, \, \, \, \, \, \, \, \, \, \, \,           (l\, =\, e,\, \mu)
\label{eq:finsta}
\end{equation}
This results in a clean four lepton final state signature accompanied by a large
amount of missing transverse energy.\\
In the following we will briefly discuss the selection criteria that allow to
suppress the Standard Model and supersymmetric backgrounds and can lead to a clearly visible signal.
Next, various methods to reconstruct the Higgs mass are proposed, all of which require a precise
knowledge of the $\ne1$ mass. This sensitivity to the mass of the $\ne1$ is further quantified for the 
SPS1a benchmark scenario. Finally, it is concluded that a precise determination of the $\ne1$ mass, 
coming from a future Linear Collider is needed for an optimal reconstruction of 
the mass of the heavy Higgs bosons with the neutralino channel.
  

\subsubsection{The $H^0,A^0 \ra \ne2 \ne2$ channel at the LHC}

\begin{figure}
\begin{center}
\epsfig{file=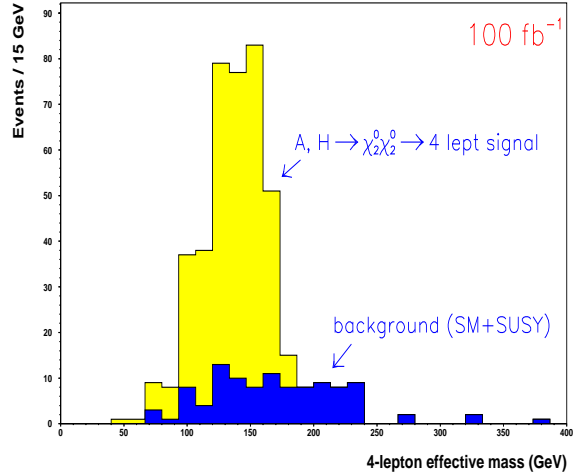,width=80mm,height=70mm}
\end{center}
\caption{Four lepton invariant mass for signal versus background (SM + SUSY) for
the SPS1a scenario where $m_A$ = 393 GeV and $\tan\beta$ = 10 (for an integrated luminosity of 100 $fb^{-1}$).
}
  \label{fig:4lept}
\end{figure}

Two categories of 
background to the Higgs signal (\ref{eq:finsta}) have to be considered: 
Standard Model processes and SUSY backgrounds. \\
The main SM backgrounds are $Z^0Z^0$ and $t\bar{t}$ production. 
They are dangerous because of their large cross sections at the LHC.
The $Z^0Z^0$ events can result in four isolated leptons but do not have no additional 
missing transverse energy. The $t\bar{t}$ background can also lead to four leptons, however
since two of the must come from a $b$ decay they will in general be soft and
non-isolated.\\
In the SUSY backgounds, one can identify the squark/gluino production, 
characterised by a large number of energetic jets and a significant amount 
of $E_T^{miss}$; the slepton and sneutrino pair production (the second one being
more dangereous because the sneutrinos can decay into two charged leptons each)
; and neutrino and chargino pair production, especially $\ne2 \ne2$ production, 
possibly mimicking the signal but with much smaller cross sections due to the 
strongly suppressed coupling of gauginos to a $Z^0$/$\gamma$ intermediate
state.\\   
One can suppress the SM and MSSM background processes, while preserving most of 
the Higgs signal, by applying the following selection criteria:
\begin{itemize}
\item two pairs of isolated leptons (electrons or muons) with 
opposite sign and same flavour are required, with transverse momentum larger than 20, 15, 
15 and 10 GeV resepctively and within $|\eta|$ $<$ 2.4. 
The isolation criterion demands that there are no
charged particles with $p_T$ $>$ 1.5 GeV in a cone of radius
$\Delta R = \sqrt{(\Delta \phi)^2+(\Delta \eta)^2}$ = 0.3 rad around each
lepton track, and that the sum of the transverse energy in the 
deposited in the electromagnetic calorimeter between 
$\Delta R$ = 0.05 and $\Delta R$ = 0.3 rad is smaller than 3 GeV.
\item a $Z^0$-veto is imposed, i.e. all dilepton pairs of opposite sign and same flavour that 
have an invariant mass in the range $m_{Z^0} \pm 10$ GeV are rejected.
\item events must have a missing transverse energy in the interval 
20 GeV $<$ $E_T^{miss}$ $<$ 150 GeV. The lower limit suppresses 
the $Z^0Z^0$ events while the upper limit reduces the squark/gluino background.
\item if needed, one can apply ceilings to the jet multiplicity and the
transverse energy of the jets to further suppress squark and gluino events if these
particles would be light (and therefore copiously produced).
\item further optimization can be done by introducing upper limits 
to the transverse momentum of the leptons and the four lepton invariant mass, 
however they depend on the specific MSSM scenario.
\end{itemize}
After such a selection procedure, a clear signal remains visible on top of the (mainly SUSY)
backgrounds,
as can be seen in Figure~\ref{fig:4lept} 
for the SPS1a benchmark scenario \cite{georg} where $m_A$ = 393 GeV and $\tan\beta$ = 10, $M_1$ = 100 GeV, $M_2$ =
192 GeV, $\mu$ = 352 GeV, $m_{\tilde{l_L}}$ = 202 GeV, $m_{\tilde{l_R}}$ = 142 GeV, $m_{\tilde{q_L}}$ = 540 GeV, 
$m_{\tilde{q_R}}$ = 520 GeV, $m_{\tilde{g}}$ = 595 GeV.
 
\subsubsection{Higgs mass reconstruction with the $H^0,A^0 \ra \ne2 \ne2$ channel}

\begin{figure}
\epsfig{file=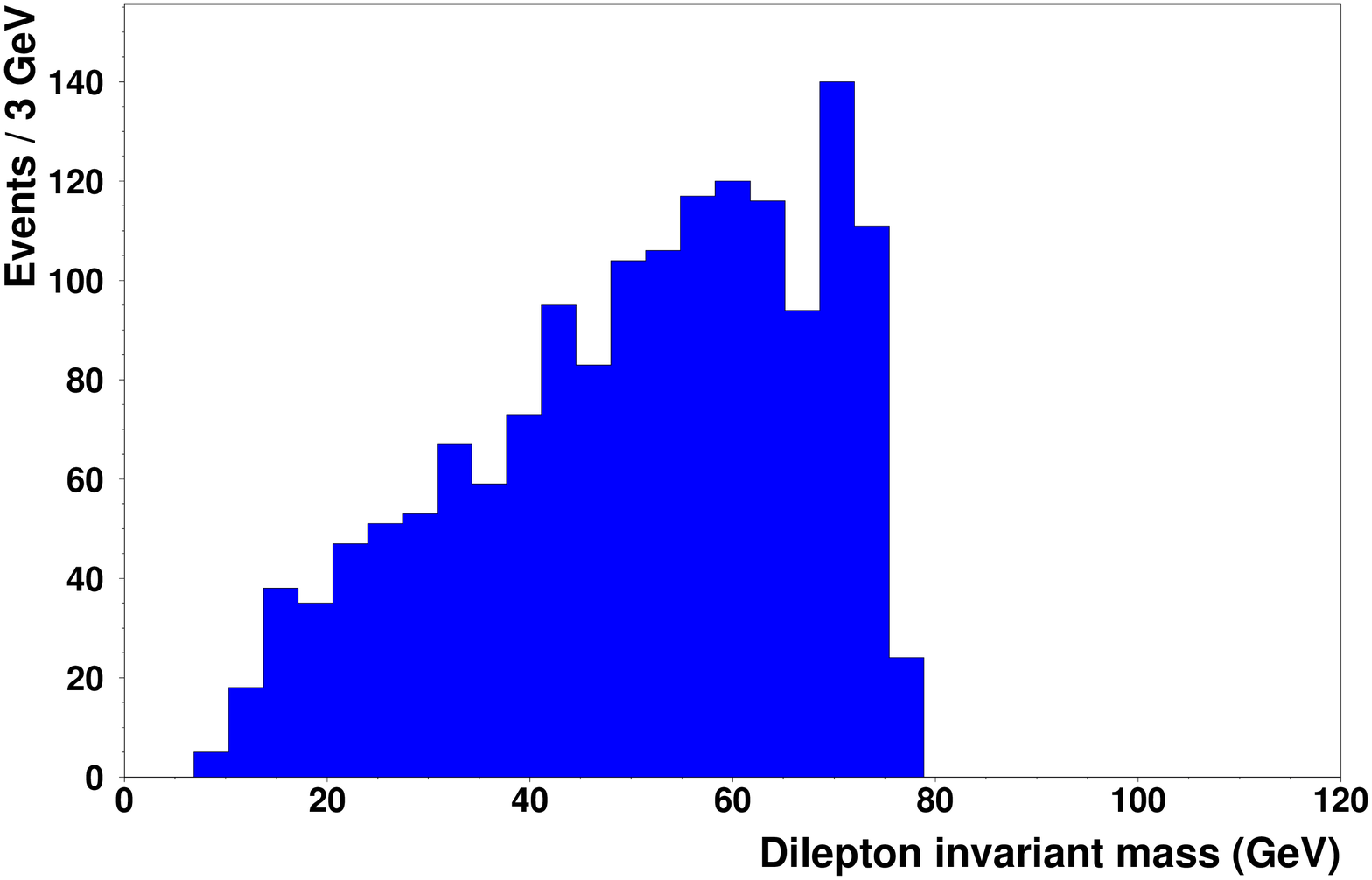,width=68mm,height=70mm}
\epsfig{file=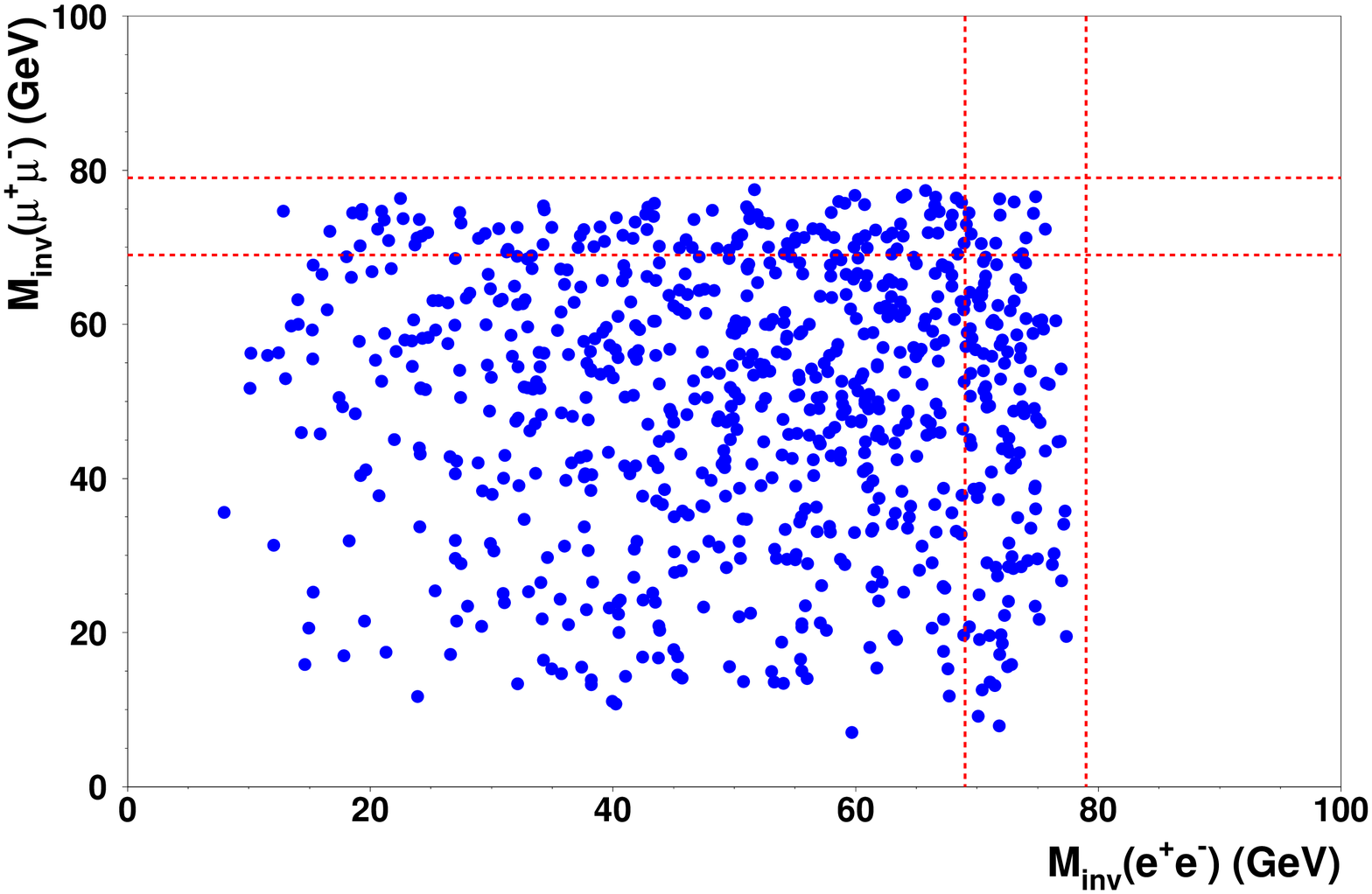,width=68mm,height=70mm}
\caption{(a) Kinematical endpoint at $m_{\ne2} - m_{\ne1}$ in the dilepton invariant
mass distribution. (b) Double kinematical edge in the di-electron versus dimuon invariant
mass distribution.} 
  \label{fig:edge1}
\end{figure}

\begin{figure}
\begin{center}
\resizebox{140mm}{80 mm}{\includegraphics{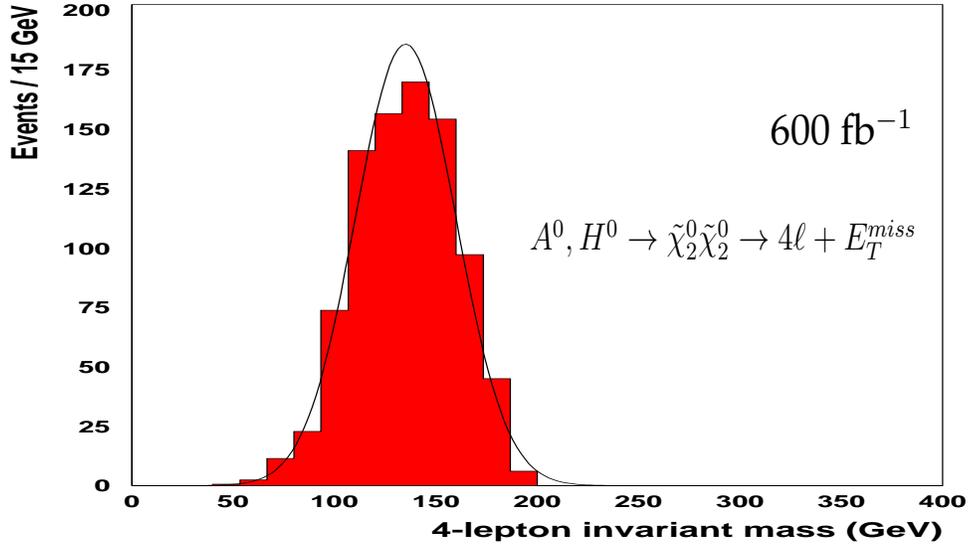}
\put(-150,380){\scalebox{1.8}[3.4]{600 fb$^{-1}$}}
\put(-280,280){\scalebox{1.4}[3.0]{$A^0, H^0 \rightarrow \tilde{\chi}^0_2 
\tilde{\chi}^0_2 \rightarrow 4\ell + E_T^{miss}$ }} }
\end{center}
\caption{Fit of the 4-lepton invariant mass to the Monte Carlo distribution for
scenario SPS1a, for a total integrated luminosity of 600 fb$^{-1}$.} 
  \label{fig:fit}
\end{figure}

\begin{figure}
\resizebox{85mm}{85 mm}{\includegraphics{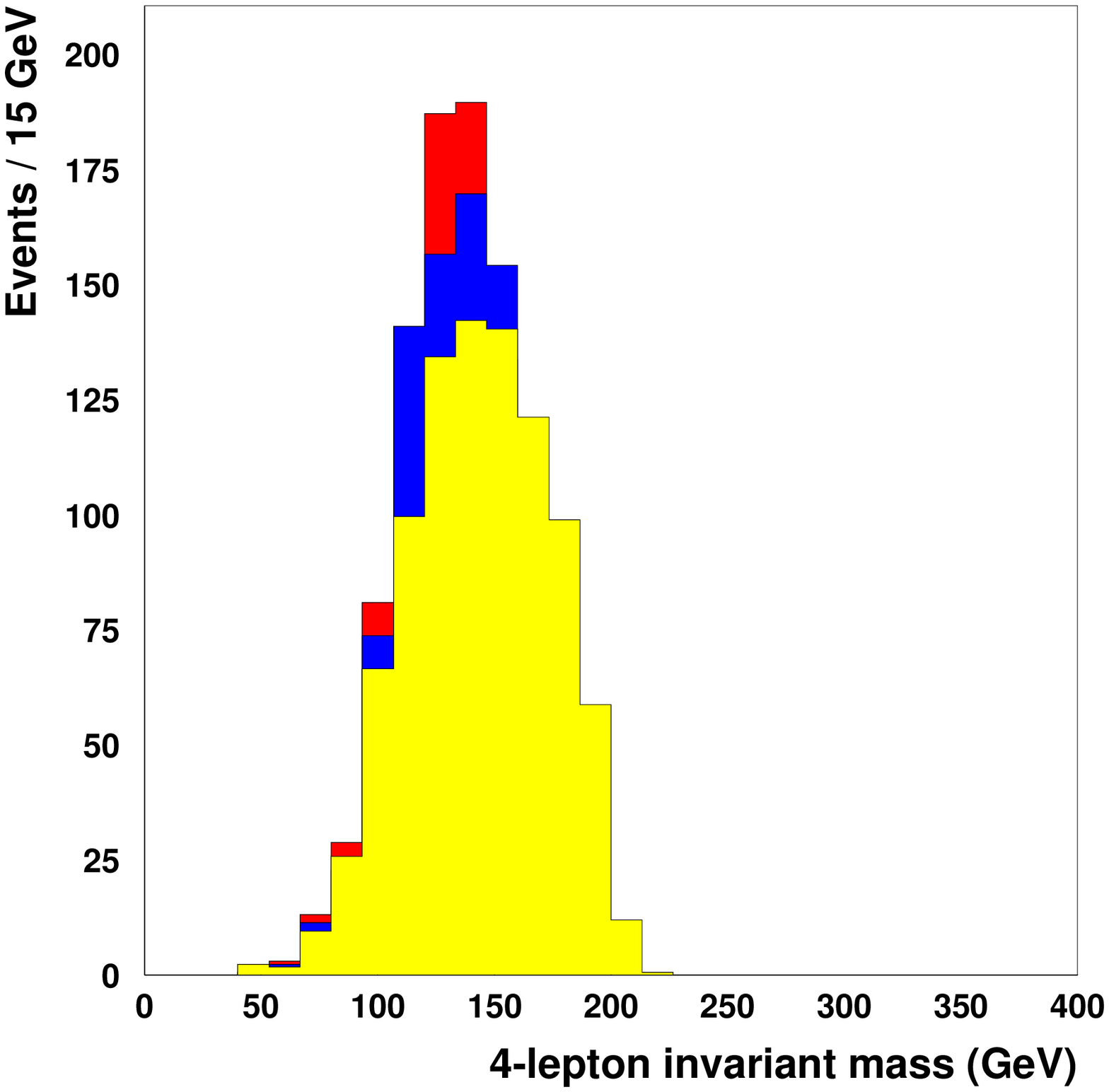}
\put(-180,150){\scalebox{2.5}[3.0]{600 fb$^{-1}$}}
\put(-330,430){\scalebox{2.5}[2.5]{$\leftarrow m_A = 373 \> {\rm GeV}$}}
\put(-300,360){\scalebox{2.5}[2.5]{$\leftarrow m_A = 393 \> {\rm GeV}$}}
\put(-290,290){\scalebox{2.5}[2.5]{$\leftarrow m_A = 413 \> {\rm GeV}$}}
}
\resizebox{85mm}{85 mm}{\includegraphics{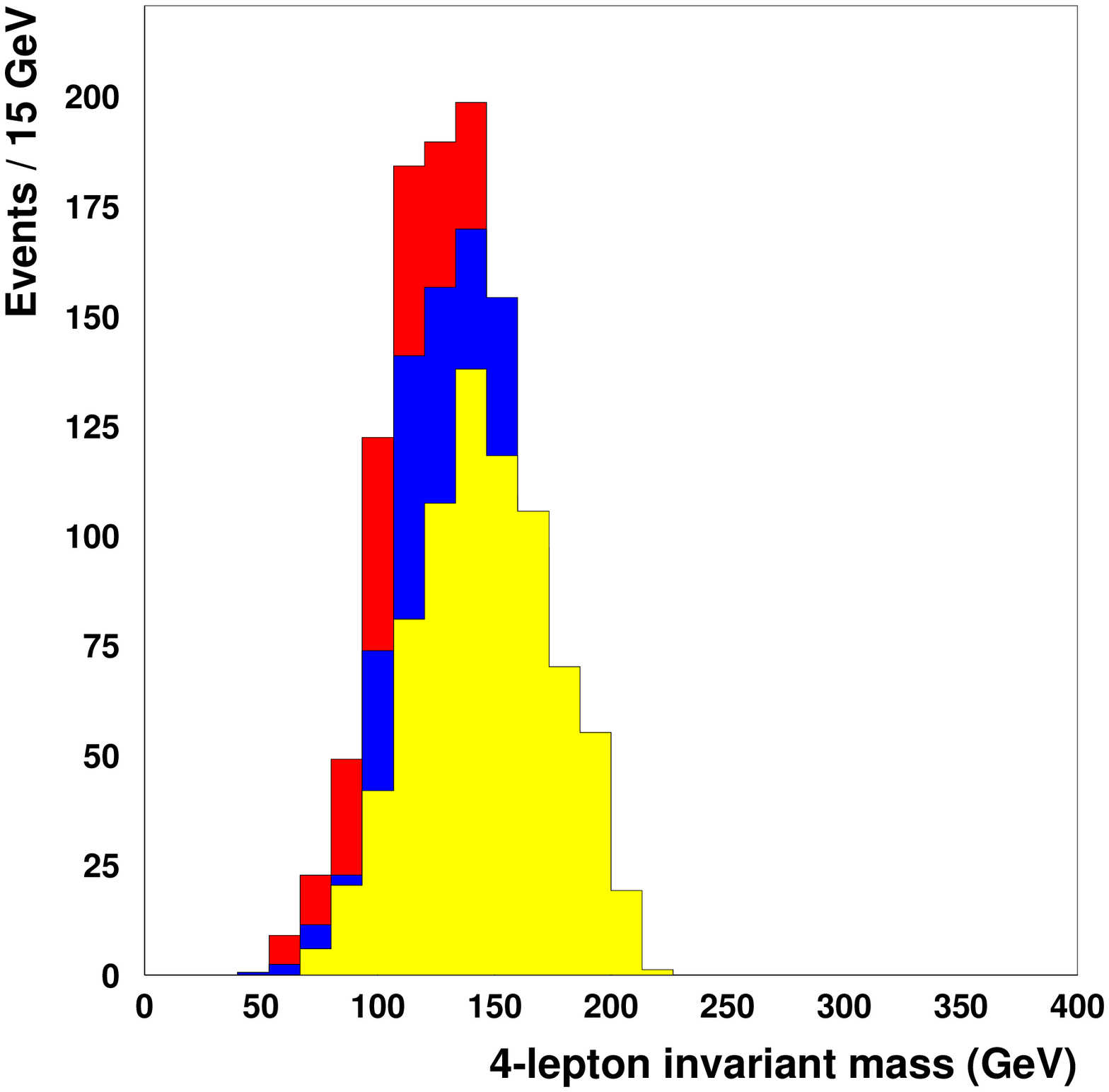}
\put(-180,150){\scalebox{2.5}[3.0]{600 fb$^{-1}$}}
\put(-330,420){\scalebox{2.5}[2.5]{$\leftarrow M_1 = 110 \> {\rm GeV}$}}
\put(-300,360){\scalebox{2.5}[2.5]{$\leftarrow M_1 = 100 \> {\rm GeV}$}}
\put(-290,300){\scalebox{2.5}[2.5]{$\leftarrow M_1 = 90 \> {\rm GeV}$}}
}
\caption{Four lepton invariant mass distribition (a) for $M_A$ = 393 $\pm$ 20 GeV; 
(b) for $M_1$ = 100 $\pm$ 10 GeV and the other parameters fixed as described in the 
text.} 
  \label{fig:varma}
\end{figure}

In order to reconstruct the mass of the $H^0,A^0$ Higgs boson we can exploit 
an interesting kinematical property of the disintegration process.
The decay of the next-to-lightest neutralino to leptons features a kinimatical
endpoint in the dilepton invariant mass spectrum near the mass difference
between the $\ne2$ and the $\ne1$, or if sleptons are intermediate in mass, 
near $\sqrt{(m_{\ne2}^2 - m_{\stilde{l}}^2)(m_{\stilde{l}}^2 - 
m_{\ne1}^2)}/m_{\stilde{l}}$. In Figure~\ref{fig:edge1}a this dilepton invariant
mass spectrum is plotted for the previously considered scenario.
Since there are two $\ne2$'s present in the Higgs decay channel, 
a double kinematical edge is visible if one selects only the events containing
two electrons and two muons and then plots the di-electron invariant mass 
versus the dimuon invariant mass. 
In Figure~\ref{fig:edge1}b these distributions are plotted 
for the previously considered scenario.
Selecting events in which both dilepton pairs are close to the kinematical edge 
 allows in principle for the direct reconstruction of the $A^0$ / $H^0$ mass (provided the 
statistics are sufficient).
By selecting dilepton pairs that have an invariant mass near the 
kinematical endpoint, one can reconstruct the four-momentum of the $\chi^0_2$:
\begin{eqnarray}
\vec{p}_{\ne2} &=&\left( 1+\frac{M_{\ne1}}{M_{ll}} \right)
\vec{p}_{ll} \nonumber\\
M_{\ne2} &=& M_{\ne1} + M_{ll} \nonumber 
\end{eqnarray}
The Higgs mass is then calculated as the invariant mass 
of the two $\ne2$'s, provided the mass of the $\ne1$ is known.  
However, as illustrated in Figure~\ref{fig:edge1}b, even in favourable scenarios
and with 600 fb$^{-1}$ of data only a small number of events survive this selection, thereby limiting the 
statistical accuracy of the Higgs mass obtained with this method. An extension of this method may 
lead to a much better stastistical significance \cite{nojiri}. \\
\\
An alternative way to determine the Higgs mass consists of fitting the
four-lepton invariant mass, which reflects the Higgs mass, with the Monte Carlo distribution
for a given $m_A$.
Figure~\ref{fig:fit} shows the fitted four lepton invariant mass for a Higgs
mass of $m_A$ = 393 GeV in scenario SPS1a. The average four-lepton invariant mass is 135 GeV.
The large discrepancy between this number and the Higgs mass is due to the
escaping $\ne1$ particles. \\
The four lepton invariant mass distribution is only mildly sensitive to the mass of 
the $A^0$ boson (the mass of the $H^0$ boson is typically a few GeV's higher).
A shift of $m_A$ by 10 (20) GeV leads to a shift of $\langle M_{llll} \rangle$ by 4 (8) GeV.
This behaviour is illustrated in Figure~\ref{fig:varma}a where the four lepton invariant
mass distributions are shown for $m_A$ = 373, 393 and 413 GeV. \\
Because of the two $\ne1$'s escape during the 
decay of the Higgs boson., the $M_{llll}$ distribution is also 
sensitive to the mass of the $\ne1$.
A shift of $m_{\ne1}$ by 5 (10) GeV leads to a shift of $\langle M_{llll} \rangle$ by 5 (15) GeV.
Figure ~\ref{fig:varma}b illustrates this dependency by showing the four
lepton invariant mass for three values of $m_{\ne1}$ = 90, 100 and 110 GeV. \\
Since the $M_{llll}$ distribution is sensitive to both $m_A$ and $m_{\ne1}$, 
the mass of the $\ne1$ needs to be known with good
precision in order to suppress this source of error in the determination of the
Higgs mass.
Assuming an ultimate LHC integrated luminosity of 600 fb$^{-1}$, the statistical error on 
$M_{llll}$ in the above scenario is $\sim$1\%. This can lead to a precision on $m_A$ of about 3\%,
provided the mass of the $\ne1$ can be determined to better than $~$1\%. 

\subsubsection{Conclusion}

Supersymmetric decay modes of the heavy MSSM Higgs bosons offer interesting possibilities 
to discover these particles in the low and intermedate $\tan\beta$ region of the MSSM parameter 
space. In particular, the $A^0, H^0 \rightarrow \tilde{\chi}^0_2 \tilde{\chi}^0_2$ decay can lead to a 
cleanly observable $4\ell + E_T^{miss}$ final state, provided neutralinos and sleptons are 
sufficiently light.
In this contribution we have investigated the importance of a precise measurement of the 
$\tilde{\chi}^0_1$ mass at a future LC for the Higgs mass reconstruction in this channel 
at the LHC.
Probably the best way to determine the mass of the $A,H$ bosons is to use the 4-lepton 
invariant mass $m_{4l}$ as an estimator for the Higgs mass, comparing its distribution
for various values of $m_A$ using Monte Carlo simulations.
However, the $m_{4l}$ distribution also depends strongly on the mass of 
the lightest neutralino $\tilde{\chi}^0_1$.
Therefore, an accurate measurement of the $\tilde{\chi}^0_1$ mass with a precision better than 
about 1\% at the LC would be needed  
for an optimal measurement of $m_A$ (i.e. with a statistical error of $~$3\% after 600 fb$^{-1}$
for the SPS1a scenario considered in this contribution) 
through the neutralino channel at the LHC.


\subsection{\label{sec:223} The neutral MSSM Higgs bosons in the
intense-coupling regime}

{\it E.~Boos V.~Bunichev, A.~Djouadi and H.J.~Schreiber}


\renewcommand{\lsim}{\raisebox{-0.13cm}{~\shortstack{$<$ \\[-0.07cm] $\sim$}}~}
\renewcommand{\gsim}{\raisebox{-0.13cm}{~\shortstack{$>$ \\[-0.07cm] $\sim$}}~}
\newcommand{\dx}{\mbox{\rm d}}
\renewcommand{\ra}{\rightarrow}
\renewcommand{\ee}{e^+e^-}
\newcommand{\tb}{\tan \beta}
\newcommand{\s}{\smallskip}
\newcommand{\nn}{\noindent}
\newcommand{\non}{\nonumber}
\renewcommand{\beq}{\begin{eqnarray}}
\renewcommand{\eeq}{\end{eqnarray}}
\newcommand{\miss}{\not\hspace*{-1.8mm}E}
\newcommand{\ct}[1]{c_{\theta_#1}}
\renewcommand{\st}[1]{s_{\theta_#1}}
\newcommand{\pn}[1]{\not{p}_#1}
\newcommand{\charpmi}[0]{\chi^\pm_i}
\newcommand{\charpmj}[0]{\chi^\pm_j}
\newcommand{\ctw}{c_W}
\newcommand{\stw}{s_W}
\newcommand{\ctwn}[1]{c^#1_W}
\newcommand{\stwn}[1]{s^#1_W}
\catcode`@=11
\def\citer{\@ifnextchar
[{\@tempswatrue\@citexr}{\@tempswafalse\@citexr[]}}
\def\@citexr[#1]#2{\if@filesw\immediate\write\@auxout{\string\citation{#2}}\fi
  \def\@citea{}\@cite{\@for\@citeb:=#2\do
    {\@citea\def\@citea{--\penalty\@m}\@ifundefined
       {b@\@citeb}{{\bf ?}\@warning
       {Citation `\@citeb' on page \thepage \space undefined}}%
\hbox{\csname b@\@citeb\endcsname}}}{#1}}
\catcode`@=12

\subsubsection{The intense-coupling regime}

In the MSSM Higgs sector, the intense-coupling regime \cite{intense,paper} is
characterized by a rather large value of $\tb$, and a pseudoscalar Higgs boson
mass that is close to the maximal (minimal) value  of the CP-even $h$ ($H$)
boson mass, $M_A \sim M_h^{\rm max}$, almost leading to a mass degeneracy of
the neutral Higgs particles, $M_h \sim M_A \sim M_H$. In the following, we will
summarize the main features of this scenario. For the numerical illustration,
we will use {\tt HDECAY} \cite{hdecay}, fix the parameter $\tb$ to the value
$\tb=30$ and choose the maximal mixing scenario, where the trilinear
Higgs--stop coupling is given by  $A_t \simeq \sqrt{6} M_S$ with the common
stop masses fixed to $M_S=1$ TeV; the other SUSY parameter will play only a
minor role. \s

Figure 1 (left) displays the masses of the MSSM Higgs bosons  as a function of
$M_A$. As  can be seen, for $M_A$ close to the maximal $h$ boson mass, which in
this case is $M_h^{\rm max}  \simeq 130$ GeV, the mass differences $M_A-M_h$
and $M_H - M_A$ are less than about 5 GeV. The $H^\pm$ boson mass, given by
$M_{H^\pm}^2 \sim M_A^2 +M_W^2$, is larger\,: in the range $M_A \lsim 140$ GeV,
one has $M_{H^\pm} \lsim 160$ GeV,  implying that charged Higgs bosons can
always be produced in top-quark decays, $t \to H^+ b$.  The couplings of the
CP-even Higgs bosons to fermions and gauge bosons  normalized to the SM Higgs
boson couplings are also shown in  Fig.~1 (right). For small $M_A$ values, the
$H$ boson has almost SM couplings, while the couplings of the $h$ boson to
$W,Z,t$ $(b)$  are suppressed (enhanced); for large $M_A$ values the roles of
$h$ and $H$ are interchanged. For medium values, $M_A \sim M_h^{\rm max}$, the
couplings of both $h$ and $H$ to gauge bosons $V=W,Z$  and top quarks are
suppressed, while the couplings to $b$ quarks  are strongly enhanced. The
normalized couplings of the CP-even Higgs particle are simply $g_{AVV}=0$ and
$g_{Abb} = 1/ g_{Att} =\tan\beta =30$.\s

These couplings determine the branching ratios of the Higgs particle, which are
shown in Fig.~2. Because the enhanced couplings, the three Higgs particle
branching ratios to $b\bar{b}$ and $\tau^+\tau^-$  are the dominant ones,
with values $\sim 90$\% and $\sim 10$\% respectively. The decays $H \to WW^*$
do not exceed the level of 10\%, even for small $M_A$ values [where $H$ is
almost SM-like] and in most of the $M_A$ range the decays $H,h \to WW^*$ are
suppressed to the level where they are not useful.  The decays into $ZZ^*$ are
one order of magnitude smaller and the decays into $\gamma \gamma$ are very
strongly suppressed for the three Higgsses and cannot be used anymore. Finally,
note that the  branching ratios for the decays into muons, $\Phi \to \mu^+
\mu^-$, are constant in the entire $M_A$ range exhibited, at the level of $3
\times 10^{-4}$. \s

Summing up the partial widths for all decays, the total decay widths of the
three Higgs particles are shown in the left-hand side of Fig.~3. As can be
seen, for $M_A \sim 130$ GeV, they are at the level of 1--2 GeV, i.e. two
orders of magnitude larger than the width of the SM Higgs boson for this value
of $\tb$ [the total width increases as $\tan^2\beta$]. The right-hand side of
the figure shows the mass bands $M_\Phi \pm \Gamma_\Phi$ and, as can be seen, 
for the above value of $M_A$, the three Higgs boson masses are overlapping.

\begin{figure}[htbp]
\begin{center}
\centerline{\psfig{file=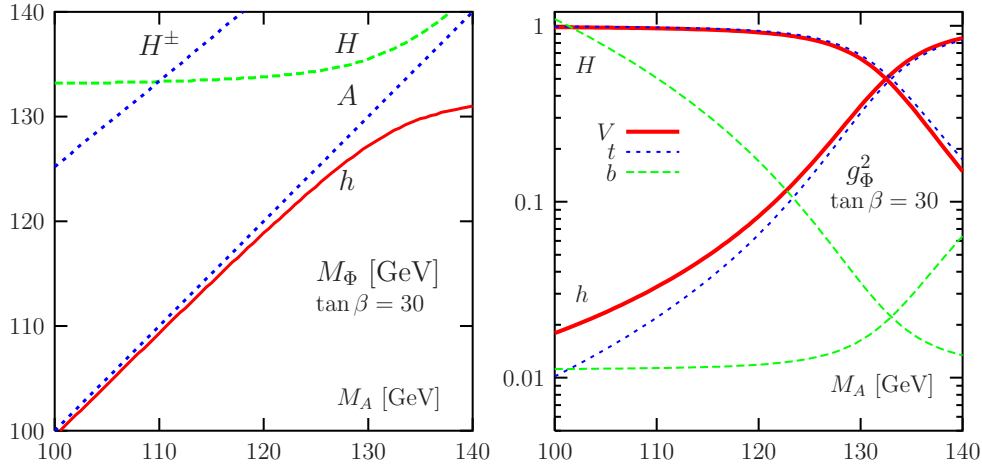,width=16cm}}
\vspace*{-15.7cm}
\caption{\it The masses of the MSSM Higgs bosons (left) and the normalized 
couplings of the CP-even Higgs bosons to vector bosons and third-generation 
quarks (right) as a function of $M_A$ and $tan \beta=30$. For the $b$-quark
couplings, the values $10 \times g_{\Phi bb}^{-2}$ are plotted.}
\end{center}
\end{figure}

\begin{figure}[htbp]
\hspace*{-1.1cm}
\begin{center}
\vspace*{-2.cm}
\centerline{\psfig{file=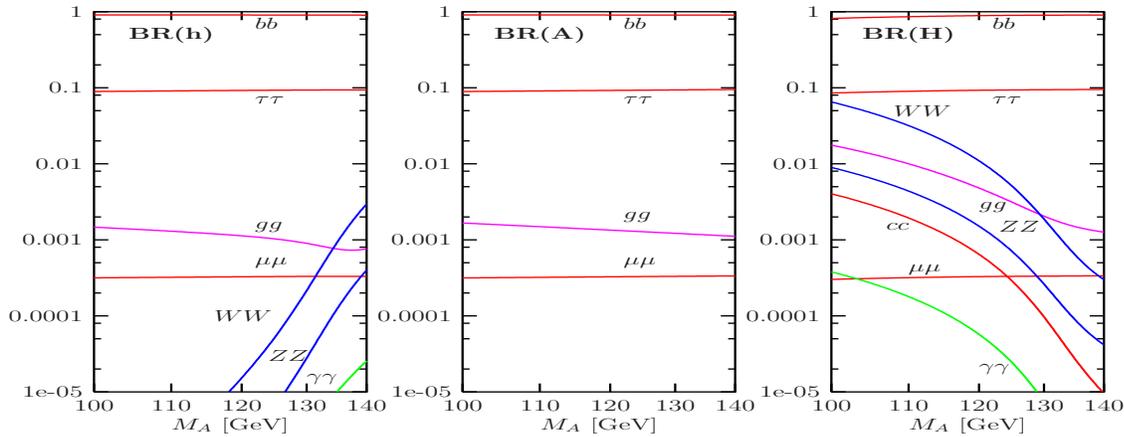,width=18cm,height=15cm}}
\vspace*{-8.1cm}
\caption{\it The branching ratios of the neutral MSSM Higgs bosons $h,A,H$ for 
the various decay modes as a function of $M_A$ and for $tan \beta=30$. }
\end{center}
\end{figure}

\begin{figure}[!ht]
\begin{center}
\centerline{\psfig{file=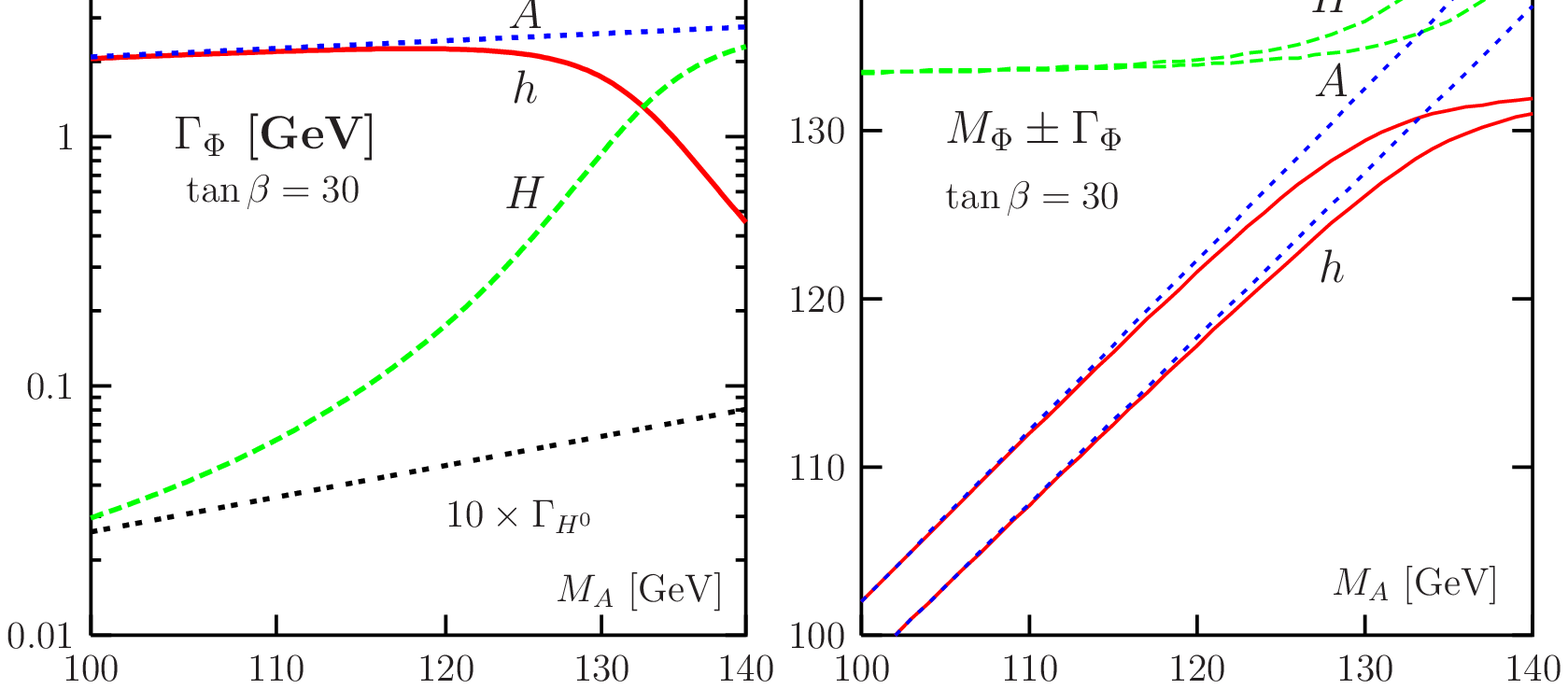,width=16cm}}
\vspace*{-15.7cm}
\caption{\it  Total decay widths $\Gamma_\Phi$ (left) and the mass bands $M_\Phi
\pm \Gamma_\Phi$ (right) for the neutral MSSM Higgs bosons as a function of 
$M_A$ and for $tan \beta=30$.}
\end{center}
\end{figure}

\subsubsection{Discrimination of the three Higgs bosons at the LHC}

The most difficult problem  we must face in the intense-coupling regime, is to
resolve between the three peaks of the neutral  Higgs bosons when their masses
are close to one another. The only decays with large branching ratios on which
we can rely are the $b\bar{b}$ and  $\tau^+ \tau^-$ modes. At the LHC, the
former has too large QCD background to be useful, while for the latter 
channel [which has been shown to be viable for  discovery] the expected
experimental resolution on the invariant mass of the  $\tau^+ \tau^-$ system is
of the order of 10 to 20 GeV, and thus clearly too large. One would then simply
observe a relatively wide resonance corresponding to $A$ and $h$ and/or $H$
production.  Since the branching ratios of the decays into $\gamma \gamma$ and
$ZZ^* \to  4\ell$ are too small, a way out is  to use the Higgs decays into
muon pairs: although the branchings ratio is rather small, BR($\Phi \to
\mu^+\mu^-) \sim 3.3 \times 10^{-4}$,  the resolution is expected to be as good
as 1 GeV, i.e. comparable to the total width,  for $M_\Phi \sim 130$ GeV. \s

Because of the strong enhancement of the Higgs couplings to bottom quarks, the
three Higgs bosons  will be produced at the LHC mainly\footnote{The  
Higgs-strahlung and vector-boson fusion  processes, as well as associated
production with top quarks, will have smaller cross sections since the Higgs
couplings to the involved particles are suppressed.} in the gluon--gluon
process, $gg \to \Phi=h,H,A \to \mu^+ \mu^-$,  which is dominantly mediated by
$b$-quark loops, and the associated production  with $b\bar{b}$ pairs,
$gg/q\bar{q} \to b\bar{b}+\Phi \to b\bar{b}+\mu^+ \mu^-$. We have generated
both the signals and backgrounds  with the program {\tt CompHEP}
\cite{comphep}.  For the backgrounds to $\mu^+\mu^-$ production, we have
included only the Drell--Yan process $pp \to  \gamma^*, Z^* \to \mu^+\mu^-$,
which is expected  to be the largest source. But for the $pp  \to \mu^+ \mu^-
b\bar{b}$ final state, however, we have included the full 4-fermion background,
which is mainly due to the process $pp \to b\bar{b} Z$ with $Z \to \mu^+
\mu^-$.  \s

The differential cross sections are shown for the scenario $M_A=125$ GeV and
$\tb=30$, which leads to $M_h=123.3$ GeV and $M_H=134.3$ GeV, as a function of 
the invariant dimuon mass in Fig.~4 (left), for  $pp (\to h,H,A) \to \mu^+
\mu^-$. As can be seen, the signal rate is fairly large but when put on top of
the huge Drell--Yan background, the signal becomes completely invisible. We
thus conclude, that already at the level of a ``theoretical simulation", the 
Higgs signal will probably be hopeless to  extract in this process for $M_H
\lsim 140$ GeV. In the right-hand side of Fig.~4, we display, again for the 
same scenario, the signal from $pp \to \mu^+\mu^- b\bar{b}$ and the complete
4-fermion SM  background  as a function of the dimuon system. The number of
signal events is an order of magnitude smaller than in the previous case, but
one can still see the two peaks, corresponding to $h/A$ and $H$ production, on
top of the background. \s

In order to perform a more realistic analysis, we have generated unweighted
events for the full 4-fermion background $pp \to \mu^+ \mu^- +b\bar{b}$ and for
the signal. With the help of the new {\tt CompHEP-PYTHIA} interface
\cite{interface},  the unweighted events have been processed through {\tt
PYTHIA 6.2} \cite{PYTHIA} for fragmentation and hadronization. To simulate
detector effects, such as acceptance, muon momentum smearing, and $b$--jet
tagging,  we take the example of the CMS detector. The details have been given
in Ref.~\cite{paper} and the main points are that: 1) the mass resolution  on
the dimuons  is about 1\%, and 2) the efficiency for $b$--jet tagging is  of
the order of 40\%. The results of the simulation for a luminosity of 100
fb$^{-1}$ are shown in Fig.~5, where the number of $\mu^+\mu^- b\bar{b}$ events
in bins of 0.25 GeV is shown as a function of the mass of the dimuon system.
The left-hand side shows the signals with and without the resolution smearing
as obtained  in the Monte-Carlo analysis, while the  figures in the right-hand
side show also the backgrounds, including the  detector effects. \s

For the point under consideration, the signal cross section for the
heavier  CP-even $H$ boson is significantly smaller than the signals from the
lighter CP-even $h$ and pseudoscalar $A$ bosons; the latter particles are too 
too close in mass to be resolved, and only one single broad peak for $h/A$ is 
clearly visible. To resolve also the peak for the $H$ boson, the   integrated
luminosity should be increased  by a factor of 3 to 4. We have also performed
the analysis for $M_A=130$ and 135 GeV. In the former case, it would be
possible to see also  the  second peak, corresponding to the $H$ boson signal
with a luminosity  of 100 fb$^{-1}$, but again the $h$ and $A$ peaks cannot be
resolved.  In the latter case, all three $h,A$ and $H$ bosons have comparable
signal rates, and the mass differences are large enough for us to hope to be
able to isolate the three different peaks, although with some difficulty. \s

\begin{figure}[!ht]
\begin{center}
\centerline{\psfig{file=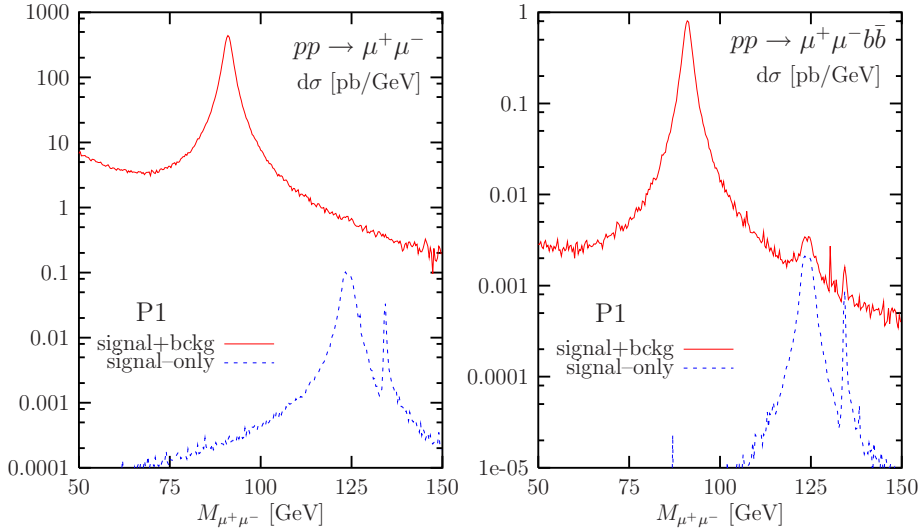,width=15cm}}
\vspace*{-13.5cm}
\caption{\it  The differential cross section in pb/GeV as a function of the
dimuon mass for the point P1, for both the signal and signal plus background 
in the processes $pp (\to \Phi) \to \mu^+ \mu^-$ (left figure) and $pp (\to 
\Phi b\bar{b}) \to \mu^+ \mu^- b\bar{b}$ (right figure).}
\end{center}
\end{figure}

\begin{figure}[!ht]
\begin{center}
\centerline{\epsfig{file=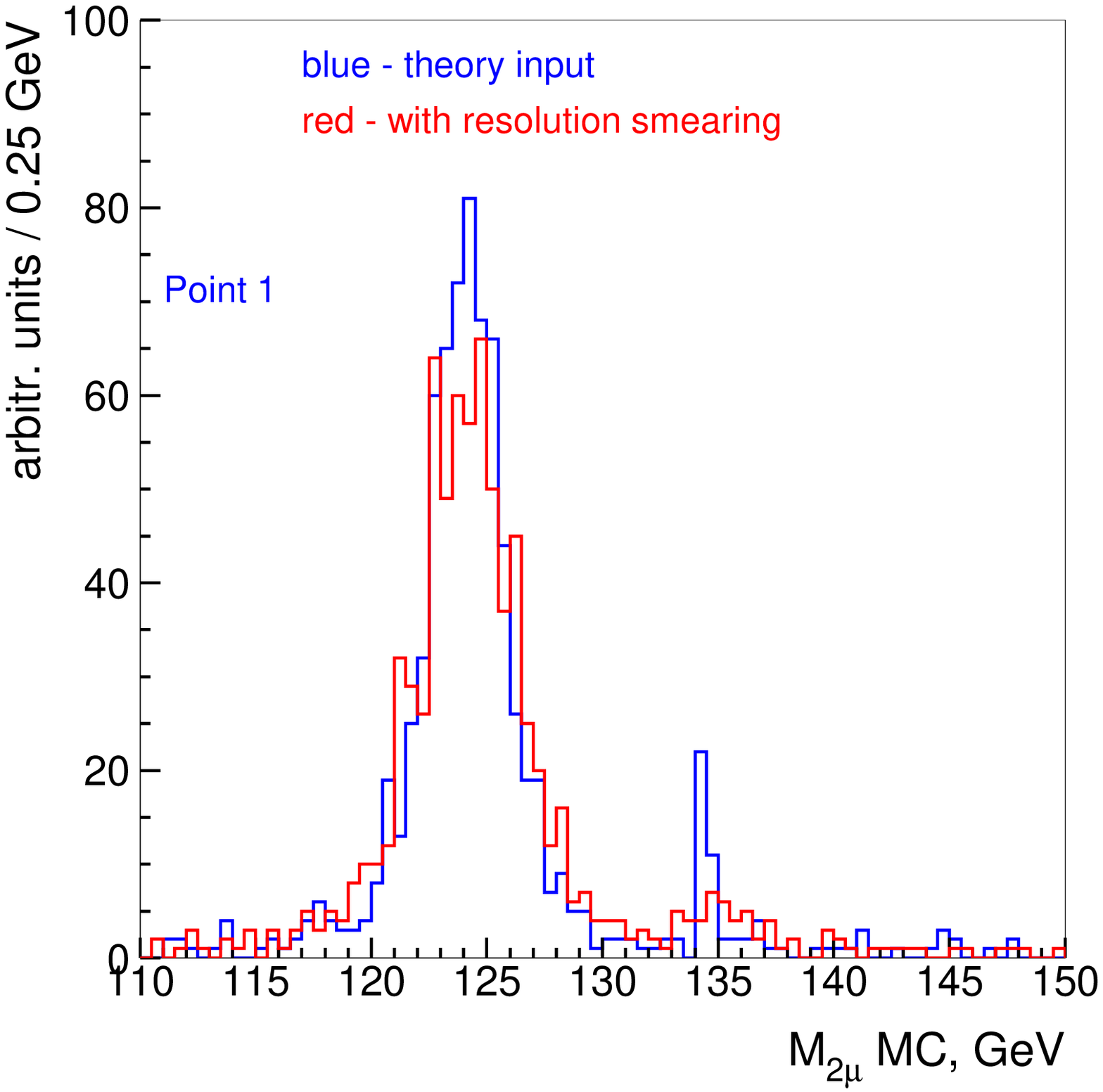,width=7cm}
\epsfig{file=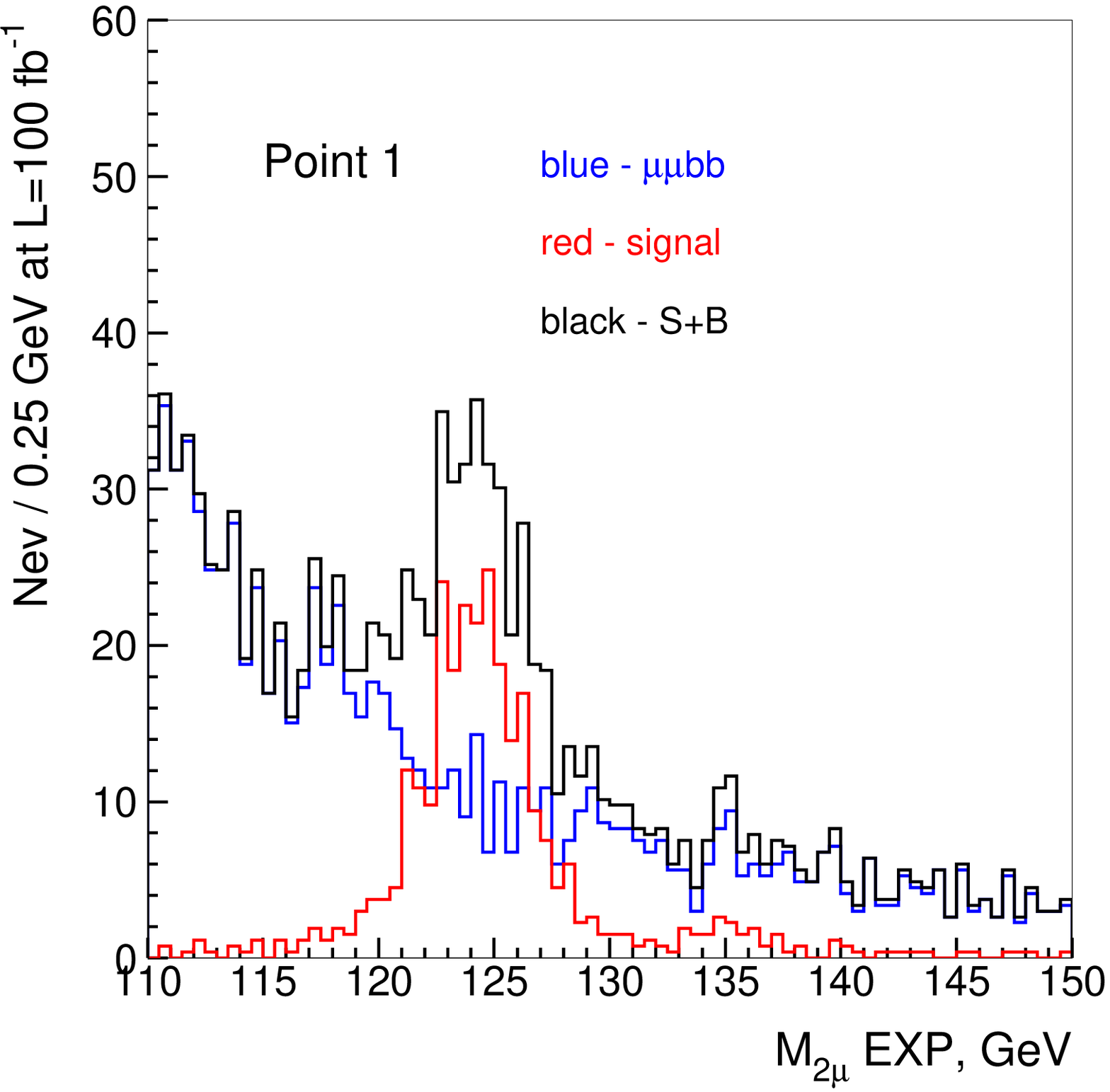,width=7cm}}
\caption{\it $\mu^+ \mu^-$ pair invariant mass distributions
for the signal before and after detector resolution smearing (left)
and for the signal and the background (right) for $M_A=125$ GeV.}
\end{center}
\end{figure}

\subsubsection{Discrimination in $e^+e^-$ collisions} 

In $\ee$ collisions, the CP-even Higgs bosons can be produced in the
bremsstrahlung, $\ee \to Z+h/H$, and  vector-boson fusion, $\ee \to \nu
\bar{\nu}+h/H$, processes. The CP-odd Higgs particle cannot be probed in these
channels since it has no couplings to gauge bosons at the tree level, but it
can be produced in association with $h$ or $H$ bosons in the process $\ee \ra A +
h/H$.  The cross sections for the bremsstrahlung and the pair production as
well as the cross sections for the production of $h$ and $H$ are mutually
complementary, coming either with a coefficient $\sin^2(\beta- \alpha)$ or
$\cos^2(\beta -\alpha)$ and one has:
\beq
\sigma(\ee \to Z+h/H) &=& \sin^2/\cos^2 (\beta-\alpha) \sigma_{\rm SM} 
\non \\ 
\sigma(\ee \to A+h/H) &=& \cos^2/\sin^2 (\beta-\alpha) \bar{\lambda} 
\sigma_{\rm SM} \non 
\eeq
where $\sigma_{\rm SM}$ is the SM Higgs cross section and $\lambda \sim 1$ for
$\sqrt{s} \gg M_A$ accounts for the $P$--wave suppression near threshold. 
Since
$\sigma_{\rm SM}$ is rather large, being of the order of 100 fb at a c.m.
energy of $\sqrt{s}=300$ GeV \footnote{Small $\sqrt{s}$ should be considered for 
this
scenario, $M_\Phi \sim 130$ GeV, since the above two processes are mediated by
$s$-channel gauge boson  exchange and the cross sections scale like $1/s$} the
production and the collective detection of the three Higgs bosons is easy
with the integrated luminosity, $\int {\cal L} \sim 1$ ab$^{-1}$, which is
expected for the TESLA machine for instance
\cite{Aguilar-Saavedra:2001rg}. 
\s  

\vspace*{1.1cm}
\begin{figure}[!ht]
\begin{center}
\centerline{\epsfig{file=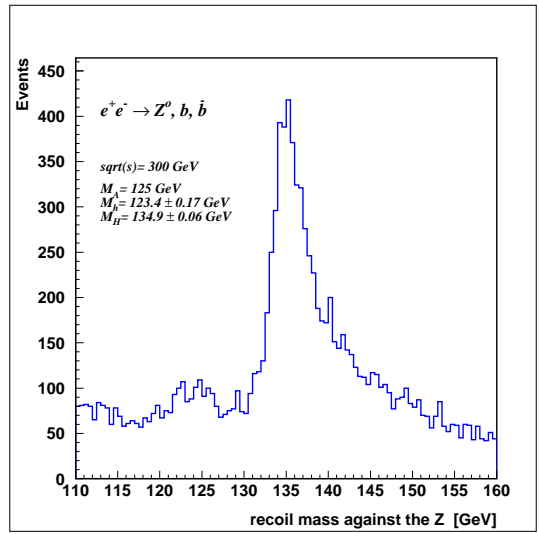,width=7cm}}
\caption{\it The recoil mass distribution for the signal and 
backgrounds including ISR and beamstrahlung for the parameter point P1}
\end{center}
\end{figure}

In $\ee$ collisions, the Higgs--strahlung processes offer a first way to 
discriminate  between the three Higgs particles, since the pseudoscalar $A$
boson is not  involved in this process. For the SM Higgs boson the 
measurement of the recoil mass in both leptonic and hadronic Z boson decay  
channels allows a very  good determination of the Higgs mass of $M_H \sim 
120$ GeV with a precision of $\Delta M_H \sim 40$ MeV
\cite{Aguilar-Saavedra:2001rg}.
In the discussed scenario when two scalar Higgs bosons h and H contribute 
with different rates the influence of the initial state radiation 
and a beamstrahlung are more important and should be carefully taken
into account. Detail simulation including all the main backgrounds
by means of the CompHEP and PYTHIA programs
and the detector effects by means of the SIMDET program \cite{SIMDET}
shows that the most promising way to measure h and H masses is to
select first the $Z b \bar{b}$ event sample and then use the recoil 
mass technique for that sample. Results of simulations for the case
of TESLA and for the MSSM parameter point P1 are shown in the Fig.~6. 
An accuracy of $\Delta M_H \sim 60$ MeV could be reached for the heavy H 
boson and of $\Delta M_H \sim 170$ MeV for the light h boson. \s

The complementary pair production channels $\ee \to A+h/H$ will allow to probe
the pseudoscalar $A$ boson. Since the $h$ and $H$ boson masses will be known, 
as discussed previously, one can concentrate on the $A$ particle and measure
its mass either via  reconstruction  in the $b\bar{b}$ and $\tau^+ \tau^-$
final states or through a scan in the threshold region. The first method has
been  discussed in Ref.~\cite{HA} for heavy Higgs bosons in the $\ee \to HA \to
4b$ final state topology at a c.m. energy $\sqrt{s}=800$ GeV and with  a
luminosity of 200 fb$^{-1}$, with the conclusion that a value $\Delta M_A/M_A 
\sim 0.2$\% can be obtained for $A$ boson masses far below the beam energy.
Since the c.m. energy that should be considered in our scenario is much
smaller, $\sqrt{s}=350$ GeV, this leads to a larger [about a factor of 5 if
there is no coupling suppression] cross section with a luminosity which is 
also larger. If we take this conservative number as a reference, one could
then measure the  pseudoscalar mass with a precision of $\Delta M_A \lsim
200$--300 MeV, which is much smaller than $M_A-M_h$ or $M_H- M_A$. \s

Higher accuracies could be obtained by measuring the $\ee \to A+h/H$ cross
sections near the respective thresholds and which rise as $\sigma \sim
\beta^3$, if one makes the analogy with slepton pair production in $\ee$
collisions, which has similar characteristics as our process. Indeed, it has
been shown in Ref.~\cite{slep} that a slepton mass of order 100 GeV can be 
measured with a precision of less than 0.1\% in a threshold scan at TESLA.  If
this holds also true for $A+h/H$ production [the cross sections are smaller but
the final states are cleaner],  this is more than enough to discriminate all
Higgs bosons in our scenario. \s

The results quoted here are from an extrapolation of SM--like searches where
only one isolated Higgs boson is produced.  A dedicated analysis,  including a
detector simulation, of the intense coupling regime in $\ee$ collisions and the
prospect for measuring the A Higgs boson mass is needed.
\cite{elsewhere}. 

\subsubsection{LHC and LC interplay} 

Once Higgs boson masses will be measured at a $\ee$ LC one can study a 
problem of a complete parameter measurement for the corresponding MSSM 
parameter point. In this sense one can use the measured masses 
and extracted branching ratios BR($\Phi \to  b\bar{b}$) at the linear collider to measure 
$glue-glue-Higgs$ couplings, branching ratios BR($\Phi \to\mu^+\mu^-$) 
and  $b-\bar{b}-Higgs$ couplings via the discussed 
processes $pp (\to \Phi) \to \mu^+ \mu^-$ and $pp (\to
\Phi b\bar{b}) \to \mu^+ \mu^- b\bar{b}$ at the LHC. Obviosly this 
can not be done at LC alone. 

\subsubsection{Conclusions}

We have shown that in the intense-coupling regime, i.e. when the $h,H$ and $A$
MSSM bosons have masses too close to the critical point $M_h^{\rm max}$ and
when the value of $\tb$ is large, the detection of the individual Higgs boson
peaks is very challenging at the LHC. It is only in the associated Higgs
production  mechanism with $b\bar{b}$ pairs, with at least one tagged $b$-jet,
and with Higgs particles decaying  into the clean muon-pair final states, that
there is a  chance of observing the three signals and resolve between  them.
This would be possible only if the Higgs  mass differences are larger than about 5
GeV. \s

In $\ee$ collisions, thanks to the clean environment and the complementarity of
the production channels, one expects the three Higgs bosons to be more easily 
separated. The Higgs--strahlung processes allow from the very beginning to
probe the $h$ and $H$ bosons and to measure their masses by studying the
recoiling $Z$ boson. The associated CP-even and CP-odd Higgs production would
then allow to probe the pseudoscalar $A$ boson, either  in the direct
reconstruction of its final decay products or via a threshold scan. 
The measured characteristics at the LC allow then to measure several others,
like glue-glue-Higgs couplings and BR($\Phi\to\mu^+\mu^-$)
at the LHC. From the other hand, in case if the Higgs masses will be not possible
to resolve completely at the LHC the information on a overall peak position
from the LHC will help to make more proper choice of the energy at LC.  

%

\subsection{\label{sec:224} Estimating the precision of a
$\tan\beta$
determination with $H/A\ra\tau^+\tau^-$ in CMS and
$\tau^+\tau^-$ fusion at a high-energy photon collider}

{\it R.~Kinnunen, S.~Lehti, F.~Moortgat, M.~M\"uhlleitner 
A.~Nikitenko and M.~Spira}

\vspace{1em}
\newcommand{\nc}{\newcommand}
\renewcommand{\lsim}{\mbox{\raisebox{-.6ex}{~$\stackrel{<}{\sim}$~}}}
\renewcommand{\gsim}{\mbox{\raisebox{-.6ex}{~$\stackrel{>}{\sim}$~}}}
\nc{\esim}{\mbox{\raisebox{-.6ex}{~$\stackrel{-}{\sim}$~}}}
\renewcommand{\beq}{\begin{equation}}
\renewcommand{\eeq}{\end{equation}}
\renewcommand{\ra}{\rightarrow}

\subsubsection{The uncertainty of the \boldmath{$\tan\beta$} measurement with 
\boldmath{$H/A\ra\tau^+\tau^-$} in CMS}

If Higgs bosons will be discovered at the LHC, the determination of
their properties will be of high relevance in order to unravel the
underlying model. In the minimal supersymmetric extension (MSSM) of the
Standard Model (SM) as well as a general type-II Two-Higgs-Doublet-Model
one of the most important parameters is $\tan\beta$, the ratio of the
two vacuum expectation values $v_{1,2}$. In this work the accuracy of
the tan$\beta$ measurement is estimated in the $H/A\ra\tau^+\tau^-$
decay channels by investigating the final states $e\mu$, $\ell\ell$ ($
\ell\ell=e\mu, ee,\mu\mu$) \cite{2lepton}, lepton+jet \cite{hljet} and
two-jets \cite{2jets}. The associated Higgs boson production cross
section $gg\rightarrow b\bar{b}H/A$ is approximately proportional to
$\tan^2\beta$ at large $\tan\beta$. Due to this feature the uncertainty
of the $\tan\beta$ measurement is only half of the uncertainty of the
rate measurement. However, due to the presence of potentially large
radiative corrections \cite{mssmrad} the extracted value has to be
considered as an effective parameter $\tan\beta_{eff}$.  The
determination of the fundamental $\tan\beta$ value requires knowledge of
the model behind.

The event rates of $H/A\ra\tau^+\tau^-$ decay channels are studied in Refs.
\cite{2lepton,2jets}. The CMS trigger efficiencies and thresholds
\cite{DA-HLT-TDR} are taken into account. Event selections to suppress
the backgrounds include lepton isolation, $\tau$ jet identification,
$\tau$ tagging with impact parameter, $b$ tagging and jet veto. The
effective $\tau^+\tau^-$ mass is reconstructed assuming that the neutrinos
are emitted collinear with the measured tau decay products. The signal
events were simulated with the following values of MSSM SUSY parameters:
$M_2$ = 200~GeV/$c^2$, $\mu$ = 300~GeV/$c^2$, $M_{\tilde{g}}$ =
800~GeV/$c^2$, $M_{\tilde{q},{\tilde{\ell}}}$ = 1 TeV/$c^2$ and
$A_{t}$ is set to 2450~GeV/$c^2$. The top mass is set to
175~GeV/$c^2$ and Higgs boson decays to SUSY particles are allowed.  We
have not included the uncertainties related to the MSSM parameters, but
kept them fixed at our chosen values.  The uncertainty of the background
estimation as well as the uncertainty of the signal selection efficiency
have not yet been taken into account in this study.  We expect, however,
that the background uncertainty and uncertainty of the signal selection
will be of the order of 5\%. The statistical errors from different
$\tau^+\tau^-$ final states are combined using the standard weighted
least-squares procedure described in \cite{ReviewOfPP}.

In addition to event rates, the accuracy of the $\tan\beta$ measurement
depends on the systematic uncertainty from the luminosity measurement
and on the theoretical uncertainty of the cross section calculation.  A
5\% uncertainty of the luminosity measurement was taken.  The
theoretical accuracy of the cross section depends on the transverse
momentum range of the two spectator quarks and reduces to 10-15\% with
the requirement of $p_{T}^{b,\bar{b}}\gsim$ 20 GeV/$c$
\cite{hep-ph/0309204,dawson}. However, since the associated $b$ jets are
very soft, reconstructing and $b$ tagging them is difficult. In order to
minimize the total measurement error, only one $b$ jet is assumed to be
$b$ tagged and a theoretical uncertainty of about 20\% is adopted
accordingly \cite{hep-ph/0309204,hep-ph/0204093}. The branching ratio
$BR(H/A\rightarrow\tau\tau)$ is approximately constant at large
$\tan\beta$, and the uncertainty of the branching ratio due to the SM
input parameters is about 3\%.

Since the value of the cross section depends on the Higgs boson mass,
the uncertainty of the mass measurement leads to an uncertainty in the
signal rate. The Higgs mass is measured using the different final
states, and the cross section uncertainties due to mass measurement errors
are combined. The mass resolution is almost constant as a function of
$M_{A}$, $\sim$ 24\% for the leptonic final states, $\sim$ 17\% for
the lepton+jet final state and $\sim$ 12\% for the hadronic final state
\cite{2lepton}. The uncertainty of the mass measurement is calculated
from the gaussian fit of the mass peak as $\sigma_{Gauss}/\sqrt{
N_{S}}$, and the error induced in the cross section
($\Delta\sigma(\Delta m)$) is estimated by varying the cross section
for Higgs masses $M_0$ and $M_0\pm\sigma_{Gauss}/\sqrt{N_{S}}$.
At the 5$\sigma$ limit where the signal statistics is lowest, the
uncertainty of the mass measurement generates 5--6\% uncertainty in the
$\tan\beta$ measurement.

Table \ref{table:errors} shows the statistical uncertainty of the
$\tan\beta$ measurement and the cross-section uncertainty due to the
mass measurement for each individual final state and for the combined
final states from $H/A\ra \tau^+\tau^-$ for 30~fb$^{-1}$.  The total
estimated uncertainty including theoretical and luminosity errors are
shown for the combined final states.  The results are shown for the
region of the ($M_{A}$, $\tan\beta$) parameter space where the
statistical significance exceeds 5$\sigma$. Close to the 5$\sigma$ limit
the statistical uncertainty is of the order of 11--12\%, but it
decreases rapidly for increasing $\tan\beta$.  As shown in the table, the
highest statistical accuracy, about 5\% for $M_{A}$ = 200 GeV/$c^2$
and $\tan\beta$ = 20, is obtained with the lepton+jet final state.
Combining other channels with the lepton+jet channel in this mass range
improves the precision only slightly.

\begin{table}[h]
\begin{small}
\fontsize{9pt}{12pt}
\centering
\vskip 0.1 in
\begin{tabular}{|c||c|c|c|c|c|c|c|c|}
\hline
\multirow{2}{2cm}{\centering \Large 30 fb$^{-1}$}
& \multicolumn{2}{|c|}{\begin{minipage}{2.5cm}
$M_{A}$~=~200~GeV/$c^2$ \\
$\tan\beta$ = 20
\end{minipage}}
& \multicolumn{2}{|c|}{\begin{minipage}{2.5cm}
$M_{A}$~=~200~GeV/$c^2$ \\
$\tan\beta$ = 30
\end{minipage}}
& \multicolumn{2}{|c|}{\begin{minipage}{2.5cm}
$M_{A}$~=~500~GeV/$c^2$ \\
$\tan\beta$ = 30
\end{minipage}}
& \multicolumn{2}{|c|}{\begin{minipage}{2.5cm}
$M_{A}$~=~500~GeV/$c^2$ \\
$\tan\beta$ = 40
\end{minipage}}\\
\cline{2-9}
 & $\Delta$stat & $\Delta\sigma(\Delta m)$ & $\Delta$stat & $\Delta\sigma(\Delta m)$
 & $\Delta$stat & $\Delta\sigma(\Delta m)$ & $\Delta$stat & $\Delta\sigma(\Delta m)$ \\
\hline
$H/A\rightarrow\tau^+\tau^-\rightarrow e\mu$    & 8.95\% & 4.82\% & 4.85\% & 3.27\% & - & - & - & -  \\
$H/A\rightarrow\tau^+\tau^-\rightarrow\ell\ell$ & 7.96\% & 3.50\% & 4.08\% & 2.37\% & - & - & - & -  \\
$H/A\rightarrow\tau^+\tau^-\rightarrow\ell j$   & 4.81\% & 2.46\% & 2.84\% & 1.65\% & - & - & 8.40\% & 4.82\% \\
$H/A\rightarrow\tau^+\tau^-\rightarrow jj$      & 13.7\% & 4.73\% & 8.25\% & 3.21\% & 12.4\% & 5.82\% & 8.45\% & 4.44\% \\
\hline
\hline
\multirow{2}{3cm}{\begin{minipage}{2.cm}\bf Combined \\ \boldmath{$e\mu+\ell j+jj$}\end{minipage}}
& 4.05\% & 1.99\% & 2.35\% & 1.34\% & 9.09\% & 4.28\% & 5.96\% & 3.26\% \\
\cline{2-9}
\cline{2-9}
& \multicolumn{2}{|c|}{$\Delta \tan\beta / \tan\beta$  }
& \multicolumn{2}{|c|}{$\Delta \tan\beta / \tan\beta$  }
& \multicolumn{2}{|c|}{$\Delta \tan\beta / \tan\beta$  }
& \multicolumn{2}{|c|}{$\Delta \tan\beta / \tan\beta$  } \\
\cline{2-9}
& \multicolumn{2}{|c|}{ 20.1\% }
& \multicolumn{2}{|c|}{ 17.7\% }
& \multicolumn{2}{|c|}{ 27.4\% }
& \multicolumn{2}{|c|}{ 23.3\% } \\
\hline
\multirow{2}{3cm}{\begin{minipage}{2.cm}\bf Combined \\ \boldmath{$\ell\ell+\ell j+jj$}\end{minipage}}
& 3.94\% & 1.85\% & 2.24\% & 1.25\% & 9.09\% & 4.28\% & 5.96\% & 3.26\% \\
\cline{2-9}
\cline{2-9}
& \multicolumn{2}{|c|}{$\Delta \tan\beta / \tan\beta$  }
& \multicolumn{2}{|c|}{$\Delta \tan\beta / \tan\beta$  }
& \multicolumn{2}{|c|}{$\Delta \tan\beta / \tan\beta$  }
& \multicolumn{2}{|c|}{$\Delta \tan\beta / \tan\beta$  } \\
\cline{2-9}
& \multicolumn{2}{|c|}{ 19.9\% }
& \multicolumn{2}{|c|}{ 17.5\% }
& \multicolumn{2}{|c|}{ 27.4\% }
& \multicolumn{2}{|c|}{ 23.3\% } \\
\hline
\end{tabular}
\end{small}
\caption{Statistical uncertainties of the $\tan\beta$ measurement and the
         uncertainties due to mass measurement for individual
         $H/A\rightarrow\tau^+\tau^-$ and combined final states in
         four ($M_{A}$,$\tan\beta$) parameter
         space point for 30 fb$^{-1}$. The total error includes statistical
         error, mass measurement error, theoretical uncertainty of the cross 
         section (20\%) and the branching ratio (3\%), and
         the luminosity uncertainty (5\%).}
\label{table:errors}
\end{table}

\begin{figure}[hbt]
  \centering
  \vskip 0.1 in
    \includegraphics[width=100mm,height=80mm]{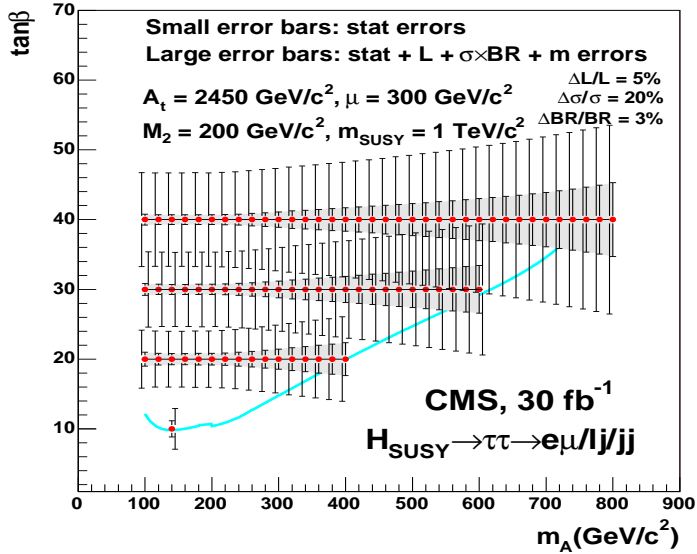}
  \caption{The uncertainty of the $\tan\beta$ measurement shown as error bars.
           The small error bars and gray area show the statistical errors only.
           The large error bars exhibit the uncertainty
           if statistical errors, the mass measurement uncertainties,
           the luminosity uncertainty (5\%)
           and the theoretical uncertainty of the production cross section (20\%)
           and the branching ratio (3\%) are
           taken into account. The solid curve corresponds to the
           5$\sigma$-discovery contour.}
  \protect\label{fig:errorbars30fb}
\end{figure}

Figure \ref{fig:errorbars30fb} shows the error on the $\tan\beta$
measurement with error bars for the combined $e\mu + \ell j + jj$
channel for 30 fb$^{-1}$ at low luminosity.  The statistical
uncertainties are depicted by the smaller error bars and gray area, the
uncertainties including the systematic errors are presented with longer
error bars. The errors are shown in the region with signal significance
larger than 5$\sigma$.  The statistical uncertainty is largest close to
the 5$\sigma$ limit, where combining the different final states improves
the accuracy most.  With one tagged b jet in the event the value of
$\tan\beta$ can be determined in the $H/A\ra\tau^+\tau^-$ decay channels
with an accuracy of better than $\sim$~35\% after collecting
30~fb$^{-1}$.

\subsubsection{The uncertainty of the \boldmath{$\tan\beta$} measurement in
\boldmath{$\tau^+\tau^-$} fusion at a photon collider}

At a high-energy photon collider light $h$ and heavy $H$ and $A$ Higgs
bosons will be produced in $\tau^+\tau^-$ fusion \begin{equation}
\gamma\gamma \to (\tau^+\tau^-) + (\tau^+\tau^-) \to \tau^+\tau^- +
h/H/A \end{equation} within the MSSM for large values of $\tan\beta$
with a rate which enables a measurement of $\tan\beta$ with high
accuracy. The production cross sections grow with the square of
$\tan\beta$ for large values of this parameter.  $\tau^+\tau^-$ fusion
at a photon collider is superior to $b\bar b$ fusion due to the larger
electric $\tau$ charge. The cross sections range between about 1 and 10
fb so that the expected event rates at a photon collider with 600 GeV
c.m. energy allow for a good statistical accuracy of the cross sections.
The size of the cross sections can be estimated by using the equivalent
particle approximation which reduces the hard process to $\tau^+\tau^-$
fusion $\tau^+\tau^-\to \Phi$ ($\Phi=h,H,A$) \cite{taufusion}.

The dominant Higgs boson decays are those into $b\bar b$ pairs, if
decays into supersymmetric particles are forbidden or rare. The
background of $\tau^+\tau^- b\bar b$ production can be sufficiently
suppressed by appropriate cuts on the minimal angle of the $\tau$
leptons and their minimal energy. By requiring the $\tau$ leptons to go
into opposite hemispheres and the invariant $b\bar b$ mass to be in a
narrow window of $\pm$0.05 $M_\Phi$ around the Higgs mass $M_\Phi$ the
background can be strongly suppressed to a negligible level for heavy
Higgs masses, i.e. 2--3 orders of magnitude below the signal processes.
This can be understood from the feature that the background is dominated
by $\tau^+\tau^-\to b\bar b$ in the equivalent particle approximation
and diffractive $\gamma\gamma\to (\tau^+\tau^-)(b\bar b)$ events, the
pairs scattering off each other by Rutherford photon exchange. The
latter can be suppressed by requiring a large invariant mass for the
$(b\bar b)$ pair and the $\tau$ leptons to go into opposite hemispheres.
The first process will mainly be diminished by the invariant $b\bar b$
mass cut.

\begin{table}[hbt]
\begin{center}
\begin{tabular}{|l||cc|}
\hline
$\Delta\tan\beta$ & \multicolumn{2}{|c|}{$M_{A}$ [GeV$/c^2$]}\\
                  &  200  &  400  \\
\hline\hline
$\Delta$stat      & 0.53  & 0.78  \\
$\Delta$tot       & 0.90  & 1.31  \\
\hline
\end{tabular}
\caption{Absolute errors on the $\tan\beta$ measurement based on $H/A$
production for $M_A=$ 200 and 400 GeV$/c^2$ at a photon collider with
c.m.~energy of 600 GeV and an integrated luminosity of 200 fb$^{-1}$,
valid for $\tan\beta> 10$. The first line labelled with $\Delta$stat
shows the statistical error, while the second line presents the total
expected accuracy after taking into account the experimental
efficiencies, i.e. $\epsilon_{b\bar b}=0.7$ and $\epsilon_{\tau^+\tau^-}
= 0.5$ \cite{deschpriv}. All errors include the sum over $H$ and $A$ Higgs
bosons. Since the background can be suppressed to a negligible level,
the absolute errors on $\tan\beta$ are independent of $\tan\beta$ at
large values.
\label{tb:dtanb}}
\end{center}
\end{table}
A first theoretical signal and background analysis leads to the results
presented in Table~\ref{tb:dtanb} for the expected accuracies of the
$\tan\beta$ measurement at a photon collider \cite{taufusion}. They
range between about 1 and 10\%. Thus, $\tau^+\tau^-$ fusion provides a
further significant observable to a global determination of $\tan\beta$
at high values.

\subsection{\label{sec:421} LHC and LC determinations of $\tan\beta$}

{\it J.~Gunion, T.~Han, J.~Jiang, A.~Sopczak}

\vspace{1em}


\def\Journal#1#2#3#4{{#1} {\bf #2}, #3 (#4)}

\def\ben{\begin{enumerate}}
\def\een{\end{enumerate}}
\def\ie{{\it i.e.}}
\def\NCA{\em Nuovo Cimento}
\def\NIM{\em Nucl. Instrum. Methods}
\def\NIMA{{\em Nucl. Instrum. Methods} A}
\def\NPB{{\em Nucl. Phys.} B}
\def\PLB{{\em Phys. Lett.}  B}
\def\PRL{\em Phys. Rev. Lett.}
\def\PRD{{\em Phys. Rev.} D}
\def\ZPC{{\em Z. Phys.} C}
\def\st{\scriptstyle}
\def\sst{\scriptscriptstyle}
\def\mco{\multicolumn}
\def\epp{\epsilon^{\prime}}
\def\vep{\varepsilon}
\def\ra{\rightarrow}
\def\to{\ra}
\def\ppg{\pi^+\pi^-\gamma}
\def\vp{{\bf p}}
\def\ko{K^0}
\def\kb{\bar{K^0}}
\def\al{\alpha}
\def\ab{\bar{\alpha}}
\def\be{\begin{equation}}
\def\ee{\end{equation}}
\def\bea{\begin{eqnarray}}
\def\eea{\end{eqnarray}}
\def\CPbar{\hbox{{\rm CP}\hskip-1.80em{/}}}

\def\tbtb{t\anti b \, \anti t b}
\def\bbbb{b\anti b b\anti b}
\def\wt{\widetilde}
\def\epem{e^+e^-}
\def\bb{b\anti{b}}
\def\qq{ q\anti{q}}
\def\bbA{\bb \ha}
\def\tanb{\tan\beta}
\def\cotb{\cot\beta}
\def\sina{\sin\alpha}
\def\cosa{\cos\alpha}
\def\sinb{\sin\beta}
\def\cosb{\cos\beta}
\def\lam{\lambda}
\def\eps{\epsilon}
\def\hl{h}
\def\hh{H}
\def\ha{A}
\def\hp{H^+}
\def\hm{H^-}
\def\hpm{H^\pm}
\def\mhpm{m_{\hpm}}
\def\anti{\overline}
\def\mhh{m_{\hh}}
\def\mha{m_{\ha}}
\def\mhl{m_{\hl}}
\def\mz{m_Z}
\def\gev{~{\rm GeV}}
\def\tev{~{\rm TeV}}
\def\fbi{~{\rm fb}^{-1}}
\def\fb{~{\rm fb}}
\def\call{{\cal L}}
\def\rts{\sqrt s}
\def\abi{~{\rm ab}^{-1}}
\def\vev#1{\langle #1 \rangle}
\def\br{{\rm BR}}
\def\gamhatot{\Gamma_{\rm tot}^{\ha}}
\def\gamhhtot{\Gamma_{\rm tot}^{\hh}}
\def\gamhpmtot{\Gamma_{\rm tot}^{\hpm}}
\def\gamres{\Gamma_{\rm res}}
\def\mt{m_t}
\def\mb{m_b}
\def\lsim{\mathrel{\raise.3ex\hbox{$<$\kern-.75em\lower1ex\hbox{$\sim$}}}}
\def\gsim{\mathrel{\raise.3ex\hbox{$>$\kern-.75em\lower1ex\hbox{$\sim$}}}}
\def\ifmath#1{\relax\ifmmode #1\else $#1$\fi}
\def\half{\ifmath{{\textstyle{1 \over 2}}}}
\def\threehalf{\ifmath{{\textstyle{3 \over 2}}}}
\def\quarter{\ifmath{{\textstyle{1 \over 4}}}}
\def\sixth{\ifmath{{\textstyle{1 \over 6}}}}
\def\third{\ifmath{{\textstyle{1 \over 3}}}}
\def\twothirds{{\textstyle{2 \over 3}}}
\def\fivethirds{{\textstyle{5 \over 3}}}
\def\fourth{\ifmath{{\textstyle{1\over 4}}}}
\def\beq{\begin{equation}}
\def\eeq{\end{equation}}
\def\bit{\begin{itemize}}
\def\eit{\end{itemize}}
\def\baselinestretch{1.0}




An important goal if SUSY is discovered 
will be a precise determination of the value of $\tanb$.
If the heavy Higgs boson, $\hh$ and $\ha$, masses are such that
their production rates are large,
Ref.~\cite{tgb} has shown that substantial sensitivity to $\tanb$
will derive from measurements, at both the LHC and a 
high-luminosity linear collider, of the $\hh$ and $\ha$ 
production processes, branching fractions and decay widths.
These are all largely
determined by the ratio of vacuum expectation values that defines
$\tanb$, and each can be very accurately measured at an LC
over a substantial range of the relevant $\tanb$ values, $1<\tanb<60$.
In particular, there are several
Higgs boson observables which are potentially able to 
provide the most precise
measurement of $\tanb$ when $\tanb$ is very large. 
In the context of the MSSM, there is a particularly 
large variety of complementary methods at the LC
that will allow an accurate determination of $\tanb$ 
when $\mha\lsim \rts/2$ so that $\epem\to\hh\ha$
pair production is kinematically allowed.
We will employ the sample case of $\mha=200\gev$ 
at a LC with $\rts=500\gev$.
[Although $\mha=200\gev$
is excluded by LEP limits in the MSSM context for some choices
of parameters, {\it e.g}. for $\tanb\lsim 3$  in the maximal-mixing 
scenario with $M_{SUSY}=1\tev$,
our results will be representative of what 
can be achieved whenever
$\rts$ is large enough for $\epem\to \hh\ha$ pair production
without much phase space suppression.
Outside the MSSM context, $\mha=200\gev$ 
is completely allowed.]
The complementarity has been demonstrated of employing: 
\begin{itemize}
\itemsep=0in
\item[a)] the $\bb\ha,\ \bb\hh \to \bbbb$ rate; 
\item[b)] the $\hh\ha\to \bbbb$ rate; 
\item[c)] a measurement of the average $\hh,\ha$ total width in $\hh\ha$ production;
\item[d)] the $\hp\hm\to \tbtb$ rate; and 
\item[e)] the total $\hpm$ width measured in $\hp\hm\to\tbtb$ production.
\end{itemize}
By combining the $\tanb$ errors from all these processes in quadrature,
we obtain the net statistical 
errors on $\tanb$ shown in Fig.~\ref{totalonly}
by the lines [solid for SUSY scenario (I) and dashed for SUSY scenario (II)],
assuming a multi-year integrated luminosity of $\call=2000\fbi$.
The scenarios are defined as:
\begin{description}
\itemsep=0.0in 
\item{(I)} $\mha=200\gev$, $m_{\wt g}=1\tev$, $\mu=M_2=250\gev$, \\ 
$m_{\wt t_L}=m_{\wt b_L}=m_{\wt t_R}=m_{\wt b_R}\equiv m_{\wt t}=1\tev$, \\ 
$A_\tau=A_b=0$, $A_t=\mu/\tanb+\sqrt6 m_{\wt t}$ (maximal mixing);
\item{(II)}  $\mha=200\gev$, 
$m_{\wt g}=350\gev$, $\mu=272\gev$, $M_2=120\gev$, \\ 
$m_{\wt t_L}=m_{\wt b_L}=356\gev$, $m_{\wt t_R}=273\gev$, 
$m_{\wt b_R}=400\gev$,\\
 $A_\tau=0$, $A_b=-672\gev$, $A_t=-369\gev$.
\end{description}
We see that, independent of the scenario, 
the Higgs sector will provide an excellent determination of $\tanb$
at small and large $\tanb$ values,
leading to an error on $\tanb$ of $10\%$ or better,
provided systematic errors can be kept below the statistical level.
If SUSY decays
of the $\hh,\ha,\hpm$ are significant [SUSY scenario (II)], 
the $\tanb$ error 
will be smaller than $13\%$ even in the more 
difficult moderate $\tanb$ range.
However, if SUSY decays are not significant
[SUSY scenario (I)] there is a limited range of moderate $\tanb$
for which the error on $\tanb$ would be large, reaching about $50\%$.
In general, it will be important to
establish the light SUSY particle spectrum
(using a combination of LHC and LC data) in
order to avoid systematic errors in the $\tanb$
determination deriving from the influence
of non-Higgs model parameters upon predictions for Higgs decays
and production rates.

Regardless of the relative magnitude of the LHC versus LC
$\tanb$ errors, the clean LC environment will provide an important and
independent measurement that will complement any LHC 
determination of $\tanb$.
Different uncertainties will be
associated with  the determination of $\tanb$ at a
 hadron and an $e^+e^-$ collider because of the different backgrounds.
Further, the LHC and LC measurements of $\tanb$ 
will be highly complementary
in that the experimental and theoretical 
systematic errors involved will be very different.

Combining all the different LC measurements
as above does not fully account for the fact
that the ``effective'' $\tanb$ value being measured in each process
is only the same at tree-level. The $\tanb$ values
measured via the $\hh\to \bb$ Yukawa coupling, the $\ha\to\bb$ Yukawa
coupling and the $\hp\to t\anti b$ Yukawa coupling could all
be influenced differently by the MSSM one-loop corrections.   
For some choices of MSSM parameters, 
the impact of MSSM radiative corrections on
interpreting these measurements can be substantial \cite{Carena:1998gk}.
However, if the masses of the SUSY particles are known, so that
the important MSSM parameters entering these radiative
corrections (other than $\tanb$) are fairly well determined, then
a uniform convention for the definition of $\tanb$ can be adopted
and, in general, an excellent determination of $\tanb$ (with
accuracy similar to that obtained via our tree-level procedures)
will be possible using the linear collider observables considered here.
Even for special SUSY parameter choices such
that one of the Yukawa couplings happens to be significantly
suppressed, the observables a)-e) would provide an excellent
opportunity for pinning down all the Yukawa couplings 
and checking the consistency of the MSSM model.

To illustrate the relation between the other MSSM model parameters
and the systematic errors in the $\tanb$ determination, we give one
example.     For $\tanb=5$, $\mha=\mhh=200\gev$,
$M_2$ large enough that there are no SUSY decays of the $\hh$ and 
$\ha$, $A_t=A_b=0$ and $m_0=1\tev$ for the 1st and 2nd generations,
one finds

\vspace*{-.1in}
\beq
\begin{array}{lll}
\br(\hh\to b\anti b)=0.61\quad\mbox{and} & \quad  \br(\ha\to b\anti b) = 0.87\quad\quad 
& {\rm for}~~m_{\tilde t}=m_{\tilde b}=500\gev, \\
\br(\hh\to b\anti b)=0.69 \quad\mbox{and}& \quad  \br(\ha\to b\anti b) = 0.88 
& {\rm for}~~m_{\tilde t}=m_{\tilde b}=1000\gev . \\
\end{array}
\eeq 
The resulting $\hh\ha\to b\anti b b\anti b$ rate differs by about 
$14\%$ between the two scenarios.  
The bottom and stop masses would need to be known to within about
$150\gev$ in order to bring the systematic uncertainty from this
source below the roughly $5\%$ statistical uncertainty shown in Fig.~\ref{totalonly}.
Dependence on the stop and sbottom sector mixing angles, 
the gluino mass and so forth are non-negligible as well. 
The precision of SM parameters, in
 particular $m_t$ and $m_b$ (the latter enters directly into
the $b\anti b$ coupling strength of the $\hh$ and $\ha$)
will also contribute to the systematic uncertainty.
Of course, because $\tanb$ 
is not a directly measurable quantity, 
other techniques for determining $\tanb$ suffer from
similar systematics issues.

Finally, it is important to note that 
the LC techniques employed for the $\tanb$ determination
discussed above can also be employed
in the case of other Higgs sector models.
For example, in the general (non-SUSY) 2HDM, if the only non-SM-like 
Higgs boson with mass below $\rts$ is the $\ha$ \cite{sec4_Chankowski:2000an}, 
then a good determination
of $\tanb$ will be possible at high $\tanb$ from the 
$\bb\ha\to\bbbb$ production rate.
Similarly, in models with more than two Higgs doublet
and/or triplet representations, the
Yukawa couplings of the Higgs bosons, and, therefore, the analogues of
the 2HDM parameter $\tanb$, 
will probably be accurately determined through Higgs
production observables in $\epem$ collisions. 

Now, we will compare the LC results summarized in Fig.~\ref{totalonly}~\cite{tgb} 
to the $\tanb$ accuracies that can be achieved at the LHC
based on $\hh,\ha,\hpm$ production and decay processes.  
First note that there is a wedge-shaped window of moderate $\tanb$
and $\mha\gsim 200\gev$ for which the $\ha$, $\hh$
and $\hpm$ are all unobservable (see, for example,
Refs.~\cite{LHCcms,LHCatlas,Gianotti:2002xx}). In this wedge,
the only Higgs boson that is detectable at the LHC is
the light SM-like  Higgs boson, $\hl$. Precision measurements of the
properties of the $\hl$ typically only provide weak sensitivity
to $\tanb$, and will not be considered here. The lower $\tanb$
bound of this moderate-$\tanb$ wedge is defined by the LEP-2
limits~\cite{lepwg}, which are at $\tanb\sim 3$ for $\mha\sim 200\gev$, falling
to $\tanb\sim 2.5$ for $\mha\gsim 250\gev$, assuming
the maximal mixing scenario [see SUSY scenario (I) defined earlier].
The upper $\tanb$ limit of the wedge is at $\tanb\sim 7$
for $\mha\sim 200\gev$ rising to $\tanb\sim 15$ at $\mha\sim 500\gev$. 
For either smaller or larger $\tanb$
values, the heavy MSSM Higgs bosons can be detected and 
their production rates and properties will provide sensitivity to $\tanb$.  

We will now summarize the 
results currently available regarding the determination of
$\tanb$ at the LHC using Higgs measurements (outside the wedge region)
assuming a luminosity of $\call=300\fbi$. The methods employed are
those proposed in \cite{Gunion:1996cn} as reanalyzed from 
a more experimental perspective by the
ATLAS and/or CMS collaborations.
The reactions that have been studied at the LHC are the following.
\ben
\item $gg\to \hh\to ZZ\to 4\ell$ \cite{LHCatlas}.

 The best accuracy that can be achieved
at low $\tanb$ is obtained from
the $\hh\to ZZ\to 4\ell$ rate. One finds $\Delta\tanb/\tanb=\pm 0.1$ 
at $\tanb=1$ rising to $>\pm 0.3$ by $\tanb=1.5$ for the sample
choice of $\mhh=300\gev$. For $\mhh<2m_Z$, $\tanb$
cannot be measured via this process. (Of course,
in the MSSM maximal mixing
scenario, such low values of $\tanb$
are unlikely in light of LEP-2 limits on $\mhl$.)

\item $gg\to \hh+gg\to \ha\to \tau^+\tau^-,\mu^+\mu^-$ and
$gg\to b\anti b \hh+b\anti b \ha\to \bb\tau^+\tau^-,\bb\mu^+\mu^-$
\cite{LHCatlas}.

At high $\tanb$ and taking $\mha=150\gev$, Fig.~19-86 of  \cite{LHCatlas}
shows that 
the $gg\to \hh\to\tau^+\tau^-$, $gg\to \ha\to\tau^+\tau^-$, and
$gg\to\bb\ha+\bb\hh\to \bb\tau^+\tau^-$ rates can, in combination,
be used to determine
$\tanb$ with an accuracy of $\pm 0.15$ at $\tanb=5$, improving
to $\pm 0.06$ at $\tanb=40$. The corresponding rates with 
$\hh,\ha\to\mu^+\mu^-$ yield a somewhat better determination
at higher $\tanb$: $\pm 0.12$ at $\tanb=10$
and $\pm 0.05$ at $\tanb=40$. 

Interpolating, using Figs.~19-86 and 19-87 from \cite{LHCatlas},
we estimate that at $\mha\sim 200\gev$ (our choice for this study)
the error on $\tanb$ based on these rates 
would be smaller than $\pm 0.1$ for $\tanb\gsim 13$,
asymptoting to $\pm 0.05$ at large $\tanb$.

The importance of including
the $gg\to \hh$, $gg\to\ha$ as well as the $gg\to \bb\hh+gg\to \bb\ha$
processes in order to obtain observable signals
for $\tanb$ values as low as 10 
in the $\mu^+\mu^-$ channels is apparent from \cite{Dawson:2002cs}.
For $\call=300\fbi$ and $\mha=200\gev$, they find that
the $\bb\mu^+\mu^-$ final states can only
be isolated for $\tanb>30$ whereas the inclusive $\mu^+\mu^-$ final
state from all production processes becomes detectable once $\tanb>10$.

\item $gg\to t\anti b \hm +\anti t b\hp$ with $\hpm\to \tau^{\pm}\nu$
\cite{Assamagan:2002ne}.

The $tb\hpm\to tb\tau\nu$ rate gives a 
fractional $\tan\beta$ uncertainty, $\Delta\tanb/\tanb$,
ranging from $\pm 0.074$ 
at $\tanb=20$ to $\pm 0.054$ at $\tanb=50$. 
 
\een
The above error estimates are purely statistical.  All three techniques
will have systematic errors deriving from imprecise knowledge of
the gluon distribution function and QCD corrections.
Background uncertainties might also enter, although not at the 
highest $\tanb$ values for the $\tanb$-enhanced processes that have
very high rates.  Additional systematic error will,
as in the LC case, derive from the need to have precise measurements
of SUSY masses and mixing angles in order to precisely relate
the $b\anti b$ Yukawa couplings of the $\hh$ and $\ha$ to the
$\tanb$ parameter. Finally, in the case of techniques [1.] and 
(for $\tanb<20-30$) [2.], the $\tanb$ sensitivity
is largely due to the loop-induced $gg\to \ha$
and $gg\to\hh$ production processes. 
Interpreting $gg$-induced  rates in terms of $\tanb$ 
requires significant knowledge of the particles, including SUSY particles,
that go into the loops responsible for the $gg\to\hh$ and $gg\to\ha$ 
couplings.

Sensitivity
to $\tanb$ deriving from direct measurements of the decay widths
has not been studied by the LHC experiments.
One can expect excellent $\tanb$ statistical 
accuracy at the higher $\tanb$ values
for which the $gg\to b\anti b \mu^+\mu^-$ signal for the $\hh$
and $\ha$ is detectable. Further, the direct width measurement
would avoid systematic errors deriving from uncertainties
in the gluon distribution function or the $gg\to \hh,\ha$ loop-induced
couplings.

We now discuss the interplay between the LHC errors and 
the LC errors for $\tanb$, 
assuming $\mha=200\gev$.  This discussion is based upon statistical
errors only. As already noted, to keep systematic errors at a level
below the statistical errors will require substantial input on
other model parameters (e.g. sparticle masses and mixings in
the MSSM context) from both the LHC and the LC.
First, consider $\tanb\leq 10$. 
As summarized above, the LHC error on $\tanb$ is $\pm 0.12$
at $\tanb\sim 10$ and at $\tanb\sim 1$, and 
the error becomes very large for $1.5\lsim \tanb\lsim 5$.
Meanwhile, the LC error from Fig.~\ref{totalonly} 
ranges from roughly $\pm0.03$ to $\pm 0.05$ for $2\lsim\tanb\lsim 5$ rising to
about $\pm 0.1$ at $\tanb\sim 10$ [in the less favorable
SUSY scenario (I)]. Therefore, for $\tanb\lsim 10$ 
the LC provides the best determination of $\tanb$ using Higgs observables
related to their Yukawa couplings.
(In the MSSM context, 
other non-Higgs LHC measurements would allow a good $\tanb$ determination
at low to moderate $\tanb$ based on other kinds of couplings.)  
In the middle range of $\tanb$ (roughly $13<\tanb<30$ at 
$\mha\sim 200\gev$), the heavy Higgs determination of
$\tanb$ at the LHC might be superior to that obtained at the LC.
This depends upon the SUSY scenario: if the heavy Higgs bosons can decay
to SUSY particles, the LC will give $\tanb$ errors that
are quite similar to those obtained at the LHC;
if the heavy Higgs bosons do not have substantial SUSY decays,
then the expected LC $\tanb$ errors are substantially larger than those
predicted for the LHC.
At large $\tanb$, the LC measurement of the heavy Higgs couplings
and the resulting $\tanb$ determination at the LC is numerically only
slightly more accurate than that obtained at the LHC.
For example, both are of order $\pm 0.05$ at $\tanb=40$.
(At this level of statistical error, the ability to reduce the
systematic error by combined LC and LHC systematic studies will be 
particularly important.) The 
statistical error comparisons are summarized in 
Table~\ref{summarytable}.
It is possible that the net LHC $\tanb$ error would
be somewhat smaller than the LC error for $\tanb\gsim 40$
if both ATLAS and CMS can each accumulate
$\call=300\fbi$ of luminosity; combining the two data sets would
presumably roughly double the statistics and decrease errors
by a factor of order $1/\sqrt 2$. Inclusion of the direct
width measurement at the LHC would also decrease the error. In any case, 
a very small statistical error on 
$\tanb$ will be achievable for all $\tanb$
by combining the results from the LC with those from the LHC. 

\begin{table}[h!]
\centering
\caption{\label{summarytable}\baselineskip 0pt 
A comparison of fractional statistical errors, 
$\Delta\tanb/\tanb$, achievable for $\call=2000\fbi$
at the LC with those expected at the LHC for $\call=300\fbi$,
assuming $\mha=200\gev$ in the MSSM. LC results are given for both 
SUSY scenarios (I) and (II),
where Higgs boson decays to SUSY particles are disallowed,
respectively allowed. LHC results are estimated
by roughly combining the determinations of $\tanb$ based on
$\hh,\ha$ production from \cite{LHCatlas}
with those using $\hpm$ production from \cite{Assamagan:2002ne},
both of which assume the standard MSSM
maximal mixing scenario. All entries are approximate.}
\medskip
\begin{tabular}{|c|c|c|c|}
\hline
 $\tanb$ range &    LHC           &       LC (case I)  &      LC (case II) \cr 
\hline
 1       &  0.12       &  0.15        &   0.1         \\
 1.5--5  &  very large &  0.03--0.05  &   0.03--0.05  \\
 10      &  0.12       &  0.1         &   0.05        \\
 13--30   &  0.05       &  0.6--0.1    &   0.05--0.1   \\
 40--60   &  0.05--0.03 &  0.05--0.025 &   0.05--0.025 \\
\hline 
\end{tabular}
\end{table}

Finally, we wish to emphasize
that the above LHC versus LC comparisons have
been made based only on $\hh,\ha,\hpm$ processes and
for the particular choice of $\mha=200\gev$ in the MSSM, 
assuming $\rts=500\gev$ for the LC. For given $\tanb$,
the LC accuracies would decline if 
$\mha>\rts/2$, since then $\hh\ha$ and $\hp\hm$ 
pair production would not be possible. The LHC accuracies will decrease
with increasing $\mha$ at fixed $\tanb$ simply as a result of
decreasing event rates. Detailed studies would be worthwhile.  

\begin{figure}[htb!]
\begin{center}
\includegraphics[width=0.7\textwidth,angle=90]{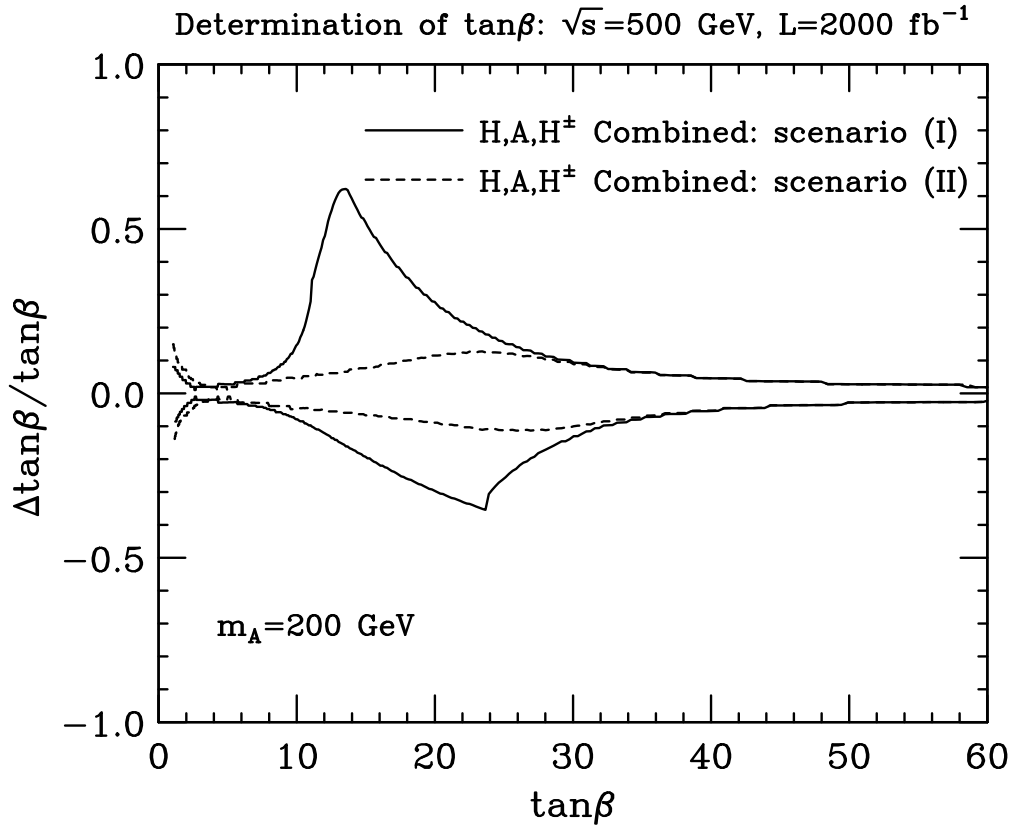}
\end{center}
\caption{\label{totalonly}\baselineskip 0pt
For the MSSM, with $\mhpm\sim \mha = 200$~GeV, and 
assuming $\call=2000\fbi$ at $\rts=500\gev$,
we plot the $1\sigma$ statistical upper and 
lower bounds, $\Delta\tanb/\tanb$,
as a function of $\tanb$ based on combining (in quadrature)
the described LC Higgs boson studies.
Results are shown for the SUSY scenarios (I) and (II) described
in the text.}
\end{figure}

%


\section{Higgs sector in non-minimal models}
%

\def\lsim{\raise0.3ex\hbox{$<$\kern-0.75em\raise-1.1ex\hbox{$\sim$}}}
\def\gsim{\raise0.3ex\hbox{$>$\kern-0.75em\raise-1.1ex\hbox{$\sim$}}}
\def\anti{\overline}
\def\beq{\begin{equation}}
\def\eeq{\end{equation}}
\def\gam{\gamma}
\def\anti{\overline}
\def\tanb{\tan\beta}
\def\br{BR}
\def\nsd#1{$N_{SD}(#1)$}
\def\what{\widehat}
\def\hpm{H^{\pm}}
\def\ie{{\it i.e.}}
\def\gev{~{\rm GeV}}
\def\stop{\tilde{t}}
\def\sto{\tilde{t}_1}
\def\stt{\tilde{t}_2}
\def\stl{\tilde{t}_L}
\def\str{\tilde{t}_R}
\def\msto{m_{\sto}}
\def\mstosq{m_{\sto}^2}
\def\mstt{m_{\stt}}
\def\msttsq{m_{\stt}^2}
\def\mt{m_t}
\def\ra{\rightarrow}
\def\neutf{\tilde{\chi}_4^0}
\def\mneutf{m_{\tilde{\chi}_4^0}}
\def\neutt{\tilde{\chi}_2^0}
\def\mneutt{m_{\tilde{\chi}_2^0}}
\def\neutth{\tilde{\chi}_3^0}
\def\mneutth{m_{\tilde{\chi}_3^0}}

In this section we discuss examples of models with a nonminimal 
Higgs sector or with a light Higgs that has
 nonstandard decays/branching 
ratios, emphasizing possible LHC-LC complementarity and 
cross talk.

The NMSSM is one of the most interesting extensions of the MSSM.  
The NMSSM Higgs sector contains one additional scalar and one
additional pseudoscalar Higgs boson beyond the MSSM Higgs states. The
MSSM Higgs states then share their couplings to fermions and gauge
bosons with the additional states.  New Higgs decay modes
also open up. A case that might make Higgs detection difficult
is that in which there is a light (CP-even) Higgs boson
which dominantly decays into two CP-odd light Higgs
bosons. If these CP-odd Higgs
bosons do not decay significantly to supersymmetric
modes, it is shown that the NMSSM Higgs discovery is
close to being guaranteed even in this case if the usually considered
LHC detection modes are supplemented by the $WW\to h \to aa$ mode,
with $aa\to b\anti b \tau^+\tau^-$.  The resulting signal would
be the only LHC signal for Higgs bosons.  Confirmation of the nature
of the signal at the LC would be vital. It is
shown that the $WW\to h\to aa$ signal, as well
as the usual $e^+e^-\to Z h\to Z aa$ signal, 
will be highly visible at the
LC due to its cleaner environment and high luminosity.
Further, the LC will be able to 
observe both the $aa\to \tau^+\tau-\to b\anti b$ and 
the $aa\to b\anti b b\anti b$ final states,
thus allowing a definitive check that the ratio
$\br(a\to \tau^+\tau^-)/\br(a\to b\anti b)$ is that expected
for a CP-odd scalar.

Next we discuss the very interesting scenario that
at the Tevatron, LHC and $e^+e^-$ Linear Collider
only one light Higgs boson will be found, with
properties as expected in the Standard Model. It
is possible to realize this SM-like scenario in a
Two-Higgs-Doublet Model both with and without
decoupling.  It is shown how precise measurements
of the Higgs boson coupling to gluons at the LHC
and to photons at a Photon Collider can allow one
to determine which scenario is realized in nature.

After this is discussed the case of a fermiophobic Higgs boson ($h_f$) 
decaying to two photons with a larger branching ratio than in the 
SM. In this case the standard production mechanisms are very suppressed
for moderate to large $\tan\beta$, both at the LC and the LHC. 
It is shown here that the search for $pp\to H^\pm h_f$ should 
substantially benefit from a previous signal at a LC in the channel
$e^+e^-\to A^0h_f$, and would provide important confirmation of any LC 
signal for $h_f$.

Should  a light  Higgs have substantial branching ratio in 'invisible' 
channels, as may well happen for a supersymmetric Higgs with mass less than 
$\sim 130$ GeV and nonuniversal gaugino masses at the high scale, 
its  search at the LHC in the standard channels may 
get compromised. Such a Higgs can be seen at an LC with ease, allowing even 
a measurement of its 'invisible' branching ratio. 
Signals at the LHC which can be searched for, so as confirm the lack
of signal at LHC in the usual channel due to these effects are discussed.

\subsection {NMSSM Higgs discovery at the LHC}
\label{sec251}
{\it U. Ellwanger, J.F. Gunion, C. Hugonie and S. Moretti}

\vspace{1em}
\def\cnone{\widetilde \chi_1^0}
\def\cpone{\widetilde \chi_1^+}
\def\cmone{\widetilde \chi_1^-}
\def\mcnone{m_{\cnone}}
\def\mcpmone{m_{\widetilde\chi_1^{\pm}}}
\def\tev{~{\rm TeV}}
\def\gev{~{\rm GeV}}
\def\mgut{M_{U}}
One of the most attractive supersymmetric models
is the Next to Minimal Supersymmetric Standard
Model (NMSSM) (see
\cite{Ellis:1988er,Ellwanger:2001iw} and
references therein) which extends the MSSM by the
introduction of just one singlet superfield, $\what
S$. When the scalar component of $\what S$
acquires a TeV scale vacuum expectation value (a
very natural result in the context of the model),
the superpotential term $\what S \what H_u \what
H_d$ generates an effective $\mu\what H_u \what
H_d$ interaction for the Higgs doublet
superfields.  Such a term is essential for
acceptable phenomenology. No other SUSY model
generates this crucial component of the
superpotential in as natural a fashion. Thus, the
phenomenological implications of the NMSSM at
future accelerators should be considered very
seriously.  One aspect of this is the fact that
the $h,H,A,\hpm$ Higgs sector of the MSSM is
extended so that there are three CP-even Higgs
bosons ($h_{1,2,3}$, $m_{h_1}<m_{h_2}<m_{h_3}$),
two CP-odd Higgs bosons ($a_{1,2}$,
$m_{a_1}<m_{a_2}$) (we assume that CP is not
violated in the Higgs sector) and a charged Higgs
pair ($h^\pm$). An important question is then the
extent to which the no-lose theorem for MSSM Higgs
boson discovery at the LHC (after LEP constraints)
is retained when going to the NMSSM; \ie\ is the
LHC guaranteed to find at least one of the
$h_{1,2,3}$, $a_{1,2}$, $h^\pm$? The first exploration
of this issue appeared in \cite{Gunion:1996fb},
with the conclusion that for substantial portions
of parameter space the LHC would be unable to
detect any of the NMSSM Higgs bosons.
Since then, there have been improvements in many
of the detection modes and the addition of
new one. These will be summarized below
and the implications reviewed.  However, these
improvements and additions do not address the 
possibly important $h\to aa$ type decays that could
suppress all other types of signals \cite{Gunion:1996fb,Dobrescu:2000jt}.

One of the key ingredients in the no-lose theorem
for MSSM Higgs boson discovery is the fact that
relations among the Higgs boson masses are such
that decays of the SM-like Higgs boson to $AA$ are
only possible if $m_A$ is quite small, a region
that is ruled out by LEP by virtue of the fact
that $Z\to hA$ pair production was not detected
despite the fact that the relevant coupling is
large for small $m_A$.  In the NMSSM, the lighter
Higgs bosons, $h_1$ or $h_2$, can be SM-like (in
particular being the only Higgs with substantial
$WW/ZZ$ coupling) without the $a_1$ necessarily
being heavy.  In addition, this situation is not
excluded by LEP searches for $e^+e^-\to Z^*\to
h_{1,2}a_1$ since, in the NMSSM, the $a_1$ can
have small $Zh_2 a_1$ ($Zh_1 a_1$) coupling when
$h_1$ ($h_2$) is SM-like. [In addition, sum rules
require that the $Zh_1 a_1$ ($Zh_2 a_1$) coupling
is small when the $h_1WW$ ($h_2WW$) couplings are
large.]  As a result, NMSSM parameters that are
not excluded by current data can be chosen so that
the $h_{1,2}$ masses are moderate in size ($\sim
100-130$~GeV) and the $h_1\to a_1a_1$ or $h_2\to
a_1a_1$ decays are dominant.  Dominance of such
decays falls outside the scope of the usual
detection modes for the SM-like MSSM $h$ on which
the MSSM no-lose LHC theorem largely relies.

In Ref.~\cite{Ellwanger:2001iw}, a partial no-lose
theorem for NMSSM Higgs boson discovery at the LHC
was established.  In particular, it was shown that
the LHC would be able to detect at least one of
the Higgs bosons (typically, one of the lighter
CP-even Higgs states) throughout the full
parameter space of the model, excluding only those
parameter choices for which there is sensitivity
to the model-dependent decays of Higgs bosons to
other Higgs bosons and/or superparticles.  Here,
we will address the question of whether or not
this no-lose theorem can be extended to those
regions of NMSSM parameter space for which Higgs
bosons can decay to other Higgs bosons.  We find
that the parameter choices such that the
``standard'' discovery modes fail {\it would}
allow Higgs boson discovery if detection of $h\to
aa$ decays is possible. (When used generically,
the symbol $h$ will now refer to $h=h_1$, $h_2$ or
$h_3$ and the symbol $a$ will refer to $a=a_1$ or
$a_2$).  Detection of $h\to aa$ will be difficult
since each $a$ will decay primarily to 
$b\anti b$ (or 2 jets if $m_a<2m_b$),
$\tau^+\tau^-$, and, possibly, $\cnone\cnone$,
yielding final states that will typically have
large backgrounds at the LHC.

In \cite{Ellwanger:2001iw} we scanned the
parameter space, removing parameter choices ruled
out by constraints from LEP on Higgs boson
production, $e^+ e^- \to Z h$ or $e^+ e^- \to h a$
\cite{LEPLEPHA}, and eliminating parameter choices
for which one Higgs boson can decay to two other
Higgs bosons or a vector boson plus a Higgs boson.
For the surviving regions of parameter space, we
estimated the statistical significances
($N_{SD}=S/\sqrt B$) for all Higgs boson detection
modes so far studied at the LHC \cite{sec25_CMS}.
These are (with $\ell=e,\mu$)

1) $g g \to h/a \to \gamma \gamma$;\par
2) associated $W h/a$ or $t \bar{t} h/a$ production with 
$\gamma \gamma\ell^{\pm}$ in the final state;\par
3) associated $t \bar{t} h/a$ production with $h/a \to b \bar{b}$;\par
4) associated $b \bar{b} h/a$ production with $h/a \to \tau^+\tau^-$;\par
5) $g g \to h \to Z Z^{(*)} \to$ 4 leptons;\par
6) $g g \to h \to W W^{(*)} \to \ell^+ \ell^- \nu \bar{\nu}$;\par
7) $W W \to h \to \tau^+ \tau^-$;\par
8) $W W \to h\to W W^{(*)}$.\par

\noindent
For an integrated luminosity of $300~{\rm
  fb}^{-1}$ at the LHC, all the surviving points
yielded $N_{SD}>10$ after combining all modes,
including the $W$-fusion modes. Thus, NMSSM Higgs
boson discovery by just one detector with
$L=300~{\rm fb}^{-1}$ is essentially guaranteed
for those portions of parameter space for which
Higgs boson decays to other Higgs bosons or
supersymmetric particles are kinematically
forbidden.

In this work, we investigate the complementary
part of the parameter space, where {\it at least
  one} Higgs boson decays to other Higgs bosons.
To be more precise, we require at least one of the
following decay modes to be kinematically allowed:
\begin{eqnarray}
& i) \ h \to h' h' \; , \quad ii) \ h \to a a \; , \quad iii) \ h \to h^\pm
h^\mp \; , \quad iv) \ h \to a Z \; , \nonumber \\
& v) \ h \to h^\pm W^\mp \; , \quad vi) \ a' \to h a \; , \quad vii) \ a \to h
Z \; , \quad viii) \ a \to h^\pm W^\mp \; .
\end{eqnarray}
After searching those regions of parameter space
for which one or more of the decays $ i) - viii)$
is allowed, we found that the only subregions for
which discovery of a Higgs boson in modes 1) -- 8)
was not possible correspond to NMSSM parameter
choices for which (a) there is a light CP-even
Higgs boson with substantial doublet content that
decays mainly to two still lighter CP-odd Higgs
states, $h\to aa$, and (b) all the other Higgs
states are either dominantly singlet-like,
implying highly suppressed production rates, or
relatively heavy, decaying to $t\anti t$, to one
of the ``difficult'' modes $i) - viii)$ or to a
pair of sparticles. In such cases, the best
opportunity for detecting at least one of the
NMSSM Higgs bosons is to employ $WW\to h$
production and develop techniques for extracting a
signal for the $h\to aa$ final state.  We
have performed a detailed simulation of 
the $aa\to b\anti b \tau^+\tau^-$ final state
and find that its detection may be possible
after accumulating $300~{\rm fb}^{-1}$ in both the
ATLAS and CMS detectors.  Further, we show that
the $WW\to h\to aa$ signal is extremely robust at
an LC.

We consider the simplest version of the NMSSM
\cite{Ellis:1988er}, where the term $\mu \widehat
H_1 \widehat H_2$ in the superpotential of the
MSSM is replaced by (we use the notation $\widehat
A$ for the superfield and $A$ for its scalar
component field)
\begin{equation}\label{2.1r}
\lambda \widehat H_1 \widehat H_2 \widehat S\ + \ \frac{\kappa}{3} \widehat S^3
\ \ ,
\end{equation}
\noindent so that the superpotential is scale invariant. 
We make no assumption on ``universal'' soft terms.
Hence, the five soft supersymmetry breaking terms
\begin{equation}\label{2.2r}
m_{H_1}^2 H_1^2\ +\ m_{H_2}^2 H_2^2\ +\ m_S^2 S^2\ +\ \lambda
A_{\lambda}H_1 H_2 S\ +\ \frac{\kappa}{3} A_{\kappa}S^3
\end{equation}
\noindent are considered as independent. 
The masses and/or couplings of sparticles will
be such that their contributions to the
loop diagrams inducing Higgs boson production by
gluon fusion and Higgs boson decay into $\gamma
\gamma$ are negligible. 
In the gaugino sector, we chose $M_2=1\tev$ (at low scales).
Assuming universal gaugino masses at the coupling
constant unification scale,
this yields $M_1\sim 500\gev$ and $M_3\sim 3\tev$.
In the squark sector, as particularly relevant
for the top squarks which
appear in the radiative corrections to the Higgs
potential, we chose the soft masses $m_Q = m_T
\equiv M_{susy}= 1$ TeV, and varied the stop
mixing parameter
\begin{equation}\label{2.4r}
X_t \equiv 2 \ \frac{A_t^2}{M_{susy}^2+m_t^2} \left ( 1 -
\frac{A_t^2}{12(M_{susy}^2+m_t^2)} \right ) \ .
\end{equation} 
\noindent As in the MSSM, 
the value $X_t = \sqrt{6}$ -- so called maximal
mixing -- maximizes the radiative corrections to
the Higgs boson masses, and we found that it leads
to the most challenging points in the parameter
space of the NMSSM.  We adopt the convention
$\lambda,\kappa > 0$, in which $\tan\beta$ can
have either sign. We require $|\mu_{\rm eff}|\ >\ 
100$~GeV; otherwise a light chargino would have
been detected at LEP. The only possibly light SUSY particle
will be the $\cnone$.  A light $\cnone$ is a frequent
characteristic of parameter choices that yield a 
light $a_1$.

We have performed a numerical scan over the free
parameters.  For each point, we computed the
masses and mixings of the CP-even and CP-odd Higgs
bosons, $h_i$ ($i=1,2,3$) and $a_j$ ($j=1,2$),
taking into account radiative corrections up to
the dominant two loop terms, as described in
\cite{Ellwanger:1999ji}.  We eliminated parameter
choices excluded by LEP
constraints~\cite{LEPLEPHA} on $e^+ e^- \to Z h_i$
and $e^+ e^- \to h_i a_j$. The latter provides an
upper bound on the $Zh_ia_j$ reduced coupling,
$R'_{ij}$, as a function of $m_{h_i}+m_{a_j}$ for
$m_{h_i} \simeq m_{a_j}$.  Finally, we calculated
$m_{h^\pm}$ and required $m_{h^\pm} > 155$~GeV, so
that $t \to h^\pm b$ would not be seen.

In order to probe the complementary part of the
parameter space as compared to the scanning of
Ref. \cite{Ellwanger:2001iw}, we required that at
least one of the decay modes $i) - viii)$ is
allowed.  For each Higgs state, we calculated all
branching ratios including those for modes $i) -
viii)$, using an adapted version of the FORTRAN
code HDECAY \cite{Djouadi:1997yw}. We then
estimated the expected statistical significances
at the LHC in all Higgs boson detection modes 1)
-- 8) by rescaling results for the SM Higgs boson
and/or the MSSM $h, H$ and/or $A$. The
rescaling factors are determined by $R_i$, $t_i$
and $b_i=\tau_i$, the ratios of the $VVh_i$,
$t\anti t h_i$ and $b\anti b h_i,\tau^+\tau^- h_i$
couplings, respectively, to those of a SM Higgs
boson.  Of course $|R_i| < 1$, but $t_i$ and $b_i$
can be larger, smaller or even differ in sign with
respect to the SM. For the CP-odd Higgs bosons,
$R_i'=0$ at tree-level; $t'_j$ and $b'_j$ are the
ratios of the $i\gamma_5$ couplings for $t\bar{t}$
and $b\bar{b}$, respectively, relative to SM-like
strength.  A detailed discussion of the procedures
for rescaling SM and MSSM simulation results for
the statistical significances in channels 1) -- 8)
will appear elsewhere.

\begin{table}[p]
\begin{center}
\footnotesize
\vspace*{-.2in}
\hspace*{-.5in}
\begin{tabular} {|l|l|l|l|l|l|l|} 
\hline
Point Number & 1 & 2 & 3 & 4 & 5 & 6  \\
\hline \hline
Bare Parameters &\multicolumn{6}{c|}{} \\
\hline
$\lambda$            & 0.2872 & 0.2124 & 0.3373 & 0.3340 & 0.4744 & 0.5212 \\
\hline
$\kappa$             & 0.5332 & 0.5647 & 0.5204 & 0.0574 & 0.0844 & 0.0010 \\
\hline
$\tan\beta$          &   2.5  &   3.5  &   5.5  &    2.5 &    2.5 & 2.5 \\
\hline
$\mu_{\rm eff}~({\rm GeV})$&    200 &    200 &    200 &    200 &    200 & 200 \\
\hline
$A_{\lambda}~({\rm GeV})$  &    100 &      0 &     50 &    500 &    500 & 500 \\
\hline
$A_{\kappa}~({\rm GeV})$   &      0 &      0 &      0 &      0 &      0 & 0 \\
\hline \hline
CP-even Higgs Boson Masses and Couplings &\multicolumn{6}{c|}{} \\
\hline \hline
$m_{h_1}$~(GeV)      &    115 &    119 &    123 &     76 &     85 &  51\\
\hline
$R_1 $               &   1.00 &   1.00 &  -1.00 &   0.08 &   0.10 &  -0.25\\
\hline
$t_1 $               &   0.99 &   1.00 &  -1.00 &   0.05 &   0.06 &  -0.29\\
\hline
$b_1 $               &   1.06 &   1.05 &  -1.03 &   0.27 &   0.37 &  0.01\\
\hline
Relative 
gg Production Rate   &   0.97 &   0.99 &   0.99 &   0.00 &   0.01 &  0.08\\
\hline
$\br(h_1\to 
b\anti b)$           &   0.02 &   0.01 &   0.01 &   0.91 &   0.91 &  0.00\\
\hline
$\br(h_1\to 
\tau^+\tau^-)$      &   0.00 &   0.00 &   0.00 &   0.08 &   0.08 &  0.00\\
\hline
$\br(h_1\to a_1 a_1)$&   0.98 &   0.99 &   0.98 &   0.00 &   0.00 &  1.00\\
\hline \hline

$m_{h_2}$~(GeV)      &    516 &    626 &    594 &    118 &    124 &  130\\
\hline
$R_2 $               &  -0.03 &  -0.01 &   0.01 &  -1.00 &  -0.99 &  -0.97\\
\hline
$t_2 $               &  -0.43 &  -0.30 &  -0.10 &  -0.99 &  -0.99 &  -0.95\\
\hline
$b_2 $               &   2.46 &  -3.48 &   3.44 &  -1.03 &  -1.00 &  -1.07\\
\hline
Relative
gg Production Rate   &   0.18 &   0.09 &   0.01 &   0.98 &   0.99 &  0.90\\
\hline
$\br(h_2\to 
b\anti b)$           &   0.01 &   0.04 &   0.04 &   0.02 &   0.01 &  0.00\\
\hline
$\br(h_2\to 
\tau^+\tau^-)$      &   0.00 &   0.01 &   0.00 &   0.00 &   0.00 &  0.00\\
\hline
$\br(h_2\to a_1 a_1)$&   0.04 &   0.02 &   0.83 &   0.97 &   0.98 &  0.96\\
\hline \hline

$m_{h_3}$~(GeV)      &    745 &   1064 &    653 &    553 &    554 &  535\\
\hline \hline

CP-odd Higgs Boson Masses and Couplings &\multicolumn{6}{c|}{} \\
\hline \hline
$m_{a_1}$~(GeV)      &     56 &      7 &     35 &     41 &     59 &  7\\
\hline
$t_1' $               &   0.05 &   0.03 &   0.01 &  -0.03 &  -0.05 &  -0.06\\
\hline
$b_1' $               &   0.29 &   0.34 &   0.44 &  -0.20 &  -0.29 &  -0.39\\
\hline
Relative
gg Production Rate   &   0.01 &   0.03 &   0.05 &   0.01 &   0.01 &  0.05\\
\hline
$\br(a_1\to 
b\anti b)$           &   0.92 &   0.00 &   0.93 &   0.92 &   0.92 &  0.00\\
\hline
$\br(a_1\to 
\tau^+\tau^-)$      &   0.08 &   0.94 &   0.07 &   0.07 &   0.08 &  0.90\\
\hline \hline

$m_{a_2}$~(GeV)      &    528 &    639 &    643 &    560 &    563 &  547\\
\hline 
Charged Higgs  
Mass (GeV)           &    528 &    640 &    643 &    561 &    559 &  539\\
\hline\hline
Most Visible of the LHC Processes 1)-8) &  2 ($h_1$) &  2 ($h_1$) &  8
                  ($h_1$) &  2 ($h_2$) &  8 ($h_2$)  &  8 ($h_2$)\\
\hline            
$N_{SD}=S/\sqrt B$ 
Significance of this process at $L=$300~${\rm fb}^{-1}$
                     &   0.48 &   0.26 &   0.55 &   0.62 &  0.53  & 0.16\\
\hline
\hline
$N_{SD}(L=300~{\rm fb}^{-1})$  for
$WW\to h\to aa\to jj \tau^+\tau^-$ at LHC & 50 &  22 &  69 &  63&  62 &  21 \\
\hline
$S(L=500~{\rm fb}^{-1})$  for
$WW\to h\to aa\to jj \tau^+\tau^-$ at LC
& 36 & 320 & 45 & 45 & 45 & 320 \\
\hline
\end{tabular}
\end{center}
\vspace*{-.2in}\caption{\label{tpoints}\footnotesize
Properties of selected scenarios that could escape detection
at the LHC. In the table, $R_i=g_{h_i VV}/g_{h_{SM} VV}$, 
$t_i=g_{h_i t\anti t}/g_{h_{SM} t\anti t}$ and $b_i=g_{h_ib\anti b}/g_{h_{SM} b\anti b}$ 
for $m_{h_{SM}}=m_{h_i}$; $t_1'$ and $b_1'$
are the $i\gam_5$ couplings of $a_1$ 
to $t\anti t$ and $b\anti b$ normalized
relative to the scalar 
$t\anti t$ and $b\anti b$ SM Higgs couplings.
We also give the $gg$ fusion production rate ratio,
$gg\to h_i/gg\to h_{SM}$, for $m_{h_{SM}}=m_{h_i}$. 
Important absolute branching
ratios are displayed. For points 2 and 6, the decays
$a_1\to jj$ ($j\neq b$) have 
$\br(a_1\to jj)\simeq 1-\br(a_1\to \tau^+\tau^-)$.
For the heavy $h_3$ and $a_2$, we give only their masses.
For all points 1 -- 6, the statistical
significances for the detection of any 
Higgs boson in any of the channels 1) --
8) are tiny; the third-to-last row gives their maximum 
together with the process number and 
the corresponding Higgs state.
The next-to-last row gives the statistical significance
of the new $WW\to h \to aa\to jj \tau^+\tau^-$
[$h=h_1$ ($h=h_2$) for points 1--3 (4--6)] LHC
signal explored here. The final row gives
the signal rate $S$
at the LC for $40<M_{jj\tau^+\tau^-}<120\gev$, where $B=0$.}
\end{table}

In our set of randomly scanned points, we selected
those for which all the statistical significances
in modes 1) -- 8) are below $5\sigma$. We obtained
a lot of points, all with similar characteristics.
Namely, in the Higgs spectrum, we always have a
very SM-like CP-even Higgs boson with a mass
between 115 and 135~GeV ({\it i.e.} above the LEP
limit), which can be either $h_1$ or $h_2$, with a
reduced coupling to the gauge bosons $R_1 \simeq
1$ or $R_2\simeq 1$, respectively. This state
decays dominantly to a pair of (very) light CP-odd
states, $a_1a_1$, with $m_{a_1}$ between 5 and
65~GeV.  The singlet component of $a_1$ cannot be
dominant if we are to have a large $h_1 \to a_1
a_1$ or $h_2\to a_1a_1$ branching ratio when the
$h_1$ or $h_2$, respectively, is the SM-like Higgs
boson.  Further, when the $h_1$ or $h_2$ is very
SM-like, one has small $Zh_1a_1$ or $Zh_2a_1$ coupling,
respectively, so that $e^+ e^- \to
h_1 a_1$ or $e^+e^-\to h_2 a_1$ associated
production places no constraint on the light
CP-odd state at LEP. We have selected six
difficult benchmark points, displayed in
Table~\ref{tpoints}.  These are such that
$a_1\to\cnone\cnone$ decays are negligible or
forbidden.  (Techniques for cases such that 
$\cnone\cnone$ decay modes are important
are under development.)
For points 1 -- 3, $h_1$ is the
SM-like CP-even state, while for points 4 -- 6 it
is $h_2$. We have selected the points so that there is
some variation in the 
$h_{1,2}$ and $a_1$ masses. The
main characteristics of the benchmark points are
displayed in Table~\ref{tpoints}. Note the large
$\br(h\to a_1 a_1)$ of the SM-like $h$ ($h=h_1$
for points 1 -- 3 and $h=h_2$ for points 4 --6).
For points 4 -- 6, with $m_{h_1}<100\gev$, the
$h_1$ is mainly singlet.  As a result, the $Zh_1a_1$
coupling is very small, implying no LEP constraints on the
$h_1$ and $a_1$ from $e^+e^-\to h_1 a_1$
production.

We note that in the case of the points 1 -- 3, the
$h_2$ would not be detectable either at the LHC or
the LC. For points 4 -- 6, the $h_1$, though
light, is singlet in nature and would not be
detectable.  Further, the $h_3$ or $a_2$ will only
be detectable for points 1 -- 6 if a super high
energy LC is eventually built so that $e^+e^-\to
Z\to h_3 a_2$ is possible.  Thus, we will focus on
searching for the SM-like $h_1$ ($h_2$) for points
1 -- 3 (4 -- 6) using the dominant $h_1(h_2)\to
a_1a_1$ decay mode.

In the case of points 2 and 6, the $a_1\to
\tau^+\tau^-$ decays are dominant. The final state
of interest will be $jj\tau^+\tau^-$, where the
$jj$ actually comes primarily from
$a_1a_1\to\tau^+\tau^-\tau^+\tau^-$ followed by jet
decays of two of the $\tau$'s: $\tau^+\tau^-\to
jj+\nu's$.  (The contribution from direct $a_1\to jj$ decays
to the $jj\tau^+\tau^-$ final state is relatively
small for points 2 and 6.)
In what follows, when we speak of
$\tau^+\tau^-$, we refer to those $\tau$'s that
are seen in the $\tau^+\tau^-\to
\ell^+\ell^-+\nu's$ final state ($\ell=e,\mu$).
  For points 1 and 3
-- 5 $\br(a_1\to b\anti b)$ is substantial.  The
relevant final state is $b\anti b \tau^+\tau^-$.
Nonetheless, we begin with a study of the backgrounds and
signals without requiring $b$-tagging.  
With our latest cuts, 
we will see that $b$-tagging is not necessary 
to overcome the apriori large Drell-Yan
$\tau^+\tau^-$+jets background.  It is eliminated
by stringent cuts for finding the highly energetic
forward / backward
jets characteristic of the $WW$ the fusion  process.
As a result, we will find good signals for all
6 of our points.

\def\cO#1{{\cal{O}}\left(#1\right)}
\def\nn {\nonumber}
\renewcommand{\ee}{\end{equation}}
\newcommand{\bn}{\begin{enumerate}}
\newcommand{\en}{\end{enumerate}}
\newcommand{\bc}{\begin{center}}
\newcommand{\ec}{\end{center}}
\newcommand{\ul}{\underline}
\newcommand{\ol}{\overline}
\newcommand{\ar}{\rightarrow}
\newcommand{\sm}{${\cal {SM}}$}
\newcommand{\as}{\alpha_s}
\newcommand{\aem}{\alpha_{em}}
\newcommand{\ycut}{y_{\mathrm{cut}}}
\newcommand{\susy}{{{SUSY}}}
\newcommand{\Dir}{\kern -6.4pt\Big{/}}
\newcommand{\Dirin}{\kern -10.4pt\Big{/}\kern 4.4pt}
\newcommand{\DDir}{\kern -10.6pt\Big{/}}
\newcommand{\DGir}{\kern -6.0pt\Big{/}}
\def\Ecm{\ifmmode{E_{\mathrm{cm}}}\else{$E_{\mathrm{cm}}$}\fi}
\def\gluino{\ifmmode{\mathaccent"7E g}\else{$\mathaccent"7E g$}\fi}
\def\photino{\ifmmode{\mathaccent"7E \gamma}\else{$\mathaccent"7E \gamma$}\fi}
\def\mgluino{\ifmmode{m_{\mathaccent"7E g}}
             \else{$m_{\mathaccent"7E g}$}\fi}
\def\taugluino{\ifmmode{\tau_{\mathaccent"7E g}}
             \else{$\tau_{\mathaccent"7E g}$}\fi}
\def\mphotino{\ifmmode{m_{\mathaccent"7E \gamma}}
             \else{$m_{\mathaccent"7E \gamma}$}\fi}
\def\ML{\ifmmode{{\mathaccent"7E M}_L}
             \else{${\mathaccent"7E M}_L$}\fi}
\def\MR{\ifmmode{{\mathaccent"7E M}_R}
             \else{${\mathaccent"7E M}_R$}\fi}

\def\lsim{\buildrel{\scriptscriptstyle <}\over{\scriptscriptstyle\sim}}
\def\gsim{\buildrel{\scriptscriptstyle >}\over{\scriptscriptstyle\sim}}
\def\Jnl #1#2#3#4 {#1 {\bf #2} (#3) #4}
\def\NPB {{\rm Nucl. Phys.} {\bf B}}
\def\PLB {{\rm Phys. Lett.}  {\bf B}}
\def\PRL {\rm Phys. Rev. Lett.}
\def\PRD {{\rm Phys. Rev.} {\bf D}}
\def\ZPC {{\rm Z. Phys.} {\bf C}}
\def\EPJC {{\rm Eur. Phys. J.} {\bf C}}
\def\Ord{\lower .7ex\hbox{$\;\stackrel{\textstyle <}{\sim}\;$}}
\def\OOrd{\lower .7ex\hbox{$\;\stackrel{\textstyle >}{\sim}\;$}}
\def\eps{\epsilon}

In principle, one could explore final states other than 
$b\anti b \tau^+\tau^-$ (or $jj\tau^+\tau^-$ for points 2
and 6). However, all other channels will be much more
problematical at the LHC. A $4b$-signal would
be burdened by a large QCD background even after
implementing $b$-tagging.  A $4j$-signal would be
completely swamped by QCD background.  Meanwhile,
the $4\tau$-channel (by which we mean that all
decay leptonically) would not allow one to
reconstruct the $h_1,h_2$ resonances.  

In the case of the $2b2\tau$ (or
$2j2\tau$) signature, we identify the $\tau$'s
through their leptonic decays to electrons
and muons. Thus, they will yield some amount of
missing (transverse) momentum, $p_{\rm miss}^T$.
This missing transverse momentum can be projected
onto the visible $e,\mu$-momenta in an attempt to
reconstruct the parent $\tau$-direction.

\noindent\underline{Results for the LHC}

Let us now focus on the $WW\to h\to aa$
channel that we believe provides 
the best hope for Higgs detection in these
difficult NMSSM cases.  (We reemphasize that the
$h_1$ [cases 1 -- 3] or $h_2$ [cases 4 -- 6] has
nearly full SM strength coupling to $WW$.)
The $b\anti b\tau^+\tau^-$ (or
$2j\tau^+\tau^-$, for points 2 and 6) final state
of relevance is complex and subject to large
backgrounds, and the $a_1$ masses of interest are
very modest in size.  In order to extract the $WW$ fusion
$2j2\tau$ NMSSM Higgs boson signature, it is crucial
to strongly exploit forward
and backward jet tagging on the light quarks
emerging after the double $W$-strahlung preceding
$WW$-fusion.  We also require two additional central
jets (from one of the $a$'s) and two opposite sign
central leptons ($\ell=e,\mu$) coming from the 
the $\tau^+\tau^-$ emerging from the decay of the other
$a$. By imposing stringent
 forward / backward jet tagging cuts, we remove the
otherwise very large background from Drell-Yan
$\tau^+\tau^-+jets$ production. 
In the end, the most important background is due
to $t\anti t$ production and decay via the purely
SM process, $gg\to t\bar t\to b\bar b W^+W^-\to
b\bar b \tau^+\tau^- + p_{\rm miss}^T,$ in
association with forward and backward jet radiation.

We have employed numerical simulations based on a version of
{\tt HERWIG v6.4}~\cite{Moretti:2002eu,Corcella:2001wc,Corcella:2000bw}
modified to allow for appropriate NMSSM couplings
and decay rates. Calorimeter emulation
was performed using the {\tt GETJET} code
\cite{GETJET}. 
Since the $a_1$ will not have been detected
  previously, we must assume a value for
  $m_{a_1}$.  In practice, it will be necessary to
  repeat the analysis for densely spaced $m_{a_1}$
  values and look for the $m_{a_1}$ choice that
  produces the best signal.
 We look among the central jets for the
  combination with invariant mass $M_{jj}$ closest
  to $m_{a_1}$. In the top plot of Fig.~\ref{MH}, we show
the $M_{jj\tau^+\tau^-}$ invariant mass distribution
obtained after cuts, but before $b$-tagging
or inclusion of $K$ factors 
--- the plot presented assumes
that we have hit on the correct $m_{a_1}$ choice.

\begin{figure}[p]
\begin{center}
\centerline{$~~~~~~~~~~~~~~~$LHC, $\sqrt{s_{pp}}=14$ TeV}
\centering\epsfig{file=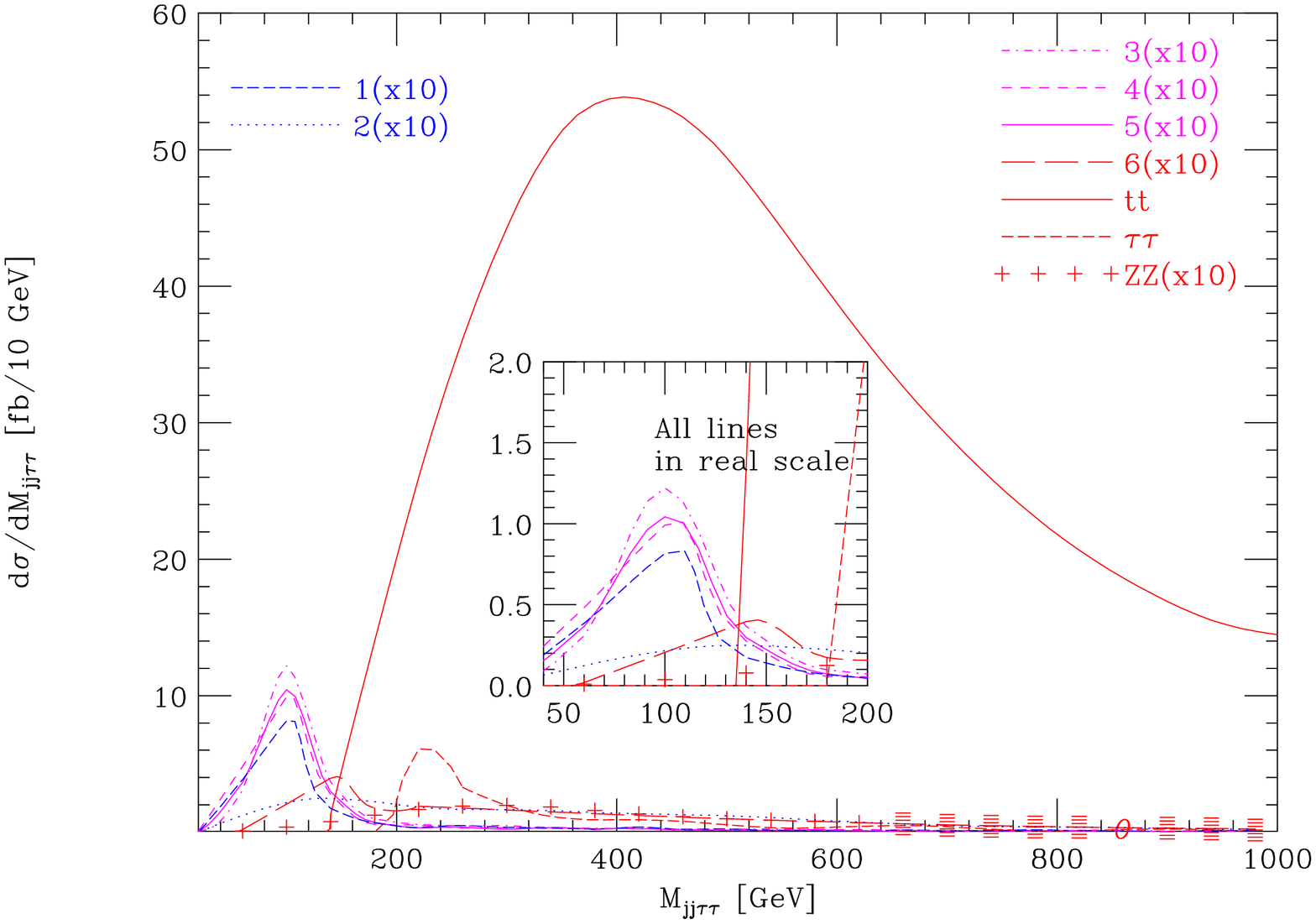,angle=90,height=8cm,width=14cm}

\vspace*{1.0truecm}

\centerline{$~~~~~~~~~~~~~~~$LC, $\sqrt{s_{e^+e^-}}=800$~GeV}
\centering\epsfig{file=sec251_AllMH_LC.ps,angle=90,height=8cm,width=14cm}
\centerline{}

\noindent
\vspace{-1.0cm}
\caption{\footnotesize We plot $d\sigma/dM_{jj\tau^+\tau^-}$ [fb/10~GeV] vs $M_{jj\tau^+\tau^-}$~[GeV]
for signals and backgrounds after basic
event selections, but before $b$ tagging. The upper (lower)
figure is that for the LHC (LC). 
In both plots, the lines corresponding to points 4 and 5
are visually indistinguishable. No $K$ factors are included.
}
\label{MH}
\end{center}
\end{figure}

The selection strategy adopted is 
a more refined (as regards
forward / backward jet tagging)
version of that summarized in \cite{ourhep}.
It is clearly
efficient in reconstructing the $h_1$ (for points
1--3) and $h_2$ (for points 4--6) masses from the
$jj \tau^+\tau^-$ system, as one can appreciate by
noting the peaks appearing in the LHC plot of
Fig.~\ref{MH} at
$M_{jj\tau^+\tau^-}\approx100$~GeV. In contrast,
the heavy Higgs resonances at $m_{h_2}$ for points
1--3 and the rather light resonances at $m_{h_1}$
for points 4--6 (recall Table~\ref{tpoints}) do
not appear, the former mainly because of the very
poor production rates and the latter due to the
fact that either the $h_1\to a_1 a_1$ decay mode
is not open (points 4, 5) or -- if it is -- the
$b$-quarks and $e/\mu$-leptons eventually emerging
from the $a_1$ decays are too soft to pass the
acceptance cuts (point 6, for which
$m_{a_1}=7$~GeV and $m_{h_1}=51$~GeV).  For all
six NMSSM setups, the Higgs resonance produces a
bump below the end of the low mass tail of the
$t\bar t$ background (see the insert in the top
frame of Fig.~\ref{MH}).  Note how small the DY
$\tau^+\tau^-$ background is after strong
forward / backward jet tagging.  Since the main
surviving background is from $t\anti t$ production,
$b$ tagging is not helpful.  For points 2 and 6,
for which the signal has no $b$'is in the final state,
anti-$b$-tagging might be useful, but has not been
considered here.

To estimate $S/\sqrt B$, we assume $L=300~{\rm
  fb}^{-1}$, a $K$ factor of 1.1 for the $WW$
fusion signal and $K$ factors of 1, 1 and 1.6 for
the DY $\tau^+\tau^-$, $ZZ$ production and $t\anti
t$ backgrounds, respectively.  (These $K$ factors
are not included in the plot of Fig.~\ref{MH}.)
We sum events over the region $40\leq
M_{jj\tau^+\tau^-}\leq 150$~GeV.  
(Had we only included masses below $130$~GeV,
we would have had no $t\anti t$ background,
and the $S/\sqrt B$ values would be enormous.  However,
we are concerned that this absence of $t\anti t$
background below $130\gev$ might be a reflection
of limited Monte Carlo statistics.  As a result
we have taken the more conservative approach of
at least including the first few bins for which
our Monte Carlo does predict some $t\anti t$ background.)

For points 1, 2, 3,
4, 5 and 6, we obtain signal rates of about $S=1636$, 702,
  2235,  2041, 2013, and  683, respectively.
The $t\anti t$+jets background rate is $B_{tt}\sim
795$. The $ZZ$ background rate is $B_{ZZ}\sim 6$.
The DY $\tau^+\tau^-$ background rate is
negligible. (We are continuing to increase our statistics
to get a fully reliable estimate.)
The resulting $N_{SD}=S/\sqrt B$ values for points 1-6
are 50,  22, 69,  63,  62, and  21, respectively.
The smaller values for points 2 and 6 are simply
a reflection of the difficulty of
isolating and reconstructing
the two jets coming from the decay of a very light $a_1$.
Overall, these preliminary results are very encouraging
and suggest that a no-lose theorem for NMSSM Higgs
detection at the LHC is close at hand.

\noindent\underline{The LC scenario}

While further examination of and refinements in the LHC analysis may
ultimately lead us to have full confidence in the viability of the
NMSSM Higgs boson signals discussed above, an enhancement at low
$M_{b\anti b\tau^+\tau^-}$ of the type shown (for some choice of
$m_{a_1}$) will nonetheless be the only evidence on which a claim of
LHC observation of Higgs bosons can be based.  
Ultimately, a means of confirmation and further study will be
critical.  Thus, it is important to summarize the prospects at the LC,
with energy up to 800~GeV, in the context of the difficult scenarios 1
--- 6 of Table~\ref{tpoints} discussed here. In the following, $h=h_1$
for points 1--3 and $h=h_2$ for points 4--6 in Table~\ref{tpoints}.

Because the $ZZh$ coupling is nearly full strength
in all cases, and because the $h$ mass is of order
100~GeV, discovery of the $h$ will be very
straightforward via $e^+e^-\to Z h$ using the
$e^+e^-\to ZX$ reconstructed $M_X$ technique which
is independent of the ``unexpected'' complexity of
the $h$ decay to $a_1a_1$. This will immediately
provide a direct measurement of the $ZZh$ coupling
with very small error~\cite{review}.  This
approach is completely independent of the final
state decay mode(s) and will thus work for all the
points 1 -- 6, including, in particular, points 2
and 6 for which the final state does not contain
$b$'s.  The next stage will be to look at rates
for the various $h$ decay final states, $F$, and
extract $BR(h\to F)=\sigma(e^+e^-\to Zh\to
ZF)/\sigma(e^+e^-\to Zh)$.  For the NMSSM points 1
and 3 -- 5, the main channels would be $F=b\anti b
b\anti b$, $F=b\anti b \tau^+\tau^-$ and
$F=\tau^+\tau^-\tau^+\tau^-$.  For points 2 and 6,
the relevant final states are
$F=4j,jj\tau^+\tau^-,\tau^+\tau^-\tau^+\tau^-$.
At the LC, a fairly accurate determination of
$BR(h\to F)$ should be possible for both sets of
the three final states $F$. This information would
allow us to determine $BR(h\to a_1 a_1)$
independently.

Here, we consider the equally (or perhaps more)
useful vector-vector fusion mode that will be
active at a LC.  At 800~GeV or above, it is the
dominant Higgs boson production channel for
CP-even Higgs bosons in the intermediate mass
range. Contrary to the case of the LHC though, the
dominant contribution (from $WW$ fusion) does not
allow for forward and backward particle tagging,
as the incoming electron and positron convert into
(anti)neutrinos, which escape detection. Although
the $ZZ$ fusion contribution would allow tagging
of forward/backward $e^-$ and $e^+$, the cross
section is a factor of 10 smaller~(see Fig. 4 of
Ref.~\cite{review}) in comparison.  
At the LC the $ZZ$ background plays a fairly
significant role. It has been simulated in our
{\tt HERWIG} and (LC-adjusted) {\tt GETJET}
numerical analysis.

At a LC, the optimal signature will thus be
different than at the LHC and a different set of
selection criteria are needed
(see \cite{ourhep} for details). 
We have chosen selection
criteria that retain both the $WW$ and the
$ZZ$ fusion Higgs boson production processes.
Basic requirements include the presence of at least 
two central jets --- 
we look among the central jets for 
the combination with invariant mass $M_{jj}$ 
closest to $m_{a_1}$. We also require
two oppositely charged central leptons ($\ell=e,\mu$). After ensuring that these are not back-to-back, 
we resolve the $p^T_{\rm miss}$ along their 
directions and reconstruct
the invariant mass $M_{\tau^+\tau^-}$.
Finally, we note that we have included Initial State
Radiation (ISR) and beam-strahlung effects, as
predicted using the {\tt HERWIG} default.  These
tend to introduce an additional unresolvable
missing longitudinal momentum, although to a much
smaller extent than do the Parton Distribution
Functions (PDFs) in hadron-hadron scattering at
the LHC. Further discussion of the details of
the cuts and simulation will be presented elsewhere. 

At the LC, the $jj\tau^+\tau^-$ background is again very
small. As a result, we do not need to employ
$b$ tagging  --- it is sufficient to simply require two 
non-forward / backward jets; these happen to be 
$b$'s for points 1 and 3 -- 5 and light quark jets for points 2 and 6.

The bottom plot of Fig.~\ref{MH}
gives the resulting $M_{jj\tau^+\tau^-}$ invariant mass 
distributions for the signal and several backgrounds. 
We note that it is not fruitful to place cuts on the 
invariant masses
$M_{jj}$ and $M_{\tau^+\tau^-}$ that exclude
$M_{jj},M_{\tau^+\tau^-}\sim m_Z$ 
in an attempt to reduce the $ZZ$
background. This is because the SM-like $h$ mass is 
typically of order $115$~GeV, \ie\ not so far from $m_Z$,
and the experimental resolutions in the two masses $M_{jj}$
and $M_{\tau^+\tau^-}$ are poor, 
either because of the large number
of hadronic tracks or the missing 
longitudinal momenta of the (anti)
neutrinos, respectively.

From Fig.~\ref{MH}, we see that the
$M_{jj\tau^+\tau^-}$ distribution reconstructed at
the LC displays resonance mass peaks (again
centered at 100~GeV) for the SM-like $h_1$ (points
1 -- 3) or $h_2$ (points 4 -- 6) that are very
clearly visible above both the $t\anti t$ and $ZZ$
backgrounds, particularly for the case of points 2
and 6 (see insert in the bottom frame of
Fig.~\ref{MH}).  Assuming $L=500~{\rm fb}^{-1}$,
the points 1,3,4,5 yield 36,45,45,45 events 
in the $40\leq M_{jj\tau^+\tau^-}\leq 120\gev$
interval where the
background is essentially zero. 
This would constitute a convincing
signal given the very small size predicted for the
background.  For points 2 and 6,
we get about 320 events over this same background-free
mass interval. The much
larger signal for these points is mainly due to two reasons.
Firstly, the overall $BR(h_1/h_2\to a_1a_1\to 2j 2\ell)$
is largest in these cases, as follows 
from the branching ratios 
in Table~\ref{tpoints} and simple combinatorics. 
Secondly, there is a kinematic difference
related to the much smaller $m_{a_1}$ in cases 2 and 6
that we now discuss. Notice that, although assigning the
entire missing transverse momentum to the
$\tau$-lepton system may seem not entirely
appropriate (given the forward/backward
(anti)neutrinos from the incoming electrons and
positrons in $WW$ fusion), this does not hamper
the ability to reconstruct the Higgs mass peaks.
However, there will be a proportion of the signal
events that tend to reproduce the overall $\sqrt{
  s_{e^+e^-}}$ value in the $M_{jj\tau^+\tau^-}$
distribution. The effect is more pronounced for
points 1 and 3--5, which is where the $a_1$ mass
is larger (see Table~\ref{tpoints}) so that most
of the hadronic tracks composing the emerging jets
easily enter the detector region. For points 2 and
6, where $m_{a_1}$ is below 10~GeV, this may often
not be true and it appears that the consequent
effect of these hadrons escaping detection is that
of counterbalancing the $p^T_{\rm{miss}}$
contributions related to the neutrinos left behind
in $WW$ fusion reactions.  For the case of the
$ZZ$ noise, in the limit of full coverage and
perfect resolution of the detector, one would have
$M_{jj\tau^+\tau^-}\equiv\sqrt{s_{e^+e^-}}$, which
explains the concentration of events with
$M_{jj\tau^+\tau^-}$ around 800~GeV. (The
``tails'' beyond $\sqrt{s_{e^+e^-}}$ are due to
the smearing of the visible tracks in our Monte
Carlo analysis.)

In summary, we have obtained a statistically very significant
LHC signal in the $jj\tau^+\tau^-$ final
state of $WW$ fusion for cases in which the NMSSM parameters 
are such that
the most SM-like of the CP-even Higgs bosons, $h$,
is relatively light and decays primarily to a pair
of CP-odd Higgs states, $h\to aa$ with $a\to b\anti b,\tau^+\tau^-$ if $m_a>2m_b$ or $a\to jj,\tau^+\tau^-$
if $m_a<2m_b$. 
 The statistical significances are (at least)
of order 50 to 70 for points with $m_a>2m_b$
and of order 20 for points with $m_a<2m_b$.
These high significances were obtained by
imposing stringent cuts requiring
highly energetic forward / backward jets
in order to isolate the $WW$ fusion signal process
from backgrounds such as DY $\tau^+\tau^-$ pair 
production. Still, this signal will
be the only evidence for Higgs bosons at the LHC. 
The LC will be absolutely essential in order to confirm that
the enhancement seen at the LHC really does
correspond to a Higgs boson. At the LC, discovery
of a light SM-like $h$ is guaranteed to be
possible in the $Zh$ final state using the recoil
mass technique.  Further, we have seen that
$WW,ZZ$ fusion production of the $h$ will also
produce a viable signal in the $jj \tau^+\tau^-$
final state (and perhaps in the $4j$ and
$\tau^+\tau^- \tau^+\tau^-$ final states as well,
although we have not examined
these~\cite{preparation}); for some parameter
choices the $jj$ will be a $b\anti b$ pair as can
be determined by $b$-tagging. Finally, we have yet
to explore the cases in which the
$a_1\to\cnone\cnone$ decay has a large branching
ratio.  Detecting a Higgs signal in such cases
will require a rather different procedure.
Work on the $WW\to h\to {\rm invisible}$ signal
is in progress~\cite{preparation}.

As we have stressed, for parameter space points of
the type we have discussed here, detection of any
of the other MSSM Higgs bosons is likely to be
impossible at the LHC and is likely to require an
LC with $\sqrt{s_{e^+e^-}}$ above the relevant
thresholds for $h'a'$ production, where $h'$ and
$a'$ are heavy CP-even and CP-odd Higgs bosons,
respectively.  

Although results for the LHC
indicate that Higgs boson discovery will be possible
for the type of situations we have
considered, it is clearly important
to refine and improve the techniques for extracting a
signal. This will almost certainly be possible 
once data is in
hand and the $t\anti t$ background can be more
completely modeled.  

Clearly, if SUSY is
discovered and $WW\to WW$ scattering is found to
be perturbative at $WW$ energies of 1 TeV (and
higher), and yet no Higgs bosons are detected in
the standard MSSM modes, a careful search for the
signal we have considered should have a high
priority.  

Finally, we should remark that the
$h\to aa$ search channel considered here in the
NMSSM framework is also highly relevant for a
general two-Higgs-doublet model, 2HDM.  It is
really quite possible that the most SM-like
CP-even Higgs boson of a 2HDM will decay primarily
to two CP-odd states.  This is possible even if
the CP-even state is quite heavy, unlike the NMSSM
cases considered here. If CP violation is introduced
in the Higgs sector, either at tree-level
or as a result of one-loop corrections (as, for example,
is possible in the MSSM),
$h\to h' h''$ decays will generally be present. The 
critical signal will be the same as
that considered here.


\subsection{An interesting NMSSM scenario at the LHC and LC}

{\it D.J.~Miller and S.~Moretti}

\subsubsection{Introduction}

The Next-to-Minimal Supersymmetric Standard Model (NMSSM) provides an
elegant solution to the $\mu$ problem of the MSSM by introducing an
extra complex scalar Higgs superfield. The extra fields have no gauge
couplings and are principally only manifest through their mixing with the
other states. This leads to scenarios where Higgs boson couplings are
reduced in comparison to the MSSM, presenting a challenge to the next
generation of colliders. In this contribution, we will examine the
phenomenology of one of these scenarios at the LHC and a future
$e^+e^-$ Linear Collider and demonstrate a synergy between the two
machines.

The NMSSM has already been discussed in Section 3.4.1 of this study,
in the context of establishing a ``no-lose'' theorem for the discovery
of at least one Higgs boson at the next generation of colliders (see
also Ref.\cite{Ellwanger:2003jt}). It was seen that for some
exceptional NMSSM parameter choices the discovery of {\it any} Higgs
boson at all will be difficult at the LHC, but for the majority of
choices at least one Higgs boson will be discovered. Here we adopt a
different philosophy and examine a ``typical'' NMSSM scenario
point. While not representative of scenarios over the entire range of
parameters, the chosen scenario is certainly not unusual and a wide
range of parameter choices will result in similar phenomenology,
differing only in numerical detail and not in general structure. This
scenario therefore presents an interesting illustrative picture of the
Higgs sector that might be waiting for us at the next generation of
colliders.

\subsubsection{The Model}

The NMSSM has the same field content as the minimal model augmented by
an additional neutral singlet superfield $\hat{S}$. Its superpotential
is given by
\begin{equation}
W=\hat{u}^c \, \mathbf{h_u} \hat{Q} \hat{H}_u  
-\hat{d}^c \, \mathbf{h_d} \hat{Q} \hat{H}_d  
-\hat{e}^c \, \mathbf{h_e} \hat{L} \hat{H}_d  
+\lambda \hat{S}(\hat{H}_u \hat{H}_d)+\frac{1}{3}\kappa\hat{S}^3,
\label{eq:superpotential} 
\end{equation}
where $\hat{H}_u$ and $\hat{H}_d$ are the usual Higgs doublet
superfields with $\hat{H}_u\hat{H}_d \equiv \hat{H}_u^+\hat{H}_d^- -
\hat{H}_u^0\hat{H}_d^0$. $\hat{Q}$ and $\hat{L}$ represent left
handed quark and lepton weak isospin doublets respectively, while
$\hat{u}^c$, $\hat{d}^c$ and $\hat{e}^c$ are the right handed quark
and lepton fields; $\mathbf{h_u}$, $\mathbf{h_d}$ and $\mathbf{h_e}$
are matrices of Yukawa couplings where family indices have been
suppressed.  The usual $\mu$-term of the MSSM, $\mu \hat{H}_u
\hat{H}_d$, has been replaced by a term coupling the new singlet field
to the usual Higgs doublets, $\lambda \hat{S} \hat{H}_u
\hat{H}_d$. When the new singlet field gains a vacuum expectation
value (VEV), an effective $\mu$-term is generated with an effective
Higgs-higgsino mass parameter given by $\mu_{\rm eff}=\lambda \langle
S \rangle$. (We adopt the notation that the superfields are denoted by
expressions with a ``hat'', while their scalar components are denoted
by the same expression without the hat.) The superpotential resulting
from this substitution (not yet Eq.(\ref{eq:superpotential})) contains
an extra symmetry --- a $U(1)$ ``Peccei-Quinn'' (PQ)
symmetry~\cite{Peccei:1977hh}, which will be broken during electroweak
symmetry breaking. As is the case when any global symmetry is
dynamically broken, this results in a massless Nambu-Goldstone boson
which is in this instance a pseudoscalar Higgs state. Since this Higgs
state has not been observed in experiment we have only two
possibilities: we must either break the Peccei-Quinn symmetry
explicitly, giving the pseudoscalar a mass and putting it out of the
kinematical reach of past experiments, or we must decouple it from the
other particles by setting $\lambda \ll 1$. Here we adopt the former
possibility\footnote{For a description of the decoupled case, see
Ref.\cite{Miller:2003hm}.} and introduce an explicit Peccei-Quinn
symmetry breaking term $\frac{1}{3}\kappa\hat{S}^3$. This results in
the superpotential given in Eq.(\ref{eq:superpotential}). We will not
elaborate on the formal details of the model here except to elucidate
our parameter choice --- for a more detailed examination of the model
see Ref.\cite{Miller:2003ay} and references therein.

At tree level, the NMSSM Higgs sector has seven parameters: the Higgs
couplings from the superpotential, $\lambda$ and $\kappa$; their two
associated soft supersymmetry breaking parameters, $A_{\lambda}$ and
$A_{\kappa}$ ; and the VEVs of the three neutral Higgs fields, which
we re-express as two ratios of VEVs, $\tan \beta=\langle H_u^0 \rangle
/ \langle H^0_d \rangle$ and $\tan \beta_s=\sqrt{2} \langle S
\rangle/v$, and the electroweak scale $v/\sqrt{2}=\sqrt{\langle H^0_u
\rangle^2+\langle H^0_d \rangle^2}$.  The scenario to be considered
here has parameters given by $\lambda=0.3$, $\kappa=0.1$, $\tan \beta
= \tan \beta_s =3$ and $A_{\kappa}=-60$~GeV. The parameter
$A_{\lambda}$ is replaced by the mass scale $M_A$ which is chosen to
be the diagonal entry of the pseudoscalar Higgs boson mass-squared
matrix that returns to the value of the physical MSSM pseudoscalar
Higgs boson mass in the MSSM limit (i.e.\ $\lambda \to 0$, $\kappa \to
0$ while keeping $\lambda/\kappa$ and $\mu_{\rm eff}$ fixed). This
choice allows the reader a more intuitive connection with the
MSSM. $M_A$ will not be fixed, but will be allowed to vary over the
physical range. Finally, we take $v=246$~GeV.

\subsubsection{The Mass Spectrum}

The Higgs mass spectrum for our parameter choice, evaluated at
one-loop precision~\cite{Kovalenko:dc}, can be seen in Fig.(\ref{fig:masses}),
\begin{figure}[htb]
\begin{center}
\includegraphics[scale=0.6]{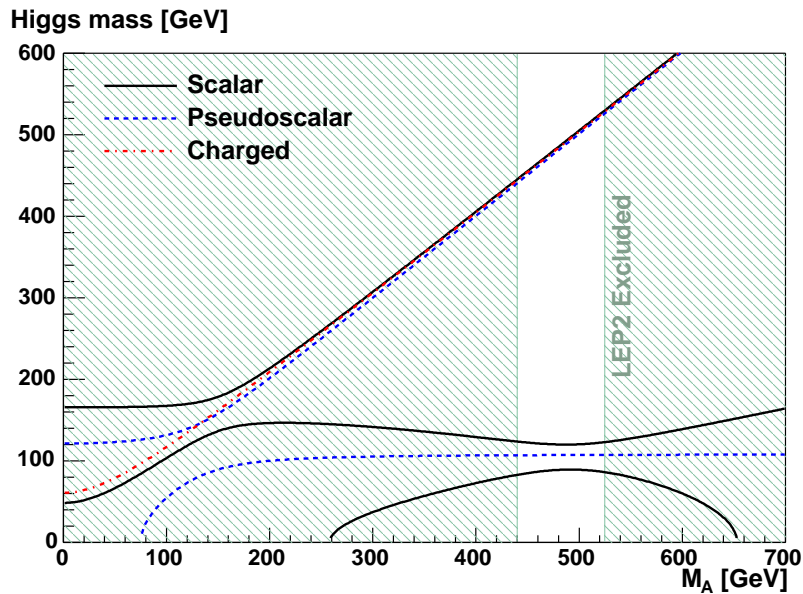}
\caption{\it 
The one-loop Higgs mass spectrum as a function of $M_A$ for
$\lambda=0.3$, $\kappa=0.1$, $\tan \beta= \tan \beta_s =3$ and
$A_{\kappa}=-60$~GeV. Also shown by the shaded area are the values of
$M_A$ that are ruled out by LEP2~\cite{sec2_Barate:2003sz} for this
parameter choice.}
\label{fig:masses}
\end{center}
\end{figure}
as a function of $M_A$. This spectrum looks remarkably like that of
the MSSM with the addition of two extra Higgs fields --- a scalar
state and a pseudoscalar state. As in the MSSM, the heavy
pseudoscalar, scalar and charged Higgs bosons all lie around the mass
scale $M_A$, while a lighter scalar state has mass around
$115$-$130$~GeV. However, in addition we see extra scalar and
pseudoscalar states with masses of order $100$~GeV and below; these
are the Higgs states which are dominated by the extra singlet degrees
of freedom.

Making an expansion in the (often) small parameters $1/\tan \beta$ and
$M_Z/M_A$ allows us to obtain simple approximate forms for the masses
of these extra singlet dominated Higgs
bosons~\cite{Miller:2003hm}. One finds that the singlet dominated
pseudoscalar Higgs fields has a mass given approximately by
\begin{equation}
M_{A_1}^2 \approx -\frac{3}{\sqrt{2}} \kappa v_s A_{\kappa}, \label{eq:mas}
\end{equation}
while the singlet dominated scalar has a mass which is maximized at 
$M_A \approx 2 \mu_{\rm eff}/\sin 2\beta$ where it is given by
\begin{equation}
M_{H_1}^2 \approx \frac{1}{2} \kappa v_s (4\kappa v_s+\sqrt{2}A_{\kappa}). \label{eq:mah}
\end{equation}
It must be stressed that these expressions are very approximate and
are not applicable over the entire parameter range; the one-loop
expressions for the masses should be used in preference, as in
Fig.(\ref{fig:masses}). However, the approximate expressions are
useful in determining the qualitative behaviour of the masses as the
parameters are varied. [Although approximate, these expressions do
surprisingly well in estimating the singlet dominated masses. For
example, for the present parameter choice they give $M_{A_1} \approx
96.2$~GeV and $M_{H_1} \approx 88.1$~GeV, which compare favourably
with the one-loop results, $107.3$~GeV and $89.5$~GeV respectively at
$M_A=495$~GeV. This is in part due to the suppression of couplings to
quarks, which reduces the impact of radiative corrections.]

In particular, the masses are strongly dependent only on the
quantities $\kappa v_s$ and $A_{\kappa}$ [and $M_A$].  The dependence
on $\kappa v_s$ (which is a measure of how strongly the PQ symmetry is
broken) is straightforward: as $\kappa v_s$ increased the masses also
increase.  Since one expects $v_s$ to be of the order of $v$ and
$\kappa$ is restricted by $\kappa^2+\lambda^2 \lesssim 0.5$ when one
insists on perturbativity up to the unification scale, it is natural
(though not mandatory) for this mass scale to be rather low, and the
extra Higgs states rather light. In contrast, the $A_{\kappa}$
contribution to the masses has opposite sign for scalar and
pseudoscalar. The dependence of the pseudoscalar mass,
Eq.(\ref{eq:mas}), on $A_{\kappa}$ indicates that $A_{\kappa}$ should
be negative, while Eq.(\ref{eq:mah}) insists that its absolute value
does not become too large. These effects are nicely summarized by the
approximate mass sum rule (at $M_A \approx 2
\mu_{\rm eff}/\sin 2\beta$):
\begin{equation}
M_{H_1}^2+\frac{1}{3}M_{A_1}^2 \approx 2 \; (\kappa v_s)^2.  \label{eq:sumrule}
\end{equation}
The overall scale for the masses is set by $\kappa v_s$, while
increasing the scalar mass leads to a decrease in the pseudoscalar
mass and vice versa.

Fig.(\ref{fig:masses}) also shows the values of $M_A$ that, for this
parameter choice, are already ruled out by LEP (the shaded
region). Although a SM Higgs boson with mass below $114.4$~GeV is now
ruled out with $95\%$ confidence by the LEP
experiments~\cite{sec2_Barate:2003sz}, lighter Higgs bosons are still
allowed if their coupling to the $Z$ boson is reduced. In the NMSSM,
since the extra singlet fields have no gauge couplings, the couplings
of the singlet dominated fields to the $Z$ boson come about only
through mixing with the neutral doublet Higgs fields. When this mixing
is small their couplings are reduced and they can escape the
Higgs-strahlung dominated LEP limits. For the LEP limits shown here we
take into account decays to both $b \bar b$~\cite{sec2_Barate:2003sz} and
$\gamma \gamma$~\cite{lhwg-2002-02}, as well as decay mode
independent searches carried out by the OPAL
detector~\cite{Abbiendi:2002qp}. As expected, the limits are dominated
by the decay $H_1 \to b \bar b$.

The dependence of the lightest Higgs boson mass on $M_A$ also makes a
prediction for the mass of the heavy states. The lightest Higgs boson
mass must be kept large enough to escape the current LEP limits.
However, since this mass decreases rapidly to either side of its
maximum (see Fig.(\ref{fig:masses}) we are forced to constrain $M_A$,
and thus the heavy Higgs boson masses, to around $M_A \approx 2
\mu_{\rm eff}/\sin 2\beta \approx \mu_{\rm eff} \tan\beta$.

There is still significant room for a rather light Higgs
bosons to be found the LHC and/or a LC. It is essential that these
light Higgs bosons be ruled out or discovered at the next generation
of colliders. In the following we will focus on the production of a
light singlet dominated scalar Higgs boson at the LHC and a LC and its
subsequent decay, but one should bear in mind that there is also
a light pseudoscalar Higgs boson which also deserves study.

\subsubsection{Branching ratios for the light scalar}

The dominant branching ratios of the lightest scalar Higgs boson are
shown in Fig.(\ref{fig:branchingratios})
\begin{figure}[htb]
\begin{center}
\includegraphics[scale=0.6]{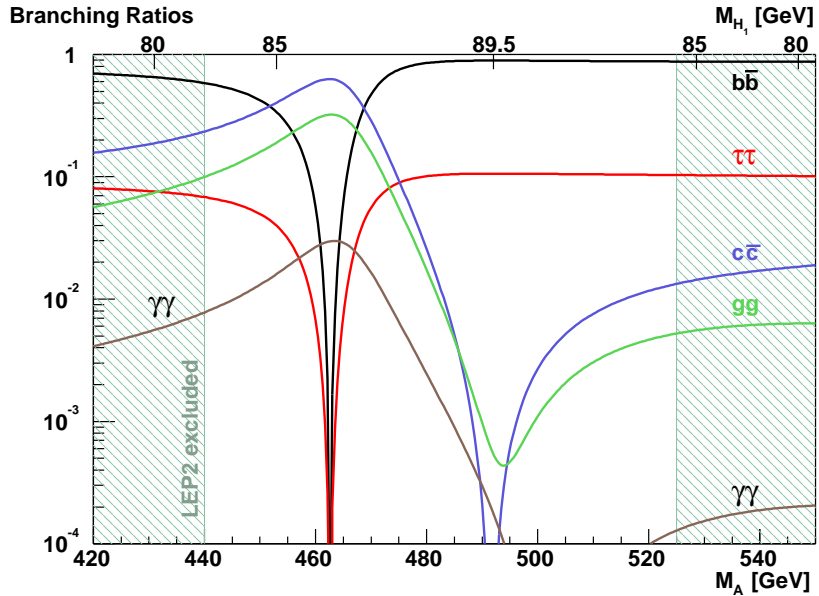}
\caption{\it 
The dominant branching ratios for the lightest scalar Higgs boson as a
function of $M_A$ for $\lambda=0.3$, $\kappa=0.1$, $\tan \beta= \tan
\beta_s =3$ and $A_{\kappa}=-60$~GeV. The complicated structure is due
to the switching off of the Higgs boson couplings to up-type and
down-type quarks and leptons.}
\label{fig:branchingratios}
\end{center}
\end{figure}
as a function of $M_A$. For a SM Higgs boson of the same mass (around
$80-90$~GeV) one would expect the dominant decays to be to bottom
quarks, $\tau$ leptons, and charm quarks, with the addition of loop
induced decays to gluons and photons. These are indeed also the
dominant decays of the singlet dominated scalar for most of the
allowed $M_A$ range, but the branching ratios now show significant
structure at approximately $463$~GeV and again at around $490$~GeV due
to the suppression of various couplings.

The couplings of the lightest Higgs scalar to up-type and
down-type quarks and leptons are given in terms of the SM Higgs
couplings by
\begin{eqnarray}
g^{\rm NMSSM}_{H_1u \bar u} &=& 
(\phantom{-} O^H_{11} \cot \beta + O^H_{21})\; g^{\rm SM}_{Hu \bar u}, \label{eq:ghuu} \\
g^{\rm NMSSM}_{H_1d \bar d} &=& 
(- O^H_{11} \tan \beta + O^H_{21})\; g^{\rm SM}_{Hd \bar d}, \label{eq:ghdd} 
\end{eqnarray}
respectively, where $O^H_{11}$ and $O^H_{21}$ are elements of the
scalar Higgs mixing matrix. The relative minus sign between terms in
Eq.(\ref{eq:ghuu}) and Eq.(\ref{eq:ghdd}) has the same origin as the
relative minus sign between the $hu \bar u$ and $hd \bar d$ couplings
in the MSSM.

The first structure seen in Fig.(\ref{fig:branchingratios}), at around
$463$~GeV, is due to the cancellation of $-O^H_{11} \tan \beta$ with
$O^H_{21}$ in Eq.(\ref{eq:ghdd}), forcing the $H_1 \to b \bar b$ and
$H_1 \to \tau^+ \tau^-$ branching ratios to vanish. As $M_A$ is
increased, $O^H_{21}$ passes smoothly through zero, eventually
canceling with $O^H_{11} \cot \beta$ in Eq.(\ref{eq:ghuu}). This
provides the structure at around $490$~GeV where the $H_1 \to c \bar
c$ branching ratio vanishes. 

The decays to $gg$ and $\gamma \gamma$ are mediated by loop diagrams
giving a more complex behaviour. $H_1 \to gg$ is dominated by top and
stop loops and consequently shows a marked decrease as the $H_1 t \bar
t$ coupling switches off; although the top-loop contribution will pass
through zero here, stop loops and bottom (s)quark loops prevent the
branching ratio from vanishing. In addition to top and bottom (s)quark
loops the $\gamma \gamma$ branching ratio is mediated by virtual $W$
bosons, charged Higgs bosons and charginos. The dominant effect is
from the $W$ bosons and the top loops and so we see a broad
suppression over the range where these couplings vanish.

\subsubsection{LHC Production}

Cross-sections for the production of the lightest scalar Higgs boson
in various channels at the LHC are shown in Fig.(\ref{fig:lhc}).
\begin{figure}[htb]
\begin{center}
\includegraphics[scale=0.6]{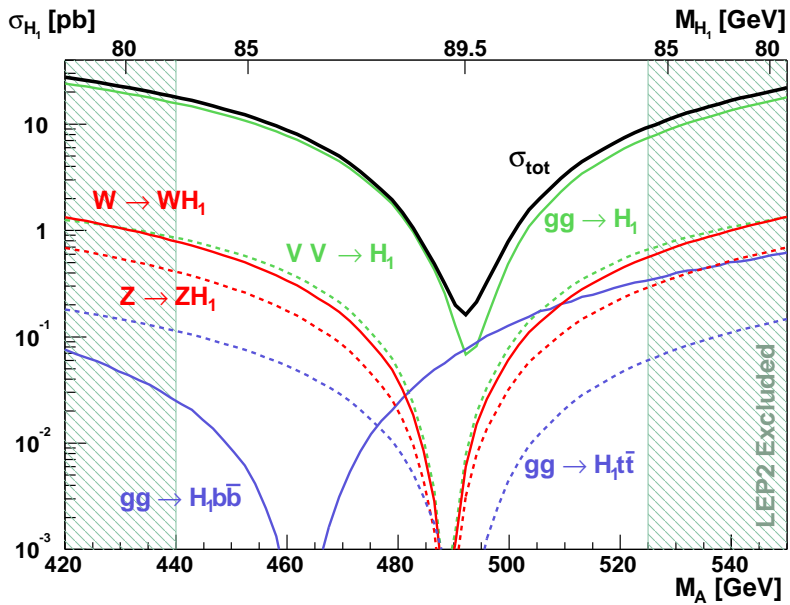}
\caption{\it 
Production cross-sections for the lightest scalar Higgs boson at the
LHC, as a function of $M_A$ for $\lambda=0.3$, $\kappa=0.1$, $\tan
\beta= \tan \beta_s =3$ and $A_{\kappa}=-60$~GeV. }
\label{fig:lhc}
\end{center}
\end{figure}
The total production cross-section is dominated by gluon-gluon fusion,
and is sizable over the entire range. Other significant production
channels are vector boson fusion ($VV \to H_1$), Higgs-strahlung ($W
\to WH_1$ and $Z \to ZH_1$) and associated production together with
top and bottom quarks ($gg \to H_1 t \bar t$ and $gg \to H_1 b \bar b$
respectively). As we saw for the branching ratios we again see
structures which are associated with the couplings of the Higgs boson
to various particles passing through zero. However, in contrast to the
earlier discussion, there are now three, rather than two, significant
values of $M_A$ where structure appears. The coupling of the Higgs
boson to a vector boson $V=W,\; Z$ with respect to the SM is given by
\begin{equation}
g^{\rm NMSSM}_{H_1VV}=O^H_{21}\; g^{\rm SM}_{H_1VV} \label{eq:ghvv},
\end{equation}
where $O^H_{21}$ is the same element appearing in
Eqs.(\ref{eq:ghuu}--\ref{eq:ghdd}), so when this mixing element
vanishes the vector boson fusion and Higgs-strahlung cross-sections
will disappear. This $M_A$ point is very close to the point where the
$H_1t \bar t$ couplings vanish because the first term on the
right-hand-side of Eq.(\ref{eq:ghuu}) is suppressed by $1/\tan \beta$.

For the lower values of $M_A$, where the Higgs decay to $b \bar b$ is
suppressed, this Higgs boson may be visible via its decay to $\gamma
\gamma$ (with a branching ratio $\gtrsim 0.1\%$ for $M_A \lesssim
480$~GeV). However, as the $\gamma \gamma$ branching ratio is turned
off at higher $M_A$, seeing this Higgs boson will become much more
challenging. Although the cross-section remains relatively large, the
Higgs boson almost always decays hadronically and the signal has a
very large QCD background. The only significant non-hadronic decay is
the Higgs decay to $\tau$-pairs with a branching fraction of
approximately $10\%$, but this also has large SM backgrounds. \\

The chosen scenario is extremely challenging for the LHC, but it is by
no means a ``worst-case scenario''. For example, increasing the value
of $\tan \beta$ would increase the separation between the $b$-quark
and vector boson switch-off points, moving the $M_A$ range with an
enhanced $H_1 \to \gamma \gamma$ branching ratio out of the allowed
region. Alternatively, increasing the value of $\kappa v_s$ slightly
would lead to a light Higgs boson sitting right on top of the
$Z$-peak, making it very difficult to disentangle from the SM
backgrounds. If the value of $\kappa v_s$ is significantly larger (and
$|A_{\kappa}|$ not too large), the singlet dominated scalar would be
heavy enough to decay to a vector boson pair, making its detection
much easier. However, if the value of $M_A$ is such that the coupling
of Eq.(\ref{eq:ghvv}) vanishes, these golden channels would be lost.

\subsubsection{LC Production}

The vanishing of the $HVV$ couplings in the region of interest is
particularly significant for a LC since the most promising production
mechanisms are vector boson fusion, e.g.\ $e^+e^- \to W^+W^-\nu \bar \nu
\to H_1 \nu \bar  \nu$, and Higgs-strahlung, $e^+e^- \to Z^* \to
ZH_1$. The cross-sections for these processes at a $\sqrt{s}=500$~GeV
LC are plotted in Fig.(\ref{fig:lc500})
\begin{figure}[htb]
\begin{center}
\includegraphics[scale=0.6]{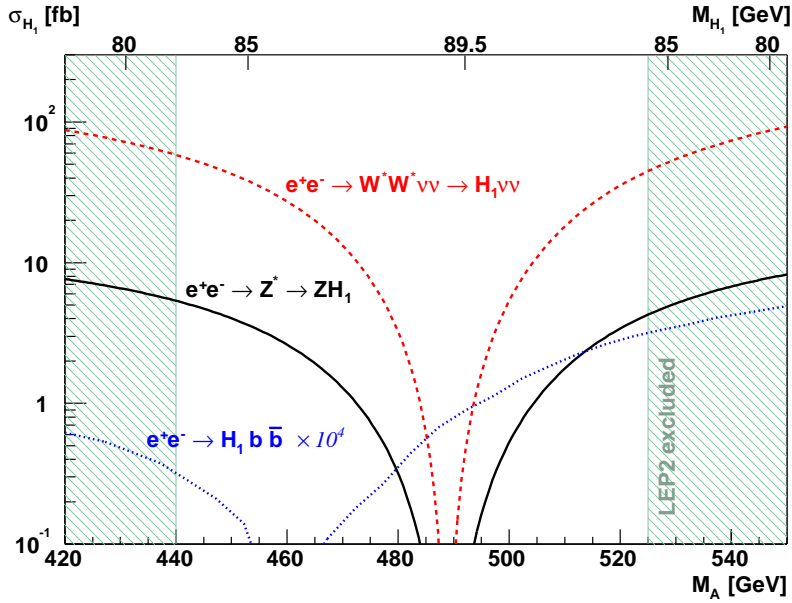}
\caption{\it 
Production cross-sections for the lightest scalar Higgs boson at a
$\sqrt{s}=500$~GeV LC, as a function of $M_A$ for $\lambda=0.3$,
$\kappa=0.1$, $\tan \beta= \tan \beta_s =3$ and $A_{\kappa}=-60$~GeV. 
The cross-section for $e^+e^- \to H_1 b \bar b$ has been multiplied 
by $10^4$.}
\label{fig:lc500}
\end{center}
\end{figure}
for our parameter choice, as a function of $M_A$, and show the
distinctive vanishing of the $H_1VV$ coupling.  Nevertheless, the
lightest scalar Higgs boson would be seen by these channels for all of
the $M_A$ range except for a small window around $490$~GeV. In
contrast to the LHC, for most of the the observable region decays
to $b \bar b$ and/or $\tau^+ \tau^-$ could be easily used due the LC's
relatively background free environment. For $M_A$ values where the
bottom and $\tau$ couplings vanish, the decays to $\gamma \gamma$ and
charm may be used instead.

It is difficult to see what production mechanism could be used to
close the remaining window around the critical point where the $HVV$
couplings vanish. Higgs production in association with a top quark
pair, $e^+e^- \to H_1 t \bar t$, is vanishingly small here because of
the proximity of the $H_1VV$ and $H_1 t \bar t$ ``turning-off'' points
(they will move even closer as $\tan \beta$ is increased). The
production in association with bottom quarks is shown in
Fig.(\ref{fig:lc500}), multiplied by a factor of $10^4$ to be visible
on the same scale. Generally, this production process has three
contributing sub-processes: Higgs-strahlung, $e^+e^- \to ZH_1$,
followed by the $Z$ decay to a bottom quark pair; Higgs pair
production, $e^+e^- \to H_1A_i$ ($i= 1,2$) followed by the
pseudoscalar decaying to bottom quarks; and bottom quark pair
production, $e^+e^- \to b \bar b$ followed by the radiation of $H_1$
off a bottom quark. The first contribution is very closely related to
the Higgs-strahlung already shown in Fig.(\ref{fig:lc500}) [simply
multiplied by the $Z \to b \bar b$ branching ratio], so contains no
new information and is {\it not} included in the $e^+e^- \to H_1 b
\bar b$ cross-section shown. The second contribution is only
kinematically allowed for the lightest pseudoscalar Higgs boson and is
vanishingly small because two small mixings are needed (neither scalar
nor pseudoscalar singlet fields have a $Z$ coupling). Therefore
the remaining process is dominated by Higgs radiation off bottom quarks,
and although this switches off at a different $M_A$ value, it is too
small to be useful because of the small bottom quark Yukawa coupling.

At a LC with $\sqrt{s}=800$~GeV, these cross-sections are modified as
shown in Fig.(\ref{fig:lc800}). 
\begin{figure}[htb]
\begin{center}
\includegraphics[scale=0.6]{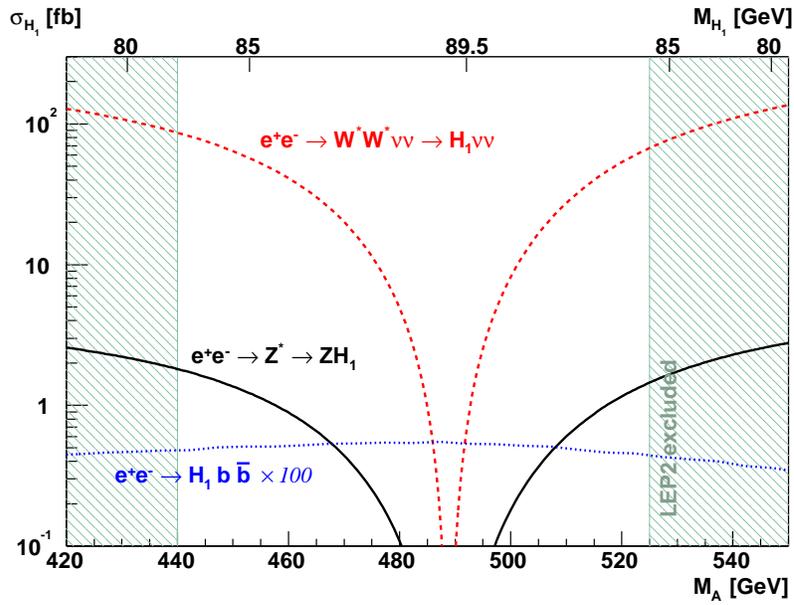}
\caption{\it 
Production cross-sections for the lightest scalar Higgs boson at a
$\sqrt{s}=800$~GeV LC, as a function of $M_A$ for $\lambda=0.3$,
$\kappa=0.1$, $\tan \beta= \tan \beta_s =3$ and $A_{\kappa}=-60$~GeV. 
The cross-section for $e^+e^- \to H_1 b \bar b$ has been multiplied 
by $100$.}
\label{fig:lc800}
\end{center}
\end{figure}
The t-channel $W$-fusion cross-section increases, while the
$s$-channel Higgs-strahlung cross-section decreases, but the overall
$M_A$ dependence remains the same, with both cross-sections vanishing
at around $490$~GeV. The $e^+e^- \to H_1 b \bar b$ associated
production cross-section has increased dramatically due to the opening
up of $e^+e^- \to H_1A_2$, which was kinematically disallowed at
$\sqrt{s}=500$~GeV. Since this new contribution contains no $H_1 b
\bar b$ coupling, the cross-section no longer vanishes at around
$460$~GeV, but unfortunately it is still too small to be of practical
use\footnote{This cross-section has been calculated under the
assumption of a fixed width (of $1$~GeV) for $A_2$, and is only
intended to present an order of magnitude estimate.}. \\

Increasing $\kappa v_s$ and thus the singlet dominated masses only
reduces the production cross-sections in line with the expectations of
a reduced phase space. If the singlet dominated scalar is heavy
enough, and $M_A$ is far enough away from its critical value, the
scalar will decay to vector bosons, making its discovery easier.

\subsubsection{Conclusions}

In this contribution we have considered a particularly challenging
NMSSM scenario, presenting masses, branching ratios and production
cross-sections at both the LHC and a future $e^+e^-$ LC. Such
scenarios have a Higgs spectrum very similar to the MSSM, i.e.\ nearly
degenerate heavy charged, scalar and pseudoscalar states and a light
Higgs boson at around $120$--$140$~GeV, supplemented by an additional
singlet dominated scalar and pseudoscalar.  We have seen that there is
still room allowed by LEP for the singlet dominated Higgs boson to be
very light, i.e.\ $\lesssim M_Z$. Despite having reasonably large
production cross-sections at the LHC, this light Higgs boson would be
difficult to see since its mainly hadronic decays cannot be easily
untangled from the SM backgrounds. At a LC, this light scalar can be
seen via vector boson fusion and Higgs-strahlung for most of the
parameter range, except for a small region where the Higgs-vector
boson coupling vanishes.  If this Higgs boson is discovered at a LC
but is missed at the LHC, LC input would be vital in providing
information for trigger and background removal when the LHC
endeavours to confirm the discovery.

We have also seen that a such a light Higgs boson may place
restrictions on the masses of the heavier Higgs bosons. For small
$\kappa v_s$, in order to avoid detection of the light scalar at LEP,
we require $M_A \approx \mu \tan \beta$. [The veracity of the
pre-condition ``small $\kappa v_s$'' may be ascertained by also
observing the singlet dominated pseudoscalar, by e.g.\ $e^+e^- \to t
\bar t A_1$, and making use of the approximate sum rule of
Eq.(\ref{eq:sumrule}).] This prediction for the heavy Higgs boson
masses would be invaluable to the LHC.

In this scenario the $H_2$, $H_3$ and $A_2$ will be present, looking
very much like the MSSM Higgs bosons $h$, $H$ and $A$ respectively
with slightly altered couplings and could be detected in the usual
way.

For heavier singlet dominated states, the position of the LHC is more
favourable, since the clean decay to vector bosons opens up [although
again, this is not useful over the entire $M_A$ range].  Also the
LHC's kinematic reach will prove useful in discovering or ruling out
very heavy singlet dominated Higgs states. On the other hand, if the
extra singlet dominated Higgs boson is found to be almost degenerate
with the lightest doublet dominated Higgs boson, LC precision may be
required to disentangle the two states.

In summary, in order to provide complete coverage over the NMSSM
parameter space, both the LHC and an $e^+e^-$ LC will be needed. Not
only can the LC probe areas where the LHC cannot, it can provide
valuable input to the LHC investigation of the NMSSM Higgs sector.


\subsection{Identifying an SM-like Higgs particle at future colliders}
{\it I.~F. ~Ginzburg,  M. Krawczyk  and P. Osland}

\vspace{1em}

\subsubsection {SM-like scenario.}
One of the great challenges at future colliders will be the SM-like scenario
that no new particle will be discovered at the Tevatron, the LHC and
electron-positron Linear Collider (LC) except the Higgs boson with partial
decay widths, for the basic channels to fundamental fermions (up- and
down-type) and vector bosons $W/Z$, as in the SM:
\begin{equation}
\bigg|\frac{\Gamma_i^{\rm exp}}{\Gamma_i^{\rm SM}}-1\bigg|
=\bigg|\left(\frac{g_i}{g_i^{\rm SM}}\right)^2-1\bigg|
\lsim\delta_i\ll 1, \text{ where }\,  i=u,d,V.
\label{widthest}
\end{equation}
New physics may still be seen via deviations of some observables from the SM
predictions.  The Higgs-boson loop couplings with photons or gluons are very
promising for such studies.  In the SM and in its extensions, all fundamental
charged particles, with mass arising from a Higgs mechanism, contribute to the
Higgs boson effective couplings with photons, and, similarly, all quarks
contribute to the Higgs boson coupling to gluons.  These couplings are absent
in the SM at tree level, and therefore, the relevant background will be
relatively low.

Results from the LHC and LC can be combined and thus facilitate in
distinguishing models. We stress here in particular the role of the
$\gamma\gamma$ mode of a linear collider (Photon Collider), where the expected
accuracy of the measurement of the two-photon width for Higgs mass equal to
120~GeV is of the order of 2\% \cite{Jikia:1999en} in comparison to the
LHC and LC with the corresponding accuracy 15--20\%
\cite{LHC,Aguilar-Saavedra:2001rg}.

Here we assume that one neutral Higgs particle has been found at the LHC
with all basic couplings, within the experimental accuracy, as expected in the
SM.  So, we have for relative couplings:
\begin{equation}
\label{Eq:chi-def}
\chi_i= \frac{g_i}{g_i^{\rm SM}},\quad {\rm with}\
\chi_i^{\rm obs}=\pm (1-\epsilon_i),\quad
\mbox{and } |\epsilon_i|\le \delta_i  \hspace{0.5cm} \text{ for } i=u,d,V.
\end{equation}
Additional constraints on these $\epsilon_i$ follow from the considered model.

We consider the Two-Higgs-Doublet Model (2HDM) with no or weak CP-violation,
using the soft $Z_2$-violation potential as given by
\cite{Diaz-Cruz:1992uw,Haber:1994mt,Branco,Ginzburg:2001ss}.  We introduce a
parameter $\mu$ via the real part of the coefficient of the bilinear mixing
term $\Re (m_{12}^2)=2\mu^2 v_1v_2/v^2$ \cite{Ginzburg:2001ss}.  For
$\mu^2\gg|\lambda_i|v^2$ the SM-like scenario with decoupling is realized.
The SM-like scenario can also be realized for $\mu<v$ with non-decoupling, in
particular for $M_{H\pm}\approx M_H\approx 700$ GeV, $M_A\approx 600$ GeV and
$M_h$=120 GeV, $\mu<0.3\, M_{H\pm}$, with $\lambda_i$ within
perturbativity and even unitarity constraints. The Yukawa interactions are of
Type~II with the basic couplings constrained by $(\chi_u
+\chi_d)\chi_V=1+\chi_u \chi_d$ (pattern relation \cite{Ginzburg:2001ss}),
valid for each neutral Higgs boson, even in the general CP-violating case.

The SM-like scenario corresponds to the observation of either $h$ or $H$. We
distinguish solutions A and B (see ref.\ \cite{Ginzburg:2001ss} and table
\ref{t1}) for the realization of an SM-like scenario for the observed Higgs
particle with mass 110 to 250 GeV. In solutions A all basic (relative)
couplings are close to 1 or all are close to $-1$. Note that only solution
A$_{h+}$ is realized in the decoupling limit.  In solutions B, one basic
coupling, $\chi_u$ or $\chi_d$, has opposite sign to the other two, and
obviously one expects here larger deviations from the SM than for solutions A.
One can see that the decay width $h(H) Z\gamma$ is less sensitive to the
discussed effects of 2HDM than the $h(H)\gamma\gamma$ one.\footnote{The
process $e\gamma\to eh(H)$ with large transverse momentum of the scattered
electron, has a sensitivity from the $h(H) Z\gamma$ vertex, with $Z$ far from
the mass shell, similar to that of $h(H)\to\gamma\gamma$ \cite{GVych}.}
\begin{table}[ht]
\begin{center}
\begin{tabular}{||c|c|c|c|c|c||}
\noalign{\vspace{-9pt}}
\hline\hline
solution&basic couplings&$|\chi_{gg}|^2$ &
$|\chi_{\gamma\gamma}|^2$ & $|\chi_{Z\gamma}|^2$\\ \hline\hline
$A_{h\pm}/A_{H_-}$
&$\chi_V\approx\chi_d\approx\chi_u\approx \pm1$
& 1.00 & 0.90 & 0.96 \\ \hline
$B_{h\pm d}/B_{H_-d}$
&$\chi_V\approx-\chi_d\approx\chi_u\approx \pm1$
& 1.28 & 0.87 & 0.96 \\ \hline
$B_{h\pm u}$
& $\chi_V\approx\chi_d\approx-\chi_u\approx \pm1$
& 1.28 & 2.28 & 1.21 \\ \hline\hline
\end{tabular}
\end{center}
\vspace*{-5mm}
\caption{\it SM-like realizations in the 2HDM~II \cite{Ginzburg:2001ss},
together with ratios of partial widths to their SM values for
$M_{h(H)}\approx120$~GeV for the case $|\chi_i|=1$, $M_{H^{\pm}}= $700~GeV
and $\mu \lesssim 0.3 M_{H^{\pm}}$.}
\vspace*{-6mm}
\label{t1}
\end{table}
\vspace*{-2mm}

\subsubsection{The two-gluon decay width.}
The two-gluon width will be measured at the LHC.  This quantity is 
for solutions A practically equal to the SM prediction (see table~1).

In the SM the contributions of $t$ and $b$ quarks partly cancel, while for
solutions $B$ they add, giving larger deviations from the SM predictions.  In
Fig.~1 (left) the solid curve corresponds to the case $\chi_u=-\chi_d=\pm 1$
for all solutions~B.  The shaded bands reflecting $1~\sigma$ experimental
uncertainties correspond to the $B_{h+u}$ case only.
\begin{figure}[htb]
\refstepcounter{figure}
\addtocounter{figure}{-1}
\begin{center}
\setlength{\unitlength}{1cm}
\begin{picture}(15.5,4.5)
\put(-0.7,-0.9){
\mbox{\epsfysize=5.5cm\epsffile{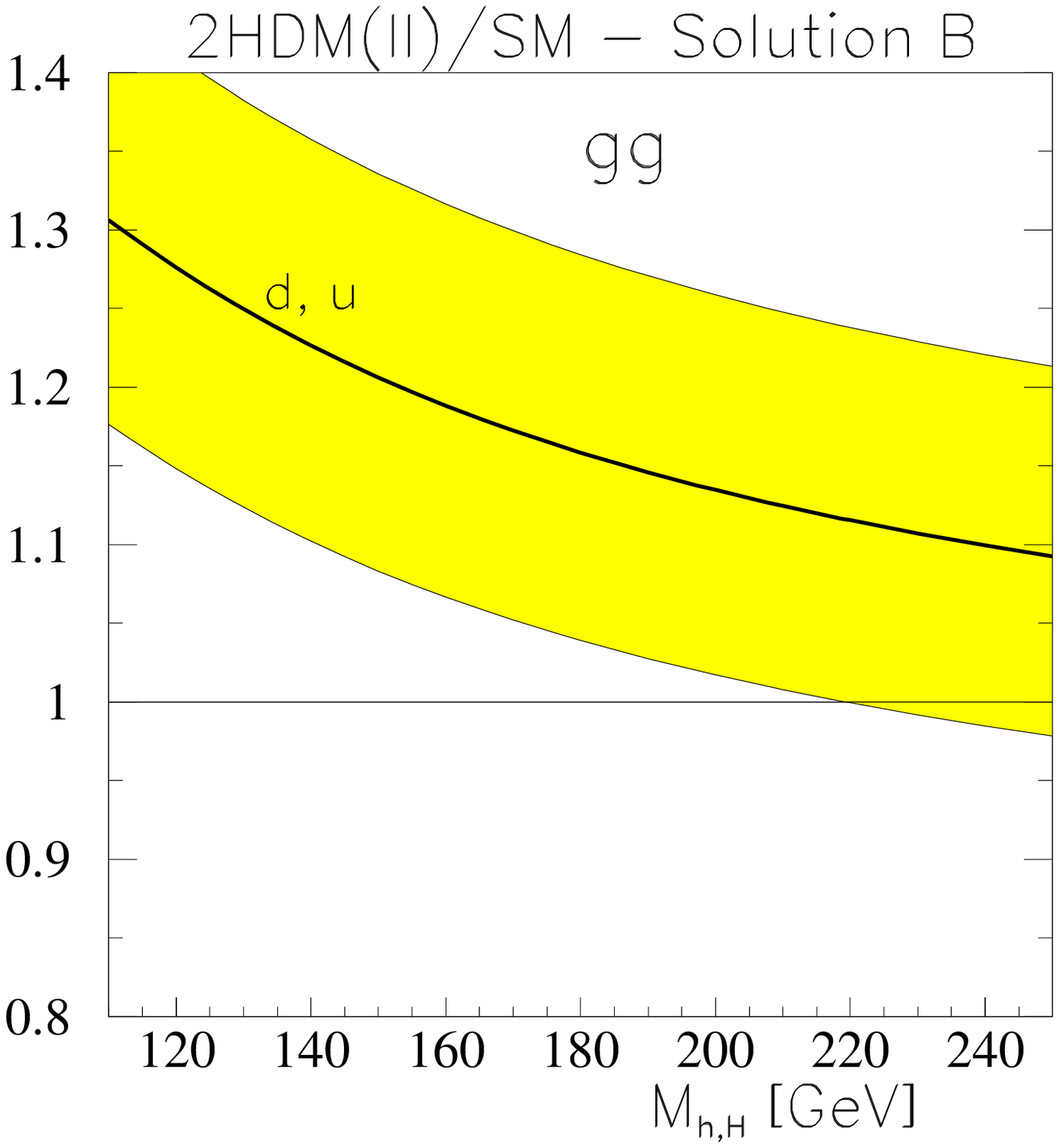}}
\mbox{\epsfysize=5.5cm\epsffile{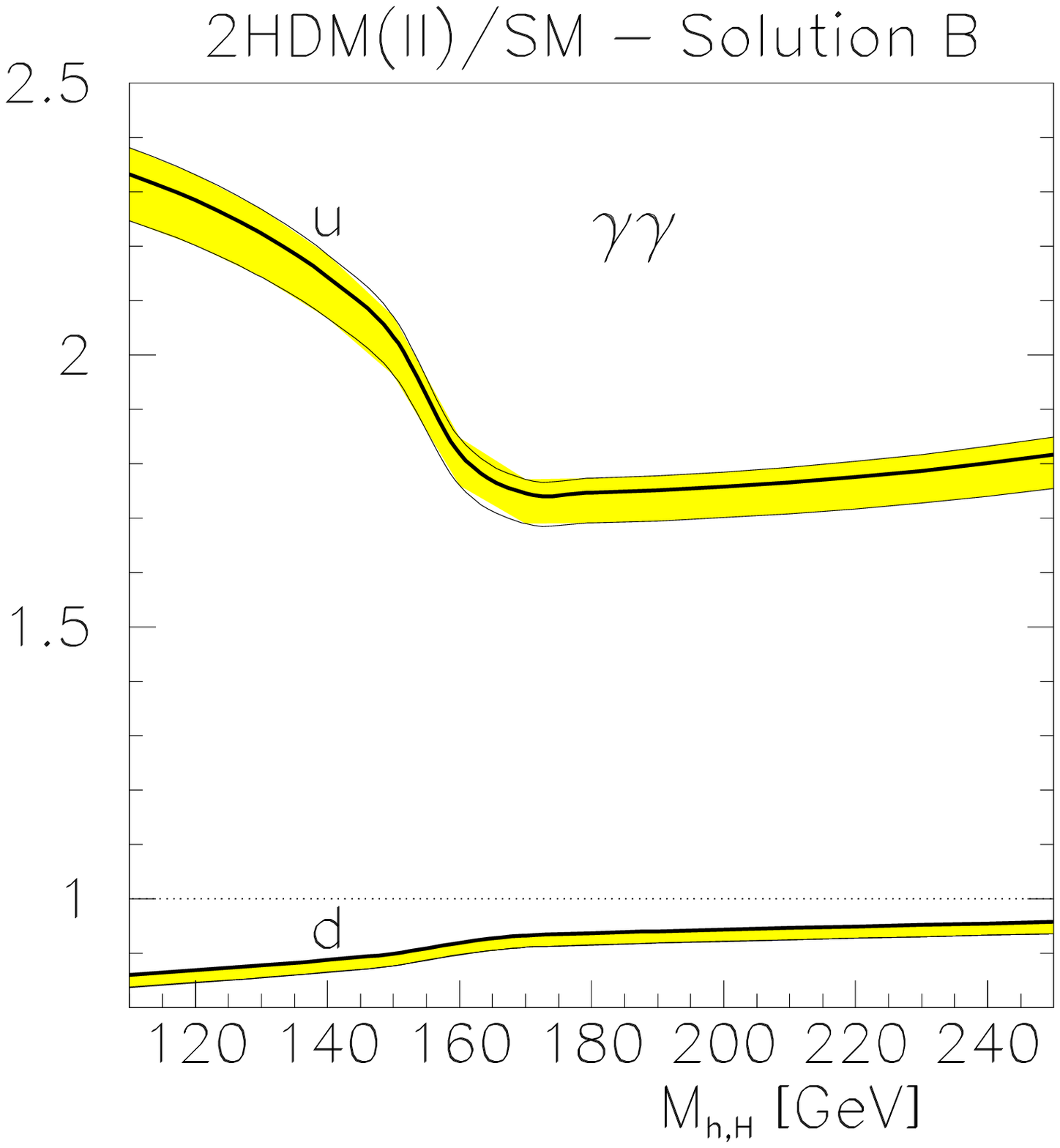}}
\mbox{\epsfysize=5.5cm\epsffile{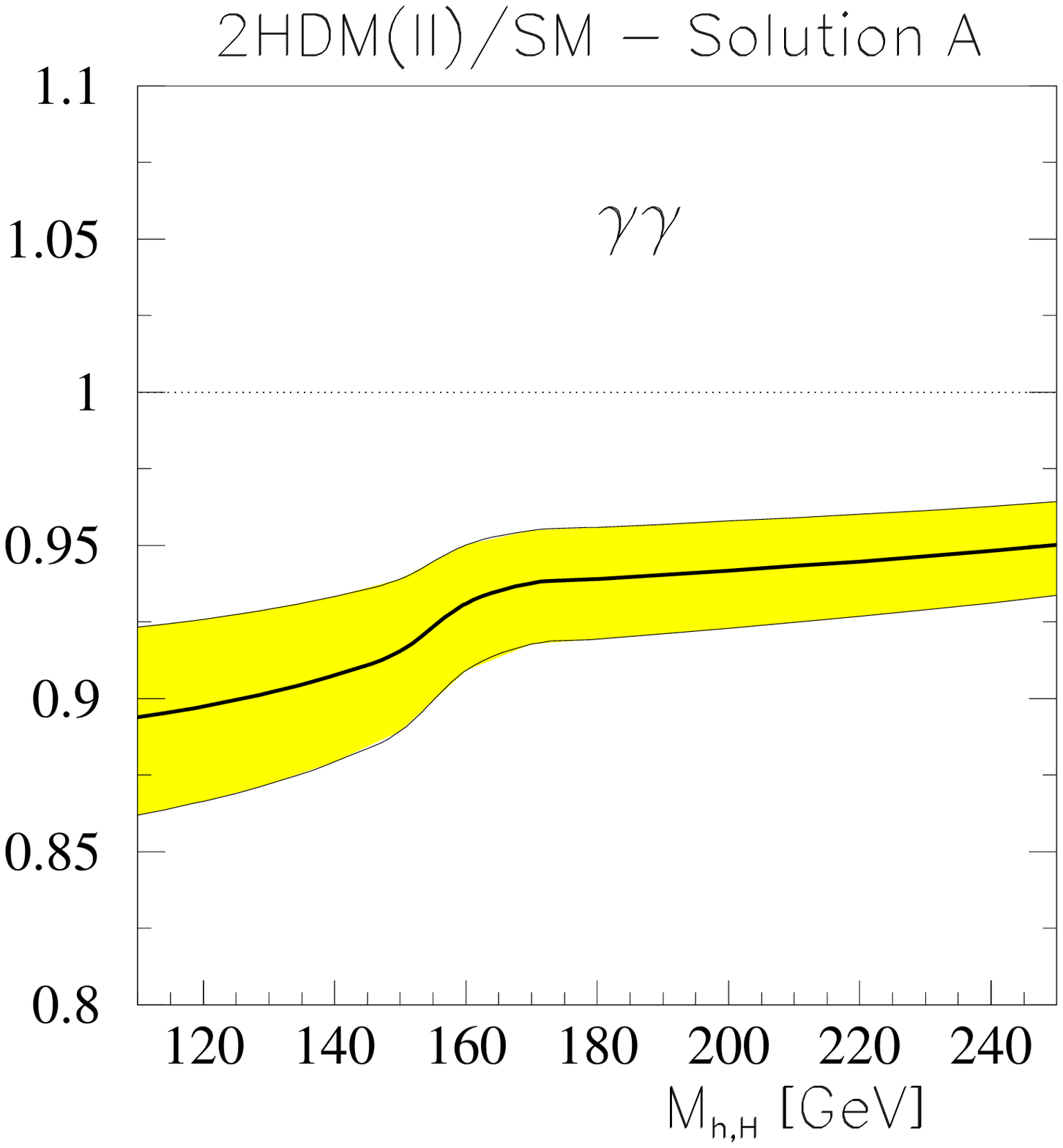}}}
\end{picture}
\vspace*{4mm}
\caption{\it Ratios of the Higgs boson decay widths in the SM-like 2HDM~II and
the SM as functions of $M_{h,H}$ for solutions B and A for $M_{H^{\pm}}$=700
GeV and $\mu \lesssim 0.3 M_H^{\pm}$.}
\vspace*{-6mm}
\end{center}
\end{figure}
\vspace*{-5mm}

\subsubsection{The two-photon decay width.}
For solutions A the widths $h (H)\to\gamma\gamma$ and $h(H)\to Z\gamma$ will
differ from the SM predictions due to the charged Higgs boson contribution,
proportional to the trilinear coupling $hH^+H^-$ ($HH^+H^-$), so we have:
\begin{equation}
\chi_{H^\pm} \equiv-\frac{vg_{h H^+H^-}}{2M_{H^\pm}^2}
=\left(1-\frac{M_h^2}{2M_{H^\pm}^2}\right)\chi_V
+\frac{M_h^2-\mu^2}{2M_{H^\pm}^2}
(\chi_u+\chi_d),
\label{hch}
\end{equation}
and similarly for $H$. The value of $\chi_{H^\pm}$ is close to $\chi_V \approx
\pm 1$ (non-decoupling) at the considered small values of
$(M_h^2/2M_{H^\pm}^2)$ for solutions A and $(\mu^2/M_{H^\pm}^2)$ for solutions
B.

For solutions $B_{h\pm d}$ the main deviation from that of the SM
is like for solution A given by the contribution of $H^{\pm}$. For
the solution $B_{h+u}$ the effect of the change of sign of the
coupling $ht\bar t$ dominates.

The ratios of the two-photon Higgs widths to the SM value are shown in Fig.~1
(see also table~1) for solutions B (center) and A (right).  The results are
presented without (solid lines) and with realistic 1~$\sigma$ uncertainties
(shaded bands) around the SM values of the measured basic couplings
\cite{Aguilar-Saavedra:2001rg}.
\vspace*{-4mm}

\subsubsection{Conclusion}
We analyse, using realistic estimates of experimental uncertainties, the
potential of future colliders to determine the nature of an SM-like Higgs
boson, SM or 2HDM, for mass 110--250 GeV.  It is crucial to combine precise
measurements of loop couplings involving gluons at LHC and photons 
at a Photon Collider, as only couplings to photons are sensitive to the
decoupling property of the model.

\subsection{Synergy of LHC, LC and PLC in testing the 2HDM~(II)}

{\it P.~Nie\.zurawski, A.F.~\.Zarnecki and  M.~Krawczyk}

\vspace{1em}

%
%

Interplay of LHC, LC and PLC in testing the 2HDM~(II) \cite{2HDM}
has been studied for  Higgs boson $h$ with mass 200 to 350~GeV,
decaying to  $W^+W^-$ and $ZZ$.
Figure~\ref{nzk:cros1}  shows
the expected rate, relative to SM,
as a function of relative (to SM) couplings to top quark, $\chi_t$, 
and vector bosons, $\chi_V$.
At LHC, cross section  sensitive mainly to  $\chi_t$
 can be measured with precision (SM case) of about 15\% 
\cite{cms_tn_95-018,cms_cr_2002-020},
while at LC, for cross section 
depending predominantly on  $\chi_V$, the precision
is $4-7$\% \cite{lc-phsm-2003-066}.
The two-photon width of the Higgs boson, 
to be measured with accuracy of $4-9$\% at PLC \cite{nzk_wwzz},
depends both on  $\chi_t$ and $\chi_V$.
At PLC also the phase of $h\rightarrow \gamma \gamma$ amplitude,
$\phi_{ \gamma \gamma}$, can be measured to $40-120$~mrad \cite{nzk_wwzz}.
By combining these measurements, couplings of the
Higgs boson can be precisely determined, as shown in Fig.~\ref{nzk:syn1}
(assumed parameter values are $\chi_V=0.9$, 
$\chi_t=-1$, $\chi_b=1$).
%

\begin{figure}[htb!]
     \epsfig{figure=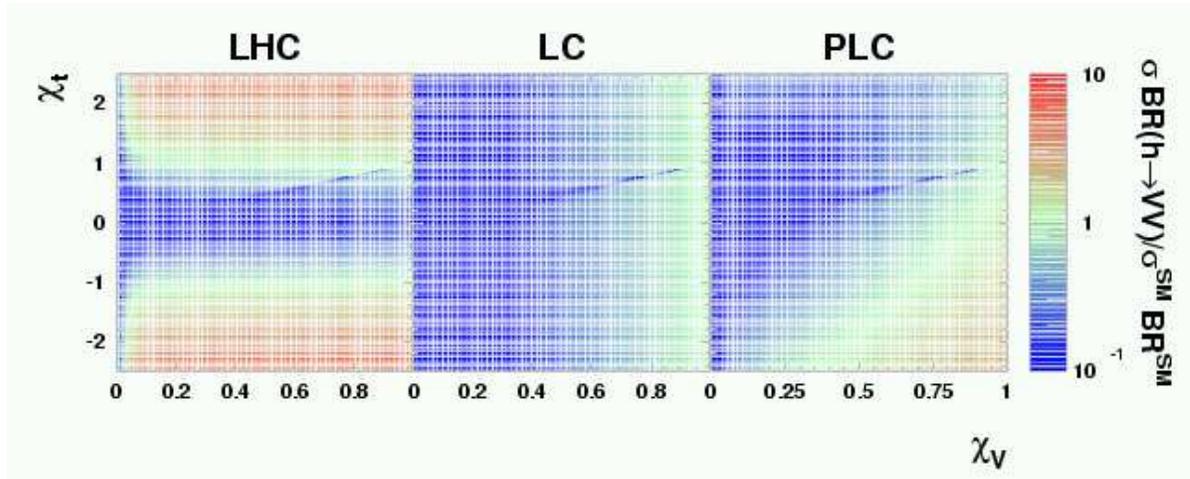,width=\textwidth,clip=}
\vspace{-1cm}
 \caption{ 
$h$ (250 GeV) production rates at LHC, LC and PLC, 
relative to SM. 
%
 } 
 \label{nzk:cros1} 
 \end{figure}


\begin{figure}[htb!]
  \begin{center}
     \epsfig{figure=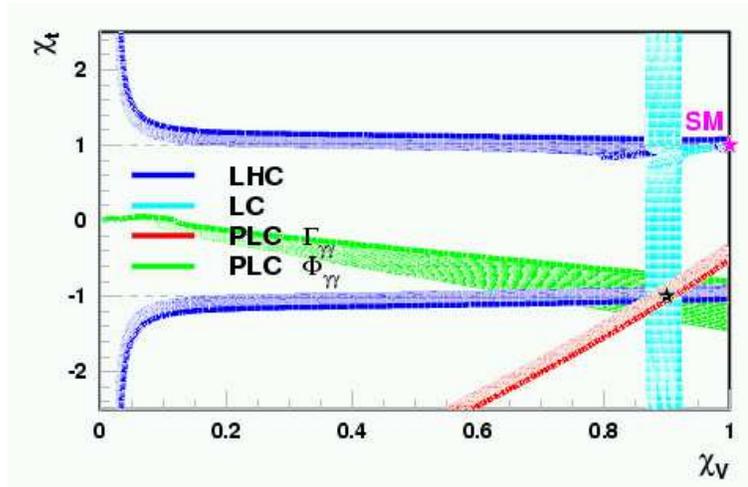,width=10cm,clip=}
  \end{center}
\vspace{-1cm}
 \caption{ 
Values of  $\chi_V$ and  $\chi_t$  determined  
from  measurements at LHC, LC and PLC, 
for assumed parameters indicated by ($\star$).
%
 }
 \label{nzk:syn1} 
 \end{figure} 

\subsection {Enhanced $h^0\to \gamma\gamma$ decays in fermiophobic Higgs
models at the LHC and LC}
{\it A.G.~Akeroyd and M.A.~D\'\i az }


\subsubsection{Introduction}

We study the production of a fermiophobic Higgs \cite{Weiler:1987an}
($h_f$) followed by $h_f\to \gamma\gamma$ decay 
at a $e^+e^-$ high energy linear collider (LC) and the 
Large Hadron Collider (LHC). A $h_f$ has very suppressed
or zero couplings to fermions and so its dominant decay
modes are to bosons, either $\gamma\gamma$ (for $m_{h_f}< 95$ GeV)
or $WW^{*}$ (for $m_{h_f}> 95$ GeV) \cite{Stange:ya}. In this paper
we shall consider $h_f\to \gamma\gamma$ decays only.
Higgs boson production 
mechanisms which depend on the Higg-fermion-fermion coupling (such 
as gluon-gluon fusion $gg\to h_f$) are very suppressed. 
Hence $h_f$ is best searched for by mechanisms which involve its
couplings to vector bosons ($V=W^\pm,Z$) and/or other Higgs bosons.
At the LHC there are two standard ways to produce $h_f$, for which
experimental simulations have been performed 
in the context of the SM Higgs ($\phi^0$). These are:

\begin{itemize}

\item[{(i)}] $pp\to W^* \to Wh_f$, $W\to l\nu$ (Higgsstrahlung)
\cite{Dubinin:1997du}

\item[{(ii)}] $pp\to qq h_f$ (Vector boson fusion) 
\cite{Rainwater:1997dg}

\end{itemize}
\noindent
At a $e^+e^-$ LC one has the following mechanisms:

\begin{itemize}

\item[{(iii)}] $e^+e^-\to h_fZ$ (Higgsstrahlung) \cite{Boos:2000bz}

\item[{(iv)}] $e^+e^-\to h_f\nu\overline\nu$ ($W$ boson fusion)
\cite{Boos:2000bz}

\end{itemize}
\noindent
At a $\gamma\gamma$ collider:

\begin{itemize}
\item[{(v)}] $\gamma\gamma\to h_f$ \cite{Asner:2002aa},
\cite{Dedes:2003cg,Ellis:2002gp}
\end{itemize}

All these mechanisms have been shown to be effective for the
SM Higgs $\phi^0$ and for $h^0$ of the MSSM 
\cite{Dedes:2003cg,Ellis:2002gp}, since both these Higgs bosons have 
substantial couplings to vector bosons. This is not
the case in a general 2 Higgs 
Doublet Model (2HDM), in which a $h_f$ may arise.
In the 2HDM the above mechanisms (i) to (v) for $h_f$ are all 
suppressed by $\sin^2(\beta-\alpha)$, which in the fermiophobic
scenario ($\alpha\to \pi/2$)
in the 2HDM (Model~I)  reduces to:
\begin{equation}
VVh_f\sim \cos^2\beta\;\;(\equiv 1/(1+\tan^2\beta)) 
\end{equation}

\begin{figure}[htb!]
\centerline{{\epsfig{file=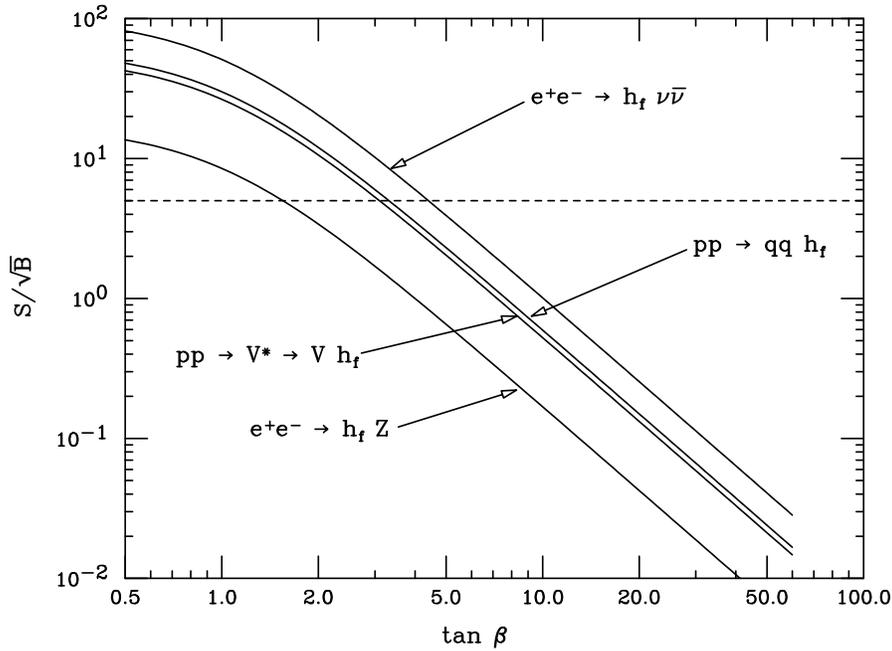,width=0.55\textwidth,angle=90}}}
\caption{$S/\protect\sqrt{B}$ as a function of $\tan\beta$ for various $h_f$
production processes.}
\label{sec253fig1}
\end{figure} 

This is a severe suppression for $\tan\beta \ge 10$ and renders
all the above mechanisms unobservable (for an earlier discussion
with just mechanism (i) see \cite{Akeroyd:1998ui}).
This is shown in Fig.~\ref{sec253fig1}, where we apply the results of the 
signal/background ($S^{\phi}/\sqrt B$) simulations
for $\phi^0\to \gamma\gamma$ to the case of a $h_f$. 
To do this we need to scale the SM Higgs signal $S^{\phi}$
by the factor BR$(h_f\to \gamma\gamma)$/BR$(\phi^0\to \gamma\gamma)$, 
and include the $\cos^2\beta$ suppression in the 
production cross--sections.
Since all the above simulations presented results for
$m_{\phi^0}=120$ GeV we will consider a $h_f$ of this mass. 
For $m_{h_f}=120$ GeV on has \cite{Mrenna:2000qh}:
\begin{equation}
BR(h_f\to \gamma\gamma)/BR(\phi^0\to \gamma\gamma)\approx 10
\end{equation}
In Fig.~\ref{sec253fig1} we plot $S/\sqrt B$ for $h_f$ as a 
function of $\tan\beta$.
We include the production mechanisms (i)-(iv) and take $m_{h_f}=120$ GeV.
We are not aware of a simulation for mechanism (v). Each curve is of the 
simple form: 
\begin{equation}
S/\sqrt B=10K_i\cos^2\beta
\end{equation}
where $K_i$ ($i=1,4$) corresponds to the SM Higgs
$S^{\phi}/\sqrt B$ for each of the mechanisms (i)-(iv) 
for the chosen luminosities (${\cal L}$) 
in Refs.\cite{Dubinin:1997du} $\to$ \cite{Boos:2000bz},
which are 50 fb$^{-1}$ for (i),(ii) and 
1000 fb$^{-1}$ for (iii),(iv), with $\sqrt s=500$ GeV.
For other choices of ${\cal L}$ the $S/\sqrt B$ scales as 
$\sqrt {\cal L}$.
One can see that all the mechanisms offer spectacular signals
($S/\sqrt B>> 5$) {\sl when} there is little suppression
in the cross-section at low $\tan\beta$. Importantly, each collider
can discover $h_f$ in two distinct channels, thus providing 
valuble confirmation of any initial signal.
However, $S/\sqrt B$ falls rapidly as $\tan\beta$ increases, and
$S/\sqrt B< 5$ at some critical value $\tan\beta_{C}$.
In Fig.~\ref{sec253fig1}, $\tan\beta_{C}$ varies between 2 and 5.
Hence unless $\tan\beta$ is fairly small a relatively light $h_f$ 
(even $m_{h_f}<< 120$ GeV) may escape detection
at both the LHC and LC. 

Fortunately there are alternative mechanisms $\sim \sin^2\beta$,
which are thus unsuppressed in the region of large $\tan\beta$
\cite{Akeroyd:2003xi}:

\begin{itemize}

\item [{(vi)}] At the LHC: $pp\to H^\pm h_f,A^0h_f$ 

\item [{(vii)}] At a LC: $e^+e^-\to A^0 h_f$ 

\end{itemize}

For $\sin^2\beta >> \cos^2\beta$ only the above channels
offer reasonable rates, and thus all three should be exploited in order
to first discover $h_f$ and thereafter provide confirmatory signals.
We are not aware of explicit signal--background
simulations for these channels. The cross--sections at Tevatron
energies for the mechanisms in (vi) were studied in \cite{Akeroyd:2003bt}.
The mechanism $pp\to H^\pm h_f$, followed by
$H^\pm\to h_fW^{*}$ (which would have a large branching ratio) 
and $W^\pm\to l^\pm\nu$ would lead
to a signature $\gamma\gamma + l$ similar to that from mechanism 
(i). Moreover, there is also the possibility of 
$\gamma\gamma\gamma\gamma$ final states \cite{Akeroyd:2003xi}.
Mechanism (vii) is usually absent in discussions of 
the MSSM Higgs bosons at a LC, due to its strong suppression
of $\cos^2(\beta-\alpha)$ for $m_A\ge m_Z$. However for a $h_f$ in 
the larger $\tan\beta$ region it provides promising rates.  

\begin{figure}[htb!]
\centerline{{\epsfig{file=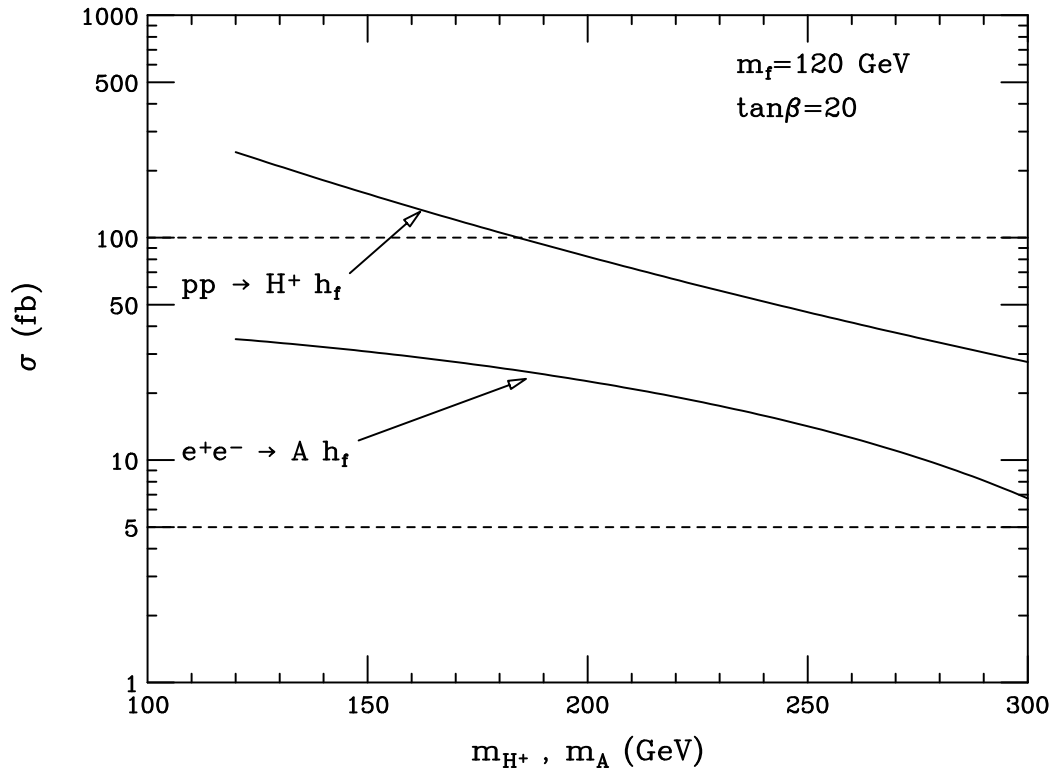,width=0.65\textwidth,angle=90}}}
\caption{$\sigma(pp\to H^\pm h_f)$ and $\sigma(e^+e^-\to A^0h_f)$ 
as a function of $m_A$ or $m_{H^\pm}$, for $m_{h_f}=120$ GeV.}
\label{sec253fig2}
\end{figure}

In Fig.~\ref{sec253fig2} we plot $\sigma(pp\to H^\pm h_f$) and 
$\sigma(e^+e^-\to A^0 h_f$)
as a function of Higgs mass $m_{H^\pm,A^0}$ for
$m_{h_f}=120$ GeV and fixing $\tan\beta=20$ 
(i.e. $\sin^2\beta\sim 1$). We do not plot 
$\sigma(pp\to A^0h_f$) which has a value approximately half that
of $\sigma(pp\to H^\pm h_f$), for $m_{H^\pm}=m_A$. 
One can see that $\sigma(e^+e^-\to A^0 h_f$) would offer a sizeable
number of events for ${\cal L}=500$ fb$^{-1}$, and backgrounds
would be smaller than for the mechanisms at the LHC.
Detection prospects for $e^+e^-\to A^0 h_f, 
h_f\to \gamma\gamma$ at larger $\tan\beta$
might be comparable to those for the Higgsstrahlung
channel $e^+e^-\to Zh_f$ at low $\tan\beta$.
It is not clear which
cross--section for $pp\to H^\pm h_f$ and $pp\to A^0 h_f$
would be observable. Detection prospects may be marginal due to the 
sizeable backgrounds at the LHC.
In this region of larger $\tan\beta$, it is more likely
that $h_f$ is detected first at a LC in the channel $e^+e^-\to A^0 h_f$,
while both $e^+e^-\to h_fZ$ and $e^+e^-\to h_f\nu\overline\nu$
are unobservable due to the aforementioned suppression of $\cos^2\beta$. 
With just one such signal for $h_f$, 
confirmatory signals at the LHC would be of great urgency.
Prior knowledge of the Higgs masses
from the LC in the observed channel $e^+e^-\to A^0 h_f$ should
facilitate the search for the analogous process $pp\to A^0 h_f$
at the LHC. In the same way, possible discovery of $H^\pm$ in 
$e^+e^-\to H^+H^-$ would aid the search for $pp\to H^\pm h_f$.
We encourage experimental simulations of the
mechanisms $e^+e^-\to A^0 h_f$ and $pp\to H^\pm h_f, A^0h_f$
followed by $h_f\to \gamma\gamma$ decay. 

Producing a $h_f$ in as many channels as possible is of utmost 
importance. It is here that the interplay of
the LC and LHC would be greatly beneficial in the studied 
region of larger $\tan\beta$.
Discovery of such a particle would strongly constrain the possible
choices of the underlying Higgs sector.

\subsection{Visible signals of an 'invisible' SUSY Higgs at the LHC} 
{\it F.~Boudjema, G.~B\'elanger, R.M.~Godbole}

\subsubsection{Introduction}
Sparticles  can affect the discovery of a Higgs of mass
$\sim 125 GeV$ in two ways. The loop effects due a light stop
of mass around that of a $t$ quark can reduce the $ggh$ coupling and hence 
the production to a level low enough to preclude the discovery of $h$ at the 
LHC \cite{Djouadi:1998az,Belanger:1999pv}. If the assumption of the 
gaugino mass unification
at the GUT scale is given up, the latest LEP and Tevatron data together with
cosmological constraints still allow the lightest supersymmetric Higgs to
have a large branching fraction into invisible 
neutralinos~\cite{Belanger:2000tg,Belanger:2001am}.

If it is the light stop that makes the Higgs 'invisible' it might still be
possible to recover the Higgs signal in the $\sto \sto^* h/ \sto \stt h$ 
channel\cite{Djouadi:1998az,Belanger:1999pv,Djouadi:1999jy}.  The large 
values of the 'invisible' branching 
fractions ($B_{invis}$) can  make the $h$  'invisible' reducing the
branching ratio into the  $\gamma \gamma$ and  $b \bar b$  channels that are
considered to be the optimal  channels to search for a $h$ at the LHC, in this
mass range in the inclusive and the associated production modes respectively.
In this case strategies for a dedicated search of a $h$ with 'invisible' 
decay products have to be devised.
It is possible to search for an 'invisible' Higgs at the LHC in the WW 
fusion channel~\cite{Eboli:2000ze,DiGirolamo:2001yv} upto an invisible 
branching ratio of $\sim 0.25$, and in its associated production with a 
$Z$\cite{Godbole:2003it,atlas2} or in the  $t \bar t$\cite{Gunion:1993jf
,Kersevan:2002zj} channel for  somewhat 
larger values of the invisible branching ratio  depending on the mass of 
the Higgs. The vector boson fusion channel is quite  promising but issues of 
QCD backgrounds  to the signal for an 'invisible'  Higgs are  yet to be 
completely resolved. 

On the other hand an LC in the first stage would be able to 
see such a  Higgs with ease, further allowing a measurement of its branching 
ratio into the 'invisible' channel. Combining this with the possible signal 
for the 'invisible' Higgs at the LHC, one can use specific features of the
sparticle phenomenology to  test the hypothesis that the 'invisibility' of 
the Higgs at the LHC is supersymmetric in origin. 

In the next section we first discuss the possible depletion in the 
inclusive Higgs signal for the case of a light stop quark. Then we 
present predicted values of the $B_{invis}$ over the supersymmetric parameter
space, while still being consistent with the data from the LEP, Tevatron as 
well as the WMAP DM constraints\cite{Bennett:2003bz}. Next we present 
special features of the neutralino phenomenology that such a scenario 
will imply at the LC as well as at the LHC.  The $\tilde \chi_1^0$ is 
a mixed gaugino-higgsino state in this case, causing an enhancement of the 
cross-section for associated production of $h$ with a pair of neutralinos
or equivalently production of $h$ in decays of $\tilde \chi_j^0$.
We point out how the associated production of the $h$ with  a pair of 
$\tilde \chi_i^0$ at the LHC and the LC will reflect this scenario as well as 
comment  how the second lightest neutral Higgs might also have 
substantial 'invisible' branching ratio in this case, thus affecting
the Higgs phenomenology further.

\subsubsection{Invisible Higgs due to Supersymmetric Effects at the LHC}

Direct decays of the lightest supersyemmetric Higgs into a pair of 
sfermions are not possible in view of the upper limit 
on $m_h$ in SUSY {\it and} the lower limits on the sparticle masses implied 
by the LEP data. As far as the loop induced $hgg$ and $h\gamma \gamma$ 
couplings are concerned, it is the light $\sto, {\tilde b}_1$ loops that 
can have the largest effect.
For large values of squark mixing  (trilinear term $A_t - \mu /\tan \beta$)
and a light stop ($\msto \sim 160 -170$) GeV, the $\sto$ loop contribution 
can interfere destructively with that of the $t$ loop. This would cause a 
reduction in $\Gamma (gg \ra h)$ and an increase in the $\Gamma (h \ra 
\gamma \gamma) $
relative to the expectations in the Standard Model. Due to the dominance 
of the latter by $W-$ loop, the  reduction in $gg \ra h$ width and hence in 
the inclusive production cross-section by light $\sto$ loop has a much more 
significant role to play. Thus in this situation a reduction 
by about a factor of 5 or more in the product of the production
cross section $\sigma (gg \ra h) $ and the $\Gamma (h \ra \gamma \gamma)$
is possible~\cite{Djouadi:1998az,Belanger:1999pv}. This is illustrated in 
\begin{figure}[htb!]
\begin{center}
\centerline{
\includegraphics*[scale=0.9]{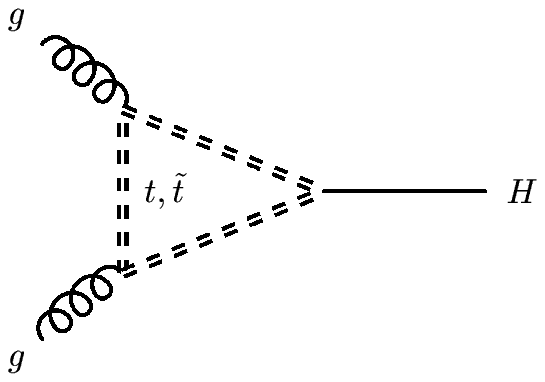}
\includegraphics[scale=1.0]{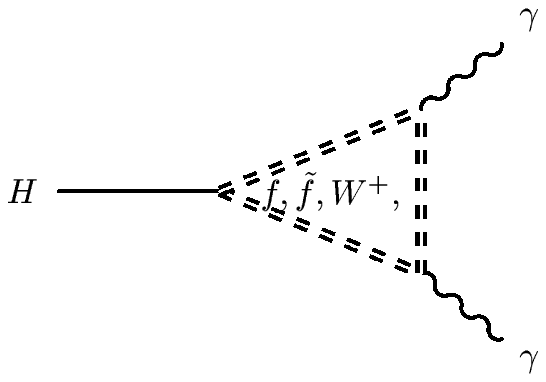}}
\includegraphics*[scale=0.95]{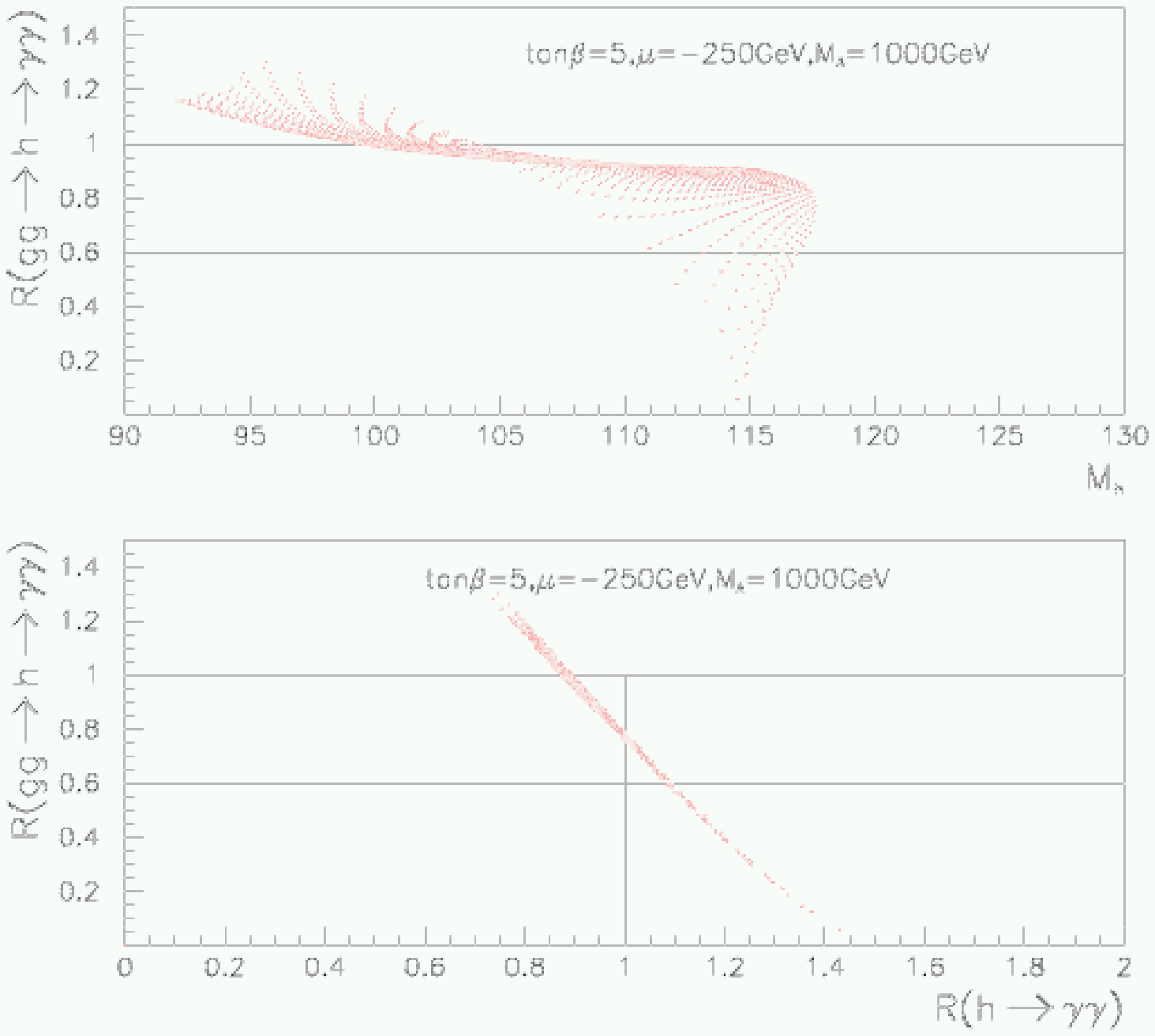}
\end{center}
\caption{\footnotesize Correlation of the ratio  $R_{gg\gamma \gamma}$ with
$R_{\gamma \gamma}$ and $m_h$ for a chosen set of values of $A_t, \mu$ and
$\tan \beta$, for light top squarks \protect\cite{Belanger:1999pv}}
\label{sec254fig1}
\end{figure}
Fig.~\ref{sec254fig1} which shows the ratio
$
R_{gg\gamma\gamma} =
\frac{\Gamma^{SUSY}(h\ra gg) \times BR^{SUSY}(h \ra \gamma \gamma)}
{\Gamma^{SM}(h\ra gg) \times BR^{SM}(h \ra \gamma \gamma)}$. 
$R_{\gamma \gamma}$ is defined similar to $R_{gg\gamma\gamma}$.
The two panels show clearly, that $R_{gg\gamma\gamma}$ can fall well
below 0.6, making the Higgs boson $h$  'invisible' in the 
inclusive channel in the $\gamma \gamma $ mode at the LHC, given the expected  
significance level of the signal as  given by the experimental studies by 
ATLAS and CMS\cite{Kinnunen:2002cr}. Two facts are worth noting though. 
We notice that large reductions in
$R_{gg\gamma\gamma}$ occur for parameter values where  $R_{\gamma \gamma}$
rises above unity thus facilitating  the search of the Higgs in the $Wh,Zh$ 
channel using the $\gamma \gamma$ decay of $h$. Further, the larger reductions
in $R_{gg\gamma\gamma}$ happen for heavier Higgses, for which the signal in 
the $\gamma \gamma$ channel remains statistically significant in spite of the 
reduction in $R_{gg\gamma\gamma}$. The reduction in  $R_{gg\gamma\gamma}$ 
becomes less pronounced with increasing values of $\tan \beta$. Further, if
the invisibility of the Higgs signal is caused by a light stop, the branching
ratio into $b \bar b$ channel is unaffected, thus keeping the prospects of the 
search is associated production mode with the decay of the $h$ in the 
$b \bar b$ channel alive.

In view of the LEP bound on the chargino mass, the effect of the sparticles 
other than the $\stop$ on the loop induced decay widths of the light higgs  
is rather small~\cite{Belanger:2000tg}. For the case of universal 
gaugino masses at high scale, the LEP bound on $\tilde \chi_1^\pm$ mass implies
a lower bound on $\tilde \chi_1^0$ mass and hence on the possible value that 
${\Gamma}_{\chi \chi} = \Gamma (h \ra \tilde \chi_1^0 \tilde \chi_1^0)$ 
can have.  However, in models with nonuniversal  gaugino masess  
at the high scale,  it is possible to have large values for 
${\cal B}_{\chi \chi} = B.R.(h \ra \tilde \chi_1^0 \tilde \chi_1^0)$
while still being consistent with the nonobservation of any effects of the
sparticles in the chargino-neutralino 
sector~\cite{Belanger:2000tg,Belanger:2001am }. For the decay to be 
kinematically possible, consistent with the LEP constraints, one needs 
$M_1/M_2$ at the EW scale to be  less than the value  of $\sim 0.5$ that 
it has in models with universal gaugino masses. Further to maximise the 
value of the $h \tilde \chi_1^0 \tilde \chi_1^0$ coupling and hence 
that of ${\Gamma}_{\chi \chi}$, one needs the LSP to be a mixture of 
gaugino and higgsino. This along with maximising the $h$ mass 
for a given choice of parameters is achieved by choosing small value of the 
higssino mass parameter $\mu$ and moderate values of $\tan \beta$.
We will see later that indeed values of ${\cal B}_{\chi \chi}$ upto 
$0.6$--$0.7$ are possible. $R_{\gamma \gamma}$ 
and  $R_{b \bar b}$ defined similarly are given approximately by 
$1.0 - {\cal B}_{\chi \chi}$. The  large values of 
${\cal B}_{\chi \chi}$  cause a reduction in the branching fraction into
the $\gamma \gamma$ channel and $b \bar{b}$ channel. Reduction in
the former affects the inclusive search, whereas the  latter is used for
the search when the Higgs is produced in association with a $W/Z/t \bar{t}$
or via the Vector Boson Fusion. Reductions in these branching fractions
then can reduce the reach of the search via these channels. Of course,  
the large values of ${\cal B}_{\chi \chi}$ give rise to new search channels 
where the Higgs decays into invisible products.

In this case the 'invisibility' of the Higgs is caused by the mixed nature 
of  Higgsino-Gaugino content of the LSP and its small mass. 
A light $\tilde\chi_1^0$  with such couplings has implications for the 
relic density of the neutralinos in the Universe, as the latter is decided by
$\sigma (\tilde \chi_1^0 \tilde \chi_1^0  \rightarrow  f^+ f^-)$. For a light
\begin{figure}[htb]
\centerline{
      \includegraphics*[scale=0.40]{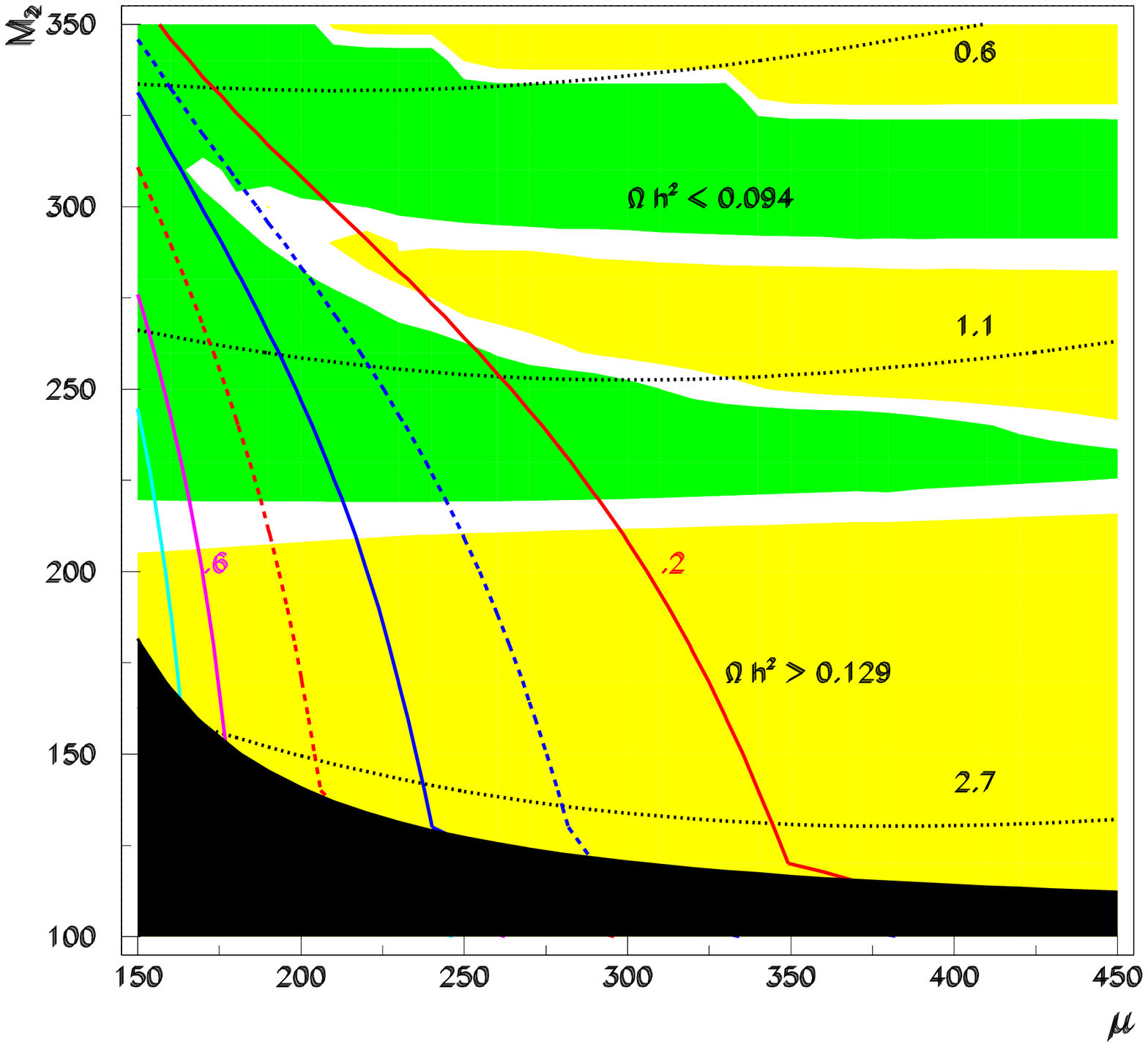}
      \includegraphics*[scale=0.40]{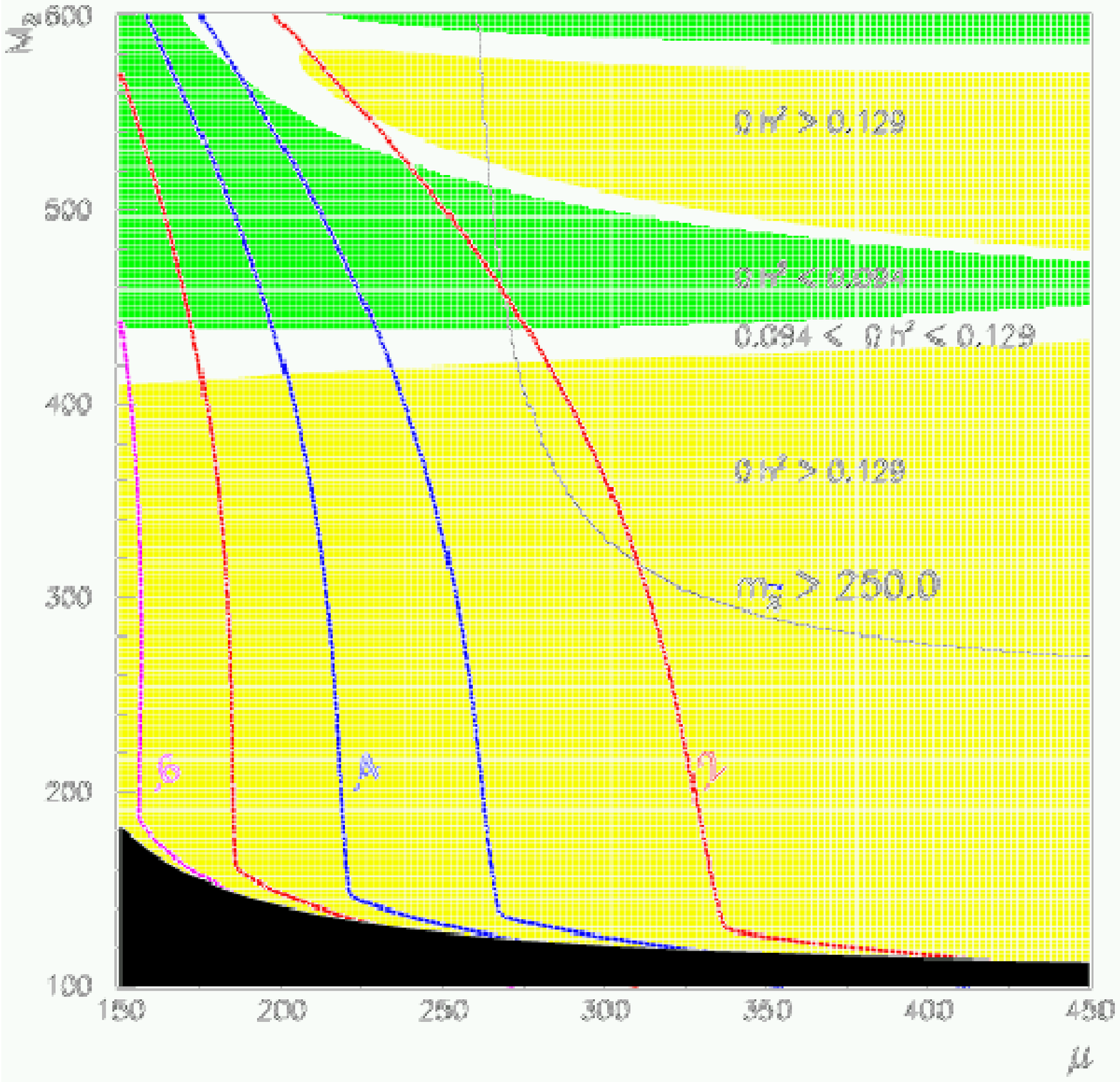}
  }
\caption{\label{sec254_fig2}\footnotesize
Left and right panel shows contours of  
$B.R. (h \rightarrow  \tilde \chi_1^0 \tilde \chi_1^0)$
for $r=M_1/M_2|_{EW}=0.2$ and $0.1$ respectively,  along with the DM and  
LEP constraints. The white region corresponds to $0.094 < \Omega h^2 < 0.129$. 
Parameter $m_0$ determining the slepton mass~\cite{Belanger:2000tg,Belanger:2001am} takes values 90  and 94 GeV for the left and right panel respectively.
$\tan \beta = 5$ and $m_h = 125$ GeV. The black region is the LEP-excluded
region.  The lightly and heavily shaded regions correspond to
$\Omega h^2 > 0.129$ and $\Omega h^2 < 0.094$ respectively.
}
\end{figure}
$\tilde \chi_1^0$, the $Z/h$-mediated $s$ channel process contributes to the
annihilation. If the $\tilde l_R$  is light  the cross-section also receives a
contribution from the $t$-channel $\tilde l_R$ exchange.
There is a clear correlation between the expected `invisible' branching
ratio for the $h$ and the relic density of the $\tilde \chi_1^0$ as the same
couplings are involved.
Left and right panels of figure~\ref{sec254_fig2} show 
results ~\cite{Belanger:2001am,
Boudjema:2002eu} Obtained for $r = M_1/M_2|_{EW} = 0.2 (0.1)$,
$\tan \beta = 5$ and $m_{\tilde l_R} \sim 100$ GeV using the
micrOMEGAs~\cite{Belanger:2001fz} program to calculate the relic density.
These show  that the requirement of an acceptable relic density does
constrain the $M_2$--$\mu$ plane quite substantially. However,
there exist large regions of this plane where  the `invisible' branching
ratio of $h$ is as large as $0.5$--$0.6$, even for `large' $(\sim 200$ GeV)
$\tilde l_R$,  consistent with the LEP constraints and with an acceptable
relic density.

\subsubsection{Implications for the sparticle phenomenology at the LHC and 
the LHC-LC interplay}
                                                                                
In a scenario where the lightest MSSM Higgs $h$ signal in the $\gamma \gamma$
or $b \bar b$ channel is invisible at the LHC due to one of the two reasons 
mentioned above, it is imperative to ask two questions : 1) whether it is 
possible to recover the Higgs signal in some other channel and 2) whether the
particular scenario responsible for making the $h$ 'invisible' can be tested at
the LHC. Of course information that even a first stage LC could provide about 
such a $h$ which is 'invisible' at the LHC will be very crucial in this 
exercise.

If it is the light $\sto$ that makes the $h$ 'invisible', the signal for the
$h$ can be recovered at the LHC in the  channel  $\sto \sto h$ and 
$\sto \stt h$ as the cross-section for this process will then be substantial. 
It may be that the Tevatron might be able to see a $\sto$ with $m_{\sto} \sim
160-170$ GeV. Feasibility of observation of a $\sto$ in this mass range, should 
$c \tilde \chi_1^0$ be its dominant decay mode, has still not been properly 
investigated. The large
QCD backgrounds at the LHC might not make it very easy. On the other hand, 
even a first phase LC will be able to probe this mass range for the $\stop$
very easily and thoroughly. 
\begin{figure*}[hbtp]
\centerline{
\includegraphics[width=8cm,height=7cm]{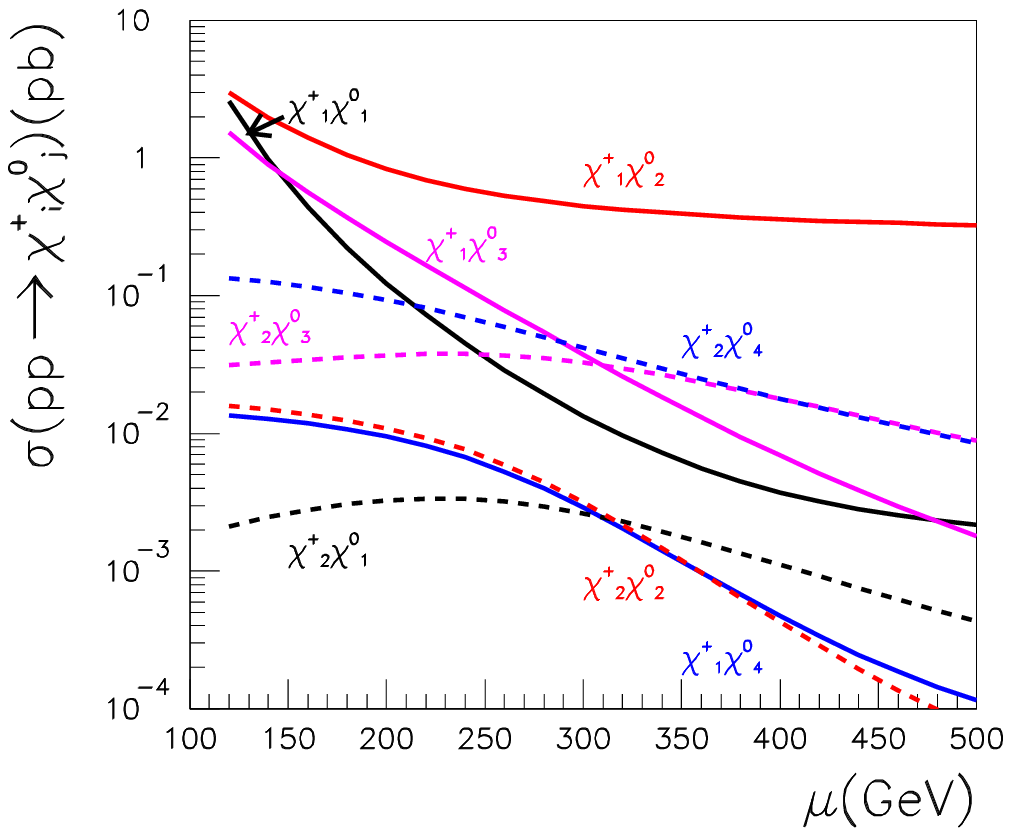}
\includegraphics[width=8cm,height=7cm]{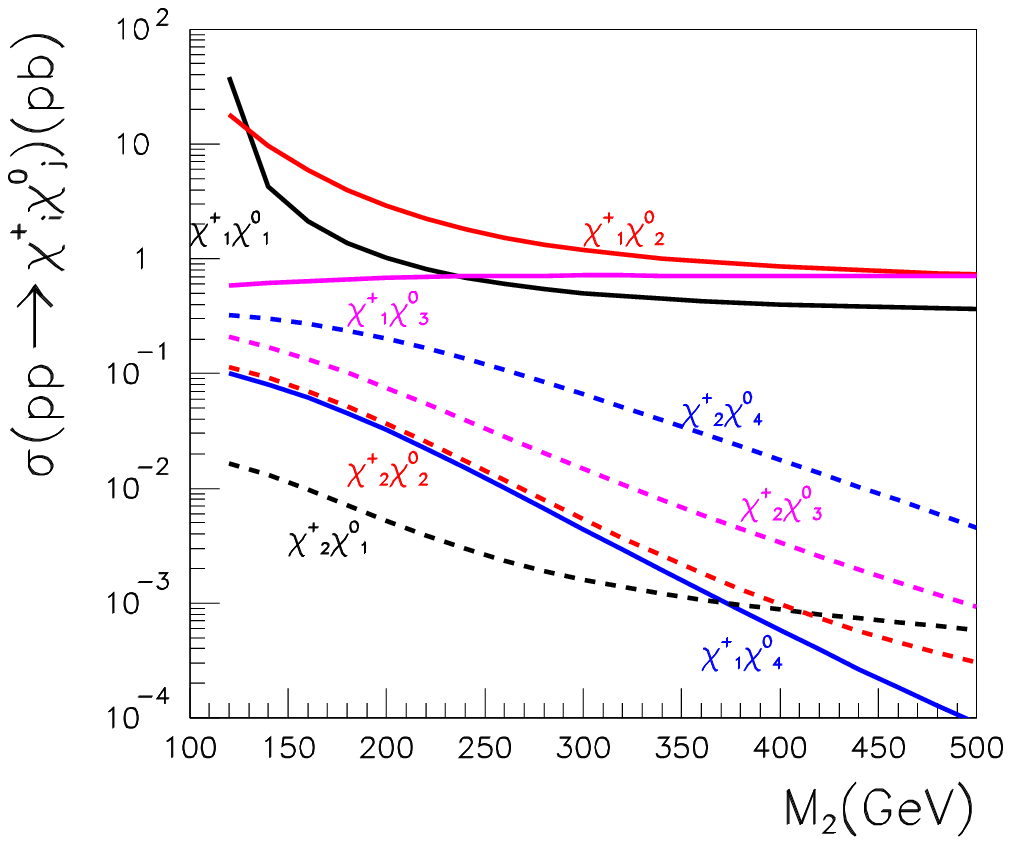}}
\caption{\label{sec254_chiprod-nounif}{\footnotesize Associated Production of
chargino and neutralino at the LHC at LO for $M_2=M_1/10$ a) as a
function of $\mu$ for $M_2=250$GeV. b) as a function of $M_2$ for
$\mu=150$GeV \/.}}
\end{figure*}
If it is the nonuniversal gaugino  masses that make the $h$ 'invisible' then 
the  smaller mass of the $\tilde \chi_1^0$ as well its mixed nature will cause
the  trilepton signal at the Tevatron to be qualitatively different from 
that expected in the universal case. Thus this scenario may be tested at 
the Tevatron via the hadronically quiet trilepton events. At the LHC 
eventhough the EW production of $\tilde \chi^\pm \tilde \chi^0$ is subdominant,
usual $b \bar b$ and $\gamma \gamma$ signatures of the Higgs,  production
of charginos and neutralino is quite substantial\footnote{Production of 
light sleptons, as constrained from cosmology in these scenarios, is on the 
other hand quite modest at the LHC.}.
Fig.~\ref{sec254_chiprod-nounif} shows that, for values of $\mu-M_2$
where $R_{\gamma \gamma}$ is below $.6$, all neutralinos and
charginos can be produced. For instance with $M_2=250$GeV, the
cross section for $\neutf \tilde \chi_2^+$ is in excess of $100$ fb while
$\neutt \tilde \chi_1^+$ is above 1 pb. In our analysis we had taken masses
of all the sfermions and heavier Higgses  to be in the TeV range. If the 
gluino/squarks are in the mass range of about $500$ GeV, the cascade 
production of $\tilde \chi^\pm \tilde \chi^0$ is then quite substantial. 
A correlation between the observations
in the chargino-neutralino sector, a possible non observation of the 
Higgs at the LHC and a signal at the LC for a Higgs with 'invisible' decay 
mode can thus unravel the issue completely.

If we now look at the (lightest) Higgs that can be produced
through cascade decays in these processes, one sees from
\begin{figure*}[hbtp]
\begin{center}
\includegraphics[width=13cm,height=8cm]{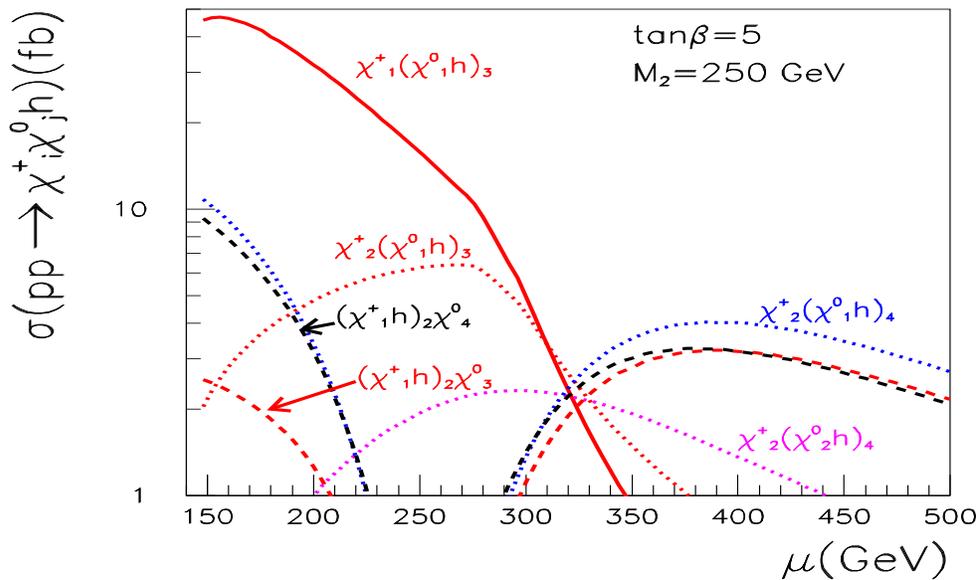}
\caption{\label{sec254_folding-nouni}{\footnotesize 
Higgs yield through charginos and
neutralinos decays as a function of $\mu$, $M_2=250$GeV,
$\tan \beta =5$, maximal mixing and $M_1=M_2/10$. The subscript for
the parentheses $(\;\;)_j$ indicates the parent neutralino or
chargino\/.}}
\end{center}
\end{figure*}
Fig.~\ref{sec254_folding-nouni} that, through essentially $\neutth$
decays, associated Higgs cross sections of about $30$fb are
possible. Nonetheless, again, it is in these regions with highest
yield that the Higgs has a large branching ratio into invisible
and would be difficult to track.  The $h$ search in neutralino decays at the
LHC, in this scenario is being investigated and should certainly provide 
clean examples of the possible LHC-LC interplay.

The implications of this scenario for the SUSY phenomenology at LC have been
recently investigated\cite{Belanger:2003wb}. It has been shown there that,
for the region of small values of $\mu$ where the Higgs can become 
'invisible' for  nonuniversal gaugino masses, for the range of nonuniversality
parameter $0.03$ -- $0.5$ (the universal case) and $\tan \beta $ values from
$10$--$50$, the  production cross-section will be at an observable level 
($> 1 fb$ at a $500$ GeV LC) for at least one of the sparticle 
pairs from among  $\tilde \chi^+ \tilde \chi^-, 
\tilde \chi_1^0 \tilde \chi_3^0, \tilde \chi_1^0 \tilde \chi_2^0$.
For smaller values of $m_0$, even the slepton signal will be large and
observable. In this case even the $\tilde \chi_1^+ \tilde \chi_1^- \gamma$
production cross-section will be appeciable.  Thus not only can an LC see an 
'invisible' Higgs easily, but it should be able to see the sparticles, thus 
testing whether SUSY is responsible for the nonobservation of the light
Higgs at the LHC. Further, measurements of the invisible decay width at
the LC, can provide useful pointers  for the sfermion and neutralino/chargino
phenomenology at the LHC. Implications for this scenario for lower masses of
the Pseudoscalar Higgs have yet to be investigated fully.


\section{A light Higgs in scenarios with extra dimensions }
\label{sec:21}

\def\lphi{\Lambda_\phi}
\def\gam{\gamma}
\def\epem{e^+e^-}
\def\br{{\mathrm BR}}
\def\mphi{M_\phi}
\def\mpl{M_{Pl}}
\def\hsm{h_{SM}}
\renewcommand{\be}{\begin{equation}}
\renewcommand{\ee}{\end{equation}}
\renewcommand{\bea}{\begin{eqnarray}}
\newcommand{\beas}{\begin{eqnarray*}}
\renewcommand{\eea}{\end{eqnarray}}
\newcommand{\eeas}{\end{eqnarray*}} 
\newcommand{\ba}{\begin{array}}
\newcommand{\ea}{\end{array}}
\newcommand{\bi}{\begin{itemize}}
\newcommand{\ei}{\end{itemize}}
\renewcommand{\ben}{\begin{enumerate}}
\renewcommand{\een}{\end{enumerate}}

In this section we discuss the detectability of a light Higgs
at the LHC, and the corresponding complementarity of a LC, for
scenarios beyond the SM based on extra dimensions.
Models with 3-branes in extra dimensions typically
imply the existence of a radion, $\phi$, that can mix with the Higgs, $h$,
thereby modifying the Higgs properties and the prospects for 
its detectability at the LHC. The presence of the $\phi$ will affect
the scope of the LHC searches. Detection of both the $\phi$ and the $h$
might be possible. We report on a study on the complementarity of the 
observation of $gg \to h$, with $h \to \gamma \gamma$
or $h\to Z^0Z^{0*}\to 4~\ell$, and 
$gg \to \phi \to Z^0Z^{0(*)} \to 4~\ell$ at the LHC 
in the context of the Randall-Sundrum 
model. The potential for determining the nature of the detected scalar(s) 
at the LHC and at an $e^+e^-$ linear collider is discussed, 
both separately and in combination.
Also the virtues of  measurements from  a 
low energy photon collider are discussed.


\subsection {On the complementarity of Higgs and radion searches at LHC
and LC}
{\it M.Battaglia,  S. De Curtis,  A. De Roeck,  D. Dominici, J.F. Gunion}

\subsubsection{Introduction}
One particularly attractive extra-dimensional model is that proposed
by Randall and Sundrum (RS)~\cite{rs}, in which there are two 3+1 dimensional
branes separated in a 5th dimension. A central prediction of
this theory is the existence of the radion,  a graviscalar which 
corresponds to fluctuations in the size of the extra 
dimension. Detection and study of the radion
will be central to the experimental probe of the RS and related
scenarios with extra dimensions. 
There is already an extensive literature on the phenomenology of the radion,
both in the absence of Higgs-radion 
mixing~\cite{Bae:2000pk,Davoudiasl:1999jd,
Cheung:2000rw,Davoudiasl:2000wi,Park:2000xp}
and in the presence 
of such a mixing~\cite{Giudice:2000av,Csaki:2000zn,Han:2001xs,Chaichian:2001rq,
Azuelos:fv,Hewett:2002nk,Dominici:2002jv}.

In this section we discuss the complementarity 
of the search for the Higgs boson and the 
radion at the LHC. 
As the Higgs-radion mixing may suppress the main discovery process 
$gg \to H \to \gamma \gamma$ for a light Higgs boson, we study the extent
to which  the 
appearance of a $gg \to \phi 
\to Z^0Z^{0(*)} \to 4~\ell$ signal ensures that LHC 
experiments will observe at least 
one of the two scalars over the full parameter phase 
space. The additional information, which could be extracted from a TeV-class 
$e^+e^-$ linear collider (LC), is also considered.
More details on the theoretical framework are given  in~\cite{Battaglia:2003gb}




\subsubsection{Radion and Higgs Boson Search Complementarity}

Here we address two issues. The first is whether there is a complementarity 
between the Higgs observability, mostly through $gg \to h \to \gamma \gamma$, 
and the 
$gg \to \phi \to Z^0Z^{0(*)} \to 4~\ell$ reaction, thus offering the LHC the 
discovery 
of at least one of the two particles over the full parameter space. The second,
 and 
related, issue,discussed in the next section,
 concerns the strategies available to understand the nature of 
the 
discovered particle. 

The couplings of the $h$ and $\phi$ to $Z^0Z^0$, $W^+W^-$ and $f\overline{f}$
 are given
relative to those of the SM Higgs boson, denoted by $H$, by:
\begin{equation}
{g_{hWW}\over g_{HWW}}={g_{hZZ}\over g_{HZZ}}={g_{hf\overline{f}}\over 
g_{H f\overline{f}}}=d+\gam b\,,\quad
{g_{\phi WW}\over g_{HWW}}={g_{\phi ZZ}\over g_{HZZ}}={g_{\phi f\overline{f}}\over g_{H f\overline{f}}}=c+\gam a\,. 
\label{sec211couplings}
\end{equation}
with $\gamma\equiv v/\lphi$ and $a,b,c,d,$ are functions of the mixing angle
defined e.g. in~\cite{Battaglia:2003gb}.
Couplings of the $h$ and $\phi$ to $\gamma\gamma$ and $gg$
receive contributions not only from the usual loop diagrams but also
from trace-anomaly couplings to $\gamma\gamma$ and
$gg$.  Thus, these couplings are not simply directly proportional to
those of the SM $H$.  Of course, in the limit of $\xi=0$, the $h$ has
the same properties as the SM Higgs boson.

The effects of the mixing of the radion with the Higgs boson have been 
studied~\cite{Dominici:2002jv} by introducing the relevant terms in the 
{\sc HDecay} 
program~\cite{Djouadi:1997yw}, which computes the Higgs couplings, including 
higher 
order QCD corrections. Couplings and widths for the radion have also been 
implemented.

Results have been obtained by comparing the product of production and
decay rates for the $h$ and $\phi$ to those expected for a light SM
$H$. The LHC sensitivity has been extracted by rescaling the results
for Higgs observability, obtained assuming SM couplings. We define 
Higgs observability as a $>5~\sigma$ excess over the SM background for
the combination of the inclusive channels: $gg \to h \to \gamma
\gamma$; $t \bar{t} h$, with $h \to b \bar{b}$ and 
$gg\to h \to Z^0Z^{0(*)} \to
4\ell$, after rescaling the SM Higgs results.
We define $\phi$ observability as a $>5~\sigma$ signal
in the $gg\to \phi \to Z^0Z^{0(*)}\to 4\ell$ channel only. We study
the results as a function of four parameters: the Higgs mass $M_h$,
the radion mass $M_{\phi}$, the scale $\lphi$ and the mixing parameter
$\xi$.

Due to the suppression, from radion mixing, of the loop-induced effective 
couplings of the $h$ (relative to the SM $H$) to 
gluon and photon pairs, the key process $gg \to h \to \gamma \gamma$ 
may fail to provide 
a significant excess over the $\gamma \gamma$ background at the LHC. 
Other modes that 
depend on the $gg$ fusion production process are suppressed too. 
For $M_\phi>M_h$, this suppression is very substantial 
for large, negative values of $\xi$. This region of significant suppression 
becomes wider at large values of $M_\phi$  and $\lphi$. 
In contrast, for $M_\phi<M_h$, the $gg\to h\to \gamma\gamma$
rate is generally only suppressed when $\xi>0$. All this is shown, in a 
quantitative way, by the contours in Figures~\ref{sec211fig:detect}, 
\ref{sec211fig:compl120} and \ref{sec211fig:compl140}.
\begin{figure}[htb!]
\begin{center}
\begin{tabular}{c c}
\hspace*{-0.85cm} 
\epsfig{file=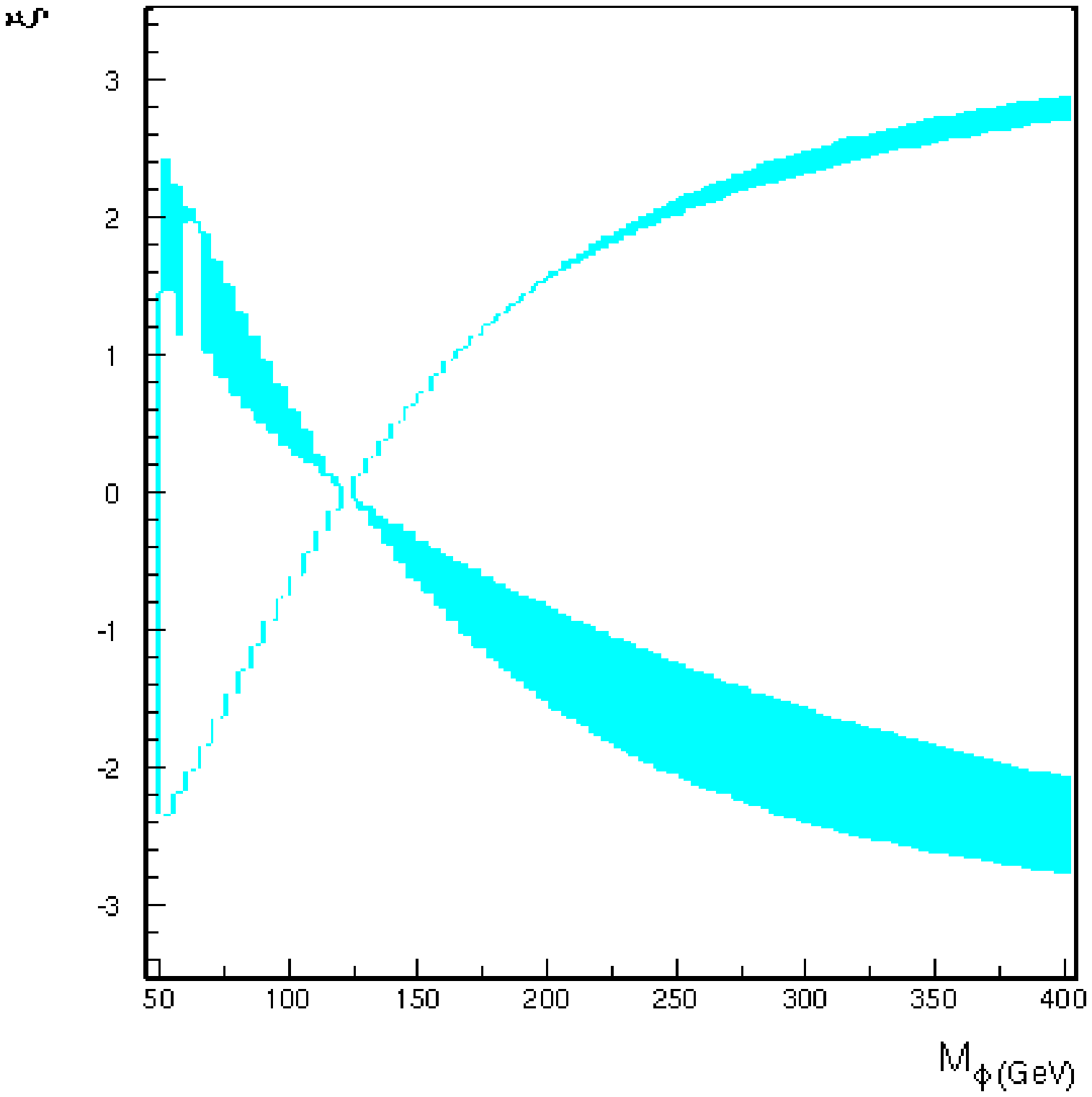,width=7.5cm,height=6.0cm} &
\hspace*{-0.7cm} 
\epsfig{file=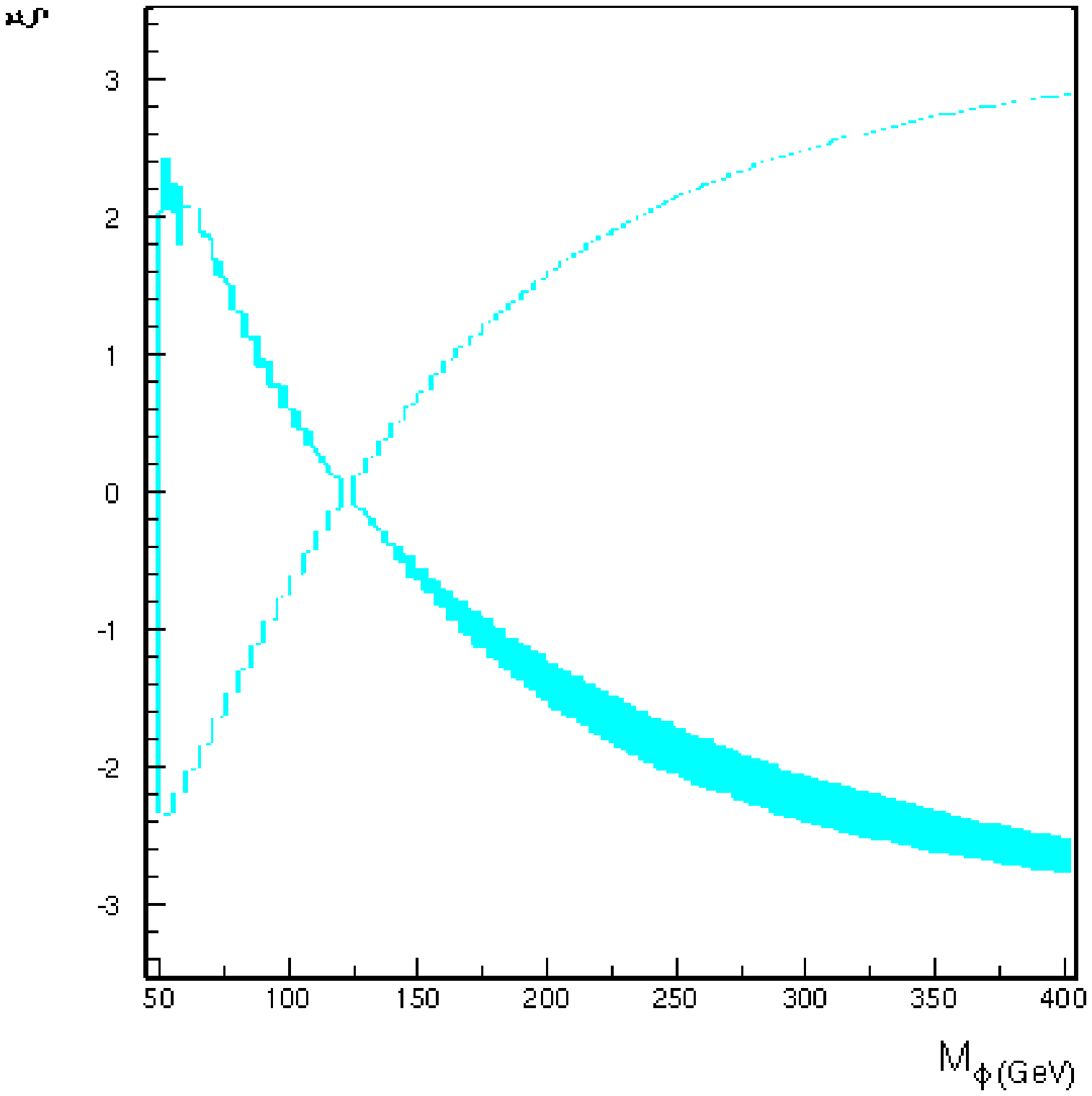,width=6.5cm,height=6.0cm} \\
\end{tabular}
\end{center}
\vspace*{-0.75cm}
\caption{\sl Regions in $(M_\phi,\xi)$ parameter space
of $h$ non-detectability (including $gg \to h\to \gamma\gamma$ and other modes)
at the LHC for one experiment and 30~fb$^{-1}$ (left) and 100~fb$^{-1}$ (right). We take $\lphi=5$ TeV and $M_h=120$ GeV.}
\label{sec211fig:detect}
\end{figure}
In all these figures, the outermost, hourglass shaped contours 
define the theoretically 
allowed region. As shown in Figure~\ref{sec211fig:detect}, 
three main regions of non-detectability may appear. 
Two are located at large values of 
$M_{\phi}$ and $|\xi|$. 
A third region appears at low $M_{\phi}$ and positive $\xi$, where
the above-noted $gg\to h\to \gamma\gamma$ suppression sets in. This latter region
becomes further expanded when $2M_\phi<M_h$ and the decay channel
$h \to \phi \phi$ opens up, thus reducing the $h \to \gamma \gamma$ 
branching ratio. Figure~\ref{sec211fig:compl120} shows how the regions
of non-detectability shrink (expand) when $M_\phi<M_h$ ($M_\phi>M_h$)
as $\lphi$ increases.
Figure~\ref{sec211fig:compl140} shows that the non-detectability 
regions shrink as $M_h$ increases from $115$ GeV to $180$ GeV. 
The detectability is increasing as additional channels, in particular 
$gg\to h \to Z^0 Z^{0*}\to 4~\ell$, become available for Higgs discovery. 
For large Higgs boson masses the region of non-detectability is reduced to a 
narrow strip along the lower $\xi$ edge. 
\begin{figure}[htb!]
\vspace*{-0.75cm}
\begin{center}
\begin{tabular}{c c c}
\hspace*{-0.85cm} 
\epsfig{file=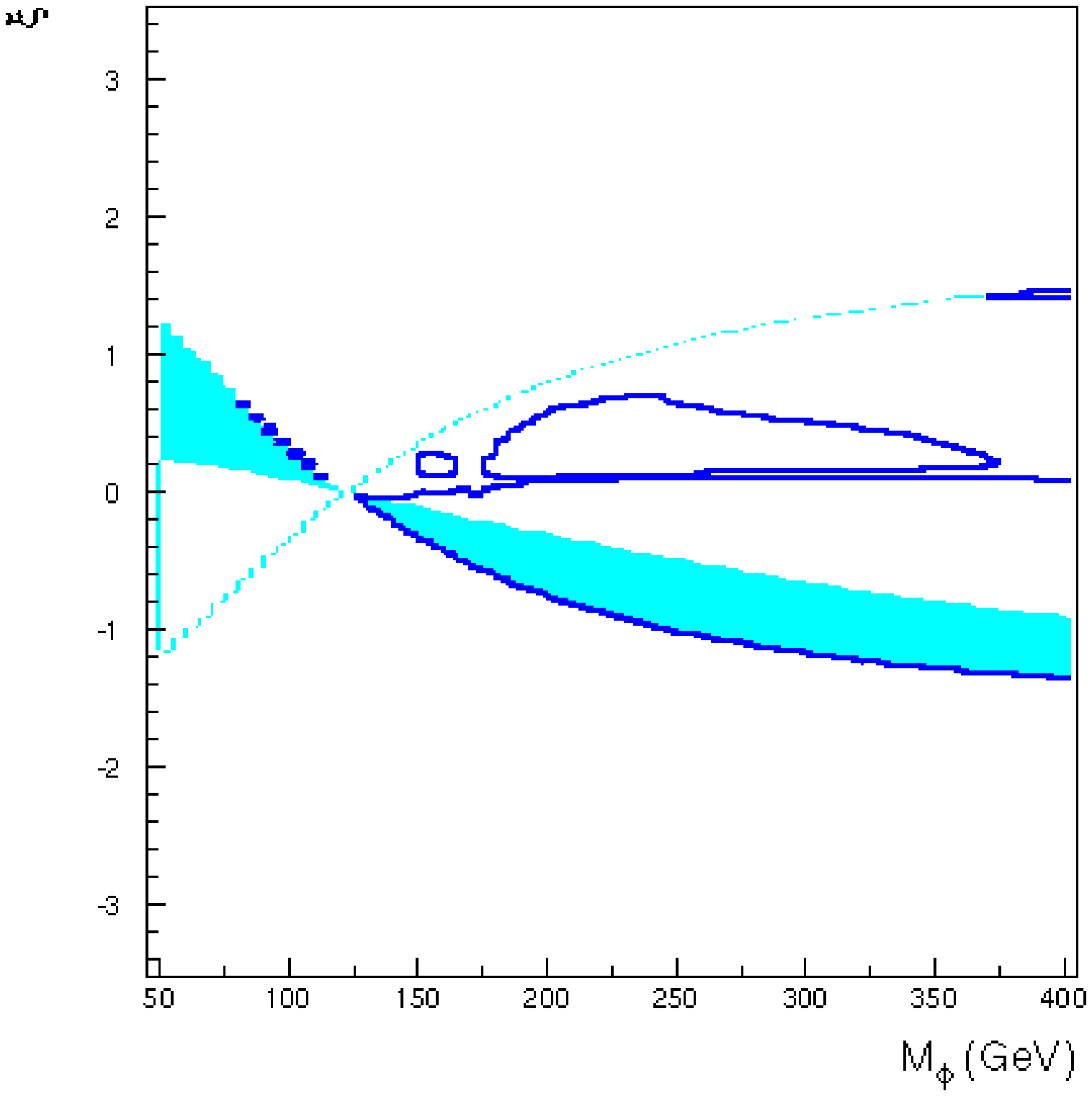,width=5.8cm,height=5.8cm} &
\hspace*{-0.7cm} 
\epsfig{file=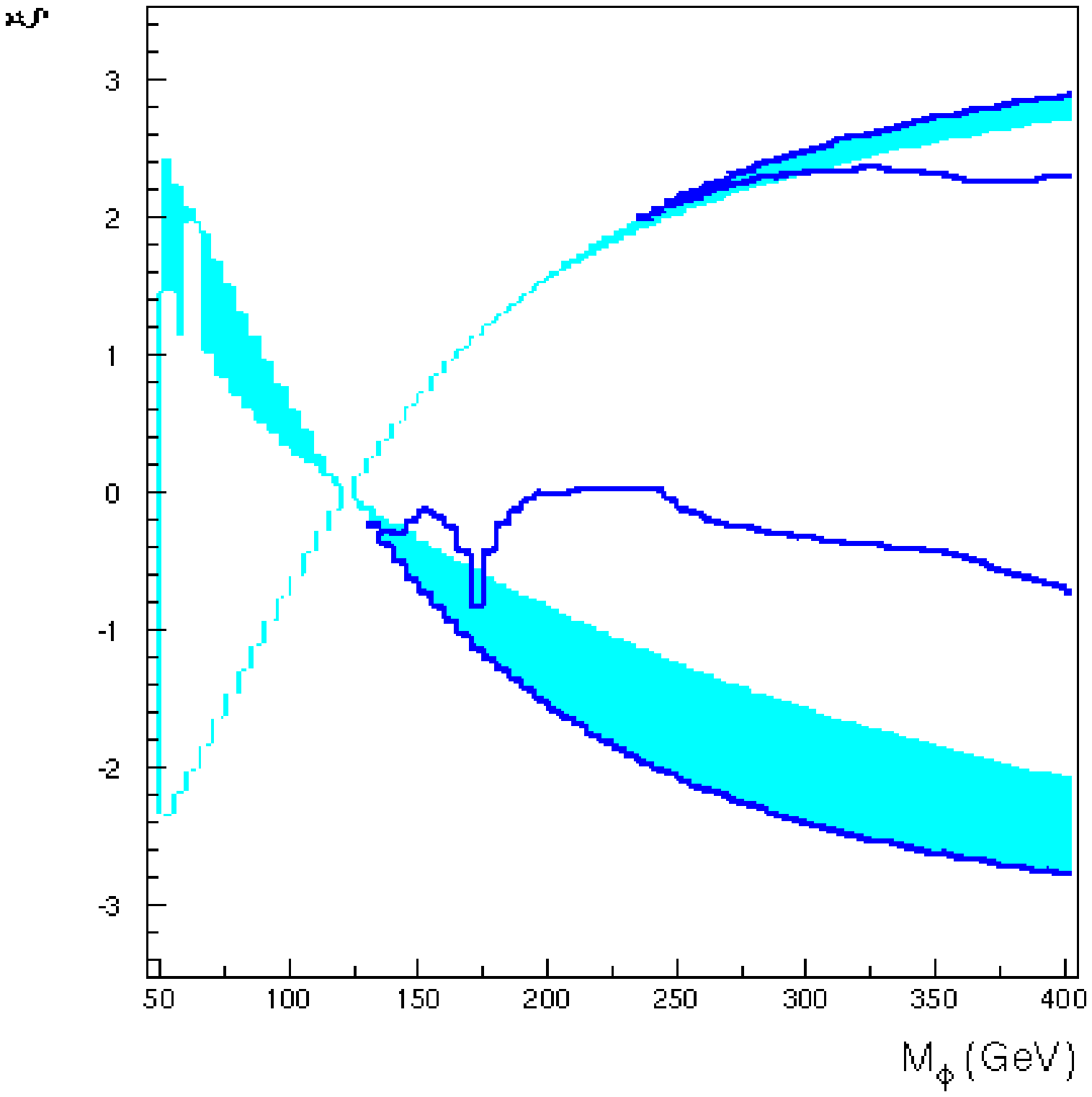,width=5.8cm,height=5.8cm} &
\hspace*{-0.7cm} 
\epsfig{file=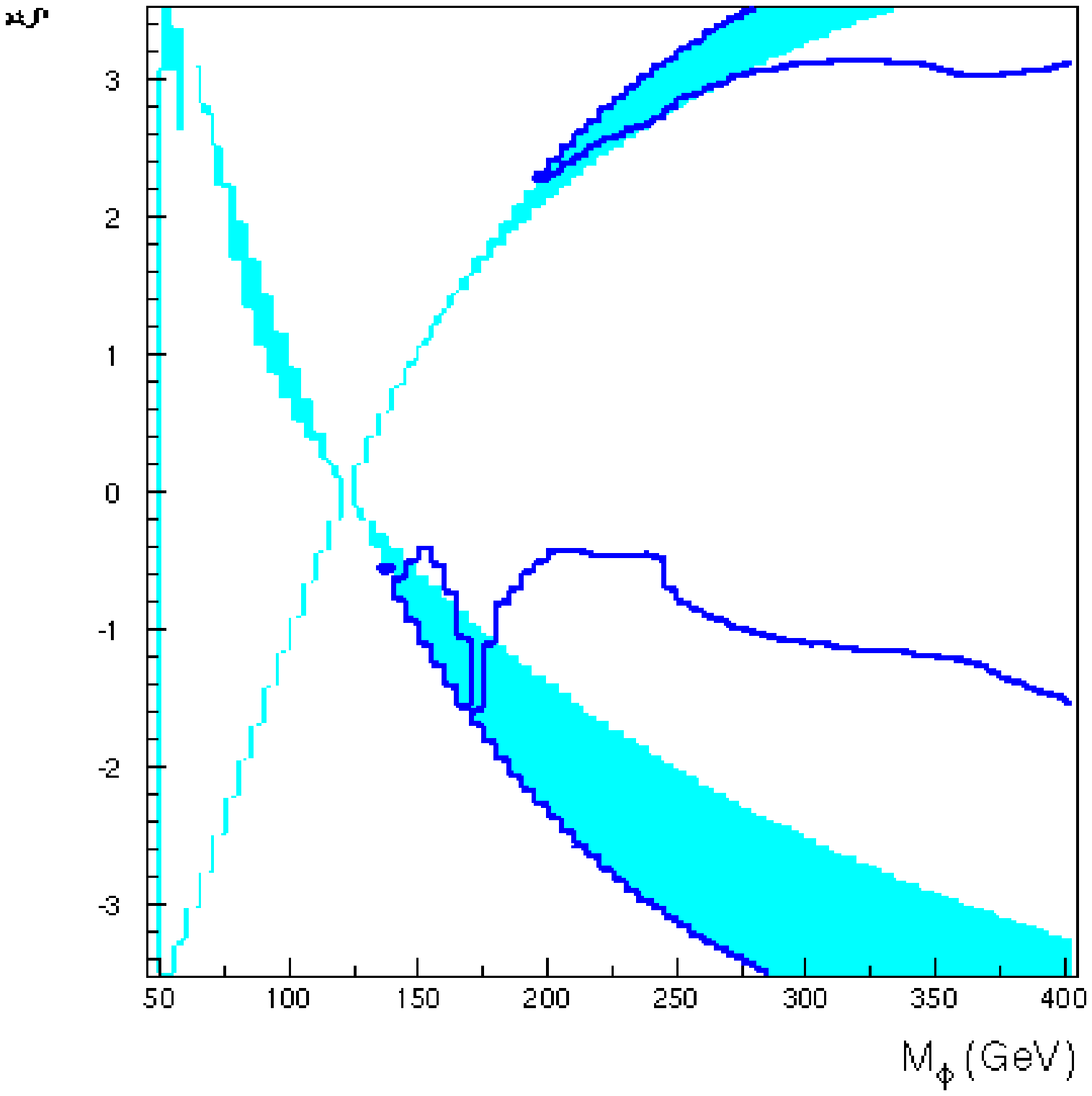,width=5.8cm,height=5.8cm} \\
\end{tabular}
\end{center}
\caption{\sl Regions in $(M_\phi,\xi)$ parameter space
of $h$ detectability (including $gg \to h\to \gamma\gamma$  and other modes)
and of $gg \to \phi \to Z^0Z^{0(*)} \to 4~\ell$ detectability 
at the LHC for one experiment and 30~fb$^{-1}$. 
The outermost, hourglass shaped contours 
define the theoretically allowed region. 
The light grey (cyan) regions show the part of 
the parameter space where the net $h$ signal significance falls below 
$5~\sigma$. The thick grey (blue) curves indicate the 
regions where the significance of the 
$gg \to \phi \to Z^0Z^{0(*)} \to 4~\ell$ signal 
exceeds $5~\sigma$. Results are presented for 
$M_h$=120~GeV and $\lphi$= 2.5~TeV (left),
5.0~TeV (center) and 7.5~TeV (right).} 
\label{sec211fig:compl120}
\end{figure}
\begin{figure}[htb!]
\begin{center}
\begin{tabular}{c c c}
\hspace*{-0.85cm} 
\epsfig{file=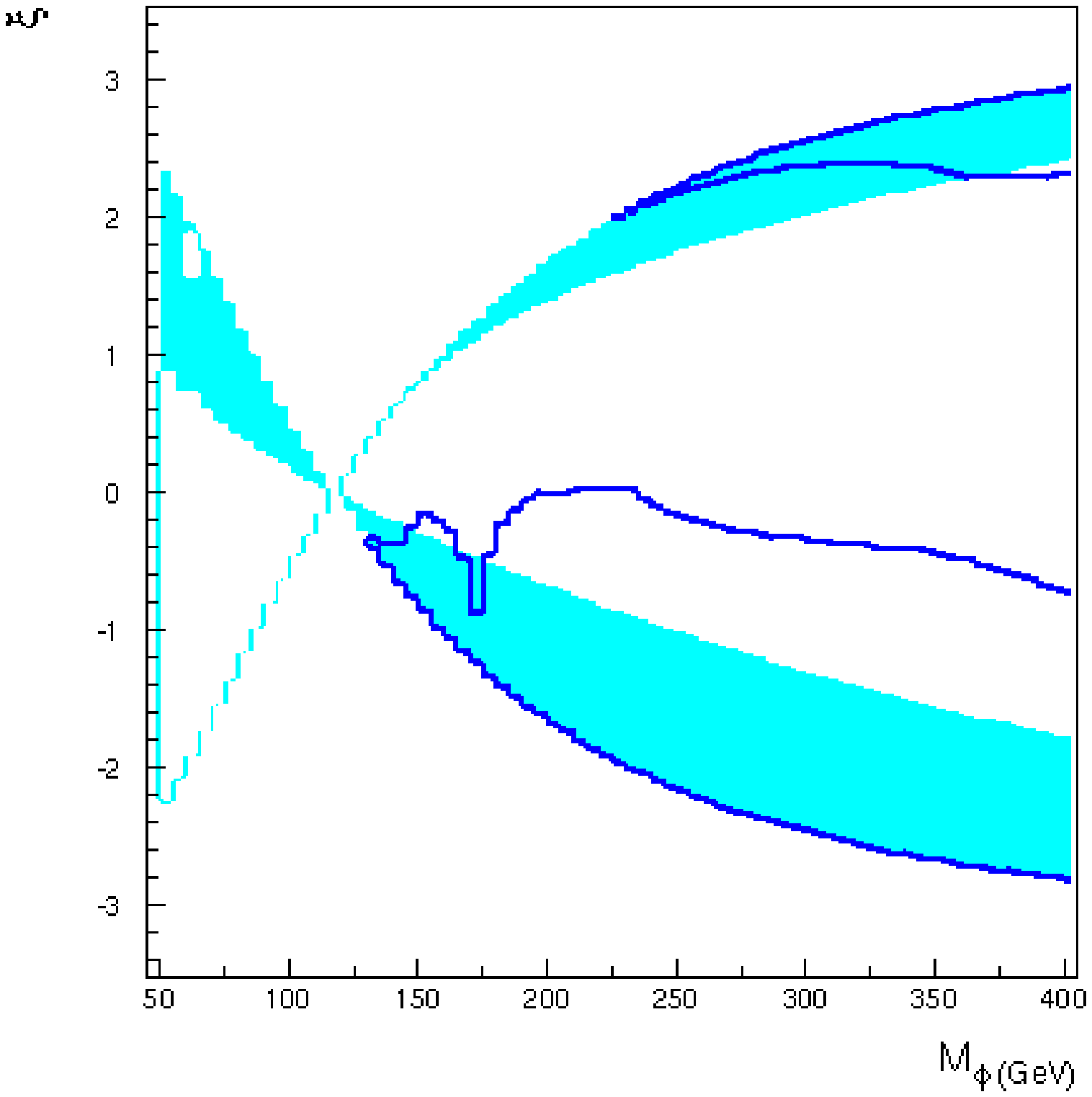,width=5.8cm,height=5.8cm} &
\hspace*{-0.7cm} 
\epsfig{file=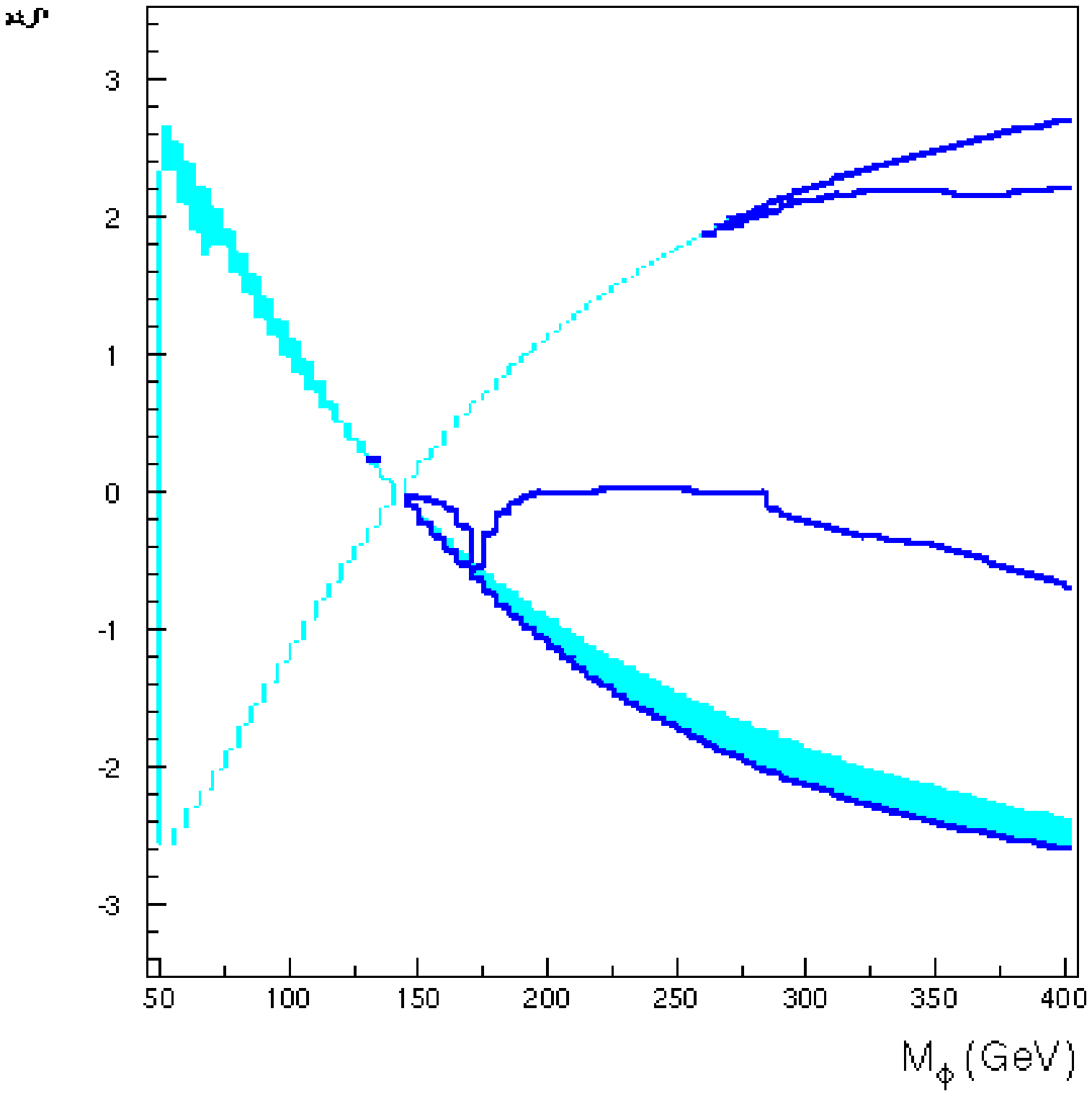,width=5.8cm,height=5.8cm} &
\hspace*{-0.7cm} 
\epsfig{file=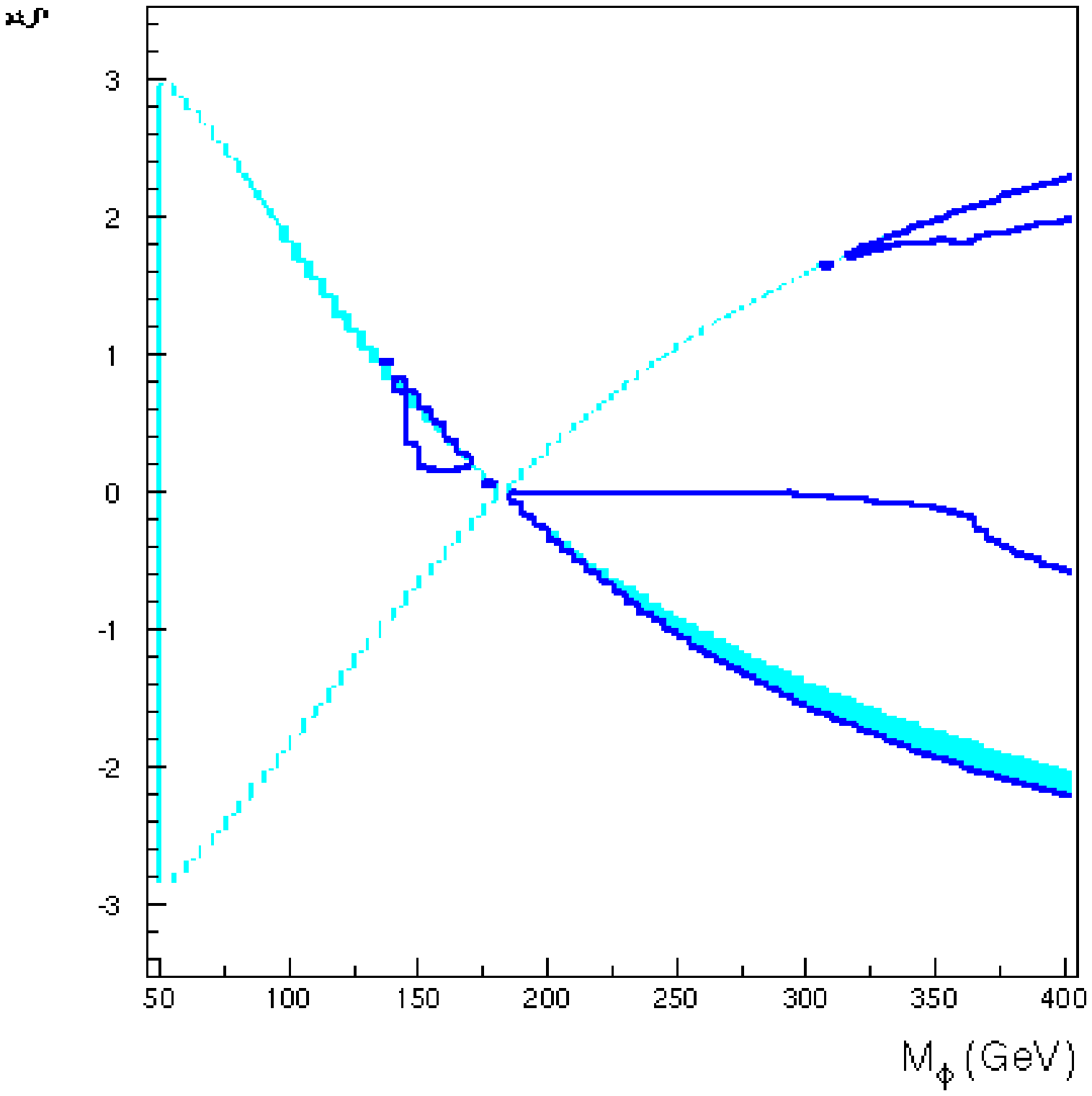,width=5.8cm,height=5.8cm} \\
\end{tabular}
\end{center}
\caption{\sl Same as Figure~\ref{sec211fig:compl120} but for $M_h$ = 115~GeV (left), 
$M_h$ = 140~GeV (center) and $M_h$ = 180~GeV (right). $\lphi$ has been fixed to 
5.0~TeV.}
\label{sec211fig:compl140}
\end{figure}

Figure~\ref{sec211fig:detect} shows that the non-detectability regions 
are reduced by considering a larger data set. In particular, 
we see that an integrated luminosity of 100~fb$^{-1}$ 
would remove the regions at large positive 
$\xi$ in the $\lphi=5$ and $7.5$~TeV plots of Figure~\ref{sec211fig:compl120}. 
Similarly, including the $qqh$, $h \to WW^* \to \ell \ell \nu \bar \nu$ channel 
in the list of the discovery modes removes 
the same two regions and reduces the large 
region of $h$ non-observability at negative $\xi$ values. 
 
In all these regions, a complementarity is potentially offered by the process 
$gg \to \phi \to Z^0Z^{0(*)} \to 4~\ell$, 
which becomes important for $M_{\phi} > 140$~GeV. 
At the LHC, this process would have the same event
structure as the golden SM Higgs mode 
$H \to Z^0Z^{0*} \to 4~\ell$, which has been thoroughly studied for an 
intermediate mass Higgs boson. 
By computing the  $gg \to \phi \to Z^0Z^{0(*)} \to 4~\ell$ rate relative
to that for the 
corresponding SM $H$ process and employing the LHC sensitivity curve for 
$H \to Z^0Z^{0(*)}$, the significance 
for the $\phi$ signal in the 
$4~\ell$ final state at the LHC can be extracted. 
Results are overlayed on 
Figures~\ref{sec211fig:compl120} and \ref{sec211fig:compl140}, 
assuming 30~fb$^{-1}$ of data.

Two observations are in order. 
The observability of $\phi$ production in the four lepton 
channel fills most of the gaps in $(M_h,\xi)$ parameter space
in which $h$ detection is not possible (mostly due to the suppression of 
the loop-induced  $gg\to h\to \gam\gam$ process).
The observation of at least one scalar is thus guaranteed over almost the full 
parameter space, with the following exceptions. (a) 
In the region of large positive $\xi$ with $M_{\phi} <M_h$
the $\phi$ couplings are suddenly suppressed (as opposed to the $M_\phi>M_h$
side of the hourglass) making the $gg\to\phi$ production rate too small
for $\phi$ observation in either the $\gam\gam$ or $Z^0Z^{0*}\to 4\ell$
final state. (The $Z^0Z^{0*}\to 4\ell$ mode is also phase space suppressed
for smaller $M_\phi$.) 
(b)  There is a narrow region at 
$M_{\phi} \simeq$~170~GeV due to the ramp-up of the 
$\phi \to W^+W^-$ channel, where a 
luminosity of order 100~fb$^{-1}$ is required to reach a $\ge 5~\sigma$ 
signal for $\phi \to Z^0Z^{0*}$. 
 We should also note that
the $\phi \to Z^0Z^{0}$ decay is reduced for $M_\phi>2M_h$ by the 
onset of the $\phi \to hh$ decay, which can become the main decay mode. 
The resulting $hh\to b \bar{b} b \bar{b}$ topology, 
with di-jet mass constraints, may represent a viable 
signal for the LHC in its own right, but detailed studies will be needed. 
Figures~\ref{sec211fig:compl120} and \ref{sec211fig:compl140} also exhibit regions
of $(M_\phi,\xi)$ parameter space in which {\it both} the $h$ and $\phi$
mass eigenstates will be detectable.
In these regions, the LHC will observe two scalar bosons
somewhat separated in mass with the lighter (heavier) having a non-SM-like 
rate for the the $gg$-induced $\gamma\gamma$ ($Z^0Z^0$) final state.
Additional information will be required to ascertain whether these two Higgs
bosons derive from a multi-doublet or other type of extended
Higgs sector or from the present type of model with Higgs-radion mixing.

An $e^+e^-$ LC should guarantee observation of both the $h$ and the
$\phi$ even in most of the regions within which detection of either at
the LHC might be difficult. Thus, this scenario provides an
illustration of the complementarity between the two machines in the
study of the Higgs sector.  In particular, in the region with
$M_\phi>M_h$ the $hZ^0Z^0$ coupling is enhanced relative to the SM
$HZ^0Z^0$ coupling and $h$ detection in $e^+e^-$ collisions would be
even easier than SM $H$ detection. Further, assuming that $e^+e^-$
collisions could also probe down to $\phi Z^0Z^0$ couplings of order
$g^2_{\phi ZZ}/g^2_{HZZ} \simeq 0.01$, the left panel of
Figure~\ref{sec211fig:lphi30} shows that the $\phi$ would be seen in almost
the entirety of the $M_\phi>M_h$ region, aside from a narrow cone near
$\xi\sim 0$.  In the $M_\phi<M_h$ region, the $hZ^0Z^0$ coupling is
suppressed, but only by a modest amount; $\epem\to Z^0 h$ would be
easily detected. As seen in Figure~\ref{sec211fig:lphi30}, detection of the
$\phi$ in the $M_\phi<M_h$ part of parameter space will only be
possible if $\xi$ is near the edges of the hourglass region. This
would include the large $\xi>0$ region (a) defined above. In regions
where both $\epem\to Z^0 h$ and $\epem\to Z^0\phi$ can be seen the
{\it four} measured quantities $g_{Z^0Z^0h}^2$, $g_{Z^0Z^0\phi}^2$,
$M_h$ and $M_\phi$ would significantly constrain the values of the
$\xi$ and $\lphi$ parameters of the model, often leaving only a
two-fold ambiguity in their determination.

\subsubsection{Determining the Nature of the Observed Scalar} 

The interplay between the emergence of the Higgs boson 
and of the radion graviscalar 
signals opens up the question of the identification of the 
nature of the newly observed particle(s).

After observing a new scalar at the LHC, some of its properties will be 
measured with sufficient accuracy to determine if they correspond to 
those expected for the SM $H$, i.e.\
for the minimal realization of the Higgs 
sector~\cite{Zeppenfeld:2000td,Zeppenfeld:2001qh}. 
In the presence of extra dimensions, 
further scenarios emerge. For the present discussion, we consider two 
scenarios. The first has a light Higgs boson, 
for which we take $M_h$ = 120~GeV, with 
couplings different from those predicted in the SM. 
The question here is if the anomaly
is due to an extended Higgs sector, 
such as in Supersymmetry, or rather to the mixing 
with an undetected radion. The second scenario consists of an intermediate-mass
scalar, with 180~GeV $< M <$ 300~GeV, observed alone. 
An important issue would then be the question of whether the 
observed particle is the SM-like Higgs boson or a radion, 
with the Higgs particle left undetected. This scenario is quite likely
at large negative $\xi$ and large $M_\phi$ --- 
see Figures \ref{sec211fig:compl120} and \ref{sec211fig:compl140}.

In the first scenario, the issue is the interpretation of discrepancies in the measured 
Higgs couplings to gauge bosons and fermions. These effects increase with $|\xi|$, 
$1/\lphi$ and $M_h/M_{\phi}$. The LHC is expected to measure some ratios of these 
couplings~\cite{Zeppenfeld:2001qh}. In the case of the SM $H$, the ratio 
$g_{HZZ}/g_{HWW}$ can be determined with a relative accuracy 
of 15\% to 8\% for 120~GeV $< M_H <$ 180~GeV, 
while the ratio $g_{H\tau\tau}/g_{HWW}$ and that of the 
effective coupling to photons, $g_{H\gamma\gamma}^{effective}/g_{HWW}$ can be 
determined to 6\% to 10\% for 120~GeV $< M_h <$ 150~GeV. 
Now, the Higgs-radion mixing would induce the same shifts in 
the direct couplings $g_{hWW}$, $g_{hZZ}$ and $g_{h\bar ff}$,
all being given by $d+\gamma b$ times the corresponding $H$ couplings ---
see Eq.~(\ref{sec211couplings}).
Although this factor depends on the $\lphi$, $M_{\phi}$ and $\xi$ parameters,
ratios of couplings would remain unperturbed and correspond to those expected 
in the SM. Since the LHC measures mostly ratios of couplings,
the presence of Higgs-radion mixing could easily be missed.
One window of sensitivity to the mixing would be offered by the 
combination $g_{h\gamma\gamma}^{effective}/g_{hWW}$. 
But the mixing effects are expected
to be limited to relative variation of $\pm 5\%$ w.r.t. the SM predictions. 
Hence, the 
LHC anticipated accuracy corresponds to deviations of one unit of $\sigma$, 
or less, except for a small region at $\lphi \simeq 1$~TeV.
Larger deviations are expected for the absolute rates~\cite{Dominici:2002jv},
especially for the $gg\to h \to \gam\gam$ channel which can be dramatically
enhanced or suppressed relative to the $gg\to H \to \gam\gam$ 
prediction for larger $\xi$ values due to the large changes in
the $gg\to h$ coupling relative to the $gg\to H$ coupling. 
Of course, to detect these deviations it is necessary to control
systematic uncertainties for the absolute $\gam\gam$ rate. 
All the above remarks would also apply to distinguishing between the light 
Higgs of supersymmetry, which would be SM-like
assuming an approximate decoupling limit, and the $h$ of the Higgs-radion
scenario. In a non-decoupling two-doublet model, the light Higgs
couplings to up-type and 
down-type fermions can be modified differently with respect to those of
the SM $H$, and LHC measurements of coupling ratios
would detect this difference.

A TeV-class LC has the capability of measuring the {\it absolute} coupling
strengths to all fermions separately. For the SM $H$, accuracies of 
order 1\%-5\% for the couplings are achieved. Further, a 
determination of the total $H$ 
width to 4\% - 6\% accuracy  
 is possible. 
These capabilities are 
important for the scenario we propose since there would be 
enough measurements 
and sufficient accuracy to detect Higgs-radion  mixing for 
moderate to large $\xi$ values~\cite{Rizzo:2002za}. This 
is shown in Figure~\ref{sec211fig:lc} 
by the additional contours, which indicate the regions 
where the discrepancy with the SM predictions 
for the Higgs couplings to pairs of $b$ 
quarks and $W$ bosons exceeds 2.5~$\sigma$.

It is worth emphasizing that one of the basic predictions of the model
is that the $W^+W^-$, $Z^0Z^0$ and $f\overline{f}$ couplings of the $h$
should all be changed by exactly the same factor. The above discussion
shows that it will be possible to check this at a basic level at the
LHC (in that ratios of branching ratios should be the same as in the
SM) and with some precision at the LC.

We note that the {\it combination} of the direct observation of 
$\phi \to Z^0Z^{0*}$ at the 
LHC and the precision measurements of the Higgs properties at a $e^+e^-$ LC 
will extend our ability to distinguish between the Higgs-radion mixing
scenario and the SM $H$ scenario 
to a large portion of the regions where at the LHC
only the $h$ or only the $\phi$ is detected and determining
that the observed boson is not the SM $H$ is difficult. 
Finally, we reemphasize
the fact that $h$ will be detected in $\epem\to Z^0 h$ throughout
the entire $(M_\phi,\xi)$ parameter space and that
$\epem\to Z^0\phi$ can be detected in all
but the region exemplified, for $M_h=120$ GeV and $\lphi=5$ TeV, 
in the left panel of Figure~\ref{sec211fig:lphi30}. This,
in particular, guarantees the 
observability of the $\phi$ in the low $M_{\phi}$, large $\xi>0$ region that 
is most difficult for the LHC.

\begin{figure}
\begin{center}
\begin{tabular}{c c}
\hspace*{-0.75cm} 
\epsfig{file=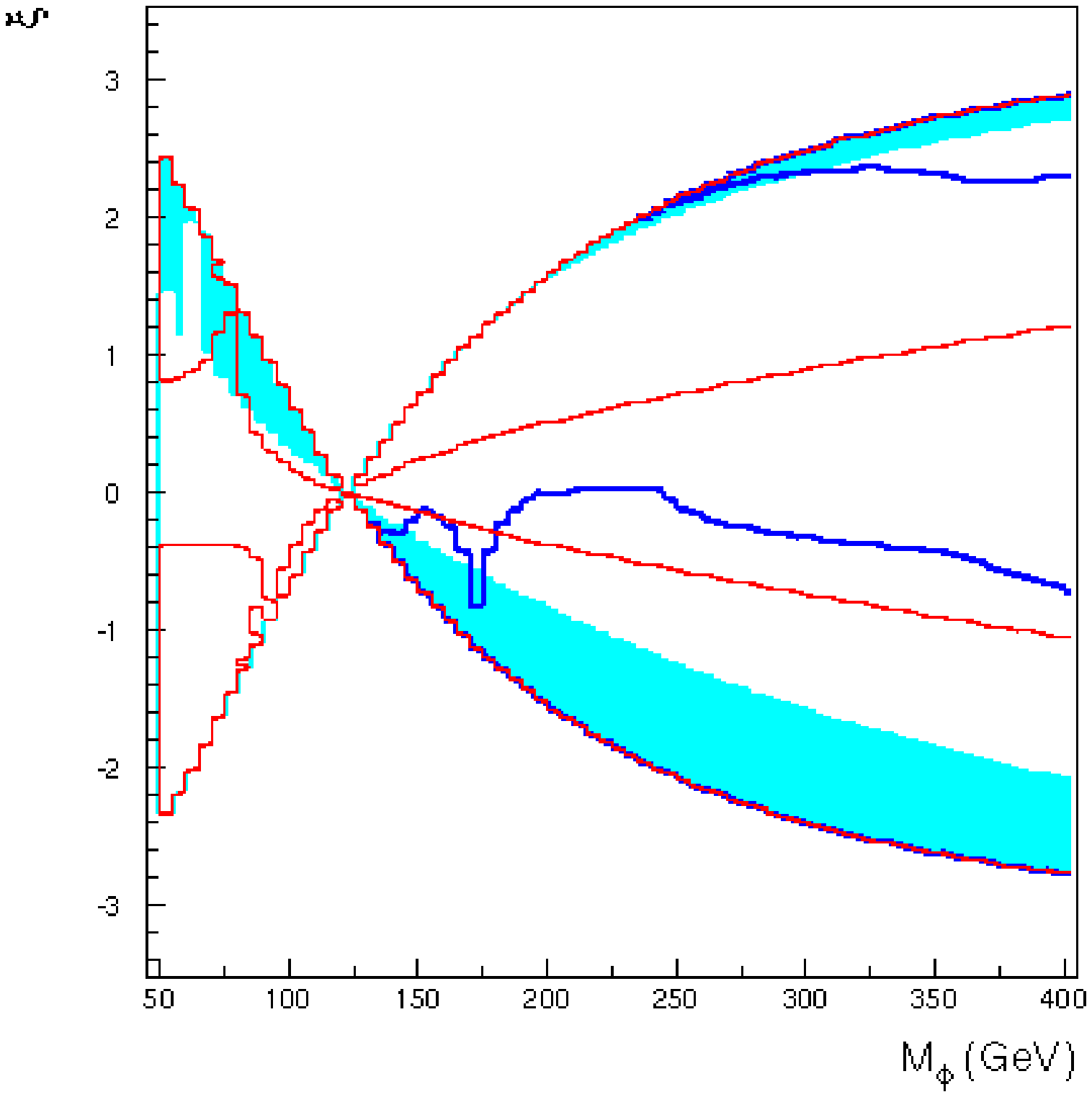,width=7.5cm,height=7.0cm} &
\epsfig{file=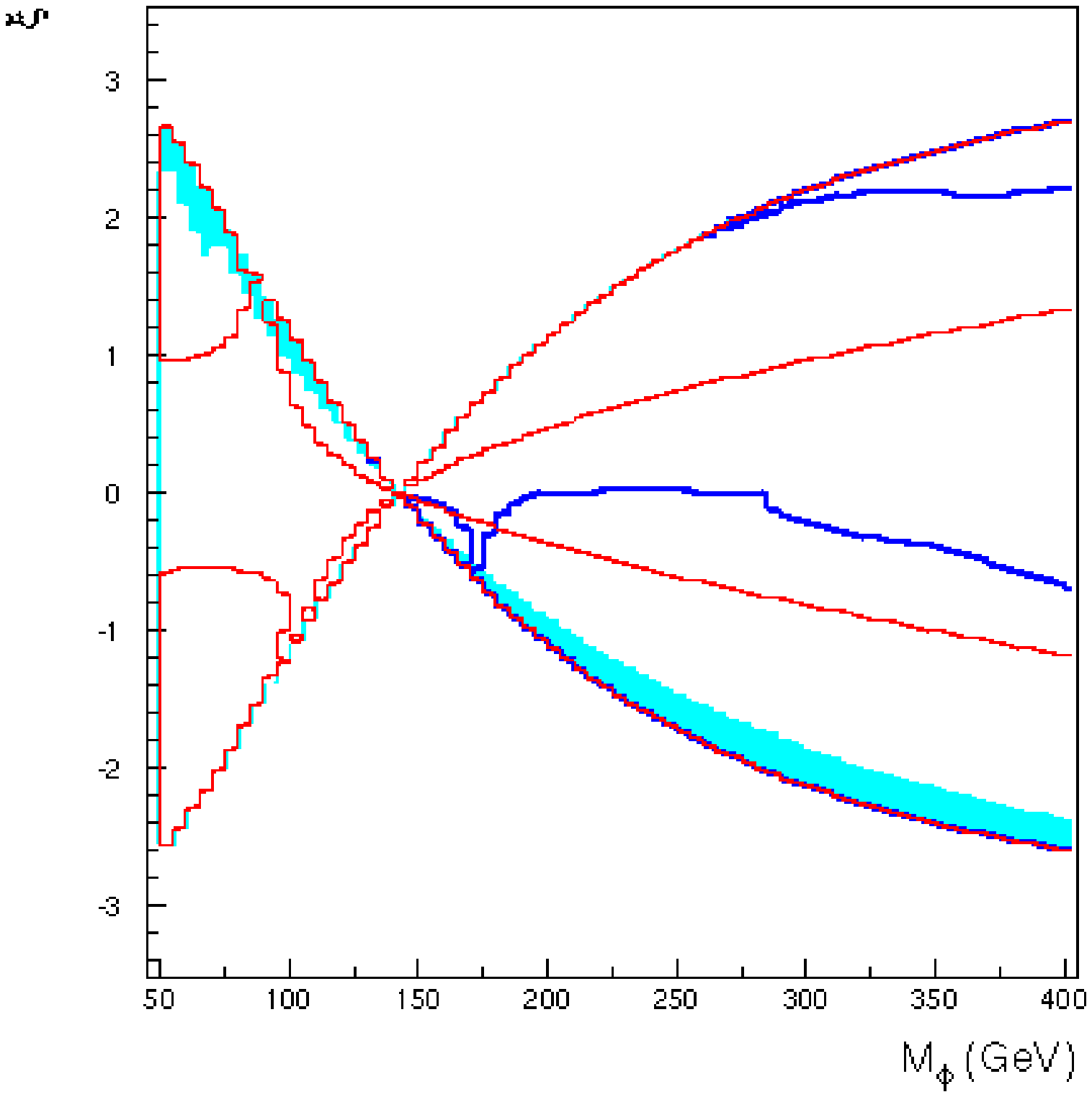,width=7.5cm,height=7.0cm} \\
\end{tabular}
\end{center}
\caption{\sl Same as Figures~\ref{sec211fig:compl120}
and \ref{sec211fig:compl140} for $M_h$ = 120~GeV (left), 
140~GeV (right) and $\lphi$ = 5~TeV with added contours, 
indicated by the medium grey (red) curves, showing 
the regions where the LC measurements of the $h$ couplings to 
$b \bar{b}$ and $W^+W^-$ would provide a $>2.5~\sigma$ 
evidence for the radion mixing effect.}
\label{sec211fig:lc}
\end{figure}

If, at the LHC, an intermediate mass scalar is observed alone, 
its non-SM-like nature can, in some cases, be determined 
through measurement of its production yield and its couplings. In 
particular, in the region at large, 
negative $\xi$ values where $\phi$ production is visible
whereas $h$ production is not, the yield of $Z^0Z^0 \to 4~\ell$ from 
$\phi$ decay can differ by a factor of 2 or more from that expected
for a SM $H$ (depending upon the value of $M_\phi$ --- see Figure~13
of Ref.~~\cite{Dominici:2002jv}). 
For $M_\phi < 2~M_h$ the deviations arise from the substantial
differences between the $gg\to \phi$ coupling and the $gg\to H$
coupling. For $M_\phi>2M_h$, this rate is also sensitive to $\br(\phi\to hh)$.
Defining $R\equiv \br(\phi\to Z^0Z^{0(*)})/\br(H\to Z^0Z^{0(*)} )$, one
finds $R>0.9$ for $\mphi<2M_h$.
Such a small deviation would not have a big
impact compared to the possibly large
deviations of $gg\to h/gg\to H$ relative to unity.
However, past the threshold for $\phi \to hh$ decays, the $Z^0Z^0$ branching 
fraction is significantly affected; for example, $R<0.7$
for a substantial portion of the $|\xi|<1.5$ part of
the $\mphi>2M_h$ region when $M_h=120$ GeV
and $\lphi=5$ TeV.  The combination of a 
reduced $Z^0Z^0 \to 4~\ell$ rate and the possibility 
to observe $\phi \to hh$ decays, 
ensures that the LHC could positively identify the existence of 
the radion in the 
region $M_{\phi} >2M_h$, $\xi \ne 0$.

\begin{figure}[htb!]
\begin{center}
\hspace*{-0.85cm}
\begin{tabular}{c c} 
\epsfig{file=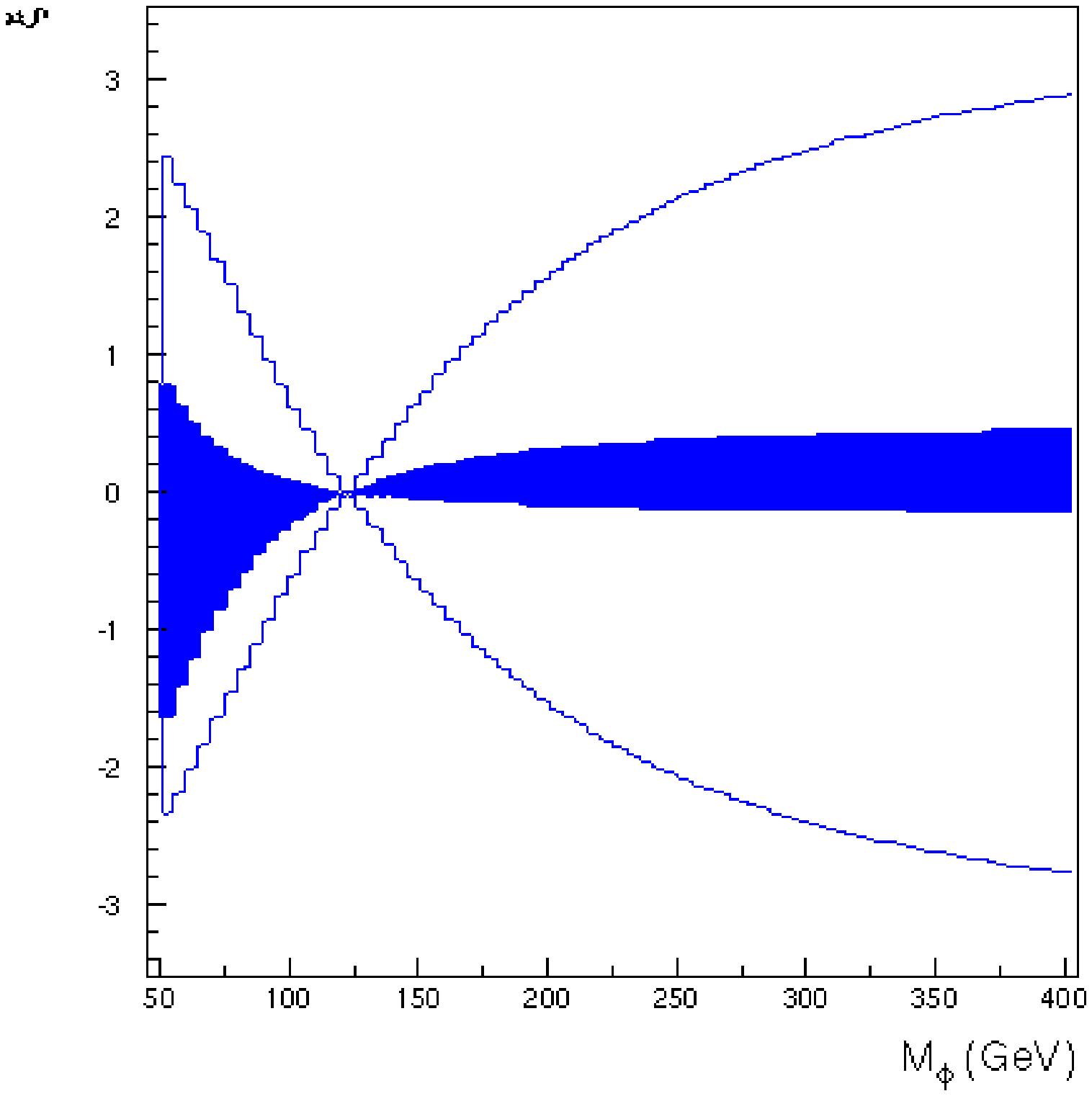,width=7.5cm,height=7.0cm} &
\epsfig{file=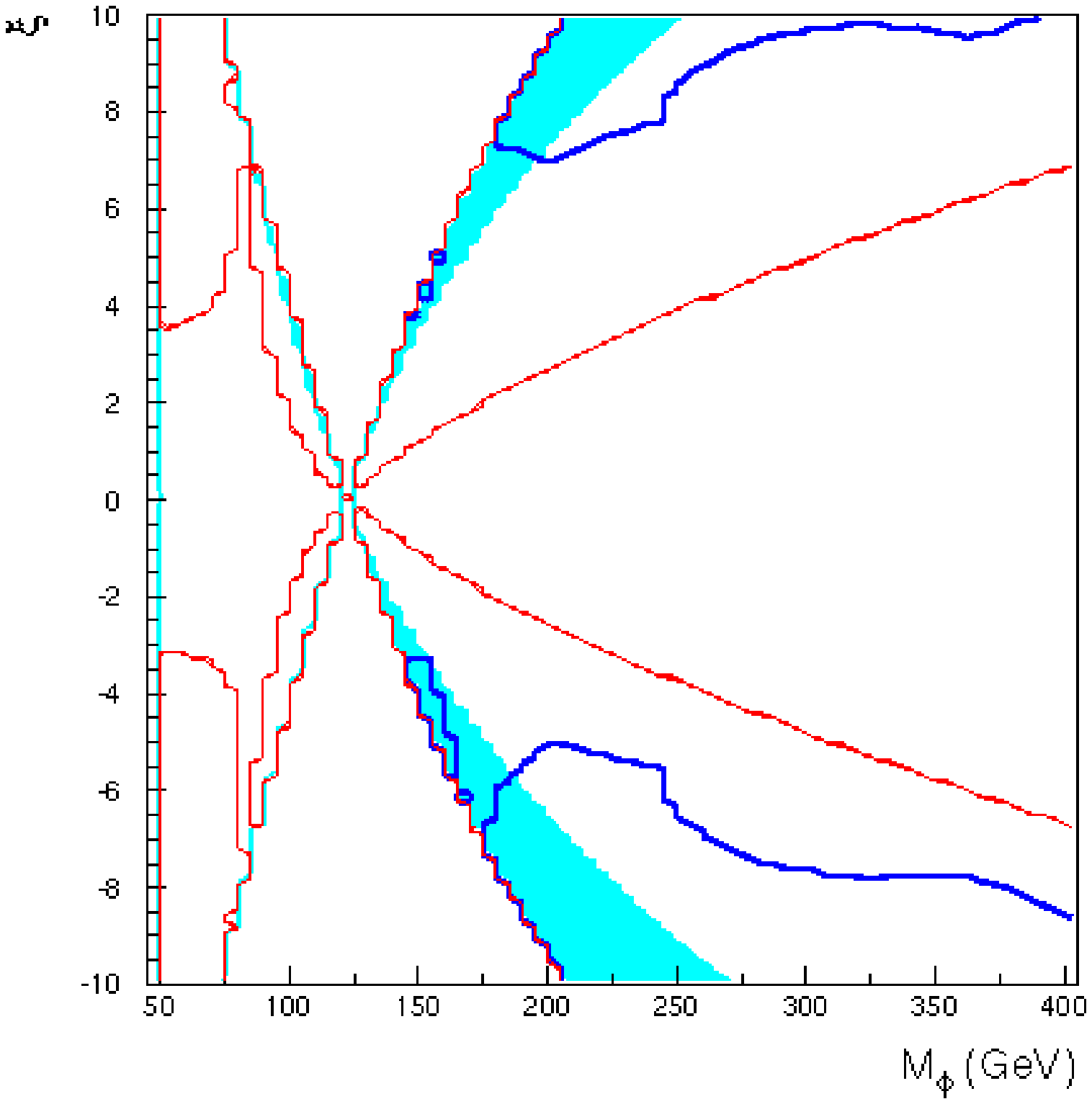,width=7.5cm,height=7.0cm} \\
\end{tabular}
\end{center}
\caption{\sl Left: Contour in $(\mphi,\xi)$ parameter space with 
$g_{\phi ZZ}^2/g_{H ZZ}^2<0.01$ indicated by the dark region, 
for $M_h$ = 120~GeV and $\lphi=5$ TeV.
Right:  Same as Figure~\ref{sec211fig:lc}, for $M_h$ = 120~GeV and $\lphi$=30~TeV.}
\label{sec211fig:lphi30}
\end{figure}

Finally, we should note that the distinctive signature of KK
graviton excitation production at the
LHC~\cite{Davoudiasl:2000wi,Davoudiasl:1999jd} will be easily observed for a
substantial range of $\lphi$.  This will not only serve as a warning
to look for a possibly mixed Higgs-radion sector but will also allow
us to determine $\lphi$ from the measurements of $m_1$ and $m_0/\mpl$.
Note that 
$m_1$, the mass of the first KK graviton excitation, given by 
\begin{equation}
m_1=x_1 {m_0\over\mpl} {\lphi\over\sqrt 6}\,
\label{sec211m1form}
\end{equation}
where  ${m_0}$ is the curvature parameter and  $x_1$ is the first
zero of the Bessel function $J_1$ ($x_1\sim 3.8$).
 $m_0/\mpl$ can be determined from the KK
excitation profile.  The 95\% CL limit for detecting the first KK
excitation, with an integrated luminosity of 100~fb$^{-1}$, is given in terms 
of $m_0/\mpl$ by $m_1({\rm
  TeV})=6.6+2\ln_{10}\left({m_0\over\mpl}\right)$ \cite{Davoudiasl:1999jd}.
Using Eq.~(\ref{sec211m1form}), we find that the signal for the first KK
excitation will be below the 95\% CL for $\lphi>{\sqrt 6\over
  x_1}\left({\mpl\over m_0}\right) \left[6.6+2\ln_{10}\left({m_0\over
      \mpl}\right)\right]$ TeV.  For example, for $m_0/\mpl=0.1$ this
corresponds to $\lphi\ge 30$ TeV, which is also consistent with
precision electroweak constraints~\cite{toharia}.  In this case, the
Higgs-radion sector becomes absolutely crucial for revealing the RS
scenario. This is illustrated in the right panel of
Figure~\ref{sec211fig:lphi30} where we show that Higgs-radion phenomenology
can be explored at the LHC for a large section of parameter space
when $\lphi=30$ TeV.

\subsection{Radions at a photon collider}
{\it 
D. Asner, S. Asztalos, A. De~Roeck, S. Heinemeyer, J. Gronberg,
 J. Gunion, H. Logan, V.~Martin, M. Szleper, 
M. Velasco}

\vspace{1em}

In this section, we demonstrate the important complementarity of
a photon collider (PC) for 
probing the Higgs-radion sector of the Randall-Sundrum
(RS) model \cite{rs}. In \cite{Asner:2001vh} a
%
SM Higgs boson with $m_{H}=115$ GeV was examined.
After the cuts, one obtains per year about $S=3280$ and $B=1660$
in the $\gam\gam\to H\to b\overline{b}$ channel, corresponding to
$S/\sqrt B \sim 80$!

We will assume that these numbers do not change significantly
for a Higgs mass of $120$ GeV.
After mixing, the $S$ rate for the $h$ 
will be rescaled relative to that for the the SM $H$.  Of course, $B$
will not change.
The rescaling is shown in Fig.~\ref{sec211gagatobb_mh120}.
The $S$ for the $\phi$ can also be obtained by rescaling
if $\mphi\sim 115$ GeV.
For $\mphi<120$ GeV, the $\phi\to b\overline{b}$ channel will
continue to be the most relevant for $\phi$ discovery,
but studies have not yet been performed to obtain the $S$
and $B$ rates for low masses.

\begin{figure}[htb!]
\begin{center}
\includegraphics[height=5.5in,width=3.5in,angle=90]{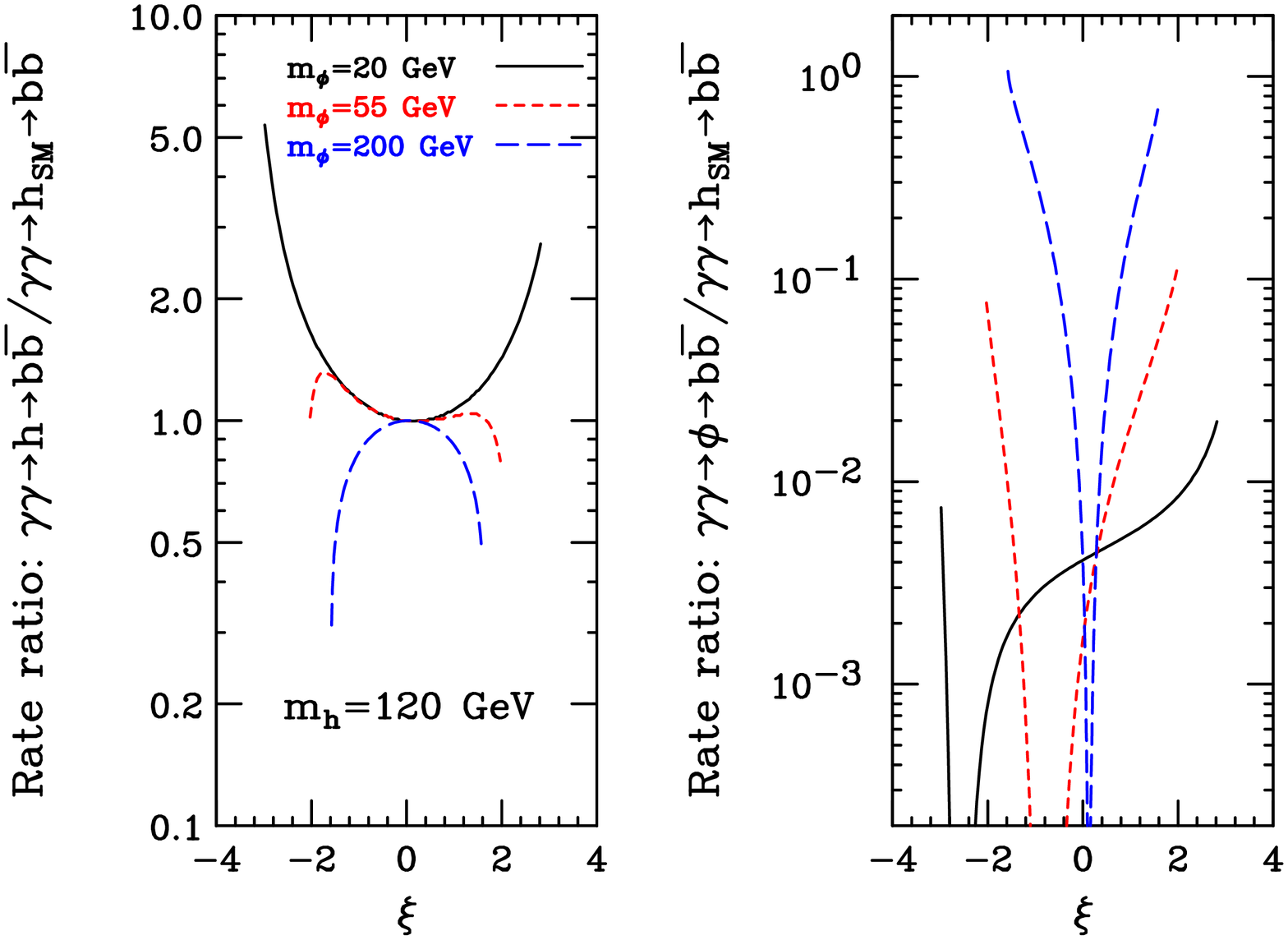}
\end{center}
\caption{The rates for $\gam\gam\to h \to b\overline{b}$
and $\gam\gam\to\phi\to b\overline{b}$ relative
to the corresponding rate for a SM Higgs boson of the same mass.
Results are shown
for $m_h=120 $ GeV and $\lphi=5$ TeV as functions of $\xi$
for $\mphi=20$, $55$ and $200$ GeV.
\label{sec211gagatobb_mh120}
}
\end{figure}
Observe that for $\mphi<m_H$ we have either little change or enhancement, whereas
significant suppression of the $gg\to H \to \gam\gam$ rate
was possible in this case for positive $\xi$.
  Also note that for $\mphi>m_H$ and large $\xi<0$
(where the LHC signal for the $h$ is marginal) there is much less suppression 
of $\gam\gam\to H\to b\overline{b}$ than for $gg\to H\to\gam\gam$ ---
at most a factor of 2 vs a factor of 8 (at $\mphi=200$ GeV).
This is no problem for the PC since $S/\sqrt B\sim \frac{1}{2} 80 \sim 40$
is still a very strong signal.
{In fact, we can afford a reduction by a factor
of $16$ before we hit the $5\sigma$ level!}  
Thus, {\it the $\gam\gam$
collider will allow $h$ discovery (for $m_H=120$) 
throughout the entire hourglass shown in Fig.~\ref{sec211fig:detect}},
which is something the LHC cannot absolutely do.

Using the factor of $16$ mentioned above
it is apparent that the $\phi$ with $\mphi<120$ GeV is very likely to
elude discovery at the $\gam\gam$ collider. (Recall that
it also eludes discovery at the LHC for this region.)
The only exceptions to this statement occur at the very largest $|\xi|$
values for $\mphi\geq 55$ GeV where $S_\phi>S_H/16$.

Of course, we need to have signal and background results
after cuts for these lower masses to know if the factor of 16
is actually the correct factor to use.
To get the best signal to background ratio we would want to lower
the machine energy and readjust cuts and so forth.  
This study should be done.
For the $\mphi>m_H$ region, we will need results for
the $WW$ and $ZZ$ modes that are under study.


Overall, the PC is more than competitive with the LHC for $h$ discovery.
In particular,
the PC can see the $h$ where the LHC signal will be marginal (i.e.
at the largest theoretically allowed $\xi$ values). Of course,
the marginal LHC regions are not very big for full $L$.
Perhaps even more interesting is the fact that
there is a big part of the hourglass where the $h$ will be seen 
at both colliders. When the LHC achieves $L>100 $ fb$^{-1}$, this comprises
most of the hourglass, shown in
\ref{sec211fig:detect}. Simultaneous observation of the $h$ at
the two different colliders will greatly increase our knowledge about the $h$
since the two rates measure different things.
The LHC rate in the $\gam\gam$ final state measures
$\Gamma(h\to gg)\Gamma(h\to \gam\gam)/\Gamma_{\rm tot}^h$ while
the PC rate in the $b\overline{b}$ final state determines
$\Gamma(h\to \gam\gam)\Gamma(h\to b\overline{b})/\Gamma_{\rm tot}^h$.
Consequently, the ratio of the rates gives us
$
 {\Gamma(H\to gg)\over \Gamma(H\to b\overline{b})}\,,
$
in terms of which we may compute
\begin{equation}
R_{hgg}\equiv \left[{\Gamma(H\to gg)\over \Gamma(H\to b\overline{b})}\right]
 \left[{\Gamma(H\to gg)\over \Gamma(H\to b\overline{b})}\right]^{-1}_{SM}\,.
\end{equation}
This is a {\it very} interesting number since it directly probes
for the presence of the anomalous $gg h$ coupling. 
In particular, 
$R_{hgg}=1$ if the only contributions to $\Gamma(H\to gg)$ come
from quark loops and all quark couplings scale in the same way.
A plot of $R_{hgg}$ as a function of $\xi$ for $M_H=120$ GeV, $\lphi=5$ TeV
and $\mphi=20$, $55$ and $200$ GeV
 appears in Fig.~\ref{sec211gggagaanomalouscoup_mh120}.

\begin{figure}[htb!]
\includegraphics[height=5.5in,width=3in,angle=90]{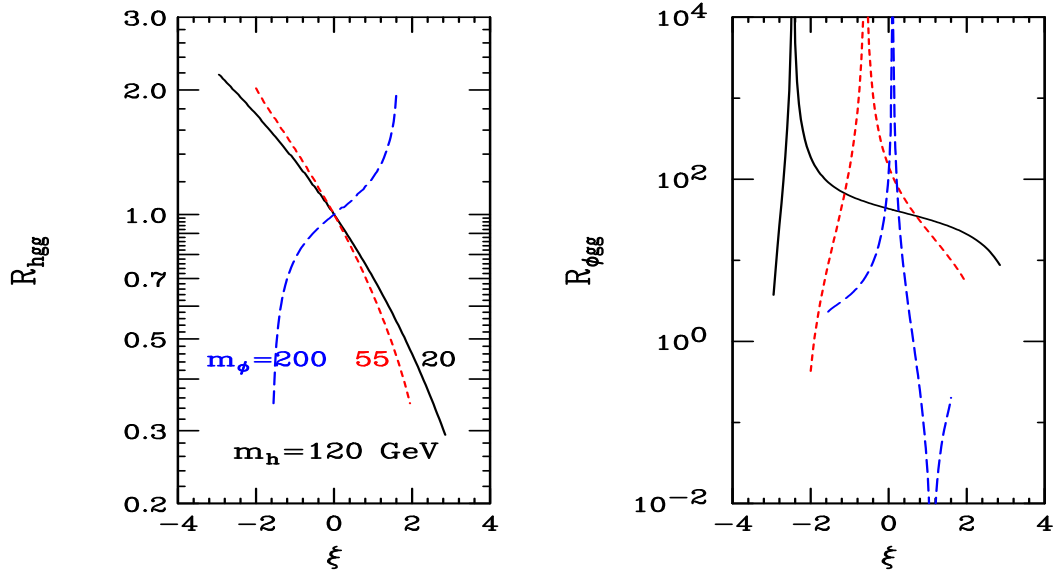}
\caption{\sl \small We plot
the ratios $R_{H gg}$ and $R_{\phi gg}$
of the $H gg$ and $\phi gg$ couplings-squared
including the anomalous contribution
to the corresponding values expected in its absence.
Results are shown
for $m_h=120 $ GeV  and $\lphi=5$ TeV as functions of $\xi$
for $\mphi=20$, $55$ and $200$ GeV.
(The same type of line is used for a given $\mphi$ in 
the right-hand figure as is used in the left-hand figure.) 
\label{sec211gggagaanomalouscoup_mh120}
}
\end{figure}

We can estimate the accuracy with which $R_{hgg}$ can be measured
as follows.  Assuming the maximal reduction of 1/2 for the
signal rate ($S$) rescaling at the $\gam\gam$ CLIC collider, we find that
$\Gamma(H\to \gam\gam)\Gamma(H\to b\overline{b})/
\Gamma_{tot}^h$ can be measured
with an accuracy of about $\sqrt{S+B}/S\sim \sqrt{3200}/1600\sim 0.035$.
The dominant error will then be from the LHC which will typically
measure $\Gamma(H\to gg)\Gamma(H\to \gam\gam)/\Gamma_{tot}^h$
with an accuracy of between $0.1$ and $0.2$ (depending on
parameter choices and available $L$).  From 
Fig.~\ref{sec211gggagaanomalouscoup_mh120}, we see that $0.2$ fractional accuracy will reveal deviations
of $R_{hgg}$ from $1$ for all but the smallest $\xi$ values.
The ability to measure $R_{hgg}$ with good accuracy 
may be the strongest reason in the Higgs context
for having the PC as well as the LHC.
Almost all non-SM Higgs theories predict $R_{hgg}\neq 1$
for one reason another, unless one is in the decoupling limit.

Depending on $L$ at the LHC, there might be a small part
of the hourglass 
(large $|\xi|$ with $\mphi>m_H$) where {\it only} the $\phi$
will be seen at the LHC and the $h$ will only be seen at the PC.
This is a nice example of complementarity between the two machines.
By having both machines we maximize the
chance of seeing both the $h$ and $\phi$.

As regards the $\phi$, we have already noted from
Fig.~\ref{sec211gagatobb_mh120} that the $b\overline{b}$ final state rate
(relevant for the $\mphi=20$ and $55$ GeV cases) will only be
detectable in the latter case (more generally for 55 GeV$ <
\mphi<2 m_W$), and then only if $|\xi|$ is as large as theoretically
allowed.  If $\gam\gam\to \phi\to b\overline{b}$ can be observed,
Fig.~\ref{sec211gggagaanomalouscoup_mh120} shows that a large deviation for
$R_{\phi gg}$ relative to the value predicted for a SM H of the
same mass is typical (but not guaranteed).  
For $\mphi>2 m_W$, $\br(\phi\to b\overline{b})$ will
be very small and detection of $\gam\gam\to \phi \to b\overline{b}$ will
not be possible.  We are currently studying $\gam\gam\to \phi\to
WW,ZZ$ final states in order to assess possibilities at larger
$\mphi$.

Overall, there is a strong  case for the PC in the
RS model context, especially if a Higgs boson is seen
at the LHC that has non-SM-like rates and other properties.

\subsection{Further scenarios}
{\it J. Gunion}

\subsubsection{Beyond Higgs-radion mixing} 



Motivations for going beyond the simple RS model
are easily found.  The one that will most
obviously have an influence on Higgs-radion mixing
physics is the fact that in the strict RS model
the radion is massless, which is to say that the
distance between the branes is not stable.  The
above studies simply introduce a mass for the
radion by hand. Several explicit mechanisms for
giving mass to the radion have been discussed.
One example is the Goldberger-Wise mechanism
\cite{Goldberger:1999uk} that relies on
introducing an additional scalar field that
propagates in the bulk. However, in the approach
of \cite{Goldberger:1999uk} the scalar field
potential employed does not yield an exact
solution to the Einstein equations. The result is
brane curvature, which might impact Higgs
phenomenology.  Possible impacts have not been
worked out.

In a more recent approach
\cite{Grzadkowski:2003fx}, the additional scalar
field is introduced in the bulk with a vacuum
profile chosen so that the RS metric (with no
curvature) is an exact solution of the Einstein
equations while at the same time the radion
becomes massive. However, there are inevitable
consequences for the Higgs sector.  One finds that
the quantum fluctuations of this additional bulk
scalar field (which include the full tower of KK
excitation fluctuations) will all mix with the
Higgs and radion. The phenomenology of the
extended Higgs-radion-KK excitation mixing matrix
has not been worked out. But, it is sure to lead
to additional freedom in the phenomenology of the
Higgs-radion sector that could possibly pose
further challenges for experimental study and an
even greater need for having both the LHC and LC
available for this study.

\subsubsection{Universal extra dimension models}
In the universal extra dimension models, all
particles propagate in the extra dimension(s).
There has been relatively little discussion of
impacts on Higgs physics in the context of these
models.  Perhaps the most important observation
made to date is that the Higgs mass can be quite
large without conflicting with precision
electroweak constraints if the extra dimension is
as large as allowed by other constraints
\cite{Appelquist:2002wb}.  If the Higgs mass is
near the upper limit of about 800 GeV, only the
LHC will be able to detect it if the LC energy is
restricted to $\sqrt s\lsim 1$~TeV.

Contributions of various KK excitation modes to
one-loop induced Higgs couplings (e.g. the
$h\gamma\gamma$ coupling) will generally be
measurable \cite{Petriello:2002uu}.  The
complexity of understanding how to relate
precision Higgs measurements to the full KK
structure will surely require both the LHC and the
LC --- probably the LHC will be needed to probe
the KK excitations while the LC will most clearly
reveal deviations from SM expectations for the
couplings most strongly influenced by the KK
modes.

\subsection{Conclusions}

In summary, for almost the entire region 
of the parameter phase space where the suppression of the Higgs signal yield 
causes the overall signal significance at the LHC to drop below 5~$\sigma$, the 
radion eigenstate $\phi$ can be observed in the $gg \to \phi \to Z^0Z^{0(*)} \to
 4~\ell$ 
process instead. An $e^+e^-$ linear collider or a low energy 
PC 
linear collider would effectively complement the LHC both 
for the Higgs observability, including the most difficult region at low 
$M_{\phi}$ 
and positive $\xi$ values, and for the detection of the radion mixing effects, 
through the precision measurements of the  Higgs particle couplings to various
 types of 
particle pairs. 

Finally, we  note
that the Higgs-radion sector is not the only means for probing
the Randall-Sundrum type of model.  The scenarios 
considered here will also yield the distinctive signature 
of KK graviton excitation production at the LHC~\cite{Davoudiasl:2000wi}. 
This easily observed signal will serve as a warning to look for
a possibly mixed Higgs-radion sector and allow to fix $\Lambda_{\phi}$.


\section{Phenomenology of the Littlest Higgs model}
\label{sec:27}

{\it H.E.~Logan}

\vspace{1em}

\renewcommand{\lsim}{\mathrel{\raise.3ex\hbox{$<$\kern-.75em\lower1ex\hbox{$\sim$}}}}
\renewcommand{\gsim}{\mathrel{\raise.3ex\hbox{$>$\kern-.75em\lower1ex\hbox{$\sim$}}}}

\noindent
{\small
The little Higgs idea is a new way to solve the little hierarchy problem
by protecting the Higgs mass from quadratically divergent one-loop 
corrections.  
We consider here the phenomenology of one particular realization
of the little Higgs idea, the ``Littlest Higgs'' model.
The Large Hadron Collider should be able to discover and measure 
some properties
of the new heavy gauge bosons, heavy vector-like partner of the top quark,
and heavy scalars, which have masses typically on the order of 
one to a few TeV.  
The linear collider should be sensitive to deviations in the precision
electroweak observables and in the triple gauge boson couplings,
and to loop effects of the new heavy particles on the Higgs boson coupling to 
photon pairs.
}

\vspace{1em}


The Standard Model (SM) of the strong and electroweak interactions has
passed stringent tests up to the highest energies accessible today.
The precision electroweak data \cite{sec2_Hagiwara:fs} point to the existence of
a light Higgs boson in the SM, with mass $m_H \lsim 200$ GeV.
The Standard Model with such a light Higgs boson can be viewed as an effective 
theory valid up to a much higher energy scale $\Lambda$, possibly all the way
up to the Planck scale. 
In particular, the precision electroweak data exclude the presence of
dimension-six operators arising from strongly coupled new physics 
below a scale $\Lambda$ of order 10 TeV \cite{Barbieri};
if new physics is to appear
below this scale, it must be weakly coupled.
However, without protection by a symmetry, the Higgs mass is quadratically
sensitive to the cutoff scale $\Lambda$ via quantum corrections, 
rendering the theory with $m_H \ll \Lambda$ rather unnatural.
For example, for $\Lambda = 10$ TeV, the ``bare'' Higgs mass-squared 
parameter must be tuned against the quadratically divergent radiative
corrections at the 1\% level.
This gap between the electroweak scale $m_H$ and the cutoff scale
$\Lambda$ is called the ``little hierarchy''.

Little Higgs models 
\cite{decon,minmoose,Littlest,SU6Sp6,KaplanSchmaltz,custodialmoose,SkibaTerning,Spencer} 
revive an old idea to keep the
Higgs boson naturally light: they make the Higgs particle a
pseudo-Nambu-Goldstone boson \cite{PNGBhiggs} of a broken global symmetry.
The new ingredient of little Higgs models is that they
are constructed in such a way that at least two
interactions are needed to explicitly break all of the global symmetry
that protects the Higgs mass.  This forbids quadratic divergences in the
Higgs mass at one-loop; the Higgs mass is then smaller than the cutoff 
scale $\Lambda$ by {\it two} loop factors, making the cutoff scale
$\Lambda \sim 10$ TeV natural and solving the little hierarchy problem.

From the bottom-up point of view, in little Higgs models
the most important quadratic divergences 
in the Higgs mass due to the top quark, gauge boson, and Higgs boson loops
are canceled by loops of new weakly-coupled 
fermions, gauge bosons, and scalars with masses around a TeV.  In 
contrast to supersymmetry, the cancellations in little Higgs models
occur between loops of particles with the {\it same} statistics.
Electroweak symmetry breaking is triggered by a
Coleman-Weinberg \cite{coleman-weinberg} potential, generated by
integrating out the heavy degrees of freedom, which also gives the Higgs
boson a mass at the electroweak scale.

The ``Littlest Higgs'' model \cite{Littlest}, which we focus on here, 
is a minimal model of this type.  
It consists of a nonlinear sigma model with a global
SU(5) symmetry which is broken down to SO(5) by a vacuum condensate 
$f \sim \Lambda/4\pi \sim$ TeV.
The gauged subgroup [SU(2)$\times$U(1)]$^2$ is broken at the
same time to its diagonal subgroup SU(2)$\times$U(1), identified
as the SM electroweak gauge group. 
The breaking of the global symmetry leads to 14 Goldstone bosons, four of
which are eaten by the broken gauge generators, leading to four 
massive vector bosons: an SU(2) triplet $Z_H$, $W^{\pm}_H$, and a 
U(1) boson $A_H$.
The ten remaining uneaten Goldstone bosons 
transform under the SM gauge group as a doublet $h$ (which becomes
the SM Higgs doublet) and a triplet $\phi$ (which gets a mass of order $f$).
A vector-like pair of colored Weyl fermions is also needed to cancel 
the divergence from the top quark loop, leading to a new heavy vector-like
quark with charge $+2/3$.

The particle content and interactions are laid out in detail in 
Ref.~\cite{LHpheno}.  
Here we summarize the features important for the collider phenomenology.
The Littlest Higgs model contains six new free parameters, which 
can be chosen as follows:

1) $\tan \theta = s/c = g_1/g_2$, where $g_{1,2}$ are the couplings of
	the two SU(2) gauge groups, with $g^{-2} = g_1^{-2} + g_2^{-2}$.

2) $\tan \theta^{\prime} = s^{\prime}/c^{\prime}
        = g_1^{\prime}/g_2^{\prime}$, where $g^{\prime}_{1,2}$ are the
	couplings of the two U(1) gauge groups, with
	$g^{\prime -2} = g_1^{\prime -2} + g_2^{\prime -2}$.

3) $f$, the symmetry breaking scale, $\mathcal{O}$(TeV).

4) $v^{\prime}$, the vacuum expectation value (vev) of the triplet 
$\phi$; $v^{\prime} < v^2/4f$.

5) $M_H$, the SM-like Higgs boson mass.

6) $M_T$, the top-partner mass (together with $m_t$ and $f$, this fixes
        the top-partner couplings up to a two-fold ambiguity).

In what follows we describe the prospects for little Higgs studies
at the LHC and the LC.  This summary is based on Refs.~\cite{LHpheno,LHloop}.

\subsection{The Little Higgs at the LHC}

\subsubsection{$Z_H$ and $W_H$}
The heavy SU(2) gauge bosons $Z_H$ and $W_H$ can be produced via Drell-Yan
at the LHC (and at the Tevatron, if they are light enough).
In the Littlest Higgs model, the SU(2) fermion doublets are chosen
to transform under the SU(2)$_1$ gauge group; their couplings
to $Z_H$ and $W_H$ are therefore proportional to $\cot\theta$,
leading to a Drell-Yan cross section proportional to $\cot^2\theta$.
In Fig.~\ref{fig:sigmaZH}(a) we show the cross section for $Z_H$ production
at the Tevatron and LHC for $\cot\theta = 1$.
\begin{figure}[htb]
\begin{center}
\resizebox{\textwidth}{!}{\includegraphics{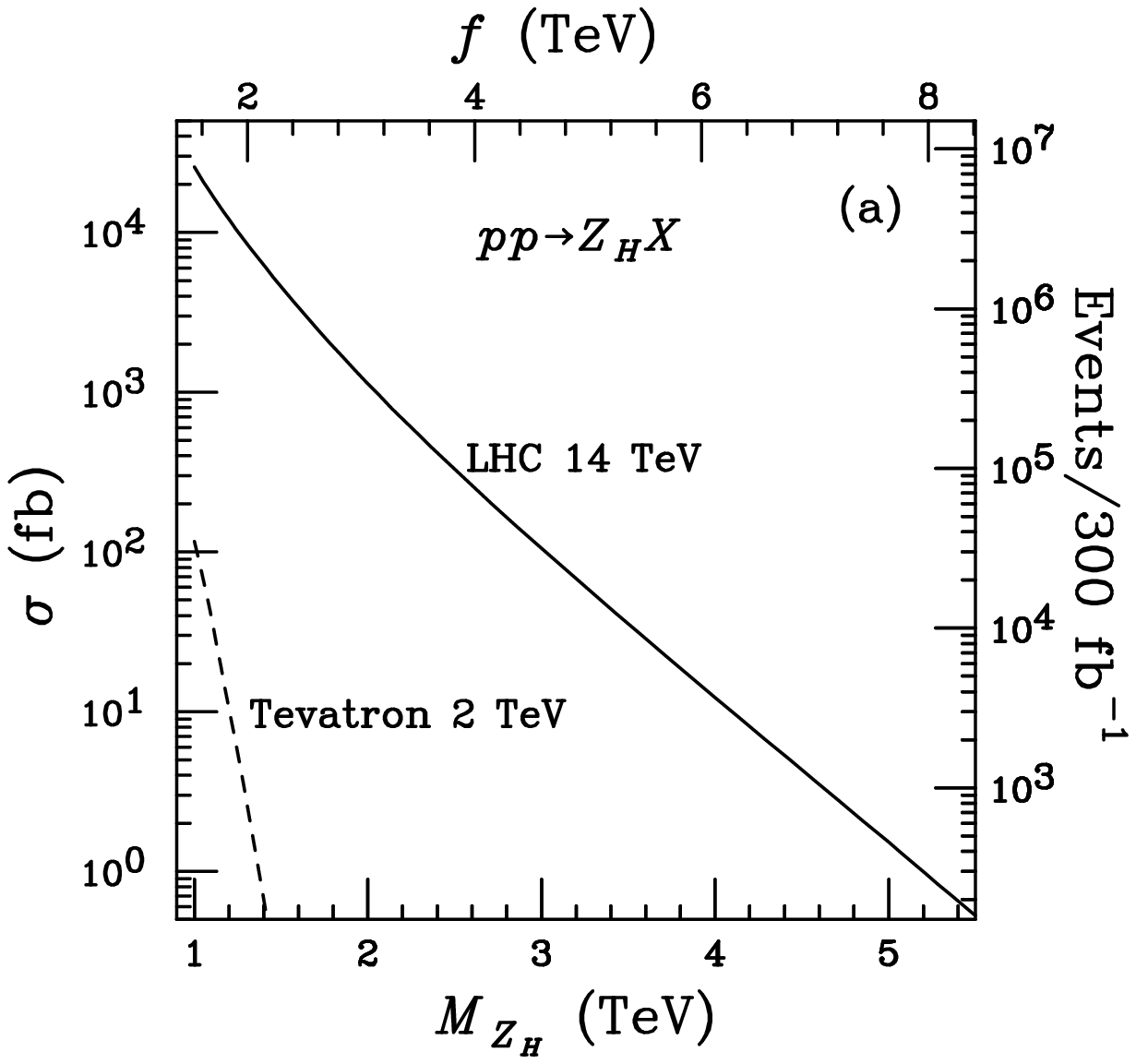}
\includegraphics{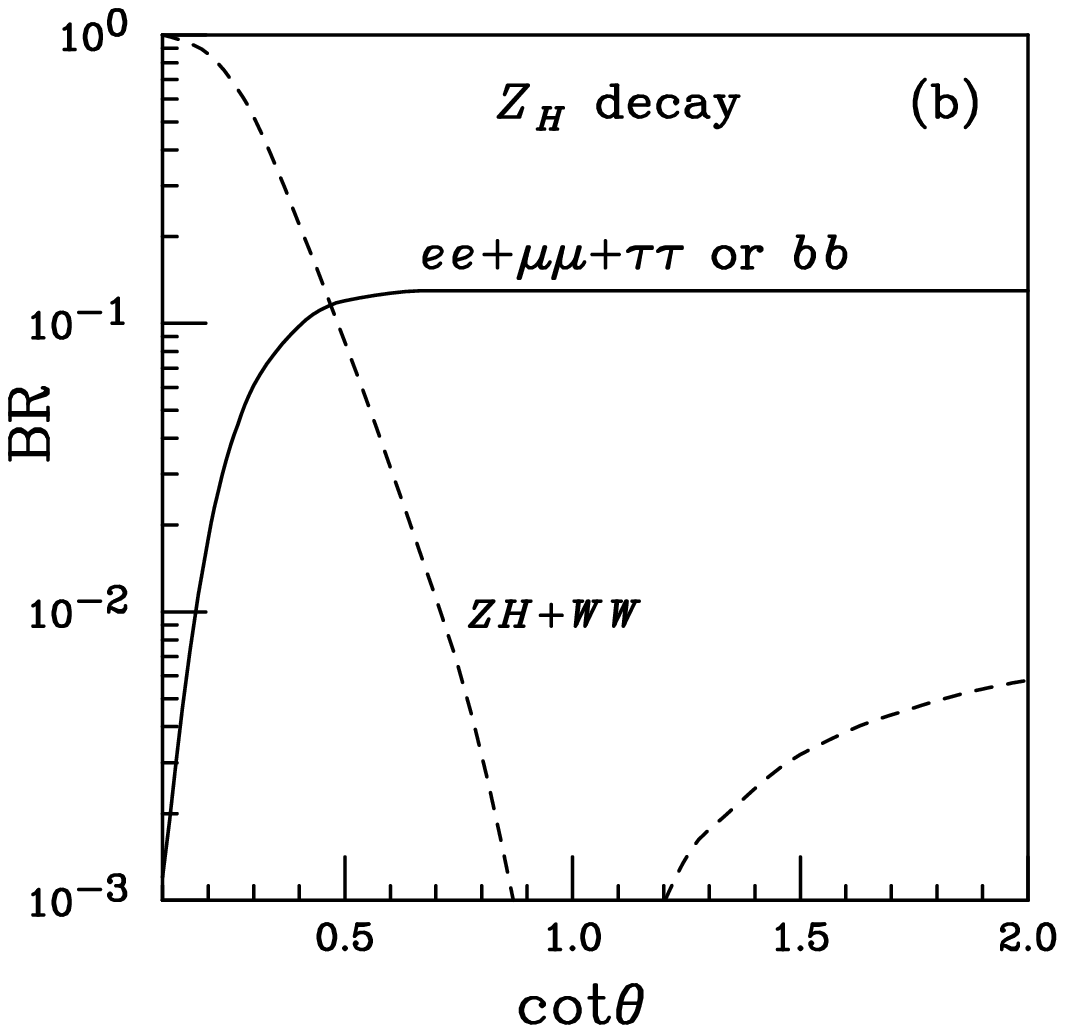}}
\end{center}
\caption{(a) Cross section for $Z_H$ production in Drell-Yan at the LHC
and Tevatron, for $\cot\theta = 1$.  From Ref.~\cite{LHpheno}.
(b) Branching ratios of $Z_H$ into SM particles as a function
of $\cot\theta$, neglecting final-state mass effects.  
}
\label{fig:sigmaZH}
\end{figure}
In the region of small $\cot\theta \simeq 0.2$, which is favored 
\cite{GrahamEW2} by the precision electroweak data, the cross section 
shown in Fig.~\ref{fig:sigmaZH}(a) must be scaled down by 
$\cot^2\theta \simeq 0.04$.  Even with this suppression factor, 
a cross section of 40 fb is expected at the LHC for $M_{Z_H} \simeq 2$ TeV,
leading to 4,000 events in 100 fb$^{-1}$ of data.
The production and decay of $Z_H$ and $W_H$ at the LHC has also been studied
in Ref.~\cite{gustavo}.

The decay branching fractions of $Z_H$ are shown in 
Fig.~\ref{fig:sigmaZH}(b).
The decays to fermion pairs follow an equipartition among the left-handed
fermion doublets.
Neglecting final-state particle masses, the branching fraction into three
flavors of charged leptons is equal to that into one flavor of quark
($\simeq 1/8$ for $\cot\theta \gsim 0.5$), due 
to the equal coupling of $Z_H$ to all SU(2) fermion doublets.
The partial widths to fermion pairs are proportional to $\cot^2\theta$.
The $Z_H$ also decays into $ZH$ and $W^+W^-$ with equal partial 
widths (again neglecting final-state mass effects).  These decays
come from the coupling of $Z_H$ to the components of the
Higgs doublet $h$, applying the
Goldstone boson equivalence theorem for the Goldstone modes eaten
by the $Z$ and $W$ bosons.
The partial widths to $ZH$ and $W^+W^-$ are proportional to 
$\cot^2 2\theta$.  
The total width of $Z_H$ depends on $\cot\theta$; for $\cot\theta \sim 0.2$
the $Z_H$ width is about 1\% of the $Z_H$ mass.

The different dependence of the bosonic and fermionic $Z_H$ decay modes 
on $\cot\theta$ offers a method to distinguish the Littlest
Higgs model from a ``big Higgs'' model with the same gauge group in which
the Higgs doublet transforms under only one of the SU(2) groups 
\cite{gustavo}, in which case the $ZH$ and $W^+W^-$ partial widths
would also be proportional to $\cot^2\theta$.

The $W_H^{\pm}$ couplings to fermion doublets are larger by a factor of
$\sqrt{2}$ than the $Z_H$ couplings; this together with the parton distribution
of the proton leads to a $W_H^{\pm}$ cross
section at the LHC about 1.5 times that of $Z_H$ \cite{gustavo}.
As for the $W_H$ decays, the branching fraction into three lepton flavors
is equal to that into one generation of quarks ($\simeq 1/4$ for $\cot\theta
\gsim 0.5$).  At low $\cot\theta$, $W_H^{\pm}$ decays predominantly into
$W^{\pm} H$ and $W^{\pm} Z$ with partial widths proportional to 
$\cot^2 2\theta$.

The general features of the production and decay of $Z_H$ and $W_H$
should extend to other little Higgs models in which the SM SU(2) gauge
group comes from the diagonal breaking of two SU(2) groups;
this is true, e.g., for the model in Ref.~\cite{SU6Sp6}.
The decays to $ZH$, $WH$ will however be modified in this model since it
contains two Higgs doublets.

\subsubsection{$A_H$}
The heavy U(1) gauge boson is the lightest new particle in the Littlest
Higgs model.  Its couplings to fermions are more model dependent than
those of the heavy SU(2) gauge bosons, since they depend on the U(1) charges
of the fermions (see Ref.~\cite{LHpheno} for details).  
Even the presence of $A_H$ is somewhat model-dependent, since
one can remove this particle from the Littlest Higgs model by gauging only
one U(1) group (hypercharge) without adding a significant amount of 
fine-tuning \cite{GrahamEW2}.  
Nevertheless, we show in Fig.~\ref{fig:sigmaAH} 
the cross section and branching ratios
of $A_H$ for the simplest anomaly-free choice of fermion U(1) charges
given in Ref.~\cite{LHpheno}. 
\begin{figure}[htb]
\begin{center}
\resizebox{\textwidth}{!}{\includegraphics{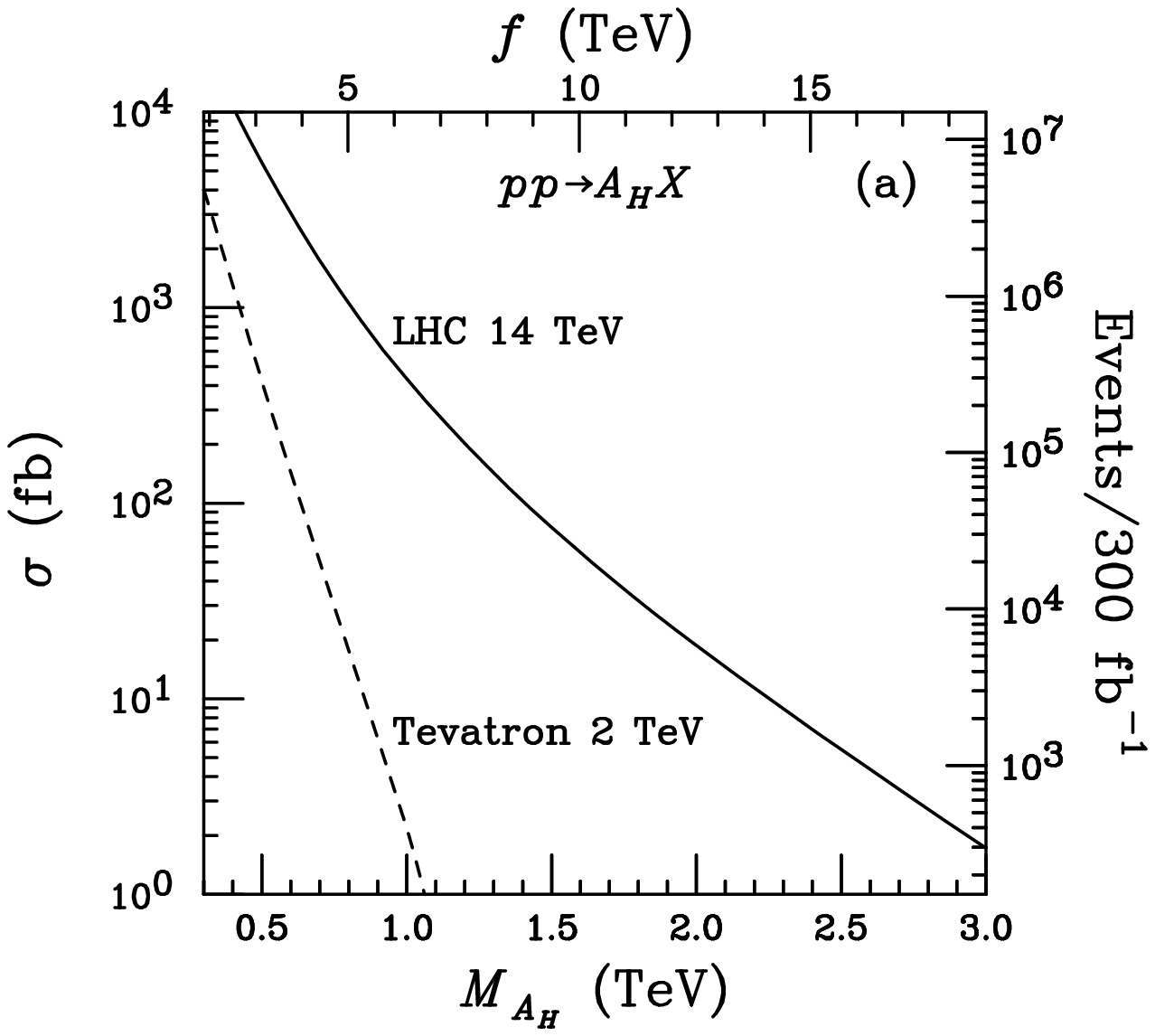}
\includegraphics{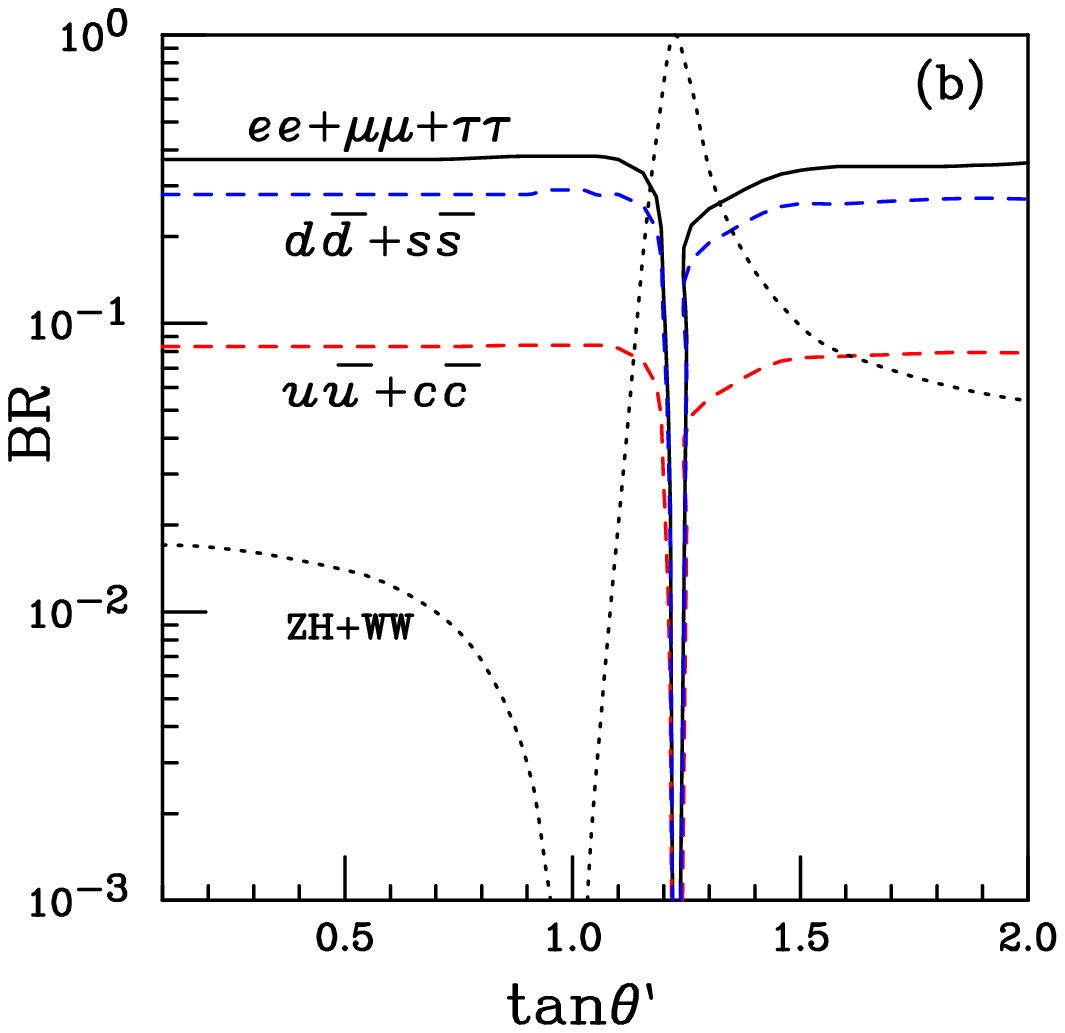}}
\end{center}
\caption{(a) Cross section for $A_H$ production in Drell-Yan at the LHC
and Tevatron, for $\cot\theta^{\prime} = 1$.  From Ref.~\cite{LHpheno}.
(b) Branching ratios of $A_H$ into fermions and $ZH+WW$ as a function
of $\tan\theta^{\prime}$, neglecting final-state mass effects.  
}
\label{fig:sigmaAH}
\end{figure}

\subsubsection{$T$}
The heavy top-partner $T$ can be pair produced via QCD interactions with
a cross section that depends only on the $T$ mass.
However, this production mode is suppressed
by phase space due to the typically high mass of the $T$.  
The single $T$ production mode, $W^+b \to T$, is dominant 
at the LHC for $M_T$ 
above about a TeV.  The cross section for single $T$ production depends 
on the $W^+bT$ coupling, which in turn depends on the amount
of mixing between $T$ and the left-handed top quark.  This coupling
is fixed in terms of $M_T$ and $f$, up to a two-fold ambiguity for
$M_T/f$ above its minimum value of $2 m_t/v$.
The cross sections are shown in Fig.~\ref{fig:toph}.
\begin{figure}[htb!]
\begin{center}
\resizebox{\textwidth}{!}{\includegraphics{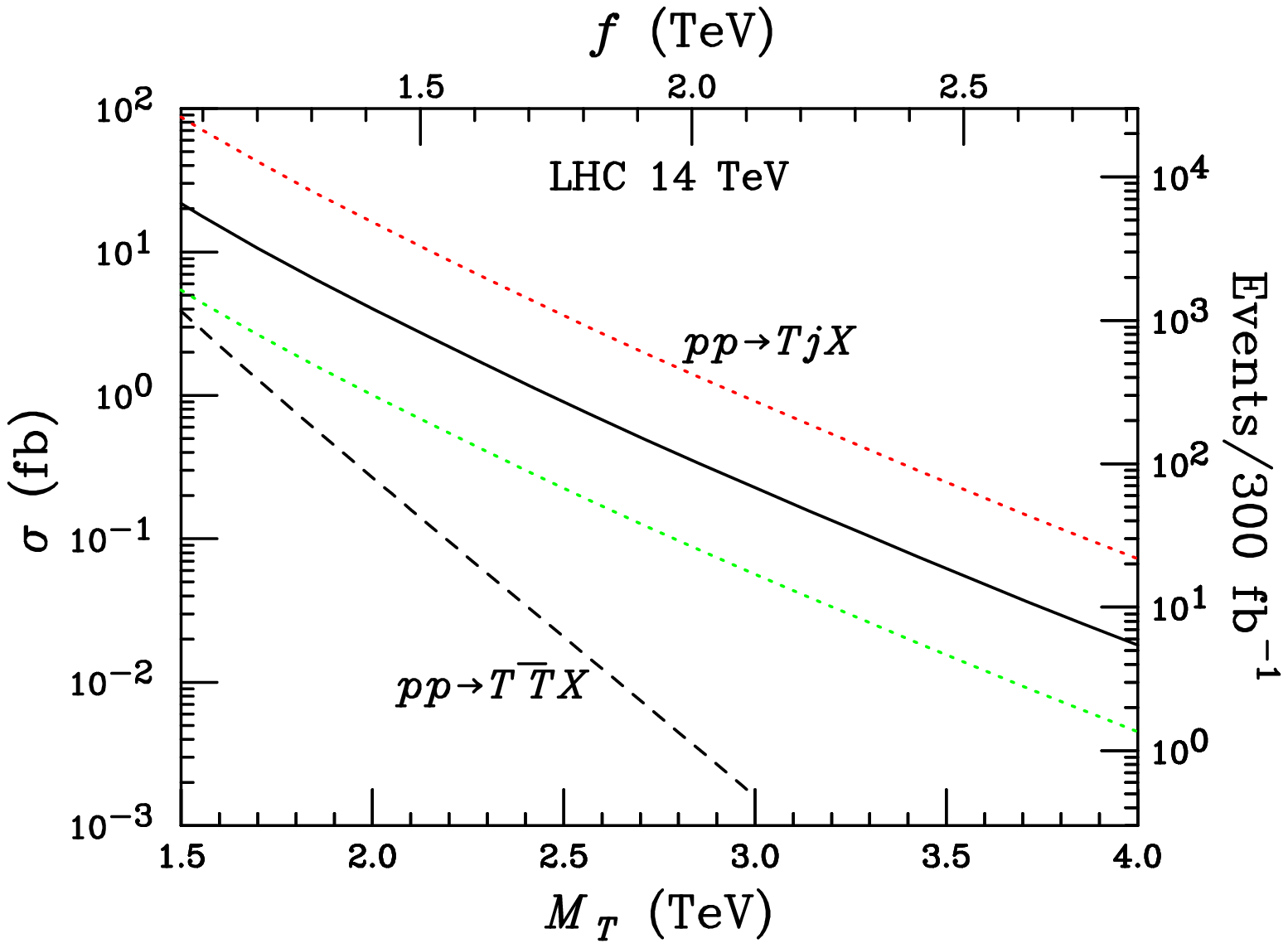}}
\end{center}
\caption{Cross sections for $T$ production at the LHC.  The single-$T$
cross section is shown for $M_T/f = 2 m_t/v$ (solid line)
and $M_T/f = 2.5 m_t/v$ (dotted lines).
The QCD pair production cross section is shown for comparison (dashed line).
The top axis shows the corresponding $f$ value for $M_T/f = 2 m_t/v$.
From Ref.~\cite{LHpheno}.}
\label{fig:toph}
\end{figure}
The top-partner $T$ decays into $tH$, $tZ$ and $bW$ with branching
fractions of 1/4, 1/4, and 1/2, respectively.  These decays come from
the coupling of $T$ to the Higgs doublet $h$ and the 3rd generation
quark doublet, applying the Goldstone boson equivalence theorem for the
Goldstone modes eaten by the $Z$ and $W$ bosons.

The structure of the top sector in
many of the other little Higgs models in the literature is quite similar
to the Littlest Higgs model,
so these general features of $T$ production and decay should carry over.  
Some models contain more than one 
top-partner \cite{SU6Sp6,KaplanSchmaltz,custodialmoose,Spencer,Anntalk}, or
contain partners for the two light generations of fermions as well
\cite{KaplanSchmaltz,SkibaTerning}; in 
these cases the phenomenology will be modified.

\subsubsection{$\Phi^{++}$}
The doubly charged component $\Phi^{++}$ of the Higgs triplet
can be singly produced through the resonant process 
$W^+W^+ \to \Phi^{++} \to W^+W^+$.  The 
cross section for this process is proportional to the square
of the triplet vev $v^{\prime}$, which must be quite small
in the Littlest Higgs model: $v^{\prime} < v^2/4f$.  This
may make resonant $\Phi^{++}$ production difficult to see 
due to lack of rate.  The doubly
charged Higgs boson could also be found via pair production from 
photon or $Z$ exchange, if it is not too heavy.
The doubly charged Higgs boson can in principle decay to a pair of like-sign
charged leptons via the dimension-four operator $L\Phi L$, offering a 
more distinctive signature than the decay into a pair of like-sign $W$ 
bosons; however, the coupling is highly model 
dependent and care must be taken to avoid generating too large a neutrino
mass from the triplet vev.

\subsection{The Little Higgs at a Linear Collider}

\subsubsection{Electroweak precision measurements}
The Littlest Higgs model introduces corrections to the precision electroweak
observables, which have been studied in 
Refs.~\cite{GrahamEW2,GrahamEW1,JoAnneEW}.
These corrections lead to constraints on the model parameter space
and a lower bound on the scale $f$ from existing electroweak data.
The constraints come from $Z$ pole data from LEP and SLD, 
low-energy neutrino-nucleon scattering, atomic parity violation,
and the $W$ boson mass measurement
from LEP-II and the Tevatron.
Together, these measurements probe contributions from the exchange of virtual
heavy gauge bosons between fermion pairs, the mixing of the heavy
SU(2) and U(1) gauge bosons with the $Z$ boson 
that modifies the $Z$ couplings to fermions,
and a shift in the ratio of the masses of the $W$ and $Z$ due both to mixing 
of the $W$ and $Z$ bosons with the heavy gauge bosons and to the nonzero 
triplet vev.

A linear collider will achieve high-precision measurements of the top
quark and Higgs boson masses, which are important inputs to the SM
electroweak fit.  LC improvement in the $W$ boson mass measurement,
together with an order-of-magnitude improvement in the measurements of 
the $Z$-pole observables at a ``Giga-$Z$'' machine,
should turn up a deviation from the SM fit due to the little Higgs model
contributions.

\subsubsection{Triple gauge boson couplings}
The $WWZ$ triple gauge boson coupling in the Littlest
Higgs model is modified from its SM form due to the modification of
$G_F$ by $W_H$ exchange \cite{LHpheno}:
\begin{equation}
g_1^Z = \kappa_Z =
        1 + \frac{1}{\cos 2 \theta_W} \left\{
        \frac{v^2}{8f^2} \left[ -4 c^2 s^2
        + 5 (c^{\prime 2}-s^{\prime 2})^2 \right]
        - \frac{2 v^{\prime 2}}{v^2} \right\},
\end{equation}
where the form-factors are defined according to \cite{WWV}
\begin{equation}
\mathcal{L}_{WWV} = i g_{WWV} \left[
        g_1^V (W^+_{\mu\nu} W^{- \mu} - W^{+ \mu} W^-_{\mu\nu} ) V^{\nu}
        + \kappa_V W^+_{\mu} W^-_{\nu} V^{\mu\nu}
        + \frac{\lambda_V}{m_W^2} W_{\mu}^{+\nu} W_{\nu}^{-\rho} V_{\rho}^{\mu}
        \right].
\end{equation}
At present, the constraints from the $WWZ$ coupling are weak compared
to those from electroweak precision measurements.
However, at a future linear collider, a precision of $10^{-3}-10^{-4}$ 
on $g_1^Z$ and $\kappa_Z$ should be achievable; this would be sensitive 
to $f\sim (15-50)v\sim 3.5-12$ TeV for generic values of $c$, $c^{\prime}$
and $v^{\prime}$.
Unfortunately for this measurement, the region of parameter space that
loosens the electroweak precision bound on $f$
(small $c$ and $v^{\prime}$ and $c^{\prime} \simeq s^{\prime}$) 
also suppresses the little Higgs contribution to $g_1^Z$ and $\kappa_Z$.

\subsubsection{Loop-induced Higgs boson decays}
The decay partial widths of the Higgs boson into gluon pairs or photon
pairs are modified in the Littlest Higgs model by the new heavy particles 
running in the loop and by the shifts in the Higgs couplings
to the SM $W$ boson and top quark \cite{LHloop}.
These modifications of the Higgs couplings to gluon or photon pairs
scale like $1/f^2$, and thus decouple at high $f$ scales.
The range of partial widths for given $f$ values accessible
by varying the other model parameters are shown in Fig.~\ref{fig:PWblob}.
\begin{figure}[htb!]
\begin{center}
\resizebox{\textwidth}{!}{
	\rotatebox{270}{\includegraphics{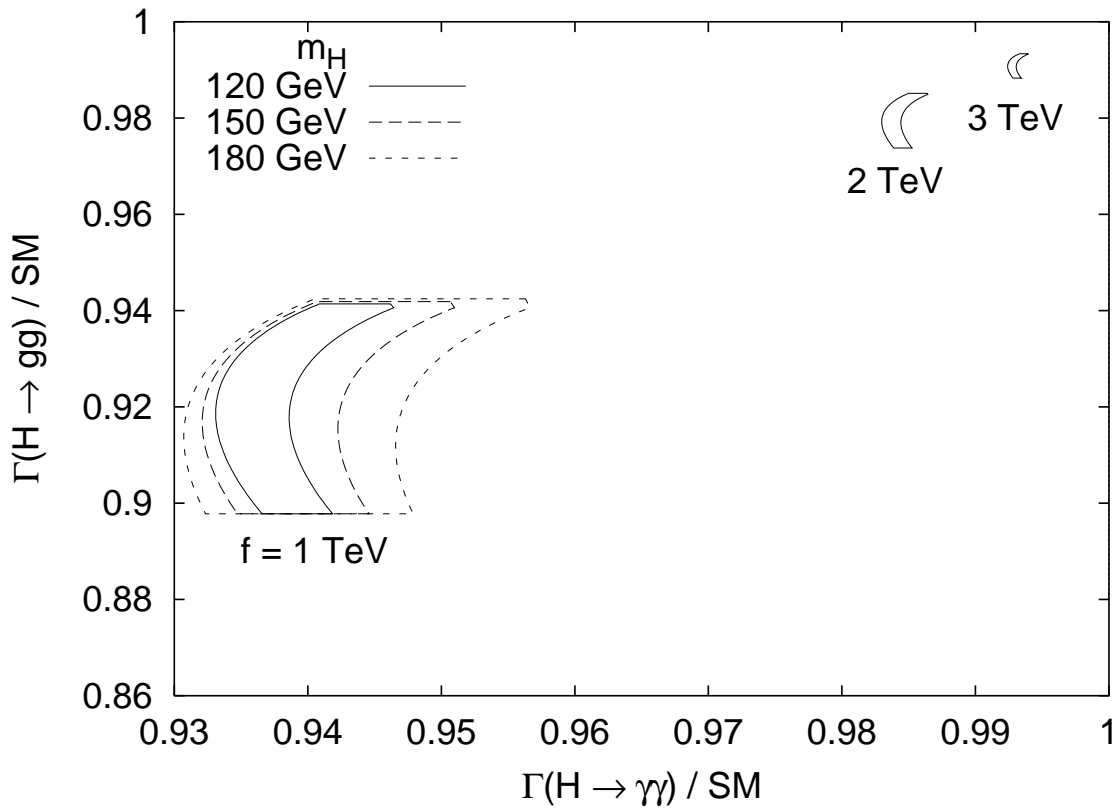}}}
\end{center}
\caption{Range of values of $\Gamma(H \to gg)$ versus 
$\Gamma(H \to \gamma\gamma)$ accessible in the Littlest Higgs model
normalized to the SM value, for $m_H = 120$, 150, 180 GeV and 
$f = 1$, 2, 3 TeV.  From Ref.~\cite{LHloop}.}
\label{fig:PWblob}
\end{figure}

Are these corrections observable?  For $f \geq 1$ TeV, the correction
to $\Gamma(H \to gg)$ is always less than 10\%.  This is already smaller
than the remaining SM theoretical uncertainty on the gluon fusion cross 
section due to uncalculated higher-order QCD corrections \cite{QCDggh}.
For the partial width to photons, the situation is more promising because
the QCD corrections are well under control.  At the LHC, the 
$H \to \gamma\gamma$ decay rate can be measured to 15--20\% \cite{Dieter}; 
this probes $f < 600$ GeV at $1\sigma$.
A linear $e^+e^-$ collider has only comparable precision since the
$H \to \gamma\gamma$ branching ratio measurement is limited by statistics
\cite{LCbook,Aguilar-Saavedra:2001rg,Boos}.
The most promising measurement would be done at a photon collider,
where the $\gamma\gamma \to H \to b \bar b$ rate can be measured 
to about 2\% \cite{gammagamma} for a Higgs boson with mass around
115--120 GeV.  (The uncertainty rises to 10\% for $m_H = 160$ GeV.)
Combining this with a measurement of the branching ratio
of $H \to b \bar b$ to about 1.5--2\% at the $e^+e^-$ collider 
\cite{LCbook,Aguilar-Saavedra:2001rg,ACFA}
allows the extraction of $\Gamma(H \to \gamma\gamma)$ with a precision 
of about 3\%.  
Such a measurement would be sensitive to $f < 1.5$ TeV at the $1\sigma$
level, or $f < 1.1$ TeV at the $2\sigma$ level.  A $5\sigma$ deviation is
possible for $f < 700$ GeV.  
For comparison, the electroweak precision constraints require
$f \gsim 1$ TeV in the Littlest Higgs model \cite{GrahamEW2}.

The biggest model dependence in the loop-induced Higgs decays in little
Higgs models comes from the content of the Higgs sector at the electroweak
scale.  In models with only one light Higgs doublet, our general conclusions
should hold, up to factors related to the multiplicity and detailed 
couplings of the new heavy particles.  However, many little Higgs models
\cite{minmoose,SU6Sp6,KaplanSchmaltz,custodialmoose,SkibaTerning}
contain two light Higgs doublets.  In this case, mixing between the 
two neutral CP-even Higgs particles and the contribution of a
relatively light charged Higgs boson running in the loop can lead to
large deviations in the couplings of the lightest Higgs boson to gluon
or photon pairs, swamping the effects from the heavy states.


%


                                                                                
\section{Exotic scenarios}
\label{sec:28}

{\it J.~Gunion}

\vspace{1em}
\def\bit{\begin{itemize}}
\def\eit{\end{itemize}}
\def\anti{\overline}
\def\beq{\begin{equation}}
\def\eeq{\end{equation}}
\def\bea{\begin{eqalign}}
\def\eea{\end{eqalign}}
\def\gev{~{\rm GeV}}
\def\mev{~{\rm MeV}}
\def\lsim{\mathrel{\raise.3ex\hbox{$<$\kern-.75em\lower1ex\hbox{$\sim$}}}}
\def\gsim{\mathrel{\raise.3ex\hbox{$>$\kern-.75em\lower1ex\hbox{$\sim$}}}}
\def\rts{\sqrt s}
\def\hp{H^+}
\def\hm{H^-}

In this section, we focus on a few additional situations in which the
available experimental signals for Higgs physics could be difficult to
detect at the LHC or the LC  and the extent to which these two
machines (including also the photon-collider $\gam$C option) will
complement one another. Whether the scenarios considered should all be
termed ``exotic'' is not clear.  Some of the models of this type
involve substantial extensions of more standard models. Other relevant
scenarios arise simply by virtue of choosing particular, theoretically
motivated parameters and/or boundary conditions within the context of
very attractive and fairly simple models.  The examples are chosen to
highlight how the LC can usually provide a clear signal for Higgs
physics in models for which the LHC cannot or vice versa, or to
illustrate how a first signal at one type of machine can be clarified
and studied in greater detail at the other collider.  In particular,
even if a scalar particle is detectable at the LHC, a full precision
study of its properties will typically require a high luminosity LC.
Additional examples of relevance for the LHC/LC complementarity, such
as the NMSSM, radion-Higgs mixing and so forth are given more detailed
treatment in earlier sections of this report.

\subsection{The CP-conserving MSSM in the decoupling limit}

\def\hl{h}
\def\mhl{m_{\hl}}
\def\hh{H}
\def\ha{A}
\def\hpm{H^\pm}
\def\mhh{m_{\hh}}
\def\mha{m_{\ha}}
\def\mhpm{m_{\hpm}}
\def\tanb{\tan\beta}
\def\cnone{\widetilde \chi^0_1}
\def\mcnone{m_{\cnone}}
\def\epem{e^+e^-}
\def\mhsm{m_{h_{SM}}}


A very probable MSSM scenario is one in which the
the $\hh,\ha,\hpm$ are fairly heavy (and rather
degenerate) while the $\hl$ is light and has very
SM properties.  As reviewed in earlier sections,
it may be possible at the LHC and LC to detect
deviations in the properties of the $\hl$ that
reveal that it is not precisely SM-like and that
provide some indication for the presence of the
more complicated MSSM Higgs sector. However, if
the $(\mha,\tanb)$ parameter space point is in the
LHC wedge of moderate $\tanb$ and $\mha\gsim
300\gev$,  
direct detection of the $\hh,\ha,\hpm$ may not be
possible at the LHC.  In addition, their detection
at an LC with $\rts<600\gev$ will also not be
possible in this same wedge region
\cite{Grzadkowski:1999wj}. This is because pair
production, $e^+e^-\to \hh\ha,\hp\hm$, is
kinematically inaccessible while the Yukawa
radiation processes, $e^+e^-\to b\anti b
\hh,b\anti b\ha,t\anti t\hh,t\anti t\ha$ all have
a very low rate for moderate $\tanb$.

\begin{figure}[!h]
\begin{center}
\hspace*{.3in}\includegraphics[width=4.5in,height=3.3in]{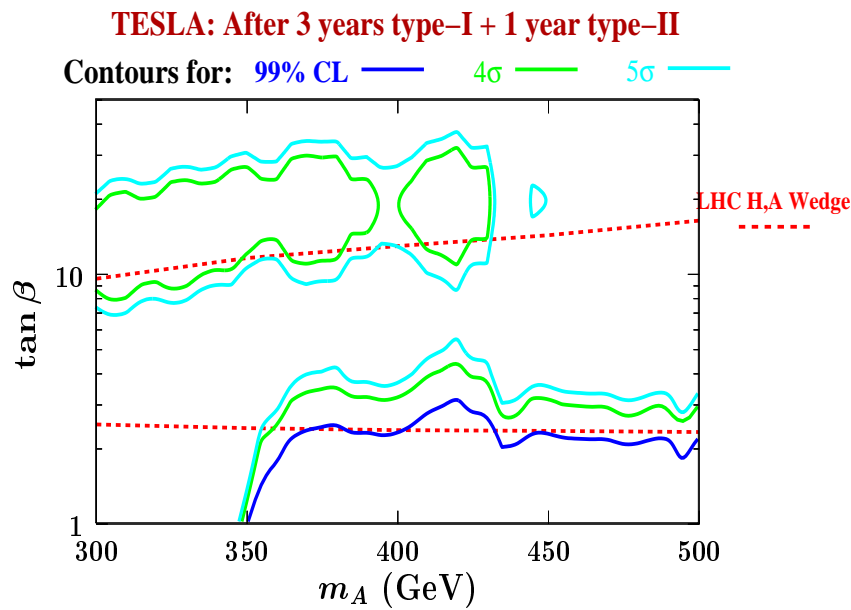}
\caption{Contours for $4\sigma$ and $5\sigma$ 
  discovery and 99\% CL exclusion in the $\gam\gam\to \hh,\ha\to b\anti b$ channel
 after 1 year of TESLA $\gam\gam$
  operation in the laser photon-electron polarization configuration II
  (designed for $\hh,\ha$ masses near the maximum reach point of
  $\mha,\mhh\sim 0.8\sqrt s$) and 3 years of operation in polarization
  configuration I (designed to maximize sensitivity to moderate
  $\hh,\ha$ masses).  For details, see
  \cite{Asner:2001ia,Asner:2003hz}. The dashed lines show the LHC
    wedge region where $\hh,\ha$ detection is not possible; the LC wedge
is even larger for $\rts\lsim 800\gev$.}
\label{asneretalplot}
\end{center}
\end{figure}

However, there are options at the LC that would
allow $\hh$, $\ha$ and $\hpm$ detection.  First,
the LC energy will eventually be upgraded to
$\gsim 1\tev$, allowing $\hh\ha$ and $\hp\hm$ pair
production up to masses $\gsim 500\gev$. Second,
it has been shown \cite{Asner:2001ia,Asner:2003hz}
(see also
\cite{Muhlleitner:2001kw,Niezurawski:2003ir}) that
the photon-collider ($\gam$C) option at a $\sqrt
s=600\gev$ LC has an excellent chance of allowing
direct discovery of the $\hh$ and $\ha$ for masses
up to about $500\gev$ in the channel $\gam\gam\to
\hh,\ha\to b\anti b$.  A plot from
\cite{Asner:2003hz} is presented in
Fig.~\ref{asneretalplot}.  This plot shows that
after 4 years of operation of a TESLA-based
$\gam$C it would be possible to detect the
$\hh,\ha$ signal precisely in the wedge of
$(\mha,\tanb)$ parameter space where their
detection would not be possible at either the LHC
or LC. The only region within the wedge for which
the $b\anti b$ final state signal is weak is the
part of the wedge with the lowest $\tanb$ values.
Here, the $t\anti t$ decays of the $\hh$ and $\ha$
become important and the $t\anti t$ final state
would be likely to provide a strong signal.  This
channel has not yet been studied in detail.

\subsection{The CP-conserving 
2HDM with the only light Higgs boson being pseudoscalar --
a special non-decoupling limit}

The easiest way to satisfy precision electroweak
constraints in the two-Higgs-doublet model (2HDM)
context is if the Higgs potential parameters are
chosen so that the decoupling limit applies. In
the decoupling limit, the lightest CP-even Higgs
boson, the $\hl$, is SM-like by virtue of all the
other Higgs bosons (the $\hh,\ha,\hpm$) being
heavy.  However, this is not the only way to
achieve consistency with the precision electroweak
data.  As shown in
\cite{Chankowski:2000an,Gunion:2000abGunion:2001vi} (see also
\cite{Gunion:2002zf}), it is possible to choose
2HDM parameters in such a way that the only light
Higgs boson is the $\ha$, all the other Higgs
bosons ($\hl$, $\hh$ and $\hpm$) having masses of
order a TeV.  The $\hl$ in this kind of model will
be SM-like and will give sizable $\Delta T<0$ and
$\Delta S>0$ contributions that would lead to
inconsistency with the precision constraints.
However, for parameters such that there is a small
non-zero $\mhpm-\mhh>0$ mass difference (an
isospin splitting) there will be a large $\Delta
T>0$ that will more than compensate the large
$\Delta T<0$ contribution from the heavy $\hl$.  This is easily
understood algebraically. When the $\ha$ is relatively light
while the $\hl$ is heavy and SM-like, one finds
\beq
   \Delta \rho=\frac{\alpha}{16 \pi m_W^2 c_W^2}\left\{\frac{c_W^2}{s_W^2}
   \frac{m_{H^\pm}^2-m_{H^0}^2}{2}-3m_W^2\left[\log\frac{m_{h^0}^2}{m_W^2}
   +\frac{1}{6}+\frac{1}{s_W^2}\log\frac{m_W^2}{m_Z^2}\right]\right\}\nonumber
\label{drhonew}
\end{equation}
The large negative contribution from the heavy SM-like $\hl$ arising
from the 2nd term in brackets is compensated by an even larger
positive contribution from the first term proportional to
$(\mhpm-\mhh)(\mhpm+\mhh)$. 
The resulting predictions of the 2HDM give postive shifts of
${\mathcal O}(0.1)$ to both $S$ and $T$ (assuming $U = 0$). 
%
%
%
In this scenario, the LHC would discover a $\sim 500\gev - 1\tev$
SM-like $\hl$  (for example, in the  $\hl\to ZZ\to
4\ell$ channel) instead of a light CP-even Higgs boson (with SM-like $WW,ZZ$
couplings) as apparently needed to satisfy precision electroweak
constraints.  The current precision electroweak constraints would
then imply that additional contributions to $S$ and/or $T$ are required,
but their source and nature would be obscure. For example, instead of
the 2HDM scenario considered here, the negative $\Delta T$ and
positive $\Delta S$ from the heavy SM-like $\hl$ could equally well be compensated
by new physics yielding a sizable $\Delta S <0$ contribution with
small $\Delta T$ (that would yield a net $S,T$ prediction in the lower
left corner of the ellipse) or new physics giving $\Delta T>0$ and $\Delta S<0$
contributions (that would return the net $S,T$ prediction to near the
center of the ellipse). 

Clarification of the situation would be 
difficult. In the 2HDM scenario being considered, and
assuming a moderate $\tanb$ value, the $\ha$ would
not be directly detectable at either the LHC or the LC ---
it falls into the wedge region described in the
previous subsection.  The $\hl\to \ha\ha$ decay, if
allowed kinematically, would typically have a reasonable
branching ratio despite the presence of the on-shell $ZZ,WW$
decay modes. Its detection would be very important to
unraveling the situation. Of course, the $(S,T)$
prediction remains within the 90\% CL ellipse even
if $\mha$ is as large as $500\gev$, for which $\hl\to\ha\ha$ decays
would be forbidden and direct detection of the $\ha$ (as well as
of the $\hpm$ and $\hh$) in $\epem$
collisions would require  a much higher energy LC.
A $\gam$C might play a crucial role.  Detection of
the $\ha$ in the $\gam\gam\to\ha\to b\anti b$
channel would generally be possible for $\mha\lsim
0.8 \sqrt s$, i.e. for $\mha\lsim 800\gev$ for
$\sqrt s\lsim 1\tev$. See \cite{Asner:2001ia} for details.

Considerable clarification would
result from a Giga-$Z$ run at the future LC (combined  with a  $WW$
threshold scan sufficient to obtain $\Delta m_W=7 MeV$). 
Given the LHC
measurement of $\mhl$, one could determine with considerable accuracy
the additional $\Delta T$ and/or $\Delta S$ from the additional new
physics with a (correlated) $1\sigma$ error of order $\delta\Delta
T\sim\delta\Delta S\sim \pm 0.05$. Some scenarios would be excluded,
but many possibilities would remain, since one could not be sure that
the observed deviation was the result
of the presence of undetected Higgs bosons or some of the other types
of new physics discussed in earlier sections of this report.

\subsection{Maximally-mixed and ``Continuum'' Higgs models}

We first describe the general type of model we wish to consider
in this section and some of the related experimental considerations.
We then use the continuum Higgs model as a particular example.

In unconstrained CP-conserving two-doublet
models away from the decoupling limit, the scalars
can mix strongly with one another and have similar
masses and many couplings possibilities.  In CP-conserving
models with more than two-doublets and/or doublets
plus one or more complex singlets there will be
mixing among the pseudoscalars as well as among
the scalars.  In CP-violating Higgs sectors (including
the MSSM two-doublet Higgs sector with CP-violation from loop
corrections), the mixing possibilities
will be even greater. Such mixing typically results in a sharing
of the $WW/ZZ$ coupling among the CP-even (or all the CP-mixed)
Higgs bosons, thereby substantially reducing key 
high-mass-resolution LHC
signals such as $gg\to h \to ZZ^{(*)}\to 4\ell$ and especially $gg\to h \to
\gam\gam$.  At the same time, the Higgs bosons can have masses that
differ by an 
amount of order the experimental resolution in other critical channels
that do not rely on the $WW/ZZ$ coupling, such as $gg\to t \anti t h$ with
$h\to b\anti b$ or $h\to \tau^+\tau^-$. Such reduced and overlapping
signals will be much more difficult to separate from background than
in the SM Higgs case. In addition, heavier Higgs bosons can decay
to lighter ones, further complicating the search possibilities.
  
Even in the absence of $h$ decays to other Higgs bosons, the $WW\to h
\to \tau^+\tau^-$ detection channel will take a ``double-hit''.
First, the production rate for each $h$ would be suppressed due to
reduced $WWh$ coupling. Second, the poor mass resolution in the final
state would mean that signals for several different $h$'s (separated
in mass by, say, $10\gev$) will overlap and make peak detection
impossible.  Instead, one must try to determine the presence of a
broad excess in the $M_{\tau\tau}$ distribution.

Because of these possibilities, it is often fairly easy to find model
parameters such that the LHC will have difficulty detecting the Higgs
boson signals.  The power of the LC is its ability to look for
$e^+e^-\to Z h$ in the inclusive $e^+e^-\to ZX$ missing-mass channel
approach by simply looking for a bump in the reconstructed $M_X$.
Even if the signals from different Higgs bosons overlap somewhat and
their strength is maximally shared, the excess in the $M_X$
distribution will be apparent at the LC.  And, of course, the
inclusive $M_X$ peak or broad excess is independent of how the Higgs
bosons decay.

Another general point is that even if the LHC does not find a direct
signal for a set of relatively light but strongly mixed and
overlapping Higgs bosons we will know that they (or some alternative
source of electroweak symmetry breaking) are present below the TeV
scale by virtue of the fact that $WW\to WW$ scattering measurements
will be consistent with perturbative expectations.

We now very briefly review a particular model
in which the above considerations have been shown
to be relevant.
This model \cite{Espinosa:1998xj} was explicitly
constructed as a worst case scenario for Higgs
detection.  The idea is to imagine a large number
of doublet and/or singlet Higgs fields with
complicated self interactions.  In general, the
Higgs sector could be CP-violating.  The worst
case arises if these many Higgs bosons are spaced
in mass at intervals slightly less than the mass
resolution in $M_{b\anti b}$, $M_{\tau^+\tau^-}$
and $M_X$. They will mix with one another and the
heavier ones will decay to the lighter ones.  In
general, they will share the $WW/ZZ$ coupling
strength.  Using continuum limit notation, the
only constraints are 
\beq \int
dm_hK(m_h)=1\,,\quad\quad \int d m_h K (m_h)
g_{ZZh}^2 m_h^2\lsim (200\gev)^2 
\end{equation}
where $g_{ZZh}^2=K(m_h)g_{ZZh_{SM}}^2$.  The latter
constraint relies on either assuming that these
Higgs bosons are entirely responsible for
explaining the precision electroweak data (a
constraint that can be avoided if there is
substantial isospin splitting between the charged
Higgs and scalar Higgs of the model) or if one
demands perturbativity of the model up to the
Planck scale.

The result is very substantial diminution of all
the standard LHC signals. 
In particular, the high resolution $gg\to h \to \gamma\gamma$ and
$gg\to h \to ZZ\to
4\ell$ final states have very low production rate
for any one of the $h$'s due to the sharing of the
$WW/ZZ$ coupling-squared. Other channels with more
limited resolution will not exhibit separated mass
peaks.  There will only be a spread out signal
that must be detected as a broad excess in some
type of mass distribution such as $M_{b\anti b}$
or $M_{\tau^+\tau^-}$ or $M_{b\anti
  b\tau^+\tau^-}$ (the latter being relevant for
heavier Higgs bosons that decay to a pair of
lighter Higgs bosons). The recent study of
\cite{Alves:2003vp}, which claims that a signal
can be seen in the $WW\to \sum_i h_i \to WW\to
2\ell2\nu$ channel, neglects the possibility of
Higgs decays to much lighter Higgs bosons (that
have very weak $WW/ZZ$ coupling and do not
contribute to the above sum rules).  Allowing for
this possibility, it would seem impossible to
guarantee an observable LHC signal for the
continuum Higgs scenario.

As discussed in \cite{Espinosa:1998xj}, the broad
$M_X$ excess in the $e^+e^-\to ZX$ channel that
would arise in this model {\it will} be detectable
with enough ($L\gsim 100\div 200~{\rm fb}^{-1}$)
integrated luminosity at the LC.  Further, with $L\sim 500~{\rm
  fb}^{-1}$ it will be possible to determine the extent
of the excess in bins with size of order $10~{\rm GeV}$
as well as to examine the dominant final states in each such bin.

\subsection{Higgsless models}

We simply mention the recently proposed Higgsless
model \cite{Csaki:2003zu} in which boundary
conditions on a brane in a warped 5th dimension
are responsible for electroweak symmetry breaking.
The unitarity of $WW$ scattering is maintained so
long as the KK excitations of the $W$ and $Z$ are
not much above the TeV scale and therefore
accessible to direct production at the LHC. These
higher KK excitations also work together in such a
way as to avoid tree-level violation of the
$\rho=1$ constraint and to make only a small
contribution to the $S$ parameter.

The exact nature of the LHC signals for the $W$
and $Z$ excitations has not been worked out.
However, it is clear that understanding what is
happening in $WW$ scattering and electroweak
symmetry breaking will require detecting all the
relevant excitations and determining details of
their couplings to one another.  This will be a
challenging task.

An LC (with $\sqrt s\lsim 1~{\rm TeV}$) might have
some difficulty studying all the relevant KK
excitations.  Thus, this is a case in which the
LHC might be superior for understanding
electroweak symmetry breaking. Still, the LC
measurements of the $M_X$ distribution would be
very revealing in that no excess (at masses below
a few hundred GeV) would be observed and one could
then be more certain that the KK resonances
observed at the LHC were indeed the entire story.




\chapter{Strong Electroweak Symmetry Breaking}
\label{chapter:strongewsymmbreak}

Editors: {\it T.~Barklow, K.~M\"onig}

\vspace{1em}

{\it T.~Barklow, S.~Boogert, G.~Cerminara,
W.~Kilian, A.~Krokhotine, K.~M\"onig, A.F.~Osorio
}

\vspace{1em}

\renewcommand{\vev}[1]{\langle #1 \rangle}
\renewcommand{\tr}[1]{\operatorname{tr}\left[#1\right]}
\newcommand{\unit}{\boldsymbol{1}}
\renewcommand{\hc}{\text{h.c.}}
\newcommand{\vW}{\mathbf{W}}
\newcommand{\vB}{\mathbf{B}}
\newcommand{\LL}{\mathcal{L}}

\newcommand{\q}[2]{{\ensuremath #1\,#2}}
\renewcommand{\qq}[3]{{\ensuremath #1\times 10^{#2}\,#3}}
\newcommand{\ql}[2]{{\ensuremath 10^{#1}\,#2}}

\newcommand{\eV}{{\ensuremath\rm eV}}
\newcommand{\keV}{{\ensuremath\rm keV}}
\newcommand{\MeV}{{\ensuremath\rm MeV}}
\newcommand{\GeV}{{\ensuremath\rm GeV}}
\newcommand{\TeV}{{\ensuremath\rm TeV}}
\newcommand{\pb}{{\ensuremath\rm pb}}
\renewcommand{\fb}{{\ensuremath\rm fb}}
\renewcommand{\ab}{{\ensuremath\rm ab}}

\makeatletter
\def\fmslash{\@ifnextchar[{\fmsl@sh}{\fmsl@sh[0mu]}}
\def\fmsl@sh[#1]#2{%
  \mathchoice
    {\@fmsl@sh\displaystyle{#1}{#2}}%
    {\@fmsl@sh\textstyle{#1}{#2}}%
    {\@fmsl@sh\scriptstyle{#1}{#2}}%
    {\@fmsl@sh\scriptscriptstyle{#1}{#2}}}
\def\@fmsl@sh#1#2#3{\m@th\ooalign{$\hfil#1\mkern#2/\hfil$\crcr$#1#3$}}
\makeatother


\noindent{\small 
If no light Higgs boson exists, quasi-elastic scattering processes of
W and Z bosons at high energies provide a direct probe of the dynamics
of EWSB.  The amplitudes can be measured in 6-fermion processes both
at LHC and at the LC.  The two colliders are sensitive to different
scattering channels and yield complementary information.  Notably,
detailed measurements of cross sections and angular distributions at
the LC will be crucial for making full use of the LHC data.  The
high-energy region where resonances may appear can be accessed at LHC
only.  A thorough understanding of the sub-TeV data of the LC and the
LHC combined will be essential for disentangling such new states.

This note collects basic facts about the phenomenology of vector boson
scattering processes at LHC and the LC and summarises representative
physics studies that have been finished or are under way.
}


\section{Introduction}

The experimental data which were obtained in recent years have
established the validity of the description of electroweak
interactions by a spontaneously broken gauge theory.  The masses of
fermions and electroweak gauge bosons are generated by the Higgs
mechanism~\cite{Higgs}, which involves the condensation of a scalar
multiplet with non-vanishing hypercharge and weak isospin quantum
numbers.  This includes the Goldstone bosons associated with the
spontaneous breaking of the electroweak $SU(2)_L\times U(1)_Y$
symmetry, which are identified with the longitudinal degrees of
freedom of the massive $W$ and $Z$ gauge bosons.

The Standard Model (SM)~\cite{GSW} is the simplest theory of
electroweak interaction that is complete in the sense that its
predictions can be extrapolated up to energies far beyond the
electroweak scale,
\begin{equation}
  v = (\sqrt2 G_F)^{-1/2} = \q{246}{\GeV},
\end{equation}
where $G_F$ is the Fermi constant.  The Higgs mechanism identifies $v$
with the vacuum expectation value of the Higgs field.  In the SM,
fluctuations around this expectation value are associated with a
scalar particle, the Higgs boson.  This hypothesis is consistent with
the electroweak precision data~\cite{PDG02}.

Nevertheless, the absence of a light Higgs boson is a logical
possibility, supported by the fact that the Higgs has not been
observed at LEP.  Actually, the best fit of the SM to the data
predicts its mass to be below $\q{100}{\GeV}$, which is already
excluded by direct searches~\cite{Chanowitz}.

If no Higgs boson exists, the Higgs mechanism is still valid formally,
but the electroweak symmetry cannot be linearly realized on the
multiplet of Goldstone bosons.  As a consequence, the model is
non-renormalisable, and scattering amplitudes of Goldstone bosons are
unbounded at high energies.  The predictivity of extrapolations is
limited to energies below the scale
\begin{equation}
  \Lambda = 4\pi v = \q{3.1}{\TeV},
\end{equation}
where the tree-level amplitudes saturate perturbative unitarity.  This
scale provides a cutoff to the low-energy effective theory.
Higher-dimensional operators are expected to contribute corrections of
order $v^2/\Lambda^2=1/(4\pi)^2$ to low-energy observables and could
mimic the existence of a light Higgs boson.

In fact, the lowest threshold where without the Higgs boson a
tree-level scattering amplitude would violate unitarity (if naively
extrapolated) is already at $\q{1.2}{\TeV}$~\cite{Uni}.  Thus, new
experiments which probe amplitudes in the ${\TeV}$ range with
sufficient precision will observe dynamics which cannot be inferred
from our present knowledge about electroweak interactions.  While
Goldstone bosons by themselves are unphysical degrees of freedom,
their identification with the longitudinal components of $W$ and $Z$
bosons makes this new dynamics observable in the scattering amplitudes
of vector bosons:
\begin{equation}\label{proc}
  W_L W_L \to W_L W_L, \quad
  W_L W_L \to Z_L Z_L, \quad\ldots
\end{equation}

In this no-Higgs scenario, it is not clear what physics we can expect
in the $\TeV$ range.  The simplest model of strong electroweak
symmetry breaking is minimal technicolour~\cite{dynEWSB}, where the
para\-digm of QCD is transferred to electroweak interactions, assuming
the existence of technifermions which are confined at the scale
$\Lambda$.  In this model, a strong vector resonance (the technirho)
is predicted in $WW$ scattering, analogous to pion scattering in the
$\GeV$ range.  However, technicolour does not account for fermion
masses and thus must be extended by additional interactions, and it
does not predict the correct sign and magnitude of the shifts in the
low-energy precision observables.  More realistic models involve more
complicated spectra and scale patterns and differ in their statements
about $\TeV$-scale phenomenology~\cite{TCmodels}.

The LHC and the Linear Collider both have the capabilities to shed
light onto this new sector of high-energy physics by measurements of
the vector boson scattering amplitudes~(\ref{proc}).  The two
facilities are complementary in several respects.  Clearly, the LHC
with its c.m.\ energy of $\q{14}{\TeV}$ is able in principle to
collect data beyond the unitarity saturation threshold.  However, the
fast falloff of the relevant cross section limits its actual reach for
the processes we are considering here, and the signals are
contaminated by a large background of SM processes.  Backgrounds are
much less severe at an $e^+e^-$ collider.  There, the machine itself
limits the energy reach, but the high luminosity anticipated for a
Linear Collider makes it more than comparable in sensitivity in the
range below $\q{1}{\TeV}$, where the new strong interactions determine
the precise pattern of the rise of vector boson scattering amplitudes
with energy.

At present, no theoretical or experimental analysis is available that
compares and combines the capabilities of both machines in the context
of strongly interacting $W$ and $Z$ bosons.  This will be a goal of
the ongoing LHC/LC workshop.  The purpose of this note is to review the
theoretical and phenomenological background, to discuss the principal
properties of the relevant processes at both colliders, and to
summarise the studies that have been completed or are under way, on a
common basis.

\section{Low-energy effective theory}

Below the cutoff $\Lambda$ there is a generic effective-theory
description of electroweak interactions that relies only on the
established facts about symmetries~\cite{EWChPT}.  The electroweak
Goldstone bosons $w^a$ ($a=1,2,3$) are used for the parameterisation
of a unitary $2\times 2$ matrix~$\Sigma$,
\begin{equation}
  \Sigma(x) = \exp\left( -\frac{i}{v}w^a(x)\,\tau^a\right),
\end{equation}
where $\tau^a$ are the Pauli matrices.  The $\Sigma$ field is
normalised to unit vacuum expectation value,
\begin{equation}
  \vev{\Sigma} = \unit,
\end{equation}
and serves as the Higgs field in the electroweak effective Lagrangian,
\begin{equation}\label{chpt}
\begin{split}
  \LL &=  
   - \tfrac12 \tr{\vW_{\mu\nu} \vW^{\mu\nu}} 
   - \tfrac12 \tr{\vB_{\mu\nu} \vB^{\mu\nu}}
  -\tfrac{v^2}{4}\tr{V_\mu V^\mu}
              +\Delta\rho\,\tfrac{v^2}{8}\tr{T V_\mu}\tr{T V^\mu}
  \\ &\quad
   + \LL_{\rm fermion} + \LL_{\rm gauge-fixing} + \LL_{\rm ghost}
  .
\end{split}
\end{equation}
The additional building blocks of this Lagrangian are the left- and
right-handed fermions, the weak and hypercharge gauge bosons $W^a$
($a=1,2,3$) and $B$ with
\begin{equation}
  \vW_\mu = W_\mu^a\tfrac{\tau^a}{2}, \qquad
  \vB_\mu = B_\mu \tfrac{\tau^3}{2},
\end{equation}
and the corresponding field strengths $\vW_{\mu\nu}$ and
$\vB_{\mu\nu}$.  The derived fields $V_\mu$ and $T$ are given by
\begin{equation}
  V_\mu = \Sigma (D_\mu\Sigma)^\dagger =  -ig\vW_\mu + ig'B_\mu + \ldots
\quad\text{and}\quad
  T = \Sigma \tau^3 \Sigma^\dagger = \tau^3 + \ldots,
\end{equation}
where $D_\mu$ is the gauge-covariant derivative.  In physical terms,
$V_\mu$ represents the longitudinal gauge bosons which are identified
with the Goldstone bosons in observable scattering amplitudes.  The
omitted terms involve Goldstone fields and their derivatives.  These
vanish in the unitary gauge, where Fadeev-Popov ghost and Goldstone
fields are set to zero and the $\Sigma$ field is set to unity.

{}From the structure of the generic Lagrangian~(\ref{chpt}) one can
infer the leading term proportional to $E^2/v^2$ in the rise of the
longitudinal vector boson scattering amplitudes.  This is called the
Low-Energy Theorem (LET)~\cite{LET}.  Transversal gauge bosons provide
contributions that are constant in energy and suppressed by the gauge
couplings, but enhanced by the larger number of degrees of freedom.
For $\Delta\rho=0$, the LET is given by
\begin{align}
  A(W^-_LW^-_L\to W^-_LW^-_L) &= -\frac{s}{v^2}\label{LET-wwww}\\
  A(W^+_LW^-_L\to W^+_LW^-_L) &= -\frac{u}{v^2}\\
  A(W^+_LW^-_L\to Z_LZ_L) &= \frac{s}{v^2} \\
  A(Z_LZ_L\to Z_LZ_L) &= 0 \label{LET-zzzz}
\end{align}
Subleading corrections (proportional to $E^4/v^4$) are parameterised by
the additional operators $\LL_1$ to~$\LL_{11}$~\cite{EWChPT}:
\begin{align}
  \LL_1 &= \alpha_1 
         gg'\tr{\Sigma\vB_{\mu\nu}\Sigma^\dagger\vW^{\mu\nu}} \label{L1}\\
  \LL_2 &= i\alpha_2 
         g'\tr{\Sigma\vB_{\mu\nu}\Sigma^\dagger[V^\mu,V^\nu]} \label{L2}\\
  \LL_3 &= i\alpha_3 g\tr{\vW_{\mu\nu}[V^\mu,V^\nu]} \label{L3}\\
  \LL_4 &= \alpha_4(\tr{V_\mu V_\nu})^2 \label{L4}\\
  \LL_5 &= \alpha_5(\tr{V_\mu V^\mu})^2 \label{L5}\\
  \LL_6 &= \alpha_6 \tr{V_\mu V_\nu} \tr{TV^\mu} \tr{TV^\nu} \label{L6}\\
  \LL_7 &= \alpha_7 \tr{V_\mu V^\mu} \tr{TV_\nu} \tr{TV^\nu} \label{L7}\\
  \LL_8 &= {\tfrac14}\alpha_8 g^2 (\tr{T\vW_{\mu\nu}})^2\label{L8}\\
  \LL_9 &= {\tfrac{i}{2}}\alpha_9 g
                 \tr{T\vW_{\mu\nu}}\tr{T[V^\mu,V^\nu]} \label{L9}\\
  \LL_{10} &= \tfrac12\alpha_{10} (\tr{TV_\mu}\tr{TV_\nu})^2\label{L10}\\
  \LL_{11} &= \alpha_{11}g\epsilon^{\mu\nu\rho\lambda}
                \tr{TV_\mu}\tr{V_\nu \vW_{\rho\lambda}}\label{L11}
\end{align}
In this list, we have omitted terms which involve fermions or violate
CP invariance.

Although in the Higgs-less scenario the point $\alpha_i\equiv 0$ has
no special properties, the couplings $\alpha_1$ to $\alpha_{11}$ are
often referred to as \emph{anomalous couplings}. Their values are
expected to be of the order $v^2/\Lambda^2=1/16\pi^2$ or
larger~\cite{NDA}.  Two of them ($\alpha_1$ and $\alpha_8$) have been
constrained at LEP1.  Together with $\Delta\rho$ in~(\ref{chpt}), they
are equivalent to the set of electroweak precision observables
$S,T,U$~\cite{STU}.  Four additional parameters ($\alpha_{2,3,9,11}$)
modify the interactions of one transversal with two longitudinal gauge
bosons and thus are equivalent to the triple-gauge couplings
$\kappa_\gamma,\kappa_Z,g_{1}^Z,g_{5}^Z$~\cite{TGC}.  
In a strong-interaction scenario, these parameters are given by the
leading Taylor coefficients of the W/Z form factors.
Finally, the
parameters $\alpha_{4,5,6,7,10}$ describe independent deviations in
the four-point interactions of longitudinal gauge bosons and thus
determine the quasielastic $W$ and $Z$ scattering amplitudes.

\section{Beyond the threshold}

As stated in the Introduction, the extrapolated tree-level scattering
amplitudes of Goldstone bosons violate unitarity in the high-energy
range above $\q{1.2}{\TeV}$.  Depending on the actual values of the
$\alpha_i$ parameters, the limits may be lower or higher.  In any
case, the low-energy effective theory does not predict scattering
amplitudes beyond a certain threshold.

Given the fact that even in the well-known QCD case strong
interactions are poorly understood, we have no calculational methods
which would provide us with reliable quantitative predictions in that
range.  Therefore, phenomenological models are used to give results
that are at least internally consistent. By comparing different models
the resolving power of an experiment in a certain energy range can be
estimated.  If no new degrees of freedom are introduced, unitarity in
the $2\to2$ scattering channels of Goldstone bosons is a necessary
requirement.  Therefore, models are often based on the unitarisation
of extrapolated scattering amplitudes.

Since low-energy pion scattering (QCD) exhibits resonances in several
channels, it is reasonable to expect similar effects for new strong
interactions in the electroweak sector.  This is implemented by the
Pad\'e unitarisation model~\cite{Pade}, which is based on the
assumption that each scattering channel is dominated by a single
resonance (or dip) in the cross section.  By contrast, the $K$-matrix
model~\cite{Kmatrix} exhibits saturation of the Goldstone boson
scattering amplitudes without resonances.  Alternatives and
refinements of these scenarios are also considered in the studies
mentioned below.

It should be stressed that such models serve as test cases for
sensitivity estimates in a region where otherwise no prediction would
be possible.  Once real data are available, signals of new strong
interactions will be extracted from data without model assumptions.
Apart from effects in Goldstone scattering, models of strong
electroweak symmetry breaking typically predict additional new
particles with properties that can be studied in various
ways~\cite{TCmodels}.  However, since no particular model is preferred
(such as the MSSM in the weakly-interacting case), this is not easily
taken into account in a generic analysis.  In the studies discussed in
the present paper, no degrees of freedom beyond the known particles
are included.

\section{Processes at the LHC and the LC}

Since vector bosons are unstable particles, they are accessible only
by their couplings to fermion pairs and photons.  Thus, at tree level,
a process, which depends on massive four-boson interaction amplitudes,
typically involves the production of six fermions in the final state.
There are two cases:
\begin{enumerate}
\item
  Vector boson scattering:
  \begin{equation}\label{VBS}
    ff \to ff+V^*V^* \to ff+VV \to 6f
  \end{equation}
\item
  Three-boson production:
  \begin{equation}\label{TBP}
    ff \to V^* \to VVV \to 6f
  \end{equation}
\end{enumerate}
Furthermore, at one-loop order, four-boson interactions modify the
imaginary part of vector boson pair production (rescattering):
\begin{equation}
  ff \to V^* \to V^*V^* \to VV \to 4f
\end{equation}

In the analysis of the processes listed above, we can make use of the
generic properties of Goldstone-boson scattering amplitudes.  Since
the relative impact of the anomalous couplings rises with the energy of
the $2\to 2$ scattering, the strongest effect occurs at large pair
invariant masses of the vector bosons.  Using angular correlations of
the decay fermions, one can enhance the fraction of longitudinally
polarised vector bosons in the event sample.  Finally, forward
scattering of vector bosons is dominated by transversal vector boson
exchange which does not involve the symmetry-breaking sector, hence
large-angle scattering is most sensitive to anomalous four-boson
interactions.

\subsection{Vector-boson scattering at the LHC}

At the LHC, four channels of vector boson scattering~(\ref{VBS}) are
accessible:
\begin{align}
  pp &\to jj+W^+W^- \label{jjWW}\\
  pp &\to jj+W^\pm W^\pm \\
  pp &\to jj+ZZ \\
  pp &\to jj+W^\pm Z \label{jjWZ}
\end{align}

In each case, detection is easiest for leptonic decays of the vector
boson.  Obviously, $ZZ\to\ell^+\ell^-\ell^+\ell^-$ is the \emph{golden
channel} since the vector boson pair can be completely reconstructed
from the final state.  In the leptonic $W$ decay $W\to\ell\nu$, the
neutrino prohibits an unambiguous determination of the invariant mass
of the vector boson pair.  Semileptonic modes (one vector boson decays
hadronically, one leptonically) are more difficult due to the QCD
background, but have a larger branching ratio.  Finally, in the
fully hadronic modes the QCD background is a major problem.

A necessary ingredient for background reduction is the tagging of the
two extra jets in~(\ref{jjWW}--\ref{jjWZ}), which are found in the
forward region.  This is due to the fact that at high energies, where
the $W$ and $Z$ masses can be neglected, they are emitted from quarks
essentially on-shell by the splitting processes $q\to qZ$ and $q\to
q'W$ which are peaked in the collinear region.  Generically, the forward
jets retain a large fraction of the original parton energy, while the
transverse momentum is of the order $M_W/2$.

Since there are five independent anomalous couplings which affect the
four-boson interactions at subleading order, one would need at least
five measurements for their determination.  At the LHC, this cannot be
done using cross sections alone, since in the subprocesses
\begin{gather}
  W^+W^-/ZZ\to W^+W^- \quad\text{and}\quad W^+W^-/ZZ \to ZZ
\end{gather}
the initial state cannot be traced.

An obvious advantage of the LHC is the accessibility of the
high-energy range beyond $\q{1}{\TeV}$.  For a rough estimate, one can
regard vector boson scattering as the result of the splittings $p\to
q$ and $q\to W/Z$, which allows for an effective subprocess energy
up to about $\q{2}{\TeV}$.  Whether this range can actually be
exploited depends on the presence of resonances.  As a guideline, one
can take the searches for the Higgs boson in the high-mass region,
which for $M_H=\q{1}{\TeV}$ is merely a broad resonance in vector
boson scattering.  While narrow resonances are likely to be observed,
the measurement of a structureless scattering amplitude becomes
difficult in this range.

\subsection{Vector-boson scattering at the Linear Collider}

At an $e^+e^-$ collider, the relevant processes are
\begin{align}
  e^+e^- &\to \nu\bar\nu W^+W^- \quad\text{and}\quad \\
  e^+e^- &\to \nu\bar\nu ZZ,
\end{align}
which contain the subprocesses
\begin{align}
  W^+W^- \to W^+W^- \quad\text{and}\quad
  W^+W^- \to ZZ,
\end{align}
respectively.  The signatures are characterised by a large missing
invariant mass due to the two neutrinos.  The $W$ and $Z$ final states
can be detected and measured in all decay channels (except
$Z\to\nu\bar\nu$), hence no reduction due to low branching ratios applies.

The difficult backgrounds are
\begin{align}
  e^+e^- &\to e^\pm \nu W^\mp Z \quad\text{and}\quad 
  e^+e^- \to e^+e^- W^+W^-,
\end{align}
which are induced by photons and thus have a cross section enhanced by
large logarithms $\ln s/m_e^2$.  Clearly, in the hadronic modes the
separation of $W$ and $Z$ is important both for disentangling the
signals and for reducing the background.  To achieve this, the jet
pair invariant mass resolution has to be better than the mass
difference $M_Z-M_W$.  Furthermore, it is essential to detect and veto
forward-going electrons.

For the Linear Collider designs which are currently considered, the
energy is initially limited to about $\q{1}{\TeV}$. There is no
particular reason to expect a resonance structure in that range (with
the possible exception of a physical Higgs boson).  However, a precise
analysis of the rise of the scattering processes in the sub-threshold
region will significantly add to any information on the high-energy
region obtainable at LHC.

By measuring the cross sections of the two processes mentioned above,
two linear combinations of the five parameters $\alpha_{4,5,6,7,10}$
can be determined.  To further separate their contributions, one can
exploit angular distributions or investigate additional scattering
processes.  These are
\begin{align}
  e^+e^- &\to e^\pm \nu W^\mp Z, \\
  e^+e^- &\to e^+e^- W^+W^-,\\
  e^+e^- &\to e^+e^- ZZ,
\end{align}
which have the subprocesses
\begin{equation}
  W^\mp Z \to W^\mp Z, \quad
  ZZ \to W^+W^-, \quad\text{and}\quad
  ZZ \to ZZ,
\end{equation}
respectively.  To reduce the large contribution of photon-induced
irreducible background for the first two cases, the electron(s) in the
final state have to be observed away from the very forward region.
Unfortunately, the $eeZ$ coupling is small, so that $ZZ$-induced
processes have a cross section an order of magnitude smaller than
$WW$-induced ones.

Finally, the process
\begin{align}
  e^-e^- &\to \nu\nu W^-W^-
\end{align}
reveals an additional channel of $WW$ scattering, which requires
running the machine in $e^-e^-$ mode.

For all $W$-induced processes, the signal can be further enhanced by a
factor of up to four if the initial beams are polarised.




\subsection{Three-boson production}

While the vector-boson scattering processes discussed above have a
cross section that logarithmically increases with energy, the cross
section of three-boson production processes falls off like $1/s$.
This limits their usefulness to subprocess energies in the lower range
where the cross section of the fusion processes is still small.

At the LHC, four channels are present:
\begin{align}
  pp &\to qq\to W^+W^-W^\pm,\quad W^+W^-Z,\quad W^\pm ZZ,\quad ZZZ,
\end{align}
where the latter channel is additionally suppressed,
cf.~(\ref{LET-zzzz}).  At the Linear Collider there are only two
processes,
\begin{align}
  e^+e^- &\to W^+W^-Z, \quad ZZZ.
\end{align}
Note that the Higgs-strahlung mode of Higgs boson production with
$H\to W^+W^-$ or $H\to ZZ$ decay is a special case of this, where a
low-lying resonance (the Higgs boson) is present in the spectrum.

\subsection{Rescattering}

The process
\begin{equation}
  e^+e^- \to W^+W^- \to 4f
\end{equation}
can be investigated with great precision at a Linear Collider.
Therefore, it is conceivable to extract the imaginary part of the
amplitude, which is the result of the rescattering $W^+W^-\to W^+W^-$.
To leading order, this imaginary part can be derived
from~(\ref{LET-wwww}), while at subleading order anomalous
contributions come into play.  In this process, $WW$ rescattering takes
place at the full collider energy.

It should be noted that the real part of the $WW$ pair production
amplitude is also a sensitive probe of a strongly-interacting symmetry
breaking sector.  By making full use of polarisation and angular
distributions at a Linear Collider, the real and imaginary parts of
all individual form factors can be disentangled~\cite{Diehl}, which in
the present context amounts to a precise measurement of the parameters
$\alpha_{2,3,9,11}$.  Together with results on vector boson pair
production at the LHC, these data have to be taken as an input for the
analysis of vector boson scattering both at LHC and at the LC.  This
is important since the parameters $\alpha_{3,9,11}$ also affect the
quartic vector boson couplings and therefore must be known if an
unambiguous determination of all independent parameters in the
low-energy effective Lagrangian is intended.

\section{Results}
\subsection{Approximations}

Several studies have been done in the past, both for the LC and the LHC.
Since vector-boson scattering is embedded in complicated six-fermion
processes where for a given final state thousands of Feynman graphs
contribute, approximations which simplify the problem are welcome.
Many of the studies listed below use one or more of the following:
\begin{enumerate}
\item
  Equivalence Theorem (ET)~\cite{ET}.  The scattering of longitudinal
  vector bosons is replaced by the scattering of Goldstone bosons.
  This is important for the implementation of unitarisation models in
  the high-energy region, and it is justified since parametrically
  the corrections (partly due to transversal gauge bosons) are
  suppressed by factors of $M_W^2/\hat s$ and $g^2$.  However, this
  suppression is typically overcome by the multiplicity of transversal
  polarisation directions and the larger structure functions of
  transversal gauge bosons.
\item
  Effective $W$ approximation (EWA)~\cite{EWA}.  The process is
  separated into the splitting of the initial fermions into vector
  bosons and the subsequent scattering, i.e.
  \begin{equation}
    f\to f'+V \quad\text{and}\quad VV\to VV.
  \end{equation}
  This approach introduces parton distribution functions for the $W$
  and $Z$ vector bosons, which differ for longitudinal and for
  transversal polarisation.  It allows us to estimate event rates when
  the ET is used for evaluating the amplitudes.  The EWA is formally
  valid up to corrections of order $M_W^2/\hat s$ and $g^2$, where
  $\hat s$ is the invariant mass squared of the $W$ pair, and $g$ is
  the weak coupling.  Unfortunately, neither $M_W$ nor $g$ are small
  enough for this approximation to be quantitatively reliable in
  realistic applications.  This is mainly due to the large number of
  numerically important Feynman diagrams which are neglected in this
  approach.  Another problem with the EWA is the transverse momentum
  of the $W$ pair which is set to zero.  This kinematic variable is
  essential for background reduction.
\item
  Narrow-width approximation for the vector bosons in the final
  state.  The process is separated into the on-shell production of
  vector bosons and their decay, i.e.
  \begin{equation}
    ff \to ffVV \quad\text{and}\quad V\to ff.
  \end{equation}
  This approximation is justified since non-resonant electroweak
  six-fermion contributions are typically very small in the
  kinematical range of interest.  Since the angular correlations of
  fermions are important, polarisation information should be carried
  through.  However, for technical reasons this information is often
  discarded.  For a realistic simulation, the $V$ mass ($W$ or $Z$)
  must be smeared by hand, which introduces an arbitrariness in the
  results.
\item
  Leading-order approximation.  In fact, complete next-to-leading
  order (NLO) results, which will be needed for precision studies of
  real data, are not available for any of the processes we are
  discussing here.  Only for on-shell $W$ scattering the NLO
  corrections ($O(g^2/16\pi^2)$) are known~\cite{Denner}.  However,
  these results have to be embedded in the narrow-width and effective
  $W$ approximations, where other $O(g^2)$ contributions are
  neglected.  The only effect that can easily be included is the
  leading-logarithmic scale dependence of the anomalous couplings,
  which is proportional to $v^2/\Lambda^2=1/16\pi^2$, not suppressed
  by $g^2$.
\end{enumerate}
While these approximations are quite useful for studying the overall
properties of the processes under consideration and for developing
experimental strategies, for detailed numerical comparisons and, in
particular, for the analysis of real data they are insufficient.  The
use of complete calculations in Monte-Carlo generators has only begun,
and on the experimental side the inclusion of detector effects is
essential as well.  Since many of the studies lack
this level of sophistication, numerical results should be compared or
combined with care.
\subsection{Existing results}
A variety of studies has been done for LC 
\cite{Phillips,Kilian1,Kilian2,Kilian3,Rosati,menges,Diehl} and LHC
\cite{Bagger1,Bagger2,Urdiales,ATLAS-TDR,Eboli,Forshaw,sec3_SLHC}.
For the TGCs the situation is relatively clear. A fast-simulation
study basically without approximations exists which has shown that the
coupling parameters belonging to the dimension four operators
($g_1^Z, \kappa_\gamma, \kappa_Z$) can be measured at $\sqrt{s} = 800 \, \GeV$
about an order of magnitude better than at LHC, resulting in $1 \sigma$ 
errors of $\Delta \alpha_2 = 0.0004$ and  $\Delta \alpha_3 = 0.0002$
(see figure \ref{fig:a23lc}) \cite{menges,tdr_phys}.
The correlations with the dimension-six operators 
$\lambda_\gamma, \lambda_Z$ are small, making the result robust.
\begin{figure}[htbp]
\begin{center}
\includegraphics[height=10.cm]{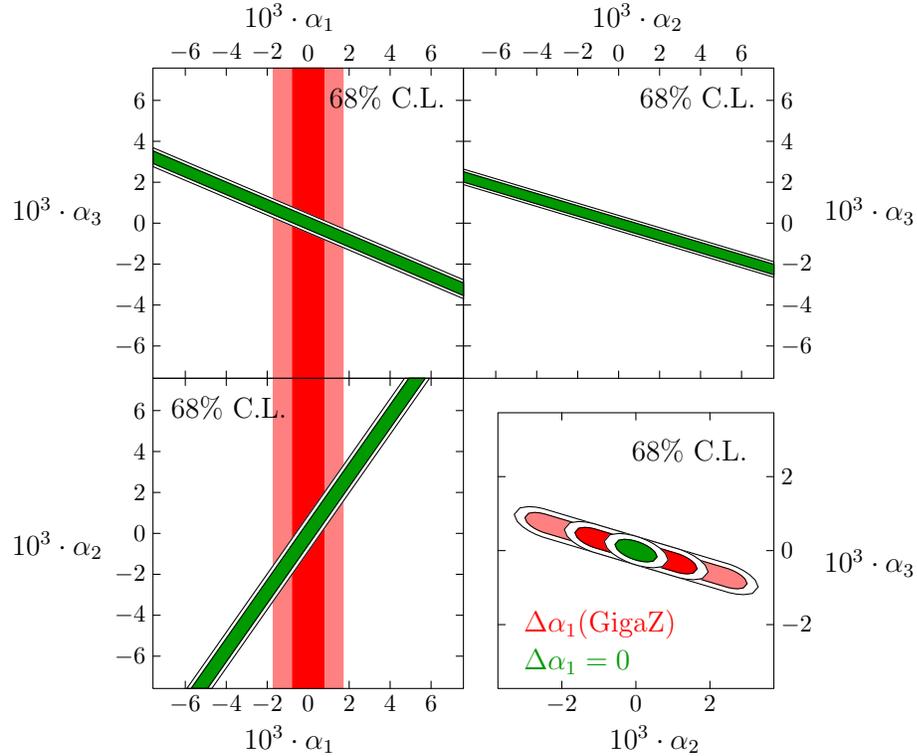}
\end{center}
\caption[Sensitivity to the effective couplings $\alpha_{1,2,3}$ from the
measurement of the triple gauge couplings at a linear collider and from
the Z-pole precision data.]
{
Sensitivity to the effective couplings $\alpha_{1,2,3}$ from the
measurement of the triple gauge couplings at a linear collider and from
the Z-pole precision data. 
For the Z-pole constraint the outer region is without and the inner region 
with the accurate $m_W$ measurement.
In the lower right plot the $\alpha_2-\alpha_3$ plane with $\alpha_1=0$ 
or with the Z-pole constraint is shown.
}
\label{fig:a23lc} 
\end{figure}

The situation is more complicated for the quartic couplings. A
complete six-fermion fast simulation study using $e^+e^-\to W^+W^- \nu\nu$
and $e^+e^-\to ZZ\nu\nu$ at $\sqrt{s} = 800 \, \GeV$ shows that 
$\alpha_4, \alpha_5$
can be measured at the linear collider 
with a precision of $0.005$ and $0.03$ assuming
$SU(2)_c$ invariance. The results of the two subprocesses are shown in
figure \ref{fig:a4a5lc} \cite{Rosati}. 
This seems to be somewhat better than what is
possible at LHC (see figure \ref{fig:a4a5lhc}) \cite{Eboli}.
The situation gets more complicated if also the $SU(2)_c$ violating couplings
$\alpha_{6,7,10}$ are allowed to differ from zero.
The process $e^+e^-\to W^+W^- \nu\nu$ contains only the subprocess 
$W^+ W^- \to W^+ W^-$ which is sensitive to $\alpha_4$ and $\alpha_5$ while
$e^+e^-\to ZZ\nu\nu$ contains only $W^+W^- \to ZZ$, 
sensitive to $\alpha_4+\alpha_6$ and $\alpha_5 +\alpha_7$.
The $\alpha_4-\alpha_5$ contour from $e^+e^-\to W^+W^- \nu\nu$ only
thus remains valid while the one from $e^+e^-\to ZZ\nu\nu$ had to be
understood as a contour in $(\alpha_4+\alpha_6)\,-\,(\alpha_5 +\alpha_7)$.

At the LHC it cannot be decided on an event by event basis if a
process is induced by $W^+ W^-$ or by $ZZ$, so that in general all
processes are sensitive to all couplings. 
This makes the LC superior to the LHC for these two processes, even if
the LHC can reach higher centre of mass energies.
However, further information can be obtained from 
$W^\pm W^\pm \rightarrow W^\pm W^\pm$ and $ZZ \rightarrow ZZ$.
The first process is sensitive to $\alpha_4$ and $\alpha_5$ only,
however with a different correlation than $W^+ W^- \to W^+ W^-$. At LC
this process can only be assessed in $e^-e^-$ running which needs
additional running time with a factor three lower luminosity where not
much else can be done. At the LHC this process is even preferred at
large $x_{Bj}$ where valence quarks dominate and easy to select with a
same sign dilepton pair in the final state.
$ZZ \rightarrow ZZ$ is sensitive to all $\alpha_i$. At LC it is
strongly suppressed due to the small Z-lepton coupling. At LHC with its
quark initial states this suppression is much weaker and $ZZ$ with at
least one leptonically decaying Z is easy to detect and the full
kinematics can be reconstructed.

With the present studies it is not possible to combine the estimated results
for LC and LHC. 
However, it has already
become clear that both colliders are able to probe quartic anomalous
couplings down to their expected values in the percent range, and
that combining the two colliders significantly reduces
the correlations between the $\alpha$ parameters.
Also the precise measurement of the triple couplings at the LC will improve
the interpretation of the quartic couplings at the LHC.

\begin{figure}[htbp]
\begin{center}
  \includegraphics[width=0.48\textwidth]{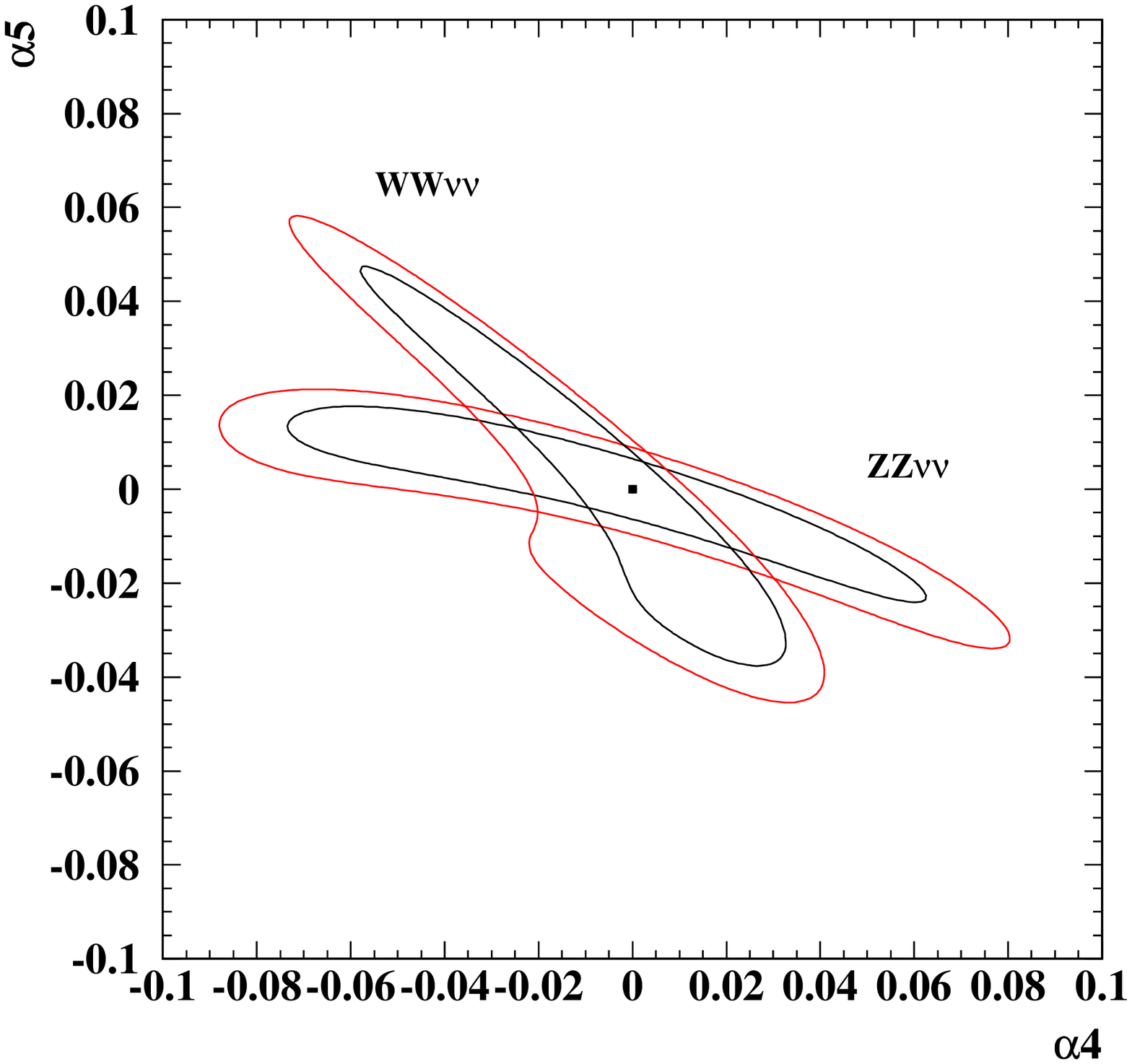}
  \includegraphics[width=0.48\textwidth]{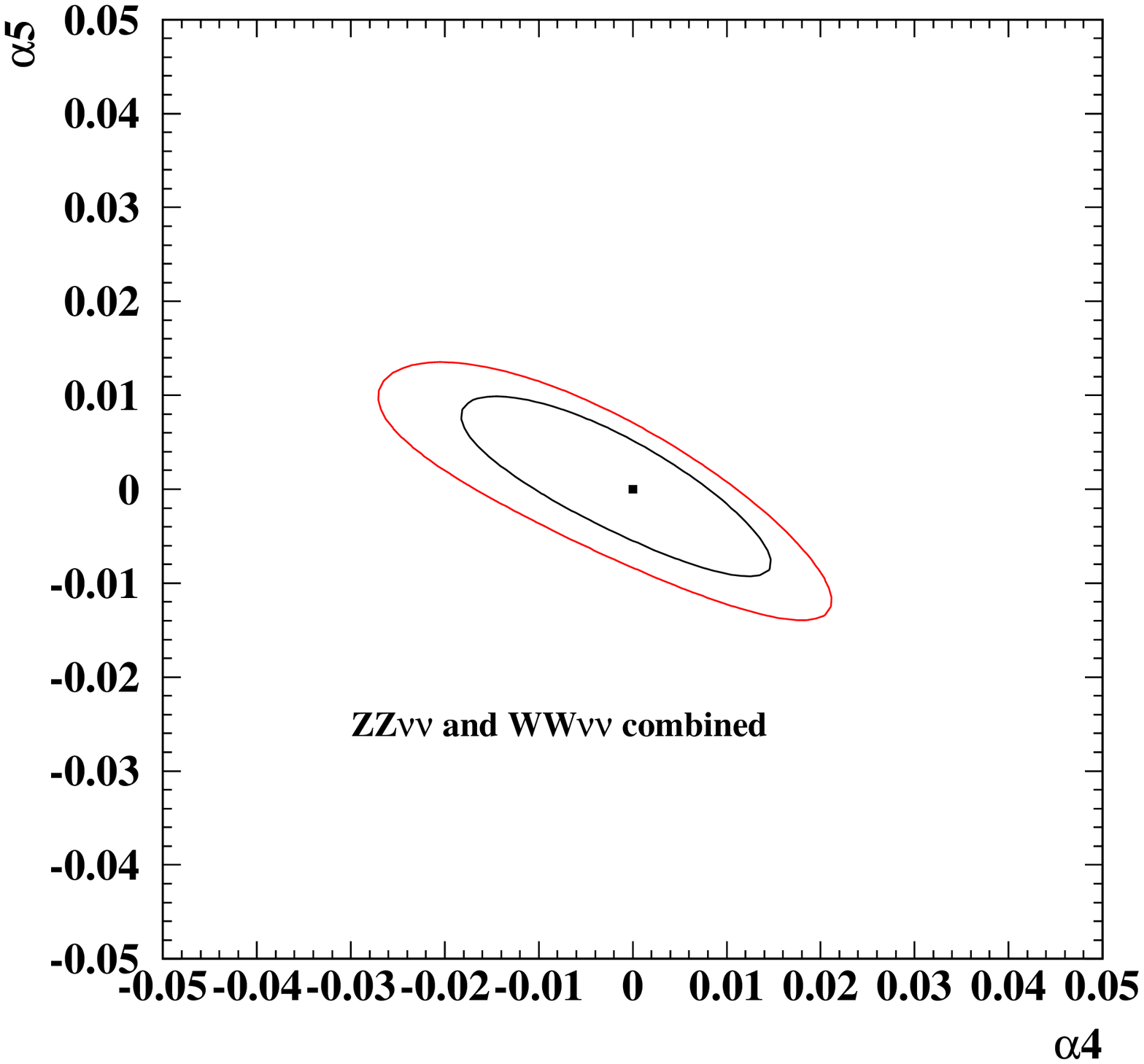}
\end{center}
\caption{
Sensitivity to $\alpha_4$ and $\alpha_5$ at a linear collider with
$\sqrt{s}=800\,\GeV$ from
$e^+ e^- \rightarrow \nu \nu W^+ W^-$ and $e^+ e^- \rightarrow \nu \nu
{ZZ}$. The inner and outer contours represent 68\% and 90\% c.l.
}
\label{fig:a4a5lc}
\end{figure}
\begin{figure}[htbp]
\begin{center}
  \includegraphics[width=0.5\textwidth,bb=16 28 492 546]{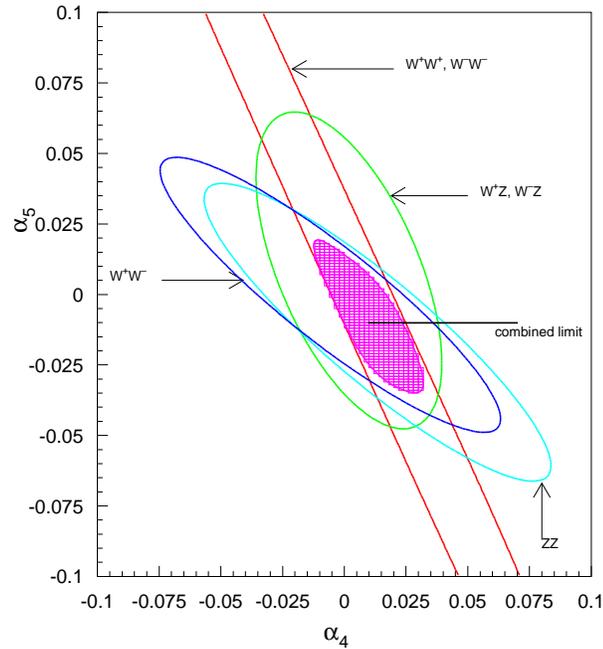}
 \end{center}
\caption{
Sensitivity to $\alpha_4$ and $\alpha_5$ (90\% c.l.) at LHC.
}
\label{fig:a4a5lhc}
\end{figure}

\subsection{Ongoing studies}

New studies are currently under way both for the LHC and the LC.
These studies are performed with contacts between the different groups from
the beginning so that a LC-LHC combination of the results should be possible.

\subsubsection{LHC}

In the CMS collaboration, a new analysis of strong $WW$ scattering is
under way~\cite{Arneodo}.  This study will use a Monte Carlo generator
which simulates all six-fermion final states in $pp$ interactions,
currently being developed in Torino~\cite{Ballestrero}.  Events will
be processed through the GEANT simulation of the CMS apparatus, and
then passed through the online selection and reconstruction programs.

While the six-fermion final state generator is being developed, the
group is currently conducting an exploratory study of $WW$ scattering
in the semileptonic mode $WW\to \mu\nu jj$.  The simulation of the
signal and backgrounds ($t\bar tX$, $Wjj$, $WX$) is based on the
PYTHIA~\cite{sec3_PYTHIA} and CompHEP~\cite{CompHEP} generators, along with
the fast simulation package of CMS~\cite{CMSJET}.  Requiring a central
high-$p_T$ muon, two central jets compatible with the $W$ mass, two
forward jets, and two rapidity gaps, the preliminary result is an
efficiency of about 15\,\% for the signal and of less than 0.2\,\% for
the backgrounds considered. A detailed investigation of the signal to
background ratio is in progress.

In the ATLAS collaboration, it is planned to extend the parton-level
study of Butterworth et al.\ \cite{Forshaw} by an experimental analysis
based on the ATLAS detector simulation.

\subsubsection{Linear Collider}

Two new studies of $WW$ scattering at the Linear Collider~\cite{Osorio,Predrag}
proceeds along similar lines.
In addition to the backgrounds considered
in~\cite{Rosati}, the $\gamma\gamma$ production of top and $W$ pairs
is taken into account. Also the processes 
$e^- e^- \rightarrow W^- W^-$, $e^+ e^- \rightarrow ZZ e^+ e^-$
and $e^+ e^- \rightarrow WZ e \nu$ will be considered.

Also a study of $WWZ$ and $ZZZ$ production in $e^+e^-$ has
started \cite{beyer}. 
Also this study will use Whizard \cite{WHIZARD} and the fast
simulation program SIMDET \cite{sec3_SIMDET}. Both studies will be combined
in a common fit for $\alpha_{4,5,6,7,10}$.

\section{Scenarios where the LHC sees resonances}

A LC can help interpret resonances that are seen in vector-boson
scattering at the LHC.  The LHC should be able to detect directly 
scalar and vector resonances with masses up to about $2\,\TeV$
\cite{Forshaw}.  Vector resonances with masses in this
range induce anomalous values for $\alpha_2$ and $\alpha_3$ which can
readily be measured at a $500\,\GeV$ LC\cite{Abe:2001nq}.  In analogy with
the pair-production of charged pions in $e^+e^-$ annihilation near the
$\rho$ resonance, a form factor $F_T$ can be used to describe the
effects of vector resonances on the pair production of longitudinally
polarised $W$ bosons in $e^+e^-$ annihilation.  The real part of
$F_T-1$ is proportional to $\alpha_2+\alpha_3$, and is related well
below threshold to the masses of the vector resonances by\cite{Barklow:2001is}
\begin{equation}
      F_T=1+s\sum_k {\frac{a_k}{ M_k^2}}\, ,
\end{equation}
where $\sqrt{s}$ is the $e^+e-$ centre of mass
energy, $M_k$ are the masses of the vector resonances ordered by 
$M_1<M_2<...$, and $a_k$ are their relative
weights with $\sum a_k=1$.  In ordinary QCD with vector meson dominance,
$a_1\approx 1$.

A $500\,\GeV$ LC could measure the mass of a  $1\,\TeV\ (1.5\,\TeV)$ vector
resonance with an accuracy of $,\GeV$ ($27\,\GeV$) by measuring
$\alpha_2+\alpha_3$ under the assumption $a_1=1$.  A measurement of $\alpha_2+\alpha_3$ which was
consistent with the mass of the LHC resonance would provide evidence
for spin-1.  By measuring both the real and imaginary parts
of $F_T$ a $500\,\GeV$ LC would also provide a 
measurement of the vector resonance width $\Gamma_1$ with an error of $19\,\GeV\ (90\,\GeV)$
for a $1\,\TeV\ (1.5\,\TeV)$ vector
resonance.

It is also possible to drop the assumption $a_1=1$ and measure $a_1$ directly.
Measurements  of the real and imaginary parts of $F_T$ at two different
centre of mass energies can be used to extract $a_1$ along with the mass and width of the first vector resonance, assuming 
$M_1\ll M_2$.  With  $250\,\fb^{-1}$ luminosity at
$\sqrt{s}=0.35\,\TeV$ and $250\,\fb^{-1}$ luminosity at $\sqrt{s}=0.5\,\TeV$, 
the mass, width and weight parameter $a_1$ of a 1~TeV vector resonance 
can be measured with accuracies of 101~GeV, 26~GeV, and 26\%, respectively.

At $\sqrt{s}=1\,\TeV$ a LC would be sitting on top of a  $1 \,\TeV$ vector 
resonance 
and would measure the mass and width with a statistical accuracy of 0.08~GeV
and 0.02~GeV, respectively, assuming $a_1=1$ and a luminosity of
$1000\,\fb^{-1}$.  Again, $a_1$ can be measured by combining results from two
different centre of mass energies. Assuming $500\,\fb^{-1}$ luminosity at
$\sqrt{s}=0.5\,\TeV$ and $1000\,\fb^{-1}$ at $\sqrt{s}=1\,\TeV$, the mass,
width, and weight parameter $a_1$ of a 1~TeV vector resonance can be measured
with accuracies of 0.08~GeV, 0.8~GeV, and 1.4\%, respectively.  If the vector
resonance has a mass of 1.5~TeV the same three parameters can be measured With
accuracies of 44~GeV, 16~GeV, and 10\%, respectively, assuming the same
combination of LC energies and luminosities.

If the resonance which was being produced at the LHC were a scalar,
then the $500\,\GeV$ LC would measure a value for $\alpha_2+\alpha_3$
which was inconsistent with vector resonance production. Hence the
$500\,\GeV$ LC result could rule out a spin-1 interpretation.  A
$0.8-1.0\,\TeV$ LC could also estimate the mass of a scalar resonance
well above $1\,\TeV$ by measuring the enhancement in the vector-boson
scattering cross-section due to a scalar resonance.  The leading order
correction to the Low Energy Theorem (LET) cross-section $\sigma_{\rm
LET}$ for $WW$scattering due to a scalar resonance of mass $M_0$ is given by 
\[
\sigma(M_0)=\left(1+\frac{8}{3}\frac{s}{M_0^2}\right)\sigma_{\rm LET}
\]
where $\sqrt{s}$ is the $WW$ centre of mass energy\cite{Barklow:1997nf}.
With $1\,\ab^{-1}$ at $\sqrt{s}=1\,\TeV$ for example, a LC could measure
the mass of a $1.5\,\TeV$ scalar resonance with a 17\% accuracy
\cite{Barklow:2001is}.

\begin{figure}[htb!]
\begin{center}
\epsfig{file=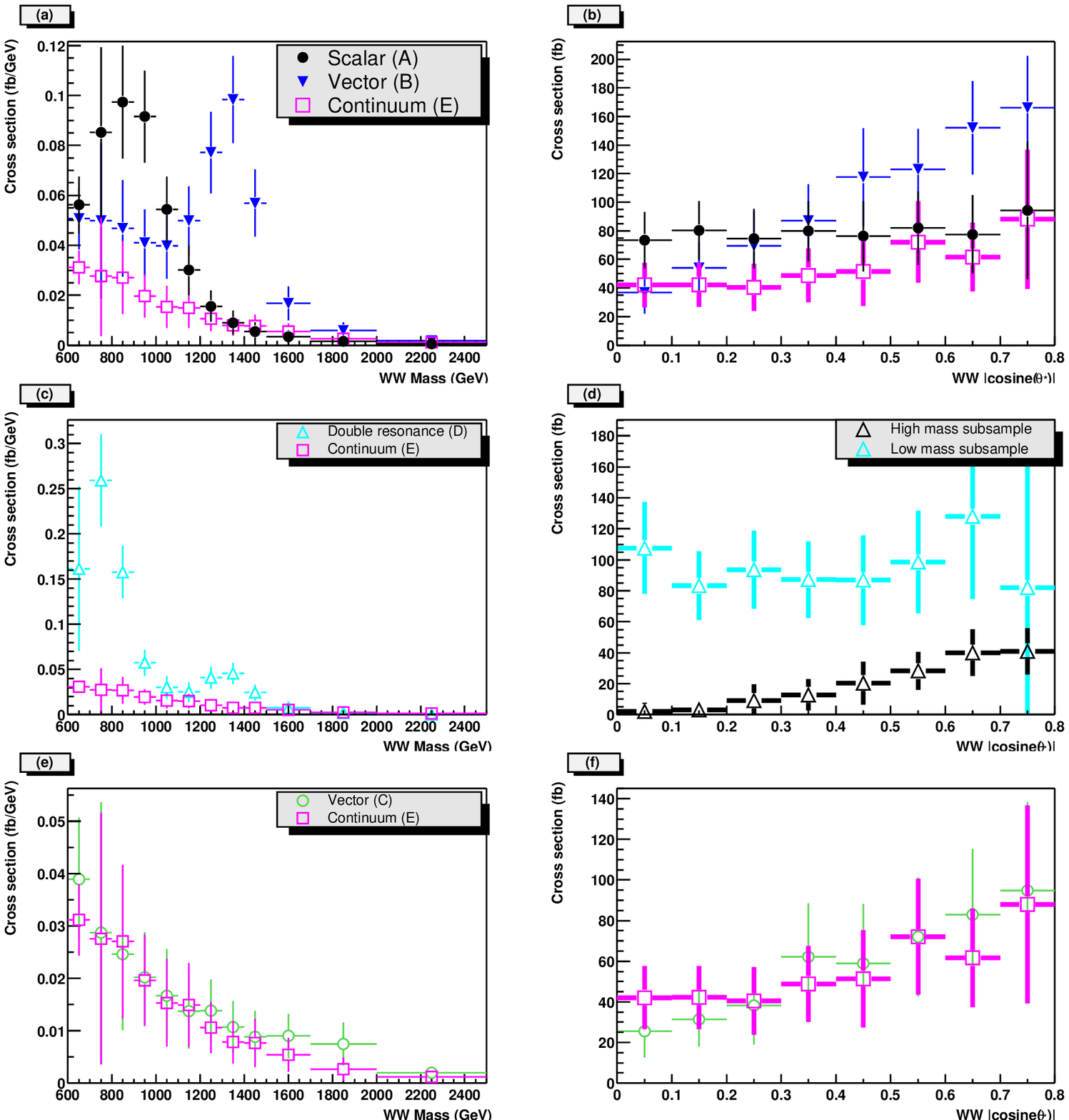,width=15.5cm,clip} 
\end{center}
\caption{Differential cross section measurements at LHC assuming
$100 \,\fb^{-1}$ of luminosity and $\sqrt{s}=14$~TeV: (left)
$d\sigma /dM_{WW}$ and (right) $d\sigma /d|\cos \theta ^{*}|$. 
The green circles are measurements assuming a single
1.9 TeV vector resonance, while the red squares are measurements assuming a
model without resonances.}
\label{sec3fig:res1900}
\end{figure}

It is instructive to study in more detail a specific example.
Figure~\ref{sec3fig:res1900} shows the precision with which the LHC
can measure the $WW$ scattering differential cross sections $d\sigma
/dM_{WW}$ and $d\sigma /d|\cos \theta ^{*}|$, where $M_{WW}$ is the
$WW$ mass and $\cos \theta ^{*}$ is the cosine of the $W$ scattering
angle in the $WW$ rest frame~\cite{Forshaw}.  Results for two $WW$
scattering models are shown: a model with a single 1.9~TeV vector
resonance and a model with no resonances.
The precision with which the LHC can measure the mass, width, and peak
cross section for a 1.9~TeV vector resonance can be obtained by
fitting the $d\sigma /dM_{WW}$ distribution of
Figure~\ref{sec3fig:res1900} to the sum of a polynomial and a Gaussian
function.  In order to obtain a lower bound on resonance parameter
precision, it is assumed that the polynomial is fixed by sideband
measurements, and that the only unknowns are the height, position, and
width of the Gaussian function.  The height of the Gaussian will be
proportional to $a_1^2$, so that the Gaussian fit can be parameterised
in terms of $a_1$, $M_1$, and $\Gamma_1$.  The wide tilted green
ellipse in the right hand side of Figure~\ref{fig:ftandellipses} is
the projection of the 3-dimensional covariance ellipse for such a fit
projected onto the $M_1\ - \ a_1$ plane.
If the constraint $a_1=1$ is imposed, the 1$\sigma$ error band on $M_1$ 
is given by the tall vertical blue band at $a_1=1$. 

\begin{figure}[htb!] 
   \begin{tabular}{lr}
     \includegraphics[angle=-90,origin=c,width=7.6cm,clip=]{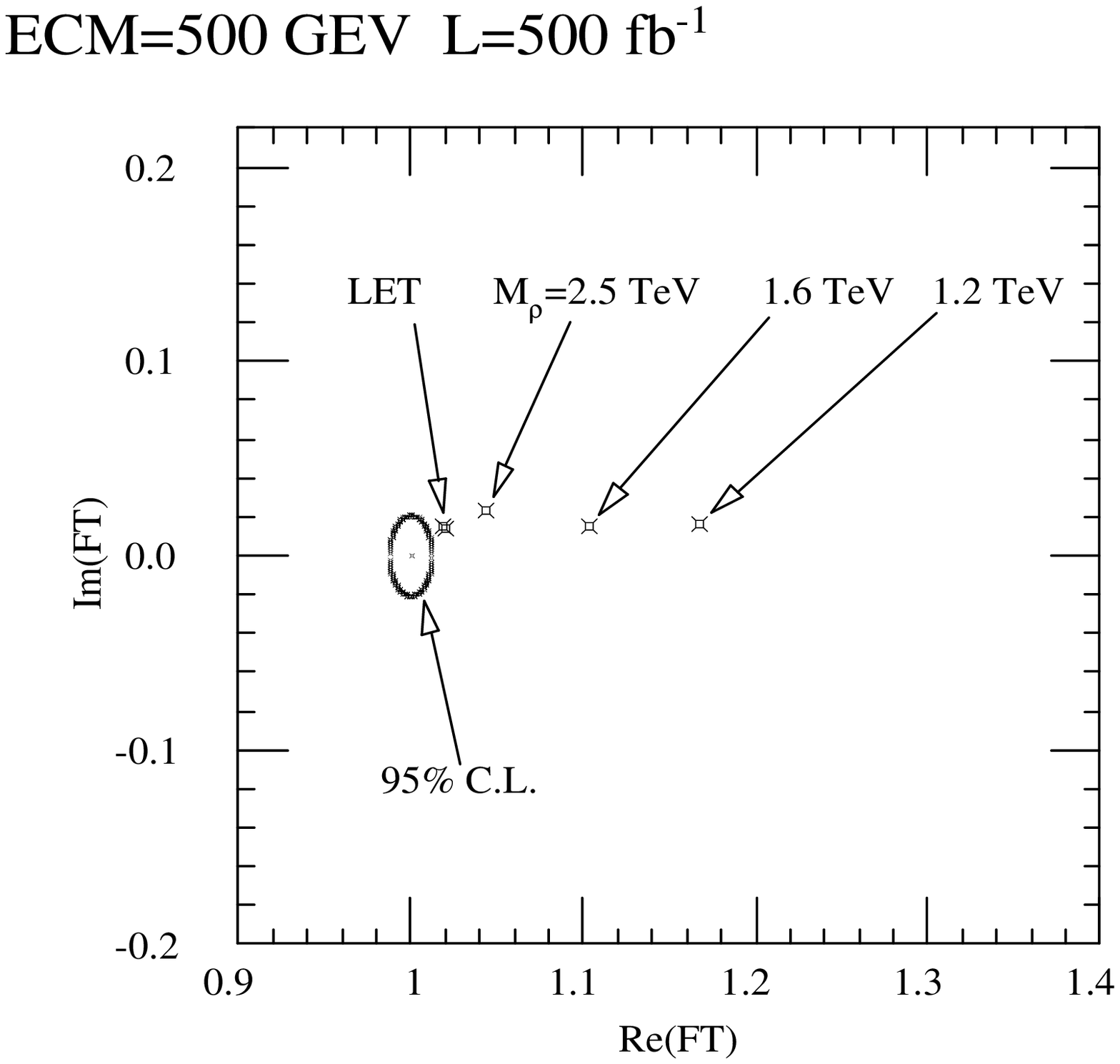}&
    \includegraphics[clip=,width=8cm]{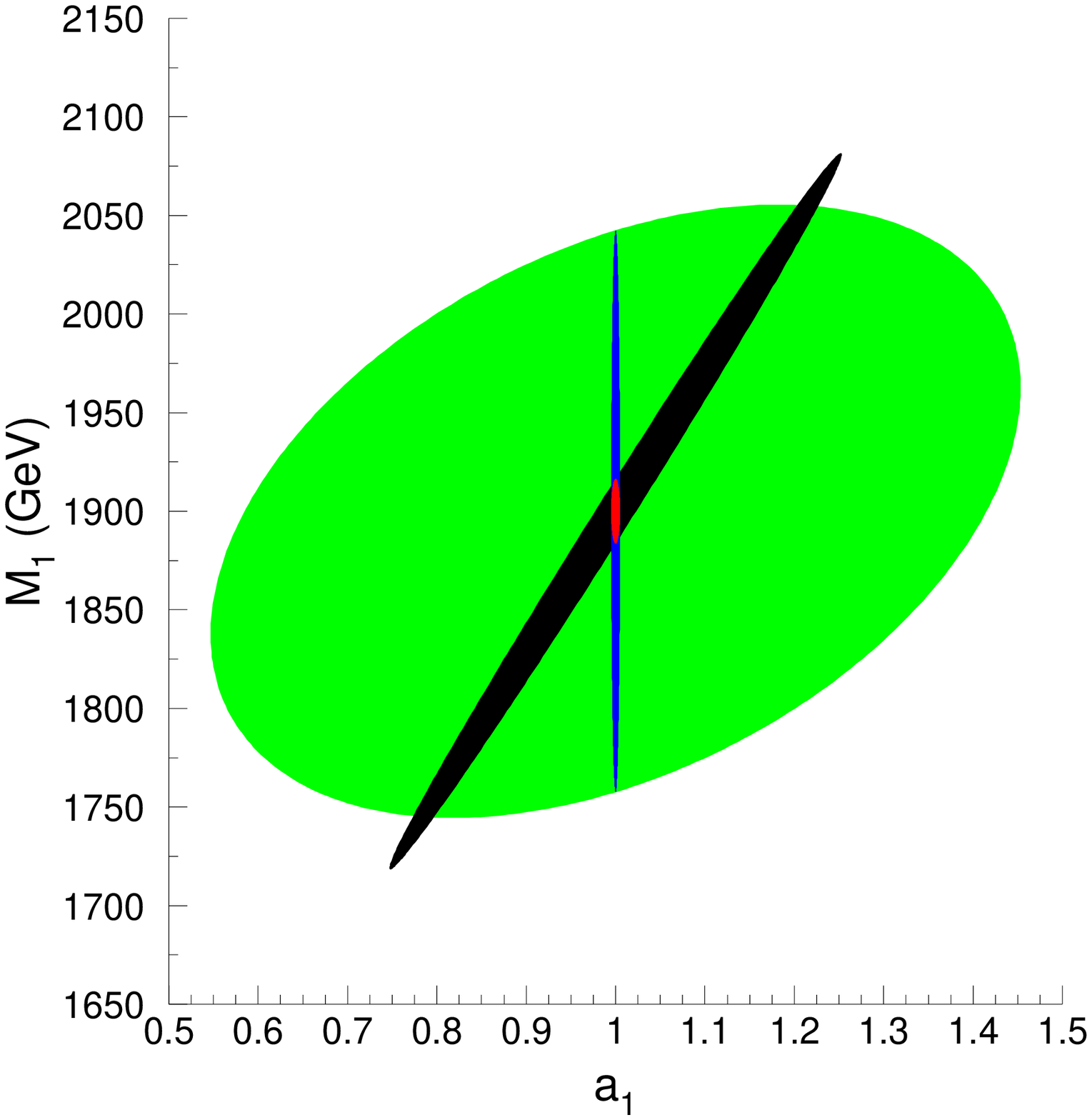}
   \end{tabular}
\caption{Left: The 95\% C.L. contour for $F_T$ measured at a LC with 
$\sqrt{s}=500$~GeV
and $500\,\fb^{-1}$, along with values of $F_T$ for various vector resonance
masses, assuming $a_1=1$ .
Right: Three dimensional covariance ellipses for
fits of $a_1,M_1,\Gamma_1$ projected onto the $M_1\ -\ a_1$ plane.  The wide
tilted green ellipse (narrow tilted black ellipse) is obtained from LHC (LC)
measurements.  If it is assumed that $a_1=1$ then the 1$\sigma$ error for
$M_1$ from the LHC (LC) is given by the tall blue (short red) vertical band at
$a_1=1$.}
\label{fig:ftandellipses}
\end{figure}

The LC can help measure the properties of a 1.9~TeV vector resonance.
The left hand side of Figure~\ref{fig:ftandellipses} shows the
accuracy with which a $\sqrt{s}=500$~GeV LC with $500\,\fb^{-1}$ can
measure the form factor $F_T$, along with the predicted values of
$F_T$ for various vector resonance masses assuming $a_1=1$.  The
signal from a 1.9~TeV vector resonance would have a significance of
15$\sigma$, and so a $\sqrt{s}=500$~GeV LC would establish the
presence of a vector resonance.  A three-parameter fit of $a_1$,
$M_1$, and $\Gamma_1$ can be obtained by combining the LC measurement
of $F_T$ at $\sqrt{s}=500$~GeV with a LC measurement of $F_T$ at
$\sqrt{s}=1000$~GeV assuming a luminosity of $1000\,\fb^{-1}$.  The
projection of the covariance ellipse for this fit onto the $M_1\ - \
a_1$ plane is given by the narrow tilted black ellipse in the right
hand side of Figure~\ref{fig:ftandellipses}.  If the constraint
$a_1=1$ is imposed then the 1$\sigma$ error band on $M_1$ from LC
measurements is given by the short vertical red band at $a_1=1$.

The LHC is sensitive to vector resonances up to a mass of $1.5\,\GeV$
\cite{ATLAS-TDR}. The
LC can distinguish the Standard Model from the Low Energy Theorem
(LET) prediction with 3.7 standard deviations at $\sqrt{s} = 500 \, \GeV$
(${\cal L} =  500 \fb^{-1}$) and with 10.7 standard
deviations at $\sqrt{s} = 1000 \, \GeV$ (${\cal L} =  1000 \, \fb^{-1}$).
The LET prediction is given by the limit of infinite mass. The mass
sensitivity is $2.5 \, \TeV$ and $4.1 \, \TeV$  for $\sqrt{s} = 500 \, \GeV$
and $\sqrt{s} = 1000 \, \GeV$, where the sensitivity is defines as
the mass which can be separated with $3 \sigma$ from the LET.

\section{Conclusions}

While it is tempting to numerically combine the results of the
existing LHC and LC studies of strong $WW$ scattering, this would be
premature due to the mixed status of the theoretical and experimental
approximations.  For a meaningful comparison, it is essential that
full six-fermion matrix elements are consistently used together with
realistic detector simulations.  In particular, the effective-$W$
approximation which has been adopted for most of the LHC studies
misses the electroweak radiation of transversally polarised $W$ and
$Z$ bosons, which constitutes a numerically important background that
needs to be taken into account.

Nevertheless, one can already draw the conclusion that in the
low-energy range it will possible to measure anomalous couplings down
to the natural scale $1/16\pi^2$.  A combination of LC and LHC data
will considerably increase the resolving power of the LHC.
Furthermore, the direct sensitivity of the LHC to resonances in the
range above $\q{1}{\TeV}$ can be fully exploited only if LC data on
the cross section rise in the sub-TeV region are available.




\chapter{Supersymmetric Models}
\label{chapter:susy}

Editors: {\it K.~Desch, K.~Kawagoe, M.M.~Nojiri, G.~Polesello}

\section{Measurement of supersymmetric particle masses, mixings and
couplings at LHC and LC}
\label{sec:41}

The precise measurement of the masses of the largest possible set
of supersymmetric particles is the most important input to the reconstruction
of the supersymmetric theory, in particular of the SUSY breaking mechanism.
At the LHC, the dominant production mechanism is pair production of gluinos
or squarks and associated production of a gluino and a squark. For these
processes, SUSY particle
masses have to be calculated from the reconstruction of long decay chains
which end in the LSP. Invariant masses can only be estimated from the
endpoints (edges) of invariant mass spectra. In particular the LSP mass
is only slightly constrained. This uncertainty propagates into the errors
of the heavier SUSY particle masses.

At a LC, the colour--neutral part of the SUSY particle spectrum 
can be reconstructed
with high precision if it is kinematically accessible. In particular, the
LSP mass can be reconstructed with a precision 
significantly better than 1~GeV. If this 
and further LC measurements are input into the LHC mass fits, significant
improvement on the masses which can only be accessed by the LHC (in particular
squarks and gluinos) can be achieved.

The following studies focus on an analysis of the SUSY benchmark point SPS 1a,
a typical mSugra scenario, but further model points are investigated as well
in Section~\ref{sec:412}. 
The main features of the SPS~1a benchmark scenario are summarised in
Section~\ref{sec:4100} for later reference. 
Section~\ref{sec:411} outlines simulations for the
reconstruction of a large set of SUSY particles in ATLAS. Section~\ref{sec:412}
describes the reconstruction of squarks and gluinos in CMS. 
In Section~\ref{sec:410} SUSY 
searches at the LC are discussed and the experimental accuracies achievable 
in the SPS~1a scenario are analysed.
The
influence of LC mass measurements on the ATLAS and CMS analyses is 
discussed in detail in Section~\ref{sec:414a}. 
In Section~\ref{sec:414} it is demonstrated that LC predictions can be
crucial for guiding the LHC search for the heaviest neutralino. 
In this way the heaviest neutralino can be identified at the LHC
and its mass can be measured with high precision. Feeding this
information back into the LC analysis leads to an improved accuracy in
the determination of the SUSY parameters in the neutralino and chargino
sector.
Section~\ref{sec:413} focusses on stop and sbottom reconstruction at the
LHC using input from the LC. 

So far only few detailed case studies for SUSY parameter determinations
at LHC and LC exist. Further LC information such as decay branching
ratios is likely to be helpful for an extraction of SUSY particle
couplings at the LHC. While the studies performed so far have mostly
been restricted to the SPS~1a benchmark point,
which is a favourable scenario both for LHC and LC, examples of other
possible scenarios should be investigated as well. It can be expected
that also in this case important synergy effects will arise from the 
LHC / LC interplay.

\subsection{\label{sec:4100}
The SPS~1a benchmark scenario}

{\it H.-U.~Martyn and G.~Weiglein}

\newcommand{\Mgl}{M_{\tilde{g}}}

\def\isajet  {{\sc Isajet}}
\newcommand{\fig}[1]{fig.~\ref{#1}}
\newcommand{\tab}[1]{table~\ref{#1}}
\def\lc     {{\sc Lc}}
\def\lhc    {{\sc Lhc}}
\newcommand{\hdick}{\noalign{\hrule height1.4pt}}
\def\ti    {\tilde}
\def\sf    {{\ti f}}
\def\sq    {{\ti q}}
\def\str   {{\ti t}_R}
\def\stl   {{\ti t}_L}
\def\st    {{\ti t}}
\def\sb    {{\ti b}}
\def\tb    {\bar{\ti t}}
\def\bb    {\bar{\ti b}}
\def\stau  {{\ti\tau}}
\def\staum {{\ti\tau}^-}
\def\staup {{\ti\tau}^+}
\def\snu   {{\ti\nu}}
\def\sell  {{\ti\ell}}
\def\sll   {{\ti\ell}}
\def\slxx    {{\ti\ell}}
\def\cx    {\ti {\chi}}
\def\ch    {\ti {\chi}}
\def\cpm   {\ti {\chi}^\pm}
\def\cpl    {\ti {\chi}^+}
\def\cm    {\ti {\chi}^-}
\def\cone  {\ti \chi^-_1}
\def\ctwo  {\ti \chi^-_2}
\def\nt    {\ti {\chi}^0}
\def\none  {\ti \chi^0_1}
\def\ntwo  {\ti \chi^0_2}
\def\nthre {\ti \chi^0_3}
\def\nfour {\ti \chi^0_4}
\def\sg    {\ti g}
\def\sG    {\ti G}
\def\sq    {\ti q}
\def\qr    {\ti q_R}
\def\ql    {\ti q_L}
\def\ur    {\ti u_R}
\def\ul    {\ti u_L}
\def\dr    {\ti d_R}
\def\dl    {\ti d_L}

\def\smu   {{\ti\mu}}
\def\smul  {{\ti\mu}_L}
\def\smur  {{\ti\mu}_R}
\def\smulm {{\ti\mu}^-_L}
\def\smurm {{\ti\mu}^-_R}
\def\smulp {{\ti\mu}^+_L}
\def\smurp {{\ti\mu}^+_R}
\def\snm   {{\ti\nu}_\mu}
\def\se    {{\ti e}}
\def\sel   {{\ti e}_L}
\def\ser   {{\ti e}_R}
\def\selm  {{\ti e}^-_L}
\def\serm  {{\ti e}^-_R}
\def\selp  {{\ti e}^+_L}
\def\serp  {{\ti e}^+_R}
\def\sne  {{\ti\nu}_e}
\def\el   {{\ti e}^-_L}
\def\er   {{\ti e}^-_R}
\def\sll  {{\ti l}_L}
\def\slr  {{\ti l}_R}
\def\snl  {{\ti\nu}_l}
\def\snu  {{\ti\nu}}

\def\stauone  {{\ti\tau}^-_1}
\def\stautwo  {{\ti\tau}^-_2}
\def\snt      {{\ti\nu}_\tau}

\def\anue {{\bar\nu}_e}
\def\anum {{\bar\nu}_\mu}
\def\anut {{\bar\nu}_\tau}

\def\cth   {\cos\theta}
\def\sth   {\sin\theta}
\def\tsf   {\theta_{\ti f}}
\def\tst   {\theta_{\ti t}}
\def\tsb   {\theta_{\ti b}}
\def\tstau {\theta_{\ti\tau}}
\def\csf   {\cos\theta_{\ti f}}
\def\cst   {\cos\theta_{\ti t}}
\def\csb   {\cos\theta_{\ti b}}
\def\cstau {\cos\theta_{\ti\tau}}

\newcommand{\msf}[1]   {m_{\ti f_{#1} }}
\newcommand{\mst}[1]   {m_{\ti t_{#1} }}
\renewcommand{\msb}[1]   {m_{\ti b_{#1} }}
\newcommand{\mstau}[1] {m_{\ti \tau_{#1} }}
\newcommand{\mnt}[1]   {m_{\ti \chi^0_{#1} }}
\newcommand{\mch}[1]   {m_{\ti \chi^+_{#1} }}
\newcommand{\msnu}     {m_{\ti \nu}}
\newcommand{\msg}      {m_{\ti g}}

\def \Eslash {E \kern-.75em\slash }
\def \Mslash {M \kern-.5em\slash }
\newcommand{\sla}[1]{\not{\! {#1}}} 
\newcommand{\rpv}{\slash\hspace{-2.5mm}{R}_{p}}
\newcommand{\dmchi}{\Delta m_{\tilde\chi_1}}
\newcommand{\cB } {{\cal B}}

\subsubsection{Introduction}

In the unconstrained version of the Minimal Supersymmetric extension of
the Standard Model (MSSM) no particular Supersymmetry (SUSY) breaking
mechanism is assumed, but rather a parametrisation of all possible soft
SUSY breaking terms is used.
This leads to more than a hundred parameters
(masses, mixing angles, phases) in this model in addition to the ones of
the Standard Model. For performing detailed simulations of experimental
signatures within detectors of high-energy physics experiments it is
clearly not practicable to scan over a multi-dimensional parameter
space. One thus often concentrates on certain ``typical'' benchmark scenarios.

The ``Snowmass Points and Slopes'' (SPS)~\cite{sec4_Allanach:2002nj}
are a set of benchmark points
and parameter lines in the MSSM parameter space
corresponding to different scenarios in the search
for Supersymmetry at present and future
experiments. The SPS~1a reference point is a
``typical'' parameter point of the minimal supergravity (mSUGRA)
scenario. It gives rise to a particle spectrum where
many states are accessible both at the LHC and the LC, 
corresponding to a rather favourable scenario for phenomenology at LHC
and LC. The SPS~1a benchmark scenario has been studied with 
detailed experimental simulations at both colliders. 

Since for no other parameter point in the MSSM a similar amount of
information about the experimental capabilities of both the LHC and the
LC is available, most analyses performed in the context of the LHC / LC
Study Group have focussed on this particular parameter point. It should
be kept in mind, however, that the interplay between LHC and LC could be
qualitatively very different in different regions of the MSSM parameter
space. In order to allow a quantitative assessment of the LHC / LC
interplay also for other parameter regions, more experimental
simulations for LHC and LC are required.

\subsubsection{Definition of the SPS~1a benchmark point}

The SPS benchmark points are defined in terms of low-energy
MSSM parameters. The benchmark values for the SPS~1a point are
the following (all mass parameters are given 
in GeV)\cite{sec4_Allanach:2002nj,Weiglein:2003cb}. 
The gluino mass $\Mgl$, the Supersymmetric Higgs mass parameter $\mu$,
the mass of the CP-odd Higgs boson $M_A$, the ratio of the vacuum
expectation values of the two Higgs doublets $\tan\beta$, and the
electroweak gaugino mass parameters $M_1$ and $M_2$ have the 
values
\begin{equation}
\Mgl = 595.2, \quad \mu = 352.4, \quad M_A = 393.6, \quad
\tan\beta =  10, \quad M_1 = 99.1, \quad M_2 = 192.7.
\end{equation}
The soft SUSY-breaking parameters in the diagonal entries of the squark
and slepton mass matrices have been chosen to be the same for the first
and second generation. They have the values
(these parameters are approximately equal to the
sfermion masses; the off-diagonal entries have been neglected for the
first two generations; the index $i$ in $M_{\tilde qi_L}$
refers to the generation)
\begin{equation}
M_{\tilde q1_L} = M_{\tilde q2_L} = 539.9, \quad
M_{\tilde{d}_R} = 519.5, \quad
M_{\tilde{u}_R} = 521.7, \quad
M_{\tilde{e}_L} = 196.6, \quad
M_{\tilde{e}_R} = 136.2 .
\end{equation}
The soft SUSY-breaking parameters in the diagonal entries of the squark
and
slepton mass matrices of the third generation have the values
\begin{equation}
M_{\tilde q3_L} = 495.9 , \quad
M_{\tilde{b}_R} = 516.9, \quad
M_{\tilde{t}_R} = 424.8, \quad
M_{\tilde{\tau}_L} = 195.8, \quad
M_{\tilde{\tau}_R} = 133.6,
\end{equation}
while the trilinear couplings of the third generation read
\begin{equation}
A_t = -510.0, \quad
A_b = -772.7, \quad
A_{\tau} = -254.2.
\end{equation}
All mass parameters for the benchmark point SPS~1a
are to be understood as defined in the
$\overline{\rm DR}$ scheme at the scale $Q = 453.6$~GeV. The value of
the
top-quark mass for all SPS benchmarks is chosen to be $\mt = 175$~GeV.

As mentioned above, these low-energy parameters correspond to a ``typical''\\
mSUGRA point with an intermediate value of $\tan\beta$. 
In order to obtain these low-energy parameters from the high-scale
parameters $m_0$, $m_{1/2}$, $A_0$ of the mSUGRA scenario a particular 
code had to be chosen. For the SPS benchmarks this was version 7.58 of
the program {\sl ISAJET}~\cite{sec4_Baer:1999sp}. Once the low-energy
parameters have been fixed, this choice is no longer relevant. The
mSUGRA parameters used for generating the SPS~1a benchmark values are
\begin{equation}
m_0 = 100 \GeV, \quad m_{1/2} = 250 \GeV, \quad A_0 = -100 \GeV,
\quad \tan\beta = 10, \quad \mu > 0 .
\end{equation}

\subsubsection{Particle spectrum and decay modes}

\begin{figure}[htb!]
  \setlength{\unitlength}{1mm}
  \begin{center}
  \begin{picture}(120,120)(0,0)
    \put(0,-5){
      \put(108,107){{\bf SPS 1a}}
      \put(0,10){\epsfig{file=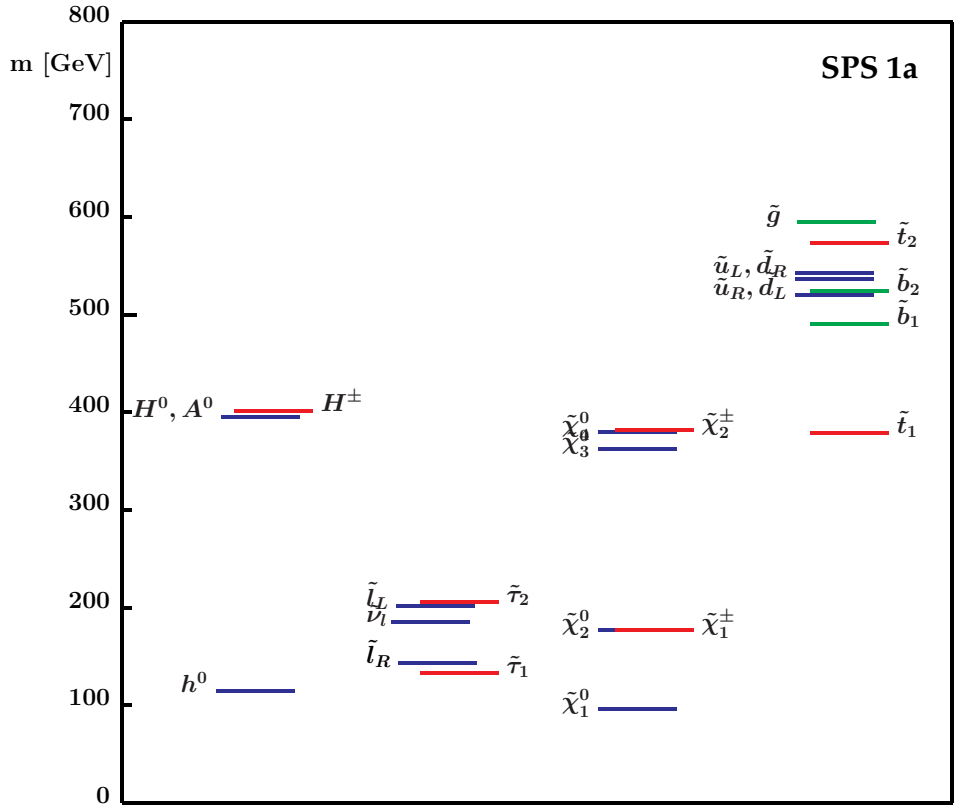,width=.8\textwidth}}
    }
  \end{picture}
  \end{center}
  \caption{The mass spectrum corresponding to the SPS~1a benchmark
   scenario (from Ref.~\cite{Ghodbane:2002kg}).}
  \label{sec4100_sps1aspectrum}
\end{figure}

While the low-energy MSSM parameters have been fixed as benchmarks by
definition, the particle spectra, branching ratios, etc.\ for the SPS
points should be calculated with an appropriate program according to the
specific requirements of the analysis that is being performed. For
simplicity, the spectrum shown in Fig.~\ref{sec4100_sps1aspectrum} and the 
branching ratios of the SUSY particles and the Higgs bosons 
listed in Tab.~\ref{sec4100_slmodes}--\ref{sec4100_higgsmodes}, see
Ref.~\cite{Ghodbane:2002kg}, have 
been obtained with {\sl ISAJET}~7.58, 
although parts of this code are not state-of-the-art. The resulting
spectra and branching ratios have been widely used for exploring the
physics potential of LHC and LC.

\begin{table}[htb!] \centering
   \begin{tabular}{|l|c|c|c|}
   \hdick & & &  \\[-2.ex]
   $\slxx$ & $m~[{\rm GeV}]$ & \ decay \ & ${\cal B}$ 
   \\[.5ex] \hdick
   $\ser    $ &  143.0 & $\nt_1    e^-               $ & 1.000 \\
   \hline
   $\sel    $ &  202.1 & $\nt_1    e^-               $ & 0.490 \\ 
              &        & $\nt_2    e^-               $ & 0.187 \\
              &        & $\cm_1    \nu_e             $ & 0.323 \\
   \hline
   $\sne    $ &  186.0 & $\nt_1    \nu_e             $ & 0.885 \\
              &        & $\nt_2    \nu_e             $ & 0.031 \\
              &        & $\cpl_1    e^-               $ & 0.083 \\
   \hdick
   $\smur   $ &  143.0 & $\nt_1    \mu^-             $ & 1.000 \\
   \hline
   $\smul   $ &  202.1 & $\nt_1    \mu^-             $ & 0.490 \\
              &        & $\nt_2    \mu^-             $ & 0.187 \\
              &        & $\cm_1    \nu_\mu           $ & 0.323 \\
   \hline
   $\snm    $ &  186.0 & $\nt_1    \nu_\mu           $ & 0.885 \\
              &        & $\nt_2    \nu_\mu           $ & 0.031 \\
              &        & $\cpl_1    \mu^-             $ & 0.083 \\
   \hdick
   $\stau_1 $ &  133.2 & $\nt_1    \tau^-            $ & 1.000 \\
   \hline
   $\stau_2 $ &  206.1 & $\nt_1    \tau^-            $ & 0.526 \\
              &        & $\nt_2    \tau^-            $ & 0.174 \\
              &        & $\cm_1    \nu_\tau          $ & 0.300 \\
   \hline
   $\snt    $ &  185.1 & $\nt_1    \nu_\tau          $ & 0.906 \\
              &        & $\cpl_1    \tau^-            $ & 0.067 \\
   \hline
   \end{tabular}
   \caption{Slepton masses and significant branching ratios
     ($\cB>3\%$) in SPS~1a
   (from Ref.~\cite{Ghodbane:2002kg}).}
   \label{sec4100_slmodes}
 \end{table}

\
\begin{table}[htb!] \centering
   \begin{tabular}{|l|c|c|c|}
   \hdick & & &  \\[-2.ex]
   $\cx$ & $m~[\GeV]$ & \ decay \ & ${\cal B}$ 
   \\[.5ex] \hdick
   $\nt_1   $ &   96.1 & $                           $ &    \\
   \hline
   $\nt_2   $ &  176.8 & $\ser^\pm e^\mp             $ & 0.062 \\
              &        & $\smur^\pm \mu^\mp          $ & 0.062 \\
              &        & $\stau_1^\pm \tau^\mp       $ & 0.874 \\
   \hline
   $\nt_3   $ &  358.8 & $\cx_1^\pm W^\mp            $ & 0.596 \\
              &        & $\nt_1    Z^0               $ & 0.108 \\
              &        & $\nt_2    Z^0               $ & 0.215 \\
   \hline
   $\nt_4   $ &  377.8 & $\cx_1^\pm W^\mp            $ & 0.526 \\
              &        & $\nt_1    h^0               $ & 0.064 \\
              &        & $\nt_2    h^0               $ & 0.134 \\
   \hline
   \end{tabular}
   \hspace{1cm}
   \begin{tabular}{|l|c|c|c|}
   \hdick & & &  \\[-2.ex]
   $\cx$ & $m~[\GeV]$ & \ decay \ & ${\cal B}$ 
   \\[.5ex] \hdick
   $\cpl_1   $ &  176.4 & $\staup_1 \nu_\tau          $ & 0.979 \\
   \hline   
   $\cpl_2   $ &  378.2 & $\nt_1    W^+               $ & 0.064 \\
              &        & $\selp    \nu_e             $ & 0.052 \\
              &        & $\smulp   \nu_\mu           $ & 0.052 \\
              &        & $\staup_2 \nu_\tau          $ & 0.056 \\
              &        & $\cpl_1    Z^0               $ & 0.244 \\
              &        & $\cpl_1    h^0               $ & 0.170 \\
   \hline
   \multicolumn{4}{c}{} \\
   \multicolumn{4}{c}{} \\
   \multicolumn{4}{c}{} \\
   \end{tabular} 
   \caption{Neutralino and chargino masses and significant branching ratios
     ($\cB>3\%$) in SPS~1a 
   (from Ref.~\cite{Ghodbane:2002kg}).}
   \label{sec4100_chimodes}
 \end{table}

The SPS~1a scenario yields a sparticle spectrum, see
Fig.~\ref{sec4100_sps1aspectrum}, of which many states are accessible
both at LHC and LC. Experimentally important and challenging, however,
are the $\tau$-rich neutralino and chargino decays, see
Tab.~\ref{sec4100_chimodes}. This is a 
generic feature of SUSY scenarios with
intermediate or large values of $\tan\beta$ (the parameter space at
smaller values of $\tan\beta$ is severely constrained by the exclusion
bounds from the LEP Higgs searches~\cite{Barate:2003sz,:2001xx}). A
non-negligible mixing leads to a significant mass
splitting between the two staus so that the lighter stau becomes the
lightest slepton. Neutralinos and charginos therefore decay
predominantly into staus and taus, which is experimentally more
challenging than the dilepton signal resulting for instance from the
decay of the second lightest neutralino into the lightest neutralino and a
pair of leptons of the first or the second generation.
This effect becomes more pronounced for larger values of $\tan\beta$.

 \begin{table}[htb!] \centering
   \begin{tabular}{|l|c|c|c|}
   \hdick & & &  \\[-2.ex]
   $\sq$ & $m~[\GeV]$ & \ decay \ & ${\cal B}$ 
   \\[.5ex] \hdick
   $\st_1   $ &  379.1 & $\nt_1    t                 $ & 0.179 \\
              &        & $\nt_2    t                 $ & 0.095 \\
              &        & $\cpl_1    b                 $ & 0.726 \\
   \hline
   $\st_2   $ &  574.7 & $\cpl_1    b                 $ & 0.206 \\
              &        & $\cpl_2    b                 $ & 0.216 \\
              &        & $Z^0      \st_1             $ & 0.225 \\
              &        & $h^0      \st_1             $ & 0.042 \\
              &        & $\nt_1    t                 $ & 0.030 \\
              &        & $\nt_2    t                 $ & 0.080 \\
              &        & $\nt_3    t                 $ & 0.033 \\
              &        & $\nt_4    t                 $ & 0.166 \\
   \hline
   \end{tabular}
   \hspace{1cm}
   \begin{tabular}{|l|c|c|c|}
   \hdick & & &  \\[-2.ex]
   $\sq$ & $m~[\GeV]$ & \ decay \ & ${\cal B}$ 
   \\[.5ex] \hdick
   $\sb_1   $ &  491.9 & $\nt_1    b                 $ & 0.062 \\
              &        & $\nt_2    b                 $ & 0.362 \\
              &        & $\cm_1    t                 $ & 0.428 \\
              &        & $W^-      \st_1             $ & 0.133 \\
   \hline
   $\sb_2   $ &  524.6 & $\nt_1    b                 $ & 0.148 \\
              &        & $\nt_2    b                 $ & 0.171 \\
              &        & $\nt_3    b                 $ & 0.053 \\
              &        & $\nt_4    b                 $ & 0.072 \\
              &        & $\cm_1    t                 $ & 0.213 \\
              &        & $W^-      \st_1             $ & 0.344 \\
   \hline
   \multicolumn{4}{c}{ } \\
   \end{tabular}
   \caption{Stop and sbottom masses and significant
     branching ratios ($\cB>3\%$) in SPS~1a 
   (from Ref.~\cite{Ghodbane:2002kg}).}
   \label{sec4100_sqmodes}
 \end{table}

Concerning the compatibility of the benchmark scenario with external 
constraints, severe restrictions on the MSSM parameter space arise from
the requirement that the lightest SUSY particle should give rise to an
acceptable dark matter density. The SPS~1a parameter values give rise to
a dark matter density in what used to be the ``bulk'' region of the 
allowed mSUGRA parameter space. Taking into account the recent precision
data from the WMAP Collaboration~\cite{sec4_Bennett:2003bz,Spergel:2003cb}, 
the dark matter density corresponding to the SPS~1a point is slightly
outside the allowed region. However, shifting the low-energy MSSM parameters
of the SPS~1a point in order to make them fully compatible with the most
recent WMAP bound would hardly affect the collider phenomenology of the
SPS~1a scenario. It should furthermore be mentioned in this 
context that allowing a small amount of R-parity violation in the model would
leave the collider phenomenology essentially unchanged, while having a 
drastic impact on the constraints from dark matter relic abundance.
The SPS~1a benchmark scenario is in satisfactory agreement with all
other experimental constraints.
  
 \begin{table}[htb!] \centering
   \begin{tabular}{|l|c|c|c|}
   \hdick & & &   \\[-2.ex]
   Higgs & $m~[\GeV]$ & \ decay \ & ${\cal B}$ 
   \\[.5ex] \hdick
   $h^0     $ &  114.0 & $\tau^-   \tau^+            $ & 0.051 \\
              &        & $b        \bar b            $ & 0.847 \\
              &        & $c        \bar c            $ & 0.035 \\
   \hline
   $H^0     $ &  394.1 & $\tau^-   \tau^+            $ & 0.059 \\
              &        & $b        \bar b            $ & 0.807 \\
              &        & $t        \bar t            $ & 0.031 \\
              &        & $\nt_1    \nt_2             $ & 0.034 \\
   \hline
   $A^0     $ &  393.6 & $\tau^-   \tau^+            $ & 0.049 \\
              &        & $b        \bar b            $ & 0.681 \\
              &        & $t        \bar t            $ & 0.092 \\
              &        & $\nt_1    \nt_2             $ & 0.065 \\
              &        & $\nt_2    \nt_2             $ & 0.058 \\
   \hline
   $H^+     $ &  401.8 & $\nu_\tau \tau^+            $ & 0.077 \\
              &        & $t        \bar b            $ & 0.770 \\
              &        & $\cpl_1    \nt_1             $ & 0.130 \\
   \hline
   \end{tabular}
   \caption{Higgs masses and significant branching ratios ($\cB>3\%$) 
            in SPS~1a (from Ref.~\cite{Ghodbane:2002kg}).}
   \label{sec4100_higgsmodes}
 \end{table}

\clearpage

\subsection{\label{sec:411}
A detailed analysis of the measurement of SUSY masses
with the ATLAS detector at the LHC}

{\it B.K.~Gjelsten, J.~Hisano, K.~Kawagoe, E.~Lytken, D.~Miller,
M.M.~Nojiri, P.~Osland, G.~Polesello}

\vspace{1em}
\def\mg{m_{\tilde{g}}^2}
\def\mq{m_{\tilde{q}_L}^2}
\def\mT{m_{\tilde{\chi}_2^0}^2}
\def\ml{m_{\tilde{l}_R}^2}
\def\mlfour{m_{\tilde{l}_R}^4}
\def\mO{m_{\tilde{\chi}_1^0}^2}
\def\threshold{{\rm thres}}
\def\edge{{\rm edge}}
\def\max{{\rm max}}
\def\min{{\rm min}}
\def\MT2{M_{T2}}
\def\qn{q_{\rm n}}
\def\qf{q_{\rm f}}
\def\ln{\ell_{\rm n}}
\def\lf{\ell_{\rm f}}

\def\GeV{{\rm GeV}}

\newcommand{\newc}{\newcommand}
\renewcommand{\beq}{\begin{eqnarray}}
\renewcommand{\eeq}{\end{eqnarray}}
\newc{\dqu}{\delta_{qu}}
\newc{\dqd}{\delta_{qd}}
\renewcommand{\non}{\nonumber}
\newc{\noi}{\noindent}
\def\sg{$\tilde g$}
\def\sq{$\tilde q$}
\def\ptmiss{$P_{T}^{miss}$}
\def\Etmiss{E_{T}^{miss}}
\def\Pt{$p_{T}$}
\def\chione{\tilde \chi_1^0}
\def\chitwo{\tilde \chi_2^0}
\def\chithree{\tilde \chi_3^0}
\def\chifour{\tilde \chi_4^0}
\def\chii{\tilde \chi_i^0}
\def\chiipm{\tilde \chi_i^\pm}
\def\chionepm{\tilde \chi_1^\pm}
\def\chionemp{\tilde \chi_1^\mp}
\def\chionep{\tilde \chi_1^+}
\def\chitwopm{\tilde \chi_2^\pm}
\def\mgl{${\tilde{g}}$}
\def\msq{${\tilde{q}}$}
\def\ttau{\tilde\tau}
\def\mA{${{A}}$}
\def\mh{${{h}}$}
\def\mH{${{H}}$}
\def\m0{${0}$}
\def\mhf{${1/2}$}
\def\tanb{tan${\beta}$}
\def\tg{{\tilde g}}
\def\tq{{\tilde q}}
\def\tu{{\tilde u}}
\def\td{{\tilde d}}
\def\ttop{{\tilde t}}
\def\tb{{\tilde b}}
\def\tchi{{\tilde\chi}}
\def\tl{{\tilde\ell}}
\def\sel{{\tilde e_L}}
\def\smul{{\tilde \mu_L}}
\def\sne{{\tilde\nu_e}}
\def\snm{{\tilde\nu_\mu}}
\def\lsp{{\tilde\chi_1^0}}
\def\GeV{{\rm GeV}}
\def\TeV{{\rm TeV}}
\def\Meff{{\rm eff}}
\def\wt{\widetilde}
\def\sgn{\mathop{\rm sgn}}
\def \sm {Standard Model }
\def \susy {supersymmetry }
\def \susyq {supersymmetric }
\def \mssm {minimal supersymmetric standard model }
\def \sugra {supergravity }
\def \fc {flavour changing }
\def \fcnc {flavour changing neutral current }
\def \brs {branching ratios }
\def \br {branching ratio }
\def\L {\Lambda }
\def\l {\lambda }
\def\ka {\kappa }
\def \t {\theta }
\def \vt {\vartheta }
\def\a {\alpha }
\def\dh {\partial }
\def \d {\delta }
\def \D {\Delta }
\def \bq {\bar q }
\def \bQ {\bar Q }
\def \g {\gamma }
\def \G {\Gamma }
\def \O {\Omega }
\def \b {\beta }
\def \S {\Sigma }
\def \s {\sigma }
\def \e {\epsilon }
\def \ud {{1 \over 2} }
\def \ut {{1 \over 3} }
\def \taud {{3 \over  2 } }
\def \bea {\begin{equation} }
\def \eea {\end{equation} }
\def \cc {coupling constant }
\def \ccs {coupling constants }
\def \tchi  {{\tilde \chi} }
\def \Eslash {E \kern-.9em\slash }
\def \pslash {p \kern-.5em\slash }
\def \kslash {k \kern-.5em\slash }
\def\tg{{\tilde g}}
\def\tq{{\tilde q}}
\def\tb{{\tilde b}}
\def\tchi{{\tilde\chi}}
\def\tl{{\tilde\ell}}
\def\tn{{\tilde\nu}}
\def\lsp{{\tilde\chi_1^0}}
\def\z2{{\tilde\chi_2^0}}
\def\cz{{\tilde\chi_3^0}}
\def\n4{{\tilde\chi_4^0}}
\def\ww{{\tilde\chi_1^{\pm}}}
\def\w2{{\tilde\chi_2^{\pm}}}
\def\GeV{{\rm GeV}}
\def\TeV{{\rm TeV}}
\def\Meff{M_{\rm eff}}
\def\wt{\widetilde}
\def\sgn{\mathop{\rm sgn}}
\def\mttwo{M_{T2}}
\def\mttwomax{{M^\rmax_{T2}}}
\def\ptlepTwo{{p_T^{l_2}}}
\def\PtlepOne{{{\bf p}_T^{l_1}}}
\def\PtlepTwo{{{\bf p}_T^{l_2}}}
\def\Qt{{\bf q}_T}
\def\pt{{p_T}}
\def\Pt{{{\bf p}_T}}
\def\ptlep{{p_T^{l}}}
\def\Ptlep{{{\bf p}_T^{l}}}
\def\slepton{\sparticle{l}}
\newc{\sparticle}[1]{{\tilde{{#1}}}}
\def\guess{\chi}
\def\slashchar#1{\setbox0=\hbox{$#1$}           
   \dimen0=\wd0                                 
   \setbox1=\hbox{/} \dimen1=\wd1               
   \ifdim\dimen0>\dimen1                        
      \rlap{\hbox to \dimen0{\hfil/\hfil}}      
      #1                                        
   \else                                        
      \rlap{\hbox to \dimen1{\hfil$#1$\hfil}}   
      /                                         
   \fi}
\def\etmiss{\slashchar{E}_T}
\newcommand{\pmiss}{{{\slashchar{p}}}}
\newcommand{\Ptmiss}{{{\slashchar{{\bf p}}}}_T}
\newcommand{\slptwo}{{{\slashchar{{\bf p}}}}}

\def\dofig#1#2{\epsfxsize=#1\centerline{\epsfbox{#2}}}
\def\dofigs#1#2#3{\centerline{\epsfxsize=#1\epsfbox{#2}%
   \hfil\epsfxsize=#1\epsfbox{#3}}}


%
\def\bentarrow{\:\raisebox{1.3ex}{\rlap{$\vert$}}\!\rightarrow}
\def\longbent{\:\raisebox{3.5ex}{\rlap{$\vert$}}\raisebox{1.3ex}%
	{\rlap{$\vert$}}\!\rightarrow}
\def\onedk#1#2{
	\begin{equation}
	\begin{array}{l}
	 #1 \\
	 \bentarrow #2
	\end{array}
	\end{equation}
		}
\def\dk#1#2#3{
	\begin{equation}
	\begin{array}{r c l}
	#1 & \rightarrow & #2 \\
	 & & \bentarrow #3
	\end{array}
	\end{equation}
		}
\def\dksl#1#2#3#4{
	\begin{equation}
	\begin{array}{r c l }
	#1 & \rightarrow & #2  \\
	 & & \bentarrow #3 \\
         & & \phantom{\bentarrow \;}  \bentarrow #4 
	\end{array}
        \label{eqq}
	\end{equation}
		}
\def\dkp#1#2#3#4{
	\begin{equation}
	\begin{array}{r c l}
	#1 & \rightarrow & #2#3 \\
	 & & \phantom{\; #2}\bentarrow #4
	\end{array}
	\end{equation}
		}
\def\bothdk#1#2#3#4#5{
	\begin{equation}
	\begin{array}{r c l}
	#1 & \rightarrow & #2#3 \\
	 & & \:\raisebox{1.3ex}{\rlap{$\vert$}}\raisebox{-0.5ex}{$\vert$}%
	\phantom{#2}\!\bentarrow #4 \\
	 & & \bentarrow #5
	\end{array}
	\end{equation}
		}
%
%
\catcode`@=11 
\def \gsim{\mathrel{\mathpalette\@versim>}}
\def \lsim{\mathrel{\mathpalette\@versim<}}
\def \@versim#1#2{\lower0.4ex\vbox{\baselineskip\z@skip\lineskip\z@skip
     \lineskiplimit\z@\ialign{$\m@th#1\hfil##\hfil$%
     \crcr#2\crcr\sim\crcr}}}
\catcode`@=12 
%


\noindent{\small
We present a series of exclusive analyses which can be performed
at the LHC for mSUGRA Point SPS~1a. The aim is to evaluate the 
precision with which sparticle masses can be evaluated at the LHC,
and how the various measurements are interrelated.

The analyses are then used to demonstrate how information
from a LC can be used to improve the LHC's mass measurement.
}

\subsubsection{Introduction}
We here describe a series of 
analyses performed on Point SPS~1a \cite{sec4_Allanach:2002nj}, 
characterized by the SUGRA parameters
\begin{gather}
m_0=100~\text{GeV}, \qquad
m_{1/2}=250~\text{GeV}, \nonumber \\
\tan\beta=10, \qquad
A=-100~\text{GeV}, \qquad
\mu>0,
\end{gather}
with the aim of providing 
an input to the studies which evaluate the complementarity
of the LHC and of the LC.
\subsubsection{\boldmath Analysis of kinematic edges involving 
$\tilde{\chi}_2^0\to\tilde{l}_Rl$}
\label{sec:edge}
In this scenario the total SUSY cross section is rather large,
and at the LHC squarks and gluinos will be produced abundantly.
The gluino is the heaviest particle and decays to a squark and 
a quark. The squarks decay to neutralinos and charginos,
which in turn decay to sleptons or lighter neutralinos and 
charginos. The sleptons then decay into the LSP, $\tilde{\chi}_1^0$.

\begin{figure}
\refstepcounter{figure}
\label{Fig:chain}
\addtocounter{figure}{-1}
\begin{center}
\setlength{\unitlength}{1cm}
\begin{picture}(10.0,3.2)
\put(0,0){
\mbox{\epsfysize=3.8cm
\epsffile{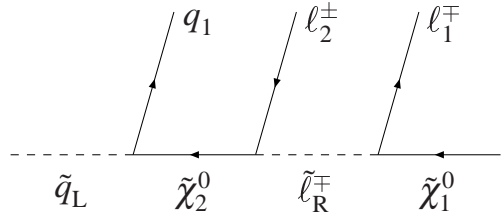}}}
\end{picture}
\vspace*{-4mm}
\caption{A squark decay chain for SPS~1a, where 
$m_{\tilde q_{\rm L}}\in\{491.9,537.2,543.0\}~\text{GeV}$,
$m_{\tilde \chi_2^0}=176.82~\text{GeV}$,
$m_{\tilde \ell_{\rm R}}=142.97~\text{GeV}$,
$m_{\tilde \chi_1^0}=96.05~\text{GeV}$.}
\end{center}
\end{figure}

Since the LSP will escape detection, it is not a straightforward 
task to reconstruct SUSY events. A possible approach is to
use kinematic edges \cite{TDR,sec4_Allanach:2002nj}.
Particularly interesting is the decay 
$\tilde{\chi}_2^0\to\tilde{l}_Rl\to l^+l^-\tilde{\chi}_1^0$.
The two leptons in the final state provide a natural trigger, 
and the energy resolution is high.

While right-handed squarks decay directly to the LSP, 
due to the bino-like nature of the $\chi_1^0$ at SPS~1a,
left-handed squarks decay to $\tilde{\chi}_2^0$ with a branching
ratio $\sim 32\%$. 
We will in this analysis look at the decay chain shown in Fig.~\ref{Fig:chain},
\begin{eqnarray}
\tilde{q}_L
\underset{\sim 32\%}\longrightarrow q\tilde{\chi}_2^0 
\underset{12.1\%}{\longrightarrow} q l_2^\pm \tilde{l}_R^\mp 
\underset{100\%}{\longrightarrow} q l_2^\pm l_1^\mp \tilde{\chi}_1^0
\end{eqnarray}
where $\tilde{q}_L$ can be $\tilde{d}_L$, $\tilde{u}_L$, 
$\tilde{b}_2$ or $\tilde{b}_1$.
The first two have very similar masses, $m_{\tilde{d}_L}=543.0~\text{GeV}$
and $m_{\tilde{u}_L}=537.2~\text{GeV}$,
and will in this analysis be grouped together and referred to as 
$\tilde{q}_L$.
For the fraction of the chain in Eq.~(\ref{Fig:chain}) which starts with 
a sbottom, $\tilde{b}_1$ is responsible for 78\%, leaving us 
insensitive to the contribution from $\tilde{b}_2$. 
Decay chains involving the stop are not considered.
The production cross section of the relevant squarks and their 
branching fractions to $\tilde{\chi}_2^0$ are
\begin{gather}
\nonumber
\sigma(\tilde{q}_L)=33\ \text{pb},\quad 
BR(\tilde{q}_L\to q\tilde{\chi}_2^0)=31.4\%
\\
\sigma(\tilde{b}_1)=7.6\ \text{pb},\quad
BR(\tilde{b}_1\to b\tilde{\chi}_2^0)=35.5\%
\end{gather}
with many of the squarks coming from gluino decay.
The stau is the lightest slepton, so it has the largest branching
ratio. In this analysis we will however only use final states 
with electrons and muons.

We shall discuss the precision that can be achieved in the
determination of this spectrum at the LHC, reconstructing it
back to the squark mass, from measurements of various
kinematic edges and thresholds of subsets of decay products.
In particular, we shall determine how this precision can be improved
with input from a Linear Collider \cite{sec4_lctdrs}.

\paragraph{Kinematics}

The invariant masses of various subsets of particles can be determined
from kinematical edges and thresholds, as discussed in
\cite{Allanach:2000kt},
\begin{eqnarray}
\label{Formula1}
\bigl(m_{ll}^2\bigr)^\edge 
&=& \frac{\bigl(\mT-\ml\bigr)\bigl(\ml-\mO\bigr)}{\ml}
\\
\bigl(m_{q ll}^2\bigr)^\edge 
&=& \frac{\bigl(\mq-\mT\bigr)\bigl(\mT-\mO\bigr)}{\mT}
\\
\bigl(m_{q l}^2\bigr)^\edge_\min 
&=& \frac{\bigl(\mq-\mT\bigr)\bigl(\mT-\ml\bigr)}{\mT}
\\
\bigl(m_{q l}^2\bigr)^\edge_\max 
&=& \frac{\bigl(\mq-\mT\bigr)\bigl(\ml-\mO\bigr)}{\ml}
\label{Formula4}
\\
\bigl(m_{q ll}^2\bigr)^\threshold 
\nonumber
&=&[(\mq+\mT)(\mT-\ml)(\ml-\mO)\\
\nonumber
&&-(\mq-\mT)\sqrt{(\mT+\ml)^2(\ml+\mO)^2-16\mT\mlfour\mO}\\
&&+ 2\ml(\mq-\mT)(\mT-\mO)]/(4\ml\mT)
\label{Formula5}
\end{eqnarray}
where ``min'' and ``max'' refer to minimising and maximising
w.r.t.\ the choice of lepton.
Furthermore ``thres'' refers to the threshold in the subset
of the $m_{qll}$ distribution for which the angle between the
two lepton momenta (in the slepton rest frame) exceeds $\pi/2$,
which corresponds to $m_{ll}^\edge/\sqrt{2}<m_{ll}<m_{ll}^\edge$.


\paragraph{Monte Carlo simulations}
In order to assess quantitatively the precision that can be achieved,
we have performed Monte Carlo simulations of SUSY production at SPS~1a,
using the {\tt PYTHIA 6.2} program \cite{sec4_PYTHIA}
and passing the particles through the {\tt ATLFAST} detector simulation
\cite{ATLFAST} before reconstructing invariant masses.
We have used a sample corresponding to 100~$\text{fb}^{-1}$, 
one year at design luminosity.
The results documented in this section have been found to 
be in agreement with results obtained using {\tt HERWIG} \cite{HERWIG}.

The cuts used to isolate the chain were the following
\begin{itemize}
\item At least four jets, the hardest three satisfying:\\
 $p_{T,1}>150\ \GeV,\quad p_{T,2}>100\ \GeV,\quad p_{T,3}>50\ \GeV$.
\item $M_{\rm eff}
\equiv E_{T,\rm miss}+p_{T,1}+p_{T,2}+p_{T,3}+p_{T,4} > 600\ \GeV$
\item $E_{T,\rm miss}>\max(100\ \GeV, 0.2 M_{\rm eff})$
\item Two isolated Opposite--Sign Same--Flavour (OS-SF) leptons (not $\tau$) 
satisfying $p_T(l)>20\ \GeV$ and $p_T(l)>10\ \GeV$.
\end{itemize}

The basic signature of our decay chain are two OS-SF leptons.
Two such leptons can also be produced in other processes.
If the two leptons are independent of each other, one would
expect equal amounts of OS-SF leptons and Opposite--Sign 
Opposite--Flavour (OS-OF) leptons. 
Their distributions should also be identical. 
This allows us to remove the background OS-SF contribution by 
subtracting the OS-OF events.

In addition to the two OS-SF leptons, our signal event will typically have
considerable missing $E_T$ and two very hard jets, one from the decay of the
squark in the chain we try to reconstruct, one from the decay of the squark in
the other chain.

The only Standard Model process to have all the features of our signal event,
is $t\bar t$ production where both $W$'s decay leptonically.  However, with
some help from the underlying event, pile-up and detector effects, other
processes might also result in the signatures above.  Together with $t\bar t$,
we therefore considered the following PYTHIA processes: QCD, $Z/W$+jet,
$ZZ/ZW/WW$.

The QCD processes are cut away by the requirement of two leptons and of
considerable missing $E_T$.  For the processes involving $Z$ and $W$ the
requirement of high hadronic activity together with missing $E_T$, removes
nearly all events.  The only Standard Model background to survive the rather
hard cuts listed above, is a small fraction of $t\bar t$ events.  However, the
rate of $W^+W^-$ (from the decay of $t$ and $\bar t$) going to $e^\pm\mu^\mp$
is identical to that going to $e^+ e^- / \mu^+\mu^-$, so with the subtraction
of OS-OF events the $t\bar t$ sample gives no net contribution to the mass
distribution, only some minor contribution to the fluctuations.

In Fig.~\ref{FIGmll} the invariant mass of the two leptons for events passing
the cuts is plotted.
The Standard Model background is clearly negligible.
The real background to our decay chain consists of other SUSY processes,
and as is illustrated in Fig.~\ref{FIGmll},
these are effectively removed by the OS-OF subtraction.
This subtraction is also included for the invariant mass distribution 
in Fig.~\ref{FIGmqlletc}.

\begin{figure}
\begin{center}
\includegraphics[scale=.35]{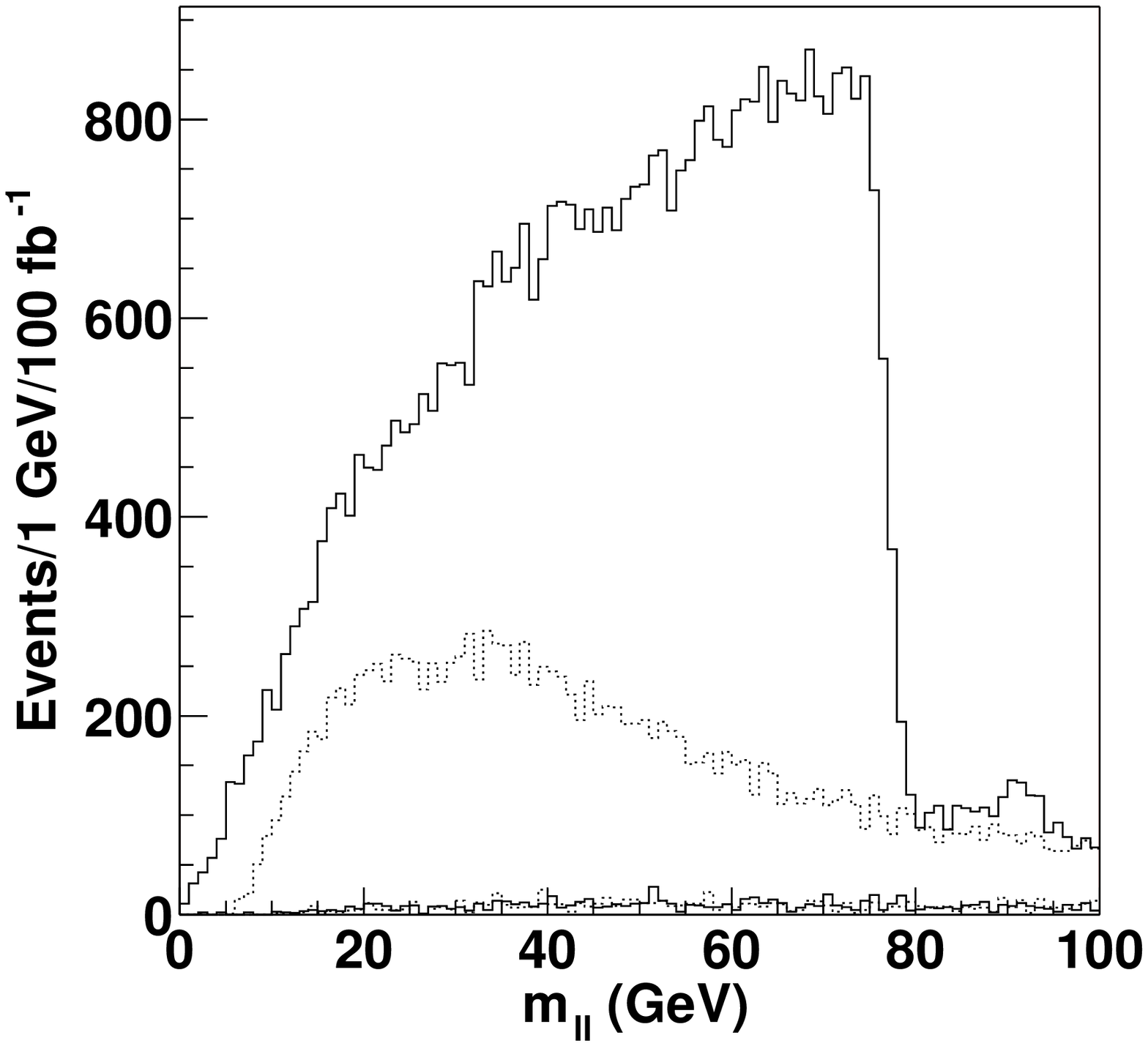}
\includegraphics[scale=.35]{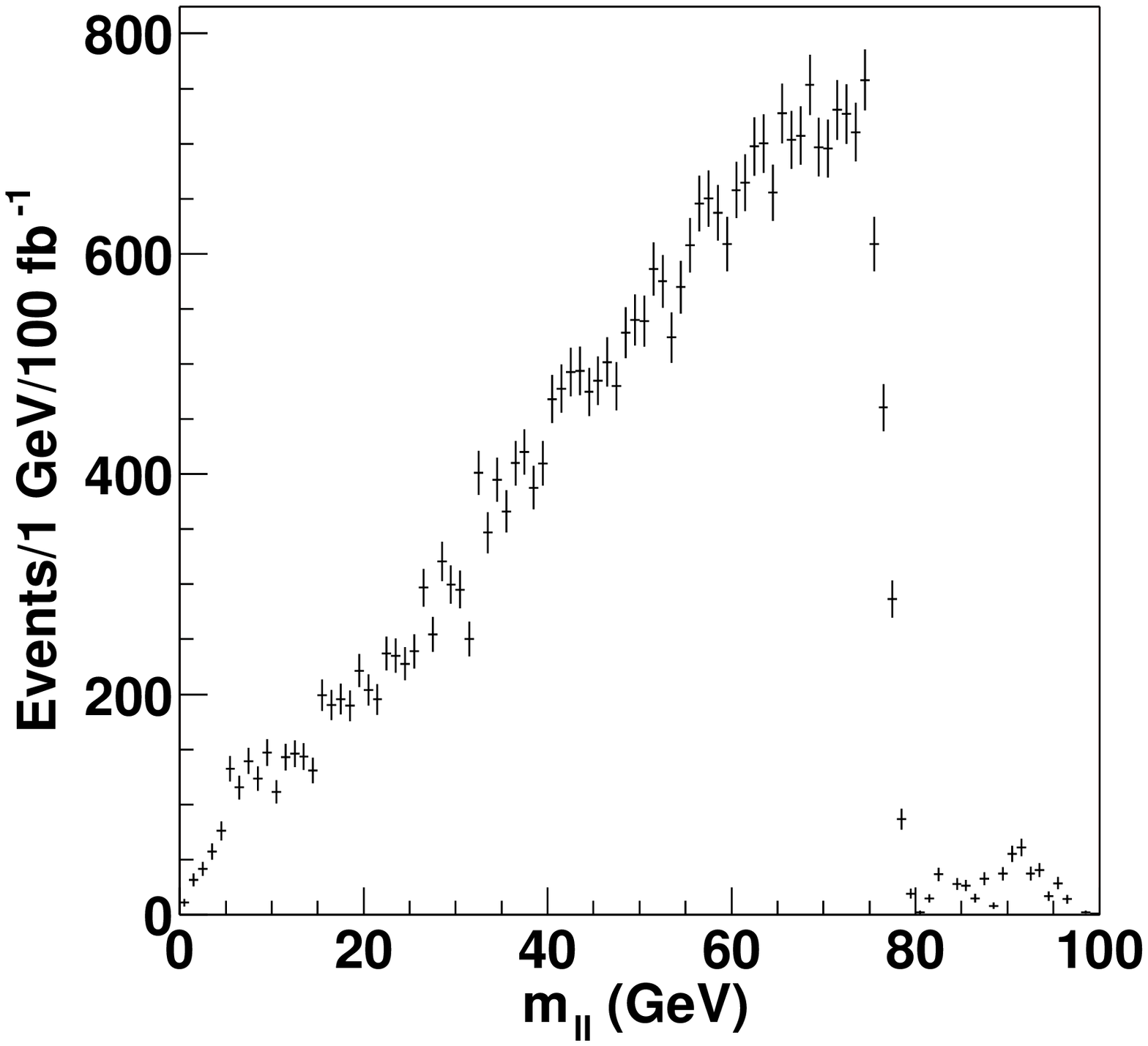}
\caption{Effect of subtracting background leptons, 
for $\int{\cal L}dt = 100~\text{fb}^{-1}$.  Left: Solid: OS-SF, Dotted: OS-OF,
Two upper curves: SUSY+SM, two lower curves: SM alone; Right: OS-SF$-$OS-OF. The
triangular shape of the theoretical expectation is reproduced.
\label{FIGmll}}
\end{center}
\end{figure}

\begin{figure}
\refstepcounter{figure}
\label{FIGmqlletc}
\addtocounter{figure}{-1}
\begin{center}
\setlength{\unitlength}{1cm}
\begin{picture}(10.0,20)
\put(-2,13){
\mbox{\epsfysize=7.20cm\epsffile{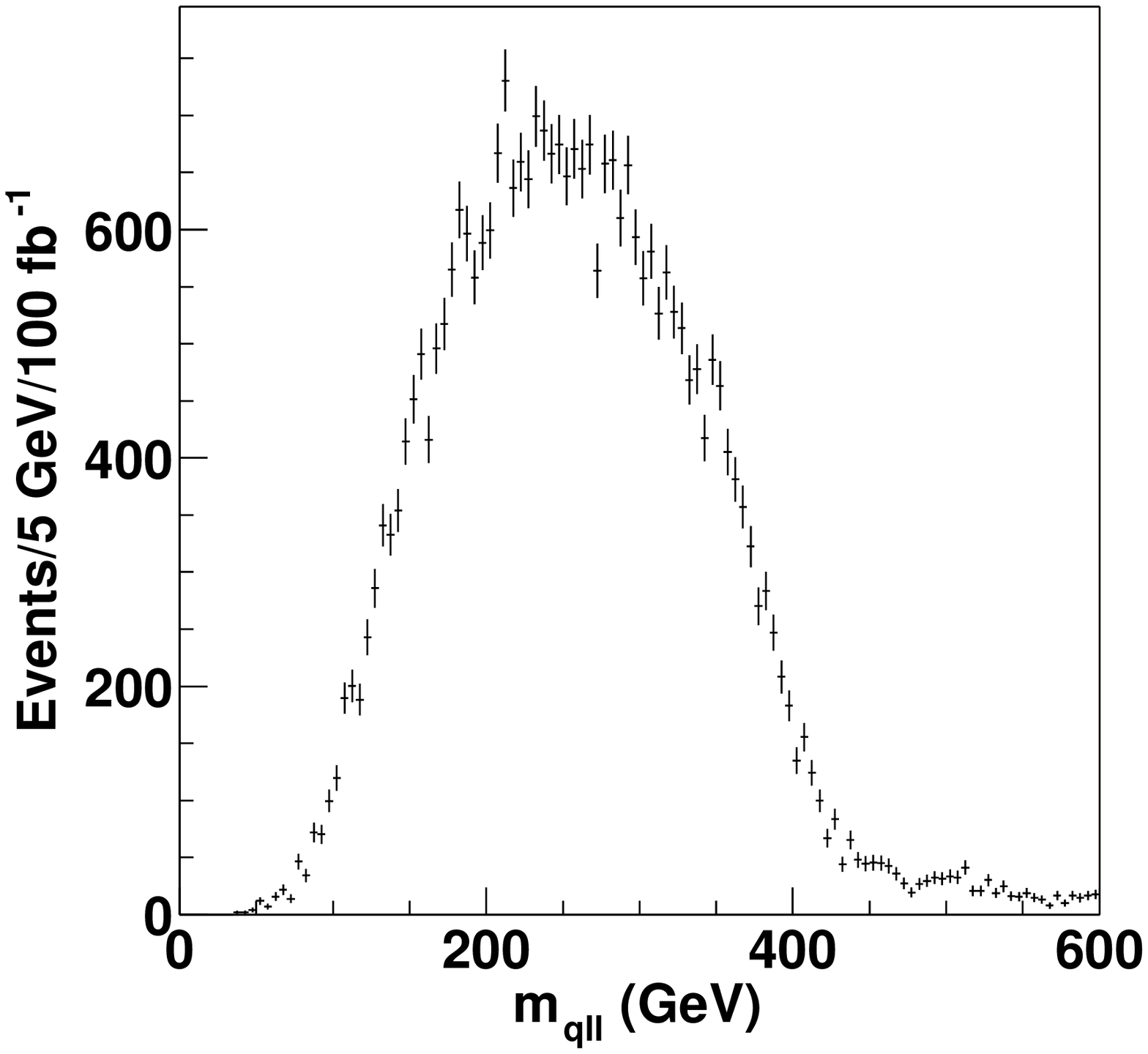}}
\mbox{\epsfysize=7.20cm\epsffile{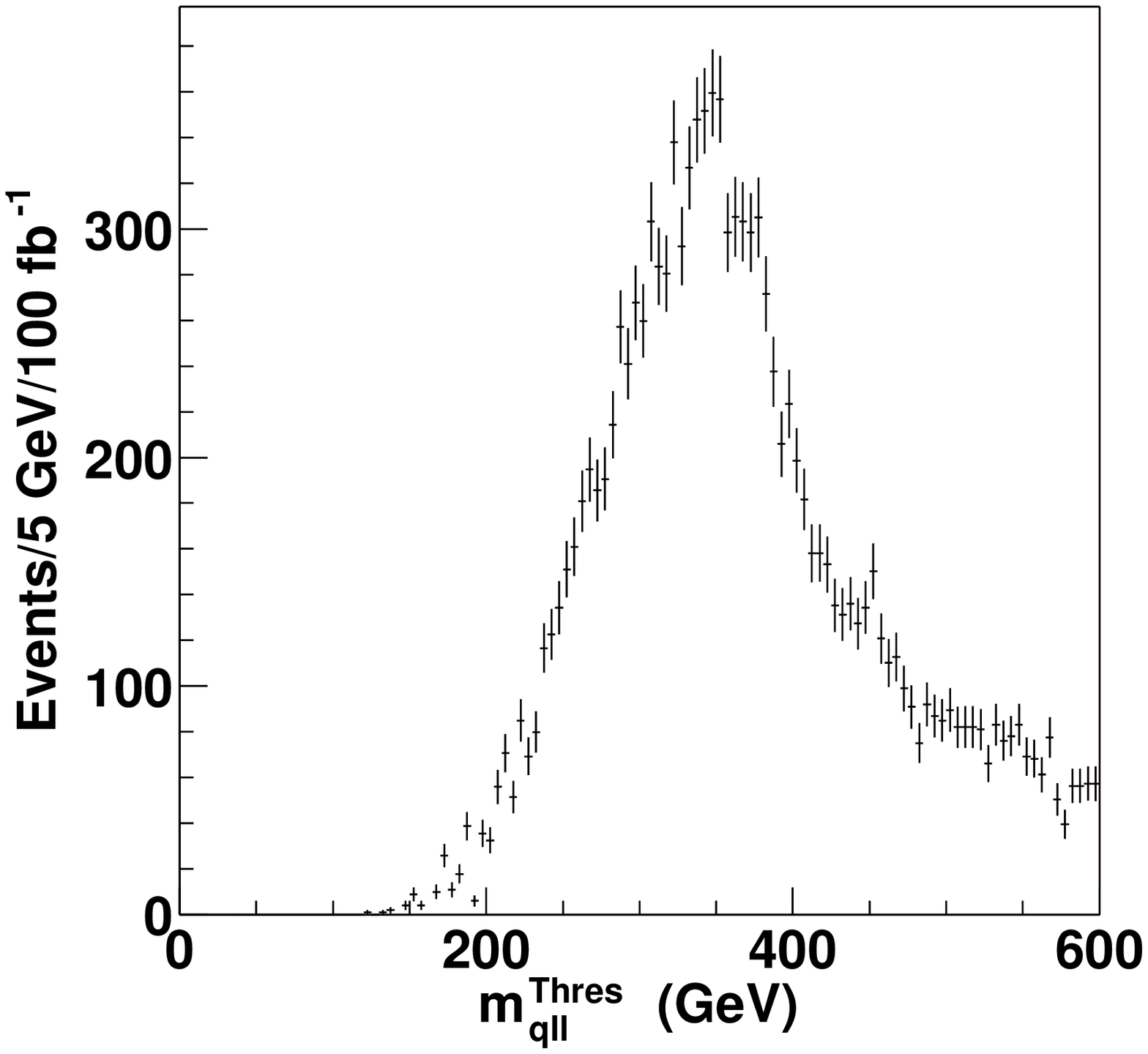}}}
\put(-2,6.5){
\mbox{\epsfysize=7.20cm\epsffile{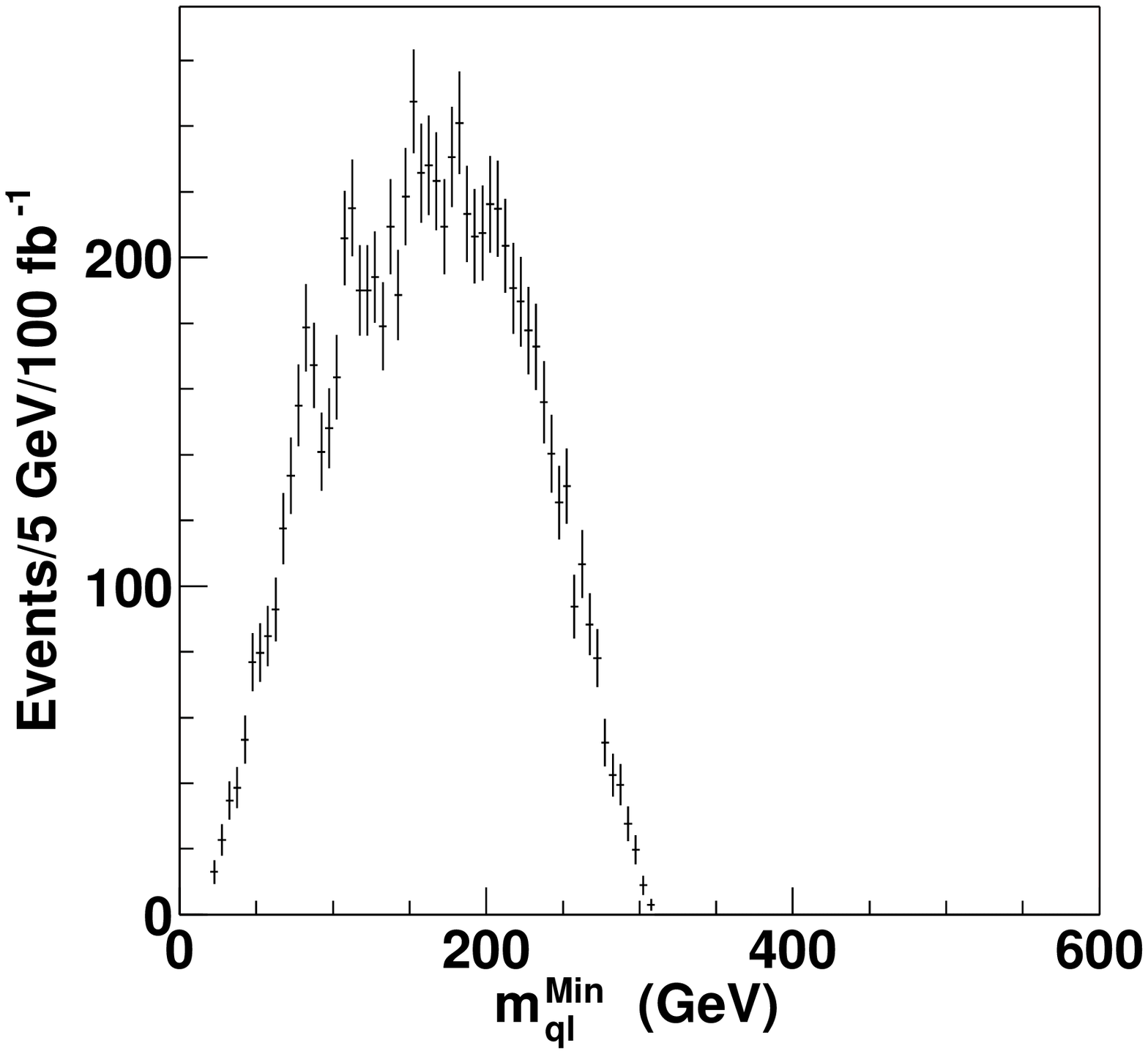}}
\mbox{\epsfysize=7.20cm\epsffile{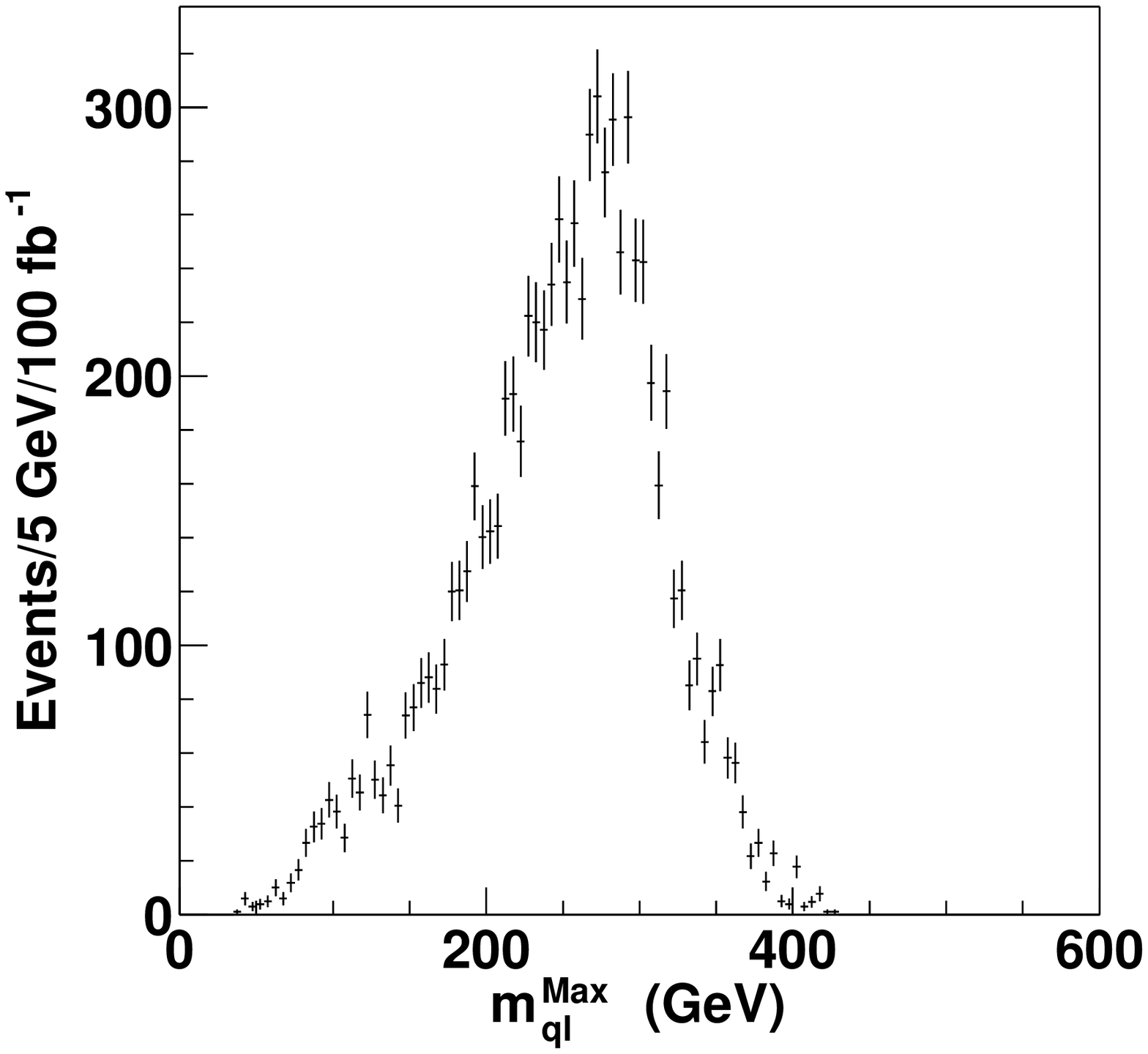}}}
\put(1.5,0){
\mbox{\epsfysize=7.20cm\epsffile{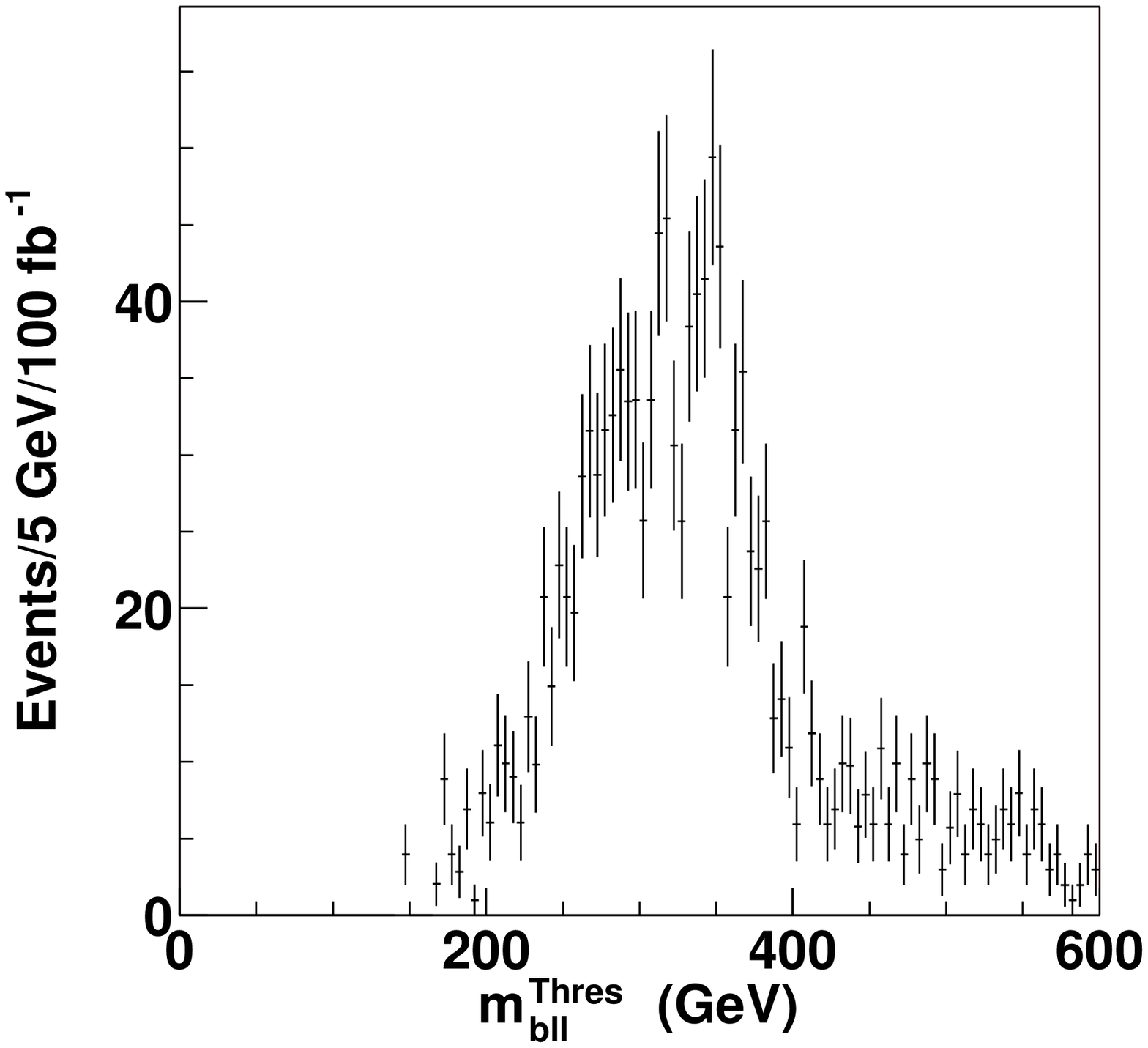}}}
\end{picture}
\vspace*{-4mm}
\caption{Invariant mass distributions with kinematical endpoints,
for $\int {\cal L}dt = 100~\text{fb}^{-1}$.}
\end{center}
\end{figure}

The edge value for $m_{ll}$, Fig.~\ref{FIGmll}, is very accurately determined
by fitting it to a triangular shape with Gaussian smearing.  For the other
distributions, Fig.~\ref{FIGmqlletc}, the end points are found with a naive
linear fit. This method is known not to be optimal. In fact, by changing the
binning or the range fitted, the fit values may change by typically a few GeV.
For a more realistic situation one would need to investigate more thoroughly
how the theoretical distributions are distorted by sparticle widths, by the
detector resolution and the cuts applied.  As already discussed in
\cite{Allanach:2000kt}, it is not possible at the level of the present studies
to perform a detailed estimate of the corresponding systematic errors,
therefore we include only the statistical errors from the fitting procedure
for each edge.  In addition one has the systematic error on the energy
scale. We use the ATLAS benchmark values, 0.1\% for leptons and 1\% for
jets. The resulting values for the endpoints and the corresponding errors are
given in table \ref{TABedges}.

\begin{table}[htp]
\begin{center}
\caption{Endpoint values found from fitting the edges in Fig.~\ref{FIGmll}
and Fig.~\ref{FIGmqlletc}, for 100~$\text{fb}^{-1}$.}
\label{TABedges}
\begin{tabular}{cccccc}
\hline\hline
Edge  &Nominal Value& Fit Value & Syst. Error & Statistical \\
  &   &  &Energy Scale & Error \\
\hline
$m(ll)^\edge$ & 77.077& 77.024 &0.08& 0.05 \\
$m(qll)^\edge$& 431.1 & 431.3 &4.3&2.4 \\
$m(ql)^\edge_\min$& 302.1 & 300.8& 3.0& 1.5 \\
$m(ql)^\edge_\max$& 380.3& 379.4& 3.8 & 1.8 \\
$m(qll)^\threshold$ &203.0 &204.6& 2.0 &2.8 \\
$m(bll)^\threshold$ &183.1 &181.1& 1.8 &6.3 \\
\hline
\end{tabular}
\end{center}
\end{table}

\subsubsection{Gluino mass measurement}
In the considered point the gluino decays 
through $\tg\rightarrow\tq q$, where $\tq$ is any 
squark flavour, except $\ttop_2$, for which the decay
$\tg\rightarrow\ttop_2t$ is kinematically forbidden.
Thus the reconstruction of the gluino can be attempted
adding a quark to an identified  $\tq_L$ decay chain.
A particularly favorable situation happens with the 
$b$ squarks, for which the decay chain includes two $b$-jets
which can be tagged, thereby reducing the combinatorial
background. In addition to the cuts given in section \ref{sec:edge}
we require:
\begin{itemize}
\item
the invariant mass of the OS-SF lepton pair should 
be larger than 65~GeV, and lower than the edge shown in 
Fig.~\ref{FIGmll}.
\item
Exactly two jets tagged as $b$, of which one
must be among the two leading jets in the event, 
and the second one must have $p_T>50$~GeV.
\end{itemize}
Using a technique described e.g.\ in \cite{TDR},
the $\tchi^0_2$ momentum can be approximated by the
expression:
$$
\vec{p}(\tchi^0_2)=\left( 1-\frac{m(\lsp)}{m(\ell\ell)}\right)\vec{p}_{\ell\ell}
$$
where $\vec{p}_{\ell\ell}$ is the vector of the sum of the 
momenta of the two leptons. 
This approximation works reasonably well for the Point under study
as the $\lsp$ momentum is close to zero 
in the $\tchi^0_2$ rest frame near the 
end-point. For situations with $\tl_R$ close in mass either to 
$\tchi^0_2$ or $\lsp$, this assumption is not justified anymore.
If the masses of the $\lsp$ and of the $\tchi^0_2$
are known, one can thus calculate the $\tb$ mass as
$m(\tchi^0_2 b)$, and the $\tg$ mass as $m(\tchi^0_2 bb)$.
The lower cut on $m_{\ell\ell}$ is chosen such as to 
to optimize the precision on the $\tchi^0_2$ momentum, while 
retaining a reasonable statistics for the analysis.
We plot in Fig~\ref{fig:2dmb} the flavour-subtracted 
distribution of $m(\tchi^0_2 b)$ versus 
$m(\tchi^0_2 bb)$, for both $b$ jets, assuming the nominal
values for $m(\lsp)$ and $m(\tchi^0_2)$.
\begin{figure}[htb]
\dofig{0.6\textwidth}{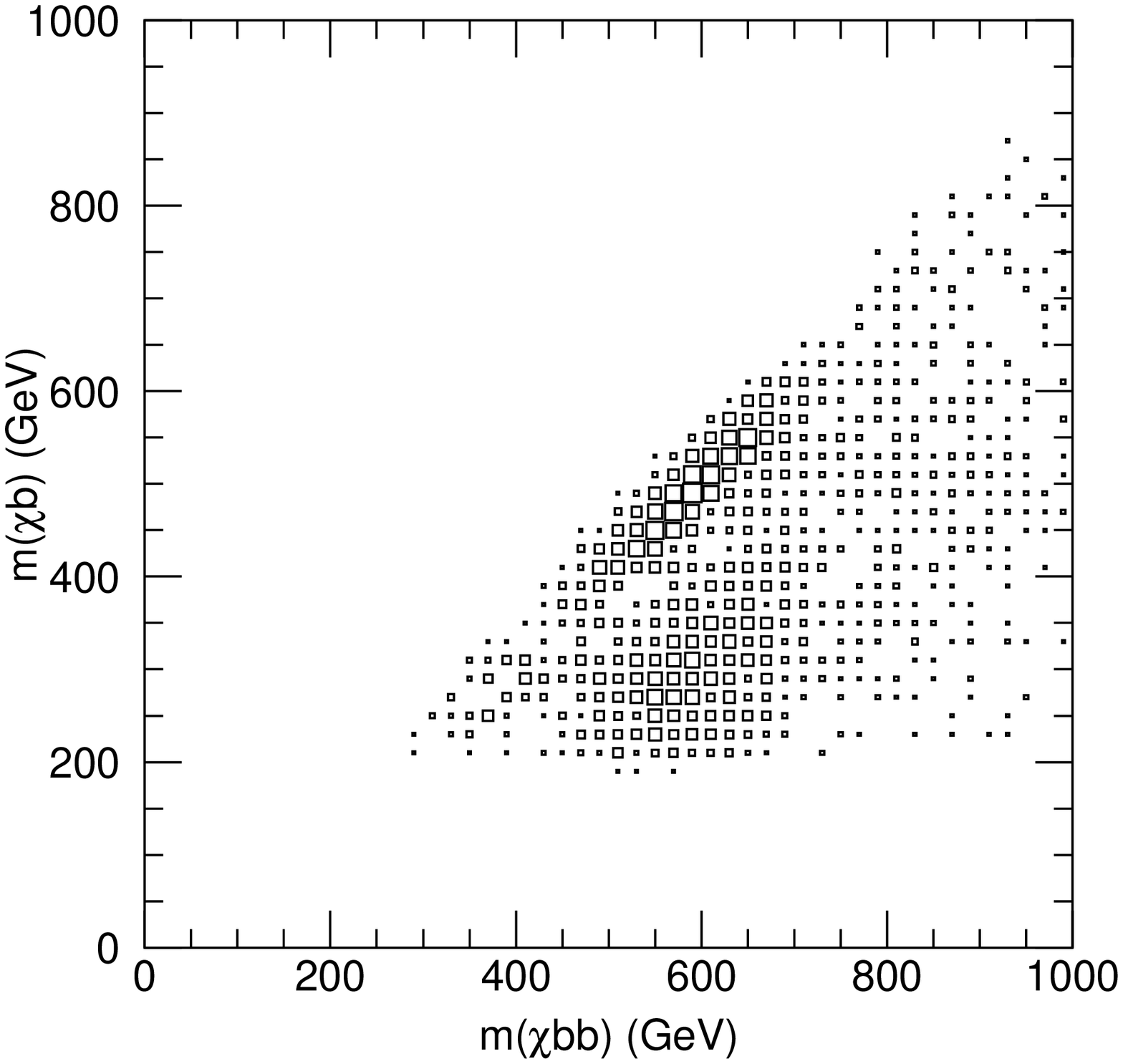}
\caption{ Distribution of $m(\tchi^0_2b)$ versus $m(\tchi^0_2bb)$ 
for events passing the selections. }
\label{fig:2dmb}
\end{figure}
Two well-separated regions appear in the plot, of which 
one corresponds to the correct $\tchi^0_2b$ pairing for the
reconstruction of the $\tb$, and shows a strong correlation 
between the $\tg$ and the $\tb$ mass.
The second region corresponds to the situation
in which $m(\tchi^0_2 b)$ is calculated taking the $b$-jet from 
the $\tg\rightarrow b\tb$ decay.
\begin{figure}
\dofig{0.7\textwidth}{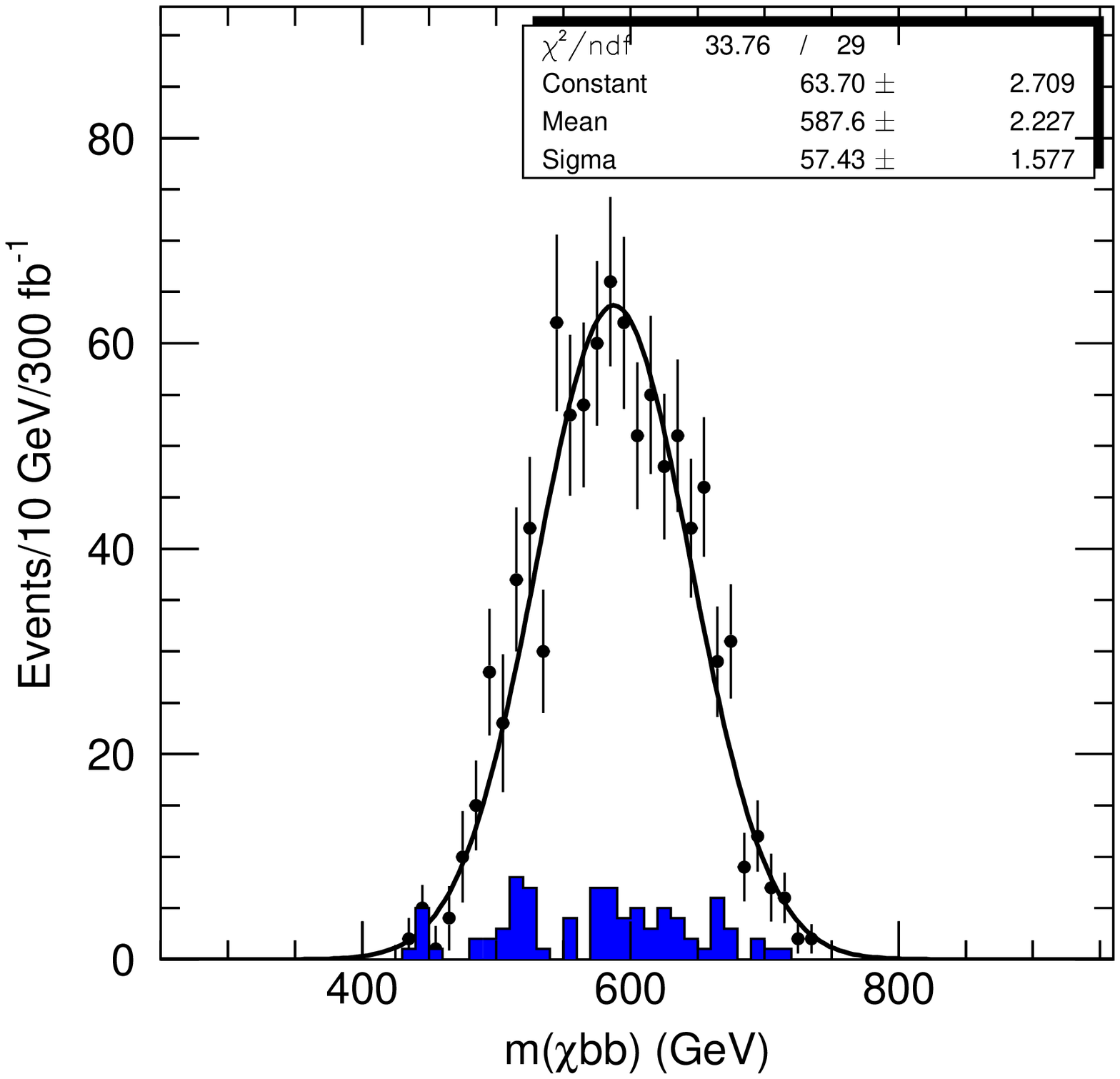}
\vskip -1cm
\caption{$m(\tchi^0_2 bb)$ after all cuts. The residual 
SUSY background is shown in blue. Superimposed is a gaussian fit. 
The distribution is shown for an integrated statistics of 300 fb$^{-1}$}
\label{fig:mgluino}
\dofig{0.7\textwidth}{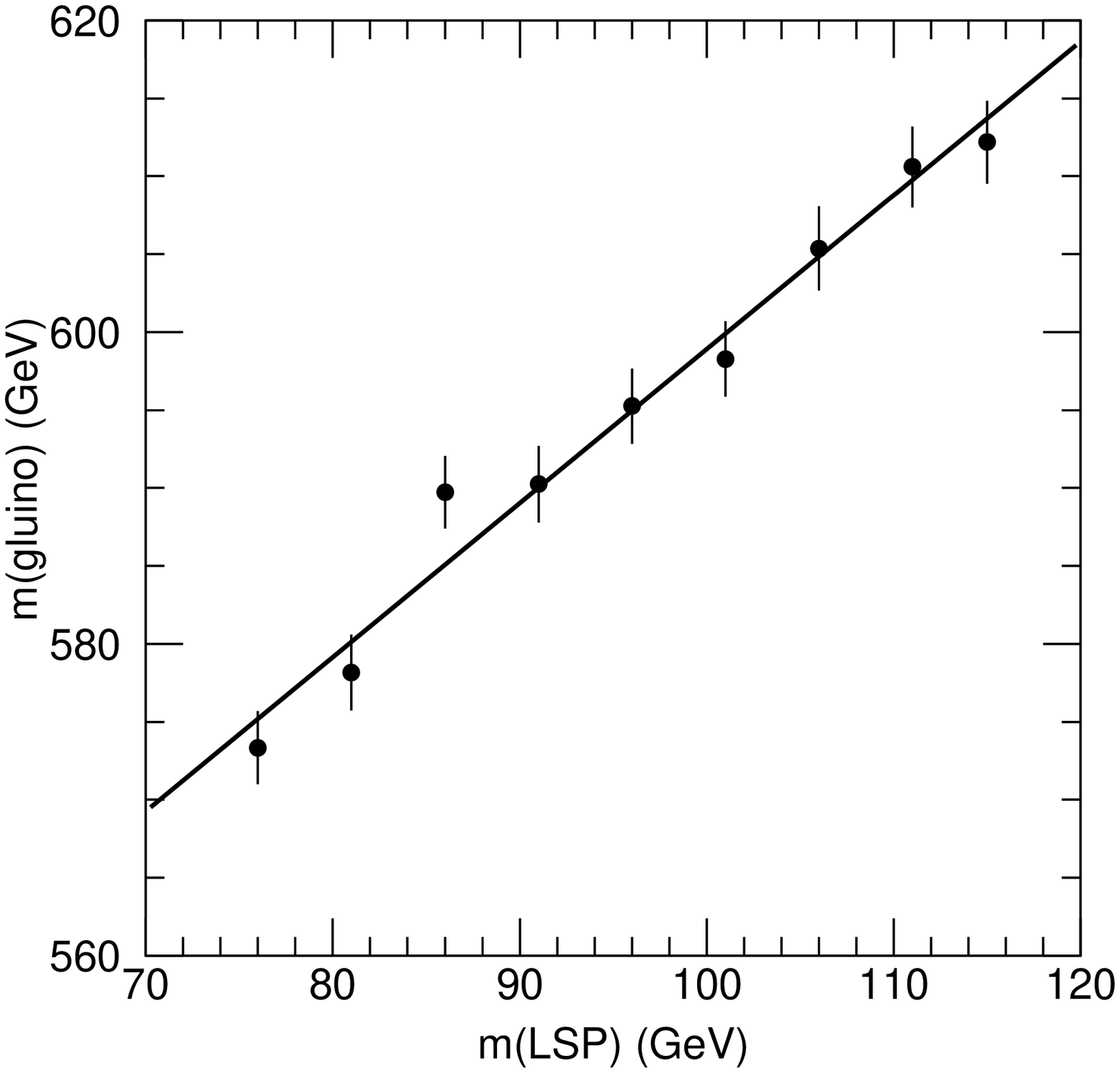}
\vskip -1cm
\caption{\em Estimated $m(\tg)$ as a function of the $m(\lsp)$ assumed
as input of the fit.}
\label{fig:gludep}
\end{figure}
We select  the interesting
region on the 2-dimensional plot by requiring $380<m(\tchi^0_2 b)<600$~GeV,
and $m(\tchi^0_2 bb)-m(\tchi^0_2 b)>150$~GeV.
The main residual background consists where the cascade 
identified by OS-SF the lepton pair originates from a squark 
of the first four generations, 
and the leading $b$ is part of a different cascade. 
We suppress this background
by requiring that the invariant mass 
of the $\tchi^0_2$ with the leading jet not tagged
as $b$ is outside of the interval 400 GeV to 600 GeV. 
The  $m(\tchi^0_2 bb)$ after these cuts is shown in Fig.~\ref{fig:mgluino}.
Superimposed in blue is the residual background.
The width of the distribution 
is dominated by the $\tchi^0_2$ momentum mismeasurement.
The statistical uncertainty on the peak position is $\sim 4$~GeV 
for 100~fb$^{-1}$ and  $\sim 2.5$~GeV for 300~fb$^{-1}$,
and the central value is $\sim 10$~GeV lower than the nominal
$\tg$ mass.
The displacement of the fit value from the nominal value is 
related to an underestimate of the energy of part of 
the $b$ jets.\par
For this analysis we assume that both $\lsp$ and  $\tchi^0_2$ 
would be measured with the technique described in the previous section.
As already discussed above, this results in a strong correlation 
between the measured $\lsp$ and  $\tchi^0_2$ masses
which can be parametrized as:
$$
m(\tchi^0_2)=82.85+0.977\times m(\lsp)
$$
Therefore, to evaluate the dependence of the measured gluino 
mass on the assumed $\lsp$ and  $\tchi^0_2$ masses,
we varied only the $\lsp$ mass between 76 and 116 GeV, and the 
$\tchi^0_2$ mass was taken from the above parametrization.
For each value of $m(\lsp)$ we performed a full gluino mass evaluation.
The results are  shown in Fig.~\ref{fig:gludep}. There is a clear 
linear dependence of the estimated gluino mass on the $m(\lsp)$
which can be parametrized as:
$$
m(\tg)=500+0.988\times m(\lsp)
$$\par
The further step in the analysis is the measurement of the mass
of the sbottom quark.
The gluino decays both to $\tb_1$ and $\tb_2$,
the selected events are therefore a mixture of the two decay channels.
We first select events for which only one of the two 
possible $m(\chi^0_2b)$ combinations
passes the selection cuts. As clearly seen from Fig.~\ref{fig:2dmb},
the spread due to the approximation on the $\tchi^0_2$
momentum measurement affects in the same way the gluino and sbottom mass
measurement, and it can to a large extent be factored
out by studying the difference  $m(\tchi^0_2bb)-m(\tchi^0_2b)$.\par
\begin{figure}
\dofig{0.65\textwidth}{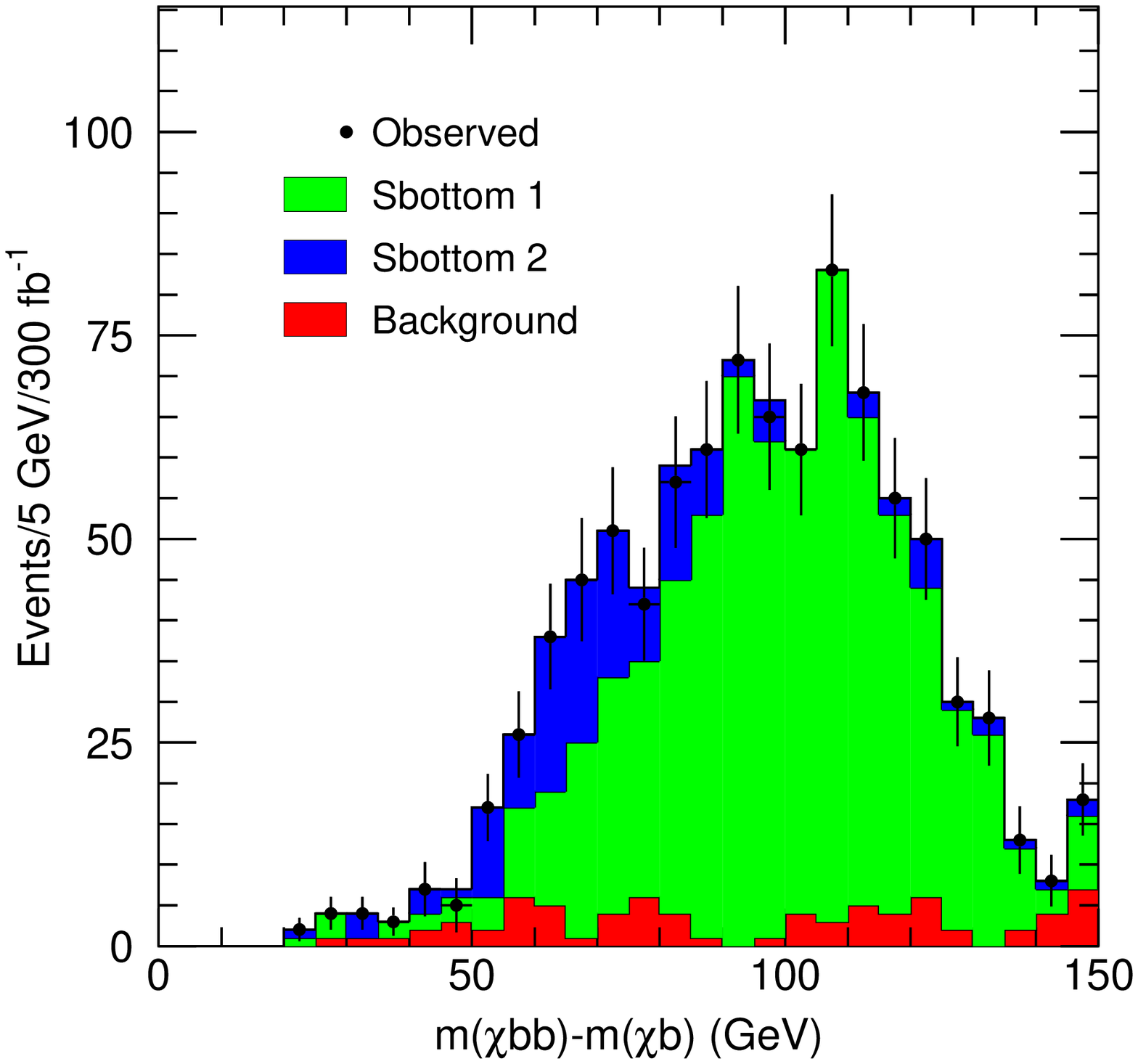}
\vskip -1cm
\caption{Distribution of  $m(\tchi^0_2bb)-m(\tchi^0_2b)$ 
for an integrated luminosity of 300~fb$^{-1}$ (points with error
bars). Superimposed are: $\tg\rightarrow\tb_1b$ (green), 
$\tg\rightarrow\tb_2b$ (blue), and background (red). }
\vskip -0.5cm
\label{fig:sbot}
\dofig{0.70\textwidth}{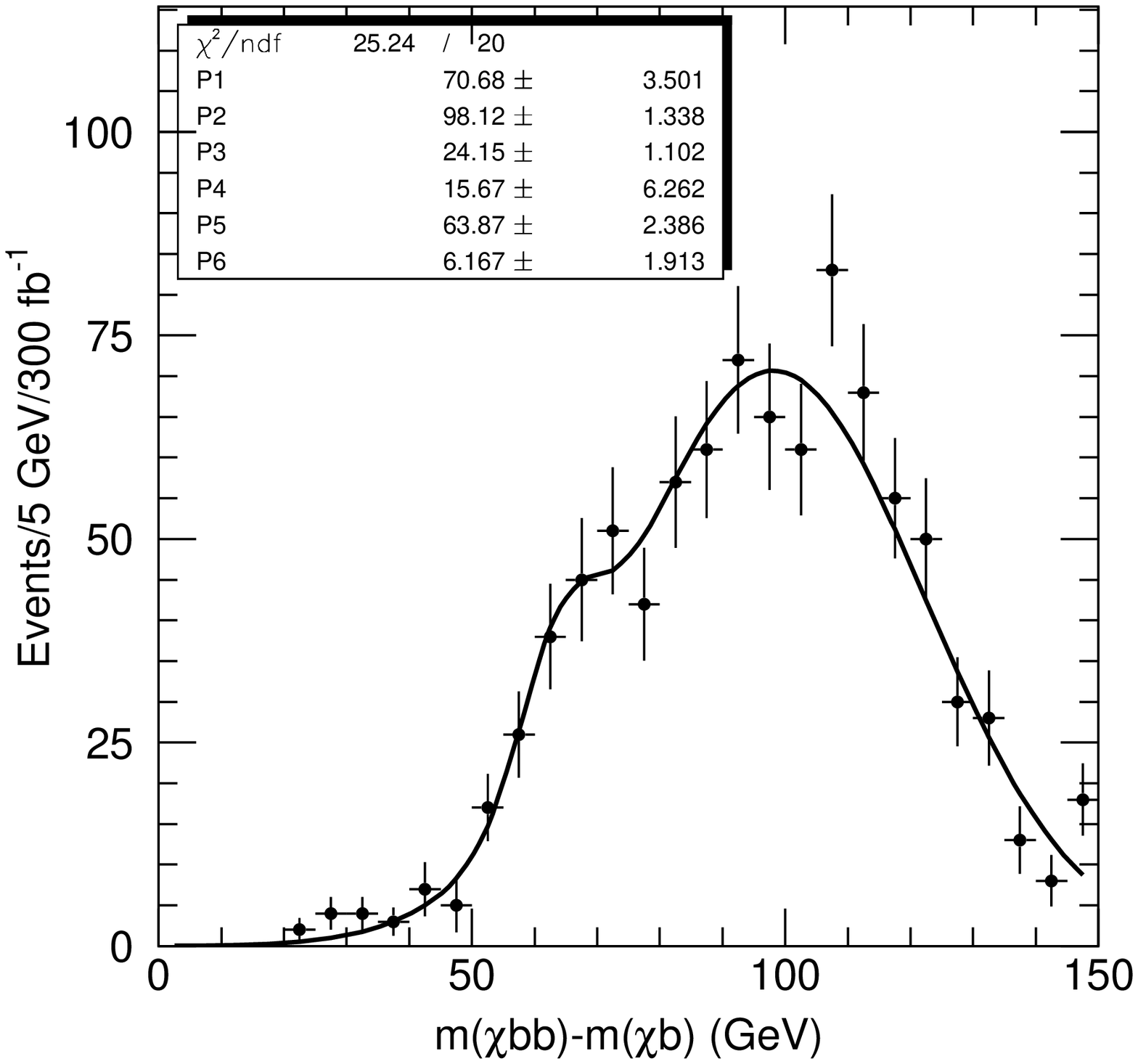}
\vskip -1cm
\caption{Distribution of  $m(\tchi^0_2bb)-m(\tchi^0_2b)$
for an integrated luminosity of 300~fb$^{-1}$. Superimposed
is the fit performed assuming the sum of two gaussian 
distributions.}
\label{fig:sbotfit}
\end{figure}
The resulting distribution is shown in Fig.~\ref{fig:sbot} for
an integrated luminosity of 300~fb$^{-1}$. The observed events, after
lepton flavour subtraction, are shown as black dots with error bars. 
In color we show the different contributions:
$\tg\rightarrow\tb_1b$ (green), $\tg\rightarrow\tb_2b$ (blue),
and background (red). The distribution is wider than could be expected
from a single resonance. However, the dominant distribution
for $\tg\rightarrow\tb_1b$ has a non-gaussian shape, because 
of the mismeasurement of part of the $b$-jets. In order to 
experimentally distinguish the two peaks, a detailed understanding 
of the response function of the detector to $b$-jets is needed.
For the present work, assuming the presence of two peaks, we naively
fit the observed distribution to the sum of two gaussian functions. 
The results are shown in Fig.~\ref{fig:sbotfit}, where   
two peaks separated by approximately the expected  $\sim35$~GeV 
are found by the fit. The statistical errors from the fit are $\sim1.5$~GeV
for $\tb_1$ and $\sim 2.5$~GeV  for $\tb_2$. 
This result is obtained under the assumption 
that two peaks do indeed exists, which must be 
demonstrated by careful experimental analysis. We have verified that 
the measured gluino-sbottom  mass difference has a negligible dependence 
on the assumed $\lsp$ mass.\par
Under the assumption that the shape of the response of the ATLAS detector to 
b-jets can be accurately modeled, it is also possible to estimate 
the relative population of $\tb_1$ and $\tb_2$ decays in the
distributions of Fig~\ref{fig:sbot} through a likelihood
fit and thus extract a measurement of 
the quantity: 
\begin{equation}
\frac{BR(\tg\rightarrow\tb_2b)\times BR(\tb_2\rightarrow\tchi^0_2 b)}
{BR(\tg\rightarrow\tb_1b)\times BR(\tb_1\rightarrow\tchi^0_2 b)}
\label{eq:brat}
\end{equation}
An approximate estimate of the achievable statistical error 
on this measurement can be obtained by considering the number 
of events in a mass interval chosen to maximize the
$\tb_2$ signal. The number of $\tb_1$ decays in that
interval can be estimated from the $\tb_1$ decay distribution
normalized to the peak at $\sim$100 GeV, and subtracted,
thus yielding the number of $\tb_2$ decays. 
The total number of flavour-subtracted 
events in the mass interval  50-80 GeV for 300~fb$^{-1}$
is 913, of which 707 are from $\tb_1$ decays, 73 from background,
and 141 from $\tb_2$ decays. We assume a 100\% error on the background 
estimate, and we neglect the systematic error on the modeling
of the detector response and on the evaluation of
the correction factor necessary to take into account the fact that
the analysis efficiency is different for $\tb_1$ and $\tb_2$ decays.
Under these hypotheses, the quantity in Equation~\ref{eq:brat}, 
can be measured to be 0.257+-0.078.\par 
For a lower statistics of 100~fb$^{-1}$, it would probably be impossible
to demonstrate the presence of two peaks, and the average value
of the distribution provides a measurement 
of the quantity:
$$
\frac{m(\tb_1)\times BR(\tg\rightarrow\tb_1b)\times 
BR(\tb_1\rightarrow\tchi^0_2)+
m(\tb_2)\times BR(\tg\rightarrow\tb_2b)\times BR(\tb_2\rightarrow\tchi^0_2)}
{BR(\tg\rightarrow\tb_1b)\times BR(\tb_1\rightarrow\tchi^0_2)+
BR(\tg\rightarrow\tb_2b)\times BR(\tb_2\rightarrow\tchi^0_2)}
$$
with a statistical error of $\sim 1.7$ GeV.



\subsubsection{Direct Slepton Production}

\newcommand{\slepr}{ $ \tilde \ell_R $ }
\newcommand{\slepl}{ $ \tilde \ell_L $ }

Sleptons (here taken to mean only selectrons and smuons) are produced by
s-channel Z$^*$ exchange with a cross section of 91 fb at SPS~1a.
The relatively high m$_{1/2}$ at this point makes for quite a large splitting
in the slepton sector. The mass of the right-handed and left-handed sleptons are 142.9 GeV
and 201.9 GeV respectively.

At SPS~1a \slepr always decays to $\lsp$ and a lepton, whereas the probability
for this decay from \slepl is 45\%. The two other decay modes for \slepl 
(to $\chitwo  + \ell$ and $\chionepm + \nu$) are
not considered in this study. The key signature for direct di-slepton production will
therefore be 2 opposite-sign same-flavour leptons, $\etmiss$ from the escaping
$\lsp$'s and no jets. Because of the low cross section for the signal we did this study in a high-luminosity environment so in reality a
small jet activity is to be expected.

The backgrounds are dilepton production from WW (most signal like), $t\bar t$ (most
events passing cuts due to the large cross section), WZ (important because of 
SF excess), and several SUSY backgrounds: chargino-neutralino production,
decays from squarks and gluinos, $\tilde \tau_1 \tilde \tau_1$, $ \tilde
\nu \tilde \nu$, and $ \tilde \nu \tilde \ell_L$.
Both signal and background were generated with {\tt PYTHIA 6.210} 
\cite{sec4_PYTHIA} and the WZ$^*$ cross checked
with {\tt HERWIG} \cite{HERWIG}. The detector simulation was 
{\tt ATLFAST} \cite{ATLFAST}, a fast simulation package for ATLAS. 

No direct mass measurement is possible due to the escaping $\lsp$'s but the
mass can still be estimated using the $M_{T2}$ variable, suggested by the
Cambridge group~\cite{lester}: 

\[
        M_{T2}^2 = \min_{\not p_1 + \not p_2 = \not p_T} \left[ \max \left\{
        m_T^2(p_T^{\ell_1},\not p_1),m_T^2(p_T^{\ell_2},\not p_2) \right\} \right]  
\]
where $\not \! p_1$ and $ \not \! p_2$ are the unknown momenta of the
neutralinos. The maximum of this distribution is a function of the mass
difference between the decaying particle and the mass of the invisible
particle (assuming the leptons are massless).


The $M_{T2}$ edge for \slepr is right on top of the edges for the SM
backgrounds but \slepl is more isolated. Since the \slepr is produced
abundantly from neutralino decay at SPS~1a it is much more realistic to measure its mass there than from direct production. 

The result after applying hard cuts and then  opposite-flavour subtraction is shown in figure~\ref{fig:sf-of}. 

\begin{figure}
\begin{center}

\mbox{\epsfig{figure=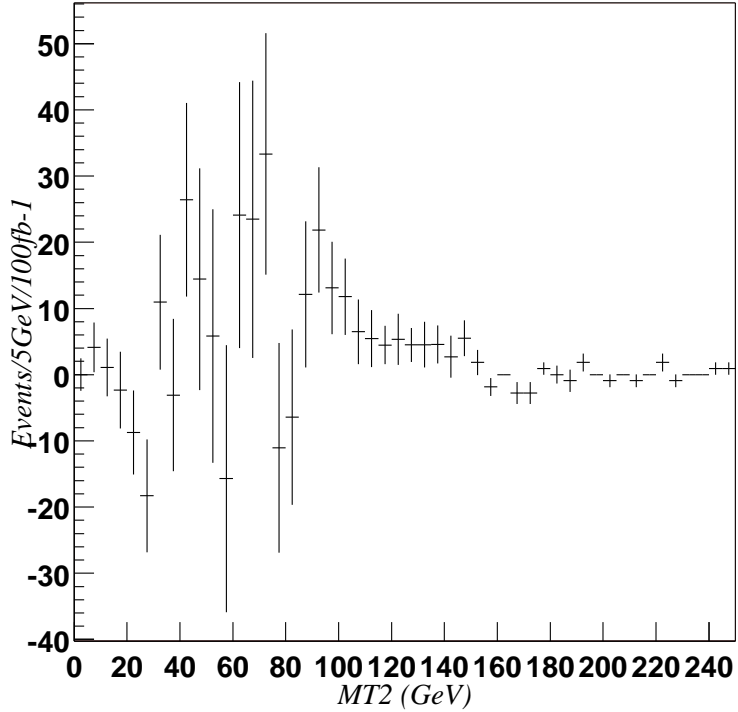,width=0.7\textwidth}}
\caption{Subtracting opposite-flavour events passing the cuts from the
  passing same-flavour events. The scale is 100 fb$^{-1}$.}
\label{fig:sf-of}
\end{center}

\end{figure}

Finally, to estimate what \slepl mass this corresponds to and how well we could do
at the LHC a set of experiments were performed. For each experiment a set of
signal SPS~1a events are picked at random and the resulting histogram from
background subtraction is compared to a sample of ultra-high statistics
signal distributions with varying \slepl masses, all scaled to 100
fb$^{-1}$. The mass is then taken as the value with lowest $\chi^2_N$.
10000 such experiments were performed.

There are very few signal events; both the signal and the backgrounds,
in particular the WW background, fluctuate and this makes 
it hard to do a precise mass measurement. 
Most experiments fall within 3 GeV of the expected value, but 
the distribution shows significant non-gaussian tails. 
From a gaussian fit to the distribution of results we quote 
a statistical uncertainty of 2.8~GeV
on the determination of the position of the  M$_{T2}$ edge.\par
A test where the Pythia WW background was replaced by a WW sample 
from Herwig was also performed. 
The result was very similar but with a somewhat higher uncertainty
on the edge position.




\subsubsection{Heavy gaugino analysis} 
%
We present here a brief overview of the full analysis 
described in \cite{gpchi04}.\\
Profiting from the high statistics sample of squarks
which will be collected at the LHC for the considered point,
we search for the rare decays 
of the squarks into the heavier gauginos. \\
\begin{table}
\begin{center}
\caption{Branching ratios for $\tu_L$ and $\td_L$ at SPS~1a, from
ISAJET 7.58 \cite{ISAJET}.\label{tab:bra}}
\begin{tabular}{|l|c|l|c|}
\hline
Decay & BR (\%) & Decay & BR (\%) \\
\hline
$\tu_L\to\tchi^0_1 u$  & 0.6 & $\td_L\to\tchi^0_1 d$ & 2.4 \\
$\tu_L\to\tchi^0_2 u$  & 31.8 & $\td_L\to\tchi^0_2 d$ & 31.0 \\
$\tu_L\to\tchi^0_3 u$  & 0.09 & $\td_L\to\tchi^0_3 d$ & 0.15 \\
$\tu_L\to\tchi^0_4 u$  & 1.0 & $\td_L\to\tchi^0_4 d$ & 4.0 \\
$\tu_L\to\tchi^+_1 d$  & 65.3 & $\td_L\to\tchi^-_1 u$ & 60.9 \\
$\tu_L\to\tchi^+_2 d$  & 1.23 & $\td_L\to\tchi^-_2 u$ & 4.06 \\
\hline
\end{tabular}
\end{center}
\end{table}
\indent
The branching fractions of the squarks into gauginos
for Point SPS~1a are given in Table~\ref{tab:bra}.
The decays into $\tilde\chi^0_4$ and
$\tchi^+_2$ dominate the ones into $\tilde\chi^0_3$.
In fact, after diagonalization of the 
mixing matrix, whereas 
the $\tilde\chi^0_3$ is almost exclusively higgsino,
the $\tilde\chi^0_4$ has typically some gaugino admixture, yielding 
a significant coupling to the $\tilde q_L$, and a BR of a few percent. 
The same is true for the $\tchi^+_2$.\\
\indent
A similar signature as the one from the decay 
$\tchi^0_2\rightarrow\tl_R l$ studied in the previous
sections can be exploited for the decay of the 
heavier gauginos. The uncorrelated lepton pairs can still be subtracted
using lepton flavour correlation, and the background from 
$\tchi^0_2$ decays is eliminated by simply considering lepton-lepton
invariant masses above the $\tchi^0_2$ kinematic edge.\\
\indent
Three decay chains for $\tchi^0_4$ and one decay chain for $\tchi^\pm_2$ 
do provide a signal with correlated lepton flavour:
\begin{displaymath}
\begin{array}{ r l l}
\tq_L\rightarrow & \tchi^0_4  & q                          \\
          &  \bentarrow &  \tilde \ell_{R}^{\pm}~~ \ell^{\mp}   \\
          &             &  \bentarrow \lsp  ~~ \ell^{\pm}~~~~~~~~~~[D1]\\
\end{array}
\end{displaymath}
\begin{displaymath}
\begin{array}{ r l l}
\tq_L\rightarrow & \tchi^0_4  & q                          \\
          &  \bentarrow &  \tilde \ell_{L}^{\pm}~~ \ell^{\mp}   \\
          &             &  \bentarrow \lsp  ~~ \ell^{\pm}~~~~~~~~~~[D2] \\
          &             &  \bentarrow \tilde\chi^0_2 ~~\ell^{\pm}~~~~~~~~~~[D3]\\
\end{array}
\end{displaymath}
\begin{displaymath}
\begin{array}{ r l l}
\tq_L\rightarrow & \tchi^{\pm}_2 & q^{\prime}                      \\
          &  \bentarrow &  \tn_{\ell}~~ \ell^{\pm}   \\
          &             &  \bentarrow \tchi^{\pm}_1~~ \ell^{\mp}~~~~~~~~~~[D4] \\
\end{array}
\end{displaymath}
The shape of the two-lepton invariant mass for each of the 
[D1]-[D4] chains displays an edge, which, for 
each decay $p1\rightarrow p2\ell^\pm\rightarrow p3\ell^\pm\ell^\mp$
is at the position $m_{l^+l^-}^{max}$ given by the expression:
$$
m_{l^+l^-}^{max} = m_{p1} \sqrt{1-\frac{m^2_{p2}}{m^2_{p1}}}
\sqrt{1-\frac{m^2_{p3}}{m^2_{p2}}}
$$
This is illustrated in Figure~\ref{fig:tota}.
The experimentally observed shape is the sum of the four shapes
from the decays [D1] to [D4], shown in different colors in the 
figure, 
weighted by the relative production rate. We observe 
that for Point SPS~1a the outermost edge is from decay [D2].
\begin{figure}
\dofigs{0.5\textwidth}{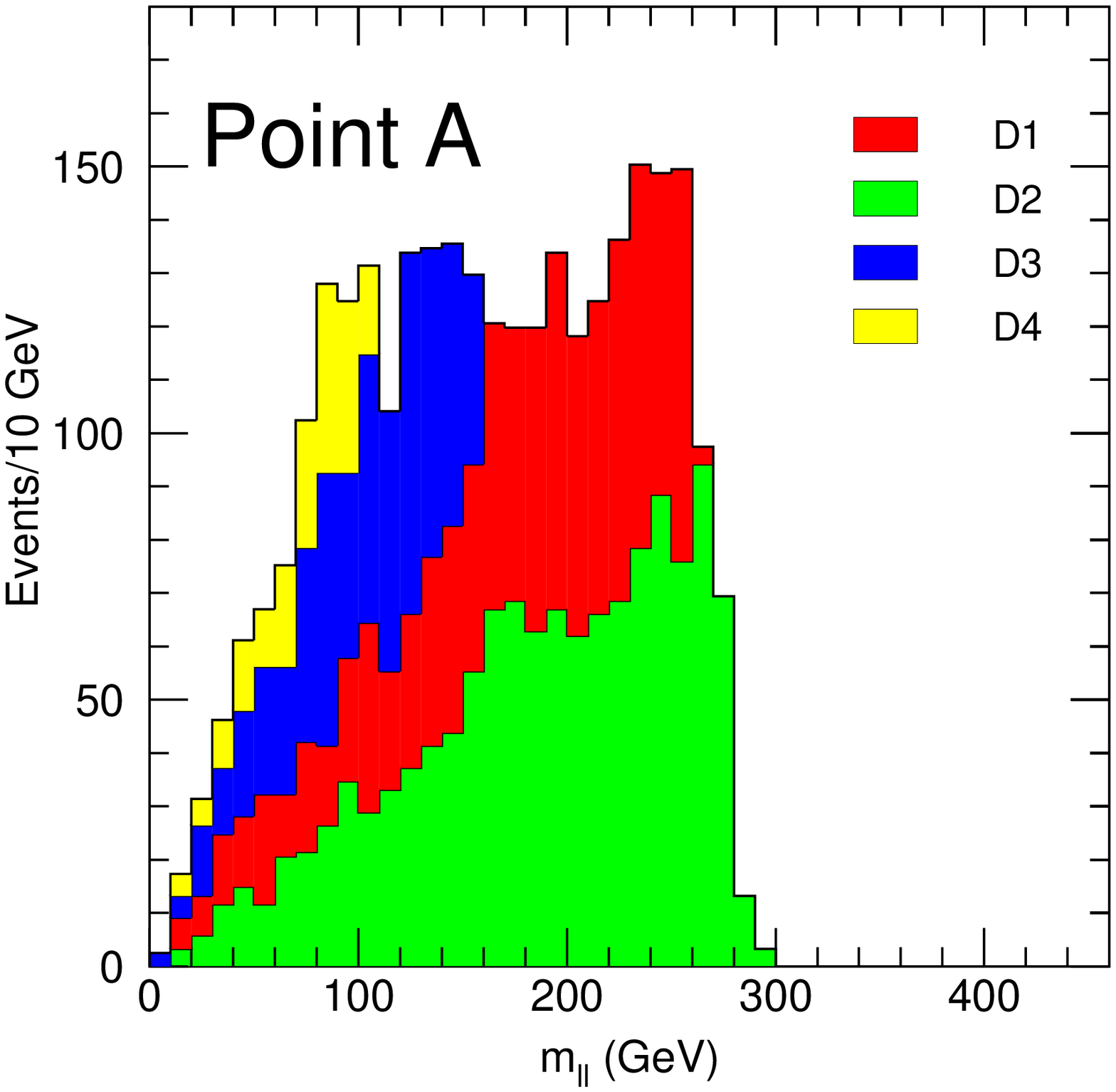}{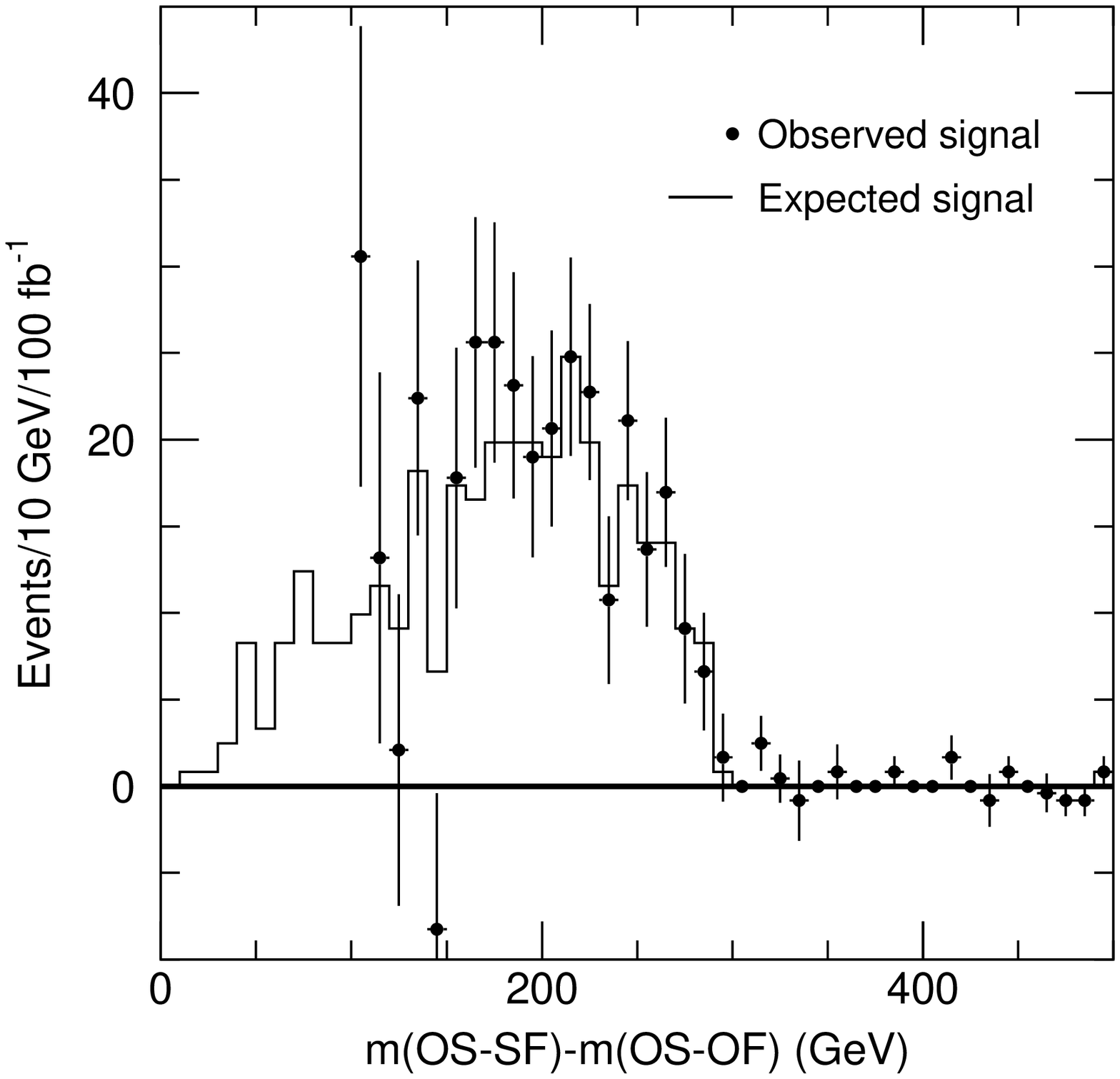}
\caption{  
Point SPS~1a. Left: distribution of $m_{\ell\ell}$ before cuts for 
the decay chains [D1] (red), [D2] (green), [D3] (blue) and
[D4] (yellow). Right: reconstructed
$m_{\ell\ell}$ spectrum after cuts and after subtracting
opposite-flavour lepton pairs from same-flavour lepton pairs.
The black points show the result of the subtraction, with
the statistical errors for 100~fb$^{-1}$. The full line is the spectrum
for events containing the desired decay chains.
}
\label{fig:tota}
\end{figure}

In order to reduce the Standard Model background 
below the SUSY background from uncorrelated gauginos decays,
we require:
\begin{itemize}
\item
Exactly two isolated opposite-sign same-flavour leptons with \\
\mbox{$P_T(1)>20$~GeV}, \mbox{$P_T(2)>10$~GeV}
\item
\ptmiss $>$ 100~GeV, at least four jets, $P_T(j1)>150$~GeV, $P_T(j2)>100$~GeV,
$M_{eff}>600$~GeV.
\item
$M_{T2} > 80$~GeV, where $M_{T2}$ is a special case of a variable discussed in \cite{lester}. 
\item
$m_{\ell^+\ell^-}>100$~GeV
\end{itemize}
\indent
The following backgrounds were considered: $\bar tt$, $Z$+jets, $W$+jets,
and produced with the PYTHIA \cite{sec4_PYTHIA} generator.\\
After these cuts, the SM background is $\sim$70 events, of 
which 42 are from $\bar tt$ and 28 from $Z$+jets for an 
integrated luminosity of 100~fb$^{-1}$. This is well
below the SUSY background which, for the SPS~1a point
is  $\sim$250 events. \\
The efficiency for the signal after cuts is $\sim$10\%.\\
\indent
The $m_{\ell^+\ell^-}$ spectra after subtraction are shown 
superimposed to the expected signal shape in the 
right side of Figure~\ref{fig:tota}. 
In Table~\ref{tab:res} we give
the numbers of signal and background events inside
the mass bin  150--290 GeV which maximizes the significance.\par 
\begin{table}
\begin{center}
\caption{Results after cuts for Point SPS~1a. We show in 
column 2 the number of signal events, in 3 the statistical significance,
in 4  the total background, in 5 the SM background, and in 6 the invariant mass
interval chosen to optimize the significance. The assumed statistics is 
100~fb$^{-1}$.}
\label{tab:res}
\begin{tabular}{|l|c|c|c|c|c|}
\hline
Point & $N_{ev}$ Signal  & Significance & Total bck. & SM bck. & Interval (GeV) \\   
\hline 
SPS~1a  &   259.1  $\pm$     21.1  &    12.3  &    92.3  &    27.1  &   150--290\\
\hline 
\end{tabular}
\end{center}
\end{table}
The invariant mass distribution after background
subtraction can be used to extract a measurement of the relevant
lepton-lepton edge.
The statistical precision of the edge measurement has been 
determined through a set of Monte Carlo experiments, each 
corresponding to an integrated luminosity of 100~fb$^{-1}$.
For each experiment the edge position was evaluated using a sliding window 
algorithm.
From this approach, the statistical error
on the $\tchi^0_4$ edge was evaluated to be $\sim4$~GeV for 
an integrated luminosity of 100~fb$^{-1}$. 

\subsubsection{Measurement of $\ttau_1$ mass}
\label{sec:411_tau}
The dominant decay mode of the $\tchi^0_2$ in Point SPS~1a
is the decay to the pair $\ttau_1\tau$, which is
favoured over the decay to $\tl_R$ due to the relatively 
high value of $\tan\beta$. 
These decays can be identified in the ATLAS 
detector through the tagging of the hadronic decays
of the $\tau$ which present a characteristic pattern in 
the detector. The graph showing the rejection 
power on quark jets as a function of the $\tau$ efficiency,
from \cite{TDR} is shown in Fig.~\ref{fig:taurej}.
A typical figure is a rejection of a factor 100 on QCD jets
for a $\tau$ efficiency of 50\%.
\begin{figure}
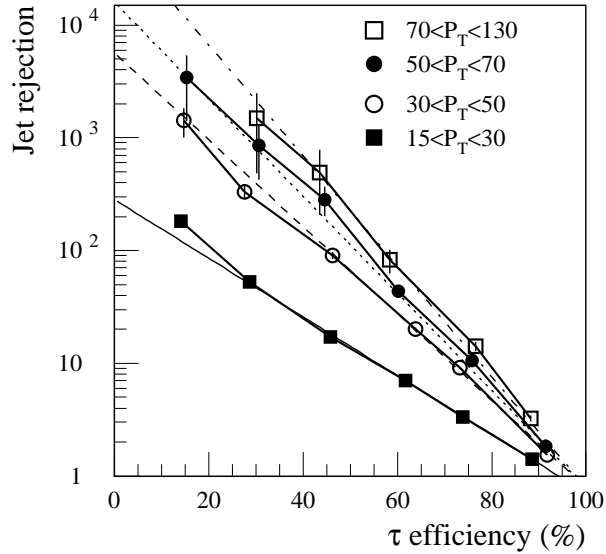

\dofig{0.5\textwidth}{sec411_fig931.epsi}
\caption{Rejection factor of QCD jets as a function of
the $\tau$ efficiency for different intervals of $\tau$
transverse momentum.}
\label{fig:taurej}
\end{figure}
The rejection values quoted are referred to low luminosity,
and we perform the analysis on an integrated luminosity 
of 30~fb$^{-1}$.\par
The first step is the reduction of the Standard Model background
to a manageable level with 
hard cuts on jet multiplicity and $\Etmiss$.
We require therefore:
\begin{itemize}
\item
At least 4 jets: $p_T>100, 50, 50, 50$~GeV
\item
$\Etmiss>max(100~{\mathrm GeV}, 0.2\times M_{eff})$
\item
$M_{eff}>500$~GeV
\item
two jets tagged as $\tau$ with $P_T(\tau_1)>30$~GeV,
and $P_T(\tau_2)>25$~GeV, with opposite charge.
\end{itemize}
Backgrounds from QCD jets, W+jets, Z+jets, $\bar tt$ were
considered and generated with PYTHIA. The use of a parton-shower 
Monte Carlo to estimate multi-jet backgrounds is known 
to underestimate the backgrounds, and the present analysis
should be considered a preliminary study, waiting for a more detailed study 
involving specialised generators. 
After the requirement on tau-tagged jets is 
applied, the dominant backgrounds are $Z$+jets and $\bar tt$.
The total SM background is reduced to about 1/10 of the SUSY backgrounds
in the $\tau\tau$ edge region and the conclusions of the analysis
would only marginally be affected by a significant increase in background.\par
The invariant mass distribution for the two $\tau$ candidates is
shown in Fig.~\ref{fig:tausub}. The background from QCD jets misidentified 
as $\tau$ jets will have random sign assignment, therefore the 
distribution of the same-sign $\tau$-candidate pairs, also shown 
in Fig.~\ref{fig:tausub}, will give a good description of the dominant 
SUSY background, constituted by a $\tau$ from $\tchi^\pm_1$ decay and a 
misidentified
jet, and of SM background. Part of the background with two real $\tau$
coming from two $\tchi^\pm_1$ will also have same-sign $\tau$, as the bulk
of SUSY production in Point SPS~1a is $\tg\tg$ and $\tg\tq$ production.\par
We can thus subtract the same-sign $\tau$ pairs from the opposite-sign
$\tau$ pairs. The result is shown in Fig.~\ref{fig:pltau} where the
resulting distribution is shown as the full points with error bars.
The distribution shown as a full line are the events containing two
$\tau$ from $\tchi^0_2$ decays, and the grey area shows the events from 
two uncorrelated $\tchi^\pm_1$ decays. The observed distribution
has a clear structure with an end-point corresponding to the edge
of the invariant mass distribution of the undecayed $\tau$, shown
as a dashed line in Fig.~\ref{fig:pltau}.
\begin{figure}
\dofig{0.6\textwidth}{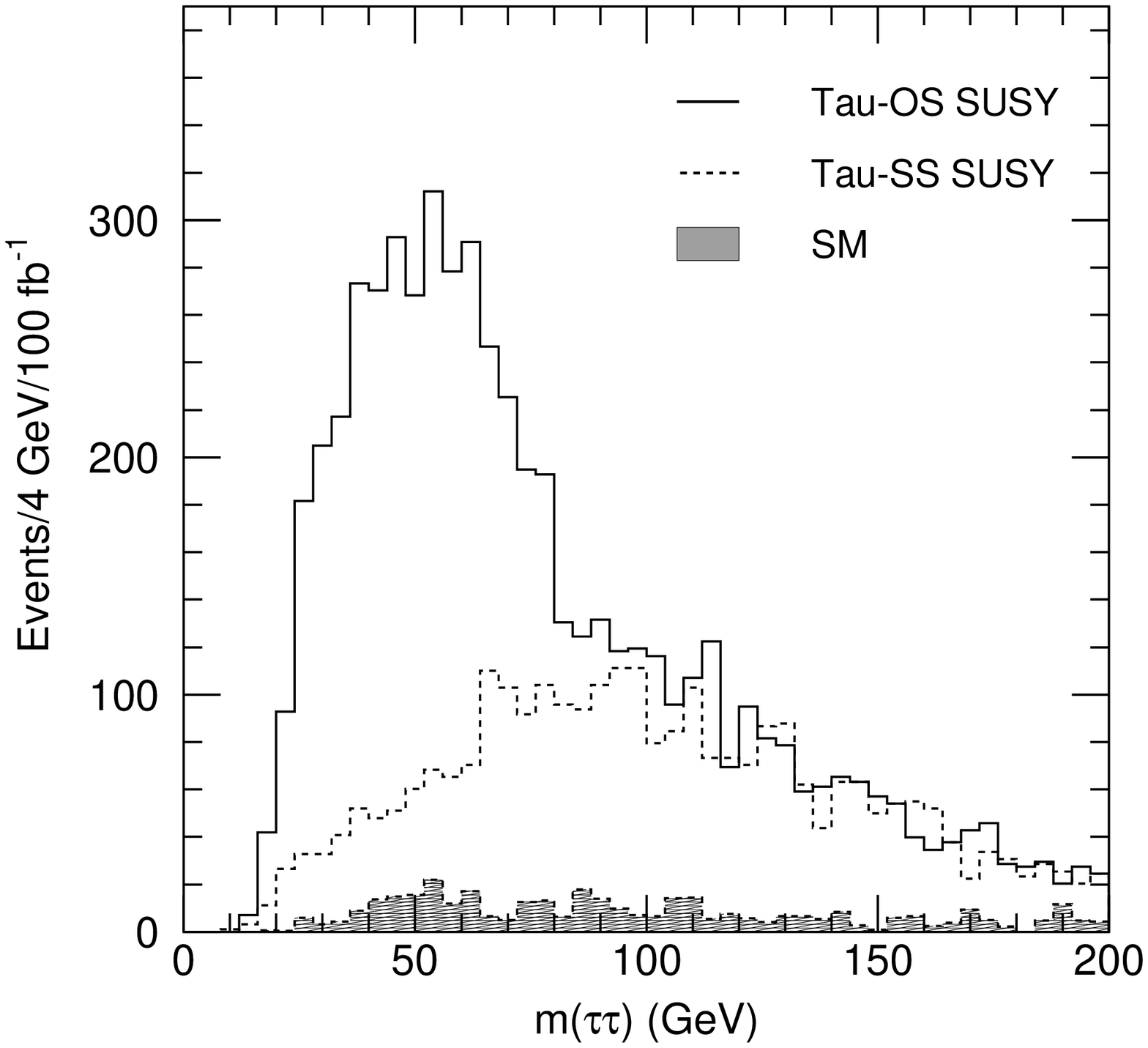}
\caption{ Invariant mass distribution of: opposite sign $\tau$-tagged
jets (full line); same-sign $\tau$-tagged jets (dashed line), 
SM background (grey). The integrated luminosity is 30~fb$^{-1}$. }
\label{fig:tausub}
\dofig{0.6\textwidth}{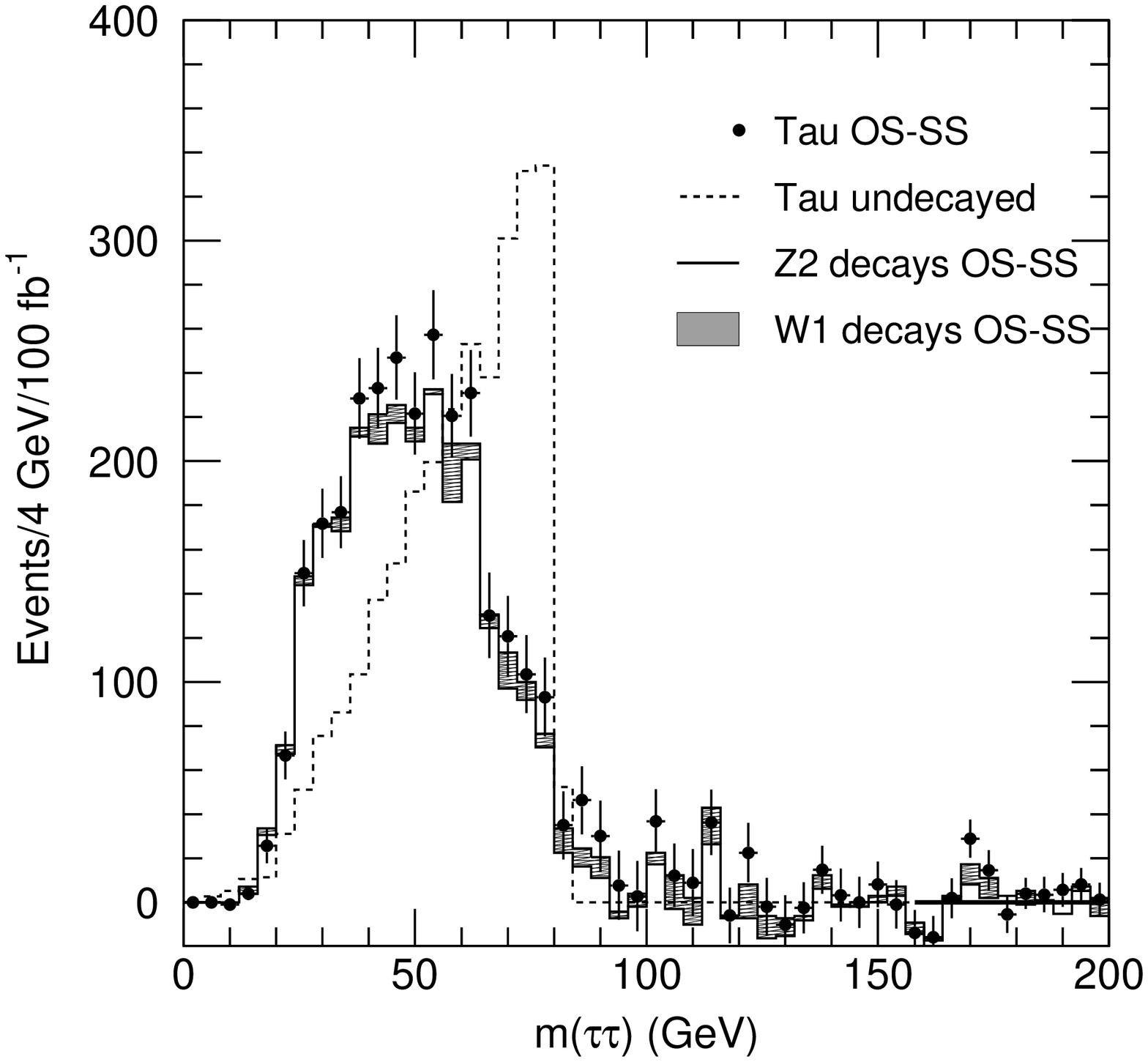}
\caption{ Invariant mass distribution of $\tau$-tagged jets after
subtraction of same-sign background (full circles). Superimposed are shown the
distributions for: $\tchi^0_2\rightarrow\ttau_1\tau$ signal (full line),
$\tchi^\pm_1$ decays (grey), undecayed $\tau$'s from $\tchi^0_2$ 
(dashed line).}
\label{fig:pltau}.
\end{figure}

Although the distribution in Fig.~\ref{fig:pltau} does not have a sharp edge,
it is nonetheless sensitive to the edge position. This has been verified, 
by producing the same distributions as for the SPS~1a point, but with 
values of the $\ttau_1$ mass such as to displace the edge position
down by five and ten GeV. On a statistical basis two distributions
displaced by five GeV can be distinguished. The identification 
of the actual end-point of the $\tau\tau$ mass distribution
requires however a detailed simulation study, outside the
scope of this analysis. We conservatively
quote a  systematic uncertainty of 5 GeV on the determination
of the edge position.

\subsubsection{Measurement of $q_R$ mass}
In mSUGRA models the $\lsp$ is essentially a bino, and the
$\tchi^0_2$ a Wino. Therefore the $\tq_R$ which has zero SU(2)
charge decays with almost 100\% BR into the corresponding
quark and the $\lsp$.\par 
In the case of SPS~1a, where the squarks are lighter
than the gluinos by only a few tens of GeV, the bulk of the 
SUSY production is given by gluino production.
The signature of events where both gluinos decay into two $\tq_R$
is thus the presence of two low $P_T$ jets from the $\tg\rightarrow q\tq_R$
decays, two high  $P_T$ jets from the $\tq_R$ decay, and $\Etmiss$. 
We apply the same cuts as for the 
hadronic analysis discussed above, and in addition we require:
\begin{itemize}
\item
Leading jet with $P_T>300$~GeV 
\item
At most 4 jets with $P_T>50$~GeV 
\item
Veto leptons, $\tau$-jets, $b$-jets 
\end{itemize}
The second and third cut are aimed at reducing the background 
from $\tq_L$ and third generation squarks. 
For the events containing two $\tq_R$ decaying to $q\lsp$,
at the end of the cascade decays there are 
two particles with the same mass, both decaying to a jet and 
a $\lsp$. A very useful variable in this situation is the 
Cambridge $\MT2$ variable calculated on the two leading jets.
We show in Fig~\ref{fig:sqr} the distribution of $\MT2$
for the SUSY events, and for the background (in red), for
a $m(\lsp)=96$~GeV. A clear edge structure can be seen, 
which approximately coincides with the $\tq_R$ mass.
The considerations on SM backgrounds given in \ref{sec:411_tau} 
apply here as well, from Fig~\ref{fig:sqr} the detectability  of an
edge structure would be guaranteed even in the presence 
of a much higher SM background.\par
At this point 65\% of the events contain two $\tq_R$, 
30\% only one $\tq_R$, and 5\% no $\tq_R$.
The bulk of the surviving SUSY background are events with one $\tq_L$ produced,
decaying either to $q\tchi^0_2$ or to $\tchi^\pm_1$.
For the SUSY background events 
the resulting edge shape
is at a lower value than for the signal,
as on the $\tq_L$ side the mass difference between the squark and the
corresponding decay gaugino is smaller than for the $\tq_R$.\par
In order to estimate the edge position, we parametrize 
the $\MT2$ shape
on a simulated sample of pure $\tq_R$ decays, and we perform a fit to
the distribution of
Fig~\ref{fig:sqr}. The resulting fit is shown in Fig.~\ref{fig:sqrfit}.
\begin{figure}[htbp]
\begin{center}
\dofig{0.6\textwidth}{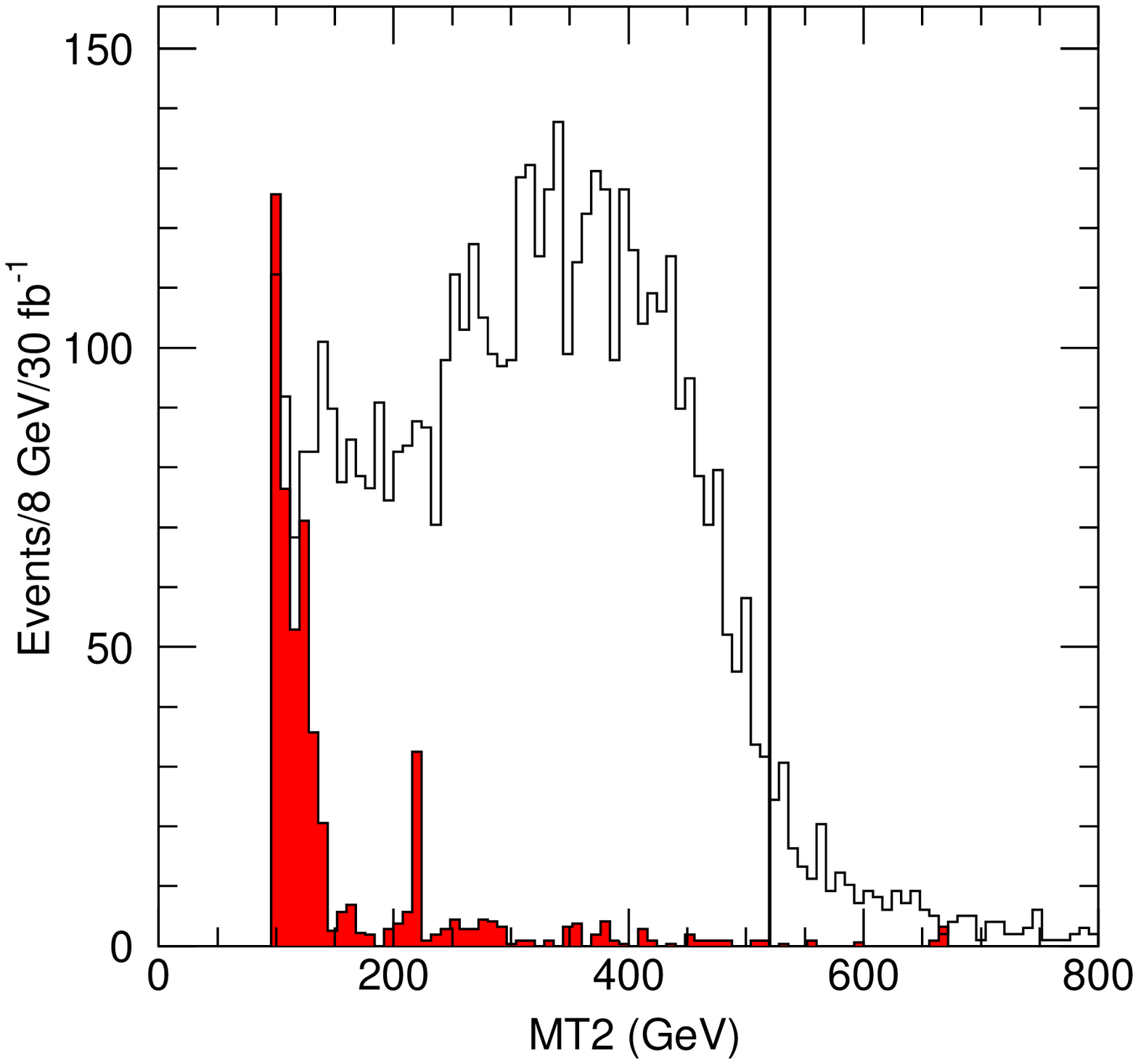}
\vskip -1cm
\caption{ Distribution of $\MT2$ for the events passing the cuts.
In red is shown the Standard Model background. The integrated statistics
in the plot is 30~fb$^{-1}$.}
\label{fig:sqr}
\end{center}
\begin{center}
\dofig{0.6\textwidth}{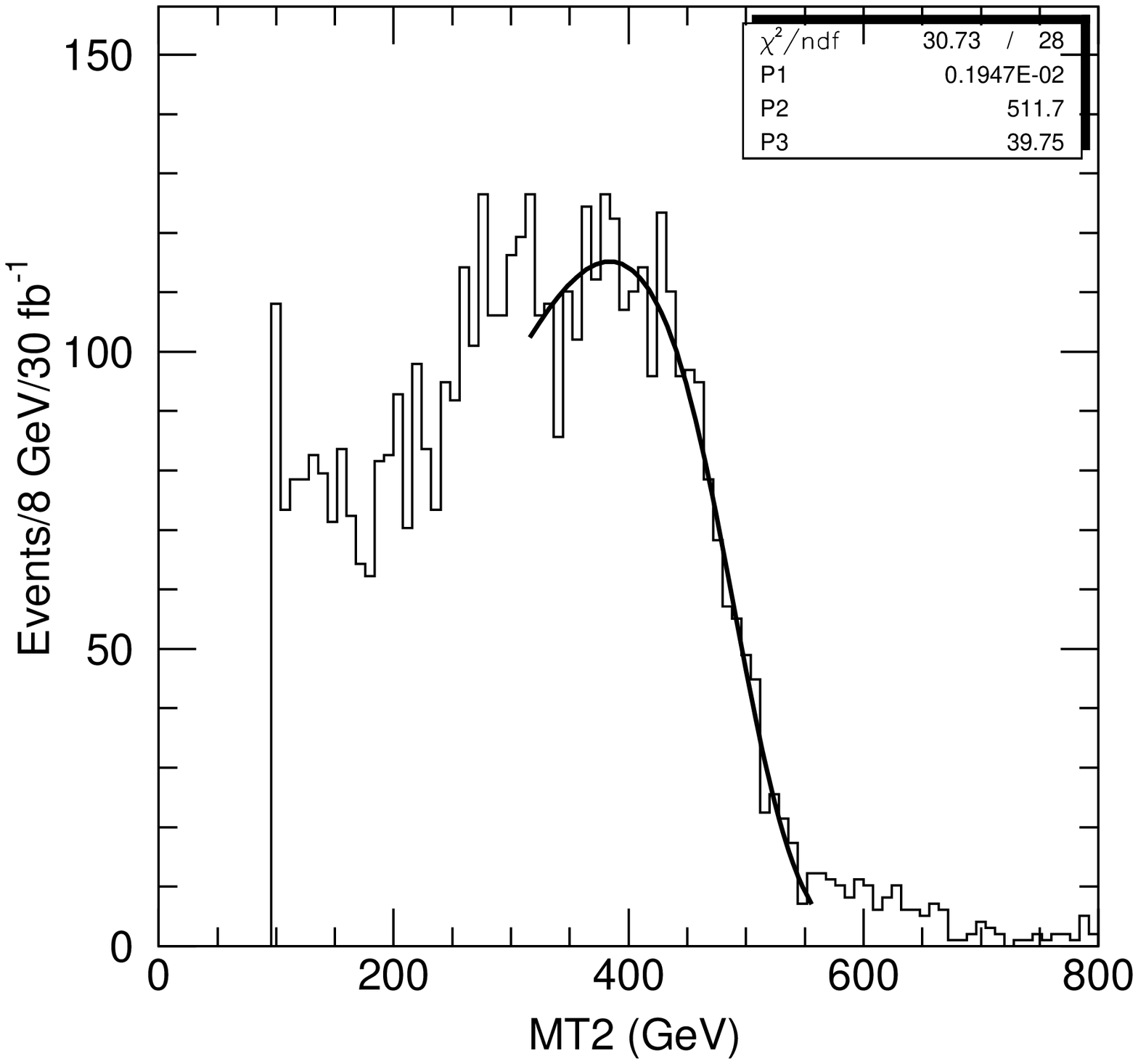}
\vskip -1cm
\caption{ Distribution of $\MT2$ for events passing the cuts. Superimposed
is the fit described in the text.}
\label{fig:sqrfit}
\end{center}
\end{figure}
The statistical error on the edge position is $\pm 3.6$~GeV for an 
integrated luminosity of 30~fb$^{-1}$. 
The position of the fitted edge is $\sim$10~GeV lower than the 
nominal one, as the presence of a significant background
with a lower edge distorts the $\MT2$ shape. 
The 10~GeV discrepancy can be taken as an approximate
estimate of the systematic uncertainty on the evaluation
of the $\tq_R$ mass through the edge fitting. As already 
discussed for the $\tq_L$ evaluation, this value of the uncertainty
is quite conservative, since in the real experiment, given that most 
of the involved sparticle masses will be known, it will be possible 
to estimate the background shape and account for it in the fit.
The $\MT2$ evaluation relies on the knowledge of $m(\lsp)$. For small
variations of $m(\lsp)$, the $\MT2$ measurement amounts to 
measuring $m(\tq_R)-m(\lsp)$, which should be taken as the output
result of this analysis.

\renewcommand{\lsim}{\buildrel<\over{_\sim}}
\newcommand{\Br}{{\rm BR}}       
\newcommand{\mtbfit}{M_{tb}^{\rm fit}}
\newcommand{\mtbw}{M_{tb}^{\rm w}}
\newcommand{\mtb}{m_{tb}}
\newcommand{\mtbmax}{m_{tb}^{\rm max}}
\newcommand{\stp}{{\tilde{t}_1}}
\newcommand{\mstp}{{m_{\tilde{t}_1}}}
\newcommand{\stps}{{\tilde{t}_2}}
\newcommand{\sbt}{{\tilde{b}_1}}
\newcommand{\sbti}{{\tilde{b}_i}}
\newcommand{\msbt}{m_{\tilde{b}_1}}
\newcommand{\sbts}{{\tilde{b}_2}}
\newcommand{\msbts}{m_{\tilde{b}_2}}
\newcommand{\glu}{{\tilde{g}}}
\newcommand{\mglu}{m_{\tilde{g}}}
\newcommand{\chapm}{{\tilde{\chi}^{\pm}_1}}
\newcommand{\cha}{{\tilde{\chi}^-_1}}
\newcommand{\chap}{{\tilde{\chi}^+_1}}
\newcommand{\chaspm}{{\tilde{\chi}^{\pm}_2}}
\newcommand{\chas}{{\tilde{\chi}^-_2}}
\newcommand{\chaps}{{\tilde{\chi}^+_2}}
\newcommand{\neu}{{\tilde{\chi}^0_1}}
\newcommand{\neus}{{\tilde{\chi}^0_2}}
\newcommand{\neuth}{{\tilde{\chi}^0_3}}
\newcommand{\neufr}{{\tilde{\chi}^0_4}}
\newcommand{\tw}{\tilde{W}}
\newcommand{\nedge}{N_{\rm edge}}
\newcommand{\Bredge}{{\rm BR(edge)}}  
\newcommand{\nall}{N_{\rm all}}
\newcommand{\nfit}{N_{\rm fit}}
\newcommand{\nprod}{N_{\rm prod}}


\subsubsection{Measurements in the stop sector}
At the LHC,
we may be able to access the nature of the stop and sbottom 
if they are lighter than the gluino ($\tilde{g}$). 
This is because they  copiously arise from the gluino decay. 
The relevant decay modes for $\tilde{b}_i$ ($i=1,2$), $\tilde{t}_1$,
to charginos $\tilde{\chi}_j^{\pm}$ ($j=1,2$) or neutralinos 
$\tilde{\chi}^0_j$ ($j=1,2,3,4$) are  
listed below (indices to distinguish a particle and its anti-particle 
is suppressed unless otherwise stated),
\begin{eqnarray}
{\rm (I)}_j&~&\glu\rightarrow b\sbt
\rightarrow bb\tilde{\chi}^0_j
\ (\rightarrow bbl^+l^-\tilde{\chi}^0_1),\cr
{\rm (II)}_j&~&\glu\rightarrow t\stp \rightarrow tt\tilde{\chi}^0_j,\cr
{\rm (III)}_j&~&\glu\rightarrow t\stp \rightarrow 
tb\tilde{\chi}^{\pm}_j,\cr
{\rm (III)}_{ij}&~&\glu\rightarrow
b \tilde{b}_i \rightarrow  b W\stp \rightarrow 
bbW\tilde{\chi}^{\pm}_j,\cr
{\rm (IV)}_{ij}&~&\glu\rightarrow b \tilde{b}_i\rightarrow tb 
\tilde{\chi}^{\pm}_j.
\label{gluinodecay}
\end{eqnarray}
In the previous literature~\cite{TDR,Hinchliffe:1996iu}, the lighter 
sbottom $\tilde{b}_1$
is often studied through the mode (I)$_2$, namely the
$bb\tilde{\chi}^0_2\rightarrow bbl^+l^-\tilde{\chi}^0_1$
channel. This mode is important when the second lightest neutralino 
$\tilde{\chi}^0_2$ has
substantial branching ratios into leptons. 
The measurement of sbottom and gluino masses 
from the study of these decay chains is discussed above.

In Refs. \cite{Hisano:2002xq,Hisano:2003qu} we proposed to measure the
edge position of the $m_{tb}$ distribution for the modes (III)$_{1}$
and (IV)$_{11}$, where $m_{tb}$ is the invariant mass of a top-bottom
($tb$) system.  The decay modes are expected to be dominant in the
minimal supergravity model (MSUGRA), as the branching ratios
$\Br(\sbt (\stp) \rightarrow t (b) \tilde{\chi}_1^{\pm})$ could be as
large as 60\%.  We focused on the reconstruction of hadronic decays of
the top quark, because the $m_{tb}$ distribution of the decay makes a
clear ``edge'' in this case.  The parton level $m_{tb}$ distributions
for the modes (III)$_j$ and (IV)$_{ij}$ are expressed as functions of
$\mglu$, $m_{\tilde{t}_1}$, $m_{\tilde{b}_i}$, and the chargino mass
$m_{\tilde{\chi}_j^{\pm}}$: $d\Gamma/dm_{tb}\propto m_{tb}$. 

The events containing $tb$ are selected by requiring the following 
conditions in addition to the standard SUSY cuts: 1) Two 
and only two $b$-jets. 2) Jet pairs consistent with a hadronic 
$W$ boson decay, $\vert m_{jj}-m_W\vert < 15$~GeV.  3) 
The invariant mass of the jet pair and one of the $b$-jets, $m_{bjj}$,
satisfies $\vert m_{bjj}-m_t \vert<30$~GeV. 
The events after the selection contain misreconstructed events.
We use a $W$ sideband method to 
estimate the background distribution due to misreconstructed events. 
Monte Carlo simulations show that the distribution of the signal modes 
(III) and (IV) after subtracting the background is very 
close to the parton level distribution.  The distribution  
is then fitted by a simple fitting function 
described with the end point $\mtbfit$, the edge height $h$ per $\Delta m$
bin, and a smearing parameter. We assume that the signal distribution 
is sitting on a linearly decreasing background near the edge, 
which is also determined by the fit.

The edge position (end point) of the $m_{tb}$ distribution $M_{tb}$ for
the modes (III)$_j$ and (IV)$_{ij}$ are written as follows;
\begin{eqnarray}
&&M_{tb}^2({\rm III})_j =  m_t^2
+\frac{m_\stp^2-m^2_{\tilde{\chi}^{\pm}_j}}{2 m_\stp^2}
\left\{
(m_\glu^2-m_{\stp}^2-m_t^2)
\right.
\nonumber\\
&&
\left.
+
\sqrt{
(m_\glu^2-(m_{\stp}-m_t)^2)
(m_\glu^2-(m_{\stp}+m_t)^2)
}
\right\},\nonumber
\label{tb_stop}
\\
&&M_{tb}^2({\rm IV})_{ij}=  m_t^2
+\frac{m_\glu^2-m_{\tilde{b}_i}^2}{2 m_{\tilde{b}_i}^2}
\left\{
(m_{\tilde{b}_i}^2-m_{\tilde{\chi}^{\pm}_j}^2+m_t^2)
\right.
\nonumber\\
&&
\left.
+
\sqrt{
(m_{\tilde{b}_i}^2-(m_{\tilde{\chi}^{\pm}_j}-m_t)^2)
(m_{\tilde{b}_i}^2-(m_{\tilde{\chi}^{\pm}_j}+m_t)^2)
}
\right\}.
\label{tb_sbot}
\end{eqnarray}
In some model parameters
$M_{tb}({\rm III})_1$ is very close to $M_{tb}({\rm IV})_{11}$.  When
they are experimentally indistinguishable, it is convenient to define
a weighted mean of the end points;
\begin{eqnarray}
\mtbw&=&\frac{\Br({\rm III}) M_{tb}({\rm III})_1+
\Br({\rm IV})_{11}M_{tb}({\rm IV})_{11}}
{\Br({\rm III})+\Br({\rm IV})_{11}},
\cr
\Br({\rm III}) & \equiv & 
\Br({\rm III})_1 +\Br({\rm III})_{11}+ \Br({\rm III})_{21}.
\label{mtbw}
\end{eqnarray}
As the final states $bbW$ from the decay chain  
$\tilde{g}\rightarrow b\tilde{b}_i\rightarrow bW\tilde{t}_1
\rightarrow bbW\tilde{\chi}^{\pm}_1$
(mode (III)$_{i1}$) could have an irreducible 
contribution to the $tb$ final state,  they are included in
the definition of $\mtbw$. 

\begin{figure}[ht]
\centerline{\psfig{file=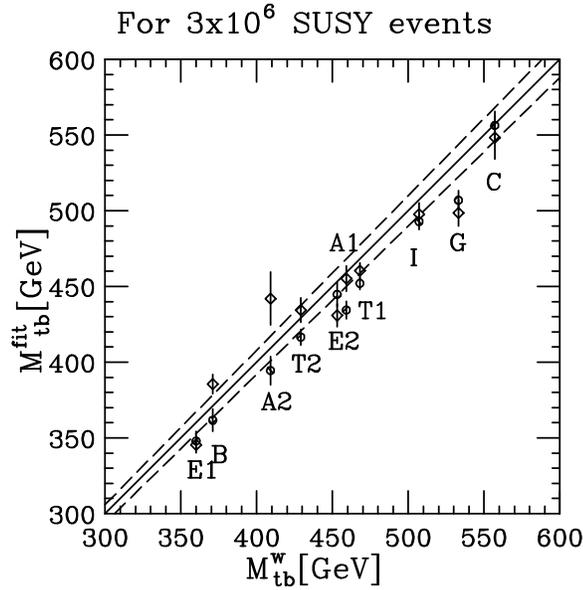,width=3in}}
\caption{ 
Relation between 
$M^{\rm w}_{tb}$ and 
$M^{\rm fit}_{tb}$ for the sample points. The solid line corresponds to  
$M^{\rm w}_{tb}= M^{\rm fit}_{tb}$ and dashed lines 
to $M^{\rm w}_{tb}(1\pm 0.02)= M^{\rm fit}_{tb}$. Bars with a diamond
and a circle correspond to  PYTHIA  and HERWIG samples, respectively. 
From \cite{Hisano:2003qu}}
\label{endpoint}
\end{figure}

We demonstrate the viability of the method by studying 
the relation between $\mtbw$ and $\mtbfit$ for several 
model points, decribed in detail in \cite{Hisano:2003qu}.
The results corresponding to a generated statistics of 
$3\times 10^6$ SUSY events for each model point 
are shown in Fig. \ref{endpoint}. The fitted value
$\mtbfit$ increases linearly with the weighted  end point $\mtbw$. 
The  $\mtbfit$ tends to be lower 
than $\mtbw$, which is the effect of
particles missed outside the jet cones.  This  is similar to what is 
found in previous studies and should be corrected 
by more careful study including the modification of 
jet definition.

The $\mtbfit$ contains the 
information on $\tilde{t}_1$ as can be seen in Eq.(\ref{mtbw}). 
Given the very precise  electroweak SUSY parameter 
measurements  and  $\tilde{g}$ and $\tilde{b}$ mass measurements with
LHC/LC, the remaining dominant uncertainty are in 
the stop mass $m_{\tilde{t}_1}$, and 
the stop and sbottom mixing angles $\theta_t$ and $\theta_b$. 
The results of the 
end point fit at SPS1 for the $3\times 10^6$ Monte Carlo  events 
are $\mtbfit=363.9\pm 4.8$~GeV and 
$h=267.3.2\pm 20.8$  ($\Delta m=10$~GeV). We show the $m_{tb}$ distribution 
and fitting curve in Fig. \ref{sps1}.

\begin{figure}
\centerline{
\psfig{file=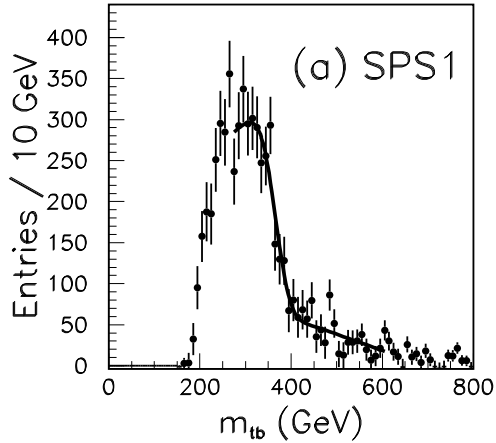,width=3in}
}
\caption{The $m_{tb}$ distributions at
 SPS1. The fit curves are also shown. From \cite{Hisano:2003qu}.}
\label{sps1}
\end{figure}

We will now discuss the relation between the edge height $h$
and the number of
reconstructed $tb$ events.  The 
number of the  reconstructed ``edge'' events
$\nedge$ arising from the decay
chains (III) and (IV) may be estimated from $M_{tb}^{\rm fit}$ and  $h$
per bin size $\Delta m$ as follows,
\begin{equation}
\nedge   \sim \nfit = 
\frac{h}{2}\left(\frac{m_t}{M^{\rm fit}_{tb}}  + 1\right)\times 
\frac{M^{\rm fit}_{tb}-m_{t}}{\Delta m }.
\label{nsig}\end{equation}
This formula is obtained by assuming the parton level distribution,
and equating the minimum of the $m_{tb}$ distribution from the decay
chain (III) or (IV) to $m_t$. The consistency between 
$\nedge\sim\nfit$ is checked by using 
the generator information in Ref. \cite{Hisano:2003qu}.

In the MSUGRA model, 
the decay modes which involve $W$ bosons (modes (II), (III) and 
(IV)) often dominate  the gluino decays to $bbX$. 
Because the events with $W$ bosons should remain after 
the $W$ sideband subtraction, 
the reconstruction efficiency $\epsilon_{tb}$ is expected to be 
similar for these decay modes. 
Thus, if the contributions from the stop
or sbottom pair productions are negligible, the numbers of events with
two bottom quarks are given approximately as
\begin{eqnarray}
\nfit &\sim&  {\epsilon_{tb}}\Bredge 
 \left[2 N(\tilde{g}\tilde{g}) 
\left(1-\Br(\tilde{g}\rightarrow bbX)\right)+ N(\tilde{g}\tilde{q})
+ N(\tilde{g}\tilde{q}^*)\right],
\cr
\nall &\sim& {\epsilon_{tb}}\Br(\tilde{g}\rightarrow bbX)
\left[
2N(\tilde{g}\tilde{g}) 
\left(1-\Br(\tilde{g}\rightarrow bbX)\right)
+ N(\tilde{g}\tilde{q})+ N(\tilde{g}\tilde{q}^*)
\right], 
\end{eqnarray}
where
\begin{eqnarray}\label{nprod}
\Bredge&\equiv & \Br({\rm III})_1 + \Br({\rm III})_{11}
 + \Br({\rm III})_{21}  + \Br({\rm IV})_{11},
\end{eqnarray}
and  $\Br(\tilde{g}\to bbX)$ is the branching ratio of the gluino
decaying into stop or sbottom, thus having two bottom quarks in the
final state.  
Therefore  $\Bredge$ $/\Br(\tilde{g}\rightarrow bbX)\sim \nfit/\nall$
is expected. 

\begin{figure}
\centerline{\psfig{file=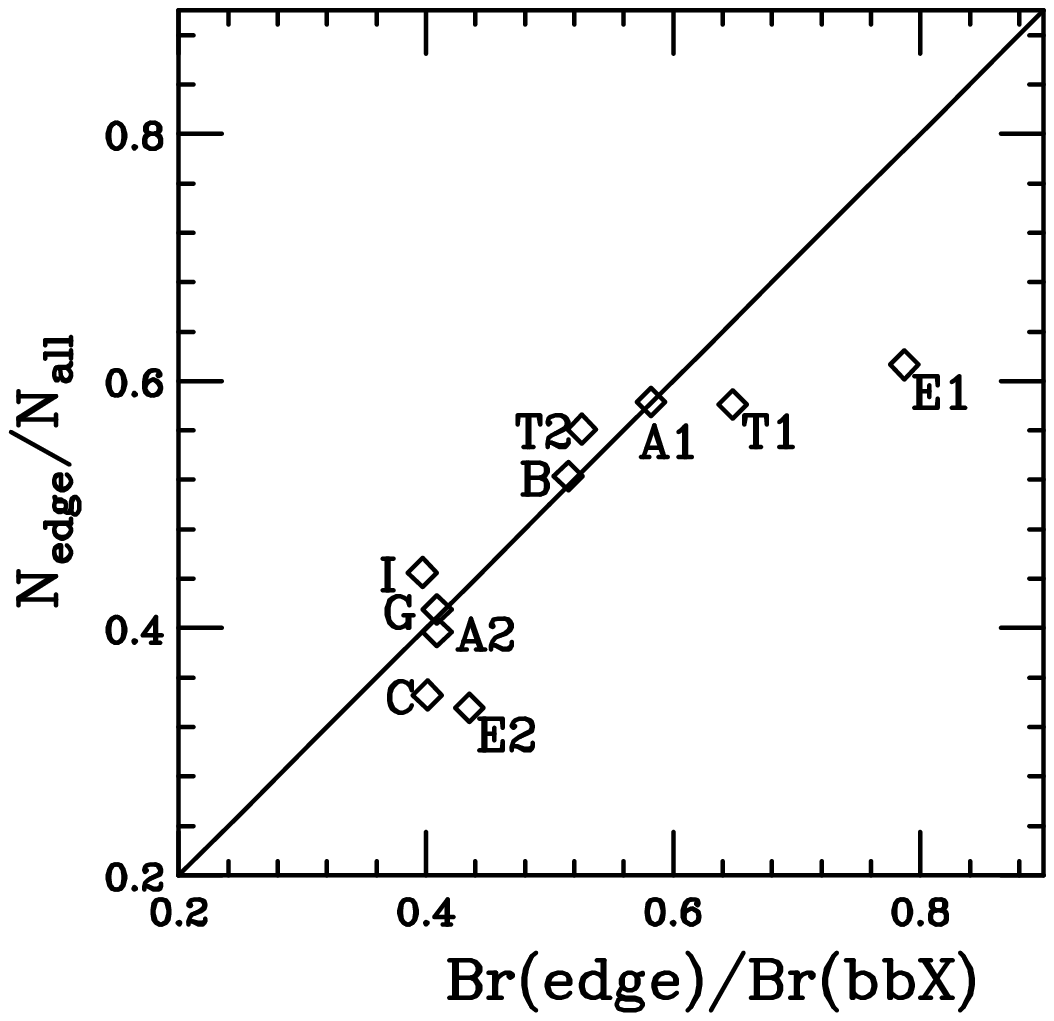,width=3in}}
\caption{Relation between
$\nedge/\nall$ and $\Bredge/\Br(\tilde{g}\rightarrow bbX)$.}
\label{ratio}
\end{figure}

In Fig.~\ref{ratio}, we plot the ratio
$\nedge/\nall$ as a function of $\Bredge$ $/\Br(\tilde{g}\rightarrow
bbX)$\footnote{ Here we plot $\nedge/\nall$ instead of $\nfit/\nall$, because
$\nfit\sim \nedge$, and the statistical
fluctuation of $\nedge$ is small ($\sim$ 2\%) with a help of the
generator information. }. The points  tend to be on the expected line 
$\nedge/\nall= \Bredge /\Br(\tilde{g}\rightarrow bbX)$. 
Some points in the plots are away from the line: The point
``C'' is off because the chargino has large branching ratios into
leptons. 
At the point ``T1'', the stop mass is
particularly light and 
$\tilde{t}_1\tilde{t}^{*}_1$ productions contributes to $\nall$. The
points ``E1'' and ``E2'' are significantly off because the first and the
second generation squarks dominantly decay into the gluino, and the 
events containing two bottom quarks are not dominant.
The description of the points are found in \cite{Hisano:2003qu}.
These exceptional cases will be easily distinguished by looking 
into the data from LHC/LC.

At the SPS1 point, we find $\nfit=3742.2\pm 291.2$ and 
 $\nall=5987.5$. This means 
$\nfit/\nall= 0.62\pm0.05$, while $\Bredge$
$/\Br(\tilde{g}\rightarrow bb X)=0.55$ is the input.

\subsubsection{Conclusions}
We have reviewed the present level of understanding of the
perspectives for SUSY measurements at the LHC with the ATLAS detector.
To this effect, we have performed an extensive and detailed study on 
the benchmark mSUGRA model SPS1a.\par
The measurement strategy is based on the identification of unique
patterns of decay products which characterise exclusive sparticle 
decay chains. For each of these decays the invariant mass distributions
of the observed decay products exhibit thresholds and end-point structures.
The kinematic structures can in turn be expressed as a function of 
the masses of the involved sparticles. By applying this procedure
to Point SPS1a, we have demonstrated that the measurements possible
at the LHC will cover most of the masses of the SUSY particles.
Information on sparticle couplings is also provided by some analyses 
e.g.  from detailed studies of the stop-sbottom sector. More studies
are however needed in order to assess the full power of the LHC 
in this field.

\subsection{\label{sec:412} 
Squark and gluino reconstruction with CMS at LHC}

{\it M.~Chiorboli, A.~Tricomi}

\vspace{1em}

\def\Journal#1#2#3#4{{#1} {\bf #2}, (#3) #4}

\def\NCA{\em Nuovo Cimento}
\def\NIM{\em Nucl. Instrum. Methods}
\def\NIMA{{\em Nucl. Instrum. Methods} A}
\def\NPB{{\em Nucl. Phys.} B}
\def\PLB{{\em Phys. Lett.}  B}
\def\PRL{\em Phys. Rev. Lett.}
\def\PRD{{\em Phys. Rev.} D}
\def\PR{\em Phys. Rev.}
\def\ZPC{{\em Z. Phys.} C}
\def\PRE{\em Phys. Rev.}
\def\MPLA{{\em Mod. Phys. Lett.} A}

\def\st{\scriptstyle}
\def\sst{\scriptscriptstyle}
\def\mco{\multicolumn}
\def\epp{\epsilon^{\prime}}
\def\vep{\varepsilon}
\def\ra{\rightarrow}
\def\ppg{\pi^+\pi^-\gamma}
\def\vp{{\bf p}}
\def\ko{K^0}
\def\kb{\bar{K^0}}
\def\mhp{$m_{H^{\pm}}$}
\def\hp{$H^\pm$}
\def\htn{$H^\pm\rightarrow\tau\nu$}
\def \ev    {\rm eV}
\def \kev   {\rm KeV}
\def \kevc  {\rm KeV/c}
\def \kevcc {\rm KeV/c$^2$}
\def \mev   {\rm MeV}
\def \mevc  {\rm MeV/c}
\def \mevcc {\rm MeV/c$^2$}
\def \gev   {\rm GeV}
\def \gevc  {\rm GeV/c}
\def \gevcc {\rm GeV/c$^2$}
\def \pb    {${~\mathrm{pb}}^{-1}$}
\def \Mw    {M$_{\mathrm{W}}$}
\def \Mz    {M$_{\mathrm{Z}}$}
\def \Mt    {M$_{top}$}
\def \Mh    {M$_{\mathrm{Higgs}}$}

\def \ups  {$\Upsilon(4S)$}
\def \WW   {${\mathrm W}^+ {\mathrm W}^-$}
\def \W   {${\mathrm W}^+$}
\def \w   {${\mathrm W}^-$}
\def \ee   {${\mathrm e}^+{\mathrm e}^-$}
\def \mm   {$\mu^+\mu^-$}
\def \tautau {$\tau^+\tau^-$}
\def \ttbar   {${\mathrm t}\overline {\mathrm t}$}
\def \bb   {${\mathrm b}\overline {\mathrm b}$}
\def \cc   {${\mathrm c}\overline {\mathrm c}$}
\def \dd   {${\mathrm d}\overline {\mathrm d}$}
\def \uu   {${\mathrm u}\overline {\mathrm u}$}
\def \ssbar   {${\mathrm s}\overline {\mathrm s}$}
\def \qq   {${\mathrm q}\overline {\mathrm q}$}
\def \ff   {${\mathrm f}\overline {\mathrm f}$}
\def \cms  {centre--of--mass} 
\def \J   {${\mathrm J}/ \psi$}
\def \Bs  {${\mathrm B}_{\mathrm s}$}
%
\def \CERN   {{\sc CERN}}
\def \CDF  {{\sc CDF}}
\def \dzero {{\sc{D}${\mathsc {\emptyset}}$}}
\def \LEP    {{\sc LEP}}
\def \Tevatron {{\sc Tevatron}}
\def \CMS    {{\sc CMS}}
\def \ATLAS     {{\sc ATLAS}}
\def \LHC  {{\sc LHC}}
\def \QCD {{\sc QCD}}
\def\al{\alpha}
\def\ab{\bar{\alpha}}
\def\be{\begin{equation}}
\def\ee{\end{equation}}
\def\bea{\begin{eqnarray}}
\def\eea{\end{eqnarray}}
\def\CPbar{\hbox{{\rm CP}\hskip-1.80em{/}}}

\def\Nuno {\tilde{\chi}_1^0}
\def\Ndue {\tilde{\chi}_2^0}
\def\Cuno {\tilde{\chi}_1^{\pm}}
\def\Cdue {\tilde{\chi}_2^{\pm}}
\def \gluino {\tilde{g}}
\def \sbottom {\tilde{b}}
\def \squark {\tilde{q}}
\def \slepton {\tilde{\ell}}
\def \sleptonpm {\tilde{\ell}^{\pm}}
\def \stau {{\tilde{\tau}}}
\def \sneutrino {\tilde{\nu}}
\def \gbb {${\gluino\to\sbottom b}$}

\newenvironment{2figures}[1]{\begin{figure}[#1] 
  \begin{center}
    \begin{tabular}{p{0.49\textwidth}p{.49\textwidth}} }
 {  \end{tabular}
  \end{center} 
 \end{figure}
}


\noindent{\small
In this paper simulation studies performed to understand the capability of the 
CMS detector at the LHC to reconstruct
strongly interacting supersymmetric particles are presented. Sbottom and gluino 
mass peaks are reconstructed through the
 $\tilde{g} \rightarrow \tilde{b}b \rightarrow \tilde{\chi}_2^0 b b \rightarrow 
\tilde{\ell}^{\pm} \ell^{\mp} bb \rightarrow \tilde{\chi}_1^0 \ell^{\pm} \ell^{\mp} bb$ 
decay chain, exploiting the characteristic dilepton edge of the
$\tilde{\chi}_2^0$ decays. The same technique has also been used to reconstruct 
squarks and gluinos  through the
$\tilde{g} \rightarrow \tilde{q}q \rightarrow \tilde{\chi}_2^0 q q \rightarrow 
\tilde{\ell}^{\pm} \ell^{\mp} qq \rightarrow \tilde{\chi}_1^0 \ell^{\pm} \ell^{\mp} qq$ 
process. 
Mass resolutions lower than 10\% are achieved for all the reconstructed 
strongly interacting particles already with an integrated luminosity of 
60~fb$^{-1}$, assuming the $\tilde{\chi}_1^0$ mass to be known. 
An estimate of the $\sigma \times BR$ of the involved processes is also given.
Emphasis is given to the dependence of the reconstruction method  
on the mass of the $\tilde{\chi}_1^0$, 
in order to estimate the possible contribution from a LC measurement.
}



\subsubsection{Introduction}
\label{intro}

One of the main purposes of the next generation colliders is to search for the Physics 
beyond the Standard Model. The discovery of superpatners of
ordinary particles, as expected in Supersymmetric extension of SM
(SUSY)~\cite{susy}, would be a proof of the existence of new
physics. If supersymmetry exists at the electroweak scale, it could
hardly escape detection at LHC. Thanks, in fact, to the centre of
mass energy of 14 TeV, which will be available at LHC, it will be
possible to extend the searches of SUSY particles up to masses of
2.5 -- 3 TeV. SUSY, if it exists, is expected to reveal itself at LHC
via excess of multijet+E$^{miss}_{\mathrm T}$+(multilepton) final
states compared to SM expectations~\cite{baer}. Determining masses of
supersymmetric particles, however, is more difficult. 
A Linear Collider could in this sense be very useful to complement measurement 
performed at LHC, increasing the degree of our knowledge of the SUSY sector.  
  The main goal of this paper, is to show the potential of the 
CMS detector~\cite{cms} to reconstruct SUSY particles and how a Linear Collider 
could improve on these measurements.

\subsubsection{Strongly interacting sparticle reconstruction}

In this section we present the results of a
new study aimed at the reconstruction of the strongly interacting
gluinos, sbottoms and squarks. In order to perform this mass reconstruction,
two different decay chains 
$\gluino\to\sbottom b, \, \sbottom\to\Ndue b,\,
\Ndue\to\slepton^\pm \ell^\mp\to\Nuno\ell^+\ell^-$ and 
$\gluino\to\squark q, \, \squark\to\Ndue q,\,
\Ndue\to\slepton^\pm \ell^\mp\to\Nuno\ell^+\ell^-$, where $\ell=e,\mu$, 
have been considered.  In the first decay chain two b-jets, two same flavour
and opposite charge isolated leptons and large missing transverse
momentum due to the escaping $\Nuno$ are produced, while the second decay chain presents 
the same topology except for the presence of two non b-jets. 
In both cases, the reconstructions are performed
starting from the $\Ndue \rightarrow {\ell}^+{\ell}^- \Nuno$ decay,
with ${\ell} = e,\mu$.  Leptons from the
$\Ndue$ decay exhibit a peculiar $\ell^+\ell^-$ invariant mass
distribution with a sharp edge, as shown in Fig.~\ref{lepedge} and 
Fig.~\ref{edge_etm_ell}.  If
$m_{\Ndue}<m_{\slepton}+m_\ell$ the $\Ndue$ decay would be a three
body decay mediated by a virtual slepton and the edge would be placed
at $m_{\Ndue}-m_{\Nuno}$.  In the opposite case, when
$m_{\Ndue}>m_{\slepton}+m_\ell$, the neutralino decay is a two body
decay and the edge would be placed at
\begin{equation}
M_{\ell^+\ell^-}^{max} = \frac{\sqrt{ \left( m_{\Ndue}^2 - m_{\slepton}^2 \right) \left( m_{\slepton}^2 
- m_{\Nuno}^2 \right)}}{m_{\slepton}}
\label{eq:edge}
\end{equation}

The analysis has been performed in a mSUGRA scenario, considering three 
different benchmark points, the so called point \\
\begin{itemize}
\item B ($m_{1/2}=250,\,m_0=100,\,\tan\beta=10,\,\mu>0\,{\mathrm and}\, A_0=0$), 
\item G ($m_{1/2}=375,\,m_0=120,\,\tan\beta=20,\,\mu>0\,{\mathrm and}\, A_0=0$), 
\item I ($m_{1/2}=350,\,m_0=180,\,\tan\beta=35,\,\mu>0\,{\mathrm and}\, A_0=0$),
\end{itemize}
of ref.~\cite{marco}. Point B is very similar to point SPS1a of ref.~\cite{sps}, the only difference being in the
value of the $A_0$ parameter. All the 
chosen benchmark points are characterized both by 
relative low value for $m_0$ and $m_{1/2}$ 
(high production cross section for strongly interacting sparticles) 
and different values of $\tan\beta$. Indeed,  the 
branching ratios of  $\Ndue\to\slepton^\pm \ell^\mp\to\Nuno\ell^+\ell^-\,(\ell=e,\mu)$
decay is
 strongly dependent on the $\tan\beta$ parameter. This effect is of fundamental 
importance for our analysis. 

The signal events are generated using PYTHIA 6.152~\cite{pythia} with 
 ISASUGRA 7.51~\cite{isajet} input parameters, 
whereas background events (\ttbar, Z+jets, W+jets and QCD jets) are
generated with PYTHIA 6.152~\cite{pythia}. The detector response has
been evaluated using the fast MC package CMSJET~\cite{cmsjet}. The
study has been realized for several different integrated luminosities. \\
Table~\ref{tab:spectra} summarizes the SUSY spectra at Point B, G and I, 
while in Table~\ref{tab.sigma} the total SUSY cross-sections at the three 
different benchmark points are shown together with the branching ratios for the 
last decay chain 
($\Ndue\rightarrow \tilde{\ell}_R^{\pm} \ell^{\mp} \rightarrow \Nuno\ell^{\pm}\ell^{\mp}$). As can be seen from this Table, while the cross-section 
and the ${\mathcal BR}$ are quite large at point B, at point G 
(intermediate $\tan\beta$ region) the situation become worse, 
while at point I (large $\tan\beta$ region) the ${\mathcal BR}$ 
for the interesting decay chain becomes even negligible. In the next sections 
we will see how these effects are reflected in the reconstruction performances. 
\begin{table}[htb]
\begin{small}
\begin{center}
\begin{tabular}{|c|r|r|r|}
      \hline \hline
      Particle                  &     \multicolumn{3}{c|}{Mass (GeV)} \\
      \hline
                                &   Point B  &  Point G  &  Point I  \\ 
      \hline \hline
      $ \tilde{u}_L         $   &   537.0  &	773.9	 &  738.4  \\
      $ \tilde{d}_L         $   &   542.8  &	778.0	 &  742.7  \\
      $ \tilde{c}_L         $   &   537.0  &	773.9	 &  738.4  \\
      $ \tilde{s}_L         $   &   542.8  &	778.0	 &  742.7  \\
      $ \tilde{t}_1         $   &   392.9  &	587.2	 &  555.4  \\
      $ \tilde{b}_1         $   &   496.0  &	701.9	 &  640.3  \\
      $ \tilde{u}_R         $   &   519.1  &	747.9	 &  714.9  \\
      $ \tilde{d}_R         $   &   520.9  &	745.8	 &  713.2  \\
      $ \tilde{c}_R         $   &   519.1  &	747.9	 &  714.9  \\
      $ \tilde{s}_R         $   &   520.9  &	745.8	 &  713.2  \\
      $ \tilde{t}_2         $   &   575.9  &	778.6	 &  736.0  \\
      $ \tilde{b}_2         $   &   524.0  &	748.4	 &  713.3  \\
      $ \tilde{e}_R         $   &   136.2  &	183.2	 &  221.3  \\
      $ \tilde{\mu}_R       $   &   136.2  &	183.2	 &  221.3  \\
      $ \tilde{\tau}_2      $   &   200.3  &	285.4	 &  304.8  \\
      $ \tilde{e}_L         $   &   196.6  &	278.9	 &  295.7  \\
      $ \tilde{\nu}_{e_L}   $   &   179.8  &	267.1	 &  284.6  \\
      $ \tilde{\mu}_L       $   &   196.6  &	278.9	 &  295.7  \\
      $ \tilde{\nu}_{\mu_L} $   &   179.8  &	267.1	 &  284.6  \\
      $ \tilde{\tau}_1      $   &   127.6  &	154.1	 &  143.7  \\
      $ \tilde{\nu}_{\tau_L}$   &   179.0  &	263.4	 &  271.2  \\
      $ \tilde{g}           $   &   595.1  &	860.8	 &  809.8  \\
      $ \tilde{\chi}_1^0    $   &    95.6  &	150.0	 &  139.8  \\
      $ \tilde{\chi}_2^0    $   &   174.7  &	277.1	 &  257.9  \\
      $ \tilde{\chi}_1^{\pm}$   &   173.8  &	276.8	 &  257.6  \\
      $ \tilde{\chi}_3^0    $   &   339.9  &	477.9	 &  446.7  \\
      $ \tilde{\chi}_4^0    $   &   361.0  &	493.6	 &  462.2  \\
      $ \tilde{\chi}_2^{\pm}$   &   361.6  &	494.3	 &  463.3  \\
      \hline	\hline	    
\end{tabular}	    
\end{center}
\end{small}
    \caption{Spectra at point B, G and I as given by {\tt PYTHIA 6.152} with input parameters taken from {\tt ISASUGRA 7.51}.}
    \label{tab:spectra}
\end{table}

\begin{table}[htb]
\begin{center}
\begin{tabular}{|c|r|r|r|}
\hline \hline
	   		     &	Point B & Point G & Point I \\
\hline
$\sigma_{\mathrm SUSY} (pb)$ &	57.77   & 8.25    & 10.14      \\
\hline
${\mathcal BR}(\Ndue\rightarrow \tilde{\ell}_R^{\pm} \ell^{\mp} \rightarrow \Nuno
\ell^{\pm}\ell^{\mp}) (\%)$  &  16.44    & 2.26    & 0.25 \\
\hline \hline
\end{tabular}
\end{center}
\caption{Total SUSY cross section and ${\mathcal BR}(\Ndue\rightarrow \tilde{\ell}_R^{\pm} \ell^{\mp} \rightarrow \Nuno\ell^{\pm}\ell^{\mp})$ for the three different benchmark 
points analysed.}
\label{tab.sigma}
\end{table}

\subsubsection{Sbottom and gluino reconstruction}

In order to perform the sbottom and gluino reconstruction, through the decay chain
$\gluino\to\sbottom b, \, \sbottom\to\Ndue b,\,
\Ndue\to\slepton^\pm \ell^\mp\to\Nuno\ell^+\ell^-$, events with
at least 2 same flavour opposite sign (SFOS) isolated leptons having
$p_T > 15$~GeV and $|\eta|<2.4$, corresponding to the acceptance of
the muon system, and at least 2 jets tagged as $b$-jets, having $p_T >
20$~GeV and $|\eta|<2.4$, are selected.

The $\sbottom$ reconstruction proceeds in two steps. First the $\Ndue
\rightarrow \tilde{\ell}_R^{\pm} \ell^{\mp} \rightarrow \Nuno
\ell^{\pm}\ell^{\mp}$ decay chain is considered.  As mentioned before,
this decay is characterized by a sharp end-point in the dilepton
invariant mass distribution.  In Fig.~\ref{lepedge} the SFOS 
dilepton pair invariant mass distribution is shown for
SUSY events superimposed over the SM background. The $t\bar{t}$
component, which represents the main background, gives a wide
distribution, while the $Z$+jets channel is visible for the $Z$ peak
which lies quite close to the end-point of the SUSY distribution.
\begin{2figures}{hbt!}
  \resizebox{\linewidth}{0.5\textwidth}{\includegraphics{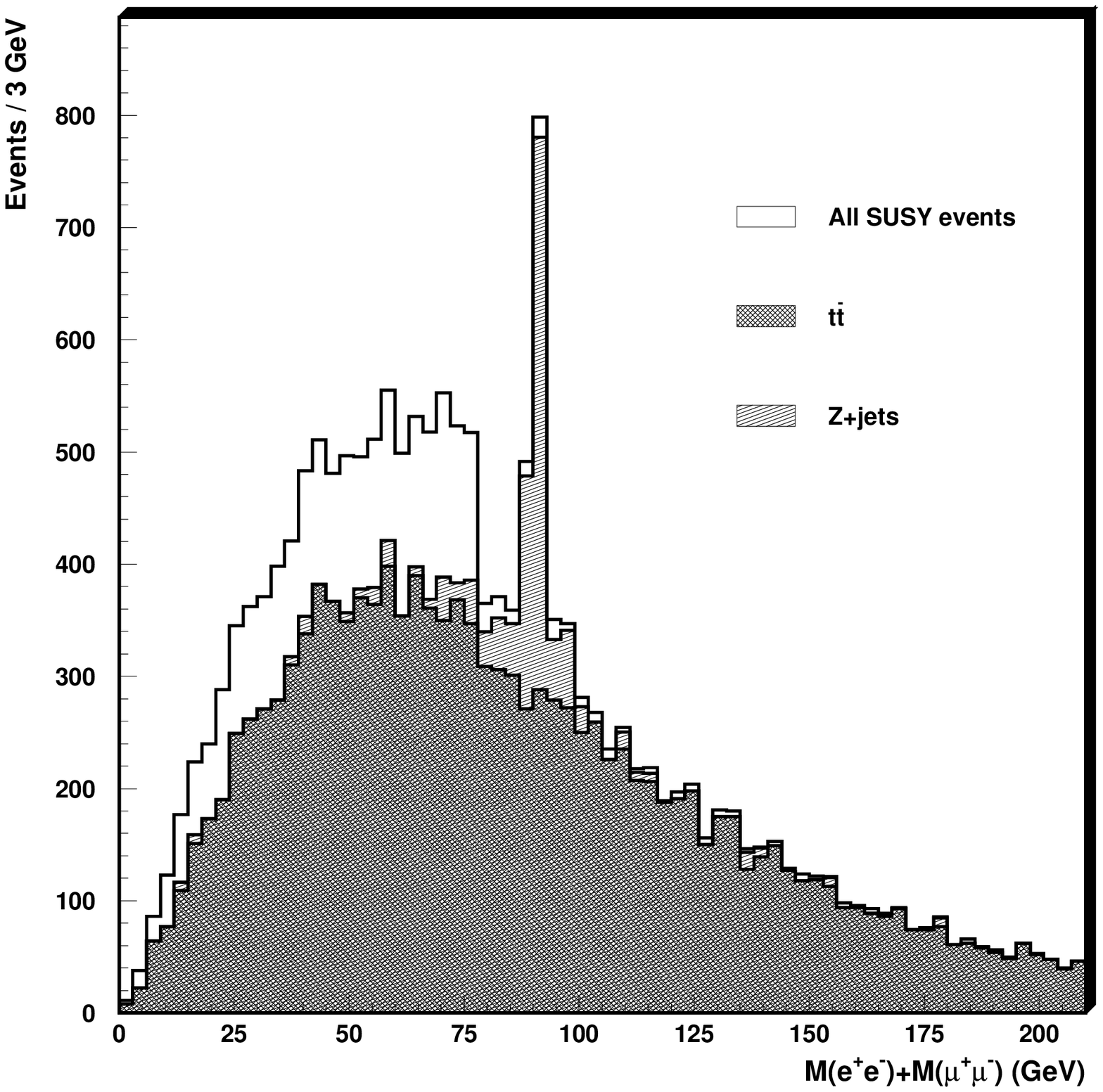}} &
  \resizebox{\linewidth}{0.5\textwidth}{\includegraphics{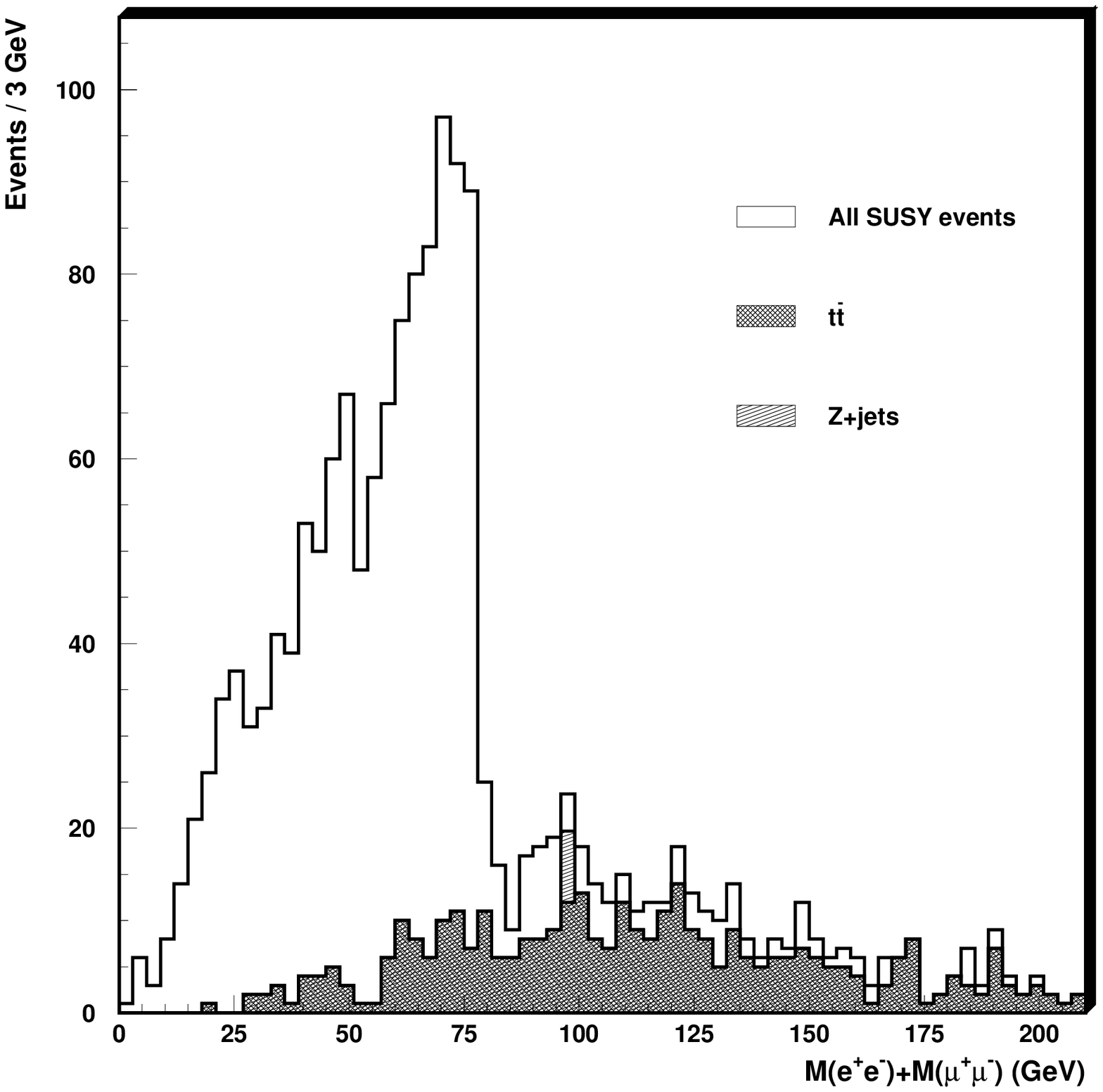}} \\
  \caption{Invariant mass distribution of same flavour opposite sign isolated leptons for SUSY events, superimposed on the
SM background, for an integrated luminosity of 10~fb$^{-1}$. The contributions of $t\bar{t}$ and $Z$+jets events are shown.\hfill\break}
\label{lepedge} &
  \caption{Same as in Fig.~\ref{lepedge} with  $E_T^{miss} > 150$~GeV and $E_{ll} > 100$~GeV cuts.\hfill\break}
\label{edge_etm_ell}
\end{2figures}
To perform the reconstruction a precise knowledge of the edge is
necessary. In order to reduce the SM background contribution, the high
missing energy content of SUSY events has been exploited.  A cut on
$E_T^{miss}>150$ GeV permits to drastically reduce the SM
background. This, combined with a cut on the dilepton energy,
$E_{\ell\ell}>100$ GeV, which suppresses other SUSY background
sources, gives a very clean dilepton edge, as can be seen in
Fig.~\ref{edge_etm_ell}.  In order to extract the value of the
end-point, a fit with a jacobian function can be performed on the
clean $M(e^+e^-) + M(\mu^+\mu^-) - M(e^+\mu^-) - M(\mu^+e^-)$
distribution, which, according to Eq.~\ref{eq:edge}, returns the value
$M_{\ell^+\ell^-}^{max} = (77.59 \pm 0.01)\,\textrm{GeV}$  for an integrated 
luminosity of 300~fb$^{-1}$.



To reconstruct the sbottom, opposite charge leptons in a window of
about 15 GeV around the edge are selected.  This requirement allows to
select a kinematical condition in which the leptons are emitted
back-to-back in the $\Ndue$ rest frame. In
this condition the $\Ndue$ momentum is
reconstructed through the relation:
\begin{equation}
\vec{p}_{\Ndue} = \left( 1 + \frac{m_{\Nuno}}{M_{\ell^+\ell^-}} \right) \vec{p}_{\ell^+\ell^-}.
\label{eq2}
\end{equation}

At this stage of reconstruction, we use the generated value for 
$m(\Nuno)$.

The $\Ndue$ momentum is then summed with the momentum of the highest
$E_T$ b-tagged jet and the $\sbottom$ is hence reconstructed. To
reduce combinatorial background coming from wrong b jets association,
further kinematical cuts have been used. All the details of the reconstruction 
procedure can be found in~\cite{susynote}. As shown in
Fig.~\ref{masspeak}a, with an integrated luminosity of 300~fb$^{-1}$, a
well visible sbottom mass peak, with a resolution well below than 10\%,
can be reconstructed for point B.   
The result of the fit, $M({\sbottom})=497\pm 2,\, \sigma=36\pm 3$~GeV, 
is in good agreement with the generated values of the two sbottoms
($\sbottom_1, \sbottom_2$), $M({\sbottom_1})=496\,{\mathrm
GeV},M({\sbottom_2})=524\,\mathrm{GeV}$. Indeed, the result of the fit 
should be considered as the superposition of the two 
contributions coming from $\sbottom_1$ and $\sbottom_2$, so we should 
compare it with the mean of the two masses weighted by the $\sigma\times BR$'s:
\begin{equation*}
\bar{M}(\sbottom) = \frac{M(\sbottom_1) \cdot \sigma \times BR(\sbottom_1) + M(\sbottom_2) \cdot \sigma \times BR(\sbottom_2)}{\sigma \times BR(\sbottom_1) + \sigma \times BR(\sbottom_2)} = 503.9 \; \textrm{GeV} \nonumber
\end{equation*}
resulting in good agreement. Unfortunately, however, the separation of the two sbottom 
contributions seems to be unaccessible even at very high luminosities due to the fact that 
the detector resolution is larger than the mass difference between the two sbottoms. However, 
with the ultimate luminosity of 300~fb$^{-1}$, reachable at the end of the LHC running period,  
it is possible to perform a double gaussian fit on the sbottom mass distribution. 
In Fig. \ref{fig:sbottomseparation} the two gaussian superimposed to the sbottom mass peak are 
shown. The results of the fit return $M(\sbottom_1)=487\pm 7$ GeV, 
$M(\sbottom_2)=530\pm 19$ GeV in agreement with the generated value. It is worth noticing that 
the ratio of the  coefficient of the two gaussians, $k_1/k_2=2.5$, is in good agreement 
with the ratio of the $\sigma\times{\mathcal BR}$ for the two sbottom states, 
$\sigma\times{\mathcal BR}(\sbottom_1)/\sigma\times{\mathcal BR}(\sbottom_2)=2.54$. 

\begin{figure}[htb!]
\vfill \begin{minipage}{.49\linewidth}
\begin{center}
\mbox{\epsfig{figure=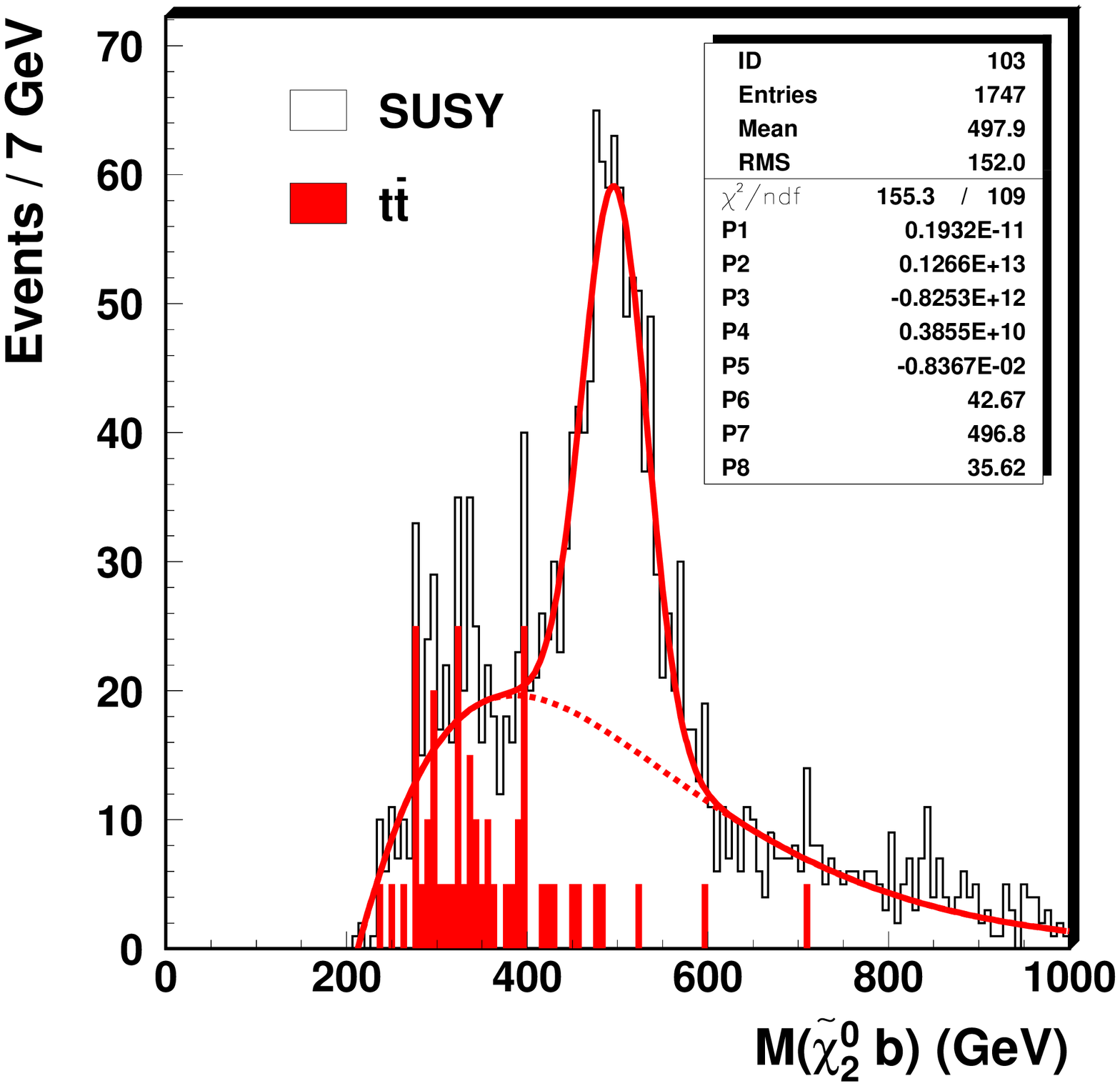,width=0.9\linewidth,clip=}}
\end{center}
\end{minipage}\hfill
\begin{minipage}{.49\linewidth}
\begin{center}
\mbox{\epsfig{figure=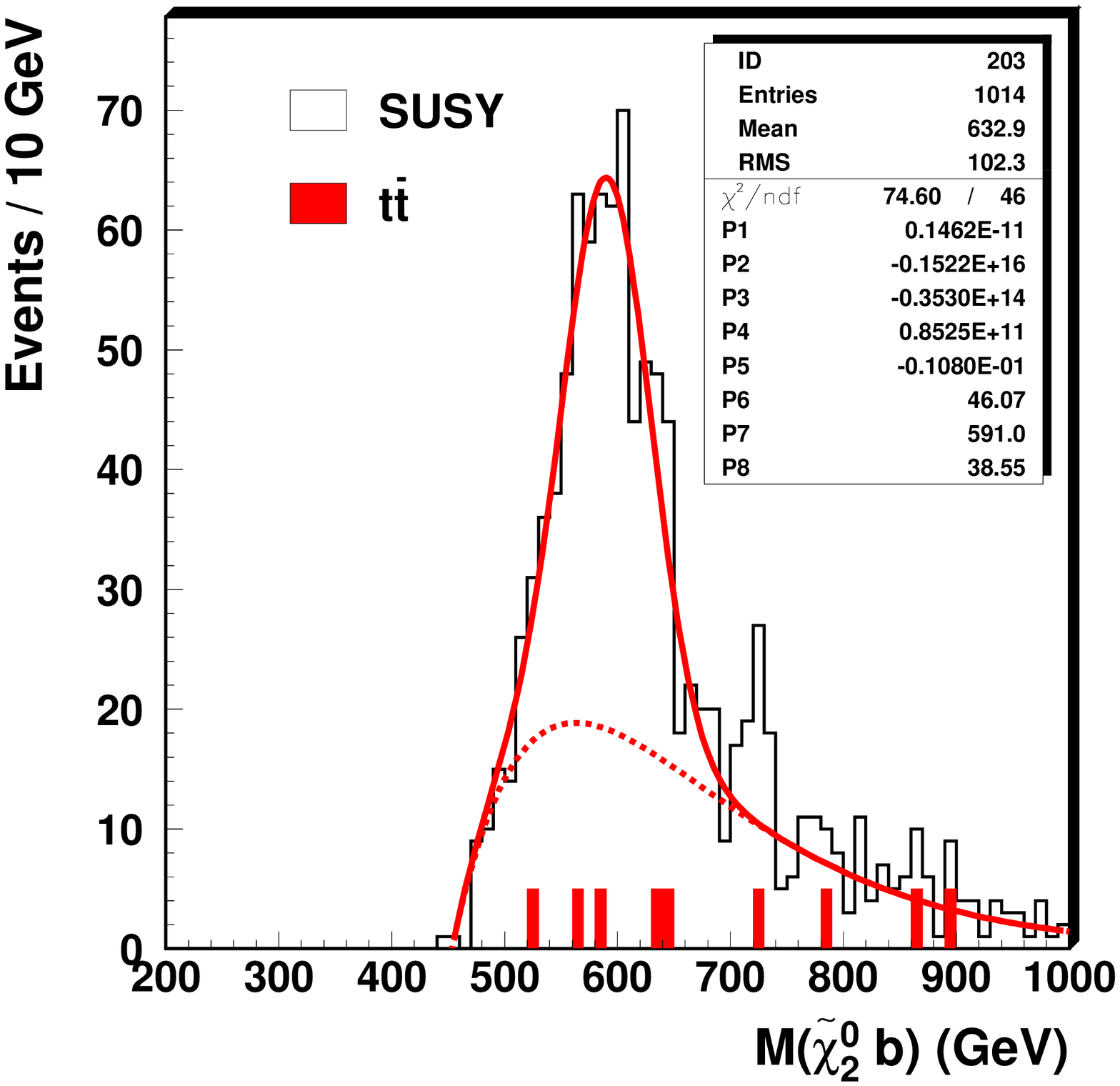,width=0.9\linewidth,clip=}}
\end{center}
\end{minipage}
\caption{(a): $M(\Ndue b)$; (b): $M(\Ndue bb)$ mass distributions for an integrated luminosity of 300~fb$^{-1}$. 
In both plots results for mSUGRA point B are presented. Events 
in the mass window $65 \,\mathrm{GeV}<\- M_{\ell^+\ell^-}\-< 80\,\mathrm{ GeV}$ 
with $E_T^{miss}>150\, \mathrm{ GeV}, 
E_{\ell\ell}> 100\, \mathrm{ GeV}$ and $E_T^b> 250\, \mathrm{ GeV}$ are considered.\hfill\break}
\label{masspeak}
\end{figure}

The gluino is reconstructed from the sbottom and the
b-tagged jet closest in angle. As shown in Fig.~\ref{masspeak}b, a resolution 
better than $10\%$ is achieved also in this case and the fitted mass value,
$M_{\gluino}=591\pm 3,\,\sigma=39\pm 3$~GeV, is in agreement with the generated value, 
$M_{\gluino}=595$~GeV. 

\begin{figure}[htb]
\begin{center}
\resizebox{0.6\textwidth}{!}{\includegraphics{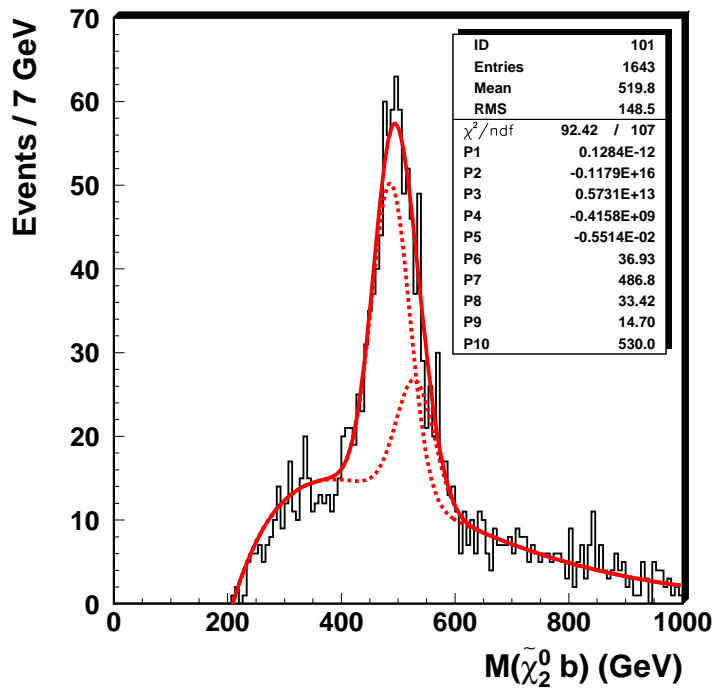}}
\end{center}
\caption{$M(\Ndue b)$ mass distribution for an integrated 
luminosity of 300~fb$^{-1}$, superimposed with a double gaussian fit.}
\label{fig:sbottomseparation}
\end{figure}

All the results shown so far are derived for point B and for an
integrated luminosity of 300~fb$^{-1}$; it is however worth to notice that the analysis at point B can
be performed also with lower integrated luminosities: the squark mass peak is visible even with an integrated luminosity of
1~fb$^{-1}$; sbottom and gluino can be seen with 10~fb$^{-1}$.  The same kind of analysis was
repeated also for point G. In this case the higher value of $\tan\beta$ reflects into higher
branching ratio for the decay $\Ndue \rightarrow \tau^+\tau^- \Nuno$
and to a lower signal $\Ndue\to\slepton^\pm
\ell^\mp\to\Nuno\ell^+\ell^-, \,\ell=e,\mu$ (see Table~\ref{tab.sigma}). 
In order to
reconstruct a clean mass peak for sbottom and gluino, not only the
cuts should be tightened but also a larger integrated luminosity is
needed. Only with an integrated luminosity of 300~fb$^{-1}$ it is possible 
to distinguish an edge in the dilepton distribution and to 
reconstruct the two mass peaks ($\sbottom$, $\gluino$) 
and perform the fits, nonetheless in this case 
a worse mass resolution is obtained.   
 An attempt was made also to repeat 
the analysis at point I of ref.~\cite{marco}, which is characterized by 
a still higher value of $\tan\beta$ ($\tan\beta=35$), but for that point, 
since the BR($\Ndue\to\slepton^\pm\ell^\mp\to\Nuno\ell^+\ell^-$) is almost negligible,  
even with an integrated luminosity of 300~fb$^{-1}$, it is not possible 
to distinghuish the dilepton edge. Hence no reconstruction of sbottoms and gluinos 
in this channel is possible. Further studies are in progress to understand the 
capability of reconstruction looking at the $\tau\tau$ final state. \\
In Table~\ref{tab:results.sb} the results obtained for the reconstructed 
masses and resolutions at point B and G, 
in the hypothesis of a known $\Nuno$ mass and 
at different integrated luminosity, are summarized.

\begin{table}[htb]
\begin{center}
\begin{tabular}{|l|c|c|c|c|c|c|c|}
\hline
\hline
  &     &  M$(\sbottom)$ &  $\sigma(\sbottom)$ & M$(\gluino)$ &  $\sigma(\gluino)$ &  M$(\gluino)$-M$(\sbottom)$ &  $\sigma(\gluino$-$\sbottom)$ \\
\hline
\hline
Point B & 10 fb$^{-1}$   & $500 \pm 7$ &   $42 \pm 5$   &   $594 \pm 7$  &     $42 \pm 7$   &   $92 \pm 3$     &   $17 \pm 4$       \\

        & 60 fb$^{-1}$   & $502 \pm 4$ & $41 \pm 4$   &   $592 \pm 4$   &     $46 \pm 3$    &   $88 \pm 2$     &   $20 \pm 2$      \\

        & 300 fb$^{-1}$  & $497 \pm 2$ & $36 \pm 3$   &   $591 \pm 3$   &     $39 \pm 3$    &   $90 \pm 2$     &   $23 \pm 2$      \\
\hline
\hline
Point G & 300 fb$^{-1}$  & $720 \pm 26$ &  $81 \pm 18$  &   $851 \pm 40$  &     $130 \pm 43$   &  $127 \pm 10$   &   $48 \pm 11$      \\
\hline
\hline
\end{tabular}
\caption{Sbottom and gluino mass resolution. All the results are expressed in GeV. \hfill\break}
\label{tab:results.sb}
\end{center}
\end{table}

\paragraph{Evaluation of the $\sigma\times BR$}

Antother interesting information that can be extracted by this analysis is the achievable 
precision on the $\sigma \times BR$ of the chain. This can be evaluated just counting  
events in the peaks corresponding to all the signal process in which a sbottom or a  
gluino is produced. \\
Indeed, the precision on the $\sigma \times BR$ measurement, is directly related to the 
number of observed events in the peak, $N_{obs}$, through 
\begin{equation}
\delta(\sigma \times BR) = \frac{\delta(N_{obs})}{{\mathcal L}_{int} \cdot \epsilon}. 
\end{equation}

For an integrated luminosity of 10 fb$^{-1}$, in the sbottom peak,
taking into account the region between $\bar{M} - 2.5\sigma$ and
$\bar{M} + 2.5\sigma$, and subtracting the events under the background
curve, we count 102 events. This means that the $\sigma \times BR$ for
events in which the sbottom is directly produced or arises from a
gluino, and then decays into the whole decay chain, can be measured
with an error of about 10\%. In the gluino peak, with the same
procedure, 59 events are counted, which correspond to an error of 13\%
on the measurement of the gluino $\sigma \times BR$.  These results
can be improved with larger collected statistics: in the sbottom decay chain 
a precision of the order of $\sim 4.5\%$ can be achieved at 60 fb$^{-1}$ and 
of the order of $\sim 1.9\%$ at 300 fb$^{-1}$. All the previous results have been obtained 
selecting events in a fixed $M_{\ell\ell}$ window ($65 < M_{\ell\ell} < 80$) 
around the edge. 
It is however important to notice that even at 10 fb$^{-1}$ of integrated
luminosity, CMS will be able to give a preliminary estimate not only
of the masses of sparticles like sbottoms and gluinos, but also of the
$\sigma \times BR$ of their production processes and consequent
decays.

\subsubsection{Squark and gluino reconstruction}

The supersymmetric partners of the light quarks ($\tilde{u}$,
$\tilde{d}$, $\tilde{c}$ and $\tilde{s}$) can be reconstructed with a
similar procedure to the sbottom reconstruction, exploiting the decay
chain $\gluino\to\squark q, \, \squark\to\Ndue q,\,
\Ndue\to\slepton^\pm \ell^\mp\to\Nuno\ell^+\ell^-$, where
$\ell=e,\mu$, which is identical to the chain considered in the case
of the sbottom, apart from the emission of a light quark $q$ instead
of a bottom in the decay $\squark \rightarrow \Ndue q$.  Non $b$-jets
have to be identified, and the $b$-tagging capability of the CMS
detector has hence to be used in order to veto the presence of
$b$-jets and to perform an anti $b$-tagging. \\
Events are selected requiring:
\begin{itemize}
\item at least two same flavour opposite sign (SFOS) isolated leptons,
with $p_T > 15$~GeV and $|\eta|<2.4$; as in the previous sections, for
leptons we mean only electrons and muons;
\item at least two jets, tagged as non $b$-jets, with $p_T > 20$~GeV
and $|\eta|<2.4$;
\item no $b$-jets.
\end{itemize}
 
The reconstruction procedure starts, as in the sbottom case, from the
identification of the dilepton edge and hence events in the window
65~GeV~$< M_{\ell\ell} <$~80~GeV are selected and associated to the
most energetic jet to get the squark invariant mass distribution. In
order to reduce the SM background a cut in the missing energy is
imposed, as in the previous reconstruction and also similar
kinematical cuts are used to reduce the SUSY combinatorial
background. A detailed description of all the cuts imposed can be
found in ref.~\cite{tesimax}.
Given the very high $\sigma \times BR$ for squarks going into the
$\Ndue$-dilepton chain, which is about four times larger than for the
sbottom chain, it is plausible to perform the reconstruction with an
integrated luminosity lower than in the sbottom case. 
In ref.~\cite{susynote} it has been shown  that squark reconstruction will be 
possible already with an integrated luminosity of
1~fb$^{-1}$, corresponding to the first two or three months of life of
the Large Hadron Collider.
The treatment of the combinatorial background is more complicated than
for sbottom since in this case the  contribution $\squark\gluino$ 
and $\squark\squark$ processes cannot be neglected. However,
similarly to the sbottom case, the use of kinematical cuts allows 
a good combinatorial reduction. Figure~\ref{fig:sq_fit_10fb} shows the squark peak for
$E_{j1} > 300$~GeV, for a sample corresponding to 300~fb$^{-1}$ of
integrated luminosity.
The measured values are:
\begin{subequations}
\begin{eqnarray}
M(\Ndue q)         & = & 536 \pm 1 \ \textrm{GeV} \nonumber \\
\sigma[M(\Ndue q)] & = & 31 \pm 1 \ \textrm{GeV}  \nonumber 
\end{eqnarray}
\end{subequations}
Due to the Wino nature of $\Ndue$, the decay $\squark\to\Ndue$
occurs almost uniquely for the left squarks, since the right
squarks decay through $\squark\to\Nuno q$ with a BR larger than 0.99.
The measured value has hence to be compared with the nominal values of the
left squark, which are
\begin{eqnarray}
M(\tilde{d}_L) = M(\tilde{s}_L) = 542.8 \ \textrm{GeV} \nonumber\\
M(\tilde{u}_L) = M(\tilde{c}_L) = 537.0 \ \textrm{GeV} \nonumber
\end{eqnarray}

The resolution of the $\sigma \times BR$ of the entire squark
production and decay process can be inferred by the number of events
in the peak, which is 18048 at 300~fb$^{-1}$, leading to a value:
\begin{equation}
\frac{\delta(\sigma \times BR)}{\sigma \times BR} = 0.7\%
\end{equation}

\begin{figure}[hbt]
\begin{center}
  \resizebox{0.6\linewidth}{!}{\includegraphics{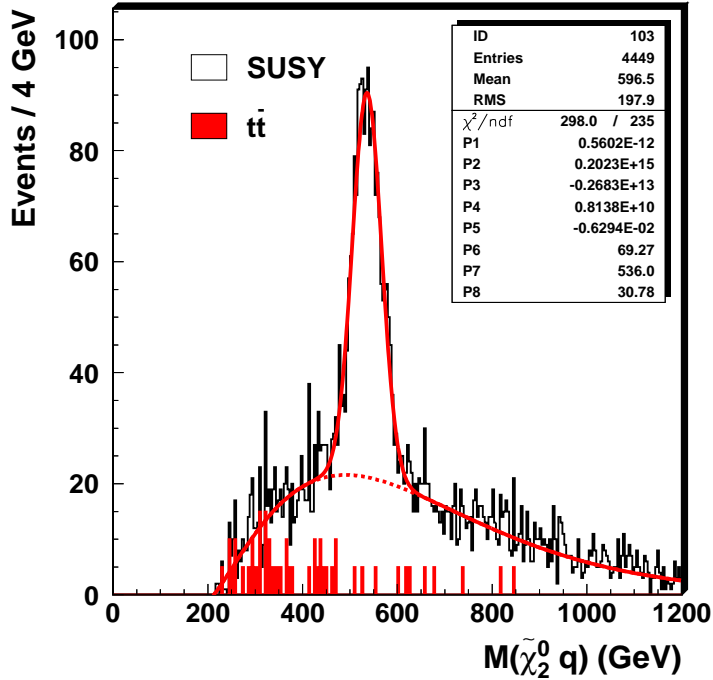}} 
\end{center}
  \caption{Invariant mass of the system $\Ndue q$ for an integrated luminosity of 300~fb$^{-1}$. Events are selected having
65~GeV$ < M_{\ell\ell} < $~80~GeV, $E_T^{miss} > 100$~GeV and $E_{j1} > 300$~GeV. A fit is performed with a gaussian superimposed over
a polynomial to take into account the combinatorial plus Standard Model background.}
  \label{fig:sq_fit_10fb} 
\end{figure}

Gluino can be reconstructed going back in the decay chain with a
procedure similar to the one used for the sbottom: after removing the
most energetic jet from the list of the available jets, the one
closest in angle to the reconstructed squark is associated to it. 
Figure~\ref{fig:glu_fit_sq} shows the final gluino peak for
an integrated luminosity of 300~fb$^{-1}$, 
after all the kinematical cuts described in ref.~\cite{tesimax} are applied. 
\begin{figure}[t!]
  \begin{center}
      \resizebox{0.6\textwidth}{!}{\includegraphics{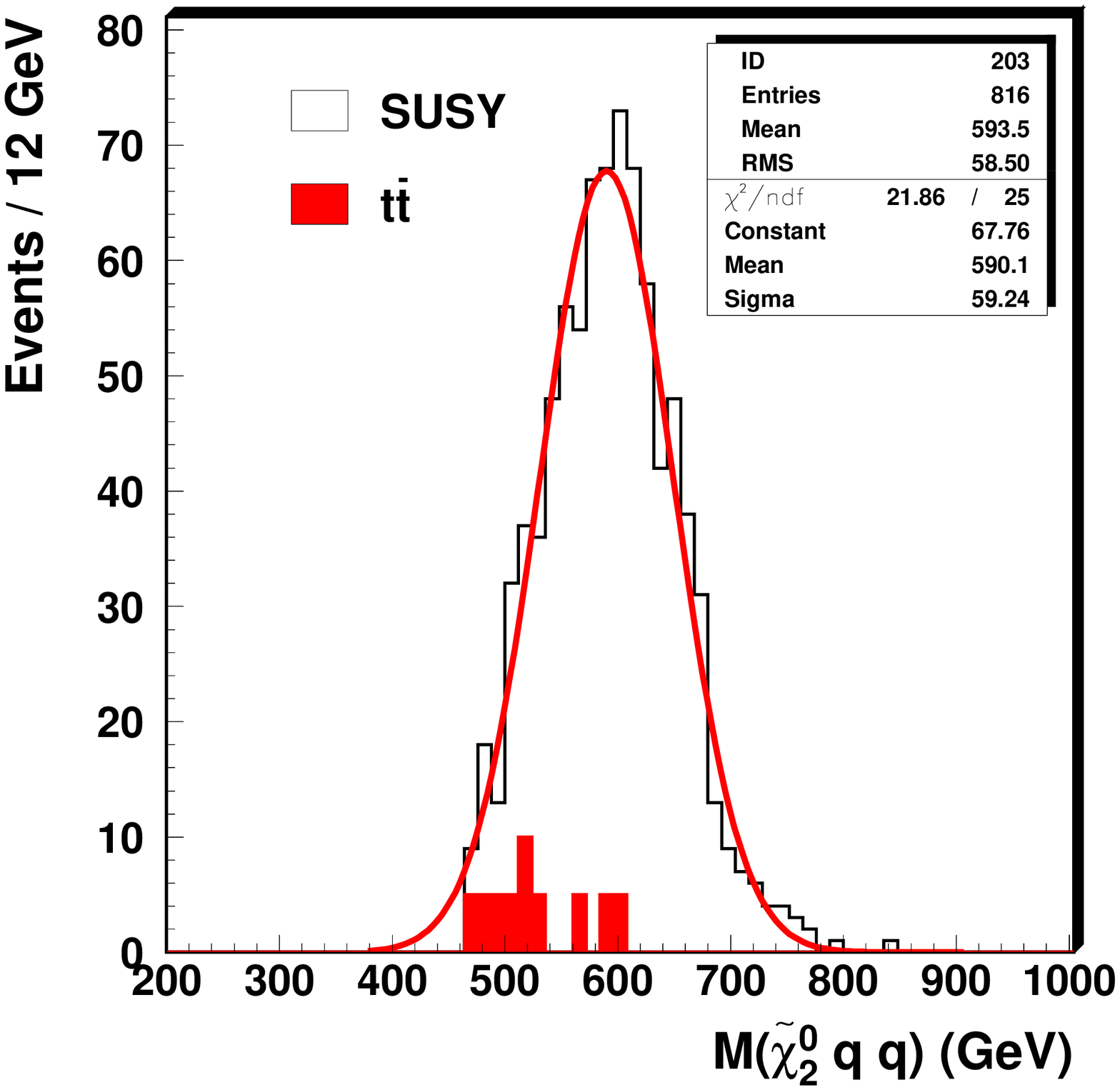}}
    \caption{Invariant mass of the system $\Ndue q q$ for events having 65~GeV$ < M_{\ell\ell} < $~80~GeV, $E_T^{miss} > 50$~GeV,
$E_{j1} > 300$~GeV, 390~GeV$ < M(\Ndue q) < $~690~GeV and $E_{j2} < 60$~GeV. The integrated luminosity is 300~fb$^{-1}$.}
    \label{fig:glu_fit_sq}
  \end{center}
\end{figure}
The gaussian fit gives:
\begin{subequations}
\begin{eqnarray}
M(\Ndue qq)         & = & 590 \pm 2 \ \textrm{GeV} \nonumber \\
\sigma[M(\Ndue qq)] & = & 59 \pm 2 \ \textrm{GeV}  \nonumber
\end{eqnarray}
\end{subequations}
\noindent which is in agreement with the mass value of the generated
gluino: $M(\gluino) = 595.1$~GeV.  Although the achieved resolution is
worse than the one of the gluino reconstructed into the sbottom chain,
this result is remarkable since it has been obtained in a totally
indipendent way, and can eventually be combined with that in order 
to have a better estimate of the gluino mass.

In Table~\ref{tab:results.sq} the results obtained for the squark and gluino reconstruction 
for several different luminosities and different benchmark points are summarized. 
Also for the squark decay chain the study has been repeated at points G and I. Given the
larger statistics, the results are slightly improved: the end-point of the dilepton mass distribution can be
seen at point G even with 10~fb$^{-1}$. Nonetheless, the peak reconstructions are possible
only with a large integrated luminosity, as reported in Table~\ref{tab:results.sq}. No reconstruction
is possible at point I.

\begin{table}[hb!]
\begin{center}
\begin{tabular}{|l|c|c|c|c|c|c|c|}
\hline
\hline
  &     &  M$(\squark)$ &  $\sigma(\squark)$ & M$(\gluino)$ &  $\sigma(\gluino)$ &  M$(\gluino)$-M$(\squark)$ &  $\sigma(\gluino$-$\squark)$ \\
\hline
\hline
Point B & 10 fb$^{-1}$   & $535 \pm 3$ &   $57 \pm 3$   &   $592 \pm 7$  &     $75 \pm 5$   &   $57 \pm 3$     &   $9 \pm 3$       \\

        & 60 fb$^{-1}$   & $532 \pm 2$ & $36 \pm 1$   &   $595 \pm 2$   &     $59 \pm 2$    &   $47 \pm 2$     &   $16 \pm 5$      \\

        & 300 fb$^{-1}$  & $536 \pm 1$ & $31 \pm 1$   &   $590 \pm 2$   &     $59 \pm 2$    &   $44 \pm 2$     &   $11 \pm 2$      \\
\hline
\hline
Point G & 300 fb$^{-1}$  & $774 \pm 9$ &  $84 \pm 9$  &   $853 \pm 11$  &     $126 \pm 11$   &  $82 \pm 3$   &   $35 \pm 3$      \\
\hline
\hline
\end{tabular}
\caption{Squark and gluino mass resolution. All the results are expressed in GeV. \hfill\break}
\label{tab:results.sq}
\end{center}
\end{table}

\subsubsection{Neutralino mass}
\label{neutralino}

All the results shown in the previous sections are obtained in the 
hypothesis of a known $\Nuno$ mass. 
In a realistic scenario, however, CMS will not be able to detect
$\Nuno$, this being a weakly interacting particle which escapes the
detector, and other strategies have to been devoloped.
One possibility is the one already 
exploited in ref.~\cite{atlas.tdr, atlas.endpoint}, which makes use of several 
different end-points, in order to constraint the mass of $\Nuno$. 
However, this can be easily done in a favourable scenario, like the 
one at point B, while this could be critical if SUSY reveal itself in a 
scenario like the one at point G where the end-points are difficult to 
select or even worse in the case of point I. 

It is worth noticing, that as both $M(\sbottom)$ and $M(\gluino)$
depend on the $\Nuno$ mass, their difference $M(\gluino)-M(\sbottom)$
is on the contrary independent on $M(\Nuno)$. As shown in
Fig.~\ref{massdiff}, CMS will be able to measure this difference 
with an error of
few percents, independent of any assumption on the sparticle spectrum.

\begin{figure}[h!]
\begin{center}
\mbox{\epsfig{figure=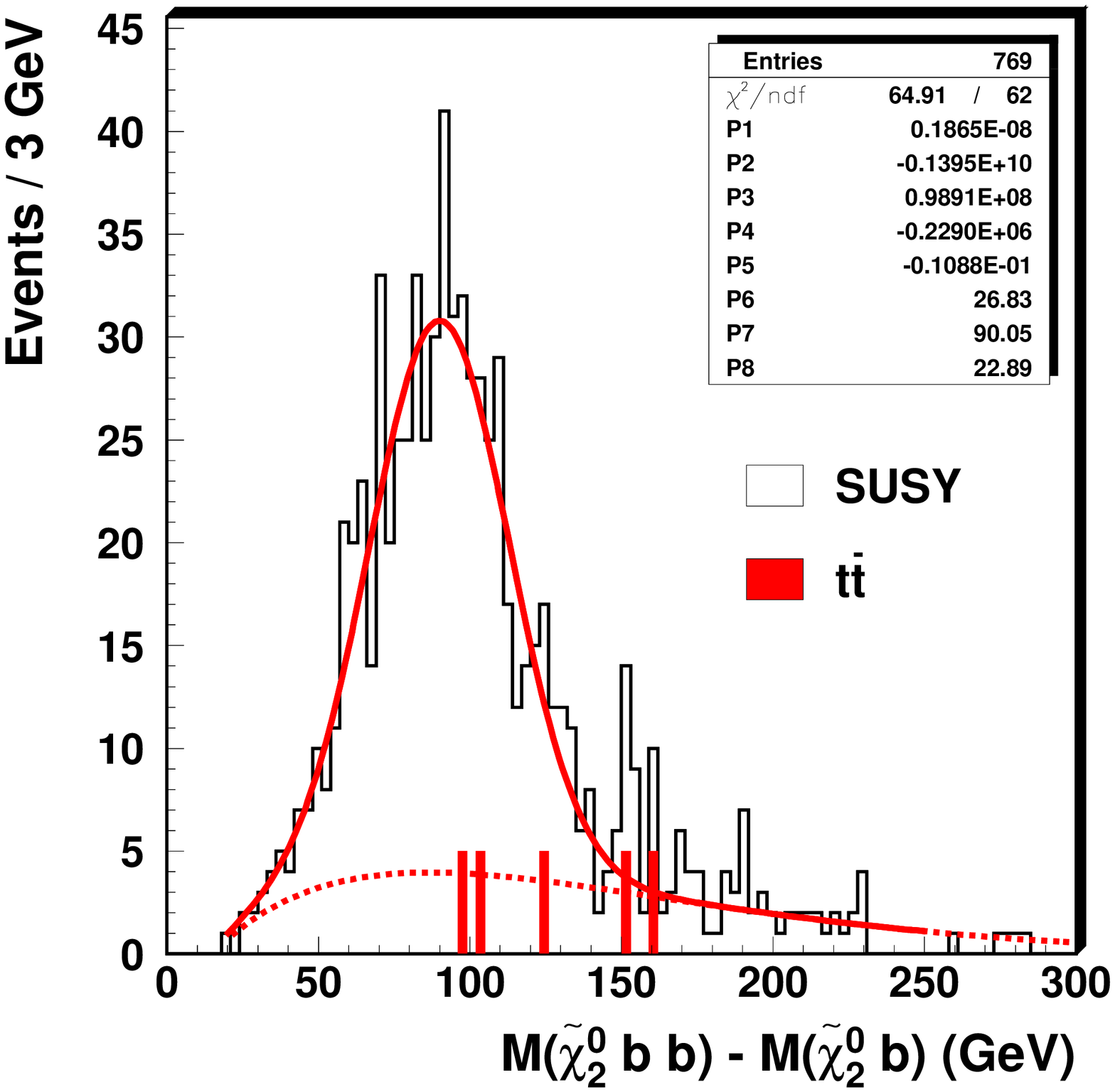,width=0.5\linewidth,clip=}}
\caption{$M(\Ndue bb)-M(\Ndue b)$ distribution for the same events as in Fig.~\ref{masspeak}.\hfill\break}
\label{massdiff}
\end{center}
\end{figure}

In Table~\ref{tab:results.sb} and \ref{tab:results.sq} also the achieved resolution in the 
 $M(\gluino)-M(\sbottom)$ for the different benchmark points and at several different 
integrated luminosities are shown.

Of course, a precise $M(\Nuno)$ measurement from a Linear Collider 
could be used as input in our measurements eliminating the biggest 
source of systematic uncertainties. To evaluate the dependence of 
sbottom, squark and gluino mass measurement on the accuracy of 
the $\Nuno$ mass knowledge, the reconstruction procedure has been 
repeated for different $\Nuno$ mass values. The dependence of 
$M(\sbottom )$ and $M(\gluino )$ (sbottom decay chain) and of 
$M(\squark )$ and $M(\gluino )$ (squark decay chain) on 
$M(\Nuno )$ is shown in Fig.~\ref{fig.nuno.sb} and in Fig.~\ref{fig.nuno.sq},  
respectively. All the masses of the reconstructed sparticles show a linear 
dependence. Performing a linear fit, we can deduce:
\begin{align}
\Delta M(\Ndue b)  & = (1.60\pm 0.03)\Delta M(\Nuno) \nonumber \\
\Delta M(\Ndue bb) & = (1.62\pm 0.05)\Delta M(\Nuno) \nonumber \\
\Delta M(\Ndue q)  & = (1.70\pm 0.01)\Delta M(\Nuno) \nonumber \\
\Delta M(\Ndue qq) & = (1.68\pm 0.07)\Delta M(\Nuno) \nonumber 
\end{align}
In the case of the sbottom, for instance, in order to have an uncertainty 
less than the statistical error achieved at 300~fb$^{-1}$, we should have 
$\Delta M(\Nuno)<1.25$ GeV. The Linear Collider should ensure an accuracy on 
the $\Nuno$ mass of the order of $\sim 1\%$ at point B. 

This is a conservative estimate of the error on the squark, sbottom and gluino mass
peaks due to the $M(\Nuno)$ uncertainty. It has in fact been evaluated living
$M(\Ndue)$ unchanged while changing $M(\Nuno)$, in a hypotesis of complet
independence between sparticles mass values. Realistically, this error should decrease
taking into account the correlation between $M(\Ndue)$ and $M(\Nuno)$.

\begin{figure}[hb!]
\begin{center}
    \resizebox{0.8\textwidth}{!}{\includegraphics{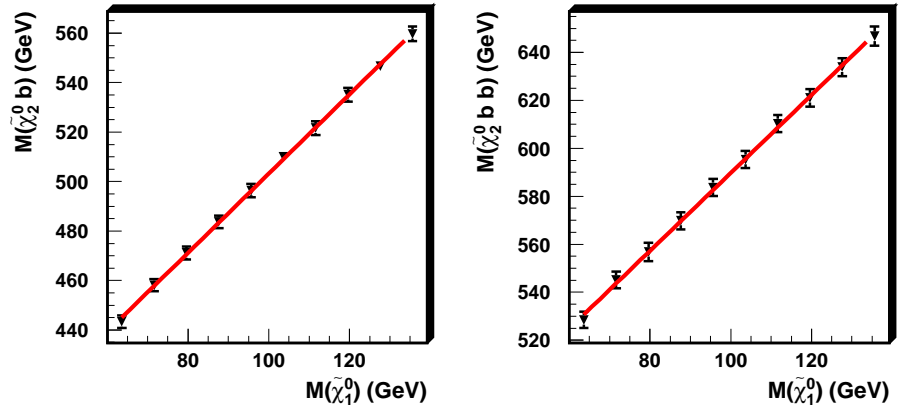}}
    \caption{Dependence of sbottom and gluino masses (sbottom decay chain) vs 
$M(\Nuno)$.}
    \label{fig.nuno.sb}
  \end{center}
\end{figure}
\begin{figure}[hb!]
\begin{center}
    \resizebox{0.8\textwidth}{!}{\includegraphics{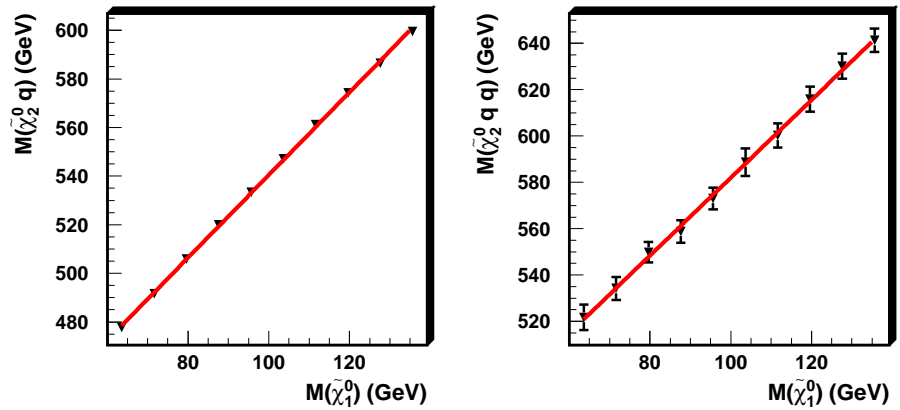}}
    \caption{Dependence of squark and gluino masses (squark decay chain) vs 
$M(\Nuno)$.}
    \label{fig.nuno.sq}
  \end{center}
\end{figure}

\subsubsection{Conclusions}

If SUSY exists at the EW scale, the CMS detector will be able to
discover it in a very large range of mSUGRA parameters. 
With the ultimate high luminosity of
300~fb$^{-1}$, strongly interacting sparticles could be discovered up
to masses of 2.5 -- 3 TeV.

Altough sparticle reconstruction is more difficult, new analyses have
shown that in many cases it will be possible to make exclusive
reconstructions. This is the case, for istance, of the decay
$\gluino\to\sbottom b$ and $\gluino\to\squark q$ which allow 
to reconstruct sbottom, squark and gluino masses. 
Resolutions better than 10\% will be attainable in the
low $\tan\beta$ region, already after the first year of data
taking. In a favourable SUSY scenario, not only the mass of 
strongly interacting SUSY particles can be measured 
but also an estimate of the $\sigma\times{\mathcal BR}$ 
of their production processes and consequent
decays will be possible. The combination of LHC/LC measurements 
will help us to reach a deeper knowledge of the SUSY sector. 
Detailed studies are going on in order to improve the present analysis and  
to evaluate the CMS capability to reconstruct SUSY sparticles.

%


\clearpage

\subsection{\label{sec:410}
Measurement of sparticle masses in mSUGRA scenario SPS~1a at a Linear
Collider}

{\it H.-U.~Martyn}
                                                                                
\setlength{\unitlength}{1mm}

\newcommand{\ord}{{\cal O}}
\renewcommand{\cO}{{\cal O}}
\renewcommand{\L}{{\cal L}}
\newcommand{\Oa}{\mathswitch{{\cal{O}}(\alpha)}}

\renewcommand{\hdick}{\noalign{\hrule height1.4pt}}
\renewcommand{\eV}  {\mathrm{eV}}
\renewcommand{\MeV} {\mathrm{MeV}}
\renewcommand{\GeV} {\mathrm{GeV}}
\renewcommand{\TeV} {\mathrm{TeV}}
\renewcommand{\fb}  {\mathrm{fb}}
\renewcommand{\fbi} {\mathrm{fb}^{-1}}
\renewcommand{\ab}  {\mathrm{ab}}
\renewcommand{\abi} {\mathrm{ab}^{-1}}

\newcommand{\sek} {\mathrm{s}}
\newcommand{\cL } {{\cal L}}
\newcommand{\cP } {{\cal P}}

\def\susy    {{\sc Susy}}
\def\susygen {{\sc Susygen}}
\def\pythia  {{\sc Pythia}}
\def\suspect {{\tt SuSpect}}

\def\tesla  {{\sc Tesla}}
\def\clic   {{\sc Clic}}
\def\nlc    {{\sc Nlc}}

\def\ee{e^+e^-}

\renewcommand{\nn}{\nonumber}
\renewcommand{\beq}{\begin{equation}}
\renewcommand{\eeq}{\end{equation}}
\renewcommand{\bea}{\begin{eqnarray}}
\renewcommand{\eea}{\end{eqnarray}}
\newcommand{\eq}[1]{eq.~(\ref{#1})}
\renewcommand{\fig}[1]{fig.~\ref{#1}}
\renewcommand{\tab}[1]{table~\ref{#1}}

\subsubsection{Introduction}

If low energy supersymmetry will be discovered at the \lhc \ its gross
features may be revealed, in particular for the coloured squark and
gluino sector. However, a Linear Collider will be indispensible
in order to
provide complementary information, in particular in the slepton and
neutralino/chargino sector, and to scrutinise the characteristics of
the underlying \susy \ structure.
High precision \lc\ experiments have to
\begin{itemize}
  \item
    measure the masses, decay widths, production cross sections,
    mixing angles, etc., of the new particles,
  \item
    prove that each particle can be associated  to its superpartner
    with the expected spin and parity, gauge quantum numbers and couplings,
  \item
    reconstruct the low energy \susy\ breaking parameters which would allow one to
    uncover the fundamental theory  and finally  extrapolate 
    its parameters to high (GUT, Planck) scales.
\end{itemize}

A concurrent operation of the \lhc\ and \lc\ would provide
answers to these elementary topics already after a few years of
running.
A nice example to support their complementarity is the SPS~1a
benchmark point~\cite{sec4_Allanach:2002nj,Ghodbane:2002kg}. 
With the recent progress in cavity
development \tesla\ energies of 1~TeV appear achievable, thus the
complete slepton and neutralino/chargino spectra as well as the light
stop $\st_1$ would be accessible.  
A peculiarity of this scenario is the large $\tan\beta = 10$ leading to
(incompletely measurable) 
multi $\tau$ final states from $\nt_i$, $\cpm_i$ cascade decays.
Therefore charged slepton and sneutrino production are very important
to measure the masses of the LSP $\nt_1$ and the light chargino $\cpm_1$.
The advantage of the \lc\ is to explore the spectrum
in a bottom-up approach, i.e. selecting particular channels by
the appropriate choice of energy and beam polarisations,
while suppressing background reactions.

First studies of SPS~1a masses were based on
extrapolations and estimates from low $\tan\beta$ scenarios~\cite{grannis}.
Meanwhile more reliable simulations have become available.
They typicaly assume integrated luminosities of $\cL = 250 - 500~\fbi$
at $\sqrt{s} = 500~\GeV$ to be accumulated within a reasonable run time
of one to two years. The \lc\ luminosity is expected to scale
linearly with energy. 
Beam polarisations of $\cP_{e^-} = \pm0.8$ and
 $\cP_{e^+} = \pm0.6$ are assumed.
For the determination of sparticle properties of SPS~1a other than masses
see~\cite{kalinowski}.

\subsubsection{Sleptons} 

Scalar leptons  are produced in pairs
\bea
   \ee & \to & \sell^+_i\sell^-_j, \, \snu_\ell\,\bar\snu_\ell  \hfill
        \qquad\qquad\qquad \ell = e,\, \mu,\, \tau \ {\rm \ and \ } \
	    [i,j = L,R \ {\rm \ or \ } \ 1,2]
   \label{slproduction}
\eea
via $s$-channel $\gamma/Z$ exchange and $t$-channel $\cx$ exchange for
the first generation.
The $L,\, R$ states can be determined using beam polarisation, e.g.
$\sell_R\sell_R$ production is much larger for
right-handed $e^-_R$ than for left-handed $e^-_L$ electrons;
positron polarisation further enhances the effect.
The isotropic two-body decays
\bea
   \sell^- & \to & \ell^-\nt_i \ ,     \label{sldecay}
   \\
   \snu_\ell & \to & \ell^-\ch^+_i     \label{snudecay}
\eea
allow for a clean identification and lead
to a uniform lepton energy spectrum. 
The minimum and maximum (`endpoint') energies 
\begin{eqnarray}  
  E_{+/-} & = &
        \frac{\sqrt{s}}{4} 
        \left ( 1 - \frac{m_{\cx}^2}{m_{\slxx}^2} \right ) 
        \, (1 \pm \beta) \ ,  \label{eminmax}
	\\[.5ex]
        m_{\tilde{l}} & = &
        \frac{\sqrt{s}}{E_{-}+E_{+}}\,\sqrt{E_{-}\, E_{+}} \ , \label{emin}
	\\[.5ex]
        m_{\cx} & = & m_{\tilde{l}} \,
          \sqrt{1 - \frac{E_{-}+E_{+}}{\sqrt{s}/2}} \label{emax}
\end{eqnarray}
can be used for an accurate determination of
the masses of the primary slepton and the secondary neutralino/chargino.

\paragraph{Charged slepton production in continuum}

Examples of mass measurements using the lepton energy spectra of
$e^+_L e^-_R\to\smur\smur$ and $\ser\ser$ production at $\sqrt{s}=400~\GeV$
are shown in fig.~\ref{slr_spectra}~\cite{martyn}.
With a moderate luminosity of $\cL = 200~\fbi$
the masses can be determined with (highly correlated) errors of
$\delta m_{\smur} \simeq \delta m_{\nt_1}\simeq 0.2~\GeV$, respectively
$\delta m_{\ser} \simeq \delta m_{\nt_1}\simeq 0.1~\GeV$.
\begin{figure}[htb] \centering
  \mbox{\hspace{-5mm}
  \epsfig{file=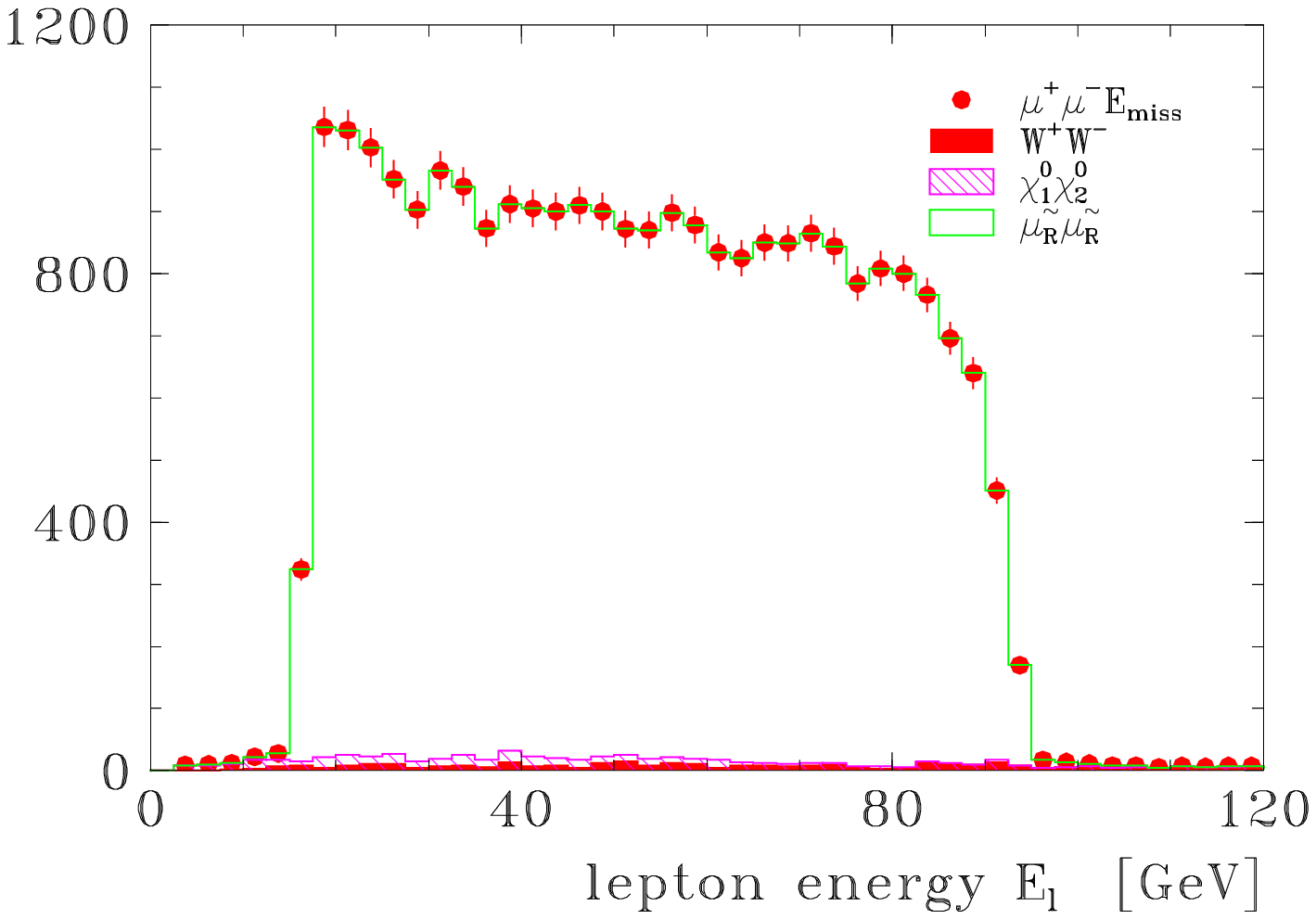,width=.52\textwidth} 
  \epsfig{file=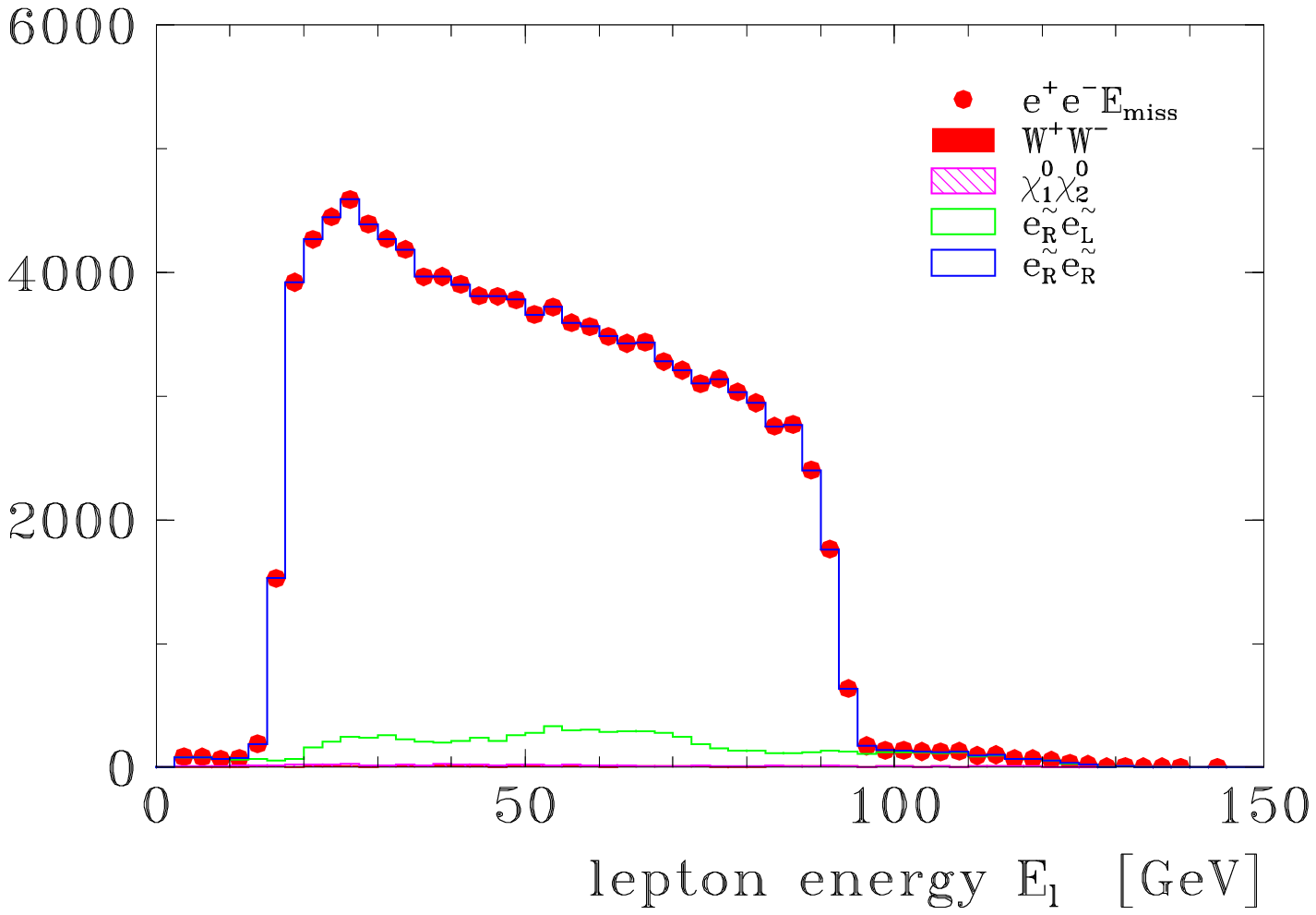,width=.52\textwidth} }
  \caption{Energy spectra of 
    $E_\mu$   from the reaction 
    $e^+_L e^-_R\to \smurp  \, \smurm   
    \to \mu^+ \nt_1 \,\, \mu^-\nt_1 $ (left) and
    $E_e$ from the reaction 
     $e^+_L e^-_R\to \serp  \, \serm   
    \to e^+ \nt_1 \,\, e^-\nt_1 $ (right),SPS~1a
     at $\sqrt{s}=400\;\GeV$ and $\cL=200\;\fbi$}
  \label{slr_spectra}
\end{figure}
A simultaneous analysis of $\ser\ser$, $\ser\sel$ and $\sel\sel$
production 
makes use of the different energy distributions of the final electrons
and positrons~\cite{nauenberg,dima}. 
The symmetric background is eliminated by a double subtraction of
$e^-$ and $e^+$ energy spectra and opposite electron beam polarisations.
This essentially results in a clean $\ser\sel$ sample where the
endpoints from $\ser$ and $\sel$ decays are easily measurable.
Assuming $\sqrt{s}=500~\GeV$ and $\cL=2\cdot500~\fbi$,
both selectron masses can be determined with an accuracy of
$\delta m_{\ser,\,\sel} \simeq 0.8~\GeV$.

In the production of
$e^+_Le^-_R\to\stau_1^+\stau_1^-\to\tau^+\nt_1\,\,\tau^-\nt_1$
the final $\tau$ leptons are incompletely measured, thus spoiling the
flat energy distribution of eq.~(\ref{eminmax}). However, the hadronic
decays $\tau\to \rho\nu_\tau, \ 3\pi\nu_\tau$ are still sensitive
to the primary $\stau_1$ mass~\cite{martyn}.
From the  $E_{3\pi}$ energy spectrum,
shown  in \fig{stau1_mass},
the expected uncertainty is $\delta m_{\stau_1} = 0.3~\GeV$,
assuming the $\nt_1$ mass to be known.
The heavier state $\stau_2$ is much more problematic to identify and
measure its mass. So far no simulations for continuum production exist.

\begin{figure}[htb]
\setlength{\unitlength}{1mm}  
  \begin{minipage}[t]{.5\textwidth}  
\begin{picture}(150,58)     
  \put(0,-5){ 
    \put(0,0){
      \epsfig{file=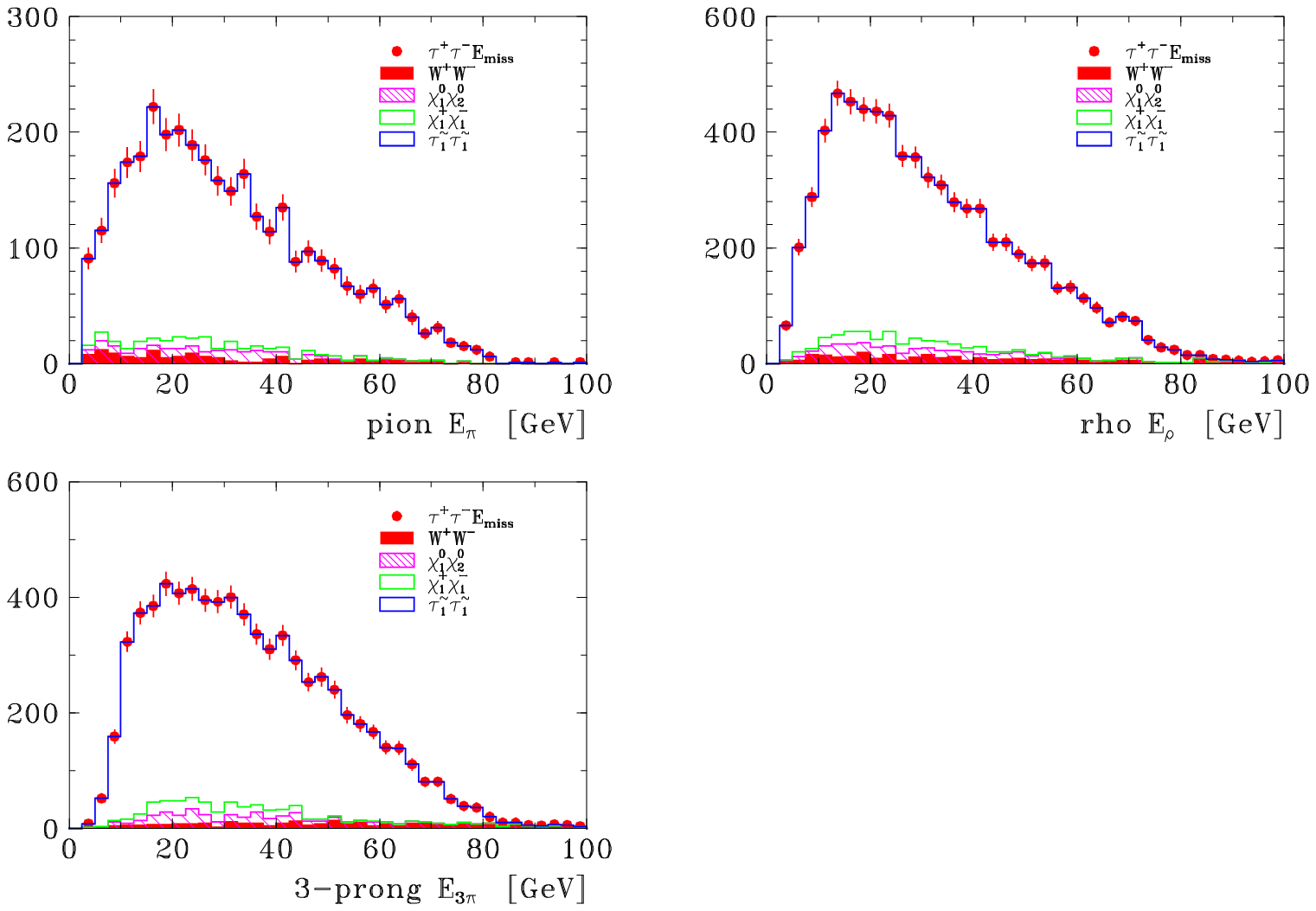,%
	bbllx=85pt,bblly=230pt,bburx=300pt,bbury=390pt,clip=,%
	width=1.04\textwidth} }   
      \put(16,52){$\tau\to 3\pi\nu$} 
    }
\end{picture}
\end{minipage} \hfill
\begin{minipage}{.48\textwidth}\vspace{-40mm} 
  \caption{Hadron energy spectrum $E_{3\pi}$ of
    $\tau\to3\pi\nu_\tau$ decays from the reaction
    $e^+_Le^-_R\to\stau_1^+\stau_1^1\to\tau^+\nt_1\,\,\tau^-\nt_1$,
    SPS~1a at $\sqrt{s}=400\,\GeV$ and
    $\cL=200\,\fbi$}
  \label{stau1_mass}
\end{minipage}
\end{figure}

\paragraph{Sneutrino production}

Sneutrinos are very difficult to detect because most of their decays
are invisible into a neutrino and a neutralino.
Only $\sim 8\% $ can be identified via the decay  (\ref{snudecay})
into the corresponding charged lepton and a chargino,
which subsequently decays via 
$\cx^\pm_1 \to \stau \nu \to \tau \nu \, \nt_1$. 
In practice only $\sne\sne$ production with the additional
t-channel chargino exchange has a large enough cross section to be observable.
A detection and measurement of $\snm$ and $\snt$ appears 
extremely challenging, if not hopeless.
The reaction
$e^+ e^-_L \to \sne\sne \to \nu_e \nt \, e^\pm \cx^\mp_1 
           \to e^\pm \tau^\mp\, \Eslash$ \ and $\tau\to \mu\nu\nu$
at $\sqrt{s}=500~\GeV$ has been studied~\cite{nauenberg}.
The energy spectrum of the primary electron can be used to
determine the electron-sneutrino mass to 
$\delta m_{\sne} = 1.2~\GeV$ and the chargino mass to
$\delta m_{\cx_1^\pm} = 1.4~\GeV$.

\paragraph{Threshold scans} 

Masses of accuracy $\cO(0.1~\GeV)$
can be obtained by scanning the excitation curve close to production
threshold.
In $e^+e^-$ annihilation  
slepton pairs $\sell^+_i\sell^-_i$, except $\ser\sel$, are produced in a
P-wave state with a characteristic rise of the cross section 
$\sigma_{\sell^+\sell^-}\sim \beta^3$, 
where $\beta=\sqrt{1-4\,m^2_\sell/s}$. 
On the other hand in $e^-e^-$ collisions $\ser^-\ser^-$ and 
$\sel^-\sel^-$ pairs are produced in a S-wave state with a steeper
rise of $\sigma_{\se^-\se^-}\sim \beta$.
Thus, the shape of the cross section carries
information on the mass and the quantum numbers.
The anticipated precision requires to take the finite width $\Gamma_{\sell}$
and higher order corrections into account~\cite{freitas}.
Examples of SPS~1a simulations within this frame 
are shown in \fig{scans}. 
Using polarised beams and $\cL=50~\fbi$ a (highly correlated)
two-parameter fit gives
$\delta m_{\ser} = 0.20~\GeV$ and $\delta\Gamma_{\ser}=0.25~\GeV$;
the resolution deteriorates by a factor of $\sim 2$ for 
$\smur\smur$ production.
For $e^-_Re^-_R\to\ser\ser$ the gain in resolution 
is substantial, yielding
$\delta m_{\ser} = 0.050~\GeV$ and $\delta\Gamma_{\ser}=0.045~\GeV$
with only a tenth of the luminosity, compared to $\ee$ beams.
But notice that a \lc\ in $e^-e^-$ mode is expected to provide much lower
luminosity, typically reduced by a factor of seven.
The precision on masses can be considerably improved if one is able fix
the sparticle width, e.g. by assuming model calculations.

\begin{figure}[htb]
\begin{picture}(150,40)
    \put(-8,0){
      \epsfig{file=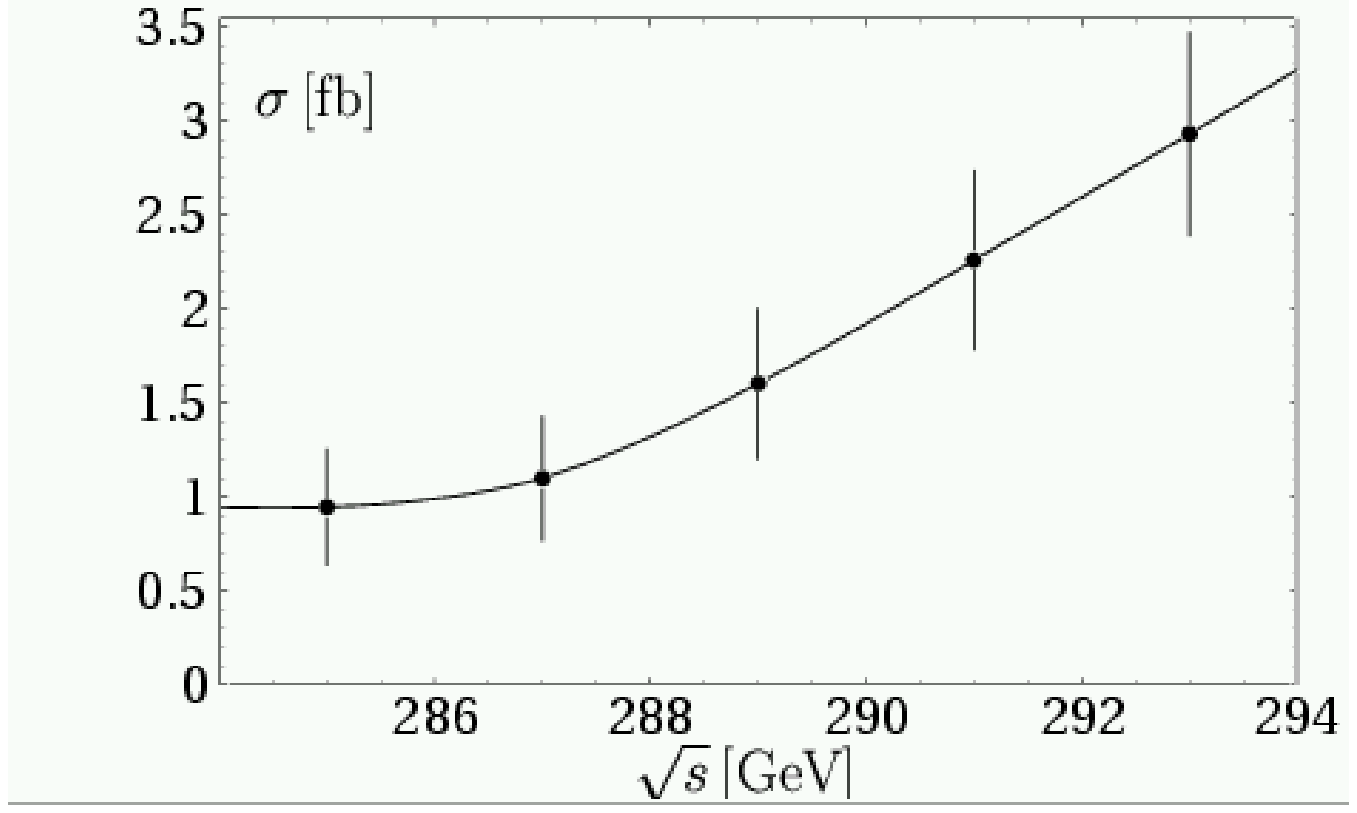,angle=0,width=.36\textwidth} 
      \epsfig{file=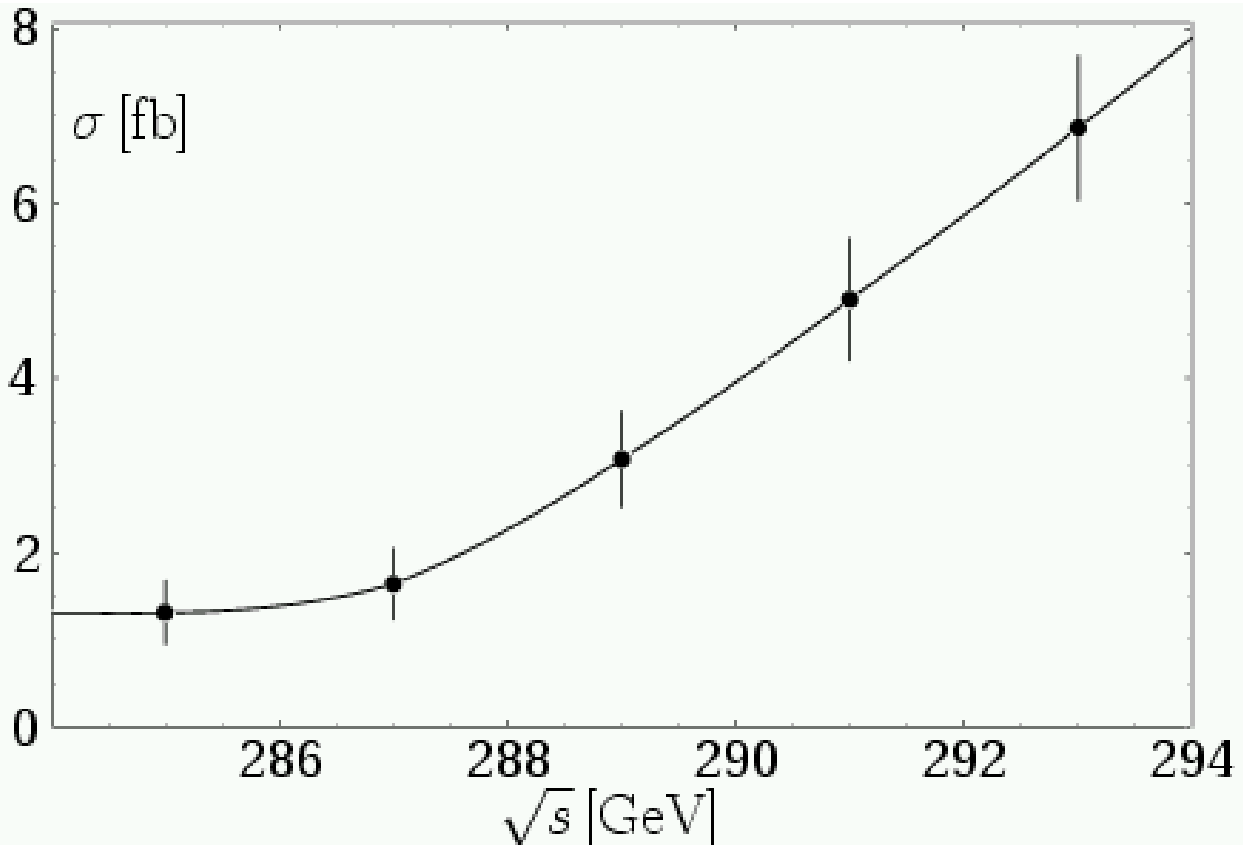,angle=0,width=.34\textwidth} 
      \epsfig{file=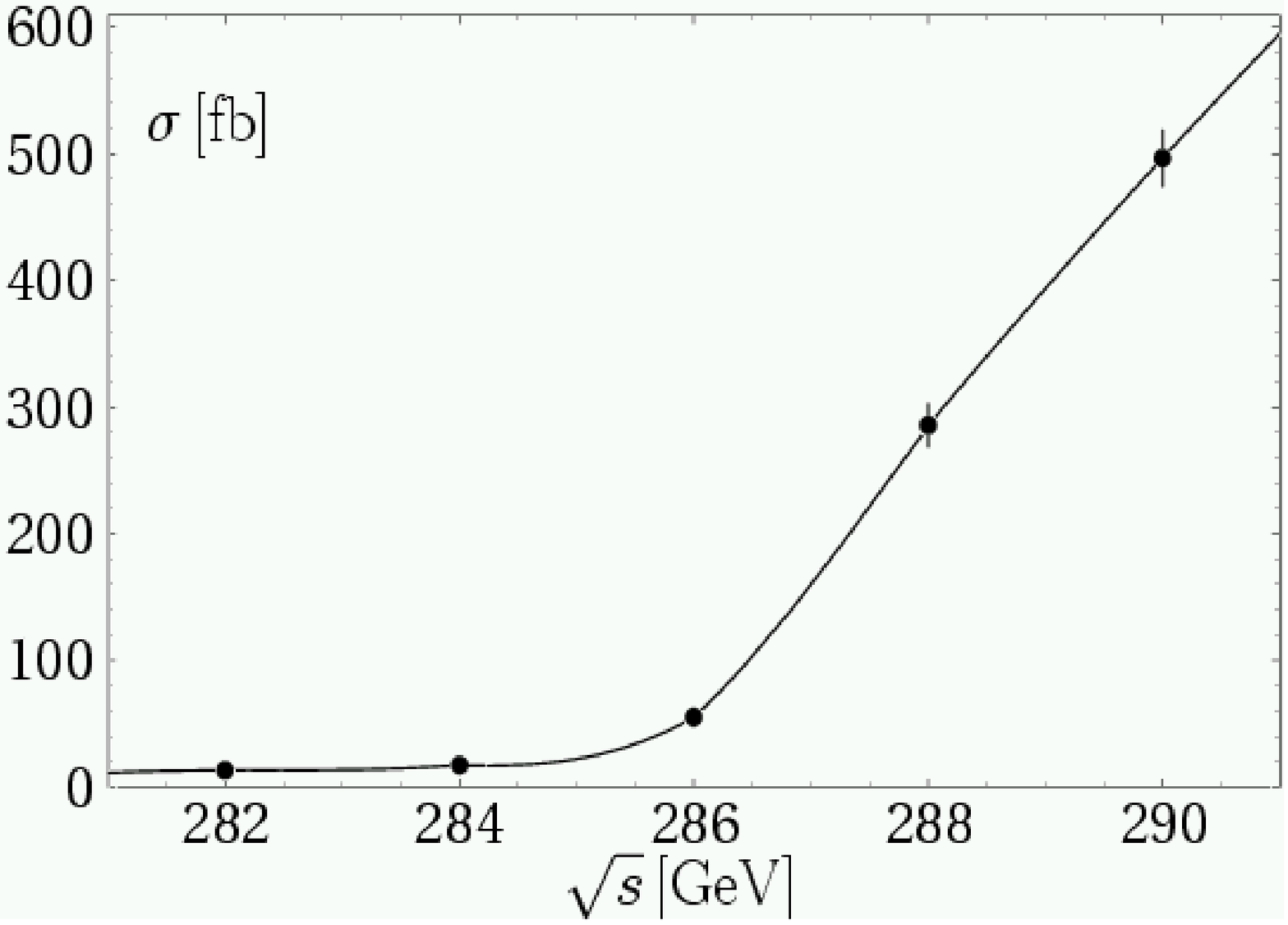,angle=0,width=.34\textwidth}  }
    \put(2,35){ {\footnotesize\small
        $e^+_Le^-_R\to\smur\smur$ \ \ \
      \footnotesize $10~\fbi$/point}   }
    \put(58,35){ {\footnotesize\small
        $e^+_Le^-_R\to\ser\ser$ \ \ \ 
      \footnotesize $10~\fbi$/point}    }
    \put(115,35){ {\footnotesize\small
        $e^-_Re^-_R\to\ser\ser$ \ \ \
      \footnotesize $1~\fbi$/point} }
\end{picture}
\caption{Cross sections at threshold for the reactions
  $e^+_Le^-_R\to\smur\smur$, $e^+_Le^-_R\to\ser\ser$ and 
  $e^-_Re^-_R\to\ser\ser$ for SPS~1a scenario
  including background~\cite{freitas}.
  Error bars correspond to a luminosity
  of $10~\fbi$ resp. $1~\fbi$ per point}
\label{scans}
\end{figure}

Taking the superior mass precision from the $\ser^-\ser^-$ scan and
combining it with the energy spectrum of  $\ser^+\ser^-$ production in
the continuum, see \fig{slr_spectra}, one can constrain the neutralino
mass to $\delta m_{\nt_1} = 0.05~\GeV$ or better. Such an accurate LSP
mass would have immediate consequences for other processes, 
e.g. the construction of the kinematically allowed minimum mass 
$m_{\rm min}(\sell)$~\cite{feng}
yielding for the smuon a resolution of 
$\delta m_{\smur} < 0.05~\GeV$~\cite{martyn}.

\subsubsection{Charginos and neutralinos}

Charginos and neutralinos are produced in pairs
\begin{eqnarray}
    e^+e^- & \to & \cpl_i \cm_j  \qquad \qquad \qquad [i,j = 1,2] 
    \label{scproduction} \\[.5ex] 
           & \to & \nt_{i} \nt_{j}  \qquad \qquad \qquad \ \ 
                           [i,j = 1, \ldots ,4] 
    \label{snproduction}
\end{eqnarray}
via $s$-channel $\gamma/Z$ exchange and $t$-channel $\se$ or $\sne$
exchange. 
Beam polarisations are important to study the $\cx$ properties
and couplings, e.g. by manipulating the $\sne$ exchange contribution.
Charginos and neutralinos decay into their lighter partners
and gauge bosons or sfermion-fermion pairs 
\begin{eqnarray}
  \cx_i & \to &  Z / W\,\cx_j \ ,
  \label{sxdecay} \\[.5ex]
  \cx^\pm_1 & \to & \stau_1\nu_\tau \to \tau\nu_\tau\, \nt_1 \ ,
  \label{scdecay} \\[.5ex]
  \nt_2 &   \to & \sell \ell \to \ell\ell\, \nt_1 \ .
  \label{sndecay}
\end{eqnarray}

\paragraph{Chargino production}

Light charginos are being detected via the process
$e^+_Re^-_L\to\cx_1^+\cx_1^1\to\tau^+\nu_\tau\nt_1\,\,\tau^-\nu_\tau\nt_1$.
The observed final state particles are the same as in the production
of $\stau_1\stau_1$ and $\nt_2\nt_1 \to \tau\tau\nt_1\nt_1$, a severe
background which may be partially removed by topological cuts.
The chargino mass may be reconstructed from the energy spectra of
hadronic $\tau$ decays. In a model similar to SPS~1a a simulation at
$\sqrt{s} = 400~\GeV$ and $\cL=200~\fbi$ yields a mass uncertainty of
$\delta m_{\cpm_1} = 1.5~\GeV$~\cite{kato}. 

Alternatively the cross section at threshold can be scanned
which rises fairly
steeply as $\sigma_{\cx\cx} \sim \beta$, characteristic for the 
chargino's  spin $1/2$.
The excitation curve is shown in \fig{c11_mass} and provides a
chargino mass determination accurate to $\delta m_{\cpm} = 0.55~\GeV$.

\begin{figure}[htb]
\setlength{\unitlength}{1mm}  
  \begin{minipage}[t]{.5\textwidth}  
\begin{picture}(150,70)     
  \put(0,-5){ 
    \epsfig{file=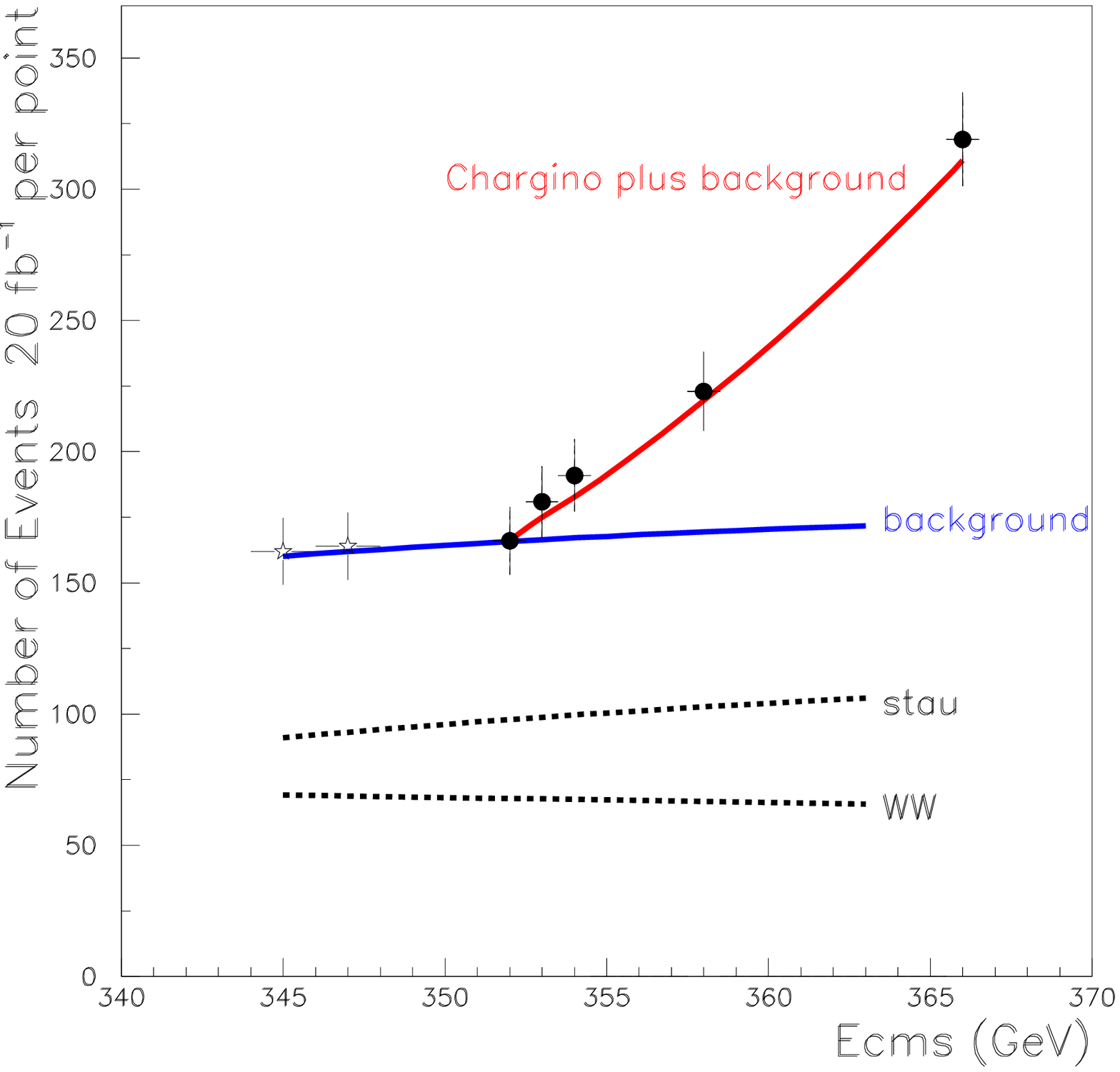,width=1.\textwidth} 
    }
\end{picture}
\end{minipage} \hfill
\begin{minipage}{.48\textwidth}\vspace{-40mm} 
  \caption{Threshold scan of 
    $e^+_Re^-_L\to\cx_1^+\cx_1^1\to\tau^+\nu_\tau\nt_1\,\,\tau^-\nu_\tau\nt_1$
    for SPS~1a assuming $\cL=100\,\fbi$}
  \label{c11_mass}
\end{minipage}
\end{figure}

The heavy chargino may be produced via $\ee\to \cx^\pm_1\cx^\mp_2$ and
identified via the bosonic two-body decays 
$\cx^\pm_2 \to Z \cx^\pm_1, \ W\nt_1$. From the energy distributions of
the reconstructed $Z,\, W$ a resolution of 
$\delta m_{\cpm_2}\sim 3~\GeV$ is estimated. 
Complete simulations of the process don't yet exist.

\paragraph{Neutralino production}

The continuum production of neutralino $\nt_2$ has been studied in
the reaction $\ee\to \nt_1\nt_2$ followed the dominant decay mode
$\nt_2\to \tau\tau\nt_1$~\cite{sec410_desch}. Both $\tau$'s are required to
be emitted in the same hemisphere, but the \susy\ background is still
very large. The shape of the $m_{\tau\tau}$ effective mass
distribution is sensitive to the neutralino mass within
$\delta m_{\nt_2} \sim 2~\GeV$.

More promising is a threshold scan of 
$e^+_Re^-_L\to \nt_2\nt_2 \to 4\,\tau + 2\,\nt_1$, which has very low
background. Assuming $\cL=100~\fbi$ one achieves a precision of
$\delta m_{\nt_1} = 1.2~\GeV$.

The simulation of inclusive cascade decays 
$\nt_2 \to e^+e^-\nt_1,\, \mu^+\mu^-\nt_1$, where the di-lepton masses
are sensitive to the $\nt_2-\nt_1$ mass difference has not yet been
tried. These chains are important in the \lhc\ 
analysess~\cite{lhc_atlas, lhc_cms},
but they are difficult to be observed at a \lc\, 
since the decay modes
are  an order of magnitude less frequent and 
the signature competes with a large
signal from $\se\se$, $\smu\smu$ production.

High mass neutralinos $\nt_3$, $\nt_4$ can be detected via the
bosonic $Z,\,W$ decays (\ref{sxdecay}). 
First studies at $\sqrt{s}=750~\GeV$ show that for high luminosity
the $Z$ energy spectra can be used to reconstruct the heavy neutralino
masses with resolutions of a 
few GeV~\cite{nauenberg}.

\subsubsection{Stop production}

Light stop production may become accessible if the \lc\ can be
operated at $\sqrt{s} = 1~\TeV$. The observation will be based on the
production and decay sequence
$e^+e^- \to \st_1\st_1 \to b \cx^+_1\,\bar{b} \cx^-_1
        \to b \tau^+\nu\nt_1 \, \bar{b} \tau^+\nu\nt_1$,
i.e. the signature are 2 $b$-jets + 2 $\tau$'s in the final state.
The analysis techniques are similar to those in the slepton sector. 
One can use the energy spectrum of the $b$-jets and exploit the $b\,b$
correlations to construct the minimum kinematically allowed mass 
$m_{\rm min}(\st)$ assuming $m_{\cpm}$ to be known~\cite{feng}.
With a luminosity of $1000~\fbi$ 
the rate will be sufficient to achieve a mass resolution of
$\delta m_{\st_1} = 2~\GeV$.

\begin{table} \centering
\begin{tabular}{|c|c|c|l|}
\hline
               & $m~[\GeV]$ & $\Delta m~[\GeV]$ & Comments\\ \hline
$\tilde{\chi}^\pm_1$ & 176.4           & 0.55      & simulation threshold scan ,
                                                     100 fb$^{-1}$ \\
$\tilde{\chi}^\pm_2$ & 378.2           & 3         & estimate
    $\cx^\pm_1\cx^\mp_2$, spectra $\cx^\pm_2 \to Z \cx^\pm_1,\, W \nt_1$
               \\ 
\hline
$\tilde{\chi}^0_1$   &  96.1           & 0.05      & combination of all methods \\
$\tilde{\chi}^0_2$   & 176.8           & 1.2       & simulation threshold scan 
         $\tilde{\chi}^0_2\tilde{\chi}^0_2$,             100 fb$^{-1}$ \\
$\tilde{\chi}^0_3$   & 358.8           & 3 -- 5    & spectra
       $\nt_3\to Z \nt_{1,2}$, \ $\nt_2\nt_3, \nt_3\nt_4$, 750 GeV, $>1000~\fbi$ \\
$\tilde{\chi}^0_4$   & 377.8           & 3 -- 5    & spectra
      $\nt_4\to W \cx^\pm_1$, \  $\nt_2\nt_4, \nt_3\nt_4$,  750 GeV, $>1000~\fbi$ \\
\hline
$\tilde{e}_R$        & 143.0           & 0.05      & $e^-e^-$ threshold scan,
                                                     10 fb$^{-1}$ \\
$\tilde{e}_L$        & 202.1           & 0.2       & $e^-e^-$ threshold scan 
                                                     20 fb$^{-1}$ \\
$\tilde{\nu}_e$      & 186.0           & 1.2       & simulation 
                                                     energy spectrum, 500 GeV,
                                                     500 fb$^{-1}$ \\
$\tilde{\mu}_R$      & 143.0           & 0.2       & simulation
						     energy spectrum, 400 GeV,
                                                     200 fb$^{-1}$ \\
$\tilde{\mu}_L$      & 202.1           & 0.5       & estimate threshold scan,
                                                  100 fb$^{-1}$ \cite{grannis} \\
$\tilde{\tau}_1$     & 133.2           & 0.3       & simulation 
                                        energy spectra, 400 GeV, 200 fb$^{-1}$ \\
$\tilde{\tau}_2$     & 206.1           & 1.1       & estimate threshold scan,
                                            60 fb$^{-1}$  \cite{grannis} \\ 
\hline
$\tilde{t}_1$        & 379.1           & 2         & estimate 
                    $b$-jet spectrum, $m_{\rm min}(\st)$, 1TeV, 1000 fb$^{-1}$ \\
\hline
\end{tabular} 
\caption{Sparticle masses and their expected precisions in Linear
  Collider experiments, 
  SPS~1a mSUGRA scenario.
\label{sps1_results}
}
\end{table}

\subsubsection{Summary}

The results of the SPS~1a sparticle mass studies are summarised in
table~\ref{sps1_results}, where the best values expected
from either production in the continuum or threshold scans are quoted.
For most sparticles they are based on realistic Monte Carlo and
detector simulations and reasonable assumptions on the \lc\ performance.
Only the heavy $\nt$, $\cpm$ and $\st_1$ states rely on some plausible
estimates. 
Typical accuracies in the per cent to per mil range can be expected at
a Linear Collider.
These precision measurements serve as input to explore \susy\
scenarios in a model independent way~\cite{susypar}.
It should be pointed out once more that \lc\ experiments provide
much more valuable information, such as accurate values on mixing
angles, couplings and quantum numbers.

The final goal would be to perform a combined analysis of all
available experimental information 
-- including masses, cross sections, branching ratios, etc. from the
\lc\ and \lhc~\cite{lhc_atlas,lhc_cms} --
in order to arrive at a high precision determination of the \susy\
Lagrange parameters and to extrapolate them to high scales aiming at
reconstructing the fundamental parameters and the mechanism of
supersymmetry breaking.
This requires also that theoretical calculations have to be
developed to the same accuracy as the anticipated experiments. 
Such a programme is in progress~\cite{spa}.

\subsection{\label{sec:414a} Sparticle mass measurements from LHC
analyses and combination with LC results}

{\it M.~Chiorboli, A.~De Roeck, B.K.~Gjelsten, 
K.~Kawagoe, E.~Lytken, D.~Miller,
P.~Osland, G.~Polesello and A.~Tricomi}

\vspace{1em}
The analyses briefly described in the previous sections
provide a series of measurements of kinematic quantities 
which are directly related to sparticle masses and branching
fractions through simple algebraic formulae.\par
Some of the measurements, performed in the framework of both ATLAS
and CMS, can be used to extract a direct measurement of
the masses of the sparticles.
Some others, in particular the analyses in the stop/sbottom
sectors provide the measurements of complex quantities 
which are used in other sections (see Sec.~\ref{sec:413}) to strongly
constrain the parameters of the MSSM.
We concentrate in this section on the information of masses
which can be obtained from the LHC analyses, and on how the 
information from LHC and LC can be combined.

In order to have a single consistent set, we base ourselves
here only on the ATLAS analyses, which address a broader 
range of signatures than the CMS ones. We have however checked
that the CMS results, once different analysis assumptions
are correctly taken into account, give results consistent 
with the ATLAS ones.

We summarize in Table~\ref{tab:summes} all of the used
measurements, with the statistical error corresponding to 
the ultimate integrated luminosity for the LHC of 300~fb$^{-1}$.
The central values of the measured quantities are calculated
from the mass spectrum of ISASUSY 7.58. \par
For all measurements, a systematic error from the uncertainty 
on the energy scale in the detector is given, corresponding to 
1\% for the measurements involving hadronic jets, and 0.1\% for 
purely leptonic measurements. For the case of $m_{\tau\tau}^{max}$
and $m(\tq_R)$, the errors given in the first column
(respectively 5 and 10 GeV) 
are very conservative estimates of the systematic uncertainty 
on the precision with which the observed
structures can be related to the corresponding physical quantities.
More detailed studies are needed for a firmer estimate.\par
\begin{table}[htb]
\begin{center}
\caption{Summary table of the SUSY measurements which can be performed
at the LHC with the ATLAS detector. The statistical errors are given 
for the ultimate integrated luminosity of 300~$fb^{-1}$. The uncertainty
in the energy scale results in an error of 1\% for measurements 
including jets, and of 0.1\% for purely leptonic mesurements.}
\label{tab:summes}
\vskip 0.2cm
\begin{tabular}{|l|c|c|c|c|}
\hline
& & \multicolumn{3}{c|}{Errors}\\
Variable & Value (GeV) & Stat. (GeV) & Scale (GeV) & Total \\
\hline
\hline
$m_{\ell\ell}^{max}$            &    77.07 &    0.03 &    0.08 &    0.08\\
$m_{\ell\ell q}^{max}$          &   428.5 &    1.4 &    4.3 &    4.5\\
$m_{\ell q}^{low}$              &   300.3 &    0.9 &    3.0 &    3.1\\
$m_{\ell q}^{high}$             &   378.0 &    1.0 &    3.8 &    3.9\\
$m_{\ell\ell q}^{min}$          &   201.9 &    1.6 &    2.0 &    2.6\\
$m_{\ell\ell b}^{min}$          &   183.1 &    3.6 &    1.8 &    4.1\\
$m(\ell_L)-m(\lsp)$             &   106.1 &    1.6 &    0.1 &    1.6\\
$m_{\ell\ell}^{max}(\tchi^0_4)$ &   280.9 &    2.3 &    0.3 &    2.3\\
$m_{\tau\tau}^{max}$            &    80.6 &    5.0 &    0.8    &    5.1\\
$m(\tg)-0.99\times m(\lsp)$     &   500.0 &    2.3 &    6.0    &    6.4\\
$m(\tq_R)-m(\lsp)$              &   424.2 &   10.0 &    4.2    &   10.9\\
$m(\tg)-m(\tb_1)$               &   103.3 &    1.5 &    1.0    &    1.8\\
$m(\tg)-m(\tb_2)$               &    70.6 &    2.5 &    0.7    &    2.6\\
\hline
\end{tabular}
\end{center}
\end{table}
The values in the table can either be used stand-alone in order to 
extract a consistent set of mass measurement from the LHC,
or combined with equivalent measurements from the LC in order to 
obtain a global picture of the SUSY mass spectra.\par
The available measurements are naturally divided into two classes:
\begin{itemize}
\item
a set of six edge measurements, each involving typically three 
among the masses of $\tilde{q}_L$, $\tilde{\chi}_2^0$,
$\tilde{l}_R$ and $\tilde{\chi}_1^0$
\item
a set of measurements involving the mass of an additional 
sparticle and one or more of the sparticles involved in the edge measurement.
\end{itemize}
A two-step strategy can therefore be used in order to calculate
the SUSY mass spectrum from the LHC data alone.
The first step consists in solving the system of equations 
(\ref{Formula1})--(\ref{Formula5})
for $\tilde{q}_L$, $\tilde{b}_1$, $\tilde{\chi}_2^0$,
$\tilde{l}_R$ and $\tilde{\chi}_1^0$. 
The second step consists, for each additional particle, in calculating
explicitly its mass using as an input the mass values for the
lighter sparticles calculated using the edge measurements.\par

The system of equations involving the edge measurements 
is solved numerically, by finding the set of mass values
 which minimizes the $\chi^2$ function,
\begin{eqnarray}
\chi^2=\sum_j\chi_j^2=\sum_j
\left[\frac{E^{\rm theory}_j(\vec m)-E^{\rm exp}_j}
{\sigma^{\rm exp}_j } \right]^2 .
\end{eqnarray}

Here $E^{\rm theory}_j(\vec m)$ is the theoretical value for the 
masses $\vec m=\{m_{\tilde{q}_L}, m_{\tilde{b}_1},
m_{\tilde{\chi}_2^0}, m_{\tilde{l}_R}, m_{\tilde{\chi}_1^0} \}$, 
$E^{\rm exp}_j$ is the measurement of edge number $j$, and
$\sigma^{\rm exp}_j$ is the estimated error of the edge measurement.

In order to assess the precision with which ATLAS can determine the
sparticle masses from the procedure of fitting kinematic edges, we generate
a large number of LHC ``experiments''.  For each ``experiment'' $i$ we
construct a set of edge measurements $E^i_j$ from the estimated errors in the
following way
\begin{equation}
E^i_j = E^{\rm nom}_j 
+ a^i_j\sigma^{\rm fit}_j 
+ b^i \sigma^{Escale}_j ,
\end{equation}
where $E^{\rm nom}_j$ is the nominal value for edge number $j$, 
$\sigma^{\rm fit}_j$ is the combined error on the fit value and 
$\sigma^{Escale}_j$ is the energy scale error. Within each ``experiment''
$b^i$ and $a^i_j$ are picked from Gaussian distributions of mean value 0 
and width 1. 
The resulting masses have a near Gaussian distribution around the
nominal values with RMS deviations given in Table~\ref{tab:summes}.

Since the analytic formulae for the edge positions,
Eqs.~(\ref{Formula1})--(\ref{Formula5}), consist mainly of mass {\it
differences} there are strong correlations between the masses.  Due to the
very accurate determination of the $m_{ll}$ edge, see Eq.~(\ref{Formula1}) and
Table~\ref{TABedges}, the two neutralinos and the slepton are very
interdependent. If one of them is overestimated, so are the other two. This is
illustrated in Fig.~\ref{masscorr} (left panel) for $\tilde{\chi}_1^0$ and
$\tilde{l}_R$.  The correlation between $\tilde{\chi}_1^0$ and
$\tilde{\chi}_2^0$ or $\tilde{l}_R$ and $\tilde{\chi}_2^0$ are similar.  In
contrast, the correlation between $\tilde{q}_L$ and the lighter particles is
less severe, but still significant, due to the less precise measurements of
the edge positions involving a jet. This can be seen in the center panel
of Fig.~\ref{masscorr}.  Finally, the rather imprecise determination of edges
involving a b-quark, leads to the $\tilde{b}_1$ having only a mild correlation
with the lighter masses, as seen in the right-hand panel of 
Fig.~\ref{masscorr}.

\begin{figure}[htp]
\begin{center}
\includegraphics[scale=.25]{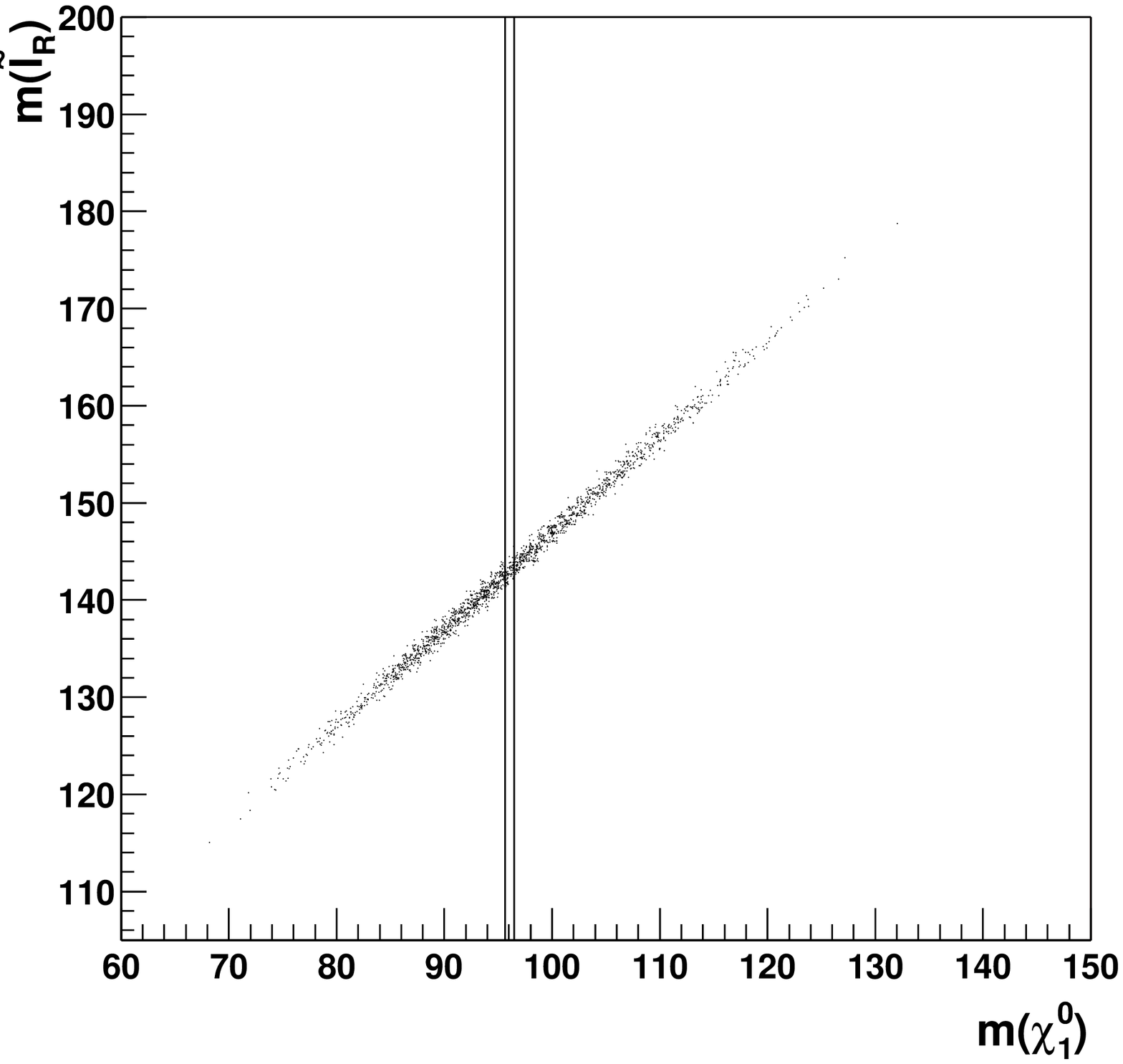}
\includegraphics[scale=.25]{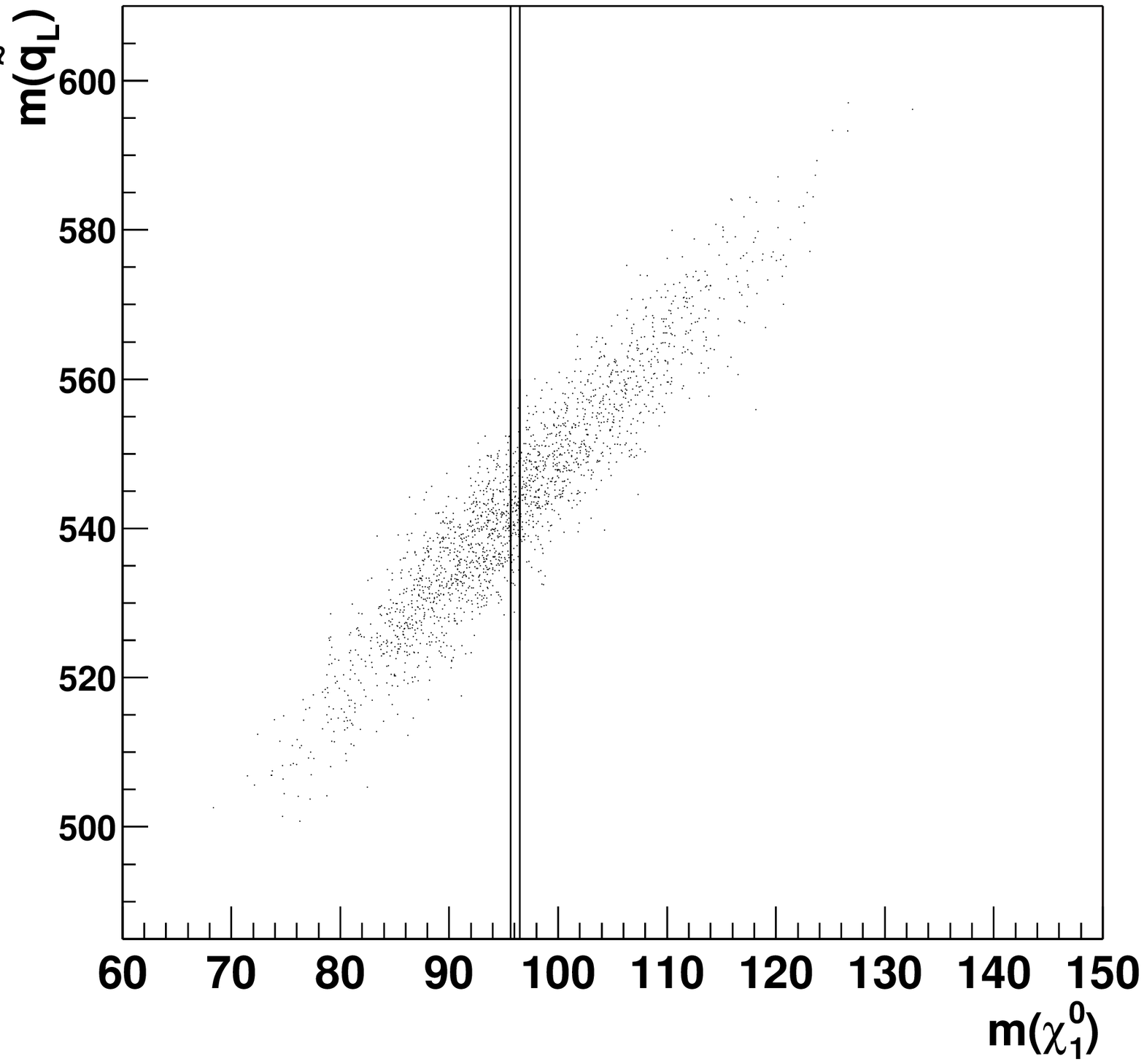}
\includegraphics[scale=.25]{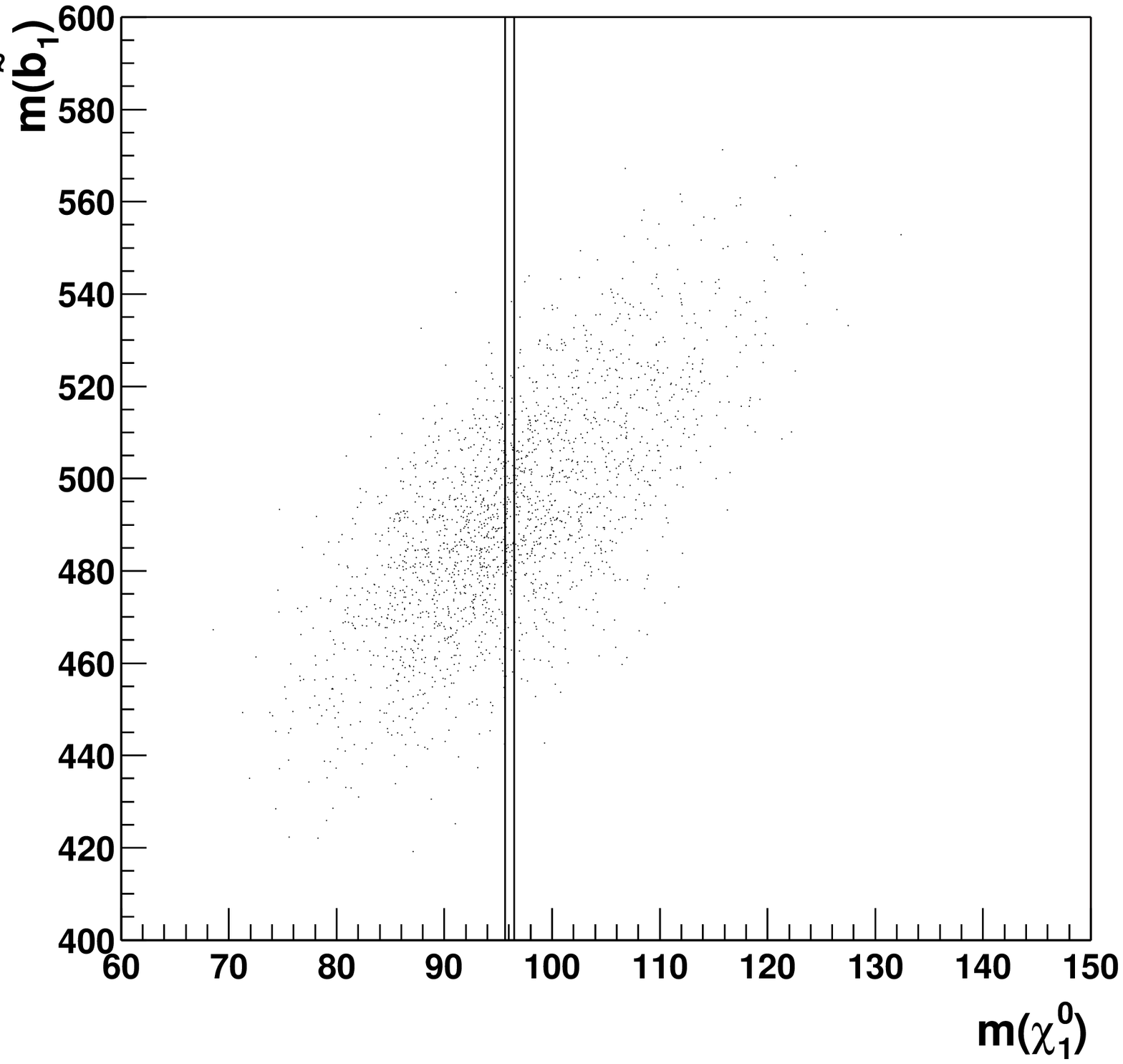}
\caption{Mass correlation plots. Dots: LHC alone.
Vertical bands: Fixing $m_{\tilde{\chi}_1^0}$ to within
$\pm2\sigma$ with LC input ($\sigma=0.2\%$).}
\label{masscorr}
\end{center}
\end{figure}
Using the mass measurements  for the lighter neutralinos 
obtained  through the edge analysis, model-independent 
mass measurements can be extracted from the measurements 
in Table~\ref{tab:summes}.\par
The resulting uncertainties
are given in Table~\ref{sec414a_TABLHCLC1}, for an integrated 
luminosity of 300~fb$^{-1}$. The measurement of $m(\tb_1)$ obtained
with the edges is superseded by the one obtained
through the study of the gluino decay chain.
For the $\ttau_1$ and $\tq_R$ masses the edge measurements
given in Table~\ref{tab:summes} include a large systematic error.
This is a rough and conservative estimate of the actual 
uncertainty, due to the fact that the analysis concerned
are very preliminary. We expect this uncertainty to be 
significantly reduced by more detailed analyses.
We quote for the mass measurement extracted from these 
analyses an interval of uncertainties, where the 
upper value corresponds to using the systematic error in full,
and the lower value to putting the systematic uncertainty to zero.
The analyses at the LHC are not able to discriminate the 
production of squarks of the first four flavours, therefore 
the uncertainty is quoted for generic $\tq_R$ and $\tq_L$.
The values of the masses are derived from kinematic measurements,
not from direct measurement. Therefore, as shown clearly in 
 Fig.~\ref{masscorr} they are strongly correlated. Therefore any analysis
trying to extract a global fit from the LHC measurements should start
from the actual measurements rather than from the values of the masses shown.
Most of the sparticle measurements shown rely on knowing the values
of the lighter sparticles, which are accessible at the LC. 
The combined result for LHC and LC was therefore obtained by 
giving as an input to the mass calculation program the LC values for the
masses of the sleptons and of the lighter neutralinos, with the errors quoted
in Table~\ref{sps1_results}. The results are shown in the second column 
of Table~\ref{sec414a_TABLHCLC1}. One can notice that the improvement on the
measurements of the masses of squarks and gluinos are moderate, as
the LHC measurements are dominated by the 1\% error on the jet energy scale.
Achieving this level of uncertainty on the jet scale will require
very detailed studies on the part of the LHC collaborations, and it
is unlikely that the uncertainty could be reduced below the 1\% level.
As an academic 
exercise we evaluated how the mass measurement uncertainty 
would change if one respectively assumes 0.5\% jet energy scale error 
and no jet energy scale error.
Note that by using the sparticle masses determined at a LC
the correlations between the masses determined at the LHC 
can be strongly reduced, as demonstrated in Fig.~\ref{masscorr}.
The vertical bands show the 2$\sigma$ precision with which the mass of
$\tilde{\chi}_1^0$ can be determined at a LC. E.g., using  
$m_{\tilde{\chi}_1^0}$ from the LC 
basically fixes the mass of the $\tilde{l}_R$ determined from the LHC data.

\begin{table}[htp]
\begin{center}
\caption{The RMS values of the mass distribution in the case of 
the LHC alone, and combined with measurements from the LC.
The results from Table~\ref{tab:summes} and 
Table~\ref{sps1_results} are used as an input. 
All numbers in GeV.} \vspace{4pt}
\label{sec414a_TABLHCLC1}
\begin{tabular}{cccc}
\hline\hline
 & LHC & LHC+LC  \\
\hline
$\Delta m_{\tilde{\chi}_1^0}$& 4.8 & 0.05 (LC input) \\
$\Delta m_{\tilde{\chi}_2^0}$& 4.7 & 0.08 \\
$\Delta m_{\tchi^0_4}$ & 5.1 & 2.23 \\
$\Delta m_{\tilde{l}_R}$ & 4.8 & 0.05 (LC input)  \\
$\Delta m_{\tl_L}$ & 5.0 & 0.2 (LC input) \\
$\Delta m_{\tau_1}$ & 5-8 & 0.3 (LC input) \\
$\Delta m_{\tilde{q}_L}$ & 8.7  & 4.9  \\
$\Delta m_{\tq_R}$ & 7-12 & 5-11  \\
$\Delta m_{\tb_1}$ & 7.5 & 5.7  \\
$\Delta m_{\tb_2}$ & 7.9 & 6.2  \\
$\Delta m_{\tg}$ & 8.0 & 6.5  \\
\hline
\end{tabular}
\end{center}
\end{table}
\begin{table}[htp]
\begin{center}
\caption{The RMS values of the mass distribution in the case of
the LHC alone, and combined with measurements from the LC, under
the assumption that an uncertainty on the jet energy scale
of 0.5\% can be achieved at the LHC.
The results from Table~\ref{tab:summes} and
Table~\ref{sps1_results} are used as an input.
All numbers in GeV.} \vspace{4pt}
\label{TABLHCLC2}
\begin{tabular}{cccc}
\hline\hline
 &  LHC (0.5\% jet scale)  & LHC (0.5\% jet scale) + LC\\
\hline
$\Delta m_{\tilde{q}_L}$ &  7.8 & 2.6  \\
$\Delta m_{\tb_1}$ &  6.0  & 3.7 \\
$\Delta m_{\tb_2}$ &  6.4  & 4.3 \\
$\Delta m_{\tg}$ &  6.0 &  3.7  \\
\hline
\end{tabular}
\end{center}
\end{table}
\begin{table}[htp]
\begin{center}
\caption{The RMS values of the mass distribution in the case of
the LHC alone, and combined with measurements from the LC, 
for the case where a vanishing uncertainty on the jet energy scale at
the LHC is assumed.
The results from Table~\ref{tab:summes} and
Table~\ref{sps1_results} are used as an input.
All numbers in GeV.} \vspace{4pt}
\label{TABLHCLC3}
\begin{tabular}{cccc}
\hline\hline
 &  LHC (0\% jet scale)  & LHC (0\% jet scale) + LC\\
\hline
$\Delta m_{\tilde{q}_L}$ &  7.4 & 0.8  \\
$\Delta m_{\tb_1}$ &  5.4  & 2.8 \\
$\Delta m_{\tb_2}$ &  5.8  & 3.4 \\
$\Delta m_{\tg}$ &  5.2 &  2.3  \\
\hline
\end{tabular}
\end{center}
\end{table}


\subsection{\label{sec:414} Susy parameter determination in combined
analyses at LHC/LC}

{\it K.~Desch, J.~Kalinowski, G.~Moortgat-Pick, M.M.~Nojiri and
G.~Polesello}

\vspace{1em}
\def\nn {\nonumber}

\def\ti    {\tilde}

\def\sel   {{\ti e}_L}
\def\ser   {{\ti e}_R}
\def\sne  {{\ti\nu}_e}
\def\el   {{\ti e}^-_L}
\def\er   {{\ti e}^-_R}
\def\nt   {{\ti \chi}^0}

\def\stauone  {{\ti\tau}_1}
\def\stautwo  {{\ti\tau}_2}
\def\snt      {{\ti\nu}_\tau}

\def\anue {{\bar\nu}_e}
\def\anum {{\bar\nu}_\mu}
\def\anut {{\bar\nu}_\tau}

\def\cL   {\cos\Phi_L}
\def\sL   {\sin\Phi_L}
\def\ctL  {{\cos2\Phi_L}}
\def\stL  {{\sin2\Phi_L}}
\def\cR   {\cos\Phi_R}
\def\sR   {\sin\Phi_R}
\def\ctR   {\cos2\Phi_R}
\def\stR   {\sin2\Phi_R}

\newcommand{\mse}[1]   {m_{\ti e_{#1} }}
\renewcommand{\mstau}[1] {m_{\ti \tau_{#1} }}
\renewcommand{\mnt}[1]   {m_{\ti \chi^0_{#1} }}
\renewcommand{\mch}[1]   {m_{\ti \chi^\pm_{#1} }}
\newcommand{\msnue}     {m_{\ti \nu_e}}

\renewcommand{\gsim}{\;\raisebox{-0.9ex}
           {$\textstyle\stackrel{\textstyle>}{\sim}$}\;}

\renewcommand{\lsim}{\;\raisebox{-0.9ex}{$\textstyle\stackrel{\textstyle<}
          {\sim}$}\;}

\renewcommand{\eq}[1]{eq.~(\ref{#1})}
\renewcommand{\fig}[1]{fig.~\ref{#1}}
\renewcommand{\tab}[1]{table~\ref{#1}}


\noindent {\small
We demonstrate how the interplay of a future $e^+e^-$ LC at its first
stage with $\sqrt{s} \lsim 500$~GeV and of the LHC could lead to a
precise determination of the fundamental SUSY parameters in the
gaugino/higgsino sector without assuming a specific supersymmetry
breaking scheme.  We demonstrate this for the benchmark scenario
SPS1a. Taking into account realistic errors for the masses and cross
sections measured at the LC with polarised beams, including errors
coming from polarisation measurements, masses of the heavier states
can be predicted.  These can provide significant guidance in the
interpretation of dilepton spectrum endpoints leading to reliable mass
measurements at the LHC. These mass measurements are then used to
improve the determination of the fundamental SUSY parameters. The
results clearly demonstrate the complementarity of the LHC and LC, and
the benefit from the joint analyses of their data.  
}


\subsubsection{Introduction}
Supersymmetry (SUSY) is one of the best motivated extensions of the
Standard Model (SM). However, since SUSY has to be broken even the 
minimal version, the unconstrained MSSM, has 105 new parameters. SUSY
analyses at future experiments, at the LHC and at a future Linear
Collider (LC) \cite{ams03}, will have
to focus on the determination of these parameters in as
model-independent a way as possible. 

With so many new parameters clear strategies will be needed in
analysing the experimental data  \cite{parameters}. 
An interesting possibility 
to resolve the new physics is to start with the 
gaugino/higgsino particles which are expected to be among 
the lightest SUSY particles. 
In the unconstrained MSSM  this sector depends only    
on 4 parameters at tree level: $M_1$, $M_2$, $\mu$ and $\tan\beta$ -- the U(1)
and SU(2) gaugino masses, the higgsino mass parameter and the ratio of
the vacuum expectations of the two Higgs fields, respectively.   

Some strategies have been worked out for the determination at the tree
level the parameters $M_1$, $M_2$, $\mu$, $\tan\beta$ 
even if only the light gaugino/higgsino particles, $\tilde{\chi}^0_1$,
$\tilde{\chi}^0_2$ and $\tilde{\chi}^\pm_1$ were kinematically
accessible at the first stage of the LC \cite{ckmz}.  In this report
we demonstrate how such an LC analysis could be strengthened if in
addition some information on the mass of the heaviest neutralino from
the LHC is available \cite{dkmnpJHEP}.  
We consider three scenarios: (i) stand alone LC
data, (ii) when the LC data are supplemented by the heavy neutralino
mass estimated from the LHC data, and (iii) joint analysis of the LC
and LHC data. In particular mass predictions of the heavier states
from the LC analysis are possible which lead to an improved
possibility to correctly interpret LHC measurements of these
particles.  The results in the last scenario will clearly demonstrate
the essentiality of the LHC and the LC and the benefit from the joint
analysis of their data.

In order to work out this hand-in-hand LHC+LC analysis for determining
the tree-level SUSY parameters, we assume that only the first
phase of a LC with a tunable energy up to $\sqrt{s}=500$~GeV 
would overlap with the LHC running. Furthermore, we assume an 
integrated luminosity of  $\int {\cal L} \sim 500$~fb$^{-1}$
and polarised beams with $P(e^-)=\pm 80\%$, $P(e^+)=\pm 60\%$. 
In the following $\sigma_L$ will refer to cross sections obtained
with $P(e^-)=- 80\%$, $P(e^+)= + 60\%$, and $\sigma_R$ with $P(e^-)=+
80\%$, $P(e^+)= - 60\%$. 
We restrict ourselves to 
the CP conserving chargino/neutralino sector and take the SPS1a as a
working benchmark \cite{sec4_Allanach:2002nj,Ghodbane:2002kg}; the inclusion of 
CP violating phases will be considered elsewhere. While the
electroweak scale parameters of the SPS1a scenario are derived from a
mSUGRA model, we do not impose any mSUGRA relations anywhere in this
analysis. Thus the analysis and the results are qualitatively valid
for any MSSM scenario with a similar mass spectrum.

Before presenting our results on the parameter determination, 
we first briefly recapitulate the main features of chargino and neutralino
sectors and sketch our strategy.

\subsubsection{ The gaugino/higgsino  sector}

\noindent\underline{a) Chargino sector}\\[.3em]
The mass matrix of the charged gaugino $\ti W^\pm$ and higgsino $\ti
H^\pm$   
is given by\footnote{One should note 
the difference between our convention of taking $\ti \chi^-$ 
as ``particles'' and e.g. the
convention of \cite{Haber-Kane}.}
\begin{eqnarray}
{\cal M}_C = \left(\begin{array}{cc}
     M_2              & \sqrt{2} m_W\cos\beta \\[2mm]
\sqrt{2} m_W \sin\beta & \mu\,
             \end{array}\right) \label{eq_charmat}
\end{eqnarray}
As a consequence of possible field redefinitions, the parameter
$M_2$ can be chosen
real and positive.
The two charginos $\tilde{\chi}^\pm_{1,2}$ are mixtures
of the charged SU(2) gauginos and higgsinos.
Since the mass matrix ${\cal M}_C$ is not symmetric,
two
different
unitary matrices acting on the left-- and right--chiral
$(\tilde{W},\tilde{H})_{L,R}$ two--component
states
\begin{eqnarray}
       \left(\begin{array}{c}
             \tilde{\chi}^-_1 \\
             \tilde{\chi}^-_2
             \end{array}\right)_{L,R} =
U_{L,R}\left(\begin{array}{c}
             \tilde{W}^- \\
             \tilde{H}^-
             \end{array}\right)_{L,R} 
\end{eqnarray}
define charginos as mass eigenstates. 
For real ${\cal M}_C$ 
the unitary matrices $U_L$ and $U_R$ can be parameterised as
\begin{eqnarray}
 U_{L,R}=\left(\begin{array}{cc}
             \cos\Phi_{L,R} & \sin\Phi_{L,R} \\
            -\sin\Phi_{L,R} & \cos\Phi_{L,R}
             \end{array}\right) 
\end{eqnarray}
The mass eigenvalues $m^2_{\tilde{\chi}^\pm_{1,2}}$ and the mixing
angles are given by
\begin{eqnarray}
 m^2_{\tilde{\chi}^\pm_{1,2}}
  & =&\frac{1}{2}(M^2_2+\mu^2+2m^2_W\mp \Delta_C)\nonumber \\
%
%
\cos 2\phi_{L,R}&=&-(M_2^2-\mu^2\mp 2m^2_W\cos 2\beta)/\Delta_C
   \nonumber
\end{eqnarray}
where 
$\Delta_C=[(M^2_2-\mu^2)^2+4m^4_W\cos^2 2\beta
              +4m^2_W(M^2_2+\mu^2)+8m^2_WM_2\mu
               \sin2\beta]^{1/2}$.

The $e^+e^-\to\tilde{\chi}^{\pm}_i \tilde{\chi}^{\mp}_j$ 
production processes occur
via the s-channel $\gamma$, $Z^0$ and the t-channel $\tilde{\nu}_e$
exchange.
Since the two mixing  angles $\Phi_{L,R}$ enter the couplings
in the $\ti \chi \ti\chi Z$ 
and  $e\ti\chi\ti\nu_e$ vertices, 
the chargino production cross sections 
$\sigma^\pm\{ij\}=\sigma(e^+e^-\to\tilde{\chi}^{\pm}_i
\tilde{\chi}^{\mp}_j)$ are  
bilinear functions of $\cos 2 \Phi_{L,R}$ \cite{CDGKSZ} and can be
written as 
\begin{equation}
\sigma^\pm\{ij\}=c_1 \cos^2 2 \Phi_L
+ c_2 \cos 2 \Phi_L + c_3  \cos^2 2 \Phi_R + c_4 \cos 2 \Phi_R
+ c_5 \cos 2 \Phi_L \cos 2 \Phi_R+c_6
\label{eq_sig11}
\end{equation}
We derived the coefficients $c_1,\ldots,c_6$
for the lightest chargino pair production cross section, see
Appendix a).

\vspace{.2cm}
\noindent\underline{b) Neutralino sector}\\[.3em]
The neutralino mixing matrix in the 
$\{\tilde{\gamma}, \tilde{Z}^0, \tilde{H}^0_1, \tilde{H}^0_2\}$
basis is given by
\begin{eqnarray}
{\cal M}_N= \left(\begin{array}{cccc}
  M_1 \cos^2_W+M_2 \sin^2_W & (M_2-M_1) \sin_W \cos_W & 0  & 0 \\[2mm]
  (M_2-M_1) \sin_W \cos_W  & M_1 \sin^2_W+M_2 \cos^2_W &   m_Z  & 0\\[2mm]
0 & m_Z &       \mu \sin{2 \beta}       &     -\mu \cos{2 \beta} \\[2mm]
0 & 0 &     -\mu \cos{2 \beta}      &       -\mu \sin{2 \beta}
\end{array}\right)\
\label{eq_neutmat}
\end{eqnarray}
The neutralino eigenvectors and their masses 
are obtained with the 4$\times$4 diagonalisation
matrix $N$:
\begin{equation}
N^{*} {\cal M}_N N^{\dagger}={\sf diag}\{m_{\tilde{\chi}^0_1},\ldots,
m_{\tilde{\chi}^0_4}\}
\label{eq_neutev}
\end{equation}

The parameter $M_1$ can only be determined from the neutralino sector.
The characteristic equation of the mass matrix squared, 
${\cal M}_N {\cal M}^{\dagger}_N$, can be written
as a quadratic equation for the parameter $M_1$:
\begin{equation}
x_i M_1^2+y_i M_1-z_i=0,\quad\mbox{for}\quad i=1,2,3,4 \label{eq_m1}
\end{equation}
where $x_i$, $y_i$, $z_i$ are given by:
\begin{eqnarray}
x_i&=&- m_{\tilde{\chi}^0_i}^6+a_{41} m_{\tilde{\chi}^0_i}^4
-a_{21} m_{\tilde{\chi}^0_i}^2+a_{01},\label{eq_m1-1}\\
y_i&=&a_{42} m_{\tilde{\chi}^0_i}^4-a_{22} m_{\tilde{\chi}^0_i}^2+a_{02},
\label{eq_m1-2}
\\
z_i&=&m_{\tilde{\chi}^0_i}^8-a_{63} m_{\tilde{\chi}^0_i}^6
+a_{43} m_{\tilde{\chi}^0_i}^4-a_{23} m_{\tilde{\chi}^0_i}^2+a_{03},
\label{eq_m1-3}
\end{eqnarray}
The coefficients $a_{kl}$, ($k=0,2,4,6$, $l=1,2,3$), being    
invariants of the matrix ${\cal M}_N {\cal M}_N^T$, can be
expressed as  functions of $M_2$, $\mu$ and $\tan\beta$. Their explicit
form is given in the Appendix b).

The $e^+e^-\to\tilde{\chi}^{0}_i \tilde{\chi}^{0}_j$ 
production processes occur
via the s-channel $Z^0$ and the t- and u-channel  $\tilde{e}_L$
and $\tilde{e}_R$ exchanges.
Since the neutralino mixing matrix $N$  is parameterised in general by
6 angles, the analytic expressions for the production cross sections
are more involved. Their explicit form can be found in \cite{ckmz}.

As one can see from eq.~(\ref{eq_m1}) for each neutralino mass
$m_{\tilde{\chi}^0_i}$ one gets two solutions for $M_1$. 
In principle, a measurement
of two neutralino masses and/or the cross section resolves this
ambiguity.  
However, one has to remember that the mass eigenvalues
show different 
sensitivity to the parameter $M_1$, depending on their
gaugino/higgsino composition.  In
our scenario, the mass of the lightest neutralino $m_{\tilde{\chi}^0_1}$ 
depends
strongly on $M_1$ if $M_1$ is in the range $-183$~GeV$<M_1<180$~GeV, 
while the others are roughly insensitive, see
Fig.~\ref{fig_mass-ms}. 
In a general MSSM, where $M_1$ and $M_2$ are independent
free parameters, this feature can completely change. We demonstrate this
in Fig.~\ref{fig_mass-ms}, where the $M_1/M_2$ GUT relation is relaxed.
We choose $M_1$ as a free parameter and all other parameters as in the 
SPS1a reference point. It can clearly be seen that the LSP becomes nearly
independent but the heavier neutralinos become more sensitive to $M_1$
with larger and larger $|M_1|$ \cite{Moortgat-Pick:1999gp}.

\begin{figure}[htb!]
\setlength{\unitlength}{1cm}
\begin{center}
{\epsfig{file=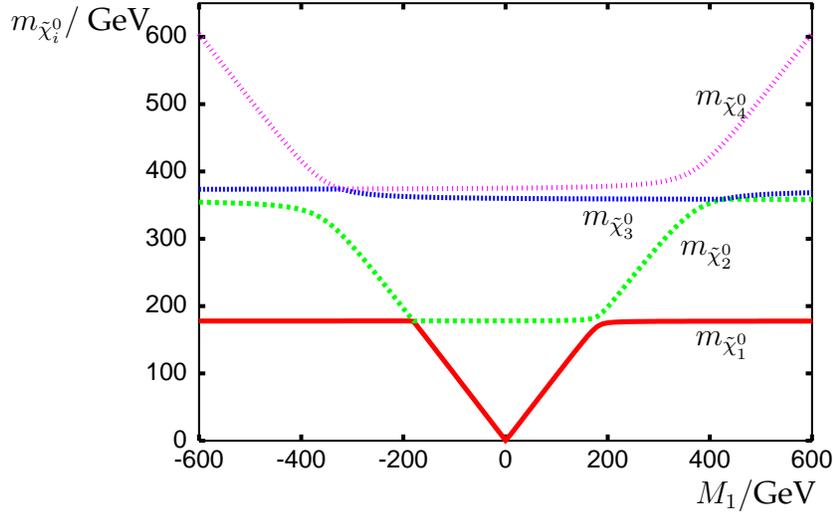,scale=.75}}
\put(-2,-.3){$M_1/$GeV}
\put(-11.1,6){$m_{\tilde{\chi}^0_i}/$~GeV}
\put(-2,1.8){$\mnt{1}$}
\put(-2.2,3){$\mnt{2}$}
\put(-3.5,3.4){$\mnt{3}$}
\put(-2,5){$\mnt{4}$}
\end{center}
\caption{\it $M_1$ dependence of the neutralino mass eigenvalues
$m_{\tilde{\chi}^0_i}$, $i=1,\ldots,4$ with $M_2$, $\mu$ and
$\tan\beta$ as in the reference scenario SPS1a. 
\label{fig_mass-ms}}
\end{figure}

\noindent\underline{c) The strategy}\\[.3em]
At the initial phase of future $e^+e^-$ linear--collider operations
with polarised beams, the
collision energy may only be sufficient to reach the production 
thresholds of the light
chargino $\tilde{\chi}^\pm_1$ and the two lightest neutralinos
$\tilde{\chi}^0_1,\, \tilde{\chi}^0_2$.
From the analysis of this restricted system, nevertheless the entire 
tree level structure of the gaugino/higgsino sector can be unraveled
in analytical form 
in CP--invariant theories as follows~\cite{CDGKSZ,ckmz}.  

It is clear from eq.(\ref{eq_sig11}) that 
by analysing the $\tilde{\chi}^+_1\tilde{\chi}^-_1$ production cross
sections with polarised beams, $\sigma^\pm_L\{11\}$ and
$\sigma^\pm_R\{11\}$, 
the chargino mixing angles $\cos2\Phi_L$ and
$\cos2\Phi_R$ can be determined \cite{CDGKSZ}. 
Any two contours, $\sigma^\pm_L\{11\}$ and $\sigma^\pm_R\{11\}$ 
for example, 
will cross at least at one point in the plane between $-1 \leq \ctL,
\ctR \leq +1$, if the chargino and sneutrino masses are known
and the SUSY Yukawa coupling is identified with the gauge coupling.
However, the contours, being of second order, may cross up to four
times.  The ambiguity can be resolved by measuring the
transverse\footnote{The measurement of the transverse cross section
involves the azimuthal production angle $\Phi$ of the charginos.  At
very high energies their angle coincides with the azimuthal angle of
the chargino decay products. With decreasing energy, however, the
angles differ and the measurement of the transverse cross section is
diluted.}  cross section $\sigma^\pm_T\{11\}$, or measuring
$\sigma^\pm_L\{11\}$ and $\sigma^\pm_R\{11\}$ at different beam
energies. We have chosen the latter solution.

In the CP conserving case studied in this paper the SUSY parameters $M_2$,
$\mu$ and $\tan\beta$ can be determined  from the 
chargino mass $\mch{1}$ and the mixing angles $\ctL$, 
$\ctR$~\cite{CDGKSZ}.  It is convenient to define
\begin{eqnarray}
p&=&\pm \left| \frac{\sin 2 \Phi_L+\sin 2 \Phi_R}{\cos 2 \Phi_L
-\cos 2 \Phi_R} \right| \label{eq_p}\\
q&=&\frac{1}{p} \frac{\cos 2 \Phi_L+\cos 2 \Phi_R}{\cos 2 \Phi_L
-\cos 2 \Phi_R} \label{eq_q}
\end{eqnarray}
Since the $\ctL$ and $\ctR$ are derived from 
$\tilde{\chi}^+_1 \tilde{\chi}^-_1$ cross sections,
the relative sign of $\sin 2\Phi_L$, $\sin 2 \Phi_R$ is not
determined and both possibilities in eqn.(\ref{eq_p}), (\ref{eq_q})
have to be considered.
From $p,q$, the SUSY parameters are 
determined as follows
($r^2=m^2_{\tilde{\chi}^{\pm}_1}/m^2_W$): 
\begin{eqnarray}
M_2&=&\frac{m_W}{\sqrt{2}}\left[(p+q)\sin\beta-
(p-q)\cos\beta\right] \label{eq_m2}\\
\mu&=&\frac{m_W}{\sqrt{2}}\left[(p-q)\sin\beta-(p+q)\cos\beta\right]
\label{eq_mu}\\
\tan\beta&=&
\left[\frac{p^2-q^2\pm \sqrt{r^2
(p^2+q^2+2-r^2)}} 
{(\sqrt{1+p^2}-\sqrt{1+q^2})^2- 2 r^2}\right]^\eta
\label{eq_tb1}
\end{eqnarray}
where $\eta=1$ for $\ctR > \ctL$, and $\eta=-1$
otherwise.  
The parameters $M_2$, $\mu$ are uniquely fixed if $\tan\beta$ is chosen 
properly. Since $\tan\beta$ is invariant
under simultaneous change of the signs of $p,q$, the definition 
$M_2 > 0$ can be exploited to remove this overall sign ambiguity.

The remaining parameter $M_1$ can be obtained from the neutralino
data \cite{ckmz}.  
The characteristic equation for the neutralino mass eigenvalues
\eq{eq_m1} is quadratic in $M_1$ if $M_2$, $\mu$ and $\tan\beta$
are already predetermined in the chargino sector.  In principle, two
neutralino masses are then sufficient to derive $M_1$, as explained in 
the previous section. 
The cross
sections $\sigma_{L,R}^0\{12\}$ and $\sigma_{L,R}^0\{22\}$ for
production of $\nt_1\nt_2$ and $\nt_2\nt_2$ neutralino
pairs\footnote{The lightest neutralino--pair production cannot be
observed. Alternatively, one can try to exploit photon tagging in the
reaction $e^+e^- \rightarrow \gamma \tilde{\chi}^0_1\tilde{\chi}^0_1$
\cite{photon}.}  with polarised beams can serve as a consistency check
of the derived parameters.

In practice the above procedure may be much more involved due to
finite experimental errors of mass and cross section measurements,
uncertainties from  sneutrino and selectron masses which enter the cross
section expressions, errors on beam polarisation measurement, etc. In
addition, depending on the benchmark scenario, some physical
quantities in the light chargino/neutralino system may turn to be 
essentially
insensitive to some parameters. For example, as seen in
\fig{fig_mass-ms},  
the first two neutralino masses are insensitive to $M_1$ 
if $M_1 \gg M_2,\,\mu$. Additional information from the LHC
on heavy states, if available, can therefore be of great value in
constraining the SUSY parameters.

Our strategy can be applied only at the tree level. Radiative
corrections, which in the electroweak sector can be ${\cal O}$(10\%),  
inevitably bring all SUSY parameters
together \cite{radiative}. Nevertheless, tree level analyses should
provide in a  relatively 
model-independent way good estimates of SUSY parameters, which can be
further refined by including iteratively 
radiative corrections in an overall fit
to experimental data.  

\subsubsection{SUSY parameters from the LC data}

\noindent\underline{a) Experimental input  at the LC}\\[.3em]

In this paper we take the unconstrained MSSM and
adopt the SPS1a scenario defined at the electroweak
scale \cite{sec4_Allanach:2002nj,Ghodbane:2002kg}. The  relevant SUSY 
parameters are 
\begin{eqnarray}
M_1=99.13~{\mbox GeV}, \quad M_2=192.7~{\mbox GeV}, \quad 
\mu=352.4~{\mbox GeV}, \quad \tan\beta=10
\end{eqnarray}
with no GUT or mSUGRA relations assumed.
The resulting chargino and neutralino masses, together with the slepton
masses of the first generation, are given in \tab{tab_mass_LC}.

\begin{table}[htb!]
\begin{center}
\begin{tabular}{|c|cc|cccc|ccc|}
\hline
& ${\tilde{\chi}^{\pm}_1}$ & ${\tilde{\chi}^{\pm}_2}$ & 
${\tilde{\chi}^0_1}$ & ${\tilde{\chi}^0_2}$ &${\tilde{\chi}^0_3}$ &
${\tilde{\chi}^0_4}$ & ${\tilde{e}_R}$ &  ${\tilde{e}_L}$ &
${\tilde{\nu}_e}$ \\
\hline
mass & 
176.03 & 378.50 & 96.17 & 176.59 & 358.81 & 377.87 & 143.0 & 202.1 & 186.0 \\
error & 
0.55   &        & 0.05  & 1.2    &        &        & 0.05 & 0.2   & 0.7   \\
\hline 
\end{tabular}
\caption{\it Chargino, neutralino and slepton  masses in SPS1a, and 
the simulated experimental errors at the LC 
\cite{Martyn,Ball}. 
It is assumed that the heavy
chargino and neutralinos are not observed at the first phase of the
LC operating at $\sqrt{s}\le 500$~GeV. 
[All quantities are in GeV.]
\label{tab_mass_LC}}
\end{center} 
\end{table}

In this scenario (which has a rather high $\tan\beta$ value) the
 $\tilde{\chi}^+_1$ and $\tilde{\chi}^0_2$ decay dominantly
into $\tilde{\tau}$ producing the signal similar to that of stau pair
production. Therefore the $\tilde{\tau}$ mass and mixing angle are also
important for the study of the chargino and neutralino sectors. The mass
and mixing angle can be determined as 
$m_{\tilde{\tau}_1}=133.2\pm 0.30$ 
GeV and
$\cos 2 \theta_{\tau}=-0.84\pm 0.04$, 
and the stau-pair production cross section ranges from 43
fb to 138~fb depending on the beam polarisation, 
see \cite{stau,Martyn} for details of the stau
parameter measurements. We
assume that the contamination of stau production events can be
subtracted from the chargino and neutralino production. Below we
included the statistical error to our analysis but we did not include
the systematic errors. 

\vspace{.2cm}
\noindent\underline{b) Chargino Sector}\\[.1em]

As observables we use the light chargino mass and polarised cross
sections 
$\sigma^\pm_L\{11\}$ and $\sigma^\pm_R\{11\}$ at $\sqrt{s}=500$~GeV and 
$\sqrt{s}=400$~GeV. The light charginos $\ti \chi^\pm_1$ 
decay almost exclusively to
$\ti \tau^\pm_1 \nu_\tau$ followed by $\ti \tau^\pm_1 \to \tau^\pm \ti
\chi^0_1$. The signature for the $\ti{\chi}^\pm_1\ti{\chi}^\mp_1$
production would be two tau jets in opposite hemispheres
plus missing energy. 

The experimental errors that we assume and take into account
are:
\begin{itemize}
\item
The measurement of the chargino mass has  
been simulated for our reference point
and the expected error is 0.55~GeV,
\tab{tab_mass_LC}.
\item With $\int {\cal L}=500$~fb$^{-1}$
at the LC,  we assume 100~fb$^{-1}$ per each polarisation configuration
and we take into account 1$\sigma$ statistical error.
\item Since the chargino production is sensitive to $m_{\tilde{\nu}_e}$, 
we include its experimental error of 0.7~GeV.
\item The measurement of the beam polarisation with an uncertainty
of $\Delta P(e^{\pm})/P(e^ {\pm})=0.5\%$ is assumed. This error is  
conservative; discussions to reach errors smaller than  
$0.25\%$ are underway \cite{Power}.
\end{itemize}
The errors on the production cross sections induced by the above
uncertainties, as well as the total errors (obtained by adding
individual errors in quadrature), are listed in 
\tab{tab_sig11}. 
We assume 100\% efficiency for the chargino cross sections 
due to a lack of realistic simulations.

\begin{table}[htb!]
\begin{tabular}{|l|cc|cc|}
\hline
\phantom{++++}$\sqrt{s}$ &\multicolumn{2}{|c|}{400~GeV} &
\multicolumn{2}{|c|}{500~GeV} \\ 
($P(e^-)$, $P(e^+)$)
&$(-80\%,+60\%)$ &$(+80\%,-60\%)$ &
 $(-80\%,+60\%)$ & $(+80\%,-60\%)$\\ \hline
$\sigma(e^+e^-\to\tilde{\chi}^+_1\tilde{\chi}^-_1)$
& 215.84 & 6.38 & 504.87 &  15.07\\ \hline
$\delta\sigma_{\mbox{stat}}$ 
& 1.47 & 0.25 & 2.25 & 0.39\\ 
$\delta\sigma_{P(e^-)}$ &  
0.48 & 0.12 & 1.12 & 0.28 \\ 
$\delta\sigma_{P(e^+)}$ &
0.40 & 0.04 & 0.95 & 0.10 \\ 
$\delta\sigma_{m_{\tilde{\chi}^{\pm}_1}}$ &
7.09 & 0.20 & 4.27 & 0.12\\
$\delta\sigma_{m_{\tilde{\nu}_e}}$  & 
0.22 & 0.01 & 1.57 & 0.04 \\ \hline 
$\delta\sigma_{\mbox{total}}$ & 7.27 & 0.35 & 5.28 & 0.51 \\ \hline
\end{tabular}
\caption{\it Cross sections 
$\sigma_{L,R}^\pm\{11\}= 
\sigma_{L,R}(e^+ e^-\to \tilde{\chi}^+_1\tilde{\chi}^-_1)$ 
with polarised beams $P(e^-)=\mp 80\%$, $P(e^+)=\pm 60\%$
at $\sqrt{s}=400$ and 500~GeV and assumed errors (in fb) corresponding to 
 100~fb$^{-1}$ for each polarisation configuration. \label{tab_sig11}}
\end{table}

Now we can exploit the \eq{eq_sig11} and draw  
$\cos 2 \Phi_R
=f(\cos 2 \Phi_L,\sigma^\pm _{L,R}\{11\})$ 
consistent with the predicted cross sections
within the mentioned error bars, as  shown in \fig{fig_mix}. 
With the $\sqrt{s}=500$ GeV data alone two possible regions in the
plane are selected. With the help of the $\sigma^\pm_L\{11\}$ at
$\sqrt{s}=400$~GeV ($\sigma^\pm_R\{11\}$ is small and does not provide
further constraints) the
ambiguity is removed and  the mixing angles are limited  
within the range
\begin{eqnarray}
\cos 2 \Phi_L&=& [0.62,0.72] \label{eq_c2lrange}\\
\cos 2 \Phi_R&=& [0.87,0.91] \label{eq_c2rrange}
\end{eqnarray}

\begin{figure}[htb!]
\setlength{\unitlength}{1cm}
\begin{center}
{\epsfig{file=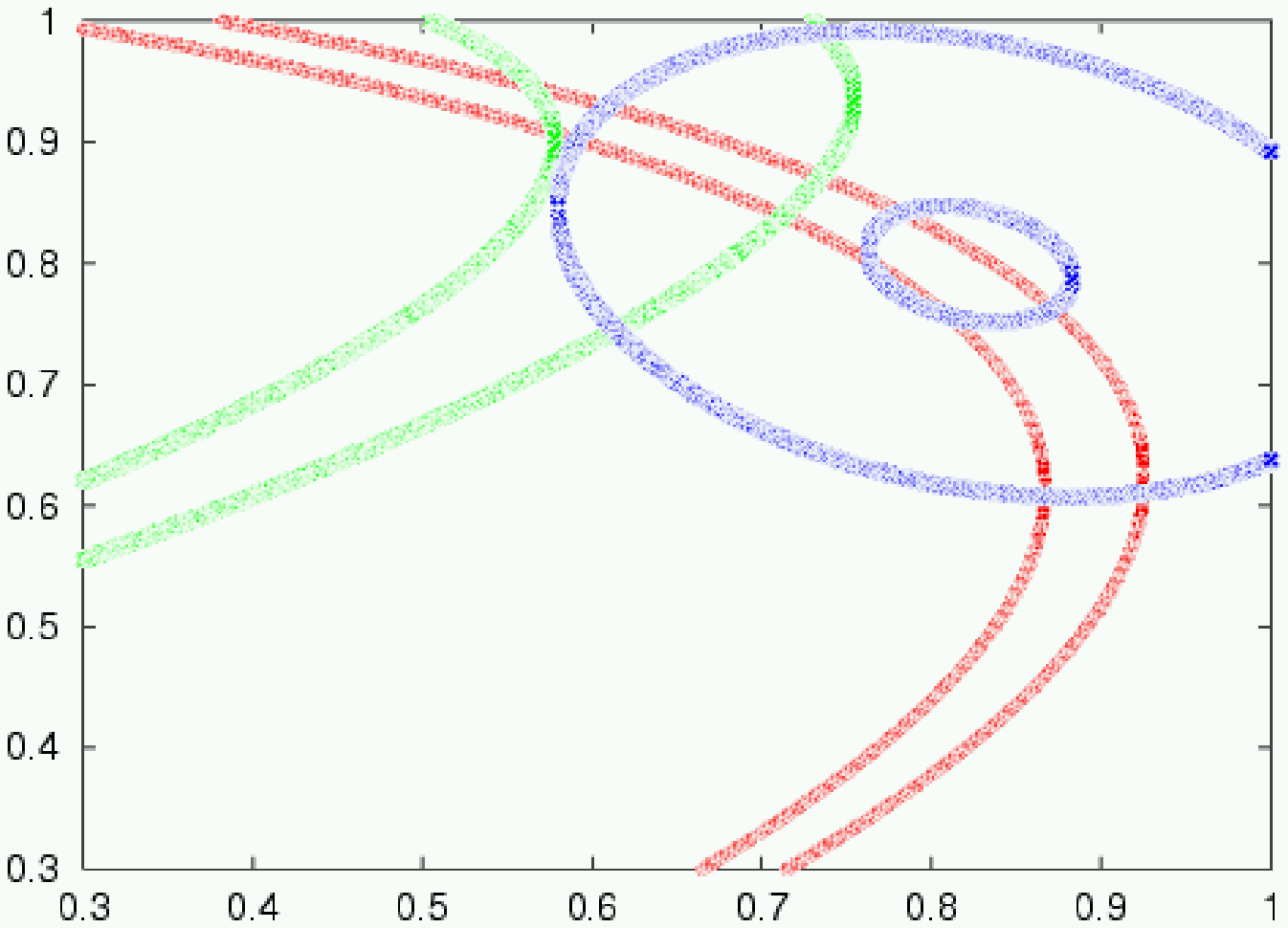,scale=.65}}
\put(-1.6,-.3){$\cos 2 \Phi_L$}
\put(-10.5,6.2){$\cos 2 \Phi_R$}
\put(-8.2,5.5){$\sigma_L^{\pm}(500)$}
\put(-8.2,2.3){$\sigma_L^{\pm}(400)$}
\put(-5,2.9){$\sigma_R^{\pm}(500)$}
\end{center}
\caption{\it $\cos 2 \Phi_R$ as a function of $\cos 2 \Phi_L$ 
for $\sigma^\pm_L\{11\}$ at $\sqrt{s}=500$~GeV (red),
and 400 GeV (green) and  $\sigma^\pm_R\{11\}$ at $\sqrt{s}=500$~GeV (blue)
within the error bounds (theo+exp) as given in \tab{tab_sig11}.
\label{fig_mix}}
\end{figure}

Although $\ctL,\ctR$ are determined rather precisely 
at a few per-cent accuracy, an
attempt to exploit eqns.~(\ref{eq_m2})-(\ref{eq_tb1}) shows that  
$M_2$ is reconstructed within 10 GeV, $\mu$ within 40 GeV, and
essentially no limit on $\tan\beta$ is obtained (we get
$\tan\beta>6$).
The main reason for this result is a relatively large error of  the
light chargino mass measurement due to the $\ti\chi^+_1\to \ti\chi^0_1
\tau^+\nu_\tau$ decay mode. Several methods exploiting other
sectors of the MSSM have been proposed to measure $\tan\beta$ in the
high $\tan\beta$ regime \cite{hightb,stau}. In the following we will 
exploit the neutralino sector (with 
eqns.~(\ref{eq_c2lrange}), (\ref{eq_c2rrange}) as the
allowed ranges for the chargino mixing angles) to improve constraints 
on $M_2$, $\mu$ and $\tan\beta$, and to determine $M_1$.  

\vspace{.5cm}
\noindent\underline{c) Neutralino Sector:}\\[.3em]

\begin{sloppypar}
As observables we use the two light neutralino masses and polarised
cross sections $\sigma^0_{L,R}\{12\}$ and $\sigma^0_{L,R}\{22\}$ at
$\sqrt{s}=400$~GeV and $\sqrt{s}=500$~GeV.  Although the production of
$\tilde{\chi}^0_1 \tilde{\chi}^0_3$ and $\tilde{\chi}^0_1
\tilde{\chi}^0_4$ pairs is kinematically accessible at
$\sqrt{s}=500$~GeV in the chosen reference point SPS1a, 
the production rates are small and the heavy states $\nt_3$
and $\nt_4$ decay via cascades to many particles.  
This feature is quite common in large parts of the MSSM parameter
space when the gaugino and higgsino mass parameters are not too close. 
Therefore in our analysis we 
take into account the experimental information available only from 
the production of the light neutralino
pairs. 
\end{sloppypar}

\begin{table}[htb!]
\begin{tabular}{|l|cc|cc|}
\hline \phantom{++++} $\sqrt{s}$ 
&\multicolumn{2}{|c|}{400~ GeV} &
\multicolumn{2}{|c|}{500~GeV} \\ 
($P(e^-)$, $P(e^+)$)
&$(-80\%,+60\%)$ &$(+80\%,-60\%)$ &
 $(-80\%,+60\%)$ & $(+80\%,-60\%)$\\ \hline
$\sigma(e^+e^-\to \tilde{\chi}^0_1\tilde{\chi}^0_2)$
& 148.38 & 20.06 & 168.42 &  20.81\\ \hline
$\delta\sigma_{\mbox{stat}}$ 
& 2.92 & 1.55 & 3.47 & 1.55\\ 
$\delta\sigma_{\mbox{bg}}$
& 0.44 & 0.02 & 0.31 & 0.03\\ 
$\delta\sigma_{P(e^-)}$ &  
0.32 & 0.05 & 0.37 & 0.06 \\ 
$\delta\sigma_{P(e^+)}$ &
0.28 & 0.001 & 0.31 & 0.01 \\ 
$\delta\sigma_{m_{\tilde{\chi}^{\pm}_1}}$ &
0.21 & 0.30 & 0.16 & 0.26\\
$\delta\sigma_{m_{\tilde{e}_L}}$  & 
0.20 & 0.01 & 0.17 & 0.01 \\ 
$\delta\sigma_{m_{\tilde{e}_R}}$  & 
0.00 & 0.01 & 0.00 & 0.01 \\
\hline
$\delta\sigma_{\mbox{total}}$ & 3.0 & 1.58 & 3.52 & 1.57 \\ \hline
\end{tabular}
\caption{\it Cross sections 
$\sigma^0_{L,R}\{12\}=
\sigma_{L,R}(e^+ e^-\to \tilde{\chi}^0_1\tilde{\chi}^0_2)$
with polarised beams $P(e^-)=\mp 80\%$, $P(e^+)=\pm 60\%$
at $\sqrt{s}=400$ and 500~GeV,   and assumed 
errors  (in fb) corresponding to 
 100~fb$^{-1}$ for each polarisation configuration.\label{tab_sig12}}
\vspace{3mm}
\begin{tabular}{|l|cc|cc|}
\hline \phantom{++++} 
&\multicolumn{2}{|c|}{400~GeV} &
\multicolumn{2}{|c|}{500~GeV} \\ 
($P(e^-)$, $P(e^+)$)
&$(-80\%,+60\%)$ &$(+80\%,-60\%)$ &
 $(-80\%,+60\%)$ & $(+80\%,-60\%)$\\ \hline
$\sigma(e^+e^-\to\tilde{\chi}^0_2\tilde{\chi}^0_2)$
& 85.84 & 2.42 & 217.24 &  6.10\\ \hline
$\delta\sigma_{\mbox{stat}}$ 
& 2.4 & 0.4 & 3.8 & 0.6\\ 
$\delta\sigma_{P(e^-)}$ &  
0.19 & 0.05 & 0.48 & 0.12 \\ 
$\delta\sigma_{P(e^+)}$ &
0.16 & 0.02 & 0.41 & 0.05 \\ 
$\delta\sigma_{m_{\tilde{\chi}^{\pm}_1}}$ &
2.67 & 0.08 & 1.90 & 0.05\\
$\delta\sigma_{m_{\tilde{e}_L}}$  & 
0.15 & 0.004 & 0.28 & 0.01 \\ 
$\delta\sigma_{m_{\tilde{e}_R}}$  & 
0.00 & 0.00 & 0.00 & 0.00 \\
\hline
$\delta\sigma_{\mbox{total}}$ & 3.6 & 0.41 & 4.3 & 0.62 \\ \hline
\end{tabular}
\caption{\it Cross sections 
$\sigma^0_{L,R}\{22\}=
\sigma_{L,R}(e^+ e^-\to \tilde{\chi}^0_2\tilde{\chi}^0_2)$
with polarised beams $P(e^-)=\mp 80\%$, $P(e^+)=\pm 60\%$
at $\sqrt{s}=400$ and 500~GeV, and assumed 
errors (in fb) corresponding to 
 100~fb$^{-1}$ for each polarisation configuration. \label{tab_sig22}}
\end{table}

The neutralino $\ti{\chi}^0_2$ decays into $\ti{\tau}^\pm_1 \tau^\mp$
with almost 90\%, followed by the $\ti{\tau}^\pm_1 \to \tau^\pm
\ti{\chi}^0_1$. Therefore the final states for the
$\ti{\chi}^\pm_1\ti{\chi}^\mp_1$ and $\ti{\chi}^0_1\ti{\chi}^0_2$ are
the same (2$\tau$ + missing energy), however with different topology.
While for the charginos, the $\tau$'s tend to be in opposite 
hemispheres with rather
large invariant mass, in the $\ti{\chi}^0_1\ti{\chi}^0_2$ process both
$\tau$'s, coming from the $ \ti{\chi}^0_2$ decay, would be more often 
in the same
hemisphere with smaller invariant mass. This feature allows to 
separate the processes to some extent
exploiting e.g.~a cut on the opening angle
between the two jets of the  $\tau's$. However, in the case
of $\ti{\chi}^0_1\ti{\chi}^0_2$, significant background from
$\ti{\chi}^\pm_1\ti{\chi}^\mp_1$ and $\ti{\tau}^\pm_1 \ti{\tau}^\mp_1$
remains.

We estimate the statistical error on $\sigma(e^+e^-\to
\ti{\chi}^0_1\ti{\chi}^0_2)$ based on the experimental simulation
presented in~\cite{Ball}. This simulation was performed at $\sqrt{s}
=500 $ GeV for unpolarised beams yielding an efficiency of 25\%. We 
extrapolate the statistical
errors at different $\sqrt{s}$ and different polarisations as
$\delta\sigma / \sigma = \sqrt{S+B}/S$ where we calculate the number
of signal (S) and background (B) events from the cross sections and
the integrated luminosity (100 fb$^{-1}$) assuming the same efficiency
as achieved for the unpolarised case. Since the cross sections for the
SUSY background processes are also known only with some uncertainty,
we account for this uncertainty in the background subtraction by
adding an additional systematic error ($\delta\sigma_{\mbox{bg}}$).

For the process $\tilde{\chi}^0_2 \tilde{\chi}^0_2 \to \tau^+\tau^-
\tau^+\tau^- \tilde{\chi}^0_1 \tilde{\chi}^0_1$
no detailed simulation exists. From the $\tau$-tagging efficiency
achieved in the $\ti{\chi}^0_1\ti{\chi}^0_2$ channel,
we assume that this final state can be reconstructed with an efficiency
of 15\% with negligible background. This is justified since no major
SUSY background is expected for the 4-$\tau$ final state,
$\mbox{BR}(\ti\nu_\tau\to\tau^+\tau^-\tilde{\chi}^0_1)^2$ is only  0.5\%.
SM backgrounds arise mainly from Z pair production and are small.

For both processes we account in addition for polarisation uncertainties
and uncertainties in the cross section predictions from the errors on
the chargino and selectron masses. Note that we implicitly assume that
the branching ratio $\ti{\chi}^0_2 \to \tau^+\tau^-\ti{\chi}^0_1$
is known, which is a simplifaction. A full analysis will have to
take into account the parameter dependence of this branching ratio
in addition, since it cannot be measured directly. 


The neutralino cross sections  depend on 
$M_1$, $M_2$, $\mu$, $\tan\beta$ and on slepton masses. We prefer to
express $M_2$, $\mu$, $\tan\beta$ 
 in terms 
of $m_{\tilde{\chi}^{\pm}_1}$ and the mixing angles $\ctL,\ctR$. 
Then we consider  neutralino 
cross sections as  functions of unknown  
$M_1, \ctL,\ctR$ with uncertainties  due to statistics and 
experimental errors on beam polarisations,    
$m_{\tilde{\chi}^{\pm}_1}, m_{\tilde{e}_L}$ and  $m_{\tilde{e}_R}$ 
included (in quadrature) in the total error, see~\tab{tab_sig12}
and \tab{tab_sig22}. 

\vspace{.2cm}
\noindent\underline{d) Results}\\[.3em]
We perform a $\Delta \chi^2$ test defined as
\begin{equation}
\Delta \chi^2  =\sum_i |\frac{O_i -\bar O_i}{\delta O_i}|^2
\label{eq_chi2}
\end{equation}
The sum over physical observables $O_i$  includes
$m_{\nt_1},m_{\nt_2}$ and 
neutralino production cross sections
$ \sigma^0_{L,R}\{12\},\sigma^0_{L,R}\{22\}$ measured
at both energies of 400 and 500 GeV. 
The $\Delta \chi^2$ is a function of unknown $M_1,\ctL,\ctR$ with 
$\ctL,\ctR$ restricted to the ranges given in
eqns.~(\ref{eq_c2lrange}),(\ref{eq_c2rrange}) as 
predetermined from the chargino sector.
$\bar O_i$ stands for the physical observables 
taken at the input values of all
parameters, and $\delta O_i$ are the corresponding errors.

\begin{figure}[htb!]
\setlength{\unitlength}{1cm}
\begin{center}
{\epsfig{file=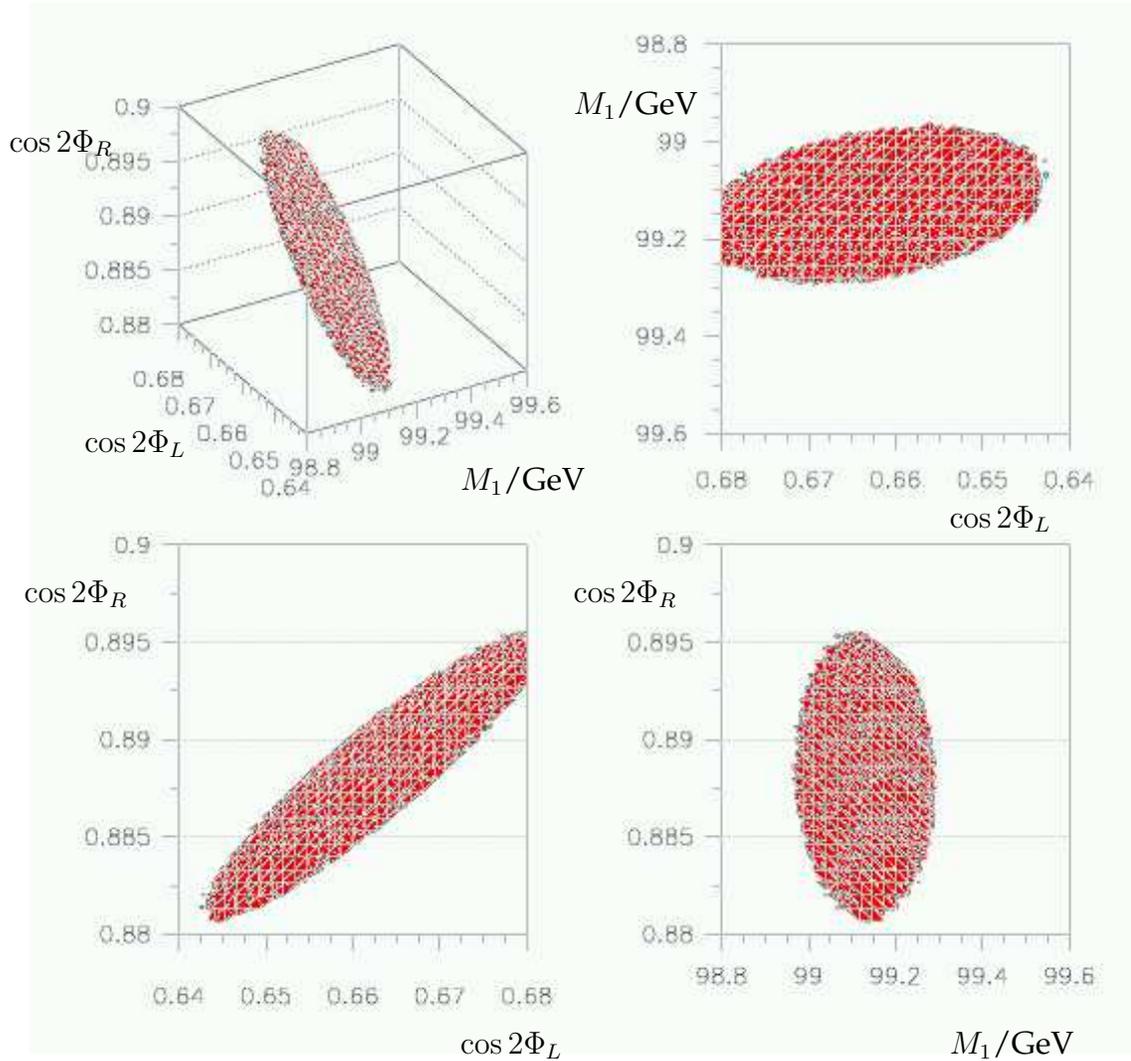,scale=0.92}}
\put(-9.,0.){$\cos 2 \Phi_L$}
\put(-2.5,0.){$M_1$/GeV}
\put(-14.8,6.){$\cos2\Phi_R$}
\put(-7.5,6.){$\cos2\Phi_R$}
\put(-2.5,7.){$\cos 2 \Phi_L$}
\put(-7.5,12.5){$M_1$/GeV}
\put(-9.,7.5){$M_1$/GeV}
\put(-15.,12.){$\cos2\Phi_R$}
\put(-14,8.){$\cos 2 \Phi_L$}
\end{center}
\caption{\it 
The $\Delta \chi^2=1$ contour in the $M_1,\ctL,\ctR$ parameter space, and
its three 2dim projections, derived from the LC data.}
\label{fig_4cygLC}
\end{figure}

In \fig{fig_4cygLC} the contour of $\Delta \chi^2=1$ is shown in the
$M_1,\ctL,\ctR$ parameter space along with 
its three 2dim projections. The projection of the contours onto the
axes determines 1$\sigma$ errors for each parameter.

Values obtained for $M_1,\ctL,\ctR$ together with $\mch{1}$ can be
inverted to derive the fundamental parameters $M_2$, $\mu$ and
$\tan\beta$. At the same time masses of heavy chargino and neutralinos
are predicted. As can be seen
in \tab{tab_lc}, in the SPS1a scenario the parameters $M_1$ and $M_2$ are 
determined at the level of a few per-mil, $\mu$ is reconstructed
within a few per-cent, while  $\tan\beta$ is found with an error of
order 15\%.

\begin{figure}[htb!]
\begin{center}
\setlength{\unitlength}{0.8cm}
{\epsfig{file=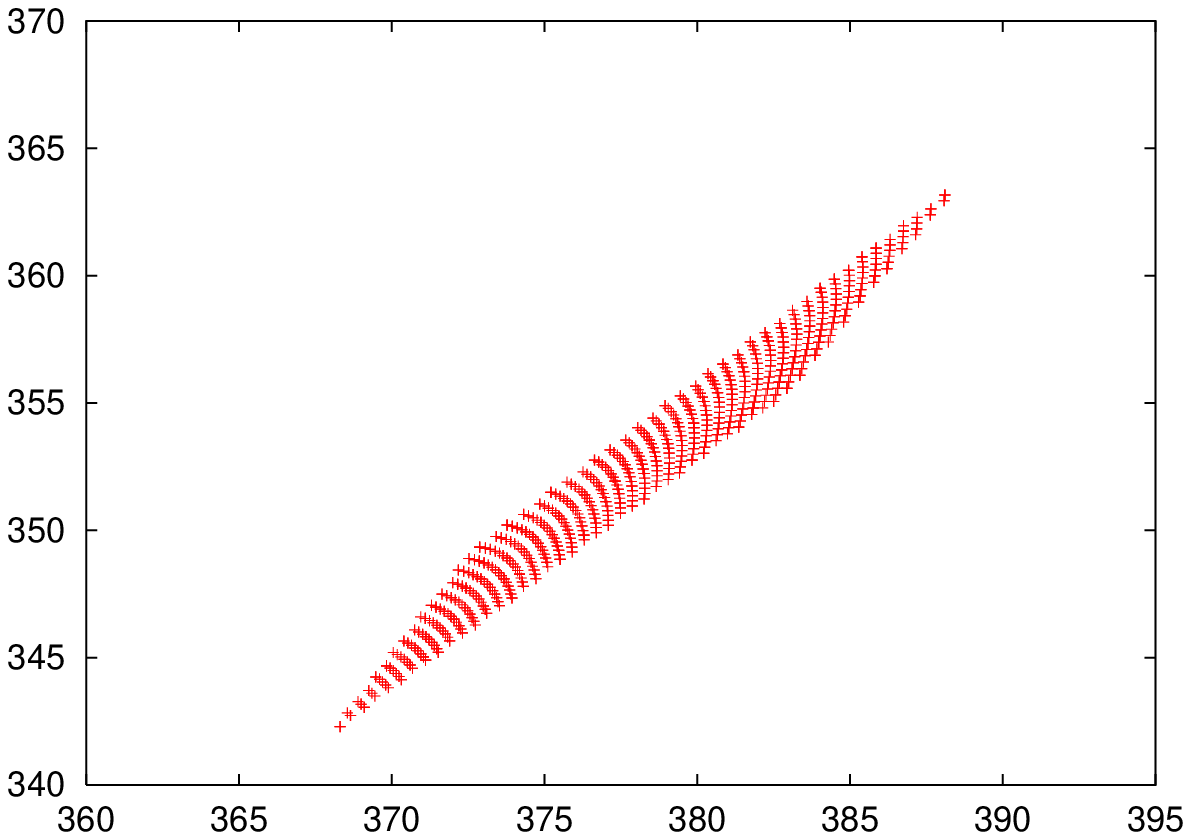,scale=.5}~~~~~~
  {\epsfig{file=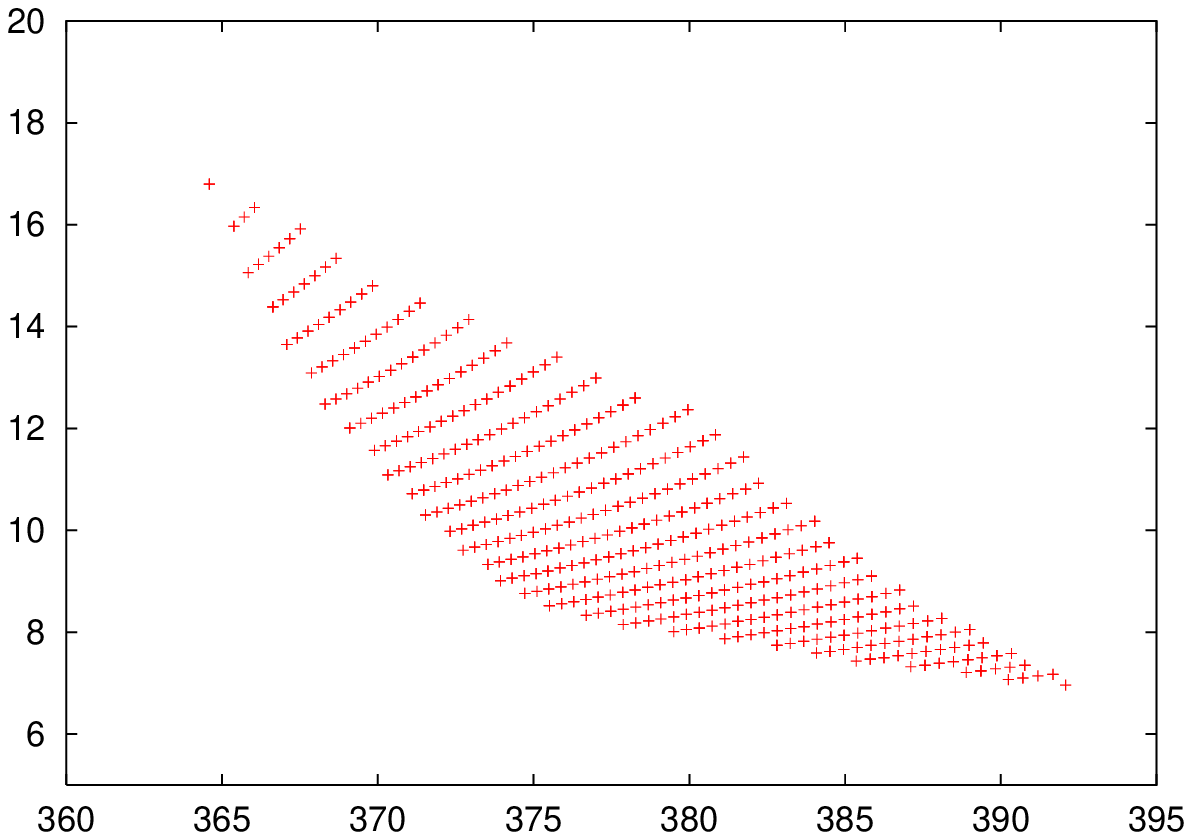,scale=.5}}}  
\put(-11.,-.3){\small $m_{\tilde{\chi}^0_4}/$GeV}
\put(-2,-.3){\small $m_{\tilde{\chi}^0_4}/$GeV}
\put(-18.3,5.){\small $\mu/$~GeV}
\put(-8.9,5.){\small $\tan\beta$}
\end{center}\par\vspace{-.5cm}
\caption{\it The correlation between predicted values of 
$\mu$ and $m_{\tilde{\chi}^0_4}$ (left panel) 
and the allowed range of 
$\tan\beta$ and 
$m_{\tilde{\chi}^0_4}$ (right panel)  from the analysis of the LC data.  
\label{fig-muchi04}}
\end{figure}

The errors on the predicted masses of the 
heavy chargino/neutralinos,  which in our
SPS1a scenario  are predominantly higgsinos, are strongly correlated
with the error of $\mu$;   the left panel of 
\fig{fig-muchi04} shows the correlation
between $\mu$ and $m_{\tilde{\chi}^0_4}$. In the right panel of this
figure a weaker correlation is
observed between $\tan\beta$ and $m_{\tilde{\chi}^0_4}$  (or between
$\tan\beta$ and  $\mu$).    Therefore,  by providing 
$m_{\tilde{\chi}^0_4}$ from endpoint measurements \cite{lhc_atlas},  
the LHC could considerably help to get a better
accuracy on $\mu$. At the same time a better determination of 
$\tan\beta$ can be expected.

\begin{table}[htb!]
\begin{tabular}{|cccc|ccc|}
\hline
\multicolumn{4}{|c|}{SUSY Parameters}&
\multicolumn{3}{|c|}{Mass Predictions}\\
$M_1$ & $M_2$ & $\mu$ & $\tan\beta$ & $m_{\tilde{\chi}^{\pm}_2}$ &
$m_{\tilde{\chi}^0_3}$ & $m_{\tilde{\chi}^0_4}$ \\[2mm] \hline
$99.1\pm 0.2$ & $192.7\pm 0.6$ & $352.8\pm 8.9$ & $10.3\pm 1.5$
& $378.8 \pm 7.8$ & $359.2\pm 8.6$ & $378.2 \pm 8.1$\\ \hline
\end{tabular}
\caption{\it SUSY parameters with 1$\sigma$ errors 
derived from the  analysis of  the LC data
  collected at the first phase of operation. 
Shown are also the predictions for the heavier chargino/neutralino masses.
\label{tab_lc}}
\end{table}

\subsubsection{Combined strategy for the LHC and LC}

\noindent\underline{a) LC data supplemented by ${\boldmath \mnt{4}}$  from
the LHC}\\[.5em]
The LHC experiments will be able to measure the masses of several
sparticles, as described in detail for the SPS1a point in
\cite{lhc_atlas}.
In particular, the LHC will provide a first measurement of 
the masses of $\nt_1$, $\nt_2$ and $\nt_4$.
The measurements of $\nt_2$ and $\nt_4$ 
are achieved through the study of the processes:
\begin{equation}
\tilde\chi^0_i\rightarrow\tilde\ell\ell\rightarrow\ell\ell\nt_1
\end{equation}
where the index $i$ can be either 2 or 4.
The invariant mass of the two leptons in the final state shows 
an abrupt edge, which can be expressed in terms of the
masses of the relevant sparticles as
\begin{equation}
m_{l^+l^-}^{\mathrm{max}} = m_{\nt_i} \sqrt{1-\frac{m^2_{\tilde\ell}}{m^2_{\nt_i}}}
\sqrt{1-\frac{m^2_{\nt_1}}{m^2_{\tilde\ell}}}
\label{eq_edge}
\end{equation}
If one only uses the LHC information, the achievable precision on
$m_{\nt_2}$ and $m_{\nt_4}$ will be respectively of 4.5 and 5.1 GeV
for an integrated luminosity of 300~fb$^{-1}$, see 
figure~\ref{fig_sigloga}. The correct interpretation of the largest
observed $m_{l^+l^-}^{\mathrm{max}}$ as originating from
$\tilde{\chi}^0_4$ decays can be facilitated by the prediction of
$m_{\tilde{\chi}^0_4}$ from the LC measurements in the MSSM context.\par 
In the case of the $\nt_4$, 
which  in the considered  scenario  is mainly higgsino, this information
can be exploited at the LC to constrain the parameter $\mu$  with a
better precision. 
If we include this improved precision on $\mnt{4}$ 
in the $\Delta\chi^2$ test of \eq{eq_chi2}, the resulting 
$\Delta\chi^2=1 $ contours get modified  
and the achievable precision is improved, as shown in  \tab{tab_lc2}. 
\begin{figure}[htb!]
\setlength{\unitlength}{1cm}
\begin{center}
{\epsfig{file=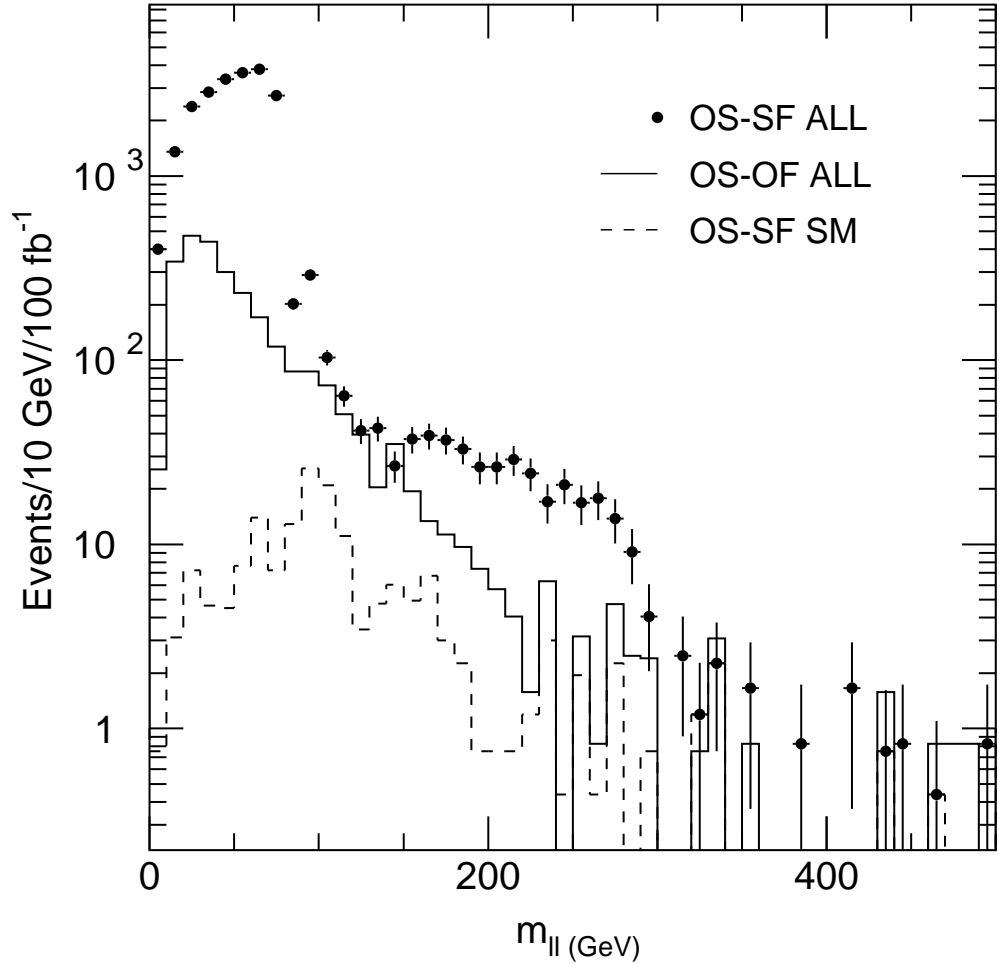,scale=.75}}
\end{center}
\caption{\it 
Invariant mass spectrum respectively for: Opposite-Sign Same-Flavour 
(OS-SF) leptons total (full dots), Opposite-Sign Opposite-Flavour 
(OS-OF) leptons total (solid line), Opposite-Sign Same-Flavour
leptons in the SM (dashed line). The signals of 
$\tilde{\chi}^0_2$, $\tilde{\chi}^0_4$ consist of OS-SF leptons 
\cite{lhc_atlas}.}
\label{fig_sigloga}
\end{figure}

\begin{table}[htb!]
\begin{tabular}{|cccc|cc|}
\hline
\multicolumn{4}{|c|}{SUSY Parameters}&
\multicolumn{2}{|c|}{Mass Predictions}\\
$M_1$ & $M_2$ & $\mu$ & $\tan\beta$ & $m_{\tilde{\chi}^{\pm}_2}$ &
$m_{\tilde{\chi}^0_3}$ \\ \hline
$99.1\pm 0.2$ & $192.7\pm 0.5$ & $352.4\pm 4.5$ & $ 10.2 \pm 0.9$
& $378.5 \pm 4.1$ & $358.8\pm 4.1$\\ \hline
\end{tabular}
\caption{\it SUSY parameters with 1$\sigma$ errors 
derived from the  analysis of  the LC data
  collected at the first phase of operation and
with   $\delta\mnt{4}=5.1$~GeV from the LHC.
Shown are also the predictions for the  masses of $\ti\chi^\pm_2$ and
  $\ti\chi^0_3$.
\label{tab_lc2}}
\end{table}

\vspace{.2cm}
\noindent\underline{b) Joint analysis of the LC and LHC  data}\\[.1em]

From the consideration of \eq{eq_edge}, one can see 
that the uncertainty on the LHC measurement of $\mnt{2}$ and $\mnt{4}$
depends both on the experimental error on the position of
$m_{l^+l^-}^{\mathrm{max}}$, and on the uncertainty on $\mnt{1}$ and 
$m_{\tilde\ell}$. The latter uncertainty, which for both 
masses is of 4.8~GeV, turns out to be the dominant contribution.
A much higher precision can thus be achieved by inserting in 
\eq{eq_edge} the values for $\mnt{1}$, $m_{\tilde e_R}$  and 
$m_{\tilde e_L}$ which are measured at the LC with 
precisions respectively of 0.05, 0.05 and 0.2  GeV,  \tab{tab_mass_LC}. 

\begin{figure}[htb!]
\setlength{\unitlength}{1cm}
\begin{center}
{\epsfig{file=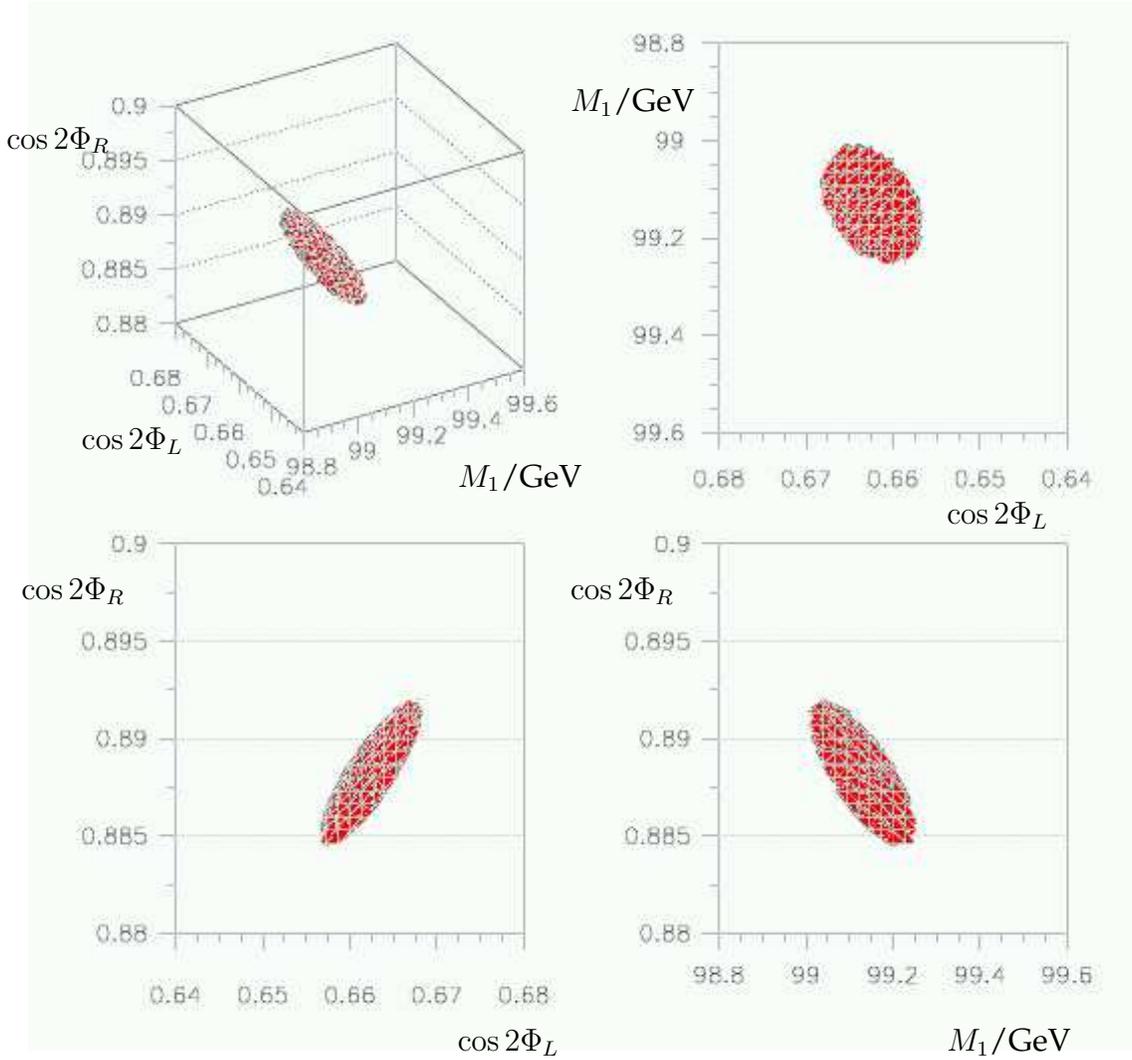,scale=.92}}
\put(-9.,0.){$\cos 2 \Phi_L$}
\put(-2.5,0.){$M_1$/GeV}
\put(-14.8,6.){$\cos2\Phi_R$}
\put(-7.5,6.){$\cos2\Phi_R$}
\put(-2.5,7.){$\cos 2 \Phi_L$}
\put(-7.5,12.5){$M_1$/GeV}
\put(-9.,7.5){$M_1$/GeV}
\put(-15.,12.){$\cos2\Phi_R$}
\put(-14,8.){$\cos 2 \Phi_L$}
\end{center}
\caption{\it The $\Delta \chi^2=1$ contour in the $M_1,\ctL,\ctR$ 
parameter space, and
its three 2dim projections, derived from the joint analysis of the LC
data and LHC data.}
\label{fig_4cygLHCLC}
\end{figure}

With this input the precisions on the LHC+LC measurements of 
$m_{\tilde{\chi}^0_2}$ and $m_{\tilde{\chi}^0_4}$ 
become: $\delta m_{\tilde{\chi}^0_2}=0.08$~GeV and  
$\delta m_{\tilde{\chi}^{0}_4}=2.23$~GeV.

\begin{table}[htb!]
\begin{tabular}{|cccc|cc|}\hline
\multicolumn{4}{|c|}{SUSY Parameters}&
\multicolumn{2}{|c|}{Mass Predictions}\\
$M_1$ & $M_2$ & $\mu$ & $\tan\beta$ & $m_{\tilde{\chi}^{\pm}_2}$ &
$m_{\tilde{\chi}^0_3}$ \\ \hline
$99.1\pm 0.1$ & $192.7\pm 0.3$ & $352.4\pm 2.1$ & $ 10.2 \pm 0.6$
& $378.5 \pm 2.0$ & $358.8\pm 2.1$\\ \hline
\end{tabular}
\caption{\it SUSY parameters  with 1$\sigma$ errors
derived from the combined analysis of the LHC and LC data
with $\delta m_{\tilde{\chi}^0_2}=0.08$~GeV and 
$\delta m_{\tilde{\chi}^0_4}=2.23$~GeV derived from the LHC when using 
the LC input of $\delta m_{\tilde{\chi}^0_1}=0.05$~GeV.
\label{tab_lhclc}}
\end{table}

From the results of the $\Delta\chi^2$ test one can calculate  the 
improvement in accuracy 
for the derived parameters by imposing the new mass constraints. The
final results are shown in figure\ref{fig_4cygLHCLC} and  
\tab{tab_lhclc}.  The accuracy for the parameters $\mu$ and
particularly $\tan\beta$ is much better, as could be
expected from  
\fig{fig-muchi04}, where the
allowed range of $\mu$ and $\tan\beta$ 
from the LC analysis is  considerably reduced once 
the measured mass $m_{\tilde{\chi}^0_4}$ at the LHC
is taken into account. In particular, the precision on $\tan\beta$ 
becomes better than from  other SUSY sectors
\cite{hightb,stau}.

\subsubsection{Summary }
We have studied the prospects for the determination of the parameters
which govern the chargino/neutralino sector of a general MSSM. We
focus on the situation where only the lightest states
($\tilde{\chi}^0_1$, $\tilde{\chi}^0_2$, $\tilde{\chi}^\pm_1$) are
accessible at the first stage of a LC.
For a specific example of the MSSM, the SPS1a scenario  
with a rather high $\tan\beta=10$, we show 
how the combination of the results from 
the two accelerators, LHC and LC,  allows
a precise determination of  the fundamental 
SUSY parameters. The analysis has been performed within the general
frame of the unconstrained MSSM.   
Our strategy does not rely on any 
particular relations among the fundamental parameters,  like the GUT or 
mSUGRA relations, and therefore  is applicable for arbitrary  
MSSM parameters which lead to a phenomenology similar to the one studied.

Measuring with high precision
the masses of the expected lightest SUSY particles $\tilde{\chi}^0_1$,
$\tilde{\chi}^0_2$, $\tilde{\chi}^{\pm}_1$ and their cross sections at 
the LC, and 
taking into account simulated mass measurement errors and
corresponding uncertainties for the theoretical predictions, we could
determine the fundamental SUSY parameters $M_1$, $M_2$, $\mu$ 
at tree level for the SPS1a point within a
few percent, while  $\tan\beta$ is estimated within $\sim$ 15 \%.
The masses of  heavier chargino and neutralinos can also be predicted
at a level of a few percent and are used as input for the LHC analysis. 
The use of polarised beams at the  LC
is decisive for deriving unique solutions.

If the LC analysis is supplemented with the LHC measurement of the
heavy neutralino mass, the errors on $\mu$ and $\tan\beta$ can be
improved. However, the best results are obtained when first  
the LSP and slepton masses from the LC
are fed to the LHC analyses to get a precise experimental determination 
of the $\nt_2$ and $\nt_4$ masses, 
which  in turn are  injected back to the analysis of the
chargino/neutralino LC data. The combined strategy 
will provide  in particular a precise measurement of the 
$\nt_4$ mass, the  $\mu$ parameter   
with an accuracy at the $\le O(1\%)$ level, and 
an error on $\tan\beta$ of the order of $\le$ 10\%,  reaching a
stage where radiative corrections become relevant in the electroweak
sector and which will have to be taken into account in  future fits 
 \cite{radiative}.

\subsubsection*{Appendix: Useful expressions for the 
gaugino/higgsino sector}
{\bf a)} For the lightest chargino pair production, $\sigma^\pm\{11\} =\sigma(
e^-(p_1) e^+(p_2)\to \tilde{\chi}^+_1(p_3)\tilde{\chi}^-_1(p_4))$, 
the coefficients 
$c_1,\ldots,c_6$ in \eq{eq_sig11} are given by:
\begin{eqnarray}
c_1&=& \int_C |Z|^2 
  \{c_{LR} L^2 f_2+c_{RL} R^2 f_1\} \nn 
\\
c_2&=&\int_C |Z|^2 \{c_{LR} L^2 (1-4L)(2 f_2+f_3)+c_{RL} R^2(1-4R)
(2f_1+f_3)\}\nonumber\\
&&-\int_C G\tilde{N} 4
\{c_{LR} L (2 f_2+f_3)+c_{RL}R(2 f_1+f_3)\}
-\int_C Re(Z)\tilde{N} c_{LR} L f_3
\nn 
\\
c_3&=& \int_C |Z|^2 (c_{LR} L^2 f_1+c_{RL} R^2 f_2)
-\int_C Z\tilde{N} 2 c_{LR} L f_1
+\int_C \tilde{N}^2 c_{LR} f_1 \nn 
\\
c_4&=&\int_C  |Z|^2 (1-4L)\{ c_{LR} L^2 (2 f_1+f_3)
+c_{RL} R^2 (2 f_2+f_3)\}+\int_C \tilde{N}^2 2 c_{LR} f_1
\nonumber\\
&&
+\int_C Re(Z)\tilde{N} c_{LR} L \{-4f_1 -f_3+4L(2f_1+f_3)\}
+\int_C G\tilde{N} 4 c_{LR} (2 f_1+f_3)
\nonumber \\
&&
-\int_C G Re(Z) 4 \{c_{LR} L (2 f_1+f_3)
+c_{RL} R (2 f_2+f_3)\} \nn 
\\
c_5&=& \int_C |Z|^2 (c_{LR} L^2+c_{RL} R^2)f_3
-\int_C Re(Z)\tilde{N} c_{LR} L f_3 \nn 
\\
c_6&=&\int_C  |Z|^2 \{c_{LR} L^2 (1-8L)+c_{RL} R^2 (1-8L)
+16 L^2 (c_{LR} L^2+c_{RL} R^2)\}(f_1+f_2+f_3)
\nonumber\\
&&-\int_C Re(Z)\tilde{N} c_{LR} L (1-4L) (2 f_1+f_3)
+\int_C G^2 (c_{LR}+c_{RL}) (f_1+f_2+f_3)
\nonumber\\
&& -\int_C Re(Z)G 8 \{c_{RL} R + c_{LR} L (1-4L)\}
(f_1+f_2+f_3)
+\int_C \tilde{N}^2 c_{LR} f_1
 \nonumber\\
&&
+\int_C G \tilde{N} 4 c_{LR} (2 f_1+f_3) \nn
\end{eqnarray}
where 
$\int_C=\frac{q_{\tilde{\chi}}}{E_b^3}\frac{1}{2 \pi}\int {\rm d}\cos\theta$, 
$L=-\frac{1}{2}+\sin^2\theta_W$, $R=\sin^2\theta_W$, and  
\begin{eqnarray*}
G&=&e^2/s,\quad 
Z=g^2/\cos^2\theta_W(s-m_Z^2+i m_Z \Gamma_Z),\quad
\tilde{N}=g^2/(t-m_{\tilde{\nu}_e}^2) \nn 
\end{eqnarray*}
denote the $\gamma$, $Z$ and $\ti\nu_e$ propagators,  
\begin{eqnarray*}
c_{LR}&=&(1-P(e^-))(1+P(e^+)),\quad
c_{RL}=(1+P(e^-))(1-P(e^+))
\end{eqnarray*}
are the beam polarisation factors, and  
\begin{eqnarray*}
f_1=(p_1 p_4)(p_2 p_3), \quad 
f_2=(p_1 p_3)(p_2 p_4), \quad 
f_3=s \, m^2_{\tilde{\chi}^\pm_i}  /2
\end{eqnarray*}
are the pure kinematic
coefficients.

{\bf b)} The coefficients 
$a_{kl}$ ($k=0,2,4,6$, $l=1,2,3$), which appear in
eqns.~(\ref{eq_m1-1}),(\ref{eq_m1-2}) and (\ref{eq_m1-3}), are   
invariants of the matrix ${\cal M}_N {\cal M}_N^T$. They can be
expressed as  functions of
the parameters 
$M_2$, $\mu$, $\tan\beta$ 
in the  following way:
\begin{eqnarray*}
a_{63}&=&M_2^2+2 (\mu^2+m_Z^2)\label{eq_inv1}\\
a_{41}&=&M_2^2+2 (\mu^2+m_Z^2 \cos^2\theta_W) \label{eq_inv2}\\
a_{42}&=&-2 \mu m_Z^2 \sin2 \beta \sin^2\theta_W\label{eq_inv3}\\
a_{43}&=&2 \mu^2 M_2^2+(\mu^2+m_Z^2)^2-2 m_Z^2\mu M_2\sin2 \beta\cos^2\theta_W
+2 m_Z^2 M_2^2 \sin^2\theta_W \label{eq_inv4}\\
a_{21}&=& \mu^4+2 \mu^2 M_2^2+2 m_Z^2 \mu^2 \cos^2\theta_W
+m_Z^4 \cos^2\theta_W-2 m_Z^2 M_2 \mu \sin2 \beta \cos^2\theta_W
\label{eq_inv5}\\
a_{22} &=&2 [ m_Z^4 M_2 \sin^2\theta_W \cos^2\theta_W
- m_Z^2 \mu^3 \sin^2\theta_W \sin2 \beta
-m_Z^2 \mu M_2^2 \sin^2\theta_W \sin2\beta] \label{eq_inv6}\\
a_{23}&=&\mu^4 M_2^2+m_Z^4 \mu^2 \sin^2 2\beta
+2 m_Z^2 \mu^2 M_2^2 \sin^2\theta_W
-2 m_Z^2 M_2 \mu^3\cos^2\theta_W\sin2\beta+m_Z^4 M_2^2\sin^4\theta_W
\nonumber
\label{eq_inv7}\\
a_{01}&=& \mu^4 M_2^2+m_Z^4 \mu^2 \cos^4\theta_W\sin^2 2\beta-2 m_Z^2\mu^3 M_2
\cos^2\theta_W \sin2 \beta\label{eq_inv8}\\
a_{02}&=&2 m_Z^4 \mu^2 M_2 \sin^2\theta_W \cos^2\theta_W \sin^2 2 \beta
-2 m_Z^2 \mu^3 M_2^2\sin^2\theta_W \sin2 \beta \label{eq_inv9}\\
a_{03}&=& m_Z^4 \mu^2 M_2^2 \sin^4\theta_W \sin^2 2 \beta
\end{eqnarray*}



\subsection{\label{sec:413} Determination of stop and sbottom sector by 
LHC and LC}

{\it J.~Hisano, K.~Kawagoe and M.M.~Nojiri}

\vspace{1em}

\noindent{\small
The information on electroweak SUSY parameters from LC would  
significantly improve resolution of masses and mixing angles 
of stop and sbottom particles produced at LHC. 
An example is shown for a Snowmass point SPS~1a, when the 
measurement of mixing angle is possible by combining LHC 
and LC data.     
}

\vspace{1em}



\subsubsection{Fits to stop and sbottom masses and mixings with 
LC inputs }

In this section, we describe the  fits to the stop and sbottom 
masses and mixings at SPS~1 using  LHC/LC data.
At this point,  $\tilde{b}_1$, 
$\tilde{b}_2\sim 500$~GeV, therefore direct production of $\tilde{b}$ at
LC is not possible without significant extension of the  energy. 
Therefore we consider  the fits of the mixing angle 
using LHC data and LC data at  $\sqrt{s}=500$~GeV.  

Note that  all sleptons, the 
lighter chargino $\tilde{\chi}^{\pm}_1$, and the lightest 
and the second lightest neutralinos $\tilde{\chi}^0_1$ and 
$\tilde{\chi}^0_2$ are within the reach of early stage of the LC
at $\sqrt{s}=500$~GeV.
The LC measures the accessible sparticles masses precisely
\cite{tesla}. Furthermore,  
Chargino and neutralino production cross sections and the 
first generation slepton production cross sections  are  
sensitive to the gaugino masses $M_1$ and $M_2$.
The expected errors 
at $\sqrt{s}=500$~GeV are given in \cite{guti}. 

The LC measurements  improve the 
mass resolution of strongly interacting sparticles at LHC \cite{giacomo}.
Errors of sparticles masses are estimated as 
$\Delta (m_{\tilde{g}}-m_{\tilde{\chi}^0_1})=2.5$~GeV, 
$\Delta(m_{\tilde{g}}- m_{\tilde{b}_1})=1.5$~GeV
and $\Delta(m_{\tilde{g}}- m_{\tilde{b}_2})=2.5$~GeV at SPS~1a
in \cite{giacomo} for  $\int{\cal L}dt=300$~fb$^{-1}$.
LHC also improve the resolution of weak SUSY parameters.
Statistics of $\tilde{\chi}^0_2$ produced from $\tilde{q}$ decay is
huge. In addition, 
$\tilde{\chi}^0_4(m_{\tilde{\chi}^0_4}= 378$~GeV)  arises 
occasionally  from squark decays. 
The same flavor and opposite sign lepton pair arises from the cascade decay 
 $\tilde{\chi}^0_2\rightarrow  \tilde{l}l
\rightarrow ll \tilde{\chi}^0_1$   and 
$\tilde{\chi}^0_4\rightarrow \tilde{l}l \rightarrow
 ll\tilde{\chi}^0_1$.  The end points 
of $m_{ll}$ distribution arising 
would be  measured precisely at LHC. 
The errors of the end points are $\Delta m_{ll}=0.08(2.3)$~GeV
for $\tilde{\chi}^0_2(\tilde{\chi}^0_4)$ respectively 
\cite{giacomo}. This is translated into the $m_{\tilde{\chi}^0_4}$ 
error $\Delta m_{\tilde{\chi}^0_4}=2.23$~GeV
which  corresponds 
to  $\Delta\mu=2.1$~GeV. 
This is an example that  the combination of 
LHC and LC fits determines $M_1, M_2, \mu, \tan\beta$ better.

Given the precise determination of weak SUSY parameters, one can 
access the structure of  sbottom and stop mass matrices. 
The mass matrices are  
controlled by the five  parameters, $m^2_{\tilde{Q}_{L3}}$, 
$m^2_{\tilde{b}_R}$, $m^2_{\tilde{t}_R}$,  $m^2_{LR}(\tilde{b})$ and 
$m^2_{LR}(\tilde{t})$.  They may be parameterized  masses and mixings 
of sbottom and stop; $m_{\tilde{b}_1}, m_{\tilde{b}_2}, \theta_b$,  $m_{\tilde{t}_1}$ 
and $\theta_{t}$. Among those parameters, 
$m_{\tilde{b}_1}$ and $m_{\tilde{b}_2}$ can  be measured 
through the peak position of 
$m(bb\tilde{\chi}^0_2)-m({bb\tilde{\chi}^0_2})$ distributions. 
On the other hand, LC measurement determine the  possible decay chains
of  $\tilde{t}$ and 
$\tilde{b}$ to the weak SUSY particles, such as the decays 
into $\tilde{\chi}^0_i$ or  $\tilde{\chi}^{\pm}_i$, and they 
further decays into $\tilde{\nu}$, $\tilde{l}$ and so on. 
Therefore $\tilde{t}$ and 
$\tilde{b}$ decay branching ratios are regarded as 
the functions of the parameters of stop and 
sbottom sector. 

The $\theta_b$ and $\theta_t$ dependent quantities 
that can be measured at LHC are discussed in the
previous section, 
\begin{itemize}
\item $\overline{\BR}(\tilde{b})\equiv $
$BR(\tilde{g}\rightarrow b \tilde{b}_2\rightarrow bb \tilde{\chi}^0_2)
/$$BR(\tilde{g}\rightarrow b \tilde{b}_1\rightarrow bb
\tilde{\chi}^0_2$)
\item The weighted end point $M^{\rm w}_{tb}$ of the cascade decays  
(III)$_1$ $\tilde{g}\rightarrow t\tilde{t}\rightarrow
tb\tilde{\chi}^+_1$ and (IV)$_{i1}$ 
$\tilde{g}\rightarrow b\tilde{b}_i\rightarrow
tb\tilde{\chi}^+_1$.

\item $\overline{\BR}(\tilde{t})\equiv$ 
 $BR(edge)/$$BR(\tilde{g}\rightarrow bbX$). 
\end{itemize}

We list the input values and estimated errors on those quantities 
in Table~\ref{table:new}, see the previous section for the details. 
Errors on the weak SUSY breaking parameters 
$M_1, M_2, \mu$ and $\tan\beta$ can be ignored compared to the errors 
on those observable in the following discussions.
 
Stop and sbottom mixing angles $\theta_t$ and $\theta_b$ 
may be determined from 
$\overline{BR}(\tilde{b})$ and $\overline{BR}(\tilde{t})$
measurement. 
Through  LHC/LC measurements, we knows $\tilde{\chi}^0_2\sim \tilde{W}$. 
Because $\tilde{W}$ couples only to the $\tilde{b}_L$, 
$BR(\tilde{b}_2\rightarrow b {\tilde{\chi}^0_2})/BR(\tilde{b}_2
\rightarrow  b {\tilde{\chi}^0_2})$ becomes a sensitive function of 
$\theta_b$. As the phase space of $\tilde{g}\rightarrow\tilde{b}_i$ 
is strongly constrained by gluino and sbottom mass measurements, the 
measured $\overline{BR}(\tilde{b})$ can be  translated immediately 
into the constraint on $BR(\tilde{b}_2\rightarrow b\tilde{\chi}^0_2)/
BR(\tilde{b}_2\rightarrow b \tilde{\chi}^0_2)$, thus $\theta_b$. 

On the other hand $\overline{BR}(\tilde{t})$ depends mainly on $\theta_t$. 
Here, the gluino branching ratio into  the edge mode is 
consisted by the contributions 
from (III)$_1$ $\tilde{g}\rightarrow 
t\tilde{t}_1\rightarrow tb\tilde{\chi}^+_1$ and 
(IV)$_{i1}$ 
$\tilde{g}\rightarrow b\tilde{b}_i\rightarrow tb\tilde{\chi}^+_1$
and (III)$_{ij}$  $\tilde{g}\rightarrow b\tilde{b}_1$
$\rightarrow bW \tilde{t}_1 \rightarrow bbW \tilde{\chi}^+_1$. 
$BR(\tilde{t}_1\rightarrow b\chi^+_1)$  
is  sensitive to  $\theta_t$ as wino like $\tilde{\chi}^+_1$ 
couples only to the $\tilde{t}_L$. 
The branching ratio of $\tilde{b}_i$ depends on $\theta_b$  
while the sum of $\tilde{b}_1$ and
$\tilde{b}_2$ contributions tends to be stable if both
$\tilde{g}\rightarrow \tilde{b}_1$ 
and $\tilde{g}\rightarrow \tilde{b}_2$ are  open. 
Altogether, one can measure $\theta_t$ from $\overline{BR}(\tilde{t})$,
and $\theta_b$ from $\overline{BR}(\tilde{b})$. 

Finally   $M^{\rm w}_{tb}$ 
is sensitive to $m_{\tilde{t}}$  as the end point depends on the 
stop mass. It also depends on the $\tilde{b}_i$ masses, 
and  $\theta_t$ and $\theta_b$ because the end point is expressed 
as the weighted average of $M_{tb}$(III)$_1$ and $M_{tb}$(IV)$_{11}$ 
Therefore the stop mass determination is possible only when 
the mixing angle is constrained.

\begin{table}
\begin{center}
\begin{tabular}{|l||l|l|}
\hline
& input & error \cr
\hline
$m_{\tilde{g}}-m_{\tilde{b}_1}$ & 103.3~GeV & 1.5~GeV\cr
$m_{\tilde{g}}-m_{\tilde{b}_2}$ & 70.6~GeV &2.5~GeV\cr
$M^{w}_{tb}$  & 370.9~GeV& 4.8(100 fb$^{-1}$ w/o sys)\cr
$\overline{\BR}(\tilde{b})$  & 0.252& 0.078 \cr
$\overline{\BR}(\tilde{t})$  & 0.583& 0.05(100 fb$^{-1}$ w/o sys) \cr
\hline
\end{tabular}
\end{center}
\caption{Measurements used for this fit. 
\label{table:new}} 
\end{table}

\begin{figure}[thb!]
\includegraphics[width=7cm]{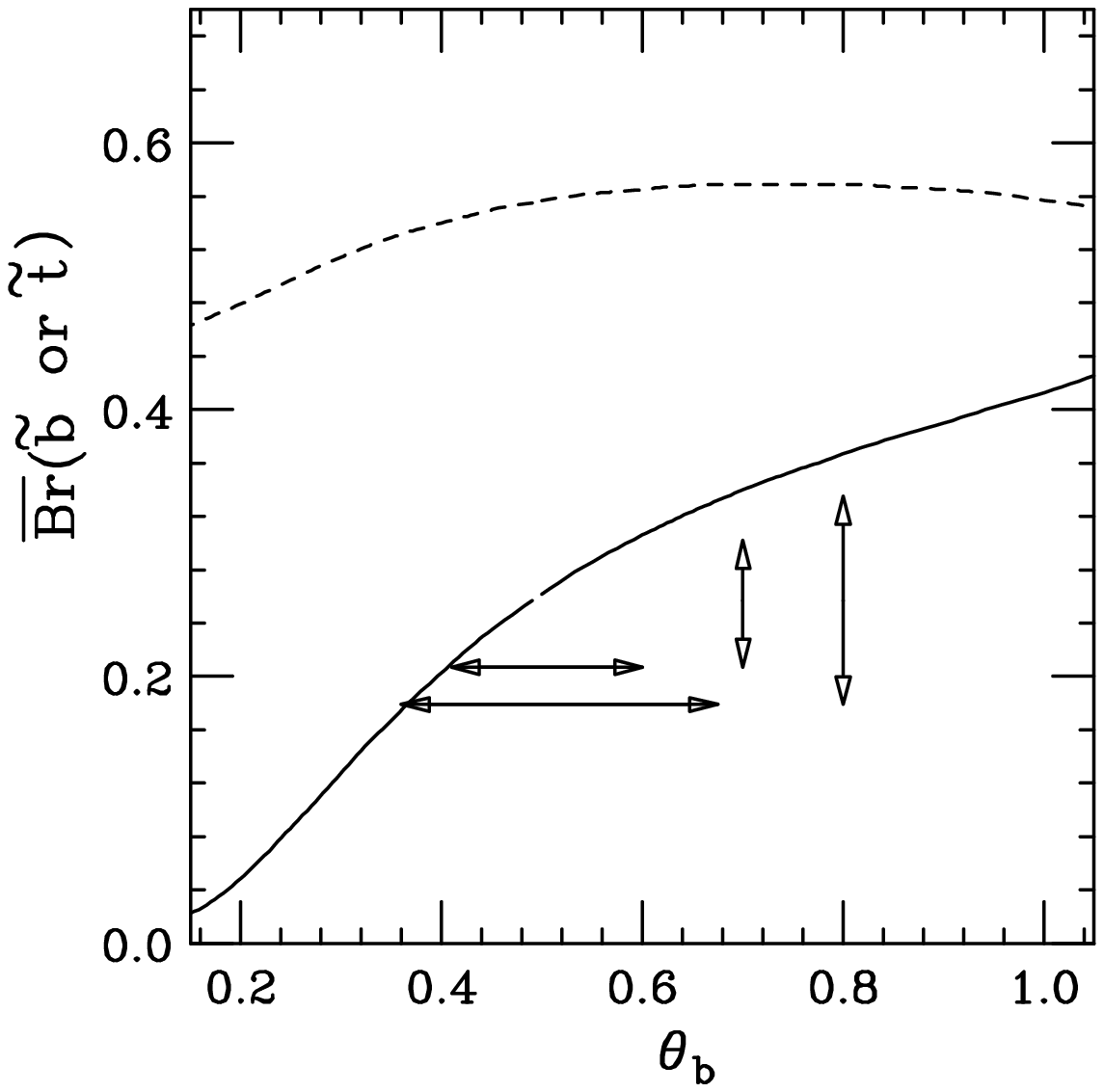}
\hskip .7cm
\includegraphics[width=8cm]{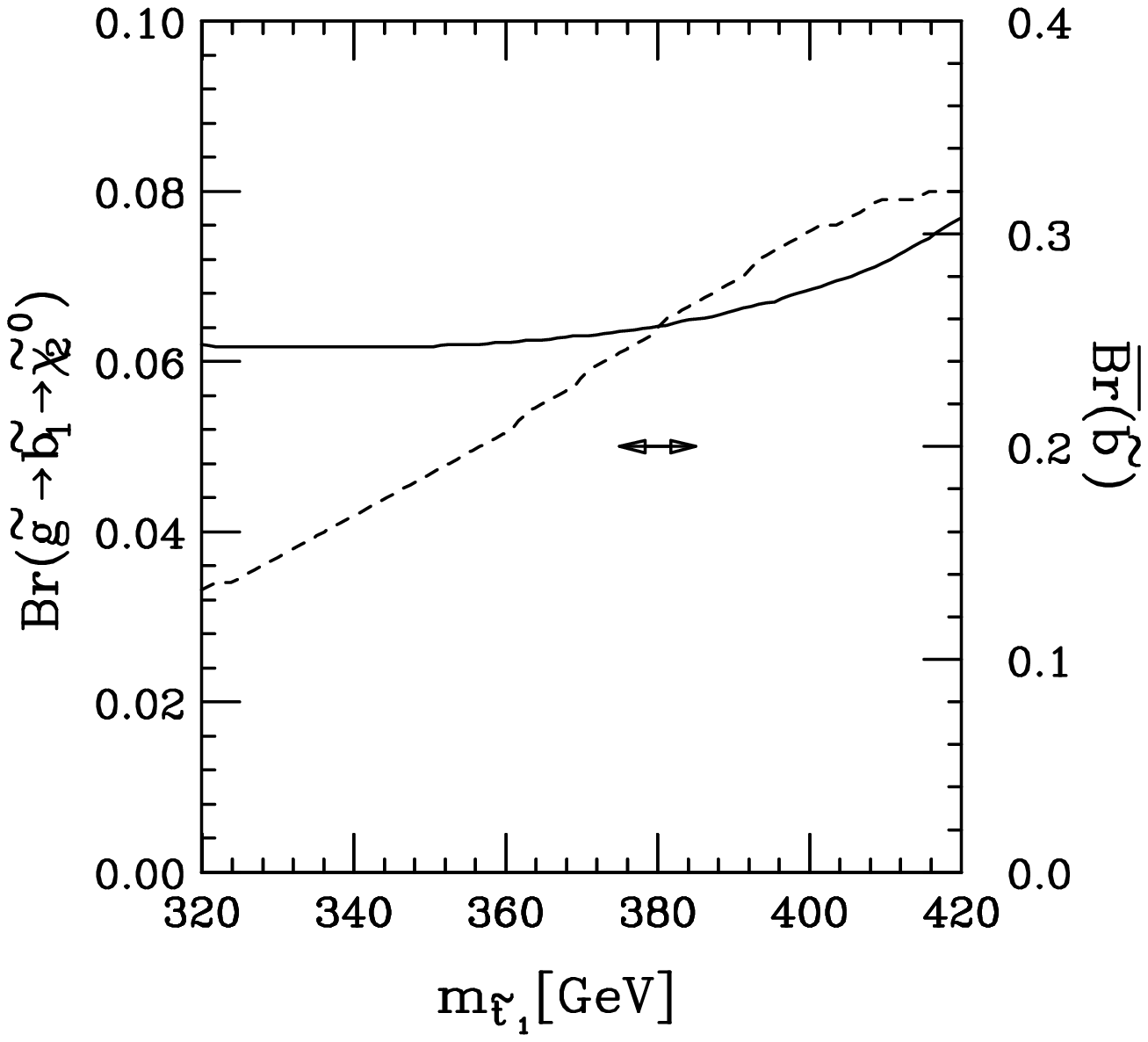}
\begin{center} (a)\hskip 8cm (b)
\end{center}
\caption{
a) $\overline{\BR}(\tilde{b})$ (solid line) and 
$\overline{\BR}(\tilde{t})$ (dotted line) as a function of $\theta_b$. 
Errors for $\overline{\BR}(\tilde{b})$ with/without systematical 
uncertainty are also shown by arrows.  See text. The other parameters 
are fixed to the inputs.
b) $\BR(\tilde{g}\rightarrow b\tilde{b}\rightarrow
bb\tilde{\chi}^0_2)$ (solid line) and $\overline{\BR}(\tilde{b})$
(dotted line) as functions of $m_{\tilde{t}}$. $\theta_t$,
$\theta_b$, $m_{\tilde{b}_1}$ and
 $m_{\tilde{b}_2}$  are fixed to SPS~1 inputs. 
An arrow shows the 1-$\sigma$ arrowed region for $m_{\tilde{t}_1}$ for 
fixed $\theta_t$. 
\label{sbottom_stop}
}
\end{figure}

We now illustrate the idea described above step by step through
theoretical calculation.  In Fig.~\ref{sbottom_stop} a), the solid line
shows $\overline{\BR}(\tilde{b})$ as a function of $\theta_{b}$ while
fixing other sparticle masses and mixings.  It increases monotonically
as $\theta_b$ increases from 0 to $\pi/2$.  This is because the
$\tilde{b}_L$ component of $\tilde{b}_2$ is proportional to $\sin
\theta_{b}$, while the wino-like $\tilde{\chi}^0_2$ couples only to
the left handed sparticles.  The statistical errors of
$\overline{\BR}(\tilde{b})$, with (without) systematic uncertainty and
the corresponding errors of $\theta_b$, $\Delta
\overline{\BR}(\tilde{b})=0.078$~$(0.045)$ and $\Delta\theta_b=0.157$
$(0.095)$, are shown by the long (short) vertical and horizontal
arrows, respectively.

We note that 
$\BR(\tilde{g}\rightarrow b \tilde{b}_i \rightarrow bb\tilde{\chi}^0_2$)
depends on the stop mass $m_{\tilde{t}_1}$ and the mixing
$\theta_t$ in addition to the sbottom mixing angle $\theta_b$. 
In Fig.~\ref{sbottom_stop}
Here 
solid and dotted lines show $\overline{\BR}(\tilde{b})$
and $BR(\tilde{g}\rightarrow b \tilde{b}_1 \rightarrow
bb\tilde{\chi}^0_2)$, respectively,  as functions of $m_{\tilde{t}_1}$.
The  stop mixing angle is fixed as $\theta_t=0.96$.
We see that the branching ratio has 
strong dependence on the stop mass. This is  because 
the decay $\tilde{b}_i\rightarrow W\tilde{t}_1$ dominates 
the sbottom decay width
if  $m_{\tilde{b}_i}\gg m_{\tilde{t}_1}$. 
On the other hand, the stop mass dependence is canceled in 
$\overline{\BR}(\tilde{b})$ in  Fig.~\ref{sbottom_stop} b).  
$\overline{\BR}(\tilde{b})$ is therefore independent 
to the stop sector and useful to extract the sbottom mixing 
angle. 
 In addition, the uncertainty of the production cross sections 
and acceptances may be canceled in the ratio.

\begin{figure}[thb!]
\begin{center}
\includegraphics[width=8cm]{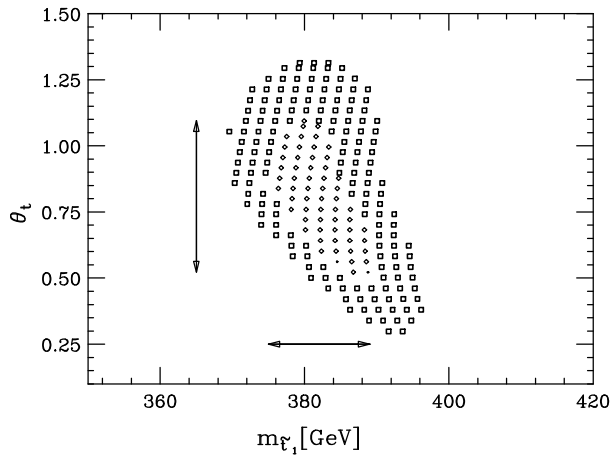}
\end{center}
\caption{Expected 1-$\sigma$ and 2-$\sigma$ errors in a $m_{\tilde{t}_1}$
and $\theta_t$ plane from 
$\overline{\BR}(\tilde{t}_1)$ and $\mtbw$ measurements.
Arrows show the 1-$\sigma$ errors of $\theta_t$ and $m_{\tilde{t}_1}$. }
\label{stop}
\end{figure}

We now discuss the constraint on 
$m_{\tilde{t}_1}$ and $\theta_t$. In Fig. \ref{stop}, 
we fix $m_{\tilde{b}_1}$, 
$m_{\tilde{b}_2}$ and $\theta_b$ to the input values, 
and scan the parameter 
space of $m_{\tilde{t}_1}$ and $\theta_{t}$. 
We define  a $\Delta\chi^2$-like function $\Delta\bar{\chi}^2$, 
\begin{equation}
\Delta\bar{\chi}^2\equiv \left(\frac{\mtbw-370.9{\rm GeV}}{4.8 {\rm GeV}}\right)^2
+\left(\frac{\overline{\BR}(\tilde{t})-0.583}{0.05}\right)^2
+\left(\frac{\overline{\BR}(\tilde{b})-0.252}{0.078}\right)^2.
\end{equation} 
In Figure \ref{stop}, the diamonds (squares) show the parameter points 
where $\Delta\chi^2<1(4)$, respectively, 
$\Delta m_{\tilde{t}_1} \sim 7$~GeV and $\Delta\theta_{t}=0.287$. 
The deviation of $\overline{\BR}(\tilde{b})$ is very small over 
the scanned region because we fixed $\theta_b$. 
Note that, the $\theta_b$ dependence of 
$\overline{\BR}(\tilde{t})$  is very weak in 
in Fig.~\ref{sbottom_stop}(a) (a dotted line)  within the allowed 
region of $\theta_b$. 

In this contribution we have shown that sbottom and stop mixing angle 
may be extracted by measuring the ratios of the branching ratios at LHC,  
provided  precise determination of the neutralino mass matrix
using LHC and LC data. The measurement of the mixing angle 
is essential to determine the sbottom mass matrix model independently,  
and the information can  be extracted only when both LHC and LC 
data is available.


%
%
                                                                                


\section{Global fits in the MSSM}

\subsection{\label{sec:424} SFITTER: SUSY parameter analysis at LHC and
LC}

{\it R.~Lafaye, T.~Plehn and D.~Zerwas}

\vspace{1em}
{\small
\noindent
SFITTER is a new analysis tool to determine supersymmetric model parameters
from collider measurements. Using the set of supersymmetric mass
measurements at the LC and at the LHC we show how both colliders
probe different sectors of the MSSM Lagrangian. This observation
is a strong motivation to move from a parameter fit assuming a
certain model to the unconstrained weak-scale MSSM Lagrangian. We
argue how the technical challenges can be dealt with in a combined
fit/grid approach with full correlations.
}

\subsubsection{Introduction}

While the Standard Model describes all
available high energy physics experiments, it still has to be regarded
as an effective theory, valid at the weak scale. New physics are
expected to appear at the TeV energy scale. The supersymmetric
extension~(\cite{Wess:1974tw}) of the Standard Model is a well motivated extension providing us with
a description of
physics that can be extended consistently up to the unification scale.

If supersymmetry or any other high-scale extension of the Standard
Model is discovered, it will be crucial to determine its
fundamental high-scale parameters from weak-scale
measurements~\cite{Blair:2002pg}. The LHC and future Linear
Colliders will provide us with a wealth of
measurements~\cite{ATLASLCTDRS,Choi:2000ta}, which
due to their complexity require proper treatment to unravel the
corresponding high-scale physics. Even in the general weak-scale
minimal supersymmetric extension of the standard model (MSSM~\cite{Fayet:1974jb}) 
without any unification or SUSY breaking assumptions some of
the measurements of masses and couplings are not 
independent measurements; moreover, linking supersymmetric
particle masses to weak-scale SUSY parameters involves non-trivial
mixing to mass eigenstates in essentially every sector of the
theory. On top of that, for example in gravity mediated SUSY
breaking scenarios (mSUGRA/cMSSM) a given weak-scale SUSY
parameter will always be sensitive to several high-scale
parameters which contribute through renormalization group running.
Therefore, a fit of the model parameters using all experimental
information available will lead to the best sensitivity and make
the most efficient use of the information available.\smallskip

\begin{table}[htb]
\begin{small} \begin{center}
\begin{tabular}{|l|cccc||l|cccc|}
\hline
 & $m_{\rm SPS1a}$ & LHC & LC & LHC+LC &
 & $m_{\rm SPS1a}$ & LHC & LC & LHC+LC\\
\hline
\hline
$h$  & 111.6 & 0.25 & 0.05 & 0.05 &
$H$  & 399.6 &      & 1.5  & 1.5  \\
$A$  & 399.1 &      & 1.5  & 1.5  &
$H+$ & 407.1 &      & 1.5  & 1.5  \\
\hline
$\chi_1^0$ & 97.03 & 4.8 & 0.05  & 0.05 &
$\chi_2^0$ & 182.9 & 4.7 & 1.2   & 0.08 \\ 
$\chi_3^0$ & 349.2 &     & 4.0   & 4.0  &
$\chi_4^0$ & 370.3 & 5.1 & 4.0   & 2.3 \\
$\chi^\pm_1$ & 182.3  & & 0.55 & 0.55 &
$\chi^\pm_2$ & 370.6  & & 3.0  & 3.0 \\
\hline
$\tilde{g}$ &  615.7 & 8.0 &  & 6.5 & & & & & \\
\hline
$\tilde{t}_1$ & 411.8 &     &  2.0  & 2.0 & & & & & \\
$\tilde{b}_1$ & 520.8 & 7.5 &       & 5.7 &
$\tilde{b}_2$ & 550.4 & 7.9 &       & 6.2 \\
\hline
$\tilde{u}_1$ &  551.0 & 19.0 & & 16.0 &
$\tilde{u}_2$ &  570.8 & 17.4 & &  9.8 \\
$\tilde{d}_1$ &  549.9 & 19.0 & & 16.0 &
$\tilde{d}_2$ &  576.4 & 17.4 & &  9.8 \\
$\tilde{s}_1$ &  549.9 & 19.0 & & 16.0 &
$\tilde{s}_2$ &  576.4 & 17.4 & &  9.8 \\
$\tilde{c}_1$ &  551.0 & 19.0 & & 16.0 &
$\tilde{c}_2$ &  570.8 & 17.4 & &  9.8 \\
\hline
$\tilde{e}_1$    & 144.9    & 4.8 & 0.05 & 0.05 &
$\tilde{e}_2$    & 204.2    & 5.0 & 0.2  & 0.2  \\
$\tilde{\mu}_1$  & 144.9    & 4.8 & 0.2  & 0.2  &
$\tilde{\mu}_2$  & 204.2    & 5.0 & 0.5  & 0.5  \\
$\tilde{\tau}_1$ & 135.5    & 6.5 & 0.3  & 0.3  &
$\tilde{\tau}_2$ & 207.9    &     & 1.1  & 1.1  \\
$\tilde{\nu}_e$  & 188.2    &     & 1.2  & 1.2  & & & & & \\
\hline
\end{tabular}
\end{center} \end{small} \vspace*{-3mm}
\caption{Errors for the mass determination in SPS1a, taken from~\cite{masses:polesello}. 
Shown are the
nominal parameter values and the error for the LHC alone, the LC
alone, and a combined LHC+LC analysis. All values are given in GeV.}
\label{tab:mass-errors}
\end{table}

In a fit, the allowed parameter space might not be sampled
completely. To avoid boundaries imposed by non-physical parameter
points, which can confine the fit to a `wrong' parameter region,
combining the fit with an initial evaluation of a
multi-dimens\-ion\-al grid is the optimal approach.

In the general MSSM the weak-scale parameters can vastly
outnumber the collider measurements, so that a complete parameter fit
is not possible and one has to limit oneself to a consistent subset of
parameters. In SFITTER both grid and fit are realised and can be
combined.  This way, one can ultimately eliminate all dependence on the
starting point of the parameter determination.
SFITTER also includes a general correlation matrix and the
option to exclude parameters of the model from the fit/grid by fixing
them to a value. Additionally, SFITTER
includes the option to apply a Gaussian smearing to all observables
before they enter the fit/grid in order to simulate realistically
experimental measurements.
In this preliminary study, however, correlations and systematic
uncertainties are neglected and the central values are used for the
measurements.

Currently, SFITTER uses the predictions for the supersymmetric masses
provided by SUSPECT~\cite{Djouadi:2002ze}, but the conventions of the
SUSY Les Houches accord~\cite{Skands:2003cj} allow us to interface
other programs.  The branching ratios and $e^+e^-$ production cross
sections are provided by MSMlib~\cite{msmlib}, which has been used
extensively at LEP and cross checked with
Ref.~\cite{Barger:2001nu}. The next-to-leading order hadron collider
cross sections are computed using
PROSPINO~\cite{Beenakker:1996ch}. The fitting program uses the MINUIT
package~\cite{James:1975dr}. The determination of $\chi^2$ includes a
general correlation matrix between measurements. In its next version
SFITTER will be interfaced with the improved branching fraction
determination of SDECAY~\cite{Muhlleitner:2003vg}, as well as
alternative renormalization group codes like
SoftSUSY~\cite{Allanach:2001kg}, ISAJET~\cite{Baer:2003mg} or
SPHENO~\cite{Porod:2003um}.

\subsubsection{mSUGRA/cMSSM Parameter Determination}

\begin{table}[htb]
\begin{center} \begin{small}
\begin{tabular}{|l|rr|rr|rr|rr|}
\hline
            & SPS1a & StartFit & LHC & $\Delta_{\rm LHC}$ & LC & $\Delta_{\rm LC}$ & LHC+LC & $\Delta_{\rm LHC+LC}$ \\
\hline
$m_0$       & 100 & 500 & 100.03 & 4.0  & 100.03 & 0.09 & 100.04 & 0.08 \\
$m_{1/2}$   & 250 & 500 & 249.95 & 1.8  & 250.02 & 0.13 & 250.01 & 0.11 \\
$\tan\beta$ &  10 &  50 &   9.87 & 1.3  &   9.98 & 0.14 &   9.98 & 0.14 \\
$A_0$       &-100 &   0 & -99.29 & 31.8 & -98.26 & 4.43 & -98.25 & 4.13 \\
\hline
\end{tabular}
\end{small} \end{center} \vspace*{-3mm}
\caption{Summary of the mSUGRA fits in SPS1a: true values, starting
values, fit values and absolute errors from the fit. As in SPS1a we
fix $\mu>0$. The mass values of the fits are based on
Tab.~\ref{tab:mass-errors}.}
\label{tab:msugra_fit}
\end{table}

Assuming that SUSY breaking is mediated by gravitational
interactions (mSUGRA/cMSSM) we fit four universal high-scale
parameters to a toy set of collider measurements: the universal
scalar and gaugino masses, $m_0$, $m_{1/2}$, the trilinear
coupling $A_0$ and the ratio of the Higgs vacuum expectation
values, $\tan\beta$. The sign of the Higgsino mass parameter $\mu$
is a discrete parameter and therefore fixed. In contrast to an
earlier study~\cite{Lafaye:2003} we assume the set of mass
measurement at the LHC and at the LC, shown in
Tab.~\ref{tab:mass-errors}.  The central value for our assumed
data set corresponds to the SUSY parameter point
SPS1a~\cite{sec4_Allanach:2002nj,Ghodbane:2002kg}, 
as computed by SUSPECT. As mentioned
in the introduction correlations, systematic errors and
theoretical errors are neglected. As the central (true) values are
used as measurements in order to study the errors on the
determination of the parameters, the $\chi^2$ values are not
meaningful and therefore are not quoted.

The starting points for the mSUGRA parameters are fixed to the mean of
the lower and upper limit (typically 1~TeV/c$^2$) of the allowed parameter range,
{\sl i.e.} they are not
necessarily close to the true SPS1a values.  The result of the fit is
shown in Tab.~\ref{tab:msugra_fit}. All true parameter values are
reconstructed well within the quoted errors, in spite of starting
values relatively far away. The measurements of $m_0$ and $m_{1/2}$
are very precise, while the sensitivity of the masses on $\tan\beta$
and $A_0$ is significantly weaker. The results for the LHC alone are
generally an order of magnitude less precise than those for the LC,
and this qualitative difference is expected to become even more pronounced
once we properly include systematical errors.

Because the data set is fit assuming mSUGRA as a unification
scenario the absence of measurements of most of the strongly
interacting particles, in particular the gluino, does not have a
strong impact on the precision of the LC determination. Therefore
the results for the combined measurements LHC+LC show only a small
improvement.

Assuming an uncorrelated data set, the correlations between the
different high-scale SUSY parameters which we obtain from the fit are
given in Tab.~\ref{tab:msugra_corr}. We can understand the correlation
matrix step by step~\cite{Drees:1995hj}: first, the universal gaugino
mass $m_{1/2}$ can be extracted very precisely from the physical
gaugino masses. The determination of the universal scalar mass $m_0$
is dominated by the weak-scale scalar particle spectrum, but in
particular the squark masses are also strongly dependent on the
universal gaugino mass, because of mixing effects in the
renormalization group running. Hence, a strong correlation between the
$m_0$ and $m_{1/2}$ occurs. The universal trilinear coupling $A_0$ can
be measured through the third generation weak-scale mass parameters
$A_{b,t,\tau}$. However, the $A_{b,t,\tau}$ which appear for example
in the off-diagonal elements of the scalar mass matrices, also depend
on $m_0$ and $m_{1/2}$, so that $A_0$ is strongly correlated with
$m_0$ and $m_{1/2}$. At this point one should stress that the
determination of $A_0$ is likely to be dominated by $A_t$ as it
appears in the calculation of the lightest Higgs mass $m_h$. After
taking into account the current theoretical error of 3~GeV on
$m_h$~\cite{schweinlein} we expect the determination of $A_0$ to
suffer significantly. The experimental errors therefore 
can be considered a call for an improvement of the 
theoretical error.\medskip

\begin{table}[htb]
\begin{center} \begin{small}
\begin{tabular}{|c|rrrr|}
\hline
             &  $m_0$  &  $m_{1/2}$ & $\tan\beta$ &   $A_0$    \\
\hline
 $m_0$       & 1.000 &   -0.555 &  0.160    &  -0.324  \\
 $m_{1/2}$   &       &    1.000 &  -0.219   &   0.617  \\
 $\tan\beta$ &       &          &  1.000    &   0.307  \\
 $A_0$       &       &          &           &   1.000  \\
\hline
\end{tabular}
\end{small} \end{center} \vspace*{-3mm}
\caption[]{The (symmetric) correlation matrix for the mSUGRA fit
given in Tab.\ref{tab:msugra_fit} with data set LHC+LC.}
\label{tab:msugra_corr}
\end{table}

In general, $\tan\beta$ can be determined in three sectors of the
supersymmetric
spectrum: all four Higgs masses, and for large values of $m_A$ in
particular the light CP even Higgs mass $m_h$ depend on $\tan\beta$.
The mixing between gauginos and Higgsinos in the neutralino/chargino
sector is governed by $\tan\beta$.  Finally, the stop mixing is
governed by $\mu/\tan\beta$, while the sbottom and stau mixing depends
on $\mu \tan\beta$. The correlation of $\tan\beta$ with the other
model parameters reflects the relative impact of these three
sectors. In an earlier analysis we assumed a uniform error of 0.5\% on
all mass measurements~\cite{Lafaye:2003} and saw that in this case
$\tan\beta$ is determined through stau mixing, which in turn means
that it shows very little correlation with $m_{1/2}$.

For the more realistic scenario in Tab.~\ref{tab:mass-errors} the
outcome is the following: the relative errors for the light Higgs mass
and for the light neutralino masses at the LC are tiny. The relevant
parameter in the Higgs sector is the light stop mass, which is
governed by $m_{1/2}$; similarly the gaugino mass $m_{1/2}$ which
fixes the light neutralino and chargino masses does not depend strongly
on $\tan\beta$. The slepton sector introduces a strong correlation between
$m_0$ and $m_{1/2}$. The resulting correlation matrix is shown in
in Tab.~\ref{tab:msugra_corr}. The
results obtained with SFITTER are in agreement with expectation.

\subsubsection{General MSSM Parameter Determination}

\begin{table}[htb]
\begin{center} \begin{small}
\begin{tabular}{|l|rrr||l|rrr|}
\hline
   & AfterGrid  & AfterFit   & SPS1a & & AfterGrid  & AfterFit   & SPS1a \\
\hline
$\tan\beta$          &       100  &      10.02$\pm$3.4     &       10 &
 $M_{\tilde{u}_R}$   &      532.1 &     532.1$\pm$2.8      &    532.1 \\
$M_1$                &       100  &     102.2$\pm$0.74     &    102.2 &
 $M_{\tilde{d}_R}$   &      529.3 &     529.3$\pm$2.8      &    529.3 \\
$M_2$                &       200  &     191.79$\pm$1.9     &    191.8 &
 $M_{\tilde{c}_R}$   &      532.1 &     532.1$\pm$2.8      &    532.1 \\
$M_3$                &      589.4 &     589.4$\pm$7.0      &    589.4 &
 $M_{\tilde{s}_R}$   &      529.3 &     529.3$\pm$2.8      &    529.3 \\
$\mu$                &       300  &     344.3$\pm$1.3      &    344.3 &
 $M_{\tilde{t}_R}$   &     420.2  &     420.08$\pm$13.3    &    420.2 \\
$m_A$                &     399.35 &     399.1$\pm$1.2      &    399.1 &
 $M_{\tilde{b}_R}$   &     525.6  &     525.5$\pm$10.1     &    525.6 \\
$M_{\tilde{e}_R}$    &     138.2  &     138.2$\pm$0.76     &    138.2 &
 $M_{\tilde{q}1_L}$  &     553.7  &     553.7$\pm$2.1      &    553.7 \\
$M_{\tilde{\mu}_R}$  &     138.2  &     138.2$\pm$0.76     &    138.2 &
 $M_{\tilde{q}2_L}$  &     553.7  &     553.7$\pm$2.1      &    553.7 \\
$M_{\tilde{\tau}_R}$ &     135.5  &     135.48$\pm$2.3     &    135.5 &
 $M_{\tilde{q}3_L}$  &     501.3  &     501.42$\pm$10.     &    501.3 \\
$M_{\tilde{e}_L}$    &     198.7  &     198.7$\pm$0.68     &    198.7 &
 $A_\tau$            &    -253.5  &    -244.7$\pm$1428     &   -253.5 \\
$M_{\tilde{\mu}_L}$  &     198.7  &     198.7$\pm$0.68     &    198.7 &
 $A_t$               &    -504.9  &    -504.62$\pm$27.     &   -504.9 \\
$M_{\tilde{\tau}_L}$ &     197.8  &     197.81$\pm$0.92    &    197.8 &
 $A_b$               &    -797.99 &    -825.2$\pm$2494     &   -799.4 \\
\hline
\end{tabular}
\end{small} \end{center} \vspace*{-3mm}
\caption{Result for the general MSSM parameter determination in SPS1a
using the toy sample of all MSSM particle masses with a universal
error of 0.5\%. Shown are the nominal parameter values, the result
after the grid and the final result. All masses are given in GeV.}
\label{tab:mssm1}
\end{table}

In this study, the unconstrained weak-scale MSSM is described by
24 parameters in addition to the standard model parameters. 
The parameters are listed in Tab.~\ref{tab:mssm1}: $\tan\beta$ as in
mSUGRA, plus three soft SUSY breaking gaugino masses $M_i$, the
Higgsino mass parameter $\mu$, the pseudoscalar Higgs mass $m_A$, the
soft SUSY breaking masses for the right sfermions, $M_{\tilde{f}_R}$,
the corresponding masses for the left doublet sfermions,
$M_{\tilde{f}_L}$ and finally the trilinear couplings of the third
generation sfermions $A_{t,b,\tau}$.

\vspace{1em}
\noindent
\underline{Toy model with all masses}

For testing purposes, we first consider a toy data set which includes
all supersymmetric particle masses. The universal error on all mass
measurements is set to 0.5\%.

In any MSSM spectrum, in first approximation, the parameters
$M_1$, $M_2$, $\mu$ and $\tan\beta$ determine the neutralino and
chargino masses and couplings. We exploit this feature to
illustrate the option to use a grid before starting the fit. The
starting values of the parameters other than $M_1$, $M_2$, $\mu$
and $\tan\beta$ are set to their nominal values, this study is
thus less general than the one of mSUGRA. The $\chi^2$ is then
minimized on a grid using the six chargino and neutralino masses
as measurements to determine the four parameters $M_1$, $M_2$,
$\mu$ and $\tan\beta$. The step size of the grid is 10 for
$\tan\beta$ and 100~GeV for the mass parameters.  After the
minimization, the four parameters obtained from grid minimization
are fixed and all remaining parameters are fitted. In a final run
all model parameters are released and fitted. The results after
the grid (including the complementary fit), after the final fit 
and the nominal values are shown in
Tab.~\ref{tab:mssm1}. The smearing option has not been applied.
However, the errors on the fitted values (once the fit converges)
should not be sensitive to these shortcomings.

The final fit indeed converges to the
correct central values within its error. The central values of the fit
are in good agreement with generated values, except for the trilinear
coupling $A_{b,\tau}$. The problem is using only mass
measurements to determine the three entries in a (symmetric) scalar
mass matrix: in the light slepton sector there are three masses, left and
right scalars plus the sneutrino, so the system is in principle calculable.
In the third generation squark sector we have three
independent diagonal entries per generation and two off-diagonal
entries. But the number of mass measurements is only four, therefore
the system is underdetermined in first order. The off-diagonal entry in the mass
matrix for down type scalars includes a term $A_{b,\tau}$ and
an additional term $\mu\tan\beta$. Even for very moderate values of
$\tan\beta$ the extraction of $A_{b,\tau}$ requires precise knowledge
of $\tan\beta$. The use of branching ratios and cross section
measurements (with polarised beams) which carry information about the scalar mixing angles
should significantly improve the determination of $A_{t,b,\tau}$.

\vspace{1em}
\noindent
\underline{Toy model with LHC-LC mass measurements}

\begin{table}[htb]
\begin{center} \begin{small}
\begin{tabular}{|l|rrrr|}
\hline
       & LHC & LC & LHC+LC    & SPS1a \\
\hline
$\tan\beta$          &      10.22$\pm$9.1   &    10.26$\pm$0.3  &   10.06$\pm$0.2   &       10 \\
$M_1$                &     102.45$\pm$5.3   &   102.32$\pm$0.1  &  102.23$\pm$0.1   &    102.2 \\
$M_2$                &      191.8$\pm$7.3   &   192.52$\pm$0.7  &  191.79$\pm$0.2   &    191.8 \\
$M_3$                &     578.67$\pm$15    &     fixed 500     &  588.05$\pm$11    &    589.4 \\
$M_{\tilde{\tau}_L}$ &       fixed 500      &   197.68$\pm$1.2  &  199.25$\pm$1.1   &    197.8 \\
$M_{\tilde{\tau}_R}$ &     129.03$\pm$6.9   &   135.66$\pm$0.3  &  133.35$\pm$0.6   &    135.5 \\
$M_{\tilde{\mu}_L}$  &      198.7$\pm$5.1   &    198.7$\pm$0.5  &   198.7$\pm$0.5   &    198.7 \\
$M_{\tilde{\mu}_R}$  &      138.2$\pm$5.0   &    138.2$\pm$0.2  &   138.2$\pm$0.2   &    138.2 \\
$M_{\tilde{e}_L}$    &      198.7$\pm$5.1   &    198.7$\pm$0.2  &   198.7$\pm$0.2   &    198.7 \\
$M_{\tilde{e}_R}$    &      138.2$\pm$5.0   &    138.2$\pm$0.05 &   138.2$\pm$0.05  &    138.2 \\
$M_{\tilde{q}3_L}$   &      498.3$\pm$110   &    497.6$\pm$4.4  &   521.9$\pm$39    &    501.3 \\
$M_{\tilde{t}_R}$    &       fixed 500      &      420$\pm$2.1  &  411.73$\pm$12    &    420.2 \\
$M_{\tilde{b}_R}$    &     522.26$\pm$113   &     fixed 500     &  504.35$\pm$61    &    525.6 \\
$M_{\tilde{q}2_L}$   &     550.72$\pm$13    &     fixed 500     &  553.31$\pm$5.5   &    553.7 \\
$M_{\tilde{c}_R}$    &     529.02$\pm$20    &     fixed 500     &  531.70$\pm$15    &    532.1 \\
$M_{\tilde{s}_R}$    &     526.21$\pm$20    &     fixed 500     &  528.90$\pm$15    &    529.3 \\
$M_{\tilde{q}1_L}$   &     550.72$\pm$13    &     fixed 500     &  553.32$\pm$6.5   &    553.7 \\
$M_{\tilde{u}_R}$    &     528.91$\pm$20    &     fixed 500     &  531.70$\pm$15    &    532.1 \\
$M_{\tilde{d}_R}$    &      526.2$\pm$20    &     fixed 500     &  528.90$\pm$15    &    529.3 \\
$A_\tau$             &       fixed 0        &   -202.4$\pm$89.5 &  352.1$\pm$171    &   -253.5 \\
$A_t$                &     -507.8$\pm$91    &  -501.95$\pm$2.7   & -505.24$\pm$3.3  &   -504.9 \\
$A_b$                &    -784.7$\pm$35603 &     fixed 0       &  -977$\pm$12467    &   -799.4 \\
$m_A$                &       fixed 500      &    399.1$\pm$0.9  &   399.1$\pm$0.8   &    399.1 \\
$\mu$                &     345.21$\pm$7.3   &   344.34$\pm$2.3  &   344.36$\pm$1.0  &    344.3 \\
\hline
\end{tabular}
\end{small} \end{center} \vspace*{-3mm}
\caption[]{Result for the general MSSM parameter determination in
SPS1a using the mass measurements given in
Tab.~\ref{tab:mass-errors}. Shown are the nominal parameter values and
the result after fits to the different data sets. All masses are given
in GeV.}
\label{tab:mssm2}
\end{table}

In the study of the three data sets LHC, LC, and LHC+LC in the MSSM, a fit
was performed for the data sets LHC and LC, whereas for LHC+LC additionally
the GRID was used for $M_1$, $M_2$, $\mu$ and $\tan\beta$ with the five chargino
and neutralino masses. The starting points were chosen to be the true values
(with the exception of the parameters used in the grid).
In order to obtain a solvable system,
for the LHC data set $m_A$, $M_{\tilde{t}_R}$, $M_{\tilde{\tau}_L}$, $A_\tau$ were
fixed. For the LC data set the first and second generation squark soft SUSY breaking
masses, the gluino mass $M_3$, $M_{\tilde{b}_R}$ and $A_b$ were fixed.
These parameters were chosen
on the basis of the measurements available in Tab.~\ref{tab:mass-errors}.
The values to which these parameters were fixed is not expected to influence the final
result of the fit. The results for the two data sets are shown in Tab.~\ref{tab:mssm2}.

Note that the general rule that the LHC is not sensitive to weakly interacting
particle masses is not entirely true: while the LHC has the
advantage of measuring the squark and gluino masses, the first and second
generation slepton mass parameters are also determined with a precision of the order of
percent. The results in Tab.~\ref{tab:mssm2}
show that the LHC alone is well capable of determining for example all
gaugino mass parameters as well as most of the scalar mass
parameters.

The situation at the LC is slightly different. Only marginal
information on the squark sector available at the LC. The
measurement of $A_t$ from the Higgs sector should be taken with a
grain of salt (theoretical error on the lightest Higgs mass).
Adding the stau mixing angle to the set of LC measurements will
improve the determination of $A_\tau$. However, the measurements
of the parameters, in particular slepton and gaugino parameters
are far more precise than at the LHC.

For the LHC+LC data set, a sufficient number of mass measurements is available,
so that no parameters need to be fixed. The superiority of the combination
of the measurements at the two colliders is obvious from this observation and
from Tab.~\ref{tab:mssm2}: The LHC contributes to reduce the error in the weak
sector ($M_2$) and the LC in the strongly interacting sector (third generation squarks).
Even more important: of 13 parameters undetermined by either the LHC or the LC,
11 are determined with good precision in the combination.
For $A_\tau$, $A_b$, we expect an improvement with the use of branching ratios and cross
section measurements.

A complete measurement of all parameters at the weak
scale is particularly important if one wants to probe unification scenarios
which link subsectors of the parameter space which are
independent at the weak scale.
An advanced tool like SFITTER can extract the information to probe
supersymmetry breaking scenarios from any set of measurements,
provided the set is sufficient to overconstrain the model parameters.

\subsubsection{Conclusions}

SFITTER is a new program to determine supersymmetric parameters from
measurements. The parameters can be extracted either using a fit, a
multi-dimensional grid, or a combination of the two. Correlations
between measurements can be specified and are taken into
account. While it is relatively easy to fit a fixed model with very few
parameters for example at a high scale to a set of collider
measurements, the determination of the complete set of weak-scale MSSM
model parameters requires this more advanced tool.
A mSUGRA inspired fit does not include the full complexity and power of
the combined LHC and LC data compared to the measurements at either
collider alone.
The results from SFITTER in the MSSM with the three data sets
show that only the combination of measurements of both the LHC and the LC
offers a complete picture of the MSSM model parameters
in a reasonably model independent framework.


%


\subsection{\label{sec:424b} Fittino: A global fit of the MSSM
parameters}

{\it P.~Bechtle, K.~Desch and P.~Wienemann}

\vspace{1em}

{\small
\noindent
If SUSY is realized, the pattern of its breaking can be generally
expressed in terms of the soft SUSY breaking Lagrangian. The program
Fittino extracts the parameters of the MSSM Lagrangian from simulated
measurements at LHC and the LC in a global fit. No prior knowledge of
the parameters is assumed. Tree-level relations between observables
and SUSY parameters are used to obtain start values for the fit.
Without the information from all sectors of the theory this fit does
not converge. Therefore both the almost complete spectrum at LHC and
the precise measurements of the lighter SUSY particles at the LC is
crucial. An example fit is performed for SPS1a, assuming unification
in the first two generations, flavor-diagonal couplings and absence of
CP-violating phases. As a result of the fit, a full error matrix of
the parameters is obtained.  }

\subsubsection{Introduction}
As for the program SFITTER, the aim of Fittino is the determination of
the parameters of the MSSM. It is implemented in C++ and focusses on
the determination of the parameters of the soft SUSY breaking
Lagrangian ${\cal L}_{\text{soft}}$, obeying the following
principles:
\begin{itemize}
\item No {\it a priori} knowledge of SUSY parameters is assumed (but
  can be imposed if desired by the user).
\item All measurements from future colliders could be used.
\item All correlations among parameters and all influences of
  loop-induced effects, where parameters of one sector affect
  observables of other sectors of the theory, are taken into account,
  as far as they are implemented into the program which provides the
  theoretical predictions.
\end{itemize}
In this way  an unbiased global fit is obtained. No attempt to extract
SUSY parameters at the GUT scale is made, since the evolution of the 
parameters and their determination at the low scale factorizes. The result
of Fittino with the full errors of the low energy SUSY parameters can therefore
be used later to extrapolate to the GUT scale.

However, all 105 possible parameters of ${\cal L}_{\text{MSSM}}$ cannot be 
determined simultaneously. Therefore, assumptions on the structure of
${\cal L}_{\text{MSSM}}$ are made. All phases are set to 0, no mixing
between generations is assumed and the mixing within the first two generations 
is set to 0. Thus the number of free parameters in the SUSY breaking sector is
reduced to 24. Further assumptions can be specified by the user.
Observables used in the fit can be
\begin{itemize}
\item Masses, limits on masses of unobserved particles
\item Widths
\item Cross-Sections
\item Branching ratios
\item Edges in mass spectra
\end{itemize}
Correlations among observables and both experimental and theoretical
errors can be supplied by the user.  Both SM and MSSM observables can
be used in the fit. Parametric uncertainties of SUSY observables can
be taken into account by fitting the relevant SM parameters
simultaneously with the MSSM parameters. The prediction of the MSSM
observables for a given set of parameters is obtained from SPheno
\cite{Porod:2003um}. The communication with SPheno is realized via
the SUSY Les Houches Accord \cite{Skands:2003cj} (SLHA). Other SUSY generators
or spectrum calculators can be included via SLHA. MINUIT \cite{Minuit}
is used for the fitting process.

In the following, we describe the principles of Fittino in more detail, followed
by example fits based on SPS1a.

\subsubsection{MSSM Parameter Determination}
\subsubsection*{General Principles of Fits with Fittino}
The full MSSM parameter space in Fittino, consisting of maximally 24 MSSM parameters
plus SM parameters, cannot be scanned completely, neither in a fit nor in a grid approach.
Therefore, in order to find the true parameters in a fit by minimizing a $\chi^2$ 
function, it is essential to begin with reasonable start values, allowing for a smooth
transition to the true minimum. As default, no {\it a priori} knowledge of the parameters
can be used in a realistic attempt of a fit, since in a real measurement no information
on true parameters will be available either. 

\begin{figure}[t]
\begin{center}
\epsfig{file=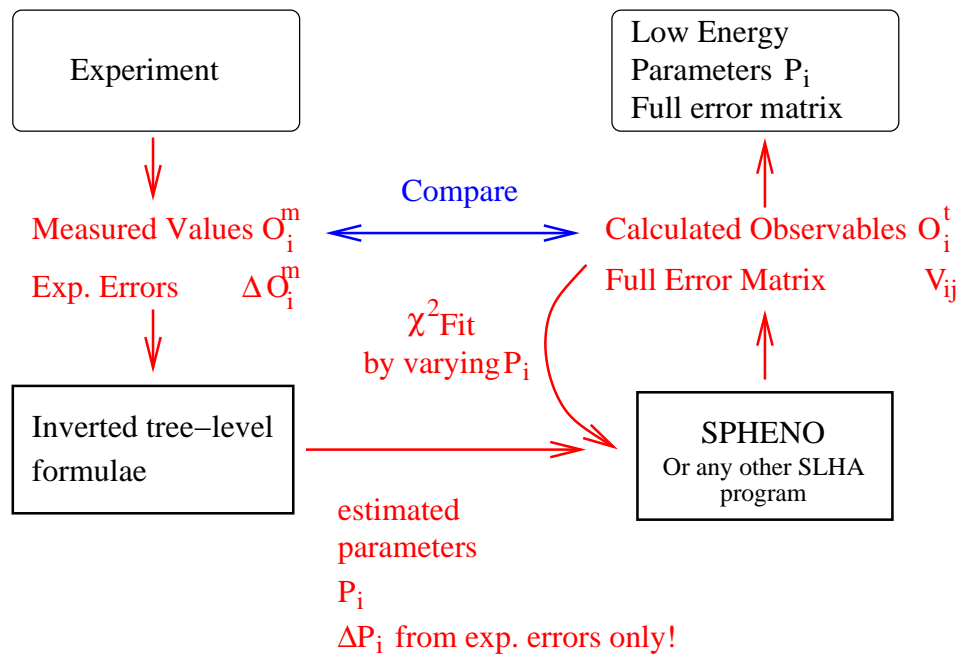, width=0.8\textwidth}
\end{center}
\caption{Iterative fit procedure}\label{fig:iterative}
\end{figure}

\begin{table}
\begin{center}
{\footnotesize
\begin{tabular}{l c c}
\hline \multicolumn{1}{c}{Measurement} & \multicolumn{1}{c}{Value} &
    \multicolumn{1}{c}{Uncertainty} \\ \hline
$m_{\text{Z}}$       &     \text{91.1187 GeV} & \text{0.0021 GeV}\\ %
$m_{\text{W}}$       &     \text{80.3382 GeV} & \text{0.039  GeV}\\ %
$m_{\text{c}}$       &     \text{1.2     GeV} & \text{0.2    GeV}\\ %
$m_{\text{b}}$       &     \text{4.2     GeV} & \text{0.5    GeV}\\ %
$m_{\text{t}}$       &     \text{174.3   GeV} & \text{0.3    GeV}\\ %
$m_{\tau}$           &     \text{1.77699   GeV} & \text{0.00029    GeV}\\ %
$\alpha_s$           &     \text{0.1172}      & \text{0.0002}\\
$G_F$                &     \text{1.16639$\cdot$10$^{\text{-5}}$ GeV$^{-2}$} & \text{1$\cdot$10$^{\text{-11}}$ GeV$^{-2}$}\\
$1/\alpha$           &     \text{127.934}     & \text{0.027}\\
$\sin^2 \theta_W$    &     \text{0.23113}     & \text{0.00015}\\
$m_{\text{h}^0}$     &     \text{110.2 GeV} & \text{0.5 GeV  }\\ 
$m_{\text{H}^0}$     &     \text{400.8 GeV} & \text{1.3 GeV   }\\ 
$m_{\text{A}^0}$     &     \text{399.8 GeV} & \text{1.3 GeV   }\\ 
$m_{\text{H}^{\pm}}$ &     \text{407.7 GeV} & \text{1.1 GeV   }\\ 
$m_{\tilde{\text{u}}_L}$ &     \text{583.5 GeV} & \text{9.8 GeV   }\\ %
$m_{\tilde{\text{u}}_R}$ &     \text{566.5 GeV} & \text{23.6 GeV  }\\ %
$m_{\tilde{\text{d}}_L}$ &  \text{586.7 GeV} & \text{9.8 GeV   }\\ %
$m_{\tilde{\text{d}}_R}$ &  \text{566.3 GeV} & \text{23.6 GeV  }\\ %
$m_{\tilde{\text{c}}_L}$ &     \text{583.6 GeV} & \text{9.8 GeV   }\\ 
$m_{\tilde{\text{c}}_R}$ &     \text{566.5 GeV} & \text{23.6 GeV  }\\ 
$m_{\tilde{\text{s}}_L}$ &     \text{586.7 GeV} & \text{9.8 GeV   }\\ %
$m_{\tilde{\text{s}}_R}$ &     \text{566.3 GeV} & \text{23.6 GeV  }\\ %
$m_{\tilde{\text{t}}_R}$ &     \text{417.5 GeV} & \text{2.0 GeV   }\\ 
$m_{\tilde{\text{b}}_R}$ &     \text{532.1 GeV} & \text{5.7  GeV  }\\ 
$m_{\tilde{\text{b}}_L}$ &     \text{565.6 GeV} & \text{6.2  GeV  }\\ 
$m_{\tilde{\nu}_{\text{e}L}}$ &     \text{192.3 GeV} & \text{0.7 GeV   }\\ 
$m_{\tilde{\text{e}}_L}$ &     \text{208.0 GeV} & \text{0.2 GeV   }\\ 
$m_{\tilde{\text{e}}_R}$ &     \text{143.91 GeV} & \text{0.05 GeV  }\\ 
$m_{\tilde{\mu}_L}$      &     \text{208.0 GeV} & \text{0.5 GeV   }\\ 
$m_{\tilde{\mu}_R}$      &     \text{143.9 GeV} & \text{0.2 GeV   }\\ 
$m_{\tilde{\tau}_R}$     &     \text{134.3  GeV} & \text{0.3 GeV   }\\ 
$m_{\tilde{\tau}_L}$     &     \text{211.8 GeV} & \text{1.1 GeV   }\\ 
$m_{\tilde{\text{g}}}$          &     \text{630.4  GeV} & \text{6.4 GeV   }\\ 

$m_{\tilde{\chi}_1^0}$   &     \text{95.74 GeV} & \text{0.05 GeV  }\\ 
$m_{\tilde{\chi}_2^0}$   &     \text{182.40 GeV} & \text{0.08 GeV  }\\ 
$m_{\tilde{\chi}_1^{\pm}}$ &     \text{180.46 GeV} & \text{0.55 GeV  }\\ 
$m_{\tilde{\chi}_2^{\pm}}$ &     \text{380.0 GeV} & \text{3.0 GeV   }\\ 

$\sigma$ ( $\text{e}^+\text{e}^- \rightarrow \tilde{\chi}_1^0 \tilde{\chi}_2^0$,
$\sqrt{s} = 500$ GeV, $P_{\text{e}^-} = 0.8$, $P_{\text{e}^+} = 0.6$ )  &  \text{22.7 fb}&  \text{2.0  fb}  \\ 
$\sigma$ ( $\text{e}^+\text{e}^- \rightarrow \tilde{\chi}_2^0 \tilde{\chi}_2^0$, 
$\sqrt{s} = 500$ GeV, $P_{\text{e}^-} = 0.8$, $P_{\text{e}^+} = 0.6$ )  &  \text{19.5 fb}&  \text{2.0  fb}  \\ 
$\sigma$ ( $\text{e}^+\text{e}^- \rightarrow \tilde{\text{e}}_L \tilde{\text{e}}_L$,
$\sqrt{s} = 500$ GeV, $P_{\text{e}^-} = 0.8$, $P_{\text{e}^+} = 0.6$)  &  \text{205.0 fb}&  \text{4.0  fb}  \\ 
$\sigma$ ( $\text{e}^+\text{e}^- \rightarrow \tilde{\mu}_L \tilde{\mu}_L$,
$\sqrt{s} = 500$ GeV, $P_{\text{e}^-} = 0.8$, $P_{\text{e}^+} = 0.6$)              &  \text{36.8   fb}&  \text{4.0  fb}  \\ 
$\sigma$ ( $\text{e}^+\text{e}^- \rightarrow \tilde{\tau}_1 \tilde{\tau}_1$,
$\sqrt{s} = 500$ GeV, $P_{\text{e}^-} = 0.8$, $P_{\text{e}^+} = 0.6$)             &  \text{39.1 fb}&  \text{4.0  fb}  \\ 
$\sigma$ ( $\text{e}^+\text{e}^- \rightarrow \tilde{\chi}_1^{\pm} \tilde{\chi}_1^{\mp}$,
$\sqrt{s} = 500$ GeV, $P_{\text{e}^-} = 0.8$, $P_{\text{e}^+} = 0.6$)      &  \text{46.7 fb}&  \text{1.0  fb}  \\ 
$\sigma$ ( $\text{e}^+\text{e}^- \rightarrow$ Z $\text{h}^0$,
$\sqrt{s} = 500$ GeV, $P_{\text{e}^-} = 0.8$, $P_{\text{e}^+} = 0.6$ )                   &  \text{11.13 fb}&  \text{0.21 fb}  \\ 
$\sigma$ ( $\text{e}^+\text{e}^- \rightarrow \tilde{\chi}_1^{\pm} \tilde{\chi}_1^{\mp}$,
$\sqrt{s} = 500$ GeV, $P_{\text{e}^-} = -0.8$, $P_{\text{e}^+} = -0.6$ )    &  \text{104.8 fb}&  \text{3.5  fb}  \\ 
$\sigma$ ( $\text{e}^+\text{e}^- \rightarrow \tilde{\chi}_1^0 \tilde{\chi}_2^0$,
$\sqrt{s} = 500$ GeV, $P_{\text{e}^-} = -0.8$, $P_{\text{e}^+} = -0.6$)&  \text{43.9 fb}&  \text{2.0  fb}  \\ 
$\sigma$ ( $\text{e}^+\text{e}^- \rightarrow \tilde{\chi}_2^0 \tilde{\chi}_2^0$,
$\sqrt{s} = 500$ GeV, $P_{\text{e}^-} = -0.8$, $P_{\text{e}^+} = -0.6$)&  \text{43.8 fb}&  \text{2.0  fb}  \\ 
$\sigma$ ( $\text{e}^+\text{e}^- \rightarrow \tilde{\text{e}}_L \tilde{\text{e}}_L$,
$\sqrt{s} = 500$ GeV, $P_{\text{e}^-} = -0.8$, $P_{\text{e}^+} = -0.6$)&  \text{97.4 fb}&  \text{4.0  fb}  \\ 
\hline
\end{tabular}
}
\end{center}
\end{table}

\begin{table}
\begin{center}
{\footnotesize
\begin{tabular}{l c c}
\hline \multicolumn{1}{c}{Measurement} & \multicolumn{1}{c}{Value} &
    \multicolumn{1}{c}{Uncertainty} \\ \hline

$\sigma$ ( $\text{e}^+\text{e}^- \rightarrow \tilde{\text{e}}_L \tilde{\text{e}}_R$,
$\sqrt{s} = 500$ GeV, $P_{\text{e}^-} = -0.8$, $P_{\text{e}^+} = -0.6$)&  \text{223.7 fb}&  \text{4.0  fb}  \\ 
$\sigma$ ( $\text{e}^+\text{e}^- \rightarrow \tilde{\text{e}}_R \tilde{\text{e}}_R$,
$\sqrt{s} = 500$ GeV, $P_{\text{e}^-} = -0.8$, $P_{\text{e}^+} = -0.6$)&  \text{29.0 fb}&  \text{2.0  fb}  \\ 
$\sigma$ ( $\text{e}^+\text{e}^- \rightarrow \tilde{\mu}_L \tilde{\mu}_L$,
$\sqrt{s} = 500$ GeV, $P_{\text{e}^-} = -0.8$, $P_{\text{e}^+} = -0.6$)            &  \text{22.7 fb}&  \text{2.0  fb}  \\ 
$\sigma$ ( $\text{e}^+\text{e}^- \rightarrow \tilde{\tau}_1 \tilde{\tau}_1$,
$\sqrt{s} = 500$ GeV, $P_{\text{e}^-} = -0.8$, $P_{\text{e}^+} = -0.6$)           &  \text{25.7 fb}&  \text{2.0  fb}  \\ 

BR ( $\text{h}^0 \rightarrow \text{b}\bar{\text{b}}$ )                       &  \text{0.82 } & \text{0.01} \\ 
BR ( $\text{h}^0 \rightarrow \text{c}\bar{\text{c}}$)                        &  \text{0.04} & \text{0.01} \\ 
BR ( $\text{h}^0 \rightarrow \tau^+ \tau^-$ )                            &  \text{0.14 } & \text{0.01}  \\ 
\hline
\end{tabular}
}
\end{center}
\caption{Simulated measurements at LHC and a 0.5 and 1~TeV LC. For the cross sections,
the corresponding center-of-mass energy and the electron and positron polarization are given.}
\label{tab:measurements}
\end{table}
The program Fittino uses an iterative procedure to determine the 
parameters. It is displayed in Fig.~\ref{fig:iterative}. 
In a first step, the SUSY parameters are estimated using tree-level relations as
follows:
\begin{enumerate}
\item $\mu,m_A,\tan\beta,M_1,M_2,M_3$ are determined from gaugino and
  Higgs sector observables using formulae from \cite{sec4_Desch:2003vw}.
  In order to extract these parameters, information from chargino
  cross-sections is needed, which enters in form of the chargino
  mixing angles $\cos2\phi_L$ and $\cos2\phi_R$.  These
  pseudo-observables are only used for the determination of the start
  values, no use is made of them for the fit.
\item $A_{\text{t}},A_{\text{b}},M_Q,M_U,M_D$ are determined from the squark sector masses,
  using formulae from \cite{Porod:Thesis}. No mixing in the third generation is assumed to get
  the start values.
\item $A_{\tau},M_L,M_E$ are determined from the slepton sector masses,
  using formulae from \cite{Porod:Thesis}. No mixing in the third generation is assumed to get
  the start values.
\end{enumerate}

Instead of the trilinear couplings $A_{\text{t}}$, $A_{\text{b}}$ and
$A_{\tau}$ the following sfermion mixing parameters
\begin{eqnarray}
    X_{\text{t}} & = & A_{\text{t}} - \mu / \tan \beta\nonumber\\
    X_{\text{b}} & = & A_{\text{b}} - \mu \cdot \tan \beta\nonumber\\
    X_{\tau} & = & A_{\tau} - \mu \cdot \tan \beta\nonumber
\end{eqnarray}
are used in the fit. This is done in order to reduce the
correlations between the parameters $\mu$, $\tan \beta$ and
$A_{\text{t}}$, $A_{\text{b}}$ and $A_{\tau}$. The tree-level
initialization is done such that $A_{\text{t,b,}\tau} = 0$.

Since the mixing in the third generation has not been taken correctly
into account on tree-level, the parameters
$X_{\text{t}},X_{\text{b}},X_{\tau}$ tend to be badly initialized. A
global fit with these starting values would most likely not converge.
Therefore next the estimates from the slepton sector are improved by
fitting only the slepton parameters $X_{\tau},M_L,M_E$ to the
observables from the slepton sector, i.~e.  slepton masses and
cross-sections. Observables not directly related to the slepton sector
can degrade the fit result, since parameters of other sectors are
likely to be still wrong. In such a case a parameter of the slepton
sector will be pulled into a wrong direction, in order to compensate
for the wrong parameters of other sectors. All parameters not from the
slepton sector are fixed to their tree-level estimates obtained in the
previous step. In this fit with reduced number of dimensions MINUIT
can handle the correlations among the parameters better than in a
global fit with all parameters free.
\begin{table}[t]
\begin{center}
\begin{tabular}{r l D{.}{.}{-1} D{.}{.}{-1}}
\hline
No. & Parameter & \multicolumn{1}{c}{Start Value (GeV)} & \multicolumn{1}{c}{Start Uncertainty (GeV)} \\
\hline
1  & $\tan \beta$             & 10.0 & 12.8 \\    
2  & $\mu$                    & 354.4 & 2.6 \\         
3  & $m_{\tilde{\text{e}}_R}$ & 150.2 & 15.0 \\
4  & $m_{\tilde{\tau}_R}$     & 141.0 & 14.1 \\     
5  & $m_{\tilde{\text{e}}_L}$ & 202.7 & 20.3 \\
6  & $m_{\tilde{\tau}_L}$     & 206.6 & 20.7 \\     
7  & $X_{\text{t}}$           & -35  & 100 \\        
8  & $X_{\tau}$               & -3558  & 100 \\        
9  & $m_{\tilde{\text{d}}_R}$ & 567 & 57 \\    
10  & $m_{\tilde{\text{b}}_R}$ & 566 & 57\\  
11 & $m_{\tilde{\text{u}}_R}$ & 567 & 57 \\     
12 & $m_{\tilde{\text{t}}_R}$ & 381 & 38 \\    
13 & $m_{\tilde{\text{u}}_L}$ & 581 & 58 \\     
14 & $m_{\tilde{\text{t}}_L}$ & 575 & 58 \\    
15 & $M_1$                    & 99.08 & 0.58\\        
16 & $M_2$                    & 195.08 & 0.73\\        
17 & $M_3$                    & 630.5 & 6.4 \\        
18 & $m_{\text{A}^0}$         & 399.8 & 1.3     \\    
19 & $m_{\text{t}}$           & 174.3   & 0.3     \\
\hline
\end{tabular}
\end{center}
\caption{Estimates of the MSSM parameters, obtained from
tree-level relations.}\label{tab:tree_results}
\end{table}

Then the third generation squark parameters are improved by only
fitting $X_{\text{t}}$, $X_{\text{b}}$, $M_Q$, $M_U$, $M_D$ to the
observables of the squark sector, masses and cross-sections. All other
parameters are fixed to their previous values.

After this step still the correlations among $\tan\beta$ and the third
generation slepton and squark parameters are not optimally modelled.
Therefore another intermediate step ist introduced, where
$\tan\beta,X_{\text{t}},X_{\text{b}},X_{\tau}$ and
$m_{\tilde{\text{t}}_{L,R}}$ are fitted to all observables and all
other parameters are fixed to their present values.

After this, all MSSM parameters are released and a global fit is done,
using the method MINIMIZE in MINUIT. During the MSSM parameter fit
Standard Model parameters have been kept fixed. If requested by the
user, an additional fit step can be done which also releases SM
parameters.  Provided the fit has converged, a subsequent MINOS error
analysis is performed, yielding asymmetrical uncertainties, the full
correlation matrix and 2D fit contours.


\subsubsection*{The SPS1a Fit}

\begin{table}[t]
\begin{center}
\begin{tabular}{r l D{.}{.}{-1} D{.}{.}{-1} D{.}{.}{-1}}
\hline
No. & Parameter & \multicolumn{1}{c}{SPS1a Value (GeV)}
    & \multicolumn{1}{c}{Fit Value (GeV)} & \multicolumn{1}{c}{Uncertainty (GeV)} \\
\hline 
1 & $A_{\tau}$     &  -250.77 & -250.82 & 65.81    \\
2 & $A_{\text{b}}$ &  -855.06 & -854.66 & 1269.16  \\
3 & $A_{\text{t}}$ &  -506.39 & -506.35 & 2.02     \\
\hline
\end{tabular}
\end{center}
\caption{Fit result for a fit where only the trilinear couplings
$A_{\mathrm{t}},A_{\mathrm{b}},A_{\tau}$ are fitted, with all other parameters 
fixed to their SPS1a values. No sensitivity is obtained for $A_{\mathrm{b}}$.}\label{tab:atau}
\end{table}

The approach described above has been tested for the scenario SPS1a
\cite{Allanach:2002nj}. For this scenario, a set of hypothetical measurements
at LHC, a \mbox{500 GeV} and a \mbox{1 TeV} Linear Collider has been
collected in Table~\ref{tab:measurements}. In this example a
theoretical error on the prediction of the lightest Higgs boson mass
of 500 MeV is included. All observables from
Table~\ref{tab:measurements} were selected as input to fit the 24
parameters of the unconstrained MSSM and the top quark mass.  In order
to reduce the number of fit parameters and to simplify the fit
procedure, unification of the first two generations has been assumed.
Table \ref{tab:mssm1} shows that this is a reasonable assumption which
does not imply any loss of accurarcy.

The parameter $X_{\text{b}}$ has been fixed to -4000 GeV. The reason
for this is summarized in Table \ref{tab:atau}.  It contains the
fitted trilinear couplings $A_{\tau}$, $A_{\text{t}}$ and
$A_{\text{b}}$ from a fit where all other parameters from Table
\ref{tab:fittedparameters} are fixed to their SPS1a values.  Evidently
the selected observables have no sensitivity to $A_{\text{b}}$. Fixing
$X_{\text{b}}$ is a passable way to circumvent this insensitivity.

Taking these simplifications into account, 18 MSSM parameters and the
top quark mass remain. All fitted parameters are listed in Table
\ref{tab:fittedparameters} including the predicted SPS1a values, the
fitted values and the uncertainties from the fit. All fitted
parameters except the third generation squark sector agree well with
the generated values. The discrepancy in the third generation squark
sector stems from the fact that $X_{\text{b}}$ is fixed to an
approximate value which does not coincide with its SPS1a value.  The
inclusion of a theoretical error of 500 MeV on the lightest Higgs mass
increases the uncertainty on $\tan \beta$ by a factor of 2.

Albeit the fitted third generation squark parameters do not match
their SPS1a values, the obtained $\chi^2$ of the fit is only 0.043.
This shows the importance to fit all parameters simultaneously, since
parameters fixed to wrong values tend to distort the whole spectrum of
the fitted parameters to compensate for the wrongly fixed parameters.
As it is obvious from Table \ref{tab:fittedparameters}, the systematic
distortion can be much larger than the $1\sigma$ uncertainties of the
parameters.


\begin{table}[htb!]
\begin{center}
\begin{tabular}{r l D{.}{.}{-1} D{.}{.}{-1} D{.}{.}{-1}}
\hline
No. & Parameter & \multicolumn{1}{c}{Generated Value}
    & \multicolumn{1}{c}{Fitted Value (GeV)}
    & \multicolumn{1}{c}{Uncertainty (GeV)} \\ \hline
   1  & $\tan \beta$             &  10.0    &  9.92    & 1.43      \\
   2  & $\mu$                    &  358.6 &  358.6  & 4.4      \\
   3  & $X_{\tau}$               & -3836.8   &  -3769  & 648     \\
   4  & $m_{\tilde{\text{e}}_R} = m_{\tilde{\mu}_R}$ 
                                 &  135.76  &  135.86  &  0.13    \\
   5  & $m_{\tilde{\tau}_R}$     &  134.6   &  133.1    & 3.0      \\
   6  & $m_{\tilde{\text{e}}_L} = m_{\tilde{\mu}_L}$ 
                                 &  195.2   &   195.19  &  0.20     \\
   7  & $m_{\tilde{\tau}_L}$     &  194.4   &  194.5   & 1.9     \\
   8  & $X_{\text{t}}$           & -506.4 & -508.7  &   35.8    \\
   9  & $m_{\tilde{\text{d}}_R}= m_{\tilde{\text{s}}_R}$ 
                                 &  528.1 &  528.2   &  15.9     \\
  10  & $m_{\tilde{\text{b}}_R}$ &  524.8 &  495.6   &   6.6     \\
  11  & $m_{\tilde{\text{u}}_R} = m_{\tilde{\text{c}}_R}$ 
                                 &  530.3 &  530.3   &  10.2      \\
  12  & $m_{\tilde{\text{t}}_R}$ &  424.3 &  412.8   & 8.0      \\
  13  & $m_{\tilde{\text{u}}_L} = m_{\tilde{\text{c}}_L}$ 
                                 &  548.7 &  548.7   & 5.2      \\
  14  & $m_{\tilde{\text{t}}_L}$ &  500.0 &  529.8   &  7.3      \\
  15  & $M_1$                    &  101.809 & 101.84    & 0.22     \\
  16  & $M_2$                    &  191.76  &  191.66   & 0.71      \\
  17  & $M_3$                    &  588.8   &  588.4   & 7.8       \\
  18  & $m_{\text{A}^0}$         &  399.77  &  399.78   & 0.73      \\
  19  & $m_{\text{t}}$           &  174.3   &  174.30   & 0.30      \\ \hline
      & $X_{\text{b}}$           & -4441.1  & -4000     & \text{fixed}  \\
      & $m_{\text{b}}$           & 4.2      & 4.2       & \text{fixed} \\
      & $m_{\text{c}}$           & 1.2      & 1.2       & \text{fixed} \\ \hline
\end{tabular}
\end{center}
\caption{Fit result for a global fit of SM and MSSM parameters
to the observables listed in Table~\ref{tab:measurements}. The
obtained total $\chi^2$ of the fit amounts to 0.043.}
\label{tab:fittedparameters}
\end{table}

\begin{table}[htb!]
\begin{center}
\begin{tabular}{l D{.}{.}{-1} l D{.}{.}{-1}} \hline
    Parameter     & \multicolumn{1}{c}{$\Delta \chi^2_{\text{total}}$} & Contributing Observable &
                    \multicolumn{1}{c}{Rel.~Contrib. (\%)} \\ \hline
    $\tan \beta$  & 75.4 & $ m_{\chi_2^0}$  & 71.4\\
                  &       & $m_{\chi_1^0}$     & 22.5\\
                  &       & $m_{\chi_1^{\pm}}$ & 1.6 \\
    $\mu$         & 21.7  & $m_{\chi_2^0}$     & 81.1\\
                  &       & $m_{\chi_1^0}$     &  9.0\\
                  &       & $m_{\chi_2^{\pm}}$ &  7.6\\
    $M_1$         & 17.2  & $m_{\chi_1^0}$     & 99.78\\
                  &       & $\sigma$ ( $\text{e}^+\text{e}^- \rightarrow \tilde{\text{e}}_L
                            \tilde{\text{e}}_L$, $500$ GeV, $0.8$, $0.6$)
                                               &  0.13\\
                  &       & $\sigma$ ( $\text{e}^+\text{e}^- \rightarrow \tilde{\text{e}}_L
                            \tilde{\text{e}}_L$, $500$ GeV, $-0.8$, $-0.6$)
                                               & 0.03\\
    $M_2$         & 68.4  & $m_{\chi_2^0}$     & 96.9\\
                  &       & $m_{\chi_1^{\pm}}$ & 2.1 \\
                  &       & $\sigma$ ( $\text{e}^+\text{e}^- \rightarrow \tilde{\chi}_1^{\pm}
                            \tilde{\chi}_1^{\mp}$, $500$ GeV, $0.8$, $0.6$)
                                               &  0.6 \\
    $M_3$         & 1.1   & $m_{\tilde{\text{g}}}$     & 91.4 \\
                  &       & $m_{\tilde{\text{t}}_R}$ & 6.0 \\
                  &       & $m_{\tilde{\text{b}}_R}$ & 0.7 \\ \hline
\end{tabular}
\end{center}
\caption{The relative contribution of the three most important observables to the
determination of the parameters $\tan \beta$, $\mu$, $M_1$, $M_2$ and $M_3$ and
the total $\Delta \chi^2_{\text{total}}$ obtained if the parameter is changed
by $\pm 1 \sigma$.}
\label{tab:contributions}
\end{table}





Improving the determination of the trilinear couplings and a release
of the $X_{\text{b}} = -4000$~GeV constraint requires the
inclusion of additional, more sensitive observables. Good candidates
are probably stop and stau polarization measurements. Unfortunately
such predictions are not yet available in SPheno.

In addition the contribution of the various observables to the
determination of a given parameter has been studied by looking at
$\Delta \chi^2$ for the individual observables if the parameter is
changed by $\pm 1 \sigma$. As an example, Table \ref{tab:contributions} summarizes
the three most important contributions to $\tan \beta$, $\mu$, $M_1$,
$M_2$ and $M_3$.

It has also been tried to carry out fits using only a reduced set of
measurements as input such as they are available in "LHC only" and "LC
only" scenarios. The "LHC only" case shows a quite comprehensive
particle spectrum but only poor information on the slepton sector is
avaiable. In contrast to that the "LC only" scenario is characterized
by an incomplete MSSM particle spectrum (no information from squark
sector) but a very precise slepton and gaugino sector. Even when 
poorly determined parameters are fixed (as in Table~\ref{tab:mssm2}), no converging
fits are obtained. This fact shows the importance of combined LHC and
LC data analyses.

\subsubsection{Conclusions}

Fittino is a program to determine the MSSM parameters from a global
fit to measurements at the LHC and a future Linear Collider. No prior
knowledge of any of the parameters is needed.  To get reasonable start
values for the fit, tree-level formulae are used to relate the
observables to the SUSY parameters. First results obtained with this
powerful tool clearly reveal the benefit from combining the LHC and LC
results. The comprehensive SUSY particle spectrum accessible at LHC
and the precise measurements of the lightest SUSY particles at a
Linear Collider are crucial to get a converging fit. Attempts to fit
only individual sectors of the theory are unsuccessful, if no prior knowledge
of any parameter is assumed.  This fact
shows the fruitfulness of combining measurements of both machines.


\section{SUSY and Dark Matter}

\subsection{\label{sec:442}
Reach of LHC and LC in Dark Matter allowed regions of the mSUGRA model}

{\it H.\ Baer, A.\ Belyaev, T.\ Krupovnickas and X.\ Tata}

\vspace{1em}



%
Recently, the WMAP collaboration has analyzed the anisotropies in
the cosmic microwave background radiation. The analysis leads to
a determination that the universe is comprised of $\sim 5\%$ baryons, 
$\sim 25\%$ cold dark matter, and $\sim 70\%$ dark energy. 
In particular, the cold dark matter density is determined to be
$\Omega_{CDM} h^2=0.1126^{+0.0161}_{-0.0181}$ (at $2\sigma$ level), 
where $\Omega_{CDM} =\rho_{CDM} /\rho_c$ is the density of cold dark matter 
in the universe, scaled to the critical closure density $\rho_c$, and
$h=0.71^{+0.04}_{-0.03}$ is the scaled Hubble constant.

$R$-parity conserving supersymmetric models provide good candidates
for CDM particles in the universe. In this section, we assume the
lightest neutralino is the CDM particle. We  work within the mSUGRA model, 
although our qualitative conclusions apply rather more broadly than this
framework might
suggest. Sparticle masses and mixings are determined within the mSUGRA
model by specifying the parameter set
\begin{equation}
m_0,\ m_{1/2},\ A_0, \tan\beta\ {\rm and}\ sign(\mu ) .
\end{equation}
We use the Isajet 7.69 program\cite{isajet769} for calculating 
sparticle masses and mixings.

The relic density of neutralinos can be calculated by solving the Boltzman
equation for a Friedmann-Robertson-Walker universe. The IsaReD 
program determines the neutralino relic density via relativistic thermal
averaging of all relevant neutralino annihilation and co-annihilation processes
in the early universe\cite{isared}. We  adopt the upper bound
$\Omega_{CDM}h^2<0.129$ as a robust limit on neutralino dark matter; 
the lower limit from WMAP need not apply in the case of mixed dark matter 
scenarios. Several regions of mSUGRA parameter space have been found to
be consistent with WMAP constraints.
\begin{itemize}
\item The bulk region at low $m_0$ and low $m_{1/2}$, where
neutralino annihilation via $t$-channel slepton exchange is dominant.
\item The stau co-annihilation region at low $m_0$ where 
$m_{\tilde{\tau_1}}\simeq m_{\tilde\chi^0_1}$.
\item The hyperbolic branch/focus point region (HB/FP) at large $m_0$,
where $|\mu |$ becomes small and the $\tilde{\chi}^0_1$ becomes
partially higgsino-like.
\item The $A$-annihilation funnel at large $\tan\beta$, where 
$2m_{\tilde{\chi}^0_1}\simeq m_A,\ m_H$, and neutralino annihilation 
takes place through the broad $s$-channel heavy higgs $A$ and $H$ resonances.
\end{itemize}
In addition, there is a narrow region at low $m_{1/2}$ where
neutralino annihilation can occur through the light Higgs $h$ resonance, and
an additional stop-neutralino co-annihilation region exists for
particular $A_0$ values.

The reach of the CERN LHC for SUSY in the mSUGRA model has been 
calculated in Ref. \cite{lhcreach} assuming 100 fb$^{-1}$ of
integrated luminosity. Briefly, sparticle pair production events
were generated for many mSUGRA model parameter choices in the 
$m_0\ vs.\ m_{1/2}$ plane for various $\tan\beta$ values. 
A fast LHC detector simulation (CMSJET) was used, and cuts were imposed to
extract signal rates in a variety of multilepton plus multijet plus
missing transverse energy channels. Backgrounds were also calculated
from a variety of QCD and vector boson production processes. A large
set of selection cuts were used to give some optimization over 
broad regions of parameter space. It was required to have at least a
$5\sigma$ signal over background, with at least 10 signal events 
in the sample. 

The reach of the CERN LHC is shown in Fig. \ref{sec442_fig1}
for the case of $\tan\beta =10$, $\mu >0$, $A_0=0$ and $m_t=175$ GeV.
The dark shaded (red) regions are disallowed by lack of 
radiative electroweak symmetry
breaking (REWSB) (right hand side) or presence of a stau LSP (left hand side).
The light gray (yellow) region is excluded by LEP2 chargino searches 
($m_{\tilde{\chi}^+}>103.5$ GeV), while the region below the yellow 
contour gives $m_h<114.4$ GeV, in contradiction of LEP2 SM Higgs searches
(here, the SUSY $h$ Higgs boson is essentially SM-like). The 
medium gray (green) regions
have $\Omega_{CDM}h^2<0.129$, and are {\it allowed} by WMAP. The broad
HB/FP region is seen on the right-hand side, while the stau co-annihilation 
region is shown on the left-hand side. At the edge of the LEP2 excluded 
region is the light Higgs annihilation corridor. The reach of the 
Fermilab Tevatron via the trilepton channel is also shown\cite{tev}, 
assuming a $5\sigma$ signal over background for 10 fb$^{-1}$. 
The reach of the CERN LHC
for 100 fb$^{-1}$ of integrated luminosity is shown by the contour labeled
``LHC''. It extends from $m_{1/2}\sim 1400$ GeV (corresponding to
a value of $m_{\tilde g}\sim 3$ TeV) on the left-hand side, to
$m_{1/2}\sim 700$ GeV (corresponding to $m_{\tilde g}\sim 1.8$ TeV) 
on the right-hand side. In particular, for this value of $\tan\beta$, the
LHC reach covers the entire stau co-annihilation region, plus the
low $m_{1/2}$ portion of the HB/FP region. The outer limit of the reach
contour is mainly determined by events in the $E_T^{miss}+$ jets channel,
which arises from gluino and squark pair production, followed by hadronic 
cascade decays.

\begin{figure} 
     \begin{center}
\epsfxsize=9cm\epsffile{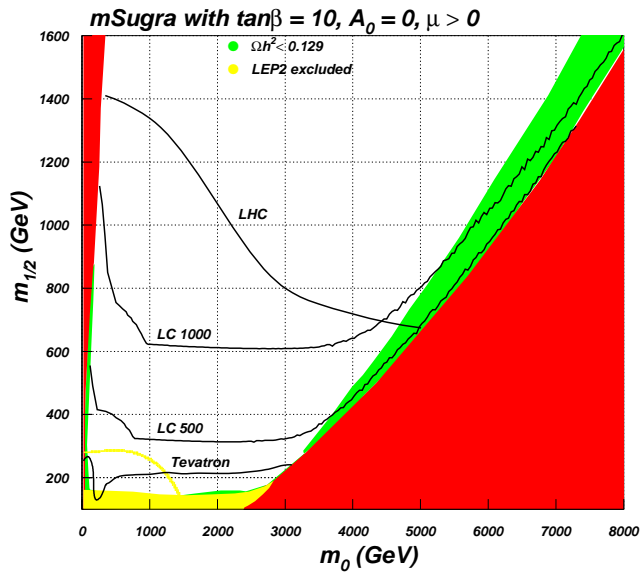} 
     \caption{ \label{sec442_fig1}Parameter space of mSUGRA model for $\tan\beta =10$,
$A_0=0$ and $\mu >0$, showing the reach of the Fermilab Tevatron, 
the CERN LHC 
and a 0.5 and 1 TeV linear $e^+e^-$ collider for supersymmetry discovery.} 
     \end{center}
\end{figure}

\begin{figure} 
     \begin{center}
\epsfxsize=9cm\epsffile{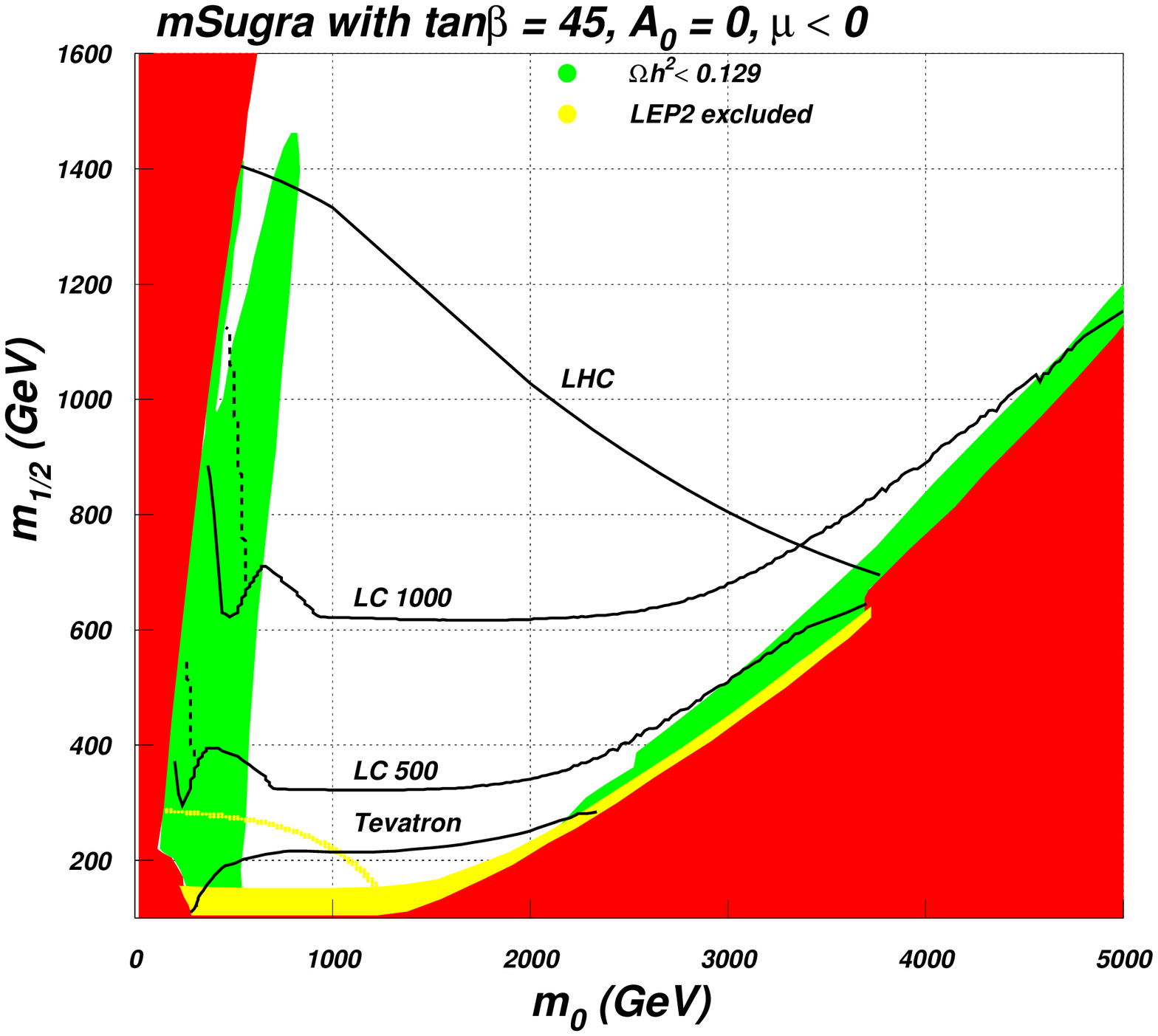}
     \caption{ \label{sec442_fig2}Parameter space of mSUGRA model for $\tan\beta =45$,
$A_0=0$ and $\mu <0$, showing the reach of the Fermilab Tevatron, 
the CERN LHC 
and a 0.5 and 1 TeV linear $e^+e^-$ collider for supersymmetry discovery.} 
    
     \end{center}
\end{figure}

     \begin{figure}[ht] 
     \begin{center}
     \vspace*{.2cm}
     \epsffile{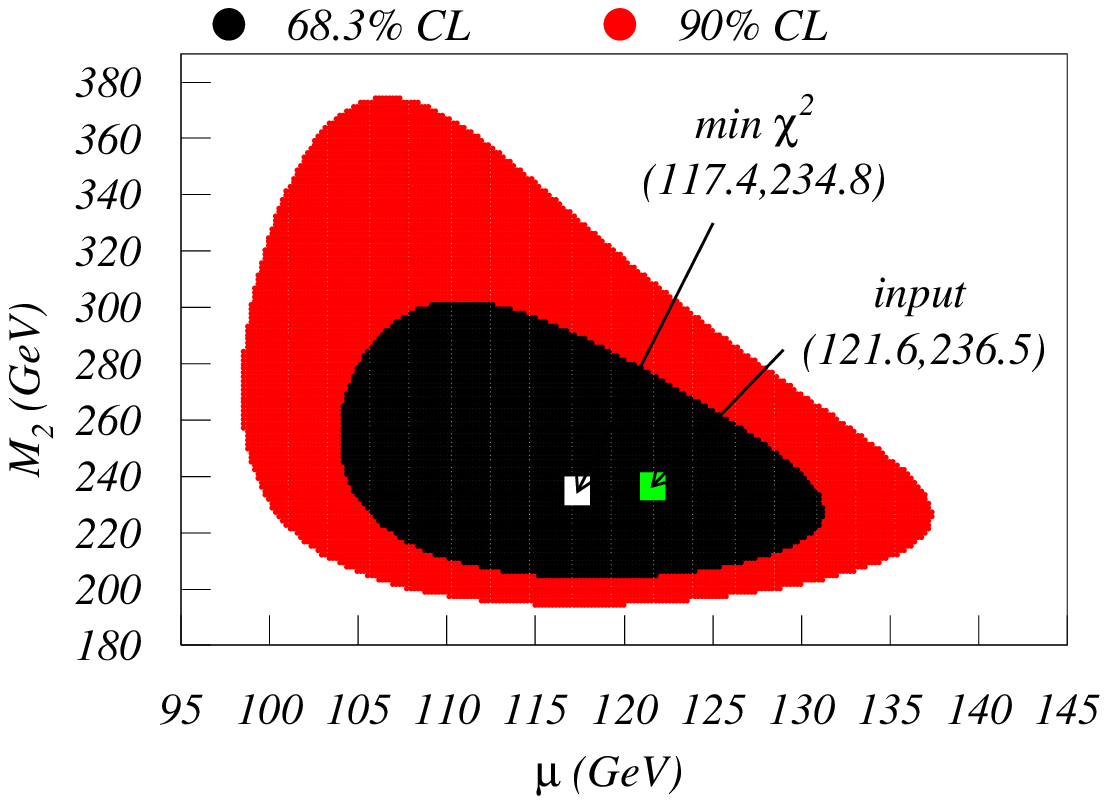}
     \end{center}
     \caption{ \label{sec442_fig3}Determination of SUSY parameters from 
examining chargino pair production at a $\sqrt{s}=0.5$ TeV LC,
for the HB/FP mSUGRA point listed in the text.}
    
     \end{figure}

We also show in the plot the reach of a $\sqrt{s}=500$ and 1000 GeV
LC, assuming 100 fb$^{-1}$ of integrated luminosity\cite{nlc}. 
Events were generated
using Isajet 7.69, and compared against various SM backgrounds.
The left-most portion of the reach contour arises where selectron and smuon
pair production are visible, while the main portion (flat with $m_{1/2}$)
arises due to chargino pair searches. An additional reach is gained between
these two regions by searching for $e^+e^-\rightarrow \tilde{\chi}_2^0
\tilde{\chi}_1^0$ production, followed by 
$\tilde{\chi}_2^0\rightarrow \tilde{\chi}_1^0 b\bar{b}$ decay. 
In addition, in Ref.~\cite{staubkt}, additional reach can be gained by 
searching for stau pair events, 
although two photon backgrounds must be accounted for, 
due to the low energy release in the stau co-annihilation region.

While a 500 GeV LC can cover only a portion of the stau co-annihilation
region, a 1 TeV LC can cover the entire region, at least for this value
of $\tan\beta$. As one moves into the HB/FP region, the LC retains
a significant reach for SUSY, which in fact extends {\it beyond} that
of the CERN LHC! It is significant that this additional reach occurs in a 
DM allowed region of parameter space. In the HB/FP region, the
superpotential $\mu$ parameter becomes small, and the lightest chargino 
and neutralino become increasingly light, with increased higgsino content. 
In fact, the 
decreasing mass gap between $\tilde{\chi}_1^+$ and $\tilde{\chi}_1^0$
makes chargino pair searches difficult at a LC using conventional cuts
because there is so little visible energy release from the chargino
decays. In Ref. \cite{nlc,staubkt}, 
we advocated cuts that pick out low energy release signal events
from SM background, and allow a LC reach for chargino pairs 
essentially up to the kinematic limit for their production.
In this case, it is important to fully account for $\gamma\gamma\to f\bar{f}$
backgrounds, where $f$ is a SM fermion.

In Fig. \ref{sec442_fig2}, we show a similar reach plot, but this time for
$\tan\beta =45$ and $\mu <0$. In this case, the broad DM $A$-annihilation
funnel has appeared on the left-hand side of parameter space. It can be 
seen that the LHC can cover most of the $A$ annihilation funnel, although
a somewhat higher integrated luminosity might be needed to cover it 
completely. Also, the stau co-annihilation region has increased to cover
higher $m_{1/2}$ values, and now extends beyond the LC1000 reach.
Still higher values of $\tan\beta$ push the allowed stau co-annihilation
region somewhat beyond the reach of even the CERN LHC. Meanwhile, the HB/FP
region is qualitatively insensitive to $\tan\beta$ values
ranging from 10-50, and a 1 TeV LC can still probe much of this region, 
well beyond what can be accessed at the LHC.

The study of Ref. \cite{nlc} also examined a case study in the HB/FP region
with parameters $m_0=2500$ GeV, $m_{1/2}=300$ GeV, $A_0=0$, $\tan\beta = 30$,
$\mu >0$ and $m_t=175$ GeV, {\it i.e.} in the HB/FP region. 
In this case, chargino pair events were selected from the 
$1\ell+$jets $+E_T^{miss}$ channel, and the dijet mass distribution was used
to extract the value of $m_{\tilde{\chi}_1^+}$ and $m_{\tilde{\chi}_1^0}$
at the 10\% level. The mass resolution is somewhat worse than 
previous case studies in the literature because the charginos undergo
three-body rather than two-body decays, and no sharp edges in energy
distributions are possible. Nonetheless, the measured value of
chargino and neutralino mass, along with a measure of the total
chargino pair
cross section, was enough to determine the SUSY parameters $M_2$ and $\mu$
to 10-20\% precision. The results, shown in Fig. \ref{sec442_fig3}, 
demonstrate that $\mu <M_2$, which points to a $\tilde{\chi}_1^+$ and
$\tilde{\chi}_1^0$ which are higgsino/gaugino mixtures, as is
characteristic of the HB/FP region. 

In conclusion,
we would like to stress that the CERN LHC and an $e^+e^-$
LC are highly complementary to each other in
exploring the dark matter allowed parameter space of the mSUGRA model.
LHC covers the  
stau co-annihilation region (completely for $\tan\beta <40$)
as well as the $H,\ A$ funnel region
(much of which is  typically beyond 
the maximum reach of a LC).
However only the lower part  of the HB/FP region can be covered by the LHC.
On the other hand, as we have demonstrated, LCs 
can probe much of the {\it upper} part of the HB/FP region
with the new proposed cuts.
Therefore, the combination of the LHC and a TeV scale 
LC can cover almost the entire  parameter space of the mSUGRA scenario.

\subsection{\label{sec:441}
Impact of the LHC and LC on the accuracy of the predicted Dark Matter
relic density
}

{\it B.~Allanach, G.~B\'elanger, F.~Boudjema and A.~Pukhov}

\vspace{1em}
\renewcommand{\ra}{\rightarrow}
\renewcommand{\mhf}{M_{1/2}}
\newcommand{\neuto}{\tilde{\chi}_1^0}
\renewcommand{\neutt}{\tilde{\chi}_2^0}
\renewcommand{\stau}{\tilde{\tau}_1}
\newcommand{\mneuto}{m_{\tilde{\chi}_1^0}}
\renewcommand{\ser}{\tilde{e}_R}
\renewcommand{\smur}{\tilde{\mu}_R}
\newcommand{\tgb}{\tan \beta}
\newcommand{\mser}{m_{\ser}}
\newcommand{\msmur}{m_{\smur}}
\renewcommand{\mneutth}{m_{\tilde{\chi}_3^0}}
\newcommand{\beqn}{\begin{eqnarray}}
\renewcommand{\eeqn}{\end{eqnarray}}

\noindent{\small
We investigate how well the relic density of dark matter can be
predicted in typical mSUGRA scenarios without the assumption of
mSUGRA when analysing data.
 We determine the parameters to which the relic
density is most sensitive and quantify the collider accuracy
needed to match the accuracy of WMAP and PLANCK.
The inclusion of experimental information from a future linear
collider facility will be essential for all viable regions of
parameter space.
}

\vspace{1em}

One of the attractive features of the minimal supersymmetric
standard model (MSSM) is that it provides a natural candidate for
cold dark matter, the neutralino, $\neuto$. With cosmology
entering the era of precision measurements and the next colliders
aiming at discovering and constraining supersymmetry some crucial
cross breeding is emerging. Already, assuming the standard
cosmology, the measurement of the relic density of dark matter has
been used to put strong constraints on the supersymmetric
model~\cite{Baer:2003}.
For example, WMAP~\cite{Spergel:2003}, which at
$2\sigma$ constrains the relic density in the range $.094<\Omega
h^2<.129$, effectively reduces the dimensionality of the MSSM
parameter space by one. The accuracy of this constraint will
increase with future data from the PLANCK satellite, which expects
precision on $\Omega h^2$ at the 2$\%$
level~\cite{Balbi:2003en}.

 In this paper, we look at the
problem from the inverse perspective and examine what is required from
collider data in order to get a precise prediction for $\Omega h^2$, which
could then be used to test the cosmology. For this we
 assume that supersymmetric particles will be produced and measured
at the LHC and the future linear collider (LC) and
 that enough information~\cite{Allanach:2001} will be present
in order to discriminate between various models of supersymmetry
breaking. When we have identified a successful model of SUSY
breaking, an accurate prediction of $\Omega h^2$ will allow us to
test the cosmological assumptions that go into its prediction.

In order to predict $\Omega h^2$ in the MSSM one generally needs
to know all the underlying parameters of the model, since that
determines the available annihilation channels for neutralinos in
the early universe. In the context of mSUGRA however, the
scenarios allowed by WMAP are rather fine-tuned and one can
concentrate only on a handful of relevant observables. The point
is that in mSUGRA the neutralino LSP  happens to be, for
practically all cases, an almost pure bino. This bino annihilates
mainly into leptons through the right-handed sleptons but this
mechanism is not efficient enough to satisfy the  newest WMAP
data.  Then only three acceptable  scenarios remain:
coannihilation, rapid annihilation through Higgs exchange or a
Higgsino LSP. Our study will cover these three  rather constrained
regions in the mSUGRA parameter space.

We will investigate the accuracy required on a few of the most relevant collider
 observables in order to control the uncertainty on the predicted
$\Omega h^2$.  These observables could be either physical masses
or Lagrangian parameters such as the chargino mixing parameter
$\mu$ or $\tan\beta$. We will refer throughout to two benchmarks
on accuracy: those that produce a 10$\%$ change in $\Omega h^2$
(``WMAP accuracy'') and those which change $\Omega h^2$ by 2$\%$
(``PLANCK accuracy''). We only quote precisions on input
parameters obtained by varying only one parameter at a time.

Although  we mention some results obtained with the mSUGRA scenario,
we concentrate mainly on what we call the  perturbed mSUGRA scenario (PmSUGRA).
In this scenario we  pick a parameter point in mSUGRA derived from the
high-scale parameters   $m_0,\mhf,A_0$ then  we examine the impact on $\Omega h^2$
of a parameter change while {\it assuming the more general MSSM only}.
In practice we use an iterative procedure to determine the
 fractional change $a=|\Delta p/p|$ in an
input parameter $p$  that will result in a fractional change
$r=\Delta \Omega / \Omega=10\%$ for WMAP accuracy.  An estimate of
the PLANCK accuracy can be obtained by dividing that of WMAP by a
factor of 5, for the interesting range of accuracies not exceeding
$50\%$, say. Here we do not address the feasibility of any
measurements, but rather only identify which measurements are
needed and with what precision. Preliminary investigations into
the effects of uncertainties in the predictions of the sparticle
mass spectrum on $\Omega h^2$ were presented in
Refs.~\cite{Allanach:2004jh} and more details can be found in
Ref.~\cite{Allanach_long}.
 The relic density is computed with {\tt
micrOMEGAs1.3}~\cite{Belanger:2004yn} with the supersymmetric
spectrum provided by {\tt SOFTSUSY1.8.7}~\cite{Allanach:2001kg}
and the interface between the two by the {\em SUSY Les Houches
Accord}~\cite{Skands:2003cj}.

\subsubsection{Coannihilation}

In mSUGRA at small $M_0$ there exists a region with almost
degenerate $\stau-\neuto$. In this region the LSP is almost purely
bino. The contribution of coannihilation channels  is essential in
bringing the relic density in the desired range.
 In computing the effective
annihilation cross section, coannihilation processes are
suppressed by a Boltzmann factor $\propto exp^{-\Delta M/T_f}$
where $\Delta M$ is the mass difference between the NLSP and the
LSP and $T_f$ the decoupling temperature. One expects $\Omega h^2$
to be very sensitive to this mass difference. Following Ref.
\cite{Battaglia:2003ab}, we take a slope ''S1" in parameter space with
$\mu>0,\tan\beta=10$, $A_0=0$ and \beqn
M_0 =   5.84615 + 0.1764 * M_{1/2} + 1.9780 \times 10^{-5} * M_{1/2}^2
\eeqn Here
masses are given in GeV units. Along this slope
the relic density is in rough agreement with the WMAP range ($\mhf=350-950$~GeV). The
coannihilation processes involving $\neuto-\stau$    dominate at
low masses  when $\Delta M\approx 10$ GeV. As one increases the
LSP mass  $\Delta M$ decreases to 300MeV, $\stau\stau$ channels
as well as coannihilation with
 selectrons and smuons become important. In fact the latter contribute up to 40\% of the
effective annihilation cross section toward the upper end of the
slope.

The physical parameters that need to be measured precisely then
are those entering the dominant  channels. The most relevant
physical parameters   include  the masses of all the light
sleptons, in particular the $\stau$,  the mass of the neutralino
LSP and the mixing in the $\stau$ sector. The latter  enters the
$\stau$ coupling to gauge bosons or to neutralinos. In addition
one needs also the couplings of the LSP. These involve the mixing
matrix of neutralinos. To estimate the sensitivity of $\Omega h^2$
on a given parameter we keep all others fixed. For example, for
 the mass difference we have, for each point on the slope,  varied the mass of the $\stau$
 while keeping all other parameters fixed.
  Fig. \ref{fig:coan} shows that the mass difference must be measured within
slightly less than 1 GeV. This dependence can directly be related to the fact that the
 coannihilation channels crucially depend on the Boltzmann factor.
The mixing angle  $\cos 2\theta_\tau$ must be measured within 0.55
for light staus to about 0.1  for heavy staus when its value is
almost 1.  Finally we also compute the accuracy on the overall
scale, defined as the accuracy required on $\mneuto$ once we keep
$\Delta M$ constant. To determine the accuracy we have varied the
parameter $M_1$ leaving all other parameters of the neutralino
sector constant and  have changed all slepton masses by the same
amount that $\mneuto$ was shifted.  We find that the required
accuracy ranges from 15\% for low masses to 5\% for
$\mneuto\approx 400GeV$. Although the overall scale is an
important  parameter, the precision required is not nearly as
important as for the mass difference. Finally, the selectron and
smuon masses need to be measured to 1.5\% to achieve WMAP
precision, see Fig.\ref{fig:coan}b. This is obtained by varying
the $\mser=\msmur$  by the same amount while keeping all other
parameters fixed.
 We have also checked that the relic density is
 not very sensitive to $\tan\beta$ and $\mu$  once we assume  constant $\Delta
 M$.

\begin{figure}{
 \unitlength=1.1in
 \vspace{-.5cm}
\begin{picture}(2.1,2.6)
\put(0.5,0){\epsfig{file=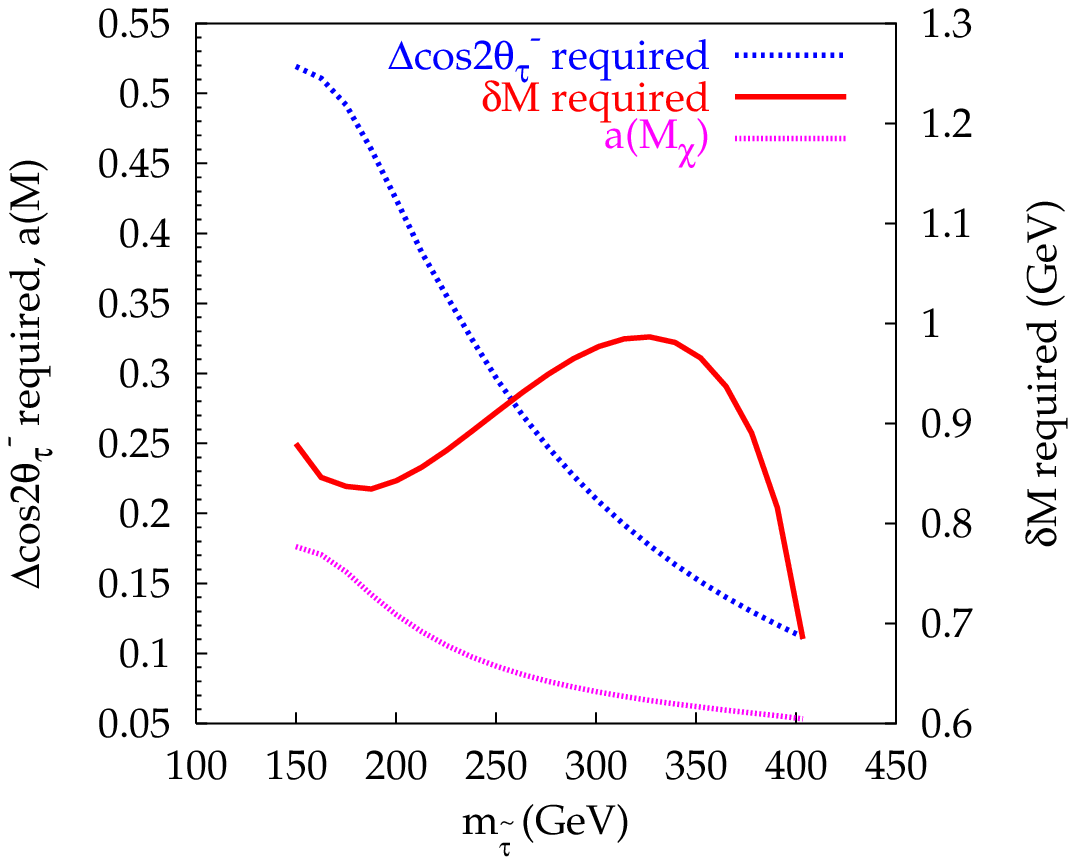,
width=8.5cm}}
\end{picture}
\begin{picture}(4,2.6)
\put(1.3,0){\epsfig{file=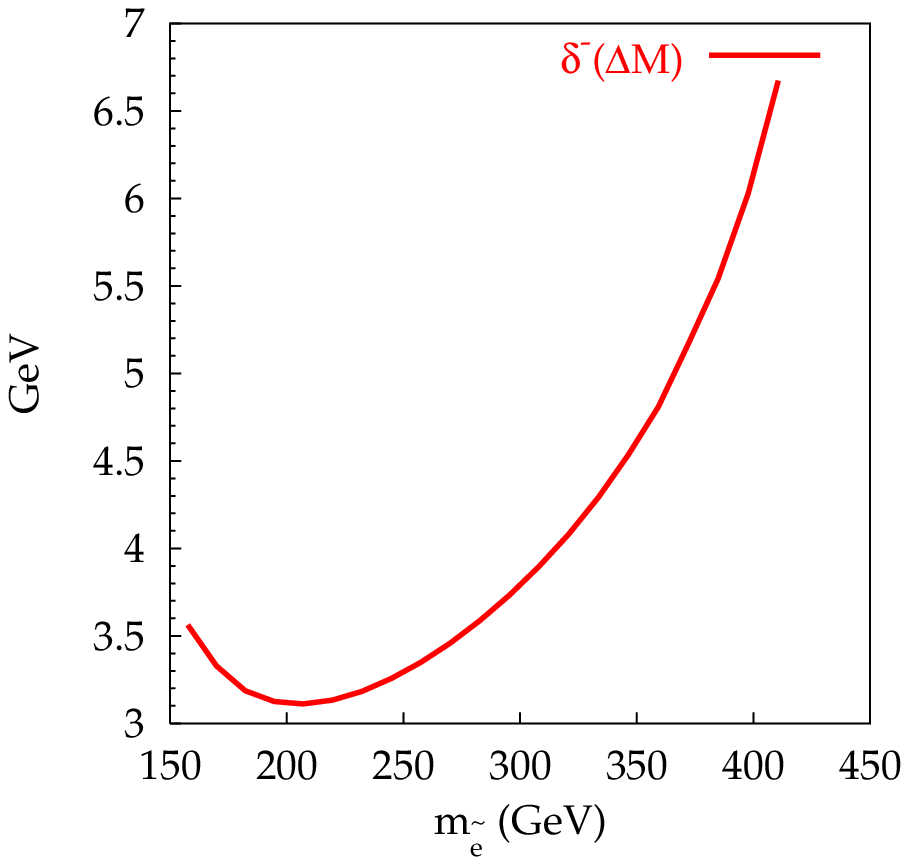,
width=8.5cm}}
\end{picture}
\caption{ \it \small (a) Required accuracy on  the
$\stau$-$\neuto$ mass difference, $\cos 2\theta_{\tau}$ and
$\mneuto$ in order to achieve WMAP precision. The latter  is
performed by keeping $\Delta M$ constant.
 (b) Required accuracy on $\Delta M=\mser-\mneuto$ assuming $\mser=\msmur$. The
abscissa range corresponds to $\mhf=350-950$ GeV.}
\label{fig:coan}}
\end{figure}
\normalsize

These results indicate that once the LHC has established
compatibility with the coannihilation  scenario one will need  a
linear collider to measure precisely both the mass of the LSP as
well as  $\Delta M$. Recent simulations of the coannihilation
region indicate that the relevant parameters can be measured with
the required accuracy to meet WMAP (and maybe even PLANCK)
precision if one is in the region with large enough $\Delta M$
\cite{Bambade}. Note that this is the region relevant for LC500 .
For increasingly smaller $\Delta M$ associated with higher masses
the situation is more problematic. On the other hand if one
assumes mSUGRA, that is that  combined fits from the LHC data
agree with the mSUGRA predictions, then $m_0, M_{1/2}$ and $A_0$
can be constrained from observables involving other particles than
just those relevant for the co-annihilation region. If the
theoretical predictions are all under control, this means that  we
could predict the relic density without having accurate
information on the stau mass or on $\Delta M$. However the
precision required on $m_0$ and $\mhf$ to match WMAP accuracy
ranges from around $1-3\%$. It remains to be seen whether such an
accuracy can be reached considering that from previous analyses
mostly done in the bulk (low $M_0-\mhf$) region, the precision
ranged from $1\%$ to $10\%$ depending on the point in parameter
space
\cite{sec4_ATLAS-TDR}.

\subsubsection{Higgs funnel}
Rapid and efficient annihilation can occur through the Higgs
resonance, this is the funnel region.  In fact because of the
Majorana nature of the neutralino the resonant enhancement
proceeds through a pseudoscalar Higgs boson. In this situation
$A\ra b \bar b$ is by far dominant at high $\tgb$, with some
contribution from $\tau \tau$. In our discussion of the Higgs
funnel we will take  $\tan \beta=50, A_0=0, \mu>0$. We
parameterise the funnel region through the slope S2, defined as a
cubic:
\begin{equation}
m_0 =  814.88 -2.20022\mhf + 3.30904 \times 10^{-3} \mhf^2
-1.05066 \times 10^{-6} \mhf^3
\end{equation}
where masses are in units of GeV.
The important physical
parameters are $M_A, \Gamma_A, \mneuto$ as well as the parameters
that enter the  vertices $Ab\bar b$  as well as $\neuto \neuto A$.
The latter   is controlled by $\mu$ which for all purposes can be
equated with $\mneutth$.

Fig. ~\ref{fig:funExp} shows that both the neutralino mass and the
pseudoscalar mass must be measured very accurately, from $2-0.2\%$
depending upon the position on the slope. More precision is needed
for the heavy spectrum. To compute the accuracy on the parameter
$\mneuto$, we in fact vary the parameter $M_1$. These accuracies
on the masses can also be expressed as an accuracy on the resonant
parameter, $2\mneuto -M_A$ which is around $5\%$.  To compute  the
accuracy on the parameter $\mu$, we simply change the value of
$\mu$ at the Lagrangian level given at the weak scale. This will
directly change  the LSP coupling to the pseudoscalar Higgs,  the
resulting  change in  the value of the LSP mass  will be very
small.  We see in Fig.~\ref{fig:funExp} that one needs an accuracy
of about $5\%$ on $\mu$ for a $10\%$ WMAP precision which
corresponds to the fact that $\Omega h^2 \propto \mu^2$.  The
accuracy on the total width is also around 10\%.   It is important
to note that once we fix the following parameters from experiment,
$M_A,
\Gamma_A,
\mneuto,\mu (\mneutth)$, the $\tgb$ dependence is very mild.

\begin{figure}{
 \unitlength=1.1in
\begin{picture}(0.2,2.6)
\put(0.1,0){\epsfig{file=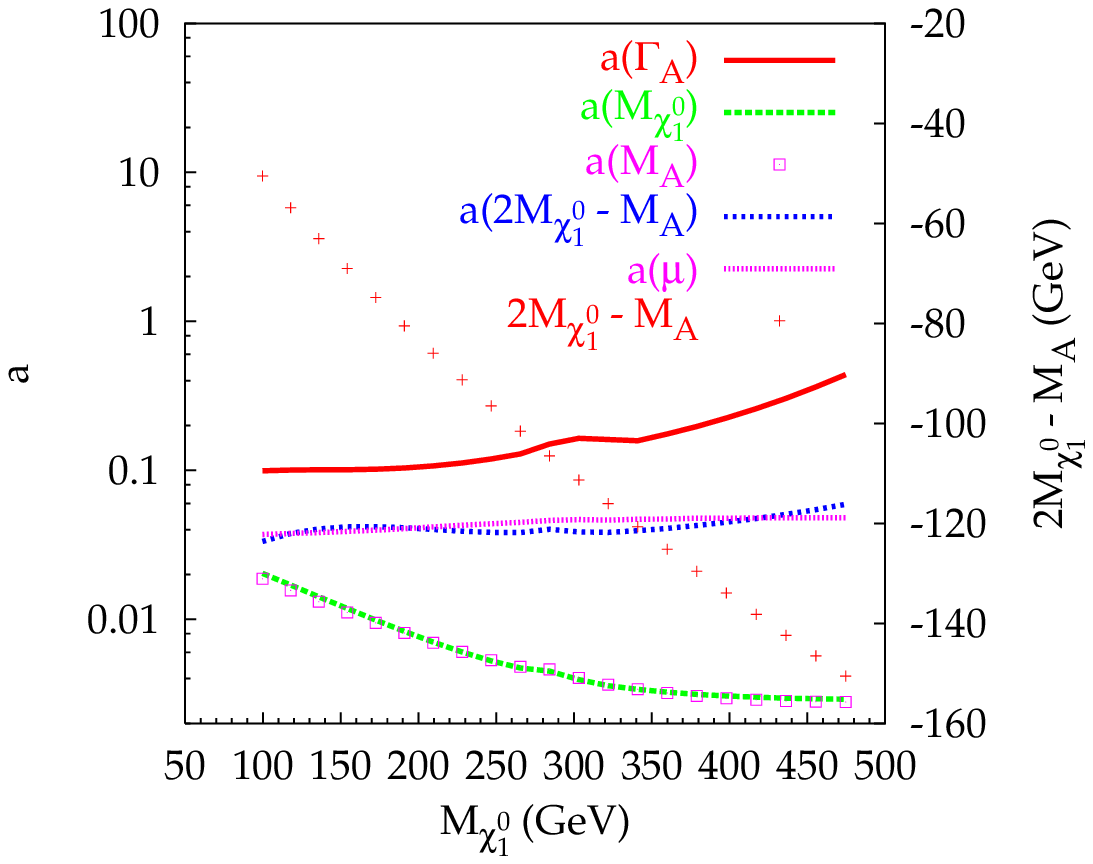, width=9.cm}}
\end{picture}
\begin{picture}(4.8,2.6)
\put(2.8,0){\epsfig{file=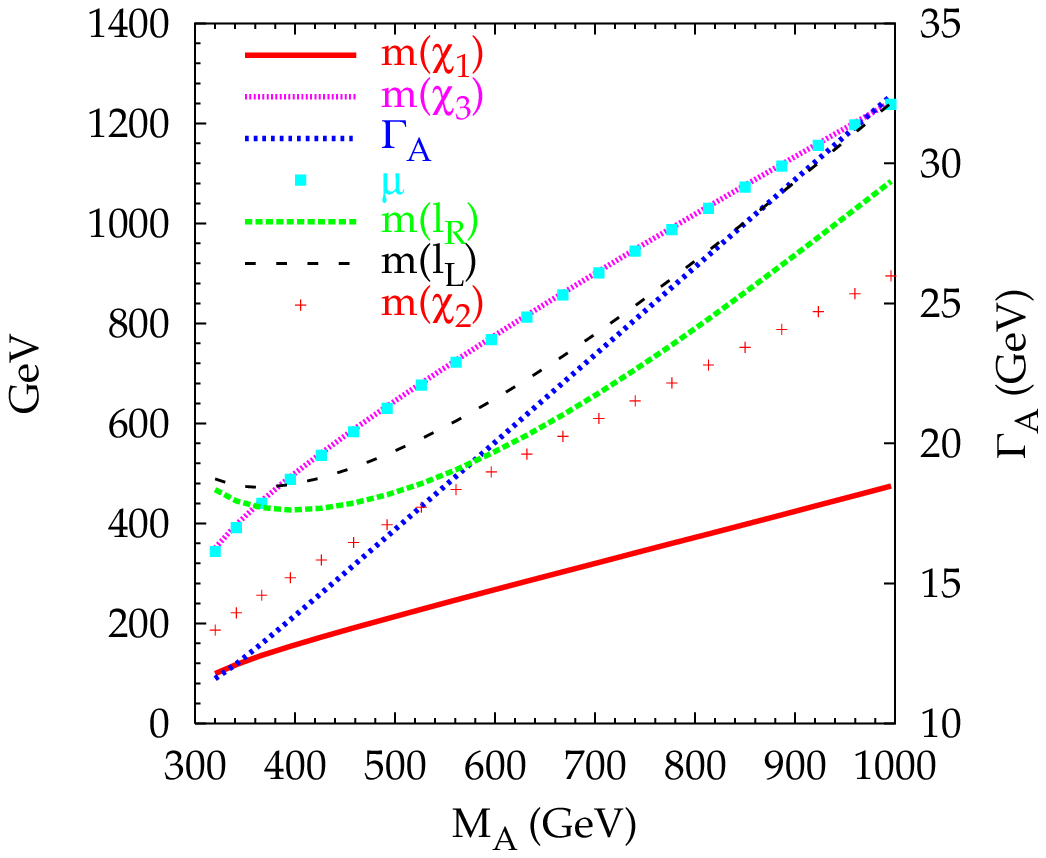, width=9.cm}}
\end{picture}
\caption{\it \small(a) Mass spectrum of the relevant particles in
the funnel region. (b) Accuracies $a$ required to achieve WMAP
precision along slope S2
  in the PmSUGRA scenario.   The scale for the  quantity  $2\mneuto-M_A$
  is displayed on the right hand axis.
  The range of the abscissa corresponds to
  $\mhf=250-1100$~GeV. }
  \label{fig:funExp}}
\end{figure}

\normalsize
 An important remark is that the funnel region features
a rather heavy spectrum, see Fig.~\ref{fig:funExp}b. In the lower part of the
slope the pseudoscalar can  be produced  at a  $500$GeV machine.
In this region,  the LHC can also measure to the needed accuracy
the mass of the pseudoscalar if the
 $A\ra \mu^+\mu^-$ channel\cite{sec4_ATLAS-TDR} can be used.
 At a linear collider, the most accessible process is the associated
production of $\neutt$, which could serve as  a good measurement
of $\mneuto$. The $\gamma \gamma$ option of a LC could bring
important constraint upon the mass and couplings of the $A$ as
well as its width. Combined with a determination of $\mu$ from the
LHC, through $\mneutth$  one could reconstruct the parameter space
that defines the funnel region.

\subsubsection{Focus point}

The  focus point region corresponds
to high values of $M_0$, near the boundary of viable electroweak
symmetry breaking and where the value of $\mu$ drops rapidly.  The LSP has a significant
 Higgsino fraction which means enhanced  couplings to the Z and the Higgses.
 Annihiliation into fermion or gauge boson pairs dominate.
Although coannihilation with heavier neutralinos/charginos can
occur these  coannihilations should not be too efficient otherwise
the relic density is less than what is measured by WMAP. We will
take $\tan \beta=50, \mu>0, A_0=0$ and define a slope S3 where the
relic density is compatible with WMAP:
\begin{equation}
m_0 = 3019.85 +  2.6928M_{1/2} -1.01648
\times 10^{-4} \left(M_{1/2}\right)^2.
\end{equation}
with $\mhf$ in the range $440-1100$GeV. Along this slope,  we have
$M_1<\mu<M_2$ and the  Higgsino component of the LSP is about
$25\%$. The relevant parameters for computing the relic density
are the weak scale values of the neutralino mass matrix, in
particular $M_1$ and $\mu$ since thes determines the Higgsino
component. $\tgb$ enters in the contribution of the $b \bar b$
annihilation and in the neutralino couplings. Some dependence on
$M_2$ and $M_A$ is also expected.

\begin{figure}
 \unitlength=1.1in
\begin{picture}(2.5,2.6)
\put(0.8,0){\epsfig{file=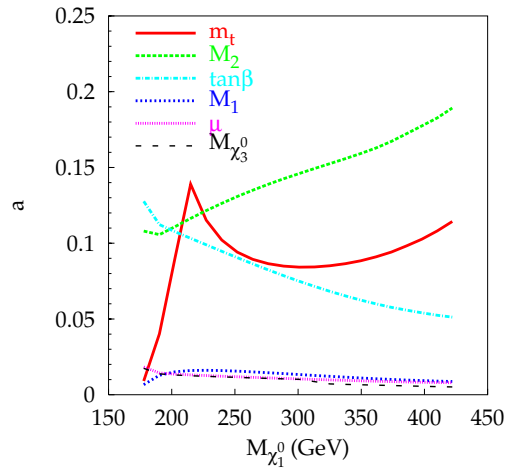, width=9.cm}}
\end{picture}
\caption{\small \it Required fractional accuracy upon MSSM
parameters in the PmSUGRA scenario along   slope S3. The abscissas
correspond to $\mhf=440-1000$ GeV.}\label{fig:foc}
\end{figure}

 Fig.~\ref{fig:foc} displays the required accuracies
$a$ for several parameters  to match the WMAP accuracy.
  The sensitivities of $\mu,M_1$ require fractional precisions of 1$\%$.
 This rather demanding accuracy originates
more from the couplings of the neutralinos to the Golsdtone and
the $A$ boson than from the neutralino mass dependence. As we show
in Fig.~\ref{fig:foc} the accuracies on $M_1$ and $\mu$ can be
converted into accuracies on $\mneuto$ and $\mneutth$.  The needed
accuracy on $M_2$ is  an order of magnitude worse, though still
relevant. Therefore we see that if one can reconstruct the
neutralino mass matrix this scenario can be very much constrained.
More problematic though is the needed accuracy on $\tan\beta$
which ranges from $5-10\%$. We also show in Fig.~\ref{fig:foc} the
accuracy needed on $m_t$,
 which at  the 10\% level is very mild. A stronger dependence is found only
 near the top threshold.
This is a major improvement over the mSUGRA case where to achieve
WMAP precision requires knowing $m_t$ to 20 MeV. The expected
precision  on $m_t$ at a linear collider when combining
theoretical and experimental uncertainties  is only 100MeV,  thus
preventing a precise prediction of $\Omega h^2$ in this scenario.
Finally we do not need to know $M_A$ very accurately,
the accuracy needed varies from more than 100
may not be accessible at LHC, it would not be possible to get even
a rough estimate of $M_A$.

This scenario is rather difficult to handle at the LHC because the
 sfermions and the pseudoscalar are too heavy to be
accessible. Furthermore the charginos and neutralinos cannot be
measured easily. On the other hand, a LC has the possibility to
measure precisely the neutralino mass matrix provided there is
enough energy to produce the neutralinos \cite{Choi:2001}. However  a realistic
study of the achievable accuracy in this scenario at a linear
collider still needs to be performed as well as a study on how to
benefit from a combination of LHC and LC data. In particular the value of
$\tan\beta$ might be very difficult to pin down, although some information
could be extracted from the light Higgs mass.

\subsubsection {Conclusion}

The relic density calculation often involves a large number of
processes. The precise knowledge of the cross-sections necessary
to make an accurate prediction of the relic density of dark matter
rests on a precise knowledge of the physical parameters of the
MSSM. Nevertheless  we have shown that within the scenarios that
are favoured by WMAP, only a few parameters are needed to be
measured with very high precision. Furthermore, using the collider
data rather than relying on some theoretical prejudice,
considerably improves the precision of the prediction of $\Omega
h^2$. Nevertheless the precision required from colliders to match
the WMAP accuracy is rather demanding. The LHC will be able to
determine roughly which  scenario one is in as well as provide
measurements of heavier particles (pseudoscalar, heavy
neutralinos) that can be essential  for an accurate prediction of
the relic density, for example in the Higgs funnel or the focus
point regions. However to make a prediction of $\Omega h^2$ that
matches  the WMAP accuracy, or even more so the PLANCK accuracy,
one absolutely needs the high precision achievable at a linear
collider.

\section{Further SUSY scenarios}

\subsection{\label{sec:422}
Non-decoupling effect in
sfermion-chargino/neutralino couplings
}
                                                                                
{\it J.~Guasch, W.~Hollik and J.~Sol\`a}

\vspace{1em}

\renewcommand{\GeV}{\mbox{ GeV}} 
\renewcommand{\TeV}{\mbox{ TeV}}
 
\newcommand{\stopp}{\ensuremath{\tilde t}} 
\renewcommand{\sbottom}{\ensuremath{\tilde b}}  
\renewcommand{\stau}{\ensuremath{\tilde \tau}} 
\newcommand{\cplus}{\ensuremath{\chi^+}} 
\newcommand{\cmin}{\ensuremath{\chi^-}} 
\newcommand{\neut}{\ensuremath{\chi^0}} 

\renewcommand{\tb}{\ensuremath{\tan\beta}} 
 
\renewcommand{\mw}{\ensuremath{M_W}} 
\newcommand{\mws}{\ensuremath{M^2_W}} 
 
\newcommand{\mxs}{\ensuremath{M^2_X}}
\newcommand{\mx}{\ensuremath{M_X}}

\renewcommand{\mA}{\ensuremath{M_{A^0}}} 
 
\newcommand{\mi}{\ensuremath{M_i}} 
\newcommand{\mis}{\ensuremath{M^2_i}}

\newcommand{\mwf}{{\mw^4}}

\newcommand{\cbta}{c_\beta}
\newcommand{\ctbt}{c_{2\beta}}
\newcommand{\cfbt}{c_{4\beta}}
\newcommand{\sbta}{s_\beta}
\newcommand{\stbt}{s_{2\beta}}

\newcommand{\sws}{s_W^2}

\newcommand{\VChaio}{V_{i1}}
\newcommand{\VChait}{V_{i2}}

\newcommand{\UChaio}{U_{i1}}
\newcommand{\UChait}{U_{i2}}

\newcommand{\mchasomenysmchast}{\left(M_1^2-M_2^2\right)}
\newcommand{\commonnumfactor}{\frac{\alpha}{4\,\pi\,\sws}}

\hyphenation{char-gi-no neu-tra-li-no char-gi-nos neu-tra-li-nos
  sfer-mion sfer-mions} 

\noindent{\small
We analyze the radiative effects induced by a heavy squark sector in
the lepton-slepton-chargino/neutralino coupling. 
These effects are known to grow as the logarithm of the heavy squark
mass. We concentrate on a scenario where sleptons and (some)
charginos/neutralinos are light enough to be produced at an $e^+e^-$
Linear Collider, whereas squarks are heavy and can only be produced at
the LHC. We conclude that the radiative effects of squarks are larger
than the expected accuracy of the coupling measurements at a LC. A
knowledge of the squark mass scale is necessary to provide a precise
prediction for slepton-chargino/neutralino observables, but a moderate
accuracy in the squark parameters is sufficient. These effects can be
treated introducing chargino/neutralino effective coupling matrices.
}



\subsubsection{Introduction}
The main aim of a high energy $e^+e^-$ Linear Collider (LC) is to perform
very high precision measurements of the elementary particles properties
at a level better than 1\%, in order to discriminate
different models of particle interactions. In order to be able to
compare these measurements with the underlying theory at this high level
of accuracy, it is not sufficient to relate the different observables
using lowest order in perturbation theory: the computation of 
radiative corrections (strong and electroweak) form a necessary ingredient
of this program.

In this note we concentrate on the properties of supersymmetric (SUSY)
particles~\cite{Nilles:1983ge,Haber:1985rc,Lahanas:1987uc,Ferrara87}. It is known that the radiative
corrections to some SUSY observables develop radiative corrections which
do not decouple if one takes some of the SUSY masses to be
large~\cite{Guasch:2002ez,Guasch:2002qa,Nojiri:1996fp,Cheng:1997sq,Cheng:1997vy,Djouadi:1997wt,Katz:1998br,Hikasa:1995bw}. Furthermore, the observables which
exhibit this non-decoupling behaviour are the ones that probe the SUSY
nature of the particles under study. Therefore the following situation
could happen: some new physics is observed at the LHC and LC, which is
roughly consistent with the predictions from SUSY models, however at the
LC only \textit{some} of the new particles are visible, but, in order to
predict their production cross-sections and branching ratios at a
sufficient level of accuracy the properties of the heavy SUSY particles
(visible at the LHC) are needed. This situation is, in fact, natural in
SUSY models, where the strongly interacting SUSY particles usually are
predicted to be much heavier than the weakly interacting ones (see
e.g.~\cite{sec4_Allanach:2002nj}). In this note we want to address this issue, asking
ourselves: how large could be the effects of the heavy particles; how
much accuracy on their masses is needed in order to make sufficiently
precise predictions for the light ones; and whether there is some way
of avoiding the uncertainty introduced by the heavy particles.

For the sake of simplicity we will work in the following scenario:
scalar-quarks (squarks) and the gluino are heavy, and beyond the reach
of the LC, whereas scalar-leptons (sleptons) and some
charginos/neutralinos are light, and can be studied at the LC. Then, we
study the radiative effects of squarks in chargino/neutralino-lepton-slepton
interactions, computing the value of the radiative corrections induced
by quarks/squarks, and comparing them with the corrections from other
sectors of the model. We will analyze these effects in the partial decay
widths of selectrons and the electron-sneutrino into charginos and
neutralinos. 

\subsubsection{Theoretical introduction: non-decoupling effects and effective
  coupling matrices} 

We concentrate in the analysis of the sfermion decays as discussed
thoroughly in
Refs.~\cite{Guasch:2002ez,Guasch:2001kz,Guasch:2002qa,Guasch:1998as}\footnote{The
  corresponding FORTRAN codes are available from~\cite{Program}}. In the radiative
corrections of the partial decay widths
\begin{equation}
\Gamma(\tilde{f} \to f' \chi)\,\,,
\label{eq:gammadef}
\end{equation}
$\tilde{f}$ being a sfermion, $f'$ a Standard Model (SM) fermion, and $\chi$
a chargino or neutralino, non-decoupling effects appear, and the
radiative corrections grow as the logarithm of the largest SUSY mass of
the model~\cite{Guasch:2002ez,Guasch:2002qa}. These non-decoupling effects reflect the
fact that SUSY is (softly) broken, and the Yukawa couplings of the
charginos/neutralinos are no longer equal to the gauge bosons and Higgs
bosons couplings beyond leading order. 

A part of these corrections can be encoded in a set of counterterm
expressions which appear in all observables. Concretely they can be
written as a shift to the $U$, $V$ and $N$ matrices that diagonalize the
chargino and neutralino mass matrices\footnote{In this note we follow
  the notation and conventions of Ref.~\cite{Guasch:2002ez}. Note, in
  particular, that $m_{\tilde{f}_1}>m_{\tilde{f}_2}$.}:
\begin{equation}
  \label{eq:effective1}
    \tilde{U}=U+\Delta U \ \ , \ \ 
    \tilde{V}=V+\Delta V  \ \ ,  \ \  \tilde{N}=N+\Delta N  \ \ , 
\end{equation}
where $\Delta U$, $\Delta V$, and $\Delta N$ are a certain combination
of counterterms, which can be computed using self-energies of gauge
bosons, Higgs bosons, charginos and neutralinos. 

Unfortunately the full
contributions to the expressions~(\ref{eq:effective1}) are
divergent. The only consistent subset of corrections which makes all the
expressions in~(\ref{eq:effective1}) finite is the subset of fermion and
sfermion 
loops {contributing to the self-energies of the gauge bosons,
 Higgs bosons, charginos and neutralinos}. With this restriction, we can
define  \textit{effective coupling   matrices} 
\begin{equation}
  \label{eq:effectivegen}
  U^{eff}=U+\Delta U^{(f)} \ \ , \ \ 
  V^{eff}=V+\Delta V^{(f)}  \ \ ,  \ \  N^{eff}=N+\Delta N^{(f)}  \ \ , 
\end{equation}
where $\Delta U^{(f)}$, $\Delta V^{(f)}$, $\Delta N^{(f)}$ are given by
the expressions 
(\ref{eq:effective1}) taking into account only loops of fermions and
sfermions. We will refer to these corrections as \textit{universal
  corrections}. They are the equivalent of the
  \textit{super-oblique corrections} of Ref.\cite{Katz:1998br}.

As an example, we have computed analytically
the electron-selectron contributions to the $\Delta U$ and $\Delta V$
matrices~(\ref{eq:effective1}), assuming zero mixing angle in the
selectron sector ($\theta_e=0$), we have identified the leading terms in the
approximation $m_{\tilde e_i}, m_{\tilde \nu}\gg (\mw,\mi) \gg m_e$, and
 analytically canceled the divergences and the  renormalization scale
 dependent terms;
finally, we have kept only the terms logarithmic in the {slepton}
masses. The result reads as follows:
\begin{eqnarray}
\Delta\UChaio^{(e)}&=& \commonnumfactor\log\left(\frac{M^2_{\tilde e_L}}{\mxs}\right)\,\bigg[
 \frac{\UChaio^3}{6} - 
  \UChait \frac{\sqrt{2}\,\mw\,(M\,\cbta +
  \mu\,\sbta)}{3\,(M^2-\mu^2)\,\mchasomenysmchast^2} 
\left(M^4 - M^2\,\mu^2 + \right.\nonumber\\
&&\left. + 3\,M^2\,\mws + 
     \mu^2\,\mws + \mwf + \mwf\,\cfbt 
    + (\mu^2-M^2)\,\mis + 
     4\,M\,\mu\,\mws\,\stbt\right)\,\bigg]\,\,,
\nonumber\\
\Delta\UChait^{(e)}&=&\commonnumfactor\log\left(\frac{M^2_{\tilde e_L}}{\mxs}\right)\,
\UChaio\,\frac{\mw\,(M\,\cbta + \mu\,\sbta)}
 {3\,\sqrt{2}\,(M^2-\mu^2)\,\mchasomenysmchast^2} \times \nonumber\\
  &\times&  \left((M^2-\mu^2)^2 
+ 4\,M^2\,\mws + 4\,\mu^2\,\mws + 2\,\mwf + 
   2\,\mwf\,\cfbt + 8\,M\,\mu\,\mws\,\stbt\right)\,\,,
\nonumber\\
\Delta\VChaio^{(e)}&=&\commonnumfactor\log\left(\frac{M^2_{\tilde e_L}}{\mxs}\right)\,\bigg[
\frac{\VChaio^3}{6}
- 
 \VChait \frac{\sqrt{2}\,\mw\,(\mu\,\cbta +
  M\,\sbta)}{3\,(M^2-\mu^2)\,\mchasomenysmchast^2} 
\left(M^4 - M^2\,\mu^2 +\right.\nonumber\\
&&\left. + 3\,M^2\,\mws + 
     \mu^2\,\mws + \mwf + \mwf\,\cfbt 
 + (\mu^2-M^2)\,\mis + 
     4\,M\,\mu\,\mws\,\stbt\right)\bigg]\,\,,
\nonumber\\
\Delta\VChait^{(e)}&=&\commonnumfactor\log\left(\frac{M^2_{\tilde e_L}}{\mxs}\right)\,\VChaio\,
\frac{\mw\,(\mu\,\cbta + M\,\sbta)}
 {3\,\sqrt{2}\,(M^2-\mu^2)\,\mchasomenysmchast^2} \times \nonumber\\
  &\times&  \left((M^2-\mu^2)^2  
   + 4\,M^2\,\mws + 4\,\mu^2\,\mws + 2\,\mwf + 
   2\,\mwf\,\cfbt + 8\,M\,\mu\,\mws\,\stbt\right),\nonumber\\
\label{eq:logterms}
\end{eqnarray}
$M^2_{\tilde e_L}$ being the soft-SUSY-breaking mass of the
$(\tilde{e}_L,\tilde{\nu})$ doublet,
whereas $\mx$ is a SM mass.

In this way the effects of the (heavy) squarks can be encoded in the
definition of \textit{effective couplings} in the
lepton-slepton-chargino/neutralino interactions.

In the following we will separate three kind of corrections: the
universal corrections induced by the quark-squark particles
$U^{eff(q)}$, $V^{eff(q)}$, $N^{eff(q)}$; the universal contributions
induced by leptons-sleptons  $U^{eff(l)}$, $V^{eff(l)}$, $N^{eff(l)}$,
and the non-universal contributions.

\subsubsection{Numerical Analysis}

For the numerical analysis we choose some default \textit{typical} set
of input parameters. We will concentrate in the parameter set given by
point 1a of the \textit{Snowmass Points and
  Slopes} (SPS)~\cite{sec4_Allanach:2002nj}\footnote{The spectrum and tree-level branching
  ratios for the several SPSs can be found e.g. in
Ref.~\cite{Ghodbane:2002kg}.}. For completeness we give here the
soft-SUSY-breaking parameters:
\begin{equation}
\begin{array}{l} 
\tb= 10, 
 \mA= 393.6 \GeV,  \mu= 352.4\GeV,
 M= 192.7\GeV,
 M'=99.1\GeV\,\,,
\\
 M_{\{\tilde{d},\tilde{s}\}_L}= 539.9 \GeV,
 M_{\tilde{b}_L}=495.9\GeV,
  M_{\{\tilde{d},\tilde{s}\}_R}=519.5\GeV,
  M_{\tilde{b}_R}=516.9\GeV,\\
 A_{\{d,s\}}= 3524\GeV,
 A_b=-772.7\GeV,\\
 M_{\{\tilde{u},\tilde{c}\}_R}= 521.7\GeV,
 M_{\tilde{t}_R}= 424.8\GeV,
 A_{u,c}= 35.24 \GeV,
 A_t=-510\GeV, \\
 M_{\{\tilde{e},\tilde{\mu}\}_L} =196.6\GeV,
 M_{\tilde{\tau}_L}=195.8\GeV,
 M_{\{\tilde{e},\tilde{\mu}\}_R} =136.2\GeV,
 M_{\tilde{\tau}_R}=133.6\GeV,\\
 A_{e,\mu}= 3524\GeV,
 A_\tau=-254.2\GeV,
\end{array}
\label{eq:sps1a}
\end{equation}
where the soft-SUSY-breaking trilinear couplings of the first and second
generation sfermions have been chosen such that the non-diagonal
elements of the sfermion mass matrix are zero.

However, a note of caution should be given, our
computation is performed in the On-shell renormalization scheme, whereas
the SPS parameters are given in the $\overline{DR}$ renormalization
scheme, and one should make a scheme conversion of the
parameters, this conversion is beyond the scope of the present work. In
this note we are interested only in establishing whether the effects of heavy
particles are important, and therefore we are only interested in
obtaining a suitable SUSY spectrum, therefore we treat the given
numerical parameters of SPS 1a as On-shell SUSY 
parameters~\footnote{Of course, once we will be analyzing the real LC
  data, the $\overline{DR}$-On-shell conversion will need to be made in order
to extract the fundamental soft-SUSY-breaking parameters.}.

\begin{table}[tb]
\begin{center}
\begin{tabular}[c]{|c|c|c|c|c|c|c|c|}
\hline
&$\Gamma^{tree}$ [GeV] & $\delta\Gamma^{(q)}/\Gamma$ & $\delta\Gamma^{(l)}/\Gamma$
& $\delta\Gamma^{no-uni}/\Gamma$ & $\delta\Gamma/\Gamma$ \\ \hline
$\tilde{e}_1\to e^-\chi^0_1$ & 0.110 & 0.043 & 0.032 & 
-0.002 & 0.073\\ \hline
$\tilde{e}_1\to e^-\chi^0_2$ & 0.047 & 0.030 & 0.034 & 
-0.012 & 0.051\\ \hline
$\tilde{e}_1\to \nu_e\chi^-_1$ & 0.081 & 0.026 & 0.033 & 
0.006 & 0.065\\ \hline
$\tilde{e}_2\to e^-\chi^0_1$ & 0.194 & 0.052 & 0.034 & 
0.000 & 0.086\\ \hline
$\tilde{\nu}_e\to \nu_e\chi^0_1$ & 0.140 & 0.059 & 0.035 & 
-0.005 & 0.089\\ \hline
$\tilde{\nu}_e\to \nu_e\chi^0_2$ & 0.006 & 0.018 & 0.033 & 
-0.014 & 0.036\\ \hline
$\tilde{\nu}_e\to e^-\chi^+_1$ & 0.016 & 0.024 & 0.033 & 
0.002 & 0.059\\ \hline
\end{tabular}
\end{center}
\caption{Tree-level partial decay widths and relative corrections for
  the selectron and sneutrino decays into charginos and neutralinos for
  SPS 1a.\label{tab:gammas}}
\end{table}

In Table~\ref{tab:gammas} we show the partial decay widths of selectrons
into charginos/neutralinos for SPS 1a. We show: the tree-level partial
widths $\Gamma^{tree}$; 
the relative corrections induced by quarks-squarks $\delta\Gamma^{(q)}/\Gamma$; 
the relative corrections induced by the lepton-slepton universal
contributions~(\ref{eq:effectivegen}) $\delta\Gamma^{(l)}/\Gamma$; 
the process-dependent non-universal contributions
$\delta\Gamma^{no-uni}/\Gamma$; and the total corrections
$\delta\Gamma/\Gamma$.

The universal corrections $\delta\Gamma^{(q)}/\Gamma$ and
$\delta\Gamma^{(l)}/\Gamma$ in Table~\ref{tab:gammas} represent a
correction that will be present whenever a
fermion-sfermion-chargino/neutralino coupling enters a given
observable. The correction $\delta\Gamma^{no-uni}/\Gamma$ represents the
process-dependent part. For the presented observables the non-universal
corrections turn out to be quite small, but this is not necessarily always
the case. From the values of Table~\ref{tab:gammas} it is clear that the
corrections of the quark-squark sector are as large as the corrections
from the (light) lepton-slepton sector, for the presented observables they
amount to a $2-6\%$ relative correction, depending on the particular decay
channel. 

For SPS 1a the squark mass scale is around 500\GeV, however the
corrections grow logarithmically with the squark mass scale. In
Fig.~\ref{fig:UniQmsquark} we show the relative corrections induced by the
quark-squark sector ($\delta\Gamma^{(q)}/\Gamma$) in the observables of
Table~\ref{tab:gammas} as a function of a common value for all
soft-SUSY-breaking squark mass parameters in~(\ref{eq:sps1a}), in a range where the squarks
are accessible at the LHC. The several lines in
Fig.~\ref{fig:UniQmsquark} are neatly grouped together: the upper lines
correspond to the lightest neutralino ($\neut_1$) which is
\textit{bino}-like, whereas the lower lines correspond to the second
neutralino and lightest chargino ($\neut_2$, $\chi^\pm_1$), which are
\textit{wino}-like. Since the coefficient of the logarithm in the
universal corrections~(\ref{eq:effectivegen}) is
proportional to the corresponding gauge coupling, the behaviour of the
corrections is different between the two kinds of gauginos, but similar
for different gauginos of the same kind. We see that for a
\textit{bino}-like neutralino the corrections undergo an absolute shift
of less than 2\% (from 4.5\% to 6.5\% in the channel
$\tilde{e}_1\to e^-\neut_1$) by changing the squark mass scale from
$500\GeV$ 
to 3\TeV. For a \textit{wino}-like gaugino the shift is much larger,
being up to 4\% in the case under study (from 2\% to 6\% in the
$\tilde{\nu}_e\to\nu_e\neut_2$ channel). We conclude, therefore, that a
certain knowledge of the squark masses is necessary in order to provide
a theoretical prediction with an uncertainty below 1\%, but only a
rough knowledge of the scale is necessary.

\begin{figure}[tbp]
\centerline{\resizebox{7cm}{!}{\includegraphics{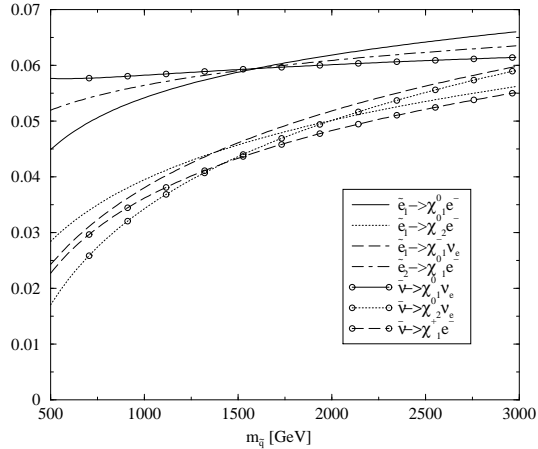}}}
\caption{Squark contributions to the radiative corrections of the
  partial decay widths of sleptons ($\delta\Gamma^{(q)}/\Gamma$) for SPS
  1a as a function of a common soft-SUSY-breaking mass parameter for all
  squarks.\label{fig:UniQmsquark}}
\end{figure} 

As explained previously these corrections admit a description in terms
of effective coupling matrices. In Fig.~\ref{fig:EffQmsq} we show the
relative finite shifts induced by the quark-squark sector in the
effective coupling matrices as a function of a common soft-SUSY-breaking
squark mass parameter. The tree-level values for the mixing matrices are:
\begin{equation}
  \label{eq:treeUVN}
\begin{array}{l}
U=\begin{pmatrix}   
  0.91 &     0.41 \cr
  -0.41 &    0.91 
\end{pmatrix}
          \, , \, \\
V=\begin{pmatrix}
  0.97 &      0.24 \cr
  -0.24&   0.97 
\end{pmatrix} \, , \, 
\end{array}
N=\begin{pmatrix}
  -0.99        &
  -0.10      &
  -0.06 i &
  0.11      \cr
  0.06 &
  -0.94    &
  0.09 i&
  -0.32           \cr
  -0.15       &
  0.28        &
  0.69 i&
  -0.64      \cr
  0.05    &
      -0.16      &
      0.71 i&
      0.68   
    \end{pmatrix}
    \,\,.
\end{equation}
\begin{figure}[t]
\begin{tabular}{cc}
\resizebox{6.5cm}{!}{\includegraphics{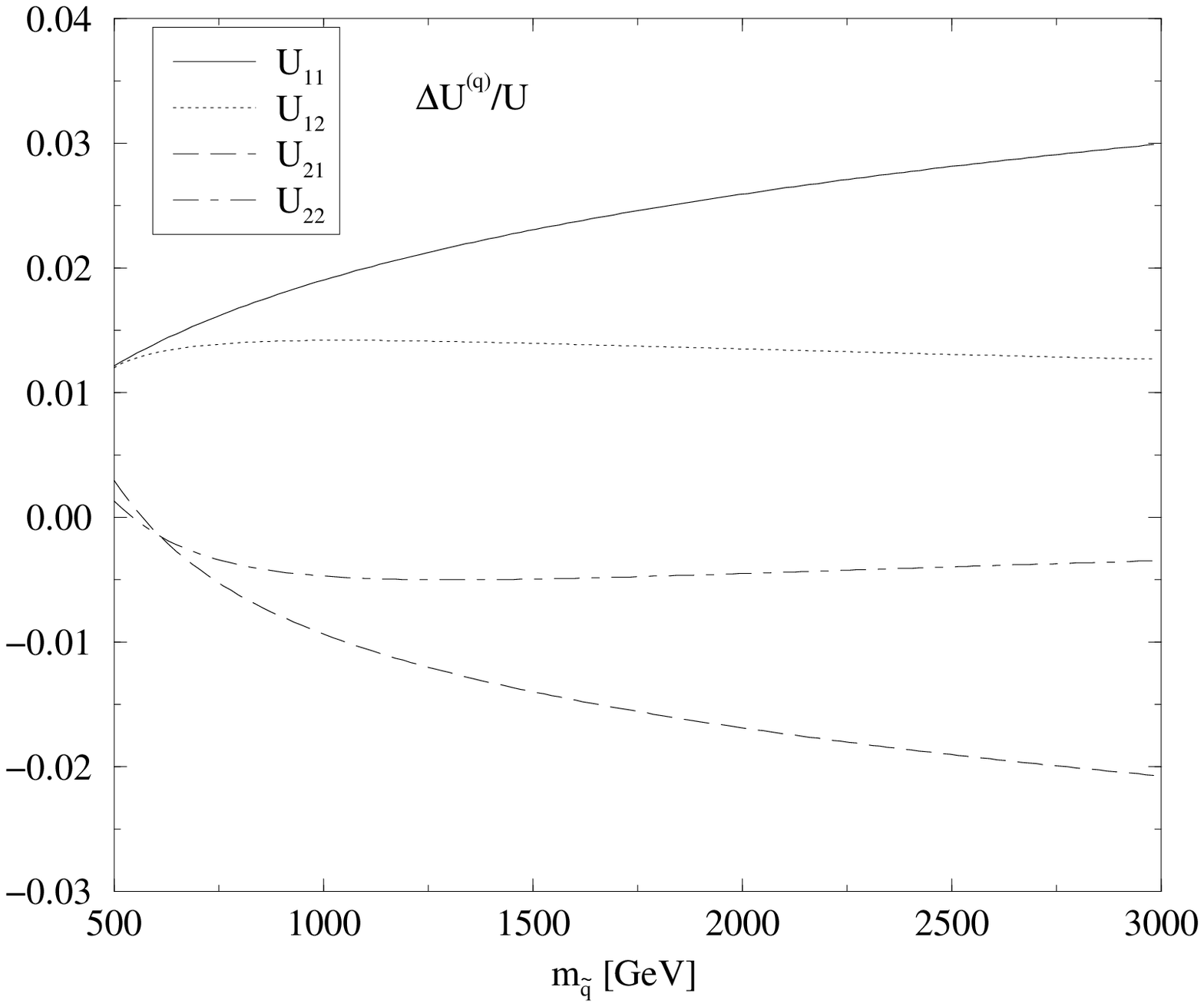}} &
\resizebox{6.5cm}{!}{\includegraphics{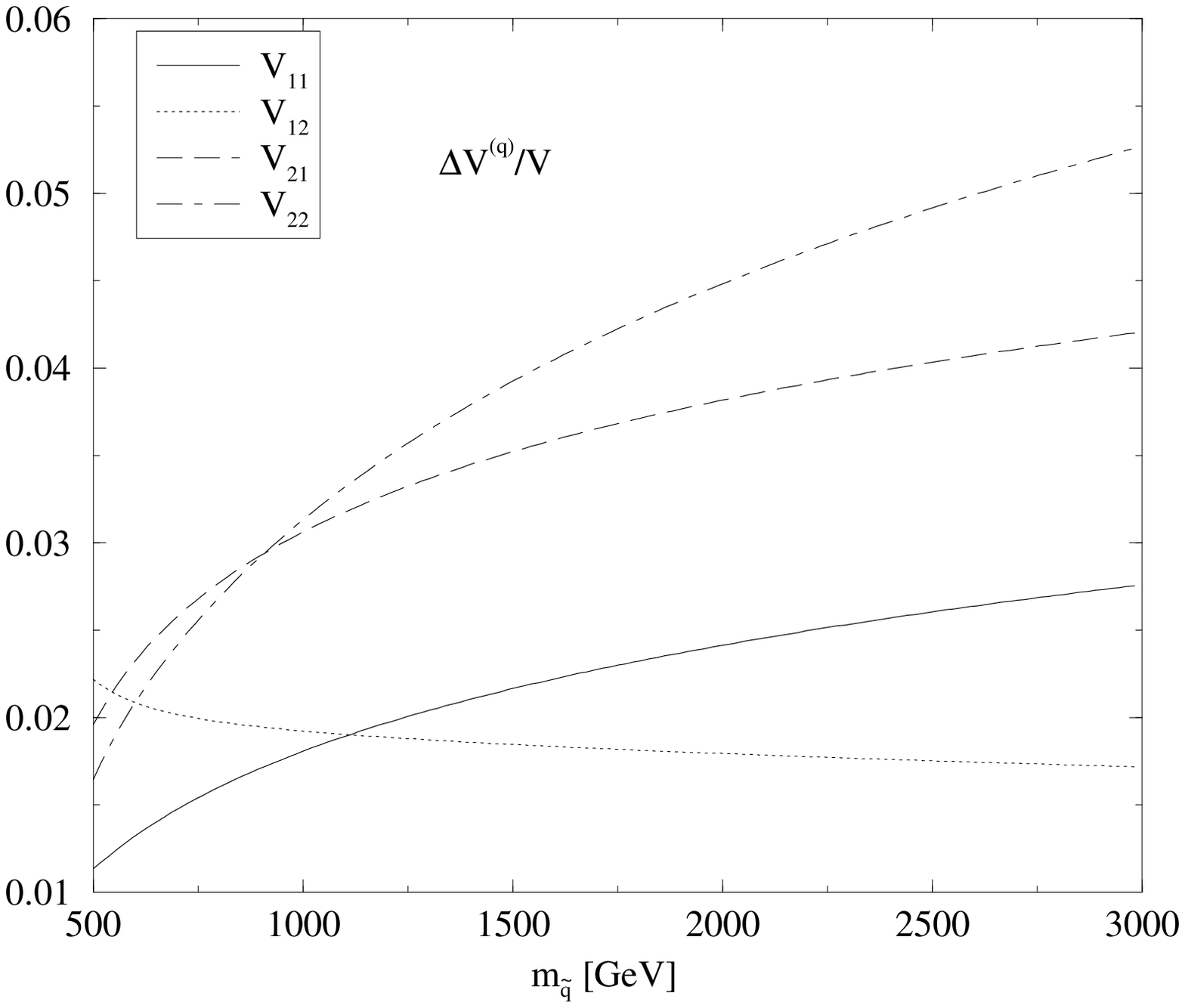}} \\
(a) & (b) \\
\resizebox{6.5cm}{!}{\includegraphics{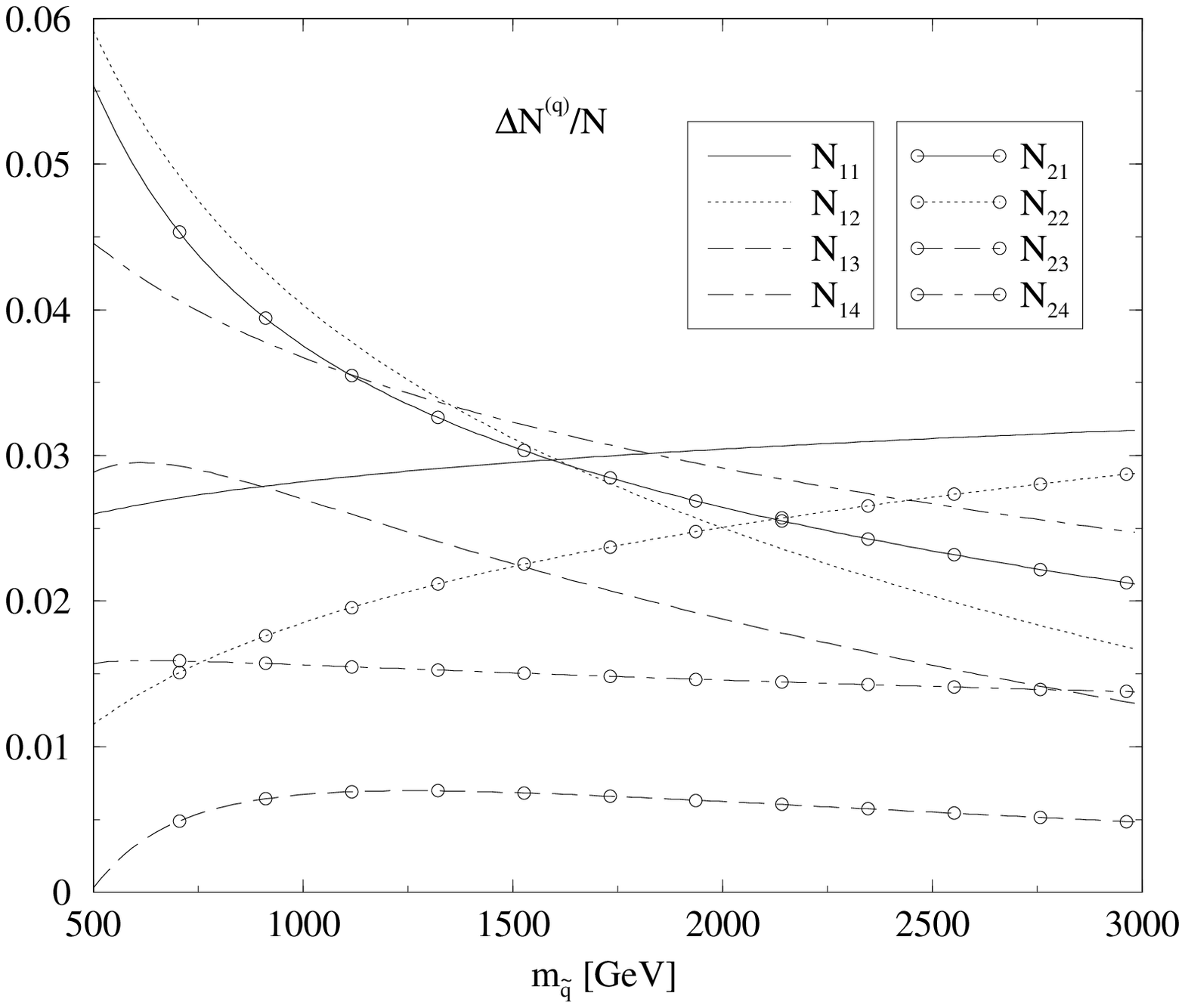}}  &
\resizebox{6.5cm}{!}{\includegraphics{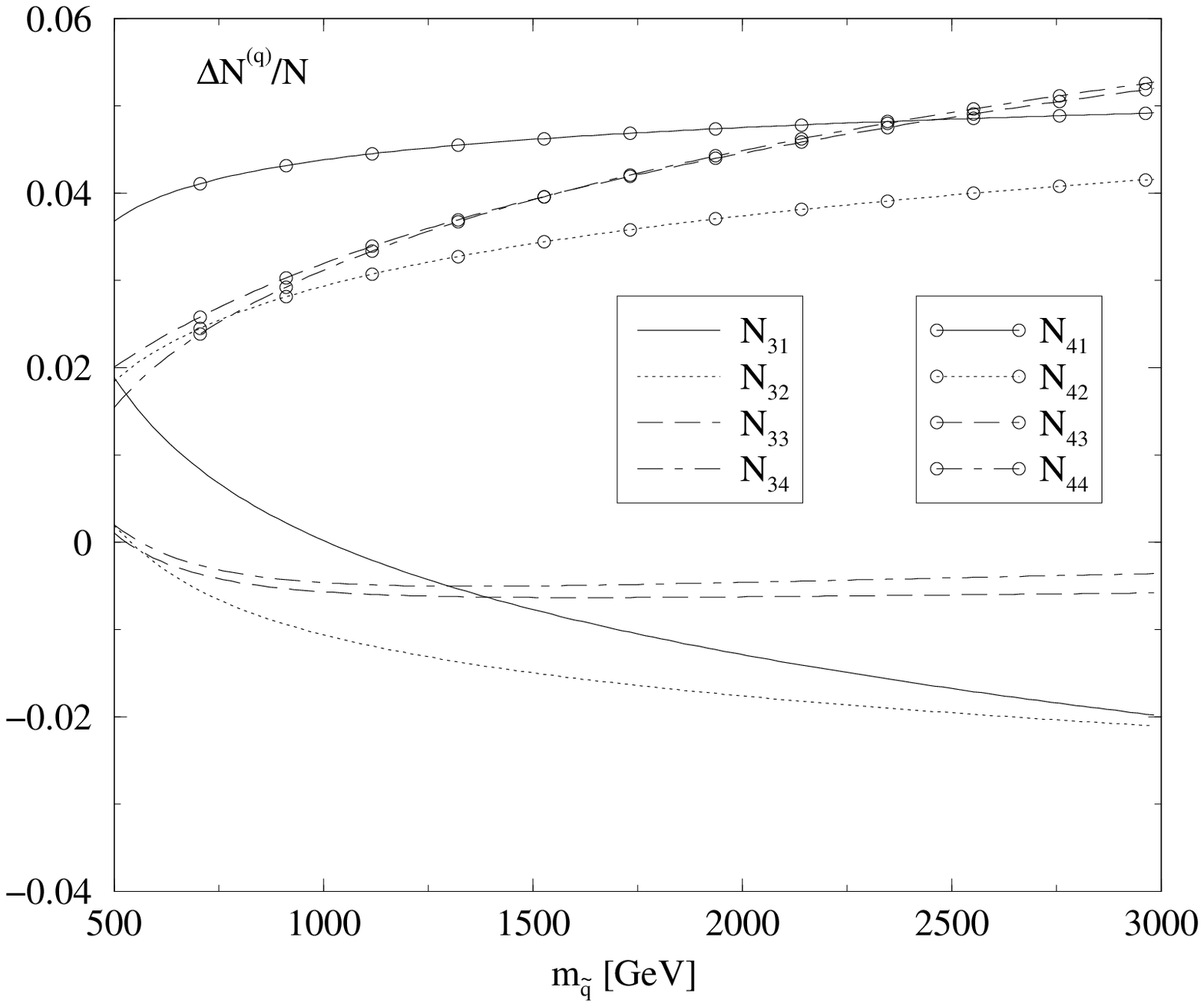}} \\
(c) & (d) 
\end{tabular}
\caption{Quark-squark contributions to the effective chargino/neutralino
  mixing coupling matrices~(\ref{eq:effectivegen}) as a function of a
  common squark mass parameter for SPS 1a.\label{fig:EffQmsq}}
\end{figure}
One can perform a one-to-one matching of Fig.~\ref{fig:EffQmsq} with
Fig.~\ref{fig:UniQmsquark}. By neglecting the small electron-higgsino
couplings we obtain:
\begin{equation}
\begin{array}{rcl}
\delta\Gamma^{(q)}(\tilde{e}_L\to e^-\neut_\alpha)/\Gamma&=&2 \frac{\Delta
N^{(q)}_{\alpha 2} - Y_L t_w \Delta N^{(q)}_{\alpha 1} }
{N_{\alpha 2} - Y_L t_w  N_{\alpha 1} },\nonumber\\
\delta\Gamma^{(q)}(\tilde{e}_L\to\nu_e\cmin_1)/\Gamma&=&2 \Delta U^{(q)}_{11}/U_{11}, \nonumber\\
\delta\Gamma^{(q)}(\tilde{e}_R\to e^-\neut_\alpha)/\Gamma&=&2 \Delta
N^{(q)}_{\alpha 1}/N_{\alpha 1} ,\nonumber\\
\end{array}
\begin{array}{rcl}
\delta\Gamma^{(q)}(\tilde{\nu}_e\to e^-\cplus_1)/\Gamma&=&2 \Delta V^{(q)}_{11}/V_{11}, \nonumber\\
\delta\Gamma^{(q)}(\tilde{\nu}_e\to \nu_e\neut_\alpha)/\Gamma&=&2  \frac{\Delta
N^{(q)}_{\alpha 2} + Y_L t_w \Delta N^{(q)}_{\alpha 1} }
{N_{\alpha 2} + Y_L t_w  N_{\alpha 1} } ;
\end{array}
\end{equation}
and $\tilde{e}_{\{L,R\}}=\tilde e_{\{1,2\}}$ for the
case~(\ref{eq:sps1a}) under 
study. We see in Fig.~\ref{fig:EffQmsq} variations up
to 5\% in the coupling matrices, which would translate to variations
up to 10\% in the observables.

\begin{figure}[tbp]
\begin{tabular}{cc}
\resizebox{7cm}{!}{\includegraphics{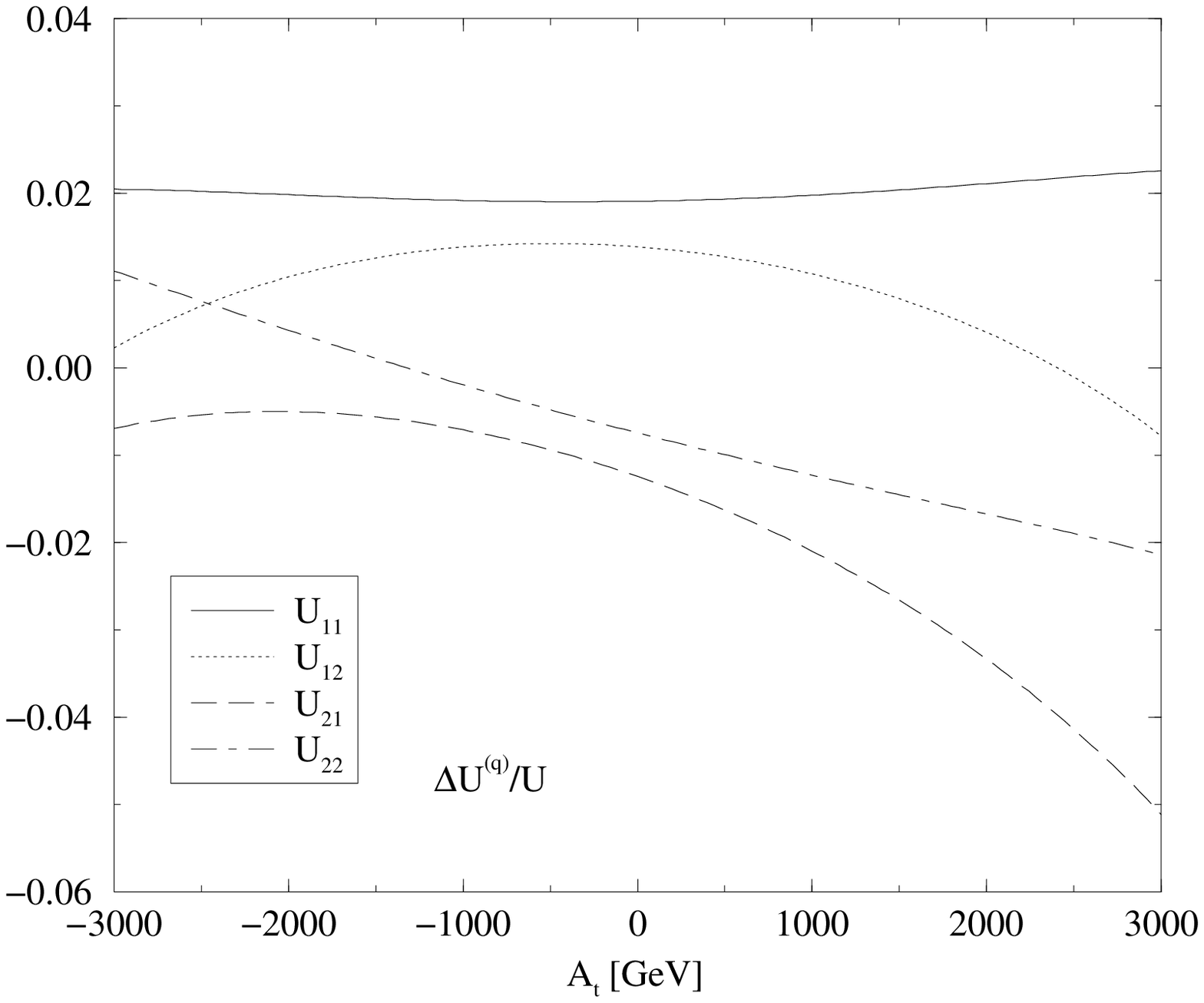}} &
\resizebox{7cm}{!}{\includegraphics{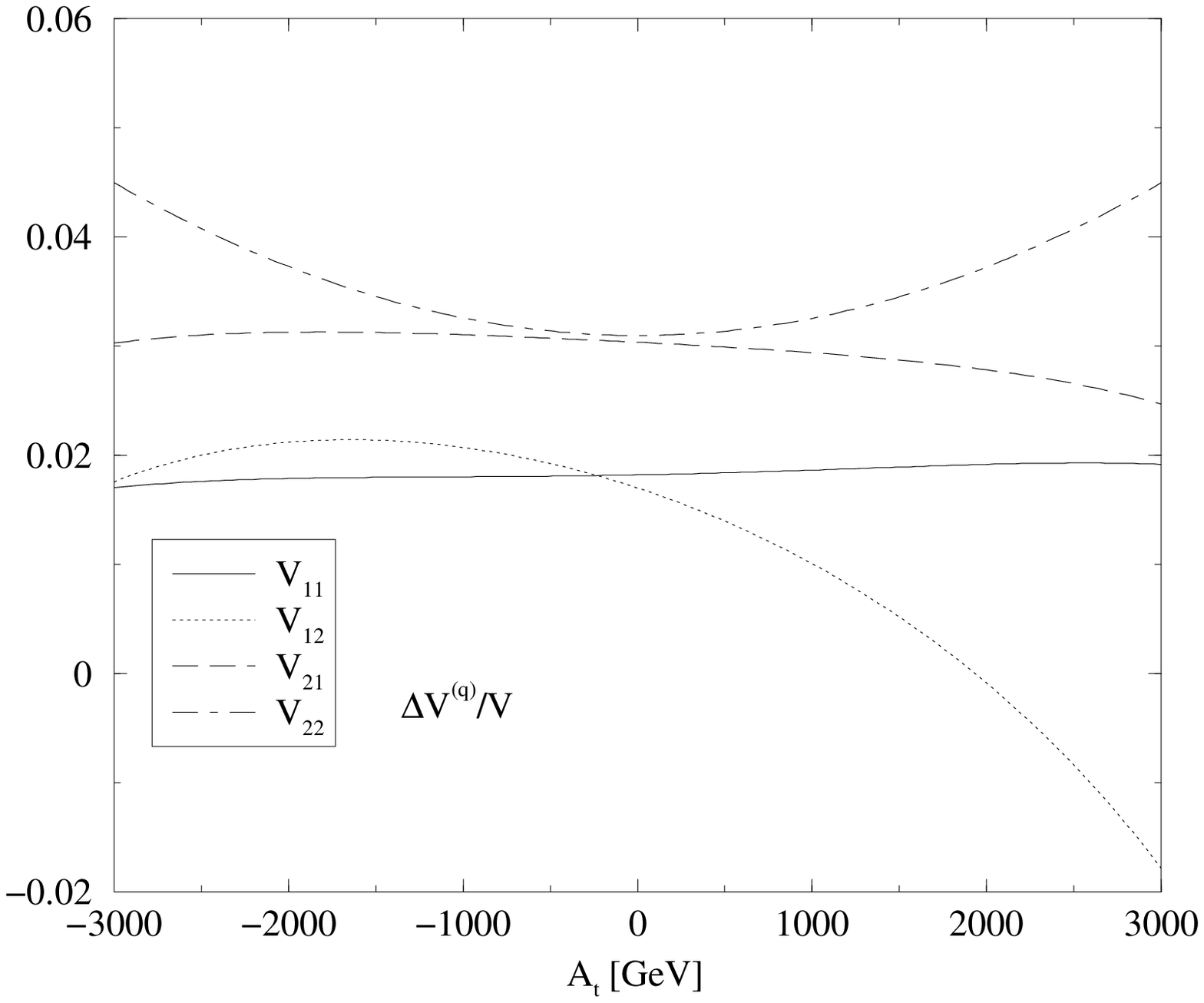}} \\
(a) & (b)
\end{tabular}
\caption{Variation of the chargino effective coupling
  matrices~(\ref{eq:effectivegen}) as a function of the top-squark
  soft-SUSY-breaking trilinear coupling $A_t$ for SPS 1a, but with a
  squark mass scale of 1\TeV. \label{fig:UVAt}}
\end{figure}

We are also interested in the variation with the soft-SUSY-breaking squark
trilinear coupling $A_q$. For the first and second generation squarks
the variation is negligible. The corrections show some variation with
$A_b$, but it is well below the 1\% level. In Fig.~\ref{fig:UVAt} we
show the variation of the chargino effective couplings with $A_t$. In
this figure we have chosen a squark mass scale of 1\TeV. Since $A_t$
enters the computation of the physical top-squark masses, choosing 
a \textit{light} squark mass scale ($\sim 500\GeV$) would produce light
physical top-squark masses ($\sim 100\GeV$) for certain values of
$A_t$. In that case one would find large variations in the
corrections which are due to the presence of light top-squark particles,
and not to the trilinear coupling \textit{per se}. Furthermore, these
light top-squark particles could be produced at the LC, and their
properties precisely measured. In this figure we see large variations of
the corrections (up to 4\%), mainly in the \textit{higgsino}
components of the charginos ($U_{i2}$, $V_{i2}$). Therefore these
corrections are mainly relevant for the couplings of third generation
sfermions ($\stau,\sbottom,\stopp$). Again, a precise knowledge of $A_t$
is not necessary to provide a prediction with sufficient precision, but
a rough knowledge of the scale and sign is needed.

We have performed a study to gauge the importance of the knowledge of
the squark mass scale. We have performed an analysis of the uncertainty
in the prediction for the partial decay widths when only part of the
spectrum is known, under two conditions: 
\begin{itemize}
\item First, including data only from the LC,  the slepton,
  chargino/neutralino and lightest stop spectrum is known. We perform a
  scan in the MSSM parameter space, for the squark spectrum between
  500\GeV and 5\TeV.
\item Second, using additionally data from the LHC, with a known squark
  mass spectrum.  
\end{itemize}
The spectrum determination at the LC/LHC+LC is taken from
Table~\ref{tab:mass-errors}.
The results are shown in Table~\ref{sec422:tab:LHCLC}. Here we show the
range of possible values in  the squark universal corrections
($\delta\Gamma^{(q)}/\Gamma$) due  to the unknown spectrum, and the
corresponding uncertainty in the prediction of the partial decay
widths. Using only LC data the uncertainty in the partial decay widths prediction is
at the percent level for all possible decays. When we restrict the
squark spectrum to the values provided by the LHC, the uncertainty
decreases significantly, in most of the channels by a factor five, or
even a factor ten. However, in some special cases, 
namely  
$\Gamma(\tilde{e}_1\to e^-\chi^0_1)$ and 
$\Gamma(\tilde{\nu}_e\to \nu_e\chi^0_1)$, the uncertainty decreases only
a factor two. 
\begin{table}
\begin{center}
\begin{tabular}{|c||c|c||c|c|c|c|c|c|c|c|c|c|c|c|c|}
\hline
& \multicolumn{2}{c||}{LC} & \multicolumn{2}{c|}{LHC+LC} \\\hline
Process & $\delta\Gamma^{(q)}/\Gamma$ & uncert (\%) &
$\delta\Gamma^{(q)}/\Gamma$ & uncert (\%) \\\hline
$\tilde{e}_1\to e^-\chi^0_1$ &
0.040 -- 0.076 & 3.6 &
0.042 -- 0.064 & 2.3
\\\hline
$\tilde{e}_1\to e^-\chi^0_2$ &
0.029 -- 0.075 & 4.6 &
0.028 -- 0.030 & 0.3
\\\hline
$\tilde{e}_1\to \nu_e\chi^-_1$ &
0.026 -- 0.082 & 5.6 &
0.024 -- 0.031 & 0.6
\\\hline
$\tilde{e}_2\to e^-\chi^0_1$ &
0.052 -- 0.063 & 1.1 &
0.052 -- 0.054 & 0.2
\\\hline
$\tilde{\nu}_e\to \nu_e\chi^0_1$ &
0.041 -- 0.075 & 3.4 &
0.045 -- 0.060 & 1.5
\\\hline
$\tilde{\nu}_e\to \nu_e\chi^0_2$ &
0.016 -- 0.080 & 6.4 &
0.015 -- 0.032 & 1.7
\\\hline
$\tilde{\nu}_e\to e^-\chi^+_1$ &
0.023 -- 0.067 & 4.4 &
0.022 -- 0.027 & 0.5
\\\hline
\end{tabular}
\end{center}
\caption{Uncertainty in the prediction of the partial decay widths of
  selectrons and sneutrinos assuming that only LC data is available, and
   combining LHC+LC data.\label{sec422:tab:LHCLC}} 
\end{table}

\subsubsection{Conclusions}

In SUSY models non-decoupling effects appear. These effects are due to
two kinds of splittings among the particle masses: a splitting between a
particle and its SUSY partner (given by the soft-SUSY-breaking masses);
and a splitting among the SUSY particles themselves. In this situation
the radiative corrections grow with the logarithm of the largest SUSY
particle of the model. In this scenario some of the particles
(presumably strongly interacting particles) are heavy, and can only be produced
at the LHC, whereas another set of particles (selectrons, lightest
charginos/neutralinos) can be studied at the LC, and their properties
measured with a precision better than 1\%. 

In order to provide a prediction at the same level of accuracy, one
needs a knowledge of the squark masses (and $A_t$) obtained from the LHC
measurements, but a high precision measurement of the squark parameters
is not necessary. 

The effects of squarks can be taken into account by the use of effective
coupling matrices in the chargino/neutralino sector. These effects can
be extracted from LC data, by finding the finite difference between the
mixing matrices obtained from the chargino/neutralino masses, and the
mixing matrices obtained from the couplings analysis.

Of course, to reach the high level of accuracy needed at the LC the
complete one-loop corrections to the observables under study is needed,
but the effective coupling matrices form a necessary and universal
subset of these corrections.




\subsection{\label{sec:423} Correlations of flavour and collider physics
within supersymmetry}

{\it T.~Hurth and W.~Porod}

\vspace{1em}
{\small
\noindent
Until now the focus within the direct search for supersymmetry has been on
mainly flavour diagonal observables. Recently lepton flavour violating
signals at future electron positron colliders have been studied. There is 
now the opportunity to analyze relations between collider observables and 
low-energy observables in the hadronic sector.
In a first work in this direction, we study flavour violation in the 
squark decays of the second and third generation taking into account
results from B-physics, in particular from the rare decay 
$b \to s \gamma$.
We show that correlations between various squark decay modes can be used 
to get more precise information on various flavour violating parameters.
}



\subsubsection{Sources of Flavour Violation}

Within the Minimal Supersymmetric Standard Model (MSSM) there are two
new sources of flavour changing neutral currents, namely new
contributions which are induced through the quark mixing like in the
SM and generic supersymmetric contributions through the squark
mixing. In contrast to the Standard Model (SM), the structure of the
unconstrained MSSM does not explain the suppression of FCNC processes
which is observed in experiments; this is the essence of the
well-known supersymmetric flavour problem.  Flavour changing neutral
current (FCNC) processes therefore yield important (indirect)
information on the construction of supersymmetric extensions of the SM
and can contribute to the question of which mechanism ultimately
breaks supersymmetry. The experimental measurements of the rates for
these processes, or the upper limits set on them, impose in general a
reduction of the size of parameters in the soft supersymmetry-breaking
terms.

To understand the sources of flavour violation that may be present in
supersymmetric models, in addition to those enclosed in the CKM matrix K,
one has to consider the contributions to the squark mass matrices
\begin{equation}
{\cal M}_f^2 \equiv  \left( \begin{array}{cc}
  M^2_{\,f,\,LL} +F_{f\,LL} +D_{f\,LL}           & 
                 M_{\,f,\,LR}^2 + F_{f\,LR} 
                                                     \\[1.01ex]
 \left(M_{\,f,\,LR}^{2}\right)^{\dagger} + F_{f\,RL} &
             \ \ M^2_{\,f,\,RR} + F_{f\,RR} +D_{f\,RR}                
 \end{array} \right) \,,
\label{massmatrixd}
\end{equation}
where $f$ stands for up- or down-type squarks.
The matrices $M_{u,LL}$ and $M_{d,LL}$  are related by
$SU(2)_L$ gauge invariance. In the super-CKM basis, where the quark mass matrices are diagonal 
and the squarks are rotated in parallel to their superpartners, the relation
reads as 
$K^\dagger M^2_{u,LL} K =  M^2_{d,LL} = M^2_Q$. In this basis the F-terms
$F_{f\,LL}$, $F_{f\,RL}$, $F_{f\,RR}$ as well as the D-terms $D_{f\,LL}$
and  $D_{f\,RR}$ are diagonal. All the additional flavour structure of the
squark sector is encoded in the soft SUSY breaking terms $M^2_Q$,
$M^2_{\,f,\,RR}$ (=$M^2_U$ for $f=u$ and $M^2_D$ for $f=d$) and 
$M{\,f,\,LR}^2$ (=$v_u (A^u)^*$ for $f=u$ and $v_d (A^d)^*$ for $f=d$).
Note, that the $A$-matrices are in general non-hermitian.

 These additional
flavour structures induce flavour violating couplings to the neutral
gauginos and higgsinos in the mass eigenbasis which give rise to
additional contributions to observables in the $K$ and $B$ meson sector.
At present, new physics contributions to $s \to d$ and $b\to d$ 
transitions are strongly constrained.
In particular,  the transitions between first- and second-generation 
quarks, namely FCNC processes in  the $K$ system, are 
the most formidable tools to shape viable supersymmetric flavour models. 
As was recently emphasized again \cite{Ciuchini:2002uv}, 
most of the phenomena involving  $b \rightarrow s$ transitions 
are still largely unexplored and leave open the possibility of large 
new physics effects in spite of the strong bound of the
famous $B \rightarrow X_s \gamma$ decay which 
still gives the most stringent bounds in this sector. 
 Nevertheless, additional experimental 
information from the $B \rightarrow X_s \ell^+ \ell^-$ decay at the
B factories and new results on the $B_s - \bar B_s$ 
mixing  at the Tevatron might change this situation in the near future.
Within the present analysis, we take the present phenomenological 
situation into account by  setting the off-diagonal elements with an 
index 1 to zero.  
Regarding the $b \rightarrow s$ transitions, we restrict ourselves 
on the most powerful constraint from the  decay $B \rightarrow X_s \gamma$ 
only.

Two further remarks are in order: 
Within a phenomenological analysis of the constraints on the
flavour violating parameters in supersymmetric models with the most 
general soft terms in the squark mass matrices, we prefer to use  
the mass eigenstate formalism which remains valid 
(in contrast to the mass insertion approximation) 
when the intergenerational mixing elements are not small.
Moreover, a consistent analysis of the bounds  should 
also include interference effects between the various contributions, 
namely the interplay between the various sources of flavour violation 
and the interference effects  of SM, gluino, chargino, neutralino 
and charged Higgs boson contributions.  
In \cite{Besmer:2001cj} such an  analysis was performed for the example 
of the rare decay  $B \rightarrow X_s \gamma$ and new bounds on simple
combinations of elements of the soft part of the squark mass matrices are
found to be, in general, one order of magnitude weaker that the bound 
on the single off-diagonal elements $m_{LR,23}$ which was derived in previous 
work \cite{Gabbiani:1996hi,Hagelin:1992tc}, where
any kind of interference effects were neglected.

\subsubsection{Squark decays}

Squarks can decay into quarks of all generations of quarks once the
most general form the squark mass matrix is considered. The most
important decays modes for the example under study are:
\begin{eqnarray}
\tilde u_i &\to& u_j \tilde \chi^0_k \, , \,  d_j \tilde \chi^+_l \\
\tilde d_i &\to& d_j \tilde \chi^0_k \, , \,  u_j \tilde \chi^-_l 
\end{eqnarray}
with $i=1,..,6$, $j=1,2,3$, $k=1,..4$ and $l=1,2$.
These decays are controlled by the same mixing matrices as the
contributions to $b \to s \gamma$. As this decay mode restricts
the size of some of the elements, the questions arises to which extent
flavour violating squark decays are also restricted. We will show below
that flavour violating decay modes are hardly constrained by present days
data. 

We will take the so--called Snowmass point SPS\#1a as a specific example
which is specified by $m_0=100$~GeV, $m_{1/2}=250$~GeV, $A_0=-100$~GeV,
$\tan\beta=10$ and $\rm{sign}(\mu)=1$. At the electroweak scale
one gets the following data: $M_2=192$~GeV, $\mu=351$~GeV, $m_{H^+}=396$~GeV,
$m_{\tilde g}=594$~GeV, $m_{\tilde t_1}=400$~GeV, $m_{\tilde t_2}=590$~GeV,
$m_{\tilde q_R} \simeq 550$~GeV, and $m_{\tilde q_L} \simeq 570$~GeV.
We have used the program SPheno \cite{Porod:2003um} for the calculation.
In the following we will concentrate on the mixing between the second
and third generation. As a specific example we have added a  set of 
flavour violating parameters
given in Table~\ref{tab:par}; the resulting up-squark masses in GeV are
in ascending order: 408, 510, 529, 542, 558 and 627.
This point is a random one out of 1000 points 
fulfilling the $b\to s \gamma$ constraint.  For the calculation of
BR$(b\to s \gamma)$ we have used  the formulas given in 
ref.~\cite{Borzumati:1999qt}. Note, that for SPS\#1a both, the chargino as
well as the gluino loops, are important for the calculation of
BR$(b\to s \gamma)$. Therefore, there is an interplay between the flavour
structure of the down-type squarks and of the one of the up-type squarks.

In what follows we will concentrate on up-type squarks. However, we want
to note that also down-type squarks as well the gluino large flavour
violating decay modes\footnote{Strictly speaking  one should use the
expression 'generation violating decay modes' in this context.}.
The corresponding branching ratios into charginos and neutralinos are given 
in  Table~\ref{tab:br}. In addition the following branching ratios are larger
than 1\%: BR($\tilde u_3 \to \tilde u_1 Z$)=2.6\%,
 BR($\tilde u_3 \to \tilde u_1 h^0$)=1.2\%,
 BR($\tilde u_6 \to \tilde g c$) = 4\%,
 BR($\tilde u_6 \to \tilde d_1 W$)=2\%,
 BR($\tilde u_6 \to \tilde u_1 h^0$)=4.9\% and
  BR($\tilde u_6 \to \tilde u_2 Z$)=1.8\%.

\begin{table}
\begin{center}
\begin{tabular}{|ccc|cccc|}
\hline
$M^2_{Q,23}$ & $M^2_{D,23}$ & $M^2_{U,23}$ & $v_u A^u_{23}$ 
 & $v_u A^u_{32}$  & $v_d A^d_{23}$ & $v_d A^d_{32}$ \\ \hline
 47066 & 9399 & 46465 & 23896 & -44763 & 14470 & 15701 \\
\hline
\end{tabular}
\end{center}
\caption{Flavour violating parameters in GeV$^2$  which are
             added to the SPS\#1a point. The
         corresponding BR$(b\to s \gamma)$ is $4 \cdot 10^{-4}$.}
\label{tab:par}
\end{table}

\begin{table}
\begin{center}
\begin{tabular}{|c|cc|cc|cc|cc|cc|cc|}
\hline
          & $\tilde \chi^0_1 c$ &  $\tilde \chi^0_1 t$ & 
              $\tilde \chi^0_2 c$ &  $\tilde \chi^0_2 t$ & 
              $\tilde \chi^0_3 c$ &  $\tilde \chi^0_3 t$ &
              $\tilde \chi^0_4 c$ &  $\tilde \chi^0_4 t$ &
              $\tilde \chi^+_1 s$ &  $\tilde \chi^+_1 b$ & 
              $\tilde \chi^+_2 s$ &  $\tilde \chi^+_2 b$ \\ 
$\tilde u_1$ & 4.7                & 18 &
               5.2                &  9.6 & 
               $6 \, \, 10^{-3}$  & 0    &
               0.02               & 0    &
              11.3                & 46.4   &
               $2 \,\, 10^{-3}$  & 4.7 \\  
$\tilde u_2$ & 19.6               & 1.1  &
               0.4                & 17.5 &
               $2 \, \, 10^{-2} $ & 0  &
               $6 \, \, 10^{-2} $ & 0 &
               0.5                & 57.5 &
               $3 \, \, 10^{-3} $ &  2.9 \\
$\tilde u_3$ & 7.3                & 3.7  &
               20                 & 1.4  &
              $6 \, \, 10^{-2} $ & 0    &
               0.6                & 0    &
               40.3               & 3.1  &
               1                  & 18.5 \\
$\tilde u_6$ & 5.7                &  0.4  &
              11.1                &  5.3  &
               $4 \, \, 10^{-2} $ &  5.7  &
              0.6                 & 13.2  &
             22.9                 & 13.1  &
             0.6                  &  8.0 \\ 
\hline
\end{tabular}
\end{center}
\caption{Branching ratios (in \%) of u-type squarks for the point specified in 
         Table~\ref{tab:par}}
\label{tab:br}
\end{table}

It is clear from Table~\ref{tab:br} that all four particles have large
flavour changing decay modes. This clearly has an impact on the edge
variables, for example, the ones involving the second lightest
neutralino: $m^{max}_{llq}$, $m^{min}_{llq}$, $m^{low}_{lq}$, and 
$m^{high}_{lq}$ \cite{Giacomo}.  
In the studies for SPS\#1a  it has been assumed up to now
that the squarks under consideration have approximately the same mass
within a few percent. In this example the masses of the squarks 
range from 408 GeV up to 627 GeV. In particular $\tilde u_1$ and $\tilde u_6$
will give rise to additional structures in the lepton and jet distributions.
In such a case a refined analysis will be necessary to decide whether this
additional structure are caused by background, new particles or flavour
changing decay modes. Here it will be of clear advantage if a linear
collider could measure the branching of the lightest squark(s) to see
if there are sizable flavour violating decays in the squark sector.

In ref.~\cite{Hisano:2003qu} several variables have been proposed 
for extracting information on stops and sbottoms in gluino decays.
One class of these variables considers final states containing
$b \tilde \chi^+_1$. In our example, three $u$-type squarks 
contribute with branching
ratios larger than 10\%, in contrast to the 
assumption that only the two stops 
contribute.
As a consequence we expect that  additional structures
will be present in the corresponding observables. Moreover, we expect
also in this case that a combination of LHC and LC will be useful in
the exploration of these structures.

In conclusion, we have seen, that large flavour changing decays of squarks
are consistent with present days data from Tevatron and the B factories.
In this note we have concentrated on the decays of up-type squark. 
These decays will lead to additional structures in the lepton and jet
distributions which are used to determine the edge variables proposed
for the LHC. A linear with sufficient energy can in principle measure
the branching ratios of the lightest up- and/or down-type squark
proving the hypothesis of large flavour violation in the squark sector.
This information can then be put back in the analysis of the LHC data.




\subsection{\label{sec:423b} Supersymmetric lepton flavour violation at
LHC and LC}

{\it F.~Deppisch, J.~Kalinowski, H.~P\"as, A.~Redelbach and R.~R\"uckl}

\vspace{1em}
\noindent{\small
In supersymmetric extensions of the Standard Model, 
the Yukawa and/or mass terms of the heavy neutrinos can generate
lepton flavour violating slepton mass terms. These new
supersymmetric 
sources of lepton flavour violation may both enhance the rates of 
charged
lepton flavour violating processes, 
$l_\alpha \to l_\beta \gamma$, and   
generate   
distinct final states, like $l_\beta l_\alpha + {\rm jets} + {E\!\!\!/}_T$, 
at future colliders. 
First, we discuss the sensitivity of future $e^+e^-$ colliders 
to the SLFV independently of the lepton flavour violating mechanism.  
Second, we study lepton flavour violating slepton pair production and decay 
at a future $e^+e^-$ linear collider in the context of the 
seesaw mechanism in mSUGRA post-LEP benchmark scenarios. 
We investigate the correlations of these signals with the corresponding 
lepton flavour violating rare decays $l_{\alpha} \rightarrow l_{\beta} 
\gamma$, and show that these correlations 
are particularly suited for probing the origin of lepton flavour violation.
%
%

}

\subsubsection{Introduction}
Neutrino oscillations imply the violation of individual lepton flavours
and raise the interesting possibility of observing 
lepton flavour violation in processes with charged leptons, such
as $\mu\rightarrow e\gamma$ or $\tau\rightarrow\mu\gamma$. 
In the Standard Model these processes are
strongly suppressed due to small neutrino masses. In the
supersymmetric extension of the Standard Model, however,  
the situation may be quite different. 
For example,  the slepton mass matrices need
not simultaneously be diagonalized with the lepton mass matrices.  
When 
sleptons are rotated to the mass eigenstate basis, the slepton mass
diagonalization matrices $W_{i\alpha}$ enter the chargino and neutralino
couplings
\begin{eqnarray}
\tilde{e}_{i} (W^*_{\tilde{l}})_{i\alpha} \bar{e}_{\alpha} \tilde{\chi}^0 
+\tilde{\nu}_{i} (W^*_{\tilde{\nu}})_{i\alpha} \bar{e}_{\alpha} \tilde{\chi}^- 
+\ldots 
\end{eqnarray}
and mix lepton flavour (Latin and Greek subscripts 
refer to the mass-eigenstate and flavour 
basis, respectively). 
Contributions from virtual slepton exchanges can therefore
enhance the rates of rare decays like $\mu\to e\gamma$. 
Furthermore, once superpartners are discovered, the supersymmetric lepton
flavour violation (SLFV) can also be searched for directly at future
colliders where the signal will come  from the production
of real sleptons (either directly or from chain decays of other
sparticles), followed by their subsequent decays. 
Searches for SLFV at colliders have a number of
advantages: superpartners can be 
produced with large cross-sections, flavour violation in the  production and 
decay
of sleptons occurs at tree level 
and therefore
is suppresed only by powers of  $\Delta
m_{\tilde{l}}/\Gamma_{\tilde{l}}$ \cite{feng2} in contrast to the
$\Delta m_{\tilde{l}}/m_{\tilde{l}}$ suppression in radiative lepton decays,
where SLFV occurs at one-loop (\cite{Deppisch:2002vz} and references therein). 
Generally, respecting
the present bounds on rare
lepton decays, large SLFV signals are possible both at the LHC
\cite{sec423b_LHC} and at $e^+e^-$ colliders
\cite{feng2,ACFH205,nojiri2,GKR,PM,Deppisch:2003wt}. This suggests that    
in some cases the LHC and future $e^+e^-$ colliders may provide 
competitive tools to search for and
explore supersymmetric lepton flavour violation.

In this note we first discuss the sensitivity at future $e^+e^-$ colliders 
to SLFV independently of the lepton flavour violating mechanism.  
The simulation has been performed assuming
a simplified situation with a pure 2-3 intergeneration mixing
between $\tilde\nu_\mu$ and $\tilde\nu_\tau$, and ignoring any mixings
with $\tilde\nu_e$. In the analysis 
the mixing angle $\tilde{\theta}_{23}$ and $\Delta \tilde{m}_{23} =
|m_{\tilde\nu_2} - m_{\tilde\nu_3}|$ have been taken as free, independent
parameters \cite{GKR}. 

In the second part, SLFV generated by the seesaw mechanism is
considered. The heavy right-handed Majorana neutrinos give rise not
only to light neutrino masses but also to mixing of different slepton
flavours due to the effects of the heavy neutrinos
on the renormalization-group running of
the slepton masses.  The implications of recent neutrino measurements
on this mixing are investigated. Moreover we  emphasize the
complementarity of the radiative decays $l_\alpha \rightarrow
l_\beta \gamma$
and  the 
specific 
lepton flavour violating processes 
$e^{\pm}e^-\rightarrow l_{\beta}^{\pm} 
l_{\alpha}^- \tilde{\chi}_b^0 \tilde{\chi}_a^0$ 
involving slepton pair production and subsequent decay  
\cite{Deppisch:2003wt}.

\subsubsection{Sensitivity at future $e^+e^-$ colliders to SLFV
\label{sfcs}}

In discussing the SLFV collider signals at future colliders, one 
has to distinguish two cases in which an
oscillation of lepton flavour can occur: in processes with 
slepton pair production and in processes with 
single slepton production, which differ in the interference
of the intermediate sleptons \cite{feng2}. 
Slepton pair production is the dominant 
mechanism at lepton colliders, but it may also occur at hadron
colliders via the Drell-Yan process.
Single  sleptons  may be produced in
cascade decays of heavier non-leptonic superparticles. Such processes are
particularly important for
hadron colliders,  but they may also be relevant for lepton colliders where
a single slepton can be the decay product of a chargino or neutralino. 

The amplitudes for pair production,   
$ \bar f\,f\to \tilde{l}^+_i \, \tilde{l}^-_i
\to l^+_\alpha \, X\, l^-_\beta\, Y $, 
and single production,  $f\,f'\to l^+_\alpha \, X\,  \tilde{l}^-_i
\to l^+_\alpha \, X\, l^-_\beta\, Y$, read, e.g., 
\begin{eqnarray}
&&{\cal M}^{\rm pair}_{\alpha\beta}= \sum_i {\cal M}^{\rm pair}_P  \frac{i}
{q^2-\tilde{m}^2_i+i\tilde{m}_i\Gamma_i} W_{i\alpha} {\cal M}^+_D  \frac{i}
{p^2-\tilde{m}^2_i+i\tilde{m}_i\Gamma_i} W^*_{i\beta} {\cal M}^-_D 
\mbox{~~~ (s-channel)~~~} \nonumber \\
&& \label{amplseins}
\\
&&{\cal M}^{\rm sin}_{\alpha\beta}= \sum_i {\cal M}^{\rm sin}_P W_{i\alpha} \frac{i}
{q^2-\tilde{m}^2_i+i\tilde{m}_i\Gamma_i} W^*_{i\beta} {\cal M}^-_D
\label{amplszwo}
\end{eqnarray}
where ${\cal M}_P$ and ${\cal M}_D$ are the respective production   
and decay
amplitudes for sleptons in the absence of SLFV, and 
$W_{i\alpha}$ stands for the lepton flavour mixing matrix element.

For nearly degenerate
in mass and narrow sleptons, $\Delta \tilde{m}_{ij} \ll \tilde{m}$
and
$\tilde{m}\overline{\Gamma}_{ij} 
\simeq (\tilde{m}_i\Gamma_i+\tilde{m}_j\Gamma_j)/2\ll
\tilde{m}^2$,  the products of slepton 
propagators  can be simplified as follows
\begin{eqnarray}
\frac{i}{q^2-\tilde{m}^2_i+i\tilde{m}_i\Gamma_i}\;\frac{-i}{q^2- 
\tilde{m}^2_j-i\tilde{m}_j\Gamma_j} 
 \sim  \frac{1}{1+i\, \Delta \tilde{m}_{ij}/ \overline{\Gamma}_{ij}}\;
 \frac{\pi}{\tilde{m}\overline{\Gamma}_{ij}} 
\; \delta(q^2-\tilde{m}^2).
\label{propags}
\end{eqnarray}
Then, in the case of 
2-3 intergeneration  mixing, the cross-sections for the above processes
(\ref{amplseins}, \ref{amplszwo}), 
take a particularly simple form \cite{jk01}:     
\begin{eqnarray}
&&\sigma^{\rm pair}_{\alpha\beta}=\chi_{23}(3-4 \chi_{23})  
\sin^2 2\tilde{\theta}_{23} \;
\sigma(\bar f\,f\to \tilde{l}^+_\alpha \,
\tilde{l}^-_\alpha) 
Br(\tilde{l}^+_\alpha \to l^+_\alpha \, X)  
Br(\tilde{l}^-_\alpha \to  l^-_\alpha\, Y) \nonumber \\
&& \label{crosspair}\\ 
&&\sigma^{\rm sin}_{\alpha\beta}=\chi_{23} \sin^2 2\tilde{\theta}_{23}\;
\sigma(f\,f'\to l^+_\alpha \, X\, \tilde{l}^-_\alpha)
Br( \tilde{l}^-_\alpha \to l^-_\alpha\, Y) 
\end{eqnarray}
where  $\sigma(f\,f'\to l^+_\alpha \, X\, \tilde{l}^-_\alpha)$, 
$\sigma(\bar f\,f\to \tilde{l}^+_\alpha \,
\tilde{l}^-_\alpha)$ and $Br( \tilde{l}^\pm_\alpha \to l^\pm_\alpha X)$ 
are the corresponding  cross-sections 
and branching ratios in the absence of flavour violation.
The slepton flavour violating mixing effects are encoded in   
\begin{eqnarray}
\chi_{23} = \frac{x_{23}^2}{2(1+x_{23}^2)} \qquad {\mbox {\rm and}} \qquad 
\sin^2 2\tilde{\theta}_{23}
\end{eqnarray}
where $x_{23} = \Delta \tilde{m}_{23}/\overline{\Gamma}_{23}$. 
In the limit $x_{23}\gg 1$, $\chi_{23} $ approaches 1/2, the
interference can be neglected and the cross-sections behave as $\sigma
\sim \sin^2 2\tilde{\theta}_{23}$ . In the opposite case, the interference
suppresses the flavour changing processes, and $\sigma \sim (\Delta
\tilde{m}_{23} 
\sin2\tilde{\theta}_{23})^2$.

To assess the sensitivity of a 500 GeV  $e^+e^-$ linear collider 
to the SLFV, the following processes have been analysed
\begin{eqnarray}
e^+e^- & \rightarrow &
\tilde{\nu}_i\tilde{\nu}^c_j  \rightarrow  \tau^\pm\mu^\mp \tilde{\chi}^+_1
\tilde{\chi}^-_1 \label{snulfv}\\
e^+e^- & \rightarrow &
\tilde\chi^+_2\tilde{\chi}^-_1   \rightarrow  \tau^\pm\mu^\mp 
 \tilde\chi^+_1\tilde\chi^-_1  \label{charlfv}\\
e^+e^- & \rightarrow &
\tilde\chi^0_2\tilde{\chi}^0_1   \rightarrow  \tau^\pm\mu^\mp 
 \tilde\chi^0_1\tilde\chi^0_1  \label{neutlfv}
\end{eqnarray} 
Here $\tilde{\chi}^\pm_1 \rightarrow \tilde{\chi}^0_1 f\bar{f}'$, and
$\tilde{\chi}^0_1$ escapes detection.  The signature of SLFV  would
be $\tau^{\pm}\mu^{\mp}+
\mbox{4 jets}+ {E\!\!\!/}_T$,  $\tau^{\pm}\mu^{\mp}+ 
\ell + \mbox{2 jets}+ {E\!\!\!/}_T$, or $\tau^{\pm}\mu^{\mp}+ {E\!\!\!/}_T$, 
depending on the hadronic or leptonic
$\tilde{\chi}^\pm_1$ decay mode. The purely leptonic decay modes are
overwhelmed by background. In particular, the neutralino pair
production process (\ref{neutlfv}), 
which could still be open if the second chargino and sleptons were too
heavy for (\ref{snulfv}) and (\ref{charlfv}), 
is difficult to extract from background. On
the other hand, with charginos decaying hadronically, the signal 
$\tau^{\pm}\mu^{\mp}+ 4\mbox{ jets}+ {E\!\!\!/}_T$ comes from both 
processes (\ref{snulfv}) and (\ref{charlfv})    
and is SM-background free. The flavour-conserving processes analogous
to (\ref{snulfv}) and (\ref{charlfv}), but with two $\tau$'s in the final
state where  one of the $\tau$'s
decays leptonically to $\mu$, contribute to the
background. On the other hand, if 
jets are allowed to overlap,    
an important SM background to the final states with $\tau^\pm\mu^\mp + \ge 
\mbox{3 jets}+ {E\!\!\!/}_T$ comes from 
$e^+ e^- \rightarrow t \bar t g$.

The simulation of the signal and background has been performed
for one of the MSSM representative points chosen for
detailed case studies at the ECFA/DESY Workshop \cite{tesla}: a 
mSUGRA scenario defined by $m_0=100$ GeV, $M_{1/2}=200$ GeV, $A_0=0$
GeV, $\tan\beta=3$ and ${\rm sgn}(\mu)=+$.  A simple parton level
simulation has been performed with a number of kinematic cuts listed
in \cite{GKR}.   
For the processes (\ref{snulfv}) and (\ref{charlfv}) we
find after cuts the following cross-sections,
$\chi_{23}(3-4 \chi_{23}) \sin^2 2\tilde{\theta}_{23} \times 0.51$ fb and
$\chi_{23} \sin^2 2\tilde{\theta}_{23} \times 0.13$ fb,
respectively, while the background amounts to 0.28 fb.

\begin{figure}
\hspace*{1.5cm} $\Delta \tilde{m}_{23}$/GeV 
\vspace{-1cm}
 \begin{center}
\epsfig{figure=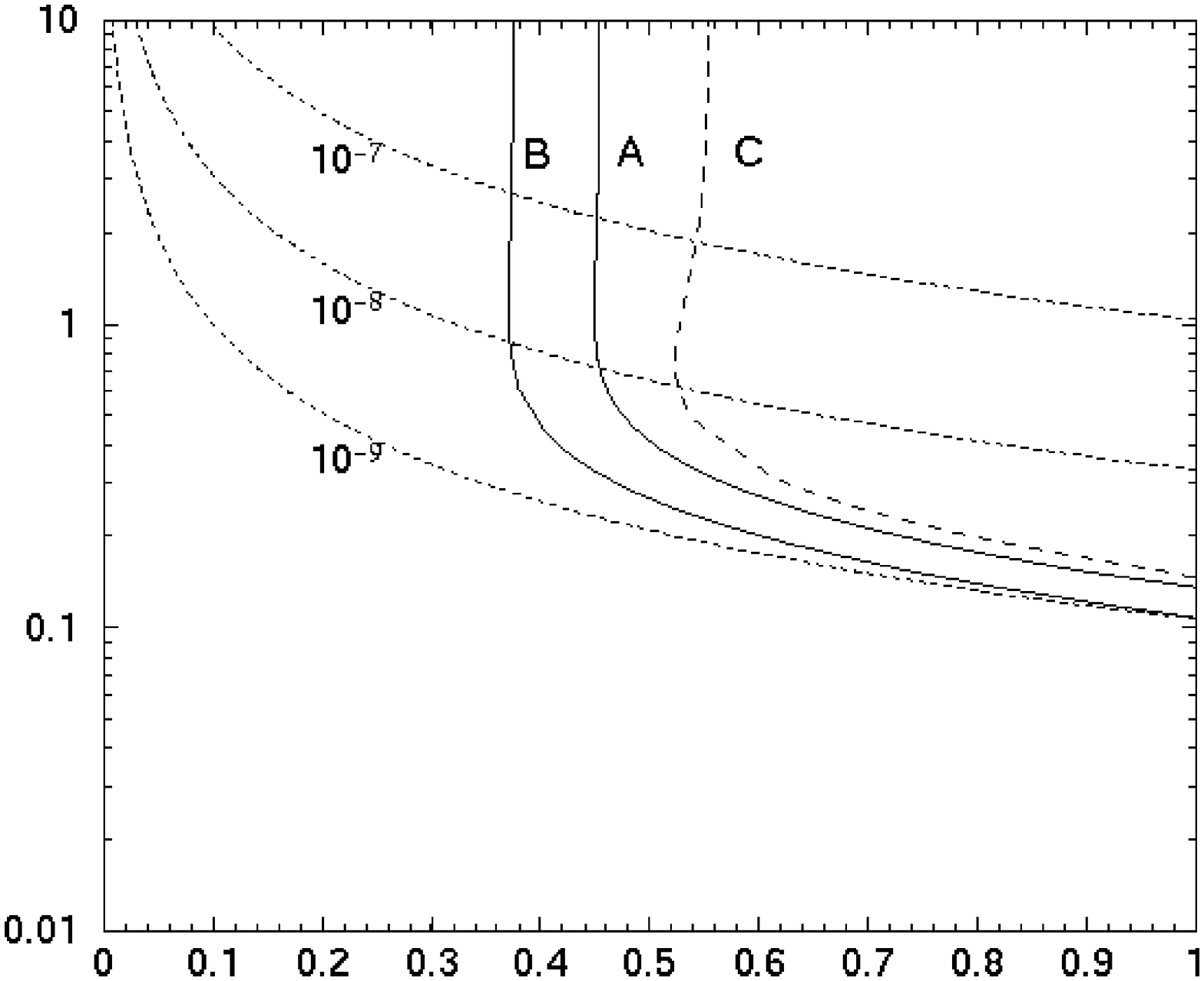,width=8cm,height=7cm}\\
$\sin2\tilde{\theta}_{23}$
 \end{center}
\caption{
Various 3$\sigma$ significance contours in 
the $\Delta \tilde{m}_{23} -\sin2\tilde{\theta}_{23}$ plane,
for the SUSY point mentioned in the text.
The contours A and B show the integrated signals
(\protect{\ref{snulfv}}--\protect{\ref{charlfv}}) 
at $\sqrt{s}=$500 GeV and for 500 fb$^{-1}$ and 1000 fb$^{-1}$, 
respectively.
The contour C shows the $\tilde\nu \tilde\nu^c$ contribution separately 
for 500 fb$^{-1}$ \cite{GKR}. The dotted lines indicate contours for
$Br(\tau\to\mu\gamma)$=10$^{-7}$, 10$^{-8}$ and 10$^{-9}$ \cite{jk02}.}
\label{fig_jk}
\end{figure}

In Fig.~\ref{fig_jk} the significance is given by $\sigma_d =
\frac{S}{\sqrt{S+B}}$ where S and B are the numbers of signal and
background events, respectively, for a given
luminosity. Shown is the
region (to the right of the curves) in the $\Delta \tilde{m}_{23} -
\sin2\tilde{\theta}_{23}$ plane that can be explored or ruled out at a
3$\sigma$ level at a linear collider of energy 500 GeV for the given
integrated luminosity. The contour A is for 500 fb$^{-1}$ and B
for 1000 fb$^{-1}$. For comparison, the boundary C shows the reach in
the process $\tilde{\nu}_i\tilde{\nu}^c_i$ alone (previously studied
in \cite{feng2,nojiri2}) using our cuts and assuming a luminosity of 500
fb$^{-1}$. The chargino contribution increases the sensitivity range
to $\sin 2\tilde{\theta}_{23}$ by 10-20\%, while the sensitivity to $\Delta
\tilde{m}_{{23}}$ does not change appreciably.

In the same figure, the contour lines for constant branching ratios of
$\tau\to \mu\gamma$ are shown for comparison \cite{jk02}. 
In the limit of small mass
splitting, $Br(\tau\to\mu\gamma)$ can be calculated in the flavour
basis using the mass
insertion technique \cite{gabbiani}.   
In our 2-3 intergeneration mixing scenario the  
radiative process $\tau\to\mu\gamma$   
constrains the combination of parameters    
\begin{equation}
\delta_{\mu\tau}= \sin2\tilde{\theta}_{23} \Delta
\tilde{m}_{23}/{\tilde m}. 
\end{equation}
The contours in Fig.~\ref{fig_jk} 
have been obtained from the approximate formula
of Ref.\cite{FNS}, normalized to the current experimental limit,
\begin{equation}
Br(\tau\to\mu\gamma)\sim 1.1 \times 10^{-6}
\left(\frac{\delta_{\mu\tau}}{1.4}\right)^2
\left(\frac{100 \mbox{
GeV}}{ {\tilde m}}\right)^4 \label{max}.
\end{equation}
This approximation 
only provides an order of magnitude estimate of the upper limit
for the supersymmetric contribution to the radiative lepton decay.  
The exact result, which is sensitive to the details of mass
spectra and mixings, can in fact be much smaller due to cancellations
among different contributions \cite{Deppisch:2002vz}.  
Fig.~\ref{fig_jk} demonstrates  that information from 
slepton production and decay could be competitive to the radiative lepton
decays.
In particular a LC  
can help to explore the 
small $\Delta \tilde{m}_{23}$ region.
It should be stressed, though, that in a given model for lepton flavour 
violation also the correlation with $\mu \to e \gamma$ has to be considered
\cite{Deppisch:2003wt},
which in many cases can yield a more severe bound, as discussed in the 
next section.


\subsubsection{Case study for the supersymmetric seesaw model}

As a definite and realistic example for SLFV we consider
the seesaw mechanism in mSU\-GRA models.
In supersymmetric theories with heavy right-handed Majorana neutrinos, 
the seesaw mechanism \cite{seesaw}
can give rise to light neutrino masses at or below 
the sub-eV scale. Furthermore, the massive neutrinos affect the 
renormalization group running of the slepton masses, 
generating flavour off-diagonal terms in the mass matrix.
These in turn lead to SLFV in scattering processes at high energies 
and in rare decays. For illustration
of the potential and complementarity of such SLFV searches we
focus on the LC processes $e^{\pm}e^-\rightarrow l_{\beta}^{\pm} 
l_{\alpha}^- \tilde{\chi}_b^0 \tilde{\chi}_a^0$ 
involving slepton pair production and subsequent decay, and
on the corresponding
radiative decay $l_\alpha \to l_\beta \gamma$. In particular,
in an early ATLAS note \cite{serin}
$\tau \to \mu \gamma$ is estimated to 
be observable at the LHC for a branching ratio of order $10^{-7}$.
However, the limit one can reasonably expect may be an order
of magnitude better \cite{denegri}. 

For our study we use the mSUGRA benchmark scenarios proposed in 
\cite{Battaglia:2001zp} for LC studies, concentrating on those
which predict charged left-handed sleptons that are light enough 
to be pair-produced at the center-of-mass energy $\sqrt{s}=500$~GeV.
Furthermore, we implement the seesaw mechanism assuming degenerate
Majorana masses for the right-handed neutrinos and constrain the
neutrino Yukawa couplings by the measured masses and mixings of
the light neutrinos. 
Further sources of SLFV exist in other models such as
GUTs \cite{guts}. However, no
realistic three generation case study of effects 
for collider processes has been 
performed so far, so that we restrict the discussion to the minimal seesaw 
model, here.

\vspace{1em}
\noindent
\underline{Supersymmetric seesaw mechanism}

If three right-handed neutrino singlet fields $\nu_R$
are added to the MSSM particle content, one has  
the additional terms \cite{Casas:2001sr}
\begin{equation}
W_\nu = -\frac{1}{2}\nu_R^{cT} M \nu_R^c + \nu_R^{cT} Y_\nu L \cdot H_2
\label{suppot4}
\end{equation}
in the superpotential.
Here, \(Y_\nu\) is the matrix of neutrino Yukawa couplings, 
$M$ is the right-handed neutrino Majorana mass matrix, and
$L$ and $H_2$ denote the left-handed 
lepton and hypercharge +1/2 Higgs doublets, respectively. 
At energies much below the mass scale $M_R$
of the right-handed neutrinos, 
$W_{\nu}$ leads to the following
mass matrix for the light neutrinos:
\begin{equation}\label{eqn:SeeSawFormula}
M_\nu = Y_\nu^T M^{-1} Y_\nu (v \sin\beta )^2.
\end{equation}
From that the light neutrino masses $m_1$, $m_2$, $m_3$
are obtained after diagonalization by the unitary MNS matrix \(U\).
The basis is chosen such that the matrices of the charged lepton 
Yukawa couplings and Majorana masses are diagonal, which is always 
possible. 

Furthermore, the heavy neutrino mass eigenstates give rise 
to virtual corrections to the slepton mass matrix 
that are responsible for lepton flavour violating processes.
More specifically, in the mSUGRA models considered, 
the mass matrix of the charged sleptons is given by 
\begin{eqnarray}
 m_{\tilde l}^2=\left(
    \begin{array}{cc}
        m_{\tilde l_L}^2    & (m_{\tilde l_{LR}}^{2})^\dagger \\
        m_{\tilde l_{LR}}^2 & m_{\tilde l_R}^2
    \end{array}
      \right)
\label{mslept}
\end{eqnarray}
with
\begin{eqnarray*}
  (m^2_{\tilde{l}_L})_{ij}     \!\!\!&=&\!\!\! (m_{L}^2)_{ij} 
+ \delta_{ij}\bigg(m_{l_i}^2 
+ m_Z^2 
\cos 2\beta \left(-\frac{1}{2}+\sin^2\theta_W \right)\bigg) 
\label{mlcharged} \\
  (m^2_{\tilde{l}_{R}})_{ij}     
\!\!\!&=&\!\!\! (m_{R}^2)_{ij} 
+ \delta_{ij}(m_{l_i}^2 - m_Z^2 \cos 2\beta 
\sin^2\theta_W) \label{mrcharged} \\
 (m^{2}_{\tilde{l}_{LR}})_{ij} 
\!\!\!&=&\!\!\! A_{ij}v\cos\beta-\delta_{ij}m_{l_i}\mu\tan\beta.
\end{eqnarray*}
When $m^2_{\tilde{l}}$
is evolved from the GUT scale \(M_X\) to the electroweak scale 
characteristic for the experiments, one obtains
\begin{eqnarray}
m_{L}^2\!\!\!&=&\!\!\!m_0^2\mathbf{1} + (\delta m_{L}^2)_{\textrm{\tiny MSSM}} 
+ \delta m_{L}^2 \label{left_handed_SSB} \\
m_{R}^2\!\!\!&=&\!\!\!m_0^2\mathbf{1} + (\delta m_{R}^2)_{\textrm{\tiny MSSM}} 
+ \delta m_{R}^2 \label{right_handed_SSB}\\
A\!\!\!&=&\!\!\!A_0 Y_l+\delta A_{\textrm{\tiny MSSM}}+\delta A \label{A_SSB},
\end{eqnarray}
where $m_{0}$ is the common soft SUSY-breaking scalar mass and $A_{0}$ the 
common trilinear coupling. The terms 
\((\delta m_{L,R}^2)_{\textrm{\tiny MSSM}}\) and 
\(\delta A_{\textrm{\tiny MSSM}}\) are well-known flavour-diagonal MSSM
corrections. In addition, the evolution generates the
off-diagonal terms $\delta m^2_{L,R}$ and $\delta A$
which, in leading-log approximation and for degenerate 
right-handed Majorana masses \(M_i=M_R,i=1,2,3\),
are given by \cite{Hisano:1998fj}
\begin{eqnarray}\label{eq:rnrges}
  \delta m_{L}^2 \!\!\!&=&\!\!\! -\frac{1}{8 \pi^2}(3m_0^2+A_0^2)(Y_\nu^\dag Y_\nu) 
\ln\left(\frac{M_X}{M_R}\right) \label{left_handed_SSB2}\\
  \delta m_{R}^2\!\!\! &=&\!\!\! 0  \\
  \delta A\!\!\! &=&\!\!\! -\frac{3 A_0}{16\pi^2}(Y_l Y_\nu^\dag Y_\nu) 
\ln\left(\frac{M_X}{M_R}\right).
\end{eqnarray}

In order to determine the product $Y_\nu^\dagger Y_\nu$ 
of the neutrino Yukawa coupling matrix entering these corrections, 
one uses the expression   
\begin{eqnarray}\label{eqn:y}
 Y_\nu  = \frac{\sqrt{M_R}}{v\sin\beta} R \cdot 
\textrm{diag}(\sqrt{m_1},\sqrt{m_2},\sqrt{m_3}) \cdot  U^\dagger,
\end{eqnarray}
which follows from \(U^T M_\nu U = \textrm{diag}(m_1,m_2,m_3)\) and 
(\ref{eqn:SeeSawFormula}) \cite{Casas:2001sr}.
Here, \(R\) is an unknown complex orthogonal matrix parametrizing
the ambiguity in the relation of Yukawa coupling and mass matrices. 
In the following we will assume \(R\) to be real which suffices for
the present purpose. In this case, 
\(R\) drops out from the product $Y_\nu^\dagger Y_\nu$,
\begin{eqnarray}\label{eqn:yy}
 Y_\nu^\dagger Y_\nu = \frac{M_R}{v^2\sin^2\beta} U \cdot
\textrm{diag}(m_1,m_2,m_3) \cdot U^\dagger.
\end{eqnarray}
Using existing neutrino data on the mass squared differences  
and the mixing matrix \(U\) together with bounds and assumptions 
on the absolute mass scale
one can calculate 
$Y_\nu^\dagger Y_\nu$. The only free parameter is the Majorana 
mass scale $M_R$. The result is 
then evolved
to the unification scale $M_X$ and
used as an input in the renormalization group corrections 
(\ref{eq:rnrges}) to the slepton mass matrix.
Finally, diagonalization of (\ref{mslept}) yields the 
slepton mass eigenvalues 
\(\tilde m_i\) and eigenstates \(\tilde{l}_i\) (\(i=1,2,...6)\).
  
\vspace{1em}
\noindent
\underline{Lepton flavour violating processes}

The flavour off-diagonal elements (\ref{eq:rnrges})
in  \(m_{\tilde l}^2\) ($\delta A=0$ in the mSUGRA
scenarios of \cite{Battaglia:2001zp})
induce, among other SLFV effects,
the processes $e^+e^- \to \tilde{l}^+_j \tilde{l}_i^-\to 
l^+_{\beta}l_{\alpha}^-\tilde{\chi}^0_b\tilde{\chi}^0_a$,
where SLFV can occur in the production and decay vertices.
The helicity amplitudes for the 
pair production of \(\tilde{l}_j^+\) and \(\tilde{l}_{i}^-\), 
and the corresponding decay amplitudes 
are given explicitly in \cite{Deppisch:2003wt}. 
In the approximation (\ref{crosspair}) for $\sigma^{\rm pair}_{\alpha \beta}$
one finds
\begin{eqnarray}
\sigma^{\rm pair}_{\alpha\beta}\propto 
\frac{|(\delta{m}_L)^2_{\alpha\beta}|^2}{
\tilde{m}^2 \Gamma^2} \;
\sigma(\bar f\,f\to \tilde{l}^+_\alpha \,
\tilde{l}^-_\alpha) 
Br(\tilde{l}^+_\alpha \to l^+_\alpha \, \tilde{\chi_0})  
Br(\tilde{l}^-_\alpha \to  l^-_\alpha\, \tilde{\chi_0})
\label{full_M_squared}
\end{eqnarray}
In the numerical evaluation no slepton degeneracy has been assumed
as in (\ref{propags}), and the amplitude 
for the complete \(2 \to 4\) processes
is summed coherently over the intermediate
slepton mass eigenstates. 

Similary, the terms (\ref{eq:rnrges}) are responsible 
for SLFV radiative decays \(l_\alpha\rightarrow l_\beta \gamma\)
induced by photon-penguin type diagrams with 
charginos / sneutrinos or neutralinos / charged sleptons in the loop.
Again schematically, the
decay rates are given by \cite{Casas:2001sr,Hisano:1998fj}
\begin{equation}
\Gamma(l_\alpha \rightarrow l_\beta \gamma) 
\propto \alpha^3 m_{l_\alpha}^5 
\frac{|(\delta m_L)^2_{\alpha \beta}|^2}{\tilde{m}^8} 
\tan^2 \beta,
\end{equation}
where $\tilde m$ stands for the relevant sparticle masses in the loop.

\begin{figure}[t]
\centering
\includegraphics[clip,scale=0.7]{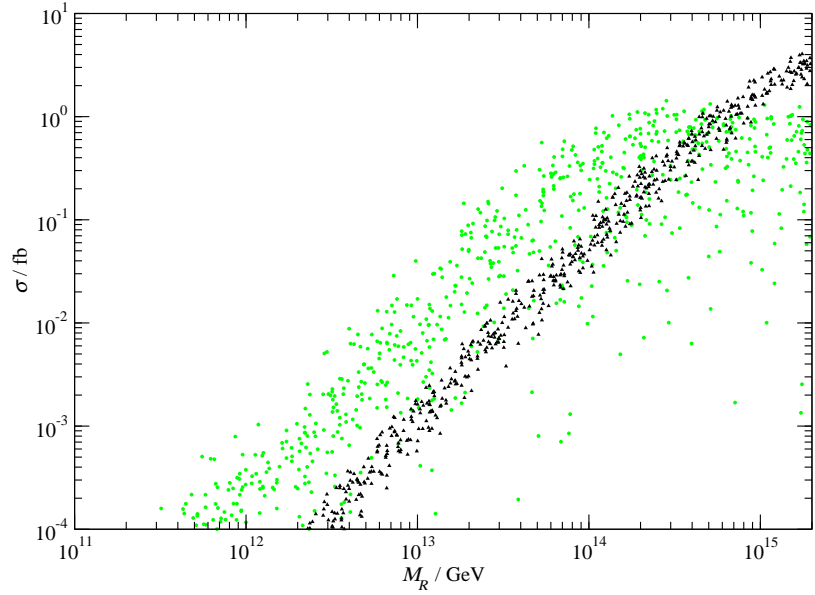}
     \caption{Cross-sections at \(\sqrt{s}=500\) GeV
              for \(e^+e^- \to \mu^+e^- +2\tilde\chi_1^0\) (circles)
              and  \(e^+e^- \to \tau^+\mu^- +  2\tilde\chi_1^0\) (triangles) 
              in scenario B.}
     \label{fig:ep}
\end{figure}

\begin{figure}[t]
\centering
\includegraphics[clip,scale=0.83]{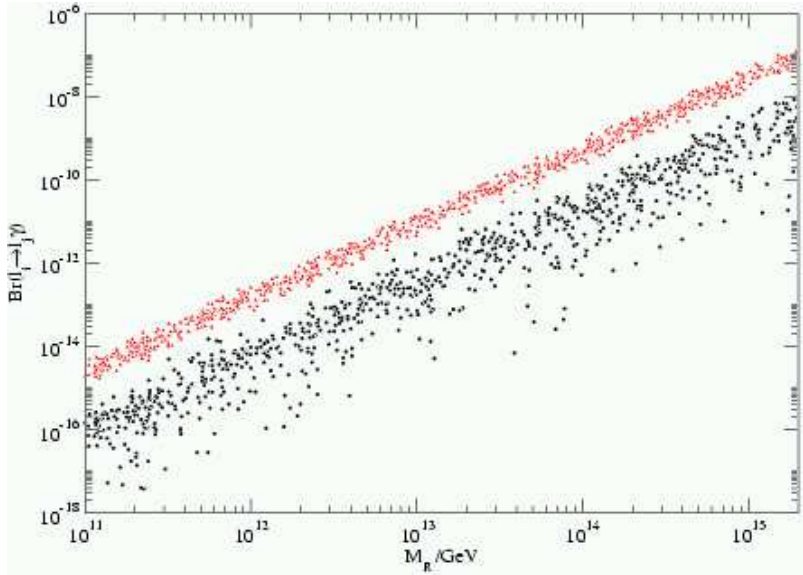}
     \caption{Branching ratios $Br(\tau \to \mu \gamma)$ (upper)
and $Br(\mu \to e \gamma)$ (lower)
in scenario B.}
     \label{fig:mue-taumu}
\end{figure}

\begin{table}[h!]
\begin{center}
\begin{tabular}{|c|c|c|c||c|c|c|}\hline
Scenario & $m_{1/2}$/GeV  & $m_{0}$/GeV & $\tan\beta$ & 
${\tilde m}_{6}$/GeV &${\tilde \Gamma}_6$/GeV & 
$m_{\tilde{\chi}_1^0}$/GeV  \\ \hline\hline
B & 250  & 100  & 10  & 208 & 0.32 & 98  \\ \hline
C & 400  & 90   & 10  & 292 & 0.22 & 164  \\ \hline
G & 375  & 120  & 20  & 292 & 0.41 & 154  \\ \hline
I & 350  & 180  & 35  & 313 & 1.03 & 143  \\ \hline
\end{tabular}
\end{center} 
\caption{\label{mSUGRAscen} Parameters of selected 
mSUGRA benchmark scenarios (from \protect{\cite{Battaglia:2001zp}}). 
The sign of \(\mu\) is chosen to be positive and \(A_0\) is set to 
zero. Given are also the mass and total width of the heaviest
charged slepton and the mass of the lightest neutralino.}
\end{table}

\vspace{1em}
\noindent
\underline{Signals and background}

Among the mSUGRA benchmark scenarios proposed in 
\cite{Battaglia:2001zp} for LC studies,
the models B, C, G, and 
I (see Tab.\ref{mSUGRAscen}) 
predict left-handed sleptons which can be pair-produced 
at \(e^+e^-\) colliders \(\sqrt{s}=500\div800\)~GeV cms energies. 
In the following we will confine ourselves 
to these models.

Most likely, at the time when a linear collider will 
be in operation, more precise measurements of the
neutrino parameters will be available than today.
In order to simulate the expected improvement, 
we take the central values of the mass squared 
differences $\Delta m^2_{ij}=|m_i^2-m_j^2|$
and mixing angles $\theta_{ij}$ from a global fit to existing 
data \cite{Gonzalez-Garcia:2000sq} with errors that indicate 
the anticipated 90 \% C.L. intervals of running and proposed 
experiments as further explained in \cite{Deppisch:2002vz}: 
\begin{eqnarray}
&&\tan^2\theta_{23}=1.40^{+1.37}_{-0.66},~~
\tan^2\theta_{13}=0.005^{+0.001}_{-0.005},~~
\tan^2\theta_{12}=0.36^{+0.35}_{-0.16}, \label{nupar1}\\
&&\Delta m_{12}^2=3.30^{+0.3}_{-0.3}\cdot 10^{-5}\textrm{ eV}^2 ,~~
\Delta m_{23}^2=3.10^{+1.0}_{-1.0}\cdot 10^{-3}\textrm{ eV}^2.
\end{eqnarray}
Furthermore, for the lightest neutrino we assume the mass range
\(m_1 \approx 0-0.03\)~eV, which at the lower end corresponds to
the case of a hierarchical spectrum. Towards the upper end, it approaches
the degenerate case.

\begin{figure}[t]
\centering
\includegraphics[clip,scale=0.83]{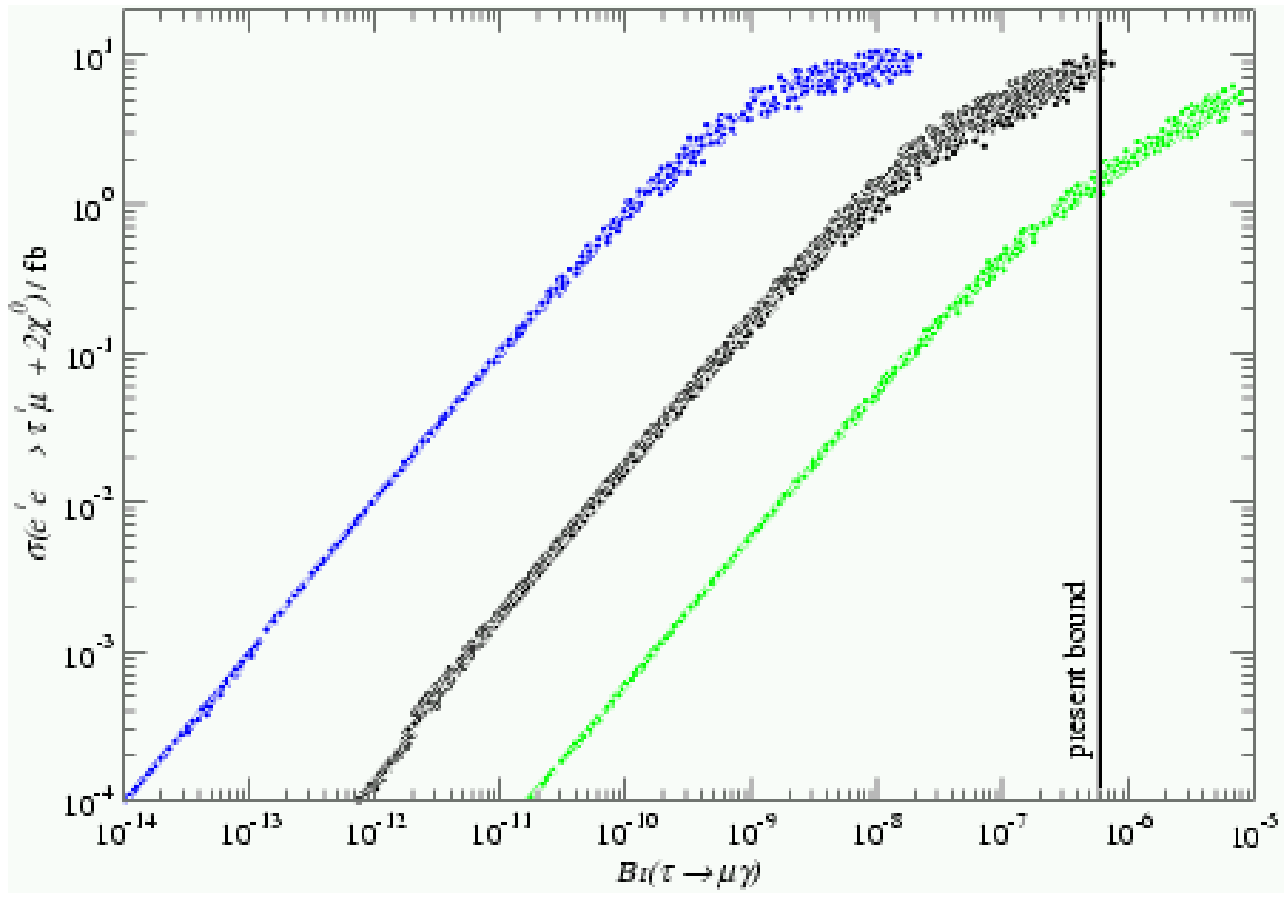}
     \caption{Correlation of \(\sigma(e^+e^- \to \tau^+\mu^- +2\tilde\chi_1^0)\)
              at \(\sqrt{s}=800\) GeV  
              with \(Br(\tau\to \mu\gamma)\) in scenario (from left to right) 
              C, G (open circles), B and I.}
     \label{fig:mutau_lowhigh}
\end{figure}

\begin{figure}[t]
\centering
\includegraphics[clip,scale=0.7]{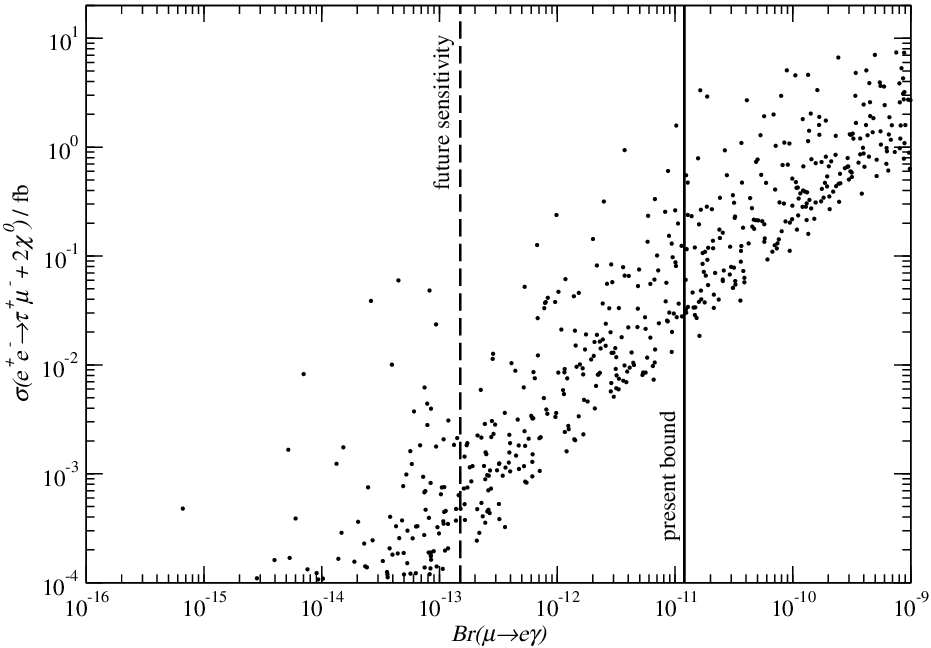}
     \caption{
Correlation of \(\sigma(e^+e^- \to \tau^+\mu^- +2\tilde\chi_1^0)\)
              at \(\sqrt{s}=800\) GeV  
              with \(Br(\mu\to e\gamma)\) in scenario B.}
     \label{fig:mutau800}
\end{figure}


In Fig.~\ref{fig:ep}, the cross-sections for 
\(e^+e^-\rightarrow \mu^+ e^- +2\tilde{\chi}^0_1\) and 
\(e^+e^-\rightarrow \tau^+ \mu^- +2\tilde{\chi}^0_1\) 
are plotted for model B. 
The channel \(\tau^+ e^- +2\tilde{\chi}_1^0\) is not
shown since it is strongly suppressed by the small mixing angle 
\(\theta_{13}\), and therefore more difficult to observe.
As can be seen, for a sufficiently large Majorana mass scale 
the SLFV cross-sections can reach several fb. 
The spread of the predictions reflects the uncertainties in the neutrino 
data.

The Standard Model background mainly comes from $W$-pair production, 
$W$ production with $t$-channel photon exchange, and $\tau$-pair production.
A 10 degree beam pipe cut and   
cuts on the lepton energy and missing energy reduce the SM background 
cross-sections to less than 30 ~fb for \((\mu e)\) final states and less
than 10 ~fb for \((\tau \mu)\) final states. If one requires a signal
to background ratio, $S/\sqrt{S+ B} = 3$,
and assumes a typical signal cross-section
of 0.1~fb, one can afford a background of about 1~fb. 
Here an integrated luminosity of 1000~fb$^{-1}$ has been assumed.
Whether or not the background process estimate above 
can be further suppressed to this level by applying  
selectron selection cuts, for example, on the acoplanarity, 
lepton polar angle and missing transverse momentum has to be studied in 
dedicated simulations. 
For lepton flavour conserving processes it has been shown
that the SM background 
to slepton pair production can be reduced to about 2-3~fb
at $\sqrt{s}=500$~GeV \cite{Becker:1993fw}.
 
The MSSM background is dominated by
chargino/slepton production with
a total cross-section of
0.2-5~fb and 2-7~fb for \((\mu e )\) 
and \((\tau \mu)\) final states, 
respectively, depending on the SUSY scenario
and the collider energy.
The MSSM
background in the ($\tau e$) channel can also
contribute to the \(\mu e\)
channel via the decay $\tau \rightarrow \mu \nu_{\mu} {\nu}_{\tau}$.
If $\tilde{\tau}_1$ and $\tilde{\chi}^+_1$ are very light, like in scenarios B 
and I, this background can be as large as 20~fb.
However, such events typically contain two neutrinos in addition to the
two LSPs which are also present in the signal events. Thus, after $\tau$ decay 
one has altogether
six invisible particles instead of two, which may allow to discriminate 
the signal 
in $\mu^+e^-+\;/\!\!\!\!E$ also from this potentially 
dangerous MSSM background 
by cutting on various distributions.
But also here one needs a dedicated simulation study, in order to make more 
definite statements.

The corresponding branching ratios,
$Br(\mu \to e \gamma)$ and $Br(\tau \to \mu \gamma)$, 
in model B are displayed in 
Fig.~\ref{fig:mue-taumu}
\cite{Deppisch:2002vz}. 
One sees that
a positive signal for $\mu \to e \gamma$ at 
the minimum branching 
ratio observable in the new PSI experiment,
$Br(\mu \to e \gamma)\simeq 10^{-13}$ \cite{PSI}
would imply a value of 
$M_R$ between $2\cdot 10^{12}$~GeV and $2\cdot 10^{13}$~GeV.
In comparison to $\mu \to e \gamma$ the channel
\(\tau\to\mu\gamma\)
is less affected by the neutrino uncertainties. 
If the sensitivity goal \(Br(\tau\to\mu\gamma)=10^{-8}\) \cite{denegri}
at the LHC is reached
one could probe $M_R=10^{15}$~GeV.

Particularly
interesting and useful are the 
correlations between SLFV in radiative decays and 
slepton pair production. Such a correlation is illustrated in 
Fig.~\ref{fig:mutau_lowhigh} for 
\(e^+e^-\rightarrow \tau^+\mu^- +2\tilde{\chi}_1^0\)
and \(Br(\tau\to \mu \gamma)\).
One sees that the neutrino uncertainties 
drop out, 
while the sensitivity to the mSUGRA parameters remains.
An observation of $\tau \to \mu \gamma$ with the branching ratio
$10^{-8}$ at the LHC would be compatible   
with a cross-section of order 10 fb for 
$e^+e^- \to \sum_{i,j}\tilde{l}_j^+\tilde{l}^-_i\to 
\tau^+ \mu^- +2\tilde{\chi}^0_1$, at least in model C. 
However, there are also correlations of different flavor channels.
This is illustrated 
in Fig.~\ref{fig:mutau800}, where the correlation of 
\(e^+e^-\rightarrow \tau^+\mu^- +2\tilde{\chi}_1^0\)
and \(\mu\to e \gamma\) is shown. Despite of the uncertainties 
from the neutrino sector, 
already the present experimental bound 
\(Br(\mu\to e \gamma)<1.2 \cdot 10^{-11}\)
yields a stronger constraint on $\sigma(e^+e^- \to
\tau^+ \mu^- +2 \tilde\chi_1^0)$
than the one obtained
from Fig.~\ref{fig:mutau_lowhigh}, making 
cross-sections larger than a few $10^{-1}$~fb  
at \(\sqrt{s}=800\)~GeV very unlikely in model B.
If this scenario is correct,
non-observation of 
$\mu \rightarrow e \gamma$
at the new PSI experiment will exclude
the observability of this channel at a LC.
As a final remark we stress that in the channel $e^+e^- \to
\mu^+ e^- +2 \tilde\chi_1^0$ cross-sections of
1~fb are compatible with the present bounds, while no signal at the 
future PSI sensitivity would constrain this channel to less than 0.1~fb. 
However we want to emphasize again that these statements are very model 
dependent, and much bigger cross-sections are possible in general, as shown 
in section \ref{sfcs}.


\subsubsection{Summary and outlook}
If superpartners are discovered at future colliders, 
we advocate the search for SUSY lepton flavour violation 
as a high priority topic of the experimental programme. 
At a LC, the most favourable signals are expected to come from
the production and decay of sleptons and charginos. 
Considering only LFV in the $\mu-\tau$ sector, 
a case motivated by the large atmospheric neutrino mixing 
but more difficult to detect than LFV in the $e-\mu$ sector 
due to the presence of decaying taus,
we have shown that the LC measurements may be complementary to 
searches for the radiative $\tau$ decay at the LHC. 
For example, a measurement  of 
$Br(\tau \to \mu \gamma) = 10^{-8}$ at the LHC  
combined with the SLFV signal at a LC would point to 
$\sin2\tilde\theta_{23}\geq 0.4$ and $\Delta\tilde m_{23}\simeq 0.3 - 1$ 
GeV. 
 
In the context of the SUSY seesaw
mechanism of neutrino mass generation,  
correlations between SLFV in radiative decays and 
slepton pair production have been found particularly
interesting. For instance, in a given MSSM scenario the
measurement of $\tau \rightarrow \mu \gamma$ at the LHC
would imply a definite cross section for  
$e^+e^- \to \tau^+ \mu^- +2 \tilde\chi_1^0$ at the LC.
Assuming a reasonable set of MSSM benchmark scenarios and 
$Br(\tau \to \mu \gamma) = 10^{-8}$ and using the 
present neutrino data, one predicts
$\sigma(e^+e^- \to \tau^+ \mu^- +2 \tilde\chi_1^0)$
in the range 0.05 to 10 fb. However, the
non-observation of $\mu \to e \gamma$ with
a branching ratio of about $10^{-13}$ 
at the new PSI experiment would exclude
the observability of $\sigma(e^+e^- \to
\tau^+ \mu^- +2 \tilde\chi_1^0)$ at a linear collider.
While the former correlation involving the same lepton flavours 
is insensitive to the uncertainties in the neutrino 
data (Fig. \ref{fig:mutau_lowhigh}), 
the latter correlation is somewhat smeared out 
(Fig. \ref{fig:mutau800}). 
However, both types of correlations remain sensitive to the mSUGRA 
parameters and, hence, provide very useful
tools for probing the origin of lepton flavour violation.
 
The complementarity of the LHC and LC (and of low-energy experiments) 
in the context of lepton
flavour violation is far from being exhausted by the present study. 
Quantitative analyses of
the impact of precise mass measurements at the LC on identifying the LFV
decay chains at the LHC (and vice-versa) and other important features  
call for detailed Monte Carlo simulations
which should be undertaken in the next round of the LHC/LC studies.


%

\subsection{\label{sec:425} Detection difficulties for MSSM and other
SUSY models for special boundary conditions}

{\it J.~Gunion}

\vspace{1em}
\def\bit{\begin{itemize}}
\def\eit{\end{itemize}}
\def\anti{\overline}
\def\beq{\begin{equation}}
\def\eeq{\end{equation}}
\def\bea{\begin{eqalign}}
\def\eea{\end{eqalign}}
\def\gev{~{\rm GeV}}
\def\mev{~{\rm MeV}}
\def\lsim{\mathrel{\raise.3ex\hbox{$<$\kern-.75em\lower1ex\hbox{$\sim$}}}}
\def\gsim{\mathrel{\raise.3ex\hbox{$>$\kern-.75em\lower1ex\hbox{$\sim$}}}}

\def\etmiss{/ \hskip-7pt E_T}
\def\emiss{/ \hskip-7pt E}
\def\etmin{/ \hskip-7pt E_T^{\rm min}}
\def\etjet{E_T^{\rm jet}}
\def\ptmiss{/ \hskip-7pt p_T}
\def\mslash{/ \hskip-7pt M}
\def\rslash{/ \hskip-7pt R}
\def\susyslash{\susy\hskip-24pt/\hskip19pt}
\def\mmissl{M_{miss-\ell}}
\def\mhalf{m_{1/2}}
\def\wtil{\widetilde}
\def\what{\widehat}
\def\wt{\wtil}
\def\gl{\wt g}
\def\mgl{m_{\gl}}
\def\wmp{W^{\mp}}
\def\chitil{\wt\chi}
\def\cnone{\wt\chi^0_1}
\def\cnonestar{\wt\chi_1^{0\star}}
\def\cntwo{\wt\chi^0_2}
\def\cnthree{\wt\chi^0_3}
\def\cnfour{\wt\chi^0_4}
\def\snu{\wt\nu}
\def\snul{\wt\nu_L}
\def\msnul{m_{\snul}}

\def\snue{\wt\nu_e}
\def\snuel{\wt\nu_{e\,L}}
\def\msnuel{m_{\snul}}

\def\snubar{\ov{\snu}}
\def\msnu{m_{\snu}}
\def\mcnone{m_{\cnone}}
\def\mcntwo{m_{\cntwo}}
\def\mcnthree{m_{\cnthree}}
\def\mcnfour{m_{\cnfour}}

\def\wt{\widetilde}
\def\wh{\widehat}
\def\cpone{\wt \chi^+_1}
\def\cmone{\wt \chi^-_1}
\def\cpmone{\wt \chi^{\pm}_1}
\def\cmpone{\wt \chi^{\mp}_1}
\def\mcpone{m_{\cpone}}
\def\mcpmone{m_{\cpmone}}

\def\cptwo{\wt \chi^+_2}
\def\cmtwo{\wt \chi^-_2}
\def\cpmtwo{\wt \chi^{\pm}_2}
\def\mcptwo{m_{\cptwo}}
\def\mcpmtwo{m_{\cpmtwo}}
\def\stautwo{\wt \tau_2}
\def\mstautwo{m_{\stauone}}

\def\dmchi{\Delta m_{\tilde\chi_1}}
\def\lampr{\lam^\prime}
\def\lampp{\lam^{\prime\prime}}

\subsubsection{Anomaly-mediated boundary conditions ---
the degenerate wino-LSP scenario}

One interesting limiting case of {\sc Susy} boundary conditions 
is loop-induced gaugino masses proportional to the respective 
renormalization group beta functions, as in Anomaly-Mediated
{\sc Susy} breaking (AMSB) and the simplest moduli-dominated {\sc Susy}
breaking models. 
In such models, the mass difference $\dmchi\equiv \mcpmone-\mcnone$
is most typically very small.
The phenomenology of $\cpmone$ production and decay has been studied
extensively in \cite{cdg,gm}, and also in \cite{randall,wells}.
For $\dmchi< m_\pi$, as achievable in some models
even after radiative corrections, the $\cpmone$
may exit the detector,
leading to an easily detected long-lived
heavily-ionizing track ({\sc Lhit}), or decay to a
soft, but visible, $e$ or $\mu$ track in the
vertex detector and inner part of the tracker
yielding a disappearing isolated track ({\sc
  Dit}). For $\dmchi$ above, but close to $m_\pi$,
the $\cpmone$ decay to a soft $\pi^\pm$ may still
have sufficient path length (including sufficient
$\beta$) to leave a distinctive track in the
vertex detector while also having low enough
$\beta$ (e.g. $\beta<0.8$) to be highly-ionizing,
yielding a ``{\sc Stub}'' track; the decay $\pi$
need not be visible.  For somewhat larger
$\dmchi$, the {\sc Stub} is not visible, but the
soft $\pi^\pm$ will be detected and may have a
high impact parameter ({\sc Hip}).  For still
larger $\dmchi$, the $\pi^\pm$ will simply be
soft, but readily detected. Once $\dmchi\gsim
2-3\gev$, the $\cpmone$ decay products have
significant energy. The most typical range for
$\dmchi$ in the models studied is
$\dmchi\in[200\mev,2\gev]$.
 
Assuming that only the $\cnone$ and $\cpmone$ are light, the
primary LC
{\sc Susy} processes are
$\cpone\cmone$, $\gam\cpone\cmone$, and $\wmp\cnone\cpmone$ production.
Observable signatures are:\\
$\bullet$~~$\dmchi<m_\pi$:
the $\cpmone$ yields a `stable particle' {\sc Lhit} 
and/or {\sc Dit} track and is easily detected: $\cpone\cmone$ production
will be easily seen. \\
$\bullet$~~$m_\pi<\dmchi<200\mev$:
the $\cpmone$ decay yields a {\sc Stub} track. Assuming small $\gam\gam$
background, no trigger would be needed; 
we denote such a signal by {\sc Snt}.\\
$\bullet$~~$200\mev<\dmchi<2-3\gev$:
the $\cpmone$ decay yields a soft, possibly {\sc Hip} ($\dmchi<1\gev$), 
$\pi$ track.
Backgrounds from $\gam\gam$-induced interactions are large.
One must tag $\cpone\cmone$ production using 
$\epem\to\gam\cpone\cmone$ \cite{ghoshetal}
or employ the more kinematically
limited $\wmp\cnone\cpmone$ final state.  \\
$\bullet$~~$\dmchi>2-3\gev$: the 
$\cpmone$ decay products are sufficiently energetic that
$\gam\gam$ induced backgrounds can be rejected to the extent
necessary for mSUGRA-like mode detection of
direct $\epem\to\cpone\cmone$ production.\\
LEP experience suggests that these various
signatures are all viable in their respective
domains of validity for an appropriately designed
detector.  Current LEP2 ($\rts\sim 200\gev$)
analyzes exclude $\mcpmone< \rts/2$ in the
`stable' and `standard' regions of $\dmchi$ and
$\mcpmone<80\gev$ or so, assuming the $\wtil\nu_e$
is relatively heavy, in the $m_\pi\leq \dmchi\leq
2\gev$ region \cite{delphi,l3}.  The backgrounds
to the `stable' and $\gamma$-tag+soft-$\pi$
signals are very small, and we assume they remain
so at higher $\rts$ with simple cuts.
(For the $\gam$-tag+soft-$\pi$ signal, we require
$p_T^\gamma>10\gev$,
$10^\circ\leq\theta_\gamma\leq 170^\circ$.)  The
appropriate {\sc Susy} discovery mode(s) at a
$\rts=600\gev$ LC are shown as a function of
$\dmchi$ and $\mcpmone$ in Fig.~\ref{regions}
\cite{gm2}.  We assume that 10 events are adequate
to establish a signal.  The rapid turn-on of
$\cpone\cmone$ production allows one to probe
almost to the kinematic limit [$\mcpmone\sim
\rts/2$ or $(\rts-p_T^\gamma)/2$], although the
totally iron-clad {\sc Snt} and {\sc Hip}
signatures disappear at the largest $\mcpmone$
values due to inadequate boost.
The {\sc Snt} signature is somewhat suppressed at
smaller $\mcpmone$ as more of the $\cpmone$'s are
boosted to $\beta>0.8$.
The discovery reach of the {\sc Snt} and {\sc Hip}
channels is increased only slightly for
$L=50\fbi\to 1\abi$.  Since
$\dmchi\in[200\mev,2\gev]$ is typical of models
with loop-dominated gaugino masses, the
$\gam$-tag signals are of great importance.

\begin{figure}[!ht]
\begin{center}
\vskip -.3in
  \includegraphics[height=12cm]{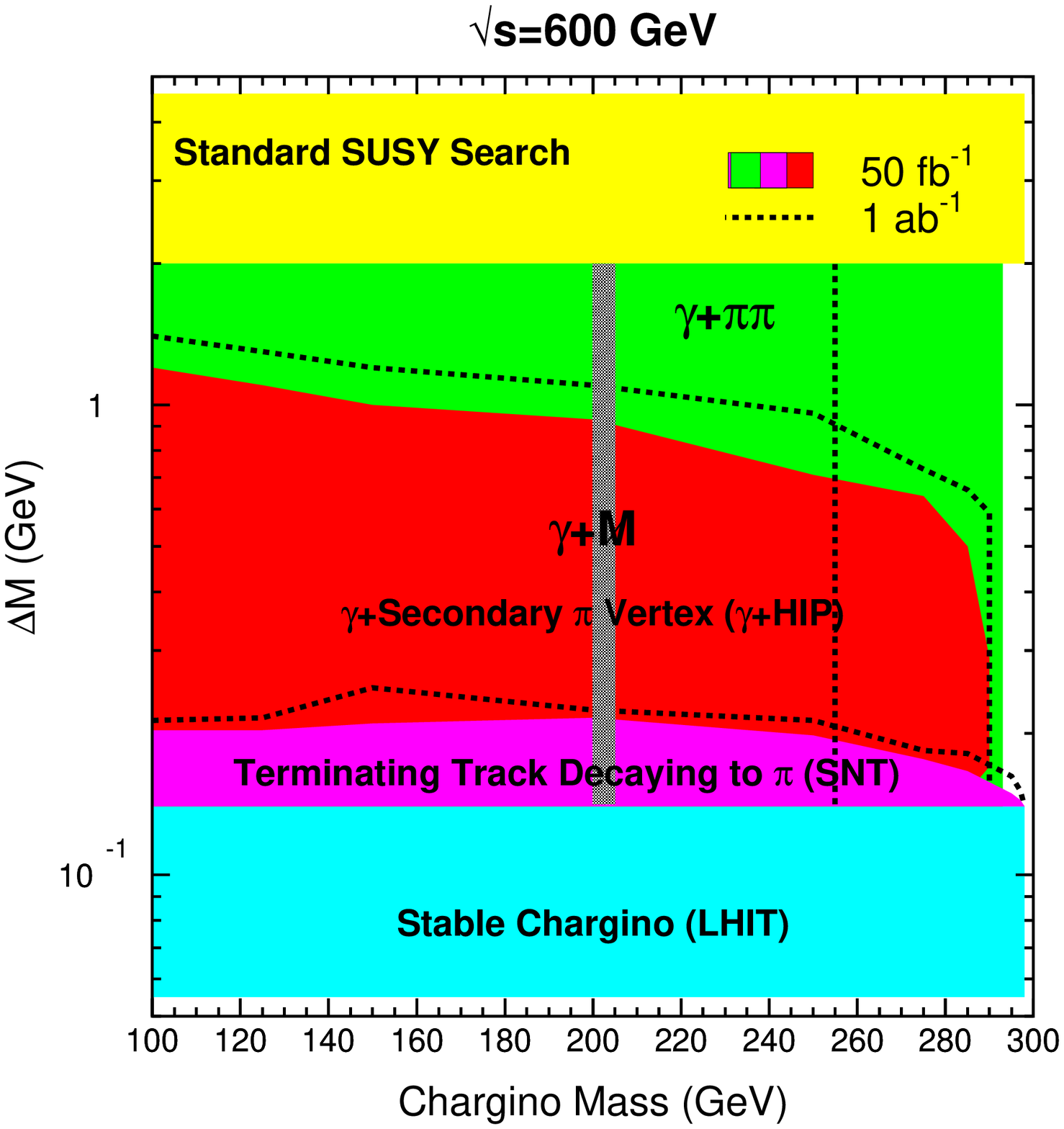}
\vskip -.2in
  \caption{Discovery reach for the various
modes described in the text as a function of $\dmchi$ and $\mcpmone$ at a $\sqrt s=600\gev$ LC;
from \cite{gm2}.}
\vskip -.3in
\label{regions}
\end{center}
\end{figure}

Detector design is critical to observing the soft
pion signatures.  If the $\vec B$ field is too
strong, soft pions will curl up before they can be
detected/tracked.
Then, the only signal $\gamma\cpone\cmone$
production yields $\gam+\mslash$, which has a
large $\gamma\nu\bar\nu$ background.
Fig.~\ref{regions} illustrates the discovery
reach.  For $\mcpmone\lsim 200\gev$ ($260\gev$),
$\mslash$ cuts can give $S/B\geq 0.02$ and
$S/\sqrt B\geq 5$ for $L= 50\fbi$ ($\sim 1\abi$).
The $\ell^\mp+\mslash$ final state from
$\wmp\cnone\cpmone$ production will provide an
alternative signal with similar reach in
$\mcpmone$; the only SM background is single
$W\to\ell\nu$ production with the $\ell$ lost down
the beam pipe.

Let us now consider the prospects for exploring
the degenerate wino-LSP scenario at the Tevatron
and LHC.  At hadron colliders, typical signatures
of mSUGRA are tri--lepton events from
neutralino--chargino production, like--sign
di--leptons from gluino pair production, and
multijets$+\etmiss$ from squark and gluino
production. The tri--lepton signal from
$\cpmone\cntwo$ production and the like--sign
di--lepton signal from $\gl\gl$ production are
both suppressed when $\dmchi$ is small by the
softness of the leptons coming from the $\cpmone$
decay(s).  For the $M_2<M_1<<|\mu|$ AMSB boundary
conditions, the tri--lepton signal is further
diminished by the suppression of the
$\cpmone\cntwo$ cross section.  For small enough
$\dmchi$, the {\sc Lhit} and {\sc Dit}
`stable'-chargino signals will be viable
\cite{gm,gm2}. But, for larger $\dmchi$, direct
observation of the $\cpmone$ becomes more
challenging.  In this case, assuming that the
$\gl$ is light enough, the most obvious signal for
SUSY in degenerate models is jet(s) plus missing
energy.  In one AMSB study, it was found that
$\gl\gl$ production at the LHC will be
sufficiently abundant that jets + $\etmiss$ will
probe up to $\mgl\sim 1.7$~TeV (1 TeV) for
$L=300\fbi$ ($L=30\fbi$), assuming that squarks
and sleptons are too heavy to provide useful
signals \cite{paigewells,Baer:2000bs}.  Using
$\mgl\sim 7\mcpmone$, $\mgl\sim 1.7$ TeV is
equivalent to $\mcpmone\sim 250\gev$, the
effective reach of a $\rts=500\gev$ LC.  A more
extensive reach for the jets + $\etmiss$ channel
at the LHC, $\mgl\sim 2.1\div 2.5\tev$, was found
in \cite{barretal}.

An important question is whether direct detection
of the charginos at the LHC is possible when
$\dmchi$ is large enough that the $\cpmone$ decay
is prompt, but small enough that the (single)
charged particle from its decay (a charged pion or
lepton) is quite soft.  Direct detection might
prove crucial to unambiguously showing that the
model is truly AMSB in character.  The most
difficult backgrounds to the soft (but isolated)
$\pi^\pm$ tracks in this case are those arising
from events containing the baryons $\Sigma^+$,
$\Sigma^-,\Xi$, and $\Omega$, all of which have
decays in which a single soft pion emerges
\cite{gm}.  One might hope to remove such
backgrounds by requiring the soft pion track(s) to
be sufficiently isolated due to the fact that the
normal hadrons above are not produced in
isolation. So even if they decay to an isolated
soft pion, there will be hadrons that are nearby
in $\Delta R$ that will be visible.  In
\cite{barretal}, this and the alternative
possibility of detecting the soft isolated
$\ell^\pm$ from the chargino decay are pursued.
Aside from appropriate isolation requirements for
the leptons, a critical ingredient is the use of
the Cambridge $m_{TX}$ variables \cite{lester}
which provide sensitivity to $\mcpmone$ when a
$\cpone\cmone$ pair is produced through looking at
distributions in $m_{TX}-\mcnone$.  For this
technique, it is necessary that $\mcnone$ be known
(from kinematic endpoints) and that the bulk of
the $\cpmone$ come from the two-body decay chain
sequence $\wtil q\to q \cpmone\to \cnone e\nu_e q$
or similar. As noted, squarks are not guaranteed
to be sufficiently light to be abundantly produced
in the AMSB models. But, assuming they are,
\cite{barretal} concludes that the isolation +
$m_{TX}-\mcnone$ procedures can determine the
presence and mass of the $\cpmone$ for
$\dmchi\gsim 700\mev$.  For $\dmchi<200\mev$, the
{\sc Hit}, {\sc Dit}, {\sc Stub}, {\sc Hip},
\ldots signatures make it easy to detect the
$\cpmone$.  In the $200<\dmchi<700\gev$ part of
parameter space, which unfortunately is somewhat
preferred, techniques for direct $\cpmone$
detection have not yet been developed.  In this
part of parameter space, the LHC is needed to see
the gluino and squark signatures while the LC is
needed to directly observed the $\cpmone$ and
determine its mass.

\subsubsection{Baryonic R-parity violation coupled
with the degenerate wino-LSP scenario}

If one wishes to construct a ``worst case''
scenario for the hadron colliders, an obvious
approach is to assume: (1) a small $\dmchi$ AMSB
(or other) scenario with $200\mev\lsim \dmchi\lsim
1\gev$; and (2) baryonic R-parity violation in
which the $\cnone$ decays to (semi-soft) jets.  A
very incomplete consideration of this kind of
scenario was given for the Tevatron and the LC in
\cite{runiirpv} (see also the earlier discussion
in \cite{Amundson:1996nw}).  The summary below is
based on this reference.

First, because $200\mev\lsim \dmchi$, the
$\cpmone$ decay will be prompt and there will be
no {\sc Hit}, {\sc Dit}, {\sc Stub}, {\sc Hip},
\ldots\ type signatures.  Second, because
$\dmchi\lsim 1\gev$, the leptons (if significant
in the $\cpmone$ decays at all) will be very soft,
making it impossible to look for either the
like-sign dilepton signal from gluino or squark
pair production or the tri-lepton signal from
$\cpmone\cntwo$ production.

Other opportunities for SUSY detection depend in
detail upon the $\cnone\to 3j$ decay scenario.
The rather strong upper bound on the baryonic
R-parity violating coupling $\lampp_{usd}$ means
that one (or more) of the $ubd$, $ubs$, $csd$,
$cbd$, $cbs$ channels will most likely dominate
$\cnone$ decay.  All of these contain at least one
heavy quark.  Since SUSY pair production events
contain at least two $\cnone$'s, we can then use
double-tagging to reduce the background relative
to the signal. For all but the $cds$ decay
channel, at least two of the jets in the final
state will be $b$-jets.  If $cds$ decays dominate,
we would have two $c$-jets.  Since $c$-tagging
will have lower efficiency than $b$-tagging,
dominance of $\cnone$ decays by the $cds$ channel
results in the most difficult scenario for
supersymmetry detection.  An exception is the case
in which all the $\lampp$ are so small that the
$\cnone$ decay is not prompt.  If the $\cnone$ is
long-lived, then one reverts to the R-parity
conserving signals of the small $\dmchi$ case.  If
the $\cnone$ decays within the detector after a
substantial path length, the sudden appearance of
three jets inside the detector will provide a
clear signature for SUSY.  The most difficult case
is when the $\cnone\to cds$ decay is prompt.

In this case, since the $\cnone$ decays promptly
to jets, there will also be no missing energy
signal for gluino/squark production. In short, the
only possible signal will be an excess of jets
(some of which can be tagged with modest
efficiency) due to the two $\cnone\to cds$ decays.
In addition, in the AMSB scenario (for which
$\mgl\sim 3\mcpmone\sim \mcnone$) each event will
typically have a number of hard jets (from the
$\gl\to q'\anti q \cpmone$ decays) accompanied by
six jets from the $\cnone$ decays.  However, since
some of the jets from $\cnone$ decay will be quite
soft (unless $\mcnone$ is quite large) the excess
of jets over standard QCD will not be that
pronounced and $c$ tagging will not be very
efficient.  To date, there is no convincing study
indicating that SUSY detection at the Tevatron or
LHC will be possible for this very difficult
scenario when $\mgl$ is sufficiently large that
the $\gl\gl$ production rate is not large.  In
other models with small $\dmchi$, $\mgl$ can be
quite a bit smaller and even of order $\mcpmone$.
In this case, one loses the extra jets from $\gl$
decay, but the $\gl\gl$ production rate is quite
large and its detection at the LHC is fairly
likely after cuts requiring a large number of jets
and several $c$ tags.

Thus, depending upon the exact model, the LC could
be absolutely crucial for SUSY detection.  In
fact, detection of $e^+e^-\to \cnone\cnone\to 6j$
would be completely straightforward at the LC and
would provide a dramatic confirmation of SUSY as
well as baryonic R-parity violation.  In addition,
the $e^+e^-\to \cpone\cmone\to \ell^+\ell^- +6j$
signal would almost certainly be observable and
allow an accurate determination of $\dmchi$.  Once
$\mcnone\sim\mcpmone$ is known from the LC data,
it might be possible to return to the LHC data and
detect the gluino/squark signals that would have
remained hidden without this knowledge.

If baryonic R-parity violation is present, a very
important goal will be to measure the relevant
$\lampp$.  There are two possibilities at a hadron
collider for directly determining $\lampp$.  If
$\lampp$ is large, RPV-induced single squark
production cross sections are also typically
substantial, and, if a signal can be isolated, the
cross section size gives a measure of $\lampp$. If
$\lampp$ is small, the decay path length for the
$\cnone$ might be observable and would again
provide a measure of $\lampp$. However, rough
estimates indicate that there is a region of
intermediate $\lampp$ at higher $m_{\wtil q_R}$
for which neither $c\tau(\cnone)$ nor
$\sigma(\wtil q_R)$ will be measurable.  In this
region, determination of $\lampp$ would only be
possible if an RPV decay mode of the $\cpmone$ is
competitive with its standard SUSY decay modes and
these can be separated from one another. The
relative size of the branching ratios would then
provide a measure of $\lampp$.  At an $\epem$
collider, if $\lampp$ is small it could again be
measured via the $\cnone$ decay length or the
relative branching ratio for RPV decays versus
normal SUSY decays.  If $\lampp$ is large, these
techniques would not be available and in addition
there are no sources of quarks or antiquarks as
needed for squark production via baryonic RPV
couplings.

In short, there is a complicated model-dependent
interplay between the abilities of the LHC and LC
in models with small $\dmchi$.  If there is no
sign of normal SUSY events at the LHC, models
having baryonic R-parity violation and small
$\dmchi$ would provide a possible explanation that
might only be tested at the LC.

\subsubsection{The gluino-LSP scenario}

It is possible to write down SUSY models in which the
gluino is the LSP \cite{nonuniv,cdg,raby}. 
Other motivations for a light gluino include
the ability to unify couplings for $\alpha_s\sim 0.12$
\cite{lrshif}
and a decreased level of fine tuning \cite{Kane:1998im}.

For a gluino-LSP, the detection of SUSY is quite
different than in the usual $\cnone$ LSP
scenarios.  Techniques have been developed
\cite{cdg,Raby:1998xr} for detecting $\gl\gl$ pair
production at the Tevatron and LHC.  They involve
signatures related to a very heavy, possibly
neutral, strongly interacting object moving
through detectors composed of much lighter
objects. By a careful examination of these
signatures, it is possible to conclude
\cite{cdg,Raby:1998xr} (by combining Run I CDF/D0
data with $Z$ pole data from LEP) that $\mgl\lsim
130\gev$ is excluded except, possibly, for a small
interval of $\mgl\sim 30\gev$.  The same studies
indicate that the range of sensitivity to $\mgl$
will increase dramatically at the LHC, with
detection being possible up to $\mgl\sim 1\tev$.

In contrast, the discovery reach of the LC will be
strictly limited by $\sqrt s$.  By definition of
the scenario, the absolutely minimal energy
required is $\sqrt s>2\mgl$.  But, $e^+e^-\to
\gl\gl$ is mediated by one-loop-induced
$\gam\gl\gl$ and $Z\gl\gl$ couplings where the
loop contains one or two squarks.  The virtual
photon exchange diagrams vanish if (for any given
flavor $q$) we have $m_{\wt q_L}=m_{\wt q_R}$. The
virtual $Z$ exchange diagrams vanish if there is
mass degeneracy for each quark isospin doublet,
i.e. if $m_d=m_u$, \ldots, and if there is mass
degeneracy also in each squark isospin doublet,
i.e. if $m_{\wt d_1}=m_{\wt d_2}=m_{\wt
  u_1}=m_{\wt u_2}$, \ldots. Thus, given that for
the first two families the quarks have small and
not very different masses while FCNC
considerations require rather degenerate squarks,
only loops involving top and bottom squarks and
quarks contribute significantly.  Still, despite
the fact that $m_t>> m_b$, if there is only small
splitting between the $\wt t_L$ and $\wt t_R$ and
if the squark mixing is also small, gluino pair
production in $e^+e^-$ collisions will be hard to
observe, even with $L=1000\fbi$
\cite{Berge:2003tb,Berge:2002ev}.

It is also interesting to consider cross sections
for $\gl\gl$ production at a $\gam$C
\cite{Berge:2003cj}.  The process $\gam\gam\to
\gl\gl$ is mediated by box diagrams containing
one, two or three squarks.  Other contributions to
$\gl\gl$ production arise from resolved photons.
For a given value of $\mgl$, the optimum machine
energy (giving the largest cross section) is
typically $\sqrt s\sim 3\mgl$.  For this optimal
choice, $\sigma(e^-e^-\to \gam\gam\to \gl\gl)$
falls rapidly with increasing squark mass and with
decreasing $\mgl$.  As an example, if $\mgl\sim
200\gev$ and $\sqrt{s_{ee}}\sim 600\gev$ one finds
$\sigma(e^-e^-\to \gam\gam\to \gl\gl)\sim 0.1 \fb$
($0.005\fb$) for a general squark mass scale of
$800\gev$ ($1.5\tev$).  For $\mgl\sim 150\gev$ and
the optimal choice of $\sqrt{s_{ee}}\sim 500\gev$,
the cross sections for these two squark mass
scales will be about a factor of three smaller.
Thus, there is no guarantee that $\gl\gl$
production will be seen at a $\gam$C facility.

A final LC possibility is the tree-level process
$e^+e^-\to q\anti q \gl\gl$ (see, for example,
\cite{cdg}). However, this has a very limited
kinematic reach.  For example, $\sum_q
\sigma(e^+e^-\to q\anti q \gl\gl)<0.01\fb$ for
$\mgl>140\gev$ at a $\rts=500\gev$ LC.

In short, while there are many scenarios for which
an LC with $\sqrt s\lsim 1\tev$ and/or the
associated $\gam$C will allow $\gl$ detection,
there are also many scenarios for which they will
not.

Thus, if the gluino is the LSP and if squarks are
considerably heavier (as typical for existing
models), it is conceivable that only the LHC would
be capable of detecting SUSY until the $\sqrt s$
of the LC is increased well above the initial
$\sim 500\gev$ first stage value.






%
%
%

\section{Determination of mSUGRA parameters and discrimination between 
SUSY breaking scenarios}

%
%

\subsection{\label{sec:432} Complementarity of LHC and Linear Collider
measurements of slepton and lighter neutralino masses}

{\it D.R.~Tovey}

\vspace{1em}
\newcommand{\ETM}{E_T^{\mathrm{miss}}}
\newcommand{\MS}{M_{\mathrm{susy}}} 
\newcommand{\MSS}{M_{\mathrm{susy}}^{\mathrm{eff}}}
\newcommand{\MX}{M_{\chi}} 
\newcommand{\mestbar}{\overline{M_{\mathrm{est}}}}
\newcommand{\mest}{M_{\mathrm{est}}}
\newcommand{\sigsus}{\sigma_{\mathrm{susy}}}
\renewcommand{\ie}{{\it i.e.}}
\renewcommand{\eg}{{\it e.g.}}


{\small
\noindent
The mSUGRA $m_0 - m_{1/2}$ plane is mapped to identify regions in
which combination of LHC and linear collider data can significantly
improve the accuracy of lighter neutralino and charged slepton mass
measurements.
}


\subsubsection{Introduction}

It has recently become clear that measurements of the masses of
supersymmetric particles (sparticles) carried out at the LHC and at a
$\sqrt{s} \geq$ 500 GeV linear collider will in many cases be
complementary. The reason for this is that although the LHC will
provide the greater search reach for SUSY particles, only the LC will
be able to provide accurate ($\lesssim$ 1\%) measurements of the
absolute masses of produced sparticles. LHC experiments may be able to
measure combinations of masses of heavier sparticles such as
$\tilde{q}_L$ \cite{Atlas:1999atl,Lester:2000cgl}, but these can only
be used to derive accurate absolute mass values with LC input
regarding absolute masses of lighter sparticles (lighter neutralinos
or charged sleptons).

While this will be the case for many models, it is clear that there
may also be models where kinematics will require that a combination of
LC and LHC data be used to measure even the masses of the lighter
sparticles. This note seeks to identify and study such models
occurring within the mSUGRA framework.

\subsubsection{Parameter Space Scan}

In this study we have chosen to work within the minimal Supergravity
(mSUGRA) framework \cite{Alvarez:1983lag} used in many previous
studies of LHC physics capabilities \cite{Atlas:1999atl}. In order to
map out representative regions of parameter space in which LC and LHC
sparticle mass measurements are complementary a scan of $m_0 -
m_{1/2}$ parameter space was performed using {\tt isajet 7.51}
\cite{Paige:1986fep} with $A_0$ $=$ 0, tan($\beta$) $=$ 10 and $\mu$
$>$ 0. At each point in parameter space the $\sigma$.BR for LHC events
producing $\tilde{\chi}^0_2$ decaying to $\tilde{\chi}^0_1$ and two
leptons was calculated for two-body decays proceeding through an
intermediate $\tilde{l}_L$ or $\tilde{l}_R$:
\begin{eqnarray}
\tilde{\chi}^0_2 \rightarrow \tilde{l}^{+(-)}_Ll^{-(+)} \rightarrow
\tilde{\chi}^0_1 l^+l^-, \\
\tilde{\chi}^0_2 \rightarrow \tilde{l}^{+(-)}_Rl^{-(+)} \rightarrow
\tilde{\chi}^0_1 l^+l^-,
\end{eqnarray}
and three-body direct ({\it i.e.} not through $Z^0$) decays:
\begin{equation}
\tilde{\chi}^0_2 \rightarrow \tilde{\chi}^0_1  l^+l^-.
\end{equation}

Fig.~\ref{fig1} shows a plot of the mSUGRA $m_0 - m_{1/2}$ plane with
full contours indicating constant values of $\sigma$.BR = 0.02 pb for
$\tilde{\chi}^0_2$ decays direct to $\tilde{\chi}^0_1$ (black),
through an intermediate $\tilde{l}_R$ (blue) or an intermediate
$\tilde{l}_L$ (red). Also shown (dashed) are contours of constant
$m(\tilde{\chi}^0_2) + m(\tilde{\chi}^0_1)$ = 500 GeV (black),
$m(\tilde{l}_R)$ = 250 GeV (blue) and $m(\tilde{l}_L)$ = 250 GeV
(red). These dashed contours correspond to the kinematics limited
sensitivities of a $\sqrt{s}$ = 500 GeV linear collider to the $e^+e^-
\rightarrow \tilde{\chi}^0_2 \tilde{\chi}^0_1$, $e^+e^- \rightarrow
\tilde{l}_R^+\tilde{l}_R^-$ and $e^+e^- \rightarrow
\tilde{l}_L^+\tilde{l}_L^-$ discovery channels respectively. It is
expected that in the event of a SUSY signal being observed in the
$e^+e^- \rightarrow \tilde{l}_R^+\tilde{l}_R^-$ or $e^+e^- \rightarrow
\tilde{l}_L^+\tilde{l}_L^-$ channels at a LC then the masses of the
produced sparticles will be measured to high accuracy
\cite{Tesla:2001tes}. In the case where $e^+e^- \rightarrow
\tilde{\chi}^0_2 \tilde{\chi}^0_1$ is kinematically accessible then
accurate measurements of the $\tilde{\chi}^0_2$ and $\tilde{\chi}^0_1$
masses can be performed using threshold scans and dilepton invariant
mass edges provided that slepton pair production (the dominant source
of SUSY background) does not occur \cite{Martyn:2002hum}. If slepton
pair production does occur then only the threshold scan can be used,
constraining $m(\tilde{\chi}^0_2)+m(\tilde{\chi}^0_1)$.
\begin{figure}[htb!]
\begin{center}
\epsfig{file=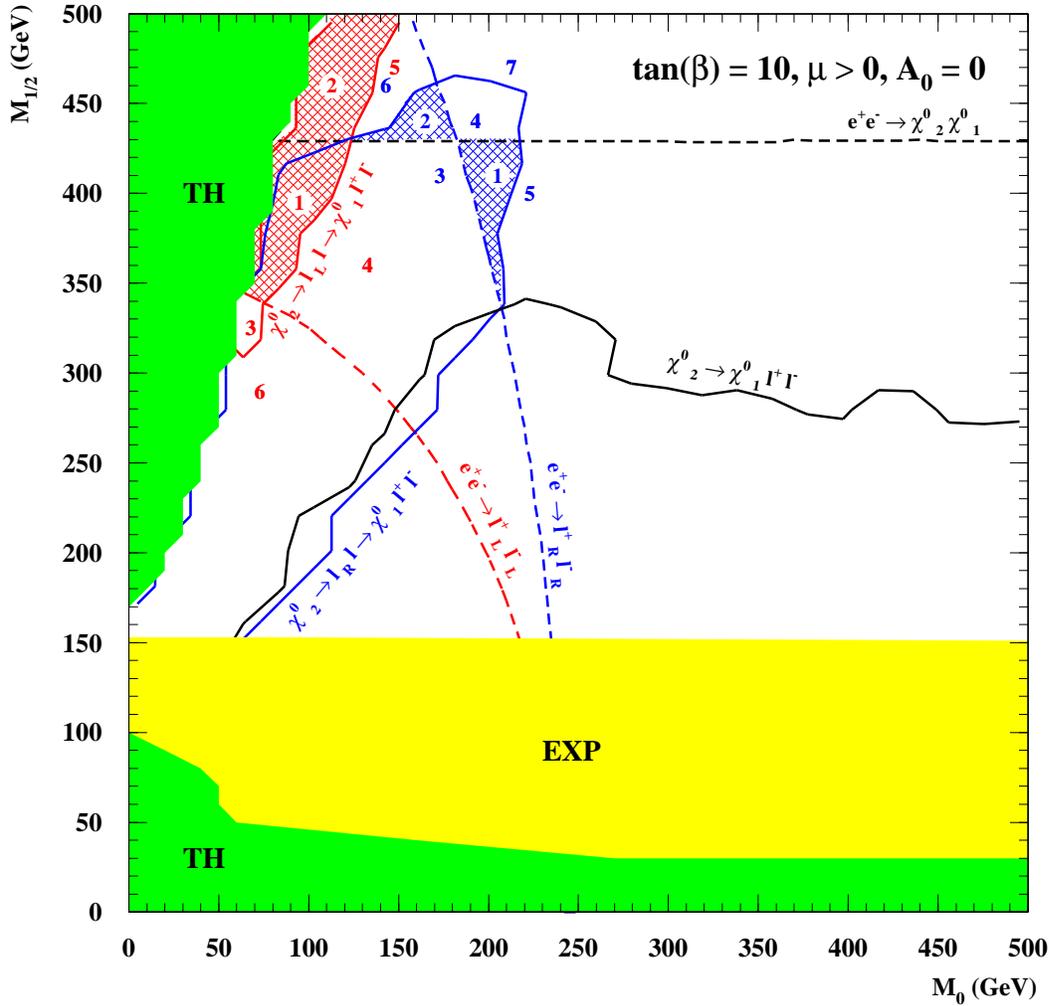,height=6.0in}
\caption{\label{fig1} {\it Contours of $\sigma$.BR = 0.02 pb for
$\tilde{\chi}^0_2$ decays in LHC events (full contours). Contours
correspond to $\tilde{\chi}^0_2$ decays direct to $\tilde{\chi}^0_1$
(black), through an intermediate $\tilde{l}_R$ (blue) or an
intermediate $\tilde{l}_L$ (red). Also shown (dashed) are contours of
constant $m(\tilde{\chi}^0_2)$ = 250 GeV (black), $m(\tilde{l}_R)$ =
250 GeV (blue) and $m(\tilde{l}_L)$ = 250 GeV (red). The full green
regions are excluded by theory, the full yellow region by experimental
bounds from LEP and elsewhere. Small blue and red numbers refer to
regions decribed in the text.}}
\end{center}
\end{figure}

Given the extreme accuracy of LC measurements we may now ask whether
there are any regions of the $m_0 - m_{1/2}$ plane in Fig.~\ref{fig1}
where LHC data can be used {\em in conjunction} with LC data to give
more precise mass bounds. Consider first the case where
$\tilde{\chi}^0_2 \rightarrow \tilde{l}^{+(-)}_Rl^{-(+)} \rightarrow
\tilde{\chi}^0_1 l^+l^-$ decays are observed at the LHC. In the
majority of the parameter space in Fig.~\ref{fig1} where this occurs
the LC will also observe $e^+e^- \rightarrow
\tilde{l}_R^+\tilde{l}_R^-$ {\em and} $e^+e^- \rightarrow
\tilde{\chi}^0_2 \tilde{\chi}^0_1$ (labelled `3' in blue). In this
case the LC will measure $m(\tilde{\chi}^0_1)$ and $m(\tilde{l}_R)$
from the dislepton channel and hence $m(\tilde{\chi}^0_2)$ from an
$e^+e^- \rightarrow \tilde{\chi}^0_2 \tilde{\chi}^0_1$ threshold scan,
with no input from the LHC. There is also a region (labelled `4')
where a LC will see nothing and here there can be no input from LC
data. In regions `1' and `2' (hatched) however there is indeed
complementarity between LHC and LC data. In region `1' the LC measures
$m(\tilde{\chi}^0_2)$ and $m(\tilde{\chi}^0_1)$ which in turn allows
the LHC to measure $m(\tilde{l}_R)$. In region `2' the LC measures
$m(\tilde{l}_R)$ and $m(\tilde{\chi}^0_1)$ enabling the LHC to measure
$m(\tilde{\chi}^0_2)$. It should be noted that a LC could use the
dilepton edge technique \cite{Atlas:1999atl} to gain independent
sensitivity to $m(\tilde{l}_R)$ in region `1', however in this case it
is not clear that there would not still be an advantage to use of the
high statistics LHC $\tilde{\chi}^0_2$ sample. In regions `5' and `6'
the LHC can measure nothing as no dilepton signature is observed
(assuming an arbitrary 0.02 pb observation threshold) and only one
channel is observed in LC data. In region `7' no dilepton measurement
channels are open either at LHC or LC (although the LHC will still
discover SUSY \cite{Tovey:2002drt}).

Consider next the case where $\tilde{\chi}^0_2 \rightarrow
\tilde{l}^{+(-)}_Ll^{-(+)} \rightarrow \tilde{\chi}^0_1 l^+l^-$ decays
are observed at the LHC (red region). In region `1' (hatched red) LHC
and LC data is complementary because the LC measures
$m(\tilde{\chi}^0_2) + m(\tilde{\chi}^0_1)$ from an $e^+e^-
\rightarrow \tilde{\chi}^0_2 \tilde{\chi}^0_1$ threshold scan
providing input to LHC measurements of $m(\tilde{l}_L)$,
$m(\tilde{\chi}^0_2)$ and $m(\tilde{\chi}^0_1)$ combinations. Note
that here constraints on $m(\tilde{l}_L)$ from LC dilepton edge
measurements would be very difficult due to dislepton SUSY
background. In region `2' (hatched red) the LC observes neither the
$e^+e^- \rightarrow \tilde{l}_L^+\tilde{l}_L^-$ channel nor the
$e^+e^- \rightarrow \tilde{\chi}^0_2 \tilde{\chi}^0_1$ channel but can
still help the LHC measure mass combinations involving
$m(\tilde{l}_L)$ and $m(\tilde{\chi}^0_2)$ by measuring
$m(\tilde{\chi}^0_1)$ using $e^+e^- \rightarrow
\tilde{l}_R^+\tilde{l}_R^-$. In region `3' the LC can measure
$m(\tilde{\chi}^0_1)$ and $m(\tilde{l}_L)$ from $e^+e^- \rightarrow
\tilde{l}_L^+\tilde{l}_L^-$ and hence $m(\tilde{\chi}^0_2)$ from a
$e^+e^- \rightarrow \tilde{\chi}^0_2 \tilde{\chi}^0_1$ threshold scan,
with no LHC input, giving no complementarity. In regions `4', `5' and
`6' the LC observes two channels, one channel or three channels
respectively but the LHC sees no $\tilde{l}_L$ mediated
$\tilde{\chi}^0_2$ decay. Again, in a limited region of parameter
space useful combined measurements significantly improving measurement
accuracy can be carried out.

Consider finally the case where $\tilde{\chi}^0_2 \rightarrow
\tilde{\chi}^0_1 l^+l^-$ decays are observed at the LHC (black). This
region lies entirely within the LC reach in the $e^+e^- \rightarrow
\tilde{\chi}^0_2 \tilde{\chi}^0_1$ channel and also the $e^+e^-
\rightarrow \tilde{\chi}^0_2 \tilde{\chi}^0_2$ and $e^+e^- \rightarrow
\tilde{\chi}^+_1 \tilde{\chi}^-_1$ channels (not shown). The LC will
therefore be able to measure both $m(\tilde{\chi}^0_2)$ and
$m(\tilde{\chi}^0_1)$ without reference to LHC data and no additional
input from the LHC will be required.

These results illustrate the scope for combined LC + LHC sparticle
mass measurements for charged sleptons and lighter neutralinos. The
regions of parameter space where such an approach is applicable are
admittedly rather limited but it must be remembered that these results
were obtained using mSUGRA models with fixed values of tan($\beta$),
$A_0$ and sign($\mu$). While it is to be expected that similar
behaviour will be exhibited for other values of these parameters
(although possibly modified at high tan($\beta$) to take into account
the large branching ratios to $\tilde{\tau}_1$ and $\tilde{\tau}_2$)
very different phenomenology may occur in less constrained SUSY
models. In these models combination of results is still likely to be
appropriate in regions where the LHC observes two-body
$\tilde{\chi}^0_2$ decay processes and the LC observes only $e^+e^-
\rightarrow \tilde{\chi}^0_2 \tilde{\chi}^0_1$ or $e^+e^- \rightarrow
\tilde{l}^+\tilde{l}^-$ (but not both) however the size of these
regions may be considerably enlarged. Further regions of interest may
also arise where the LHC observes three-body $\tilde{\chi}^0_2$ decays
and the LC observes only $e^+e^- \rightarrow \tilde{l}^+\tilde{l}^-$
thereby providing input on $m(\tilde{\chi}^0_1)$. Finally, in less
constrained models with different relationships between sparticle
({\it e.g.} gaugino) masses it may be advantageous to make use of LC
$\tilde{\chi}^0_1$ mass measurements provided by {\it e.g.} chargino
or sneutrino decays.

\subsubsection{Conclusions}

The mSUGRA $m_0 - m_{1/2}$ plane has been mapped to identify regions
in which combination of LHC and linear collider data can significantly
improve the accuracy of lighter neutralino and charged slepton mass
measurements.


%


\subsection{\label{sec:433} Discriminating SUSY breaking scenarios}

{\it B.C.~Allanach, D.~Grellscheid and F.~Quevedo}

\vspace{1em}














\noindent{\small
We approach the following questions: if supersymmetry is discovered,
how can we select among different supersymmetric extensions of the
Standard Model? What observables best distinguish the models and what is the
required discriminating accuracy?
We examine scenarios differing by the fundamental string scale and
concentrate on GUT and intermediate scale models. 
We scan over the  parameters in each scenario, finding ratios of
sparticle masses that provide the maximum discrimination between them.
The necessary accuracy for discrimination is determined in each
case. A
future linear collider could provide the necessary precision on
slepton masses, whereas combined LHC and LC information could provide 
a further check of the gluino to squark mass ratio.
}



\subsubsection{Introduction}
Supersymmetric (SUSY) phenomenology is notoriously complicated, and many
studies of experimental measurement capabilities have focused on individual
points in the parameter space (see, for example, SUSY studies in this
document). Here, we want to examine what is
required to discriminate between different SUSY breaking scenarios, when 
the parameters within each are allowed to vary (as long as they agree with
current experimental data). By this we hope
to provide {\em guaranteed} discrimination,
whatever the parameter point, provided measurement accuracies are low
enough. The results we present here are updated versions of 
the ones in ref.~\cite{Allanach:2001qe}, using a more up-to-date 
spectrum predictor: {\small \tt SOFTSUSY1.71}~\cite{Allanach:2001kg}.

Given that we need to scan over parameters in the scenarios we will consider,
a detailed empirically-based analysis is unfeasible. The existence of SUSY
backgrounds and signals depend crucially upon the part of MSSM parameter space
one examines, thus it would be necessary to perform a
separate study for each point in parameter space. A more tractable analysis is
followed: we try to find simple functions of sparticle masses that may be 
used to discriminate between some test scenarios. In this initial study, we
examine sparticle mass {\em ratios}.

The three scenarios we choose to study are inspired by type I string
models~\cite{Burgess:1998px}:
(1) String scale at the GUT scale $M_{GUT} \sim O(10^{16})$~GeV, defined by 
the scale of electroweak gauge unification $g_1(M_{GUT})=g_2(M_{GUT})$.
(2) Intermediate string scale 
($M_I = 10^{11}$~GeV) 
with extra leptons to achieve gauge coupling  
unification at $M_I$, which we will refer to as {\it early unification} (EU)
and 
 (3) intermediate string scale ($M_I = 10^{11}$~GeV)  with {\it mirage
unification}~\cite{Ibanez:1999st} at $M_I$.  
In scenarios (1) and (3), we assume that the low-energy effective field theory
corresponds to the $R$-parity conserving MSSM. In order to keep the volume of
SUSY breaking parameter space manageable, we will here look only at the
dilaton dominated limit of scenarios (1)-(3).

\subsubsection{Discriminators}

As advertised above, we use sparticle mass ratios to discriminate between
models. Ratios rather than absolute masses are used in order to factor out
dependence upon the overall SUSY mass scale. 
Fig.~\ref{fig:scans}a shows the idea: we expect each model to inhabit 
some closed 2-dimensional surface in the ratio space $\{R_1,R_2\}$. The
``distance'' vector between the two regions is defined as the vector $(
\Delta R_1, \Delta R_2)$ where $\sqrt{\Delta R_1^2+\Delta R_2^2}$, 
is minimised. It can only be defined when separation is achieved, ie where the
regions do not overlap. The distance vector is therefore the 
measurement accuracy of the ratios $R_{1,2}$ required to discriminate between
the two models {\em in all cases}. In 
\begin{figure}
\unitlength=1in
\begin{picture}(6,2.3)
\put(-0,2.2){(a)}
\put(0,0){\epsfig{file=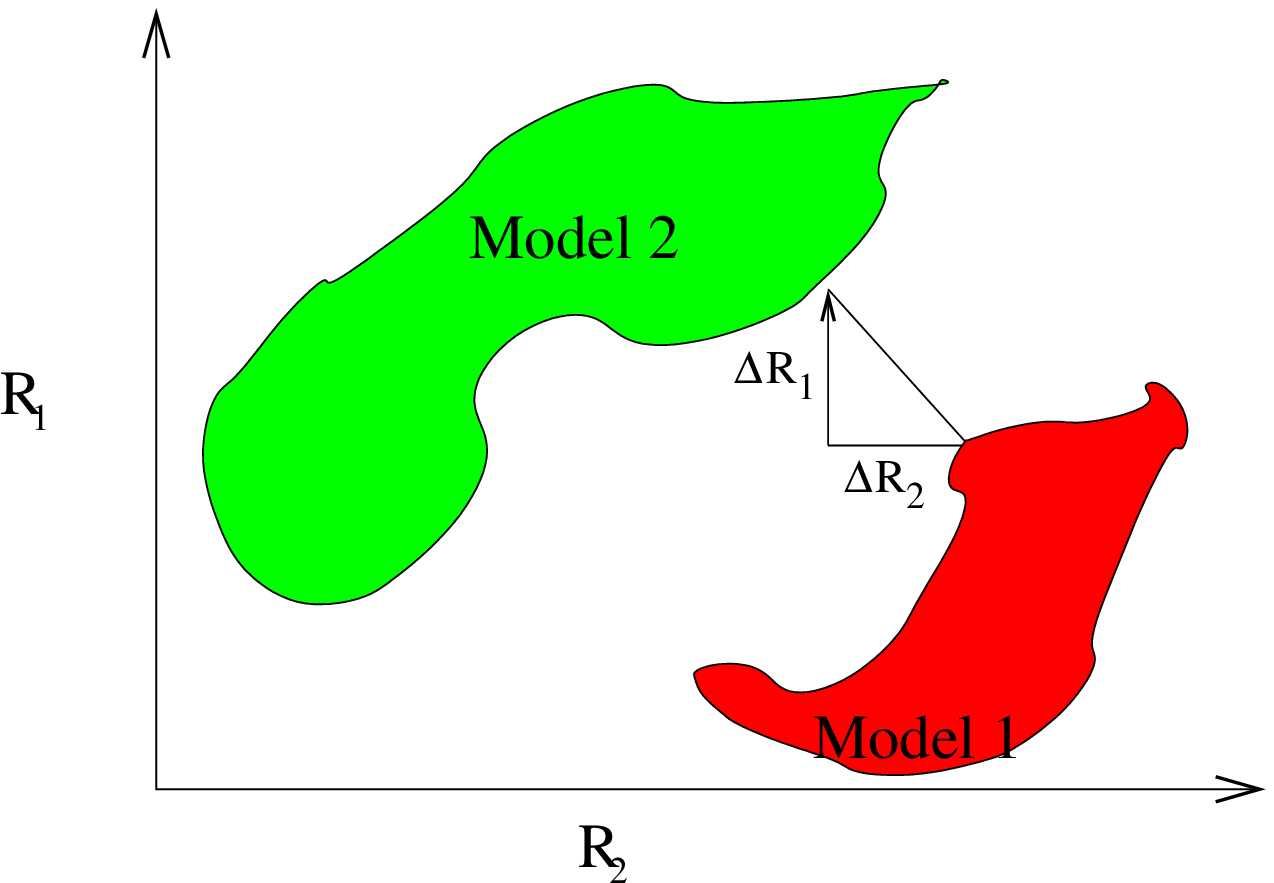, width=3in}}
\put(3,2.2){(b)}
\put(3,0.2){\epsfig{file=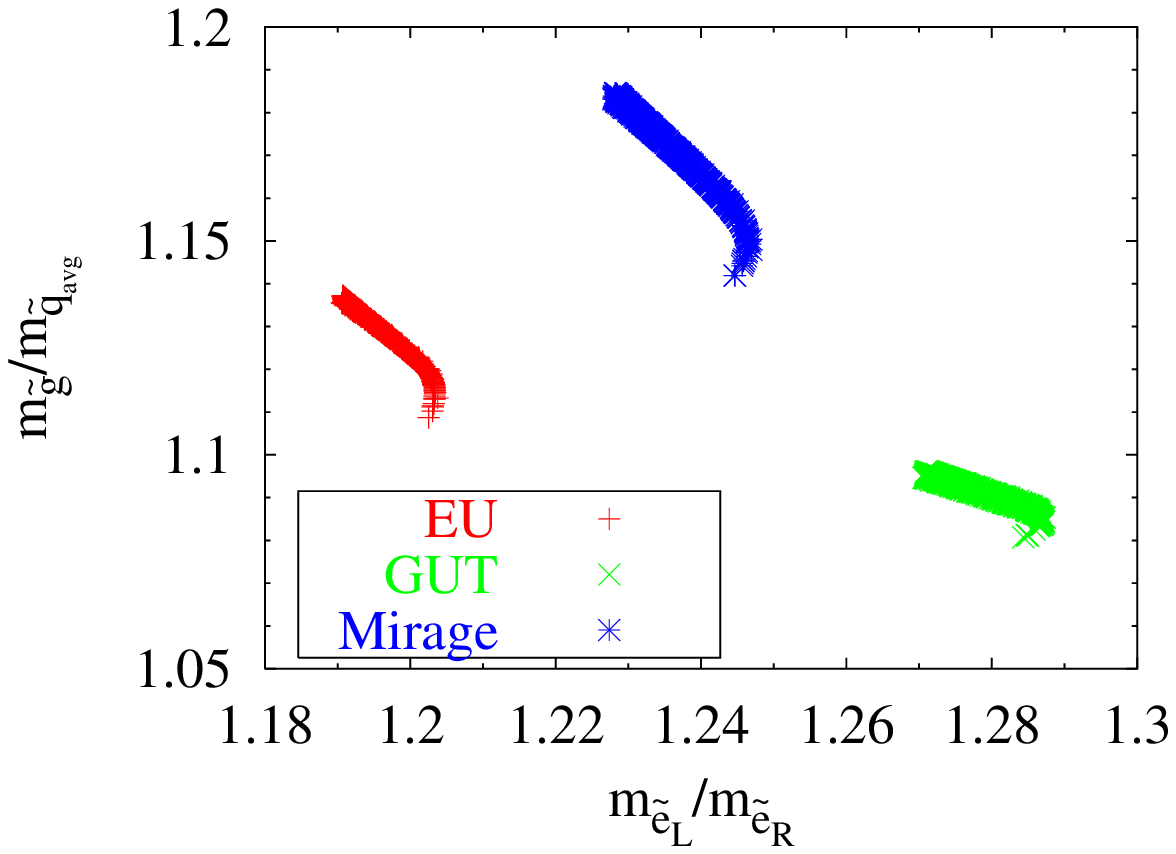, width=3in}}
\end{picture}
\caption{Discriminating ratios and parameter scans: (a) schematic, (b) results
  of the scans for the three scenarios considered.}
\label{fig:scans}
\end{figure}
practice, we populate the regions with a random 
scan over parameter space that satisfy experimental constraints and do not
have sparticles heavier than a few TeV (see
ref.~\cite{Allanach:2001qe} for details). 

We achieve separation, as shown in
figure~\ref{fig:scans}b. The distance ratios here are shown to be 
$(\Delta [m_{\tilde g}/m_{{\tilde q}_{avq}}], \Delta [m_{{\tilde e}_L}/m_{{\tilde e}_R}])=(0.03,
0.04)$ to
discriminate EU and mirage scenarios, $(0.02, 0.07)$ to discriminate the EU and
GUT scenarios and $(0.05, 0.03)$ to discriminate between the GUT and mirage
scenarios, in their respective dilaton-dominated limits.
$m_{{\tilde q}_{avg}}$ is an average over the first two generations' squark
  masses. 

\subsubsection{Conclusions}

We have examined three test SUSY breaking scenarios in the MSSM, finding
discriminating ratios of sparticle masses. 
Simple combinations of two
ratios can discriminate all three scenarios. 
The percent-level accuracies required indicate a rough level for the desired
accuracy upon sparticle masses. The combined experimental and
theoretical~\cite{Allanach:2003jw} errors should therefore not be larger than
the percent-level in order to achieve discrimination.
$3\%$ accuracies on the measurements of $m_{{\tilde e}_L}/m_{{\tilde e}_R}$
would be 
required in order to always discriminate between the scenarios. Such a
precision should be readily available at a future LC facility 
that has polarised beams. We require maximum
errors of $2\%$ on $m_{\tilde g}/m_{{\tilde q}_{avq}}$ for
discrimination. At the LHC, errors upon $m_{\tilde g}$ or $m_{{\tilde q}_{av}}$
  are expected to be of order 10$\%$, too high for our purposes. At a
  500 GeV LC 
  on the   other hand, it is difficult to produce squarks and gluinos because
  they are typically kinematically inaccessible. 
However a few-percent uncertainty upon the ratio 
might be feasible if LC information on slepton and weak gaugino masses
is fed into an LHC analysis. New techniques to study discrimination in models
which depart from dilaton domination are currently underway~\cite{future}.





\subsection{\label{sec:433a} Gravitino and goldstino at colliders}

{\it W.~Buchm\"uller, K.~Hamaguchi, M.~Ratz and T.~Yanagida}

\vspace{1em}
\renewcommand{\baselinestretch}{1.2}    
\newcommand{\re}{{\rm Re}}
\newcommand{\im}{{\rm Im}}
\renewcommand{\tr}{{\rm tr}}
\newcommand{\Tr}{{\rm Tr}}
\newcommand{\diag}{{\rm diag}}
\newcommand{\quabla}{\boldsymbol{\square}}
\newcommand{\CenterObject}[1]{\ensuremath{\vcenter{\hbox{#1}}}}
\renewcommand{\D}{\mathrm{d}}
\newcommand{\I}{\mathrm{i}}
\renewcommand{\stau}{\widetilde{\tau}}

\noindent
{\small
We consider theories with spontaneously broken global or local
supersymmetry where the pseudo-goldstino or the gravitino is the
lightest superparticle (LSP).  Assuming that the long-lived
next-to-lightest superparticle (NSP) is a charged slepton, we study
several supergravity predictions: the NSP lifetime, angular and energy
distributions in 3-body NSP decays. The characteristic couplings of
the gravitino, or goldstino, can be tested even for very small masses.
}

\subsubsection*{Introduction}

The discovery of supersymmetry at the Tevatron, the LHC or a future
Linear Collider would raise the question how supersymmetry is realized
in nature. Clearly, supersymmetry is broken. Spontaneously broken
global supersymmetry would predict the existence of a spin-1/2
goldstino ($\chi$) whereas the theoretically favoured case of local
supersymmetry requires a massive spin-3/2 gravitino ($\psi_{3/2}$).

In a recent paper \cite{BHRY} we have studied how a massive gravitino,
if it is the lightest superparticle (LSP), may be discovered in decays
of $\stau$, the scalar $\tau$ lepton, which is naturally the
next-to-lightest superparticle (NSP). The determination of gravitino
mass and spin appears feasible for gravitino masses in the range from
about $1\,\mathrm{GeV}$ to $100\,\mathrm{GeV}$.  As we shall discuss
in this note, evidence for the characteristic couplings of a
pseudo-goldstino, which corresponds to the spin-1/2 part of the gravitino, 
can be obtained even for masses much smaller than $1\,\mathrm{GeV}$.  

The gravitino mass depends on the mechanism of supersymmetry breaking.
It can be of the same order as other superparticle masses, like in
gaugino mediation~\cite{Kaplan:1999ac} or gravity
mediation~\cite{Nilles:1984ge}. But it might also be much smaller as
in gauge mediation scenarios \cite{Giudice:1998bp}. As LSP, the
gravitino is an attractive dark matter candidate~\cite{Bolz:1998ek}.

The $\stau$ NSP has generally a long lifetime because of the small,
Planck scale suppressed coupling to the gravitino LSP.  The production
of charged long-lived heavy particles at colliders is an exciting
possibility~\cite{Drees:1990yw}.  They can be directly produced in
pairs or in cascade decays of heavier superparticles.  In the context
of models with gauge mediated supersymmetry breaking the production of
slepton NSPs has previously studied for the
Tevatron~\cite{Feng:1997zr}, for the LHC~\cite{Ambrosanio:2000ik} and
for a Linear Collider~\cite{Ambrosanio:1999iu}.

The dominant $\stau$ NSP decay channel is $\stau\to\tau+\text{missing
energy}$.  In the following we shall study how to identify the
gravitino or goldstino as carrier of the missing energy. First, one
will measure the NSP lifetime. Since the gravitino couplings are fixed
by symmetry, the NSP lifetime is predicted by supergravity given the
gravitino mass, which can be inferred from kinematics. Second, one can
make use of the 3-body NSP decay $\stau\to\tau+\gamma+X$ where
$X=\psi_{3/2}$ or $X=\chi$.  The angular and energy distributions and
the polarizations of the final state photon and lepton carry the
information on the spin and couplings of gravitino or goldstino.

For gravitino masses in the range from about $10\,\mathrm{keV}$ to
$100\,\mathrm{GeV}$, the NSP is essentially stable for collider
experiments, and one has to accumulate the NSPs to study their
decay. Sufficiently slow, strongly ionizing sleptons will be stopped
within the detector. One may also be able to collect faster sleptons
in a storage ring.  For gravitino masses less than
$\mathcal{O}(10\,\mathrm{keV})$ the $\stau$ can decay inside the
detector, which may be advantageous from the experimental point of
view.

At LHC one expects $\mathcal{O}(10^6)$ NSPs per year which are mainly
produced in cascade decays of squarks and gluinos
\cite{Beenakker:1997ch}.  The NSPs are mostly produced in the forward
direction~\cite{Maki:1998ih} which should make it easier to accumulate
$\stau$s in a storage ring.  In a Linear Collider an integrated
luminosity of $500\,\mathrm{fb}^{-1}$ will yield $\mathcal{O}(10^5)$
$\stau$s
\cite{sec4_lctdrs}. Note that, in a Linear Collider, one
can also tune the velocity of the produced $\stau$s by adjusting the
$e^+e^-$ center-of-mass energy. A detailed study of the possibilities
to accumulate $\stau$ NSPs is beyond the scope of this note. In the
following we shall assume that a sufficiently large number of $\stau$s
can be produced and collected.

This study is strongly based on Ref.~\cite{BHRY}. Here, we discuss
in more detail the case of a very light gravitino, or
pseudo-goldstino, for which the $\stau$ NSP can decay inside the
detector.  Although in this case it is difficult to determine mass and
spin of the gravitino, one can still see the characteristic coupling
of the gravitino, which is essentially the goldstino coupling, via the
3-body decay $\widetilde{\tau}\to\tau+\gamma+X$ with $X=\psi_{3/2}$ or
$X=\chi$.

\subsubsection*{Planck mass from $\boldsymbol{\stau}$ decays}

The $\stau$ decay rate is dominated by the two-body decay into $\tau$
and gravitino,
\begin{eqnarray}
 \Gamma_{\stau}^\mathrm{2-body}
 & = &
 \frac{\left( m_{\stau}^2 - m_{3/2}^2 - m_{\tau}^2 \right)^4 }{
	48\pi\,m_{3/2}^2\,M_\mathrm{P}^2\,m_{\stau}^3 }\,
 \left[1-\frac{4m_{3/2}^2\,m_{\tau}^2}{
 	\left( m_{\stau}^2 - m_{3/2}^2 - m_{\tau}^2 \right)^2} \right]^{3/2}\;,
 \label{eq:2bodyDecayRateWithTau}
\end{eqnarray}
where $M_\mathrm{P}=(8\pi\, G_\mathrm{N})^{-1/2}$ denotes the reduced
Planck mass, $m_\tau=1.78\,\mathrm{GeV}$ is the $\tau$ mass,
$m_{\stau}$ is the $\stau$ mass, and $m_{3/2}$ is the gravitino
mass. For instance, $m_{\stau}=150\,\mathrm{GeV}$ and
$m_{3/2}=10\,\mathrm{keV}$ leads to a lifetime of
$\Gamma_{\stau}^{-1}\simeq 7.8\times 10^{-7}\,\mathrm{s}$, and
$m_{\stau}=150\,\mathrm{GeV}$ and $m_{3/2}=75\,\mathrm{GeV}$ results
in $\Gamma_{\stau}^{-1}\simeq 4.4\,\mathrm{y}$.

Since the decay rate depends only on two unknown masses $m_{\stau}$
and $m_{3/2}$, independently of other SUSY parameters, gauge and
Yukawa couplings, it is possible to test the prediction of the
supergravity if one can measure these masses. The mass $m_{\stau}$ of
the NSP will be measured in the process of accumulation.  Although the
outgoing gravitino is not directly measurable, its mass can also be
inferred kinematically unless it is too small,
\begin{eqnarray}
  m_{3/2}^2 &=& m_{\stau}^2 + m_\tau^2 -2m_{\stau} E_\tau\;.
\end{eqnarray}
The gravitino mass can be determined with the same accuracy as
$E_\tau$ and $m_{\stau}$, i.e.\ with an uncertainty of a few GeV.

Once the masses $m_{\stau}$ and $m_{3/2}$ are measured, one can
compare the predicted decay rate \eqref{eq:2bodyDecayRateWithTau} with
the observed decay rate, thereby testing an important supergravity
prediction.  In other words, one can determine the `supergravity
Planck scale' from the NSP decay rate which yields, up to
$\mathcal{O}(\alpha)$ corrections,
\begin{eqnarray}
 M_\mathrm{P}^2(\text{supergravity}) & = &
 \frac{\left( m_{\stau}^2 - m_{3/2}^2 - m_{\tau}^2 \right)^4 }{
	48\pi\,m_{3/2}^2\,m_{\stau}^3\, \Gamma_{\stau}}\,
 \left[1-\frac{4m_{3/2}^2\,m_{\tau}^2}{
 	\left( m_{\stau}^2 - m_{3/2}^2 - m_{\tau}^2 \right)^2} \right]^{3/2}\;.
\end{eqnarray}
The result can be compared with the Planck scale of Einstein gravity,
i.e.\ Newton's constant determined by macroscopic measurements,
$G_\mathrm{N}=6.707(10)\cdot10^{-39}\,\mathrm{GeV}^{-2}$
\cite{Hagiwara:2002fs},
\begin{eqnarray}
  M_\mathrm{P}^2(\mathrm{gravity}) &=& (8\pi\, G_\mathrm{N})^{-1} 
  \,=\, (2.436(2)\cdot 10^{18}\,\mathrm{GeV})^2\;.
\end{eqnarray}
The consistency of the microscopic and macroscopic determinations of
the Planck scale is an unequivocal test of supergravity.

Note that the measurement of the gravitino mass yields another
important quantity in supergravity, the mass scale of spontaneous
supersymmetry breaking $M_\mathrm{SUSY}\,=\,
\sqrt{\sqrt{3}M_\mathrm{P}\,m_{3/2}}$.  This is the analogue of the
Higgs vacuum expectation value $v$ in the electroweak theory, where $v
= \sqrt{2}m_W/g = (2\sqrt{2}G_\mathrm{F})^{-1/2}$.

\subsubsection*{Gravitino and goldstino versus neutralino}

If the measured decay rate and the kinematically determined mass of
the invisible particle are consistent with
Eq.~\eqref{eq:2bodyDecayRateWithTau}, one already has strong evidence
for supergravity and the gravitino LSP. To analyze the couplings of
the invisible particle, one can study the 3-body decay
$\stau\to\tau+\gamma+X$ for the gravitino $X=\psi_{3/2}$ and compare
it with the case where $X$ is a hypothetical spin-1/2 neutralino.
This is of particular importance if the mass of the invisible particle
is so small that the supergravity prediction for the NSP lifetime, 
as described in the previous section, cannot be tested.

The NSP $\stau$ is in general a linear combination of
$\stau_\mathrm{R}$ and $\stau_\mathrm{L}$, the superpartners of the
right-handed and left-handed $\tau$ leptons $\tau_\mathrm{R}$ and
$\tau_\mathrm{L}$, respectively.  The interaction of the gravitino
$\psi_{3/2}$ with scalar and fermionic $\tau$ leptons is described by
the lagrangian \cite{Wess:1992cp},
\begin{eqnarray}
 L_{3/2}
 & = & 
 -\frac{1}{\sqrt{2}M_\mathrm{P}}
 \left[
   (D_\nu\,\stau_\mathrm{R})^*\overline{\psi^\mu}\,\gamma^\nu\,\gamma_\mu\,
   P_\mathrm{R} \tau
   +
   (D_\nu\,\stau_\mathrm{R})\,\overline{\tau} P_\mathrm{L}
   \gamma_\mu\,\gamma^\nu\,\psi^\mu\right]\;,
 \label{eq:FullGravitinoLagrangian}
\end{eqnarray}
where $D_\nu\,\stau_\mathrm{R} = (\partial_\nu + \I e\, A_\nu)
\stau_\mathrm{R}$ and $A_\nu$ denotes the gauge boson.  The
interaction lagrangian of $\stau_\mathrm{L}$ has an analogous form.

As an example for the coupling of a hypothetical spin-1/2 neutralino
to $\stau$ and $\tau$, we consider the Yukawa
interaction\footnote{This interaction would arise from gauging the
anomaly free U(1) symmetry $L_{\tau}-L_{\mu}$, the difference of
$\tau$- and $\mu$-number, in the MSSM, with $\lambda$ being the
gaugino.},
\begin{equation}
 L_\mathrm{Yukawa}\,=\,
  h\left(\stau_\mathrm{R}^*\,\overline{\lambda}\,P_\mathrm{R}\,\tau
         +\stau_\mathrm{L}^*\,\overline{\lambda}\,P_\mathrm{L}\,\tau \right)
+\text{h.c.}\;.
\label{eq:LYukawa}
\end{equation}
Note that for very small coupling $h$, the $\stau$ decay rate could
accidentally be consistent with the supergravity prediction
Eq.~(\ref{eq:2bodyDecayRateWithTau}).

Also the goldstino $\chi$ has Yukawa couplings of the type given in
Eq.~(\ref{eq:LYukawa}).  The full interaction lagrangian is obtained
by performing the substitution $\psi_\mu\to \sqrt{{2\over 3}}{1\over
m_{3/2}}\partial_\mu\chi$ in the supergravity lagrangian. The 
non-derivative form of the effective lagrangian for $\chi$ is given by
\cite{Lee:1998aw},
\begin{equation}
 L_\mathrm{eff}
 \,=\,
 \frac{m_{\widetilde{\tau}}^2}{\sqrt{3}M_\mathrm{P}\,m_{3/2}}
 \left(\widetilde{\tau}_\mathrm{R}^*\,\overline{\chi}\,P_\mathrm{R}\,\tau
 +\widetilde{\tau}_\mathrm{R}\,\overline{\tau}\,P_\mathrm{L}\,\chi\right)
 -\frac{m_{\widetilde{\gamma}}}{4\sqrt{6}M_\mathrm{P}\,m_{3/2}}
 \overline{\chi}[\gamma^\mu,\gamma^\nu]\,\widetilde{\gamma}\,F_{\mu\nu}\;,
 \label{eq:GoldstinoLagrangian}
\end{equation}
where we have neglected a quartic interaction term which is irrelevant
for our discussion. In the following, we consider a massive
pseudo-goldstino $\chi$, in order to compare it with the massive
gravitino and neutralino.  Like a pseudo-Goldstone boson, the
pseudo-goldstino has goldstino couplings and a mass which explicitly
breaks global supersymmetry.

Note that the goldstino coupling to the photon supermultiplet is
proportional to the photino mass $m_{\widetilde{\gamma}}$. As a
consequence, the contribution to 3-body $\stau$-decay with
intermediate photino (cf.\ Fig.~\ref{fig:PhotinoContribution}) is not
suppressed for very large photino masses. As we shall see, this leads
to significant differences between the angular distributions for pure
Yukawa and goldstino couplings, even when $\chi$ and $\lambda$ are
very light.

\begin{figure}[t]
\begin{center}
 \subfigure[{}]{
 \CenterObject{\includegraphics[scale=1]{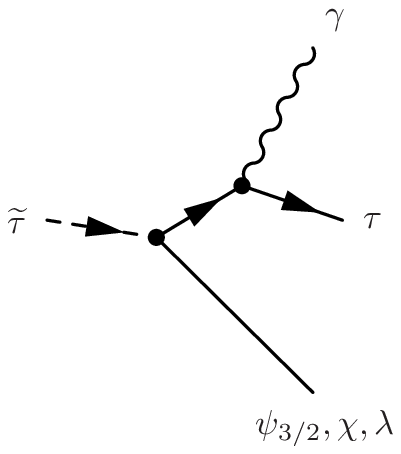}}}
\hfil
 \subfigure[{}]{
 \CenterObject{\includegraphics[scale=1]{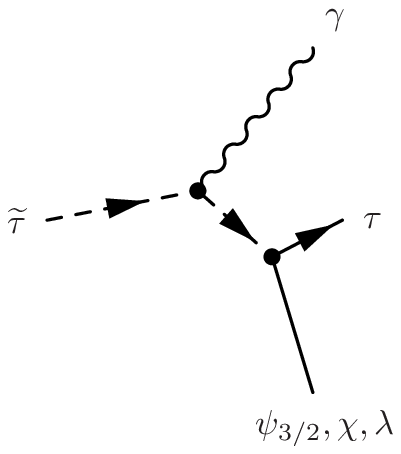}}}
 \\
 \subfigure[{}\label{fig:PhotinoContribution}]{
 \CenterObject{\includegraphics[scale=1]{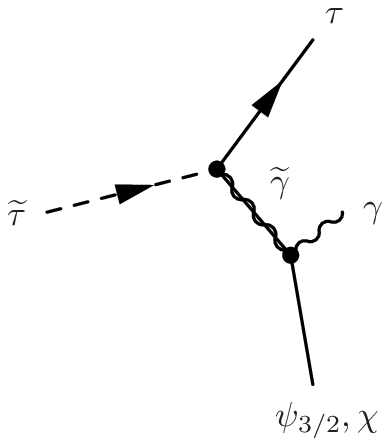}}}
 \hfil
 \subfigure[{}\label{fig:4piontVertex}]{
 \CenterObject{\includegraphics[scale=1]{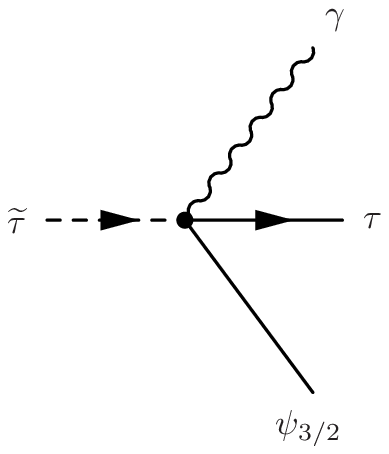}}}
\end{center} 
 \caption{Diagrams contributing to the 3-body decay
 $\widetilde{\tau}\to\tau+ \gamma+X$ where
 $X=\psi_{3/2},\chi,\lambda$. In the limit of very large
 $m_{\widetilde{\gamma}}$, diagram (c) becomes irrelevant for the
 gravitino, but it always contributes for the goldstino where it leads
 effectively to a 4-point interaction.}
 \label{fig:3bodystauR2tauRgammaGoldstino}
\end{figure}

In $\stau$ decays both, photon and $\tau$ lepton will mostly be very
energetic.  Hence the photon energy $E_{\gamma}$ and the angle
$\theta$ between $\tau$ and $\gamma$ can be well measured (cf.\
Fig.~\ref{fig:KinematicalConfiguration}).  We can then compare the
differential decay rate
\begin{equation}
 \Delta(E_\gamma ,\cos\theta)\,=\,{1\over\alpha\,\Gamma_{\widetilde{\tau}}}
 \frac{\D^2\Gamma(\widetilde{\tau}\to\tau+\gamma+X)}{\D E_\gamma\,\D \cos\theta}\;,
\end{equation}
for the gravitino LSP ($X=\psi_{3/2}$), the pseudo-goldstino
($X=\chi$) and the hypothetical neutralino ($X=\lambda$). 
Details of the calculation are given in
Ref.~\cite{BHRY}.  The differences between $\psi_{3/2}$, $\chi$ and
$\lambda$ become significant in the backward direction ($\cos\theta <
0$) as demonstrated by Fig.~\ref{fig:CompareDifferentialDecayRate}
(b)-(d), where $m_{\widetilde{\tau}}=150\,\mathrm{GeV}$ and
$m_{X}=75\,\mathrm{GeV}$ ($X=\psi_{3/2},\chi,\lambda$). The three
differential distributions are qualitatively different and should
allow to distinguish experimentally gravitino, goldstino and
neutralino.

Let us now consider the case of small $m_X$. Then  the goldstino lagrangian
\eqref{eq:GoldstinoLagrangian} effectively describes the
gravitino interactions. Therefore, one can no longer distinguish
between gravitino and goldstino in this case.  However, even for small
$m_X$ one can discriminate the gravitino or goldstino from the
neutralino. The difference between goldstino $\chi$ and neutralino
$\lambda$ stems from the photino contribution (cf.\
Fig.~\ref{fig:PhotinoContribution}) which does not decouple for large
photino mass $m_{\widetilde{\gamma}}$.  This is different from the
gravitino case where the analogous diagram becomes irrelevant in the
limit $m_{\widetilde{\gamma}}\gg m_{\stau}$ which is employed
throughout this study. 

The arising discrepancy between gravitino or goldstino and neutralino
is demonstrated by Fig.~\ref{fig:DiffDecayRateGoldstino}. It clearly
shows that even for very small masses $m_{3/2}$ and $m_\lambda$, the
differential decay rates $\Delta$ for gravitino $\psi_{3/2}$ and
neutralino $\lambda$ are distinguishable. This makes it possible to
discriminate gravitino and goldstino from a hypothetical neutralino
even for very small masses. Note that the plots of 
Fig.~\ref{fig:DiffDecayRateGoldstino} remain essentially the same as long as 
$r=m_X^2/m_{\stau}^2\ll 1$.

\begin{figure}[htb!]
\begin{center}
 \subfigure[Kinematical configuration. The arrows denote the momenta.
 \label{fig:KinematicalConfiguration}]{
 	\CenterObject{\includegraphics{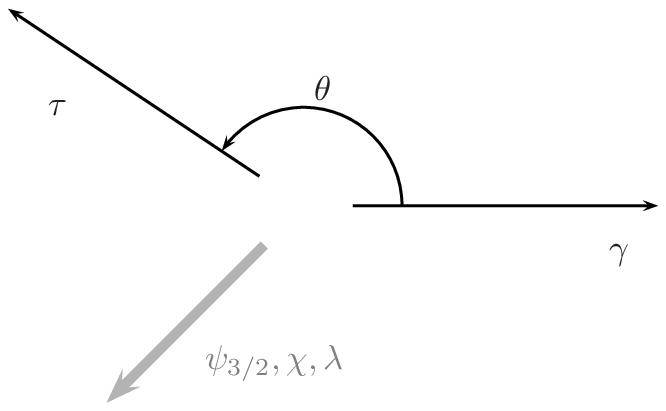}}}
 \quad
 \subfigure[Gravitino $\psi_{3/2}$]{
 	\CenterObject{\includegraphics[scale=0.85]{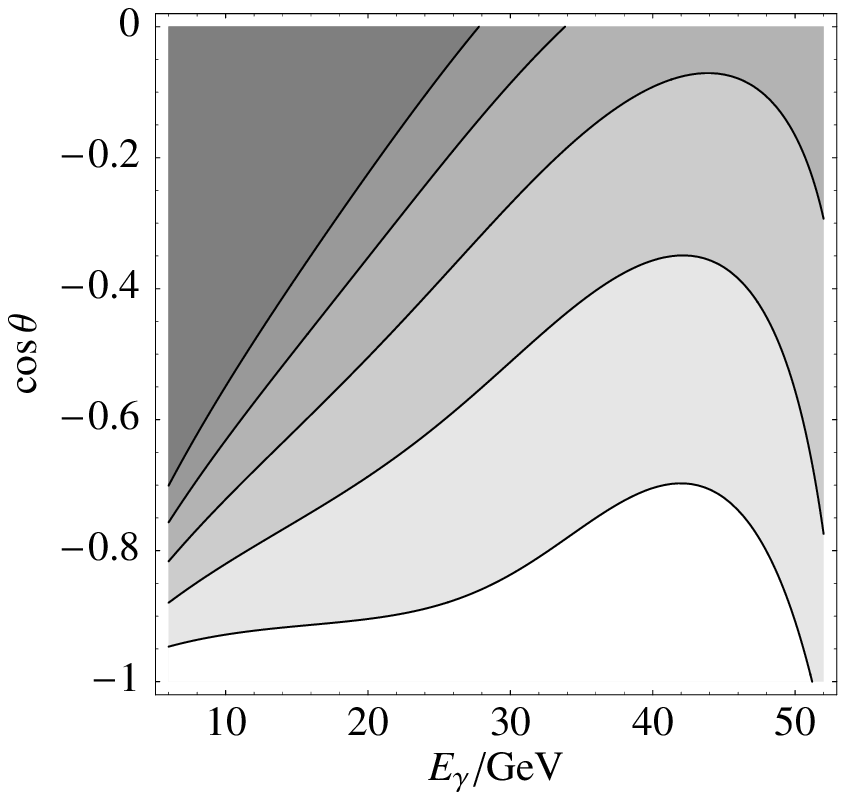}}}
 \\
 \subfigure[Pseudo-goldstino $\chi$]{
 	\CenterObject{\includegraphics[scale=0.85]{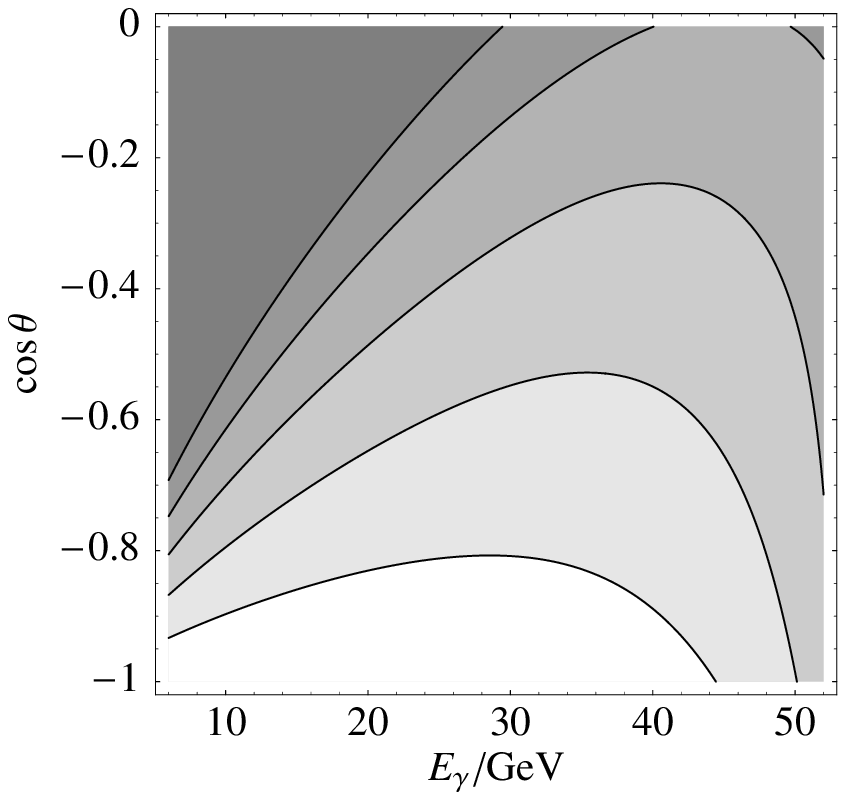}}}
 \quad
 \subfigure[Spin-1/2 neutralino $\lambda$]{
 	\CenterObject{\includegraphics[scale=0.85]{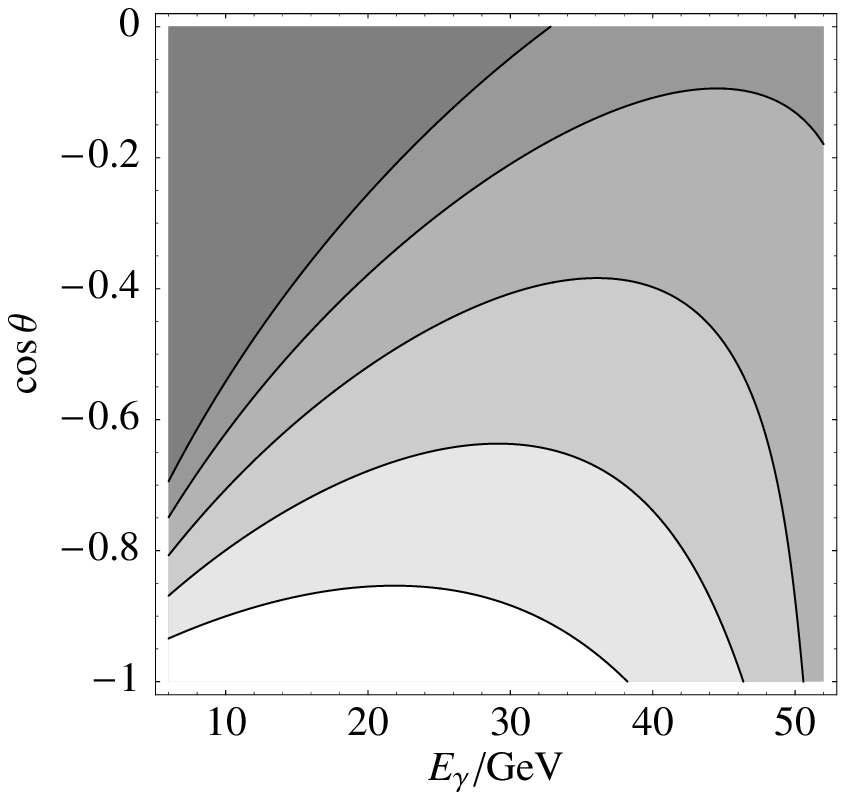}}}
\end{center}
\caption{(a) shows the kinematical configuration of the 3-body decay.
  The others are contour plots of the differential decay rates for (b)
  gravitino $\psi_{3/2}$, (c) pseudo-goldstino $\chi$ and (d)
  neutralino $\lambda$.  $m_{\widetilde{\tau}}=150\,\mathrm{GeV}$ and
  $m_{X}=75\,\mathrm{GeV}$ ($X=\psi_{3/2},\lambda,\chi$).  The
  boundaries of the different gray shaded regions (from bottom to top)
  correspond to $\Delta(E_\gamma ,\cos\theta)[\mathrm{GeV}^{-1}]
  =10^{-3}, 2\times10^{-3}, 3\times10^{-3}, 4\times10^{-3},
  5\times10^{-3}$.  Darker shading implies larger rate.}
\label{fig:CompareDifferentialDecayRate}
\end{figure}


Let us finally comment on the experimental feasibility to determine
gravitino or goldstino couplings. The angular distribution of the
3-body decay is peaked in forward direction ($\theta=0$).  Compared to
the 2-body decay, backward ($\cos\theta<0$) 3-body decays are
suppressed by $\sim 10^{-1}\times\alpha\simeq10^{-3}$. Requiring
10\dots100 events for a signal one therefore needs $10^4$ to $10^5$
$\stau$s, which appears possible at the LHC and also at a Linear
Collider according to the above discussion.

\begin{figure}[htb!]
 \begin{center}
  \subfigure[Gravitino/goldstino]{%
    \CenterObject{\includegraphics[scale=0.85]{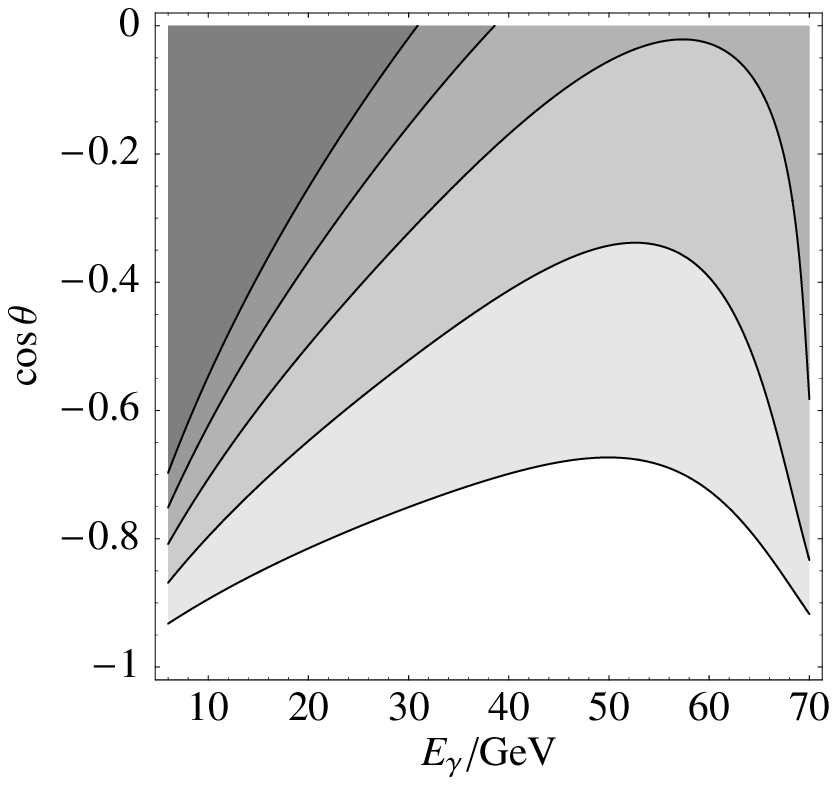}}
	}
	\hfil
   \subfigure[Neutralino]{%
    \CenterObject{\includegraphics[scale=0.85]{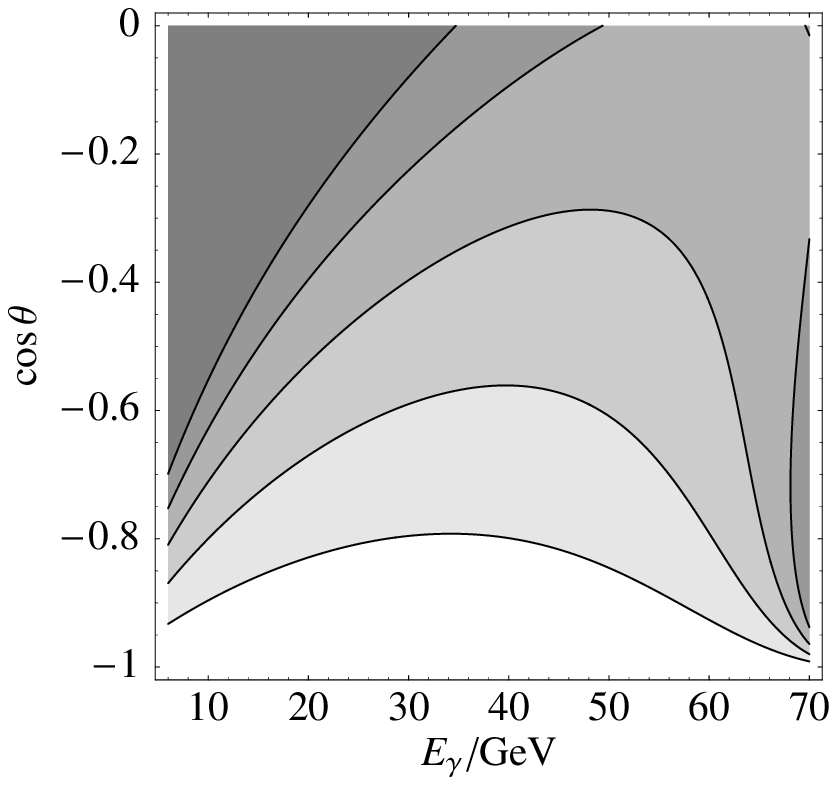}}
	}
  \end{center}
  \caption{Contour plots of the differential decay rates for (a)
  gravitino $\psi_{3/2}$ and (b) neutralino $\lambda$.
  $m_{\widetilde{\tau}}=150\,\mathrm{GeV}$, $m_{X}= 0.1~\mathrm{GeV}$
  ($X=\psi_{3/2},\lambda$). The figures remain essentially the
  same as long as $r=m_X^2/m_{\stau}^2\ll 1$. The contours have the same
  meaning as in Fig.~\ref{fig:CompareDifferentialDecayRate}.}
  \label{fig:DiffDecayRateGoldstino}
\end{figure}

\subsubsection*{Gravitino spin}

A third test of supergravity is intuitively more straightforward
though experimentally even more challenging than the previous ones. It
is again based on 3-body decays. We now take into account also the
polarizations of the visible particles, $\gamma$ and $\tau$. The main
point is obvious from Fig.~\ref{fig:TypicalSpin32} where a left-handed
photon and a right-handed $\tau$ move in opposite
directions.\footnote{For simplicity, we here restrict ourselves to the
case of a right-handed $\stau$ LSP, leaving finite left-right mixing
angles for future investigations.} Clearly, this configuration is
allowed for an invisible spin-3/2 gravitino but it is forbidden for a
spin-1/2 goldstino or neutralino.  Unfortunately, measuring the
polarizations is a difficult task.

\begin{figure}[!htb]
\begin{center}
 \subfigure[Characteristic spin-$3/2$ process. The thick arrows represent the
 	spins.\label{fig:TypicalSpin32}]{
 	\CenterObject{\includegraphics{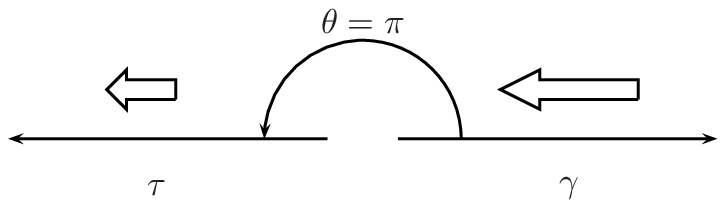}}}
 \hfil
 \subfigure[$m_X=10\,\mathrm{GeV}$.]{
 	\CenterObject{\includegraphics[scale=0.85]{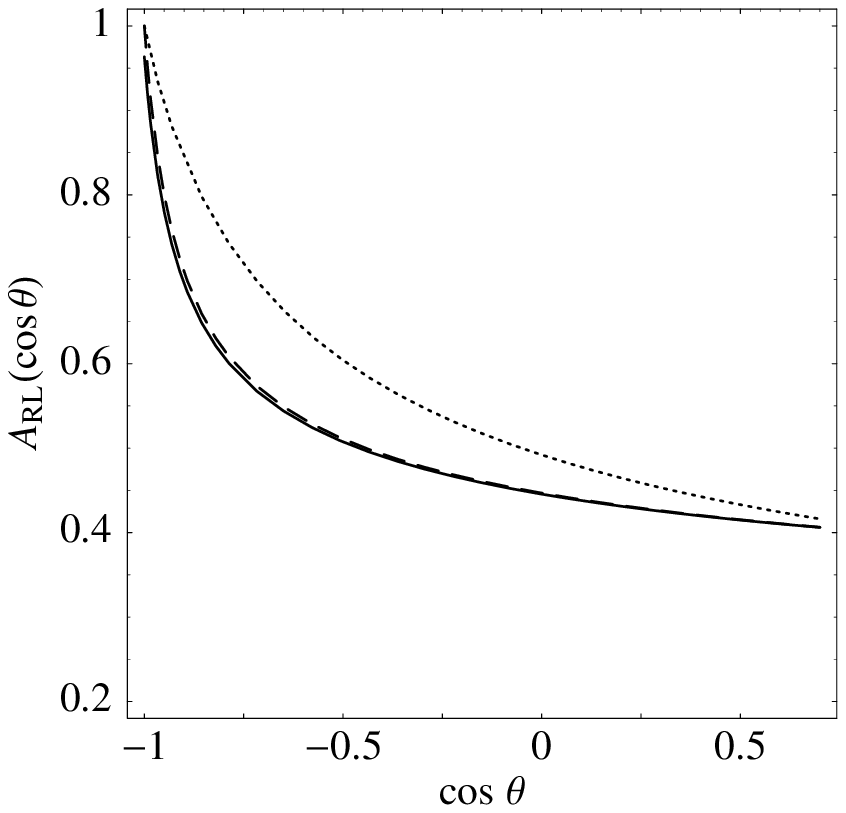}}}
 \\
 \subfigure[$m_X=30\,\mathrm{GeV}$.]{
 	\CenterObject{\includegraphics[scale=0.85]{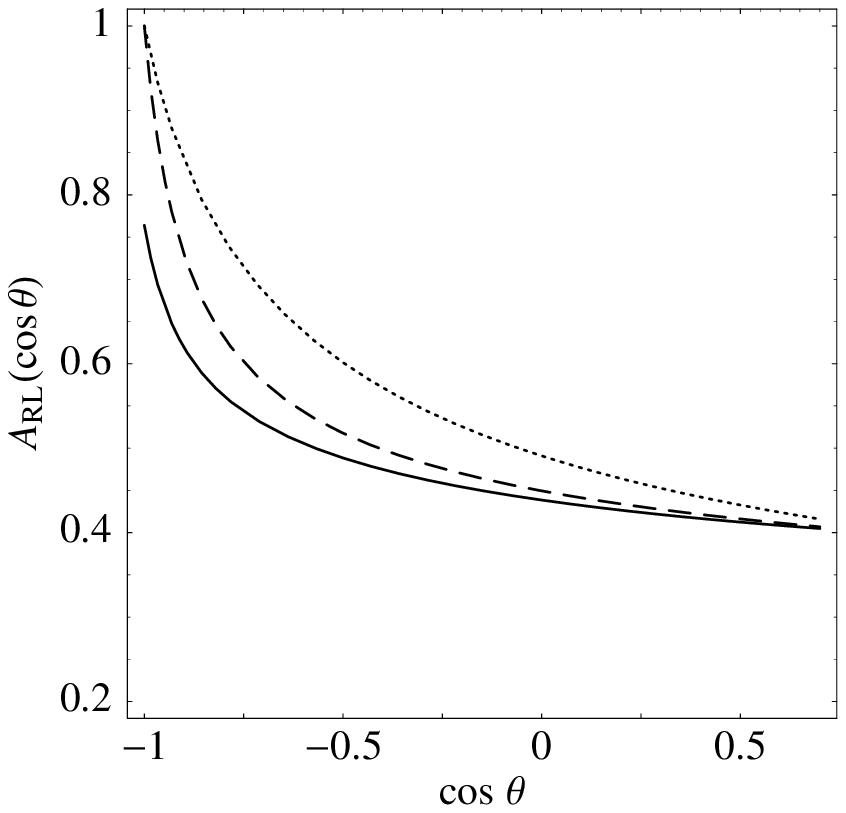}}}
 \hfil
 \subfigure[$m_X=75\,\mathrm{GeV}$.]{
 	\CenterObject{\includegraphics[scale=0.85]{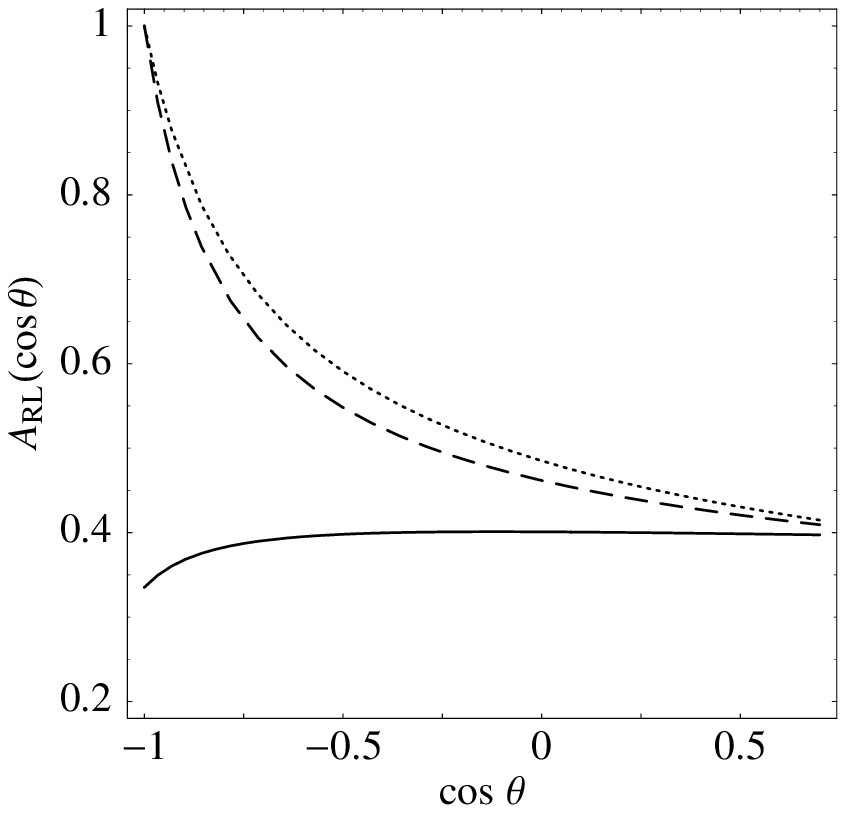}}}
\end{center} 
\caption{(a) illustrates the characteristic spin-3/2 process: 
 photon and $\tau$ lepton move in opposite directions and the spins
 add up to $3/2$, so the invisible particle also has spin $3/2$.  The
 other figures show angular asymmetries for gravitino $\psi_{3/2}$
 (solid curve), goldstino $\chi$ (dashed curve) and neutralino
 $\lambda$ (dotted curve).  $m_{\stau}=150\,\mathrm{GeV}$. The photon
 energy is larger than $10\%$ of the maximal kinematically allowed
 energy (cf.\ Ref.~\cite{BHRY}).  Note that the asymmetries only
 depend on the ratio $r=m_X^2/m_{\stau}^2$ ($X=\psi_{3/2},
 \chi,\lambda$).}
\label{fig:CompareLandR}
\end{figure}


As Fig.~\ref{fig:TypicalSpin32} illustrates, the spin of the invisible
particle influences the angular distribution of final states with
polarized photons and $\tau$ leptons.  An appropriate observable is
the angular asymmetry
\begin{equation}
 A_\mathrm{RL}(\cos\theta)\,=\,
 \frac{\displaystyle 
 \frac{\D \Gamma}{\D\cos\theta}(\stau_\mathrm{R}
 \to\tau_\mathrm{R}+\gamma_\mathrm{R}+X)
 -\frac{\D \Gamma}{\D\cos\theta}(\stau_\mathrm{R}
 \to\tau_\mathrm{R}+\gamma_\mathrm{L}+X)}{\displaystyle
 \frac{\D \Gamma}{\D\cos\theta}(\stau_\mathrm{R}
 \to\tau_\mathrm{R}+\gamma_\mathrm{R}+X)
 +\frac{\D \Gamma}{\D\cos\theta}(\stau_\mathrm{L}
 \to\tau_\mathrm{R}+\gamma_\mathrm{L}+X)}\;,
\end{equation}
where $X$ denotes gravitino ($X = \psi_{3/2}$), goldstino ($X=\chi$)
or neutralino ($X=\lambda$). Note that, as discussed before, the
photino does not decouple in the case $X=\chi$.

The three angular asymmetries are shown in Fig.~\ref{fig:CompareLandR}
for $m_{\stau}=150\,\mathrm{GeV}$ and different masses of the
invisible particle.  As expected, the decay into right-handed $\tau$
and left-handed photon at $\theta=\pi$ is forbidden for spin-1/2
invisible particles ($\chi$ and $\lambda$), whereas it is allowed for
the spin-3/2 gravitino. This is clearly visible in
Figs.~\ref{fig:CompareLandR}(c) and (d); for small gravitino masses
the goldstino component dominates the gravitino interaction as
illustrated by Fig.~\ref{fig:CompareLandR}(b). The discrepancy between
gravitino and goldstino compared to a hypothetical neutralino persists
for arbitrarily small $m_X$, which is analogous to the double
differential distribution discussed in the previous section.

\subsubsection*{Conclusions}

We have discussed how one may discover a massive gravitino, and
thereby supergravity, at the LHC or a future Linear Collider, if the
gravitino is the LSP and a charged slepton is the NSP. With the
gravitino mass inferred from kinematics, the measurement of the NSP
lifetime will test an unequivocal prediction of supergravity. The
analysis of 3-body NSP decays will reveal the couplings of the
gravitino or the goldstino. For very small masses, one can distinguish
the gravitino from the neutralino but not from the goldstino.  For
masses larger than about $1\,\mathrm{GeV}$, the determination of 
gravitino mass and spin appears feasible.
 
%


\subsection{\label{sec:434} Reconstructing supersymmetric theories by
coherent LHC / LC analyses}

{\it B.C.~Allanach, G.A.~Blair, S.~Kraml, H.-U.~Martyn, G.~Polesello, 
W.~Porod, P.M.~Zerwas}

\vspace{1em}
\renewcommand{\cha}{\tilde{\chi}}
\renewcommand{\neu}{\tilde{\chi}^0}
\def\SPHENO{SPheno\,2.2.0}
                                                                                
                                                                                
{\small
\noindent
Supersymmetry analyses will potentially be a central
area for experiments at the LHC and at a future $e^+ e^-$ linear collider.
Results from the two facilities will mutually complement and
augment each other so that
a comprehensive 
and precise picture of the supersymmetric world can be developed.
We will demonstrate in this report how coherent analyses at LHC and LC
experiments can be used to explore the breaking mechanism of supersymmetry
and to reconstruct the fundamental theory at high energies, in particular
at the grand unification scale. This will be exemplified for minimal
supergravity in detailed experimental simulations performed
for the Snowmass reference point SPS1a.
}
                                                                                

\subsubsection{Physics Base}

The roots of standard particle physics are expected to go as 
deep as the Planck length of $10^{-33}$~cm where gravity is 
intimately linked to the particle system.  A stable bridge
between the electroweak energy scale of 100~GeV and
the vastly different Planck scale of $\Lambda_{\rm PL}\sim10^{19}$~GeV,
and the (nearby) grand unification scale $\Lambda_{\rm GUT}\sim10^{16}$~GeV,
is provided by supersymmetry.  If this scenario is realized in nature,
experimental methods must be developed to shed light on the physics 
phenomena near $\Lambda_{\rm GUT}$/$\Lambda_{\rm PL}$.  Among other
potential tools, the extrapolation of supersymmetry (SUSY) 
parameters measured at 
the LHC and an e$^+$e$^-$ linear collider with high precision, can play
a central r\^ole~\cite{r1}.  A rich ensemble of gauge and Yukawa couplings,
and of gaugino/higgsino and scalar particle
masses allows the detailed study of the 
supersymmetry breaking mechanism and the reconstruction of the 
physics scenario near the GUT/PL scale.

The reconstruction of physical structures at energies more than fourteen
orders above the energies available through accelerators is
a demanding task.  Not only must a comprehensive picture
be delineated near the electroweak scale, but
the picture must be drawn, moreover, as precisely as possible to keep
the errors small enough so that they do not blow up beyond control when the
SUSY parameters are extrapolated over many orders of magnitude.
The LHC~\cite{r2} and a future e$^+$e$^-$ linear collider (LC)~\cite{r3}
are a perfect tandem for solving such a problem: {\bf (i)} While the
colored supersymmetric particles, gluinos and squarks, can be
generated with large rates for masses up to 
2 to 3~TeV at the LHC, the strength of e$^+$e$^-$ linear colliders
is the comprehensive coverage of the non-colored particles, 
charginos/neutralinos and sleptons.  If the extended Higgs spectrum
is light, the Higgs particles can be discovered and investigated at
both facilities; heavy Higgs bosons can be produced at the 
LHC in a major part of the parameter space; at an e$^+$e$^-$
collider, without restriction, for masses up to the beam energy;
{\bf (ii)} If the analyses are performed coherently, the 
accuracies in measurements of  cascade decays at LHC
and in threshold production as well as decays of supersymmetric
particles at LC complement and augment
each other {\it mutually} so that a high-precision
picture of the supersymmetric parameters at the electroweak scale
can be drawn.  Such a comprehensive and precise picture is 
necessary in order to carry out the evolution of the 
supersymmetric parameters to high scales, driven by perturbative 
loop effects that involve the entire supersymmetric particle spectrum.

Minimal supergravity (mSUGRA) provides us with a scenario within
which these general ideas can be quantified.  The  form of this
theory has been developed in great detail, creating a
platform on which semi-realistic experimental studies can be performed.
Supersymmetry is broken in mSUGRA in a hidden sector and the breaking
is transmitted to our eigenworld by gravity~\cite{Chamseddine:jx}.  This
mechanism suggests, yet does not enforce, the universality of the
soft SUSY breaking parameters -- gaugino and scalar masses, and
trilinear couplings -- at a scale that is generally identified with
the unification scale.  The (relatively) small number of
parameters renders mSUGRA a well-constrained system that suggests itself
in a natural way as a test ground for coherent experimental analyses
at LHC and LC.  The procedure will be exemplified for a specific set of
parameters, defined as SPS1a among the Snowmass reference
points~\cite{sec4_Allanach:2002nj,Ghodbane:2002kg}.

\subsubsection{Minimal Supergravity: SPS1a}

The mSUGRA Snowmass reference point SPS1a is characterised by
the following values~\cite{sec4_Allanach:2002nj,Ghodbane:2002kg}: 
\begin{eqnarray}
  \begin{array}{ll}
    M_{1/2} = 250~{\rm GeV} \qquad & M_0=100~{\rm GeV} \\
    A_0=-100~{\rm GeV} & {\rm sign}(\mu)=+\\
    \tan\beta=10 & \\
  \end{array}
\end{eqnarray}

\noindent
for the universal gaugino mass $M_{1/2}$, 
the scalar mass $M_0$, the trilinear coupling
$A_0$, the sign of the higgsino parameter $\mu$, and $\tan\beta$,
the ratio of the vacuum-expectation values of the two Higgs fields.
As the modulus of the Higgsino parameter is fixed at the
electroweak scale by requiring
radiative electroweak symmetry breaking, $\mu$ is finally given by:
\begin{equation}
  \mu = 357.4~{\rm GeV} 
\end{equation}

This reference point is compatible with the constraints
from low-energy precision data, predicting 
BR($b\rightarrow s\gamma) =2.7\cdot 10^{-4}$
and $\Delta[g_\mu-2]= 17\cdot 10^{-10}$.  The amount of cold dark matter
is, with $\Omega_\chi h^2=0.18$, on the high side but still compatible
with recent WMAP data \cite{r5} if evaluated on their own without 
reference to other experimental results;
moreover, only a slight shift in $M_0$ downwards drives the value
to the central band of the data while such a shift does not alter
any of the conclusions in this report in a significant way.

In the SPS1a scenario the squarks and gluinos
can be studied very well at the LHC while the non-colored gauginos
and sleptons can be analyzed partly at LHC and in comprehensive form
at an e$^+$e$^-$ linear collider operating at a total energy below
1 TeV with high integrated luminosity close to 1~ab$^{-1}$.

The masses can best be obtained at {\underline {LHC}} by analyzing 
edge effects in the
cascade decay spectra. The basic starting point is the identification
of the a sequence of two-body decays:
\mbox{$\tilde q_L\rightarrow\tilde\chi^0_2 q\rightarrow\tilde\ell_R\ell q
\rightarrow \tilde\chi^0_1\ell\ell q$}. This is effected  through the 
detection of an edge structure of the invariant mass 
of opposite-sign same-flavour leptons
from the $\chi^0_2$ decay in events with multi-jets and $E_T^{miss}$.
One can then measure the kinematic edges 
of the invariant mass distributions among the two leptons and the jet resulting
from the above chain, and 
thus an approximately model-independent determination 
of the masses of the involved sparticles is obtained. 
This technique was developed in Refs.~\cite{HinPai, Allanach:2000kt} 
and is worked 
out in detail for point SPS1a in Ref.~\cite{SPSLHC}.
The four sparticle masses ($\tilde q_L$, $\tilde\chi^0_2$,
$\tilde\ell_R$, and $\tilde\chi^0_1$) thus measured are used as an input
to additional analyses which rely on the knowledge
of the masses of the lighter gauginos in order 
to extract masses from the observed 
kinematic structures. Examples are the studies of 
the decay 
\mbox{$\tilde g\rightarrow\tilde b_1 b\rightarrow \tilde\chi^0_2 bb$},
where the reconstruction of the gluino and sbottom 
mass peaks relies on an approximate 
full reconstruction of the $\tilde\chi^0_2$, 
and the shorter decay chains \mbox{$\tilde q_R\rightarrow q \tilde\chi^0_1$}
and \mbox{$\tilde\chi^0_4\rightarrow\tilde\ell\ell$}, 
which require the 
knowledge of the sparticle masses downstream of the cascade.
For SPS1a the heavy Higgs bosons can also be searched for in 
the decay chain: 
$A^0 \to \tilde \chi^0_2 \tilde \chi^0_2
     \to \tilde \chi^0_1 \tilde \chi^0_1 l^+ l^- l^+ l^-$ 
\cite{moortgat}. The invariant four-lepton mass depends
sensitively on $m_{A^0}$ and $m_{\chi^0_1}$. The same holds true for $H^0$.
Note however that the main source of the neutralino final states are
$A^0$ decays, and that the two Higgs bosons $A^0$ and $H^0$ cannot 
be discriminated in this channel. 

The mass measurements obtained at the LHC are thus very 
correlated among themselves, and this correlation must be taken into
account in the fitting procedure.  Another source of correlation 
comes from the fact that in most cases the uncertainty on the
mass measurement is dominated by the systematic uncertainty on the 
hadronic energy scale of the experiment, which will affect 
all the measurements involving jets approximately by the same 
amount and in the same direction.

At {\underline {linear colliders}} very precise mass values can be extracted 
from decay spectra and threshold scans \cite{r6A,r6B,r6C}.
The excitation curves for chargino production
in S-waves~\cite{r7} rise steeply with the velocity of the particles
near the threshold and thus are very sensitive to their mass values;
the same is true for mixed-chiral selectron pairs in
$e^+e^-\to \tilde e_R^+ \tilde e_L^-$ 
and for diagonal pairs in 
$e^-e^-\to \tilde e_R^- \tilde e_R^-, \;  \tilde e_L^- \tilde e_L^-$
collisions \cite{r6C}.  
Other scalar sfermions, as well as neutralinos, 
are produced generally in P-waves, with a
somewhat less steep threshold behaviour proportional to the
third power of the velocity~\cite{r6C,r8}.  Additional information,
in particular on the lightest neutralino $\tilde{\chi}^0_1$, can
be obtained from decay spectra.

Typical mass parameters and the related measurement
errors are presented in Table~\ref{tab:massesA}.  
The column denoted ``LHC'' collects
the errors from the LHC analysis, the column ``LC'' the errors
expected from the LC operating at energies up to 1~TeV with an
integrated luminosity of $\sim1$~ab$^{-1}$.  
The error estimates are based on detector simulations for the production
of the light sleptons, $\tilde e_R$, $\tilde \mu_R$ and $\tilde \tau_1$,
in the continuum.
For the light neutralinos and the light chargino threshold scans have 
been simulated. 
Details will be given elsewhere; see also Ref.\cite{R13A}.
The expected precision of the other particle masses is taken from
Ref.\cite{r6C},
or it is obtained by scaling the LC errors from the previous 
analysis in Ref.\cite{r6A},
taking into account the fact that the 
$\tilde \chi^0 / \tilde \chi^\pm$ cascade decays
proceed dominantly via $\tau$ leptons in the reference point SPS1a,
which is experimentally  challenging.
The third column of Tab.~\ref{tab:massesA} denoted 
``LHC+LC'' presents the corresponding errors if the
experimental analyses are performed coherently, i.e. the
light particle spectrum, studied at LC with very high precision,
is used as an input set for the LHC analysis.

\renewcommand{\arraystretch}{1.1}
\begin{table}
\begin{center}
\begin{tabular}{|c||c||c|c||c|}
\hline
               & Mass, ideal & ``LHC''  &\ ``LC''\ & ``LHC+LC''
\\ \hline\hline
$\tilde{\chi}^\pm_1$ & 179.7 &          & 0.55   &  0.55  \\
$\tilde{\chi}^\pm_2$ & 382.3 &     --   & 3.0    &  3.0   \\
$\tilde{\chi}^0_1$   &  97.2 &     4.8  & 0.05   &  0.05  \\
$\tilde{\chi}^0_2$   & 180.7 &     4.7  & 1.2    &  0.08  \\
$\tilde{\chi}^0_3$   & 364.7 &          & 3-5    &  3-5   \\
$\tilde{\chi}^0_4$   & 381.9 &     5.1  & 3-5    &  2.23  \\
\hline
$\tilde{e}_R$        & 143.9 &     4.8  & 0.05   &  0.05  \\
$\tilde{e}_L$        & 207.1 &     5.0  & 0.2    &  0.2   \\
$\tilde{\nu}_e$      & 191.3 &     --   & 1.2    &  1.2   \\
$\tilde{\mu}_R$      & 143.9 &     4.8  & 0.2    &  0.2  \\
$\tilde{\mu}_L$      & 207.1 &     5.0  & 0.5    &  0.5  \\
$\tilde{\nu}_\mu$    & 191.3 &     --   &        &       \\
$\tilde{\tau}_1$     & 134.8 &     5-8  & 0.3    &  0.3  \\
$\tilde{\tau}_2$     & 210.7 &     --   & 1.1    &  1.1  \\
$\tilde{\nu}_\tau$   & 190.4 &     --   & --     &  --   \\
\hline
$\tilde{q}_R$        & 547.6 &    7-12  &    --  & 5-11  \\
$\tilde{q}_L$        & 570.6 &     8.7  &    --  &  4.9  \\
$\tilde{t}_1$        & 399.5 &          & 2.0    &  2.0  \\
$\tilde{t}_2$        & 586.3 &          &   --   &       \\
$\tilde{b}_1$        & 515.1 &     7.5  &    --  &  5.7  \\
$\tilde{b}_2$        & 547.1 &     7.9  &    --  &  6.2  \\
\hline
$\tilde{g}$          & 604.0 &     8.0  &    --  &  6.5  \\
\hline
$h^0$                & 110.8 &     0.25 & 0.05   & 0.05   \\
$H^0$                & 399.8 &          & 1.5    & 1.5   \\
$A^0$                & 399.4 &          & 1.5    & 1.5   \\
$H^{\pm}$            & 407.7 &     --   & 1.5    & 1.5   \\\hline 
\end{tabular}\\
\end{center}
\caption {{\it Accuracies for representative mass measurements
at ``LHC'' and ``LC'', and in coherent ``LHC+LC'' analyses
for the reference point SPS1a [masses in {\rm GeV}].
$\tilde q_L$ and $\tilde q_R$ represent the flavours
$q=u,d,c,s$ which cannot be distinguished at LHC.
Positions marked by bars cannot be filled either due to 
kinematical restrictions or due to small signal rates; blank positions 
could eventually be filled after significantly more investments 
in experimental simulation efforts than performed until now.
The ``LHC'' and ``LC'' errors have been derived in Ref.~\cite{SPSLHC}
and Ref.~\cite{LC-errors}, respectively, in this document.}} 
\label{tab:massesA}
\end{table}
 
Mixing parameters must be obtained from measurements of cross
sections and polarization asymmetries, 
in particular from the production of chargino pairs and
neutralino pairs~\cite{r7,r8}, both in diagonal or mixed form: 
$e^+e^- \rightarrow {\tilde{\chi}^+_i}{\tilde{\chi}^-_j}$
[$i$,$j$ = 1,2] and ${\tilde{\chi}^0_i} {\tilde{\chi}^0_j}$ 
[$i$,$j$ = 1,$\dots$,4]. 
The production cross sections for
charginos are binomials of $\cos\,2\phi_{L,R}$, the mixing angles
rotating current to mass eigenstates. Using polarized electron
and positron beams, the cosines can be determined in a model-independent
way.  [In specified models like MSSM the
analysis of the low-lying states is already sufficient for this
purpose \cite{DKMNP}.]

\renewcommand{\arraystretch}{1.1}
\begin{table}[t]
\begin{center}
\begin{tabular}{|c||c|c|}
\hline
           & Parameter, ideal   & {``LHC+LC''} errors
\\ \hline\hline
 $M_1$        & 101.66   &   0.08  \\
 $M_2$        & 191.76   &   0.25  \\
 $M_3$        & 584.9    &   3.9   \\
\hline
$\mu$         & 357.4    &   1.3   \\
\hline  
 $M^2_{L_1}$  &$3.8191 \cdot 10^4$ & 82.   \\
 $M^2_{E_1}$  &$1.8441 \cdot 10^4$ & 15.   \\
 $M^2_{Q_1}$  &$29.67 \cdot 10^4$  & $0.32\cdot 10^4$ \\
 $M^2_{U_1}$  &$27.67 \cdot 10^4$ &  $0.86 \cdot 10^4$ \\
  $M^2_{D_1}$ &$27.45 \cdot 10^4$ &  $0.80 \cdot 10^4$ \\
 $M^2_{L_3}$  &$3.7870 \cdot 10^4$&  360.  \\
 $M^2_{E_3}$  &$1.7788 \cdot 10^4$&   95.   \\
 $M^2_{Q_3}$  &$24.60 \cdot 10^4$ & $0.16 \cdot 10^4$\\
 $M^2_{U_3}$  &$17.61 \cdot 10^4$ & $0.12 \cdot 10^4$\\
 $M^2_{D_3}$  &$27.11 \cdot 10^4$ & $0.66 \cdot 10^4$\\
\hline
$M^2_{H_1} $  & \hspace*{5mm}$3.25 \cdot 10^4$ & $ 0.12 \cdot 10^4$ \\
$M^2_{H_2} $  &$-12.78 \cdot 10^4$& $0.11 \cdot 10^4$  \\
$A_t $        & $-497.$       &  9.    \\
\hline
$\tan\beta$   & 10.0              &  0.4   \\
\hline
\end{tabular}\\
\end{center}
\caption {{\it The extracted SUSY Lagrange mass and Higgs parameters 
at the electroweak scale in the reference point SPS1a
[mass units in {\rm GeV}].
}} 
\label{tab:params}
\end{table}

Based on this high-precision information, the fundamental SUSY
parameters can be extracted at low energy in analytic form. To lowest
order:
\begin{eqnarray}
\left|\mu\right|&=&M_W[\Sigma + \Delta[\cos2\phi_R+\cos2\phi_L]]^{1/2}
\nonumber\\
\mbox{sign}(\mu)&= &[ \Delta^2
                   -(M^2_2-\mu^2)^2-4m^2_W(M^2_2+\mu^2) \nonumber \\
 & &                   -4m^4_W\cos^2 2\beta]/8 m_W^2M_2|\mu|\sin2\beta 
\nonumber\\
M_2&=&M_W[\Sigma - \Delta(\cos2\phi_R+\cos2\phi_L)]^{1/2}\nonumber\\
|M_1|&=& \left[ \textstyle \sum_i m^2_{\tilde{\chi}_i^0}  
                 -M^2_2-\mu^2-2M^2_Z\right]^{1/2}
\nonumber\\
|M_3|&=&m_{\tilde{g}} \nonumber\\
\tan\beta&=&\left[\frac{1+\Delta (\cos 2\phi_R-\cos 2\phi_L)}
           {1-\Delta (\cos 2\phi_R-\cos 2\phi_L)}\right]^{1/2} 
\label{eqn:basicLE}
\end{eqnarray}
where $\Delta = (m^2_{\tilde{\chi}^\pm_2}-m^2_{\tilde{\chi}^\pm_1})/(4M^2_W)$
and 
$\Sigma =  (m^2_{\tilde{\chi}^\pm_2}+m^2_{\tilde{\chi}^\pm_1})/(2M^2_W) -1$.
The signs of $M_{1,3}$ with respect to $M_2$ follow from measurements of
the cross sections for ${\tilde{\chi}} {\tilde{\chi}}$ production and 
gluino processes. In practice one-loop corrections to the mass relations 
have been used to improve on the accuracy.  
 
The mass parameters of 
the sfermions are directly related to the
physical masses if mixing effects are negligible: 
\begin{equation}
m^2_{\tilde{f}_{L,R}}=M^2_{L,R}+m^2_f + D_{L,R} 
\end{equation}
with $D_{L} = (T_3 - e_f \sin^2 \theta_W) \cos 2 \beta \, m^2_Z$ 
and $D_{R} = e_f \sin^2 \theta_W \cos 2 \beta \, m^2_Z$ 
denoting the D-terms.  The non-trivial
mixing angles in the sfermion sector of the third generation
can be measured in a way similar to the charginos and neutralinos.  
The sfermion production cross sections for
longitudinally polarized e$^+$/e$^-$ beams are bilinear
in $\cos$/$\sin2\theta_{\tilde f}$.  The mixing angles and the two physical
sfermion masses are related to the tri-linear couplings $A_f$,
the higgsino mass parameter $\mu$ and $\tan\beta(\cot\beta)$
for down(up) type sfermions by:
\begin{equation}
A_f-\mu\tan\beta(\cot\beta)=\frac{m^2_{\tilde{f}_1}-m^2_{\tilde{f}_2}}{2 m_f}\sin2\theta_{\tilde f} ~~~~~[f:{\rm down(up)~type}]
\end{equation}
This relation gives us the opportunity to measure 
$A_f$ if $\mu$ has been determined
in the chargino sector.

Accuracies expected for the SUSY Lagrange parameters at the
electroweak scale for the reference point SPS1a are shown in
Table~\ref{tab:params}.  The errors are presented for 
the coherent ``LHC+LC'' analysis.  They have been obtained by fitting
the LHC observables and the masses of SUSY particles and Higgs bosons
accessible at a 1 TeV Linear Collider.
For the fit the programs SPheno2.2.0 \cite{Porod:2003um} and 
MINUIT96.03 \cite{James:1975dr} have been used.
The electroweak gaugino and higgs/higgsino parameters cannot be
determined individually through mass measurements at the LHC
as the limited number of observable masses leaves this sector 
in the SPS1a system under-constrained. Moreover, the Lagrange 
mass parameters in the squark sector can be determined 
from the physical squark masses with sufficient accuracy only
after the LHC mass measurements are complemented by LC measurements
in the chargino/neutralino sector; this information is necessary as 
the relation between the mass parameters is affected by large loop 
corrections.

\subsubsection{Reconstruction of the Fundamental SUSY Theory}

As summarized in the previous section, 
the minimal supergravity scenario mSUGRA is characterized
by the universal gaugino parameter $M_{1/2}$, the scalar
mass parameter $M_0$ and the trilinear coupling $A_0$, all
defined at the grand unification scale.  These parameters
are complemented by the sign of the higgs/higgsino mixing parameter
$\mu$, with the modulus determined by radiative symmetry
breaking, and the mixing angle, $\tan\beta$, in the Higgs sector.

The fundamental mSUGRA parameters at the GUT scale are
related to the low-energy parameters at the electroweak scale
by supersymmetric renormalization group 
transformations (RG)~\cite{RGE1,RGE2}
which to leading order generate the evolution for
\begin{center}
\begin{tabular}{lclr}
 gauge couplings &:&  $\alpha_i = Z_i \, \alpha_U$ & (5) \\
  gaugino mass parameters &:& $M_i = Z_i \, M_{1/2}$ & (6) \\
 scalar mass parameters &:&   $M^2_{\tilde\jmath} = M^2_0 + c_j M^2_{1/2} +
        \sum_{\beta=1}^2 c'_{j \beta} \Delta M^2_\beta$  & (7) \\
  trilinear  couplings &:&  $A_k = d_k A_0   + d'_k M_{1/2}$ & (8) 
\end{tabular}
\end{center}
\refstepcounter{equation}
\refstepcounter{equation}
\label{eq:gaugino}
\refstepcounter{equation}
\label{eq:squark} 
\refstepcounter{equation}
The index $i$ runs over the gauge groups $i=SU(3)$, $SU(2)$, $U(1)$.
To leading order, the gauge couplings, and the gaugino and scalar mass
parameters of soft--supersymmetry breaking depend on the $Z$ transporters
with 
\begin{eqnarray}
Z_i^{-1} =  1 + b_i \frac{\alpha_U}{ 4 \pi}
             \log\left(\frac{M_U}{ M_Z}\right)^2 
\end{eqnarray}
and $b[SU_3, SU_2, U_1] = -3, \, 1, \, 33 / 5$;
the scalar mass parameters depend 
also on the Yukawa couplings $h_t$, $h_b$, $h_\tau$
of the
top quark, bottom quark and $\tau$ lepton.
The coefficients $c_j$ [$j=L_l, E_l, Q_l, U_l, D_l, H_{1,2}$; $l=1,2,3$] 
for the slepton and squark doublets/singlets of generation $l$, 
and for the two Higgs doublets
are linear combinations of the evolution
coefficients $Z$; the coefficients $c'_{j \beta}$ are of order unity. 
The shifts $\Delta M^2_\beta$ are nearly zero for the first two families of 
sfermions but they can be rather large for the third family and for the 
Higgs mass
parameters, depending on the coefficients $Z$, 
the universal parameters $M^2_0$, $M_{1/2}$ and $A_0$,
and on the Yukawa couplings $h_t$, $h_b$, $h_\tau$.
The coefficients $d_k$ of the trilinear
couplings $A_k$ [$k=t,b,\tau$]  
depend on the corresponding Yukawa couplings 
and they are approximately unity for the
first two generations while being O($10^{-1}$) 
and smaller if the Yukawa couplings are
large; the coefficients $d'_k$, depending on gauge 
and Yukawa couplings, are of order unity.
Beyond the approximate solutions shown explicitly, the evolution equations 
have been solved numerically in the present analysis  to
two--loop order \cite{RGE2} and threshold effects have been
incorporated at the low scale \cite{bagger}.
The 2-loop effects as given in
Ref.~\cite{Degrassi:2001yf} have been included for
the neutral Higgs bosons and the $\mu$ parameter. 

\noindent
\underline{Gauge Coupling Unification}

Measurements of the gauge couplings at the electroweak scale
support very strongly the unification of the couplings at a scale
$M_U \simeq 2\times 10^{16}$~GeV \cite{r13A}.  
The precision, being at the per--cent level, is
surprisingly high after extrapolations over
fourteen orders of magnitude in the energy 
from the electroweak scale to the grand unification scale $M_U$. 
Conversely, the
electroweak mixing angle has been predicted in this approach at the
per--mille level. The evolution of the gauge couplings from 
low energy  to the GUT scale $M_U$ is carried out at two--loop accuracy. 
The gauge couplings $g_1$,
$g_2$, $g_3$ and the Yukawa couplings are calculated
in the $\overline{DR}$ scheme
by adopting the shifts given in Ref.\cite{bagger}.
These parameters are evolved to $M_U$ using 2--loop RGEs \cite{RGE2}. 
At 2-loop order the gauge couplings do not meet
exactly \cite{Weinberg:1980wa}, the
differences attributed to threshold effects at the unification
scale $M_U$ which leave us with an ambiguity in the definition of
$M_U$. In this report we define $M_U$ as the scale, {\it ad libitum}, 
where $\alpha_1 = \alpha_2$, denoted $\alpha_U$, in the RG evolution.
The non--zero 
difference $\alpha_3 - \alpha_U$ at this scale is then accounted
for by threshold effects
of particles with masses of order $M_U$. The quantitative evolution implies 
important constraints on the particle content at $M_U$
\cite{Ross:1992tz}.

\begin{figure*}[htb!]
\setlength{\unitlength}{1mm}
\begin{center}
\begin{picture}(160,85)
\put(-25.5,-101.5){\mbox{\epsfig{figure=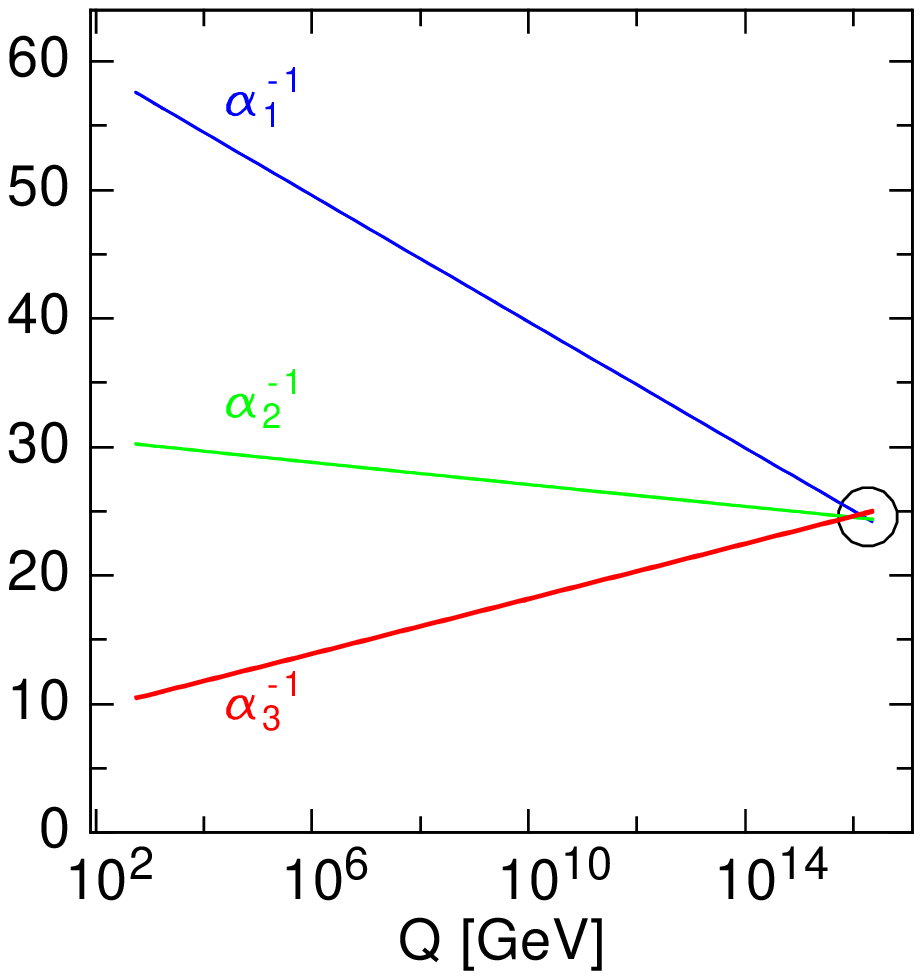,
                                   height=22.cm,width=16.cm}}}
\put(76,31.5){\mbox{\huge $\Rightarrow$}}
\put(60.5,-101.5){\mbox{\epsfig{figure=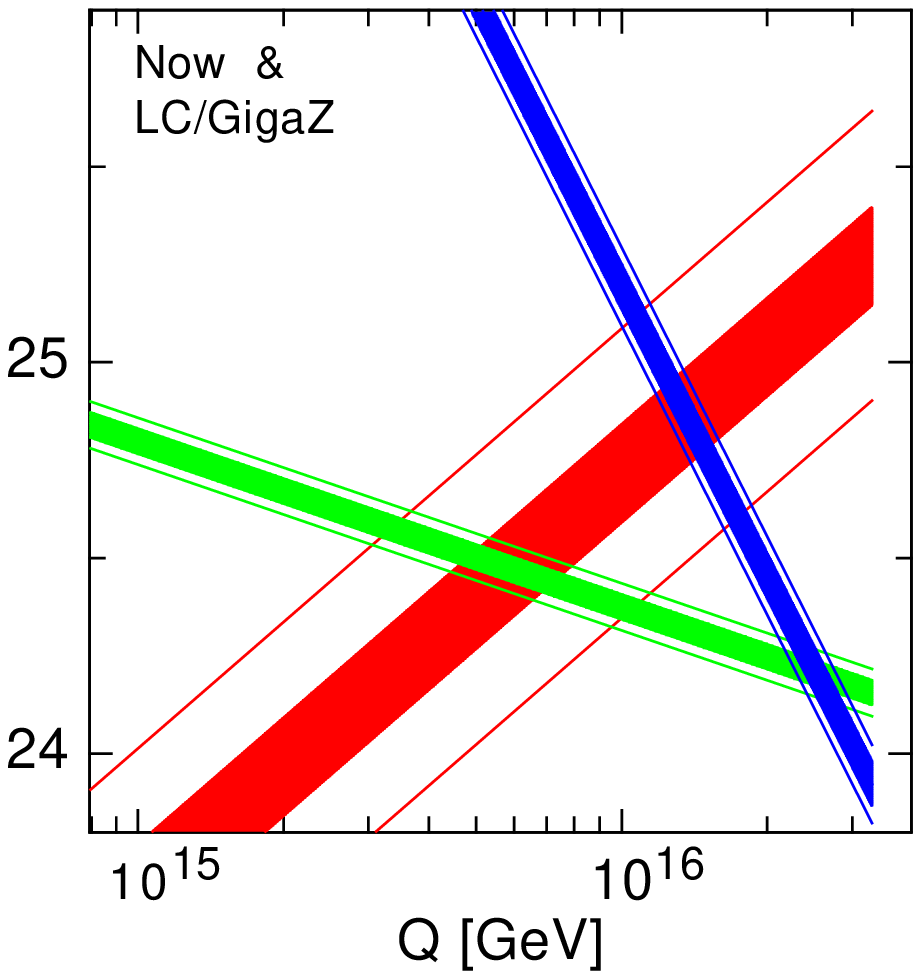,
                                   height=22.cm,width=16.cm}}}
\put(-3,80){\mbox{\bf a)}}
\put(84,80){\mbox{\bf b)}}
\end{picture}
\end{center}
\caption{{\it (a) Running of the inverse gauge couplings from low
  to high energies.
  (b) Expansion of the area around the unification point $M_U$ 
      defined by the meeting point of $\alpha_1$ with $\alpha_2$.
      The wide error bands are based on present data, and the spectrum
      of superymmetric particles from LHC measurementes within mSUGRA. 
      The narrow bands demonstrate the improvement expected by future GigaZ 
      analyses and the measurement of the complete spectrum at ``LHC+LC''.}}
\label{fig:gauge}
\end{figure*} 
\begin{table}
\begin{center}
\begin{tabular}{|c||c|c|}
\hline
 & Present/''LHC'' & GigaZ/''LHC+LC'' \\
\hline \hline
$M_U$ & $(2.53 \pm 0.06)\cdot 10^{16} \, \rm {GeV}$ & 
           $ (2.532 \pm 0.016) \cdot 10^{16} \, \rm {GeV}$ \\
$\alpha_U^{-1}$ & $  24.12 \pm 0.10 $ &  $ 24.12 \pm 0.05 $\\ \hline
$\alpha_3^{-1} - \alpha_U^{-1}$ & $0.96 \pm 0.45$ & $0.95 \pm 0.12$ \\ \hline
\end{tabular}
\end{center}
\caption{{\it Expected errors on $M_U$ and $\alpha_U$ for the mSUGRA 
 reference point, derived for the present level of accuracy and
 compared with expectations from GigaZ [supersymmetric spectrum as discussed
  in the text].  Also shown is the difference between
 $\alpha_3^{-1}$ and $\alpha_U^{-1}$ at the unification point $M_U$.}}
\label{tab:gauge}
\end{table}
Based on the set of low--energy gauge and Yukawa parameters 
$\{\alpha(m_Z)$, $\sin^2
\theta_W$, $\alpha_s(m_Z)$, $Y_t(m_Z)$, $Y_b(m_Z)$, $Y_\tau(m_Z)\}$
the evolution of the inverse couplings $\alpha_i^{-1}$ $[i=U(1)$, $SU(2)$,
$SU(3)]$ is depicted in Fig~\ref{fig:gauge}. The evolution is performed for
the mSUGRA reference point defined above. Unlike earlier analyses, the
low--energy thresholds of supersymmetric particles can be calculated
in this framework
exactly without reference to effective SUSY scales. The outer lines
in Fig.~\ref{fig:gauge}b
correspond to the present experimental accuracy of the gauge
couplings \cite{PDG}: 
$\Delta \{\alpha^{-1}(m_Z)$, $\sin^2\theta_W$, $\alpha_s(m_Z)\}$ 
$=\{ 0.03, 1.7 \cdot 10^{-4}, 3 \cdot 10^{-3} \}$, and the spectrum of
supersymmetric particles from LHC measurements complemented in the top-down
approach for mSUGRA. 
The full bands demonstrate the improvement for the absolute errors
$\{8 \cdot 10^{-3}, 10^{-5},10^{-3}  \}$ after operating GigaZ 
\cite{Monig:2001hy,Erler:2000jg} and inserting the complete spectrum
from ``LHC+LC'' measurements.  
The expected accuracies in $M_U$ and $\alpha_U$
are summarized in the values given in Tab.~\ref{tab:gauge}.
The gap between $\alpha_U$ and 
$\alpha_3$ is bridged by contributions from high scale physics.
Thus, for a typical set of SUSY parameters, the evolution of the gauge
couplings from low  to high scales leads to a precision of 1.5 per--cent
for the Grand Unification picture.

\noindent
\underline{Gaugino and Scalar Mass Parameters: Top-down Approach}

The structure of the fundamental supersymmetric theory is assumed,
in the top-down approach, to be defined
uniquely at a high scale. In mSUGRA the set of
parameters characterizing the specific form of the theory includes,
among others, the scalar masses $M_0$ and the gaugino masses $M_{1/2}$. 
These universal parameters are realized at the grand unification point $M_U$.
Evolving the parameters from the high scale down to the electroweak scale
leads to a comprehensive set of predictions for the masses, mixings and 
couplings of the physical particles. Precision measurements of these 
observables can be exploited to determine the high-scale parameters
 $M_0$, $M_{1/2}$, etc., and to perform consistency tests of the
underlying form of the theory. The small number of fundamental parameters,
altogether five in mSUGRA, gives rise to many correlations between a 
large number of experimental observables. They define a set of
{\it necessary consistency conditions} for the realization of the
specific fundamental theory in nature. \\

\noindent
{\it Interludium:}
In addition to the experimental errors, theoretical uncertainties must be
taken into account. They are generated by truncating the perturbation series
for the evolution of the fundamental parameters in the $\overline{DR}$ scheme
from the GUT scale to a low SUSY scale $\tilde M$ near the electroweak
scale, and for the relation between the parameters at this point to the
on-shell physical mass parameters, for instance. Truncating
these series in one- to two-loop approximations leads to a residual
$\tilde M$ dependence that would be absent from the exact solutions and may
therefore be interpreted as an estimate of the neglected higher-order 
effects.

\renewcommand{\arraystretch}{1.1}
\begin{table}
\begin{center}
\begin{tabular}{|c|c||c|c|}
\hline
Particle & $\Delta_{th}$~[GeV]
 & Particle & $\Delta_{th}$~[GeV] \\ 
\hline \hline
$\tilde \chi^+_1$ & 1.2 &  $\tilde q_R$ & 8.4\\
$\tilde \chi^+_2$ & 2.8 &  $\tilde q_L$ & 9.1\\ 
$\tilde \chi^0_1$ & 0.34 & $\tilde t_1$ & 4.4\\
$\tilde \chi^0_2$ & 1.1 & $\tilde t_2$ & 8.3 \\
$\tilde \chi^0_3$ & 0.6 &  $\tilde b_1$ & 7.4\\
$\tilde \chi^0_4$ & 0.3 &  $\tilde b_2$ & 8.2 \\ \hline
$\tilde e_R$ & 0.82 & $\tilde g$ & 1.2 \\ \cline{3-4}
 $\tilde e_L$ & 0.31 & $h^0$ & 1.2 \\ 
 $\tilde \nu_e$ & 0.24 & $H^0$ & 0.7 \\ 
 $\tilde \tau_1$ & 0.59 & $A^0$ & 0.7  \\
 $\tilde \tau_2$ & 0.30 & $H^+$ & 1.0 \\
 $\tilde \nu_\tau$ & 0.25 & & \\ \hline
\end{tabular}
\end{center}
\caption{{\it Theoretical errors of the SPS1a mass spectrum, 
calculated as difference between the minimal 
and the maximal value of the masses
if the scale $\tilde M$ is varied between 100 GeV and 1 TeV.}}
\label{tab:masserrorth}
\end{table}

\renewcommand{\arraystretch}{1.1}
\begin{table}
\begin{center}
\begin{tabular}{|c||cccccccc|}
\hline
 & $m_{ll}^{max}$  
 & $m_{llq}^{max}$ 
 & $m_{llq}^{min}$ 
 & $m_{lq}^{high}$ 
 & $m_{lq}^{low}$  
 & $m_{\tau\tau}^{max}$  
 & $m_{ll}^{max}(\tilde\chi^0_4)$ 
 & $m_{llb}^{min}$ \\ \hline \hline
\SPHENO & 80.64 & 454.0 & 216.8 & 397.2 & 325.6 & 83.4\; & 283.4 & 195.9 \\
$\Delta_{exp}$  
    & \hphantom{0}0.08 
    & \hphantom{00}4.5 
    & \hphantom{00}2.6 
    & \hphantom{00}3.9 
    & \hphantom{00}3.1 
    &  \hphantom{0}5.1 
    & \hphantom{00}2.3 
    & \hphantom{00}4.1 \\
$\Delta_{th}\;\,$ 
    & \hphantom{0}0.72 
    & \hphantom{00}8.1 
    & \hphantom{00}3.6 
    & \hphantom{00}7.7 
    & \hphantom{00}5.5 
    &  \hphantom{0}0.8 
    & \hphantom{00}0.7 
    & \hphantom{00}2.9 \\ \hline \hline
\end{tabular}\\[4mm]
\begin{tabular}{|c||cccccc|}
\hline 
 & $m_{\tilde q_R}-m_{\tilde\chi^0_1}$
 & $m_{\tilde l_L}-m_{\tilde\chi^0_1}$ 
 & $m_{\tilde g}-m_{\tilde b_1}$
 & $m_{\tilde g}-m_{\tilde b_2}$
 & $m_{\tilde g}-0.99\,m_{\tilde\chi^0_1}$
 & $m_{h^0}$ \\ \hline \hline
\SPHENO & 450.3 & 110.0 & 88.9 & 56.9 & 507.8 & 110.8 \\
$\Delta_{exp}$  
    & \phantom{0}10.9 
    & \phantom{00}1.6 
    & \phantom{0}1.8 
    & \phantom{0}2.6 
    & \phantom{00}6.4 
    & \phantom{000}0.25 \\
$\Delta_{th}\;\,$  
    & \phantom{00}8.1 
    & \phantom{000}0.23 
    & \phantom{0}6.8 
    &    \phantom{0}7.6 
    & \phantom{00}1.3 
    & \phantom{00}1.2 \\ \hline
\end{tabular}
\end{center}
\caption{{\it LHC observables assumed for SPS1a and their 
   experimental ($\Delta_{exp}$) and present theoretical ($\Delta_{th}$)  
   uncertainties. [All quantities in GeV].} 
\label{tab:LHCobs}}
\end{table}

We estimate these effects by varying $\tilde M$ between the electroweak
scale and 1 TeV. The theoretical uncertainties 
of the physical masses and  LHC observables derived in this way 
are listed in 
Tables~\ref{tab:masserrorth} and \ref{tab:LHCobs}, respectively. 
They are of similar size as the  differences found by comparing
the observables with different state-of-the-art 
codes for the spectra \cite{Allanach:2003jw}.
The comparison of the present theoretical
uncertainties with the experimental errors at LHC demonstrates that the
two quantities do match {\it cum grano salis} at the same size.
Since LC experiments will reduce the experimental errors roughly by an
order of magnitude, considerable theoretical efforts are needed 
in the future  to reduce $\Delta_{th}$ to a level that matches 
the expected experimental precision at  LC.  
Only then we can deepen our 
understanding of the underlying supersymmetric theory by tapping the full
experimental potential of ``LC'' and of the combined ``LHC+LC'' analyses. \\

\begin{figure*}
\setlength{\unitlength}{1mm}
\begin{center}
\begin{picture}(160,80)
\put(-10,-25){\mbox{\epsfig{figure=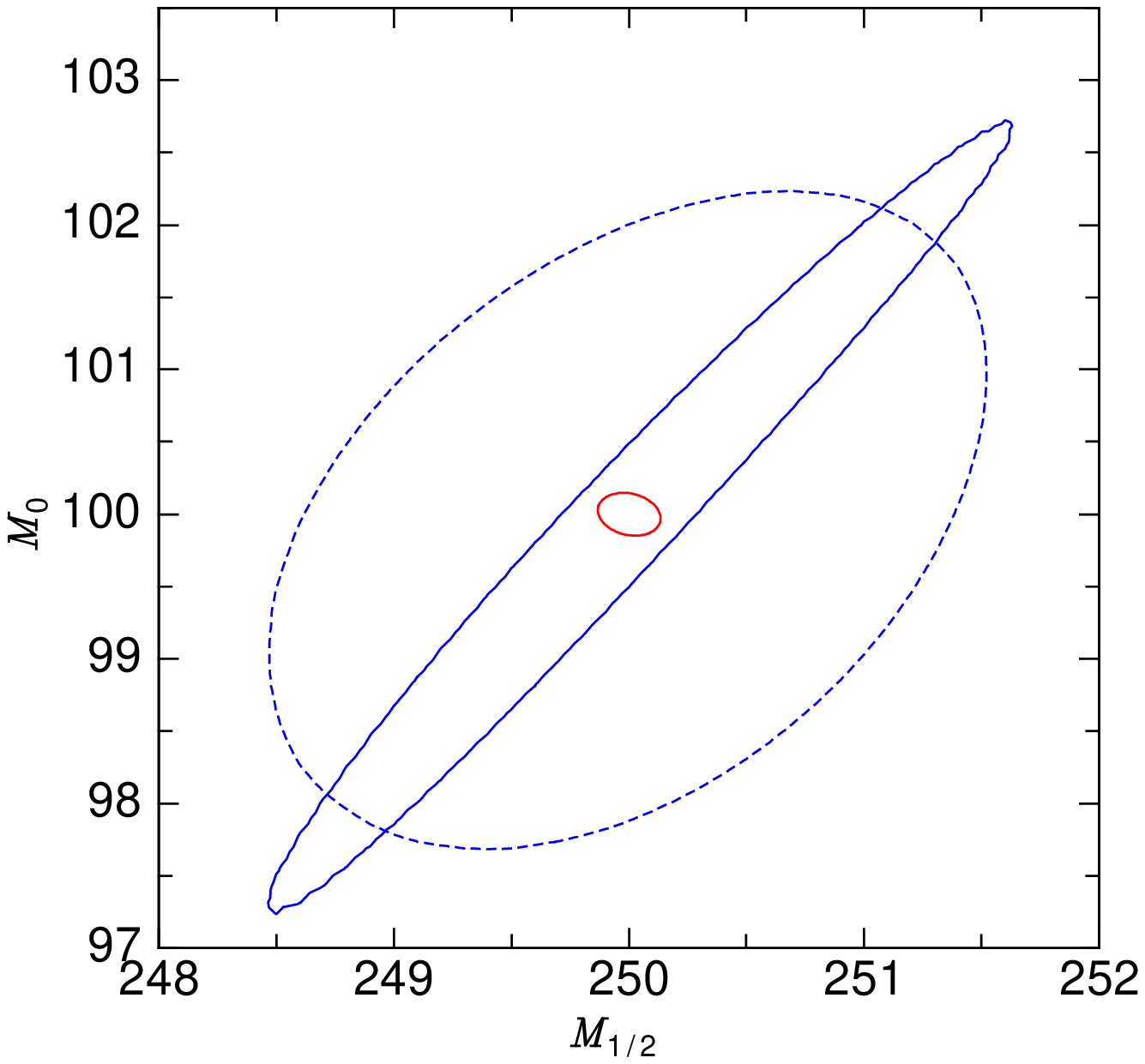,
                                   height=17.5cm,width=12.cm}}}
\put(74,-25){\mbox{\epsfig{figure=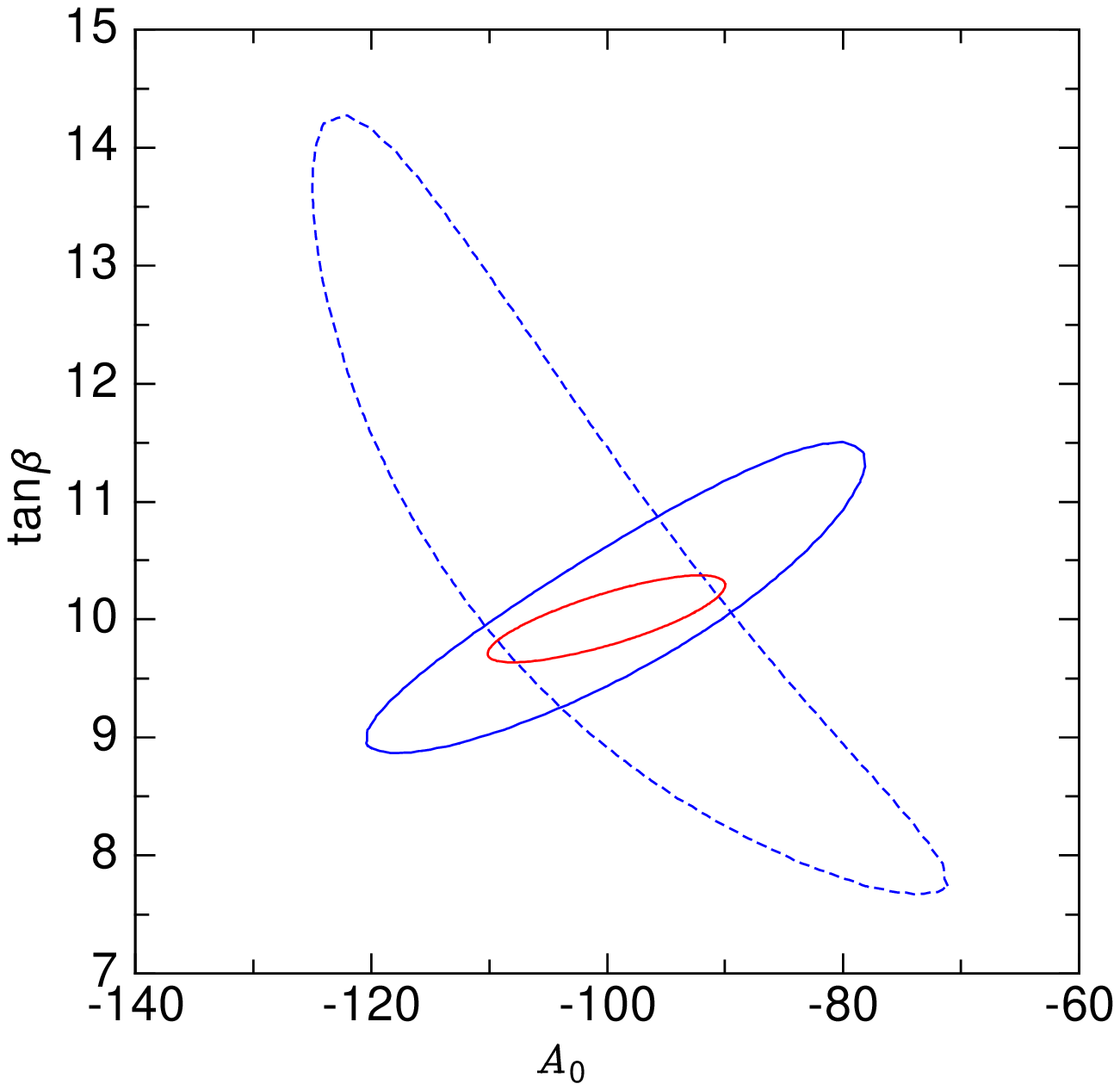,
                                   height=17.5cm,width=12.cm}}}
\put(-3,75){\mbox{\bf a)}}
\put(80,75){\mbox{\bf b)}}
\end{picture}
\end{center}
\caption{\it 1-$\sigma$ error ellipses for the mSUGRA parameters
  in the top-down approach [i.e. the contours of $\Delta\chi^2=4.7$ 
  in the $M_0-M_{1/2}$ and $\tan\beta-A_0$ planes with the respective 
  other parameters fixed to their best fit values, 
  c.f. Table~\ref{tab:topdownresults}].  
  The full blue ellipses are the results obtained from LHC measurements 
  alone while the red ones are for the combined ``LHC+LC'' analyses. 
  The dashed blue lines show the results for the ``LHC'' case including 
  today's theoretical uncertainties.}
\label{fig:topdown}
\end{figure*}

In the top-down approach, models of SUSY-breaking are tested by fitting 
their high-scale parameters to experimental data. 
The minimum $\chi^2$ of the fit gives a measure of the probability that 
the model is wrong. 
The results of such a fit of mSUGRA to anticipated ``LHC'', ``LC'', and ``LHC+LC'' 
measurements are shown in Table~\ref{tab:topdownresults} and 
Fig.~\ref{fig:topdown}. 
For the ``LHC'' case the observables in Table~\ref{tab:LHCobs} have been used, 
for the LC the masses in Table~\ref{tab:massesA} and for ``LHC+LC''
the complete information have been used. 
If mSUGRA is assumed to be the underlying supersymmetric theory, the 
universal parameters $M_{1/2}$ and $M_0$ can be determined at the LHC 
at the per--cent level. LC experiments and coherent ``LHC+LC'' analyzes 
improve the accuracy by an order of magnitude, thus allowing for much 
more powerful tests of the underlying supersymmetric theory. 
Table~\ref{tab:topdownresults} takes only experimental errors into account.
The accuracy of the present theoretical calculations matches the errors
of the ``LHC'' analysis and can thus be included in a meaningful way
in a combined experimental plus theoretical error analysis. 
Adding $\Delta_{th}$ and $\Delta_{exp}$ quadratically the
 errors of the ``LHC'' analysis increases to:
$\Delta M_{1/2}= 2.7$~GeV, $\Delta M_0= 2.9$~GeV, $\Delta A_0= 51$~GeV, and 
$\Delta\tan\beta= 5$. 
As argued above,  significant theoretical improvements by an order of
magnitude, i.e. ``the next loop'', are necessary to exploit fully
 the ``LC'' and ``LHC+LC'' potential.

The minimum $\chi^2$ of the fit to mSUGRA as in Table~\ref{tab:topdownresults}
 is indeed small, 
$\chi^2_{min}/n.d.o.f. \leq 0.34$ for ``LHC'', ``LC'', as well as ``LHC+LC''. 
When fitting instead mGMSB model parameters as an alternative 
to the same data, 
we would obtain $\chi^2_{min}/14\,d.o.f. = 68$ from LHC data alone. 
Such a result would clearly disfavour this model.

\begin{table}
\begin{center}
\begin{tabular}{|c||c|c||c|}
\hline
             & ``LHC''        & ``LC'' & ``LHC+LC'' \\ \hline \hline
$M_{1/2}$   & $250.0 \pm 2.1$ &  $250.0\pm 0.4$ & $250.0 \pm 0.2$  \\
$M_0$       & $100.0 \pm 2.8$ &  $100.0\pm 0.2$ & $100.0 \pm 0.2$  \\
$A_0$       & $-100.0\pm 34$  & $-100.0\pm 27$  & $-100.0\pm 14$   \\
$\tan\beta$ & $ 10.0 \pm 1.8$ &  $10.0 \pm 0.6$ & $10.0 \pm 0.4$  \\
\hline
\end{tabular}
\end{center}
\caption{{\it Results for the high scale parameters in the
   top-down approach including the experimental errors.}}
\label{tab:topdownresults}
\end{table}

\noindent
\underline{Gaugino and Scalar Mass Parameters: Bottom-up Approach}

In the bottom-up approach the fundamental supersymmetric theory
is reconstructed at the high scale from the available {\it corpus} of
experimental data without any theoretical prejudice. This approach exploits
the experimental information to the maximum extent possible and reflects an
undistorted picture of our understanding of the basic theory.
 
At the present level of preparation in the ``LHC'' and ``LC'' sectors, such
a program can only be carried out in coherent ``LHC+LC'' analyses  while
the separate information from either machine proves insufficient.
The results for the evolution of the mass parameters from the electroweak
scale to the GUT
scale $M_U$ are shown in Fig.~\ref{fig:sugra_LHC}.  

On the left of Fig.~\ref{fig:sugra_LHC}a
the evolution is presented for the 
gaugino parameters $M^{-1}_i$, which clearly
is under excellent control for the coherent ``LHC+LC'' analyses, 
while ``LHC'' [and ``LC''] measurements alone are insufficient for the 
model-independent reconstruction
of the parameters and the test of universality
in the $SU(3) \times SU(2) \times U(1)$ group space.
The error ellipse for the unification of the gaugino masses 
in the final analysis is depicted on the right of 
Fig.~\ref{fig:sugra_LHC}a. Technical details of the ``LHC+LC'' analysis
can be found in Ref.~\cite{r1}. 

\begin{figure*}
\setlength{\unitlength}{1mm}
\begin{center}
\begin{picture}(160,170)
\put(-4,0){\mbox{\epsfig{figure=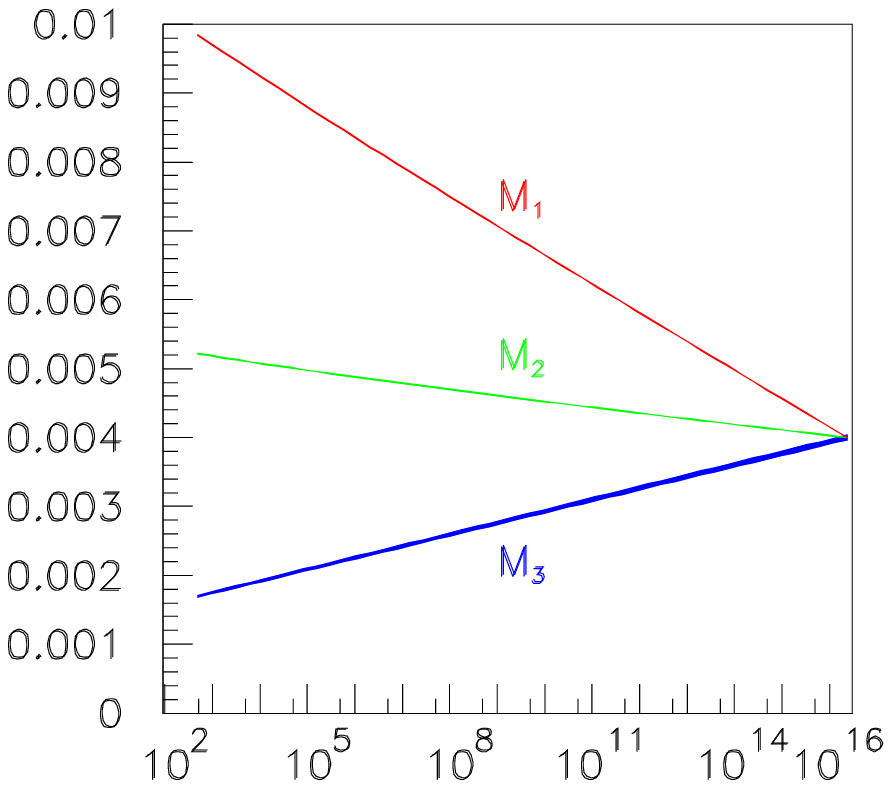,height=17cm,width=18cm}}}
\put(84.5,83.5){\mbox{\epsfig{figure=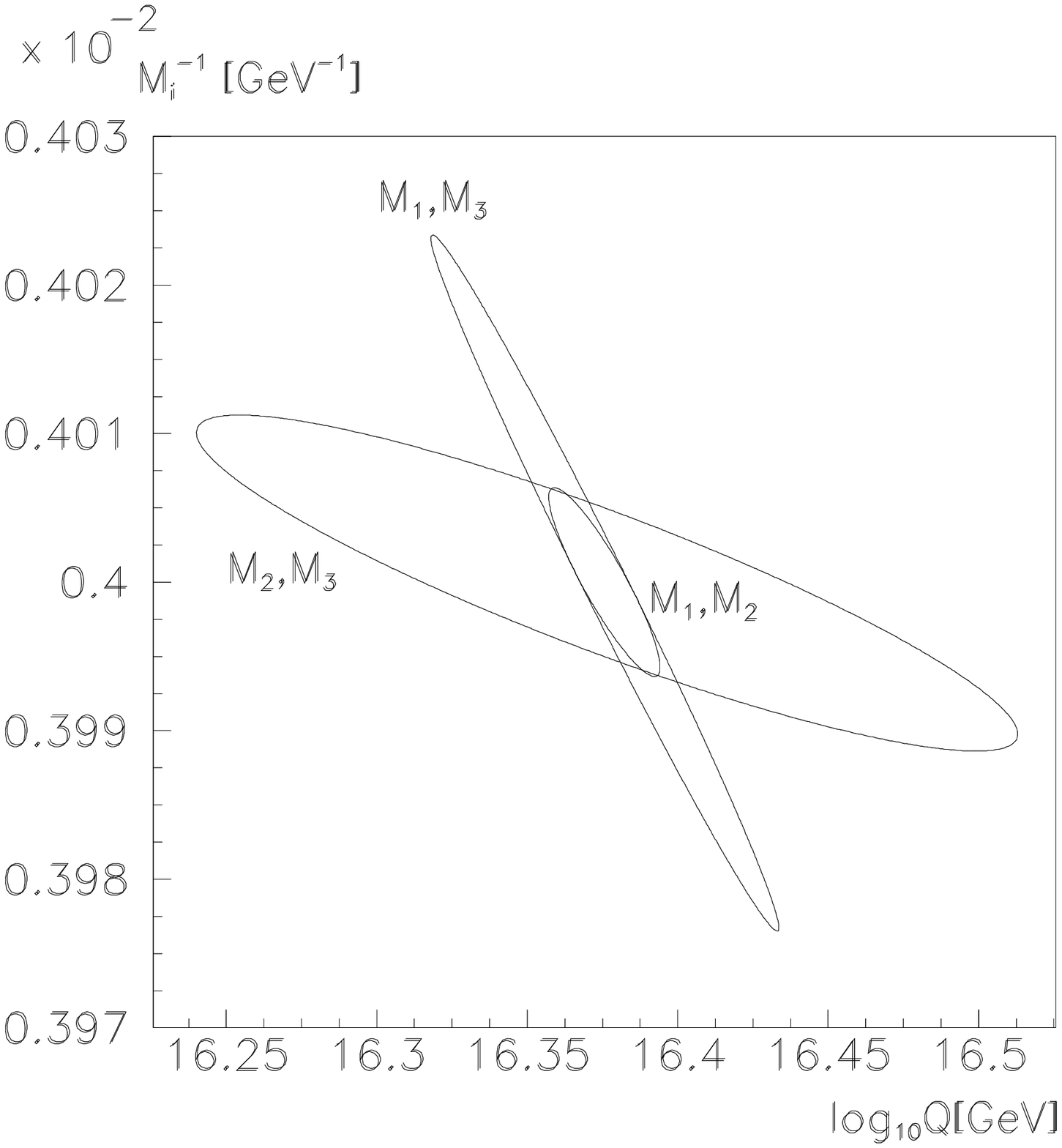
                              ,height=7.95cm,width=8.3cm}}}
\put(-4,-86){\mbox{\epsfig{figure=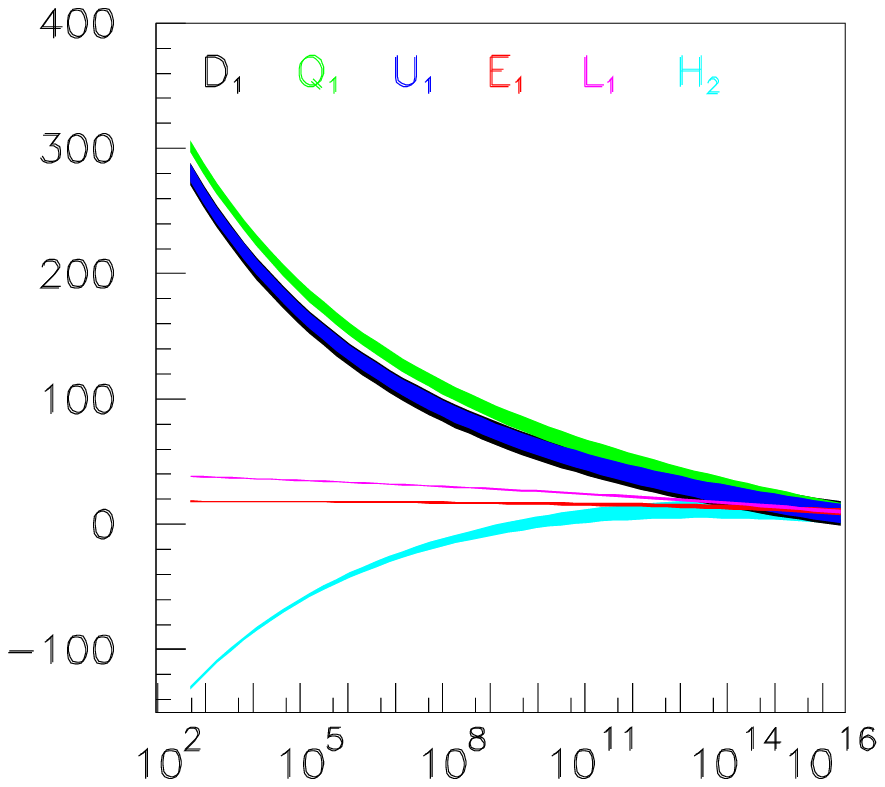,height=17cm,width=18cm}}}
\put(78,-86){\mbox{\epsfig{figure=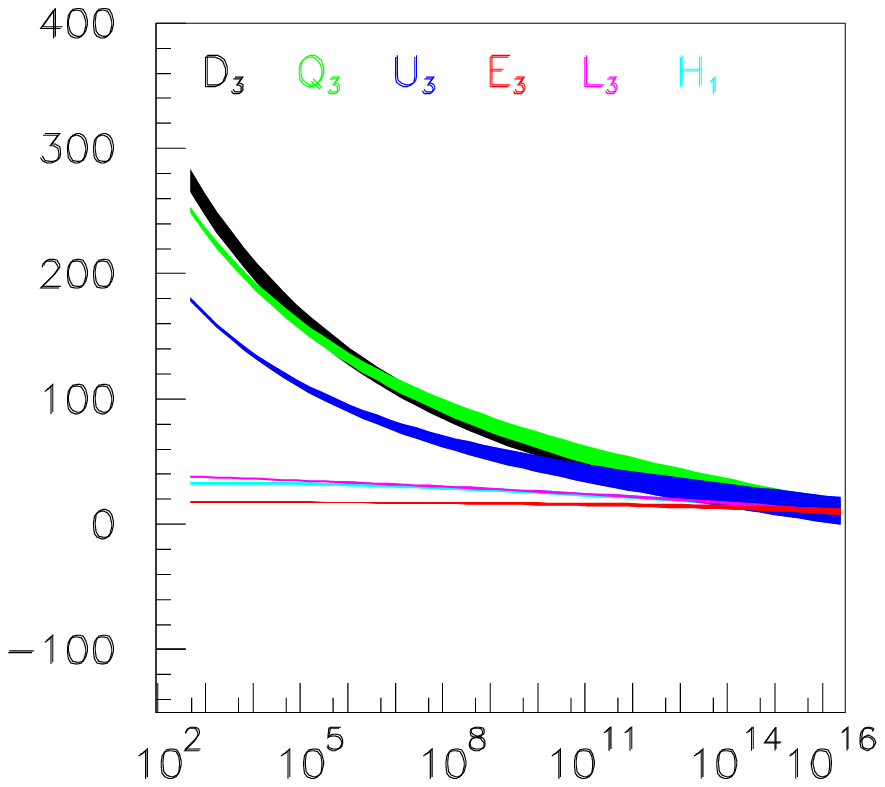,height=17cm,width=18cm}}}
\put(-1,158){\mbox{\bf (a)}}
\put(10,156){\mbox{$1/M_i$~[GeV$^{-1}$]}}
\put(60,82){\mbox{$Q$~[GeV]}}
\put(-1,73){\mbox{\bf (b)}}
\put(14,71){\mbox{$M^2_{\tilde j}$~[$10^3$ GeV$^2$]}}
\put(65,-3){\mbox{$Q$~[GeV]}}
\put(80,159){\mbox{
}}
\put(80,73){\mbox{
}}
\put(95,71){\mbox{$M^2_{\tilde j}$~[$10^3$ GeV$^2$]}}
\put(147,-3){\mbox{$Q$~[GeV]}}
\end{picture}
\end{center}
\caption{{\it  Evolution, from low to high scales, (a) of 
the gaugino mass parameters
for ``LHC+LC'' analyses
and the corresponding error ellipses of the universal GUT values;
(b) left: of the first--generation sfermion mass parameters 
(second generation, {\it dito}) and  
the Higgs mass parameter $M^2_{H_2}$; right: of the
third--generation sfermion mass parameters and
the Higgs mass parameter $M^1_{H_2}$.}   
}
\label{fig:sugra_LHC}
\end{figure*}

In the same way the evolution of the scalar mass parameters can be
studied, presented in Figs.~\ref{fig:sugra_LHC}b 
separately for the first/second
and the third generation in ``LHC+LC'' analyses.
Compared with the slepton parameters, the accuracy deteriorates
for the squark parameters, 
and for the Higgs mass parameter
$M^2_{H_2}$.
The origin of the differences between the errors for slepton and 
squark/Higgs mass parameters can be traced back to the numerical 
size of the coefficients
in Eqs.~(\ref{eq:squark}). Typical examples, 
evaluated at $Q=500$~GeV, read as follows \cite{r1}:
\begin{eqnarray}
M^2_{\tilde L_{1}} &\simeq& M_0^{2} + 0.47 M^2_{1/2} \\
M^2_{\tilde Q_{1}} &\simeq& M_0^{2} + 5.0 M^2_{1/2}  \\
M^2_{\tilde H_2} &\simeq&  -0.03 M_0^{2} - 1.34 M^2_{1/2}
           + 1.5 A_0 M_{1/2} + 0.6 A^2_0 \\
|\mu|^2 &\simeq& 0.03 M_0^{2} + 1.17 M^2_{1/2}
           - 2.0 A_0 M_{1/2} - 0.9 A^2_0
\end{eqnarray}
While the coefficients for the sleptons are of order unity, 
the coefficients $c_j$
for the squarks grow very large,  $c_j \simeq 5.0$, so that small errors
in $M^2_{1/2}$ are magnified by nearly an order of magnitude in the solution
for $M_0$. By close inspection of Eqs.(\ref{eq:squark}) for the Higgs mass
parameter it turns out that 
the formally leading 
$M^2_0$ part is nearly cancelled by the $M^2_0$ part
of $c'_{j,\beta} \Delta M_\beta^2$. Inverting Eqs.(\ref{eq:squark}) for
$M^2_0$ therefore gives rise to large errors in the Higgs case.
Extracting the trilinear parameters $A_k$ is difficult and
more refined analyses
based on sfermion cross sections and Higgs and/or sfermion decays are
necessary to determine these parameters accurately.

A representative set of the final mass values and the associated errors,
after evolution from the electroweak scale to $M_U$, are presented
in Table~\ref{tab:evolved_params}.
It appears that the joint ``LHC+LC''analysis generates
a comprehensive and detailed picture of the fundamental SUSY parameters 
at the GUT/PL scale. Significant improvements however would be welcome
in the squark sector where reduced experimental errors would refine the
picture greatly.

\renewcommand{\arraystretch}{1.1}
\begin{table}
\begin{center}
\begin{tabular}{|c||c|c|}
\hline
               & Parameter, ideal & ``LHC+LC'' errors     \\
\hline\hline
 $M_1$        &  $250.$           & 0.15   \\
 $M_2$        &  {\it ditto}     & 0.25   \\
 $M_3$        &                   & 2.3    \\
\hline\hline  
 $M_{L_1}$  &$100.$     & 6.  \\
 $M_{E_1}$  &{\it ditto}          & 12.  \\
 $M_{Q_1}$  &                   & 23.  \\
 $M_{U_1}$  &                   & 48.  \\
 $M_{L_3}$  &                   & 7.  \\
 $M_{E_3}$  &                   & 14. \\
 $M_{Q_3}$  &                   & 37. \\
 $M_{U_3}$  &                   & 58.  \\
\hline
$M_{H_1} $  &{\it ditto}          & 8.  \\
$M_{H_2} $  &                   & 41.  \\
\hline \hline
$A_t       $  &  $-100.$          & 40.                 \\
\hline
\end{tabular}\\
\end{center}
\caption {
{\it Values of the SUSY Lagrange mass parameters after extrapolation 
to the unification scale where gaugino and scalar mass parameters are 
universal in mSUGRA [mass units in {\rm GeV}].} 
} 
\label{tab:evolved_params}
\end{table}

\subsubsection{Summary}
We have shown in this brief report that in supersymmetric theories stable
extrapolations can be performed from the electroweak scale
to the grand unification scale, close to the Planck scale.
This feature has been demonstrated compellingly in the evolution
of the three gauge couplings and of the soft supersymmetry breaking
parameters, which approach universal values at the GUT scale in minimal 
supergravity.  As a detailed scenario we have adopted the Snowmass reference
point SPS1a. It turns out that the information on the mSUGRA
parameters at the GUT scale from pure ``LHC'' analyses
is too limited to allow for the reconstruction of the high-scale
theory in a model-independent way. 
The coherent ``LHC+LC'' analyses however in which the measurements 
of SUSY particle
properties at LHC and LC mutually improve each other, result in a
comprehensive and detailed picture 
of the supersymmetric particle system. In particular, the gaugino sector
and the non-colored scalar sector are under excellent control.
 
\begin{table}
\begin{center}
\begin{tabular}{|c||c|c|}
\hline
                &  Parameter, ideal    & Experimental error \\ 
\hline\hline
$M_U$           & $2.53\cdot 10^{16}$ &  $2.2  \cdot 10^{14}$       \\
$\alpha_U^{-1}$ &   24.12          &     0.05     \\ \hline
$M_\frac{1}{2}$ & 250.             & 0.2     \\
$M_0$           & 100.             & 0.2     \\
$A_0$           & -100.            & 14      \\  
\hline
$\mu$           & 357.4            & 0.4     \\
\hline
$\tan\beta $    &  10.             & 0.4      \\  
\hline
\end{tabular}
\end{center}
\caption[]{\it Comparison of the ideal parameters with the
experimental expectations 
in the combined ``LHC+LC'' analyses
for the particular mSUGRA reference 
point adopted in this report [units in {\rm GeV}].} 
\label{tab:univ_params}
\end{table}
 
Though mSUGRA has been chosen as a specific example, the methodology can
equally well be applied to left-right symmetric theories and to
superstring theories.  The analyses offer the exciting opportunity to 
determine intermediate scales in left-right symmetric theories and to
measure effective string-theory parameters near the Planck scale.

Thus, a thorough analysis of the mechanism of
supersymmetry breaking and the reconstruction of the
fundamental supersymmetric theory at the grand unification scale
has been shown possible in the high-precision high-energy experiments at
LHC and LC.  This point has been highlighted by performing a global
mSUGRA fit of the universal parameters, c.f. Tab.~\ref{tab:univ_params}
Accuracies at the level of per-cent to per-mille can be
reached, allowing us to reconstruct the structure of nature
at scales where gravity is linked with particle physics.




\chapter{Electroweak and QCD Precision Physics}
\label{chapter:ewandqcd}
Editors: {\it E.~Boos, A.~De~Roeck, S.~Heinemeyer, W.J.~Stirling}

\section{\label{sec:81} Top physics}

{\it E.~Boos, A.~Sherstnev, S.~Slabospitsky, Z.~Sullivan, S.~Weinzierl}

\vspace{1em}




\def\bq{\begin{eqnarray}}
\def\eq{\end{eqnarray}}
\def\l{\langle}
\def\r{\rangle}
\def\eps{\varepsilon}



\def \slhclum{$10^{35}\ \mathrm{cm}^{-2}\mathrm{s}^{-1}$}
\def \lhclum{$10^{34}\ \mathrm{cm}^{-2}\mathrm{s}^{-1}$}
\def\d {$} 
\def \pt   {\mbox{$p_{\scriptscriptstyle T}$}} 
\def \ifb {Feb$^{-1}$}
\renewcommand{\gev}{\mathrm{GeV}}


The top quark, with the mass slightly less than the mass of the gold
nucleus, is the heaviest elementary particle found.  The direct
top-quark mass (pole mass) measurement by the CDF~\cite{Abe:1995hr}
and D0 \cite{Abachi:1995iq} collaborations in Run I at the Tevatron
gives $M_t = 174.3\pm 5.1$~GeV~\cite{Hagiwara:fs}, which is in
spectacular agreement with the result from the electroweak (EW) data
analysis by LEP and SLC \cite{Groom:in}.  This is a well known
historical example of important influence of lepton collider results
on physics analyses at hadron colliders.

The Standard Model (SM) top quark couplings are uniquely fixed by the
principle of gauge invariance, the structure of the quark generations,
and a requirement of including the lowest dimension interaction
Lagrangian.  
Within the SM the top quark is considered as a point-like particle, albeit
its heavy mass.
The mass of the top quark is close to the electroweak
breaking scale, and the top Yukawa coupling $\lambda_{t} =
(2\sqrt{2}G_F)^{1/2}m_{t}$ is numerically very close to unity.  This
suggests that a study of the properties of the top quark might also
reveal details of the electroweak symmetry breaking mechanism.  In
particular, it has been argued that new physics might lead to
measurable deviations from the Standard Model values.

The top quark decay width, $\Gamma_t/|V_{tb}|^2 = 1.39$ GeV for the
pole mass, was calculated in the SM to second order in QCD
\cite{Chetyrkin:1999ju} and to first order in EW \cite{Denner:1990ns}
corrections including the $W$-boson and $b$-quark masses. Since the
CKM matrix element $V_{tb}$ is close to one, in the SM the
top-quark lifetime $\tau_t \approx 0.4 \times 10^{-24}$s is much
smaller then the typical time for a formation of QCD bound states
$\tau_{QCD}\approx 1/\Lambda \approx 3 \times 10^{-24}$s. 
Therefore, the top quark decays long before 
it can hadronize \cite{Bigi:1986jk} and it provides a very clean
source for fundamental information on the constituents of matter.  

The physics of the top quark at the LHC \cite{Beneke:2000hk} and LC
\cite{sec8_Aguilar-Saavedra:2001rg} has been studied in great 
detail, including in many cases a realistic simulation of the detectors.
We summarize here only the basic predictions.

\subsection{Top-quark production}

At hadron and lepton colliders, top quarks may be produced either in
pairs or singly.  The pair-production cross section, about 850 pb, is
known at the LHC to NLO level \cite{Nason:1987xz} including the
re-summation of the Sudakov logarithms (NLL)
\cite{Catani:1996yz}. About 90\% of the rate is due to gluon-gluon
collisions, while quark-antiquark collisions give the remaining
10\%. The estimated overall theoretical systematic uncertainty is
about 12\%, which would lead to about 4\% accuracy in the determination of
the top-quark mass.  However, combining various top-quark decay
channels and using various methods, the error on the top-quark mass
(dominated by systematics) is expected to reach $\sim$~0.5\%, beyond
which more data offers no obvious improvement.

A precise knowledge of the top-quark mass is necessary for many
precision observables.  The better the top-quark
mass from the LHC is known, the more appropriate the energy interval
for the threshold scan at the LC can be chosen for $t \bar t$-pair
production. Several NNLO
calculations have been performed for various definitions of the
threshold top-quark mass parameter \cite{Hoang:1998xf}. A delicate
comparison of different calculations leads to the conclusion that a
scan at the threshold is expected to reduce the error on the top mass
down to $\delta m_t \simeq 100 \,\mbox{MeV}$
\cite{sec8_Aguilar-Saavedra:2001rg}, a value not achievable at hadron
machines.

The total top-quark pair production cross section at the LC in the
continuum has been calculated to the NLO QCD \cite{Chetyrkin:1996cf}
and 1-loop EW \cite{Beenakker:1991ca} level. The cross section is
about $10^3$ times smaller than at the LHC approaching about 0.85 pb
at maximum around 390 GeV and falling down with the energy as
$1/s$. However, a very clean environment of the LC experiment and a
possibility to use beam polarization allow access to unique
information on various couplings, as we discuss later.     
In particular, an important aspect related to $t \bar{t}$-production
is the top-Higgs Yukawa coupling. This subject is considered
in details in the section~\ref{chapter:ewsymmbreak}.

Single-top-quark production provides many new avenues to physics that
cannot be observed in $t\bar t$ production.  Single-top-quark
production allows a direct measurement of the CKM matrix element
$V_{tb}$, and can provide a verification of the unitarity of the CKM
matrix.  The top quark is produced through a left-handed interaction
and is highly polarized.  Since no hadronization occurs, spin
correlations survive in the final decay products.  Hence,
single-top-quark production offers an opportunity to observe the
polarization of the top-quark at production.  Finally, measurement of
the charged-current couplings of the top quark may probe non-standard
couplings that would provide hints about new physics.  A survey of all
new-physics-scenarios which have been studied is beyond the scope of
this document and we refer the reader to the review article
\cite{Chakraborty:2003iw} and the references therein.

The single-top-quark production rate at the LHC is known to the NLO
level for all tree production mechanisms which are classified by the
virtuality of the $W$-boson involved: $t$-channel ($q^2_W < 0$)
\cite{Stelzer:1997ns}, $s$-channel ($q^2_W > 0$) \cite{Smith:1996ij},
and associated $tW$ ($q^2_W = M_W^2 $) \cite{Tait:1999cf}. The
representative diagrams for the three production mechanisms
\cite{Belyaev:1998dn} are shown in Fig.\ \ref{sec81:fig1}.
Note that the LHC is a $p p$-collider and therefore the cross sections
for $t$ and $\bar{t}$ production are not equal for the $t$ and $s$-
channel mechanisms as shown in the Table \ref{sig81_tab:stnlo}.
For the associated $W$ production channel, the cross section for 
$t$ and $\bar{t}$ production are the same,
giving a $W^-t+W^+\bar t$ cross section of about $60$~pb. 
\begin{table}
\begin{center}
\caption{\label{sig81_tab:stnlo}
NLO cross sections for $t$ and $\bar t$ production at the LHC,
$\sqrt{s} = 14$~TeV.}
\medskip
\begin{tabular}{c|c|c} \hline
 & $s$-channel & $t$-channel \\ \hline
$\sigma_{NLO}(t)$, pb      & 6.55 & 152.6  \\
$\sigma_{NLO}(\bar t)$, pb & 4.07 &  90.0  \\ \hline
\end{tabular}
\end{center}
\end{table}

\begin{figure}[htb]
\centerline{\psfig{file=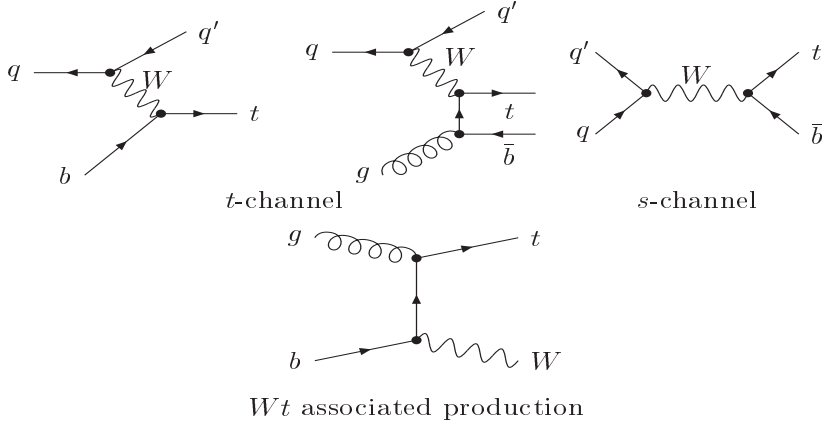,width=12.truecm,height=6.truecm,clip}}
\caption{Representative Feynman diagrams for single-top-quark
production at hadron colliders.}
\label{sec81:fig1}
\end{figure}

It was demonstrated that single-top-quark production at the LHC could
be extracted from the backgrounds, see \cite{Stelzer:1998ni,
Belyaev:1998dn, atlasphystdr, cmsewtop, Harris:2002md} for details.
In order to do that correctly, fully differential Monte Carlo programs
which include the decay of the top quark and the complete irreducible
non-resonant background have been used.

At the LC the single-top-quark production cross section was calculated
to LO at $e^+e^-$, $\gamma\gamma$ and $\gamma e$ collision modes
including various beam polarizations \cite{Boos:2001sj}. The recently
calculated NLO corrections to the single-top-quark production in
$\gamma e$ are well under control and rather small \cite{Kuhn:2003pn}.  

The single-top-quark production rate at LHC energies is about $1/3$ of
the top-pair rate, while at a 500--800 GeV LC it is smaller by a
factor of 1/80 than the corresponding pair-production rate.
Single-top-quark production in $\gamma e$ collisions is of special
interest; the rate is smaller than the top-pair rate in $e^+e^-$
only by a factor of 1/8 at 500--800 GeV energies, and it becomes the
dominant LC processes for top production at a multi-TeV LC like CLIC
as demonstrated in Fig.~\ref{sec81:figLC}.

\begin{figure}
\begin{center}   
  \epsfxsize=8cm
  \epsffile{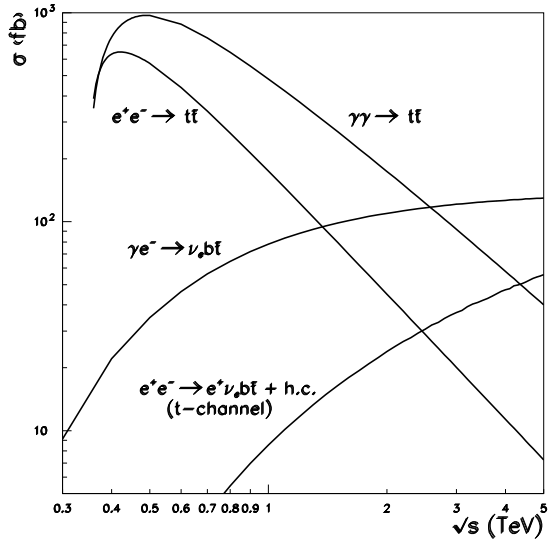}
\caption{Top-quark cross sections at the LC}
\label{sec81:figLC}
\end{center}
\end{figure}

Since single-top-quark production allows a direct measurement
of the CKM matrix element $V_{tb}$ it is worth examining the
measurement of the CKM matrix element $V_{tb}$ in more detail.
$V_{tb}$ is known indirectly from unitarity of the CKM matrix
to a very high precision \cite{Hagiwara:fs}:
\bq
\label{sec81:Vtbindirect}
\left| V_{tb} \right| & = & 0.9990 - 0.9993 \;.
\eq
In the near future there will be no way to measure $\left| V_{tb}
\right|$ directly to this precision.  Without an assumption of unitarity
the $\left| V_{tb} \right|$ element 
becomes virtually unconstrained \cite{Hagiwara:fs}:
\bq
\left| V_{tb} \right| & = & 0.08 - 0.9993 \;.
\eq

A direct measurement of $V_{tb}$ without any assumptions on the number
of generations is possible through two processes.  In single-top-quark
production the cross section at both LHC and LC is proportional to $\left| 
V_{tb} \right|^2$, and hence is a direct measure of $\left| V_{tb} 
\right|^2$. 
The partial width of the top-quark decaying into $bW$ is also 
proportional to $\left| V_{tb} \right|^2$, and at a linear collider 
one can measure indirectly the partial width by extracting the total top
quark width from the $t\bar{t}$ threshold and by measuring the 
branching of $t\rightarrow Wb$. 
The anticipated accuracy of a direct measurement of   
$\left| V_{tb} \right|$ is about $7\%$ at the LHC~\cite{atlasphystdr}
and in the $e^+e^-$ option of a linear collider
\cite{Stelzer:1998ni,Beneke:2000hk,Boos:2001sj}, and can possibly
reach $1\%$ for a polarized $\gamma e^-$ collider
\cite{Boos:2001sj}.

Note, the $|V_{tb}|$ matrix element can be measured at a LC
significantly more accurately as compared to the LHC. Therefore, 
its value could be used at LHC in
single-top analysis of the process $t$-channel $q b \rightarrow q' t$
to measure the $b$-quark distribution function in the
proton. Such a measurement is of a great importance in order to get
correctly the MSSM Higgs production rate, especially at large values of
$\tan\beta$, and gives a nice example of the potential LHC\&LC interplay.

\subsection{Spin correlations in top production and decays}

Since the top quark decays before hadronization, its spin properties
are not spoiled. Therefore spin correlations in top production and
decays is an interesting issue in top-quark physics. 
 As well known from the polarized top decays
 \cite{Czarnecki:1994pu}  the down type fermions
are the best top-quark spin analyzers. Such a unique property of
polarized top-quark decay is a consequence of the pure (V-A) structure
of the charged currents in the Standard Model.  NLO corrections do not
change this property drastically: to the leptonic $K_l$ they are very
small $-0.0015\alpha_s$ \cite{Czarnecki:1990pe}, and to quarks
$K_{d,s}$ they are about $-6\%$ \cite{Brandenburg:2002xr}.

The (V-A) structure of the top quark interaction leads to simple
top-quark spin properties for the single-top-quark production at the
LHC.  It was shown \cite{Mahlon:1996pn} that the top-quark spin in
each event should follow the direction of the down-type quark momentum
in the top-quark rest frame.  This is the direction of the initial
$\bar{d}$-quark for the $s$-channel, and the dominant direction of the
final $d$-quark for the $t$-channel single-top-quark production
processes. This follows from the above results on polarized top
decays, because these two production modes could be considered as
top-quark decays ``backward in time'' \cite{Boos:2002xw}.  Further,
the best variable to observe maximal top-quark spin correlations in
single-top-quark $s$-channel or $t$-channel production and the
subsequent decay of the top-quark is the angle between this down-type
quark direction in the production processes and the charged lepton (or
$d,s$-quark) direction from the top-quark decay in the top-quark rest
frame.  Since the $u$-quark density is the largest among the quark
densities, the observed light-quark jet in the $t$-channel is
predominately initiated by $\bar{d}$-quarks.  It is useful to define
the quantity \cite{vanderHeide:2000fx}
\bq
a & = & \frac{1}{2} \left( 1 + \cos \theta_{q\bar{l}} \right) \;,
\eq
where $\theta_{q\bar{l}}$ is the angle between the light-quark jet and
the charged lepton.
In the rest frame of the top quark the
differential cross section with respect to this variable is given by
\bq
\frac{d\sigma}{da} & = & \sigma \left( 2 P a + (1-P) \right),
\eq
where $P$ is the polarization of the top quark.
As an example, a plot for the spin correlation is shown in Fig.\ 
\ref{sec81:fig4}, based on the complete LO matrix elements (cf.\
\cite{Stelzer:1998ni} for a method to extract this signal from the
backgrounds). It was demonstrated that the top spin correlation
properties
for the single top production and decay are preserved at next-to-leading
order \cite{Campbell:2004ch, Cao:2004ky}.

\begin{figure}[htb]
\centerline{\psfig{file=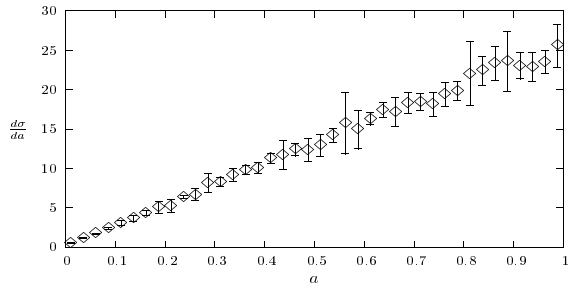,
 width=8.truecm,height=4.truecm,clip}}
\caption{ The distribution for the angular correlation $a$ for
$t$-channel production at the LHC.}
\label{sec81:fig4}
\end{figure}

\begin{figure}
\begin{center}   
  \epsfxsize=8cm
  \epsffile{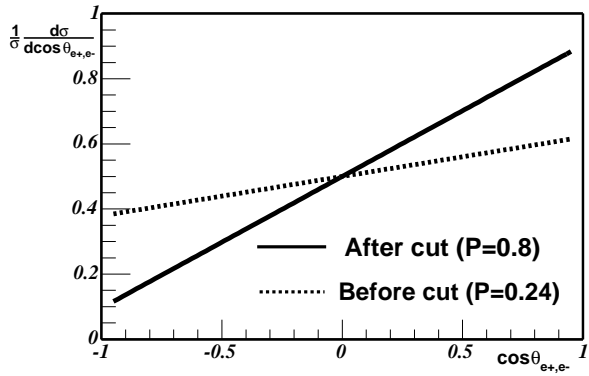}
\caption{The angular distribution for the associated $tW$ production
before and after cuts.}
\label{sec81:cut}
\end{center}
\end{figure}

Spin properties of the $t$-quark in the $tW$ production process
follows from the analogy of the $tW$ production mode to radiative
polarized top decay. Here one can find a kinematic region in which top
quarks are produced with the polarization vector preferentially close
to the direction of the charged lepton or the $d,s$-quark momentum
from the associated $W$ decay. In this kinematic region the direction
of the produced charged lepton or the $d,s$-quark should be as close
as possible to the direction of the initial gluon beam in the
top-quark rest frame \cite{Boos:2002xw} as demonstrated in Fig.\
\ref{sec81:cut}.

The latter result of the LHC process is an additional nice example of
"theoretical" LHC\&LC interplay. 
Indeed it has an immediate implication for single-top-quark
production in $\gamma e$ collisions in the process $\gamma e
\rightarrow \nu_e \bar t b$. Here the directions of the initial photon
and the electrons play the role of the gluon and lepton direction in
the $tW$ process at a hadron collider. The directions of $\gamma$ and
electron beams are close to the top-quark rest frame since the top is
moving slowly here. So one would expect the that the top-quark is strongly
polarized in the direction of the initial electron beam. Indeed, the
angular distribution for the angle between the lepton from the top
decay and the initial electron beam shows about 90\% correlation
\cite{boos-sherstnev1} (see Fig.\ \ref{sec81:figAE}). 

\begin{figure}
\begin{center}   
  \epsfxsize=8cm
  \epsfysize=5cm
  \epsffile{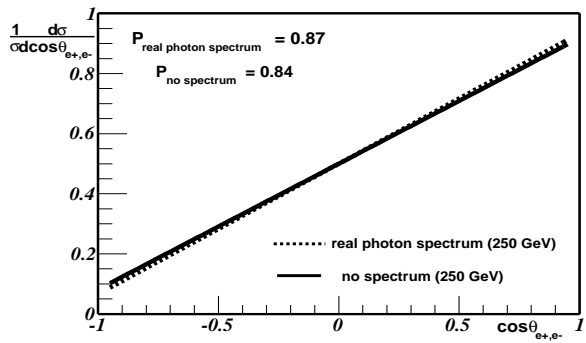}
\caption{Angular distribution for the angle between the
electron beam and muon from the top decay in single-antitop-quark
production at $\gamma e^-$ 500 GeV collider.}
\label{sec81:figAE}
\end{center}
\end{figure}


For top-quark-pair production at Linear Colliders, and in the
quark-antiquark production part at hadron colliders, specifically
important at the Tevatron, one can find a
top-quark spin-quantization axis, or in other words, a top-spin basis
(so called ``off-diagonal'' basis),
in which there will be very strong spin correlations for produced top
and anti-top quarks \cite{Parke:1996pr}. 

For the case of $e^+e^-$ collisions one can show \cite{Kiyo:2000th}
the QCD corrections lead to the known K-factors for the dominant
Up-Down spin configuration and to a very small appearing Down-Up
configuration in the ``off-diagonal basis.'' So, one may conclude
that, in contrast to the helicity basis, the QCD corrections in the
``off-diagonal'' basis affect very little the top-antitop spin
configuration. Therefore, it provides an interesting possibility to
search for anomalous top-quark interactions by looking for deviations
from the predicted SM $t\bar{t}$ spin correlations in the
``off-diagonal'' basis.  However, detailed studies in that direction,
both at the $t\bar{t}$ threshold and in the continuum regions, still
remain to be done.

The situation with $t\bar{t}$ spin correlations at the LHC is more
complex.  Here the $q\bar{q}$ collisions give the only about 10\% of
the rate.  Unfortunately, for the dominating $gg$ collision mode there
is no optimal spin basis \cite{Bernreuther:2003ga}. As a result
one needs to analyze the top-antitop spin correlation matrix in this
case. Since there is no optimal basis for the case of top pair
production at the LHC one can use either the beam or helicity basis and
compute, for example, double-angle distributions for the top and
antitop decay products: e.g.\ charged leptons
\begin{eqnarray}
\frac{1}{\sigma}
\frac{d^2\sigma(pp\rightarrow t\bar{t}X \rightarrow l^+l^- X)}
{d \cos\theta_+ \cos\theta_-} =
\frac{1}{4}(1 + B_1 \cos\theta_+ + B_2 \cos\theta_- C 
\cos\theta_+\cos\theta_-),  \label{sec81:pptt}
\end{eqnarray}
where $\theta_+, \theta_-$ are the angles of charged leptons with
respect to the top-quark spin-quantization axis in the $t(\bar{t})$
rest frame.  It was shown \cite{Bernreuther:2003ga} that the QCD
corrections are smaller for the helicity basis, and therefore this
choice of the basis looks more promising.  The problem here is how
well one can reconstruct the top-quark rest frame.  No realistic
studies including LHC detector responses have been performed so far.


\subsection{Anomalous couplings}

The search for {\em anomalous (i.e.\ non-SM) interactions} is one of
the main motivations for studying top-quark physics.  New physics can
manifest itself in two ways. Firstly, we may expect a modification of
the SM couplings ($g t \bar t$, $\gamma t \bar t$, 
$Z t \bar t$, and $tWb$ vertexes).  Secondly, the
new physics may lead to the appearance (or huge increase) of new types
of interactions (like $tH^+ b$ or anomalous Flavor Changing Neutral
Current (FCNC) -- $tgc$, $t\gamma c$, and $tZc$ interactions).
We do not know which type of new physics will be responsible for a
future deviation from the SM predictions.  However, top-quark
couplings can be parametrized in a model independent way by an
effective Lagrangian (cf.\ \cite{Beneke:2000hk}).

In practice, any $tVq$ coupling (with $V$ being any gauge boson: the
gluon, photon, or $W^{\pm}, Z$-bosons) could be parametrized with
four (complex, in general) parameters:
\begin{equation}
 t \, V \, q \, = \, \frac{g}{\sqrt{2}}
\kappa (V - A \gamma^5) \gamma^{\mu} V^{\mu}
 \, + \, \frac{g}{\sqrt{2}}
\frac{\eta}{\Lambda} \sigma^{\mu \nu} (f + i h\gamma^5)
 G_{\mu \nu},
\label{sec81:eq_an1}
\end{equation}
where $\Lambda$ is a new physics scale (we set $\Lambda = 1$~TeV);
$G_{\mu \nu} = \partial_{\mu} V_{\nu} - \partial{_\nu} V_{\mu}$ and
$V_{\mu}$ is a gauge boson, $|V|^2 + |A|^2 = |f|^2 + |h|^2 = 1$.  New
physics in the top-quark sector will be probed through anomalous top
decays as well as through anomalous production rates or channels.

Both types of couplings (the ``vector-like'' $\propto \gamma^{\mu}$
and ``tensor-like'' $\propto \sigma^{\mu \nu}$, see Eq.\
(\ref{sec81:eq_an1})) could be investigated by means of rare top
decays.  On the other hand these two types of anomalous interactions
have different behavior with respect to the energy of the subprocess
of $t$-quark production. Indeed, near the threshold region the
``vector-like'' coupling gives the main contribution, while the
``tensor-like'' coupling becomes essential at high values of $\sqrt{
\hat s } \gg m_t$. Note, that in hadronic collisions the production of
a system with the high values of $\sqrt{ \hat s }$ is suppressed due
to parton luminosities. Therefore, we may conclude very roughly that
in the production processes the ``vector-like'' couplings will be
better investigated in hadronic collisions, while a linear collider
provides the better potential to probe the ``tensor-like''
interactions. So, one expects the only combine analysis of the LHC and LC data
could provide tiny constraints on the both types of couplings.
Obviously that can not be achieved by analyzing the data from the only one type 
of a collider.

\subsubsection {Probes of anomalous $g t \bar t$ coupling}

Due to gauge invariance, only ``tensor-like'' interaction (top quark anomalous 
chromoelectric and chromomagnetic moments)
($\propto \sigma^{\mu \nu}$) could contribute to anomalous $g t \bar
t$ couplings~\cite{Beneke:2000hk}. 
 Such an anomalous coupling could be probed via the $t$-quark
production  at both LHC and LC machines.  
In hadronic collisions as it was shown in Ref.\
\cite{Rizzo:1996zt} the high-end tail of the top-quark $p_{T}$
and $M_{t \bar t}$ distributions are the observables most sensitive to
non-zero values of $\eta^g f$, with a reach for the combination
$\frac{4m_t}{\Lambda}\eta^g f$ as small as $\simeq 0.03$.  For these
values only a minor change in the total $t \bar t$ rate is expected.
The information on $\eta^g h$
could be obtained also by studying the correlation observables between
$\ell^+ \ell^-$ lepton pairs produced in the dilepton decays of $t
\bar t$ pair~\cite{Lee:1997up}. 
In $e^+e^-$ collisions such a coupling could be investigated by a measurement
of  the energy spectrum of the additional light (gluon) jet radiated 
off the top quark in the process $e^+e^- \to t\bar tg$ \cite{Rizzo:1996zt}.

From the LHC collider alone one can achieve sensitivities to 
$\frac{4m_t}{\Lambda}\eta^g$  of  order 0.03 
 with 100 $fb^{-1}$ of integrated luminosity. This is similar in magnitude 
to what can be 
obtained at a 500 GeV LC with an integrated luminosity of 50 $fb^{-1}$.
One should point out that at the LHC and at LC different kinematic
characteristics are mostly sensitive to the anomalous  moments.
Therefore one expects that the combining  analysis  of the LHC and LC data
will allow to get better accuracy in measurement of top-gluon anomalous couplings
and possibly to distinguish chromoelectric and chromomagnetic moments 
by using polarized options of LC including photon collisions.
The polarized electron-positron and photon-photon collisions are also
allow to measure uniquely the $\gamma t\bar t$ and $Z t\bar t$  anomalous
couplings \cite{Bernreuther:vd}.

\subsubsection {Search for anomalous $\bf Wtb$ couplings}
 
The $Wtb$ vertex structure can be probed and measured using either
top-quark-pair or single-top-quark production processes.  The total
$t \bar t$ rate depends very weakly on the $Wtb$ vertex structure, as
top quarks are dominantly produced on-shell~\cite{Boos:1997ud}.
However, more sensitive observables, like $C$ and $P$ asymmetries,
top-quark polarization, and spin correlations provide interesting
information. The single-top-quark production rate is directly
proportional to the square of the $Wtb$ coupling, and therefore it is
potentially very sensitive to the $Wtb$ structure 
\cite{Parke:1994dx}.  The potential to measure anomalous $Wtb$
couplings at the LHC via the production rate of single top quarks and
from kinematic distributions has been studied in several
papers~\cite{Whisnant:1997qu, Larios:1997dc,Boos:1999dd,Tait:2000sh,
atlasphystdr}. 

The potential of the hadron colliders can be compared to the potential
of a next generation $e^+e^-$ linear collider~(LC) where the best
sensitivity could be obtained in high energy $\gamma e$-collisions
\cite{Boos:1997ud, Cao:1998at}. The results of this comparison are
shown in Table~\ref{sec81:tb_twb} [in this table we present the
constraints on the following combinations of anomalous couplings:
$F_{L2}=\frac{2M_W}{\Lambda}\eta^W (-f^{W} - i h^{W})$ and
$F_{R2}=\frac{2M_W}{\Lambda}\eta^W (-f^{W} + i h^{W})$].  One may
conclude that the upgraded Tevatron will be able to perform the first
direct measurements of the structure of the $Wtb$~coupling. The LHC
with $5\%$ systematic uncertainties will improve the Tevatron limits
considerably, rivaling the reach of a high-luminosity (500 fb$^{-1}$)
500~GeV LC option. A very high energy LC with 500 fb$^{-1}$
luminosity will eventually improve the LHC limits by a factor of three
to eight, depending on the coupling under consideration.
\begin{table}[htb]
\begin{center}
  \caption{\label{sec81:tb_twb}
    Uncorrelated limits on anomalous couplings from measurements at
    different machines.}
\medskip
    \begin{tabular}{|l|lcl|lcl|}\hline
        &\multicolumn{3}{|c|}{$F_{L2}$}
        &\multicolumn{3}{|c|}{$F_{R2}$} \\\hline\hline
      Tevatron ($\Delta_{{\rm sys.}}\approx10\%$)
               & $-0.18$ &$\ldots$&$+0.55$ & $-0.24$ &$\ldots$&$+0.25$ \\
      LHC ($\Delta_{{\rm sys.}}\approx5\%$)
               & $-0.052$&$\ldots$&$+0.097$ & $-0.12$ &$\ldots$&$+0.13$ \\
      $\gamma e$ ($\sqrt{s_{e^+e^-}}=0.5\,{\rm TeV}$)
               & $-0.1$ &$\ldots$&$+0.1$ & $-0.1$ &$\ldots$&$+0.1$ \\
      $\gamma e$ ($\sqrt{s_{e^+e^-}}=2.0\,{\rm TeV}$)
               & $-0.008$&$\ldots$&$+0.035$ & $-0.016$&$\ldots$&$+0.016$ \\
      \hline
    \end{tabular} 
\end{center}  
\end{table}

\subsection {FCNC in top quark physics}

While most of the rare decays expected in the SM are beyond any
possible reach, there is a large class of theories beyond the SM where
branching fractions for decays of top quarks induced by
flavor-changing neutral currents (FCNC) could be as large as
$10^{-5}-10^{-6}$.
Present constraints on top anomalous couplings are derived from
low-energy data, direct searches of top rare decays~\cite{Abe:1997fz},
deviations from the SM prediction for $t\bar t$ production and
searches for single-top-quark production at LEP 2~\cite{fcnc-lep} and
HERA~\cite{fcnc-hera}.  A short summary of the present constraints
on FCNC couplings, recalculated in terms of ``branching ratio'' (BR)
is given in Table~\ref{sec81:tab_curfcnc}.

\begin{table}[htb]
\begin{center}
\caption{\label{sec81:tab_curfcnc} Current constraints on top-quark FCNC 
anomalous interactions presented in terms of ``branching ratios'' (here
the symbol $q$ stands for an up or charm quark). }
\medskip
\begin{tabular}{|c|c|c|c|} \hline
               &  CDF & LEP-2 & HERA  \\  \hline
 BR($t \to g       q$) & $\le 29\%$  & -- & --\\ \hline
 BR($t \to \gamma  q$) & $\le 3.2\%$ & --   & $\le 0.7\%$ \\ \hline
 BR($t \to Z       q$) & $\le 32\%$  & $\le 7.0\%$ & -- \\ \hline
\end{tabular}
\end{center}
\end{table}

At present only a few cases (like-sign top-pair production, $t \to qZ$
and $t \to q \gamma$ decays, see~\cite{atlasphystdr, Gouz:1998rk,
sec8_Gianotti:2002xx}) have been investigated with a more or less realistic
detector simulation (ATLFAST~\cite{atlfast} or CMSJET~\cite{sec8_cmsjet}).
Other investigations were done at the parton level (the final quarks
were considered as jets and a simple smearing of lepton, jet and
photon energies was applied, cf.\ \cite{Malkawi:1995dm, Tait:1996dv,
delAguila:1999v, Aguilar:2001}).

For the LHC case all three possible FCNC decays have been
investigated~\cite{atlasphystdr,sec8_Gianotti:2002xx}:
\begin{eqnarray}
 && t \to q \, V , \;\;\; {\rm where} \;\;  V=\gamma, Z, g \;\;\; {\rm and} \;\;\; 
   q=u \,\, {\rm or } \,\, c 
\end{eqnarray}
The most important background processes include the production of $t
\bar t$~pairs, $W + $~jets, $WW+WZ+ZZ$, $W + \gamma$, and single-top-quark
production.  The achievable branching ratios are summarized in
Table~\ref{sec81:tab_future}.

The anomalous FCNC top-quark interaction can be probed also in the
processes of single-top-quark production.  There are four different
subprocesses which lead to anomalous top-quark production in the final
state together with one associated jet (see Fig.~\ref{sec81:fig11}):
\begin{eqnarray}
 q \bar q \to  t \bar q, \quad g g \to t \bar q, \quad
 q q  \to t q, \quad  q g \to  t g\,.  \label{sec81:anomeq_twotwo}
\end{eqnarray} 
The major background comes from $W+2$ jets and $W+b \bar b$
production, as well as from single-top-quark production within SM 
approach~\cite{Beneke:2000hk}. 

\begin{figure}[htb]
\begin{minipage}{.48\textwidth}
\vspace*{0cm}
  \begin{center}
    \resizebox{6cm}{!}{\includegraphics{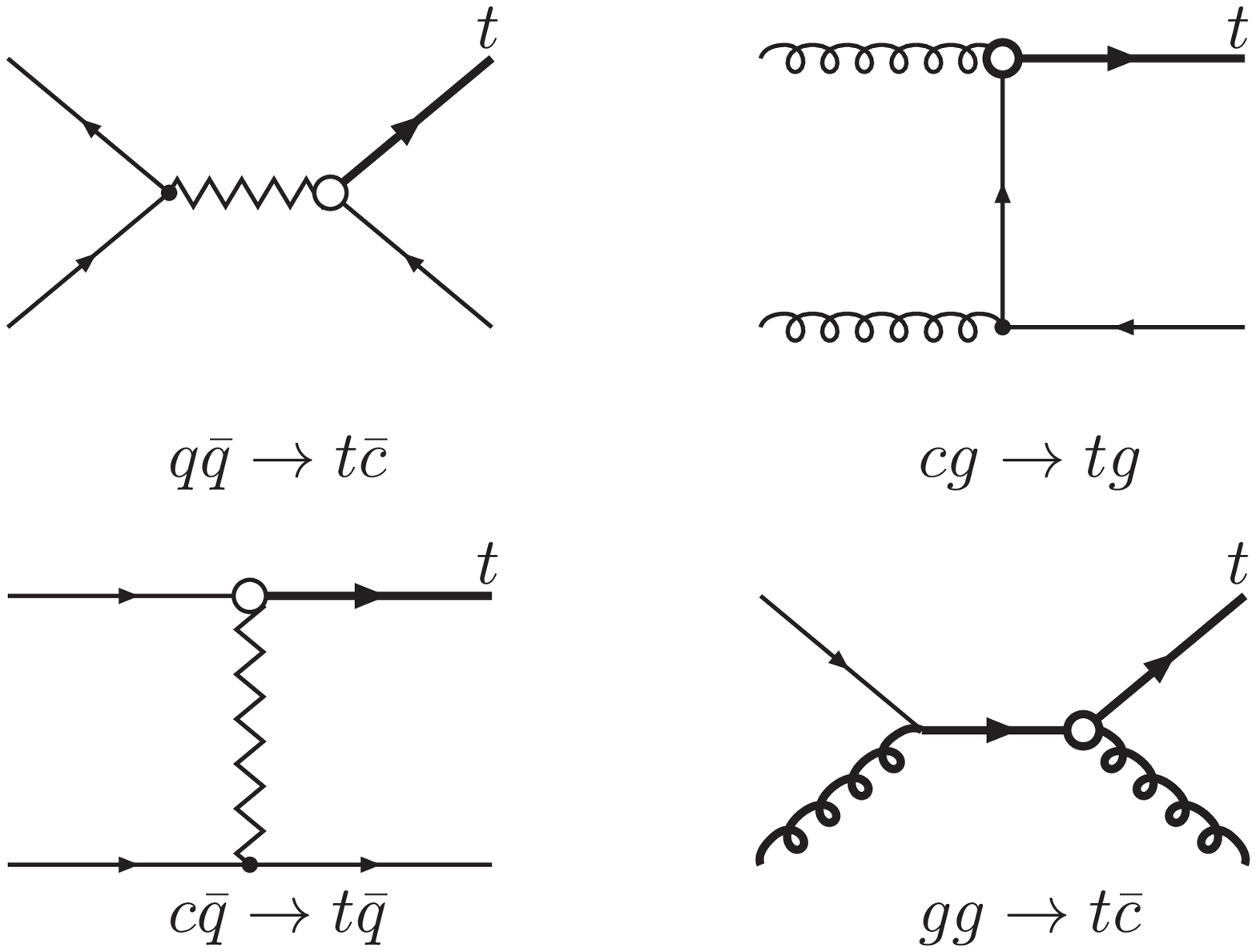}}
\vspace*{0cm} 
\caption{ $2 \to 2$ single-top-quark production.}
\label{sec81:fig11}
  \end{center}
\end{minipage}   
\hfill
\begin{minipage}{.48\textwidth}
\vspace*{+1.5cm}
  \begin{center}
    \resizebox{6cm}{!}{\includegraphics{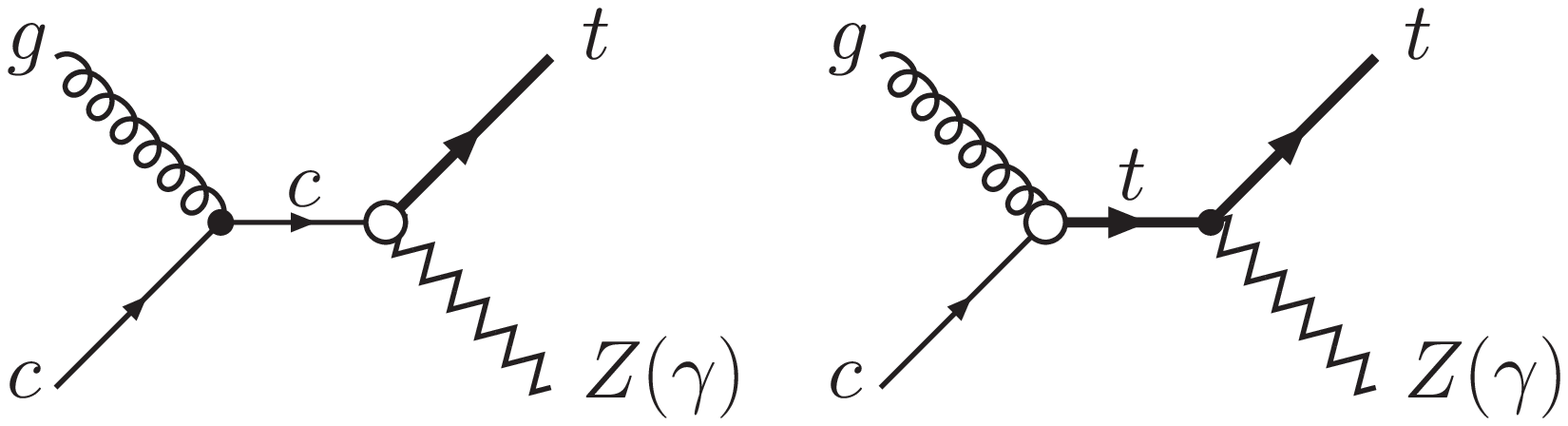}}
\vspace*{1.5cm}
     \caption{ $s$-channel diagrams for $t V$ ($V = Z, \gamma$)
      production  }
    \label{sec81:fig12}
  \end{center}
\end{minipage}
\end{figure} 

All anomalous FCNC couplings may contribute to the processes of
single-top-quark production with an associated hard photon or
$Z$-boson~\cite{delAguila:1999v}:
\begin{eqnarray*}
 q  g \to  \gamma t, \quad q g \to Z t.
\end{eqnarray*}
 The left diagram in
Fig.~\ref{sec81:fig12} corresponds to the $Z(\gamma)tq$ coupling, while  
the right one shows the top-gluon anomalous coupling (the
corresponding $t$-channel diagrams are not shown).
It has been shown that the best limits on the top-quark FCNC couplings
can be obtained from the decay channels $Zt \to \ell^+ \ell^- \, \ell
\nu b$ and $\gamma t \to \gamma \, \ell \nu b$.
The most promising way to measure the anomalous FCNC top-gluon
coupling seems to be the investigation of single-top-quark production
processes, as the search for $t \to g q$ decays would be overwhelmed
by background from QCD multi-jet events. At the same time, both
top-quark production and decay would provide comparable limits on
top-quark anomalous FCNC interactions with a photon or a $Z$-boson.
In general, the studies shown above indicate that the LHC will improve
by a factor of at least 10 the Tevatron sensitivity to top-quark FCNC
couplings~\cite{sec8_Gianotti:2002xx}.

At a linear $e^+ e^-$ collider both methods of searching for FCNC
interactions were considered (see~\cite{Aguilar:2001,
Bar-Shalom:1999iy}): $t \bar t$-pair production via SM
$\gamma/Z$-exchange with following FCNC decays $t \to q \gamma \, (Z)$:
\begin{eqnarray}
 e^+ e^- \to t \bar t, \quad {\rm with} \quad t \to q \gamma \, (Z) ,\
 \label{sec81:ee1}
\end{eqnarray} 
and single-top-quark production due to FCNC anomalous interactions
with the SM decay channel $t \to b W$: 
\begin{eqnarray}
e^+ e^- \to t  u(\bar {c}), \quad {\rm with} \quad t \to b W.
 \label{sec81:ee2}
\end{eqnarray} 


  It was shown in~\cite{Aguilar:2001}, that
the top-quark decay signal from the process~(\ref{sec81:ee1}) is
cleaner than ($t \bar q$) production~(\ref{sec81:ee2}).  On the other
hand, the cross sections for $t \bar q$-processes is larger than for $t
\bar t $-production. At the same time the processes with FCNC top-quark
decays~(\ref{sec81:ee1}) could help to determine the nature of the
couplings involved, $t\gamma q$ or $tZq$.  Also, it was shown that
beam polarization will be very useful to improve the limits from $t
\bar q$ production due to a substantial background decrease
 (see~\cite{Aguilar:2001} for details).


\begin{table}[htb!]
\begin{center}
  \caption{\label{sec81:tab_future} Future expectations on FCNC interactions
from Run~II of the Tevatron, LHC(CMS), and linear $e^+ e^-$  
collider.}
\medskip
\begin{tabular}{|l|r|l|l|l|} \hline
 & Tevatron  &\multicolumn{2}{c|}{ LHC}   & $e^+ e^-$  \\
$t \to      $ &  Run II  & decay  & production & $\sqrt{s} > 500$~GeV \\ 
\hline
$g \,  q    $ & $0.06 \%$ & $1.6 \times 10^{-3}$ & $1 \times 10^{-5}$  & -- \\ 
\hline
$\gamma \, q$ &$0.28 \%$ & $2.5 \times 10^{-5}$
         &  $3 \times 10^{-6}$  &$ 4 \times 10^{-6}$\\ \hline
$Z \, q  $ &$1.3 \%$ & $1.6 \times 10^{-4}$ 
         & $1 \times 10^{-4}$ &$2 \times 10^{-4}$  \\ \hline
\end{tabular}
\end{center}
\end{table}

Table~\ref{sec81:tab_future} presents a short comparison of LHC and LC
potentials to investigate FCNC couplings.  Branching ratios of
order $10^{-6}$ are achievable, which are of interest for some
theories beyond the Standard Model, as discussed in Ref.\
\cite{Beneke:2000hk}.  Note that, due to the limited statistics which could
be gathered in a future linear collider in the reaction $e^+ e^- \to t
\bar t$, the LHC has an advantage in the searches for rare top-quark
decays.  On the other hand, the future LC has a much smaller
background.  Therefore, both the LHC and a future LC have great potential to
discover top-quark production due to anomalous interactions.  Only for
anomalous interactions with a gluon ($tgc$ or $tgu$) will the LHC have
an evident advantage.  There are many studies devoted to top-quark
rare decays, like $t \to bWZ$, $t \to b H^+$, etc. However, almost all
these investigations were performed for the LHC option only. We refer the
reader to the most comprehensive review~\cite{Beneke:2000hk}.


It has to be stressed that different types of new interactions may
manifest itself in different ways at different colliders even for the 
same observables.
So, only  mutual LHC\&LC analysis will provide 
more definite conclusions about a type of possible new interactions.

%

\section{\label{sec:82} Electroweak precision physics}

{\it U.~Baur, S.~Heinemeyer, S.~Kraml, K.~M\"onig, W.~Porod and G.~Weiglein}

\vspace{1em}

\renewcommand{\be}{\beta}
\newcommand{\de}{\delta}
\newcommand{\De}{\Delta}
\newcommand{\sweff}{\sin^2\theta_{\mathrm{eff}}}
\newcommand{\sw}{s_W}
\newcommand{\MW}{M_W}
\newcommand{\MZ}{M_Z}
\renewcommand{\mh}{m_h}
\newcommand{\MA}{M_A}
\renewcommand{\mt}{m_t}
\newcommand{\mtexp}{m_t^{\rm exp}}
\newcommand{\Xt}{X_t}
\newcommand{\At}{A_t}
\newcommand{\Ab}{A_b}
\renewcommand{\tb}{\tan\be\;}
\renewcommand{\ML}{\left( \begin{array}{cc}}
\renewcommand{\MR}{\end{array} \right)}
\newcommand{\Stop}{\tilde{t}}
\newcommand{\StopL}{\tilde{t}_L}
\newcommand{\StopR}{\tilde{t}_R}
\newcommand{\Stope}{\tilde{t}_1}
\newcommand{\Stopz}{\tilde{t}_2}
\newcommand{\MstL}{M_{\tilde{t}_L}}
\newcommand{\MstR}{M_{\tilde{t}_R}}
\newcommand{\mste}{m_{\tilde{t}_1}}
\newcommand{\mstz}{m_{\tilde{t}_2}}
\renewcommand{\tst}{\theta_{\tilde{t}}}
\newcommand{\msbe}{m_{\tilde{b}_1}}
\newcommand{\msbz}{m_{\tilde{b}_2}}
\renewcommand{\mgl}{m_{\tilde{g}}}
\renewcommand{\lsim}{\;\raisebox{-.3em}{$\stackrel{\displaystyle <}{\sim}$}\;}
\renewcommand{\gsim}{\;\raisebox{-.3em}{$\stackrel{\displaystyle >}{\sim}$}\;}
\renewcommand{\gev}{\,\, \mathrm{GeV}}
\renewcommand{\tev}{\,\, \mathrm{TeV}}
\renewcommand{\non}{\nonumber}

In this section the electroweak precision measurements,
triple gauge boson couplings (TGC), and contact interactions are analyzed. 
Electroweak precision observables can play a role in various
models. In many physics scenarios they can provide information about
new physics scales that are too heavy to be detected directly. Several
possible applications of precision observables are discussed in
different parts of this report. Examples of the application of
electroweak precision observables are
\begin{itemize}
\item 
the search for new gauge bosons, see Sect.~\ref{subsub-zpgigaz},
\item
limits for new physics scales in little Higgs models, see
Sect.~\ref{sec:27},
\item
investigations on strong electroweak symmetry breaking, see
Sect.~\ref{chapter:strongewsymmbreak},
\item the search for heavy scales in supersymmetric theories, which
will be covered in this section.
\end{itemize}
For the electroweak precision measurements (in the context of
Supersymmetry) three example studies are presented. Physics scenarios
which are dominated by LHC measurements can be probed much better if
in addition the information from electroweak precision observables,
obtained at the LC, is used.
The triple 
gauge couplings are analyzed in view of a possible combination of the
results of the LHC and the LC. In the case that both colliders can
obtain independently about the same uncertainty, the combination of the
two experiments can result in an even higher precision. 

Concerning the physics of contact interactions a possible LHC/LC
interplay could arise as follows. 
Precision measurements of $e^+ e^- \to f \bar f$ at the LC and of
Drell-Yan production at the LHC might exhibit deviations that could be
interpreted in terms of 
contact interactions. While the LHC is sensitive to
$u \bar u \to \ell^+ \ell^-$ and to a somewhat lesser extend to
$u \bar d \to \ell^+ \ell^-$  the linear collider measures
$e^+ e^- \to q \bar q$ with unidentified quarks,
$e^+ e^- \to b \bar b$ and $e^+ e^- \rightarrow \ell^+ \ell^-$
with identified leptons. The combination of the two machines can thus
disentangle the flavour structure of the new interactions in case a
signal is seen.


\subsection{Electroweak precision measurements at the LHC and the LC}

The precision that can be achieved for electroweak precision
observables (EWPO), including $\mt$ and $\mh$ has been analyzed in
detail in Ref.~\cite{Baur:2001yp} and is reviewed in
Table~\ref{sec82:tab:ewpo}.  
The numbers listed there represent the sensitivities
that can be reached by each individual collider. 
The reach of the LC is seen to exceed that of the LHC (due to its cleaner
environment) 
for all relevant observables. It is therefore difficult to find a case of
positive LHC/LC interplay by just looking at EWPO's. This,
however, changes if SUSY scenarios are analyzed. It is conceivable
that part of the SUSY spectrum, due to the large masses of some of the
SUSY partners, can
only be determined by the LHC~\cite{sec82_Allanach:2002nj}. This may
especially be relevant in the case of scalar top quarks which play an
important role for the SUSY interpretation of EWPO's, see e.g.\
Refs.~\cite{sec8_Erler:2000jg,Heinemeyer:1998np}. In this case, the measurement of
the SUSY mass spectrum at the LHC and the LC can be combined with
precision measurements of the EWPO's to derive
constraints on the remaining unknown parameters or to perform stringent
consistency checks of the underlying model. This will be shown for some
representative scenarios in the following subsections.

\begin{table}[htb]
\renewcommand{\arraystretch}{1.5}
\begin{tabular}{|c||c||c|c|c||c|c|}
\cline{2-7} \multicolumn{1}{c||}{}
& now & Tev.\ Run~IIA & Run~IIB & LHC & ~LC~  & GigaZ \\
\hline\hline
$\de\sweff(\times 10^5)$ & 17   & 78   & 29   & 14--20 & (6)  & 1.3  \\
\hline
$\de\MW$ [MeV]           & 34   & 27   & 16   & 15   & 10   & 7      \\
\hline
$\de\mt$ [GeV]           &  5.1 &  2.7 &  1.4 &  1.0 &  0.2--0.1 & 0.1 \\
\hline
$\de\mh$ [MeV]           &  --- &  --- & ${\cal O}(2000)$ &  
200 &   50 &   50 \\
\hline
\end{tabular}
\renewcommand{\arraystretch}{1}
\caption{Current and anticipated future experimental uncertainties for 
the effective leptonic mixing angle, $\sweff$, the $W$~boson mass,
$\MW$, the top quark mass, $\mt$, and the Higgs boson mass, $\mh$. See
Ref.~\cite{Baur:2001yp} for a detailed discussion and further references.}
\label{sec82:tab:ewpo}
\end{table}


\subsection{Constraints on the parameters of the scalar top sector}
\label{sec82:subsec:MSSMstop}

Once a Higgs boson compatible with the MSSM predictions has been
discovered, the dependence of $\mh$ on the top and scalar
top~\cite{Heinemeyer:1998np,sec82_Degrassi:2002fi} sector can
be utilized to determine unknown parameters of the
$\Stop$~sector~\cite{Heinemeyer:2000jd,Heinemeyer:2003ud}.

The mass matrix relating the interaction eigenstates $\StopL$ and $\StopR$ 
to the mass eigenstates is given by
\begin{equation}
\label{sec82:stopmassmatrix}
{\cal M}^2_{\Stop} =
  \ML \MstL^2 + \mt^2 + \cos 2\be\; (\frac{1}{2} - \frac{2}{3} \sw^2) \MZ^2 &
      \mt \Xt \\
      \mt \Xt &
      \MstR^2 + \mt^2 + \frac{2}{3} \cos 2\be\; \sw^2 \MZ^2 
  \MR ~,
\end{equation}
where $\Xt$ can be decomposed as $\Xt = \At - \mu/\tb$. $\At$ denotes
the trilinear Higgs--$\Stop$ coupling. Assuming that $\tb$ and $\mu$ can
be determined from other sectors, there are three new parameters in the
mass matrix, the soft SUSY-breaking parameters $\MstL$, $\MstR$, and $\At$.
The mass eigenvalues of the scalar top states are obtained after a rotation
with the angle $\tst$,
\begin{equation} 
{\cal M}^2_{\Stop} 
\quad
\stackrel{\tst}{\longrightarrow}
\quad
\ML \mste^2 & 0 \\ 0 & \mstz^2 \MR ~.
\end{equation}

\underline{Scenario I:}
the two masses $\mste$,
$\mstz$ are determined at the LHC, but $\tilde t_1$ and $\tilde t_2$ 
are too heavy for direct 
production at the LC. In this case the direct measurement of the 
off-diagonal entry in the
$\Stop$~mass matrix Eq.~(\ref{sec82:stopmassmatrix}), i.e.\ a measurement of
$\tst$, is not possible and 
the trilinear coupling $\At$ cannot be determined. However,
measurement of $\mh$ would allow an indirect determination of
$\At$~\cite{Heinemeyer:2003ud}. This is shown 
for the benchmark scenario SPS1b~\cite{sec82_Allanach:2002nj} in
Fig.~\ref{sec82:fig:mhAt} (evaluated with 
{\em FeynHiggs}~\cite{Heinemeyer:1998yj,Frank:2002qa,wwwfeynhiggs}). 
In this scenario, a
measurement of $\MA$ and $\tb$ at the LHC is
possible~\cite{sec82_Cavalli:2002vs}. 
In Fig.~\ref{sec82:fig:mhAt} the experimental error of $\mh$ is indicated, 
whereas the theoretical error 
due to unknown higher order corrections has been neglected; see
Ref.~\cite{Heinemeyer:2003ud} for details.
Fig.~\ref{sec82:fig:mhAt} demonstrates that $\At$ can be determined with LHC
measurements alone; the allowed range is indicated by the green (light
shaded) area. However, the precise determination of 
$\mt$ at the LC improves the accuracy for
$\At$ by about a factor of three over that which can be achieved with
the LHC $\mt$ measurement. Concerning resolving the sign ambiguity of
$\At$, see Sect.~\ref{sec:221}. 

\begin{figure}[htb!]
\begin{center}
\epsfig{figure=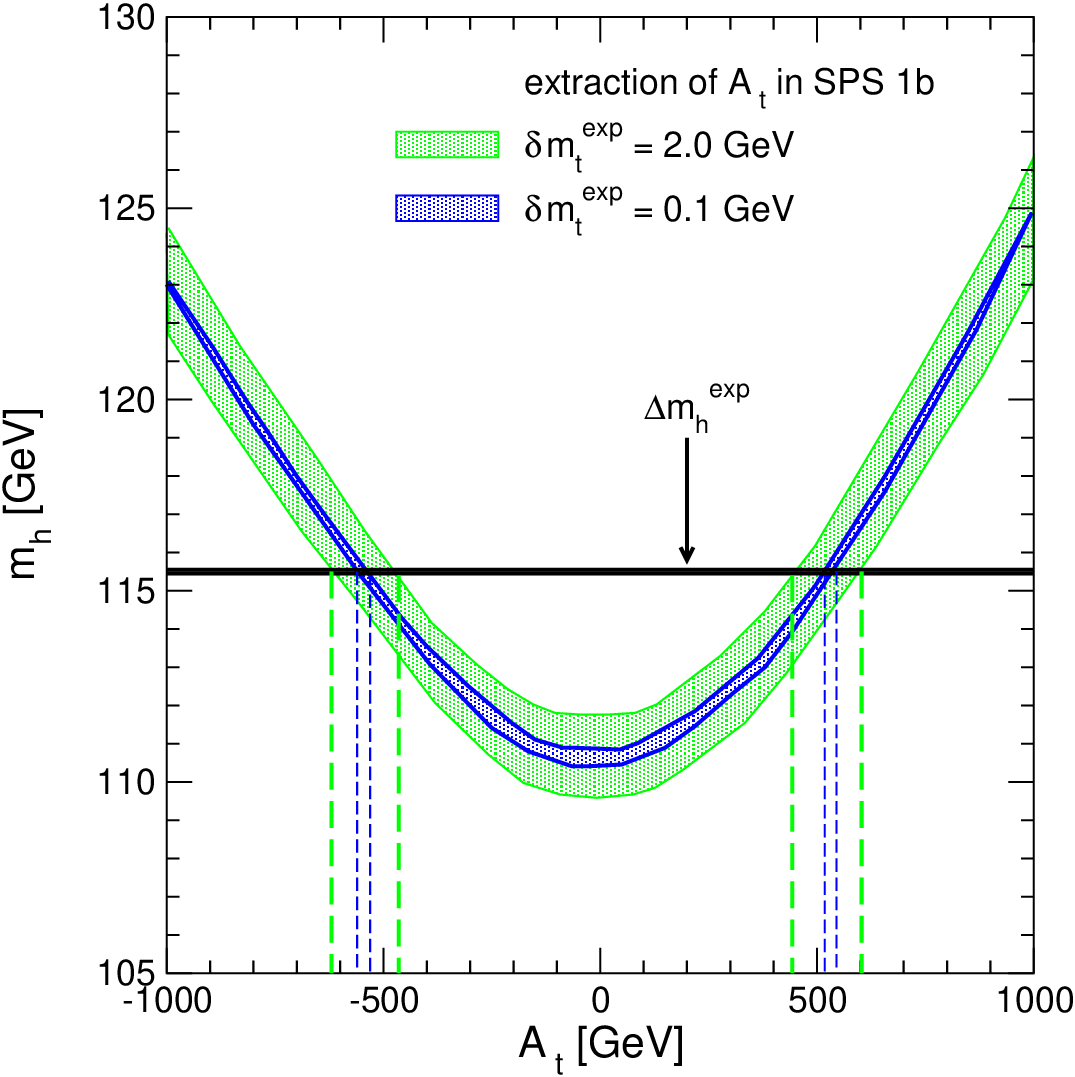, width=12cm,height=8cm}
\caption{
Indirect determination of $\At$ in the SPS1b scenario for $\tb=30$, 
$\de\mtexp = 2 \gev$ (LHC) and $\de\mtexp = 0.1 \gev$
(LC). 
The statistical uncertainty of the Higgs boson mass, $\De\mh^{\rm exp}$, is
indicated. The SUSY parameters $\mste, \mstz, \msbe, \msbz$ are assumed
to be given by the values predicted by SPS1b with an uncertainty of 5\%.
$\MA$ and $\tb$ are assumed to be determined with a precision of 10\%
and 15\%, respectively. 
}
\label{sec82:fig:mhAt}
\end{center}
\end{figure}

\underline{Scenario II:} 
in this example~\cite{sec8_Aguilar-Saavedra:2001rg,Heinemeyer:2003ud},
the lighter scalar top quark and $\MA$ are accessible at the LC, whereas
only a lower bound can be established for $\tb$. The LC 
will provide precise measurements of its mass, $\mste$, and the
$\Stop$~mixing angle, $\tst$~\cite{Berggren:1999ss,Bartl:2000kw}. On
the other hand, the heavier scalar top, 
$\Stopz$, can only be detected at the LHC. 
The measurement of $\mh$ can then be used to obtain
indirect limits on $\mstz$. Comparison of the indirectly determined
$\mstz$ with the value obtained from a direct measurement at the LHC
will provide a stringent consistency test of the MSSM.
In Fig.~\ref{sec82:fig:mhMSt2} we show the allowed region in the
$\mstz-\mh$-plane for this scenario.
Assuming that the top quark mass can be determined with a precision of
$\de\mt = 0.1 \gev$ at the LC, one finds $680 \gev \lsim
\mstz \lsim 695 \gev$. For comparison, for $\de\mt = 2 \gev$, as
expected from LHC experiments, one obtains $670 \gev \lsim \mstz \lsim
705 \gev$, i.e. the precise determination of $\mt$ at the LC reduces the
uncertainty of $\mstz$ by about a factor~3.
Again the theoretical error 
due to unknown higher order corrections to $\mh$ has been neglected; see
Ref.~\cite{Heinemeyer:2003ud} for details.

\begin{figure}[htb!]
\begin{center}
\epsfig{figure=sec82_mhMSt202b.cl.eps, width=12cm,height=7cm}
\caption{
Indirect determination of $\mstz$ from the measurement of $\mh$ for
$\de\mtexp = 2 \gev$ (LHC) and $\de\mtexp = 0.1 \gev$ (LC). 
The experimental error of the Higgs boson mass, $\de\mh^{\rm exp}$, is
indicated. The other parameters are given by
$\mste = 180 \pm 1.25 \gev$, $\cos\tst = 0.57 \pm 0.01$, 
$\MA = 257 \pm 10 \gev$, $\tb > 10$, 
$\mu = 263 \pm 1 \gev$, $\mgl = 496 \pm 10 \gev$, $\Ab = \At \pm 30\%$,
and a lower bound of $200 \gev$ has been imposed on the lighter sbottom
mass. 
}
\label{sec82:fig:mhMSt2}
\end{center}
\end{figure}
\begin{figure}[htb!]
\begin{center}
\epsfig{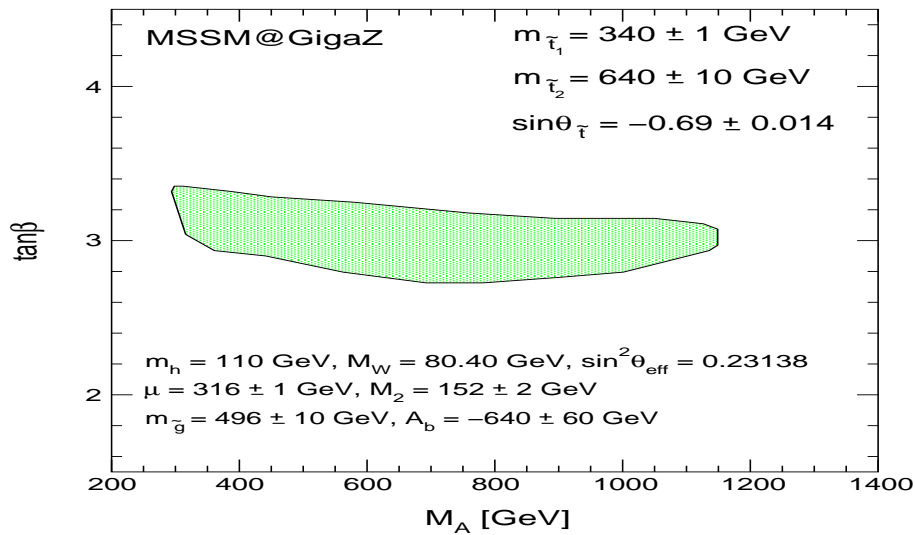}
\caption{The region in the $\MA-\tb$-plane allowed by the measurements
  of the SUSY spectrum at the LHC and the LC and by the measurements
  of $\MW$, $\sweff$ and $\mh$ at the LC.
}
\label{sec82:fig:TBMA}
\end{center}
\end{figure}


\subsection{Constraints on the parameters of the MSSM Higgs boson sector}
\label{sec82:subsec:MSSMhiggs}

In the previous scenarios the parameters of the Higgs sector were
assumed to be known. However, it may also be possible to use SUSY mass
measurements at the LHC in combination with the EWPO measurement
(especially $\MW$ and $\sweff$) at the LC to obtain information on the
(unknown) parameters of the Higgs sector~\cite{sec8_Erler:2000jg}. 

\underline{Scenario III:}
In Fig.~\ref{sec82:fig:TBMA} we show the parameter space of 
a scenario where the MSSM Higgs
mass scale, the mass of the $CP$-odd Higgs boson $\MA$, cannot be
measured directly at the LHC nor the LC (for most of the shown
parameter space). However, the combination of LHC and LC mass
measurements together with the $\MW$, $\sweff$ from GigaZ and the
$\mh$ determination at the LC (see the plot for further details) can set
an upper bound of $\MA \lsim 1200 \gev$.


\subsection{Triple gauge boson couplings}
\label{sec82:subsec:tgc}

The LHC will significantly improve the precision of the measurements of
the triple gauge boson couplings (TGC) compared to the LEP and Tevatron
results~\cite{Haywood:1999qg}. In the SM, the TGC's are
uniquely fixed by gauge invariance. Extensions to the SM, in particular
models in which the $W$~and $Z$~bosons are composite objects, often lead
to deviations from the SM predictions for the TGC's. 

Assuming electromagnetic gauge invariance, Lorentz invariance, and 
$C$-and $P$-conservation, five parameters can be used to describe the three
gauge boson vertices~\cite{Haywood:1999qg}: 
\begin{equation} 
\De g_1^Z = g_1^Z-1~, \qquad  
\De \kappa_V = \kappa_V - 1~, \qquad
 \lambda_V \qquad\qquad (V=\gamma,\,Z) ~.
\end{equation} 
In the SM, at tree level, all five parameters vanish. 

At the LHC, $W\gamma$ and $WZ$ production can be used to constrain
TGC's. $WW$ production, which is also sensitive to the weak boson
self-couplings, suffers from a large $t\bar t$ background, and,
therefore, is not considered. $W\gamma$ production probes the $WW\gamma$
couplings $\De\kappa_\gamma$ and $\lambda_\gamma$. $WZ$ production is
sensitive to the $WWZ$ couplings $\Delta g_1^Z$, $\Delta\kappa_Z$ and
$\lambda_Z$.

\begin{figure}[htb!]
\begin{center}
\includegraphics[width=12.0cm,height=8cm]{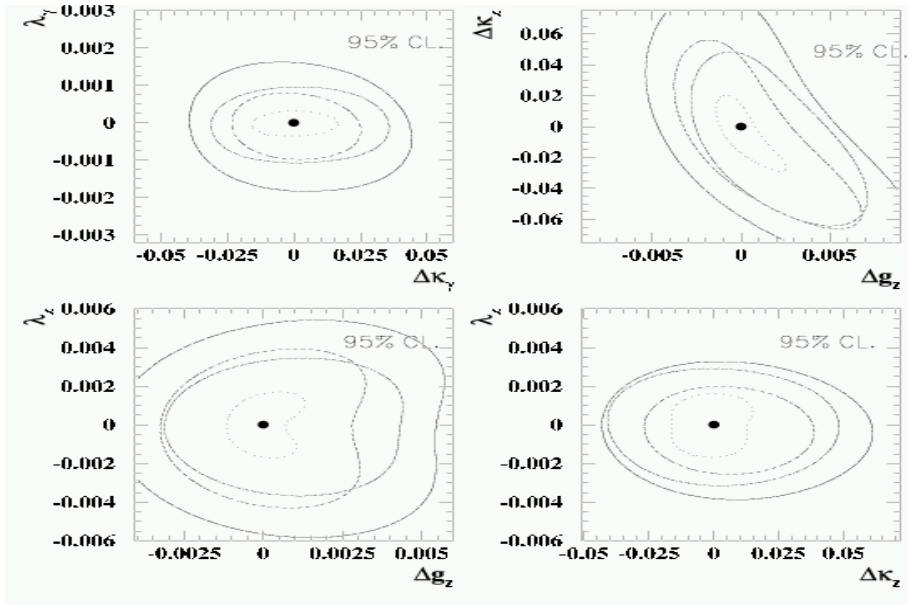}
\caption[]{\label{sec82:fig:LHC} Expected 95\% CL constraints on TGC's in
ATLAS resulting from two-parameter fits. Shown are results for
$\sqrt{s}=14 \tev$ and 100~fb$^{-1}$ (solid), $\sqrt{s}=28 \tev$ and
100~fb$^{-1}$ (dot-dash), $\sqrt{s}=14 \tev$ and 1000~fb$^{-1}$ (dash),
and $\sqrt{s}=28 \tev$ and 1000~fb$^{-1}$ (dotted). The figure has been
taken from Ref.~\cite{sec8_Gianotti:2002xx}.} 
\end{center}
\end{figure}
%
\begin{figure}[htb!]
\begin{center}
\begin{tabular}{cc}
\includegraphics[height=3.7cm,bb=2 8 542 528]{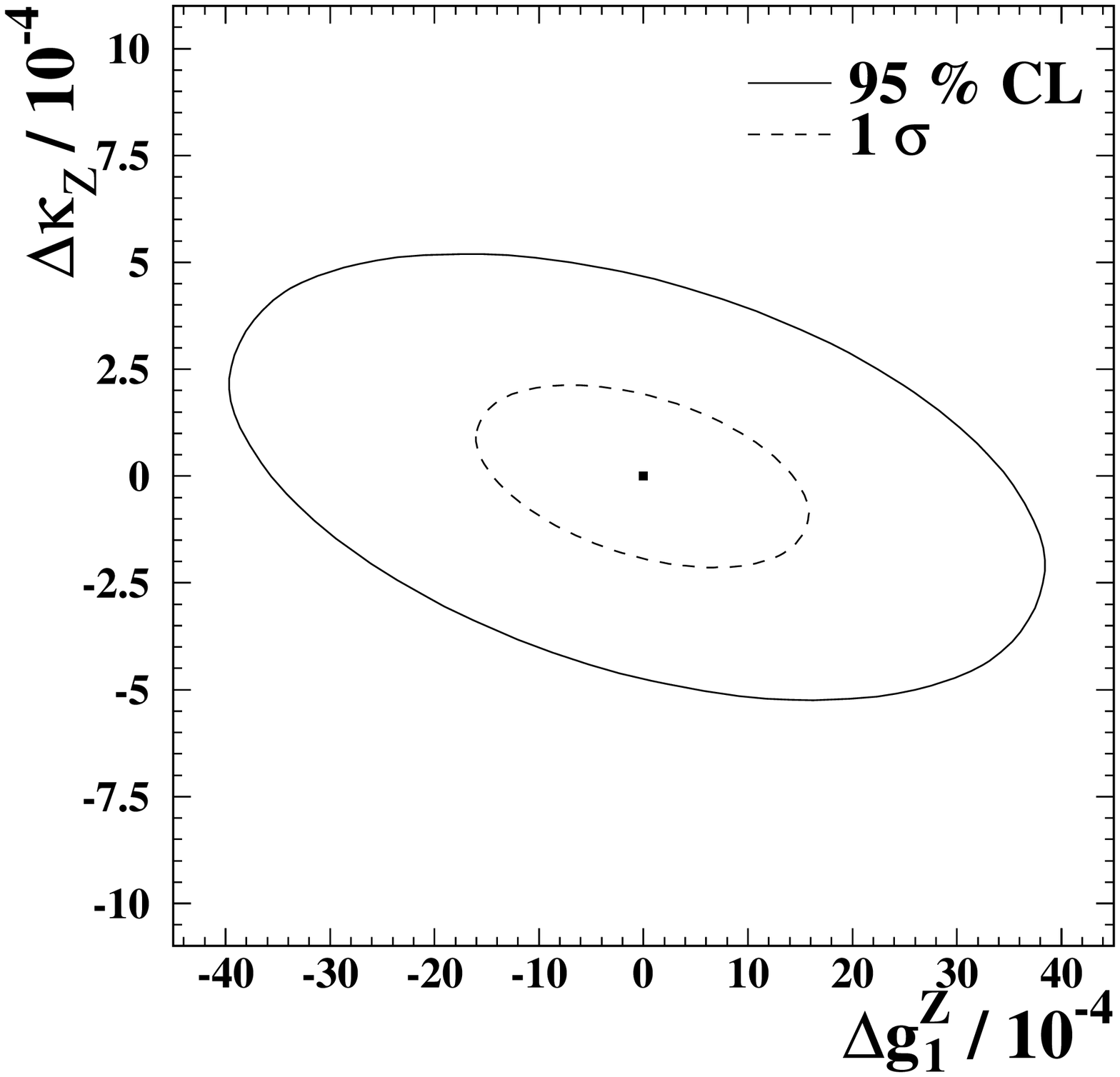}&
\includegraphics[height=3.7cm,bb=2 8 542 528]{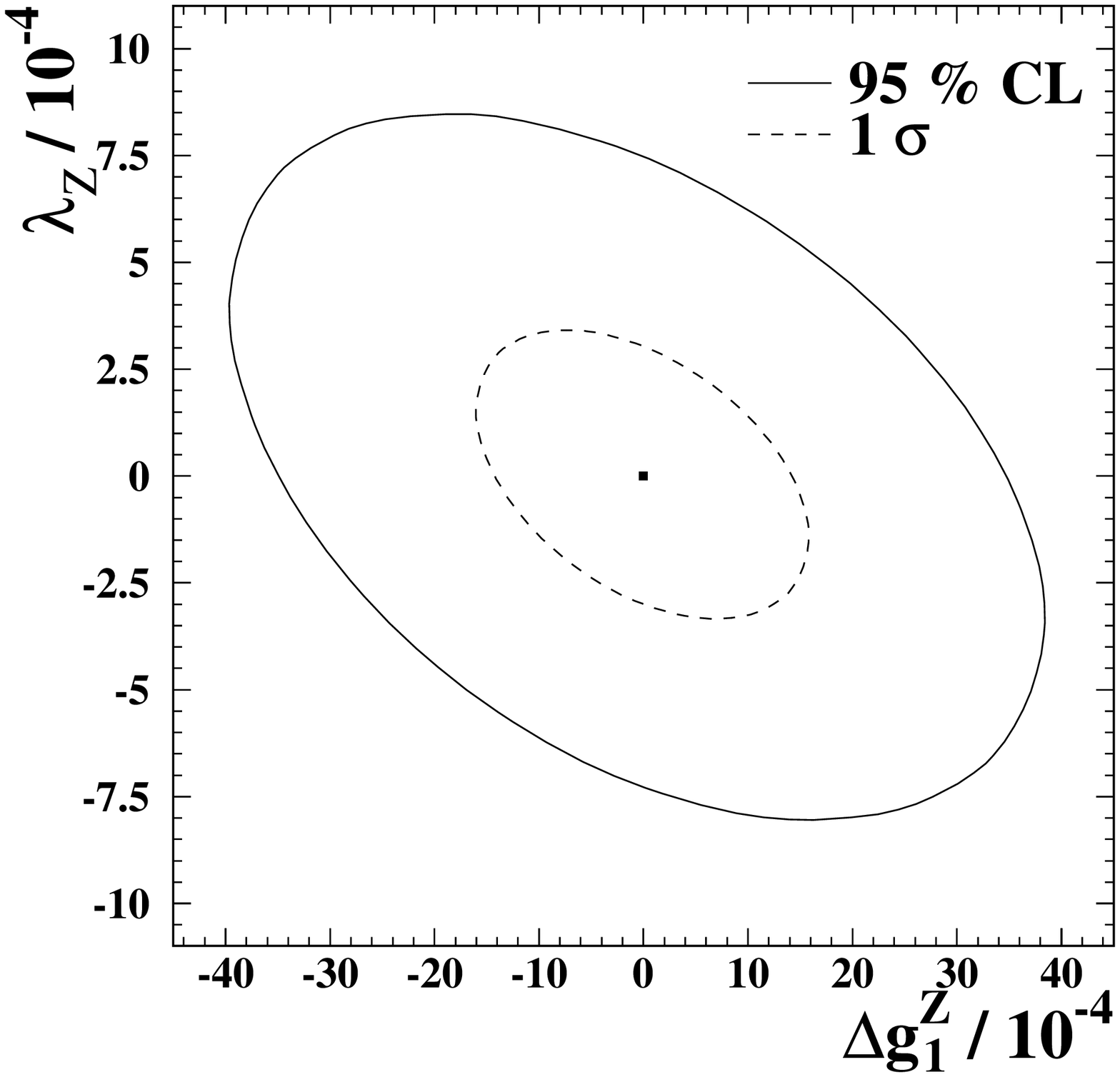}\\
\includegraphics[height=3.7cm,bb=2 8 542 528]{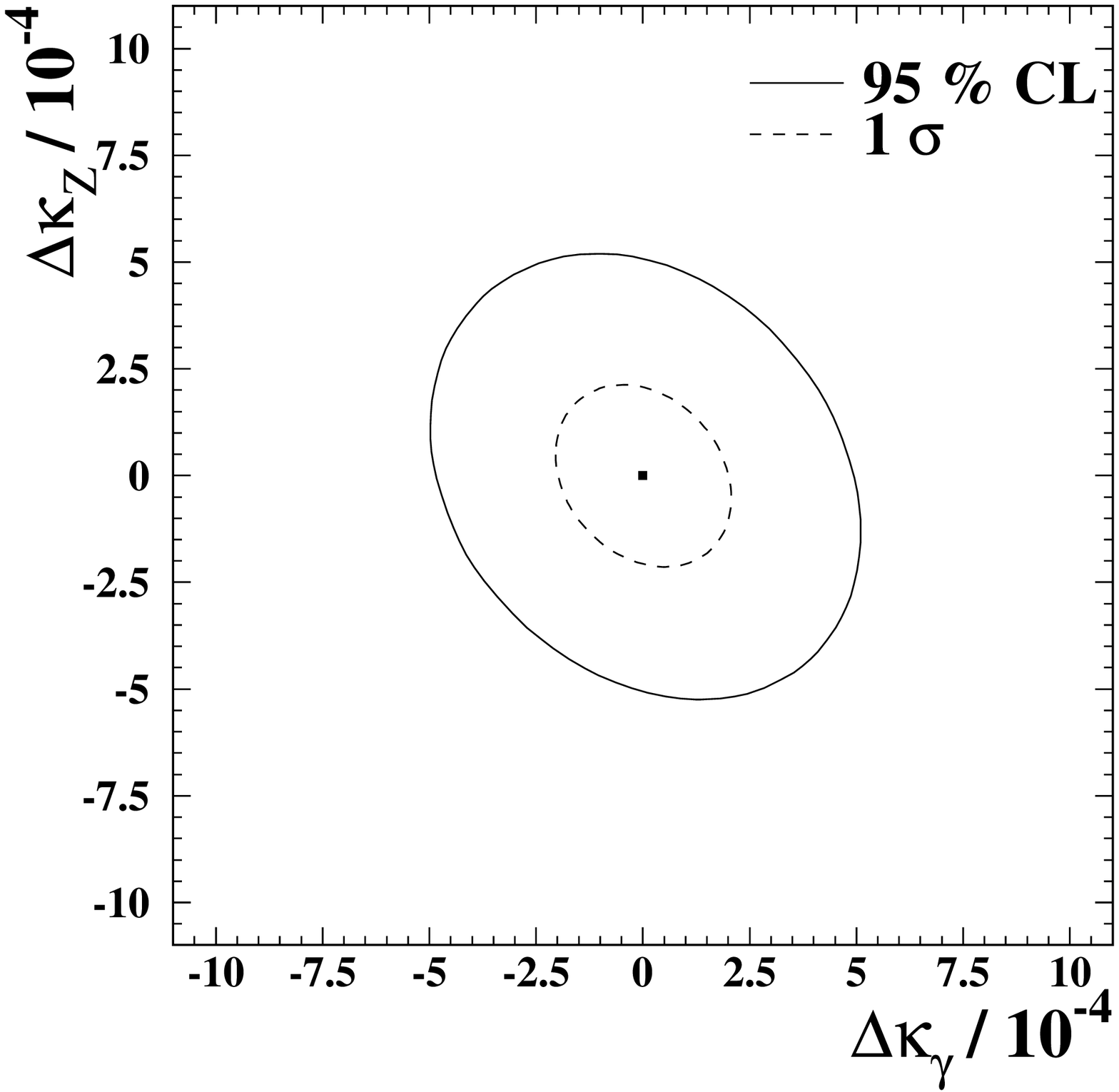}&
\includegraphics[height=3.7cm,bb=2 8 542 528]{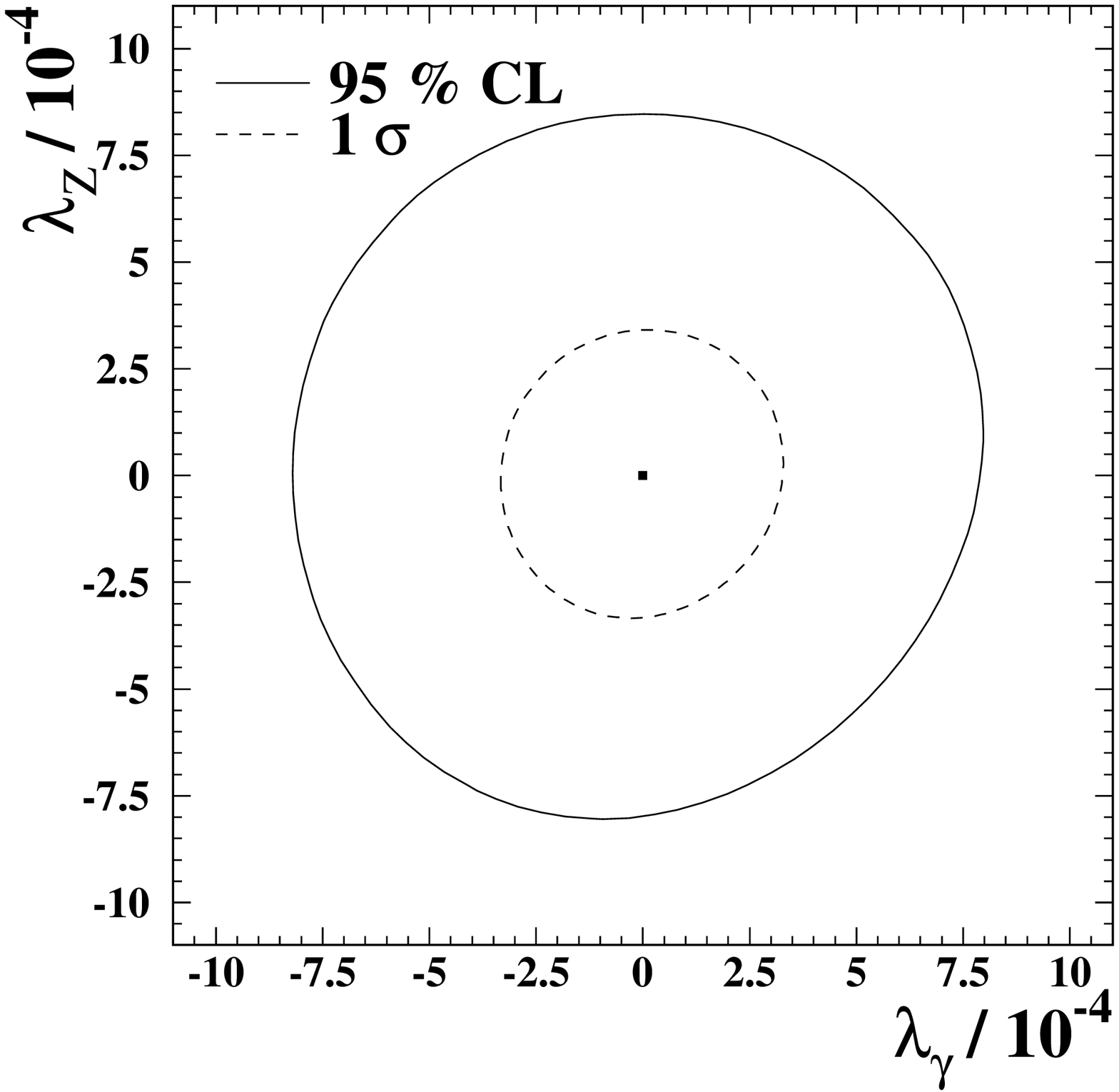}
\end{tabular}
\end{center}
\caption{$1 \sigma$ and $95\%$ CL (2D) contours for TGC's in the
5-parameter fit of $e^+e^-\to W^+W^-$
  ($\sqrt{s}=800 \gev, \, {\cal L} = 1000\,{\rm fb}^{-1},\, 
  {\cal P}_{e^-}=0.8,\,{\cal P}_{e^+}=0.6$). 
The figure is taken from Ref.~\cite{sec8_Aguilar-Saavedra:2001rg}. 
}
\label{sec82:fig:LC} 
\end{figure}

The experimental sensitivity to anomalous TGC's arises from the growth
of the non-standard contributions to the helicity amplitudes with
energy, and from modifications to angular distributions. In hadronic
collisions, one usually exploits the former by measuring the transverse
momentum distribution of the photon or $Z$ boson~\cite{Haywood:1999qg}. The
energy dependence is most pronounced for $\lambda_V$ and $\De g_1^Z$. 
Figure~\ref{sec82:fig:LHC} shows the expected 95\% confidence level
constraints on TGC's in ATLAS, resulting from two-parameter
fits~\cite{sec8_Gianotti:2002xx}. 
There are strong correlations between $\De g_1^Z$ and the other $WWZ$
couplings, $\De\kappa_Z$ and $\lambda_Z$. At the LHC, the limits for
$\lambda_V$ and $\De g_1^Z$ 
are of ${\cal O}(10^{-3})$, while the precision of $\De\kappa_V$ 
is ${\cal O}(10^{-2})$. An increase in the integrated luminosity by a
factor~10 improves the limits 
by a factor~1.3 to~2. A similar improvement is observed if the machine
energy is doubled. Doubling the energy and increasing the luminosity
yields bounds which are up to a factor~7 more stringent than those
obtained for $\sqrt{s}=14 \tev$ and 100~fb$^{-1}$.

At the LC, the most stringent bounds are obtained
from angular distributions in $W$~pair production. Sensitivity bounds
for $\sqrt{s}=800 \gev$ and an integrated luminosity of 1~ab$^{-1}$ 
are shown in Fig.~\ref{sec82:fig:LC}~\cite{sec8_Aguilar-Saavedra:2001rg}. 
For these bounds electron and positron polarization has been assumed,
${\cal P}_{e^-}=0.8,\,{\cal P}_{e^+}=0.6$. If positron polarizaation
would not be available, the bounds would be weakened by about a factor~2.
For $e^+e^-$ collisions at $\sqrt{s}=500 \gev$, but both beams
polarized, one finds bounds which are a factor~1.2 to~2 less stringent
than those shown in Fig.~\ref{sec82:fig:LC}.
All bounds, with the exception of those for $\De g_1^Z$, 
are of ${\cal O}(10^{-4})$ and thus are significantly better than
those obtained at the LHC, even with an energy and/or luminosity upgrade. 
The results for the anticipated precision of all TGC's for LEP, 
the Tevatron assuming 1~fb$^{-1}$ for each
experiment~\cite{Thurman-Keup:ka}, 
the LHC assuming 30~fb$^{-1}$ for each experiment~\cite{dobbs}
and for the LC assuming either 500~fb$^{-1}$ at $\sqrt{s} = 500 \gev$ 
or 1000~fb$^{-1}$ at $\sqrt{s} = 800 \gev$~\cite{sec8_Aguilar-Saavedra:2001rg}
is summarized in Fig.~\ref{sec82:fig:summary}.
Both, LEP and Tevatron analyses employ the relations 
$\Delta \kappa_\gamma = -\cot^2\theta_W (\Delta \kappa_Z - \Delta g_1^Z)$
and $\lambda_\gamma = \lambda_Z$ (with $\cos\theta_W = \MW/\MZ$).

\begin{figure}[htb!]
\begin{center}
\includegraphics[height=6cm]{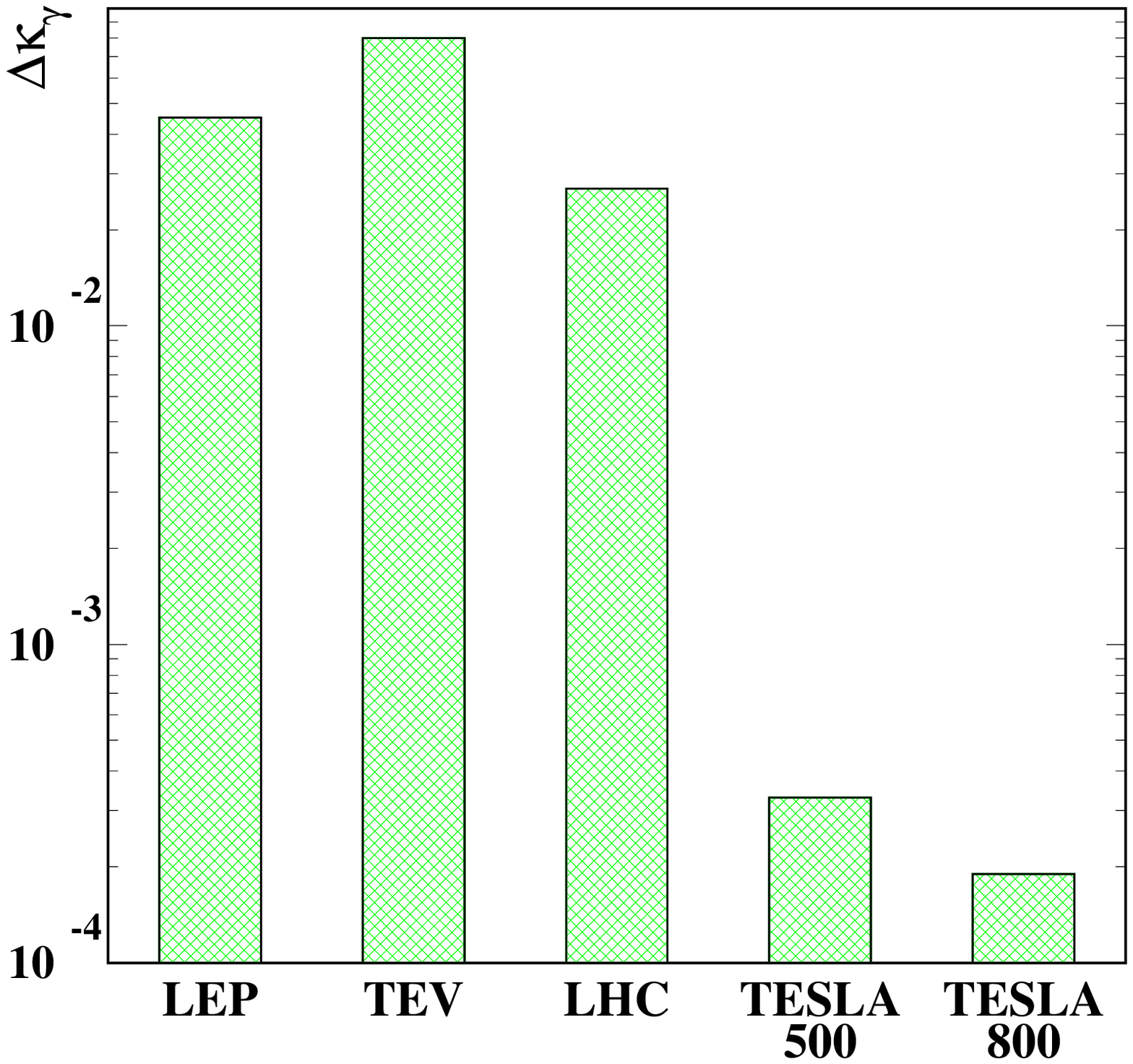}
\includegraphics[height=6cm]{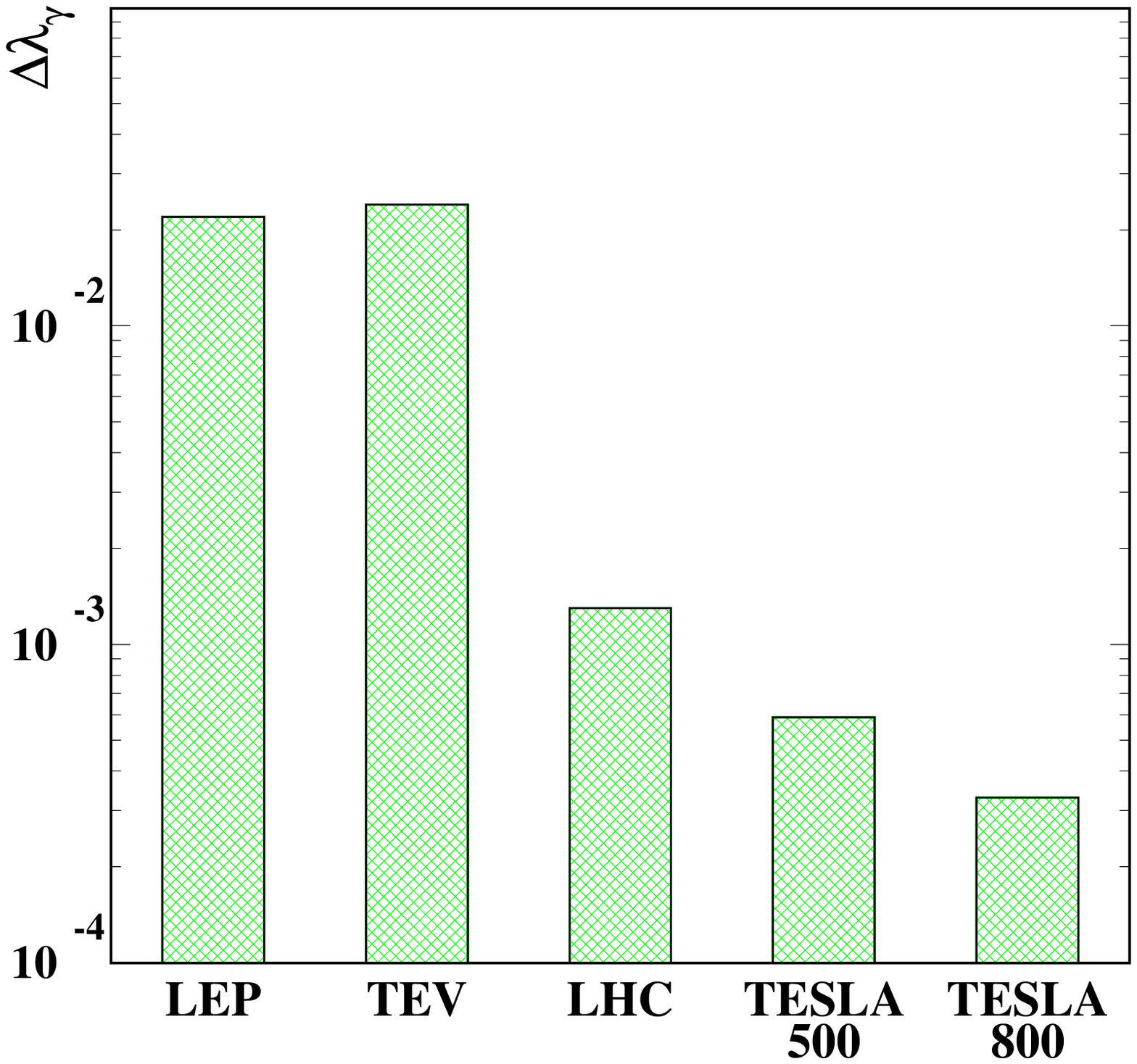}\\[-1cm]
\includegraphics[height=6cm]{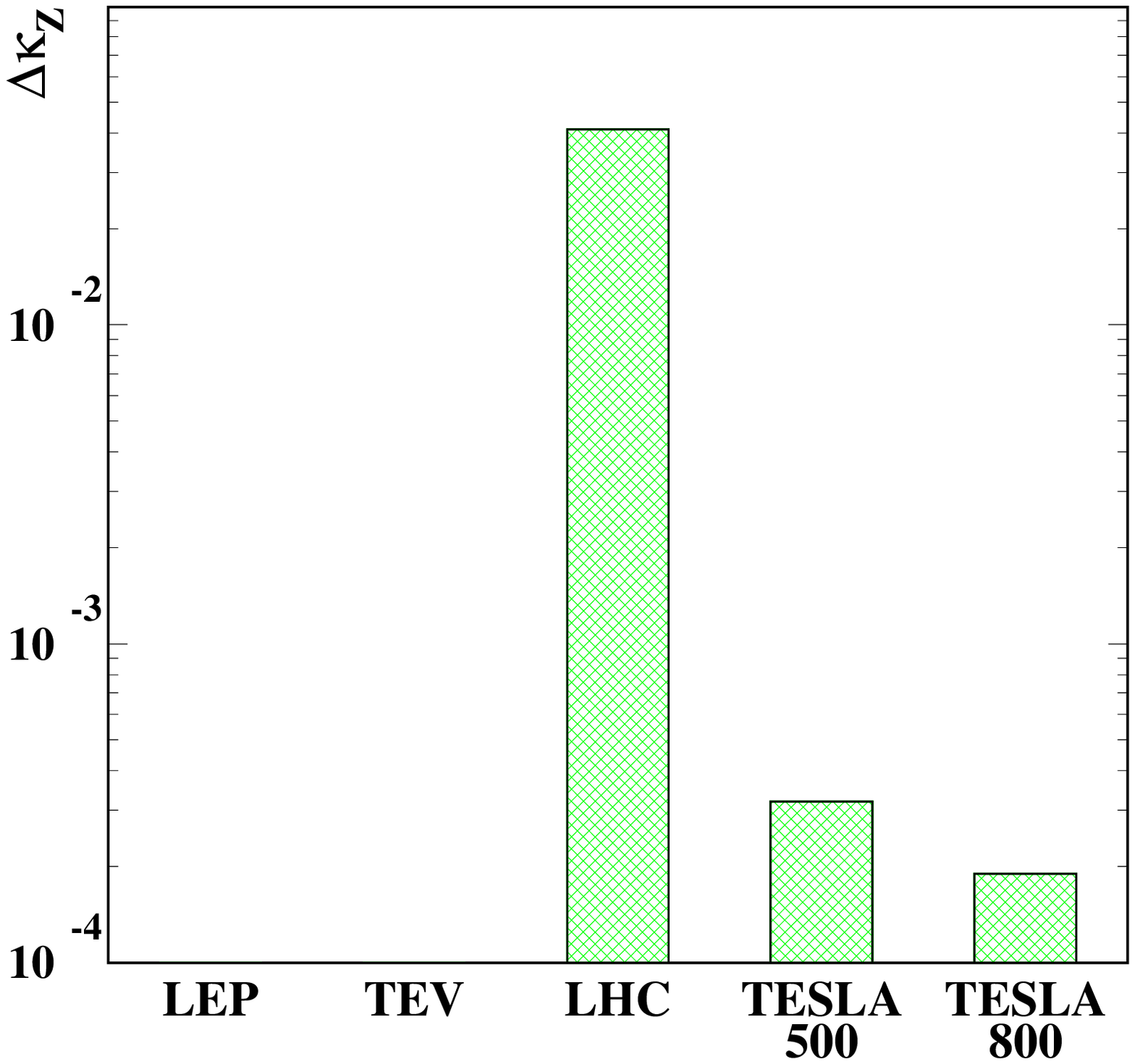}
\includegraphics[height=6cm]{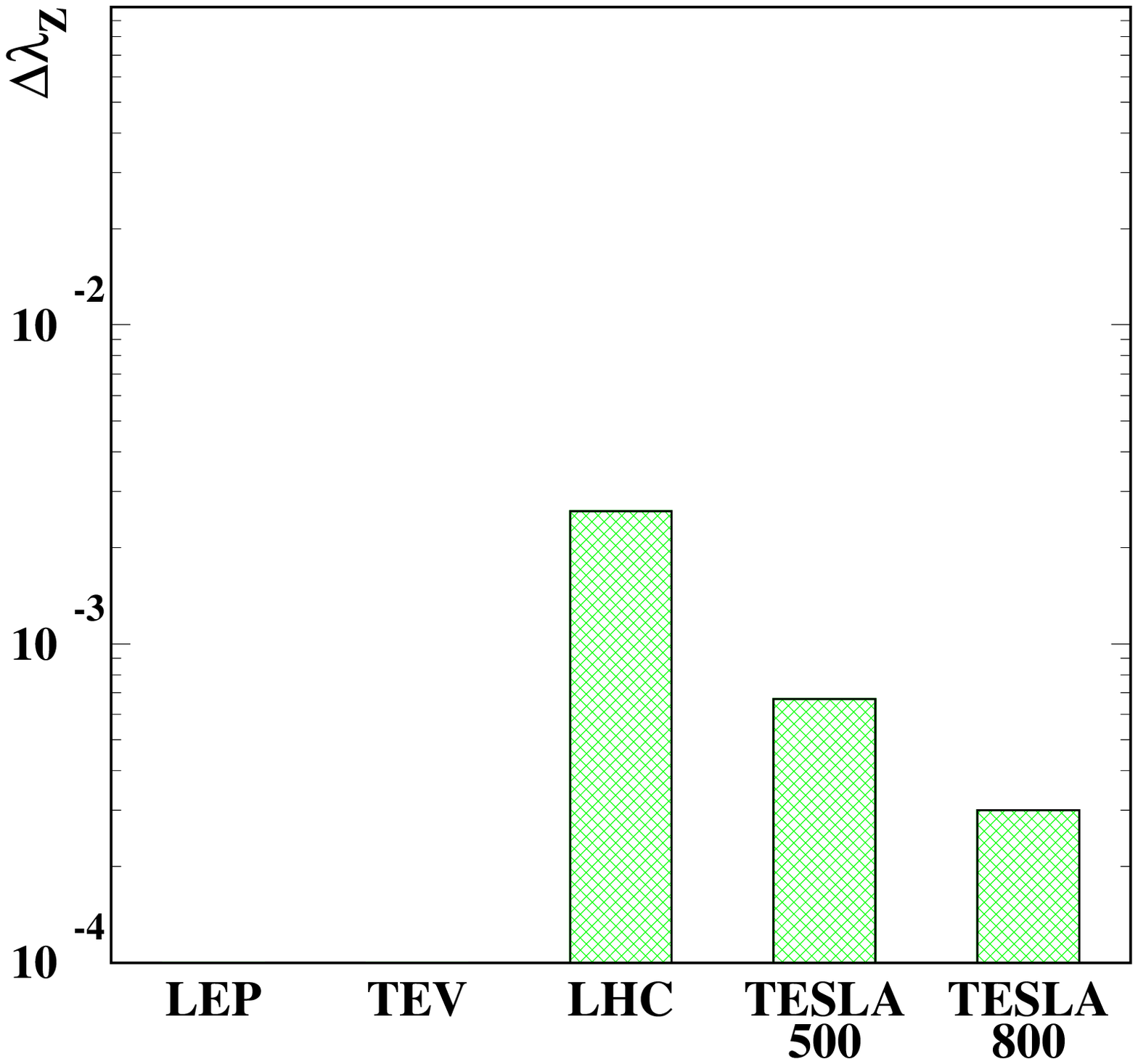}\\[-1cm]
\includegraphics[height=6cm]{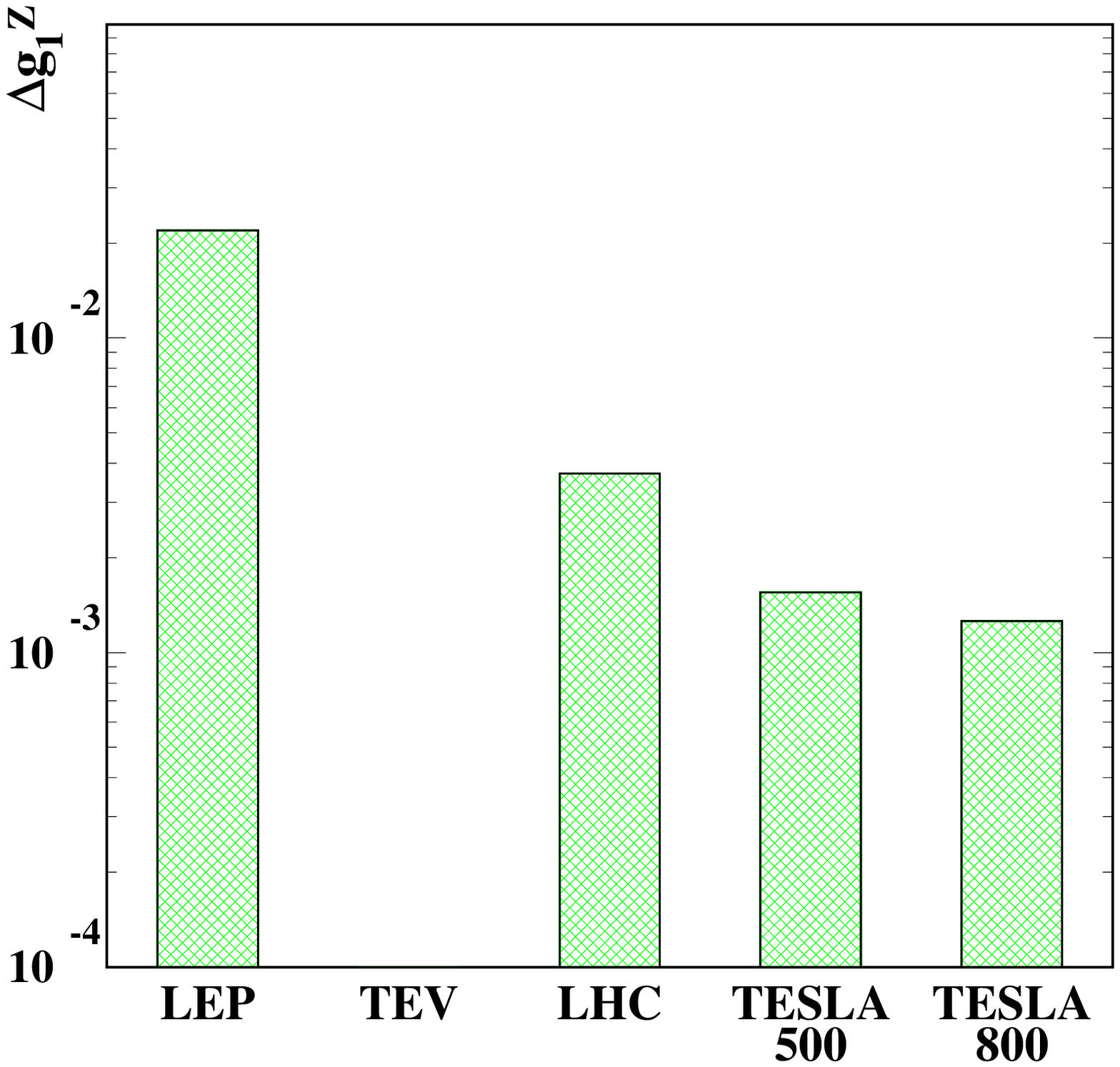}
\end{center}
\caption{Anticipated precisions for the TGC's at the
Tevatron~\cite{Thurman-Keup:ka} 
(assuming 1~fb$^{-1}$ for each experiment), the LHC~\cite{dobbs}
(assuming 30~fb$^{-1}$ for each experiment) and a
LC~\cite{sec8_Aguilar-Saavedra:2001rg} (assuming either
500~fb$^{-1}$ at $\sqrt{s} = 500 \gev$ or 1000~fb$^{-1}$ at 
$\sqrt{s} = 800 \gev$). Both, LEP and Tevatron analyses employ the relations 
$\Delta \kappa_\gamma = -\cot^2\theta_W (\Delta \kappa_Z - \Delta g_1^Z)$
and $\lambda_\gamma = \lambda_Z$.
}
\label{sec82:fig:summary} 
\end{figure}

Only for $\De g_1^Z$ would a combined measurement from LHC and LC
experiments result in a noticeable improvement of our knowledge of gauge
boson self-interactions: 
from Figs.~\ref{sec82:fig:LHC} and~\ref{sec82:fig:LC} one observes that 
$\De g_1^Z$ can be measured with similar precision independently at
the LHC and a LC. Combining LHC and LC limits for $\De g_1^Z$ may thus
result in significantly improved constraints by up to a factor of two
for this parameter. 

\smallskip
Electroweak radiative corrections within the SM to the TGC's are of order
$\sim 10^{-3}$ \cite{Argyres:1992vv} and thus
significantly larger than the expected sensitivities at the
LC. Similar  corrections can be expected in the MSSM. No
detailed study exists yet, but 
it can be expected that LHC/LC analyses similar to the one presented in
Sect~\ref{sec82:subsec:MSSMhiggs} are possible.

\smallskip
Although (with the above exception) 
TGC's can in general be determined much more precisely at a
LC, a positive interplay with the LHC may arise in specific models. For
example, in models where the electroweak symmetry is broken dynamically, 
TGC's may be related to the masses of heavy vector resonances which appear
in these models. These resonances have masses which typically are of
${\cal O}(1-2 \tev)$. 
A precise measurement of the TGC's at a 500--800~GeV LC may thus
allow for an indirect determination of the masses of such resonances, 
and provide specific information for a search for such particles at
the LHC (see Sect.~\ref{chapter:strongewsymmbreak} for more details).

One can also probe the self-couplings of the neutral gauge bosons
($ZZ\gamma$,> $Z\gamma\gamma$ and $ZZZ$ couplings) at the
LHC~\cite{babe,br} and LC~\cite{ntgclc}. In the SM, at tree level, these
couplings all vanish. The sensitivity limits which can be achieved for
these couplings at the LHC strongly depend on the form factor scale
assumed. For a form factor scale in the range of $2-3$~TeV, the bounds
which can be achieved at the LHC and LC are similar. They are about one
order of magnitude larger than what the SM predicts at the one-loop
level~\cite{babe}.


\subsection{Conclusions}

In this section the electroweak precision measurements (in the context
of supersymmetry) and triple gauge boson couplings (TGC) have been analyzed. 
Other applications of electroweak precision observables can be found in
Sects.~\ref{subsub-zpgigaz},~\ref{sec:27},~\ref{chapter:strongewsymmbreak}. 

For the electroweak precision measurements within Supersymmetry three
example studies have been performed. Often SUSY scenarios possess
scales that cannot be detected at the LHC and LC. The three examples
have shown how in this case electroweak precision observables
(e.g.\ $\MW$, $\sweff$ and $\mt$), measured most accuratly at the LC,
can help to determine these heavy scales. The interplay of observables
measured at the LHC and at the LC, together with the precision
observables can be crucial.

The triple gauge couplings have been analyzed in view of a possible
combination of the  results of the LHC and the LC. In the case of 
$\De g_1^Z$ both colliders can obtain independently about the same
uncertainty. Thus the combination of the two experiments can result in
an higher precision by up to a factor of~2.

Concerning the physics of contact interactions the LHC/LC interplay
could help to disentangle the flavour structure of the new
interactions in case a signal is seen.





\section{QCD studies}


\def\sgg{sigma_{\gamma^*\gamma^*}} 
\def\gaga{\gamma^*\gamma^*}
\def\delas{\Delta\alpha_S(M_Z^2)}


\vspace{1em}
Both the Large Hadron Collider and the Linear Collider are ideal machines
for testing Quantum Chromodynamics, and in particular for making precision measurements
related to the strong interaction. As has already been clearly demonstrated by
past and present machines (for example LEP and
 the CERN and Tevatron $p \bar p$ colliders), the two colliders are complementary
in what they can measure and with what precision.
The benchmark QCD measurements for each machine are:
\bigskip

\noindent {\it Linear Collider} \cite{sec83_TESLATDR}
\begin{itemize}
\item precision determination of $\alpha_S$ through event shape observables, multijet final
states, etc.
\item $Q^2$ evolution of $\alpha_S$ and  \lq\lq higher-twist'' (i.e. $Q^{-n}$) contributions
to hadronic observables
\item QCD effects in $t \bar t$ production, both at threshold and in the continuum
\item fragmentation functions
\item $\gamma\gamma$ physics
\begin{itemize}
\item unpolarised and polarised photon structure functions
 $F_2^\gamma(x,Q^2)$ ( $g_1^\gamma(x,Q^2)$) at low $x$ or high $Q^2$
\item $\gamma\gamma$ and $\gamma^*\gamma^*$ total cross sections as a test of 
non-perturbative and perturbative QCD (BFKL dynamics) respectively
\end{itemize}
\end{itemize}
\bigskip

\noindent{\it LHC}\cite{LHCSMReport}
\begin{itemize}
\item production of high $E_T$ jets, to determine $\alpha_S$ and pin down parton distribution 
functions (pdfs) and to calibrate backgrounds to new-physics processes
\item $W$ and $Z$ total cross sections, as a test of precision NNLO QCD predictions
\item $W$ and $Z$ transverse momentum distributions, as a test of Sudakov resummation
\item heavy flavour ($c$, $b$ and $t$) cross sections, as a test of QCD dynamics, for example
threshold resummations, production mechanisms for quarkonia, etc.
\item prompt photon and diphoton production, to measure the gluon distribution and to calibrate backgrounds
to Higgs production
\item forward jet, $W+\;$jet, heavy quark, .. production as a test of BFKL dynamics
\item diffractive processes, both \lq hard' and \lq soft' 
\end{itemize}

In terms of the {\it complementarity} of the QCD measurements at the two machines, 
one can note the following:
\begin{itemize}
\item There is a strong correlation between the gluon distribution and the strong
coupling $\alpha_S$ in the theoretical prediction for the large $E_T$ jet cross section 
at hadron colliders. A precision measurement of $\alpha_S$ at high $Q^2$ at the LC can be 
fed back into the LHC analysis to extract the gluon pdf from large $E_T$ jet and prompt
photon data with greater precision, see Section~\ref{alphasmeasurements} below.

\item Among the backgrounds to Higgs $\to \gamma\gamma$ production at the LHC, one
of the most important arises from the fragmentation of a quark or gluon into
a photon, $q, g \to \gamma + X$. A precise knowledge of these fragmentation functions
is required to control the SM background. The LC can provide such information, particularly
in the large $z$ region. This is discussed in more detail in Section~\ref{Binoth} below.

\item The tests of (small-$x$) BFKL dynamics at the two machines centre on the exchange
of a \lq perturbative pomeron' between two softly scattered hadronic systems, but
utilise different initial and final states ($\gamma^*\gamma^*\to q \bar q q \bar q$
for the LC and $gg \to gg, b \bar b b \bar b$, ... for the LHC) and therefore
have different systematics. The phenomenology of such processes is not straightforward
at either machine, and therefore information from both will probably be necessary
to provide firm evidence for BFKL effects. This is discussed in more detail 
in Section~\ref{Orr} below.

\item If supplemented with forward detectors to measure the 
scattered protons, the LHC could also study
$\gamma\gamma$ interactions. Centre-of-mass energy 
values, $\sqrt{s_{\gamma\gamma}}$, above 1 TeV could be reached.
If total $\gamma\gamma$ cross sections can be measured at these high energies, 
with sufficiently small systematic errors, these
measurements can be used to estimate more
accurately the expected background at a LC, 
particularly for a multi-TeV collider such as 
CLIC. Conversely, the measurement of
 total $\gamma\gamma$ cross sections at a LC,  in particular at
its derived photon collider, with much better precision than at the LHC 
will give improved understanding of the LHC results in 
the  $\sqrt{s_{\gamma\gamma}}$
region of overlap,  i.e. below 500~GeV.
\end{itemize}

\subsection{Measurements of $\alpha_S$}\label{alphasmeasurements}

{\it A.~De~Roeck and W.J.~Stirling}

\vspace{1em}
The strong coupling $\alpha_S$ can  in principle be measured at hadron colliders through processes involving jets,
either from `pure QCD' multijet final states or from processes like $W +$~jet production.
Difficulties arise, however, from the need for an accurate measurement of the jet (i.e. parton) energies,
which requires detailed knowledge of calorimeter response, fragmentation effects, the underlying event, etc.

A further problem is that $\alpha_S$ is inevitably correlated with the 
parton distribution functions. For example,  in the combination
$\alpha_S f(x,Q^2)$, where $f=q,g$,   at high $x$  increasing the value of  $\alpha_S$ will, via the DGLAP
evolution equations, cause a
decrease in $f$, thus compensating the effect due to the explicit factor of the coupling. In fact for this reason
 the $W+$ jet cross section at the Tevatron
is relatively insensitive to the value of the coupling constant, see for example \cite{Abachi:1996gy}.

The inclusive high $E_T$ jet cross section offers high statistics and more
sensitivity to the coupling. For example, a recent measurement by the CDF collaboration
at the Tevatron $p \bar p$ collider \cite{Affolder:2001hn} gives
\begin{equation}
\alpha_S(M_Z^2)_{\rm NLO} = 0.1178 \pm 0.0001\; \mbox{(stat)}\; {+0.0081 \atop -0.0095}
\; \mbox{(expt. syst.)}
\end{equation}
There are many contributions to the experimental systematic error, the largest being due
to the calorimeter response to jets.

The above result does not include any additional theoretical error from factorisation and 
renormalisation scale dependence --- the standard variation of these scales
is estimated to give rise to an additional (theoretical) error of approximately $\pm 0.005$ \cite{Giele:1995kb}.
There is also a
systematic error from the dependence on the input parton distribution functions.\footnote{A study of the 
impact of the pdf uncertainty on the production cross sections for Drell-Yan lepton pairs and Higgs bosons
at the LHC has been performed by Bourilkov as part of this workshop, see 
Ref.~\cite{Bourilkov:2003kk}.} Because of the
strong correlation between these and $\alpha_S$, it  is difficult to estimate this accurately
in isolation from the global pdf fit. Note that Tevatron large $E_T$ jet data are included
in both the latest MRST \cite{MRST2002} and CTEQ \cite{CTEQ6} global analyses, and have some influence on the
$\alpha_S$ value that is a by-product of such studies.\footnote{For example, the most recent
MRST value is $\alpha_S(M_Z^2) = 0.1165 \pm 0.002\; \mbox{(expt.)}\; \pm 0.003
\; \mbox{(theory)}$.} 
Combining errors in quadrature would give $\alpha_S(M_Z^2) = 0.118 \pm 0.011$ from the CDF
jet data, to be compared to
the world average (PDG, 2002) value of  $\alpha_S(M_Z^2) = 0.1172 \pm 0.002$ \cite{PDG2002}.

When considering how such a measurement would extrapolate to the LHC, one can assume that the
NNLO corrections will be known (for a recent review of progress in this area, see for example
\cite{Glover:2002gz}), and therefore that the theoretical systematic error will be much reduced.
The statistical error will also be small, and 
the systematic errors associated with the jet measurement
may improve by a factor two or so, using the channel $W\rightarrow$ jets, in 
tagged $t\overline{t}$ events.
 However, the problem with the $\alpha_S$ -- pdf correlation will
persist, indeed the jet cross section data will be used to pin down the gluon distribution
at medium and high $x$. Note that even though the measurements of $\alpha_S$ will be  made
at a very high scale $Q \sim E_T^{\rm jet}\sim {\cal O}({\rm TeV})$, this does not help improve the precision on 
the derived value of $\alpha_S(M_Z^2)$.

At the $e^+e^-$ colliders LEP and SLC, $\alpha_S$ was measured in a number of ways, including
 (i) from the $Z^0$ hadronic width
via a global precision electroweak fit  ($\alpha_S(M_Z^2) = 0.1200 \pm 0.0028\;  \pm 0.002
\; \mbox{(scale)}$, \cite{PDG2002}) (LEP1 and SLC)  and (ii) from various hadronic final state
jet and event shapes measures (Thrust, $C$ parameter, $R_3$  etc, ...) at LEP1, LEP2 and SLC. These
latter measurements are 
all based on the basic ${\cal O}(\alpha_S)$ $e^+e^- \to q \bar q g$ process, with NLO
${\cal O}(\alpha_S^2)$ perturbative QCD corrections included. An average value based on  many such measurements
from all the collaborations at LEP and SLC is $\alpha_S(M_Z^2) = 0.122 \pm 0.007$, where the 
error is totally dominated by the theoretical uncertainties associated with the choice of scale,
and the effect of hadronisation on the different quantities fitted \cite{PDG2002}.

Extrapolation of the hadronic final state $\alpha_S$ measurement to a LC is reasonably straightforward, assuming
of course similar or even improved detector capabilities.\footnote{The results for the estimated 
errors on $\alpha_S$ at a LC in this section are taken from 
\cite{sec83_TESLATDR}.} 

With luminosities  of order $10^{34}\; {\rm cm}^{-2}{\rm s}^{-1}$ or greater, and centre-of-mass
energies $\sqrt{s} = 500 - 1000$~TeV, a LC should produce enough $e^+e^- \to q \bar q (g)$
events to achieve a statistical error of order $\delas \simeq 0.001$ or better. Assuming hermetic detectors
with good calorimetry and tracking, the systematic uncertainties from detector effects should also
be controllable to this level of accuracy. The uncertainties associated with hadronisation
depend quite sensitively on the observable, but in every case are expected to decrease
with increasing energy at least as fast  as $1/Q$. An order of magnitude improvement can therefore
be expected at a LC compared with LEP1/SLC, and once again the associated uncertainty
should be well below the $\delas = 0.001$ level. 
This leaves the theoretical systematic error from unknown higher-order corrections
as the dominant source of uncertainty on $\alpha_S$, as at LEP/SLC. However, it will certainly
be the case that the NNLO corrections to all the relevant jet rates and event shape variables
will be known by the time a LC comes into operation \cite{Glover:2002gz}. The current error
on $\alpha_S$ from this source, $\delas \simeq 0.006$, is therefore likely to decrease 
significantly, and could well be at the $0.001$ level. Putting everything 
together, a reasonable expectation is 
\begin{equation}
\delas \;\vert_{\rm LC} = \; \pm 0.001\; \mbox{(expt.)} \; \pm 0.001\; \mbox{(theory)}
\end{equation}
which is (a) comparable with the current world average error, and (b) much smaller than
the likely error from any measurement at the LHC. 

Of course by running a linear collider in a GigaZ mode at the $Z$ resonance, precision $\alpha_S$
measurements can be made using the same techniques as at LEP1 and SLC. For example, the ratio
of leptonic to hadronic $Z$ decay widths could be measured with a precision of order $\delta R_l / R_l 
= \pm 0.05\%$, or maybe even slightly better, which provides a very clean determination of the strong coupling (via the perturbative
QCD corrections $[1 + \alpha_S/\pi + ...]$  to the hadronic decay widths) with an estimated
error of $\delas \approx \pm 0.001$. For a more complete discussion see
Refs.~\cite{Winter:2001av,sec8_Erler:2000jg}.

In terms of the synergy between the two machines, therefore, precision measurements
of $\alpha_S$ at a LC, made at comparable $Q$ scales,  can be input into QCD analyses 
at the LHC, thus improving the precision on the pdf measurements for example.
To see the possible quantitative effect of this, consider the contribution
to the theoretical systematic error on the Higgs ($gg \to H$) total cross section at
LHC from the uncertainty on the input gluon distribution. From Fig.~\ref{fig:2dplotlhc}, taken from 
Ref.~\cite{MRST2002}, we see that this 
is currently $\pm 3\%$ when $\alpha_S$ is allowed to vary in the fit, but reduces
to  $\pm 2\%$ when  $\alpha_S$ is fixed, for example by a high-precision 
external measurement such as could be made at the LC.

\begin{figure}[ht]
\centerline{\psfig{file=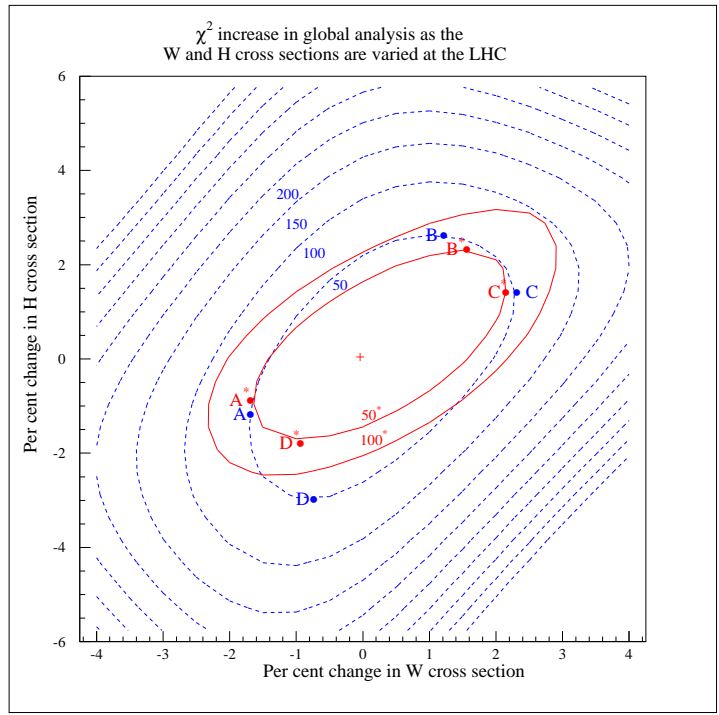,height=12cm}}
\vspace*{-2cm}
 \caption{ Contours with $\Delta\chi^2=50,100\dots$ obtained by performing
global pdf fits with the values of $\sigma_W$ and $\sigma_H$, at the LHC energy,
fixed in the neighbourhood of
their values predicted by the unconstrained MRST2001 fit. The
$\Delta\chi^2=50$ contour is taken to represent the errors on $\sigma_W$
and $\sigma_H$ (arising from the experimental errors on the data used in the
global pdf fit).  The dashed contours are obtained if $\alpha_S(M_Z^2)$ is
allowed to vary. The superimposed solid $\Delta\chi^2=50,100$ contours
are obtained if $\alpha_S(M_Z^2)$ is fixed at 0.119.}
\label{fig:2dplotlhc}
\end{figure}

\subsection{BFKL physics}\label{Orr}

{\it A.~De~Roeck, L.H.~Orr and W.J.~Stirling}

\vspace{1em}
\subsubsection{Introduction }

Many processes in QCD can be described by a fixed order expansion in the 
strong coupling constant $\alpha_S$.  In some kinematic regimes, however,  each
power of  $\alpha_S$ gets multiplied by a large logarithm (of some ratio of relevant
scales),  and fixed-order calculations must give way to leading-log
calculations in which such terms are resummed.  The BFKL equation \cite{bfkl}
resums these large logarithms when they arise from multiple (real and virtual)
gluon emissions for scattering processes in the so called high-energy limit,
for example $gg \to gg$ for $s \to \infty,\ |t| $ fixed.  In the BFKL regime, the transverse momenta of the
contributing gluons are  comparable but they are strongly ordered in rapidity.

The leading-order BFKL equation can be solved analytically, and its solutions usually result in
(parton-level) cross sections that increase as the power $\lambda$, where 
$\lambda = 4C_A\ln 2\, \alpha_s/\pi \approx 0.5$.\footnote{$\lambda$ is also
known as $\alpha_P-1$.}
For example, in dijet production at large rapidity separation $\Delta$ in hadron 
colliders \cite{muenav},
BFKL predicts for the ($qq$, $qg$ or $gg$) parton-level cross section 
$\hat\sigma \sim  e^{\lambda\Delta}$. In virtual photon - virtual photon scattering into hadrons, as
measured for example in $e^+e^-$ collisions, BFKL becomes relevant when
the electron and positron emerge with a small
scattering angle and hadrons are produced centrally \cite{Bartels:1996ke,brodskyetal}. In that case
 BFKL predicts  $\sigma_{\gamma^*\gamma^*} \sim (W^2/Q^2)^\lambda$, where  $W^2$ is the
invariant mass  of the  hadronic system (equivalently, the photon-photon
centre-of-mass energy) and $Q^2$  is the invariant mass of either photon.

\subsubsection{Experimental status and improved predictions}
The experimental status of BFKL is ambiguous at best, with existing results 
being far from definitive.  The data tend to lie between the predictions of
fixed-order QCD and analytic solutions to the BFKL equation.   This happens,
for example, for the azimuthal decorrelation in dijet production at the Fermilab
Tevatron \cite{dzeroazi} and for the virtual photon cross section at
LEP \cite{maneesh}. Similar results are found for forward jet production
in deep inelastic $ep$ collisions at
HERA \cite{Adloff:1998fa,Breitweg:1998ed}.

It is not so surprising that analytic BFKL predicts stronger effects 
than seen in data.  Analytic BFKL solutions implicitly contain sums over 
arbitrary numbers of gluons with arbitrary energies, but the kinematics are 
leading-order only.  As a result there is no kinematic cost to emit gluons, and 
energy and momentum are not conserved, and BFKL effects are
artificially enhanced.

This situation can be remedied by a Monte Carlo implementation of solutions to
the BFKL equation \cite{os,schmidt}.  In such an implementation the BFKL
equation is solved by iteration, making  the sum over gluons explicit.  Then
kinematic constraints can be implemented  directly, and conservation of energy
and momentum is restored.  This tends to lead to  a suppression of BFKL-type
effects.  The Monte Carlo approach has been applied to dijet
production at hadron colliders \cite{os,schmidt,osmore,Andersen:2001kt} leading to 
better (though still not perfect) 
agreement with the dijet azimuthal
decorrelation data at the Tevatron \cite{os}. Predictions  
for the azimuthal angle decorrelation in
 dijet production at the Tevatron and the 
LHC, as a function of the dijet rapidity difference  $\Delta y$,
are shown in Fig.~\ref{fig83:dijets}~\cite{osmore}.
The normalisation is such that $\langle \cos\Delta\phi\rangle = 1$
corresponds to jets which are back-to-back in the transverse plane,
while completely uncorrelated jets have  $\langle \cos\Delta\phi\rangle = 0$.
 Kinematic suppression of gluon emission accounts for a substantial
part of the difference between analytic (asympototic) and Monte Carlo BFKL predictions.
This kinematic suppression is however not as dramatic at the LHC, because
of its relatively higher centre-of-mass energy compared
to the jet transverse momentum threshold. As a result, improved BFKL MC
predictions tend to retain more BFKL-type behaviour because of the greater
phase space for emitting gluons \cite{osmore}.
Studies of dijet production in BFKL physics therefore look promising 
at the LHC, and such studies
will benefit from jet detection capabilities that extend far into the
forward region and allow for $p_T$ thresholds as low as possible.
\begin{figure}
\begin{center}
\mbox{\epsfig{figure=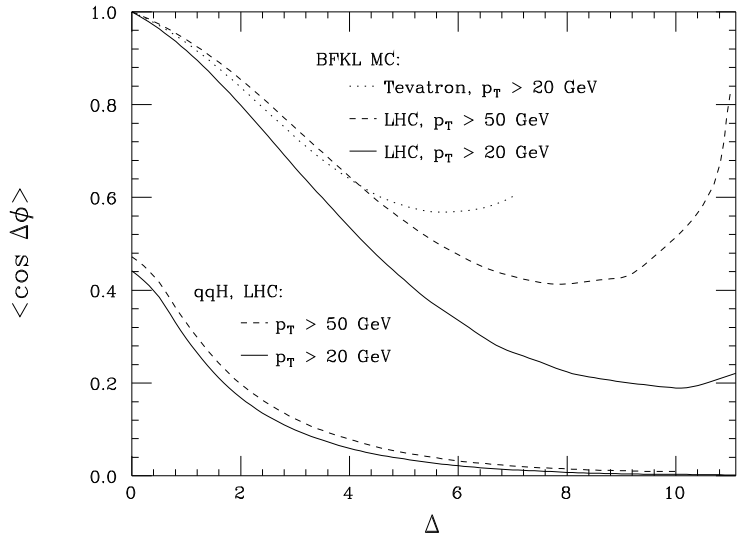,width=12cm}}
\end{center}
\caption{The azimuthal angle decorrelation in dijet production at the Tevatron 
($\sqrt{s}=1.8$~GeV) and LHC ($\sqrt{s}=14$~TeV)
as a function of dijet rapidity difference $\Delta = \Delta y$.  
The upper curves are computed 
using the improved BFKL MC with running $\alpha_s$;
they are: (i) Tevatron, $p_T>20$~GeV (dotted curve),
(ii) LHC, $p_T>20$~GeV (solid curve), and (iii) LHC, $p_T>50$~GeV
(dashed curve).  The lower curves are for dijet production in the process
$qq\to qqH$ for $p_T>20$~GeV (solid curve) and $p_T>50$~GeV
(dashed curve).}\label{fig83:dijets}
\end{figure}

Other similar studies have considered forward $W + $~jet production  at hadron colliders
\cite{Andersen:2001ja}, and  the associated jet production in dijet production 
\cite{Andersen:2003gs}.
Applications to forward jet production at
HERA  and to virtual photon scattering in $e^+e^-$ collisions 
are underway;
an update on the latter appears in the next section.

\subsubsection{$\gaga$ scattering at the Linear Collider:  a closer look}

BFKL effects can arise in $e^+e^-$ collisions via the scattering 
of virtual photons emitted from the initial $e^+$  and $e^-$.
The scattered electron and positron appear in the forward and backward
regions (\lq\lq double-tagged'' events)
with hadrons in between.  With total centre-of-mass energy $s$, 
photon virtuality $-Q^2$, and photon-photon invariant mass ($=$ invariant
mass of the final hadronic system) $W^2$, BFKL effects are expected 
in the kinematic regime where $W^2$
is large and $$s \gg Q^2 \gg \Lambda_{QCD}^2.$$  
At fixed order in QCD, the dominant process is four-quark production with
$t$-channel gluon exchange (each photon couples to a quark box; the quark
boxes are connected via the gluon).  The corresponding BFKL
contribution arises from diagrams with a gluon ladder attached to the 
exchanged $t$-channel gluon.

\begin{figure}[ht]
\centerline{\psfig{file=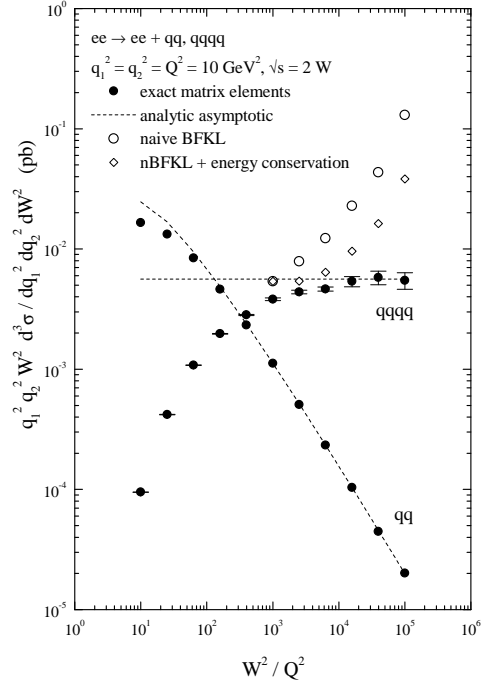,height=10cm}}
\caption{Exact (closed data points) and analytic asymptotic (dashed line) 
$e^+e^- \to e^+e^-q \bar q$  and 
$e^+e^- \to e^+e^-q \bar q q \bar q$  cross sections versus 
$W^2/Q^2$ at fixed $W^2/s = 1/4$.  Also shown:  analytic BFKL without (open 
circles) and with (open diamonds) energy conservation imposed.} 
\label{fig:compare}
\end{figure}

The relative contributions of fixed-order QCD and BFKL are most easily
understood by looking at  
\begin{equation}
W^2 Q_1^2 Q_2^2 \; \frac{d^3 \sigma}{d W^2 d Q_1^2 d Q_2^2 } 
\end{equation}
as a function of $W^2/Q^2$ for fixed $\sqrt{s}/W$.  The asymptotic regime then
corresponds to large $W^2/Q^2$.  This quantity is shown in Figure~\ref{fig:compare} for
$Q_1^2=Q_2^2=Q^2=10\ {\rm GeV}^2$ and $\sqrt{s}=2W$.  The solid points are
the QCD calculations of two-quark (\lq qq') and four-quark production (\lq qqqq'); we
see that  the latter dominates for large $W^2/Q^2$ and approaches a
constant asymptotic  value.  In contrast, the analytic BFKL result, shown with
open circles,  rises well above that of fixed-order QCD.  The diamonds show
analytic BFKL with energy conservation imposed, but not exact kinematics;
it can be interpreted as an upper limit for the Monte Carlo prediction,
the calculation of which is in progress.

It is important to note in Figure~\ref{fig:compare} that although BFKL makes a definite
leading-order prediction for the behavior of the cross section as  
a function of $W^2/Q^2$, the origin  in
$W^2/Q^2$  (i.e., where  BFKL 
meets asymptotic QCD) is  {\it not} determined in leading order.
  We have chosen $W^2/Q^2=10^3\ {\rm GeV}^2$ as a 
reasonable value where the QCD behavior is sufficiently asymptotic for 
BFKL to become relevant, but another choice might be just as reasonable.  
Only when higher order corrections are computed can the BFKL prediction
be considered unique.  
 
\begin{figure}[ht]
\centerline{\psfig{file=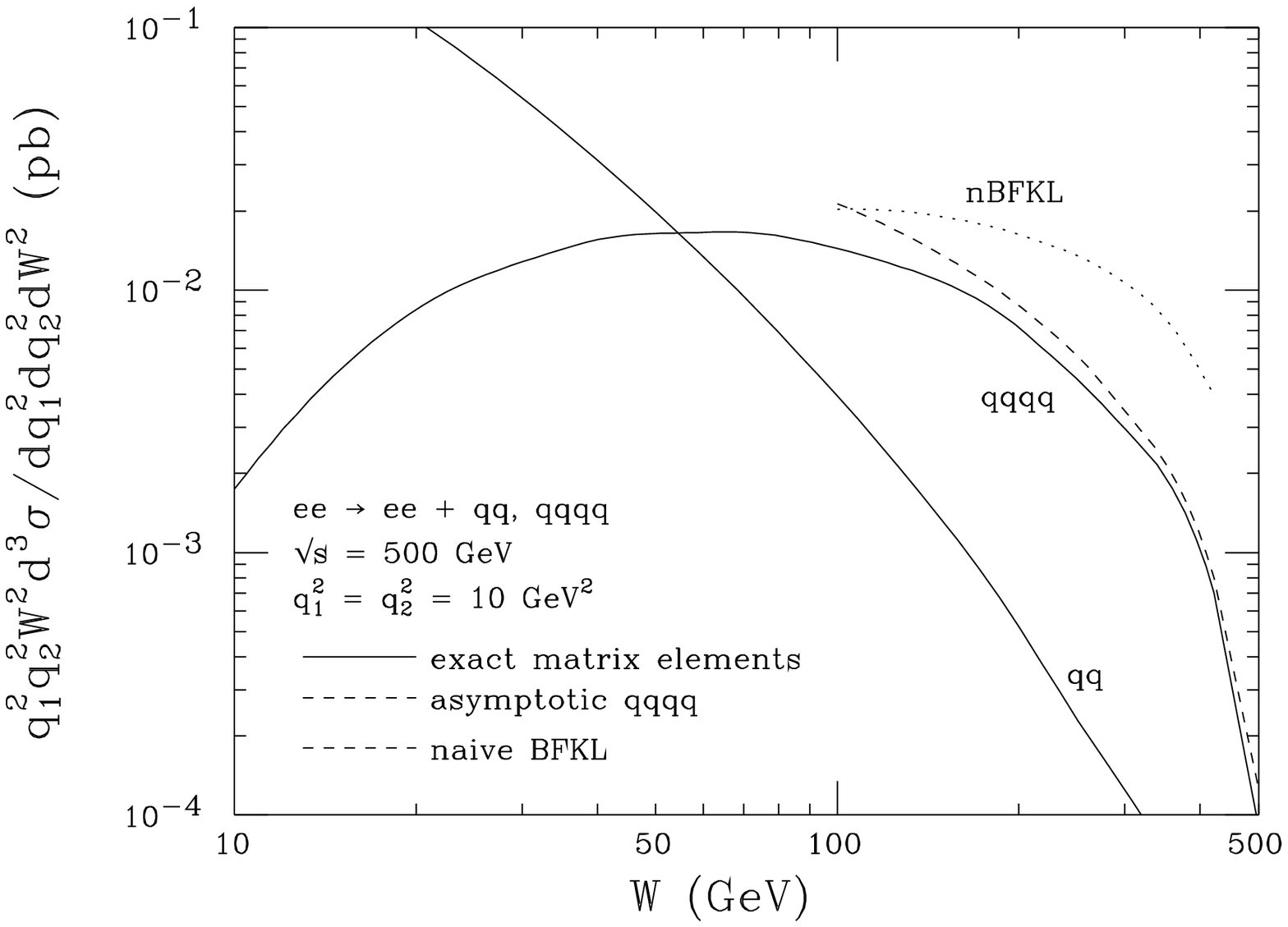,width=13cm}}
\caption{Exact (solid lines) and analytic asymptotic (dashed line) 
$e^+e^- \to e^+e^-q \bar q$  and 
$e^+e^- \to e^+e^-q \bar q q \bar q$  cross sections versus 
$W^2/Q^2$ at fixed $\sqrt{s}=500\ {\rm GeV}$.  Also shown:  analytic BFKL
(dotted line).}  \label{fig:lc}
\end{figure}

From an experimental point of view, the cross
section at fixed $\sqrt{s}$ is more directly relevant.  Figure~\ref{fig:lc} shows 
$W^2 Q_1^2 Q_2^2 \; \frac{d^3 \sigma }{ d W^2 d Q_1^2 d Q_2^2 }$
for a linear collider energy $\sqrt{s}=500\ {\rm GeV}$.  The solid lines
show the exact fixed-order QCD prediction.  The dashed line
is the asymptotic four-quark production cross section, and the 
dotted line is the analytic BFKL prediction. Now we see that all
of the curves fall off at large $W$, but the BFKL cross section lies well
above the others.

\subsubsection{Status of NLO Corrections}

It is apparent that, although it is not yet clear 
whether BFKL is necessary to describe the data in hand, leading-order analytic
BFKL is  not sufficient. In view of the experimental
precision that is likely to be achieved for the above processes at the LHC
and the LC, a next-to-leading order (strictly
speaking, next-to-leading log order) analysis will be required.  
The calculation of the NLO BFKL
kernel was completed several years ago \cite{Fadin:1998py,Ciafaloni:1998gs},  
but it is only very recently that
the formalism has been implemented \cite{Andersen:2003an} in a Monte Carlo 
framework appropriate for phenomenology at the LHC and the LC.
Although in principle the NLO contributions should affect the relevant hadron collider
and $e^+e^-$ collider cross section in a similar way, only  detailed numerical 
studies will show whether this is in fact true. The convergence of the perturbation
series is likely to depend sensitively on the kinematics of the intial and final states
that can be accessed experimentally.

\subsubsection{Other BFKL processes at a Linear Collider}

There are a number of related processes in $\gamma\gamma$ collisions
at high energy which also provide a test of BFKL dynamics. These include
$\gamma\gamma \to J/\psi J/\psi$ \cite{Kwiecinski:1999hg}, 
where the $J/\psi$ mass plays the role
of the $\gamma^*$ virtuality in $\sigma_{\gaga}$, and $\gamma\gamma \to \rho \rho$
at fixed momentum transfer $|t|$.

Inclusive polarised structure function 
measurements may show BFKL effects:
the most singular terms  in the  small $x$ resummation
on $g_1(x,Q^2)$
behave like $\alpha_s^n\ln^{2n}1/x$, compared to $\alpha_s^n\ln^{n}1/x$
in the unpolarised case. 
Thus large $\ln 1/x$ effects are expected to set in 
much more rapidly for polarised than for unpolarised structure measurements.
For
leading order calculations, including consistency constraints, 
the differences in predictions for $g_1$ 
with and without these large logarithms can be as large as a factor 
 2 to 3 for $x= 10^{-4}$ and could thus be easily measured with a few
years of data taking at a photon collider~\cite{Kwiecinski:2000yk}.

Further candidate measurements are
the processes $e^+e^- \rightarrow e^+e^-\gamma X$ and  
$\gamma\gamma \rightarrow \gamma X$~\cite{Evanson:1999kw} and 
a process similar to the 'forward jets' at HERA: $e\gamma$ 
scattering  with a forward jet produced
in the direction of the real photon~\cite{Contreras:np}.

\subsubsection{Conclusions}

In summary, BFKL physics is a complicated business.  Tests have so far been performed
in a variety of present experiments (Tevatron, HERA, LEP) and there is 
potential for the future as well (LHC, LC).  The advantage of these latter machines is
that they provided a better access to the asymptotic (high-energy) regime
where the formalism most naturally applies.
At present energies, however, comparisons between
theory and experiment are not straightforward; leading-order BFKL is apparently
insufficient, and subleading corrections such as kinematic constraints can be
very important.  The recent availability of the next-to-leading order corrections
to the BFKL kernel (and their implementation in a Monte Carlo framework suitable
for studying the processes described above) provides a way of gauging the convergence of the
perturbative approach at present and future colliders. Much detailed phenomenology
remains to be done.

\subsection{Improving the $H\to\gamma \gamma$ background prediction by 
using combined collider data}\label{Binoth}

{\it T.~Binoth}

\vspace{1em}
\subsubsection{Introduction}

Particle searches at future colliders rely on a precise
understanding of Standard Model backgrounds. In prominent 
cases like Higgs boson search in the two-photon channel at the LHC,
the background is dominated by fragmentation processes 
where photons and mesons are produced in the hadronisation of 
a jet. To reduce these backgrounds severe isolation cuts are 
imposed which lead to the situation that fragmentation models
are tested at large $z$, where $z$ is the ratio between the
transverse energy of the photon or meson and the jet.  
Fragmentation functions are up to now experimentally
well constrained  only for not too large $z$ ($\sim z<0.7$).
On the other hand predictions of the two photon 
background are very sensitive to the large $z$ region.
By measuring high $p_t$ mesons and photons at the Tevatron and
the LC, these uncertainties can be pinned down considerably.
The study is aimed to quantify the present uncertainties 
of fragmentation functions by comparing recent parametrizations. 

\subsubsection{Sensitivity of the $H\to \gamma\gamma$ background on fragmentation functions}
The LEP data favour a light Higgs boson above 114 GeV and below about 200 GeV \cite{LEPHiggs1}.
In the mass window between 80 and 140 GeV one of the most promising search channels
at the LHC is the Higgs decay into two photons, as the hadronic decays are swamped by 
huge QCD backgrounds. The corresponding background is constituted out of direct photons,
photons from fragmentation and neutral mesons, the latter being misidentified as a 
single photon in the detector.  

Recently a dedicated  calculation 
was completed which  treated all three contributions in 
next--to--leading order (NLO) in $\alpha_s$ \cite{Binoth:1999qq,Binoth:2002wa}. 
Comparison of NLO QCD  with existing di-photon \cite{Binoth:2000zt} 
and di-pion \cite{Binoth:ym} data  shows an excellent agreement  between theory
and experiment. 
\unitlength=1mm
\begin{figure}
\begin{picture}(160,75)
\put(0,0){\epsfig{file=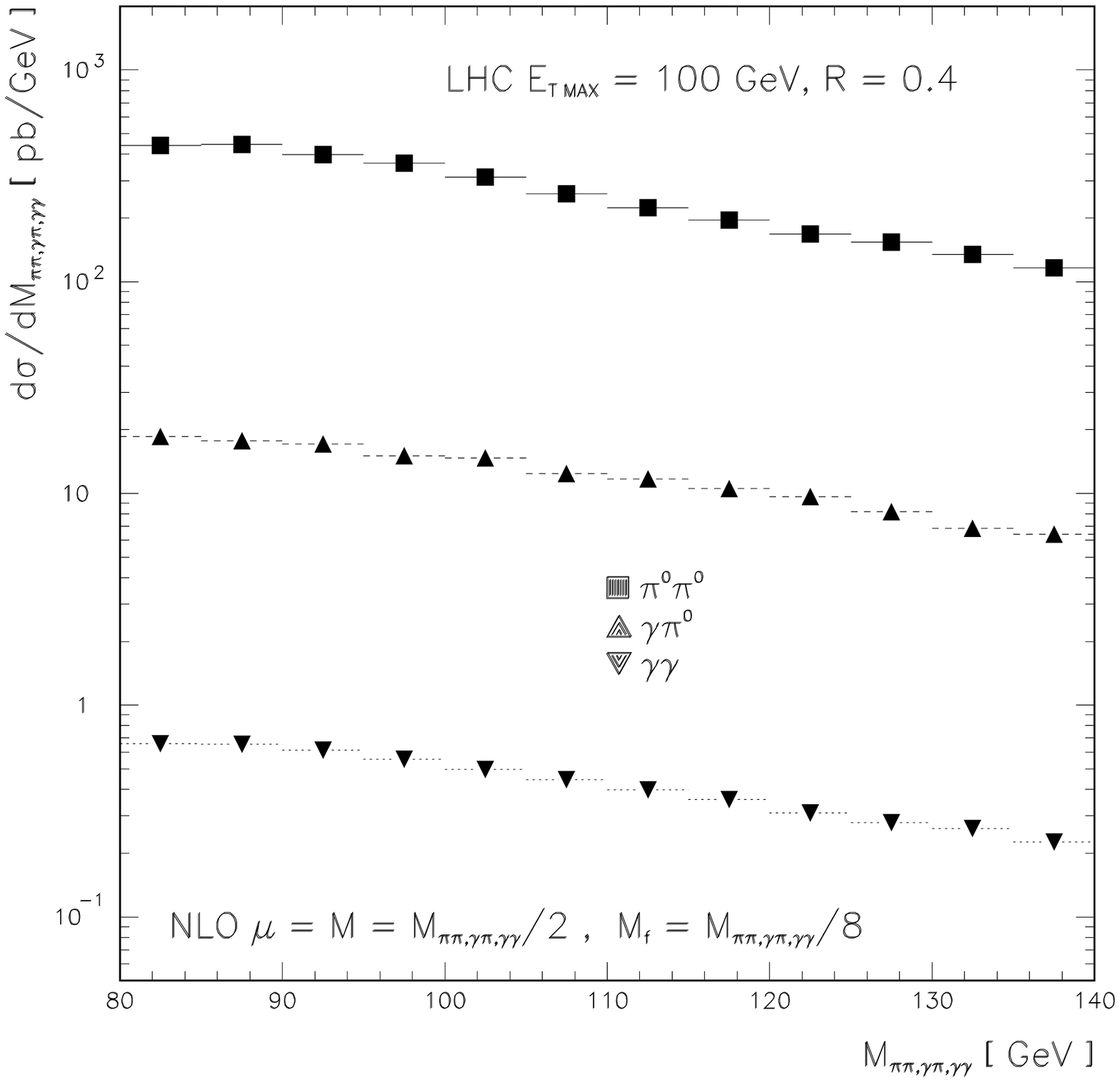,height=7.5cm}}
\put(85,0){\epsfig{file=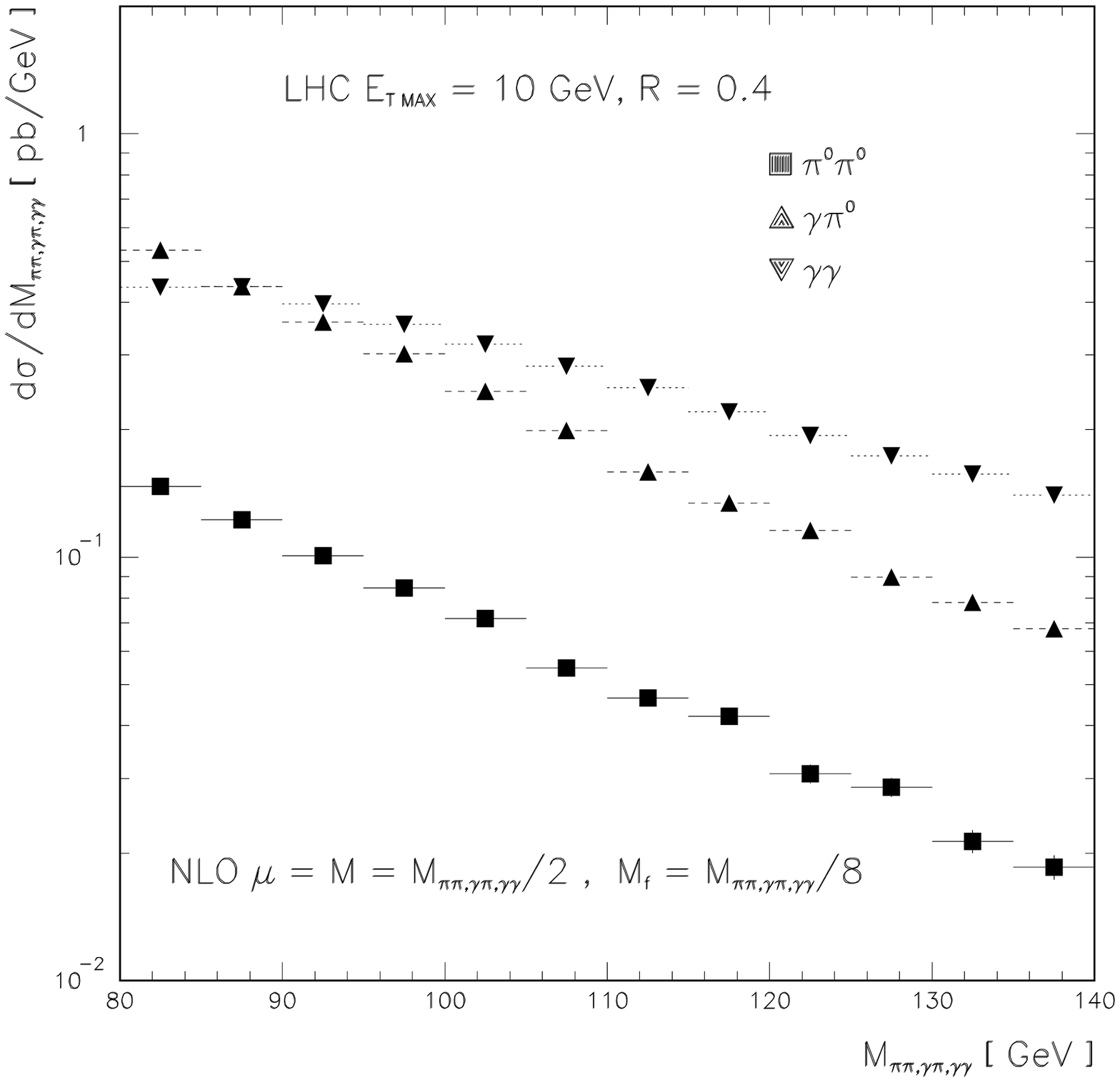,height=7.5cm}}
\end{picture}
\caption{NLO predictions for $\gamma\gamma$, $\gamma\pi^0$, $\pi^0\pi^0$ rates
at the LHC. Severe isolation (right) is needed to suppress the
pion contributions.}\label{fig_1_binoth}
\end{figure}
In Fig.~\ref{fig_1_binoth} the $\gamma\gamma$, $\gamma\pi^0$, $\pi^0\pi^0$  NLO predictions
are plotted for the invariant mass distribution of the photon/pion pairs at the 
LHC. In \cite{Bern:2002jx} it was shown that the inclusion of NLO corrections 
generally increases the discovery potential for the Higgs boson in the two photon channel.

To isolate a Higgs signal of the order 100 fb/GeV from this distribution, severe
isolation cuts have to be applied to reduce the background from fragmentation
processes. Applying isolation criteria around the photon/pion means that only  
a certain amount of transverse hadronic energy, $E_{T\,max}$, is allowed in a cone 
in rapidity and azimuthal angle space around the photon/pion, 
$R=\sqrt{(\Delta\eta)^2+(\Delta\phi)^2}$. As the minimal $p_T$ values at the LHC
are 25 and 40 GeV for the detected particles one finds for example for the 
 energy  fraction of the pion with respect to the jet:
\begin{equation}
z = \frac{E_{T\,\pi^0}}{E_{T\, jet}} > \frac{p_{T\,min}}{p_{T\,min} + E_{T\,max}} = 0.7
\end{equation}       
when typical values for  $E_{T\,max}=10$ GeV and $p_{T\,min}=25$ GeV are used.
This means that fragmentation functions are tested in a regime where they are
not restricted by experimental data.  In Fig. \ref{fig_1_binoth} (right)
one sees that the pion contributions are still sizable, but photons from fragmentation
are less important. Although additional
photon/pion selection cuts will help to reduce the pions further it has to be
stressed that  the uncertainties for the pion contributions are large.  

To quantify the uncertainty from fragmentation functions in the given prediction
three recent NLO parametrizations of the parton--to--pion fragmentation
functions were compared to each other.
\begin{figure}     
\unitlength=1mm
\begin{picture}(160,75)
\put(0,0){\epsfig{file=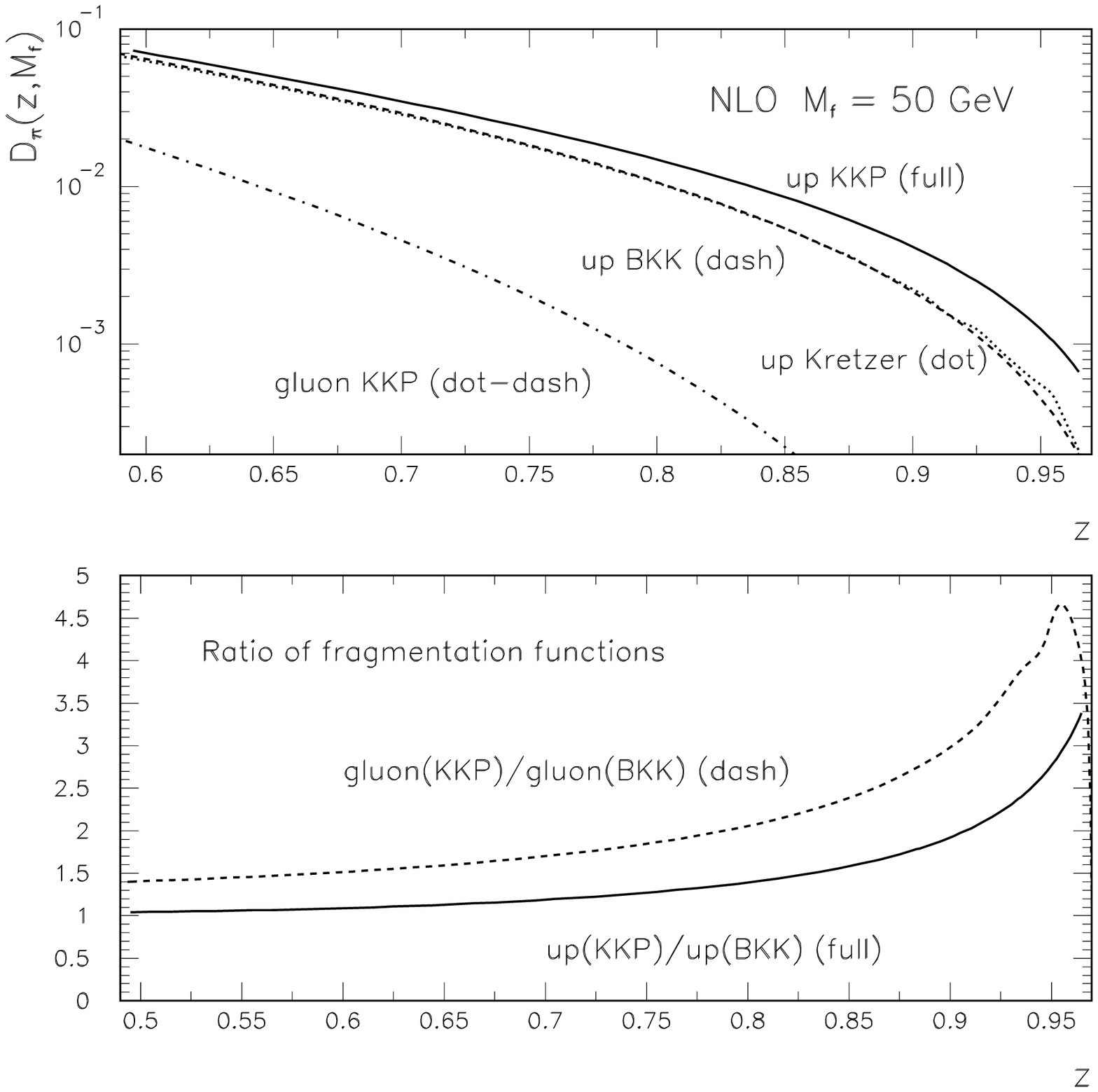,height=7.5cm}}
\put(85,0){\epsfig{file=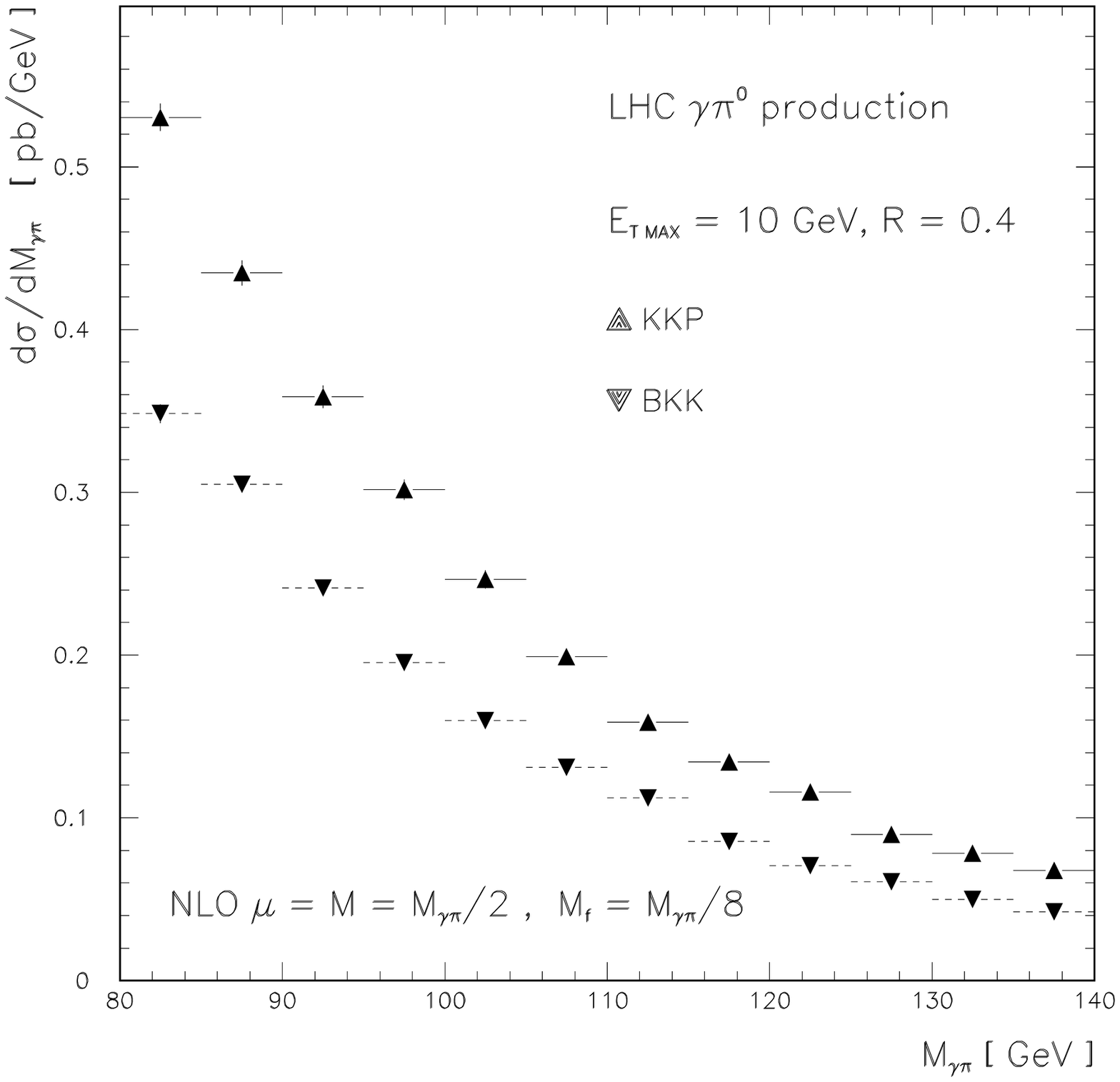,height=7.5cm}}
\end{picture}
\caption{Differences of the tails of fragmentation functions (left)
lead to large uncertainties in the predictions for $\gamma\pi^0$ rates.
Only the quark to pion fragmentation matters here.}\label{fig_2_binoth}
\end{figure}
In Fig.~\ref{fig_2_binoth} (right) one sees that the $\gamma\pi^0$ invariant mass spectrum
scales by a factor of about 1.5 when switching from the BKK \cite{Binnewies:1996df}
to the KKP \cite{Kniehl:2000fe} parametrization. 
The parametrization of Kretzer \cite{Kretzer:2000yf} is almost indistinguishable 
to the one of BKK and is not plotted separately.
The difference in the predicted rates is easily understood by looking at the ratio 
of the respective fragmentation functions $D_{q\to \pi^0}(z,Q^2)$ Fig.~\ref{fig_2_binoth} (left).
As only the tails of the distributions are probed the prediction is sensitive to the
experimentally unconstrained tails.
For the $\pi^0\pi^0$ component of the background the situation is even more dramatic as 
one pion has to have a $p_{T\, min}=40$ GeV leading to $z_{min}=0.8$. Taking
into account also usual theoretical scale uncertainties one can  not claim that 
the  $\gamma\pi^0$, $\pi^0\pi^0$ contribution of the $H\to \gamma\gamma$ background is
quantitatively well under control.      

\subsubsection{Conclusion}
The Run 2 at the Tevatron  will provide large data samples of multi-jet 
events which will enhance the experimental handles to constrain the 
fragmentation functions. However, in particular the clean data expected at a  
forthcoming LC will provide very important measurements for these studies. 
By analyzing  mesons/photons with high $p_t$ in the jets
fragmentation functions can be determined with an unprecedented accuracy.
Especially the region $z>0.7$ where data from LEP and SLD is becoming insufficient will
be probed  much more precisely. This in turn will lead to 
an improvement  of our understanding of two-photon events at the LHC.
This is not only relevant   for Higgs searches but also for other 
non-standard particles coupled to photons. 
Apart from that, as the di-photon spectrum will be measured 
very precisely at the LHC, it also
amounts to a good test of perturbative QCD in general.



%




\chapter{New Gauge Theories}
\label{chapter:newgaugeth}

Editor: {\it S.~Riemann}

\vspace{1em}
                                                                                
{\it D.~Bourilkov, M.~Dittmar, A.~Djouadi, S.~Godfrey, 
A.~Nicollerat, F.~Richard, S.~Riemann, T.~Rizzo}

\vspace{1em}

\section{Scenarios with extra gauge bosons}
New massive gauge bosons are suggested by most extensions of the Standard 
Model gauge group; neutral heavy gauge bosons Z$'$ as well as charged bosons, 
W$'$. Within a variety of models the new bosons sit just below the 
Planck scale 
or near the weak scale or they originate  in certain classes of 
theories with extra dimensions. Another recent example for a theory beyond 
the Standard Model is the 'Little Higgs' model.

For the physics program of near future colliders  those predictions with gauge
bosons  of a few hundred GeV or few TeV  are of interest. Some of the models 
also include the existence of an extended fermion sector. 

Here, the following models 
will be considered:
\begin{itemize}
\item 
The popular scenario for extra gauge bosons is based on the 
symmetry breaking schemes: 
SO(10)$\rightarrow$SU(2)$_L \times$SU(2)$_R \times$U(1) predicts 
new neutral and charged gauge bosons, 
Z$'$ and W$'$, and
E$_6\rightarrow$SU(3)$\times$SU(2)$\times$U(1)$\times$U(1)$_{Y'}$ anticipates 
neutral Z$'$ bosons.
Details about the models can be found in \cite{ref:WZp-model}. 
\item 
The 'Little Higgs' scenario \cite{ref:csaki,ref:hewettdec,ref:han,ref:LHpheno} 
is  proposed as an alternative to  SUSY and extra dimensions
to provide a solution to the hierarchy problem. 
In these models new symmetries imply the 
existence of a rich spectrum of new particles, in 
particular new gauge bosons that could be light. 
The Higgs boson appears as a pseudo-Goldstone boson and is protected by 
the global symmetry from 1-loop quadratic divergences. 
\item
KK excitations of the Standard Model gauge bosons are a natural 
prediction of models with additional dimensions~\cite{ref:KK-model}. 
The details of such models 
are described in chapter~8. 
\item
In the so called Universal Extra Dimension scenario (UED)~\cite{ref:ued} 
all SM fields
propagate in extra dimensions, there is no direct vertex involving one 
non-zero Kaluza-Klein (KK) mode. 
%
\end{itemize}

\subsection{Sensitivity to new physics models}

New interactions beyond the Standard Model can be parametrized in terms of 
effective four-fermion 
contact interactions \cite{ref:ehlq},
\begin{equation}
{\cal L}_{CI} =             \sum_{i,j = \mathrm{L,R}} \eta_{ij}
           \frac{g^2}{\Lambda^2_{ij}}
           (\bar{\mathrm{u}}_{F,i} \gamma^{\mu}\mathrm{u}_{F,i})
           (\bar{\mathit{u}}_{f,j} \gamma^{\mu}\mathit{u}_{f,j}).
\label{ci_lagr}
\end{equation}
with 
the helicity coefficients $\eta_{ij}$ and the couplings $g$ (by convention $g^2=4\pi$).  
$\Lambda$ can be interpreted as the compositeness scale, or it approximates 
the effects of exchanged new particles, $M_X \approx \Lambda$.
Contact interactions preserve the chiral symmetry and 
the Standard Model matrix element is extended by an additional contact term.

The new interactions  could arise in hadron-hadron 
collisions producing lepton or quark pairs at  the LHC   
by means of the Drell-Yan mechanism or in fermion-pair production processes at 
lepton colliders.
The parton cross section or the differential cross section, respectively, have 
the form
\begin{equation}
\frac{d \sigma}{d\cos \Omega}\propto \sum_{i,j=L,R} \rho_{ij} |A_{ij}|^2, 
\label{eq:sum-heliamps}
\end{equation}
with 
\begin{eqnarray}
A_{ij} &\sim& \displaystyle{ 
A_{ij}^{SM}+\frac{\eta_{ij}\cdot s}{\Lambda^2 }}\nonumber
\\
\rho_{LL,RR} &=&(1 + \cos \theta )^2 \label{eq:ci-ang-dep} \\
\rho_{LR,RL} &=&(1 - \cos \theta )^2 .\nonumber
\end{eqnarray}
The angular distribution functions, $\rho_{ij}$, 
describe the exchange of the spin-1 particles.
The sensitivity reaches for the new 
interactions can be tested by searching for deviations from the Standard Model 
expectations.

\subsubsection{Sensitivity to Contact Interactions at the LHC and LC}

The 
sensitivity reaches for various models of four-fermion contact interactions 
 expected for the LHC and LC 
have been determined \cite{ref:ci-lhc,ref:snow,ref:sriemann,ref:paver,ref:ci-lhc-db}. 
Some main results are summarized in Table~\ref{tab:ci-reaches} for 
qqee interactions  at the LHC as well as for eeqq and  eell interactions at the LC  
(for $\Lambda_{eeqq}(LC)$
the same strength of contact interaction to all quark flavors is assumed). 
New studies~\cite{ref:ci-lhc-db} show that 
these conservative bounds for the LHC can be improved by a factor roughly
1.5.
The sensitivity
regions for $e^+e^- \rightarrow e^+e^-$ are related to an optimistic scenario 
for systematic uncertainties; with the high statistics the systematic effects can 
be better controlled. A 'safe' realistic scenario lowers the $\Lambda_{eeee}$
bounds by a factor $\approx 2$ \cite{ref:ci-lhc-db}.  
\begin{table}[h]
\renewcommand{\arraystretch}{1.2}
 \begin{center}
\begin{tabular}{|rl|cccc|cccc|}
 \hline
 & & \multicolumn{4}{|c|}{LHC}& \multicolumn{4}{|c|}{LC}\\ 
 & &  \multicolumn{4}{|c|}{$\Lambda$ [TeV]}& \multicolumn{4}{|c|}{ $\Lambda$ [TeV]}\\
\hline 
\multicolumn{2}{|c|}{model }    &  LL &  RR &  LR &  RL &  LL &  RR &  LR &  RL         \\ \hline
eeqq:& $\Lambda_+$
          & 20.1    & 20.2    &  22.1   & 21.8    & 64    & 24    & 92    & 22           \\
 & $\Lambda_-$
          &  33.8     &33.7     & 29.2    & 29.7 & 63  & 35    &  92   &  24          \\
ee$\mu\mu$: & $\Lambda_+$
          &     &     &     &     &90     & 88    & 72    & 72           \\
& $\Lambda_-$
          &     &     &     &     &90     & 88    & 72    &  72          \\
eeee:& $\Lambda_+$ &     &     &     &  
          &44.9    &43.4 & 52.4 & 52.4  \\
& $\Lambda_-$
          &     &     &     &     & 43.5 & 42.1 &50.7 & 50.7            \\
\hline
\end{tabular}
\end{center}
\caption{The  95\% sensitivity reaches for a basic choice of contact interactions
expected for the LHC \cite{ref:ci-lhc} 
($L_{int}=100~fb^{-1}$ at 14~TeV and $\delta L$=5\%) and the  LC \cite{ref:ci-lhc-db,ref:sriemann}  
($L_{int}=1~ab^{-1}$ at 0.5~TeV and $P_{e^-}$=0.8, $P_{e^+}$=0.6).
}
\label{tab:ci-reaches}
\end{table}
With the LHC new interactions between 
quarks and leptons can be examined whereas a LC opens the window 
to new interactions between leptons only or leptons and quarks. 
In Ref. \cite{ref:usubov} also the sensitivity of LHC measurements to 
quark compositeness is derived by analysing the dijet angular distributions
and the transverse energy distribution of jets.
With an integrated luminosity of 30~(300)~$fb^{-1}$ 
a sensitivity to compositeness of quarks up to a scale of 25~(40)~TeV can be achieved.

This 
complementarity demonstrates the need to have both colliders to be sensitive to
the full spectrum of  new physics. 

\subsubsection{Distinction of new physics models}
Contact terms describe the effective new contributions 
in model-independent manner. If hints to new effects are found the
quest for their source 
could be answered with the exchange of new particles, $M_X \approx \Lambda$. 
Well known examples of these interpretations are  the exchange of 
extra gauge bosons, leptoquarks, supersymmetric particles, gravitons etc.
It will be a puzzle to deduce from bounds on effective 
interactions back to  special models and phenomena. An important step is the
check of the angular distribution:
If new  spin-1 particles are exchanged the typical 
$(1\pm\cos \theta)^2$ (Eq.~\ref{eq:ci-ang-dep}) 
behaviour will be observed and can be distinguished from an exchange of 
other particles (spin-0, spin-2).
In case of an exchange of spin-2 particles like gravitons 
the additional contribution to the 
helicity amplitudes  depend on the scattering angle \cite{ref:ratazzi-cullen,ref:ci-ed}
 (see also chapter 8) 
and Equations (\ref{eq:sum-heliamps},
\ref{eq:ci-ang-dep})
are modified:
\begin{eqnarray}
A_{LL,RR} &=& A_{LL,RR}^{SM}-\displaystyle{
\frac{\lambda \cdot s^2 }{4 \pi \alpha M_S^4}(2\cos \theta - 1)}\nonumber\\
A_{LR,RL} &=& A_{LR,RL}^{SM}-\displaystyle{
\frac{\lambda \cdot s^2 }{4 \pi \alpha M_S^4}(2\cos \theta + 1)} \label{eq:graviton-ang-dep}
\end{eqnarray}
$M_S$ is the cut-off scale (\cite{ref:hewett}) and $|\lambda|$ is of the  
order 1.
Details about the identification of this kind of new physics will be 
discussed in the next chapter.

\subsubsection{Z$'$ in the context of Contact Interactions}

Let's assume that the angular distributions indicates the exchange of a new 
spin-1 particle but the available energy is to low to produce it directly. 
This particle could be a Z$'$: 
Then a simultaneous analysis of the contact terms $\eta_{LL}^{ef}/\Lambda^2$,
$\eta_{RR}^{ef}/\Lambda^2$, $\eta_{LR}^{ef}/\Lambda^2$ and $\eta_{RL}^{ef}/\Lambda^2$
will show that the condition
\begin{equation}
\displaystyle{
\frac{\eta_{LL}^{ef}}{\Lambda^2} \frac{\eta_{RR}^{ef}}{\Lambda^2} =
 \frac{\eta_{LR}^{ef}}{\Lambda^2} \frac{\eta_{RL}^{ef}}{\Lambda^2} 
\sim 
\frac{g_L^e}{M_{Z'}}\frac{g_L^f}{M_{Z'}}\frac{g_R^e}{M_{Z'}}\frac{g_R^f}{M_{Z'}} }
\nonumber
\end{equation}
is fulfilled. 
The helicity amplitudes in Equ.~(\ref{eq:ci-ang-dep}) can be replaced by 
amplitudes including $\gamma$, Z and Z$'$ exchange:
\begin{equation}
A_{ij} = Q^e Q^f + 
  \frac{g_i^Z g_j^Z}{s_W^2 c^2_W} \frac{s}{s-M_Z^2   + iM_Z \Gamma_Z}
+ \frac{g_i^{Z'} g_j^{Z'}}{ c^2_W}\frac{s}{s-M_{Z'}^2+ iM_{Z'} \Gamma_{Z'}}
\label{eq:zp-heli-amp}
\end{equation}
Exploring the expected search reaches in terms of contact interactions shown 
in Table~\ref{tab:ci-reaches} it is evident that the sensitivity to a Z$'$ 
could be higher with a LC. But it is wideley assumed that 
new heavy vector bosons are 'light' enough to be directly produced 
at the LHC
by means of the Drell-Yan 
mechanism \cite{ref:cms-tdr,ref:atlas-tdr}, $pp \rightarrow q\bar{q} 
\rightarrow W_R^+W_R^-,~ Z_R, Z'$ and 
discovered through 
fermionic or bosonic decay modes. Most likely the energy of a LC - 
at least in the first phase of operation - will not be not sufficient 
to run  a Z$'$ factory and to measure all characteristics.
But  below a Z$'$
production threshold new gauge bosons appear 
virtually and cause  significant deviations from the Standard Model 
expectations that allow to determine  Z$'$ parameters or bounds 
on them.

With a GigaZ option the propagator term in Eq.~(\ref{eq:zp-heli-amp}) 
becomes negligible but 
the sensitivity  of  loop corrections to new bosons can be explored: Precision
measurements of $g_i^Z$, $g_j^Z$, $\sin^2 \theta_{eff}$ 
and the $\rho$-parameter will allow to deduce
the effects beyond the Standard Model and a  potential Z-Z$'$ mixing can be measured.

Recent direct  measurements 
from the Tevatron
by CDF and D0 
exclude Z$'$ masses in conventional models
below $\approx$700~GeV \cite{ref:zp-tevatron}, 
corresponding indirect limits from LEP vary between 330 GeV and 670 GeV 
for E$_6$ models and between 470 GeV and 900 GeV for left-right symmetric 
models~\cite{ref:zp-lep2}.  Extra charged bosons, W$'$, are excluded below
785 GeV \cite{ref:wp-tevatron}. 
Hence, with a linear collider operating at energies of 0.5 TeV up to about 
1~TeV it is expected to 
observe  the virtual effects of new gauge bosons.

\section{Z$'$ studies at the LHC}

\subsection{Z$'$mass reaches at the LHC}

Detailed studies have been performed on sensitivity 
reaches for new gauge bosons
at the LHC and LC 
\cite{ref:cvetic,ref:leike,ref:godfrey,ref:riemann,ref:sriemann,ref:atlas-tdr,ref:cms-tdr}.

The Z$'$ production cross section at LHC, $\sigma(pp\rightarrow Z')$, is measured
via 
the leptonic cross section
$\sigma(pp\rightarrow Z')Br(Z' \rightarrow l^+ l^-)$. 
Depending on the underlying model new particles can
influence  the total Z$'$ width. 
Therefore, it is assumed that all Z$'$ decay modes are known. 
In a recent paper \cite{ref:dittmar}
the potential of a Z$'$ search at
the LHC is updated. Based on a more realistic 
simulation the signals of Z$'$ bosons originating in various theoretical models
are analyzed and 
the discovery potential up to $m_{Z'}=5~$TeV for $\cal{L}$=100fb$^{-1}$
is reconfirmed. 
In \cite{ref:atlas-tdr} 
the ability of the process 
Z$'~\rightarrow~ j j$ for Z$'$ detection is studied: Depending on the 
model a Z$'$ observation 
up to $m_{Z'}\leq 3-4~$TeV will be possible.  
The discovery power of this channel does not reach that of the leptonic 
channel but it comes closer for high Z$'$ masses where the statistics in the 
leptonic channels is low. 

\subsection{Distinction of models at the LHC}

The measurement of the production cross section, 
well suited for 
$Z'$ detection
by observing a mass bump, 
is not sufficient to distinguish models. 
In \cite{ref:cvetic} the authors demonstrate the determination of 
Z$'$ couplings
with LHC measurements. They introduce a new parameter set  of normalized  
Z$'$ couplings  resolving the left- and right-handed
coupling contributions 
that are characteristic 
of the particular model. 
Although they neglect all 
systematic errors and assume 100\% efficiency in data reconstruction 
it is accepted that a distinction of Z$'$ models will be possible 
by measuring the cross section, the asymmetry and the rapidity 
distribution~\cite{ref:lhc-rapidity,ref:dittmar}.
Besides ambiguities this will of course
work better for light Z$'$ bosons.

%
\subsubsection{{Forward-Backward Asymmetry measurement at LHC}}
%

The more realistic study of the Z$'$ search potential  in \cite{ref:dittmar}  
devotes special attention to the discrimination 
between models. Due to the model-dependent values of the Z$'$ couplings 
to quarks and leptons the  foward-backward asymmetry, $A_{FB}^l$, is a
sensitive measure 
to distinguish models. The asymmetry cannot be measured 
directly in the symmetric collisions of LHC.
However, from the rapidity
distribution of the dilepton system the quark direction can
be obtained by assuming to be the boost direction of the $ll$ system 
with respect to the beam axis (see \cite{ref:dittmar-afbl}).
The probability to assign the correct  quark direction increases 
for larger rapidities of the dilepton system. 
With a cut on the rapidity distribution, $Y_{ll}>0.8$, 
a cleaner 
but smaller signal  sample can be  observed and  
the asymmetry obtained, see Figure~\ref{sec5_fig:dittmarafblrapid}(a).
\begin{figure}[htpb]
\begin{flushleft}
 \begin{tabular}{lr}
   \includegraphics[width=0.49\textwidth,height=0.49\textwidth]{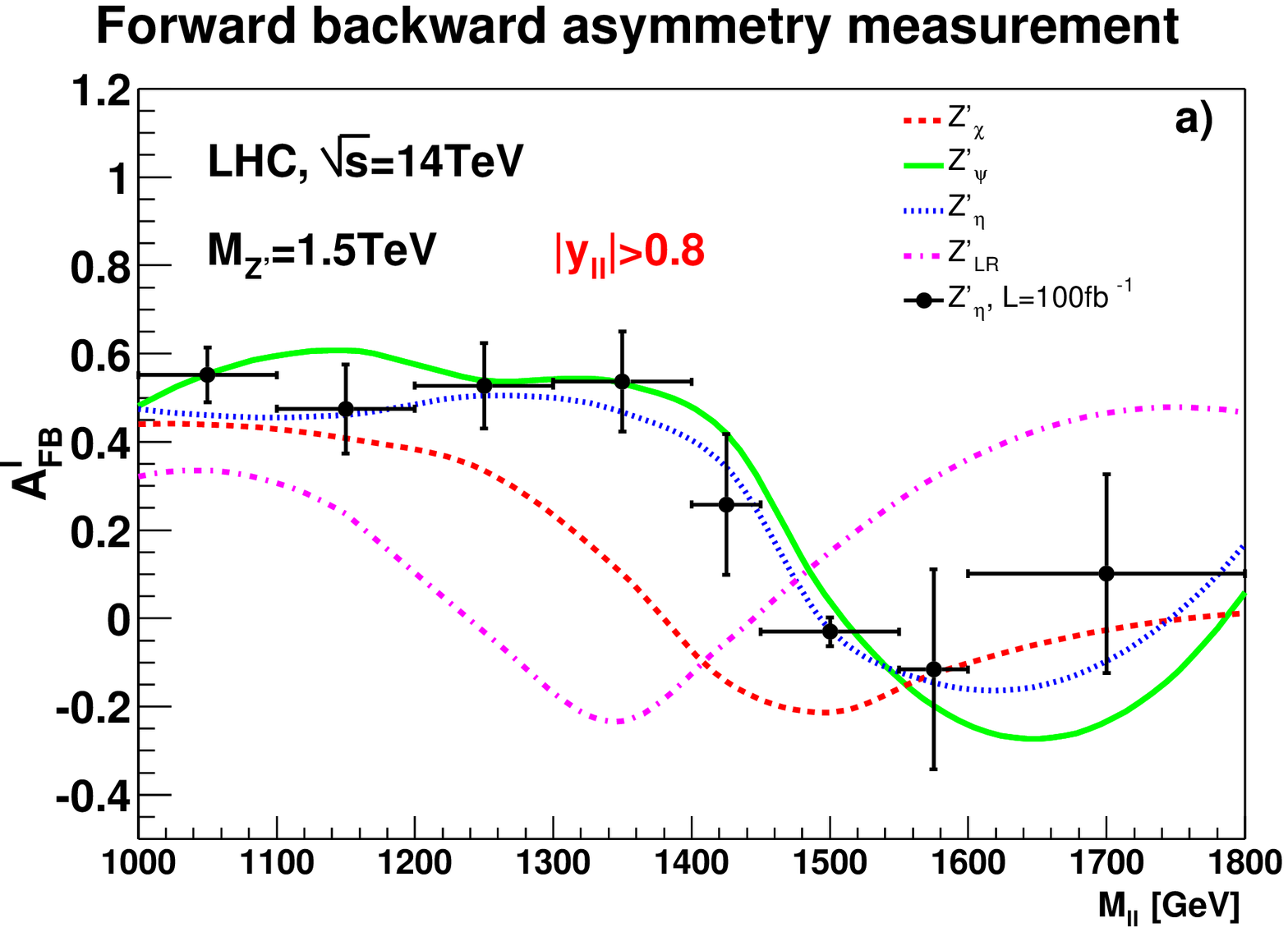}&
   \includegraphics[width=0.49\textwidth,height=0.49\textwidth]{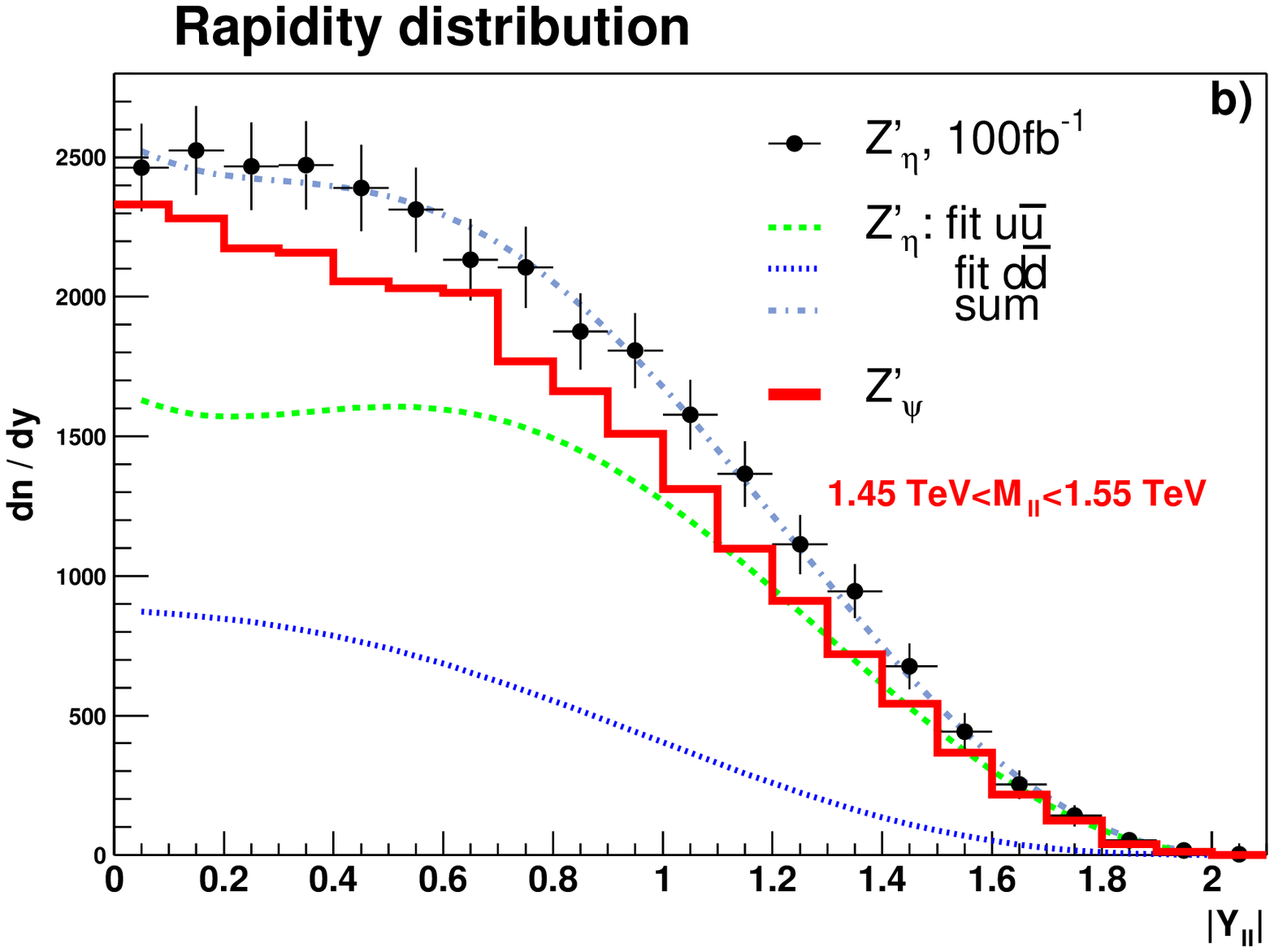}
 \end{tabular} 
\caption{$A^l_{FB}$ (a) as a function of $M_{ll}$ for several Z$'$ models.
The observable rapidity distribution for two Z$'$ models is shown in (b), 
including the fit results which determine the types of $q\bar{q}$fractions.
A simulation of the statistical errors, including random fluctuations of the 
Z$'_{\eta}$ model and with errors corresponding to a luminosity of 
100$~fb^{-1}$
has been included in both plots. For details see \cite{ref:dittmar}. 
\label{sec5_fig:dittmarafblrapid}
}
\end{flushleft}
\end{figure}

\subsubsection{{Rapidity Distribution}}

To complete the Z$'$ analysis, one
can obtain some information  about the fraction of Z$'$'s produced from
$u\bar{u}$ and $d\bar{d}$ by analyzing the $Z'$ rapidity distribution. 
Assuming that the $W^{\pm}$ and $Z$ rapidity distribution has been measured  in
detail, following the ideas given in~\cite{ref:dittmar-lhclumi}, 
relative parton
distribution functions for $u$ and $d$ quarks as well as for the corresponding 
sea quarks and antiquarks are well known. Thus, the rapidity spectra can be
calculated separately for  $u\bar{u}$ and $d\bar{d}$ as well as for  sea quark
anti-quark annihilation and for the mass region of interest to analyze the $Z'$
rapidity distribution~\cite{ref:lhc-rapidity}.  
Using these distributions 
a fit can be
performed to the $Z'$ rapidity distribution which allows to obtain the
corresponding fractions of $Z'$'s produced from $u\bar{u}$, $d\bar{d}$ as well
as for  sea quark anti-quark annihilation.  This will thus reveal how the $Z'$
couples to different quark flavors in a particular model.

\vspace{0.5cm}

\subsubsection{{Extracting the Z$'$ couplings}}

Due to the significant shape of $A_{FB}^l$ in the vicinity of a Z$'$ peak,
the Z$'$ peak region as well as the 'interference' region
provide complentary information for the analysis.
Furthermore, the analysis of the rapidity distribution gives 
additional Z$'$ information. 
Figure~\ref{sec5_fig:dittmarafblrapid}(b) shows the expected 
rapidity distribution  for the
$Z'_{\eta}$ model. A particular $Z'$ rapidity distribution is fitted using a
combination of the three pure quark-antiquark rapidity distributions.
The fit output gives the  $u\bar{u}$, $d\bar{d}$ and $sea$ quarks fraction in
the sample. In order to demonstrate the analysis power of this method  we also
show the $Z'_{\psi}$ rapidity distribution which has equal couplings to 
$u\bar{u}$ and $d\bar{d}$ quarks.

\subsection{Summary: Z$'$ search at the LHC}

Z$'$ signals like mass bumps will be observed up to $M_{Z'}=5~$TeV.
The statistical significance of $A_{FB}^l$ measurements   
will allow a distinction of 'usual' models with the LHC 
if $m_{Z'}<2-2.5~$TeV. 

However, if nothing is detected 
at the LHC the indirect searches 
at the LC 
are strongly required to check whether heavier gauge bosons 
could exist or whether they have couplings that make them invisible at the LHC.

\section{Z$'$ studies at the LC}

\subsection{Z$'$ mass reaches at the LC}

Most likely, at the LC new bosons will  be detected via Z-Z$'$ 
interference effects. Then the significance of deviations from the Standard 
Model expectations is determined by the  ratio of Z$'f\bar{f}$ couplings 
and Z$'$ mass.

Usually, the sensitivity reach of indirect searches is given at 
the 95\% C.L. assuming special models, e.g. 
E$_6$ or left-right symmetric models. To compare 
these 2$\sigma$ deviations from the Standard Model expectations with the 
direct search reaches one has to consider at least a 5$\sigma$ deviation. 
Figure~\ref{sec5_fig:zplhclc} illustrates the discovery and sensitivity reaches 
of the LHC and LC.  
The sensitivity limits at LC would improve up 
to roughly 12\% if all systematic errors could be reduced to zero. 
\begin{figure}[htp]
\begin{center}
    \includegraphics[width=0.7\textwidth,height=0.7\textwidth]{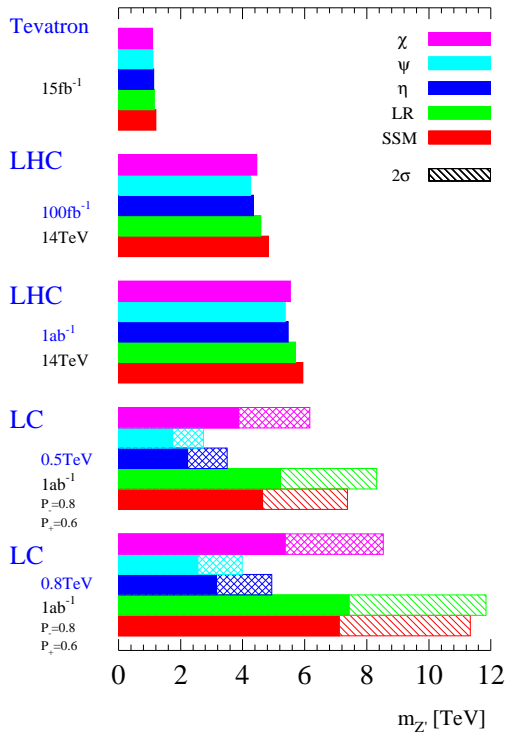}
\caption{Comparison of the Z$'$ discovery reaches at LHC \cite{ref:godfrey}
and the 2$\sigma$ and 5$\sigma$ sensitivity bounds at LC \cite{ref:sriemann}. 
\label{sec5_fig:zplhclc}
}
\end{center}
\end{figure}

\subsection{Distinction of models with the LC}

To learn about the Z$'$ properties the measurement of the 
couplings is essential. The couplings and also the behaviour of the angular 
distribution of the observables measured with polarized beams allow 
the study of the nature 
of  new gauge bosons.
First, the usual scenario is considered where the extra gauge 
bosons carry vector and axial vector couplings. 

In the ideal case the new gauge bosons are light enough to be 
found with the LHC. Then both colliders provide 
the feasibility to determine 
the Z$'$ couplings.

At the LC 
at high energies,
fermion pair production is a process with high statistics
and clear topologies. Moreover with the 
expected integrated luminosity of  
1~ab$^{-1}$ and with the possibility of polarisation of at least the 
electron beam.
If the mass of a potential Z$'$ is known from LHC 
the Z$'$ model can be resolved with a good accuracy.
Figure~\ref{sec5_fig:zpcoupl} demonstrates the powerful interplay of LHC and LC 
measurements: Leptonic final states obtained at a LC 
are analyzed with the knowledge of the mass of a potential Z$'$.
\begin{figure}[htpb]
\begin{flushleft}
\begin{tabular}{lr}
    \includegraphics[width=0.49\textwidth]{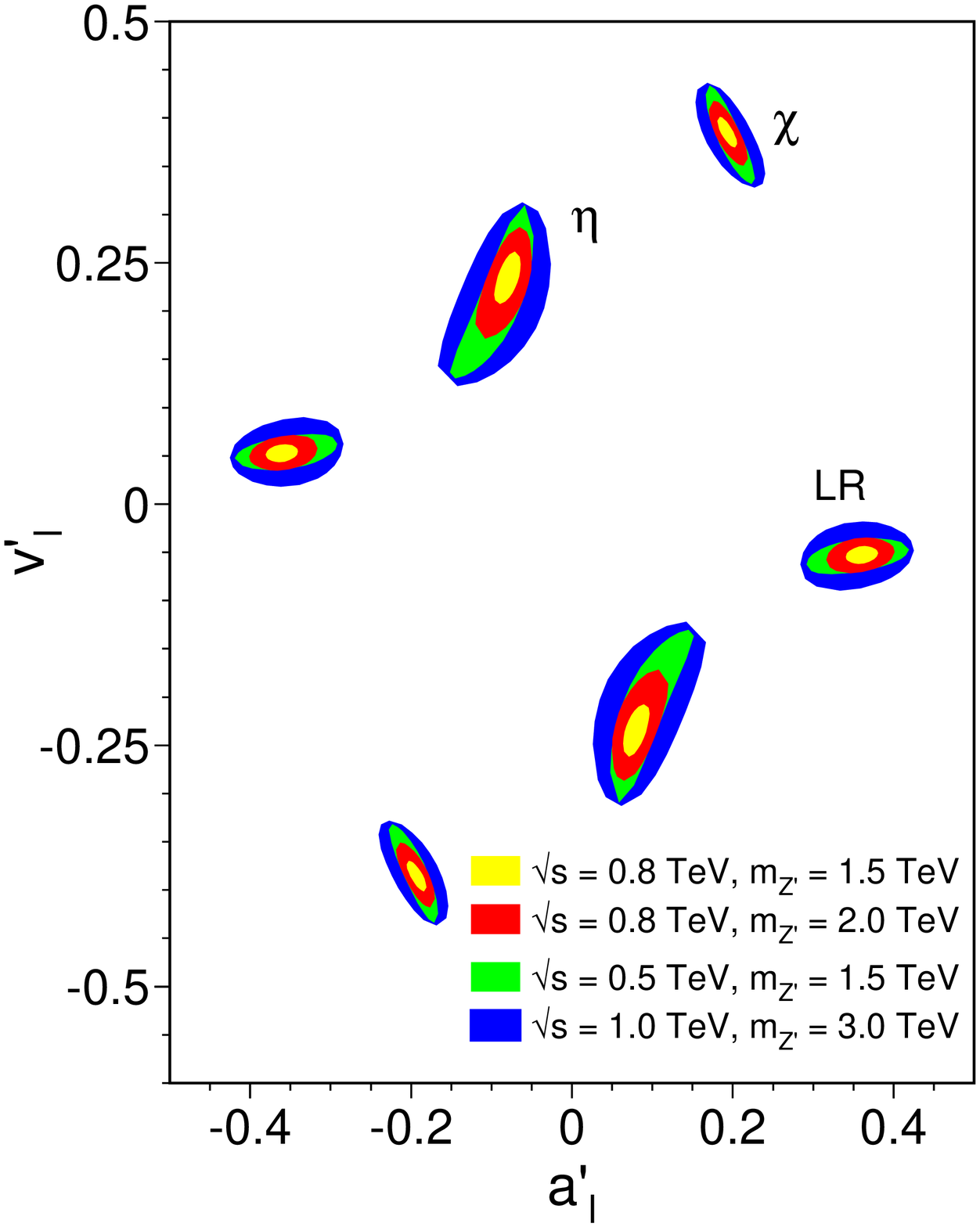}&
    \includegraphics[width=0.49\textwidth]{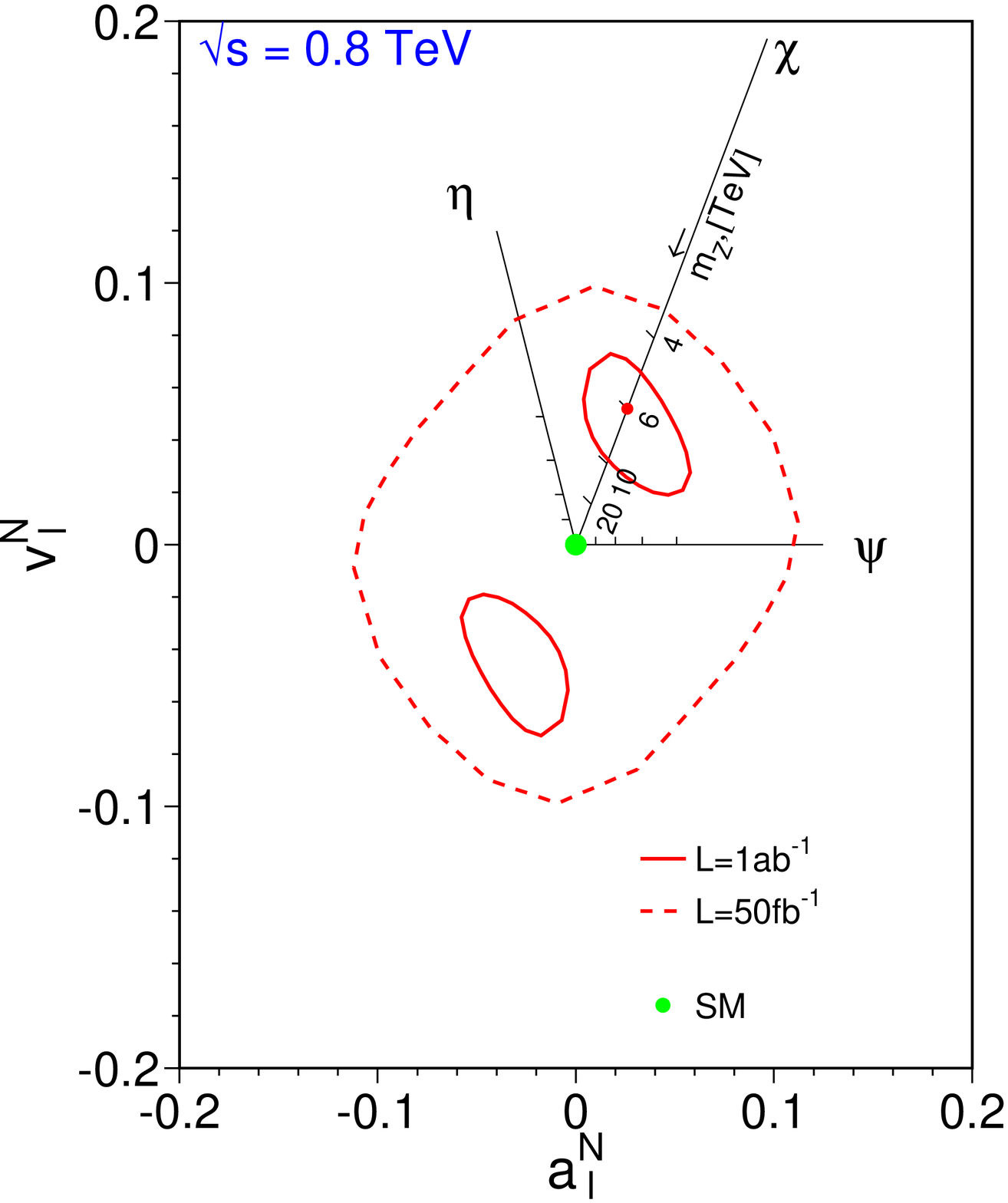}
\end{tabular}
\caption{Left figure: The 95\% C.L. contours on leptonic Z$'$ couplings, 
assuming the 
mass of the Z$'$ is measured at LHC. Right figure:
The 95\% C.L. contours on normalized leptonic 
Z$'$ couplings. 
The LC parameters are
${\cal{L}}_{int} = 1ab^{-1}, ~~\Delta {\cal{L}}_{int} = 0.2\%$,
$P_-=0.8,~~P_+=0.6,~~\Delta P_- =  \Delta P_+ = 0.5\%$,
$\Delta$sys(lept)=0.2\%.
\label{sec5_fig:zpcoupl}}
\end{flushleft}
\end{figure}

\subsubsection{{Sensitivity to Z$'$ models beyond the LHC reach }}

An interesting situation 
arises if the Z$'$ is too heavy to be detected 
at the LHC, e.g. $m_{Z'} > 5~$TeV, or if the Z$'$ couplings to quarks are 
unexpectedly negligable. If the Z$'$ is not too heavy  
(approximately 6 TeV for the usual Z$'$ models) 
there is still the 
possibility of measuring significant deviations
from the Standard Model predictions. 
Instead of  a 
direct implementation of the Z$'$ parameters 
normalized couplings have to be 
considered:
\begin{equation}
a^N_f = a'_f \displaystyle{\sqrt{\frac{s}{m_{Z'}^2-s}}};~~~~~~
v^N_f = v'_f \displaystyle{\sqrt{\frac{s}{m_{Z'}^2-s}}}.
\end{equation}
It should be remarked that these normalized couplings correspond to contact 
interactions, $\eta_{ij}^{ef}/\Lambda^2 \approx g_i^e g_j^f /M_{Z'}^2$.

An example is given in Figure~\ref{sec5_fig:zpcoupl}: 
A Z$'$ with $m_{Z'}= 6~$TeV
realised in the $\chi$ model cannot be seen at the LHC 
but can just be detected at the LC.
The Z$'$ models are  characterized by the lines 
$v^N_l = f(a^N_l)$, the distance of a special reconstructed 
point on this line allows 
the determination of the  model and the mass of the  
Z$'$ boson. 
It should be remarked that the lines, $v^N_l = f(a^N_l)$,
correspond  to the observables $P_V^l$ introduced in 
\cite{ref:lhc-rapidity,ref:cvetic}.
By measuring the normalized couplings at two different c.m.s. energies, the 
Z$'$ mass and the Z$'$ couplings, $a'_f$ and $v'_f$, can possibly be
determined. The success of this procedure depends on the c.m.s. 
energies of 
the measurements, the luminosity distributed among these energies and, 
of course, 
on the Z$'$ mass (see \cite{ref:sr-zp-reco,ref:rizzo-zp-reco}).   

Similar studies were also performed for the process 
$e^+e^- \rightarrow q \bar{q}$ 
\cite{ref:leike-riemann,ref:riemann,ref:sr-zpcoupl}.
The expected 
resolution power, based on Z$'q\bar{q}$ couplings, is less 
significant than that derived from Z$'l^+l^-$ couplings. It  depends on 
the uncertainty of the measurement of leptonic Z$'$ couplings and on
systematic errors due to quark flavour tagging and charge reconstruction.
Although the performance of the detector and the high luminosity
will provide an excellent and efficient reconstruction of the b- and c-quarks,
the resulting  Z-Z$'$ interference effects for leptonic initial and quark 
final states provide less stringent constraints on Z$'$ parameters. A global 
analysis of the measurements of all final states at and above the Z resonance
will enhance the resolution sensitivity. But a numerical evaluation of the 
expected improvement is rather model-dependent.   
The combination of LHC and LC measurements allows a comprehensive analysis
of  the new gauge boson parameters,
and becomes even more important for higher Z$'$ masses approaching
the discovery limit of the LHC.

\subsection{Z$'$ search at GigaZ} \label{subsub-zpgigaz} 
The new gauge bosons can mix with the 
standard model Z$^0$ 
and therefore modify  
electroweak  parameters. Precision measurements on the Z peak are sensitive to
these modifications.
Measurements at LEP/SLD 
result in a Z-Z$'$ 
mixing angle which is presently consistent with zero 
(see for example  \cite{ref:zp-mixing}). 
Hence,  all Z$'$ studies at energies off the Z or Z$'$ resonances  based
on Z-Z$'$ and $\gamma$-Z$'$ interference effects are less sensitive to the
 Z-Z$'$ mixing and 
can be performed with the assumption of zero Z-Z$'$ mixing.
%

In general, it should be remarked that the majority of studies with new 
gauge bosons 
assume a scenario with the weak parameter $\rho=1$ at tree level.
Other models would require new loop calculations and will lead to largely 
modified 
electroweak corrections and their relations to the Standard Model parameters. 
In \cite{ref:gluza} this is demonstrated in detail 
for the  $\Delta r( m_{top})$ 
behaviour considering Z$'$ bosons in a specific Left-Right model without the restriction $\rho=1$.
A similar procedure has to be executed in the case of UED.

%
Z-Z$'$ mixing  modifies electroweak  parameters, e.g.
$M_W$, $\sin^2 \theta_{eff}^{lept}$, Z total and partial widths.
This results in 
a very high sensitivity of precision measurements performed at  
the Z resonance to the Z-Z$'$ mixing. 
With accuracies available at GigaZ it will become possible to elucidate the 
origin 
of new gauge bosons and the symmetry breaking mechanism responsible 
for the mixing. Even in the 'zero coupling' limit 
there could remain an observable 
contribution to the $\rho$ parameter, e.g. in the case of UED the effect of 
t-b mass splitting leading to weak isospin violation in the KK spectrum.

Although the precision data from LEP/SLD \cite{ref:lep-sld} constrain
the Standard Model Higgs to be light, the contribution of a heavy Higgs could 
be compensated by a Z$'$.  
In \cite{ref:richard} it is shown that such scenarios are not very likely for 
the popular E$_6$ and LR models and they are 
already excluded by recent precision measurements. Nevertheless,
the discovery of Higgs bosons at the LHC gives a strong restriction for the 
extremely precise measurements with the GigaZ option at a LC. It is obvious 
that GigaZ results have to be compatible with LHC results, LC results at high 
energies and with the existing LEP/SLD/APV data. 
Only the consideration of all results will allow the extraction of the physics 
of symmetry breaking. Disagreements between direct measurements 
(including  non-observation of hypothetical particles) and precision 
measurements will improve the search strategies at least at the LHC. 
Examples of the resolution power for new physics with GigaZ are given in
Figures~\ref{fig:richard1} and \ref{fig:richard2} 
by demonstrating the (mis)matching of electroweak observables and new physics
depending on the Standard Model Higgs mass.
\begin{figure}[htb!]
\begin{center}
    \includegraphics[width=0.45\textwidth,height=0.6\textheight]{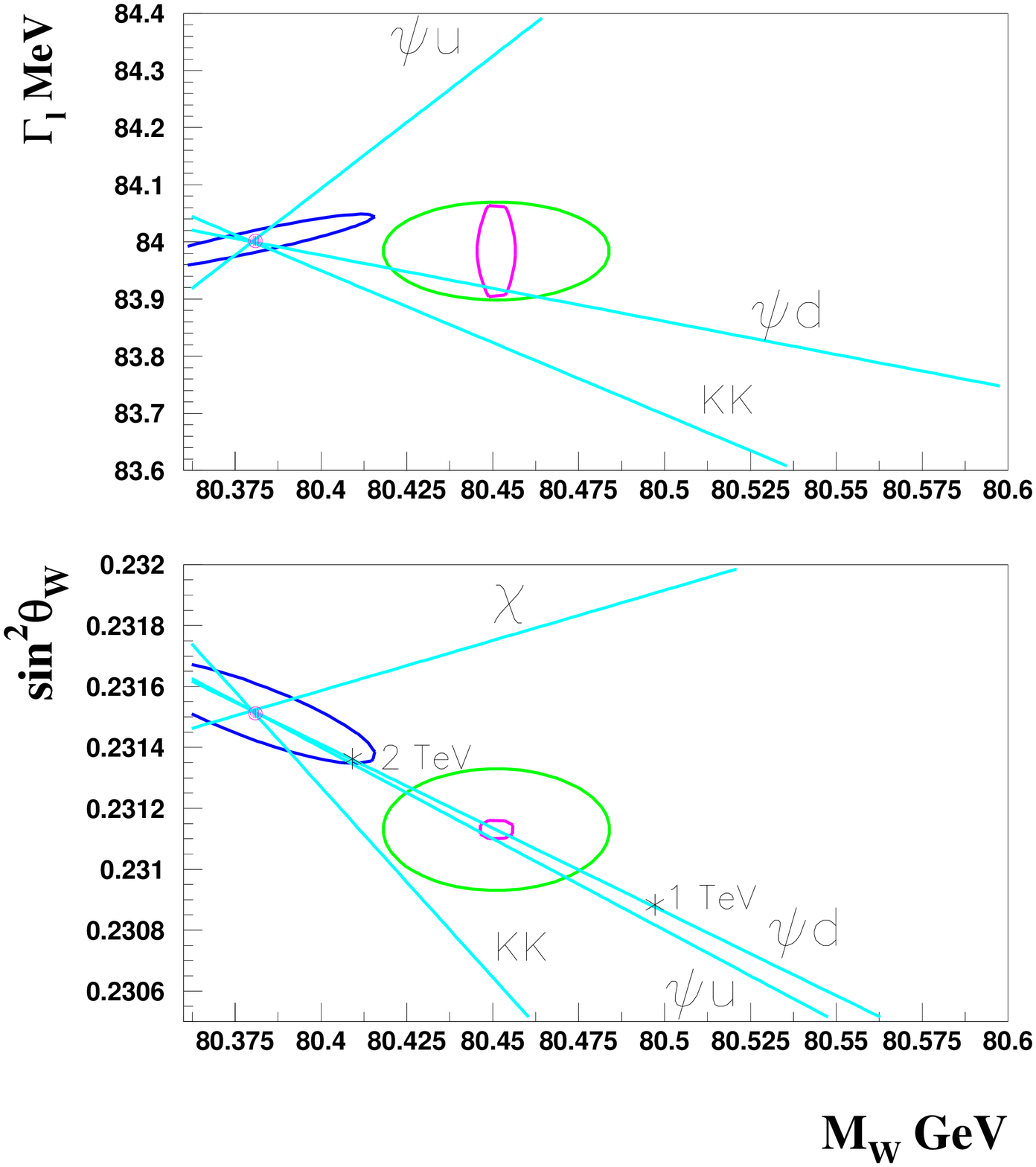}
    \includegraphics[width=0.45\textwidth,height=0.6\textheight]{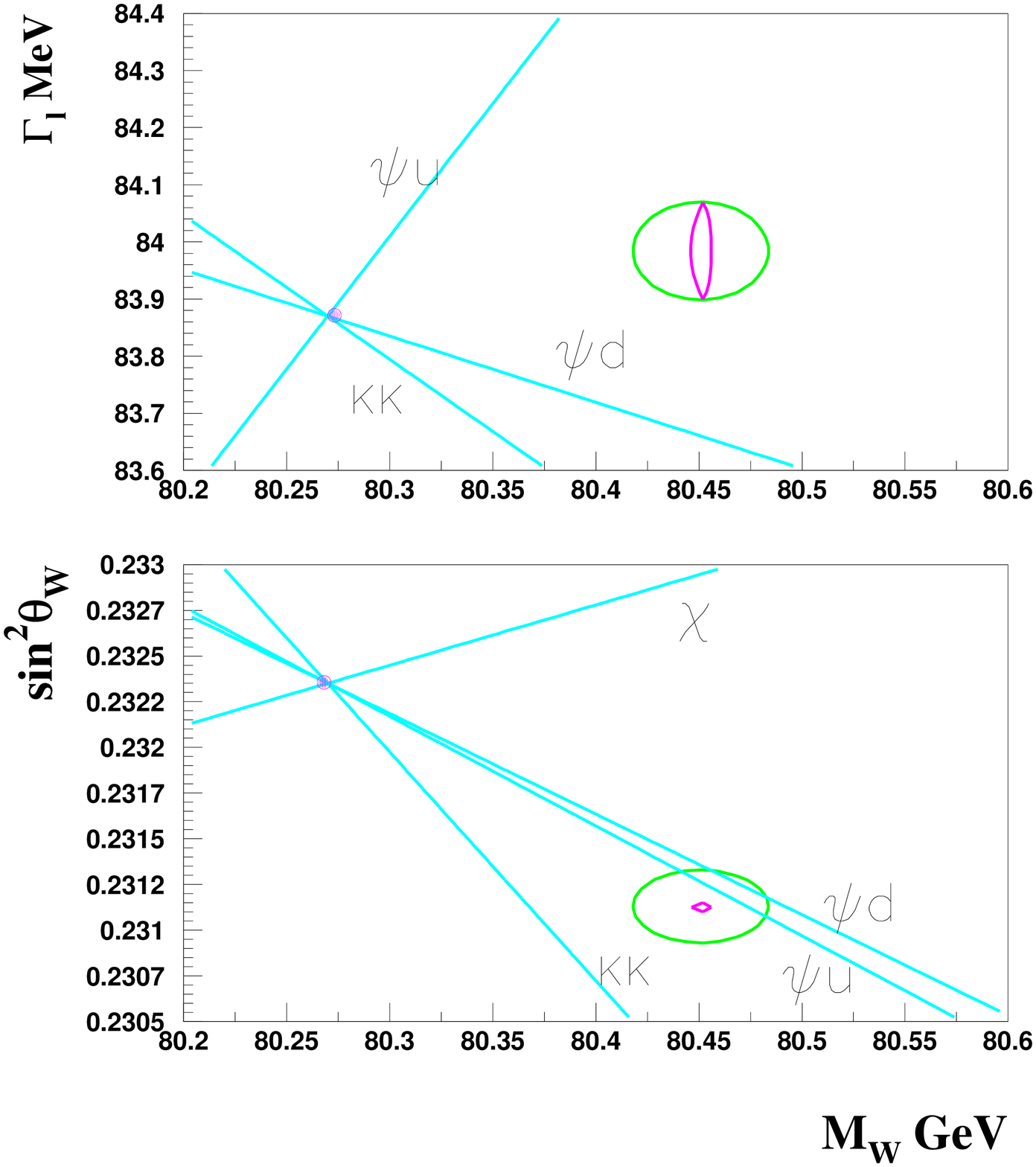}
\end{center}
\caption{ SM prediction, 
 LEP/SLD  and GigaZ expected precision 
 compared with Z$'$ models ($m_H$=115 GeV).
On the left parts of the upper plots
 are indicated the SM prediction ellipses (dark blue) 
taking into account uncertainties coming 
from the top mass and from $\alpha(M_Z)$ and assuming a 115 GeV (500 GeV)
SM Higgs boson in the upper (lower) part of the figure. 
The experimental measurements from LEP/SLD/Tevatron 
appear on the second ellipse (right, green at each plot)
 while inside these ellipses are the anticipated ellipses (purple) 
from GigaZ. For details see \cite{ref:richard}.
\label{fig:richard1}
}
\end{figure}
\begin{figure}[htb!]
\begin{center}
    \includegraphics[width=0.45\textwidth,height=0.6\textheight]{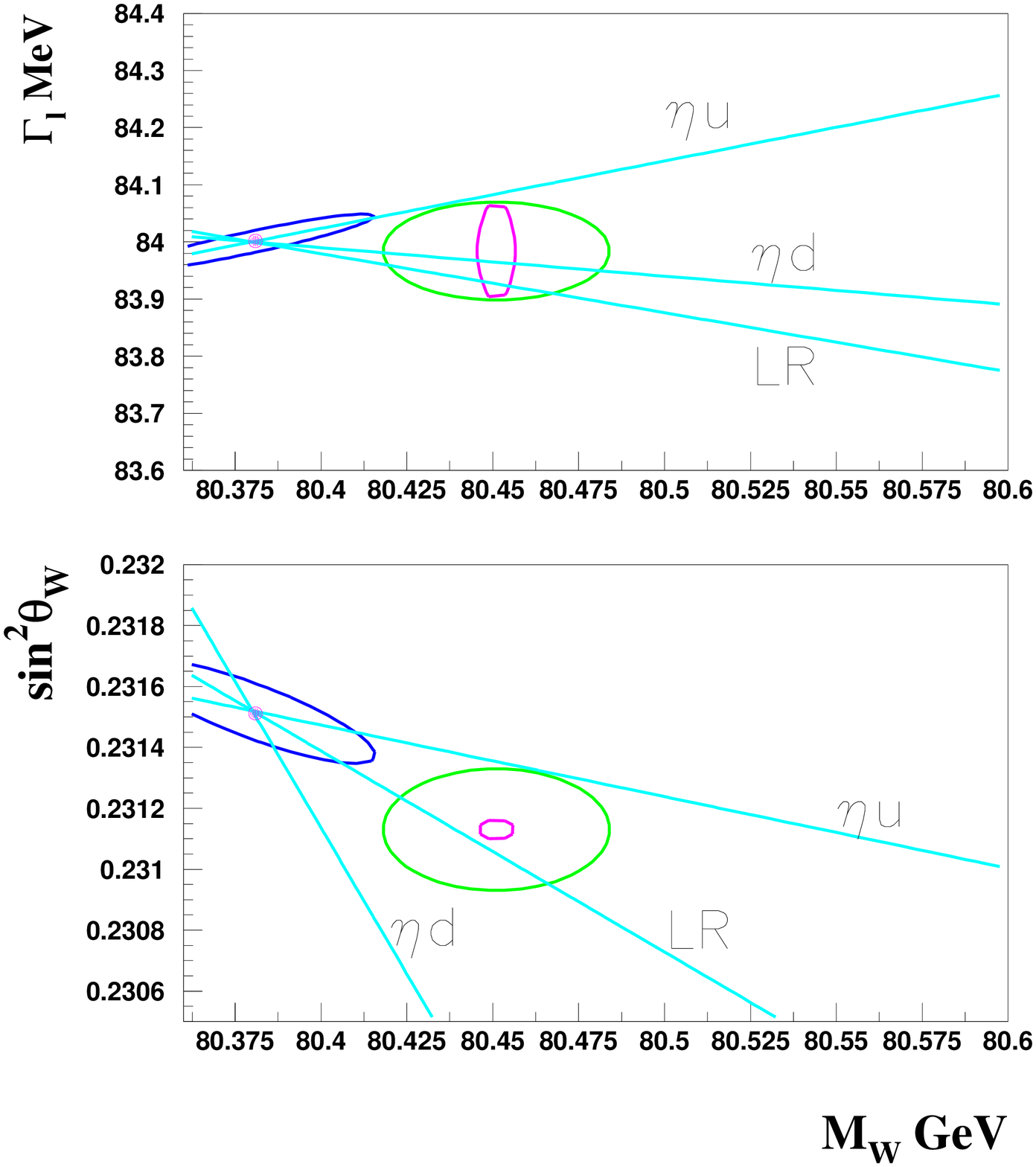}
    \includegraphics[width=0.45\textwidth,height=0.6\textheight]{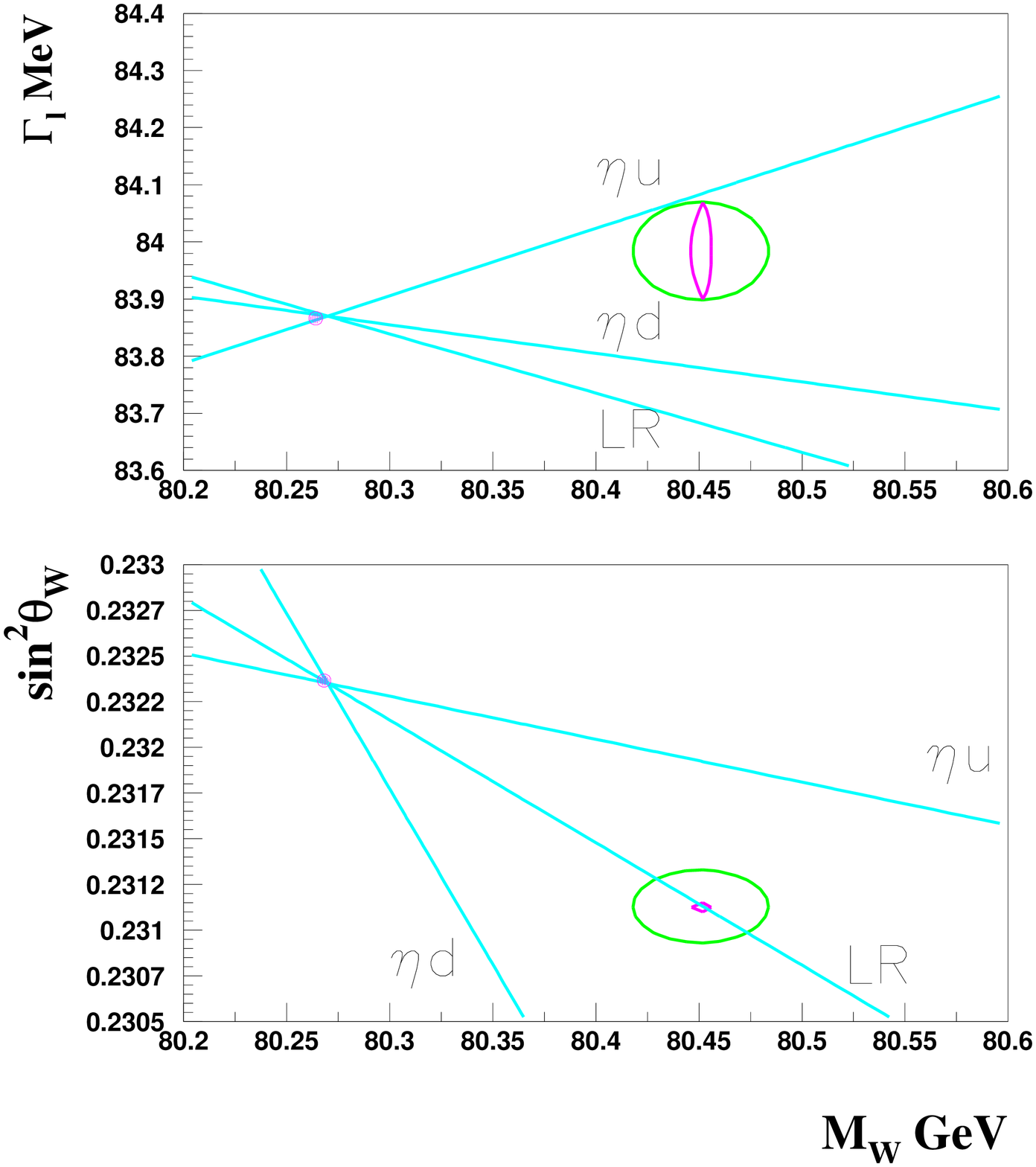}
\end{center}
\caption{ Corresponding to Figure~\ref{fig:richard1}, three Z$'$ models are 
tested against the LEP/SLD/Tevatron data 
 assuming a 115 GeV (500 GeV)
SM Higgs boson in the upper (lower) part of the figure. 
For details see \cite{ref:richard}.
\label{fig:richard2}
}
\end{figure}

Finally,
in Figure~\ref{sec5_fig:zplhclcgigaz} the mass regions covered by the LHC, LC and 
a GigaZ option at the LC for various Z$'$ models are compared and 
summarized \cite{ref:richard}, the limits for the LC are given at the 95\% C.L.
\begin{figure}[htb!]
\begin{flushleft}
    \includegraphics[width=0.9\textwidth,height=0.7\textwidth]{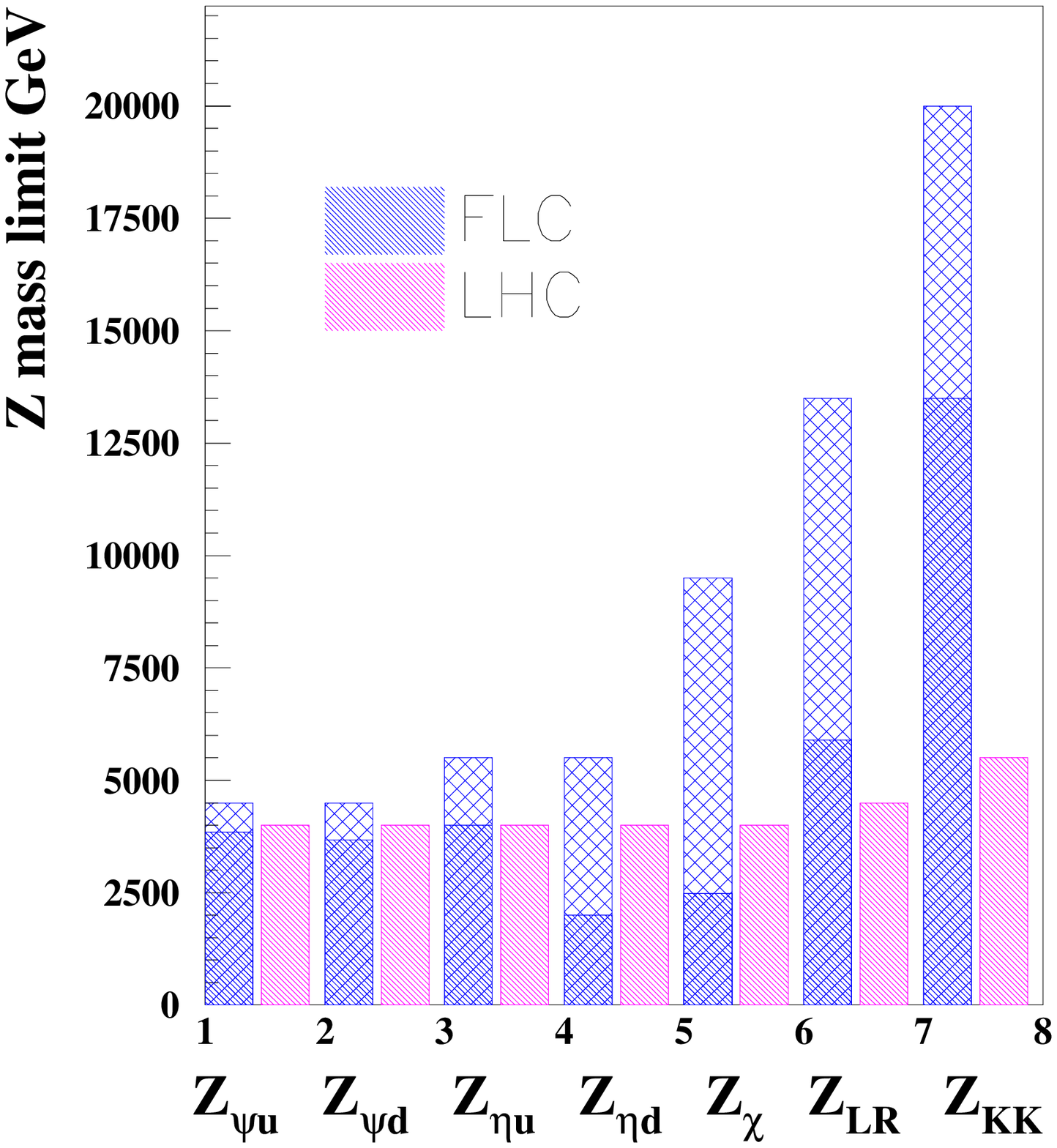}
\end{flushleft}
\caption{Mass regions covered by LHC and future LC (FLC) for a Z$'$ in various 
scenarios. For the FLC (the left tower, in blue) the heavy hatched region is 
covered by exploiting the GigaZ option (sensitive to the Z-Z$'$ mixing) 
and the high energy region (sensitive to Z-Z$'$ interference effects) 
\cite{ref:richard}.
\label{sec5_fig:zplhclcgigaz}
}
\end{figure}

\section{Z$'$ in Little Higgs models}

In 'Little Higgs' models (LHM) 
new symmetries imply the 
existence of a rich spectrum of new particles, in 
particular new gauge bosons that could be light. 
In the minimal version of LHM only a triplet and a 
singlet of gauge bosons, a triplet of Higgs bosons and a vector-like quark, 
$T$, are generated. This leads to two Z$'$: $B_H$, the U(1) singlet, and
and the triplet $Z_H$. For calculations the 'Little Higgs' formalism  
includes 3 new parameters: the mass scale $f$ and the two coupling 
ratios,  $x=g'_t/g'$ and $y=g_t/g$ \cite{ref:hewettdec}. 
Only with $t-T$ mixing significant 
contributions from the $T$ to pseudoobservables at the Z resonance 
will be expected. 

\subsection{Studies at LC}

In \cite{ref:richard} it is discussed in detail 
how Z$'$ bosons arising from LHM 
influence measurements on the Z resonance and at high energies.
It is still quite difficult to puzzle the existing precision measurements from 
LEP/SLD and the search limits from the Tevatron 
with future results from GigaZ, LC high energies and the LHC. A crucial 
point is the mass of the Higgs boson which is presently restricted to values 
below 200 GeV. Performing a fine-tuning one can find bounds on the mass scale 
$f$ as shown in Figure~\ref{fig:richard-ft}. 
(It should be remarked that 
here  the scale $f$ corresponds to the definition in 
\cite{ref:hewettdec} and therefore it is a factor $1/\sqrt{2}$ smaller 
than the $f$ used in section 3.6.)
\begin{figure}[htb!]
\begin{center} 
    \includegraphics[width=.7\textwidth]{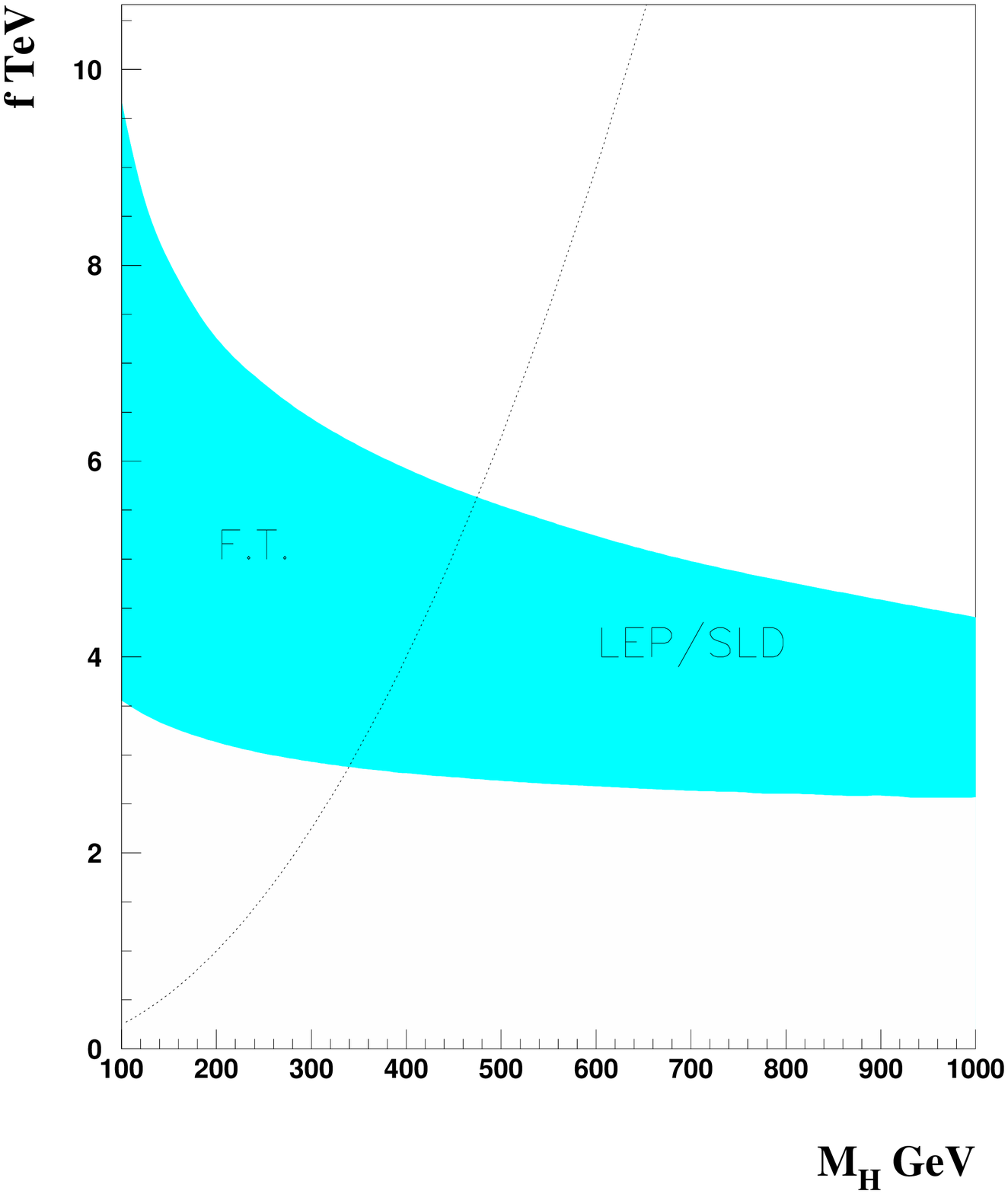}
\end{center}
\caption{Allowed regions for the mass scale parameter $f$ in the 'Little Higgs'
scenario versus the Higgs mass in the low $x$ approximation \cite{ref:richard}. 
The blue (dark) 
band is allowed by the measurements from LEP/SLD. The dotted curve indicates 
the fine-tuning limit and cuts away the LEP/SLD solution in the region marked 
F.T. The mass of B$_H$ is of the same order as $f$ if one takes $x=0.1$.
It scales like $1/x$.
\label{fig:richard-ft}
}
\end{figure}

In \cite{ref:sgodfrey} the sensitivities 
of various observables to Z$'$ bosons in the LH and KK
models are examined. Figure~\ref{fig:zp-lhm} 
shows how the modified observables influence  the $\chi^2$ distribution
and therefore it
illustrates the 
discrimination potential of the LC.  For example, the LH model gives 
rise to large LR asymmetries while the Kaluza-Klein $Z'$ does not.
The $\chi^2$ contributions  are given for 50$~fb^{-1}$ but they reflect the 
relations in general; a scaling to other luminosities and c.m.s. energies 
can be done following the well known scaling law 
$\sim (s\cdot L_{int})^{1/4}$. 
\begin{figure}[htb!]
\begin{center} 
    \includegraphics[width=.7\textwidth]{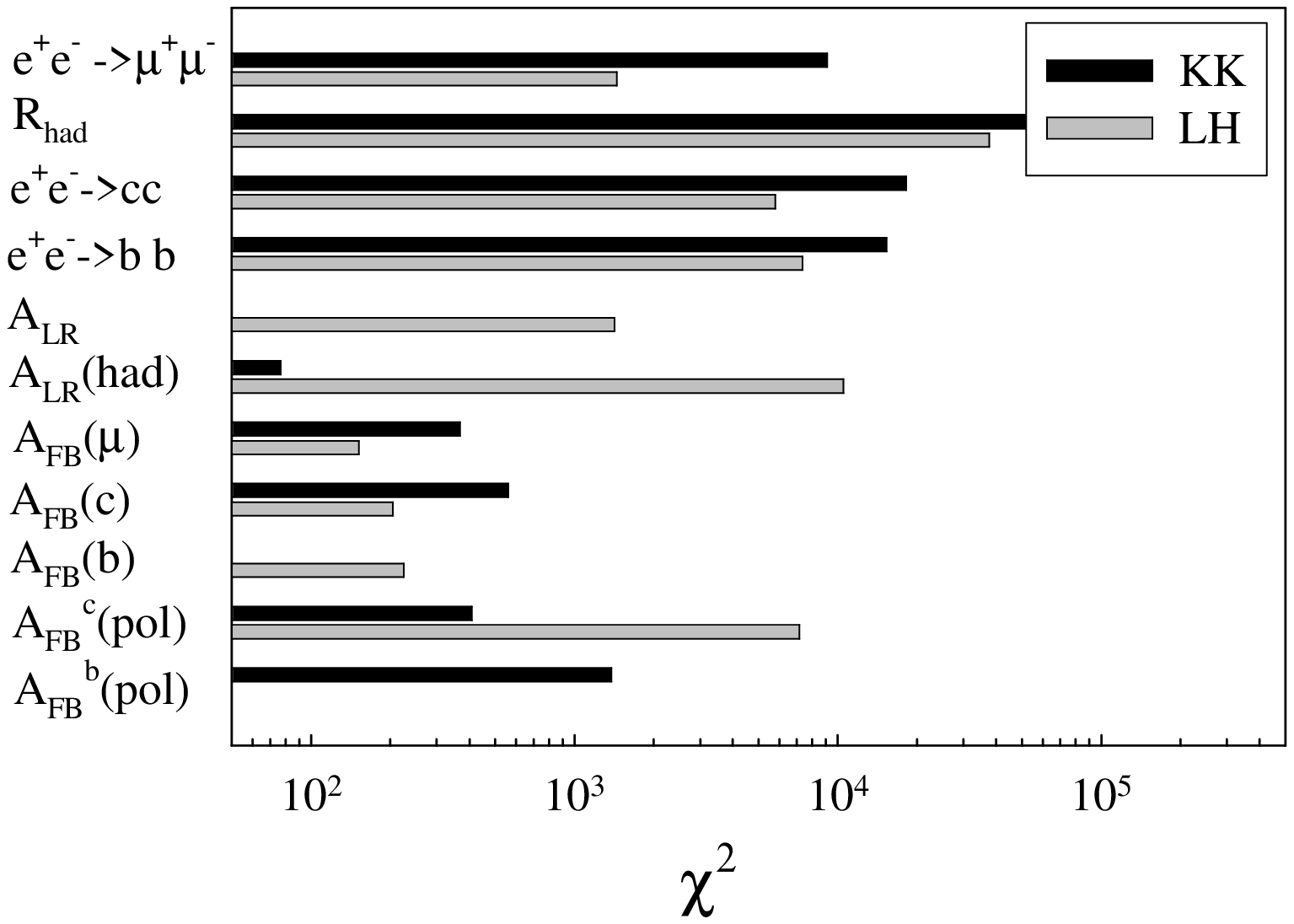}
\end{center}
\caption{The contributions to $\chi^2$ for the observables
$\sigma(e^+e^- \to \mu^+\mu^-)$, $R^{had}$, $\sigma(e^+e^- \to 
c\bar{c})$,
$\sigma(e^+e^- \to b\bar{b})$, $A_{LR}^\mu$, $A_{LR}^{had}$, $A_{FB}^c$,
$A_{FB}^b$, $A_{FB}^c(pol)$, and $A_{FB}^b(pol)$ for the Kaluza-Klein 
and Little Higgs models (see \cite{ref:sgodfrey}).  These are based on 
$\sqrt{s}=500$~GeV, L=50~fb$^{-1}$, and $M_{Z'}=2$~TeV.  The $\chi^2$ 
is based solely on the statistical error.  We assume 100\% polarization 
and do not include finite $c$ and $b$-quark detection efficiencies.
\label{fig:zp-lhm}
}
\end{figure}

\subsection{Studies at LHC}

Studies of Little Higgs Models at the LHC are described in detail in
section 3.6. 
It includes the 
heavy SU(2) gauge bosons $Z_H$ and $W_H$ that can be produced via Drell-Yan
at the LHC. 
The cross section for the  $Z_H$ production will be about 1$~fb^{-1}$  
for $M_H\approx 5$~TeV and  $\cot\theta=1$. The $W_H^{\pm}$ cross
section is expected to be   about 1.5 times that of $Z_H$ due to the
larger $W_H^{\pm}$ couplings to fermion doublets.  

\section{Charged new gauge bosons}

Limits have been placed on the existence of new charged gauge  bosons based on
indirect searches for deviations of electroweak measurements from the 
Standard Model predictions.
Limits from $\mu$ decay constrain the mass of a righthanded extra W boson,
W$'$, to $m_{W_R}>$550~GeV \cite{ref:wprime-mudecay}. 
More stringent are the results derived from $K_L-K_S$ mass splitting with
 $m_{W_R}>$1.6~TeV \cite{ref:wprime_KlKs}. 

A W$'$ search at hadron colliders considers the direct production via the  
Drell-Yan process and the subsequent decay. Present bounds from measurements at
the Tevatron collider exclude low W$'$ masses,    
$m_{W_R}>$720~GeV \cite{ref:wp-tevatron}. 
Former studies show that 
the LHC is expected to be able to detect extra charged gauge bosons up to masses of $\approx 5.9~$TeV \cite{ref:wprime-rizzo-lhc}.
In  \cite{ref:wprime-cvetic} the resolution power of experiments at the LHC is 
considered and the possible 
approaches for disentangling
the source of the new bosons are illustrated.
  
Complementary, the search for a W$'$ at e$^+$e$^-$ colliders is studied in 
\cite{ref:wp-leike-godfrey}. Based on the assumption that a W$'$ will 
be found at the LHC the process $e^+e^- \rightarrow W' \rightarrow \nu \nu \gamma$ 
is used to reconstruct the W$'$ couplings; ad-hoc systematic errors are taken 
into account.
The results demonstrate the necessary combination of LHC and LC results to 
cover all possible sources of a W$'$.

\section{Z$'$ from a Kaluza-Klein excitation}

\subsection{Discovery reach for a
Z$'$ from a Kaluza-Klein excitation}

KK excitations of the Standard Model gauge bosons are a natural 
prediction of models with additional dimensions. 
In \cite{ref:sgodfrey} the discovery limits for a Kaluza-Klein Z$'$
are derived and compared for the LC and LHC.
The discovery limits for the LHC are based on 10 events 
in the  Drell-Yan $e^+e^- + \mu^+\mu^-$ channels using the EHLQ quark 
distribution functions \cite{ref:ehlq} set 1, taking $\alpha=1/128.5$, 
$\sin^2\theta_w=0.23$, and including a 1-loop $K$-factor in the $Z'$ 
production \cite{ref:k-factor}. The
2-loop QCD radiative corrections and 1-loop QED radiative 
corrections are included in calculating the $Z'$ width.  Using different 
quark distribution functions results in a roughly 10\% variation in 
the $Z'$ cross sections \cite{ref:rizzo-kk} with the subsequent change in 
discovery limits.  
The discovery limit for the 
Little Higgs  Z$'$ 
given in Table~\ref{tab:zptable} 
is consistent with other calculations
\cite{ref:burdman,ref:han}.

At $e^+e^-$ collider searches 
the  process $e^+e^-_\lambda \to f\bar{f}$  
is used, 
$\lambda$ denotes the $e^-$ polarization; here 90\% is assumed.  There 
are numerous observables that can be used to search for the effects of 
$Z'$ bosons.
Table~\ref{tab:zptable} presents the 
95\% C.L. exclusion limits
and a comparison with  the LHC.
\begin{table} \label{tab:zptable}
\begin{center}
\begin{tabular}{l l l  l l } \hline
Collider & $\sqrt{s}$ & $L$  & KK & LHM \\
                &  &    & (TeV) & (TeV) \\ \hline
LHC $pp$ & 14 TeV & 100 fb$^{-1}$ & 6.3 & 5.1    \\
LC ($e^+e^-$) & 0.5 TeV &  1 ab$^{-1}$ & 15 & 13\\
        & 1 TeV & 1 ab$^{-1}$ & 25 & 23 \\
\hline
\end{tabular}
\caption{Discovery limits for the extra gauge bosons arising in 
theories with finite size extra dimensions and in  the Little Higgs Model 
at the LHC and the LC (\cite{ref:sgodfrey}).
}
\end{center}
\end{table}

\subsection{Distinguishing a 'conventional' and KK Z$'$}

In the simplest schemes, the 
lightest 
KK states is assumed to 
be produced  
at the LHC yet sufficiently heavy
that any higher KK excitation are not observable. 
In \cite{ref:rizzo-kk} the capabilities of the LHC (directly) and the LC 
(indirectly) are explored to distinguish
these types of states from the more conventional GUT-type Z$'$ discussed above.
It is shown  that
such a separation is straightforward at the LC while only possible under 
certain circumstances at the LHC even with a luminosity upgrade.

The model studied (see also chapter 8) 
assumes one flat extra dimension where all the fermions 
are constrained to lie at one of the two orbifold fixed points, $y=0, \pi R$,
associated with the compactification on $S^1/Z_2$ \cite{ref:4inRizzo}, where 
$R$ is the compactification radius.
Two specific cases are considered: All fermions are placed at $y=0 ~(D=0)$ or 
quarks and leptons are localized at opposite fixed points $(D=\pi R)$. Here, 
$D$ is the distance between leptons and quarks in one extra dimension.

Up to $M_c \le  7~$TeV, KK excitations are directly accessible at the LHC by 
observing a single bump in the $l^+l^-$ channel. 
Figure~\ref{fig:zp-kk-rizzo}
shows the invariant mass spectrum and the forward-backward asymmetries  
for $M_{Z'}=4~$TeV in 'conventional' models and the forward-backward 
asymmetries
for $M_c=4$~TeV for the two cases, $D=0,~ D=\pi R$.
For $D=0$ a narrow dip (destructive interference) below the KK resonance will
be obtained at $M \approx 0.55 M_c$ in the asymmetry an the invariant mass 
spectrum (see chapter 8). 
This structure with a dip position sensitive to the model 
is missing in the case $D=\pi R$ due to the constructive interference.
For either $D$ choice the cross section and asymmetry excitations are 
qualitatively different from that expected with a typical Z$'$. 
Nevertheless, it could happen that a special Z$'$ model mimics the KK cases. 
Therefore, the ability to distinguish between Z$'$ and KK scenarios 
is 
an interesting problem. 
The probabilities to identify at the LHC and at the LC
a  KK excitation as a Z$'$ for the 
two scenarios D=0 and  D=$\pi$R are summarized 
in Table~\ref{tab:prob-zp=kk-lhc}.
The separation of a Z$'$ and KK should be possible 
up to $M_c \approx 5$~TeV for sufficiently high luminosities
at the LHC.
The Z$'$-KK separation for larger values of $M_c$ is 
slightly better at a 500~GeV LC. With increasing c.m.s. energy the resolution 
power of the LC 
increases substantially compared to the LHC.  
 \begin{figure}[htb!]
 \begin{flushleft}
 \begin{tabular}{lcr}
     \includegraphics[width=0.38\textwidth,height=0.31\textwidth,angle=90]{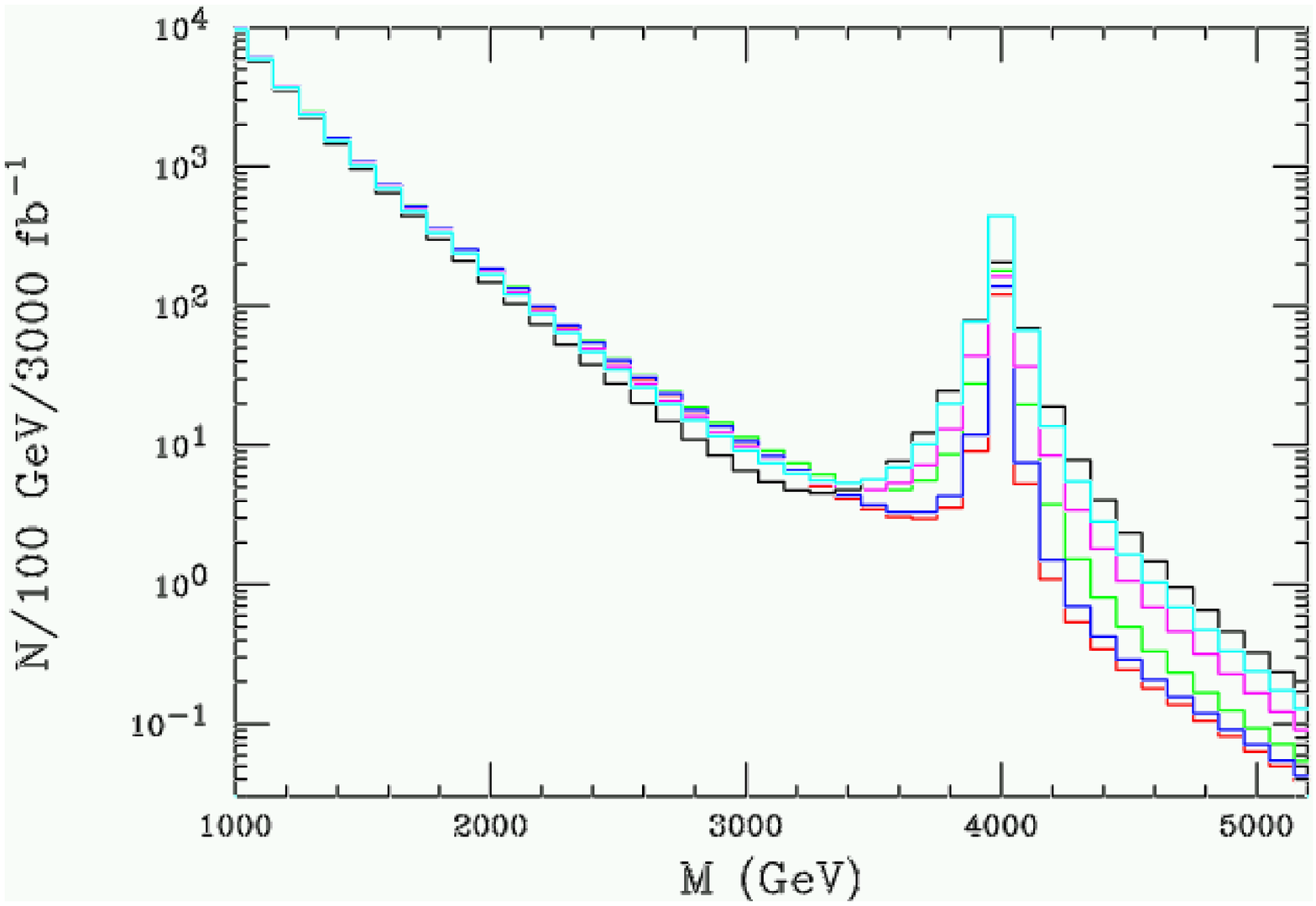}&
     \includegraphics[width=0.38\textwidth,height=0.31\textwidth,angle=90]{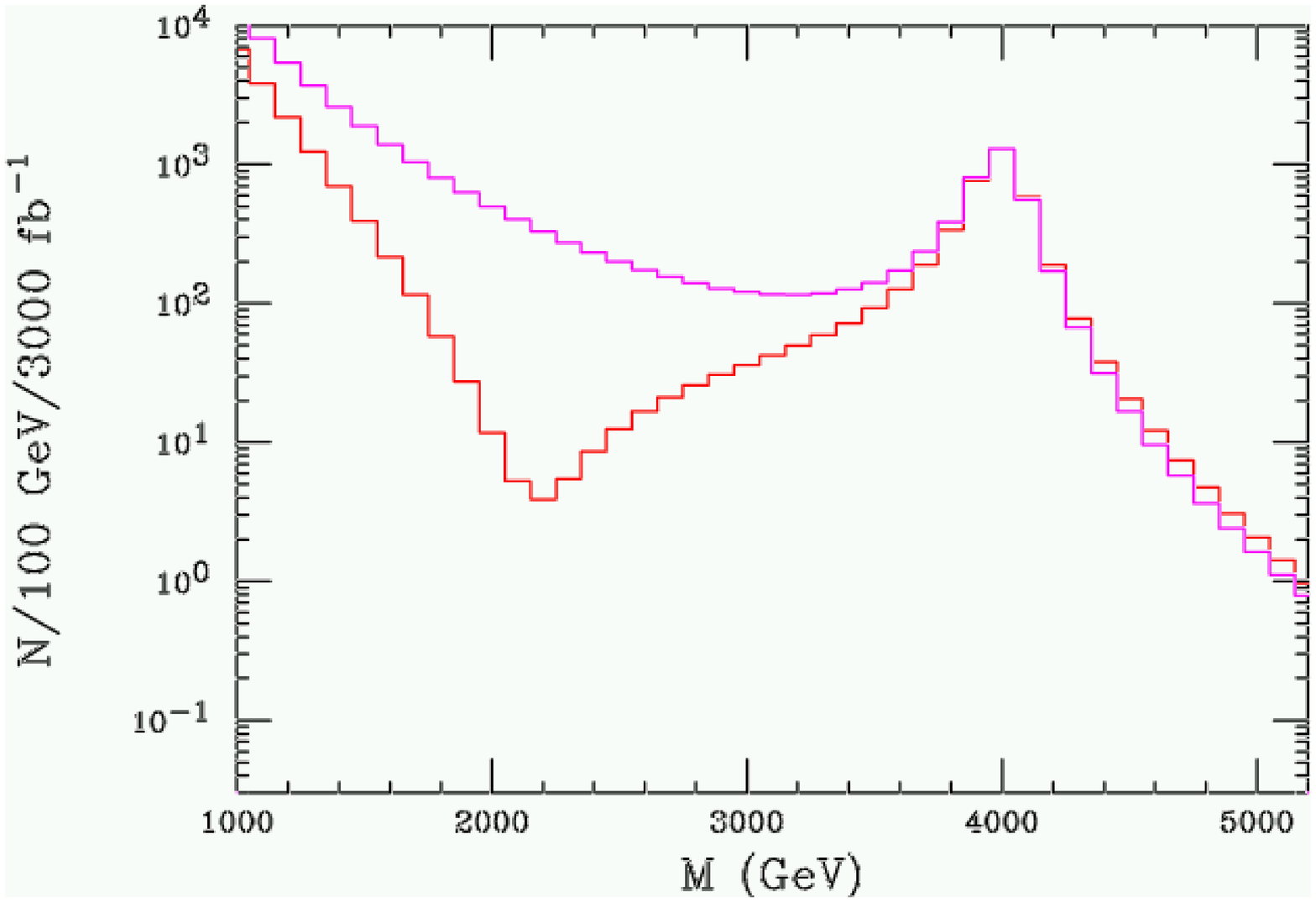}&
     \includegraphics[width=0.38\textwidth,height=0.31\textwidth,angle=90]{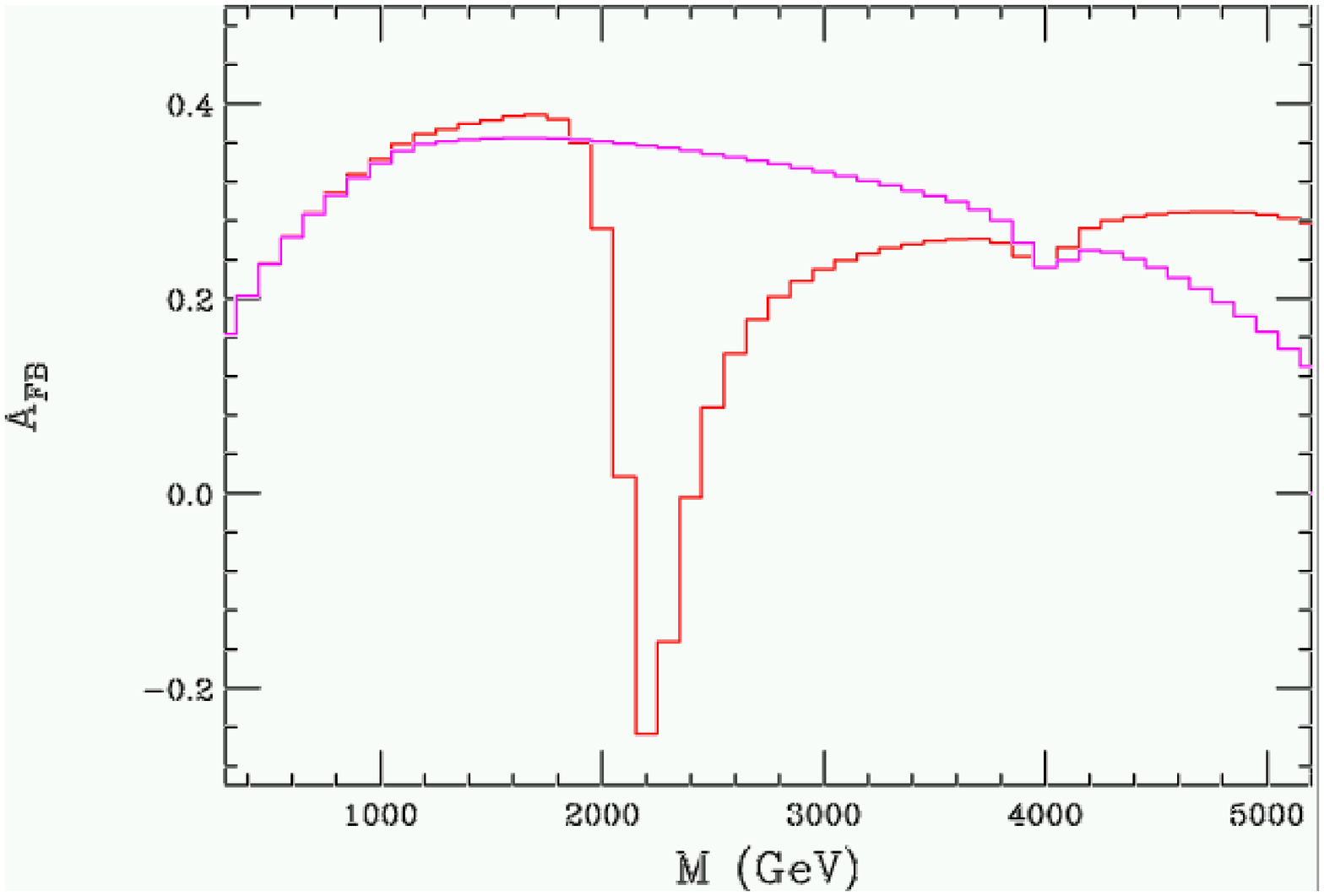}
 \end{tabular}\caption{
The lepton pair invariant mass spectrum (left) and the 
forward-backward lepton asymmetry for the production of a 4~TeV Z$'$ in 
various models (middle). The red(green, blue, magenta, cyan, black)
histograms correspond to $E_6$ model $\psi(\chi, ~\eta)$, the Left Right
Symmetric Model with $\kappa=g_R/g_L=1$, the Alternative Left Right Model and
the Sequential Standard Model, respectively. For descriptions of these
models and original references see References \cite{ref:snow},\cite{ref:leike}.
On the right: 
Forward-backward lepton asymmetry for the production of a 4 TeV KK
resonance (right) for $D=0$ (red) and $D=\pi R$ (magenta).
\label{fig:zp-kk-rizzo}
}
 \end{flushleft}
 \end{figure}
\begin{table}[htb!]
\renewcommand{\arraystretch}{1.2}
 \begin{center}
\begin{tabular}{|c|c|rr|rr|}
 \hline
 & & \multicolumn{4}{|c|}{required luminosity [$fb^{-1}$]}\\ 
\cline{3-6} 
 probability  & $M_c$ &  \multicolumn{2}{|c|}{LHC} & \multicolumn{2}{|c|}{LC} \\
of confusion  & [TeV]& $~~~~D=0$ & $~~~~D=\pi R$ &$\sqrt{s}=0.5~$TeV & $\sqrt{s}=1.0~$TeV\\ 
\hline
$5\cdot 10^{-1}$ & 4 &  30 &  800 & 190 & $<$50 \\
                 & 5 & 220 & 2500 & 500 & 130 \\
                 & 6 & $>$3000 &  & 1200& 260 \\
                 & 7 &         &  & $>$3000 &480 \\
                 & 8 &         &  &       & 820 \\
                 & 9 &         &  &       & 1300 \\
\hline
$ 10^{-5}$       & 4 &  65 &  1800 & 300 & 100 \\
                 & 5 & 450 &       & 720 & 200 \\
                 & 6 &     &       & 2400& 400 \\
                 & 7 &         &  &      &730 \\
                 & 8 &         &  &      &1250 \\
                 & 9 &         &  &      & $>$2000 \\
\hline
\end{tabular}
\end{center}
\caption{The probability for the Z$'$ hypothesis in fits to KK generated data.
Depending on the mass of the generated  KK excitation, $M_c$, 
the luminosity required at LHC and LC 
to reach a  confusion probability of $5\cdot 10^{-1}$   and of $10^{-5}$ 
is given.
For the LC the cases $D=0$ and $D=\pi R$ are here identical.}
\label{tab:prob-zp=kk-lhc}
\end{table}

\section{Summary}

With the LHC and the LC new interactions between and quarks and leptons  
will be studied. In addition, the LC is  sensitive to new phenomena in 'pure' 
electron-lepton interactions and the LHC to quark compositeness.
Depending on the model assumed and the type of fermions involved in the new
interaction a sensitivity is given up to $\approx$30 TeV at the LHC and 
roughly 100$\cdot \sqrt{s}~$ at the LC. 
The full picture of new physics demands investigations at both colliders.

If a Z$'$ exists it will be seen with the LHC up to c.m.s. energies of 5 TeV.
The reconstruction of the Z$'$ model will be possible only if $M_{Z'}<2~$TeV$ - 2.5$~TeV.

With a linear collider operating at 0.5~TeV to 1~TeV -- 
most likely below the production threshold of new gauge bosons --  
these particles will be observed indirectly.
A very good resolution of the Z$'$ models will be possible for Z$'$ bosons 
with masses below  $\approx 3 \cdot \sqrt{s}$.
But there is also a reasonable sensitivity 
to Z$'$ effects up to masses of $\approx 7\cdot \sqrt{s}-8\cdot \sqrt{s} $ that allows 
to conclude the
Z$'$ models. This 
search reach extends that of a LHC substantially.

With both together, LHC and LC, the model resolution is substantially 
improved: The knowledge of the new particle 
mass from LHC can be used for more precise distinctions of the possible new 
physics sources.

Without a LC it is more than challenging to find out the properties of new physics.
To enlighten a Confusion between models as for example in the case of KK excitations and Z$'$ bosons the LC is essential.

The clean topologies and the high luminosity at a LC allow precison measurements 
at high energies, completed by
highest precision measurements on the Z peak. 
The accuracy of these measurements extends the sensitivity widely over the 
direct search ranges of a linear collider.



\chapter{Models with Extra Dimensions}
\label{chapter:extradim}

Editor: {\it J.L.~Hewett}

\vspace{1em}

{\it M.~Battaglia, D.~Bourilkov, H.~Davoudiasl, D.~Dominici, J.F.~Gunion, 
J.L.~Hewett, T.G.~Rizzo, M.~Spiropulu, T.~Tait}

\vspace{1em}

\newcommand\Real{{\cal R \mskip-4mu \lower.1ex \hbox{\it e}\,}}
\newcommand\Imag{{\cal I \mskip-5mu \lower.1ex \hbox{\it m}\,}}
\renewcommand\ie{{\it i.e.}}
\renewcommand\eg{{\it e.g.}}
\newcommand\etc{{\it etc.}}

\newskip\zatskip \zatskip=0pt plus0pt minus0pt
\newcommand\matth{\mathsurround=0pt}
\renewcommand\lsim{\mathrel{\mathpalette\atversim<}}
\renewcommand\gsim{\mathrel{\mathpalette\atversim>}}
\def\atversim#1#2{\lower0.7ex\vbox{\baselineskip\zatskip\lineskip
\zatskip\lineskiplimit
0pt\ialign{$\matth#1\hfil##\hfil$\crcr#2\crcr\sim\crcr}}}
\def\undertext#1{$\underline{\smash{\vphantom{y}\hbox{#1}}}$}

It has been proposed that the hierarchy between the electroweak and
Planck scales may be related to the geometry of extra spatial 
dimensions.
This idea makes use of the fact that gravity has yet to be probed at
energy scales much above $10^{-3}$ eV in laboratory experiments,
admitting for the possibility that gravity behaves differently than 
expected at higher energies.  If new dimensions are indeed related to the
source of the hierarchy, then they should provide detectable signatures
in experiments at the electroweak scale.

Theoretical frameworks with extra dimensions have some general
features.  In most scenarios, our observed 3-dimensional space is
a 3-brane and is embedded in a higher $D$-dimensional
spacetime, $D=3+\delta+1$, which is known as the `bulk.'  The
$\delta$ extra spatial dimensions are orthogonal to our 3-brane.
If the additional dimensions are small enough, the Standard Model
gauge and matter fields are phenomenologically allowed to
propagate in the bulk;
otherwise they are stuck to the 3-brane.  Gravity, however, propagates
throughout the full higher dimensional volume.  Conventional wisdom
dictates that if the additional dimensions are too large then 
deviations
from Newtonian gravity would result and hence the extra dimensional
space must be compactified.  As a result of
compactification, fields propagating in the bulk expand into a
series of states known as a Kaluza-Klein (KK) tower, with the
individual KK excitations being labeled by mode numbers.  The collider
signature for the existence of additional dimensions is the observation
of a KK tower of states.  The detailed properties of the KK states
are determined by the geometry of the compactified space and their
measurement would reveal the underlying geometry of the bulk.
A review of extra dimensional models and their experimental signatures
can be found in Ref. \cite{sec6_jlhms}.

\section{Large extra dimensions}

In this scenario\cite{sec6_add} the apparent gauge hierarchy is 
generated by a large volume of the extra dimensions.  The Standard 
Model gauge and matter fields are confined to the 3-brane, while 
gravity propagates and becomes strong in the bulk.  The Planck scale 
of the effective 4-dimensional 
theory is related to the Fundamental scale where gravity becomes 
strong in the full higher dimensional spacetime, $M_D$, via Gauss' 
Law,
\begin{equation}
M^2_{\rm Pl}=V_\delta M_D^{2+\delta}\,,
\end{equation}
where $V_\delta\sim (2\pi R_c)^\delta$ is the volume of the
compactified space and $R_c$ represents the radius of the compactified
dimensions.  Taking
$M_D\sim 1$ TeV eliminates the gauge hierarchy and results in extra
dimensions of size $\sim 0.1$ mm to 1 fm for $\delta=2$ to 6.  
The states in the resulting
KK tower of gravitons are evenly spaced with masses $m_{\vec n}=
\sqrt{\vec n^2/R_c^2}$, where $\vec n$ represents the KK level
number, and couple to the wall fields with inverse
Planck strength.  The details of the KK
decomposition for the bulk gravitons and the derivation of their Feynman
rules can be found in \cite{sec6_grw,sec6_hlz}.  We note that both
gravitensor and graviscalar fields result from the KK decomposition.

\subsection{Indirect effects: graviton exchange}
\label{virt}

\renewcommand\epem{\mathrm{e^{+}e^{-}}}
\newcommand\rs{\sqrt{s}}
\renewcommand {\be}{\begin{equation}}
\renewcommand {\ee}{\end{equation}}


\newcommand {\Eqref}[1]{Equation~(\ref{#1})}
\newcommand {\Figref}[1]{Fig.~\ref{fig:#1}}
\newcommand {\Tabref}[1]{Table~\ref{tab:#1}}

One collider signature for this scenario is that of virtual
exchange of the graviton KK excitation states in all $2\to 2$
scattering processes.  This results in deviations in cross sections
and asymmetries in SM processes as well as giving rise to new
reactions which are not present at tree-level in the SM, such
as $gg\to G_n\to\ell^+\ell^-$.
In particular, as discussed in 
\cite{sec6_grw,sec6_hlz,sec6_Hewett,sec6_Shrock,sec6_Rizzo,sec6_Bourilkov:2003kj}, 
the exchange of spin-2 graviton KK states
modifies the differential cross section for fermion
pair production in $e^+e^-$ and $pp$ collisions in a unique way, 
providing clear signatures for the existence of large extra dimensions.
The exchange process is governed by the effective Lagrangian
\begin{equation}
{\cal L} = i{4\lambda\over\Lambda_H^2}T^{\mu\nu}T_{\mu\nu}\,,
\end{equation}
employing the notation of \cite{sec6_Hewett}, and is similar to a 
dimension-8 contact interaction.  $T^{\mu\nu}$ is the conserved
stress-energy tensor.
The sum over the propagators
for the full graviton KK tower is divergent for $\delta\geq 1$
and thus introduces a sensitivity to the unknown ultraviolet
physics.  Several approaches for regulating this sum may be employed.
The most model independent is the introduction of a naive 
cut-off, with the cut-off being set to $\Lambda_H$ which is of order
the fundamental scale $M_D$.  In
addition, the parameter $\lambda=\pm 1$ is usually incorporated.

\begin{table}[htb]
\renewcommand{\arraystretch}{1.20}
\caption{Sensitivity reach for extra dimensions at 95\% CL at a 
$\sqrt s = 0.5$ TeV linear collider, taking $\lambda=+1$,
as a function of the accumulated luminosity.  
From \cite{sec6_Bourilkov:2003kj}.}
\label{tab:ll-ed}
 \begin{center}
  \begin{tabular}{|r|c|c|}
   \hline
   \multicolumn{3}{|c|}{\boldmath $e^{+}e^{-} \rightarrow e^{+}e^{-}$ 
\unboldmath} \\
 & \multicolumn{1}{c}{\boldmath ``Realistic'' \unboldmath} & 
\multicolumn{1}{c|}{\boldmath Optimistic \unboldmath} \\
   \hline
Luminosity  & $\Lambda_H$        &  $\Lambda_H$         \\
$[fb^{-1}]$ & [TeV]     &  [TeV]      \\
   \hline
   \hline
    1       &     2.6      &      2.6       \\
   \hline
   10       &     3.1      &      3.5       \\
   \hline
  100       &     3.3      &      4.2       \\
   \hline
 1000       &     3.3      &      4.6       \\
   \hline
   \hline
   \multicolumn{3}{|c|}{\boldmath $e^{+}e^{-} \rightarrow 
\mu^{+}\mu^{-}$ \unboldmath} \\
 & \multicolumn{1}{c}{\boldmath ``Realistic'' \unboldmath} & 
\multicolumn{1}{c|}{\boldmath Optimistic \unboldmath} \\
   \hline
Luminosity  & $\Lambda_H$        &  $\Lambda_H$         \\
$[fb^{-1}]$ & [TeV]     &  [TeV]      \\
   \hline
   \hline
    1       &     1.6      &      1.6       \\
   \hline
   10       &     2.1      &      2.1       \\
   \hline
  100       &     2.8      &      2.8       \\
   \hline
 1000       &     3.5      &      3.5       \\
   \hline
  \end{tabular}
 \end{center}
\end{table}

\begin{figure}[!ht]
  \begin{center}
    
\resizebox{0.80\textwidth}{11.0cm}{\includegraphics{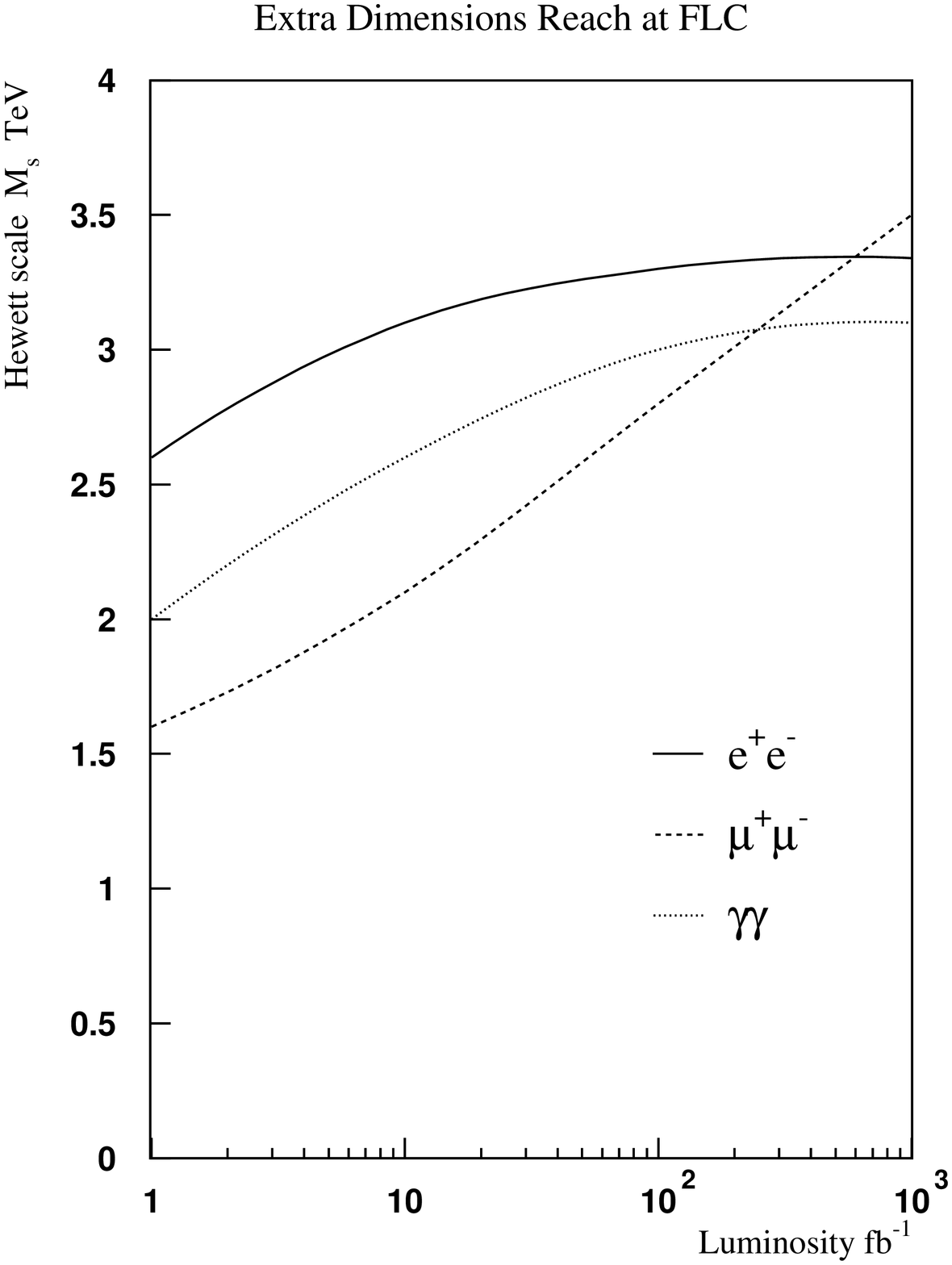}}
  \end{center}
  \caption{\em Evolution of the sensitivity reach for extra dimensions 
at a linear
           collider with \mbox{$\rs = 0.5$} TeV
           in different final states with the accumulated luminosity.
$M_S=\Lambda_H$.  From \cite{sec6_Bourilkov:2003kj}.}
  \label{fig:ed-lumi}
\end{figure}

\noindent{\bf $\bullet$ $e^{+}e^{-}$ and $\mu^{+}\mu^{-}$ Production at a 
Linear Collider}
\vspace{0.25cm}

The effects of graviton exchange at a LC with $\sqrt s=500$ GeV are 
computed here \cite{sec6_Bourilkov:2003kj}
with a semi-analytical program in the improved Born approximation,
using effective couplings.  QED effects in the initial and final states
are taken into account.  Events without substantial energy loss due
to initial state radiation are selected by a cut on the ``effective''
energy: $\sqrt s'/\sqrt s>0.85$.  With this cut, the interactions occur
close to the nominal machine energy and offer the best sensitivity
for manifestations of new physics.  
The virtual graviton effects are computed using the calculations
from~\cite{sec6_grw,sec6_Hewett,sec6_Rizzo}.  
The sensitivity to graviton KK
exchange is determined by a log likelihood fit.  Two cases are
distinguished:

1.  {\bf Realistic}:  a cross section error is composed of the
statistical error and a systematic error of $0.5\%$ coming from the
experiment, $0.2\%$ from the luminosity determination, and a
theoretical uncertainty of $0.5\%$.  The forward-backward asymmetry
error consists of the statistical error and a systematic uncertainty
of $0.002$ (absolute) for $e^+e^-$ and 0.001 (absolute) for
$\mu^+\mu^-$ final states.  The main origins of the latter are
from charge confusion of the leptons and uncertainties in the
acceptance edge determination.  Both of these effects are more
important for electrons due to the forward peak in the differential
cross section and the longer lever arm for measuring the muon
momenta.  One should stress that the forward-backward asymmetry
systematics is lower than what was achieved at LEP, and requires
a substantially improved detector.

2.  {\bf Optimistic}:  a cross section error is composed just of the
statistical error and a $0.2\%$ contribution from the luminosity
determination.  The forward-backward asymmetry error consists of the
statistical error and a systematic uncertainty which is given by the
minimum of the systematic uncertainty for the `Realistic' case and
the statistical error.  The rationale behind this is the hope that 
with increasing statistics one can better control the systematic 
effects.  In practice, this turns out to play a role only for
$e^+e^-$, as for muon pairs the statistical error of the 
forward-backward asymmetry is always larger than 0.001.

The optimization of the acceptance range is an important experimental
question.  The strong forward peak of Bhabha scattering is less
sensitive to new physics as the SM amplitudes dominate the 
interference terms.  We have investigated two regions:

$(i)$ barrel:  from $44^\circ$ to $136^\circ$, where the polar
angle is with respect to the electron beamline.

$(ii)$ barrel+backward endcap:  from $44^\circ$ to $170^\circ$, so
the region of the backward scattering is added.

The sensitivity reach from our fits \cite{sec6_Bourilkov:2003kj} 
for electron and muon pair
production are summarized
in~\Tabref{ll-ed}. They evolve from 2.6 (1.6)~TeV for electrons 
(muons) at nominal luminosity
to 3.3 (2.8)~TeV for 100~fb$^{-1}$. Here, the electron 
channel is
saturated by the systematic uncertainties, while the muon channel is 
still statistically dominated and continues to yield an improved
sensitivity for the highest luminosities.
In the Table, only the numbers for the positive interference case 
($\lambda=+1$) are shown, as
the sensitivity reach for negative interference is practically the 
same.
The results are also displayed in~\Figref{ed-lumi} and agree
with estimates from~\cite{sec6_Hewett,sec6_Rizzo,sec6_Sabine:2001}.

It is interesting to note the large difference between the
``Realistic'' and the Optimistic scenarios for the two channels: while 
the systematic errors for the electrons
start to saturate above 10~fb$^{-1}$, the muons do not show any 
saturation at all.
This is explained by the fact that for Bhabhas the sensitivity comes 
mainly from the
cross section measurement, while for muons it is dominated completely 
by the
forward-backward asymmetry.

We also investigate the optimal acceptance region for this process. 
For final state electrons, the gain in sensitivity is below 1\% from 
including
the backward endcap, so the measurement may as well be restricted to the 
barrel
region. For muons, the sensitivity comes from the asymmetry which is
best measured at lower angles, hence the results in the Table are derived 
under the
assumption that both the barrel and the two endcaps are used {\it i.e.},
from 10$^{\circ}$ to 170$^{\circ}$.
\vspace{0.25cm}

\noindent{\bf $\bullet$ $\gamma\gamma$ Production at a Linear Collider}
\vspace{0.25cm}

The production of photon pairs in $\epem$ collisions is described by
$t$- and $u$-channel QED diagrams.
The differential cross section has the following simple form
\vspace{-0.2cm}
\be
\frac{d \sigma}{d \Omega} = |e_t+e_u +  New\;Physics|^2
\ee
with
\be
\left(\frac{d \sigma}{d \Omega}\right)_{QED} = 
\frac{\alpha^2}{2s}\left[\frac{t}{u}+\frac{u}{t}\right] = 
\frac{\alpha^2}{s}\left[\frac{1 + 
\cos^2\theta}{1 - \cos^2\theta}\right]\,.
\ee
Deviations from QED typically have the form:
\vspace{-0.2cm}
\begin{eqnarray}
\label{eq11}
\frac{d \sigma}{d \Omega} & = & \left(\frac{d \sigma}{d 
\Omega}\right)_{QED} \left(1 \pm \frac{s^2\sin^2\theta}{2
({\cal L}_{\pm}^{QED})^4} \right)\,,  \\
\label{eq12}
\frac{d \sigma}{d \Omega} & = & \left(\frac{d \sigma}{d 
\Omega}\right)_{QED} \left(1 \pm {\lambda s^2\sin^2\theta\over
2\pi\alpha(\Lambda_H)^4} + ...\right)\,.
\end{eqnarray}
The QED cut-off in \Eqref{eq11} represents the basic form of possible 
modifications to quantum electrodynamics.
 \Eqref{eq12} displays the deviations specifically due to graviton 
KK exchange as calculated in \cite{sec6_grw,sec6_Agashe}.
If we ignore the contributions from higher order graviton terms 
(denoted as $...$),
the two equations predict the same angular form of modifications 
to the differential cross section.  Thus, 
it is particularly simple to compare the results from
different searches by transforming the relevant parameters;
the relation is 
$$\Lambda_H = 2.57\; {\cal L}^{QED}.$$

The sensitivity reach from our fits \cite{sec6_Bourilkov:2003kj}
is summarized in~\Tabref{gg-ed} and \Figref{ed-lumi}. 
The sensitivity
evolves from 2~TeV at nominal luminosities 
to 3~TeV for 100~fb$^{-1}$, where the process becomes
saturated by systematic effects. As in the case of
Bhabha scattering, there is no gain in sensitivity from going 
outside of the barrel
region;  in this case, the differential cross section is symmetric, 
exhibiting both a forward and a backward peak.

\begin{table}[htb]
\renewcommand{\arraystretch}{1.20}
\caption{Sensitivity reach for extra dimensions at 95\% CL at a linear 
collider
with \mbox{$\rs = 0.5$} TeV for the case $\lambda=+1$.
From \cite{sec6_Bourilkov:2003kj}.}
\label{tab:gg-ed}
 \begin{center}
  \begin{tabular}{|r|c|c|}
   \hline
   \multicolumn{3}{|c|}{\boldmath $e^{+}e^{-} \rightarrow \gamma\gamma$ 
\unboldmath} \\
 & \multicolumn{1}{c}{\boldmath ``Realistic'' \unboldmath} & 
\multicolumn{1}{c|}{\boldmath Optimistic \unboldmath} \\
   \hline
Luminosity  & $\Lambda_H$        &  $\Lambda_H$         \\
$[fb^{-1}]$ & [TeV]     &  [TeV]      \\
   \hline
   \hline
    1       &     2.0      &      2.0       \\
   \hline
   10       &     2.6      &      2.6       \\
   \hline
  100       &     3.0      &      3.4       \\
   \hline
 1000       &     3.1      &      4.1       \\
   \hline
   \hline
${\cal L}^{QED}$ 1000&  1.2      &      1.6       \\
   \hline
  \end{tabular}
 \end{center}
\end{table}

Summing over all fermion final states ($e\,,\mu\,,\tau\,,c\,,b$, 
and $t$) and examining the deviations in the total cross section,
forward-backward asymmetry, and left-right asymmetry in each case,
as well as the tau polarization asymmetry, yields the search
reach \cite{sec6_tgr_lumi} 
for $\Lambda_H$ shown in Fig. \ref{sec6_tgrmh} as a function
of integrated luminosity.  We see that for a 1 TeV LC, scales of
order $\Lambda_H\sim 8.4$ TeV can be observed with 500 fb$^{-1}$.

\begin{figure}[t]
\centerline{
\includegraphics[width=8cm,angle=90]{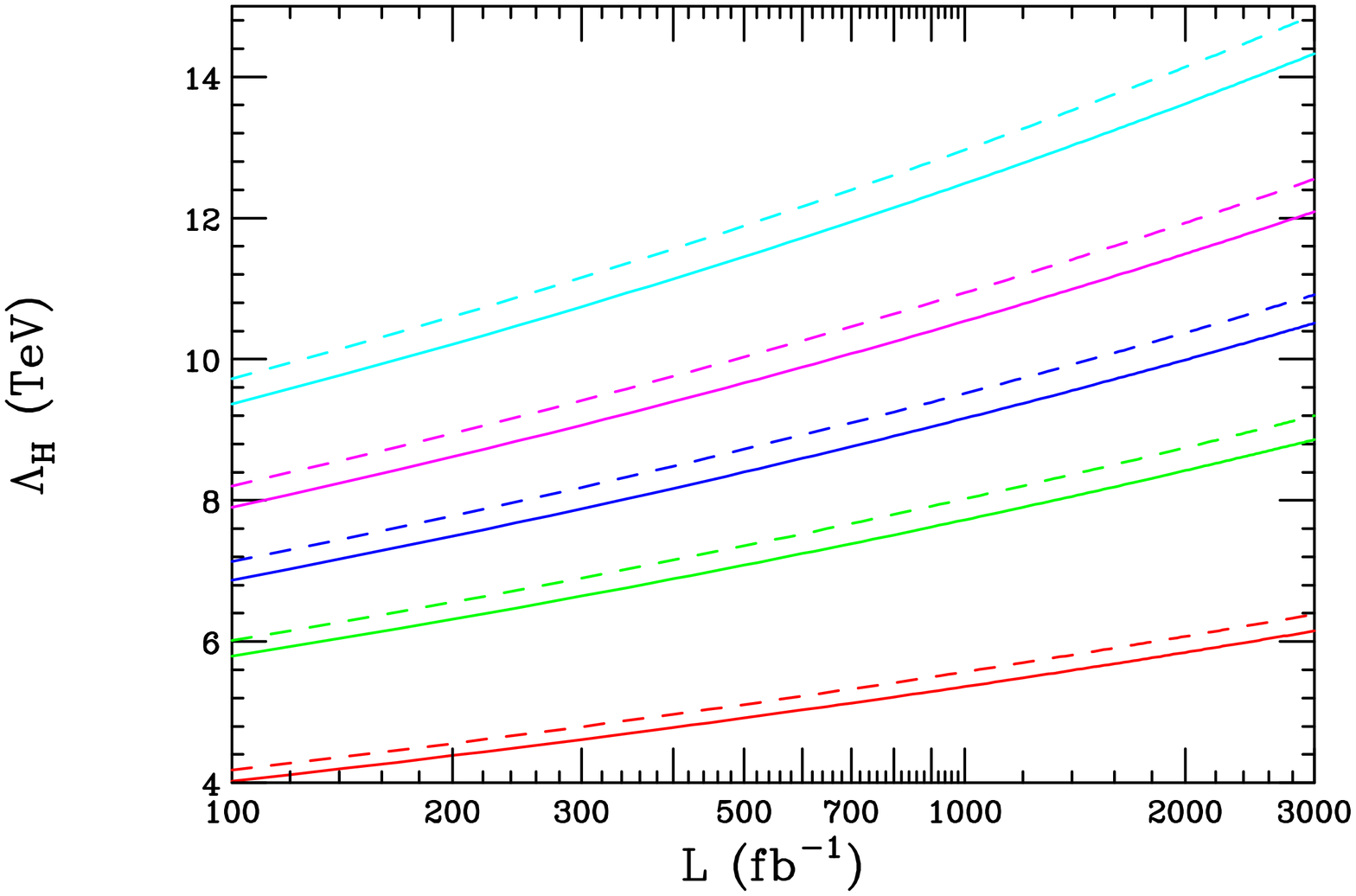}}
\vspace*{-.1cm}
\caption{95\% CL search reach for the scale associated with
the exchange of graviton KK states in $\epem\to f\bar f$.  The solid
(dashed) curves correspond to a initial state positron polarization
of 0 (60)\%.  Each pair of curves is for $\rs= 0.5\,, 0.8\,, 1.0\,,
1.2\,,$ and 1.5 TeV, from bottom to top, respectively.
From \cite{sec6_tgr_lumi}.}
\label{sec6_tgrmh}
\end{figure}

The sensitivities for virtual graviton exchange at the LHC have been 
evaluated in \cite{sec6_lhckk} and are found to be 
$\Lambda_H\leq 7.5, 7.1$ TeV for Drell-Yan and di-photon 
production, respectively.  We see that this
search region is comparable to that of a TeV-class LC.  In addition,
the spin-2 exchange also modifies the forward-backward asymmetry
in Drell-Yan production; however, the utility of this asymmetry 
in extending the search reach or for identifying the spin-2 nature
of the exchange has yet to be analyzed.

If deviations are observed at the LC or LHC from the virtual exchange 
of new particles, then it will be necessary to have techniques
available to differentiate between the possible scenarios giving
rise to the effect.  One such
method is the use of transverse polarization at the LC.
If longitudinal positron polarization is available at the LC, then 
spin rotators may be used to convert these to transversely polarized 
beams.  This allows for new asymmetries to be constructed which
are associated with the azimuthal angle formed by the directions
of the $e^\pm$ polarization and the plane of the momenta of the 
outgoing fermions in $e^+e^-\to f\bar f$.  The spin-averaged matrix
element for this process can be written as
\begin{eqnarray}
|{\cal M}|^2 & = & {1\over 4}(1-P_L^-P_L^+)(|T_+|^2+|T_-|^2)
+(P_L^--P_L^+)(|T_+|^2-|T_-|^2)\nonumber\\
& & +(2P_T^-P_T^+)
[\cos 2\phi\Real(T_+T^*_-)-\sin 2\phi\Imag(T_+T^*_-)]\,,
\end{eqnarray}
where $\phi$ is the azimuthal angle defined on an event-by-event
basis described above, $P_{L,T}^-(P_{L,T}^+)$ represent the polarizations
of the electron(positron), and $T_{+,-}$ correspond to the relevant
helicity amplitudes.
Note that the $\phi$ dependence in this expression is only 
accessible if both beams are transversely polarized.  Spin-2
exchange does not introduce any new helicity amplitudes over
those present for spin-1 exchange, but does yield
an asymmetric distribution in $\cos\theta$  unlike
the case of spin-1 exchange.  Asymmetries in the differential
cross section $d\sigma/d\cos\theta d\phi$ extend the
search reach for graviton KK exchange by more than a factor of two,
and provide an additional tool for isolating the signatures for
spin-2 exchange up to mass scale in excess of $10\sqrt s$.
These results \cite{sec6_tgr_trans} are shown
in Table \ref{sec6_tgrpol}.  If deviations due to virtual graviton
exchange were observed in, {\it e.g.}, Drell-Yan production 
at the LHC, we see that a LC with positron polarization and
$\sqrt s\geq 800$ GeV can identify
the spin-2 nature of the exchange for the entire LHC search region. 

If positron polarization is not available at the LC, then the
spin-2 nature of virtual graviton exchange can be identified
by expanding the fermion pair production cross section into
multipole moments.  Studies indicate \cite{Rizzo:2002pc} that
this technique can uniquely identify spin-2 exchange at the 
$5\sigma$ level for fundamental scales $M_D$ up to $6\sqrt s$.

\begin{table}
\centering
\begin{tabular}{|c|c|c|} \hline\hline
 $\sqrt s$ (GeV)  & Search Reach $\Lambda_H$ (TeV) & ID Reach (TeV) 
\\ \hline
500 & 10.2 & 5.4 \\
800 & 17.0 & 8.8 \\
1000 & 21.5 & 11.1 \\
1200 & 26.0 & 13.3 \\
1500 & 32.7 & 16.7 \\ \hline\hline
\end{tabular}
\caption{95\% CL search reach and identification of spin-2 exchange
from the azimuthal asymmetries discussed in the text.  From 
\cite{sec6_tgr_trans}.}
\label{sec6_tgrpol}
\end{table}

\subsection{Direct production: graviton emission}
\label{emit}

A second class of collider signals for large extra dimensions is
that of real emission of graviton states in the scattering processes
$e^+e^-\to \gamma(Z)+G_n$ and $pp\to$jet$+G_n$
\cite{sec6_grw,sec6_peskin}.  The produced graviton
behaves as if it were a massive, non-interacting stable particle
and thus appears as missing energy in the detector.  The emission
processes probe the fundamental scale $M_D$ directly.  The cross section
is computed for the production of a single massive KK excitation
and then summed over the full tower of KK states.  Since the mass
splittings in the KK tower are so small, this sum may be replaced by an 
integral weighted by the density of KK states.  Due to this integration,
the radiated graviton appears to have a continuous mass distribution;
this corresponds to the probability of emitting gravitons with 
different extra dimensional momenta.  The observables for graviton
production are then distinct from those of other physics processes
with a missing energy signature involving fixed masses for the
undetectable particles.  The expected discovery reach from this
process has been computed in \cite{sec6_teslatdr} at a 800 GeV LC
with 1000 fb$^{-1}$ of integrated luminosity and various
configurations for the beam polarization.  These results are displayed
in Table \ref{sec6_emit_tab} and include kinematic acceptance cuts,
initial state radiation, and beamsstrahlung.  In hadronic collisions,
the effective theory breaks down for some regions of the parameter
space as the parton-level center of mass energy can exceed the
value of $M_D$.  Experiments are then sensitive to the new physics
appearing above the fundamental scale that is associated with the
low-scale UV theory of 
quantum gravity.  An ATLAS simulation \cite{sec6_ianh} of
the missing transverse energy in signal and background events at
the LHC with 100 fb$^{-1}$ is presented in Fig. \ref{sec6_ian}
for various values of $M_D$ and $\delta$.  This study results in
the discovery range displayed in Table \ref{sec6_emit_tab}.  The
lower end of the range corresponds to where the ultraviolet
physics sets in and the effective 4-dimensional 
theory fails, while the upper
end represents the boundary where the signal is observable above
background.  Note that the search regions for a polarized LC
are comparable to that of the LHC.

\begin{table}
\centering
\begin{tabular}{|c|l|c|c|c|} \hline\hline
$e^+e^-\to\gamma+G_n$ & & 2 & 4 & 6 \\ \hline
LC & $P_{-,+}=0$ & 5.9 & 3.5 & 2.5 \\
LC & $P_{-}=0.8$ & 8.3 & 4.4 & 2.9 \\
LC  & $P_{-}=0.8$, $P_{+}=0.6$  & 10.4 & 5.1 & 3.3\\
\hline\hline
$ pp\to g+G_n$ & & 2 & 3 & 4\\ \hline
LHC  & & $4 - 8.9$ & $4.5-6.8$ & $5.0-5.8$ \\ \hline\hline
\end{tabular}
\caption{$95\%$ CL sensitivity to the fundamental scale $M_D$ in TeV
for different values of $\delta$,
from the emission process for various polarization configurations and
different colliders as discussed in the text.  $\sqrt s = 800 $ GeV
and 1 ab$^{-1}$ has been assumed for the LC and 100 fb$^{-1}$ for the
LHC.  Note that the LHC only
probes $M_D$ within the stated range.
From \cite{sec6_teslatdr,sec6_ianh}.}
\label{sec6_emit_tab}
\end{table}

If an emission signal is observed, one would like to determine the
values of the fundamental parameters $M_D$ and $\delta$.  The
evolution of the emission cross section with center of mass energy
in $e^+e^-$ annihilation depends quite strongly on the values
of these parameters.  This is displayed in Fig. \ref{sec6_softb},
where the cross section is normalized at 500 GeV for $M_D=5$ TeV
and $\delta=2$.  The dashed curves in the bottom panel 
also include the effects of including a finite value for the
brane tension.  A finite brane tension takes into
account the effects of a non-rigid brane and introduces the
additional parameter $\Delta$.
Measurement of this cross section at different
values of $\sqrt s$ thus determines the values of the fundamental
parameters of the theory.  We see that a long lever arm 
(corresponding to measurements at high values of $\sqrt s$)
are necessary to disentangle the effects of a finite brane
tension.

Looking at the LHC missing energy distributions displayed in  
Fig. \ref{sec6_ian}, we see that the shape and normalizations of the
curves also vary for different values of $M_D$ and $\delta$.  However,
there are uncertainties, {\it e.g.}, due to the parton densities,
associated with determining the overall normalization at the LHC.
Input from the cross section measurement at the LC for a single value of
$\sqrt s$ would help to determine the overall normalization at
the LHC.  With this information from the LC, 
the determination of the
fundamental parameters from LHC data would be improved 
\cite{sec6_albert}.  This will be quantified below.

\begin{figure}[htbp]
\centerline{
\psfig{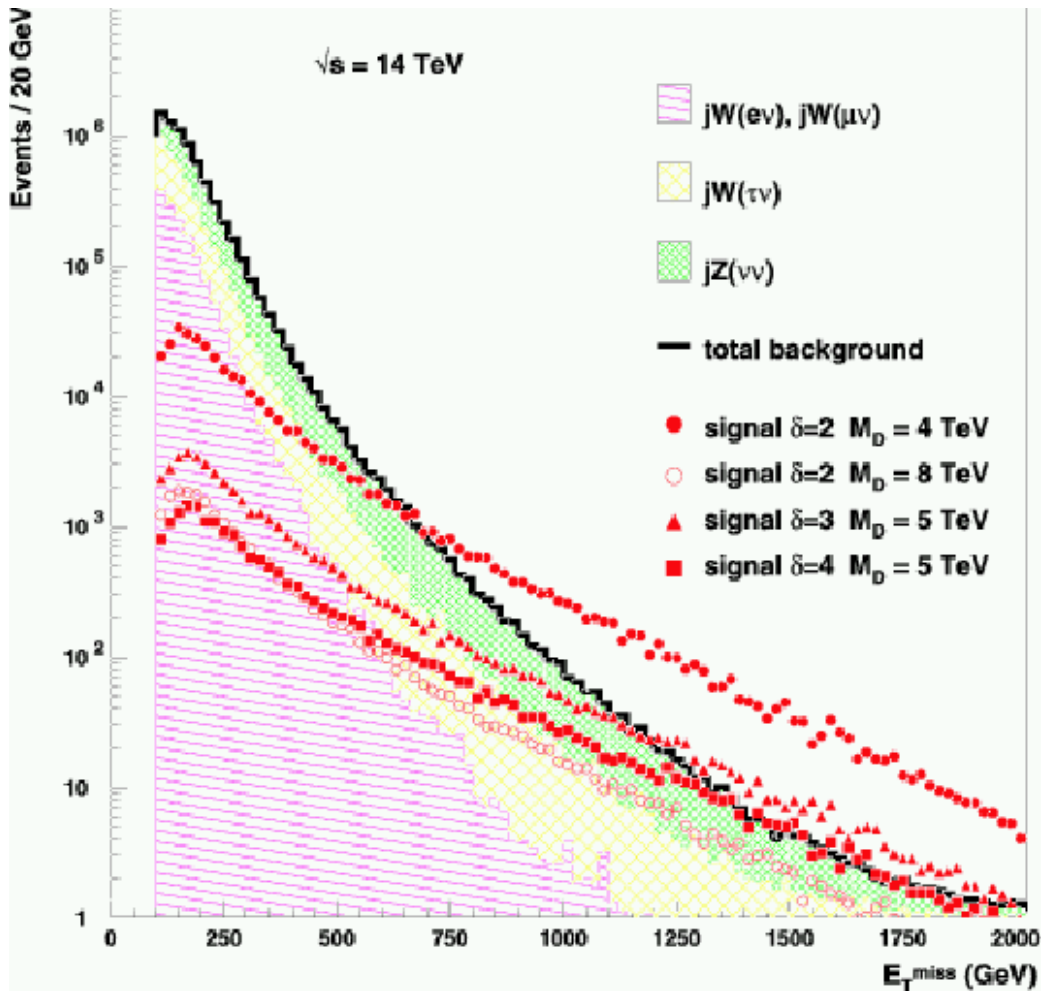}}
\vspace*{-.1cm}
\caption{Distribution of the missing transverse energy in background
events and signal events for 100 fb$^{-1}$. The contribution of the
three principal Standard Model background processes is shown as well as 
the distribution of the signal for several values of $\delta$ and $M_{D}$.
From \cite{sec6_ianh}.}
\label{sec6_ian}
\end{figure}

\begin{figure}[htbp]
\centerline{
\includegraphics[width=9cm,angle=90]{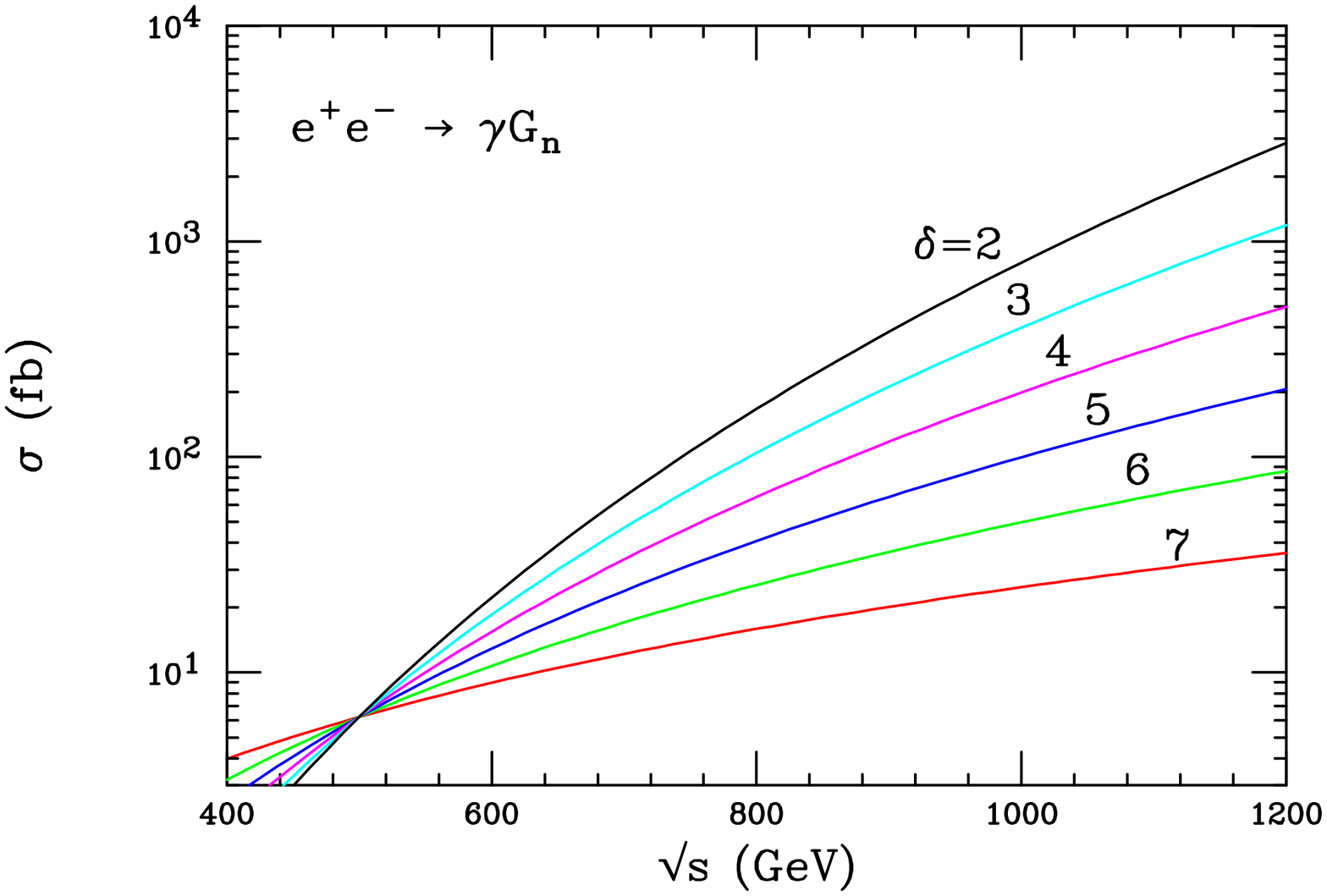}}
\vspace*{0.4cm}
\centerline{
\includegraphics[width=9cm,angle=90]{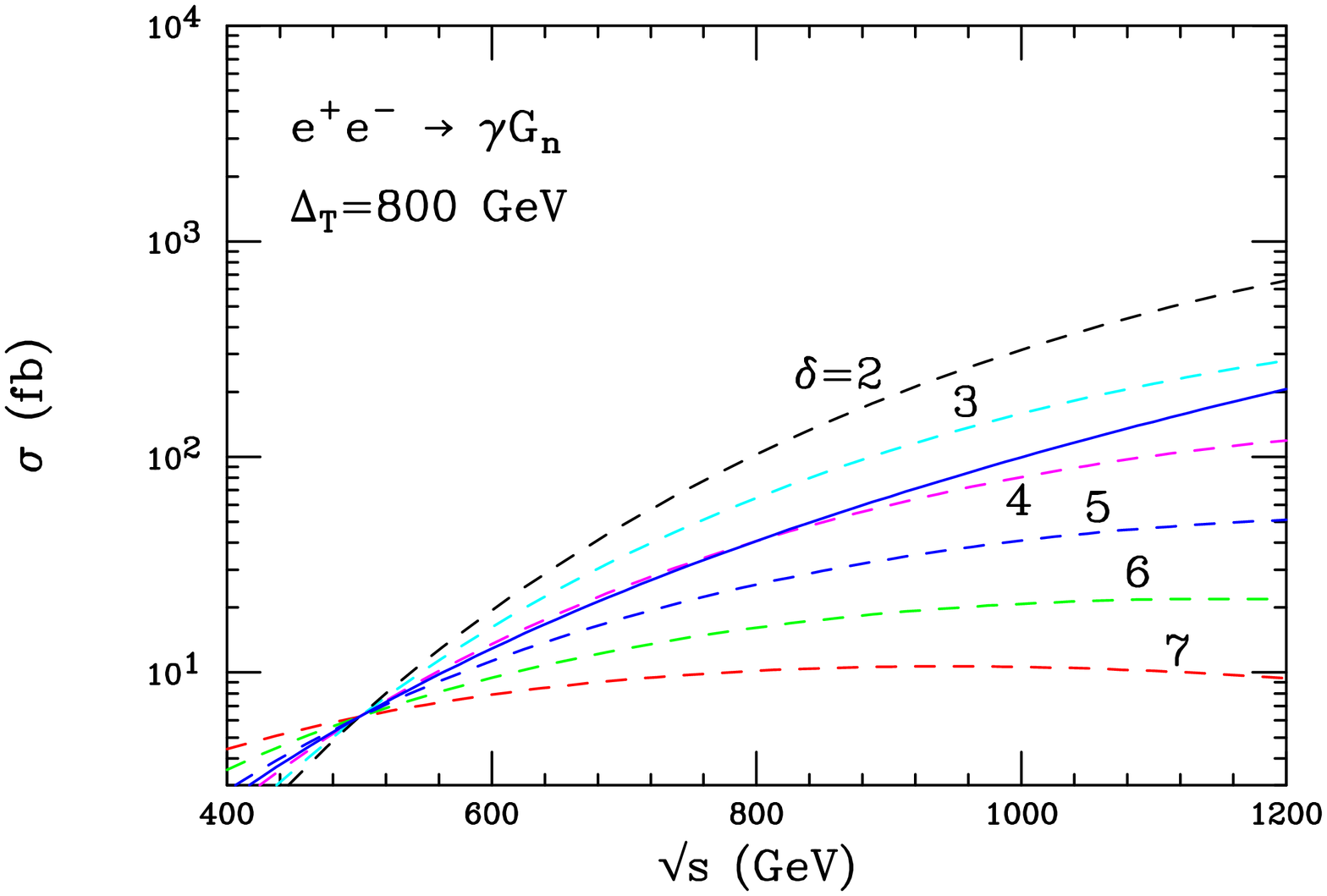}}
\vspace*{0.1cm}
\caption{Emission cross section in $e^+e^-$ annihilation as
a function of $\sqrt s$ for $\delta=2-7$ from top to bottom
on the right-hand side.  The cross sections are normalized to
$M_D=5$ TeV and $\delta=2$ at $\sqrt s$ = 500 GeV.  Top:  Brane terms
are not included.  Bottom:   The effects of finite brane tension
are included, taking
the relevant tension parameter to be $\Delta=800$ GeV.  Here, the lone
solid curve represents the case without brane term effects for $\delta=5$.
From \cite{sec6_tgrbrane}.}
\label{sec6_softb}
\end{figure}

\subsection{Graviscalar effects in Higgs production}

\newcommand\heff{h_{eff}}

Another class of signals that can arise in models with large extra dimensions
is associated with the allowed mixing between the Higgs boson 
and the graviscalar states.  Such mixing
is discussed in Sec 3.5 in this Report in the case of warped
extra dimensions.   The present scenario differs from the warped case
in that there is a KK tower of densely-packed graviscalar
states rather than a single radion.  The graviscalar KK states have
masses $m_{\vec n} = \sqrt{\vec n^2/R_c^2}$ and couple to the trace of
the stress-energy tensor with inverse Planck strength.
The presence of this graviscalar
KK tower alters the phenomenology of the Higgs boson.  For example,
instead of computing the production of a single Higgs boson, one
must now consider the production of the full set of densely spaced
mass eigenstates, all of which are mixing with one another.  Due to this
mixing, the Higgs will effectively acquire a potentially
large branching ratio to invisible final states which are
composed primarily of the graviscalars.  

The interaction between
the  complex doublet Higgs field $H$ and the 
Ricci scalar curvature $R$ of the induced 4-dimensional metric
$g_{ind}$ is given by the action
\begin{equation}
S=-\xi \int d^4 x \sqrt{g_{ind}}R(g_{ind})H^\dagger H\,.
\end{equation}
After the usual shift $H=({v+ h\over \sqrt{2}},0)$,
this interaction leads to the mixing term \cite{sec6_Giudice:2000av}
\begin{equation}
{\cal L}_{\rm mix}= -{2\sqrt 2\over \overline M_{Pl}}\xi v 
m_h^2\sqrt{{3(\delta-1)\over \delta+2}}\,  h \sum_{\vec n >0}s_{\vec n}\,.
\end{equation}
Here,  $\xi$ is a dimensionless parameter which is naturally of
order unity and
$s_{\vec n}$ represents the graviscalar KK excitations.

At colliders, one should consider the production and
decay of the coherent state $\heff= h'+\sum_{\vec n>0}s'_{\vec  n}$\,,
which is the sum of the physical eigenstates $h'$ and $s'_{\vec n}$ that
are obtained after diagonalizing the Hamiltonian. 
For a SM initial state $I$ ({\it e.g.}, 
think of $WW\to \heff$ fusion at the LHC or
$Z+\heff$ at the LC) and a SM final state $F$, the net
production cross section is given by
\begin{equation}
\sigma(I \to \heff \to F)\simeq \sigma_{SM}(I\to h \to F)
\left[{\Gamma_{h\to F}^{SM}\over 
\Gamma_h^{SM}+\Gamma_{\heff\to graviscalar}}\right]\,.
\label{sec6_xsec}
\end{equation}
Here,
\begin{eqnarray}
\Gamma_{\heff\to graviscalar}&=&
2\pi\xi^2 v^2 \frac {3(\delta -1)}
{\delta +2}
\frac {m_h^{1+\delta}}{M_D^{2+\delta}}{S_{\delta -1}}\nonumber\\
&\sim& (16\,{\rm MeV}) 20^{2-\delta } \xi^2
S_{\delta-1}\frac {3(\delta -1)}
{\delta +2} \left ( \frac {m_h}{150\, {\rm GeV}} \right )^{1+\delta}
\left ( \frac 
{3\, TeV} {M_D}\right )^{2+\delta}\,,
\label{sec6_invwidth}
\end{eqnarray}
and can be thought of an invisible width for the coherent $\heff$
state.  $S_{\delta-1}$ is the surface of a unit-radius sphere in
$\delta$ dimensions.
The net result is that the coherently summed
amplitude gives the SM Higgs cross section multiplied by
a branching ratio to the final state 
that must be computed with the inclusion of the invisible
$\heff\to graviscalar$ width.  
For graviscalar final states, the production
rate $\sigma(I\to \heff
\to graviscalar)$ is obtained by replacing $\Gamma_{h\to F}^{SM}$
with $\Gamma_{\heff\to graviscalar}$.
The branching ratio into
invisible final states, including contributions from mixing as
well as from the direct decay into graviscalars, is  
\begin{equation}
BR(\heff \to invisible)={\Gamma_{\heff\to graviscalar} \over
  \Gamma_h^{SM}+\Gamma_{\heff\to graviscalar}}\,.
\end{equation}
The rates for the usual SM Higgs decay channels
are then reduced by $1-BR(\heff\to invisible )$.
The importance of the invisible decay width is illustrated
in Figs.~\ref{sec6_mh120del2contours} and \ref{sec6_mh120del4contours} 
in the $M_D,\xi$ parameter space for $m_h=120$ GeV with $\delta=2$
and $\delta=4$, respectively. One should keep in mind
that the SM Higgs boson with mass of 120 GeV would
have a width of about 3.6 MeV.  We thus see that the graviscalar
mixing results in substantial corrections
to the expected Higgs width and a significant invisible branching fraction.

\begin{figure}[p]
\begin{center}
\includegraphics[width=7.0cm,height=6.0cm]{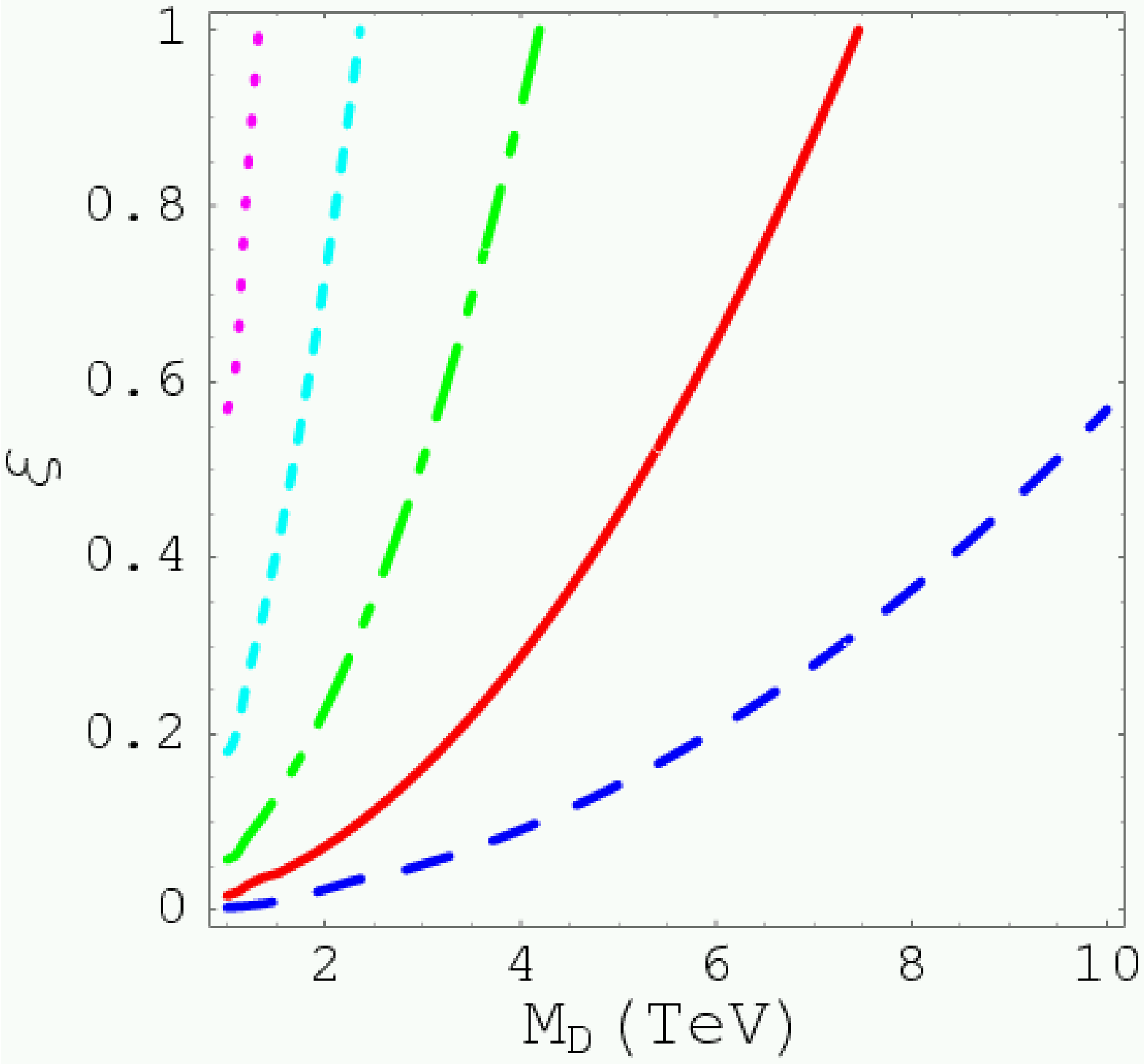}
\includegraphics[width=7.0cm,height=6.0cm]{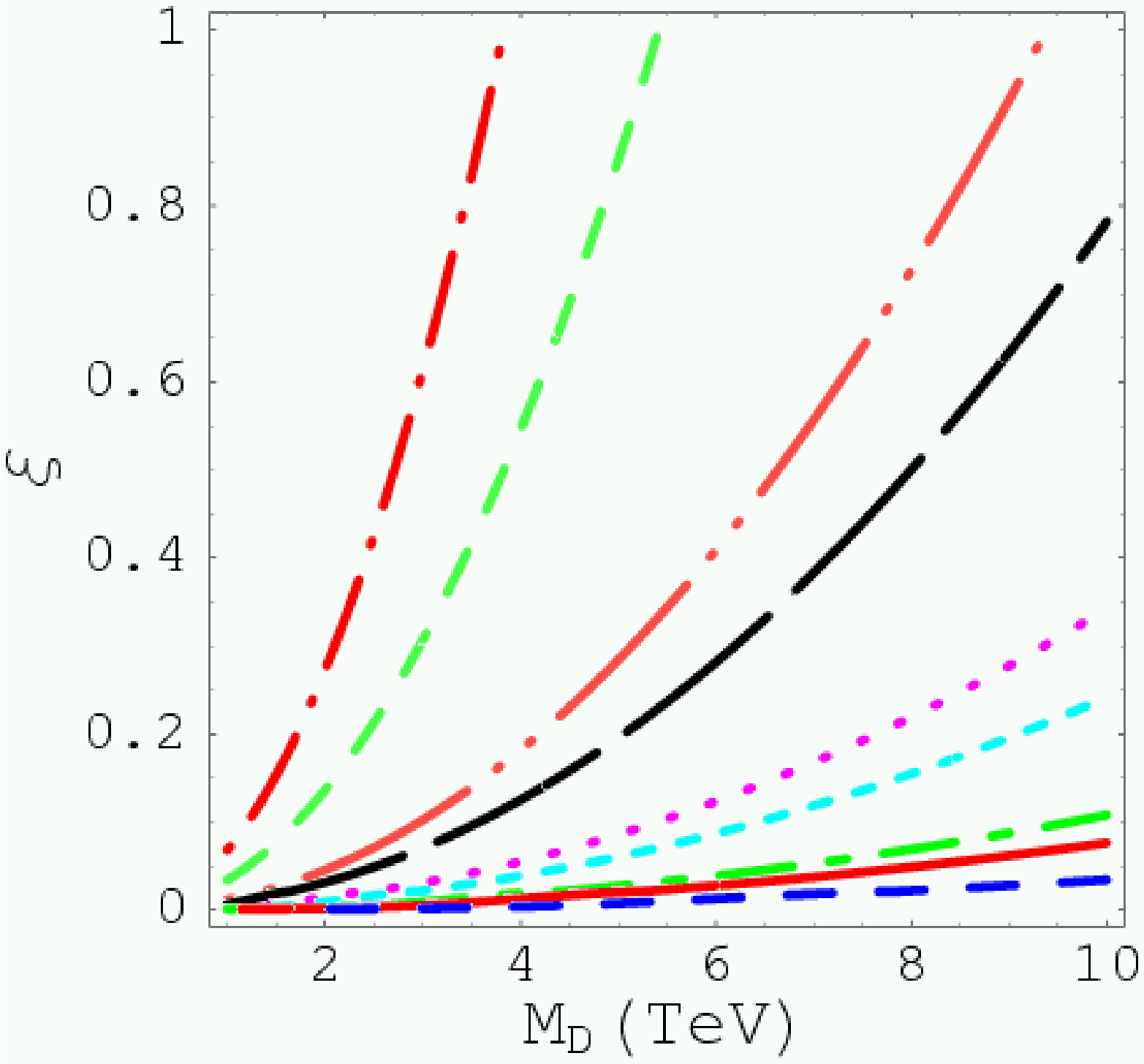}
\caption{Contours of fixed $\Gamma_{\heff\to graviscalar}$ (left) 
and fixed $BR(\heff\to invisible)$ (right) in the
$M_D$ -- $\xi$ parameter space for $m_h=120$ GeV,
taking $\delta=2$. The width contours correspond to: $0.0001$ GeV 
(large blue dashes), $0.001$ GeV (solid red line), $0.01$ GeV (green
long dash -- short dash line), $0.1$ GeV (short cyan dashes), and 
$1$ GeV (purple dots).  The $BR$ contours correspond to: $0.0001$
(large blue dashes), $0.0005$ (solid red line), $0.001$ (green
long dash -- short dash line), $0.005$ (short cyan dashes), 
$.01$ (purple dots), $.05$ (long black dashes), $0.1$ (chartreuse
long dashes with
double dots), $0.5$ (green dashes), and $0.85$ (red long dash,
short dot line at high $\xi$ and
low $M_D$).  From \cite{sec6_gunion}.
 }
\label{sec6_mh120del2contours}
\end{center}
\begin{center}
\includegraphics[width=7.0cm,height=6.0cm]{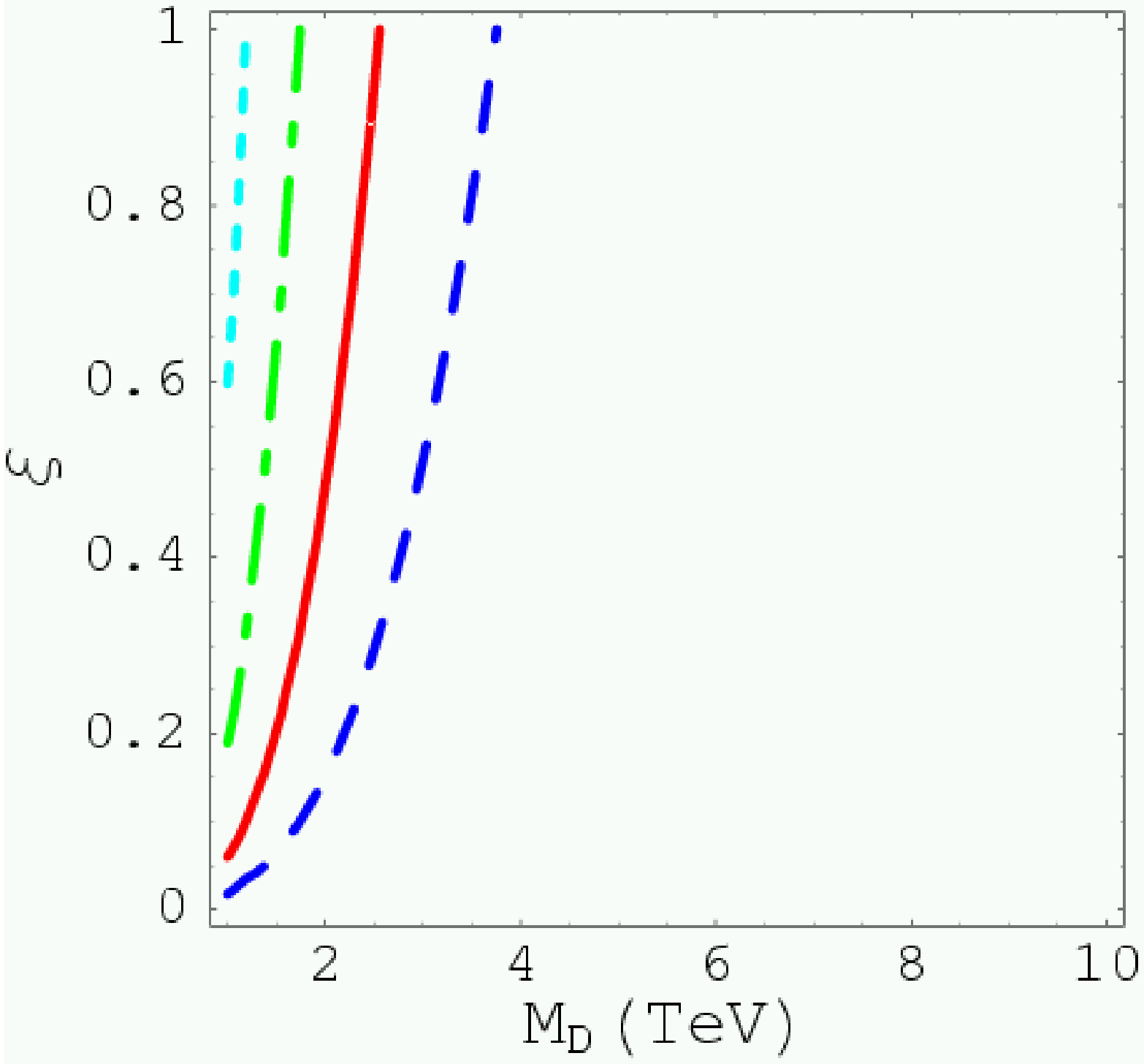}
\includegraphics[width=7.0cm,height=6.0cm]{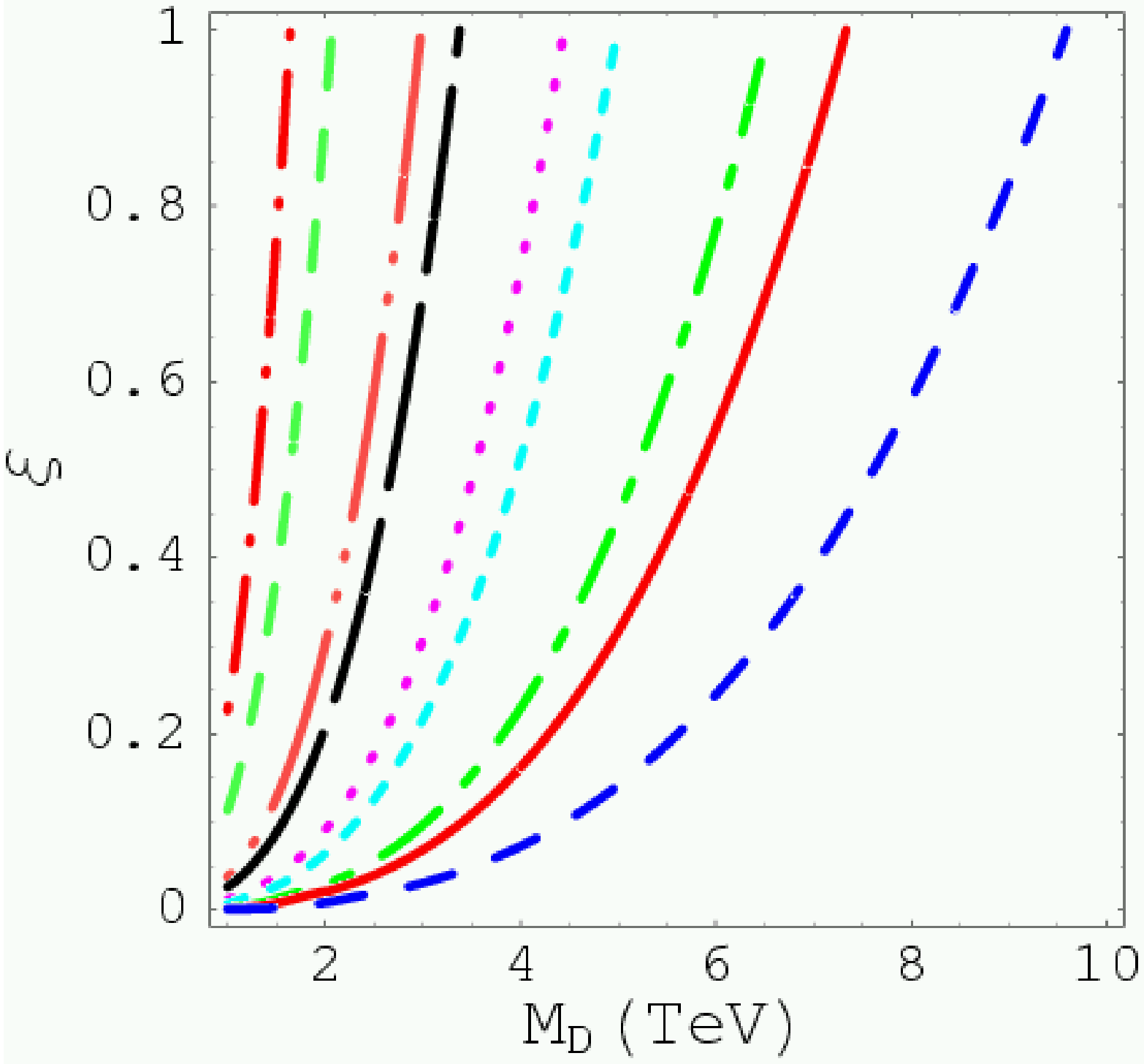}
\caption{As in Fig.~\ref{sec6_mh120del2contours} but for
  $\delta=4$.}
\label{sec6_mh120del4contours}
\end{center}
\end{figure}

Detailed studies of the SM Higgs boson signal significance, with
inclusive production, have been carried out by the 
ATLAS~\cite{sec6_atlas} and CMS~\cite{sec6_cms}  collaborations.
Here, the results of Fig.~25 in \cite{sec6_cms}, 
obtained for $L=30$ fb$^{-1}$, are employed by reducing the
SM Higgs boson signal rates in the usual visible channels by
$1-BR(h_{eff}\to invisible)$. For $L=100$ fb$^{-1}$, 
the statistical significance in each channel is simply rescaled 
by $\sqrt{100/30}$. To understand the impact of the invisible branching
ratio, we focus on $m_h=120$ GeV and perform a
full scan of the parameter space by varying $M_D$ and $\xi$ for
different values of the number of extra dimensions $\delta$. 
It is found \cite{sec6_gunion} that 
there are regions at high $\xi$ where the
significance of the Higgs boson signal in the canonical channels drops
below the 5~$\sigma$ threshold.  However, the LHC experiments will
also be sensitive to an invisibly decaying Higgs boson produced via
$WW$-fusion, with tagged forward jets. In
Ref.~\cite{sec6_cms} the results of a detailed CMS study for this mode
are given in Fig.~25 for integrated luminosity of
10~fb$^{-1}$.  These results assume a 100\% invisible branching ratio;
they are rescaled here by multiplying the signal rate by
$BR(h_{eff}\to invisible)$ and are adjusted for different
luminosities by scaling statistical significances according to
$\sqrt{L({\rm fb}^{-1})/10}$.  (The latter is probably
somewhat inaccurate 
for high-luminosity operation because of effects from pile-up.) 
The portion of the $(M_D,\xi)$ parameter
space where the $\heff $ signal can be recovered at the $5\sigma$
level through invisible decays for $L=30$ fb$^{-1}$ and 
$L=100$ fb$^{-1}$ is summarized in
Fig.~\ref{sec6_figure120}.  It is important
to observe that whenever the Higgs boson sensitivity is lost due to
the suppression of the canonical decay modes the invisible rate from 
$WW$ fusion is
large enough to still ensure detection through a dedicated analysis.

\begin{figure}[p]
\begin{center}
\begin{tabular}{c c}
\includegraphics[width=8.0cm,height=8.0cm]{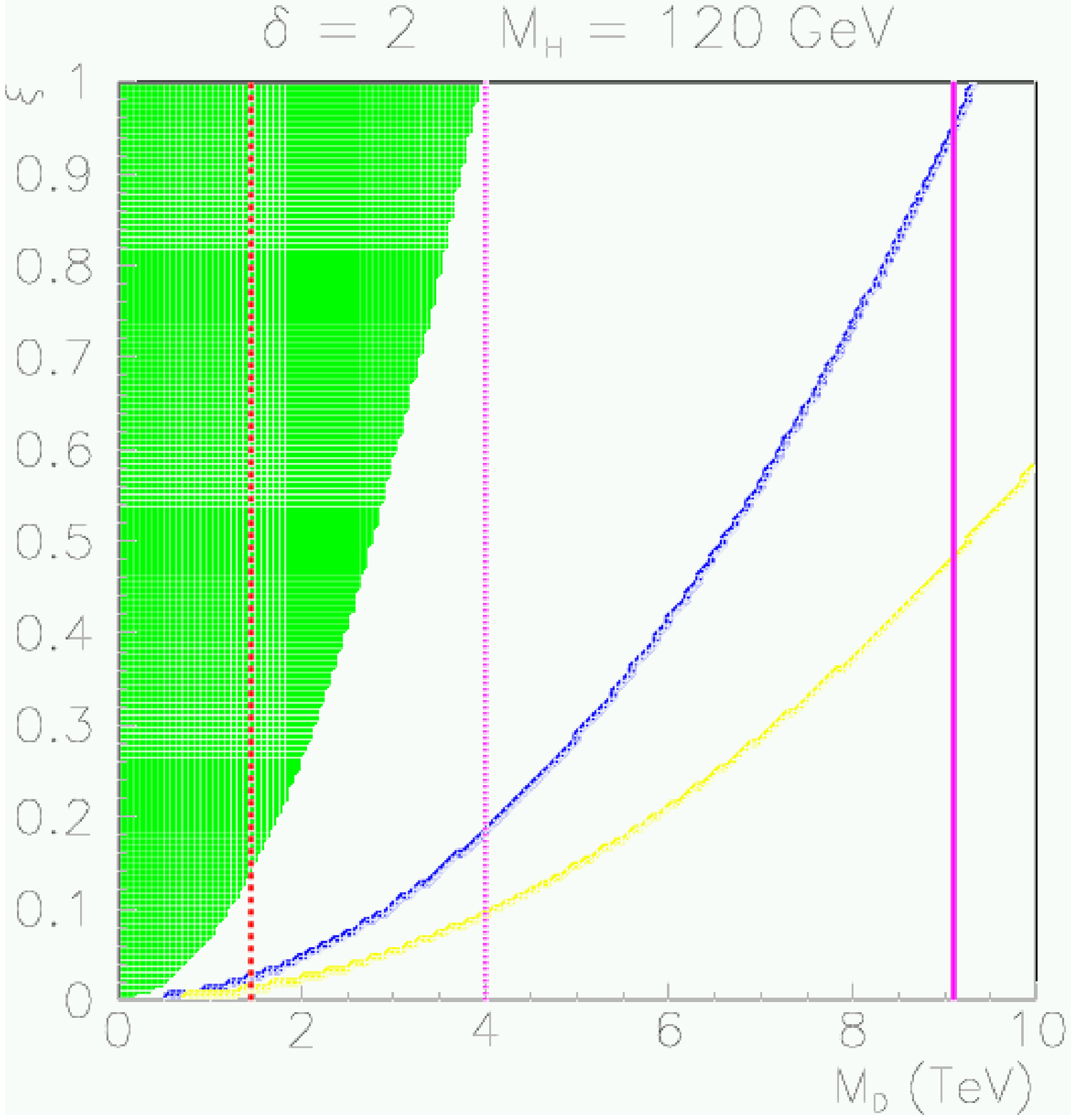} &
\includegraphics[width=8.0cm,height=8.0cm]{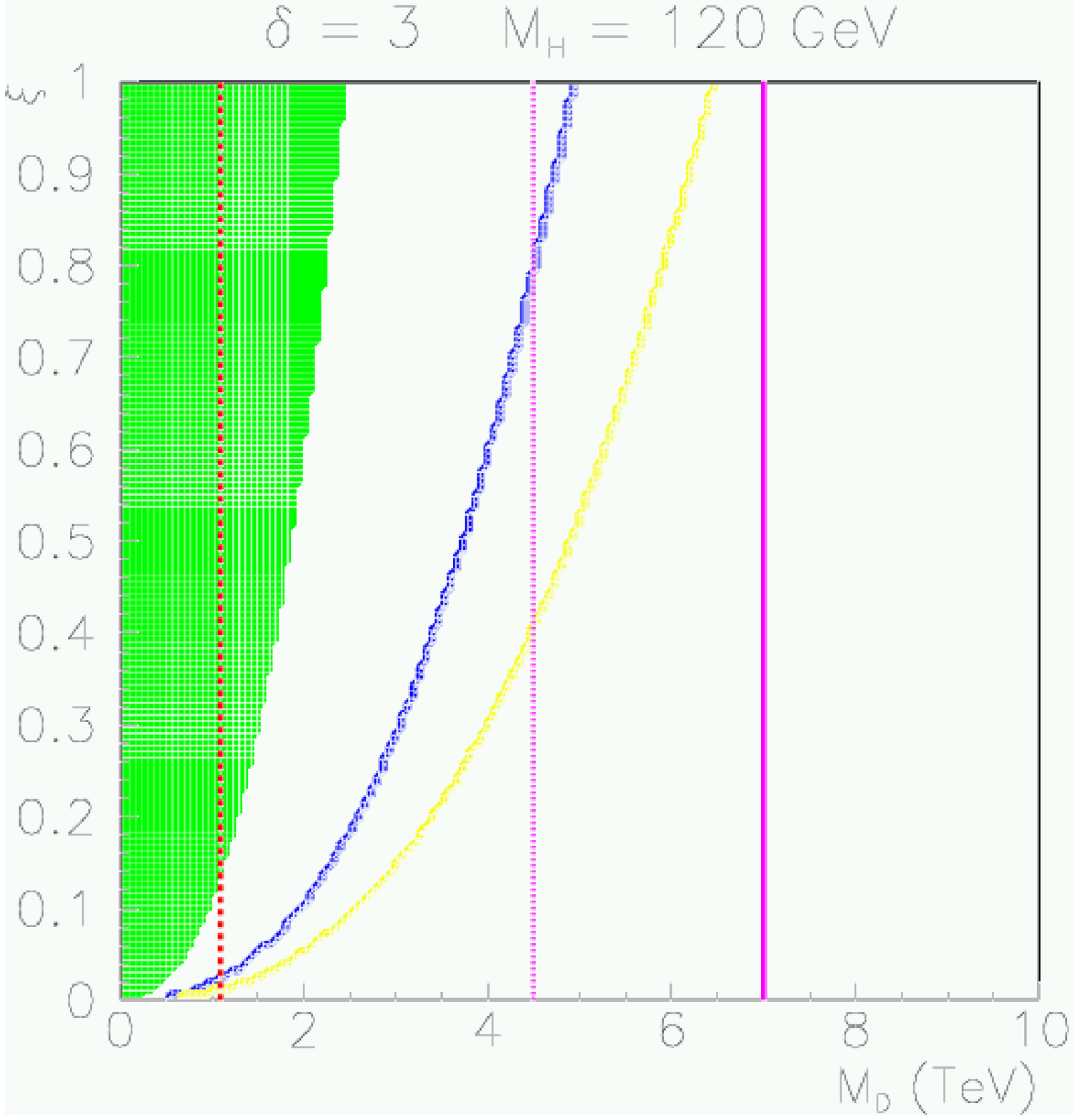}\\
\includegraphics[width=8.0cm,height=8.0cm]{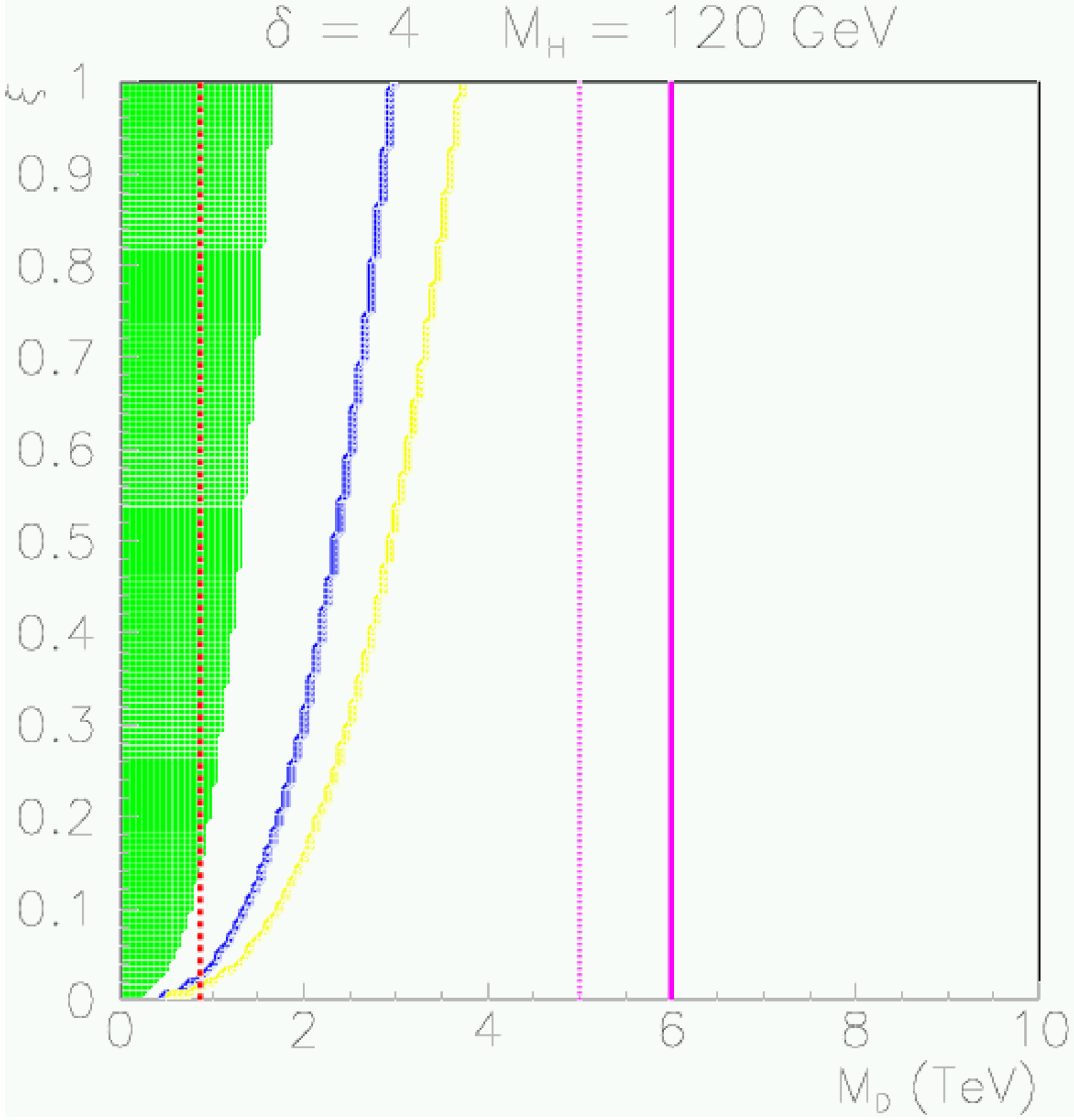}&
\includegraphics[width=8.0cm,height=8.0cm]{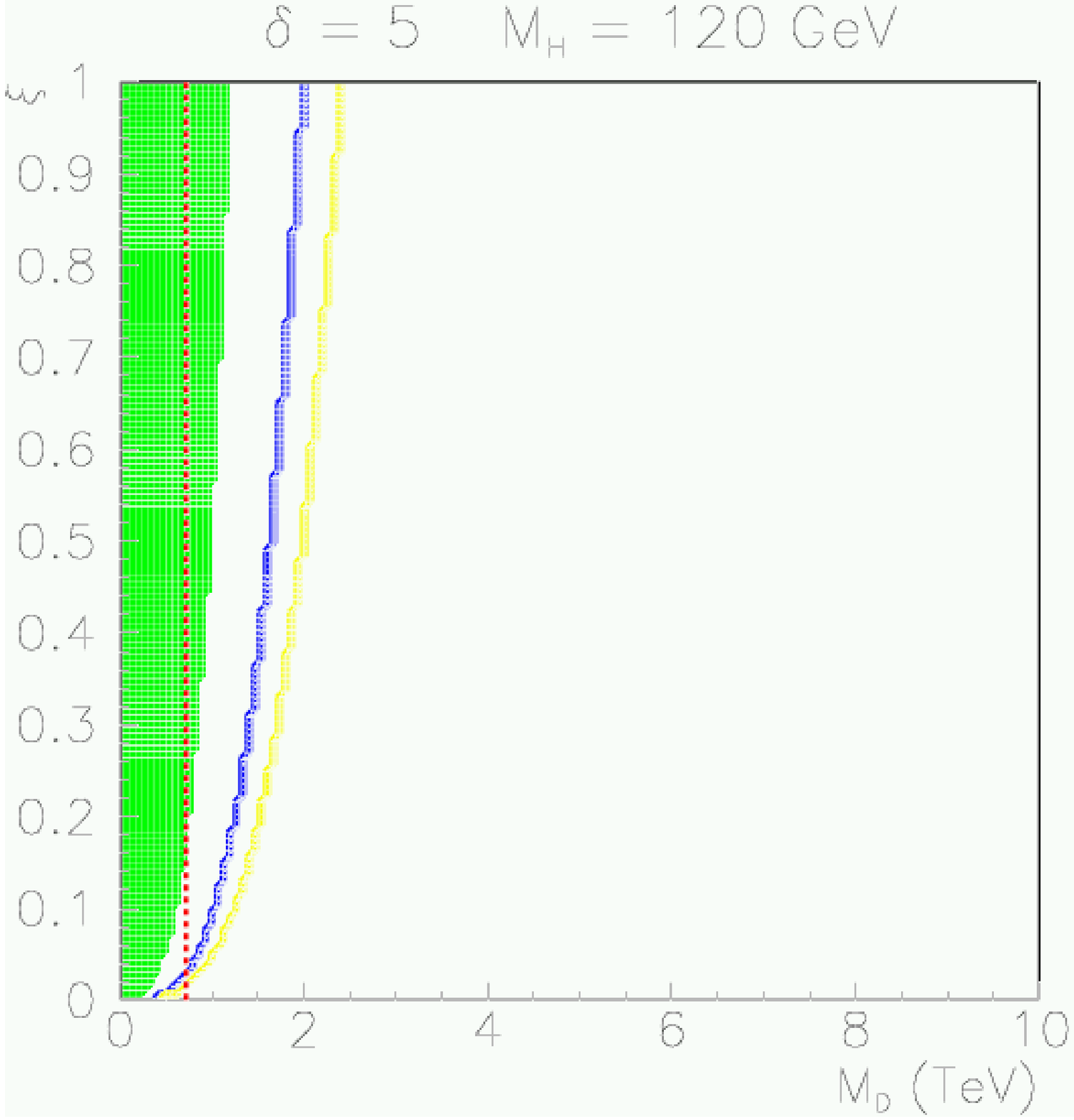}
\end{tabular}
\caption{Invisible decay width effects in the $\xi$ - $M_D$ plane for 
  $m_h$ = 120~GeV. The figures are for different values of $\delta$
  as labeled. The
  shaded regions indicate where the Higgs signal at the LHC
  drops below the 5$\sigma$ threshold in the canonical channels 
  for 100 $fb^{-1}$ of data.  The
  regions above the blue (bold) line are the parts of the parameter
  space where the LHC invisible Higgs signal in the $WW$-fusion
  channel exceeds 5$\sigma$ significance. 
  The regions above the yellow (light gray) line are the parts of the
  parameter space where the LC invisible Higgs signal will exceed
  $5\sigma$ assuming $\sqrt s=350$ GeV and $L=500$ fb$^{-1}$.
  The solid vertical line at
  the largest $M_D$ value in each panel shows the upper limit on
  $M_D$ which can be probed at the $5\sigma$ level
  by the analysis of $jets/\gamma$ with
  missing energy at the LHC.  The middle dotted vertical line 
  shows the value of $M_D$ below which the
  theoretical computation at the LHC is unreliable  (For
  $\delta=5$, there is no value of $M_D$ for which the LHC computation
  is reliable.) The dashed vertical line at the lowest $M_D$ value is
  the 95\% CL lower limit coming from combining Tevatron and LEP/LEP2 
  limits.  From \cite{sec6_gunion}.}
\label{sec6_figure120}
\end{center}
\end{figure}

As discussed in the previous section, 
the production of $jets/\gamma +$ missing energy
from graviton emission 
is also sensitive to the values of $M_D$ and $\delta$.
For comparison, the $5\sigma$ search range given in Table \ref{sec6_emit_tab} 
for $M_D$ is also displayed in Fig.~\ref{sec6_figure120}.
Also shown is the 95\% CL lower limit on $M_D$ coming from
the combination of LEP, LEP2 and Tevatron data, as summarized in
\cite{sec6_Giudice:2003tu}.  From Fig.~\ref{sec6_figure120} we observe that the
invisible Higgs decay width is predicted to probe (at $5~\sigma$)
parts of parameter space where the $jets/\gamma$ + missing energy signature
either falls below this level or is not reliably computable.

A TeV-class $e^+e^-$ linear collider will be able to see the $\heff $
Higgs signal regardless of the magnitude of the invisible branching
ratio simply by looking for a peak in the $M_X$ mass spectrum
in $e^+e^-\to ZX$ events.
As shown in \cite{sec6_schumacher}, a substantial signal
for the case where $X$ is an invisible final state
is possible down to fairly low values of $BR(\heff \to invisible)$.
We have employed the $\sqrt s =350$ GeV, $L=500$ fb$^{-1}$ results of
\cite{sec6_schumacher} to determine the portion of $(M_D,\xi)$ parameter
space for which the invisible Higgs signal will be observable at
the LC at the $5\sigma$ or better level.  This is the region
above the light gray (yellow) curves in Fig.~\ref{sec6_figure120}. 
The LC will be able to detect this signal over a
larger part of the parameter space than can the LHC.

\subsection{Determination of the model parameters}

In principle, the graviton emission signature at the LHC and LC can
be used to provide information on $M_D$ and $\delta$ and then, as
discussed above, the
Higgs signals in visible and invisible channels can be used to further
constrain these parameters and to determine $\xi$.  However, as
discussed in section \ref{emit}, the LHC
has two difficulties  regarding the use
of the $jets/\gamma +$ missing energy signal.  First, it is not possible to
measure the cross section at different center of mass energies in
a controlled fashion, and second, the effective theory breaks down when the
parton center of mass energy exceeds the value of $M_D$.
Thus, although the LHC
may well see an excess in this channel from graviton emission, its
interpretation in terms of $M_D$ and $\delta$ will be highly ambiguous.
The situation at the LC for the $\gamma/Z+$ missing energy signal is completely
different. There, the machine energy can be controlled to probe
several fixed energies (below $M_D$), and as discussed previously
(see Fig. \ref{sec6_softb}) the ratio of the cross
sections at two different energies can determine $\delta$
and the absolute normalization
of the cross sections can determine $M_D$.

With regard to the Higgs sector, a crucial first
test for this model will be
to determine if the $e^+e^-\to  ZX$ events at the LC exhibit
a resonance structure with the predicted rate for a SM Higgs
with an observed peak mass.
This can be done at about the 3\% level.  If such a peak is observed
with SM normalization,
then a determination of the parameters for large extra dimensions can be
performed.  Without the LC, there will be no
decay-mode-independent means for checking that the Higgs 
is produced with a SM-like rate.  This can only be accomplished 
by checking for consistency of the rates for visible and invisible
final states in various production modes 
with the prediction from large extra dimensions 
that the standard visible states
are reduced in rate from the SM prediction 
by the uniform factor of $[1-BR(\heff \to invisible)]$. 

The parameter determination at the LHC is computed \cite{sec6_gunion}
assuming that the production cross section for the Higgs
signal in each of the many production modes studied by ATLAS and CMS
is SM-like.  The errors on the parameter determination will be
somewhat increased if one allows for the possibility of non-SM
production rates. Thus, the results presented here 
for LHC operation alone are somewhat optimistic.  The
procedure is as follows: 

\begin{itemize}
\item The $jets/\gamma+$ missing energy signal from graviton emission
is not included in the determination of $M_D$ and
$\delta$ due to the ambiguities discussed above.
\item For the Higgs signal in visible channels,  the SM ATLAS and CMS 
results are rescaled according to $1-BR(\heff\to invisible)$.
\item For the Higgs signal in the invisible final state, 
  the detailed results of \cite{sec6_Eboli:2000ze} (used in
the CMS analysis \cite{sec6_cms}) are employed.  In this reference, 
the Higgs signal  
and background event rates are given for
the $WW\to Higgs \to invisible$ channel assuming SM production
rate and 100\% invisible branching ratio.  Here,
the signal rate is rescaled using $S_{inv}^{\heff}=BR(\heff\to
  invisible)S_{inv}^{SM}$ and  
the error in the signal rate is computed as $[\Delta S_{inv}^{\heff }
]^2=S_{inv}^{\heff}+B_{inv}$. 

\end{itemize}

The LC will be able to
improve the determination of the model parameters 
considerably with respect to the LHC alone.  In the present analysis
\cite{sec6_gunion}, 
the Higgs signals in both visible and invisible final states
as well as the $\gamma+$ missing energy signal for graviton
production have been employed. 
\begin{itemize}

\item

For the process of graviton emission, the results from 
the TESLA study \cite{sec6_teslatdr} (see Table \ref{sec6_emit_tab})  
have been used.  The
$e^+e^-\to \nu_e\bar \nu_e+\gamma$ background has been computed using
the {\tt KK}  \cite{sec6_Ward:2002qq} and 
{\tt nunugpv}~\cite{sec6_Montagna:ec}
simulation programs. Results from the two programs agree well.
It is assumed that measurements are performed at both
$\sqrt s =500\,, 1000$ GeV   with integrated
luminosities of $1000$ fb$^{-1}$ and $2000$ fb$^{-1}$,
respectively.  (Results obtained for $500$ fb$^{-1}$ and 
$1000$ fb$^{-1}$,
respectively, are not very different.)

\item For the invisible Higgs signal, the results of
  \cite{sec6_schumacher} are employed from which the fractional error for
  the measurement of any given $BR(\heff\to invisible)$ can be extracted.

\item For the visible Higgs signal, 
the best available LC errors for the 
$b\bar b$ final state are used, assuming operation
at an energy of $\sqrt s=500$ GeV with luminosity
of $1000$ fb$^{-1}$  and with
polarization.  The corresponding
SM fractional error for this channel at $m_h=120$ GeV is  $0.02$.
This SM result is adjusted assuming signal rate reduction by
the factor $1- BR(\heff\to invisible)$.

\end{itemize}

To determine how well the $M_D,\xi,\delta$ parameters can be determined,
specific input values, $M_D^0,\xi^0,\delta^0$, are chosen. 
The expected experimental errors are then computed
for each of the above observables assuming this input parameter set.
The expected $\Delta\chi^2$ for each of the above
observables is then computed for different values of $M_D,\xi,\delta$.

The $BR(\heff\to visible)$ measurement turns out to be 
important in discriminating between different models when the
invisible branching fraction is large (requiring small to moderate
$m_h$, small $M_D$, $\delta=2$ or $3$, and
substantial $\xi$).  In such a case, the visible branching fraction
can be quite small and typically varies rapidly as a function of the
model parameters (in particular $\xi$), whereas the invisible branching
fraction, although large, will vary more slowly and
will not provide as good a discrimination between different parameter
values.  Of course, if $BR(\heff\to visible)$ is so small that the
background is dominant, 
our ability to determine $\xi$, $M_D$ and $\delta$ from this measurement
deteriorates.  Complementary statements apply to the case when
$BR(\heff\to invisible)$ is small and $BR(\heff\to visible)$ is
slowly varying.  

Given the $\Delta\chi^2$ for the five measurements outlined above,
we will characterize the net discrimination between 
models by using LHC data alone, $\Delta\chi^2(LHC)$, and LHC+LC data,
$\Delta\chi^2(LHC+LC)$ (the latter being
completely dominated by the LC information), where
\begin{eqnarray}
\Delta\chi^2(LHC)&=&
\Delta\chi^2(LHC,~H_{vis})+\Delta\chi^2(LHC,~H_{inv})\nonumber\\
\Delta\chi^2(LC)&=&
\Delta\chi^2(LC,~\gamma+{\rm missing~energy})
+\Delta\chi^2(LC,~H_{inv})+\Delta\chi^2(LC,~H_{vis})
\nonumber\\
\Delta\chi^2(LHC+LC)&=&
\Delta\chi^2(LHC)+\Delta\chi^2(LC)\,.
\end{eqnarray}

Note that since $m_h$ will be very precisely measured, one can
concentrate on the ability to determine the parameters $M_D$, $\delta$
and $\xi$. Regions of parameter space corresponding to a
95\% CL determination, which for three parameters corresponds to
$\Delta\chi^2=7.82$ are presented in
Figs.~\ref{sec6_95cla}--\ref{sec6_95clb}.  Here, 
$m_h=120$ GeV and the fixed input value
$\xi^0=0.5$ are assumed.

\begin{figure}[p]
\begin{center}
\begin{tabular}{c c}
\includegraphics[width=8.5cm,height=9.5cm]{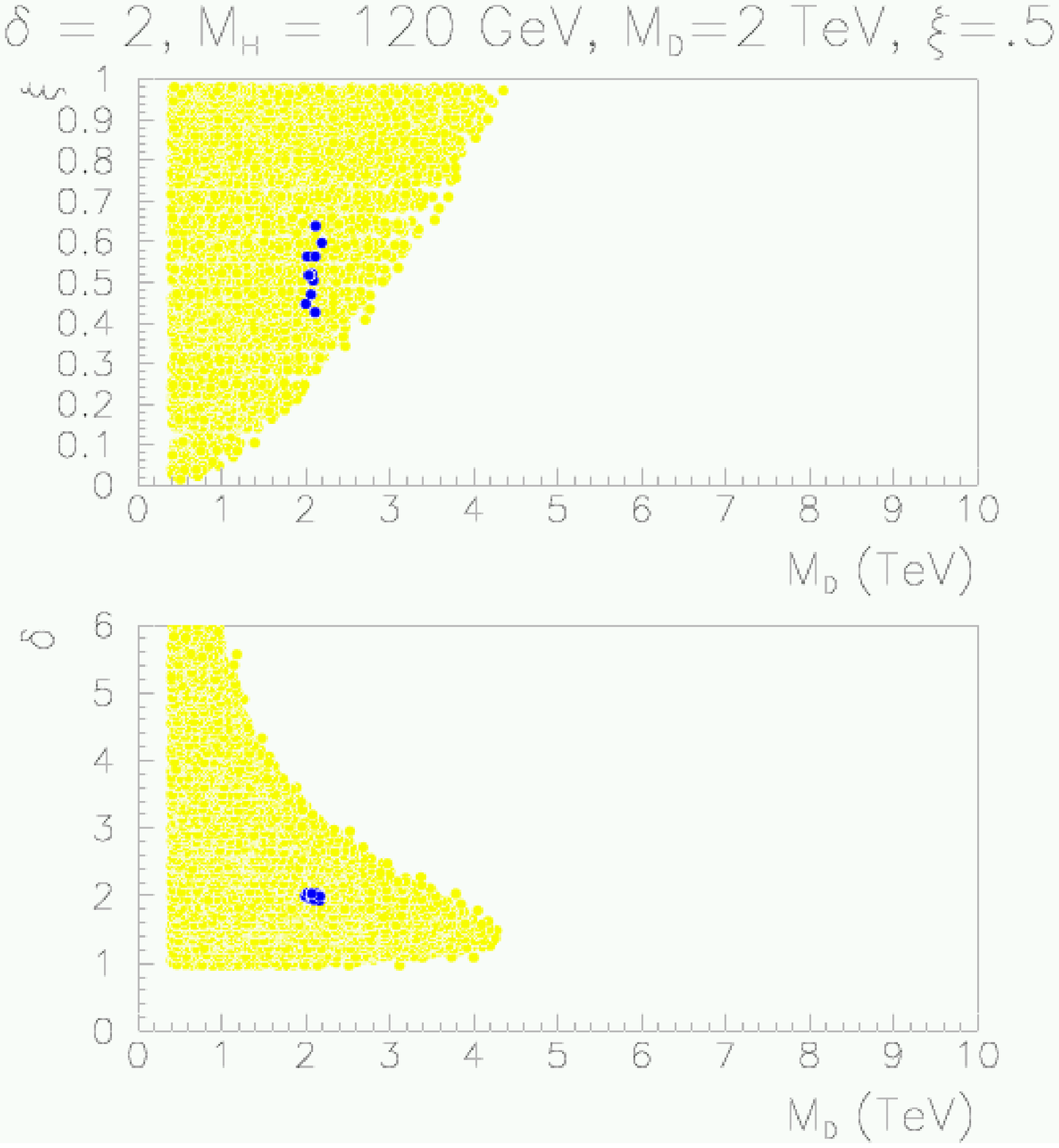} &
\includegraphics[width=8.5cm,height=9.5cm]{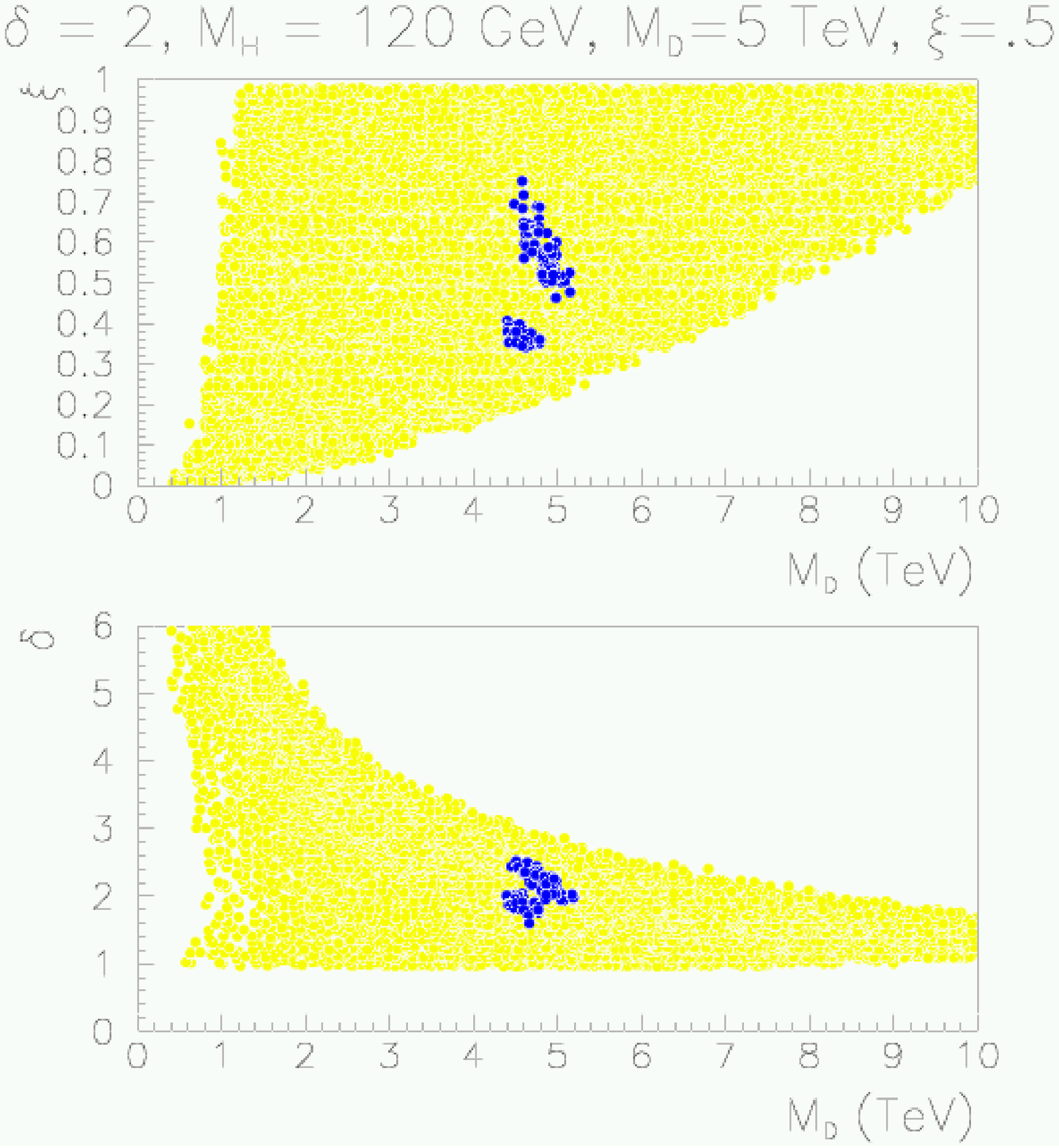}\\
\includegraphics[width=8.5cm,height=9.5cm]{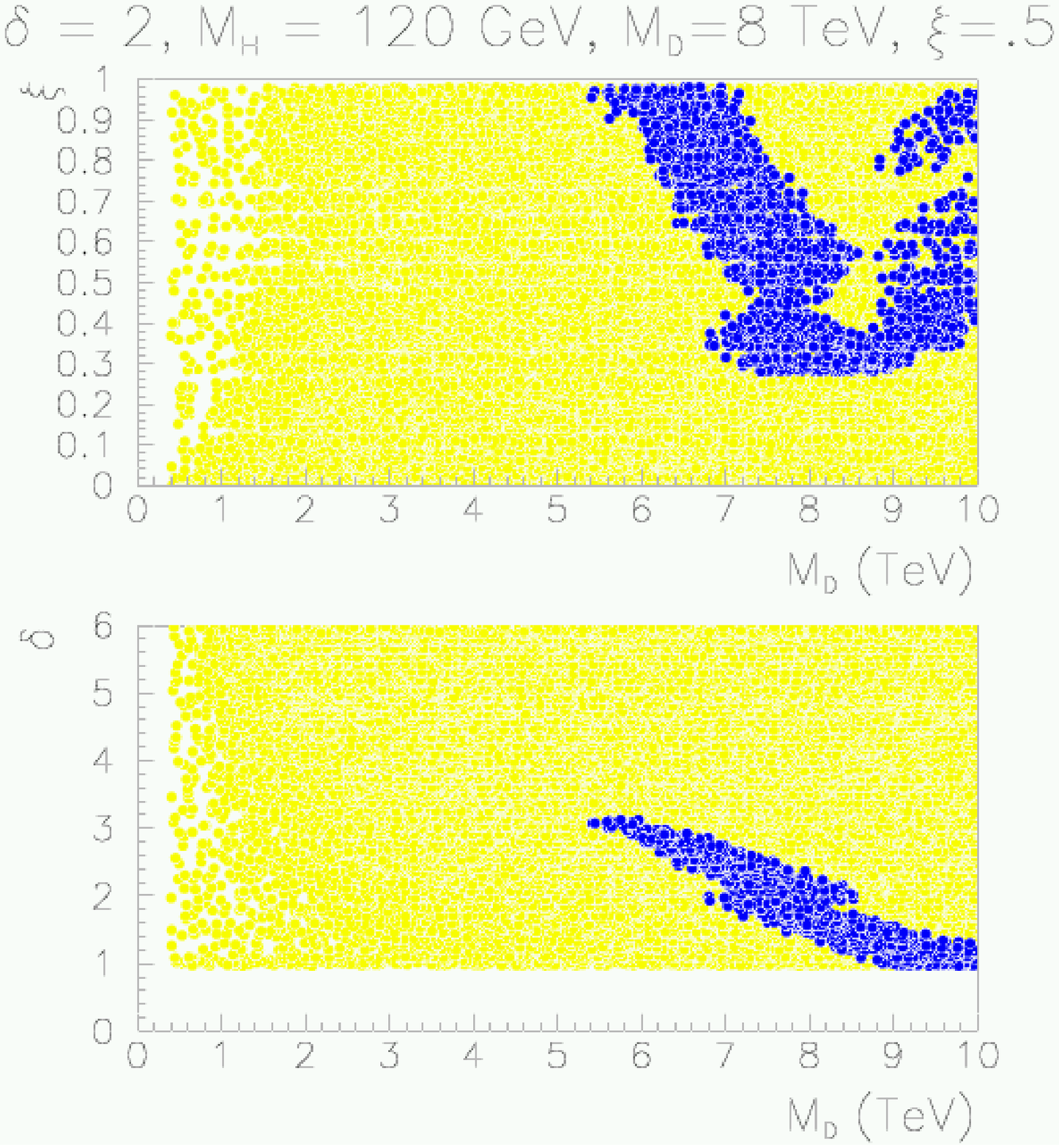}&
\includegraphics[width=8.5cm,height=9.5cm]{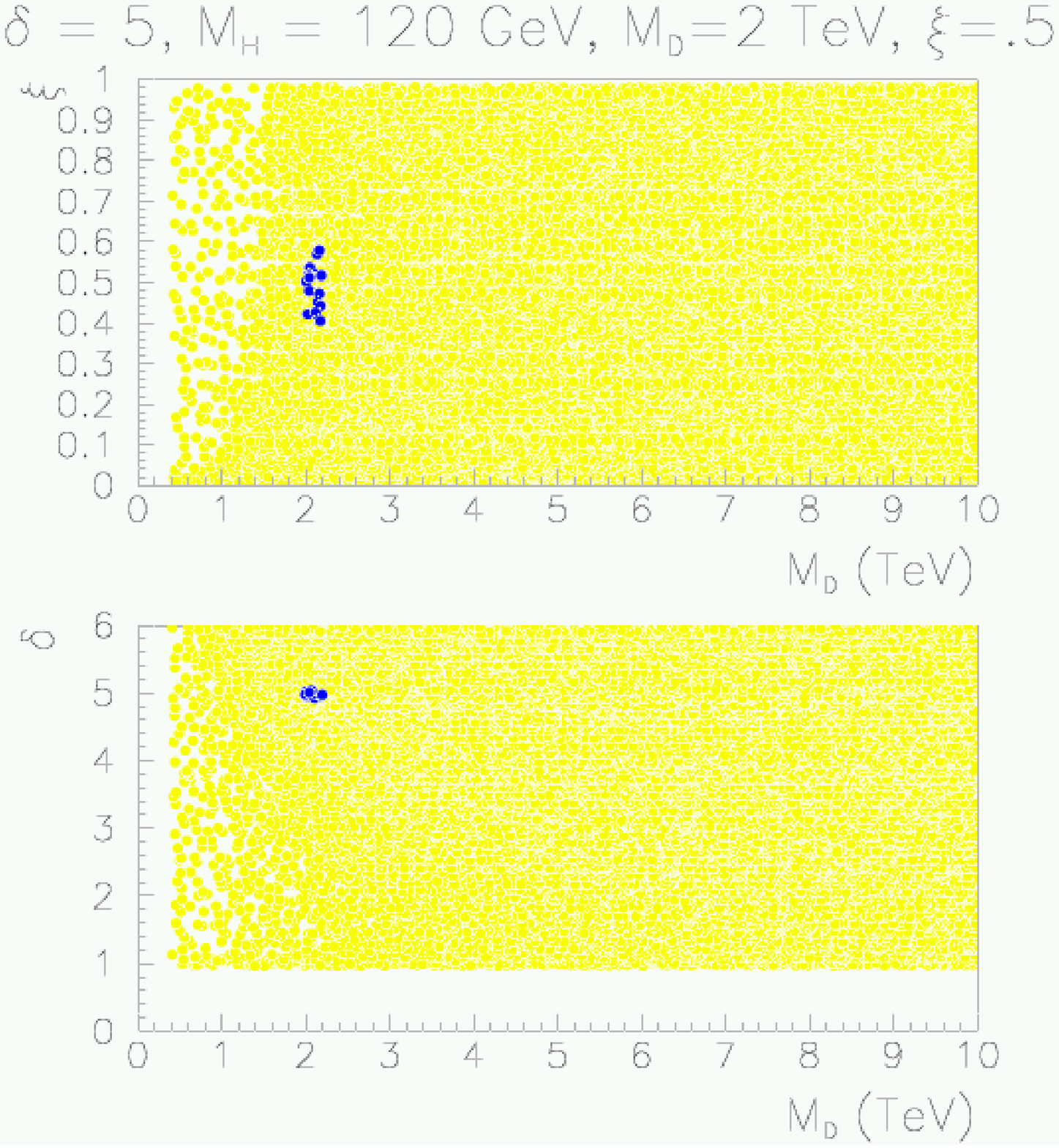}
\end{tabular}
\caption{95\% CL contours for determination of
  $M_D$, $\xi$ and $\delta$ assuming $m_{h}=120$ GeV,
input $\xi^0=0.5$ and input $\delta^0$ and $M_D^0$ values as
indicated above each pair of figures. All results are obtained
assuming $L=100$ fb$^{-1}$ Higgs measurements 
at the LHC, $\sqrt s=350$ GeV ($500$ GeV) invisible (visible) 
mode Higgs measurements
at the LC, and $\sqrt s=500$ GeV and $\sqrt s=1000$ GeV 
$\gamma+$ missing energy
measurements at the LC with $L=1000$ fb$^{-1}$ and $L=2000$ fb$^{-1}$ at the
two respective energies.  
The larger light gray (yellow) regions are the 95\% CL regions in
the $\xi,M_D$ and $\delta,M_D$ planes using only $\Delta\chi^2(LHC)$.
The smaller dark gray (blue) regions or points are the 95\% CL regions in
the $\xi,M_D$ and $\delta,M_D$ planes using $\Delta\chi^2(LHC+LC)$.
From \cite{sec6_gunion}.  }
\label{sec6_95cla}
\end{center}
\end{figure}
\begin{figure}[h]
\begin{center}
\begin{tabular}{c c}
\includegraphics[width=8.5cm,height=9.5cm]{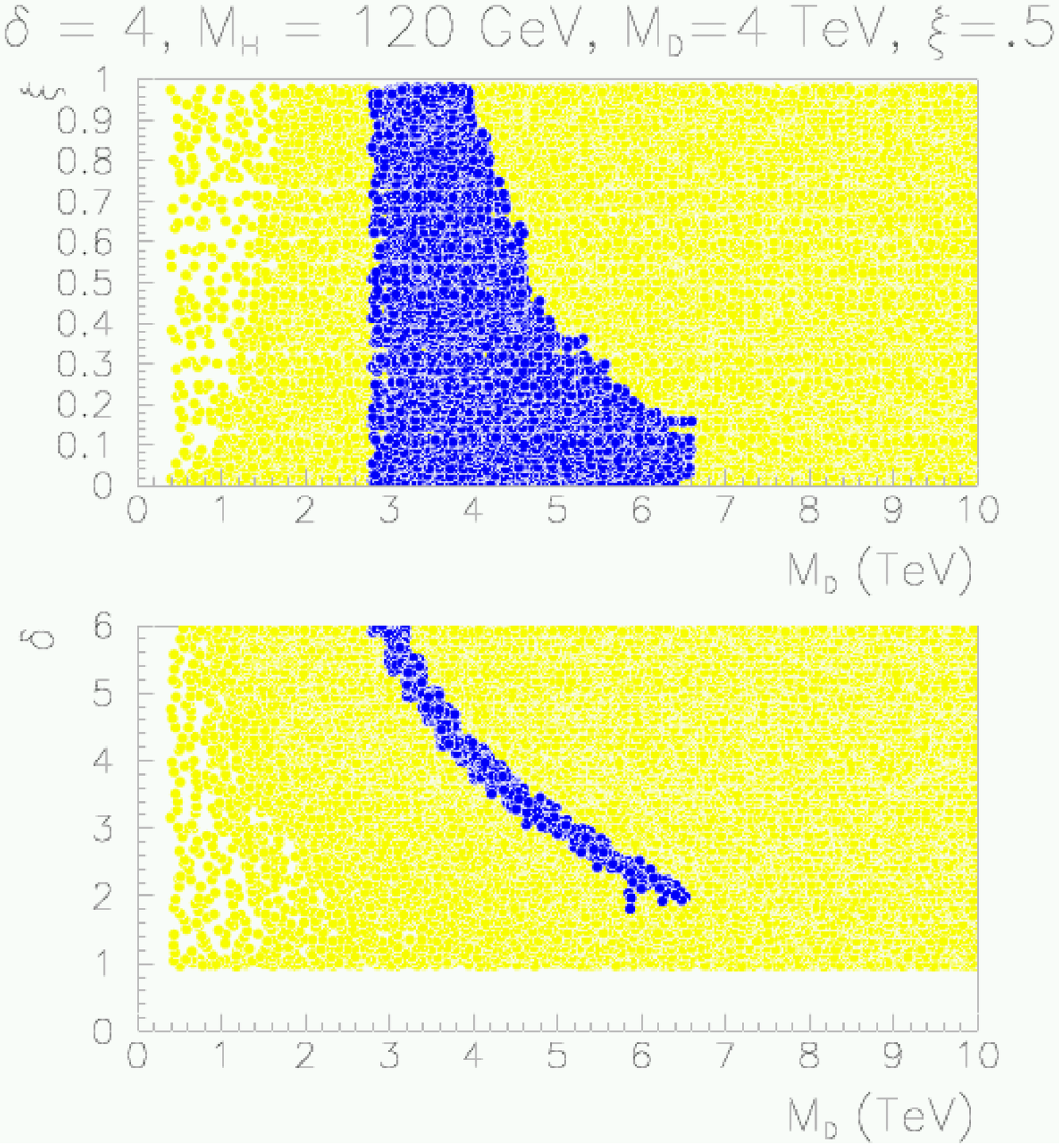} &
\includegraphics[width=8.5cm,height=9.5cm]{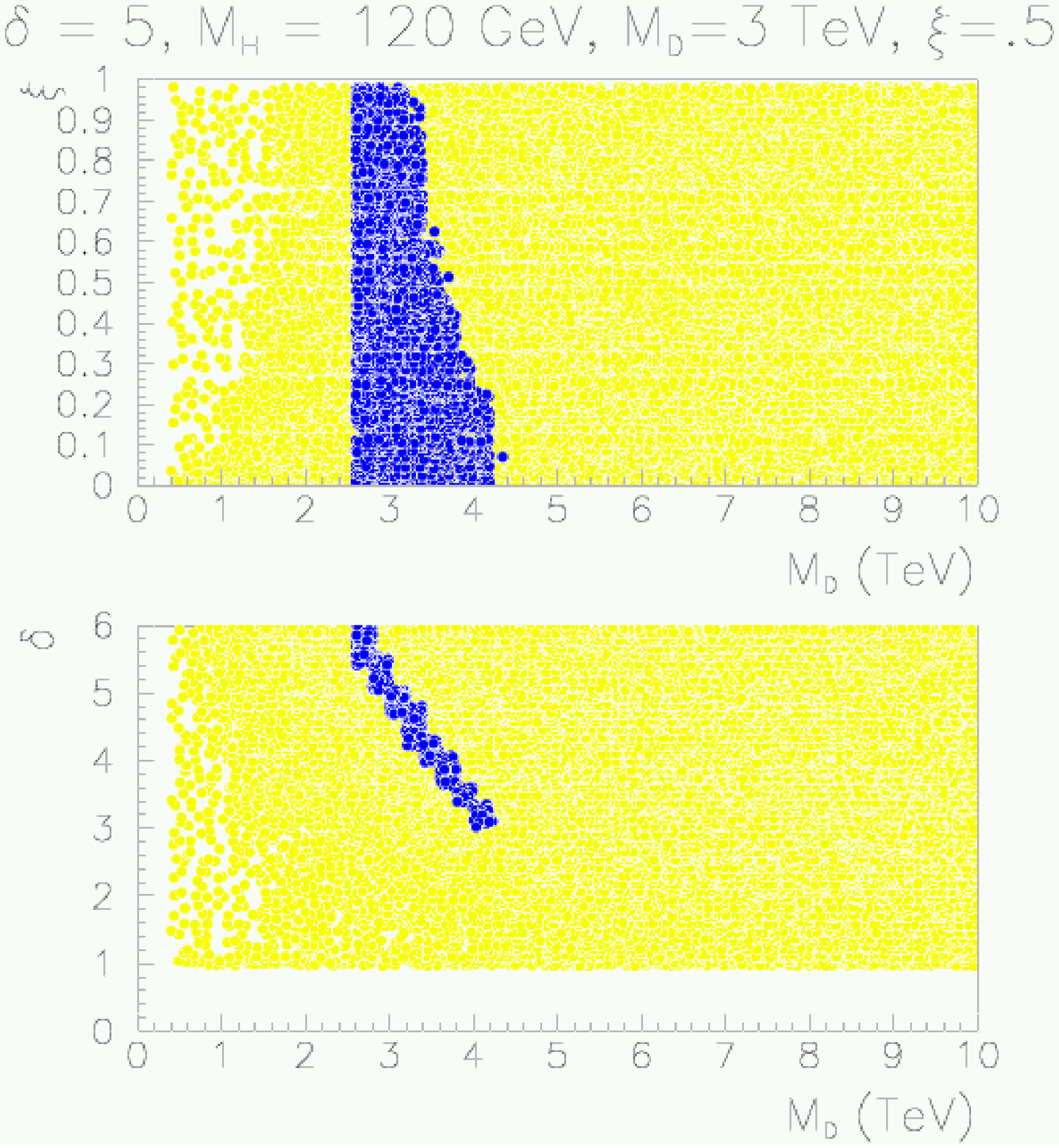}\\
\end{tabular}
\caption{As in Fig.~\ref{sec6_95cla}, but for different values of
$\delta^0$ and $M_D^0$ as indicated.
  }
\label{sec6_95clb}
\end{center}
\end{figure}

The most important point of these figures is that the ability of the
LHC alone to determine the input parameters is {\bf very} limited, whereas
by including the LC data, a quite precise $\delta$, $\xi$
and $M_D$ determination is possible when $M_D$ and $\delta$ are not
too large.   Similar results are also found (see \cite{sec6_gunion})
for the case $m_{h}>2M_W$.

\section {TeV$^{-1}$ extra dimensions}

The possibility of TeV$^{-1}$-sized extra dimensions naturally
arises in braneworld theories \cite{sec6_ant}.  By themselves, they
do not allow for a reformulation of the hierarchy problem, but
they may be incorporated into a larger structure in which this
problem is solved.  In these scenarios, the SM fields are
phenomenologically allowed to propagate in the bulk.  This presents
a wide variety of choices for model building:  (i) all, or only
some, of the SM gauge fields exist in the bulk; (ii) the Higgs
field may lie on the brane or in the bulk; (iii) the SM fermions
may be confined to the brane or to specific locales in the extra
dimensions, or they may freely reside in the bulk.  The
phenomenological consequences of this scenario strongly depend
on the location of the fermion fields.

\subsection{Gauge fields in the bulk}

We first discuss the case where the SM matter fields are rigidly fixed
to the brane and do not feel the effects of the additional
dimensions.  The simplest model of this class is the case of only
one extra TeV$^{-1}$-sized
dimension, where the fermions are constrained to lie
at one of the two orbifold fixed points, $y=0,\pi R$, associated
with the compactification on the orbifold $S^1/Z_2$, where $R$
is the radius of the compactified TeV$^{-1}$ dimension.
Two specific cases will be considered below: 
either all of the fermions are placed at 
$y=0$ ($D=0$), or
the quarks and leptons are localized at opposite fixed points ($D=\pi 
R$).
Here $D$ is the distance between the quarks and leptons in the single
extra TeV$^{-1}$ dimension.
The latter scenario
may assist in the suppression of proton decay. 

In this framework, the fermionic couplings of
the KK excitations of a given gauge field are identical to those 
of the SM,
apart from a possible sign if the fermion 
lives at the $y=\pi R$ fixed point, and an overall factor of
$\sqrt 2$. The gauge boson KK excitation masses are given to lowest 
order in
$(M_0/M_c)^2$ by the relationship 
\begin{equation}
M_n^2=(nM_c)^2+M_0^2\,,
\label{sec6_bulkmass}
\end{equation}
where $n$ labels
the KK level as usual, $M_c =1/R\sim 1$ TeV is the compactification
scale, and $M_0$ is the
zero-mode mass.  $M_0$ is obtained via spontaneous symmetry breaking 
for the cases of the
$W$ and $Z$ and vanishes for the $\gamma$ and gluon.  
Note that the KK excitations of all the gauge states
will be highly degenerate.  For example, if
$M_c=4$ TeV, the splitting between the first $Z$ and $\gamma$ KK states
is less than $\sim$ 1 GeV, which is too small to be observed at the LHC;
the two states then appear as a single resonance in the Drell-Yan
channel.

An updated analysis {\cite {sec6_bunch}}
of precision electroweak data constrains $M_c \gsim 4-5$ TeV, 
independently ($i$) of whether the Higgs field vev is mostly in the bulk 
or on the brane, or ($ii$) of which
orbifold fixed points the SM fermions are located. 
At a LC with 
$\sqrt s= 500-1000$ GeV the effects of gauge KK exchanges with masses
well in excess of the $M_c\sim 4-5$ TeV
range are easily observable via their virtual exchange
 as shown in Fig.~\ref{sec6_fig0}. This
implies that there will be sufficient `resolving power' at the LC
to determine the properties of the exchange state in analogy with the
case of extra gauge bosons discussed in the previous chapter.

The large constraint from precision data on $M_c$ implies that
the LHC experiments will at best observe only a single resonance in the
$\ell^+\ell^-$ channel.  The next set of KK states, which are 
essentially twice as heavy with $M_2\gsim 8-10$ TeV,
are too massive to be seen even with an integrated luminosity of order
$1-3$ $ab^{-1}$ {\cite {sec6_tgr}}. 
This can be seen from Fig.~\ref{sec6_figm1}.  As mentioned above, these
apparently isolated single resonance structures are, of course,
superpositions
of the individual KK excitations of both the SM $\gamma$ and $Z$.
This double excitation plus the existence of
the heavier KK tower states lead to the very unique resonance shapes 
displayed in the figure.
This figure shows that KK states up to masses somewhat in excess
of $\simeq 7$ TeV or so should
be directly observable at the LHC (or at the LHC with a luminosity upgrade) 
in a single lepton pair channel 
(See also \cite{sec6_before}).  
In addition, gluon KK states may be
observable in the dijet channel at the LHC for $M_c<15$ TeV 
\cite{sec6_leshouches04}.

\begin{figure}[htbp]
\centerline{
\includegraphics[width=9cm,angle=90]{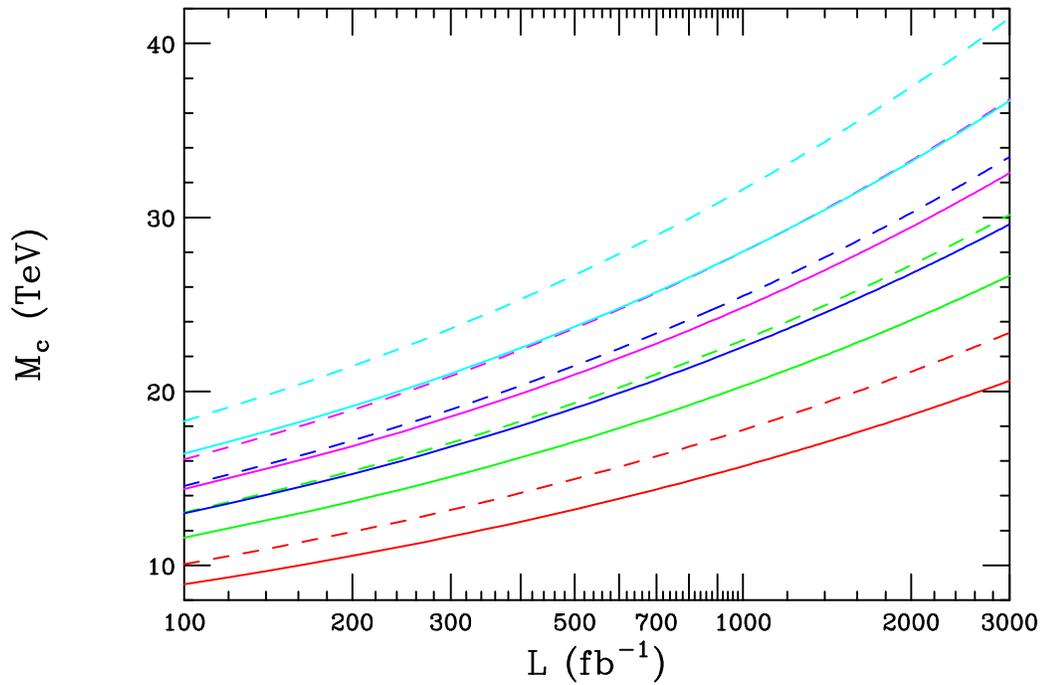}}
\vspace*{0.1cm}
\caption{$95\%$ CL bound on the scale $M_c$ from the reaction 
$e^+e^-\to f\bar f$, where
$f=\mu, \tau, c,b,t$ have been summed over, as a function of the LC 
integrated luminosity . The
solid (dashed) curves assume a positron polarization $P_+=0 (0.6)$; an
electron polarization of $80\%$ has been assumed in all cases. From
bottom to top the center of mass energy of the LC is taken to be
0.5, 0.8, 1, 1.2 and 1.5 TeV, respectively.  From \cite{sec6_tgr_lumi}.}
\label{sec6_fig0}
\end{figure}
\begin{figure}[htbp]
\centerline{
\includegraphics[width=9cm,angle=90]{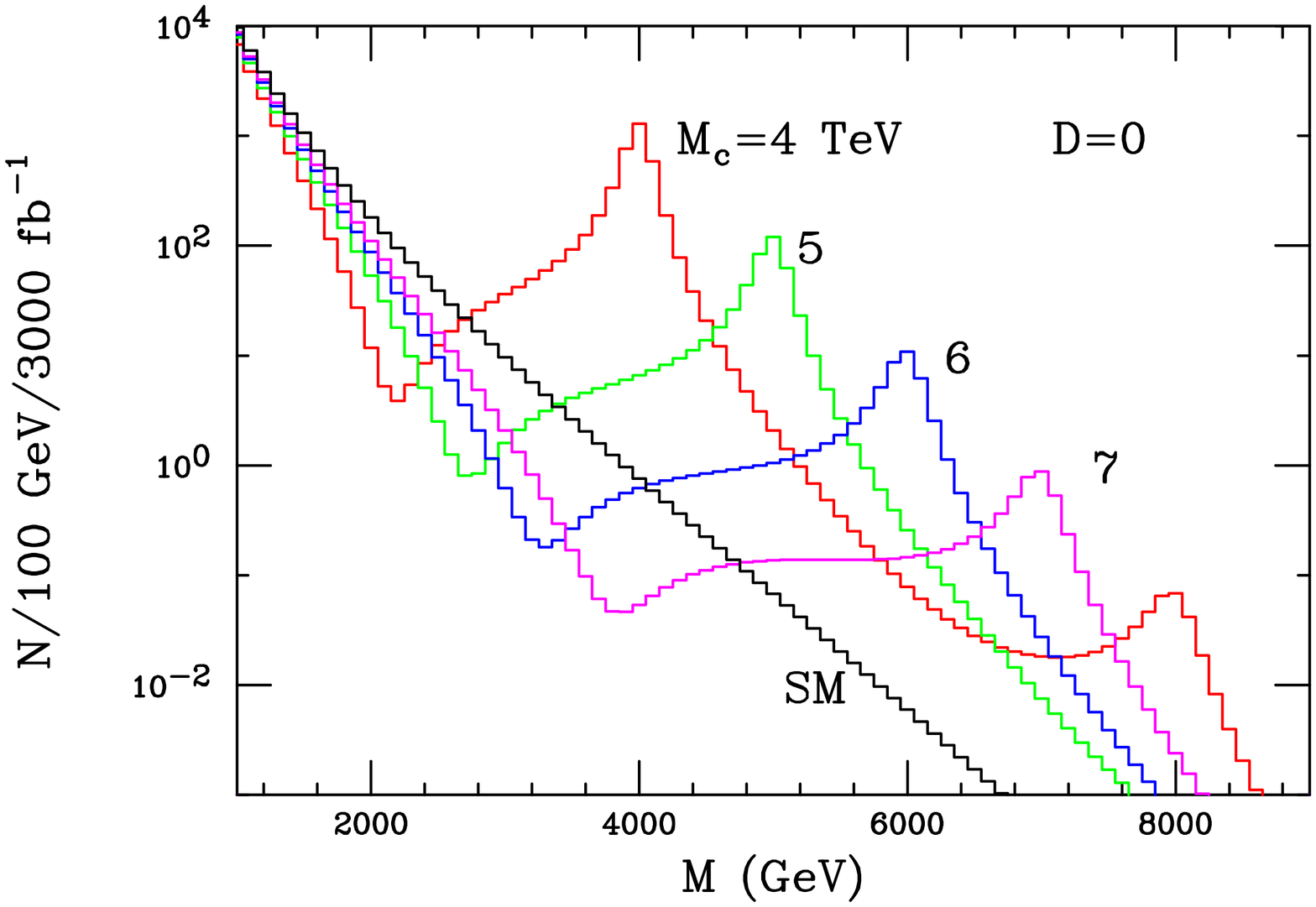}}
\vspace{0.4cm}
\centerline{
\includegraphics[width=9cm,angle=90]{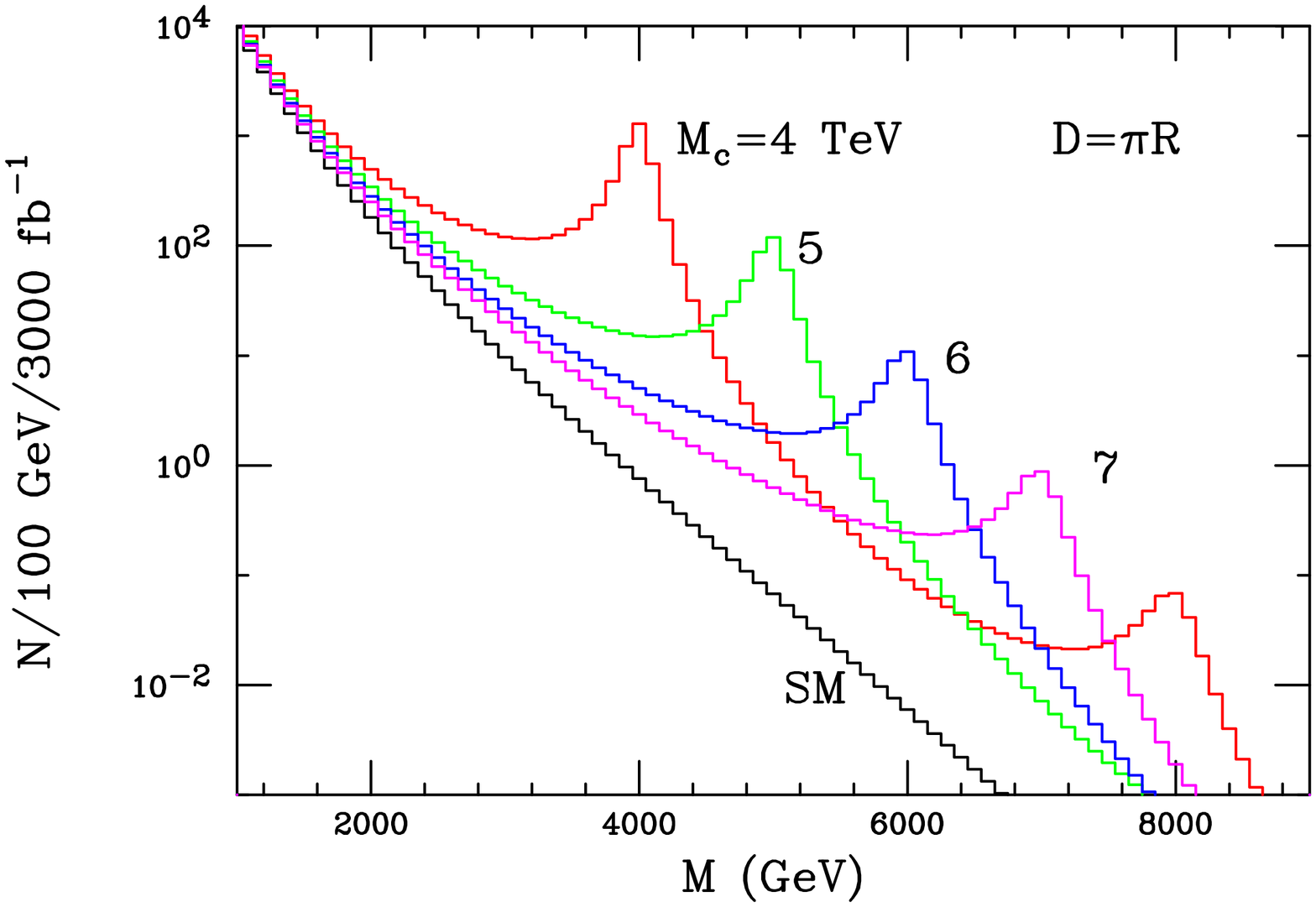}}
\vspace*{0.1cm}
\caption{Production rate in the Drell-Yan channel $pp\to e^+e^- X$
for $\gamma/Z$ KK
resonances as a function of dilepton invariant mass assuming a very 
high
luminosity LHC. A rapidity cut $|\eta_l|\leq 2.5$ has been applied to 
the
final state leptons. The red(green, blue, magenta)
histogram corresponds $M_c$=4(5, 6, 7) TeV, respectively. The black 
histogram
is the SM background. In the top panel all fermions
are assumed to lie at the $y=0$ fixed point, $D=0$, while the quarks 
and
leptons are split, $D=\pi R$,  in the lower panel.  
From \cite{sec6_tgrkk}.}
\label{sec6_figm1}
\end{figure}

Next we examine \cite{sec6_tgrkk} the identification of gauge
KK states at both the LHC and LC. 
Here we consider a time line
where the LC turns on after several years of data taking by the LHC,
at roughly the time of an LHC luminosity upgrade, and assume that the 
LHC discovers a resonance peak in the Drell-Yan invariant mass 
distribution.  Measurement
of the lepton pair angular distribution at the LHC will determine
that the resonance is spin-1,  provided sufficient
statistics are available. 
Perhaps the most straightforward interpretation 
of such a resonance would be that
of an extended gauge model{\cite {sec6_snow}}, {\it e.g.},
a GUT scenario,  which predicts the
existence of a degenerate $W'$ and $Z'$; many such models have been
explored in the
literature{\cite {sec6_models}}.  Here we address whether 
such a scenario with a
degenerate $Z'/W'$ can be distinguished from KK excitations without 
the observation of any of the higher KK excitations. 
This issue has been previously discussed
to a limited extent by several authors {\cite {sec6_before}}.

Fig.~\ref{sec6_fig2} shows a closeup of
the excitation spectra and forward-backward asymmetries, $A_{FB}$,
for KK production near the first resonance region
assuming $M_c=4$ TeV and with $D=0,\pi R$. 
Note the strong
destructive interference minimum in the cross section for the
$D=0$ case near $M\simeq 0.55M_c$ which is also reflected in the
corresponding narrow
dip in the asymmetry. This dip structure is a common feature 
in higher dimensional models ($\delta>1$) and in models with 
warped extra dimensions, with the precise location of the dip 
being sensitive to model details.
In addition, while the overall behaviour of the $D=0$ and $D=\pi R$
cases is completely different below the peak, it is almost identical
above it.  This difference in the two spectra is due solely to 
an additional factor of $(-1)^n$ appearing in the KK sum; this
arises from the placement of
the quarks and leptons at opposite fixed points. 
Lastly, note also that the
peak cross section and peak $A_{FB}$ values are nearly
identical in the two cases.

\begin{figure}[htbp]
\centerline{
\includegraphics[width=9cm,angle=90]{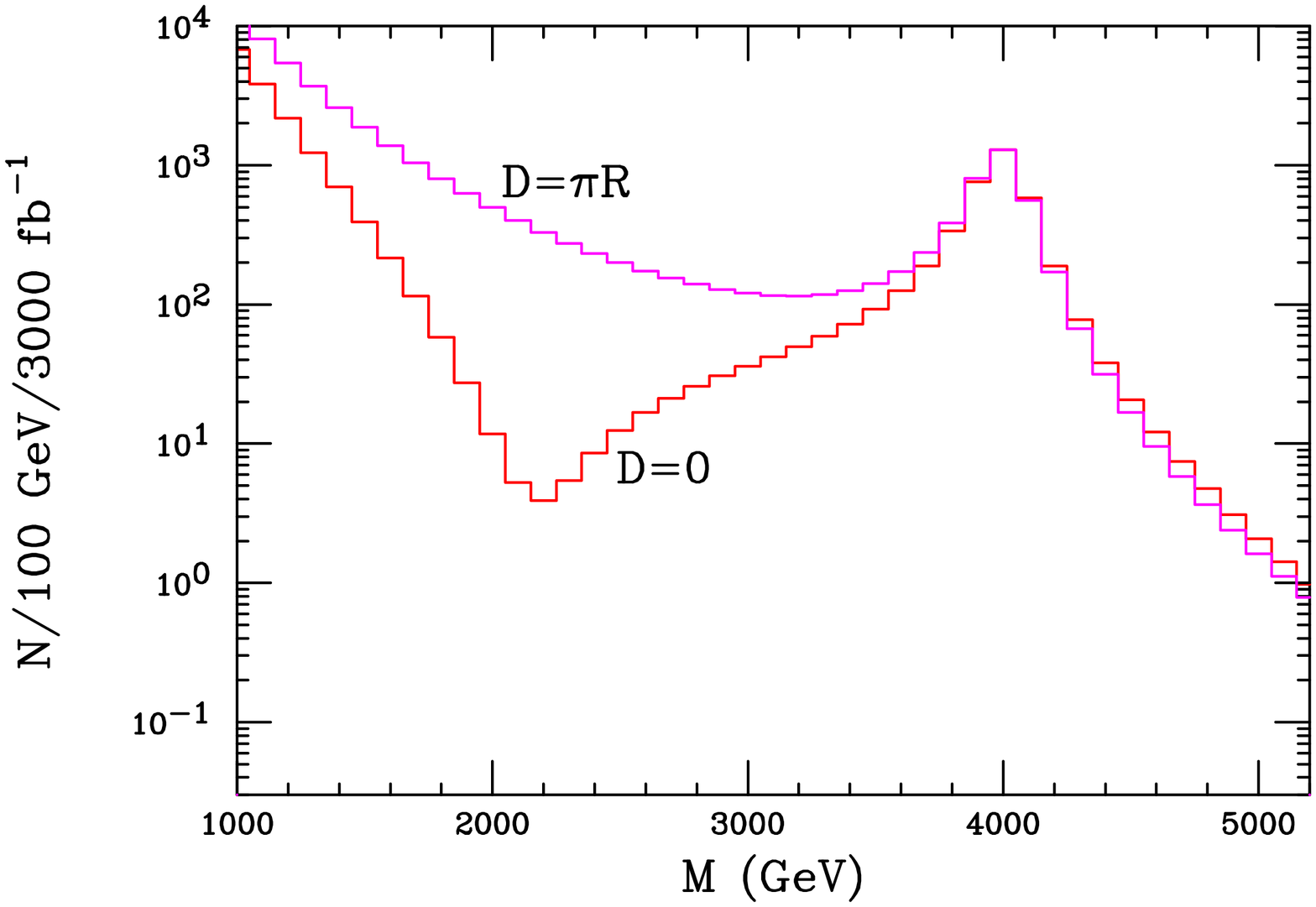}}
\vspace*{0.4cm}
\centerline{
\includegraphics[width=9cm,angle=90]{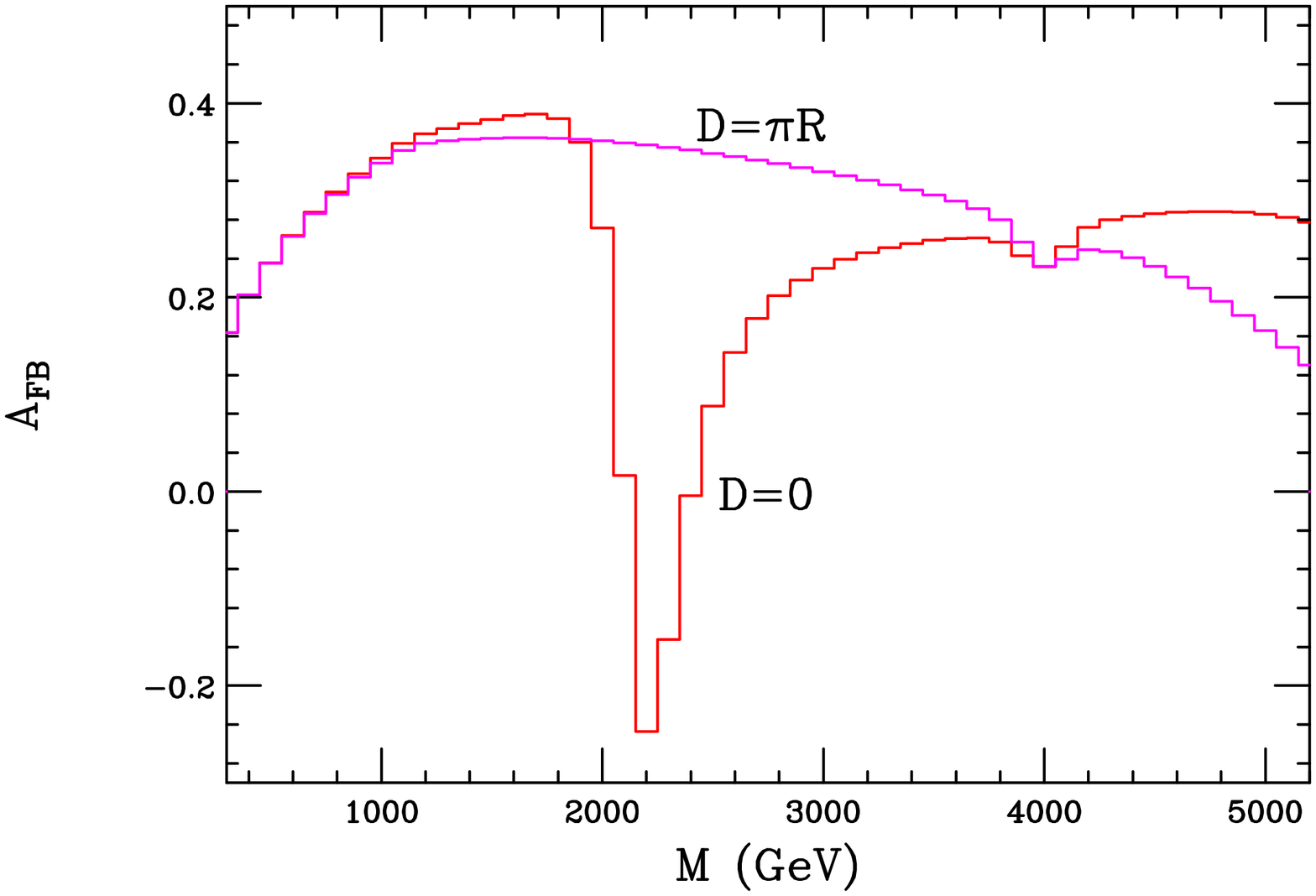}}
\vspace*{0.1cm}
\caption{A comparison of the lepton pair invariant mass spectrum and
forward-backward lepton asymmetry for the production of a 4 TeV KK
resonance for the two choices of $D$. The red(magenta) histograms are 
for
the case $D=0(\pi R)$.  From \cite{sec6_tgrkk}.}
\label{sec6_fig2}
\end{figure}

For either choice of the fermion placement,
the excitation curve and $A_{FB}$  for KK production appear to be
qualitatively  different than that for typical $Z'$
models{\cite{sec6_snow}} as discussed in the previous
chapter.  None
of the various $Z'$ models produces resonance structures 
similar to those for KK excitations.
The resonance structure
for the KK case is significantly wider and has a larger peak cross 
section than does the typical $Z'$ model, and the latter does
not have the strong destructive interference
below the resonance. (Recall, however, that the height and
width of the $Z'$ or KK resonance also depends on the set of allowed 
decay modes.)
In addition, the dip in the value of $A_{FB}$ occurs much closer to the
resonance region for the typical $Z'$ model than it does in the KK 
case.  Clearly, while the KK resonance structure 
does not resemble that of a
conventional $Z'$, without further study, one could not simply 
claim that some unusual $Z'$ model could not mimic KK production.

Next, the differences between the KK and $Z'$ scenarios are quantified
\cite{sec6_tgrkk}.  For $M_c=4 (5,~6)$ TeV
cross section `data' at the LHC is generated corresponding to dilepton 
masses in the range 250-1850 (2150, 2450) GeV in 100 GeV bins for 
both the $D=0$ and $\pi R$ cases.  Lower invariant dilepton masses are 
not useful due to the presence of the
$Z$ peak and the photon pole.  We next try to fit these cross section
distributions by making the assumption that the data
is due to the presence of a single $Z'$. For simplicity, we restrict our 
attention to
the class of $Z'$ models with generation-independent couplings and 
where the associated a new gauge group
generator commutes with weak isospin.
These conditions are satisfied, \eg,  by GUT-inspired $Z'$ models
as well as by many others {\cite {sec6_snow}}.
If these constraints hold then the $Z'$ couplings
to SM fermions can be described by only 5 independent parameters: 
the couplings to the left-handed quark and lepton doublets and the 
corresponding ones to the right-handed quarks and leptons.
These couplings are all varied independently
in order to obtain the fit to the dilepton mass distribution with the 
best $\chi^2$ per degree of freedom, and the
relevant probability (confidence level) of the fit
is obtained  using statistical errors
only.  The overall
normalization of the cross section is determined at the $Z$-pole which 
is outside of the fit region and is governed solely by SM physics.

\begin{figure}[htbp]
\centerline{
\includegraphics[width=9cm,angle=90]{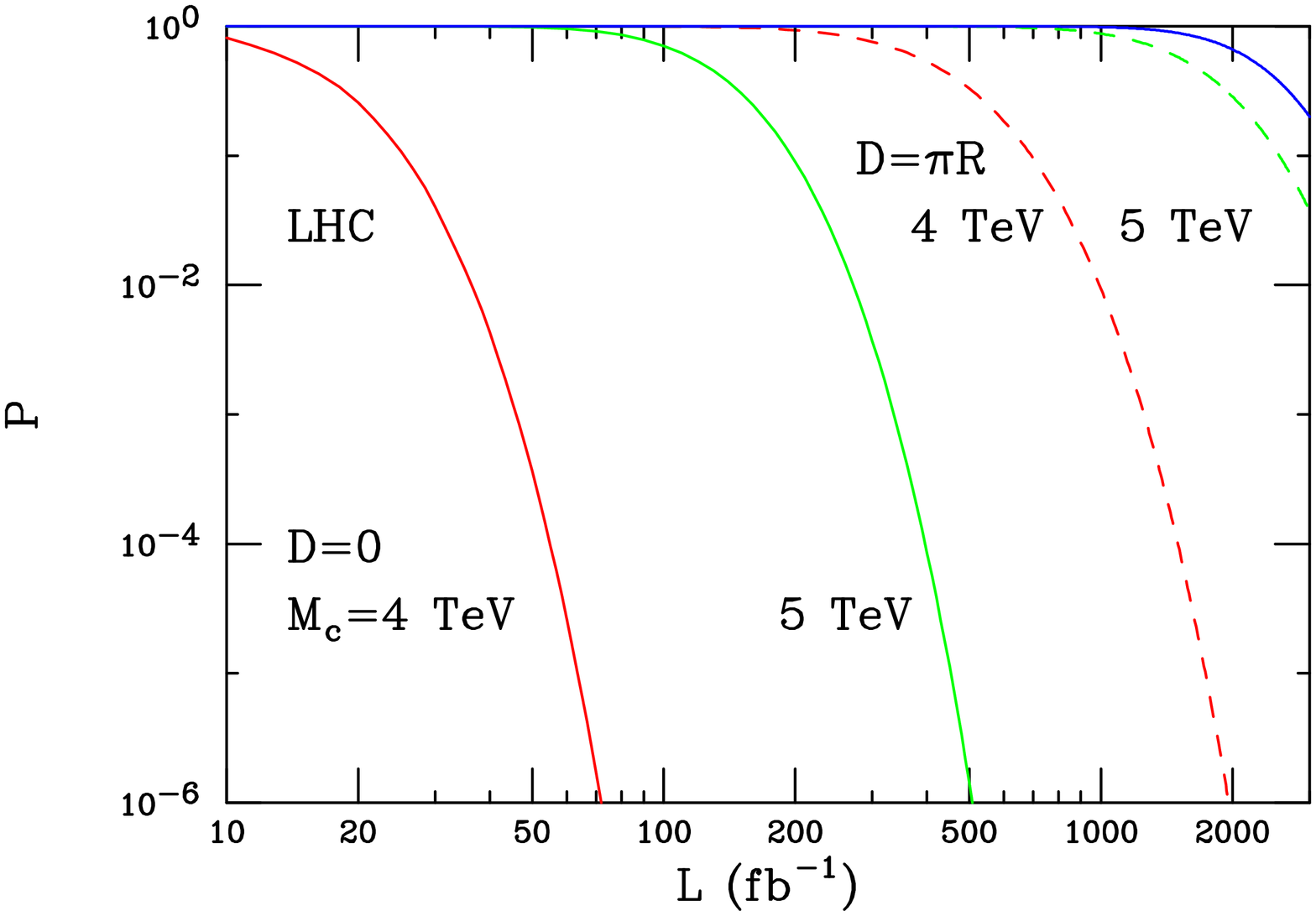}}
\vspace*{0.4cm}
\centerline{
\includegraphics[width=9cm,angle=90]{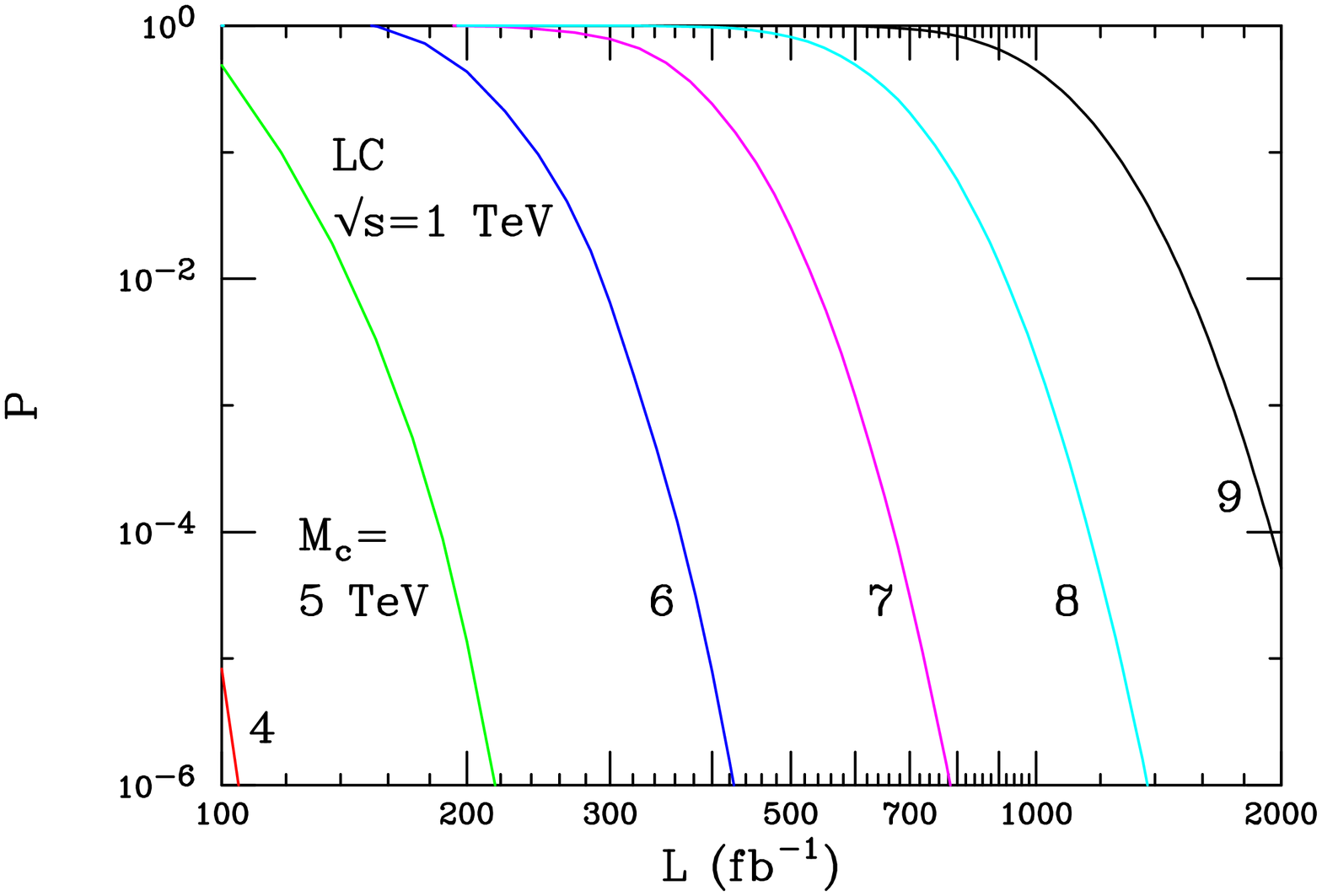}}
\vspace*{0.1cm}
\caption{Top Panel: Probability associated with the best $Z'$ fit 
hypothesis as a function of the LHC integrated luminosity for the 
cases $D=0$ (solid) and $D=\pi R$ (dashed).
Bottom Panel: Same as the above but now for a 1 TeV LC using leptonic
data only. 
In this case, the $D=0$ and $D=\pi R$ cases are identical. The
value of $M_c$ is assumed to be determined at the LHC.
From \cite{sec6_tgrkk}.}
\label{sec6_fig4}
\end{figure}

The results \cite{sec6_tgrkk} 
of performing these fits for different values of $M_c$ and the
two choices $D=0,\pi R$ are displayed in Fig.~\ref{sec6_fig4}. Explicitly, 
this shows
the best fit probability for the $Z'$ hypothesis to the KK generated
data. For example, taking the case $D=0$ with $M_c=4$ TeV we see that 
with an
integrated luminosity of order 60 fb$^{-1}$ the best fit probability is
near a few $\times 10^{-5}$. For such low probabilities one could 
certainly
claim that the KK generated `data' is not well fit by the $Z'$ 
hypothesis.
As the mass of the KK state increases, the difference in the 
production
cross section from the SM expectation is reduced and greater statistics 
are needed;  for $M_c=5$ TeV an
integrated luminosity of order 400 fb$^{-1}$ is required to get to the 
same level of rejection of the $Z'$ hypothesis. 
For the $D=\pi R$ case, we see  
that the level of confusion between KK states and $Z'$ scenarios
is potentially greater. Even for
$M_c=4$ TeV we see that only at very high integrated luminosities,
of order $\sim 1.5~ab^{-1}$,  can the KK and $Z'$ scenarios be 
distinguished at the level discussed above.
For larger KK masses this separation becomes
essentially impossible at the LHC.

The corresponding analysis \cite{sec6_tgrkk} 
for the LC is different since
actual resonances are not observed.
It is assumed that data is taken at a single value of $\sqrt s$ so 
that the mass of the KK or $Z'$ resonance obtained from the LHC
must be used as input to the analysis.
Without such input an analysis can still 
be performed provided data from at least two distinct values of 
$\sqrt s$ are available {\cite {sec6_oldt}}. In that case $M_c$ 
becomes an additional 
fit parameter. 

At the LC, a tower of gauge KK states with fixed $M_c$ are 
exchanged in the process $e^+e^-\to f\bar f$.
`Data' is generated for both
the differential cross section and the Left-Right polarization
asymmetry, $A_{LR}$ as functions of the
scattering angle $\cos \theta$ in 20 (essentially) equal sized bins,
and including the effects of ISR.
The electron beam is assumed to be $80\%$ polarized and angular 
acceptance cuts are applied.
Other detailed assumptions in performing this analysis are the same
as those employed in Ref. \cite{sec6_tgr_lumi}.  A fit is performed
to this `data' making the assumption
that the deviations from the SM are due to the exchange of a single $Z'$.
For simplicity, we concentrate on the processes 
$e^+e^- \to \mu^+\mu^-,\tau^+\tau^-$ as only
the two leptonic couplings are then involved in performing the fits. 
In this case,
the $D=0$ and $D=\pi R$ scenarios lead to identical results for 
the shifts in all observables at the LC.
Adding new final states, such as $b\bar b$ or $c\bar c$, may lead to 
potential improvements in the fit although additional parameters 
must now be introduced and
the $D=0$ and $D=\pi R$ cases would be distinct as at
the LHC. 

As in the above analysis for the LHC, the two assumed $Z'$ couplings 
to leptons are varied and the best $\chi^2/df$ is obtained for the fit. 
For the case of a $\sqrt s$ = 500 GeV LC, we find that an
integrated luminosity of 300 fb$^{-1}$ is roughly equivalent to
60 fb$^{-1}$
at the LHC for the case of $M_c=4$ TeV assuming $D=0$.
For larger values of $M_c$,
800~(2200) fb$^{-1}$ at a 500 GeV LC gives equivalent discrimination
power as 400 (7500) fb$^{-1}$ at the LHC
assuming $M_c=5~(6)$ TeV.   The results \cite{sec6_tgrkk} 
for a 1 TeV LC are displayed
in Fig.~\ref{sec6_fig4}.  Here, we see that once the LC center-of-mass 
energy increases, it is superior to the LHC in resolving power, 
but the analysis still relies upon the LHC for the input
value of $M_c$.  This figure shows results for values
of $M_c$ beyond the range which is directly observable at 
the LHC $M_c\ge 7-8$ TeV (although still assuming that the value
of $M_c$ is an input).  This would seem to imply that
by extending the present analysis to include data from at least two 
values of $\sqrt s$,~KK/$Z'$ separation may be feasible out to 
very large masses at the LC.

In summary, the power of the LC to discriminate between
models is shown to be better than that of the LHC, however, the LC analysis 
depends upon the LHC determination of the resonance mass as an input. 

\subsection{Universal extra dimensions}

In the scenario known as universal extra dimensions,  
all SM fields propagate in the bulk and branes need not be present.
The simplest model of this type contains a single extra dimension 
compactified on an $S_1/Z_2$ orbifold.
Translational invariance in the higher dimensional space is 
preserved.  This leads to the tree-level conservation of the
$\delta$-dimensional momentum of the bulk fields, which in turn implies 
that KK number is conserved at tree-level while KK parity, $(-1)^n$, 
is conserved to all orders.  Two immediate
consequences of KK number and parity conservation are that ($i$) 
at tree-level, KK excitations can no longer be produced as s-channel
resonances and can only be produced in pairs, 
and ($ii$) the lightest
KK particle (LKP) is stable and is thus a dark matter candidate
\cite{sec6_timt}.  The former results
in a substantial reduction of the sensitivity to such states in 
precision electroweak and collider data with the present bound 
on the mass of the first KK excitation being of order a few
hundred GeV for $\delta=1$ \cite{sec6_ued}.

Due to the conservation of KK parity, the phenomenology of universal
extra dimensions resembles that of supersymmetric theories with R-parity
conservation.  Every SM field has KK partners.  These KK partners carry 
a conserved quantum number, KK parity, which guarantees that the LKP
is stable.  Heavier KK modes decay via cascades to the LKP, which
escapes detection resulting in a missing energy signature.  The 
Lagrangian of the model includes both bulk terms and 
interactions which are localized
at the orbifold fixed points.  The bulk
interactions yield masses for the KK states as given in Eq. 
(\ref{sec6_bulkmass}); these are highly degenerate within each KK
level.  The boundary interactions take the form of loop induced
localized kinetic terms \cite{sec6_blkt} and are not universal for
the different SM fields.  The low scale boundary terms are
determined via high energy parameters which are evolved through 
the renormalization group.  In addition, non-local radiative
corrections \cite{sec6_cms1} also affect the KK mass spectrum.
Combined, this has the effect of removing the degeneracy between the modes
in a KK level and also affects their decay chains.  
A typical KK spectrum \cite{sec6_cms2} is displayed
in Fig. \ref{sec6_uedfig}.  The KK modes with strong interactions
receive larger corrections than those with only electroweak
interactions.  This spectrum resembles that of the
superpartners in a Minimal Supersymmetric Standard Model where
the sparticles all lie relatively close together.

Once the KK mass degeneracy is lifted, the states decay promptly.
The resulting collider phenomenology is strikingly similar to that 
of supersymmetry \cite{sec6_cms2}.  Indeed, if such a scenario were 
observed at the LHC, it is reasonable to expect that it would be
mis-identified as the discovery of supersymmetry!   Superpartner
status would most likely be assigned to the various KK states.  
Hence, techniques to distinguish the two scenarios need to be developed.

The cleanest way to identify Universal Extra Dimensions 
would be the observation 
of the second KK level.  However, due to the conservation of KK
parity, second level KK states must be pair produced with each
other (or singly at loop-level) 
and will likely be out of kinematic reach of the LHC.
The distinguishing factor for this scenario is then that the KK 
states have the same spin as their SM partners, unlike the case of
supersymmetry.  It is thus imperative to measure the spin of
the particles in the spectrum.  In principle, particle spins can
be determined at the LHC from examination of their decay
distributions.  However, given the cascade decay chains and the
size of the pair production cross sections, this is expected to
be problematic at the LHC.  At a LC, threshold cross sections can be
easily measured and their S-wave versus P-wave behaviour can be
cleanly determined.  The cross sections for 300 GeV 
muon KK state  as well as smuon pair 
production \cite{sec6_tait} as a function of the $\epem$ center 
of mass energy are displayed in Fig. \ref{sec6_tait}.  The S- and 
P-wave behaviour is clearly observable and distinguishes between 
the two spins.  If the LC determined that KK states were being
produced at the LHC, then dedicated searches for the second KK
level would be well-motivated.  Since the mass scale and signatures
would be known, the $n=2$ modes may be detectable with the LHC
luminosity upgrades.

\begin{figure}[htbp]
\centerline{
\includegraphics[width=10cm,angle=0]{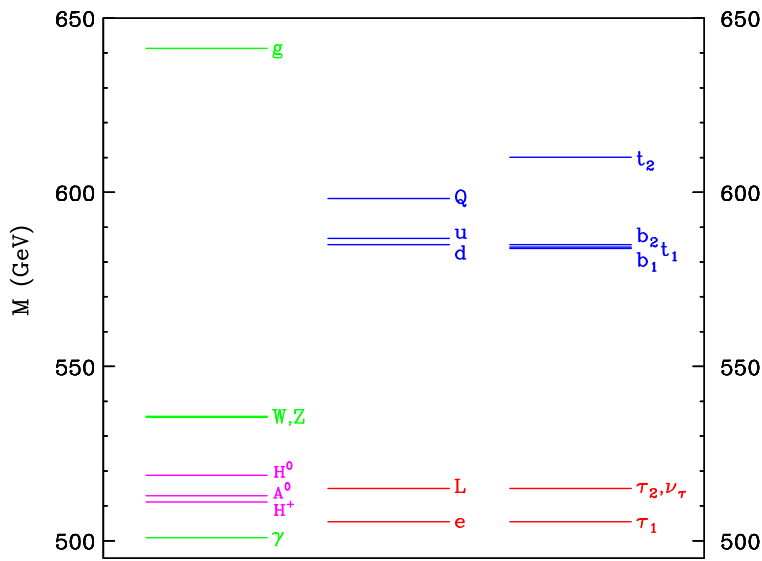}}
\vspace*{0.1cm}
\caption{Mass spectrum, including radiative corrections, of the
first level KK states in the Universal Extra Dimensions scenario,
taking $R^{-1}=500$ GeV.  From \cite{sec6_cms2}.}
\label{sec6_uedfig}
\end{figure}

\begin{figure}[htbp]
\centerline{
\includegraphics[width=9cm,angle=0]{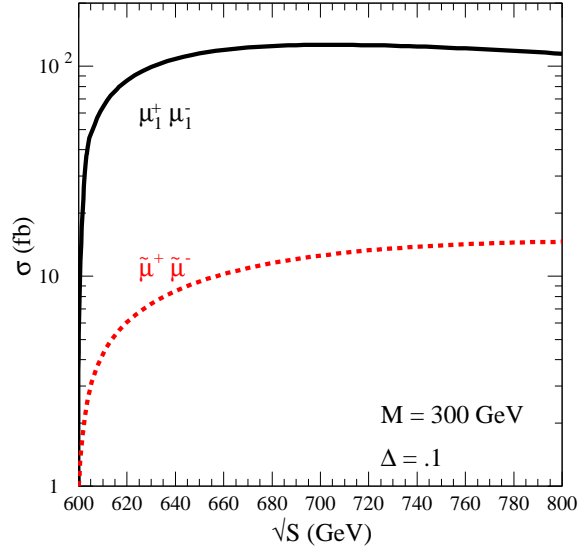}}
\vspace*{0.1cm}
\caption{Pair production of 300 GeV smuons and muon KK states
at an $\epem$ collider as a function of center of mass energy.
$\Delta$ represents the mass splitting between the smuon and LSP
as well as of the muon KK and LKP.  From \cite{sec6_tait}.}
\label{sec6_tait}
\end{figure}

\section{Warped extra dimensions}

In this scenario, known as the Randall-Sundrum (RS) model 
\cite{sec6_rs}, the hierarchy is explained in the context of a 
5-dimensional nonfactorizable background geometry with large curvature.  
This background manifold 
is a slice of anti-de Sitter ($AdS_5$) spacetime where
two 3-branes of equal and opposite tension sit at orbifold fixed
points at the boundaries of the $AdS_5$ slice.  The 5-d warped 
geometry induces a 4-d effective scale $\Lambda_\pi = e^{-kr_c\pi}
\overline M_{Pl}\sim$ TeV on one of the branes, denoted as
the TeV-brane.  Here, $k$  represents 
the 5-d curvature scale, which together with $M_5$ is 
assumed to be of order $\overline M_{Pl}$ (where $\overline M_{Pl}=
M_{Pl}/\sqrt{8\pi}$), and $\pi r_c$ is the 
length of the fifth dimension.  Consistency of the low-energy 
description requires that the curvature be bounded by $k/\overline 
M_{Pl} \lsim 0.1$.  TeV scales are naturally realized and stablized 
\cite{sec6_goldwise} on the TeV-brane provided that $kr_c\sim 10$.

\subsection{Conventional RS model}

In the simplest version of this scenario, the SM fields reside on the
TeV-brane and gravity propagates throughout the 5-d spacetime.
The 4-d phenomenology of the graviton KK tower 
is governed by two parameters, $\Lambda_\pi$
and $k/\overline M_{Pl}$.  The masses of the 4-d graviton 
states are given by $m_n=x_n\Lambda_\pi k/\overline M_{Pl}$, with $x_n$ 
being the roots of the first-order Bessel function $J_1$, 
{\it i.e.}, $J_1(x_n)=0$,
and their 
interactions with the SM fields on the TeV-brane are \cite{sec6_dhr1}
\begin{equation}
{\cal L} = -{1\over\overline M_{Pl}}T^{\mu\nu}(x)h^0_{\mu\nu}(x)
- {1\over\Lambda_\pi}T^{\mu\nu}(x)\sum_{n=1}^\infty h^{(n)}_{\mu\nu}
(x)\,,
\end{equation}
where $T^{\mu\nu}$ is the conserved stress-energy tensor.
Note that in this scenario, the coupling strength of the graviton
KK tower interactions and the mass of the first excitation are of
order of the weak scale.
The hallmark signature for this scenario is the presence of 
TeV-scale spin-2 graviton resonances at colliders.  The KK spectrum
in $e^+e^-\to\mu^+\mu^-$, taking $m_1=500$ GeV, 
is shown in Fig. \ref{sec6_rs1}.  Note
that the curvature parameter controls the width of the resonance.  
The LHC can discover these resonances in the Drell-Yan channel if 
$\Lambda_\pi < 10 $ TeV \cite{sec6_dhr1}, provided that the resonance 
width is not too narrow, and determine their spin-2 nature via the
angular distributions of the final-state lepton pairs 
\cite{sec6_allanach} if enough statistics are available.
An accurate determination of the branching fractions for the graviton 
KK decays to various final states will probe the universal 
$T^{\mu\nu}$ structure
of the couplings and verify the production of gravity.  Numerical
studies of such coupling determinations have yet to be performed, but
are likely to demonstrate the benefits of the LC even if the graviton 
KK states are kinematically inaccessible at the LC and are produced 
indirectly; this is in analogy to the $Z'$ studies discussed in the 
previous chapter.

\begin{figure}[htbp]
\centerline{
\includegraphics[width=9cm,angle=90]{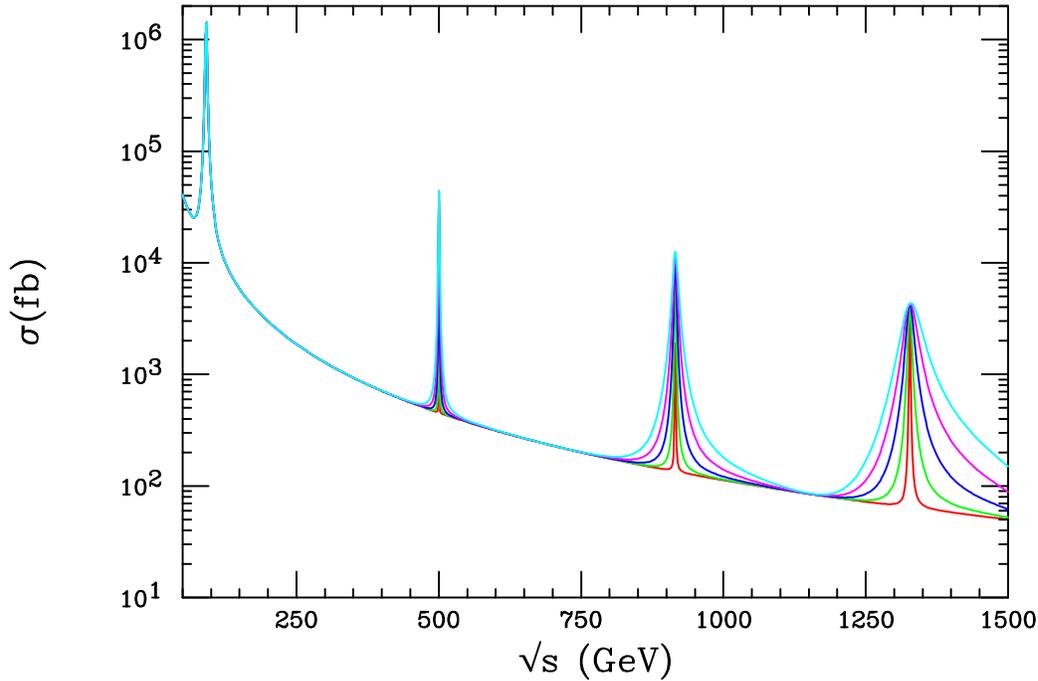}}
\vspace*{0.1cm}
\caption{The cross section for $e^+e^-\to\mu^+\mu^-$ including the
exchange of a KK tower of gravitons in the RS model with $m_1=500$
GeV.  The various curves correspond to $k/\overline M_{Pl}$ in the range
$0.01-0.1$.  From \cite{sec6_dhr1}.}
\label{sec6_rs1}
\end{figure}

If the KK gravitons are too massive to be produced directly, their
contributions to fermion pair production may still be felt via virtual
exchange.  In this case, the uncertainties associated with a cut-off
(as present in the large extra dimensions scenario) are avoided, since
there is only one additional dimension and thus the KK states may be neatly
summed.  The resulting sensitivity to the scale $\Lambda_\pi$ at
the LHC and LC is displayed in Table \ref{sec6_rscont}.  We see that
the reach of the 500 GeV LC is complementary to that of the LHC and
that a 1 TeV LC extends the discovery reach of the LHC.  This degree
of sensitivity to virtual graviton KK exchange at the LC implies that 
the KK coupling measurements discussed above should be viable.

\begin{table}
\centering
\begin{tabular}{|c|c|c|c|} \hline\hline
 & \multicolumn{3}{c|}{$k/\overline M_{Pl}$ } \\ \hline
 & 0.01 & 0.1 & 1.0\\ \hline
LC $\sqrt s=0.5$ TeV & 20.0 & 5.0 & 1.5 \\
LC $\sqrt s=1.0$ TeV & 40.0& 10.0 & 3.0 \\
LHC & 20.0 & 7.0 & 3.0 \\ \hline\hline
\end{tabular}
\caption{$95\%$ CL search reach for $\Lambda_\pi$ (in TeV) in the contact
interaction
regime taking 500 fb$^{-1}$ and 100 fb$^{-1}$ of integrated luminosity 
at the LC and LHC, respectively.  From \cite{sec6_dhr1}.}
\label{sec6_rscont}
\end{table}

\subsection{Extensions of the RS model}

\noindent {\bf $\bullet$ Extended Manifolds}
\vspace{0.25cm}

From a theoretical perspective, the RS model may be viewed as an
effective theory whose low energy features originate from a full theory
of quantum gravity, such as string theory.  One may thus expect that a
more complete version of this scenario admits the presence of
additional dimensions compactified on a manifold ${\cal M}^\delta$ of
dimension $\delta$.  From a model-building point of view, it can also
be advantageous to place at least some of the SM fields in the higher
$\delta$-dimensional space to {\it e.g.}, suppress proton decay or 
address gauge coupling unification and the flavor problem.  The presence 
of an additional manifold may reconcile string theory and several 
model-building features with the RS scenario.  

The existence of an extra manifold modifies the conventional RS 
phenomenology and collider signatures.  As a first step, the additional
manifolds $S^\delta$ with $\delta\geq 1$, representing both flat and
curved geometries, have been considered \cite{sec6_dhr6}.  In this case,
additional graviton KK states appear, corresponding to orbital
excitations of the $S^\delta$ manifold.  For the simplest scenario of
$S^1$, the RS metric is expanded to
\begin{equation}
ds^2=e^{-2kr_c\phi}\eta_{\mu\nu}dx^\mu dx^\nu +r_c^2d\phi^2+R^2d\theta^2\,,
\end{equation}
where $\theta$ parameterizes the $S^1$, and $R$ represents its radius.
The masses of the KK states
are now given by $m_{n\ell}=x_{n\ell}\Lambda_\pi k/\overline M_{Pl}$,
where the $x_{n\ell}$ are solutions of the equation
\begin{equation}
2J_\nu(x_{n\ell})+x_{n\ell}J'_\nu(x_{n\ell})=0\,,
\end{equation}
with $\nu\equiv\sqrt{4+(\ell/kR)^2}$.  The KK mode number $\ell$
corresponds to the orbital excitations, while $n$ denotes the usual RS
$AdS_5$ mode levels.  The couplings of the $m_{n\ell}$ graviton KK
states are then given by
\begin{equation}
{\cal L} = -{1\over\overline M_{Pl}}T^{\mu\nu}(x)h^0_{\mu\nu}(x)
- {1\over\Lambda_\pi}T^{\mu\nu}(x)\sum_{n=1}^\infty \xi(n\ell)
h^{(n,\ell)}_{\mu\nu}(x)\,,
\end{equation}
where $\xi(n\ell)$ depends on $k\,, R$, and $x_{n\ell}$ \cite{sec6_dhr6}.
The KK spectrum and couplings are thus modified due to the presence
of the new manifold.  

In particular, the addition of the $S^\delta$
background to the RS setup results in the emergence of a forest of
graviton KK resonances.  These originate from the orbital excitations on
the $S^\delta$ and occur in between the original RS resonances.
A representative KK spectrum is depicted as a histogram in
Drell-Yan production in the electron channel at the LHC in Fig 
\ref{sec6_LHC_s1}.  Here, $kR=1.0$ and $m_{10}=1$ TeV is assumed and
detector smearing \cite{sec6_atlas} is included.
In this case, the individual peaks at smaller masses are well
separated due to this choice of model parameters.  At larger masses
there is increased overlap among the KK states and individual
resonances may be difficult to isolate.  A detailed detector study
of this KK forest is required to determine what separation between
the states is necessary in order to isolate the resonances.  
The separation between the states decreases as the model parameters
are varied.  This is displayed in Fig. \ref{sec6_rs_s1} which shows
the cross section in $e^+e^-\to\mu^+\mu^-$ for $kR=2.0$
and $m_{10}=600$ GeV.  Note that the density of the KK
spectrum is substantially increased; in particular, 
both $n$ and $\ell$~~KK excitations are present in the same
kinematic region and the
peaks are not well separated.  In the case shown here, the spacing
between the $n=1\,, \ell=0$ and $n=1\,,\ell=1$ states is only 425 MeV,
which is comparable to the individual KK widths of $\simeq 525$ MeV.
The $\mu$ pair resolution at a LC should be sufficient to resolve 
these two states if they are within kinematic reach.  However, the
separation of these states may prove problematic at the LHC; in this
case, the LC would be needed to observe the distinct states.
\vspace{0.25cm}

\begin{figure}[htbp]
\centerline{
\includegraphics[width=9cm,angle=90]{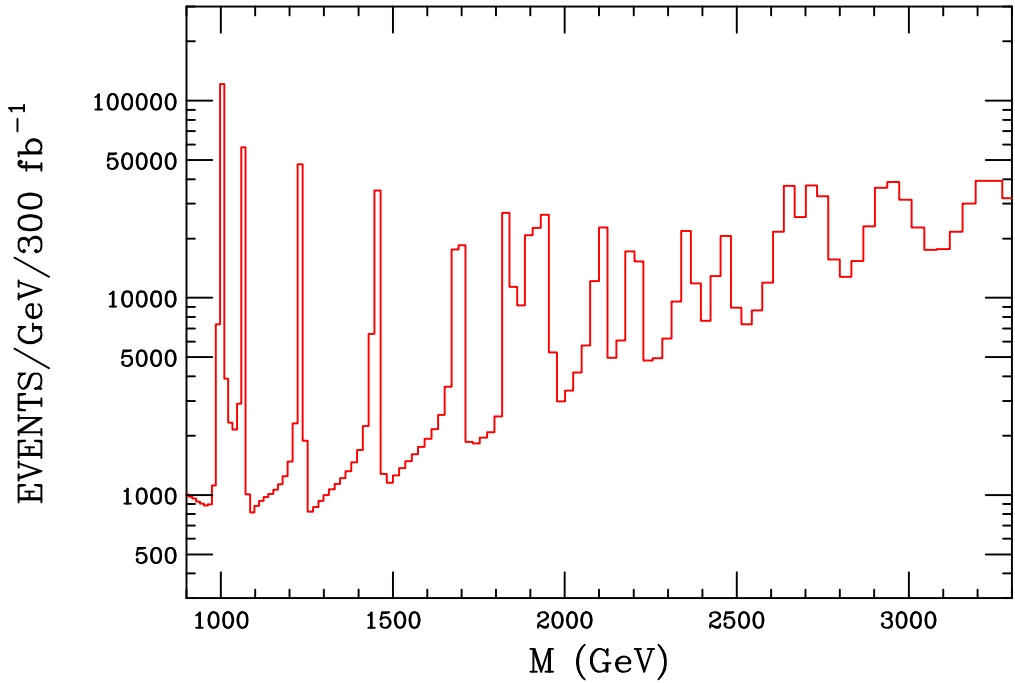}}
\vspace*{0.1cm}
\caption{Binned Drell-Yan cross section for $e^+e^-$ production at the
LHC assuming $m_{10}=1$ TeV, $k/\overline M_{Pl}=0.03$, and $kR=1.0$.
The cross section has been smeared by an electron pair mass resolution 
of $0.6\%$ as might be expected at ATLAS {\cite {sec6_atlas}}.
From \cite{sec6_dhr6}.}
\label{sec6_LHC_s1}
\end{figure}

\begin{figure}[htbp]
\centerline{
\includegraphics[width=9cm,angle=90]{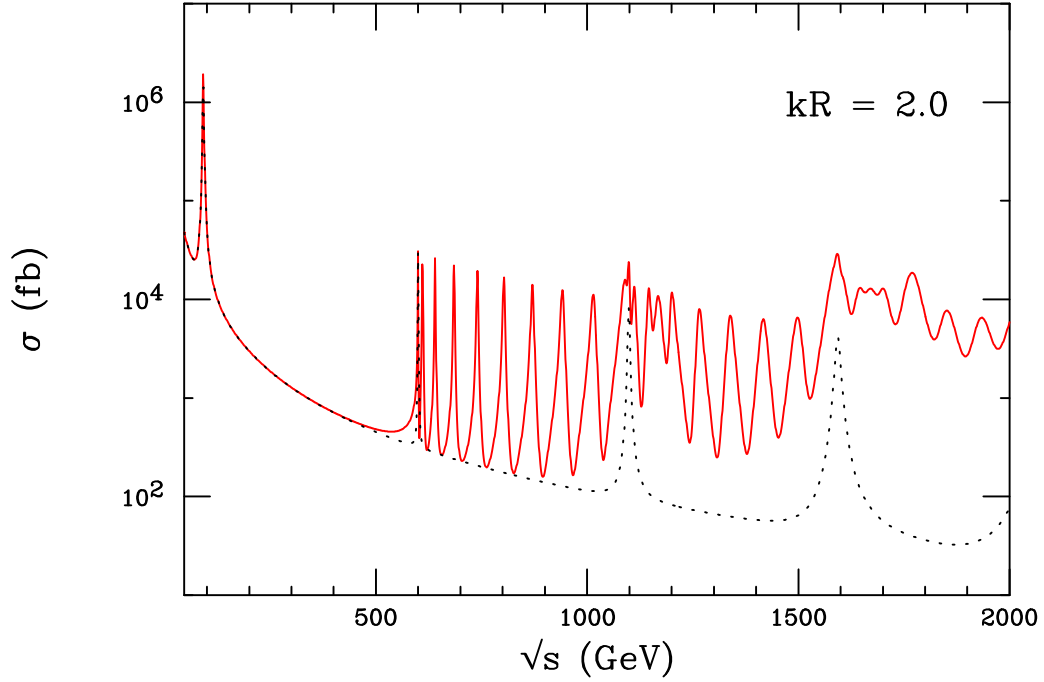}}
\vspace*{0.1cm}
\caption{The solid (red) curve corresponds to the cross section for
$e^+e^- \to \mu^+\mu^-$ when the additional
dimension is orbifolded, \ie, for $S^1/Z_2$, with $m_{10}=600$ GeV,
$k/\overline M_{Pl}=0.03$ and  $kR=2.0$. 
The result for the conventional RS model is
also displayed, corresponding to the dotted curve.  From \cite{sec6_dhr6}.}
\label{sec6_rs_s1}
\end{figure}

\noindent {\bf $\bullet$ Brane Localized Kinetic Terms}
\vspace{0.25cm}

An additional well-motivated extension to the conventional RS model
is the inclusion of brane curvature terms for the graviton.  These
terms respect all 4-d symmetries, and are expected to be present in
a 4-d theory, leading to brane localized kinetic terms for the 
graviton.  In fact, it has been demonstrated \cite{sec6_bktrs} that brane
quantum effects can generate such brane kinetic terms, and that these
terms are required as brane counter terms for bulk quantum effects.
In addition, a tree-level Higgs-curvature term is allowed in the 
brane action and can also induce brane kinetic terms for the graviton.  
There are thus many good theoretical reasons for assuming the presence 
of such terms for gravitons.  

The existence of boundary kinetic terms results in novel features and
can substantially modify the KK phenomenology of both gravitons and
bulk gauge fields \cite{sec6_bktrs}.  In the presence of these terms,
the graviton KK spectrum and couplings is again modified and alters
the low-energy 4-d phenomenology of the RS model \cite{sec6_bktrs}.  
In this case, the graviton action is now augmented by brane localized 
curvature terms on both branes.  The graviton KK
masses are again given by $m_n=x_n\Lambda_\pi k/\overline
M_{Pl}$, where the $x_n$ are now roots of the equation
\begin{equation}
J_1(x_n)-\gamma_\pi x_nJ_2(x_n)=0\,.
\end{equation}
Here, $\gamma_\pi$ represents the coefficient of the boundary term 
for the TeV-brane and is naturally of order unity.  The couplings are
modified to be
\begin{equation}
{\cal L} = -{1\over\overline M_{Pl}}T^{\mu\nu}(x)h^0_{\mu\nu}(x)
-{1\over\Lambda_\pi}T^{\mu\nu}(x)\sum_{n=1}^\infty \lambda_n
h^{(n)}_{\mu\nu}(x)\,,
\end{equation}
where $\lambda_n$ depends on the coefficient of the boundary terms
on both branes.  This results in a dramatic reduction of the graviton
KK couplings to SM fields on the TeV-brane, even for small values of
the brane kinetic term coefficients.  For coefficients of order $\sim
10$, the graviton KK couplings are reduced by a factor of $\sim 1/100$
of their value in the conventional RS model.  This clearly results in
a substantial decrease in the cross section for spin-2 graviton KK 
resonance production, which is the hallmark collider signature of 
the RS scenario, due to their extremely narrow width in this scenario.  
The resulting degradation in the graviton search reach at
the LHC is displayed in Fig \ref{sec6_lhcbt} for 100 fb$^{-1}$ of
integrated luminosity.  From this figure, it is clear that the LHC
can no longer cover all of the interesting parameter space for this
model.  For example, a first graviton KK excitation of mass 600 GeV
with $k/\overline M_{Pl}=0.01$ may still miss detection.  Very light
KK gravitons may thus escape direct detection at the Tevatron and LHC.

However, it is possible that such narrow resonances may be observed
at the LC via radiative return.  The differential cross section for
radiative return in $e^+e^-\to mu^+\mu^-\gamma$ \cite{sec6_osland}
including a 750 GeV KK state with $\gamma_\pi=-8$ and 
$k/\overline M_{Pl}=0.025$, a resonance which is invisible
at the LHC, is displayed in  
Fig. \ref{sec6_lhcbt}.  From the figure we
see that there is a sharply defined peak; the narrrow width approximation
yields a $\sim 1$ fb$^{-1}$ cross section under the peak.  Hence, the
 radiative return search technique at the LC is very 
powerful and may guide dedicated analyses at the LHC in
the search for narrow resonances.

\begin{figure}[htbp]
\centerline{
\includegraphics[width=9cm,angle=90]{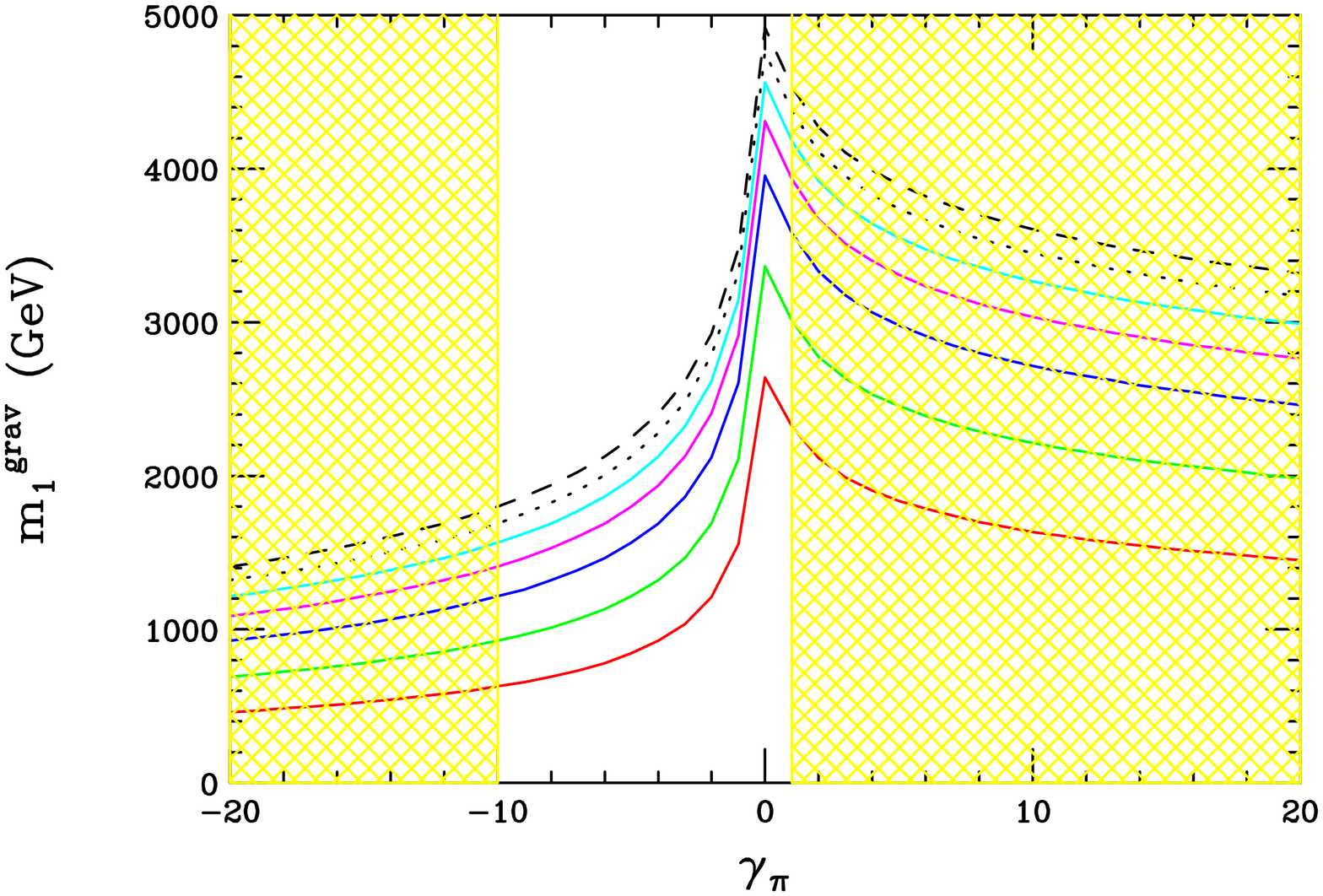}}
\vspace*{0.4cm}
\centerline{
\includegraphics[width=9cm,angle=90]{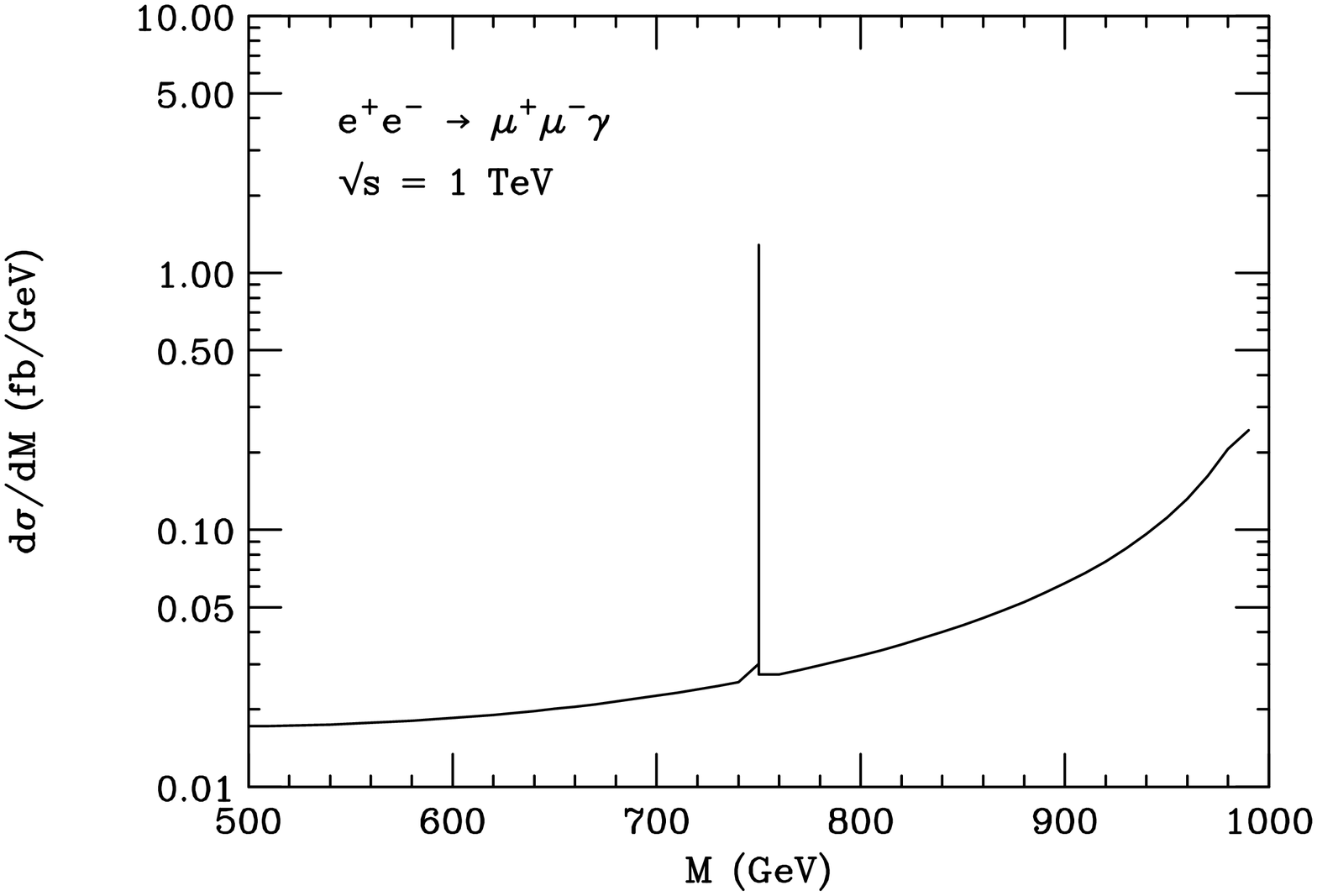}}
\vspace*{0.1cm}
\caption{Top: Search reach for the first graviton KK resonance employing the
Drell-Yan channel at the LHC with  an integrated
luminosity of 100 $fb^{-1}$ as
a function of the boundary term coefficient
$\gamma_\pi$ assuming $\gamma_0=0$. From bottom to
top on the RHS
of the plot, the curves correspond to $k/\overline M_{Pl}=0.01, 0.025, 
0.05, 0.075, 
0.10, 0.125$ and 0.15, respectively.  The unshaded region is that
allowed by naturalness considerations and the requirement of a
ghost-free radion sector.  From \cite{sec6_bktrs}  Bottom:
Differential cross section for $e^+e^-\to\mu^+\mu^-\gamma$ including
a 750 GeV graviton excitation with $k/\overline M_{Pl}=0.025$ and
boundary term $\gamma_\pi=-8$.  From \cite{sec6_jlh}.}
\label{sec6_lhcbt}
\end{figure}

\subsection{Conclusions}

Collider signatures for the presence of extra spatial dimensions
are wide and varied, depending on the geometry of the additional
dimensions.  The basic signal is the observation of a Kaluza-Klein
(KK) tower of states corresponding to a particle propagating in the
higher dimensional space-time.  The measurement of the properties
of the KK states determines the size and geometry of the extra 
dimensions.

In the scenario of large extra dimensions, where gravity alone
can propagate in the bulk, the indirect effects and direct
production of KK gravitons are both available at the LHC and at 
the LC.  For the indirect exchange of KK gravitons, the search
reach of the LC exceeds that of the LHC for $\sqrt s \gsim 800$
GeV.  Measurement of the moments of the resulting angular 
distributions at the LC can identify the spin-2 nature of the
graviton exchange for roughly half of the search reach range.
If positron polarization is available, then azimuthal asymmetries
can ($i$) extend the search for graviton exchange by a factor of 2,
probing fundamental scales of gravity up to 21 TeV for $\sqrt s=
1$ TeV with 500 fb$^{-1}$ of integrated luminosity, and ($ii$)
identify the spin-2 exchange for the entire LHC search region.
In the case of direct KK graviton production, the LHC and LC
have comparable search reaches.  However, the LHC is hampered
by theoretical ambiguities due to a break-down of the effective
theory when the parton-level center of mass energy exceeds
the fundamental scale of gravity; in this case the LHC may directly
detect effects of quantum gravity.  Measurement of direct graviton
production at two different center of mass energies at the LC
can determine the number of extra dimensions and the absolute
normalization of the cross section can determine the fundamental
scale.  Simultaneous determination of all the model parameters
has been examined in a quantitative fashion with the result
that data from the LC and LHC analyzed together substantially 
improves the accuracy of this determination over the LHC data 
taken alone.

Standard Model fields are allowed to propagate in extra dimensions
with size less than TeV$^{-1}$.  Signatures for the KK states of
the SM gauge fields mimic those for new heavy gauge bosons in
extended gauge theories.  The LHC may discover electroweak gauge 
KK states via direct production in the mass range $M_c\simeq 4-6$ 
TeV (lower masses are excluded by LEP/SLC data), while indirect
detection at the LC is possible for $M_c\lsim 20$ TeV for
$\sqrt s=1$ TeV.  Indirect detection of electroweak gauge KK
states is also possible at the LHC for $M_c\lsim 12$ TeV via a 
detailed study of the shape of the Drell-Yan lepton-pair
invariant mass distribution.  If discovered, a quantitative
analyses shows that the determination
of the mass of the first gauge KK excitation at the LHC, together
with indirect effects at the LC can be used to distinguish the
production of a KK gauge state from a new gauge field in
extended gauge sectors.

The possibility of universal extra dimensions, where all SM
fields are in the bulk, can cause confusion with the production
of supersymmetric states, since the KK spectrum and phenomenology 
resembles that of supersymmetry.  In fact, the lightest KK 
state is a Dark Matter candidate.  In this case, threshold
production of the new (s)particle at the LC can easily determine
its spin and distinguish universal extra dimensions from
supersymmetry.  Spin determination analyses are on-going for
the LHC.  

Lastly, the presence of warped extra dimensions results in the
resonance production of spin-2 gravitons.  This produces a
spectacular signature at the LHC for the conventional construction
of the Randall-Sundrum model.  However, extensions to this model,
such as the embedding in a higher dimension manifold, or the
inclusion of kinetic brane terms, result in reduced coupling strengths
and extremely narrow-width graviton KK states.  Such states may
be difficult to observe at the LHC and even very light KK states
may escape detection.  In this case, radiative return at the LC
may pinpoint the existence of these states, which can then be
confirmed by a dedicated search at the LHC.


%
%
%
%
%
%
%
%


\chapter{Conclusions and Outlook}


The Large Hadron Collider and the Linear Collider
will explore physics at the TeV scale, opening 
a new territory where ground-breaking discoveries are expected.
There is a huge variety of possible manifestations of new physics in this 
domain that have been advocated in the literature, and even more
possibilities surely exist beyond our present imagination.
The physics programme of both the LHC and the LC in exploring this
territory will be very rich. The different characteristics of the two
machines give rise to different virtues and capabilities.
The high collision energy of the LHC leads to a
large mass reach for the discovery of heavy new particles. The clean
experimental environment of the LC allows detailed studies of directly
accessible new particles and gives rise to a high
sensitivity to indirect effects of new physics.

Thus, physics at LHC and LC is complementary in many respects. 
While qualitatively this is obvious, a more quantitative investigation 
of the  interplay between LHC and LC requires detailed
information about the quantities that can be measured at the two
colliders and the prospective experimental
accuracies. Based on this input, case studies employing realistic
estimates for the achievable accuracy of both the experimental measurements 
and the theory predictions are necessary for different physics
scenarios in order to assess the synergy from the interplay of 
LHC and LC.
The LHC / LC Study Group was formed in order to tackle this task. The
present report is the first step on this way, and summarises the initial
results obtained within the LHC / LC Study Group. The results are based
on a close collaboration between the LHC and LC experimental communities 
and many theorists. 

A large variety of possible scenarios has been investigated,
including different manifestations of physics of weak and strong
electroweak symmetry breaking, supersymmetric models, new gauge
theories, models with extra dimensions and possible implications of
gravity at the TeV scale. For scenarios where detailed
experimental simulations of the possible measurements and the achievable 
accuracies are available both for LHC and LC, the LHC / LC interplay has
been investigated in a quantitative manner. In other scenarios many
promising possibilities have been pointed out where the interaction
between LHC and LC can be foreseen to have a large impact. 

Based on the results in this report important synergy from a 
concurrent running of LHC and LC has been established.
The intimate interplay of the results of the two collider facilities will
allow one to probe, much more effectively and more conclusively 
than each machine separately, the fundamental interactions of nature and the
structure of matter, space and time. 

Experience from the past shows that
the gain of knowledge often proceeds in iterative steps. Once
experimental data from both LHC and LC will be available, it is 
very likely that new questions to the LHC will arise which weren't
apparent before. The synergy between LHC and LC will
extend the physics potential of both machines. Results from both
colliders will be crucial in order to decipher the
underlying physics in the new territory that lies ahead of us and to
draw the correct conclusions about its nature. This information will be
decisive for guiding the way towards effective experimental strategies and 
dedicated searches. It will not only sharpen the goals for a subsequent 
phase of running of both LHC and LC, but will also be crucial for the
future roadmap of particle physics.

The interplay between LHC and LC is a very rich field, of which only very
little has been explored so far. The ongoing effort of the LHC and LC
physics groups in performing thorough experimental simulations for
different scenarios will enable further quantitative assessments of the 
synergy between LHC and LC. The exploratory work collected in the
present report is aimed to serve as a starting point for
future studies.


\vfill

\subsection*{Acknowledgements}

This research was supported in part 
by the
Alfred P.\ Sloan Foundation,
by the
Benoziyo center for high energy physics, 
by the 
Bundesministerium f\"ur Bildung und
Forschung (BMBF, Bonn, Germany) under
the contract numbers 05HT1WWA2, 05HT4WWA2,
by the Davis Institute for High Energy Physics,
by the
Department of Science and Technology, India, under project number 
SP/S2/K-01/2000-II,
by the
'Erwin Schr\"odinger fellowship No.~J2272'
of the `Fonds zur F\"orderung der wissenschaftlichen Forschung' of
Austria
by the 
European Community's Human
Potential Programme under contract HPRN-CT-2000-00149 Physics at
Colliders and contracts HPRN-CT-2000-00131, HPRN-CT-2000-00148, 
HPRN-CT-2000-00152,
by the
France-Berkeley fund,
by the 
German Federal Ministry of
Education and Research (BMBF) within the Framework of the
German-Israeli Project Cooperation (DIP),
by the 
Grant-in-Aid for Science Research, Ministry
of Education, Science and Culture, Japan,
No.~11207101, 13135297, 14046210, 14046225 and 14540260,
by the
KBN Grant 2 P03B 040 24 (2003-2005) and
115/E-343/SPB/DESY/P-03/DWM517/2003-2005,
by
MECYT and FEDER under project FPA2001-3598,
by the
Polish Committee for Scientific
Research, grant  no.~1~P03B~040~26 and project
no.~115/E-343/SPB/DESY/P-03/DWM517/2003-2005, 
by the RFBR grants 04-02-16476 and 04-02-17448,
by the
Swiss Nationalfonds,
by UK-PPARC,
by the 
grant of the Program ``Universities of Russia'' UR.02.03.028,
by the
U.S.\ National Science Foundation under grants No.\
ITR-0086044, 
PHY-0122557,
PHY-0139953, 
PHY-0239817,
PHY-9600155, PHY-9970703,
by the
U.S.~Department of Energy
under grants DE-AC02-76CH02000, DE-AC02-76SF00515, DE-AC03-76SF00515,
DE-FG02-91ER40674, DE-FG02-95ER40542, DE-FG02-91ER40685, DE-FG02-95ER40896, 
and by the
Wisconsin Alumni Research Foundation.
%


\end{document}